\documentclass[11pt,oneside,english]{book}

\usepackage{mathptmx}
\usepackage[T1]{fontenc}

\usepackage[latin9]{inputenc}

\usepackage[a4paper]{geometry}
\usepackage{color}
\usepackage{babel}
\usepackage{array}
\usepackage{longtable}
\usepackage{varioref}
\usepackage{float}
\usepackage{footnote}
\usepackage{amsmath}
\usepackage{amssymb}
\usepackage{graphicx}
\usepackage{setspace}
\usepackage[authoryear]{natbib}
\makeatletter

\makesavenoteenv{tabular}

\providecommand{\LyX}{L\kern-.1667em\lower.25em\hbox{Y}\kern-.125emX\@}
\providecommand{\tabularnewline}{\\}
\newcommand{\lyxdot}{.}

\usepackage{html}              
\usepackage{url}                 
\usepackage{ae,aecompl}   
\usepackage{lscape}           
\usepackage{epsfig}           
\usepackage{rotating}         
\usepackage{psfrag}           
\usepackage{graphicx}        
\usepackage[small,normal,bf,up]{caption} 

\usepackage{footnote}
\usepackage{threeparttable}
\usepackage{setspace}
\definecolor{darkgreen}{rgb}{0, 0.7, 0}
\vrefwarning                        

\usepackage{fancyhdr}                    

\usepackage{xcolor}
\usepackage{hyperref} 

\hypersetup{
    colorlinks,
    linkcolor={blue!50!black},
    citecolor={blue!50!black},
    urlcolor={blue!80!black}
}

\makeatother

\begin{document}
\frontmatter\pagestyle{fancy}

\title{{\Huge{}\vspace{-3cm}Structural and Nucleosynthetic Evolution of
Metal-poor \& Metal-free Low and Intermediate Mass Stars}\vspace{1.5cm}\\
{\huge{}Simon Wattana Campbell}\\
{\large{}BSc (Astro Hons), BA (Lang), Dip Eng (Aero)}\vspace{1.5cm}\\
{\large{}A thesis submitted for the degree of Doctor of Philosophy.}\\
{\large{}Centre for Stellar and Planetary Astrophysics,}\\
{\large{}School of Mathematical Sciences,}\\
{\large{}Monash University, Australia.}}
\date{\vspace{-1.5cm}March 2007}

\maketitle
\newpage
\begin{figure}
\begin{centering}
\includegraphics[width=0.8\columnwidth]{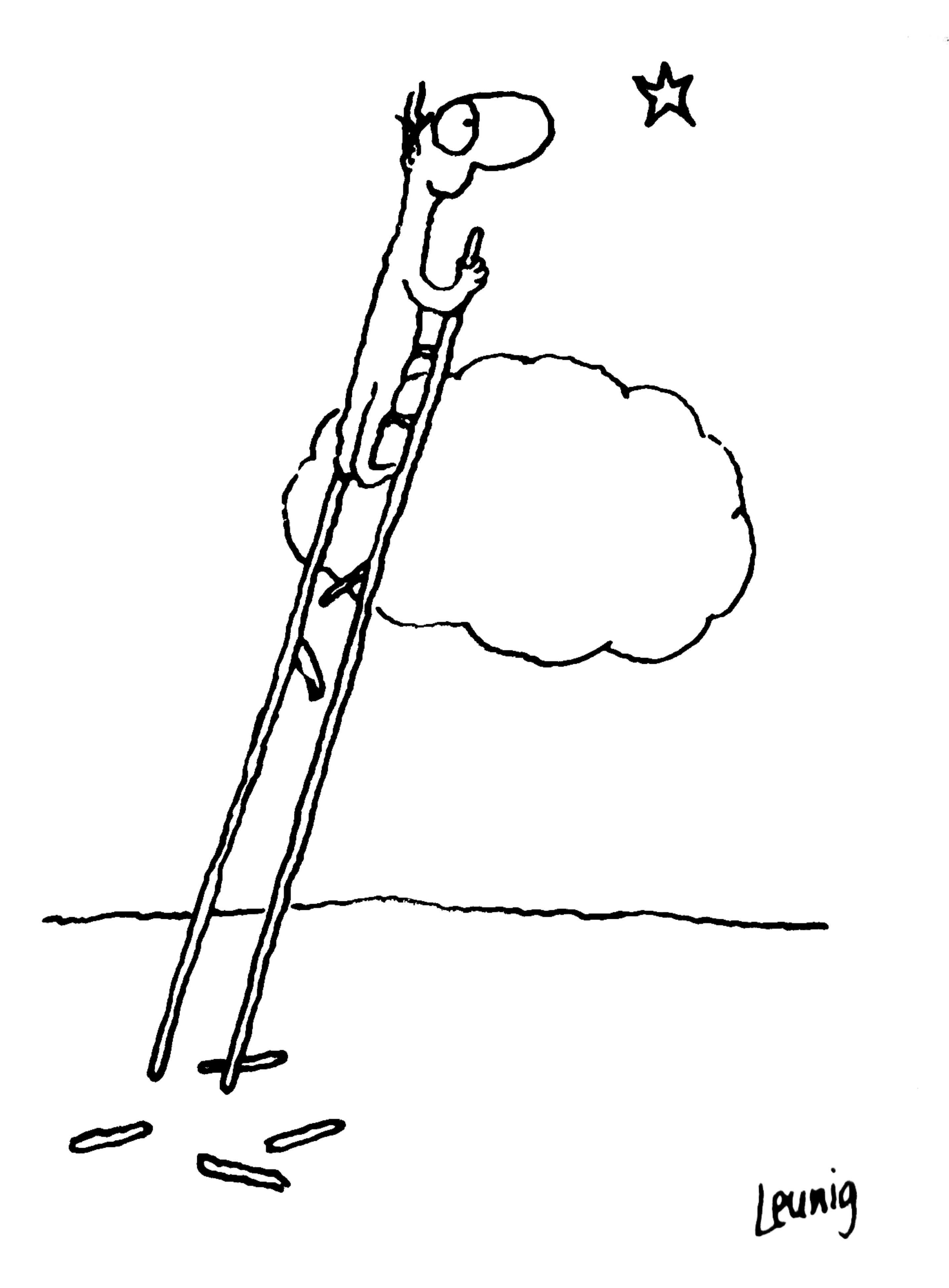}
\par\end{centering}
An apt cartoon from the Melbourne artist Michael Leunig. Reproduced
from a coffee mug given to me by Lisa Pinter. We note that the assumptions
used in the current work are (usually) more substantial than the support
for the ladder!
\end{figure}

\begin{spacing}{0.6}

\tableofcontents{}

\end{spacing}

\pagestyle{plain}

\begin{spacing}{1.0}\begin{center}
\textbf{\LARGE{}Abstract}\addcontentsline{toc}{section}{Abstract}
\par\end{center}

In this study we have investigated stellar evolution and nucleosynthesis
in the low and extremely low metallicity regime -- including models
of stars with a pure Big Bang composition (i.e. $Z=0$). The metallicity
range of the extremely metal-poor (EMP) models we calculated is $-6.5<\textrm{[Fe/H]}<-3.0$,
whilst most of our models are in the mass range $0.85<M<3.0$ M$_{\odot}$.
We have also calculated a series of models with a metallicity of $\textrm{[Fe/H]}=-1.4,$
to compare with observations of abundance patterns in Galactic globular
cluster stars.

Many of the extremely metal-poor (EMP) and $Z=0$ models experience
violent evolutionary episodes not seen at higher metallicities. We
refer to these events as `Dual Flashes' since they are characterised
by peaks in the hydrogen and helium burning luminosities occurring
at roughly the same time. Some of the material processed by these
events is later dredged up by the convective envelope, causing very
significant surface pollution. These events have been reported by
previous studies, so our results confirm their occurrence -- at least
in stellar models. 

The novelty of this study is that we have calculated the entire evolution
of the $Z=0$ and EMP models, from the ZAMS to the end of the AGB
-- including detailed nucleosynthesis. We have also calculated the
nucleosynthetic yields, which we make available in the appendices.
Although subject to many uncertainties these are, as far as we are
aware, the only yields available in this mass and metallicity range.

We find that our models predict an increased number of carbon-rich
stars at the lowest metallicities. This is mainly due to the extra
pollution provided by the Dual Flash (DF) events -- which do not
occur in higher metallicity models. This concurs well with the observations
that show the proportion of carbon-enhanced metal-poor (CEMP) stars
in the Galactic halo to be higher at lower metallicities. It is also
found that the pollution arising from the DF events is simultaneously
C- and N-rich, as also observed in the CEMP stars. This contrasts
with the pollution expected from third dredge-up at low mass, which
would be C-rich only. Furthermore, the models predict that the proportion
of CEMP stars should continue to increase at lower metallicities.
We also compare the chemical pollution arising from our models with
the detailed abundance patterns available for some of the most metal-poor
CEMP stars, and find mixed results.

In relation to our globular cluster (GC) models we find that our `standard'
AGB models have fundamental problems in explaining the GC abundance
anomalies, mainly due to the occurrence of third dredge-up (3DUP).
As an experimental test we also calculate a series of models in which
we `turn off' 3DUP. Here we find that the match with observations
is better, but problems still remain. We provide the yields for all
these GC models in the appendices. 

Finally we note that all these calculations contain many uncertainties.
These include the unknown mass-loss rates, uncertain nuclear reaction
rates, and the treatment of convection. In the case of the Dual Flash
events a further uncertainty is the possibility that full fluid dynamics
calculations are really needed to model these violent episodes.

\begin{center}
\textbf{\LARGE{}Statement}\addcontentsline{toc}{section}{Statement}
\par\end{center}

This thesis contains no material which has been accepted for the award
of any other degree or diploma in any university or other institution.
To the best of my knowledge this thesis contains no material previously
published or written by any other person except where reference is
made in the text. The length of this thesis is less than 100,000 words,
exclusive of figures, tables, appendices and bibliographies.\\
\\
\\

Simon Wattana Campbell
\begin{center}
\textbf{\LARGE{}Acknowledgements}\addcontentsline{toc}{section}{Acknowledgements}
\par\end{center}

Firstly I'd like to thank my supervisor -- John Lattanzio -- of
Monash University, Australia, for his support and friendship over
the last 4.9 years. Indeed, he certainly deserves thanks right now
because he has just read through every word of this thesis -- which,
as you can see, is no small feat! Particularly since the pace at which
I have fired chapters at him has accelerated in the last few months.
Even at this 11th hour (well, 27th for me today) he is religiously
reading the last sections. I would also like to thank him for encouraging
me with the forging of many scientific collaborations, to follow my
scientific interests, to attend conferences, and to give seminars.
I also think he deserves thanks for maintaining the Stellar Interiors
and Nucleosynthesis (SINS) group throughout the years. Our SINS meetings
were a key part of my time at Monash. Of course JL is synonymous with
port and cheese appreciation (not to mention wine and beer), for which
I am all the more educated (thanks to his mum Joy for those P \& C
nights and BBQs too!).

There are so many friends and colleagues I would like to thank at
Monash that I will probably forget someone, so please don't hold it
against me! In no particular order: Gareth Kennedy (for many interesting
conversations -- and inspiring the lecture-hopping culture), John
Mansour (for all our long chats about music, bikes, etc.), Johnny
Hitti (for those long philosophical arguments -- and music appreciation
at the NFF etc.), Hamed Moradi (for passionate discussions on religion
\& politics + organising the soccer -- and African music appreciation),
Carolyn Doherty (for putting up with my extreme mess in the office
-- and my hogging of all the Evoln notes), John McCloughan (for all
those Cinque Lire cafe trips and organising the happy hours at the
Nott), Keith Hsuan (for interesting discussions and an insight into
swing dancing), Lisa Elliott (for letting me tag along on those observing
trips \& sharing travel stories), Amanda Karakas (for the early help
with the stellar codes and her very useful thesis), Christian Lerrahn
(for some interesting email debates \& organising the soccer), Marie
Gibbon (for keeping the Global Warming issue alive), Allie Ford (for
the vego section of the Maths Dept. BBQs and letting me tag along
on the observing run), Juan Oliveros (some more sangria soon!), Charles
Morgan (for some interesting discussions), and Paul Lasky and Hannah
Schunker for resurrecting Happy Hour. Thanks also to the CSPA staff.
All of these people have contributed to helping me survive the PhD.
They have also helped put the `P' into the PhD -- through all the
great discussions/debates we've had over the years. I've been very
lucky to be in an environment so rich with nice, intelligent people.
A particular thanks to Gertrude Nayak, who helped me with getting
into honours -- if that hadn't happened I wouldn't have reached this
stage, and the secretaries (Linda, Denise, Barb \& Doris et al.).
Finally I thank Monash University for the PhD scholarship that supported
me financially for 3.5 years.

I have been fortunate enough to have gone to various conferences and
visited various universities around the world during my time at Monash
(largely thanks to John Lattanzio). In my travels I have met many
nice people, such as Onno Pols (thanks for the binary star education
even though that work did not end up in this thesis, for looking after
me in Utrecht \& of course for the Belgian beer education!), Chris
Tout (thanks for the hospitality in Cambridge -- especially those
traditional Cambridge Uni dinners), Jordi Jose (thanks for looking
after me in Barcelona), Sandro Chieffi (thanks for the help with the
high-T opacity tables), David Yong (always the fellow vegetarian at
conferences), Sergio Cristallo (mi hijo), Lionel Siess, Peter Cottrell,
Agostino Renda, Jarrod Hurley, Brad Gibson, Yeshe Fenner, Rob Izzard,
Axel Bonacic Marinovic, Liz Wylie, Kenji Bekki, and many more. A special
thanks to the organisers of the Torino workshops over the years, at
Granada (Inma \& Carlos \& Olga) and Cambridge (Maria, Richard, John
\& Ross et al.).

As most of this work was done on computers I think it fitting to thank
the people that contributed to the open-source software I have used.
All of my work has been done using Linux operating systems. A few
pieces of software I have made great use of have been: Yorick (for
most of the plots in the thesis), the GNU Image Manipulation Program
(GIMP), and \LyX{} (a \LaTeX{} frontend). I have used \LyX{} to write
this entire document -- and I can highly recommend it! Thanks to
the people that provide those services that are indispensable for
astronomers and astrophysicists -- the NASA ADS and astro-ph. I also
have to thank the excellent staff at the Australian Partnership for
Advanced Computing (APAC supercomputer) and also the Victorian version,
VPAC. Much of this work was done under the APAC grant Project Code
$g61$ (``Element Production by Intermediate Mass Red-Giant Stars'').
Of course a big thanks goes the original authors of the Stellar codes
I have used in this thesis -- Peter Wood, Rob Cannon, John Lattanzio,
Cheryl Frost, Faulkner \& Gingold. Thanks also to the IT guys for
support over the years (Jon, Declan, Sean \& Mike).

Other friends I would like to thank are the NFF crew (Dave, Claudia,
Jan, Pat, Dave \& Barb, Suki) -- for the great Easter music festivals/get-togethers
over the years. It was always a welcome change of scene. Rachel, Matt
\& Nathan for our pub dinners, Marina for having us over for dinner
so often (braving the Vego cooking) and those great African nights
full of dancing at the Espy, Mary \& Tony for the musical outings.
A big thanks to Nim and Man \& Manvina for always being great hosts
in Malaysia.

I thank my family -- Mum \& Dave for being so caring, especially
all those supportive phone calls and emails, my nomadic brothers Alex
and Jeff -- it was always nice to see you on your jaunts to Melbourne
-- hopefully we'll see more of you two when we move to Asia! Thanks
also to our amazing net-surfing, e-mailing, beach-combing Grandma,
for her moral support, all the email updates on the family -- and
lovely luncheons in Anglesea. Two more people I consider family are
Lisa's mum Maria and Lisa's brother Joe. Maria has been very supportive
over the years -- making many yummy dinners, and looking after our
various animals when we go away. Joe has been there for us too, ever
reliable and always willing to help out.

Here I would like to thank the person most central to my life --
my loving partner of 13 years, Lisa Pinter. She has supported me emotionally,
morally and practically over my many years of study. It has been a
very hard road to get the thesis finished this year. I appeared to
have had bitten off a bit more than I could chew! Nevertheless, Lisa
supported me through the months and months of 7 day weeks (sometimes
wisely steering clear of me in my PhD lair!). It was a difficult time
for her too. We seem to have both survived though -- I think the
daily coffee together at Santucci's helped with that. There is definitely
some catching up to do. I have been away in thesis land for quite
a while! 

Finally we note that the writing of this thesis has been on a laptop
in our house which sources most of its \textbf{\emph{energy from a
star!}}\footnote{We have $12\times165$ Watt solar panels on the roof.}
\begin{center}
\textbf{\LARGE{}Publications}\addcontentsline{toc}{section}{Publications}
\par\end{center}

The following papers were published during the Author's candidature:

\subsection*{Refereed}
\begin{enumerate}
\item \textbf{Origin of Abundance Inhomogeneity in Globular Clusters} \\
Bekki, K., Campbell, S. W., Lattanzio, J. C., Norris, J. E. \\
2007, MNRAS, accepted January 2007
\item \textbf{The Chemical Evolution of Helium in Globular Clusters: Implications
for the Self-Pollution Scenario} \\
Karakas, A. I., Fenner, Y., Sills, A., Campbell, S. W., Lattanzio,
J. C.\\
2006, ApJ, 652, 1240
\item \textbf{Fluorine Abundance Variations in Red Giants of the Globular
Cluster M4 and Early-Cluster Chemical Pollution}\\
Smith, V.V., Cunha, K., Ivans, I.I., Lattanzio, J.C., Campbell, S.W.,
Hinkle, K.H.\\
2005, ApJ, 633, 392S
\item \textbf{Abundance Anomalies in NGC6752 - Do AGB Stars Have a Role?}\\
Campbell, S.W., Fenner, Y., Karakas, A.I., Lattanzio, J.C., Gibson,
B.K.\\
2005, Nuclear Physics A, 758, 272
\item \textbf{The Evolution of Fluorine in Galactic Systems}\\
Renda, A., Fenner, Y., Gibson, B.K., Karakas, A.I., Lattanzio, J.C.,
Campbell, S.W., Chieffi, A., Cunha, K., Smith, V.V.\\
2005, Nuclear Physics A, 758, 324
\item \textbf{Modelling Self-Pollution of Globular Clusters from Asymptotic
Giant Branch Stars}\\
Fenner, Y., Campbell, S.W., Karakas, A.I., Lattanzio, J.C., Gibson,
B.K.\\
2004, MNRAS, 353, 789
\item \textbf{On the origin of fluorine in the Milky Way}\\
Renda, A., Fenner, Y., Gibson, B.K., Karakas, A.I., Lattanzio, J.C.,
Campbell, S.W., Chieffi, A., Cunha, K., Smith, V.V.\\
2004, MNRAS, 354, 575
\end{enumerate}

\subsection*{Conference Proceedings}
\begin{enumerate}
\item \textbf{Are There Radical Cyanogen Abundance Differences Between Galactic
Globular Cluster RGB and AGB Stars?}\\
Campbell, S.W., Lattanzio, J. C., Elliott, L. M.\\
2006, MmSAI, 77, 864
\item \textbf{Helium enhancements in globular cluster stars from Asymptotic
Giant Branch star pollution}\\
Karakas, A. I., Fenner, Y., Sills, A., Campbell, S. W., Lattanzio,
J. C.\\
2006, MmSAI, 77, 858
\item \textbf{Nucleosynthesis in Early Stars}\\
Campbell, S.W.\\
2004, in Carnegie Observatories Astrophysics Series, Vol. 4: Origin
and Evolution of the Elements, ed. McWilliam, A., Rauch, M., sympE,
7C
\item \textbf{Abundance Anomalies in NGC 6752. Are AGB Stars the Culprits?}\\
Campbell, S.W., Fenner, Y., Karakas, A.I., Lattanzio, J.C., Gibson,
B.K.\\
2004, MmSAI, 75, 735
\item \textbf{The Role of AGB Stars}\\
Lattanzio, J., Karakas, A., Campbell, S.W., Elliott, L., Chieffi,
A.\\
2004, MmSAI, 75, 322
\end{enumerate}

\end{spacing}

\mainmatter\pagestyle{fancy}
\part{INTRODUCTION}

\chapter{Setting the Scene}
\begin{quote}
``Science is a way of trying not to fool yourself. The first principle
is that you must not fool yourself, and you are the easiest person
to fool.''
\begin{flushright}
\vspace{-0.7cm}-- Richard Feynman
\par\end{flushright}

\end{quote}

\section{Preface}

The perennial questions of `What is out there?' and `Is there any
other life in the Universe?' are not solely the domains of scientists
such as astrophysicists and astronomers. They are key questions for
humanity. Interest in these questions pervades our society at large.
For astronomers and astrophysicists this is frequently evidenced in
the reactions of people to their profession. No matter what the age-group
a whole series of questions usually ensues. These questions are always
very pertinent -- and often very difficult to answer. I believe that
this great, raw, interest is a healthy sign. Given this passion for
the subject of astronomy it would be nice if our education systems
reflected this interest and offered greater opportunities for children
-- and adults -- to learn more about the Universe. Exposing people
to the amazing reality that is our Universe opens our minds and puts
our place in the Universe (and on our planet) into perspective. Furthermore,
learning how information about the Universe is gathered by scientists
is highly revealing of just how powerful a method Science is for understanding
reality.

This thesis represents a tiny step (but hopefully a useful one!) in
the process of attempting to understand the Universe. It is predicated
on many other small (and some large) steps taken by scientists --
and indeed other researchers -- before us. The exact topic is necessarily
very narrow, highly specialised. However the broader context is easily
understandable and, as noted above, of great interest to many people.
Therefore, for the non-specialists that have, for some reason, found
themselves with a copy of this highly technical monograph, we first
provide a section that gives some general information on the important
role of stars in our Universe. We also give a non-specialist introduction
to the current study towards the end of that section. A more technical
introduction is given in Section \ref{section-Intro-CurrentStudy}.

\section[Layperson's Introduction to Stars]{A Layperson's Introduction to \protect \\
the Importance of Stars}

When talking about stars the image that generally pops into people's
minds is that of the night sky filled with millions of pinpoints of
light. Practically all of the light that we see in space comes from
stars of one sort or another. Some of those stars that we see at night
are actually entire \emph{galaxies}. However these galaxies are themselves
just collections of many, many stars -- like a city is a collection
of many, many houses. The galaxies just look like tiny bright dots
because they are so far away. Indeed the galaxy that we live in --
known as the `Milky Way' -- is one of these collections of stars.
It has \emph{billions} of stars. So what are these things called stars
that seem to be everywhere in the Universe, shining brightly? The
answer to that is best seen during the \emph{day} rather than the
night. The star that shines the brightest (at least for us) is our
own Sun. It turns out that all those stars in the night sky are just
like the Sun -- huge bright balls of gas. The only reason the Sun
looks so much brighter than all the other stars is because we are
so close to it. Indeed, we are so close to it that we can feel the
heat from it -- and get sunburnt by its UV radiation. So the Sun
is our `pet' star. It also turns out that it is a pretty average star
compared to all the other stars out there. It has an average size
(which is pretty big!), an average brightness, and even an average
age. So there are lots of stars out there quite similar to our Sun.
However there are also stars out there that are much bigger and much
smaller. Some are much brighter but some are much dimmer. There is
even a whole range of colours!

\subsubsection*{The Light From Stars}

A key question that many people ask about stars is `Why do they shine?'.
This is a great question. It took researchers quite a long time to
work this out. Back in the early 1900s we had no idea. People suggested
that the Sun produces so much heat that it must be burning \emph{coal!}
Another theory was that the heat was produced by the slow collapse
of the Sun. This is possible, but it turned out that the Sun wouldn't
shine for very long in this case -- it would have shone for much
less time than the Earth has been around -- so that theory was rejected
too. About 20 years later we still didn't know how the Sun shone.
One of the famous astronomers at the time (scientists could be famous
back then!), Arthur Eddington, wrote a brief article summarising some
of the theories about how the Sun might shine (\citealt{1919Obs....42..371E}).
This is a fascinating article because it shows science at work. Eddington
had collected all the evidence about the Sun and worked out the key
characteristics of the source of energy -- but still didn't know
what it actually was. None of the explanations at the time fitted
the evidence. It turns out that the reason they couldn't work it out
was because the source of energy had not even been discovered on Earth
yet -- nuclear energy! It is impressive that -- even before nuclear
energy was discovered -- the scientists around that time were saying
that it must have been \emph{something} like that. They were calling
it `sub-atomic energy'. Since this energy must be causing the stars
to shine so brightly and for so long, it was naturally wondered if
humans could harness it, as \citet{1920Obs....43..341E} noted: 
\begin{quotation}
\emph{``If, indeed, the sub-atomic energy in the stars is being freely
used to maintain their great furnaces, it seems to bring a little
nearer to fulfillment our dream of controlling this latent power for
the well-being of the human race -- or for its suicide.'' }
\end{quotation}
In this statement he also foreshadowed one of the unfortunate spin-offs
of these scientific investigations -- the development of nuclear
weapons. Interestingly Eddington's comment was made well before nuclear
weapons were invented.

So we now know that it is \emph{nuclear energy} that powers the stars.
This energy is released in nuclear reactions, and, in particular,
nuclear \emph{fusion} reactions\emph{.} Nuclear fusion simply means
that two (sub-atomic) particles are fused together. This usually happens
in the very centre of stars -- where it is very hot and dense. The
sub-atomic particles are moving around so fast that when they collide
they stick together. Nuclear fusion has two effects. The first we
have already mentioned -- that nuclear energy is released, which
causes the star to shine. The second is a very important effect --
\emph{a new particle is formed} (out of the combination of the two
involved in the collision). This is occurring all the time in the
Sun. For example, in the Sun there is a lot of hydrogen. When some
of these hydrogen nuclei collide they eventually combine to form the
next element in the periodic table -- helium. Then, if there is a
lot of helium (and the temperature is high enough), carbon can be
formed, and so on. This is alchemy -- turning one element into another.
It appears that Nature is doing alchemy all the time, in stars. Indeed,
it turns out that almost all elements are made in stars -- even gold,
silver and uranium! The modern name for this alchemy (or `synthesis
of the elements'), is \emph{nucleosynthesis,} which is the topic of
the next section.

\subsubsection*{Stellar Alchemy: The Origin of the Elements}

Around the late 1930s the picture of stellar nucleosynthesis was beginning
to be fleshed out (based on earlier work by researchers such as George
Gammow and Ernest Rutherford). In \citeyear{1939PhRv...55..434B}
Hans Bethe published a paper called ``Energy Production in Stars''
where he proposed that the particular nuclear reactions occurring
in stars like the Sun were those that involved hydrogen being `burnt'
to helium. He proposed two sets of reactions -- the CNO cycles and
the proton-proton (p-p) chains. We now know that this is correct --
and that it is the p-p chains that provide the most energy in the
Sun. 

After about 20 more years of research (by many scientists) the famous
work called ``Synthesis of the Elements in Stars'' by \citet{1957RvMP...29..547B}
(also known as the B$^{2}$FH paper) really consolidated and clarified
our knowledge of the origin of the elements. They investigated all
the different types of nuclear reactions that lead to the production
of the elements: hydrogen burning, helium burning, neutron capture
reactions, etc. It had been discovered that \textbf{\emph{stars are
the major source of almost all elements}} -- except for hydrogen
and helium which were the main elements produced by the Big Bang.
Our picture of the origin of the elements has not changed much since
then, although there has been an explosion of research work into the
details of all the different types of stars. The current study is
one of these research works. 

An interesting consequence of the fact that almost all elements are
produced by stars is that everything on Earth must have, at one stage,
been made inside a star (apart from the H and He) -- including ourselves
and the planet itself! This realisation must have inspired that now
famous statement: \textbf{\emph{`We are all made of stardust'}}. We
note that the way that the stardust (or gas) gets out of the stars
is either through mass loss via stellar winds or violent explosions
of stars (i.e. supernovae). 

Thus it can be seen that stars are very important objects in the Universe
- they provide most of the light that we can see and also produce
almost all of the Elements that surround us!

\subsubsection*{Information from the Starlight}

One of the most amazing things about starlight is that it actually
contains a huge amount of information. People often ask how astronomers
know exactly what a star is made of, or how fast a star is rotating.
Both of these pieces of information come from the starlight itself.
The best example to explain this is our own star's light -- sunlight.
Everyone has seen a spectrum of light from the Sun -- through a crystal
or a rainbow for example. Sunlight is made up of all these colours.
The crystal (or water in the case of rainbows) spreads the sunlight
out so we can see the array of colours. It was originally thought
that the spread of colours was continuous -- that there were no breaks
in the pattern. However, in 1802 William Wollaston was looking at
one of these spectra and noticed that there were lots of little dark
lines in it. These are now called `Fraunhofer lines', after Joseph
Fraunhofer who investigated them in detail around 1814. It wasn't
for another 60 years before scientists finally figured out what these
lines were. The breakthrough came when \citet{1860AnP...186..161K}
made the discovery that every chemical element has a set of particular
colours that it absorbs (and emits). These are called \emph{spectral
lines.} \citeauthor{1860AnP...186..161K} then realised that these
were the lines that people were seeing in the Sun's spectrum. This
was a very important breakthrough. It opened up a huge door that has
enabled us to explore the Universe in great detail -- without having
to go anywhere in a spaceship. It turned out that, if we had a powerful
enough telescope, it was possible to get chemical information from
millions and millions of stars. Indeed, we are still using this method
today in exploring the Universe. It has given us a huge amount of
information that has enabled astronomers and astrophysicists to develop
theories about how the Universe has evolved. Finally we note that
the same light can be used to work out how fast a star is moving away
from or towards us (by using the Doppler shift) and also how fast
it is spinning. It can also be used to work out if the star has a
companion star (i.e. a binary star system) -- or even a planet!

\subsubsection*{Computer Simulations of Stars \& The Current Study}

With all this information being collected from the starlight we have
a lot of things that need to be explained. There are still many things
that we do not understand about the Universe. However we do know a
fair bit about many areas of physics. For example, we have a good
understanding of the effects of Gravity, we know a lot about nuclear
reactions (through experiments and theory), and we know a lot about
how gas behaves. Add to this the fact that these physical laws (`laws
of Nature') are, in general, easily described by the language of mathematics,
and it appears possible for us to put it all together and work out
what is happening inside a star. Indeed, this is the job description
of a stellar astrophysicist! In the early days of stellar astrophysics
scientists would do all the calculations \emph{by hand}. In some cases
the scientists had a whole `army' of people that would do the calculations
of stars by hand. Naturally the invention of `automatic digital computers'
came as a great boon to this area of research. As computers became
more capable around the end of the 1950s, many scientists applied
them to the problem of stellar evolution (eg. \citealt{1956MNRAS.116..515H};
\citealt{1959ApJ...129..489S}; \citealt{Hen59}). This lead to a
great increase in understanding of stars since we could now follow
the whole life-cycle of a star with a computer program. This is very
handy because stars take millions or billions of years to evolve in
reality. As computers became faster and faster so the stellar evolution
simulations became faster and more detailed. For example, to calculate
the life of the Sun up to its present age would have taken weeks in
the 1960s -- but now we can calculate it in a matter of minutes,
and with much greater precision.

This study is a study in stellar evolution, using computer codes as
just described. We look at particular types of stars -- stars that
were born just after the Big Bang. Thus they are the first generation
of stars that `lit up' the Universe. They are also the first generation
of stars that released newly-formed elements into the Universe (often
referred to as `pollution'). We also investigate the second generation
of stars. These two groups of stars are of particular interest at
the moment because astronomers have discovered a group of stars --
in our Milky Way galaxy -- that have seen very little pollution.
This means that they are extremely old -- they are `stellar relics'
from the early times of the Universe. Understanding them (through
stellar modelling) should give a big insight into the early stages
of the evolution of our Galaxy -- if not the Universe. Hopefully
our models help in some small way.

\section{Synopsis of the Study\label{section-Intro-CurrentStudy}}

In astronomy and astrophysics elements with atomic numbers greater
than that of hydrogen and helium are referred to as `metals'. The
Big Bang produced only hydrogen and helium (and trace amounts of heavy
elements), so the `metallicity' of a star is naturally a measure of
how much pollution its gas has been subjected to since the beginning
of the Universe. Since the material that pollutes the Universe is
primarily produced via nucleosynthesis inside stars, the rate of pollution
is dependent on the timescales of stellar evolution. The metallicity
of a star is thus often taken as an age indicator (although this does
not always hold due to other factors affecting the pollution rate,
eg. the local star formation rate). When we talk of metallicity in
what follows we shall use the traditional definition -- {[}Fe/H{]}\footnote{Where $\textrm{[Fe/H]}_{star}=\log\left(\frac{N_{Fe}}{N_{H}}\right)_{star}-\log\left(\frac{N_{Fe}}{N_{H}}\right)_{\odot}$. }.
We note that this has limitations as a star can be metal-rich in carbon
(even super-solar) but still extremely metal-poor in terms of Fe --
as seen in the carbon-enhanced metal-poor halo stars.

The main thrust of the current study centres on understanding some
of the recent -- and not so recent -- observations of low and extremely
low metallicity stars. The key question is: `Can we match theory with
observations of these very old stars?'. We investigate three populations
of low metallicity stars in particular, the first of which is (so
far) unobserved and therefore purely hypothetical:
\begin{enumerate}
\item Population III stars ($Z=0$)
\item Extremely metal-poor stars ($-6.5<\textrm{[Fe/H]}<-3.0$)
\item Mildly metal-poor Galactic Globular Cluster stars ($\textrm{[Fe/H]}=-1.4$)
\end{enumerate}
The mass range which we (primarily) explore is $0.85<M<3.0$ M$_{\odot}$.
At extremely low metallicities the upper limit at which the core He
flash occurs is only $\sim1.3$ M$_{\odot}$ (in terms of initial
stellar mass). In the current study we define stars above this mass
to be intermediate-mass stars, and those below to be low-mass stars.
Thus our mass range covers a variety of stars with different evolutionary
features.

Although no Pop III stars have been observed to date there are many
observations of extremely metal-poor (EMP) stars in the Galactic halo.
The observed numbers of these interesting stars has greatly increased
in the past decade or so, mainly due to surveys such as the Hamburg/ESO
Objective Prism Survey (HES, \citealt{CWR+99}) and the HK survey
(\citealt{BPS92}). As these stars are so metal poor -- reaching
down to $\textrm{[Fe/H]}=-5.4$ (\citealt{2005IAUS..228..207F}) --
they represent a possible link to Pop III. This is due to the fact
that the gas from which they formed could only have been enriched
by a small number of pollution events -- or even a single event.
Thus these stars may bear the chemical signatures of Pop III. Another
possibility is that these EMP halo stars may actually be members of
Pop III themselves. Their envelopes may have accreted heavy element
polluted material over their $\sim10$ Gyr lifespans. For these reasons
we have calculated the series of zero metallicity models (Chapters
\ref{chapter-Z0-StructEvoln} and \ref{Chapter-ZeroZ-Nucleo}), as
well as the grid of models having metallicities comparable to those
of the most metal-poor (EMP) halo stars (Chapter \ref{Chapter-HaloStarModels}).
For the initial composition of our EMP stars we mixed the ejecta from
a $Z=0$ supernova model with pristine Big Bang material, to mimic
the single pollution event theory mentioned above (see Section \ref{subSec-HaloStar-InitialComp}
for details).

Making detailed calculations of stars of such low metallicities provided
a set of stellar modelling problems that had to be resolved before
computation was possible. The key modification was the addition of
time-dependent mixing to the structural evolution code. This was needed
due to the occurrence of violent hydrogen flashes in these models
(which do not occur in higher metallicity models). During these events
the evolutionary timesteps become comparable to the mixing turnover
timescales, so instantaneous mixing is no longer a valid approximation.
These, and other code modifications, are detailed in Chapter \ref{sevmods}.
The numerical codes which we have used for our study are described
in Chapter \ref{Chapter-NumericalCodes}. 

A substantial amount of previous work has been done in relation to
the modelling of EMP and $Z=0$ stars. We compile a list of the literature
in Table \ref{table-LowZ-LitReview}, which provides a convenient
overview of the progress of the field at a glance. Although the structural
evolution for these stars has been partly investigated, some uncertainties
remain unresolved. We highlight these in the sections in which we
compare our results with those in the literature (eg. Sections \ref{subsec-m0.85z0-ComparePrevStudies},
\ref{subsec-m2z0-ComparePrevStudies} and \ref{section-SummaryAndCompare-HaloMods}).
Less work has been done in trying to ascertain what the nucleosynthetic
yields of these stars are. Indeed, full nucleosynthetic evolution
and yields of low- and intermediate-mass $Z=0$ and EMP models remains
quite an unexplored area. We report on the detailed nucleosynthesis
of two of our $Z=0$ models in Chapter \ref{Chapter-ZeroZ-Nucleo}
and for some of our EMP models in Section \ref{section-HaloStars-Nucleo}.
Thanks to high resolution spectroscopy now available on large (8-10
m) telescopes there are now detailed abundance observations for a
selection of the EMP halo stars. These observations provide multidimensional
constraints on the stellar modelling, since we need to match a range
of elemental abundances at once. These abundance patterns provide
new insights/clues to the nucleosynthesis occurring inside EMP stars.
We compare the nucleosynthetic results of our models with some of
these abundance patterns in Section \ref{Section-HaloStarModels-CompareObs}.
In particular we discuss the results of our models in terms of the
interesting discovery that the proportion of carbon-enhanced EMP stars
increases with decreasing metallicity (see eg. the review by \citealt{BC05}),
in Sections \ref{Section-HaloStarModels-CompareObs} and \ref{SubSec-CEMPchannels-Conclusion}.
Taking the structural and nucleosynthetic evolution together, these
models, although certainly full of uncertainties, represent the most
comprehensive investigation into EMP evolution and nucleosynthesis
to date. Although naturally also full of uncertainties, we have provided
the yields of all our models in Appendix \ref{APPX-YieldsLowZandZ0}.
These will be of use to chemical evolution studies of the early Universe.
Indeed, they are the only yields at these masses and metallicities
of which the Author is aware.

A large part of the Author's time was spent on an interesting low-metallicity
tangent. This is reflected by the presence of Chapter \ref{GC-chap},
in which we explore the intriguing mystery of the Galactic globular
cluster abundance anomalies. We quantitatively investigate the popular
theory that the abundance anomalies (eg. the Mg-Al and Na-O anticorrelations)
originate in AGB stars. On finding that `standard' AGB models have
fundamental problems with producing these anomalies we try another
tack in which we artificially `turn off' third dredge-up in the AGB
models (Section \ref{Section-GC-No3DUP}). These models prove to be
a better match to the observations but still fall short. Thus the
solution to the mystery remains as elusive as ever. We note that we
provide the yields for all of these models in Appendix \ref{Appx-GCYields}.

In Chapter \ref{Section-Background-LowMetallicityObsEvidence} we
give a brief overview of the various observations at low metallicity,
to put our modelling in context. 

Finally, in Chapter \ref{chapter-ConcludingRemarks}, we reflect on
the road that lead to this somewhat substantial body of work, and
give a list of future research directions. As mentioned there we shall
not be lacking of things to do!

\chapter{Background}
\begin{quotation}
``Our imagination is stretched to the utmost, not, as in fiction,
to imagine things which are not really there, but just to comprehend
those things which `are' there.\textquotedbl{}
\begin{flushright}
\vspace{-0.3cm}-- Richard Feynman
\par\end{flushright}

\end{quotation}

\section{Low Metallicity Observational Evidence\label{Section-Background-LowMetallicityObsEvidence}}

First we note that there has been no direct observation of any Population
III ($Z=0$) stars to date. This is probably not surprising since
we would expect stars that had been in existence for approximately
a Hubble time to have had their surfaces polluted to some degree via
accretion from the ISM. Observations of metal-poor objects are however
routinely made. These observations can provide indirect information
on the nature of the first generation of stars. Metal-poor objects
can be observed locally -- in old stellar populations -- or by `looking
back in time' at young objects/systems (ie. at high redshift). In
this section we first present the results of a literature search for
low metallicity objects, in graphical form. We then go on to describe
some of these low metallicity objects, grouped by their locations
(i.e. high or low redshift).

\subsection{Low Metallicity Diagram of the Universe}

\begin{sidewaysfigure}

\begin{center}
\includegraphics[width=0.9\columnwidth,keepaspectratio]{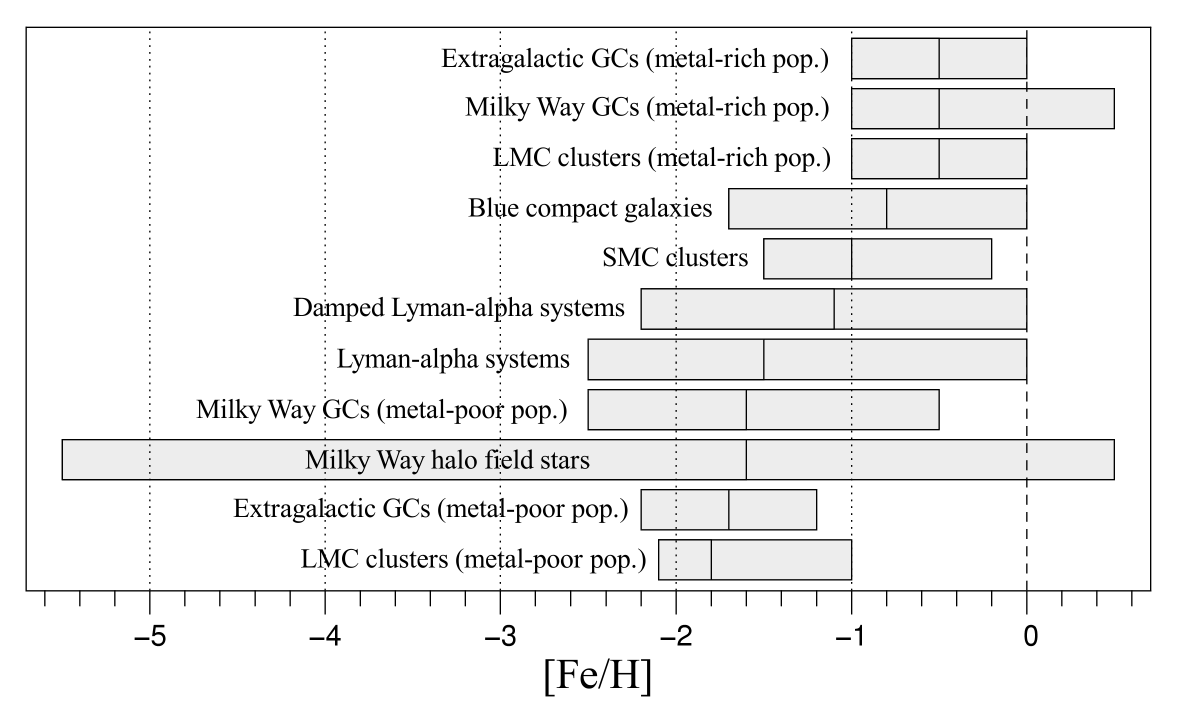}\vspace{-0.5cm}
\par\end{center}

\caption{A \emph{rough} summary of metallicity, as defined by {[}Fe/H{]}. The
object groups are plotted in order of median metallicity which is
denoted by the vertical line in each box. See text for more details
on this figure, including data sources. \label{fig-bulk-metals}}

\end{sidewaysfigure}

In Figure \ref{fig-bulk-metals} we provide a diagram that summarises
the overall metallicity picture we have of the Universe at the moment,
based on a literature search we performed during the course of the
current study. It schematically displays the ranges of metallicity
present in all low metallicity systems (of which the author is aware).
Presenting the metallicity information in this way gives an interesting
overview of the metal pollution throughout the Universe. We need to
add some caveats though. The ranges displayed are not necessarily
definitive in any of the cases -- new observations may extend the
ranges. They are also affected by observational bias. There are even
differences between results in the literature. The median values are
also somewhat uncertain but, due to the gaussian-like distributions
of most of the object groups (and the log scale), they should be more
robust than the spreads. 

In the diagram we use the traditional definition of metallicity --
{[}Fe/H{]}. As can be seen, each group actually spans at least an
order of magnitude in metallicity. The extreme case is the Galactic
halo field stars which, with the current data available, span 6 orders
of magnitude. However, the mean metallicity of this group is actually
comparable to the Galactic GCs and Damped Ly$\alpha$ systems. It
can also be seen that the populations least polluted with Fe (based
on observations to date) are the Ly$\alpha$ systems -- and the stars
that lie in the extremely metal-poor tail of the Galactic halo population
(also see Figure \ref{mdf-halo}). Iron is commonly accepted to have
two main sources -- a fast release component via type II supernovae
(SNeII) and a delayed release component via type Ia supernovae (SNe-Ia).
The delayed release ($\sim1$ to 3 Gyr) of Fe by SNe-Ia is thought
to be the reason for the {[}$\alpha$/Fe{]} gradient in the Galaxy
-- as the fresh Fe gets released, {[}$\alpha$/Fe{]} reduces to $\sim$
Solar. However this probably does not explain the extreme iron deficiency
of the field halo stars. In this case the material from which they
formed must have been subjected to a very small number of pollution
events (possibly only by a single SNII) or have been diluted to a
great degree --- or both. It is these two populations that should
have the most `direct' evidence for the nature of Population III.
The Galactic Halo stars will be most useful for comparisons with the
stellar models presented in this study (see eg. Section \vref{Section-HaloStarModels-CompareObs}
where we compare our metal-poor and $Z=0$ models with observations). 

We give some information and the data sources for the metallicity
diagram here, in order to save space in the caption:

\textbf{\emph{Details for Figure \ref{fig-bulk-metals}:}} Note that
the ends of the distributions are not necessarily definitive in any
of the cases --- new observations may extend the ranges. They are
also affected by observational bias and there are differences between
results in the literature. The median values are also somewhat uncertain
but due to the gaussian-like distributions of most of the object groups
(and the log scale) they should be more robust than the spreads. As
dust depletion of iron is significant in non-stellar sources (eg.
DLA systems), the {[}Fe/H{]} values have been increased by about 0.4
dex for these objects (which brings the values roughly in line with
the observed {[}Zn/Fe{]} abundances; see eg. \citealt{Pet04}). Also,
due to the paucity of {[}Fe/H{]} values available for Ly$\alpha$
systems, the {[}Fe/H{]} values for these systems have been derived
by scaling from the oxygen and carbon abundances assuming a scaled-Solar
metal distribution. 

\textbf{\emph{Data sources for Figure \ref{fig-bulk-metals}}} (from
bottom to top): LMC cluster {[}Fe/H{]} values are from \citet{1993ASPC...48..351O}
and \citet{OSM96} (metal-poor \& metal-rich populations), Extragalactic
GC values from \citet{BS06} (metal-poor \& metal-rich populations),
Milky Way halo field star values from \citet{2006ApJ...636..804A},
Milky Way GC values from \citet{2006AA...450..105B} and \citet{1996AJ....112.1487H}
(metal-poor \& metal-rich populations), Ly$\alpha$ forest values
from \citet{LKA01} and \citet{Pet04}, Damped Ly$\alpha$ values
from \citet{2003ApJS..147..227P} and \citet{Pet04}, SMC cluster
values from \citet{OSM96}, and Blue Compact Galaxy values from \citet{KO00}
and \citet{IT99}.

We now go on to describe some of the low metallicity groups included
in the metallicity diagram. In particular we give details on the metal-poor
Galactic halo stars and, for some perspective, the low metallicity
objects seen at high redshift. These descriptions support the large
amount of information condensed in Figure \ref{fig-bulk-metals}.

\subsection{Local Archaeology and the Metal Poor Halo Stars\label{SubSection-LowZObs-LocalUniverse}}

Low metallicity near-field objects are useful for studying the chemical
history of the early Universe. Indeed, much more detailed information
about their chemical composition can be obtained from these objects
than from the high-redshift objects because they are so close to us.
The same high-resolution instruments that are used to glean information
from high redshift objects are able to give much more detail for nearby
stars and galaxies. In the case of nearby galaxies the sample is also
not limited to very bright objects (or objects in front of bright
background objects) as it is at high redshift. The ability to sample
smaller objects (even down to individual stars) within galaxies outside
the Milky Way also contributes to the wealth of information available
for the taking. Much work has been done to try and unravel the chemical
history of the Galaxy, and information from the oldest stellar populations
in and near our Galaxy provides vital constraints on the theories.

The three most metal-poor, low-redshift groups of objects from Figure
\ref{fig-bulk-metals} are:
\begin{itemize}
\item The Milky Way Halo field stars
\item The metal-poor Globular Clusters (galactic and extragalactic)
\item The LMC clusters (low metallicity group)
\end{itemize}
The lowest metallicity objects known to date are found in the Galactic
Halo -- the extremely metal-poor Halo stars (EMP halo stars). However
the LMC clusters and metal-poor globular clusters (in the Milky Way
halo and their counterparts around other galaxies) also have metallicities
comparable to the bulk of the field Halo population. All of these
object groups provide valuable information about the early chemical
history of the Universe, and, at the low-metallicity end of their
distributions, they should provide information about the elusive Population
III. 

For the current study it is the EMP halo stars that are of most interest,
as the low-metallicity tail of their distribution represents the most
metal-poor objects known in the Universe. Indeed, \emph{the} most
metal-poor object is the EMP halo star HE 1327-2326 (\citealt{2005Natur.434..871F}).
Furthermore, much data is currently being collected for these objects,
in the form of detailed chemical compositions (eg. \citealt{2006ApJ...639..897A};
\citealt{2006AA...455..291S}; \citealt{2007AJ....133.1193B}; \citealt{2007ApJ...655..492A}).
This observational data provides stringent constraints on the theoretical
models, which we discuss later in this study. For these reasons we
now give some background information on this interesting population.

\subsubsection*{Milky Way Halo Field Stars\label{subsection-EMPHs}}

The Milky Way halo is a potential source of a great amount of information
about the low metallicity Universe. Comprising the lowest metallicity
population of our Galaxy, it contains the lowest metallicity objects
observed in the Universe to date -- with a significant number of
stars exhibiting heavy element abundances much lower than those of
Ly$\alpha$ systems at high redshift. The halo population falls into
two groups -- the field stars and the globular clusters (GCs). The
GCs are a significant low metallicity group and are the subject of
Chapter \ref{GC-chap}. Here we focus on the field star population.

\begin{figure}
\begin{centering}
\includegraphics[width=0.9\columnwidth,keepaspectratio]{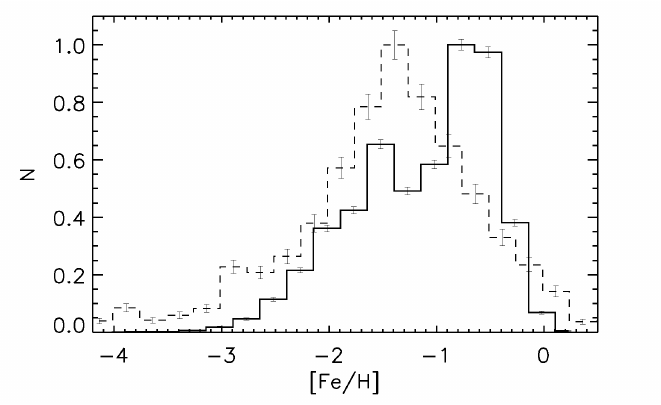}
\par\end{centering}
\caption{Metallicity versus scaled number of stars N for a region of the Milky
Way. The two groups of observations are taken at different distances
from the galactic plane: within 3 kpc (solid line) and $>8$ kpc (dashed
line). Observations are from data release 3 of the SDS. {[}From \citealt{BCN+05},
with permission{]}\label{mdf-halo}}
\end{figure}

The halo field star system is generally metal-poor but there is a
spread in {[}Fe/H{]}. \citet{BCN+05} have put together a rough metallicity
distribution function (MDF) for the halo based on the available (but
observationally biased) data. It can be seen in Figure \ref{mdf-halo}
that the halo is clearly metal-poor relative to the thick disc population
(dashed line). Future surveys with less observational bias will refine
the MDF but a significant tail of extremely metal-poor stars is evident
already. We note that the extreme tail of the MDF is where the least
polluted information about Population III should lie, as these objects
have probably only been polluted by a few (or even just one) first
generation stars. Thus the chemical composition of their envelopes
may reflect that of Pop III ejecta.

The halo field stars have the great advantage of being very close
(relative to external galaxies and DLAs for example), enabling observations
of very high resolution to be made using 8-10 m class telescopes with
spectrographs. Indeed, there has been an explosion of such observations
in the last decade as the realisation that extremely metal-poor stars
actually exist. It used to be thought that there was a lower cut-off
in the MDF at $\textrm{[Fe/H]}\sim-2.5$ -- ie. that low-mass extremely
metal-poor (or Z=0) stars never formed (eg. \citealt{Bon81}). Subsequent
surveys specifically looking at the halo population revealed lower
and lower metallicity stars, as the limiting (apparent) magnitudes
of the searches increased, indicating that \emph{low-mass very low-metallicity
stars did indeed form.} 

The two major low-metallicity halo field star surveys to date are
the HK and Hamburg/ESO surveys. Originally known as the Preston/Shectman
Survey, the HK survey is named after the method of using the H and
K calcium lines to select metal-poor stellar candidates (by eye; \citealt{BPS85},
\citeyear{BPS92}). The candidates were then targeted with follow-up
medium resolution spectroscopy to confirm if they were really Fe-poor.
To date (the data analysis is still in progress) the survey has found
$\sim1000$ halo stars with $\textrm{[Fe/H]}<-2.0$ and $\sim100$
with $\textrm{[Fe/H]}<-3.0$ (\citealt{BC05}). The more modern Hamburg/ESO
Survey (HES, originally designed as a QSO survey; \citealt{WKG+96})
is more efficient at finding low metallicity stars. It utilises an
automatic/digital selection technique and also reaches down to dimmer
stars (with a higher limiting magnitude of $B\sim17.5$ as compared
to $B\sim15.5$ in the HK survey; \citealt{CWR+99}). So far it has
uncovered $\sim200$ stars with $\textrm{[Fe/H]}<-3.0$. 

A very recent development has been the utilisation of the publicly
available data collected by the Sloan Digital Sky Survey (SDSS; \citealt{YAA+00}).
This survey was designed to collect information about QSOs and galaxies
but actually contains many Milky Way stars. \citet{BCN+05} note that
by SDSS data release 3 (SDSS DR3) there were spectrographic data for
70,000 Milky Way stars available. Analysing this wealth of data \citet{2006ApJ...636..804A}
put together a database with $T_{eff}$, $\log(g)$ and {[}Fe/H{]}
for more than 20,000 of these stars. Amongst the analysed stars were
$\sim2000$ stars with $\textrm{[Fe/H]}<-2.0$ and $\sim150$ stars
with $\textrm{[Fe/H]}<-3.0$. This dramatically increases the total
number of metal poor halo stars known, which now stands at $\sim450$
with $\textrm{[Fe/H]}\lesssim-3.0$. However, new surveys currently
underway and planned for the future will increase the sample size
by orders of magnitude. These surveys, such as the Sloan Extension
for Galactic Understanding and Exploration (SEGUE, using the same
telescope as the SDSS; \citealt{2005AAS...20714704B}), and GAIA,
the European Space Agency's 3-telescope satellite to be placed at
the L2 Earth-Sun Lagrangian point, will most likely uncover stars
of much lower metallicity than currently known -- or at least increase
the sample size at the extremely metal-poor end of the MDF. GAIA will
have a limiting magnitude of $V\sim20$ so will be able to sample
the outer halo, which may be more metal-deficient than the inner halo.
However, as \citet{BC05} note, stars of this magnitude will be very
difficult to follow up with the high resolution needed for abundance
determination (using current instruments). Also, new telescopes such
as the Chinese Large-Area Multi-Object Spectrographic Telescope (LAMOST)
will be able to perform rapid searches for metal-poor stars and also
to follow-up medium resolution spectroscopy (\citealt{2005AIPC..752..155Z}).
Thus the near future is bright for gaining low-metallicity information
from the Galactic halo stars. This should give a much deeper insight
into the nature of Population III, and the chemical evolution of the
Universe in general. 

Finally we note that higher resolution follow-up observations of EMP
stars found by the surveys gives us detailed chemical compositions
which are readily compared with theoretical stellar and chemical evolution
models. Indeed, this is what we do in Section \vref{Section-HaloStarModels-CompareObs}
where we compare our metal-poor (and $Z=0$) models with the EMP halo
star abundance observations from the literature.

\subsection{High Redshift: Information from the Young Universe}

Here we give another perspective on the low-metallicity Universe.
There are two main ways in which chemical information is gathered
about the young (high redshift) Universe. The first is by directly
analysing the light emitted by bright objects at high redshift, such
as that from quasars, Seyfert galaxies or Lyman-break galaxies. The
second is by analysing the absorption features present in the spectra
of the bright objects -- absorption that has occurred as the light
has travelled from the high-redshift objects through innumerous obstacles
(such as intergalactic gas clouds) before finally arriving at Earth. 

As the emission features of the bright objects at high redshifts show
that they are (in general) quite metal-rich, here we concentrate on
the information gathered via absorption features. 

\subsubsection*{Ancient Backlighting: Early Active Galactic Nuclei}

It is now generally agreed that quasi-stellar objects (QSOs or `quasars')
are the early Universe counterparts to current-day active galactic
nuclei (AGN). They are extremely luminous (up to $\sim$10$^{14}$
L$_{\odot}$) and the fact that some display variability on the order
of weeks implies that they must be very compact. They are thought
to be very massive accreting black holes -- the majority of their
luminosity comes from the accretion of matter rather than thermonuclear
reactions, as their spectra are dominated by non-thermal emission
(although there is often a thermal/stellar component which is now
understood to be that of the surrounding stars of the host galaxy).
As these QSOs, or galactic cores, are so luminous, and have been in
existence since at least the first 10\% of the age of the Universe,
they act as useful probes into the high redshift Universe. Indeed,
QSOs are now routinely discovered at larger and larger redshifts (eg.
in surveys such as the Sloan Digital Sky Survey (SDSS)). The highest
redshifted QSO observed to date is $J114816.64+525150.3$ which has
a redshift of 6.4, corresponding to a look-back time of $\sim$13
billion years \citep{Fea03}. 

\subsubsection*{Lyman $\alpha$ Forests}

\begin{figure}
\begin{centering}
\includegraphics[width=0.85\columnwidth,keepaspectratio]{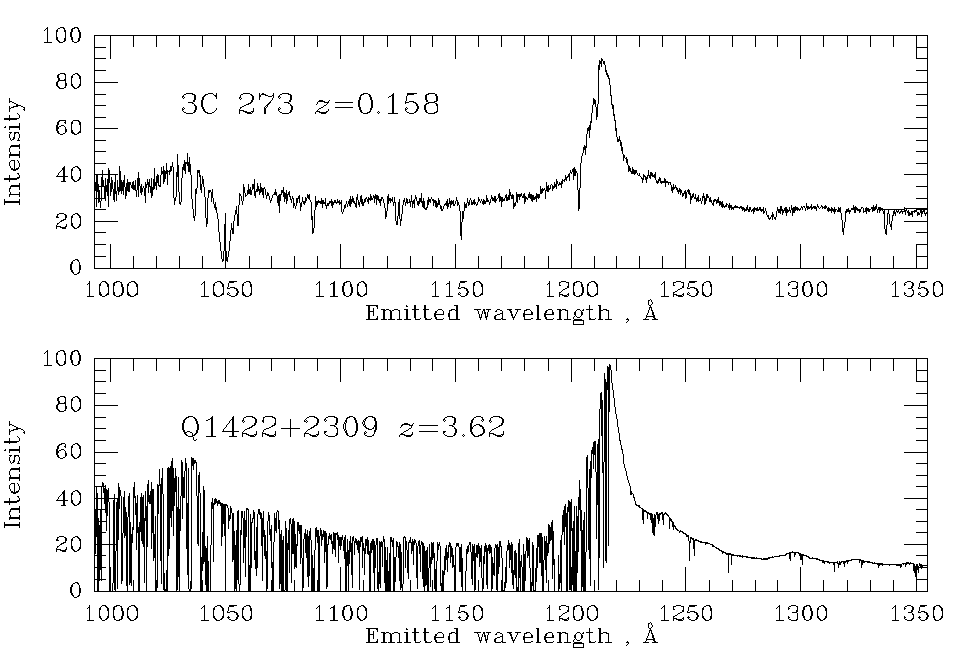}
\par\end{centering}
\caption{The spectra of two QSOs at different redshifts (from the HST and Keck
I telescopes, respectively). Both spectra have the same bandwidth
range and have been shifted so the scale is approximately in the rest
frame of the QSOs (ie. the large QSO Ly$\alpha$ emission lines are
at $\sim$1215 $\textrm{Å}$). The second peak at $\sim$1026 $\textrm{Å}$
is the QSO Lyman $\beta$ emission. Most of the absorption line features
between these peaks are due to absorbers in the line of sight (IGM,
galaxies) and are redshifted to a lesser and lesser extent as one
moves from the QSO Ly$\alpha$ emission towards shorter wavelengths.
It can be seen that there are many more absorption features at high
redshift. {[}Image courtesy of William Keel, University of Alabama,
Tuscaloosa, USA, private communication.{]}\label{fig-lyaforest}}
\end{figure}

QSOs themselves have some very interesting chemical features. For
example they exhibit strong metal emission lines (eg. O, N, Si, Fe,
\citealt{LBJ+95}) indicating that there had already been a large
amount of nucleosynthesis even by this early stage of the Universe,
at least in very dense regions. However, it is the use of QSOs to
sample the high redshift inter-galactic medium (IGM) that has provided
the most information on low metallicity systems in the early Universe.
By the time the strong continuum emission from QSOs reaches Earth,
the photons have passed through innumerous obstacles along the line
of sight. This gives rise to absorption lines against the background
continuum. Furthermore, as the QSO emission has (often) come from
the high redshift universe, it has passed through gas clouds and galaxies
located at various distances/red shifts along the line of sight. Hydrogen,
in addition to being an ubiquitous chemical species in the Universe,
also absorbs strongly in the ultraviolet (UV), notably at 1215.67
$\textrm{Å}$ -- the Lyman $\alpha$ (Ly$\alpha$) line of neutral
hydrogen (HI). Although the majority of hydrogen is actually ionised
(even in the high redshift IGM), absorption via Ly$\alpha$ is so
strong that even at low HI abundances it will absorb most of the radiation
at 1215.67 $\textrm{Å}$. As the absorbing gas clouds and galaxies
are at differing redshifts, each absorber subtracts a section of the
redshifted QSO continuum. The resultant spectrum that we observe thus
has multiple sharp HI absorption lines at various offsets to the blue,
relative to the QSO continuum, as the intervening absorbers are less
redshifted than the QSOs. This multitude of lines reflects the multitude
of Ly$\alpha$ absorbing systems that the radiation has passed through
-- and is known as the \emph{Ly$\alpha$ forest}. The Ly$\alpha$
forest was first discovered using optical telescopes. Thus the earliest
observations were actually at long look-back times as the Ly$\alpha$
lines had been shifted from the UV far into the red end of the spectrum.
It is now known that the absorbers that produce the Ly$\alpha$ forests
are of low HI column densities ($10^{12}\lesssim N(HI)\lesssim10^{16}$
cm$^{-2}$). This distinguishes them from \emph{Damped} Ly$\alpha$
systems as the absorption lines are sharp (ie. not damped). Figure
\ref{fig-lyaforest} shows examples of the Ly$\alpha$ forest at two
different redshifts. 

A new development in sampling the high redshift IGM is to use Gamma-Ray
Bursts (GRBs) as background sources instead of QSOs (\citealt{FSL+06}).
In fact all these spectrographic observations are quite a recent development
($\sim$15 years) and have been made possible by the current generation
of 10 metre class telescopes with high resolution spectrographs (eg.
HIRES on Keck I). The advent of this technology has drastically increased
the distances/redshifts at which we can gain information on chemical
abundances (see \citealt{Pet04} for a recent review), revolutionising
our understanding of the chemical evolution of the Universe. In addition
to the HI absorption there are also many metal lines observed. Ly$\alpha$
systems are the lowest metallicity systems know at high redshift,
having metallicities of $\sim10^{-2}\rightarrow10^{-3}$ times solar.
They thus represent the closest link we have to Population III at
high redshift. 

\subsubsection*{Damped Lyman $\alpha$ Systems}

Damped Lyman $\alpha$ systems (DLAs, eg. \citealt{WTS+86}) are the
higher-density analog of Ly$\alpha$ systems. Quantitatively they
have neutral hydrogen column densities: N(HI) $\gtrsim2\times10^{20}$
cm$^{-2}$. The high density causes broad absorption lines with damping
wings. The absorbers appear to be galaxies in early stages of evolution
(although this is somewhat uncertain; see eg. \citealt{Pet04} for
a review). Some have recently been associated with galaxies but it
appears some do not have detectable stellar populations. Their metallicities
are, in general, higher than Ly$\alpha$ systems but still substantially
sub-solar, while their early state of evolution is inferred from their
chemical pattern. 

\subsubsection*{Brief Summary of Pollution of the Young IGM}

The young intergalactic medium (IGM) was once thought to consist of
pristine primordial intergalactic gas clouds, unpolluted by the processes
going on in, for example, quasar host galaxies. It is now known that
much of the early IGM was polluted (through the methods just discussed
above). Depending on the cloud HI column densities, it has been shown
that up to $90\%$ of clouds at $z\sim3$ are polluted with metals
such as silicon, carbon and nitrogen (\citealt{SC96}). Indeed, Figure
1 in \citealt{SC96} shows the `C-IV forest' -- a multitude of instances
of a particular carbon line at different redshifts, analogous to the
Ly$\alpha$ forest (the C-IV doublet, at $\sim1550$ $\textrm{Å}$,
is redward of Ly$\alpha$). The metal abundances in the high-redshift
clouds have been found to be $\sim10^{-1}\rightarrow10^{-2}$ solar
(eg. \citealt{Son01}). In terms of their abundance \emph{pattern}
they are relatively high in silicon as compared to carbon -- at a
redshift of $\sim$3 the Si/C ratio is $\sim$3 times solar (\citealt{SC96}).
This overabundance of Si implies that the stellar source of the pollution
was predominantly Type II supernovae (SNeII). Interestingly, this
abundance pattern is also reminiscent of the Galactic Halo field population.
Despite this clue, the origin of the metal pollution is still unresolved.
Some authors suggest that it originates from galactic outflows --
galactic winds that, at early epochs, were strong enough to pollute
the IGM (eg. \citealt{ASS+05}), some argue that tidal stripping of
galaxies is the only way to inject so much metal into the IGM (eg.
\citealt{Gne98}), and others argue that there was a \emph{pregalactic}
population of stars that initially polluted the Universe at a redshift
$\gg6$. Recent observations that show that metal abundances are roughly
constant with redshift ($1.5<z<5.5$, \citealt{Son01}) -- within
clouds of similar density -- appear to support the hypothesis that
the metal pollution was implanted at a very early epoch. \citet{Son01}
also notes that the metallicity is so high (carbon $\sim10^{-2}$
solar) -- even at $z\sim5$ -- that the majority of the pollution
must have occurred before this epoch. It would be interesting to compare
the pollution of the young IGM with that of the EMP Galactic halo
stellar population, particularly since so much carbon is seen in both.
This is however outside the scope of the current study so we shall
investigate this at a later date.

\section{Big Bang Nucleosynthesis\label{sec-BBN}}

\subsection{Background}

\begin{figure}
\begin{centering}
\includegraphics[width=0.7\columnwidth,keepaspectratio]{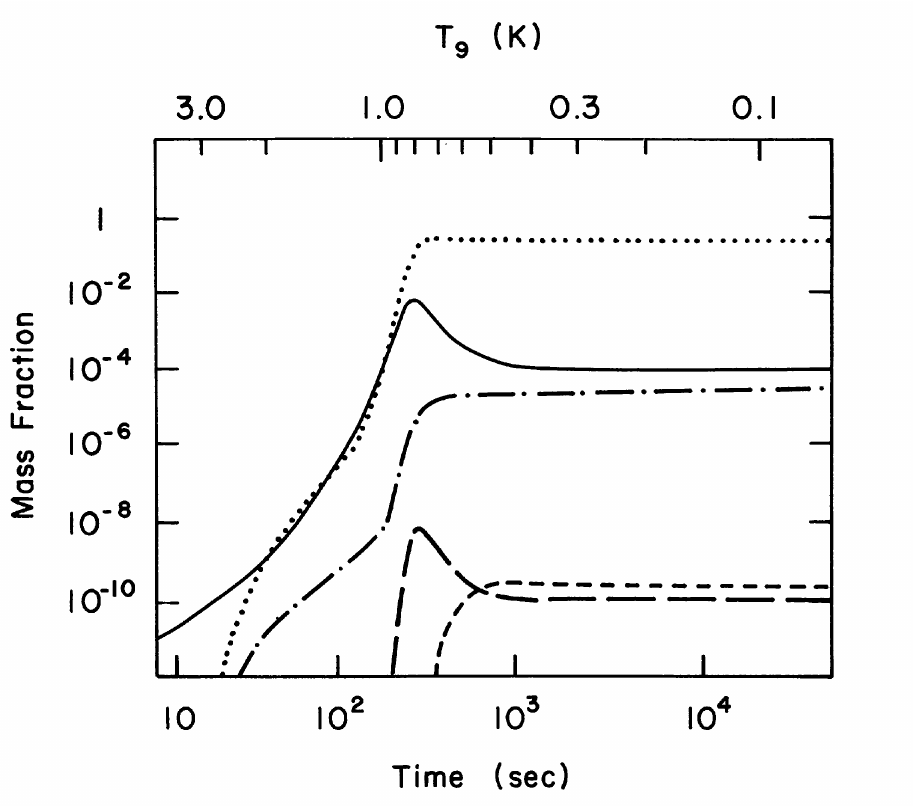}
\par\end{centering}
\caption{The abundance evolution from a Big Bang nucleosynthesis calculation.
Nuclides shown are $^{4}$He (dotted line), $^{2}$H (solid), $^{3}$He
(dash-dotted), $^{7}$Li (long dashed) and $^{7}$Be (short dashed).
This figure is from the annual review by \citet{Boes0985}. \label{bbn}}
\end{figure}

Apart from predicting the well-known Cosmic Background Radiation,
Big Bang Theory also predicts a rapid nucleosynthesis of elements
in the first few minutes of the Universe. Big Bang Nucleosynthesis
(BBN) is a mature area of research, such that the main elemental predictions
have been known since the 1960s (see eg. \citealt{Boes0985} for a
review). Figure \ref{bbn} shows the results of a typical BBN calculation.
As can be seen in the Figure, most of the nucleosynthesis happens
in a brief window of time -- when the density and temperature are
conducive to nuclear reactions. As the universe rapidly expands and
cools nuclear reactions are no longer possible and the primordial
abundance pattern is essentially 'frozen' in. It remains like this
until (local) temperatures and densities are again high enough for
nuclear reactions -- usually in stars. 

In the overall picture of the chemical evolution of the Universe it
is this primordial mix of isotopes that everything else stems from.
The first generation of stars form from this gas and their subsequent
evolution is primarily governed by this chemical make-up. Importantly,
the nucleosynthesis of fresh nuclei that occurs in these stars is
then (usually) released into the interstellar environment via winds
or supernovae explosions, polluting the ambient gas that will, in
turn, go into the formation of the next generation of stars. Thus
it is important to have the initial composition of the first generation
correct. There is some debate about the extent to which BBN occurs
in a homogeneous gas. Standard BBN (SBBN) assumes the primordial gas
is homogeneous. Inhomogeneous BBN (IBBN) theory allows for some regions
of higher density in which more advanced nucleosynthesis can occur.
Higher abundances of nuclei crucial to determining stellar structure
can be produced in IBBN (eg. $^{12}$C). Recent studies into IBBN
even suggest that \emph{very} heavy nuclei can be formed -- through
the p-process and r-process. In their IBBN simulations \citet{Mats1205}
produce heavy elements such as $^{158}$Eu and $^{90}$Mo at the mass
fraction level of $\sim10^{-8}$. However the size of the regions
in which this happens must be kept small so as to stay within the
observational evidence constraints. Indeed, current observations of
the anisotropies in the Cosmic Microwave Background Radiation (CMB)
by the Wilkinson Microwave Anisotropy Probe (WMAP) satellite suggest
that SBBN theory is sufficient to explain the primordial abundances
(\citealt{LKM06}), even though there may have been small levels of
inhomogeneity. For this reason we use abundance predictions from homogeneous/standard
BBN theory in the current study.

\subsection{Primordial Abundances for the Stellar Models}

The most abundant nuclei produced by SBBN are $^{1}$H, $^{2}$H,
$^{3}$He, $^{4}$He and $^{7}$Li. The recent observations by WMAP
have significantly constrained the predictions of SBBN. In Table \ref{bbnabunds}
we list all the starting abundances used for the stellar structure
models in this study, which are taken from the SBBN calculations of
\citet{Coc0104}. 

\begin{table}
\begin{centering}
\begin{tabular}{|c|c|}
\hline 
Nuclide & Primordial Mass Fraction\tabularnewline
\hline 
\hline 
$^{1}$H & $0.754992$ $(0.754796)$\tabularnewline
\hline 
$^{2}$H & $(1.96\times10^{-4})$\tabularnewline
\hline 
$^{3}$He & $7.85\times10^{-6}$\tabularnewline
\hline 
$^{4}$He & $0.24500$\tabularnewline
\hline 
$^{7}$Li & $(3.13\times10^{-10})$\tabularnewline
\hline 
$^{12}$C & $0.0$\tabularnewline
\hline 
$^{14}$N & $0.0$\tabularnewline
\hline 
$^{16}$O & $0.0$\tabularnewline
\hline 
\end{tabular}
\par\end{centering}
\caption{Initial abundances used in the zero-metallicity stellar models, as
given by the Standard Big Bang Nucleosynthesis calculations by \citet{Coc0104}.
This material was also used to dilute SNe ejecta to give the initial
composition of our Halo star models. Values in brackets are the abundances
used in the nucleosynthesis calculations. $^{7}$Li and $^{2}$H are
only used in the nucleosynthesis calculations, as these species are
not included in the structure code. \label{bbnabunds}}
\end{table}

Of particular interest is the primordial abundance of $^{4}$He (Y$_{P}$).
A recent review and study by \citet{Olive1204} in light of the recent
WMAP observations (from which $\eta$, the primordial baryon to photon
ratio, is inferred) finds that Y$_{P}$ gained through observations
actually has large errors associated with it. However, even taking
into account the errors, the observed Y$_{P}$ is easily within the
error bars of Y$_{P}$ given by SBBN calculations, as $\eta$ (the
only parameter in the calculations) also has an uncertainty -- despite
the very detailed observations of WMAP. On reanalysing the observational
data \citet{Olive1204} find $0.232\leq Y_{P}\leq0.258$. Using the
WMAP constraint on $\eta$, many authors have revised the SBBN calculations.
Using SBBN combined with the WMAP constraint \citet{Coc0104} finds
$Y_{P}=0.2479\pm0.0004$. \citet{Cyb0402} find $Y_{P}=0.2484_{-0.0005}^{+0.0004}$.
For the current study we adopt a value of $Y_{P}=0.245$ which is
roughly in the centre of the observational error bars found by \citet{Olive1204}
and close to the values given by SBBN + WMAP calculations. 

$^{7}Li$ has long been a contentious issue between observers and
SBBN simulators (see eg. the review by \citealt{Boes0985}). To the
current day there remains a strong discrepancy between the observations
and the SBBN calculations. Primordial $^{7}Li$ is thought to be visible
on the surface of (hot) metal-poor Galactic halo stars. The `Spite
Plateau', observed by \citet{SS82}, shows that there is a very constant
abundance of $^{7}Li$ across these stars. It is this plateau value
that is thought to be the primordial abundance, as there is such a
narrow spread between stars (of varying metallicity) that it seems
unlikely that any other process could have produced it. We would expect
other processes, such as production in SNe or AGB stars, to have produced
a much greater, stochastic, spread. SBBN calculations, which have
multiple constraints, predict a higher primordial abundance of $^{7}$Li
-- by a factor of $\sim3$. Although not a very large factor it is
still large enough to lie outside the observational errors. As the
situation is still unresolved, in the current study we choose to adopt
the SBBN prediction for $^{7}$Li. This decision is lent some weight
by the fact that SBBN predictions are supported by other observational
evidence, such as $^{2}$H (and to some extent $^{4}$He). 

\subsubsection*{Primordial Abundances of Heavier Nuclei}

If trace amounts of heavier nuclei were available in the early Universe
this may have a very significant effect on the evolution of zero metallicity
stars. In particular, if $^{12}$C is present at the (mass fraction)
level of $~10^{-12}\rightarrow10^{-10}$ then the CNO cycle will be
able to operate on the main sequence and during H-shell burning --
rather than just the p-p chains. Depending on the actual abundance
this would affect the structural evolution, as the efficiency of the
CNO cycles increases with the abundance of CNO nuclei. However it
is generally accepted that SBBN does not not produce significant amounts
of nuclei with atomic mass $>7$. For example, \citet{VCC00} include
reactions involving B, Be and C but find that the mass fractions of
B and Be produced are of the order $10^{-16}$or less. This is 3 orders
of magnitude below the lowest abundances of these elements that has
been observed (in Galactic Halo stars, see eg. \citealt{BK93}).

\section[Previous Theoretical Models of $Z=0$ and EMP Stars]{Previous Theoretical Models of $Z=0$ and Extremely Metal-Poor Stars\label{Section-PreviousModels}}

\subsection{Introduction}

Here we give an overview of the literature on the subject of zero
and very-low metallicity stellar modelling, from its beginnings to
the present day. We focus in particular on low- and intermediate-mass
stars, as this is the range of interest for the current study. No
literature survey is perfect -- we may have missed a paper or two
in our survey -- but we believe the survey does cover the vast majority
of the papers published. Importantly, the main developments over the
$\sim50$ year history of the subject are provided, through the descriptions
of the main findings of each paper. 

Table \ref{table-LowZ-LitReview} lists the studies covered in our
review. In the table we have also provided some key details of the
studies, namely the mass and metallicity ranges of the models, the
stages of evolution reached, and the helium content used. The table
thus provides a quick overview of the development of the field at
a glance. 

The whole review is organised chronologically (in the table and the
text). The text is divided into three sections. The first section
details the early developments whilst the second section details developments
in the 1980s, 1990s and early 2000s. The third section is delineated
by the start date and end date of the Author's thesis candidature
(i.e. by the time-frame of the current study). This choice highlights
the fact that much work was done in the field during this period,
as it has recently been (and still is) a `hot topic' in stellar astrophysics.
This is mainly due to the recent discovery of some extremely metal-poor
Galactic halo stars, particularly the C-rich star HE0107-5240 (\citealt{2002Natur.419..904C}). 

We also provide a fourth section that places the current study into
the context of the development of the field.

\subsection{ZAMS and Beyond: 1960s to 1980s}

\subsection*{The Early Days}

Around the same time that it was realised that the newly invented
`automatic digital computers' could be applied to the problem of stellar
evolution (eg. \citealt{1956MNRAS.116..515H}; \citealt{1959ApJ...129..489S};
\citealt{Hen59}; \citealt{1962ApJ...136..150S}), the watershed work
of \citet{1957RvMP...29..547B} (now known as the B$^{2}$FH paper)
detailed eight nucleosynthetic paths to provide weight to the argument
that stars are the the origin of most elements in the Universe. Noting
that this theory did not explain the production of hydrogen in the
Universe, \citet{1961ApJ...133..159E} constructed some potential
primordial ZAMS models composed solely of H (in the mass range $1-2000$
M$_{\odot}$). Although the only nuclear reactions included in her
models were those of the p-p chains, Ezer found that the central temperatures
reached were high enough to initiate the 3$\alpha$ process, thereby
producing primary $^{12}$C and allowing the CNO cycles to operate. 

With the maturing of Big Bang Nucleosynthesis theory (BBN, eg. \citealt{1966ApJ...146..542P};
\citealt{1973ApJ...179..343W}) and new observations (eg. \citealt{1972ApJ...173...25S})
it appeared that helium was also one of the main components of the
primordial gas, and thus stellar modellers began constructing primordial
ZAMS models with metal-free compositions containing $\sim20\%$ He
and $\sim80\%$ H (eg. \citealt{1971ApSS..14..399E}; \citealt{1974ApSS..31....3C};
\citealt{1975ApSS..35..185C}).

\citet{1972ApSS..18..226E} appears to be the first to evolve a primordial
stellar model past the main sequence, taking the evolution of a 3
M$_{\odot}$, $Z=0$ star through the core He burning stage. Ezer
found that the star spent most of its time on the blue side of the
HR diagram and produced $^{12}$C in the core towards the end of the
MS. She also found that core He ignition occurred while the star was
still on the blue side of the HR diagram, thereby preventing the star
from reaching the RGB phase. In terms of evolutionary timescales she
found that the MS lifetime was considerably longer than that of Pop
I models but also that the core He burning phase was a factor of 3
\emph{shorter}. Although the details of some evolutionary features
would change in the future, due to improved input physics, the majority
of her findings for this model are still indicative of our understanding
of $Z=0$ intermediate mass (IM) stellar evolution today.

At low masses (LM) \citet{1974ApJ...191..173W} calculated models
of metal-poor and extremely metal-poor stars in the mass range $0.65\rightarrow2.5$
M$_{\odot}$ up to and including the RGB in order to investigate their
stellar lifetimes and pulsational stability (see Table \ref{table-LowZ-LitReview}
for the metallicity range). Wagner found that the stellar lifetimes
were dependent on the metallicity $Z$ and also the H abundance. Low
metallicity models had shorter lifetimes than Pop I models if the
H content was the same, but lifetimes increased if H was increased. 

It wasn't until almost decade later that the first true $Z=0$ LM
models of advanced stages of evolution were calculated. \citet{1982AA...115L...1D}
followed the evolution of 0.9 M$_{\odot}$ and 1 M$_{\odot}$ models
up to the end of the RGB. They noticed an important feature of these
models -- that the tip of the RGB is $\sim$1 dex less luminous than
stars of Pop I or II. They suggested that this would not affect the
observability of primordial low-mass stars as the SGB/RGB lifetime
was longer due to the reliance on pp-chain energy production (we note
that the lifetime prediction is at odds with the current study's findings).
Another very important observation they made was that the core He
flash began well off centre. They go on to suggest that it is so off
centre that the He convection zone may breach the H-He discontinuity
and consequently pollute the surface with C and N. This appears to
be the first mention of this evolutionary feature in the literature.
This phenomenon would be `re-discovered' at the beginning of the 1990s,
with \citet{1990ApJ...349..580F} showing that the He convection zone
does indeed break through, causing a H-flash. We refer to this event
as the Dual Core Flash (DCF). 

Meanwhile \citet{1981ApSS..79..265E} continued the evolution of her
3 M$_{\odot}$ star onto the early AGB, finding that the shell burning
phase also proceeded at a slow rate in IMS, extending the lifetime
as compared to Pop II stars.

A few years later \citet{1984ApJ...287..749F} performed semi-analytic
calculations to determine if $Z=0$ stars would pulsate on the AGB
and hence possibly contribute to the early pollution of the Universe
(via third dredge-up (3DUP)). They predicted that stars with mass
$\lesssim4$ M$_{\odot}$ would indeed develop thermal pulses on the
AGB and consequently dredge up $^{12}$C to their surfaces. For higher
masses they predicted that thermal pulses would \emph{not} occur --
unless there had been some prior pollution of their envelopes that
had brought the $Z$ content above some critical amount ($\sim10^{-7}\rightarrow10^{-4},$
dependent on core mass). 

Later that year \citet{1984ApJ...287..745C} made a full calculation
of a 5 M$_{\odot}$ model, following it part of the way into the AGB
phase. They found that no thermal pulses or third dredge-up (3DUP)
occurred, as predicted by \citet{1984ApJ...287..749F}. They attributed
this to the fact that 3$\alpha$ reactions occur at the bottom of
the H shell in $Z=0$ models, thus the two shells are both burning
He and therefore they move outwards (in mass) at the same rate, preventing
the development of an instability. 

The 1980s continued to be a busy decade for primordial stellar evolution,
although the later stage of evolution, namely the TP-AGB, remained
unexplored, as did the core He flash in the low mass models. \citet{1985ApSS.117...95E}
extended their work by calculating the evolution of more massive IMSs
(M = $5,7,9$ M$_{\odot}$) with $Z=0$ to core He exhaustion and
\citet{1986MNRAS.220..529T} explored the mass interval $2.5>M>8.0$
M$_{\odot}$ for stars of low and extremely low $Z$ (although mainly
to ascertain the nature of metal-poor Type I supernovae). In 1987
\citeauthor{1987ApSS.136...83K} also calculated MS $Z=0$ models
of very low mass stars (M = $0.2\rightarrow0.8$ M$_{\odot}$). 

We give some information and the data sources for the literature review
table here, in order to save space in the caption:

\textbf{\emph{Details for Table \ref{table-LowZ-LitReview}:}} Mass
is given in units of M$_{\odot}$ (an arrow indicates a range of masses).
Metallicity is given as $\log(Z/Z_{\odot})$ except when $Z=0$ where
we write `zero'. Using $\log(Z/Z_{\odot})$ gives a rough approximation
to {[}Fe/H{]}, which is useful for comparisons with observational
studies. In column 5 we show the maximum stage of evolution that the
models were evolved \emph{through} (so it is an inclusive designation).
The `$\sim$' symbol is used where the models were evolved only part
of the way through the indicated phase of evolution. In the last column
we show the helium mass fraction used. {[}Abbreviations for the evolutionary
stages are: CHeB (core helium burning), DCF (dual core flash, the
H-He core flash in low mass low-Z stars, also known as HEFM in the
literature), SRGB (secondary RGB), DSF (dual shell flash, the H-He
shell flash near the start of the AGB in low- and intermediate-mass
stars, also known as HCE).{]}

\begin{sidewaystable}

\begin{center}

\scriptsize

\begin{tabular}{|c|c|c|c|c|c|}
\hline 
 Year &  Author & Mass & Metallicity & Max. Evolution & Helium\tabularnewline
\hline 
\hline 
1961 & \citet{1961ApJ...133..159E} & $1\rightarrow2000$ & zero & ZAMS & 0.0\tabularnewline
\hline 
1971 & \citeauthor{1971ApSS..14..399E} & $5\rightarrow200$ & zero & MS & 0.2\tabularnewline
\hline 
1972 & \citeauthor{1972ApSS..18..226E} & 3.0 & zero & CHeB & 0.2\tabularnewline
\hline 
1974 & \citeauthor{1974ApSS..31....3C} & $2\rightarrow20$ & zero & ZAMS & $0.0,0.23,0.3$\tabularnewline
\hline 
1974 & \citeauthor{1974ApJ...191..173W} & $0.65\rightarrow2.5$ & $-6,-4,-2$ & RGB & 0.26\tabularnewline
\hline 
1975 & \citeauthor{1975ApSS..35..185C} & $1\rightarrow100$ & zero & ZAMS & $0.0,0.2,0.4$\tabularnewline
\hline 
1981 & \citeauthor{1981ApSS..79..265E} & 3.0 & zero & EAGB & 0.2\tabularnewline
\hline 
1982 & \citeauthor{1982AA...115L...1D} & $0.9,1.0$ & zero & RGB, $\sim$DCF & 0.2\tabularnewline
\hline 
1983 & \citeauthor{1983AA...118..262G} & 0.9 & zero & MS & 0.2\tabularnewline
\hline 
1984 & \citeauthor{1984ApJ...287..745C} & 5.0 & zero & $\sim$AGB & 0.2\tabularnewline
\hline 
1985 & \citeauthor{1985ApSS.117...95E} & $5,7,9$ & zero & CHeB & 0.2\tabularnewline
\hline 
1986 & \citeauthor{1986MNRAS.220..529T} & $2.5\rightarrow8.0$ & $-6,-4,-2$ & EAGB  & 0.2\tabularnewline
\hline 
1987 & \citeauthor{1987ApSS.136...83K} & $0.2\rightarrow0.8$ & zero & MS & 0.2\tabularnewline
\hline 
1990 & \citeauthor{1990ApJ...349..580F} & 1.0 & zero & $\sim$DCF & 0.23\tabularnewline
\hline 
1990 & \citeauthor{1990ApJ...351..245H} & 1.0 & zero & DCF & 0.23\tabularnewline
\hline 
1993 & \citeauthor{1993ApJS...88..509C} & $0.7\rightarrow15$ & $-8,-4,-2$ & RGB/EAGB & 0.23\tabularnewline
\hline 
1996 & \citeauthor{1996ApJ...459..298C} & $0.7\rightarrow1.1$ & $-8,-4,-3$ & $\sim$DSF (M=0.8) & 0.23\tabularnewline
\hline 
1998 & \citeauthor{1998IAUS..189P.150F} & $0.8\rightarrow4$ & zero & RGB/AGB & 0.23\tabularnewline
\hline 
2000 & \citeauthor{2000ApJ...529L..25F} & $0.8\rightarrow4$ & zero,$-4,-2$ & DCF/DSF & 0.23\tabularnewline
\hline 
2000 & \citeauthor{2000ApJ...533..413W} & $0.8\rightarrow1.2$ & zero & RGB & 0.23\tabularnewline
\hline 
2001 & \citeauthor{2001AA...371..152M} & $0.7\rightarrow100$ & zero & $\sim$AGB (no CHeF)\footnote{The low mass stars were not evolved through the CHeF but zero age
HB branch models were constructed to enable evolution to the AGB.} & 0.23\tabularnewline
\hline 
2001 & \citeauthor{2001ApJ...554.1159C} & $4\rightarrow8$ & zero & DSF, $\sim$AGB & 0.23\tabularnewline
\hline 
2001 & \citeauthor{2001ApJ...559.1082S} & $0.8\rightarrow1.0$ & zero & DCF, SRGB, $\sim$AGB & $0.23\rightarrow0.25$\tabularnewline
\hline 
2001 & \citeauthor{2001AA...378L..25G} & 3 & zero & AGB s-process\footnote{The s-process nucleosynthesis was done using a separate nucleosynthesis
code (using a TPAGB stellar model for input physical conditions).
A (parameterised) partial mixing down of protons was added to provide
a $^{13}$C pocket.} & 0.23\tabularnewline
\hline 
2002 & \citeauthor{2002ApJ...570..329S} & $0.8\rightarrow20$ & zero & DSF, $\sim$AGB & 0.235\tabularnewline
\hline 
2002 & \citeauthor{2002AA...395...77S} & 0.8 & zero, $-3,-2$ & DCF & 0.23\tabularnewline
\hline 
2003 & \citeauthor{2003ASPC..304..318H} & 2 \& 5 & zero & DSF, $\sim$AGB & 0.23?\tabularnewline
\hline 
2004 & \citeauthor{2004ApJ...602..377I} & $1\rightarrow3$ & -2.7 & DSF, $\sim$AGB & 0.24\tabularnewline
\hline 
2004 & \citeauthor{2004ApJ...609.1035P} & $0.8\rightarrow1.5$ & zero,$-6,-5,-4$ & DCF, SRGB, $\sim$EAGB & 0.23 \& 0.27\tabularnewline
\hline 
2004 & \citeauthor{2004AA...422..217W} & 0.82 & zero, -5 & DCF, SRGB & 0.23?\tabularnewline
\hline 
2004 & \citeauthor{2004ApJ...611..476S} & $0.8\rightarrow4.0$ & zero & $\sim$DSF, $\sim$DSF, $\sim$AGB & 0.23?\tabularnewline
\hline 
2006 & Campbell (This study) & $0.8\rightarrow3+$\footnote{Some models with $M=4$ \& 5 M$_{\odot}$ were also calculated but
only part of the way through the AGB.} & zero, $-6,-5,-4,-3$ & DCF, DSF, SRGB, AGB \& Yields & 0.245\tabularnewline
\hline 
\end{tabular}\end{center}\vspace{-0.4cm}

\caption{A summary of the literature (to the best of our knowledge) for theoretical
studies of low- and intermediate-mass stars of very low- and zero-metallicity.
See text for details on this table.\label{table-LowZ-LitReview}}

\end{sidewaystable}

\normalsize

\subsection{DSF Discovery and DCF Modelling: 1990s to 2001}

It wasn't until the 1990s that the modelling of the most demanding
of the peculiar evolutionary traits of $Z=0$ stars was attempted
-- the Dual Core Flash (DCF). Although discovered almost a decade
earlier (\citealt{1982AA...115L...1D}) no work had been published
since on the topic. In their first paper on the subject \citet{1990ApJ...349..580F}
evolved a 1 M$_{\odot}$ star through to just after the peak of the
core He flash at the tip of the RGB. As reported by \citet{1982AA...115L...1D}
they also found that the core He flash occurs much further off-centre
than in more metal-rich models. This has the consequence that the
resulting He convective zone breaks through the H-He discontinuity,
mixing down protons into regions of very high temperature. Although
their calculations did not follow the mixing in this paper, the same
group modified their evolution code to include time-dependent mixing
and subsequently followed the H-flash that ensued (\citealt{1990ApJ...351..245H}).
They found that the protons were mixed down to a point where the timescale
of proton capture is much shorter than that of the mixing timescale.
Once the H luminosity exceeded that of the He flash luminosity the
convection zone split into two -- resulting in an upper H burning
convective zone and a lower He burning one. Both convection zones
contain $^{12}$C whilst the upper one also contains CNO cycle burning
products. This is of significance because they also found that the
convective envelope moves inwards after the two flashes have abated
and dredges up some of this polluted material to the surface. They
found that the initially metal-free star now had a metallicity of
$Z_{cno}\sim0.004$ and noted that the high N abundance in a newly
observed extremely metal-poor giant star (CD -38$^{\circ}245$, \citealt{1984ApJ...285..622B})
may be due to the same mechanism (although their $Z=0$ model produced
too much N). 

A few years later \citet{1993ApJS...88..509C} published a grid of
calculations of very- and extremely-metal-poor models covering a large
mass range ($0.7>M>15$ M$_{\odot}$, also see Table \ref{table-LowZ-LitReview}).
Their models were evolved up to the end of the RGB (not including
the DCF phase) for low mass stars, and through to core He exhaustion
in the more massive models. They found that the value of $Z$ remains
important in the efficiency of the CNO cycles down to $Z\sim10^{-6}$
and that all their models with $Z=10^{-10}$ succeeded in self-producing
$^{12}$C before core H exhaustion, enabling the CNO cycles to operate.
They also note that the maximum stellar mass at which the core He
flash occurs (due to degeneracy in the core) decreases substantially
with $Z$. At solar $Z$ the cut-off is $\sim2.3$ M$_{\odot}$ whilst
at $Z=10^{-10}$ they find it to be $\sim1.2$ M$_{\odot}$. As \citeauthor{1993ApJS...88..509C}
mention, this implies that very low metallicity populations may be
more blue in colour (we note however that also depends on the IMF).
The same group (\citealt{1996ApJ...459..298C}) also investigated
low mass stars ($0.7>M>1.1$ M$_{\odot}$) at similar metallicities.
They produced isochrones from their models and also found that very
few RR Lyrae variables would be expected at very low metallicity.
A model of mass $0.80$ M$_{\odot}$ and $Z=10^{-10}$ was also followed
to the onset of thermal pulses on the AGB. The core He flash was not
modelled so the DCF event was missed. Instead a ZAHB model was constructed
to follow the later evolutionary stages. It was found that the He
convective zone produced by the first (strong) AGB He shell flash
was likely to break through the H-He discontinuity, dredging down
protons to high temperatures and therefore causing a secondary (hydrogen)
flash to occur. This appears to be the first report of what we refer
to as a Dual Shell Flash (DSF). We use the term `dual' because the
event is characterised by a double peak in luminosity -- one from
the He flash and one from the H flash -- occurring quite close together
in time. They were however unable to follow the evolution further
as the code did not include time-dependent mixing. They likened the
event to the DCF event described in the previous section (see eg.
\citealt{1990ApJ...351..245H}), as the He convective zone (a result
of the normal TP-AGB He flash) penetrates the H-He discontinuity in
both cases, with similar results.

Moving back to $Z=0$ models, \citet{1998IAUS..189P.150F} brought
the overall $Z=0$ evolutionary picture into more focus. Whilst arguing
that the metal deficient dwarf carbon star G77-61 (observed by \citealt{1988AA...189..194G})
was the result of a binary star mass transfer event from a companion
population III star, they divide the peculiar evolutionary traits
of $Z=0$ stars into three distinct cases. The three cases are delineated
by stellar mass:
\begin{enumerate}
\item $M\lesssim1.1$ M$_{\odot}$: Dredge-up after the Dual Core Flash
pollutes the surface with C, N and He. Stars \emph{may} go on to experience
3DUP on the AGB, further polluting the surface. 
\item $1.2\lesssim M\lesssim4.0$ M$_{\odot}$: Dredge-up after the Dual
Shell Flash at the start of the TP-AGB pollutes the surface with C,
N and He. Stars may go on to experience 3DUP on the AGB, further polluting
the surface. 
\item $M\gtrsim4.0$ M$_{\odot}$: No DCF or DSF -- surface remains unpolluted
during entire evolution.
\end{enumerate}
Note that the terminology used here (i.e. DCF and DSF) is only used
in the current study. 

Motivated by the growing number of observations of very metal deficient
stars (eg. from the HK survey, \citealt{BPS92}), the same group (\citealt{2000ApJ...529L..25F})
extended their investigation to models of low and very low metallicity
stars. In doing this they improved on their classification scheme,
taking into account the metallicity dependence of the occurrence of
the DCF and DSF. In Figure 2 of their paper \citet{2000ApJ...529L..25F}
provide a schematic diagram in the mass-metallicity plane. It can
be seen in this figure that the phenomena of DCFs and DSFs continue
up to (relatively) metal-rich models. Case I (in which the DCF in
low mass stars leads to pollution of the surface) is however limited
to a small mass range, and, in particular, very low metallicities.
They argue that the stellar models can \emph{qualitatively} explain
the observed increased occurrence of C-rich extremely metal-poor stars
(CEMPs), especially when taking binary mass transfer into consideration.
We note that we also provide a mass-metallicity-pollution diagram
in Section \vref{section-HaloStarModelsPollutionSummary}, which summarises
the pollution results of all our $Z=0$ and EMP models. Our models
are evolved through their entire evolution (i.e. to the end of the
AGB), including detailed nucleosynthesis, so it is the results from
the yields that we present in our diagram.

Motivated by the same observations \citet{2000ApJ...533..413W} investigated
how the surface of a low-mass $Z=0$ MS or RGB star could be polluted
to result in a match to the observations . They examined two scenarios,
one where the hot p-p chain reactions were taken into account, whereby
extra $^{12}$C may have been produced, altering the evolution, and
one where the star is polluted from a nearby massive star. In the
first scenario they revealed that the hot p-p chains are unimportant
in these stars (the 3$\alpha$ reaction is very dominant in terms
of producing $^{12}$C). In the second scenario they found that the
extra metals never diffuse down to the regions in which energy generation
occurs. Interestingly \citeauthor{2000ApJ...533..413W} did \emph{not}
find that a DCF occurs in their 1 M$_{\odot}$, $Z=0$ model. This
is at odds with previous studies (eg. \citealt{1982AA...115L...1D};
\citealt{1990ApJ...351..245H}). We note however that the same group
later report that their models do experience the DCF (\citealt{2001ApJ...559.1082S};
\citealt{2004AA...422..217W}). 

In 2001 \citeauthor{2001AA...371..152M} published the largest grid
of $Z=0$ models to date. A fine grid in mass was used, with a mass
range of 0.7 M$_{\odot}$ to 100 M$_{\odot}$. Mass loss was not taken
into account. In terms of low and intermediate mass stars the evolution
was followed up till the beginning of the AGB. In two cases ($M=2.5$
and 5.0 M$_{\odot}$) the evolution was taken a short way into the
TP-AGB. They found no evidence of thermal pulses in the 2.5 M$_{\odot}$
model but did find well-developed thermal pulses in the 5 M$_{\odot}$
model. They note that these results are at odds with the semi-analytical
predictions of \citet{1984ApJ...287..749F} for the occurrence of
TPs, which predict that the 2.5 M$_{\odot}$ model should have experienced
TPs whereas the 5 M$_{\odot}$ model should not. \citeauthor{2001AA...371..152M}
also discuss the two critical values that are often used to define
the boundary between low-, intermediate- and high-mass stars, namely
$M_{HeF}$ and $M_{up}$. $M_{HeF}$ is the upper limit at which He
ignites in degenerate conditions, giving rise to a core He flash.
Stars of this mass and less are classified as LM stars. $M_{up}$
is the stellar mass at which carbon ignition occurs. Stars above this
mass are classified as massive stars. Between these two limits are
the IM stars which ignite He quiescently and also never attain core
temperatures high enough to ignite carbon. The reason we detail this
here is because these values change significantly at very low and
zero metallicities. \citeauthor{2001AA...371..152M} find (confirming
previous work by \citealt{1993ApJS...88..509C} and others) that $M_{HeF}$
drops to $\sim1.1$ M$_{\odot}$ at $Z=0$. $M_{up}$ on the other
hand exhibits interesting behaviour as (initial) metallicity decreases
-- first decreasing then increasing. They note that $M_{up}$ is
primarily dependent on the mass of the H-exhausted core at the end
of the MS. As metallicity first decreases (from solar) core masses
increase, thereby reducing $M_{up}$. However, as the p-p chains become
dominant in the cores of extremely low metallicity MS stars the (convective)
cores become smaller again, due to the relatively low temperature
dependence of the reactions. In fact, \citeauthor{2001AA...371..152M}
find that $M_{up}$ is much higher for $Z=0$ stars than for stars
of solar metallicity, such that $M_{up,Z=0}\sim7.5$ M$_{\odot}$
(see their Figure 6). In terms of surface pollution they find that
the first dredge-up in low mass stars is very shallow, only mixing
up marginal amounts of He. Second dredge-up is more efficient, significantly
increasing the surface He abundance in the mass range $2\lesssim M\lesssim8$
M$_{\odot}$ and also increasing the surface CNO abundance in the
mass range $5\lesssim M\lesssim6.5$ M$_{\odot}$. Other polluting
episodes such as the DCF, DSF and the third dredge-up on the AGB were
not taken into account.

The same year \citet{2001ApJ...554.1159C} published results on the
AGB evolution of IM stars in the mass range $4.0\lesssim M\lesssim8.0$
M$_{\odot}$ (see also \citealt{2000MmSAI..71..781D}). Contrary to
the semi-analytical predictions of \citet{1984ApJ...287..749F} they
found that all their models \emph{do} experience 3DUP. Some other
interesting results from this study include the discovery of what
we shall call `Hot third dredge-up' (H3DUP, following \citealt{2003ASPC..304..318H})
in the more massive models, and the confirmation of the DSF phenomenon
in the less massive models (which \citeauthor{2001ApJ...554.1159C}
refer to as hydrogen convective episode (HCE) H-flashes). The first
phenomenon, H3DUP, was found to occur in the mass range $6.0\lesssim M\lesssim8.0$
M$_{\odot}$. When the H-rich convective envelope extended inwards
protons were brought into regions of very high temperature (the temperature
of the H burning shell in $Z=0$ models is high due to the paucity
of CNO catalysts). This gave rise to moderate H-flashes ($L_{HF}\sim10^{6}$
L$_{\odot}$) at each H3DUP episode. They note that the reason that
3DUP occurs in these models in the first place is because the envelope
is enriched with CNO nuclei during the second dredge-up. The second
phenomenon they report appears to be what we refer to as the `Dual
Shell Flash' (also see \citealt{1996ApJ...459..298C}; \citealt{1998IAUS..189P.150F}).
\citealt{2001ApJ...554.1159C} appear to be the first to examine the
phenomenon in detail. They find a convective feature in their lower
mass models ($4.0\lesssim M\lesssim6.0$ M$_{\odot}$), which we describe
here. During the early TP-AGB the H burning shell becomes convectively
unstable, just after the maximum of the He shell flash. Initially
this H convective episode (HCE) has no tangible effect on the evolution.
However, a few TPs into the AGB they find that the HCE ingests some
$^{12}$C (a product of the He burning below) and subsequently a thermonuclear
runaway occurs. The now C and N enriched (N is produced via the CNO
cycles) material is then dredged up by the convective envelope, enriching
the surface. This event is of great importance to the further evolution
of the star as it is now a more `normal' AGB star -- the envelope
has a reasonably high metallicity ($Z_{cno}\sim10^{-6}$) enabling
efficient CNO shell burning (thus lower shell temperatures). In fact,
\citet{2001ApJ...554.1159C} find that these models go on to experience
normal 3DUP episodes from then on. The interpulse periods also lengthen
to become more `normal'. They also stress that the treatment of convective
boundaries is very important in terms of 3DUP (and the DSF). They
argue that, due to opacity discontinuities that develop in these situations,
overshoot is physically plausible and thus 3DUP and other mixing events
are likely to occur. In summary they find that $Z=0$ IMS go on to
experience normal TP-AGBs because of the various envelope polluting
episodes that occur, and these stars are thus important sources of
N and C in the early Universe.

Inspired by the theoretical predictions of 3DUP in $Z=0$ AGB stars,
\citet{2001AA...378L..25G} explored the possibility of s-process
nucleosynthesis in IM stars. They chose a 3 M$_{\odot}$ model as
a case study, using the physical profiles in a $Z=0$ TP-AGB stellar
model as input for a post-process nucleosynthesis code. By imposing
a H profile just below the maximum incursion of the convective envelope
during 3DUP they simulated the often postulated proton diffusion required
to produce a $^{13}$C pocket. They found that, despite the lack of
heavy nuclei seeds for the s-process, the neutron density was sufficient
to take the s-process from CNO and Ne seeds all the way to Pb and
Bi. In fact they found that Pb and Bi were particularly overabundant
(ie. relative to the solar distribution), such that these stars could
be called `Pop III lead stars'. 

The year 2001 was a busy one for $Z=0$ stellar modelling. \citet{2001ApJ...559.1082S}
continued the work at the low-mass end, evolving models in the mass
range ($0.8\lesssim M\lesssim1.0$ M$_{\odot}$) through the dual
core flash at the tip of the RGB. Although this had been done previously
(\citealt{1990ApJ...351..245H}; \citealt{2000ApJ...529L..25F}) they
tested the robustness of the phenomenon by varying physical parameters.
Their main result was that the DCF is quite a robust phenomenon but
it also appears somewhat dependent on the He content of the star (at
least for $M=1.0$ M$_{\odot}$). They state that the key factor is
actually the degree to which the core He flash begins off-centre.
The closer the ignition is to the H-He discontinuity the more likely
it will be that a DCF will occur. In terms of surface pollution from
this event they find similar results to earlier studies (ie. huge
amounts of C and N are dredged to the surface). They also report a
new phenomenon in low-mass primordial stars -- a \emph{second} RGB
(SRGB) and second core He flash. After the DCF and subsequent dredge-up
of CNO and He rich material the He burning shell is virtually extinct,
leaving the H shell to support the star. Thus the star again has an
RGB structure, and proceeds to move up the giant branch, until a second
(weaker) He core flash occurs. This time there is no associated H
flash and the star subsequently moves to quiescent core He burning.
Another interesting feature \citeauthor{2001ApJ...559.1082S} discovered
is that the surface temperature of the star jumps discontinuously
as the envelope is polluted with DCF material (in the sense that the
surface becomes cooler, see their Figure 1). This is due to the sudden
increase in opacity. They however note that the low temperature opacity
tables they use do not properly represent the composition of the envelope,
so the temperature jump may be somewhat inaccurate. 

\subsection[Latest Studies: 2002 to 2006]{Latest Studies -- Published During the Course of the Present Study
(2002 -- 2006)}

In 2002 \citeauthor{2002ApJ...570..329S} calculated a grid of models
with $Z=0$ and a mass range of $0.8\lesssim M\lesssim20$ M$_{\odot}$.
Interestingly they found no DCF in the low mass models, at variance
with most (if not all) previous studies. They confirmed the occurrence
of 3DUP in the IM stars but also found it to occur in lower mass stars,
such that it was operating in the mass range $1\lesssim M\lesssim7$
M$_{\odot}$. In the less massive models ($1\lesssim M\lesssim5$
M$_{\odot}$) this was due to the fact that their models experienced
the same H convective episodes (ie. the DSF) as those reported by
\citet{2001ApJ...554.1159C}, also subsequently mixing up CNO nuclei
and thus allowing the star to experience a more normal TP-AGB. In
their 7 M$_{\odot}$ model they found that it was the 2DUP episode
that enriched the envelope so much as to allow a normal TP-AGB evolution.
They also note that the small amount of overshoot they used also contributed
to the onset of 3DUP.

\citet{2002AA...395...77S} revisited the problem of explaining the
C-rich extremely metal-poor halo stars (CEMPHs) which have been observed
in increasing numbers in recent years. They evolved a series of 0.8
M$_{\odot}$ stars through the DCF (which they refer to as Helium
Flash-Induced Mixing, or HEFM). Their main conclusion was that their
models (and those of other studies, eg. \citealt{2000ApJ...529L..25F})
produce C and N rich stars, thus qualitatively matching the results,
but in quantitative terms they produce far too much C and N. In addition
to this they note that the DCF/HEFM phenomenon does not occur at the
metallicities of the currently observed CEMPHs. Thus they conclude
that these models are not a good match to the observations. 

In 2003 \citeauthor{2003ASPC..304..318H} reported his investigation
into the beginning of the AGB for $Z=0$ and very low metallicity
models of masses 2 M$_{\odot}$ and 5 M$_{\odot}$. He confirmed the
occurrence of the HCEs at the start of the TP-AGB along with the subsequent
H-flash (DSF) as protons are mixed down and C mixed up. He also found
that strong H burning was occurring during 3DUP, calling this phenomenon
`Hot Dredge-Up'. We note that this may be the same phenomenon reported
by \citet{2001ApJ...554.1159C}, as described earlier.

The year 2004 was another busy one for $Z=0$ stellar modelling. \citet{2004ApJ...602..377I}
continued the search for an explanation of the CEMP stars by exploring
the (early) TP-AGB evolution of intermediate mass stars ($1>M>3$
M$_{\odot}$). They confirmed the findings of earlier studies (eg.
\citealt{1996ApJ...459..298C}; \citealt{2000ApJ...529L..25F}; \citealt{2001ApJ...554.1159C})
on the occurrence of DSFs at the beginning of the TP-AGB. They also
found deep 3DUP in all their models which continues to enrich the
envelope with CNO nuclei after the post-DSF 3DUP (although they note
that the convective mixing algorithm they use mixes one mesh point
past the formal Schwarzschild boundary, enhancing the likelihood of
3DUP). A key finding was that the DSF appears to be a source for lithium
enrichment in extremely metal-poor models. The Li produced during
the DSF (in the H convective zone) is dredged up and enriches the
envelope in all the models (except the 3 M$_{\odot}$ star which does
not experience the DSF). In the 2.5 M$_{\odot}$ model the Li is quickly
burnt by HBB but in the lower mass models it survives and probably
goes on to pollute the interstellar environment via stellar wind mass
loss. They also suggest that the DSF may be a source of s-process
nucleosynthesis in primordial low mass stars ($M\sim$1 M$_{\odot}$).

\citet{2004ApJ...609.1035P} also attempt to explain the observations
of extremely low-metallicity C-rich stars, and in particular the star
HE0107-5240, a very popular star in 2004 (the next two papers also
refer to this star). HE0107-5240 was the most iron-poor star known
at that stage, having a metallicity of {[}Fe/H{]} = -5.3 ( \citealt{2002Natur.419..904C}).
\citeauthor{2004ApJ...609.1035P} evolved low mass models ($0.8>M>1.5$
M$_{\odot}$) of zero and very low metallicity (see Table \ref{table-LowZ-LitReview}
for the $Z$ range) through the DCF and on to the AGB. Like \citet{2001ApJ...559.1082S}
they also find a secondary RGB phase develops just after the DCF.
Lithium is dredged up after the DCF, enriching the surface, which
concurs with the findings of \citet{2004ApJ...602..377I}. By calculating
a range of masses and metallicities they find that the DCF phenomenon
is a robust feature, always occurring in models of $M\lesssim0.9$
M$_{\odot}$ with $\log(Z)\lesssim-6$. For the same mass range they
also find that even a large increase in helium content (up to $Y=0.27$)
does not prevent the occurrence of the DCF. In comparing their surface
pollution results with that of CEMPH stars they find many reasons
to conclude that these stars' strange abundance patterns are \emph{not}
due to the DCF event. One key reason is that, like previous studies,
their models produce far too much C and N. Another is that, when confronting
the theoretical model with the observations in the $\log(g),\,\log(T_{eff})$
plane, the (real) star appears to be on the early RGB -- a long way
before the DCF pollution is predicted to occur.

\citet{2004AA...422..217W} also took up the HE0107-5240 challenge.
The novelty in their study was to use ejecta from supernovae calculations
to 1) pollute the surface of a $Z=0$ model and 2) mix with Big Bang
material to make a `Pop II.5' model. As with previous authors they
found that scenario 1 resulted in too much N and C being produced,
as well as the $^{12}$C/$^{13}$C ratio being too low compared to
observations. In addition to this they find that HE0107-5240 appears
to be on the RGB, well before the core He flash (or DCF) and associated
envelope enrichment is expected to occur, concurring with the findings
of \citet{2004ApJ...609.1035P}. In scenario 2 some of these problems
are circumvented as the DCF enrichment does not occur (due to the
higher $Z$ of the model) and the supernova material was chosen so
as to match the C overabundance (and Fe abundance). The N overabundance
was then obtained via first dredge-up, as CN cycling of the initial
C occurs prior to this. A side effect of using SNe yields is that
there is also a strong overabundance of oxygen. They find that too
much O is produced for a match with HE0107-524 but also note that
other CEMPs show more O enrichment. Their conclusion is that the SN
plus primordial gas mixture models provide the best fit to the observations,
particularly because it is possible to find a combination of SNe to
mix in order to match {[}C/Fe{]} a priori. 

In the same year a third group, \citet{2004ApJ...611..476S}, also
attempted to explain the surface abundances of HE0107-5240. Their
pollution scenario is based on mass transfer from a binary companion.
The binary system is thought to be primordial (ie. true $Z=0$) at
birth and the small amount of iron is assumed to arise from accretion
(only onto the surface) from a polluted primordial gas cloud. Enrichment
in the other elements arise through the mass transfer of the faster-evolving
component of the binary (the primary component, which is now a white
dwarf). Since the primary component of the binary is a true Pop III
star it undergoes a DSF (it's mass is expected to be in the range
1.2-3.0 M$_{\odot}$), which provides the large amounts of C and N
needed, as well as the enhancements in O and Na. They also discuss
a potential s-process signature that could be used to distinguish
between Pop III and Pop II pollution material (see their Figure 6).
To explore this scenario the authors calculate the evolution of a
range of low and intermediate mass $Z=0$ stars ($0.8<M<4.0$ M$_{\odot}$).
As they did not include time-dependent mixing they did not evolve
the stars through the DCFs and DSFs, halting the evolution at the
onset of these events. Thus, in order to obtain estimates for the
pollution arising from these phenomena, \citeauthor{2004ApJ...611..476S}
utilise the results of \citet{1990ApJ...351..245H}. The key in identifying
the primary component was that they noticed that the DSF event produces
less C and N than the low-mass DCF event. As noted earlier, excess
C and N has been a consistent problem with the low-mass Pop III pollution
scenario (eg. \citealt{2002AA...395...77S}; \citealt{2004ApJ...609.1035P}).
This is the main reason they predict the mass of the primary to be
in the range 1.2-3.0 M$_{\odot}$. They also discuss the production
of O through $^{13}$C burning during the DSF and also the potential
to produce Na, Mg and Al (which can also be produced through later
3DUP episodes). Finally, it is interesting to note that they did not
find any 3DUP in their 4 M$_{\odot}$model, in contrast to earlier
studies (eg. \citealt{2001ApJ...554.1159C}; \citealt{2002ApJ...570..329S}).

\subsection{The Current Study In Context}

One of the main aims of the current study is to examine the nucleosynthesis
of primordial stars. In order to achieve this it was necessary to
first modify the structural evolution code substantially (SEV code,
see next two chapters) and the nucleosynthesis code to a lesser degree.
The next stage was to run the calculations. Owing to the peculiar
evolution of primordial stars, as evidenced by the studies described
above, there were many difficult phases to evolve through. Table \ref{table-LowZ-LitReview}
gives a clear representation of where our study fits in to the further
development of the field. We have calculated a grid of models with
a modest range of masses ($0.85\rightarrow5.0$ M$_{\odot}$) and
a wide range of metallicities (zero and $\log(Z/Z_{\odot})\approx-6,-5,-4,-3$).
Looking at Table \ref{table-LowZ-LitReview} we can see that this
grid is essentially covered by the previous studies. We note however
that whilst the range has been covered it is evident that no consistent
study has been performed across the range in question -- it requires
the addition of a few independent studies to cover the present one
(especially if the stages of evolution evolved through is taken into
account). This study provides a self-consistent set of models that
cover all phases of evolution. In addition to this there is still
a consensus lacking in some of the evolutionary traits. For example
the DCF phase is found to occur in the calculations of some authors
(eg. \citealt{1990ApJ...351..245H}; \citealt{2001ApJ...559.1082S})
but not all (\citealt{2002ApJ...570..329S}). This in itself warrants
another investigation into the evolution of low mass $Z=0$ stars.
Furthermore, to the best of our knowledge, our models are \emph{the
first calculations of the entire AGB phase at these metallicities}.
Importantly this has enabled us to calculate detailed chemical yields
for these primordial stellar models, which has also never been done
before (again, to the best of our knowledge). Finally we note that
much work was done in the field during the period of the Author's
candidature, as it has recently been (and still is) a `hot topic'
in stellar astrophysics. This is mainly due to the discovery of some
extremely metal-poor Galactic halo stars, particularly the C-rich
star HE0107-5240 (\citealt{2002Natur.419..904C}). With the recent
discovery of an even more metal-poor Halo object (HE1327-2326, \citealt{2005Natur.434..871F})
and the likelihood of more discoveries in the near future, we anticipate
the interest in models of these stars will continue.

\part{METHOD}

\chapter{Numerical Codes\label{Chapter-NumericalCodes}}
\begin{quote}
``We know very little, and yet it is astonishing that we know so
much, and still more astonishing that so little knowledge can give
us so much power.'' 
\begin{flushright}
\vspace{-0.5cm}-- Bertrand Russell (1872-1970)
\par\end{flushright}
\end{quote}

\section*{Overview}

The stellar model simulations are handled in two distinct parts --
the structural evolution and the nucleosynthetic evolution. The structural
evolution is computed first, using the Monash version of the Mount
Stromlo Stellar Structure (MSSS) code also known as `$EVOLN$'. For
clarity I will refer to this code as the SEV code (structural evolution
code) in this thesis. The SEV code includes only the nuclear reactions
important for the structure and thus provides only a small amount
of chemical evolution information. The second code, which is employed
to calculate the detailed nucleosynthesis, is known as `$DPPNS45$'.
The nucleosynthetic code (hereafter, NS code) uses the structural
evolution information output by the SEV code necessary for calculating
the chemical evolution (ie. the temperature profile, density profile,
convective velocities, etc.). Thus the NS code is a post-process code.
A benefit of having this two-step paradigm is that the structure need
only be computed once, whereas the nucleosynthesis can be re-run many
times, allowing the fast computation of models with differing reaction
rates.

As discussed in the preceding chapter, very low metallicity stellar
models (if not real stars!) are known to have some unique evolutionary
episodes. These episodes, such as the proton ingestion episodes (PIEs)
are notoriously difficult to model. Neither code was set up to handle
episodes such as these. Thus I have made modifications to both codes
in order to model these types of stars. The SEV code required the
most changes, most notably the introduction of time dependent mixing
to handle the rapid evolution during PIEs.

Both codes are described in detail in the sections below. For each
code I also discuss the modifications I made during the course of
my research.

\section{Structural Evolution Code\label{sevcode}}

This section briefly describes the structural evolution code as obtained
by the author at the beginning of the current study. Modifications
made during the course of this research are described in s\ref{sec-SEV-IncludedPhysics}.

\subsection{History of Development and Basic Assumptions}

The Mount Stromlo Stellar Structure Program was originally developed
in the late 1960s (\citealt{Fau68}). Initially designed to model
He shell burning stages in low mass stars ($M\sim0.9$ M$_{\odot}$),
it has been expanded and updated numerous times such that it is now
capable of modelling all phases of low to intermediate mass stars,
from the pre-main sequence (PMS), through the thermally pulsing AGB
phase, and on to the white dwarf (WD) cooling phase. As noted in the
PhD thesis by \citet{LATT84}, descriptions of the code are widely
scattered throughout the literature. To the best knowledge of the
author, \citet{LATT84} is the source with the most detail about the
code. In particular it describes the difference equations used and
the particular implementation of semiconvection. To provide some further
key references and outline the history of the development of the code
we list some landmark studies and associated code updates below:

\begin{spacing}{0.6}
\begin{itemize}
\item Thermal pulses in He shell burning stars (\citealt{FW72})

\begin{itemize}
\item Neutrino losses added, opacities updated
\end{itemize}
\item Hydrogen added (early 1970s. Previously only included He and C burning)
\item AGB evolution of a 0.6 M$_{\odot}$ star (\citealt{Gin74})

\begin{itemize}
\item No shell-shifting methods used.
\end{itemize}
\item AGB He shell flashes in stars of mass 0.8 to 3.0 M$_{\odot}$ (\citealt{1981ApJ...247..247W}).

\begin{itemize}
\item Entire stellar model now in Henyey matrix (ie. no envelope fitting).
\end{itemize}
\item Third dredge-up in low mass stars (\citealt{1989ApJ...344L..25L},
\citealt{1996ApJ...473..383F}).
\item Hot Bottom Burning in AGB stars (\citealt{1992PASAu..10..120L}, \citealt{Frost98}).
\end{itemize}
\end{spacing}

The version of this program obtained at the beginning of the current
study is the Monash University version, known as `EVOLN' (\citealt{LATT84};
\citealt{1986ApJ...311..708L}; \citealt{1996ApJ...473..383F}). For
clarity and in order to distinguish it from the nucleosynthesis code
I shall refer to it as the SEV code (structural evolution code) in
this thesis. 

The main assumptions and approximations used in the code are:
\begin{enumerate}
\item Spherical symmetry (ie. no rotation or magnetic fields) $\implies$one
spatial dimension.
\item Hydrostatic equilibrium $\implies$quasi-static evolution.
\item Radiation transport is treated as a diffusive process.
\item The Mixing Length Theory (MLT) is used for convection (local theory).
\item Convective zones are mixed instantaneously (ie. time-dependent mixing
not included).
\end{enumerate}
All these assumptions and approximations are widely used in other
one-dimensional stellar codes. Indeed, the first three lead (with
other physical considerations) to the standard form of the well known
stellar structure equations (see eg. \citealt{1958ses..book.....S};
\citealt{1983psen.book.....C}). Using mass as the independent variable
the equations take the form:

\begin{eqnarray}
\frac{\partial r}{\partial m} & = & \frac{1}{4\pi r^{2}\rho}\label{mass}
\end{eqnarray}

\begin{eqnarray}
\frac{\partial P}{\partial m} & = & -\frac{Gm}{4\pi r^{4}}\label{eqm}
\end{eqnarray}

\begin{eqnarray}
\frac{\partial L}{\partial m} & = & \epsilon_{nuc}-\epsilon_{\nu}-\left(\frac{\partial U}{\partial t}+P\frac{\partial V}{\partial t}\right)\label{energy}
\end{eqnarray}

\begin{eqnarray}
\frac{\partial T}{\partial m} & = & -\frac{GmT}{4\pi r^{4}P}\nabla\label{trans}
\end{eqnarray}

where the variables have the usual meanings. Equation \ref{mass}
is the mass continuity equation, equation \ref{eqm} the equation
for hydrostatic equilibrium, equation \ref{energy} the energy equation,
and equation \ref{trans} the energy transport equation. Regions in
a model are deemed convective based on the \citet{SWCH1906} criterion,
such that if the radiative temperature gradient is greater than the
adiabatic temperature gradient then the region is convectively unstable.
The standard procedure of taking $\nabla$ (the temperature gradient)
equal to $\nabla_{radiative}$ in radiative zones is used, whilst
in convective zones $\nabla$ is calculated from the MLT.

The SEV code solves the four stellar structure equations to produce
a model in hydrostatic equilibrium. The temporal advance is then based
on the chemical changes brought about over a time interval $\Delta t$
by nuclear reactions (and turbulent mixing). The nuclear reactions
alter the mean molecular weight $\mu$ which in turn affects the density
structure, and therefore the thermal structure, etc. A new hydrostatic
model is calculated with the new composition profile at the time $t+\Delta t$,
thus enabling us to follow the evolution through a series of `quasistatic'
models. 

\subsection{Overview of Numerical Method}

The core of the SEV code is based on the Henyey-matrix numerical method
(\citealt{Hen59}; \citealt{Hen64}) which is basically a Newton-Raphson
method. It uses successive iterations whereby corrections are applied
to an initial guess for the structure, the new guess fed back into
the matrix, then new corrections are calculated and applied, and so
on. This iterative method is also known as `relaxation'. The code
is deemed to have converged if the difference between the current
iteration's structure and the previous one's is below a certain threshold
(typically $\sim10^{-3}$). A recent description of the relaxation
method can be found in \citet{1994sipp.book.....H}.

The change in chemical composition, which is the driving force for
evolution, is derived from the rates of change of abundances due to
nuclear reactions and the movement of chemical species due to convective
mixing. A semi-implicit scheme is used to advance the chemical profile
to the next timestep: 

\begin{eqnarray}
X_{i}^{n} & = & X_{i}^{n-1}+\frac{\Delta t}{2}\left[\left(\frac{dX_{i}}{dt}\right)^{n}+\left(\frac{dX_{i}}{dt}\right)^{n-1}\right]\label{chemevoln}
\end{eqnarray}

where the $X_{i}$ are the chemical species, $\Delta t=t^{n}-t^{n-1}$
and $n$ and $n-1$ represent the current and previous timesteps respectively.
Thus, in addition to the converged hydrostatic model, we now have
a change in composition profile with which to calculate the next hydrostatic
model. In this way a series of hydrostatic models form the quasistatic
evolutionary sequence.

Mass is essentially the independent variable, however the SEV code
has a peculiar formalism, whereby $x$ is actually used:

\begin{eqnarray}
x & = & \left(\frac{m_{r}}{M}\right)^{\frac{1}{3}}.\label{massvar}
\end{eqnarray}

\citet{Fau68} made this choice so as to include the central point
as a normal mesh point -- as all derivatives using this variable
are finite at the centre -- and the calculations are simplified.
The dependent variables in the SEV code are: $L_{r}/L_{\odot},\,ln(T),\,r/R_{\odot}$
and $ln(P)$. 

The stellar model is divided into many mass shells (the term `shells'
is commonly used but it should be remembered that the scheme is actually
one dimensional). State variables (T, $\rho,$ etc.) are recorded
at each mesh point (some being defined between mesh points). There
are typically between 500 and 3000 mesh points used, depending on
the stage of evolution. The Lagrangian mesh is adaptive -- the points
are distributed to maintain a prescribed minimum resolution in the
state variables. Thus in regions where there are steep gradients in
physical properties the resolution is increased. Likewise, the timestep
is constrained so as not to let the state variables (or chemical abundances)
change too much between models. 

\subsection{Included Physics\label{sec-SEV-IncludedPhysics}}

\subsubsection*{Chemical Species and Nuclear Reactions}

The SEV code currently follows six chemical species: $^{\textrm{1}}$H,
$^{\textrm{3}}$He, $^{\textrm{4}}$He, $^{\textrm{12}}$C, $^{\textrm{14}}$N
and $^{\textrm{16}}$O. These are the only isotopes considered explicitly
however the rest of the isotopes involved in H, He and C burning reactions
are included \emph{implicitly}. The missing isotopes involved in the
CNO cycles are included implicitly by assuming that they are always
present in their equilibrium abundances. Temperature-dependent branching
ratios are used for isotopes with two possible exit channels. Only
following the major isotopes simplifies and speeds up model calculations
however it means that the code contains only basic nucleosynthesis
-- enough to properly follow the structural evolution -- it lacks
the detailed nucleosynthesis that is needed to confront stellar models
with the detailed observations currently available. For this reason
we utilise the post-processing nucleosynthesis code which is able
to follow an arbitrary number of elements (this code discussed in
section \ref{nscode}). The SEV code also contains one pseudo-element,
called $zother$, which represents all the elements heavier than $^{\textrm{16}}$O.
The heavy products of He and C burning are added to $zother$ for
baryon conservation.

The code currently solves for H burning (pp chains and CNO tri-cycle),
He burning (triple-alpha reaction) and C burning (which is never used
in this study). Further burning stages (Ne, O, etc.) are not included,
which imposes an upper limit to the mass and/or stage of evolution
we are able to follow certain stars. All nuclear reaction rates are
from \citet{CF88}, except the rate for the $^{\textrm{1}}$H(p, $e^{+}+\nu$)$^{\textrm{2}}$H
reaction which is from \citet{Har83}. Although the rates used are
from fairly old compilations, the aim of the SEV code is to calculate
the structure only, whereas the detailed nucleosynthesis is calculated
by the nucleosynthesis code which allows the accurate tracking of
a large number of isotopes and is constantly updated with the latest
reaction rates. We note however that recent updates to the key SEV
code rates (which have not been used in the current study, but will
be used for all future studies) have shown that these rates have not
changed substantially from that given in the \citet{CF88} compilation.

Neutrino energy losses are included following \citet{1967ApJ...150..979B}
for the pair neutrino process, photoneutrino process and plasma neutrino
process, whilst bremsstrahlung rates are from \citet{1969PhRv..180.1227F}.
Corrections due to neutral currents are taken from \citet{1976MNRAS.176....9R}
(for pair, plasma, and photoneutrino processes) and \citet{1976ApJ...210..481D}
(for bremsstrahlung). 

\subsubsection*{Opacity}

Opacity is of particular importance to stellar modelling as it determines
the heat flow through the star and thus the thermal structure, which
then feeds back onto the rest of the structure and consequently the
nuclear reaction rates. Here we describe the SEV code opacity regime
in use at the beginning of the present study. Substantial changes
were made to the opacity subroutines during the course of the current
study and are described in section \vref{opacmods}.

In the SEV code opacities for each mesh point are obtained by interpolating
in opacity tables. Tables of Rosseland mean opacities made available
by the Lawrence Livermore National Laboratory OPAL project (\citealt{1996ApJ...464..943I})
cover the bulk of the temperature and density range of low and intermediate
mass stars. We interpolate within these tables in the temperature
range: $7000\,\textrm{K}<T<5\times10^{8}\,\textrm{K}$. The OPAL tables
do not include opacities for conductive material so in regions where
the density is very high we use the analytical fit by \citet{1975ApJ...196..525I}
(appendix A of that paper) to the \citet{1969ApJS...18..297H} conductive
opacity formulae for non-relativistic electrons and those of \citet{1983ApJ...273..774I}
and \citet{1984ApJ...277..375M} for dense (liquid-metal) material
containing relativistic electrons. Opacities for solid material are
from \citet{1982ApSS..87..193R}. The subroutine that calculates these
conductive opacities was supplied by MacDonald (1992, private communication).
At temperatures below 7000 K a combination of the \citet{1970ApJS...19..261C}
and \citet{HMMJ+77} tables are used. At very low temperatures ($T\lesssim4000\,\textrm{K}$),
such as those found in the envelopes of RGB or AGB stars, molecular
opacity becomes important. As the \citet{1970ApJS...19..261C} low
temperature tables do not include molecular contributions, we add
in the opacity contributions from CN, CO, H$_{2}$O and TiO using
the formulae suggested by \citet{1989AAS...77....1B} which are based
on the molecular (and grain) opacity tables from \citet{1975ApJS...29..363A}
and \citet{1983ApJ...272..773A}. The corrections to the formulae
by \citet{1993ApJS...86..541C} are taken into account. All the opacity
tables assume a scaled-solar heavy element distribution.

\subsubsection*{Turbulent Convection and Mixing\label{evcodeCvnMixing}}

Regions in each stellar model are deemed convective based on the \citet{SWCH1906}
criterion, such that if the radiative temperature gradient is greater
than the adiabatic temperature gradient then the region is convectively
unstable. The standard procedure of taking $\nabla$ (the temperature
gradient) equal to $\nabla_{radiative}$ in radiative zones is used.
In convective zones $\nabla$ is calculated from the MLT. 

Fluid mechanics theory predicts that the onset of convection occurs
when the Rayleigh number ($Ra$) reaches some critical value $Ra_{c}$,
where $Ra_{c}$ is of the order 10$^{3}$. As conditions in stellar
convective zones give rise to extremely large Rayleigh numbers ($\sim$
10$^{20}$, \citealt{1971ARAA...9..323S}) -- and also very small
Prandtl numbers ($Pr\sim10^{-9}$, where $Pr=\frac{kinematic\ viscosity\ \nu}{thermal\ diffusivity\ \kappa}$)
-- it is currently impossible to simulate anything even close to
these conditions in the laboratory. This lack of experimental evidence
necessitates either 1) a complete hydrodynamic calculation (eg. solution
of the Navier-Stokes equations) or 2) a decent approximate theory
of convection. The first option is very computationally demanding,
especially if it is done properly in three dimensions. Some two and
three dimensional hydrodynamic simulations have been calculated but
they necessarily focus on small regions within a star (to make the
problem tractable in a computing sense), such as the convective layer
in Solar-type stars (or part thereof: \citealt{1990AA...228..155N};\citealt{1995AA...295..703S}).
A recent development is that of the DJEHUTY project (\citealt{2001AAS...198.6513D})
which now has a working three dimensional hydrodynamical code that
can simulate an entire star (given enough computing resources). However
the time it takes to evolve stars is still prohibitive in terms of
a full evolutionary sequence, despite the fact that it is being run
on the largest computing cluster in the world. For this reason the
second option has been employed since the early days of stellar evolution.
Convective approximations, mainly in the form of the widely used Mixing
Length Theory (MLT; \citealt{Pr25}; \citealt{1951ZA.....28..304B};
\citealt{1958ZA.....46..108B}), have allowed the modelling of convective
zones in stellar models to be calculated in reasonable lengths of
time. Much of the simplification of the calculations arise because
the MLT is a \emph{local} formalism, such that the properties of each
convecting mass element are given solely by the conditions at its
current location. Some attempts at including non-local effects in
stellar modelling have been made, mainly as modifications to the MLT
(eg. \citealt{1973ApJ...184..191S}; \citealt{Mad76}; \citealt{Xiong85}).
The various incarnations of the MLT are all based on the Boussinesq
approximation. One key feature of this formalism is that we have a
characteristic `mixing length' \emph{l} which is assumed to be much
shorter than other length scales in the star (eg. the pressure scale
height, $H_{P}$). However, as is well known, application of the MLT
only gives reasonable results when \emph{l $\sim H_{P}$} (see eg.
\citealt{1994sipp.book.....H} for a brief review)\emph{.} Indeed,
most authors use a mixing length greater than the pressure scale height
(eg. \emph{l $\sim1.5H_{P}$}). Despite this internal inconsistency
within the (application of) MLT, and despite the wholesale assumptions
used in constructing the MLT equations used in stellar model calculations,
the predictive success of stellar models appears to indicate that
the MLT gives a reasonable description of convection. For this reason,
and for the sake of efficient computation, most authors still use
the MLT (or variations thereof) and it is the formalism currently
employed in the SEV code. 

Another convection formalism that the author is aware of is `Full
Spectrum of Turbulence' (FST) (\citealt{1996ApJ...473..550C}; \citealt{Ven05}).
While still a local theory, and still based on the MLT, FST relaxes
the single eddy approximation of the MLT, taking into account turbulence
on all length scales. This group (Canuto, Mazzitelli, Ventura et al.)
finds substantially different results when comparing with stellar
models using the MLT. This serves as a further reminder that the treatment
of convection is far from perfect in stellar models.

In the SEV code the mixing of chemical elements within a convective
zone is assumed to happen instantaneously. This assumption is valid
for most stages of stellar evolution as the convective turnover timescale
is usually much shorter than the evolutionary timescale. Thus the
chemical profile will be easily mixed in one timestep. However, in
evolutionary stages where there are rapid changes and the timestep
is consequently lowered, this assumption is not valid. This may happen
during the core He flash in low mass stars or during proton ingestion
episodes in zero- or very low-metallicity stars. As the current study
is concerned with these cases, a time dependent mixing scheme had
to be added to the SEV code (detailed in section \vref{sevmods}).

Finally we note that the SEV code includes the method suggested by
\citet{1981ApJ...248..311W} for taking into account the change in
entropy due to the dredge up of heavy nuclei during the third dredge-up
on the AGB. The main effect of including this physics is to slightly
increase the depth of the third dredge-up. 

\subsubsection*{Overshoot\label{sub-evoln-overshoot}}

Directly related to the above discussion on convection is the phenomenon
of overshoot. Overshoot is defined as the extension of convection
beyond the formal (eg. Schwarzschild) convective boundary. In the
Schwarzschild formalism the convective boundary is defined as the
first point where the actual temperature gradient $\nabla$ is equal
to the adiabatic temperature $\nabla_{ad}$. However, a convecting
parcel of gas reaching this boundary will, by definition, still have
a small temperature excess $\Delta T$, and thus a finite acceleration.
As the MLT is strictly a local formalism the first point above the
Schwarzschild boundary has no information about the still moving parcel
below and it is designated as a radiative (non-convective) point.
If a non-local scheme were used then the parcel would be followed
through the Schwarzschild boundary into the stably stratified material
on the other side. This subadiabatic material would then cause a deceleration
of the parcel. Of course the key question here is how far the parcel
will overshoot the formal boundary before finally coming to rest,
as this will define the extent of the extra mixing. Non-local MLTs
are often formulated in order to follow this phenomenon (eg. for convective
core overshoot: \citealt{1975AA....40..303M}; \citealt{1973ApJ...184..191S}).
However, the results based on this type of method have given varying
results. Indeed, \citet{1987AA...188...49R} notes that adding non-local
prescriptions to the local MLT can lead to non-physical equations
and he highlights the huge variation in overshoot distances between
authors (from almost zero to a few times the original Schwarzschild
convective zone size). Another method to gauge the extent of overshoot
is to utilise results from hydrodynamical computations. These computations
are quite difficult and still require the use of many assumptions
(\citealt{1996AA...313..497F} provides a list). Some recent work
in this area includes that by \citet{1996AA...313..497F} and \citet{1995AA...295..703S}.
These groups investigate overshooting using two- and three-dimensional
hydrodynamic codes respectively. \citet{1996AA...313..497F} find
that the vertical velocities fall off exponentially starting from
within the formal convective boundaries. They also note that the concept
of an overshoot \emph{distance} is not useful (although commonly used)
since their results show that the extent of overshoot is very dependent
on the particular conditions at the edge of the convection zone. \citet{1995AA...295..703S}
also find this dependence (based on their stability parameter $S$).
Their results indicate that the overshoot distance is some fraction
of $H_{P}$ (the pressure scale height), which is proportional to
the velocity scale height. Very recently \citet{2006ApJ...642.1057H}
have performed 3D hydrodynamic simulations to mimic the convection
in an AGB He shell flash. Although they do not quantify the extent
of overshooting in their study, they suggest that some overshoot does
occur, but at quite a minimal level due to the physical conditions.
A reassuring finding in their study is that the velocities in the
convective regions given by the hydrodynamic code are very similar
to those obtained from the MLT. The main exception is in the exponential
drop-off of velocities \emph{before} the bottom of the convective
zone (and probably continuing just beyond the formal convective boundary). 

Despite the great work done to date in this field of multidimensional
hydrodynamic simulations of stellar convection, it must be remembered
that the field is still in its beginning stages. Many assumptions
are included even in the most detailed models (as noted by the authors
themselves), so the results must be used with caution. That said,
the beginning of a consistent picture is starting to form, consisting
of 1) there appears to be an exponential decay of velocities at the
edge of convection zones and 2) the degree of overshoot seems to be
very dependent on the local conditions. However it appears that we
still don't really have a good quantitative handle on the extent of
overshooting. Thus for now most stellar models still employ ad-hoc
(although reasoned) approaches.

The method used to account for overshoot in the SEV code involves
a `search for convective neutrality' (\citealt{1986ApJ...311..708L}).
The algorithm linearly extrapolates $\nabla_{rad}/\nabla_{ad}$ taken
from the last two convective mesh points to the next point (which
is radiative). If the extrapolated ratio is greater than unity (ie.
$\nabla_{rad}>\nabla_{ad}$) then that point is reclassified as convective.
Conversely, if the ratio is less than unity then the point remains
radiative. The algorithm only allows one radiative point to be added
to the convective zone each iteration. In using this procedure we
ensure that the last convective point is as close to neutral buoyancy
as possible. This method is applied to all convective boundaries.
We note that although the method allows for core growth and (the disputed)
`core breathing pulses' during core He burning, it does not have a
significant effect on the convective boundaries of flash-driven He
convective zones (eg. during the TPAGB). 

Finally we note that the above method for overshoot is only used in
some of the stellar models in this study. During the course of the
research the SEV code was modified to include a time-dependent hydrodynamically-based
prescription for overshoot (see section \vref{sevmods} for details).

\subsubsection*{Semiconvection\label{sub-semiconvection}}

Semiconvection in the SEV code is intimately related to overshoot.
The overshoot method described above may even be classified as a method
to include semiconvection, as it does bear some resemblance to the
semiconvection algorithm often used in stellar codes (the \citealt{1972ApJ...171..309R}
method). The Robertson and Faulkner method also searches for convective
neutrality but achieves it by adjusting the abundances in a region
just outside the formal convective boundary -- if there exists a
small convective region outside the convective core. The region between
the two fully convective zones is said to be semiconvective, and a
mixing procedure is used to mimic this physical process. 

The method used to follow semiconvection in the SEV code relies on
the previously described overshoot formalism but adds in one parameter,
$\Delta q_{scvn}=\Delta m/M$ (\citealt{LATT84}). $\Delta q_{scvn}$
prescribes the mass spacing of the mesh points around a convective
boundary. Differing values of $\Delta q_{scvn}$ give rise to differing
amounts of semiconvection/overshoot. In the absence of quantitative
knowledge about the extent of semiconvection and overshoot (apart
from inferences from some observations), $\Delta q_{scvn}$ was calibrated
with reference to models using the \citet{1972ApJ...171..309R} method
described above.

With regards to the current study this method of following semiconvection
was only used for some of the stellar models. A new formalism was
introduced, as described in section \vref{semimods}. The physical
nature of semiconvection is also discussed in more detail in that
section.

\subsubsection*{Mass loss\label{sub-MassLossDescription}}

A particularly large uncertainty arises in the later stages of stellar
evolution (in low to intermediate mass stars) when mass loss from
the surface of a star becomes significant. As yet there are no theoretical/computational
models developed from first principles to explain mass loss in cool
stars. There is the general idea though that radiation pressure exerted
on grains in the outer atmospheres of cool stars is a major factor.
However currently the only method to include mass loss in stellar
evolution calculations is to make use of empirical formulae. The empirical
formulae are based on observations of mass loss in various types of
stars. Mass loss rates are deduced by observing the Doppler shifts
of circumstellar absorption lines. The formula for RGB stars by \citet{1975MSRSL...8..369R}
is still probably the most used today:

\begin{eqnarray}
\dot{M}(M_{\odot}yr^{-1}) & = & 4\times10^{-13}\eta_{R}\frac{RL}{M}\label{reimers}
\end{eqnarray}

where R, L and M are the star's radius, luminosity and mass in solar
units. $\eta_{R}$ is the Reimers mass loss parameter and is of order
unity. As can be seen, the mass loss rate is a function of the gross
physical properties of the star and thus $\dot{M}$ evolves with the
star. Although this relation was derived from observations of RGB
stars it is often used for modelling AGB stars by varying $\eta_{R}$.
In the SEV code we use Reimers' law only on the RGB (with $\eta_{R}=0.4$)
as the mass loss rate towards the end of the AGB is known to be extremely
strong. Indeed, a \emph{superwind,} with $\dot{M}\sim10^{-5}M_{\odot}/yr$,
is needed to form planetary nebulae, which is not produced by equation
\ref{reimers}. The AGB is also phenomenologically different from
the RGB as AGB stars are known to pulsate. Observations show that
mass loss is related to the period of pulsation. Thus for the AGB
we make use of the empirical formula derived by \citet{1993ApJ...413..641V}
(hereafter VW93) which is based on observations of AGB stars:

\begin{eqnarray}
log\left(\dot{M}(M_{\odot}yr^{-1})\right) & = & -11.4+0.0125P\label{vw93}
\end{eqnarray}

where $P$ is the pulsation period in days. Assuming that AGB stars
oscillate in their fundamental mode the period can be related to the
stellar radius and mass (in solar units):

\begin{eqnarray}
log\left(P(days)\right) & = & -2.07+1.94\log R-0.9\log M\label{vw93period}
\end{eqnarray}

VW93 note that this equation gives superwind values of $\dot{{M}}$
in stars with periods of $\sim500$ days. Such a strong wind quickly
enshrouds the central star in dust and it is no longer visible at
optical wavelengths. This is observed in low mass populations at P
= 500 days but the effect does not set in in higher mass populations
(M$\sim5$ M$_{\odot}$) until P $\gtrsim750$ days. VW93 account
for this by using a different mass loss formula for more massive stars
(their equation (5)). In the SEV code we use a different method whereby
we switch to the following relation when $P>500$ days:

\begin{eqnarray}
\dot{{M}}(M_{\odot}yr^{-1}) & = & \frac{L}{cv_{exp}}\label{mdotradpress}
\end{eqnarray}

where $v_{exp}$ is the expansion velocity and $c$ is the speed of
light (both in $km\,s^{-1}$). This corresponds to a pure radiation-pressure-driven
wind although we still include a dependence on the pulsation period
by calculating the expansion velocity using equation (3) from VW93:

\begin{eqnarray}
v_{exp} & = & -13.5+0.056P\label{vexp}
\end{eqnarray}

Finally, an upper limit is set on $v_{exp}$ ($=15\,km\,s^{-1}$)
to mimic the observed radiation-pressure-driven limit. 

\subsubsection*{Equation of State}

The equation of state (EOS) describes the interdependency of the density,
temperature and pressure of a material (usually a plasma in stellar
models). Depending on the physical conditions a given material will
behave in different ways. For relativistic or electron-degenerate
gas the SEV code uses the fitting formulae of \citet{1967ApJ...150..979B}.
For fully-ionised gas we use the perfect gas equation (including radiation
pressure). For cool outer layers where the gas is only partially ionised
the Saha equation plus the method of \citet{1965ZA.....62..221B}
is applied, with radiation pressure added. 

As the various equations of state utilised in the SEV code are quite
old, \citet{2000PASA...17..284M} compared the results to a more modern
EOS (provided by the OPAL project, \citealt{1996ApJ...456..902R}).
He found that the differences were so small ($<2\%$ in pressure)
that it did not warrant the factor of four increase in computation
time to implement the OPAL EOS. However we note that the comparison
was only performed for the parameter space in which the OPAL EOS is
valid, which does not include relativistic and/or degenerate gas.

\section{Nucleosynthesis Code\label{nscode}}
\begin{quotation}
``To the extent that it is possible, it is the isotopes that keep
the theorists honest.''
\begin{flushright}
\vspace{-0.6cm}-- David Arnett 
\par\end{flushright}
\end{quotation}

\subsection{Introduction}

Modern abundance observations provide data for a large portion of
the Elements in many stars. Since the elements have varying sources
of production these observations amount to myriad constraints on stellar
models (although there are sometimes large uncertainties in the observations
that weaken the constraints). In addition to the constraints imposed
by individual elements, the abundance \emph{patterns} of stars also
provide important constraints -- and insights. Clues can be sought
in these patterns for the likely stellar sites of nucleosynthesis
via abundance ratios, correlations and anticorrelations. Indeed, it
is always very difficult to match all the observed elements in a single
star (this is also due to other complicating factors such as multiple
pollution sources). Moreover, in some cases even isotopic abundances
or ratios can be deduced from the observations. For example the $^{12}$C/$^{13}$C
ratio is routinely measured (eg. \citealt{2003ApJ...585L..45S}) and
it is possible to deduce the $^{24}$Mg:$^{25}$Mg:$^{26}$Mg ratios
in some stars (eg. \citealt{2003AA...402..985Y}). Knowledge of the
isotopic abundances provides even more stringent constraints on the
models. For example, in the globular cluster chapter of the present
study (Chapter \ref{GC-chap}) we find that the elemental abundances
of Mg in the globular cluster NGC 6752 are satisfactorily reproduced
by the models -- but the isotopic abundances show a large discrepancy
relative to the observations. Thus the isotopic observations can ``keep
the theorists honest'' -- or show up problems in the models and/or
reaction rates. We also note that (very accurate) determinations of
isotopic ratios have been made in pre-solar grains such as those in
the Murchison meteorite (see eg. \citealt{1995GeCoA..59.4029H}).
These also provide constraints on the models. 

With these points in mind it can be seen that the NS code, which currently
has a network of 74 species and 506 reactions, is an important part
of our modelling. It allows us to make detailed predictions of abundance
patterns which we can confront with the observations.

\subsection{Code Overview}

As mentioned in the introduction to this chapter our nucleosynthesis
code (NS code) is a post-processing code. During the calculation of
the structural evolution of a stellar model (with the SEV code) the
values of many structural properties are stored in a file. The NS
code then uses the these structural properties as a basis for calculating
the detailed nucleosynthetic evolution of the model. This two step
method is possible because only a few reactions (and nuclides) contribute
significantly to the structural evolution of stars of low- and intermediate
mass (as discussed in the SEV code section). The myriad minor nuclides
and their reactions supply only negligible amounts of energy and thus
do not affect the structure of the models. They do however provide
interesting and complex nucleosynthesis. 

The code is based on one developed by \citet{1993MNRAS.263..817C}
to model the extreme HBB ($rp-process$) occurring in Thorne-\.{Z}ytkow
objects (stars with degenerate neutron cores and massive H-rich envelopes).
It was used for the complex nucleosynthesis and time-dependent mixing
required for nuclear burning in those objects. At that stage it was
attached to the Eggleton stellar evolution code. The severe HBB is
the main source of energy generation in these objects so the NS code
fed back physical information to the structure code. Stellar structure
solutions were determined through iterations between the two codes.
In this way the energy from the HBB calculated by the NS code was
taken into account in the models. The code was rewritten by Lattanzio
and Cannon to take input from the Monash Mount-Stromlo Stellar Structure
code (ie. the SEV code). Unlike the \citet{1993MNRAS.263..817C} code
which used a static mesh the new NS code was written to include a
moving mesh. Since all the energetically important reactions for the
structure of low- and intermediate-mass stars were already included
in the SEV code, there was no need for feedback between these two
codes. All the (energetically) minor reactions important for nucleosynthesis
could be calculated post-process. Also, since the reactions in the
SEV code are a subset of those included in the NS code, and the NS
code calculates the energy generation from all the reactions, it was
possible to compare the energy generation given by each code as a
consistency check. Finally we note that the code is written so that
is is quite an easy task to expand the reaction network, and to alter/update
individual nuclear reaction rates. 

\subsection{Input Physics}

\subsubsection*{Inputs from the SEV Code}

Not every quasi-static model that the SEV code calculates is output
to the NS input file. Instead there is a set of selection criteria
based on changes in the stellar properties such as the rates of change
of luminosity and abundances. This reduces the time resolution during
stages in which not much nucleosynthesis is occurring (but in which
structural changes occurred), speeding up the nucleosynthesis calculations.
Physical variables supplied to the NS code for each model include
the temperature profile, density profile, abundance profiles and the
velocity profile (all versus mass). This gives the NS code enough
information to calculate the time dependent mixing (through the velocity
profile) and the nucleosynthesis (through the temperature, density
and initial abundance profiles). The total hydrogen and helium burning
luminosities from the SEV code are also provided to the NS code as
a consistency check. 

\subsubsection*{The Nuclear Network}

\begin{figure}
\begin{centering}
\includegraphics[width=0.85\columnwidth,keepaspectratio]{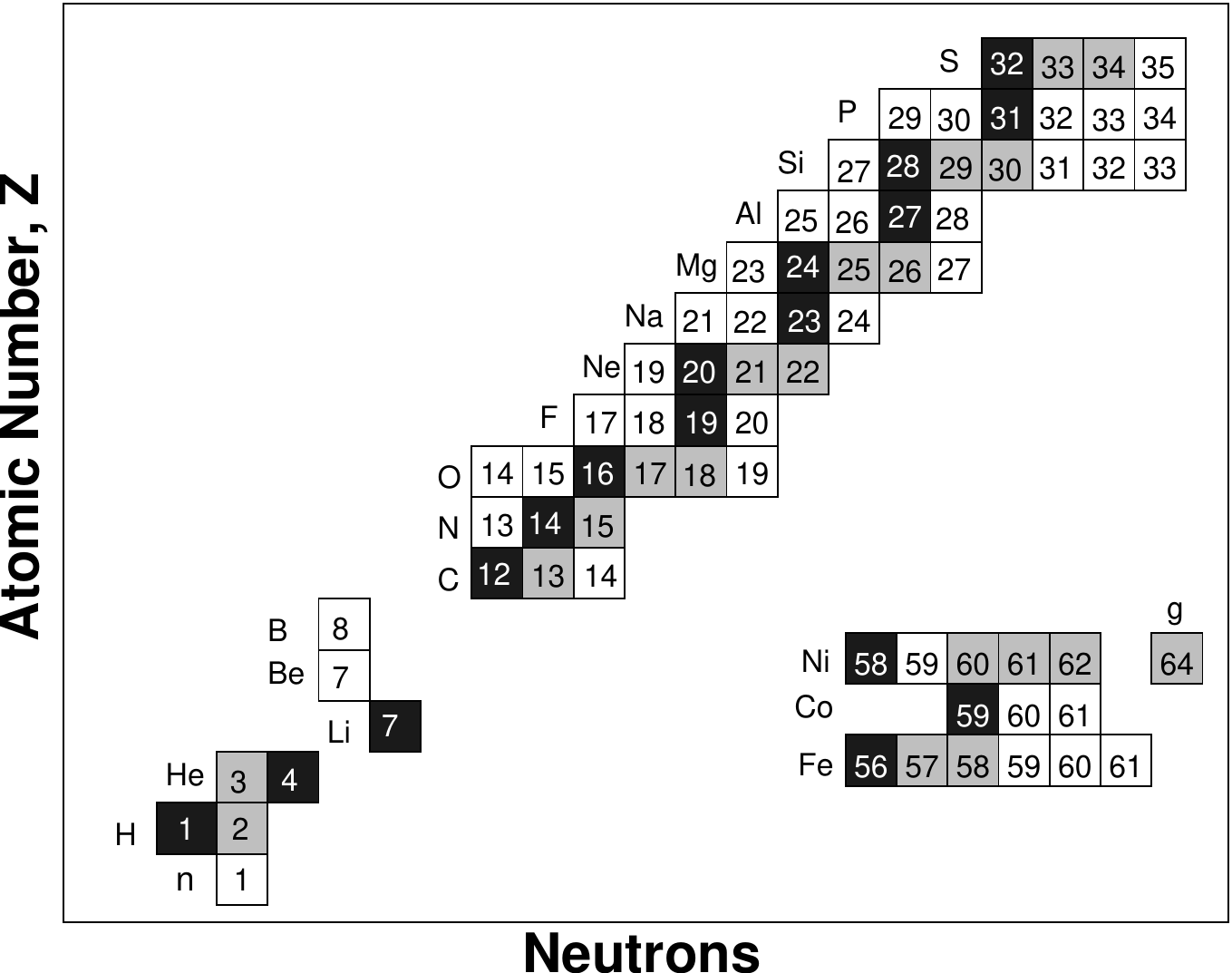}
\par\end{centering}
\caption{All the species included in the nuclear reaction network used for
the current study (except the synthetic species used for $^{26}$Al$_{m}$,
the metastable state of $^{26}$Al). Dark and light grey shading indicate
stable nuclides, whilst white indicates unstable nuclides. \label{fig-NScode-allSpecies}}
\end{figure}

In Figure \ref{fig-NScode-allSpecies} we show the 74 species that
are included in our nuclear network. The network is broken into two
sections. The first covers all species from neutron to $^{35}$S (58
species) whilst the second incorporates some iron group nuclides (Fe,
Co and Ni -- 14 species). Two extra `synthetic' species are also
included -- $^{26}$Al$_{m}$ and `$g$'. $^{26}$Al$_{m}$ represents
the metastable state of $^{26}$Al. The $g$ species is used to terminate
the network. It is produced via a synthetic decay from $^{62}$Ni
which itself is produced by neutron capture on $^{61}$Ni: $^{61}$Ni(n,$\gamma$)$^{62}$Ni$\rightarrow$$^{64}g$.
This species is given a mass of 64 to reflect that of the next stable
species formed by neutron captures, $^{64}$Ni. Following the method
of \citet{1989AA...221..161J} the cross section for the $^{61}$Ni(n,$\gamma$)$^{62}$Ni
reaction has been artificially given as an average representing that
of all the nuclides from $^{61}$Ni to $^{209}$Bi. In this way $g$
keeps track of the number of neutron captures occurring beyond the
end of our network. This can be used as a rough guide to estimate
the further s-processing that would occur amongst the heavier elements.
A similar tactic is used to account for the gap in the network between
S and Fe (ie. Cl to Mn). In this case the $^{34}$S(n,$\gamma$)$^{35}$S
reaction is given an averaged cross section (see \citealt{Lug01}
for details on the calculation of these averaged cross sections).

\subsubsection*{Nuclear Reaction Rates}

Most of the reaction rates we use in the current study come from the
Reaclib library (\citealt{Thi87}). However many of the rates have
been updated using various sources. These include: $^{22}$Ne(p,$\gamma$)$^{23}$Na
(\citealt{1995ApJ...451..298E}); $^{13}$C($\alpha$,n)$^{16}$O
(\citealt{1995AIPC..327..255D}); $^{24}$Mg($p,\gamma$)$^{25}$Al
(\citealt{DCP+99}); $^{25}$Mg($p,\gamma$)$^{26}$Al (\citealt{1996PhRvC..53..475I});
$^{26}$Mg($p,\gamma$)$^{27}$Al (\citealt{1990NuPhA.512..509I});
$^{22}$Ne($\alpha,n$)$^{25}$Mg and $^{22}$Ne($\alpha,\gamma$)$^{26}$Mg
(\citealt{1994ApJ...437..396K}). We note that this set of reaction
rates was compiled by \citet{Lug01} who lists all the updated rates
(relative to the \citealt{Thi87} compilation) in her Appendix A.
We refer to this combination of rates as our `standard' set. There
have been many new determinations of rates during the course of the
current study, but they have not been used. It is this standard set
that we use for all of our models -- unless otherwise stated.

\subsubsection*{Time-Dependent Mixing}

Time dependent mixing is important for many nucleosynthetic sites.
One classic example is lithium production through HBB in AGB stars
(see eg. \citealt{1992ApJ...392L..71S}). It will also be important
in our $Z=0$ and extremely metal-poor models, since these models
experience the dual core flash and dual shell flash events, which
occur on rapid timescales. 

\begin{figure}
\begin{centering}
\includegraphics[width=0.5\columnwidth,keepaspectratio]{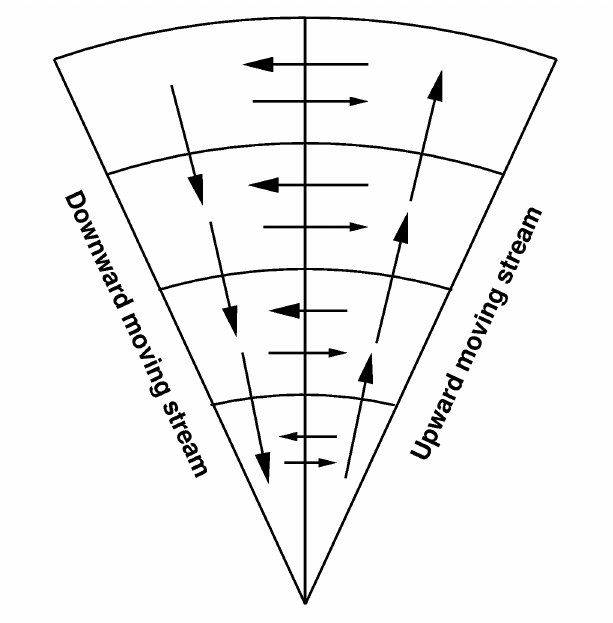}
\par\end{centering}
\caption{Schematic representation of the mixing algorithm used in the NS code
(taken from \citealt{1993MNRAS.263..817C}). Note that each cell is
linked to three other cells. Material moves between these three cells
(upwards, downwards and laterally) at rates derived from the local
values of the convective velocities and mixing lengths (which are
taken from the SEV code calculation). \label{NScode-CannonMix}}
\end{figure}

The NS code receives the run of mixing velocities and mixing lengths
(given by the mixing length theory of convection) versus mass from
the SEV code. With this information it uses a `donor cell' scheme
to model time-dependent mixing throughout the star (\citealt{1993MNRAS.263..817C}).
The algorithm allows for both upward and downward moving streams/eddies.
Thus any composition difference between upward moving eddies and the
downward moving eddies at the same radial position in the star --
which can arise if the mixing timescale is much shorter than the nuclear
reaction timescales -- is taken into account. The algorithm also
allows for lateral mixing. \citet{1993MNRAS.263..817C} notes that
if the lateral mixing is very fast then the donor-cell algorithm approximates
the diffusion equation (which is usually used for time-dependent mixing).
In Figure \ref{NScode-CannonMix} we show the mixing scheme schematically.

\subsubsection*{Other Code Details}

\begin{figure}
\begin{centering}
\includegraphics[width=0.8\columnwidth,keepaspectratio]{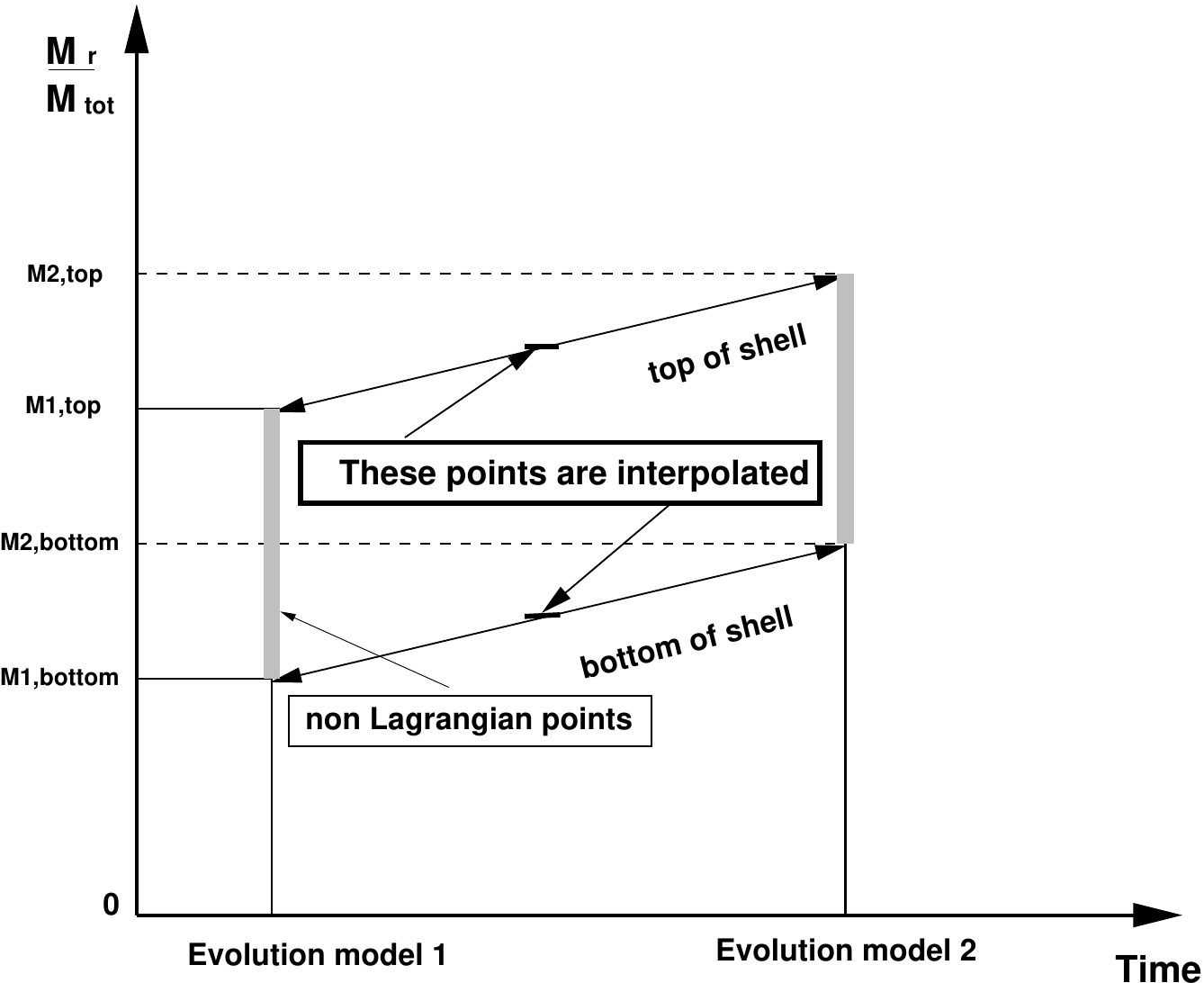}
\par\end{centering}
\caption{Schematic showing the location of two of the non-Lagrangian points
used to follow the H and He burning shells between two structural
evolution models (M$_{top}$ and M$_{bottom}$, for each model). The
mesh resolution is enhanced in and around these key points, to follow
the rapid nucleosynthesis that occurs in the shells. \label{fig-NScode-Mesh}}
\end{figure}

The mesh used in the NS code is a mix of Lagrangian and non-Lagrangian
points. The non-Lagrangian points are not Eulerian but follow the
hydrogen and helium burning shells. The NS code reads in two structural
models at a time. It then locates the H and He burning shells in each
structural model, based on the H and He abundances, and places the
non-Lagrangian points at key positions within the shells. An interpolation
is performed between the positions of the shells in the two structural
models. In this way the positions of the shells are tracked (see Figure
\ref{fig-NScode-Mesh}). This allows the mesh resolution to be selectively
enhanced around the burning shells. This is necessary because this
is where most of the nucleosynthesis occurs. The NS code then evolves
forward, using its own timesteps. When the end point of the nucleosynthetic
evolution between these two structural models is reached, the next
model is read in and the process repeated. A typical number of mesh
points in an AGB model is $\sim200$. In some cases we have however
increased this to $\sim2000$ in order to resolve small radiative
pockets between convection zones (eg. during Dual Flashes in $Z=0$
models). 

\subsection{NS Code Modifications}

As noted in the introduction to the present chapter the NS code required
much less modification than the SEV code in order to enable the modelling
of $Z=0$ and extremely metal-deficient stars. There are two reasons
for this. Firstly the NS code only calculates the nucleosynthetic
evolution, from a given set of quasi-static structural models. Supplying
it with $Z=0$ or extremely metal-poor models has no impact on its
operation, although we note that the mesh resolution needed to be
increased in some of the models due to the appearance of narrow radiative
zones between convective zones. Secondly the NS code already treated
convective mixing in a time-dependent manner, so the conditions given
by the extreme events in our new models in which the mixing and burning
timescales become comparable (ie. the Dual Flash events) were already
taken into account.

The only significant modification we made during the course of this
study was to remove the scaled-solar assumptions for the initial abundances.
This allowed the use of custom-made initial compositions such as those
of our $Z=0$ and extremely metal-poor models. 

\subsection{Yield Calculation and Synthetic Pulses\label{subsection-YieldCodeCalcsAndSynthPulses}}

Our yields are generally given in mass fraction of each species in
the total ejecta (most of them are tabulated in the appendices). Thus
the yields are always positive -- unlike some definitions that take
into account the initial composition of the models. In the tables
we also give the initial stellar mass and final mass (remnant mass),
as well as the initial compositions. Using this information it is
easy to convert our yields to any format. 

The yields are calculated by integrating the mass of each species
lost by the star over its lifetime ($\tau_{*}$):

\begin{equation}
M_{i}^{tot}=\int_{0}^{\tau_{*}}X_{i}(t)\frac{dM}{dt}dt\label{eqn-YieldCalc}
\end{equation}

where $X_{i}$ is the mass fraction of species $i$. The total mass
of each species lost to the ISM, $M_{i}^{tot}$, is then scaled with
$M_{ej}$, the total mass lost by the star, to give the mass fractions.

In some cases our models failed to converge towards the end of the
AGB. This is a common problem with stellar codes. Often there was
very little mass left in the envelope so this was just added to the
yields. However in some cases there was enough mass left that it would
not have been lost in one interpulse period. In these cases we did
a short synthetic evolution calculation of the remaining thermal pulses
(including third dredge-up and core growth) to complete the evolution,
following the method of \citet{entry-5}. Yields were then calculated
including the extra mass loss. In most models the number of thermal
pulses calculated in this way was $\lesssim8$. This represents between
$\sim1$ and $10\%$ of the total number of thermal pulses in most
of our models. Thus the synthetic pulses generally have a minor impact
on the yields.

\chapter{Structural Evolution Code Modifications\label{sevmods}}
\begin{quote}
\textquotedbl It's better to light a candle than curse the darkness.\textquotedbl{}
\begin{flushright}
\vspace{-0.4cm}-- Chinese Proverb
\par\end{flushright}

\end{quote}

\section{Time-Dependent Mixing\label{timedepmix}}

\subsection{Motivation}

As discussed in section \ref{evcodeCvnMixing}, the SEV code contains
the assumption that the timescale of mixing in a convection zone is
much smaller than that of the evolutionary timescale. This is implemented
through the instantaneous mixing approximation whereby the abundances
in any convective zone are mixed to homogeneity over the whole zone.
This is done each iteration after the change in composition from nuclear
burning has been applied. In the majority of situations involved in
the evolution of low- and intermediate-mass stars this is a valid
approximation. It reduces mixing to a very simple process, speeding
up calculations. 

However there do exist some evolutionary episodes in which physical
changes are happening at a very high rate. During these episodes the
timestep is reduced markedly in order to follow the evolution accurately.
In these situations the timescale of mixing over a convective region
can become comparable to the timestep being used. If this happens
then the instantaneous mixing approximation is not valid and the time
dependence of the mixing needs to be followed. The chemical profile
in a convective zone at the end of a timestep will not be uniform
(as it would be with instantaneous mixing) in this case, as the convective
zone will be only partially mixed. In order to model these evolutionary
episodes more properly the Author decided to modify the SEV code to
include time-dependent mixing. In what follows we describe the mixing
paradigm used as well as the practical details of the implementation
(and testing) of the new time-dependent mixing subroutine.

\subsection{The Diffusive Mixing Paradigm}

As discussed in section \ref{evcodeCvnMixing}, convection in stars
is in general highly turbulent. Thus mass (and heat) movement is happening
on all scales simultaneously -- there is a spectrum of scales, velocities,
energies involved. Unfortunately it is still not feasible to model
convection from first principles. Indeed, a full mathematical description
of convection remains one of the Holy Grails of physics. Solving the
Navier-Stokes equations is also unfeasible for evolutionary timescales
such as those of stars, as the computational demand is enormous. We
are thus left with the task of \emph{approximating} the effects of
turbulent convection (and mixing) in stars. 

Also discussed in section \ref{evcodeCvnMixing} is the Mixing Length
Theory (MLT) of convection. The MLT is the most commonly used formalism
for convection in stellar models. In practice it provides a (locally
determined) velocity and mixing length for each point in a convection
zone, amongst other physical quantities. Like many other authors we
have chosen to retain the MLT as the convection formalism for the
SEV code.

In order to simulate the \emph{mixing} in the convective zones in
a time-dependent manner we need a reasonable physical/mathematical
description of turbulent mixing. Due to the random motions in turbulent
convection (as opposed to the regular turnover structures in the non-turbulent
regime), the problem has been likened to a diffusion process whereby
random mass motions gradually smooth out the composition gradient.
Indeed, diffusion can be understood in the context of kinetic theory,
whereby random motions of particles of say, a gas, which has a composition
gradient, will over time smooth out the gradient (eg. \citealt{Gian88}).
This process, which is related to the second law of thermodynamics,
was first quantified by \citet{fick} and is known as Fick's Law:

\begin{eqnarray}
J & = & -D\frac{dC}{dx}.\label{ficks}
\end{eqnarray}

Here $J$ is the rate of transport for the quantity, say `particles'
(so units are: particles per unit area per second), $D$ the diffusion
coefficient (area$^{\textrm{2}}$ per second), $C$ the concentration
(particles per unit volume), and $x$ the distance. It can be seen
that the flux of material is dependent on its gradient in space. A
steeper gradient leads to a faster rate of redistribution. In terms
of the second law of thermodynamics, the material tends towards a
(system-wide) maximum in entropy (disorder) -- a state of equilibrium.
This happens spontaneously and without any loss of energy. The natural
phenomenon of diffusion/entropy-maximisation has been observed in
a huge number of systems. Some simple examples include: diffusion
of water into spaghetti; diffusion of perfume around a room; diffusion
of He out of a balloon. Other physical process are also well described
by diffusion, such as: heat conduction, momentum diffusion and electron
diffusion. The main difference in all of the different scenarios lies
in the diffusion coefficient $D$. The diffusion coefficient contains
the physical information about each particular system. For instance
characteristic $D$s in liquids are quite different to those of gases.
However, a standard result from kinetic theory arguments gives a simple
form of $D$, relating it to the \emph{mean free path} $l_{m}$ of
the moving material and its characteristic velocity $v$ (in the direction
of diffusion):

\begin{eqnarray}
D & = & -\frac{1}{3}vl_{m}\label{diffcoeff}
\end{eqnarray}

Thus, if we accept that turbulent mixing can be described as a diffusion
process (ie. Fickian transport), and if we have some information about
the mean free path of convecting parcels and their velocities then
we can make an order of magnitude estimation of the diffusion coefficient
in stars.

Early stellar modellers first applied the concept of diffusion to
the problem of semiconvection (eg. \citealt{1971ApJ...165..295S};
\citealt{1971BAAS....3R.394S}). It was needed in this regime of mixing
because semiconvection is thought to cause a slow redistribution of
chemical elements (see below for a discussion of semiconvection) and
thus warranted the use of time-dependent mixing. Other descriptions
of semiconvection had mimicked the essentially diffusive process of
semiconvection by adjusting chemical profiles to achieve `convective
neutrality' (ie. $\nabla_{rad}=\nabla_{ad}$, eg. \citealt{1972ApJ...171..309R}).
A year or so later the diffusive mixing approximation to turbulent
mixing was introduced into some stellar evolution codes to calculate
mixing over the entire star (eg. \citealt{1972MNRAS.156..361E}; \citealt{1974ApJ...187..555S}).
These authors both retained the MLT as the convection formalism. Of
central importance to applying the diffusion approximation is the
determination of a physically plausible diffusion coefficient. In
his indomitable style, \citet{1972MNRAS.156..361E} chose a ``na\"{i}ve
formula'' for the turbulent mixing diffusion coefficient and stated
that ``...it is not necessary to have a physically correct expression
in order to obtain sensible answers.''. However, he did construct
his formula based on order of magnitude physical estimates. \citet{1974ApJ...187..555S}
take a more pragmatic approach, taking the diffusion coefficient as:

\begin{eqnarray}
D & = & v_{conv}l_{mix}\label{sackmann}
\end{eqnarray}

where $v_{conv}$ and $l_{mix}$ are the local convective velocity
and mixing length, as calculated from the MLT. This is basically equation
\ref{diffcoeff}. As the MLT is a crude approximation to turbulent
convection, this can only be taken as an order of magnitude estimate,
and the factor of $\frac{1}{3}$ in equation \ref{diffcoeff} is irrelevant. 

Equation \ref{ficks} is actually Fick's \emph{1$^{\textrm{st}}$}
Law, describing steady state diffusion. Fick's 2$^{\textrm{nd}}$
Law describes the more common case of non-steady state diffusion.
It is known as the standard diffusion equation:

\begin{eqnarray}
\frac{dC}{dt} & = & \frac{\partial}{\partial x}\left(D\frac{\partial C}{\partial x}\right)\label{diffneqn1}
\end{eqnarray}

in the case of one spatial dimension. In the SEV code we use mass
as the independent variable and assume spherical symmetry, so the
diffusion equation becomes:

\begin{eqnarray}
\frac{dX_{i}}{dt} & = & \frac{\partial}{\partial m_{r}}\left(\sigma D\frac{\partial X_{i}}{\partial m_{r}}\right)\label{diffneqn2}
\end{eqnarray}

where

\begin{eqnarray}
\sigma & = & \left(4\pi r^{2}\rho\right)^{2}\label{sigma}
\end{eqnarray}

and we have changed to the usual stellar notation of $X_{i}$ representing
each chemical species that will be transported. Equation \ref{diffneqn2}
is the one we have applied to turbulent mixing in the SEV code. Diffusion
equations have the added advantage in that there are good numerical
methods to solve them. The next section describes the implementation
of Equation \ref{diffneqn2} in the new time-dependent-mixing subroutine.

\subsection{Choice of Implementation Method}

We begin by noting that there are a few different ways in which the
new mixing paradigm could be implemented in the SEV code. As the rate
of nuclear burning in a star is partly tied to the local composition,
and mixing alters that composition, it is sensible to say that the
mixing and burning are \emph{coupled}. Taking account of this coupling
would lead to an equation for composition change:

\begin{eqnarray}
\frac{dX_{i}}{dt} & = & \left(\frac{\partial X_{i}}{\partial t}\right)_{nuc}+\frac{\partial}{\partial m_{r}}\left(\sigma D\frac{\partial X_{i}}{\partial m_{r}}\right)\label{diffneqn3}
\end{eqnarray}

(eg. \citealt{1974ApJ...187..555S}) which is just Equation \ref{diffneqn2}
with the sources and sinks from nuclear reactions summarised in the
first term on the right. Furthermore, there will be (some) feedback
on the structure itself if the nuclear rates are significantly altered,
implying that Equation \ref{diffneqn3} should be coupled with the
equations of stellar structure. The next question is: how does this
fit in with the SEV code? The SEV code traditionally applies convective
mixing after 1) calculating the (approximate) structure and 2) applying
the changes in composition due to nuclear burning. This is done in
a series of iterations. In this context it appears we have three possible
scenarios for implementing diffusive mixing:
\begin{enumerate}
\item Keep the current method of mixing each convection zone each iteration,
giving a `semi-coupled' evolution, ie. just replace the the instantaneous
mixing with Equation \ref{diffneqn2}.
\item Apply a coupled mix and burn each iteration, so the burning and mixing
are fully coupled and the chemical changes are `semi-coupled' to the
structure, ie. replace the current method of calculating the chemical
changes due to nuclear reactions \emph{and} the instantaneous mixing
by Equation \ref{diffneqn3}.
\item Solve Equation \ref{diffneqn3} for each species simultaneously with
the structure equations in the Henyey matrix, ie. fully-couple structure
and composition change.
\end{enumerate}
At first glance, it appears that the most self-consistent method would
be that of (3). However it can be argued (Wood and Lattanzio 2006,
private communication) that an iterative solution (ie. method (1))
would give the same result as a fully coupled solution. Indeed, in
the SEV code and many other stellar codes, the nuclear burning and
mixing are always treated in an iterative fashion. The fact that the
models produced by these codes are not significantly different to
fully coupled codes (also known as fully-simultaneous codes) suggests
that the iterative method is equivalent to the fully-coupled method.
Recent work using an updated version of the fully-coupled Cambridge
University STARS code by \citet{2001MmSAI..72..299P} shows that this
is indeed the case, at least for AGB stars. They state that their
results are comparable to those from the SEV code. Wood and Lattanzio
argue that all the equations would be essentially solved together
with successive substitutions in the diffusion equations after each
Henyey (structure) iteration. Wood also notes that convergence would
probably be slower for method (1) than for an explicitly fully-coupled
method, but that the iterative method may actually be more \emph{stable}
because of the slower convergence. Stability is certainly a sought-after
characteristic for stellar evolution codes. 

The initial decision taken by the Author was to implement method (1)
and then method (3) in order to compare the results of each method
directly, without the added uncertainties of comparing different codes.
However, due to time constraints, only method (1) was completed in
time for this thesis. Given the arguments above it seems that this
should be a reasonable approach.

\subsection{Choice of Numerical Method}

Given the context just discussed above, we now describe the procedure
taken to replace the instantaneous mixing procedure with Equation
\ref{diffneqn2}.

\subsubsection*{Overview and External Testing}

The first step in adapting differential equations to numerical/computational
work is to difference the equations. There are many methods for differencing
equations, with each method having certain desirable or undesirable
qualities. The methods basically fall into three groups: explicit,
implicit and hybrid (a combination of explicit and implicit). Explicit
methods are not usually used for stellar evolution. The reason for
this is that they become unstable when using large timesteps such
as those required to model stars. The approximate point at which explicit
methods become unstable (giving unphysical random or periodic solutions)
is given by the well-known Courant condition (Equation \ref{courant}).
The benefit of explicit methods is that they are accurate when they
are stable. This contrasts with implicit methods which are unconditionally
stable with any sized timestep (ie. no Courant condition) but can
suffer from inaccuracies if the timestep is too large. Hybrid methods
combine the explicit and implicit methods and have qualities stemming
from both methods. They maintain stability for much longer timesteps
than explicit methods but do have the equivalent of a Courant condition
above which they are unstable. However they are more accurate (in
general) than implicit methods while they are stable. The reason we
describe these methods here is because we have run tests on the diffusion
subroutine using a variety of differencing methods, to check for accuracy
and stability. We tried three different differencing methods:
\begin{enumerate}
\item Our simple implicit method obtained by directly differencing Equation
\ref{diffneqn2} (see eg. \citealt{1992nrfa.book.....P}):
\begin{eqnarray}
\frac{X_{j}^{t+1}-X_{j}^{t}}{\Delta t} & = & \sigma_{j+\frac{1}{2}}^{2}D_{j+\frac{1}{2}}\left(\frac{X_{j+1}^{t+1}-2X_{j}^{t+1}+X_{j-1}^{t+1}}{\left(\Delta m\right)^{2}}\right)\label{simimplicit}
\end{eqnarray}
where $t$ represents the timestep, $j$ the mesh point and $\sigma=(4\pi\rho r^{2})$.
Note that $\Delta m$, $\rho$ and $D$ are constant in these tests.
The physical values at $j+\frac{1}{2}$ are taken as averages except
for $D$ where we follow \citet{2004AA...416.1023M}: 
\begin{eqnarray}
D_{j+\frac{1}{2}} & = & \frac{D_{j+1}D_{j}}{fD_{j+1}+(1-f)D_{j}}\label{MeyDiffcoeff}
\end{eqnarray}
 which allows the smallest diffusion coefficient to dominate between
mesh points. This formula for $D$ is used in all the following methods
also.
\item The implicit method described in \citet{2004AA...416.1023M} (they
make an approximation to reduce the second order equation to first
order) adjusted to the SEV code formalism of meshpoint numbering starting
at the centre rather than the surface:
\begin{multline}
\frac{X_{j}^{t+1}-X_{j}^{t}}{\Delta t}=-\frac{\sigma_{j-\frac{1}{2}}}{\Delta m}D_{j-\frac{1}{2}}\left(\frac{X_{j}^{t+1}-X_{j-1}^{t+1}}{r_{j}-r_{j-1}}\right)\\
+\frac{\sigma_{j+\frac{1}{2}}}{\Delta m}D_{j+\frac{1}{2}}\left(\frac{X_{j+1}^{t+1}-X_{j}^{t+1}}{r_{j+1}-r_{j}}\right)\label{MeynetImplicit}
\end{multline}
\item The Crank-Nicholson method described in \citet{2004AA...416.1023M},
also adjusted for the SEV code formalism.
\end{enumerate}
All these methods reduce to solving a set of simultaneous linear algebraic
equations. The systems of equations are all tridiagonal thus easily
solved using standard methods. As an example we show the set of equations
we solve for one of the implicit schemes (the modified version of
the Meynet et al. scheme), rearranged ready for numerical coding:

\begin{multline}
X_{j}^{t}=\left(1+\frac{[j,j-1]}{\Delta m_{j}}+\frac{[j+1,j]}{\Delta m_{j}}\right)X_{j}^{t+1}\\
-\left(\frac{[j,j-1]}{\Delta m_{j}}\right)X_{j-1}^{t+1}-\left(\frac{[j+1,j]}{\Delta m_{j}}\right)X_{j+1}^{t+1}\label{matrix1}
\end{multline}

where, for example:

\[
[j,j-1]=\Delta t\,\frac{4\pi r_{j-1/2}^{2}D_{j-1/2}^{\,}}{r_{j}-r_{j-1}}.
\]

Boundary conditions were taken such that the flux at the centre and
surface is zero:

\[
\left(\frac{\partial X_{i}}{\partial m}\right)_{cent}=\left(\frac{\partial X_{i}}{\partial m}\right)_{surf}=0
\]

which, from Equation \ref{matrix1} implies we have at the centre
(where $j=1$):

\begin{equation}
X_{1}^{t}=\left(1+\frac{[2,1]}{\Delta m_{1}}\right)X_{1}^{t+1}-\left(\frac{[2,1]}{\Delta m_{1}}\right)X_{2}^{t+1}\label{matrixCentre}
\end{equation}

and at the surface (where $j=jm$):

\begin{equation}
X_{jm}^{t}=\left(1+\frac{[jm,jm-1]}{\Delta m_{jm}}\right)X_{jm}^{t+1}-\left(\frac{[jm,jm-1]}{\Delta m_{jm}}\right)X_{jm-1}^{t+1}.\label{matrixSurface}
\end{equation}

To solve the matrix system we employed the subroutine from Numerical
Recipies called $TRIDAG$ (\citealt{1992nrfa.book.....P}). The diffusive
mixing subroutine is named \emph{diffuse} (in \emph{diffusion.f})
and is called by \emph{abund-diffn} which is a modified version of
the original SEV code subroutine \emph{abund.f}. 

In order to have some sort of reliable benchmark comparison for the
above methods, we also coded an explicit method differenced in the
following way:

\begin{eqnarray}
\frac{X_{j}^{t+1}-X_{j}^{t}}{\Delta t} & = & \sigma_{j+\frac{1}{2}}^{2}D_{j+\frac{1}{2}}\left(\frac{X_{j+1}^{t}-2X_{j}^{t}+X_{j-1}^{t}}{\left(\Delta m\right)^{2}}\right)\label{SimExplicit}
\end{eqnarray}

which is the same as Equation \ref{simimplicit} except it only uses
information at the current timestep $t$ to evaluate the values at
the next timestep $t+1$. When calculating diffusion using this method
we always maintained the timestep well below that required for stability
by the Courant condition:

\begin{eqnarray}
\Delta t & \ll & \frac{\left(\Delta r\right)^{2}}{2D}\label{courant}
\end{eqnarray}

\begin{figure}
\begin{centering}
\includegraphics[width=0.6\textwidth,keepaspectratio,angle=270]{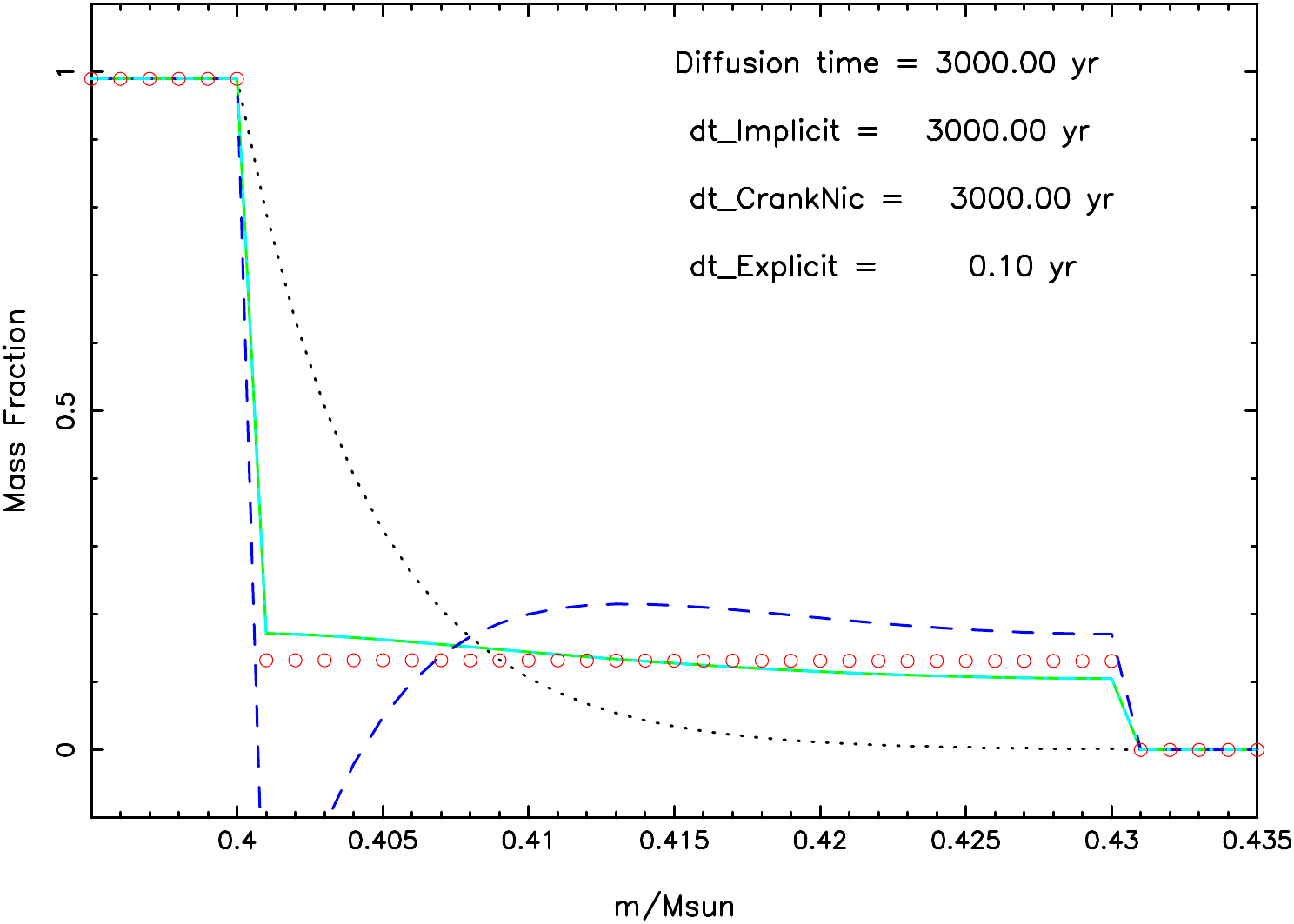}
\par\end{centering}
\caption{Initial testing of the diffusion subroutine (outside SEV code). The
initial composition profile (dotted line, black) was designed to mimic
the main product of a burning shell (eg. helium). The smoothly varying
section of the initial profile resembles a build-up of composition
after one timestep due to nuclear fusion. Normally this region would
be homogenised by the instant mixing routine. The diffusion coefficient
was taken as zero in the regions $m<0.4$ and $m>0.43$ to represent
radiative zones around a convective shell with a moderate diffusion
coefficient (one that results in the homogenisation of the region
in $<3000$ years, ie. as given by the explicit scheme with red circles).
The simple model has $M=M_{\odot}$ and is of constant density. Resulting
profiles from four differencing methods are plotted: 1) Our implicit
(solid green line), 2) Meynet et al. 2004 implicit (dash-dot cyan
line), 3) Crank-Nicholson (dashed blue line), and 4) Our \emph{explicit}
method (circles, red). The red circles also indicate the mesh spacing,
which is the same for all cases. Note that the two implicit scheme
lines overlay each other in the graph. The explicit method was given
a timestep small enough to stay under the Courant condition (0.1 yr),
while all the other methods had a single timestep of 3000 yr. All
the diffused profiles should end up going through the red circles.
It can be seen that, while all the non-explicit methods are quite
inaccurate (with the given conditions), the implicit methods give
smooth results whilst the hybrid Crank-Nicholson gives a result very
far from the explicit solution. In fact the CN method gives `non-physical'
results such that the abundance profile becomes negative in one region.
Note that the final profiles from both implicit methods are practically
identical.\label{fig-diffnTests1}}
\end{figure}

The practicality of making comparisons in this way is limited by the
timesteps for the explicit method, as they are so tiny (often $<1\textrm{yr}$).
Here we have limited our tests to an evolutionary time interval of
3000 yr, which is not unreasonable for a single timestep in a stellar
evolution calculation (depending on the evolutionary phase). We initially
tested the different differencing methods outside the SEV code, by
setting up an artificial chemical profile (single element) in a constant
density model with uniform mass spacings $\Delta m$. The diffusion
coefficient was taken as a step function with radiative (non-convective)
zones on either side of a uniformly convective shell. This setup was
designed to mimic a convective nuclear burning shell. Figure \ref{fig-diffnTests1}
displays the results of one of the tests. The main results were 1)
both the implicit and Crank-Nicholson (CN) methods had significant
errors and 2) the implicit methods gave quite smooth profiles (although
inaccurate) whilst the CN method gave an almost random profile.

\begin{figure}
\begin{centering}
\includegraphics[width=0.6\textwidth,keepaspectratio,angle=270]{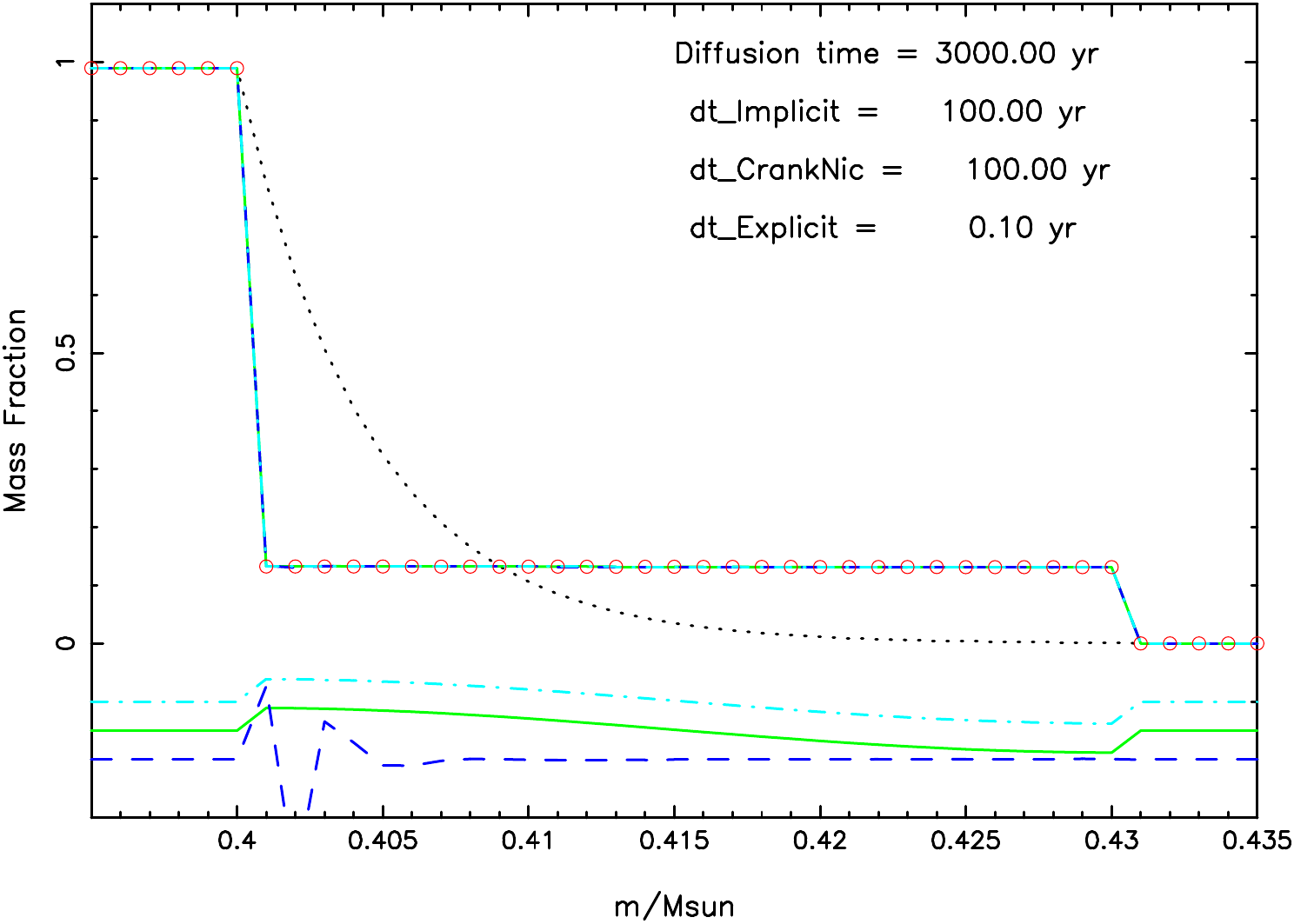}
\par\end{centering}
\caption{Same as Figure \ref{fig-diffnTests1} except the timestep for the
non-explicit methods has been reduced to 100 yr, resulting in 30 timesteps
within the evolutionary timestep of 3000 yr. The three lines at the
bottom are the differences between each non-explicit method and the
explicit method, multiplied by $10^{3}$, to give an indication of
error. It can be seen that all methods performed much better with
the smaller timestep, as all the curves are very close to the explicit
method's solution. Again, the implicit methods give a smooth profile
whilst the Crank-Nicholson is non-smooth.\label{fig-diffnTests2}}
\end{figure}

To check the dependence of accuracy on timestep we reduced the timestep
for the non-explicit methods to 100 yr. Figure \ref{fig-diffnTests2}
shows the results of this test. Reducing the timestep by an order
of magnitude and consequently taking a series of steps within the
evolutionary timestep of 3000 years increased the accuracy dramatically,
although the CN method still gave a non-smooth profile. Further tests
revealed that the CN requires a very small timestep to return smooth
results, though not as small as that needed for the explicit method.

\begin{figure}
\begin{centering}
\includegraphics[width=0.6\columnwidth,keepaspectratio,angle=270]{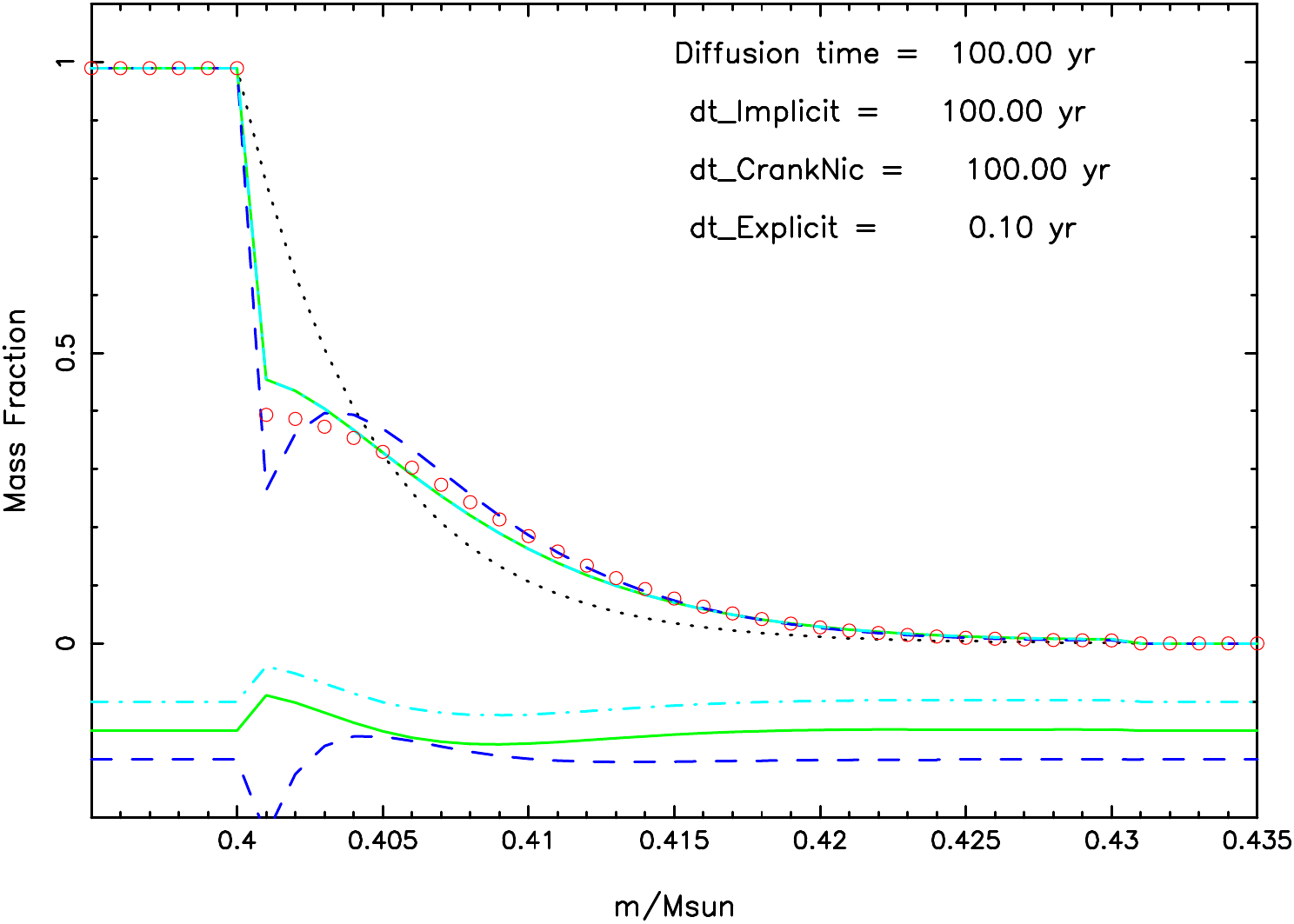}
\par\end{centering}
\caption{Same as Figure \ref{fig-diffnTests2} except that the results for
the first timestep are plotted (rather than for the final timestep).
The errors have \emph{not} been magnified. In this case the chemical
profile has not had time to homogenise over the shell, resulting in
an intermediate profile (ie. as given by the explicit case, red dots).
It can be seen that there are substantial errors in the abundance
distributions after a single timestep despite the good results seen
in Figure \ref{fig-diffnTests2} after completing all the timesteps.
This suggests that the driving source of error is the change in profile
over the timestep rather than the timestep itself, indicating that
the best approach is to limit the amount of change in each timestep.
\label{fig-diffnTests3}}
\end{figure}

Another interesting result from the testing is displayed in Figure
\ref{fig-diffnTests3}. The plot shows the resulting composition profiles
after only one sub-evolutionary timestep. Large errors can be seen
in all the results, although again the implicit method results are
smooth whilst the CN method results are disjointed. These results
are interesting because the results by the end of the timestepping
(previous plot, Figure \ref{fig-diffnTests2}) show very little error,
despite going through this initial inaccurate stage. Comparing all
three test Figures suggests that the main driver of error (in the
non-explicit methods) is not the size of the timestep but the size
of the composition jump -- ie. how much movement of material there
is during the timestep. This implies that the key issue in maintaining
accuracy in the diffusion routine is to place a limit on the amount
of material diffused in one timestep. This could also be achieved
by limiting the amount of material burnt in an evolutionary timestep.
Alternatively the evolutionary timestep could be divided into many
`diffusion timesteps', as the profile obtained in the case with small
timesteps (Figure \ref{fig-diffnTests2}) was much more accurate than
that when using a single large timestep (Figure \ref{fig-diffnTests1}).

Another important conclusion drawn from these tests was that the implicit
methods are the only practical methods to include diffusion in the
stellar evolution code. The reason for this is that it is totally
impractical to use timesteps as small as are needed for stability
by the explicit and CN methods. Although the implicit methods give
somewhat inaccurate results (although stable) we believe that this
is not a significant problem as there are many uncertainties in the
turbulent diffusive mixing paradigm itself. For instance, the diffusion
coefficient, which determines the rate of diffusion, is based on the
velocities and length scales given by the MLT (Equation \ref{diffcoeff}),
which is in itself a rough approximation of convection (see section
 \ref{evcodeCvnMixing}). Based on an investigation which mainly focused
on stability (as opposed to accuracy), \citet{2004AA...416.1023M}
also recently came to the conclusion that implicit finite difference
methods are the best to use for diffusion in stellar modelling. 

\subsubsection*{Grafting the Diffusion Routine to the SEV Code}

The vast majority of the time taken in implementing turbulent diffusive
mixing was spent installing the subroutine into the SEV code. Interfacing
the codes proved to be more difficult than expected. However the implementation
was successful in the end.

Given that the two different implicit methods give almost identical
results and that they both share the same matrix solver, we decided
to implement both schemes to enable comparisons within the SEV code.
The instantaneous mixing routine was retained, and a switch added
in the code so the operator can choose one of three mixing scenarios:
\begin{enumerate}
\item Instantaneous mixing all the time
\item Instantaneous mixing with a mixing timescale test that invokes the
diffusion routine if needed
\item Diffusive mixing all the time
\end{enumerate}
Mass conservation is a prime requirement of the diffusion routine.
In order to check this we sum the mass of each chemical species before
entering the diffusion routine:

\begin{equation}
\sum_{j=1}^{jm}\Delta m_{j}W_{j+\frac{1}{2}}\label{abundsum}
\end{equation}

where $W$ is the variable used for abundance in the SEV code. After
the profile is diffused the same sum is calculated and compared with
the original. If they differ than more than a prescribed limit (say
$\epsilon_{W}$) then the evolutionary timestep is reduced and the
model recalculated. This is quite a stringent test as the errors are
summed over \emph{every mesh point,} although it does depend on the
value taken for $\epsilon_{W}$, which we have initially chosen to
be $10^{-7}$. Initial testing showed that the method of \citet{2004AA...416.1023M}
(Equations \ref{MeynetImplicit} and \ref{matrix1}) conserves mass
better than the other method, so we have retained it as the default
method for diffusion. Mass conservation is observed to improve as
the timestep is decreased, as expected from the external testing.
In the SEV code the timestepping is controlled by parameters that
limit the changes of various quantities (eg. the luminosity, density,
etc.). Of particular relevance to the diffusion routine is the $STW$
parameter, which tracks the abundance changes in time and is used
to limit the timestep if changes are occurring too quickly. It was
found that this existing control parameter is ideal for limiting the
errors in the diffusion routine, as they appear to be dependent on
the degree of change in chemical profile. As a further check for the
degree of mass conservation needed in practice, we also calculated
the mass sums of each element before and after the nuclear burning
was calculated. It was found that the errors were sometimes as high
as $10^{-6}$. Again this is the error summed over all points in the
star. Thus we believe using $\epsilon_{W}=10^{-7}$ should be of sufficient
accuracy in practise.

\begin{figure}
\begin{centering}
\includegraphics[width=0.8\columnwidth,keepaspectratio]{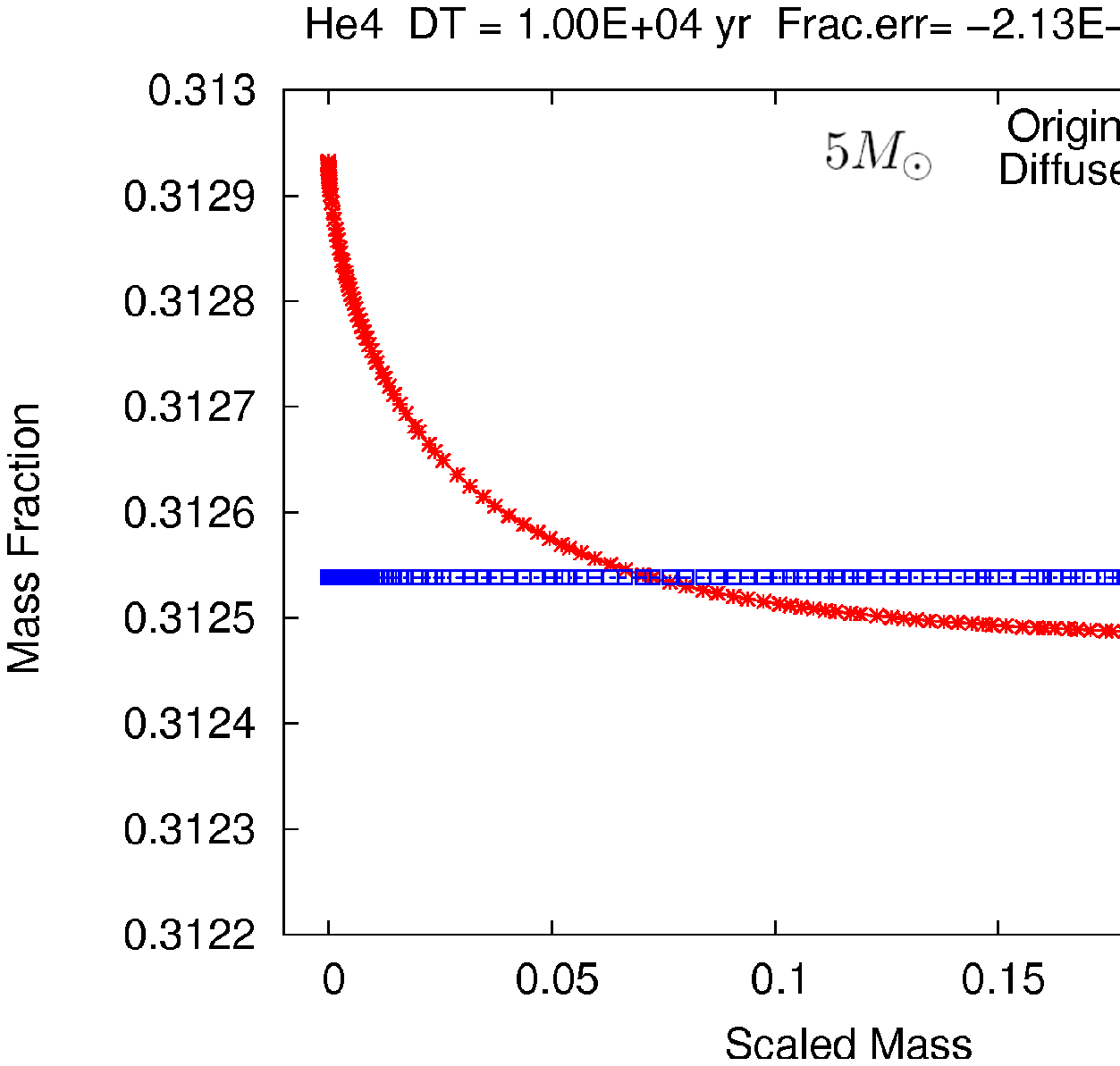}
\par\end{centering}
\caption{In situ testing of the diffusion routine. The stellar model has a
mass of 5 M$_{\odot}$ and is on the main sequence so it has a convective
core and a radiative envelope (the model started with $\textrm{Y}=0.25$).
This particular model has a timestep of $10^{4}$ yr. Plotted in red
is the $^{\textrm{4}}$He profile in the core after the nuclear burning
has been calculated (during one iteration). The blue line is the $^{4}$He
profile after applying the diffusion. It can be seen that the mixing
timescale is easily shorter than the evolutionary timestep as the
profile is completely homogenised at the end of the timestep. Also
of note is the small scale of abundance inhomogeneity before it enters
the diffusion routine. This leads to the small error in the sum of
Y ($\sim10^{-9}$). The markers on each line represent the mesh points.
\label{fig-diffuseHe4}}
\end{figure}

Figure \ref{fig-diffuseHe4} shows the diffusive mixing subroutine
in action in a real stellar model. A very simple case has been chosen
here -- a 5 M$_{\odot}$ low metallicity star on the main sequence
(MS). The $^{4}$He profile in the core is plotted after the nuclear
burn had been calculated and again after diffusion was applied. The
initial profile shows that an increase of helium mass fraction of
$\sim10^{-4}$ has been produced via hydrogen burning. After the diffusion
is applied the distribution is flat, indicating that the timescale
of mixing is much smaller than the evolutionary timestep of $10^{4}$
yr. Mass is conserved adequately in this example, deviating by $\sim10^{-9}$.

\begin{figure}
\begin{centering}
\includegraphics[width=0.8\columnwidth,keepaspectratio]{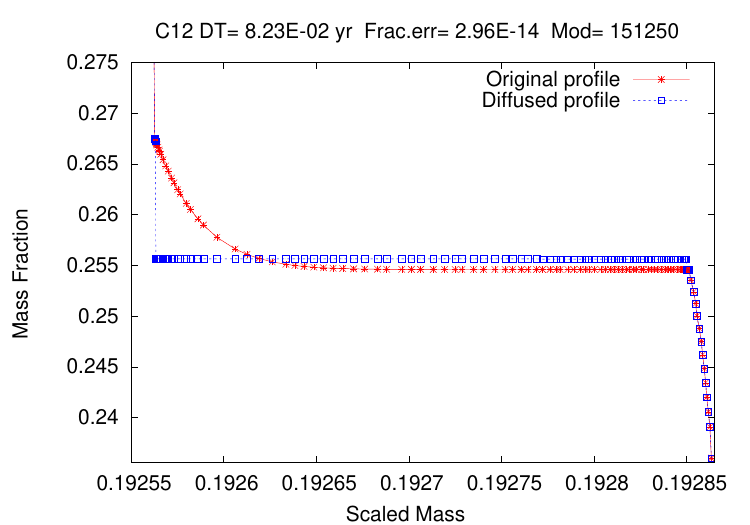}
\par\end{centering}
\caption{The same as Figure \ref{fig-diffuseHe4} but during the AGB evolution
and for $^{12}$C. The region plotted is the convective intershell
during a He flash. Again the timescale of mixing is much smaller than
the timestep ($\sim10^{-2}\textrm{ yr}$), as evidenced by the flat
composition profile after diffusion. The mass conservation is very
good ($\sim10^{-14}$). \label{fig-diffn-Intshell}}
\end{figure}

The next example (Figure \ref{fig-diffn-Intshell}) is from the same
star but later in its evolution. Here we choose a different type of
convection zone -- the convective intershell during a He flash on
the AGB. In this case helium is burning at the base of the convection
zone (creating the convective zone), mainly producing carbon. As can
be seen, the initial abundance profile is qualitatively similar to
the previous example. Despite the small timestep of $\sim10^{-2}$
yr, the mixing timescale is again much shorter than the evolutionary
timestep resulting in a flat profile. The error in this example is
quite small, being $\sim10^{-14}$ when summed over every mesh point.

\begin{figure}
\begin{centering}
\includegraphics[width=0.8\columnwidth,keepaspectratio]{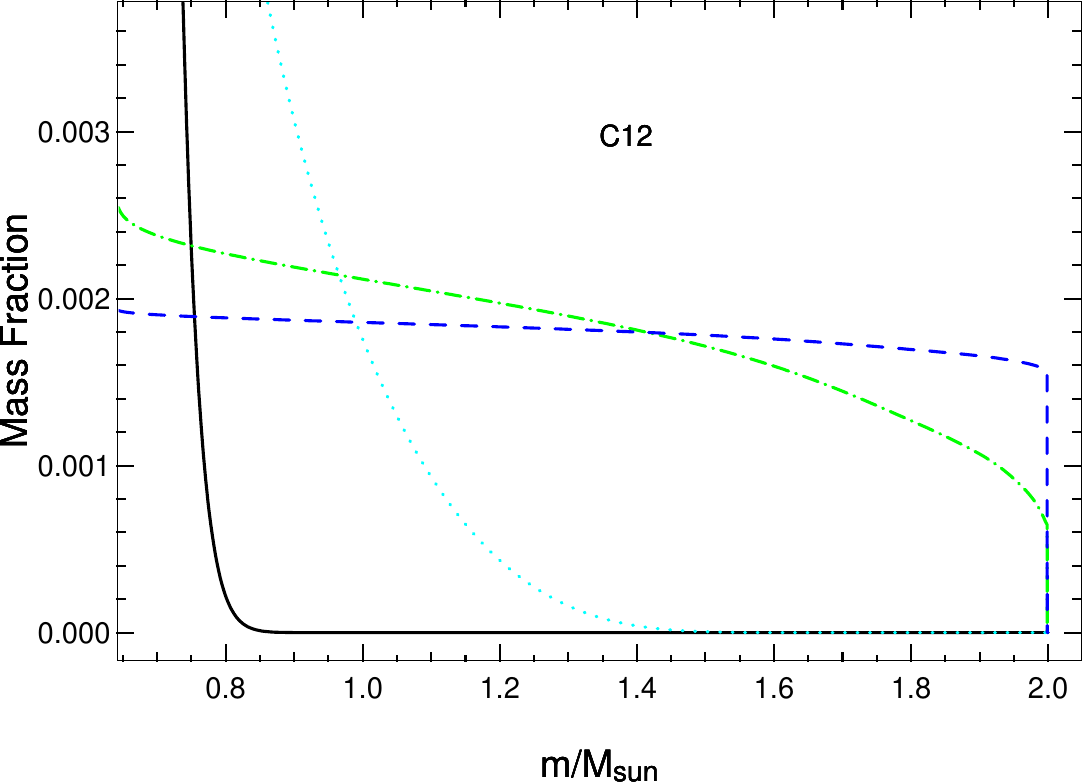}
\par\end{centering}
\caption{A series of $^{12}$C abundance profiles taken at different timesteps.
The star has a mass of 2 M$_{\odot}$ and is hyper metal-poor ($\textrm{[Fe/H]}=-5.5$).
The profiles are taken at consecutive stages during the dredge-up
of the entire intershell after a very strong H-flash near the start
of the AGB (overshoot was included in this model). The solid line
is first in time, showing that the convective envelope initially has
almost zero $^{12}$C. The dashed line is the final profile plotted,
whilst the other two are taken at intermediate timesteps. It can be
seen that in this case the mixing timescale is much \emph{longer}
than the evolutionary timestep, as it takes many timesteps to diffuse
the material out to the surface. The number of models between the
first profile and the last profile is 1000, whilst the evolutionary
timesteps were of the order $10^{-3}\textrm{ yr}$, so the time span
represented in the figure is $\sim1$ yr. \label{fig-diffn-LMAF}}
\end{figure}

Our final example is taken from one of the models calculated for this
thesis. Plotted in Figure \ref{fig-diffn-LMAF} are a series of $^{12}$C
abundance profiles taken during the dredge-up of the entire AGB intershell
in a hyper-metal-poor star of $2\,\textrm{M}_{\odot}$ (the model
included overshoot on the AGB). In this case it can be seen that the
mixing timescale is much \emph{longer} than the evolutionary timescale,
leading to incomplete mixing throughout the convective envelope. The
plot shows the gradual mixing out of $^{12}\textrm{C}$ over 1000
timesteps. The chemical profiles are reminiscent of classical diffusion
profiles. In this case instantaneous mixing would have been a poor
approximation. 

Based on the above empirical tests we conclude that the diffusion
subroutine is performing well, although, as always, there are more
tests that could be undertaken. In particular we believe it would
be possible to use the explicit differencing method on selected stages
of evolution in order to test the accuracy of the implicit method.
This would be time consuming but computers are fast enough to do this
test in a reasonable amount of time now. Another check that could
be performed would be to improve the mass conservation of the routine.
Although the routine performed very well outside the SEV code, it
does not do quite as well inside the SEV code. However we currently
have checks to keep the errors within reasonable bounds, as described
above, and the in situ empirical tests show that the errors are not
excessive in practise.

We now move on to treat two major uncertainties in stellar evolution
calculations, overshoot and semiconvection -- taking into consideration
the new mixing paradigm.

\subsection{Diffusive Overshoot\label{overshootMods}}

An overview of overshoot is given in section

\ref{sub-evoln-overshoot}. Here we describe the implementation of
a new formalism for overshoot in the SEV code. 

As mentioned in section \ref{sub-evoln-overshoot}, there are a number
of methods for including overshoot in stellar modelling. Seeing as
we now have mixing being treated time-dependently, it is a natural
progression to treat overshoot time-dependently. Indeed, the problem
is simplified in some regards as one can explicitly specify that a
certain number/range of points beyond the edge of a convective zone
have mixing velocities, and hence artificially create an overshoot
zone. However, as mentioned earlier, the problem with overshoot is
knowing how \emph{far} to overshoot. Numerical simulations of convection
(eg. \citealt{1995AA...295..703S}; \citealt{1996AA...313..497F})
have suggested that the velocity profile reduces in an exponential
fashion from the edge of a convection zone. Recently \citet{1997AA...324L..81H}
implemented the parameterised results of the \citet{1996AA...313..497F}
numerical simulations into their stellar evolution code. Since then
many authors have applied similar methods (eg. \citealt{2004AA...421L..25G};
\citealt{2005AA...431..279V}). The approach taken by \citet{1997AA...324L..81H}
makes use of the local `velocity scale height':

\begin{equation}
H_{v}=f_{OS}H_{p}\label{velScaleHeight}
\end{equation}

where $H_{p}$ is the local pressure scale height and $f_{OS}$ is
a scaling factor. This is used in equation (3) in \citet{1997AA...324L..81H}
to produce an expression for the exponentially reducing diffusion
coefficient profile:

\begin{equation}
D_{OS}=D_{0}e^{\frac{-2z}{H_{v}}}\label{Dovershoot}
\end{equation}

where $D_{0}=v_{0}H_{p}$ is the diffusion coefficient just inside
the formal convective boundary ($v_{0}$ is the corresponding velocity
given by the MLT) and $z$ is distance from the same boundary. 

We chose to use Herwig's formalism in the SEV code. Implementation
was a simple procedure, as it just required the addition of a velocity
profile described by Equation \ref{Dovershoot} onto the existing
velocity profile in each convective region. The extra mixing alters
the composition profile and then feeds back on the structure. 

\subsubsection*{Testing the Diffusive Overshoot Routine}

\begin{figure}
\begin{centering}
\includegraphics[width=0.8\columnwidth,keepaspectratio]{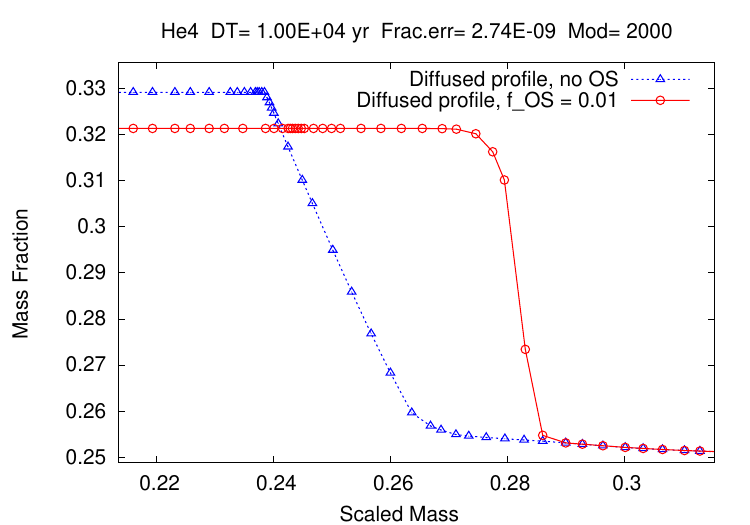}
\par\end{centering}
\caption{Comparison between models with and without diffusive overshoot. The
region shown is the edge of the convective cores in two 5 M$_{\odot}$
stars. Both stellar models were run with diffusive mixing but one
had overshoot switched on a few hundred models before the timestep
plotted. The two main effects of including diffusive overshoot can
be seen: 1) the core is larger (an increase of $\sim15\%$ in mass)
and 2) the abundance profile at the edge of the core is smoother.
\label{fig-overshoot-MS-He4}}
\end{figure}

In Figure \ref{fig-overshoot-MS-He4} we give an example of the resulting
profiles when using the new diffusive mixing overshoot procedure,
as compared to using a `hard' Schwarzschild boundary. The stellar
model used is the same as that in Figure \ref{fig-diffuseHe4}, a
MS star with a mass of 5 M$_{\odot}$. It can be seen that the abundance
profile from the run with overshoot switched on has a significantly
larger core mass (increasing from $\sim0.24$ to $\sim0.28$ M$_{\odot}$,
or $\sim15\%$). Having a greater mass in the convective core leads
to a longer MS lifetime as the star has more fuel to burn. It also
leads to a higher core mass at the end of the MS and thus affects
the later stages of evolution. 

\begin{figure}
\begin{centering}
\includegraphics[width=0.8\columnwidth,keepaspectratio]{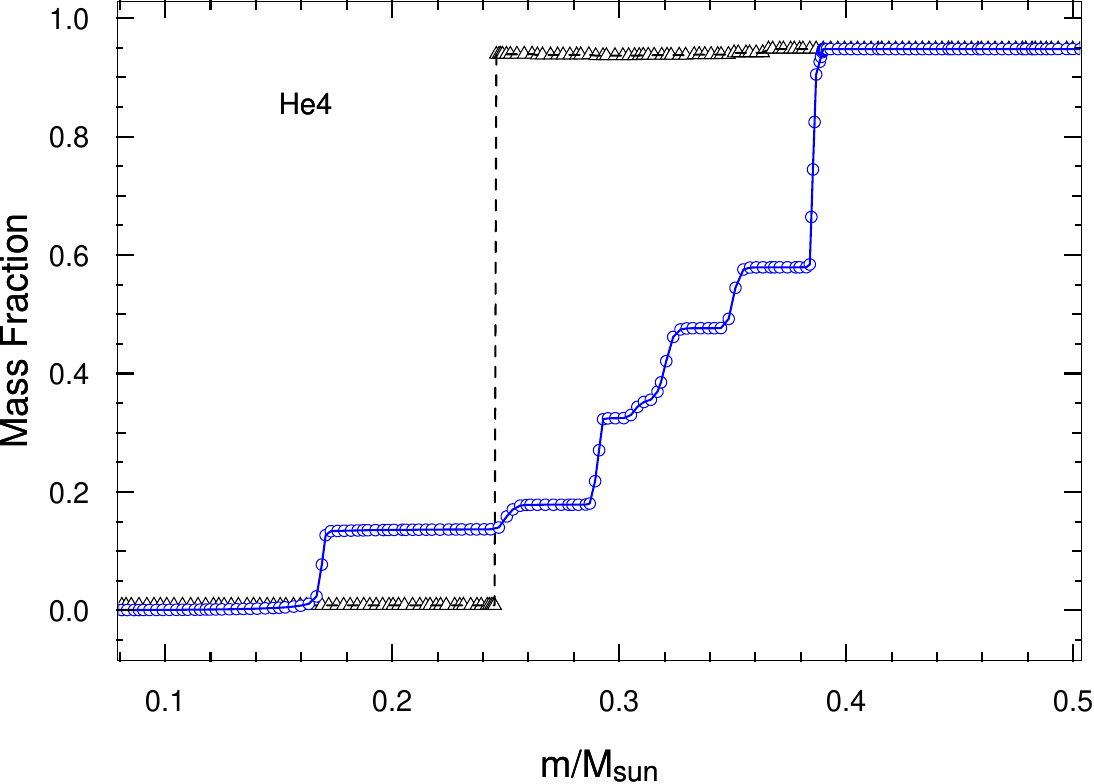}
\par\end{centering}
\caption{The $^{4}$He profiles of two 1 M$_{\odot}$ stars near the end of
core helium burning. Both were calculated using the new diffusive
mixing subroutine. One has no overshoot (dashed line, black) and one
had overshoot switched on half way through core helium burning (solid
line, blue). The overshoot parameter had a value: $f_{OS}=0.01$.
Markers represent the mesh points. It can be seen that the model without
OS retains a very strong discontinuity in the composition profile,
arising from the hard Schwarzschild boundary that did not allow any
mixing. This also means that there is a discontinuity in the opacity
profile, and hence the radiative gradient. The model that includes
OS on the other hand has an almost step-wise composition gradient
which has resulted from core breathing episodes (see text for details).
\label{fig-He4Profiles}}
\end{figure}

The other main difference is the smoothness of the profile on the
edge of the convective zone. This is formed due to the smoothly reducing
velocity profile. Without overshoot the core abundance profile has
a sharp edge. The smoothness of the abundance profile is of particular
interest in low mass stars towards the end of core helium burning.
With a strict Schwarzschild boundary there is a growing discontinuity
in the opacity at the edge of the core (due to the composition difference
between the material inside and outside the core) that becomes quite
large as the core helium depreciates. Just how rigid this boundary
is is debatable but it seems reasonable to say that it is not perfectly
impenetrable. Any mixing across the boundary (say through semiconvection)
whilst the opacity discontinuity is large would cause the regions
just outside the convection zone to also become convective due to
an increase in opacity, thereby increasing the mass of the convective
core. Through the addition of a slower velocity drop-off on the edge
of the convective core, the diffusive overshoot routine naturally
gives some slow mixing in the tail of the velocity profile. This leads
to quite different results at the start of the He shell burning stage.
Figure \ref{fig-He4Profiles} shows the situation at the end of core
helium burning with and without diffusive overshoot. The profiles
are quite different, the first being basically a single step function
as the core has not mixed out at all, and the second being a series
of steps. The profile from the overshooting model results from a series
of `core breathing pulses' in which the core repeatedly expands due
to an entrainment of extra He fuel (see eg. \citealt{1985ApJ...296..204C};
\citealt{1991ASPC...13..299S}; \citealt{2003ApJ...583..878S}). The
sudden expansion and contraction (in mass) leaves behind a series
of steps in the composition profiles. 

\begin{figure}
\begin{centering}
\includegraphics[width=90mm,keepaspectratio]{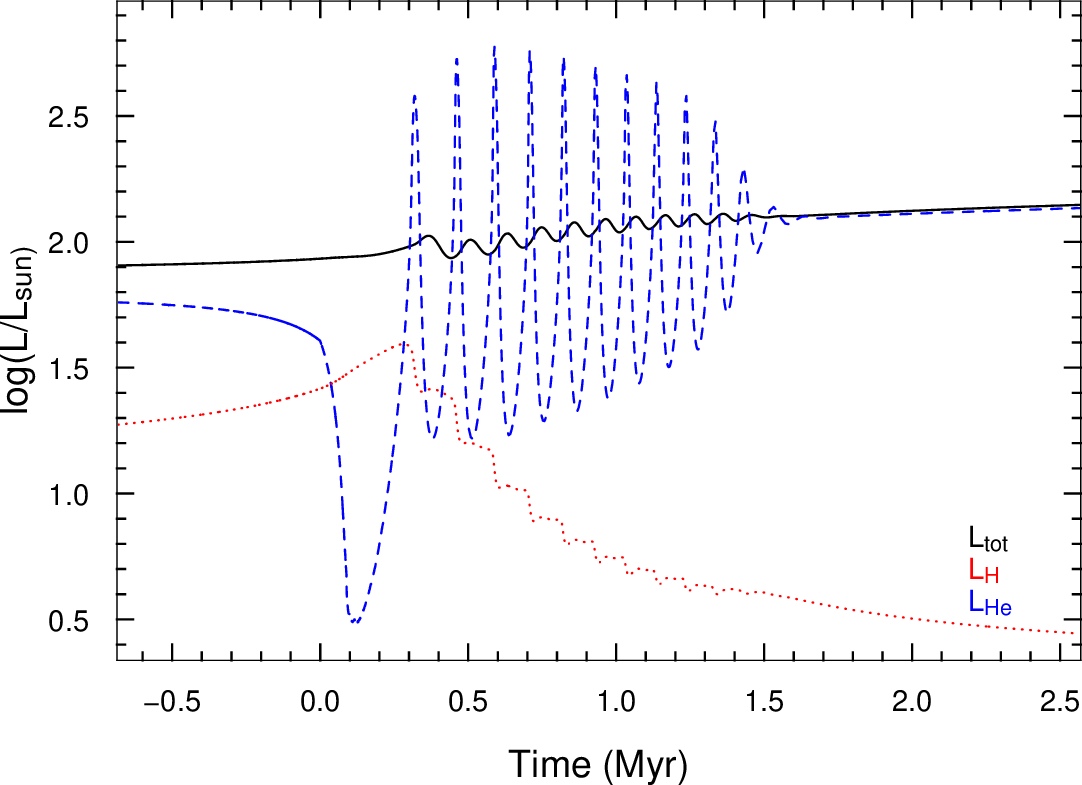}
\par\end{centering}
\caption{The subsequent evolution of the star in Figure \ref{fig-He4Profiles}
without overshoot. Plotter are the helium luminosity (dashed line),
hydrogen luminosity (dotted line) and total luminosity (solid line).
The zero point in time is taken (roughly) at the end of core He burning.
The gravonuclear instability is clearly seen in the He luminosity
evolution. \label{fig-gravoNuclearLoops}}
\end{figure}

\begin{figure}
\begin{centering}
\includegraphics[width=0.75\columnwidth,keepaspectratio]{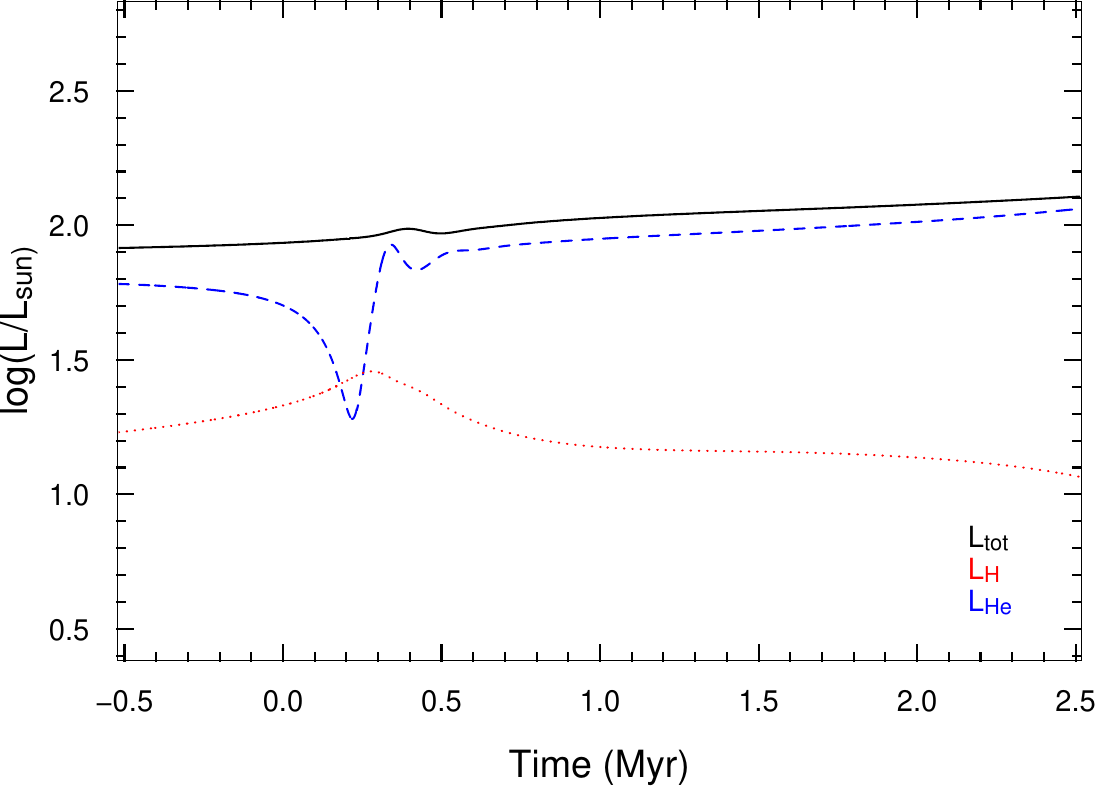}
\par\end{centering}
\caption{Same as Figure \ref{fig-gravoNuclearLoops} but for the star with
overshoot. The axes are on the same scale as Figure \ref{fig-gravoNuclearLoops}
to enable direct comparison. As can be seen the instability does not
arise in this model, as the more gradually changing He profile leads
to a thicker He burning shell (see text for more details). \label{fig-NoGravoNucleoLoops}}
\end{figure}

\begin{figure}
\begin{centering}
\includegraphics[width=0.75\columnwidth,keepaspectratio]{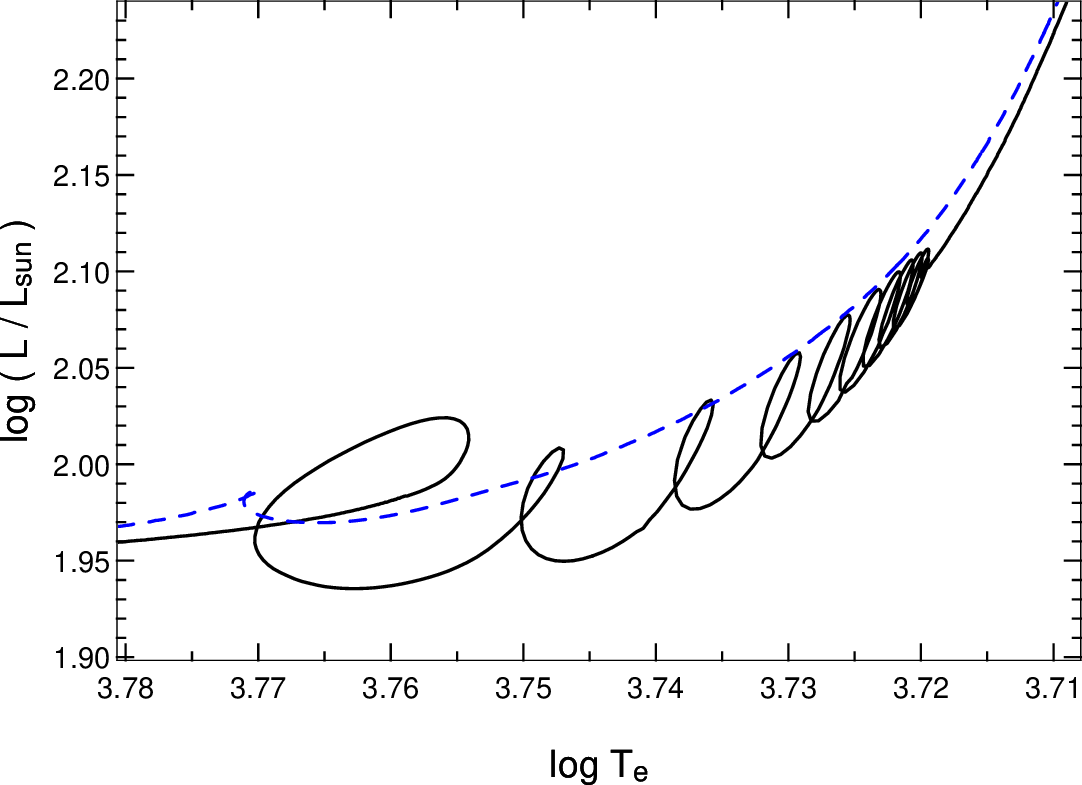}
\par\end{centering}
\caption{The same stars as in Figure \ref{fig-He4Profiles} but showing the
gravonuclear loops in the H-R plane. The dashed line is the model
with diffusive overshoot (no GNLs) whilst the solid line is the model
without overshoot. \label{fig-GNLs-hrd}}
\end{figure}

The difference in abundance profiles has a profound effect on the
subsequent helium shell burning phase after core helium burning is
completed. Figures \ref{fig-gravoNuclearLoops} and \ref{fig-NoGravoNucleoLoops}
compare the subsequent (luminosity) evolution of the stars in Figure
\ref{fig-He4Profiles}, ie. overshoot versus no overshoot. In Figure
\ref{fig-gravoNuclearLoops} (the model with no overshoot) it can
be seen that the helium luminosity drops significantly when helium
is finally exhausted in the core. Then there is a series of `pulsations'
of helium burning before the star settles into a stable He shell burning
configuration. This phenomenon has been noted before and it was \citet{1997ApJ...489..822B}
who coined the term `gravonuclear loops' (GNLs) to describe it. \citet{1997ApJ...489..822B}
studied gravonuclear loops in some detail and note that \citet{1965ApJ...142..855S}
found the same instability in a 1 M$_{\odot}$ star. \citet{1965ApJ...142..855S}
subsequently performed an analytical analysis to see if we would expect
an instability at this stage of evolution. They concluded that an
instability should indeed arise -- if the He burning shell is very
thin. This differs from the conclusion of \citet{1997ApJ...489..822B}
(based on super-metal-rich models with m $\sim$ 0.5 M$_{\odot}$)
who suggest that the driving factors for GNLs are 1) the opacity in
the envelope (in particular the iron `opacity bump' in these models,
see their Appendix for details) and 2) the mass of the envelope. They
also suggest that the instabilities are metallicity dependent, such
that it is more common in stars with high metallicities. More recently
\citet{2000LIACo..35..529S} performed some tests to see if the method
of suppressing core breathing pulses has an impact on the occurrence
of GNLs. They concluded that this was actually the prime driver for
GNLs, as it determines the nature of the helium profile at the start
of the shell helium burning stage. Methods that gave a thin burning
shell resulted in GNLs whilst methods that had a smoother profile
give a thicker He burning shell, thus avoiding the instability. This
is in line with the analysis by \citet{1965ApJ...142..855S} and is
also supported by Figures \ref{fig-He4Profiles} - \ref{fig-GNLs-hrd}
here. They also conclude that the effect is not dependent on metallicity
or the mass of the envelope, in variance with \citet{1997ApJ...489..822B}.

In Figure \ref{fig-NoGravoNucleoLoops} we show the case in which
diffusive overshoot was included. It can be seen that GNLs do not
arise, due to the less abrupt He profile and resulting thicker He
burning shell. In Figure \ref{fig-GNLs-hrd} we plot both cases in
the L-T$_{eff}$ plane, in which the loops are very evident.

\subsubsection*{How Far to Overshoot?}

We now return to the perennial question for any method of overshoot
-- how far do we extend the convective boundary beyond the formal
one? 

\begin{figure}
\begin{minipage}[b][1\totalheight][t]{0.5\columnwidth}%
\begin{center}
\includegraphics[height=40mm,keepaspectratio]{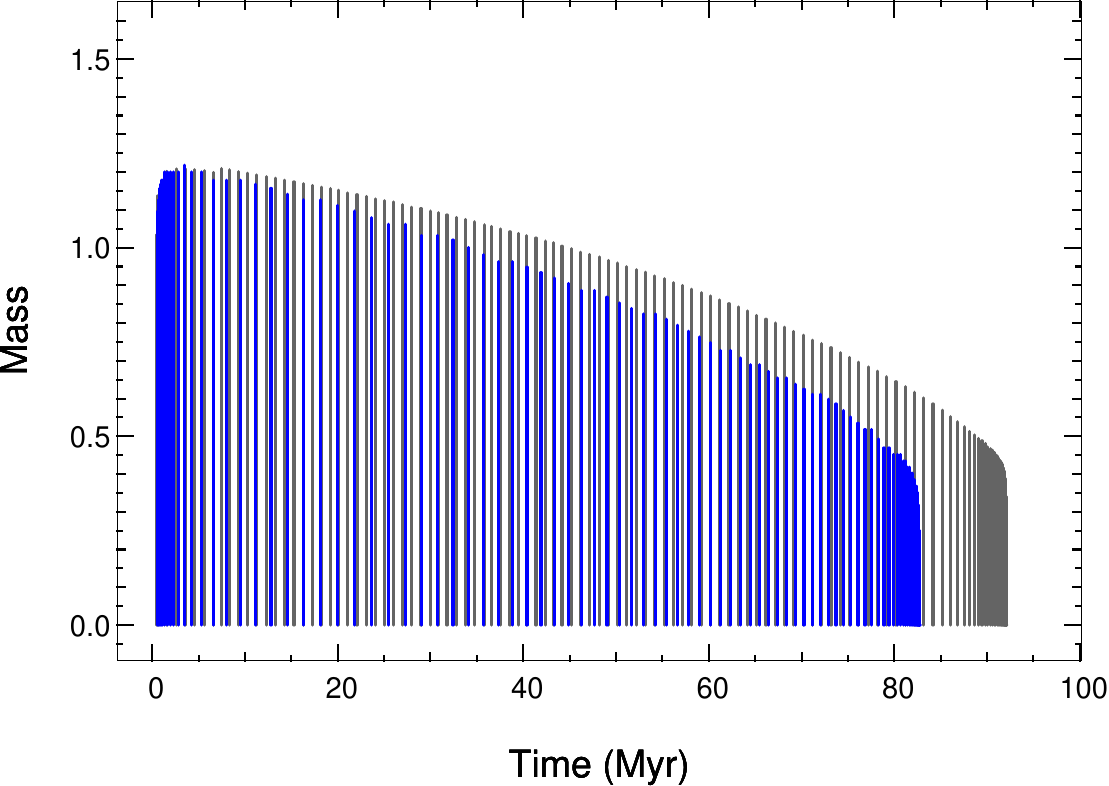}
\par\end{center}%
\end{minipage}%
\begin{minipage}[b][1\totalheight][t]{0.5\columnwidth}%
\begin{center}
\includegraphics[height=40mm,keepaspectratio]{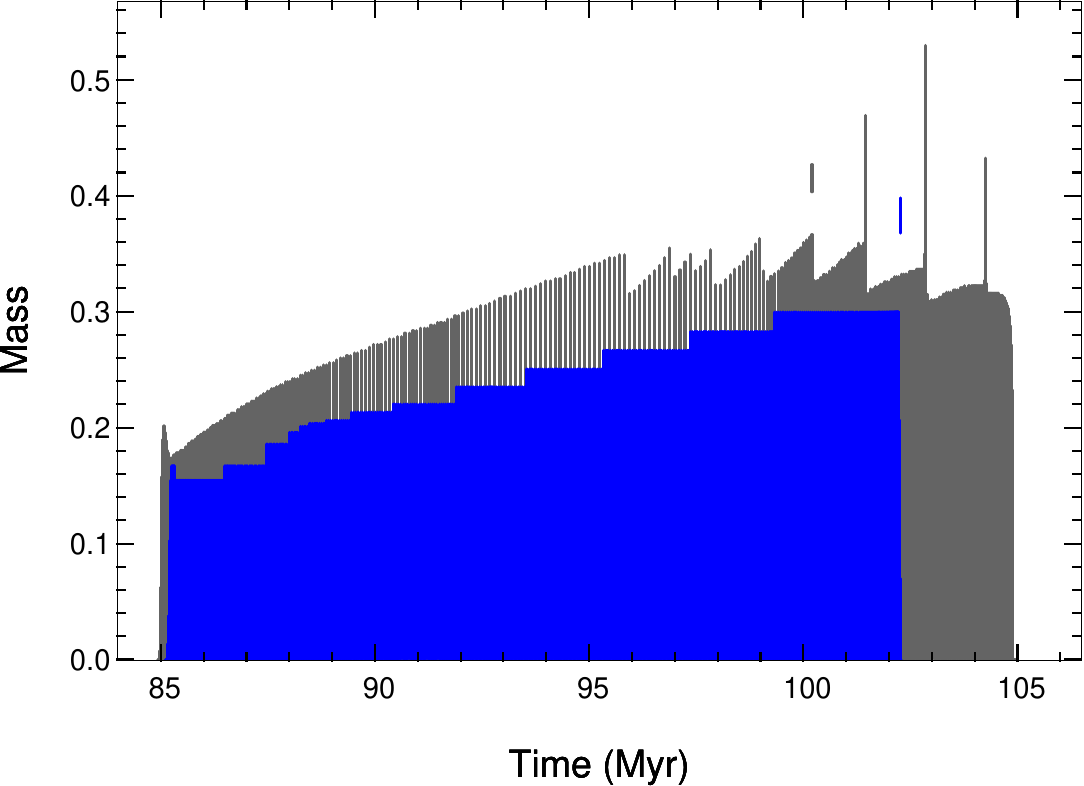}
\par\end{center}%
\end{minipage}

\caption{Comparison between the original SEV code and the version with diffusive
mixing and overshoot (and updated opacities). The vertical lines represent
convection zones (where the velocity is non-zero). The results from
the original code are in blue whilst those from the modified code
are in grey. Both stars have a mass of 5 M$_{\odot}$. The plot on
the left shows the extent of convection during the main sequence whilst
the plot on the right shows convection during core helium burning.
The time axis in the HB plot has been offset for clarity and the convection
profiles aligned for easy comparison. Both versions of the code include
overshoot but use different methods (see text for details). The diffusive
overshoot parameter, $f_{OS}$ was set at 0.01 for the diffusive mixing
case. It can be seen that this is a reasonable choice on the MS, where
the mass of the convective cores are fairly similar. The slight increase
in the diffusive mixing case leads to a slightly longer MS lifetime.
More discrepancy is noted in the HB evolution, where the lifetime
of this stage is increased by $\sim10\%$ due to a larger core throughout
the HB. Significant core breathing pulses are also present in the
diffusive mixing case.\label{fig-m5Compare}}
\end{figure}

\begin{figure}
\begin{centering}
\includegraphics[width=0.85\columnwidth,keepaspectratio]{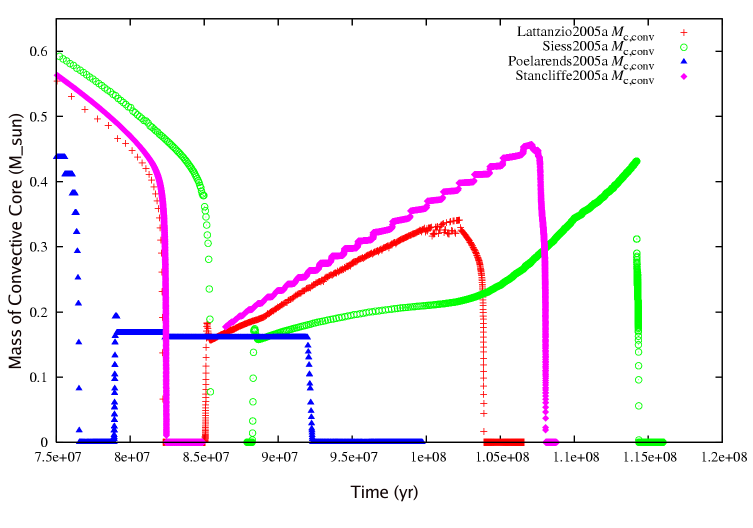}
\par\end{centering}
\caption{The evolution of the convective core of a 5 M$_{\odot}$ star on the
HB as given by four different stellar codes. This plot was generated
on the \emph{CIQuA Stellar Database} website (\citealt{2004MmSAI..75..631O}).
It can be seen that the size of the convective core varies significantly
(from $\sim0.2$ to $\sim0.5$ M$_{\odot}$). Comparing this with
Figure \ref{fig-m5Compare} above suggests that the core mass given
by the SEV code with diffusive mixing and overshoot (where $m_{core}\sim0.4$
M$_{\odot}$) is reasonable. The length of the core helium burning
stage also varies immensely (from $\sim13$ to $\sim23$ Myr). This
compares well with the modified SEV code results which give a lifetime
of $\sim20$ Myr.\label{fig-OtherCodes-HB-m5}}
\end{figure}

The degree of overshoot in the diffusive mixing overshoot method is
governed by $f_{OS}$. By varying this parameter one can have as much
or as little overshoot as one would like. This leads to different
core masses on the MS and HB and consequently affects the later stages
of stellar evolution. In order to gauge what value of $f_{OS}$ to
use as a standard we performed comparison tests between the original
SEV code (which uses instant mixing and a traditional overshooting
method) and the modified code. A series of models with a range of
values of $f_{OS}$ used for overshoot were calculated. It was found
that a value around 0.01 reproduced the core sizes and lifetimes given
by the original SEV code reasonably well. To illustrate we show the
results of the comparisons at two different stellar masses. Figure
\ref{fig-m5Compare} shows the convection zones of a 5 M$_{\odot}$
during the MS and during core helium burning. In the diffusive overshoot
case we have used $f_{OS}=0.01$ whilst we used standard overshooting
in the original SEV code case (see Section \ref{sub-evoln-overshoot}
for the standard overshoot method). The main differences are 1) the
convective cores are slightly more massive in the diffusive overshoot
case (on the MS and HB) and 2) the lifetimes are both slightly longer
in the diffusive overshoot case. This is indicative of the overshoot
parameter $f_{OS}$ being slightly too large to reproduce the original
case exactly. Another feature is that the diffusive overshoot case
exhibits some core breathing pulses, whilst the instant-mixing case
does not. How significant these differences are is difficult to ascertain,
as the core helium burning stage of evolution is notoriously uncertain.
In order to try to gauge the significance of the differences in HB
characteristics we have plotted the results from four different stellar
evolution codes in Figure \ref{fig-OtherCodes-HB-m5}. This plot was
generated on the \emph{CIQuA Stellar Database} website (\citealt{2004MmSAI..75..631O})
which is a useful resource for code comparisons. The figure highlights
the fact that the size of the convective core varies significantly
between codes (from $\sim0.2$ to $\sim0.5$ M$_{\odot}$). Comparing
this with the results from our version of the SEV code (where $m_{core}\sim0.4$
M$_{\odot}$, Figure \ref{fig-m5Compare}) we note that our results
are well within the limits of variation. The length of the core helium
burning stage also varies immensely -- by a factor of two in fact
-- from $\sim13$ to $\sim23$ Myr. Again the results from the modified
SEV code lie within the variation, giving a HB lifetime of $\sim20$
Myr. 

\begin{figure}
\begin{centering}
\includegraphics[width=0.65\columnwidth,keepaspectratio]{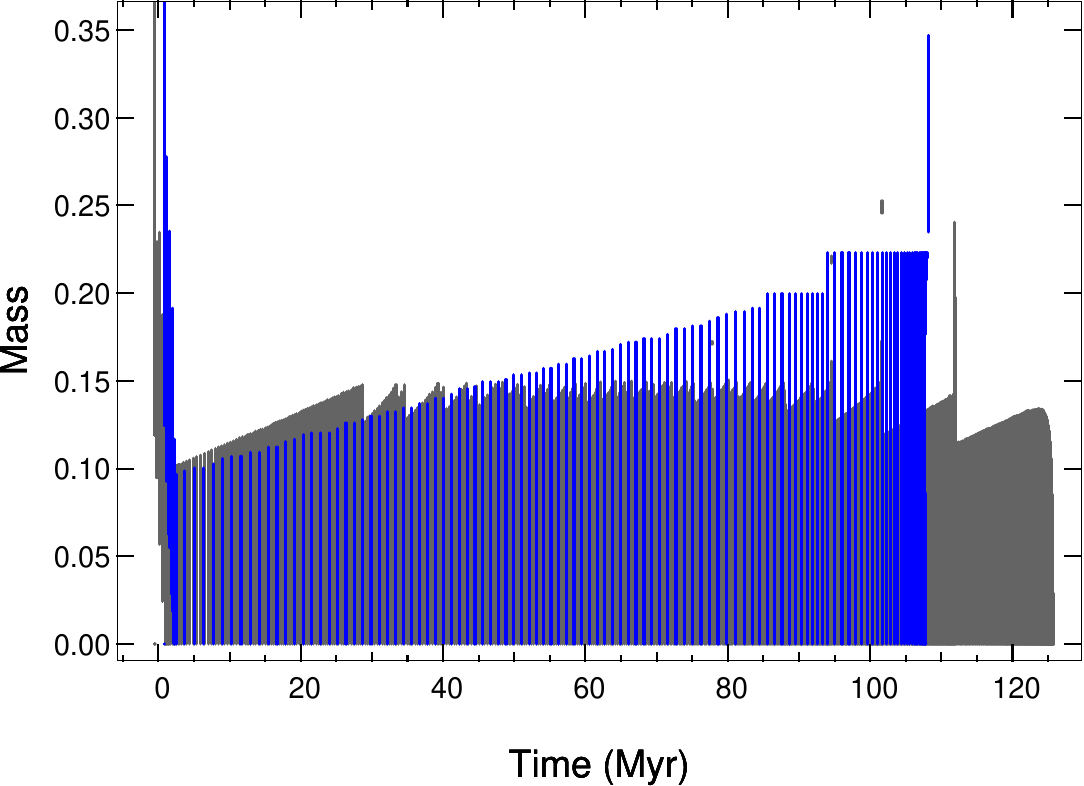}
\par\end{centering}
\caption{Same as Figure \ref{fig-m5Compare} except the stars have a mass of
1 M$_{\odot}$. Only the helium burning stage is shown as there is
no convection on the MS at this mass (the MS lifetimes only differ
by $\sim1\%$). Here we see a different effect than with the 5 M$_{\odot}$
comparison -- the original code gives a \emph{larger} core mass in
this case. However, the HB lifetime is again longer despite this.
The same $f_{OS}$ value of 0.01 was used.\label{fig-m1compareHB}}
\end{figure}

\begin{figure}
\begin{centering}
\includegraphics[width=0.8\columnwidth,keepaspectratio]{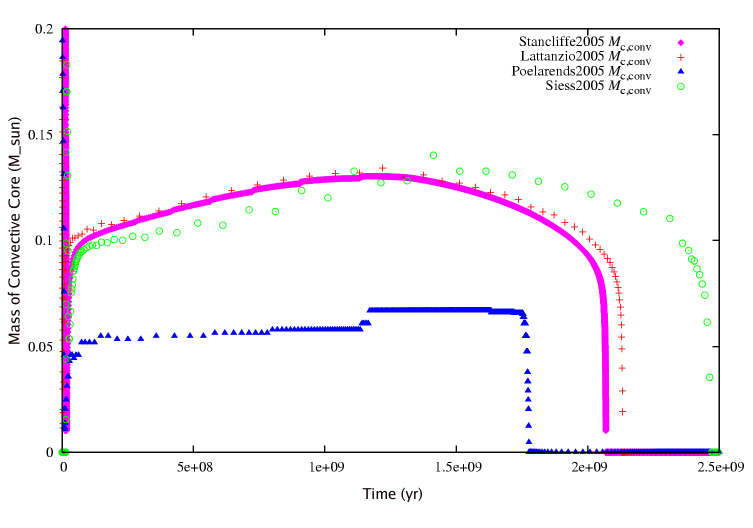}
\par\end{centering}
\caption{The evolution of the convective core of a 1.5 M$_{\odot}$ star on
the HB as given by four different stellar codes. This plot was generated
on the \emph{CIQuA Stellar Database} website (\citealt{2004MmSAI..75..631O}).
It can be seen that the size of the convective core varies significantly
(from $\sim0.07$ to $\sim0.15$ M$_{\odot}$). Although our comparison
model has a mass of 1 M$_{\odot}$, the core sizes should be similar.
Comparing these models with Figure \ref{fig-m1compareHB} above suggests
that the core mass given by the SEV code with diffusive mixing and
overshoot (where $m_{core}\sim0.15$ M$_{\odot}$) is reasonable.
The length of the core helium burning stage varies significantly but
the Lattanzio2005 code (which is a previous version of the SEV code)
results compare well with the others, suggesting that the modified
SEV code should also (given the apparent uncertainties). \label{fig-OtherCodes-HB-m1}}
\end{figure}

We continue the comparisons with Figure \ref{fig-m1compareHB}. Here
we plot the convective regions during the core helium burning phase
of two 1 M$_{\odot}$ stars -- one calculated with the original SEV
code and one with the modified (diffusive mixing) SEV code. We don't
show the MS as there is no core convective zone. We do note however
that the MS lifetimes only differ by $\sim1\%$ between the two cases.
For the diffusive mixing case we use the same $f_{OS}$ value of 0.01.
Here we see a different effect than in the 5 M$_{\odot}$ comparison
-- the original code gives a \emph{larger} core mass during the HB.
Despite this the HB lifetime is again longer, probably due to the
large core breathing pulses near core helium exhaustion. In Figure
\ref{fig-OtherCodes-HB-m1} we again try to gauge the significance
of the discrepancies by comparing with other authors. A star of mass
1 M$_{\odot}$ was not available in the \emph{CIQuA} database so we
compare with a 1.5 M$_{\odot}$ star. We expect the HB convective
core to be somewhat larger in the 1.5 M$_{\odot}$ star. The core
mass from the diffusive mixing version of the SEV code compares well
with those found with the other codes. Interestingly, the core mass
from the instant mixing version of the SEV code is somewhat larger
than that obtained from the other codes. We note that one of the comparison
codes in Figure \ref{fig-OtherCodes-HB-m1} is actually another version
of the SEV code (labelled Lattanzio2005). It uses instant mixing but
includes updated opacities. Comparing the HB lifetimes with ours in
Figure \ref{fig-m1compareHB}, it appears that our modified SEV code
compares reasonably well with the other codes.

In conclusion, given the variability of HB lifetimes found with different
stellar evolution codes evident in the plots above, it is apparent
that the core helium burning stage of low and intermediate mass stars
is significantly uncertain. Since the degree of core overshoot has
a strong influence on HB evolution this leads us to conclude that
the diffusive overshoot factor $f_{OS}$ is also somewhat uncertain.
However, we have found that a value of $\sim0.01$ gives results consistent
with other authors. The value should be calibrated to observations,
and we leave this for future work. It must also be noted here that
the degree of overshoot is not necessarily uniform throughout the
evolution of a star -- or even through different convection zones
within a star. It most likely depends on the precise physical conditions
of each convection zone (and the stable layers above or below). The
simple method used for overshoot here is most likely insufficient
to model overshoot from all types of convection zones, especially
with a single value of $f_{OS}$. Again, the most promising solution
to this problem is the increasing feasibility of performing multidimensional
hydrodynamic calculations. 

\subsubsection*{Pre-AGB Comparison in the HR Diagram}

\begin{figure}
\begin{centering}
\includegraphics[width=0.85\columnwidth,keepaspectratio]{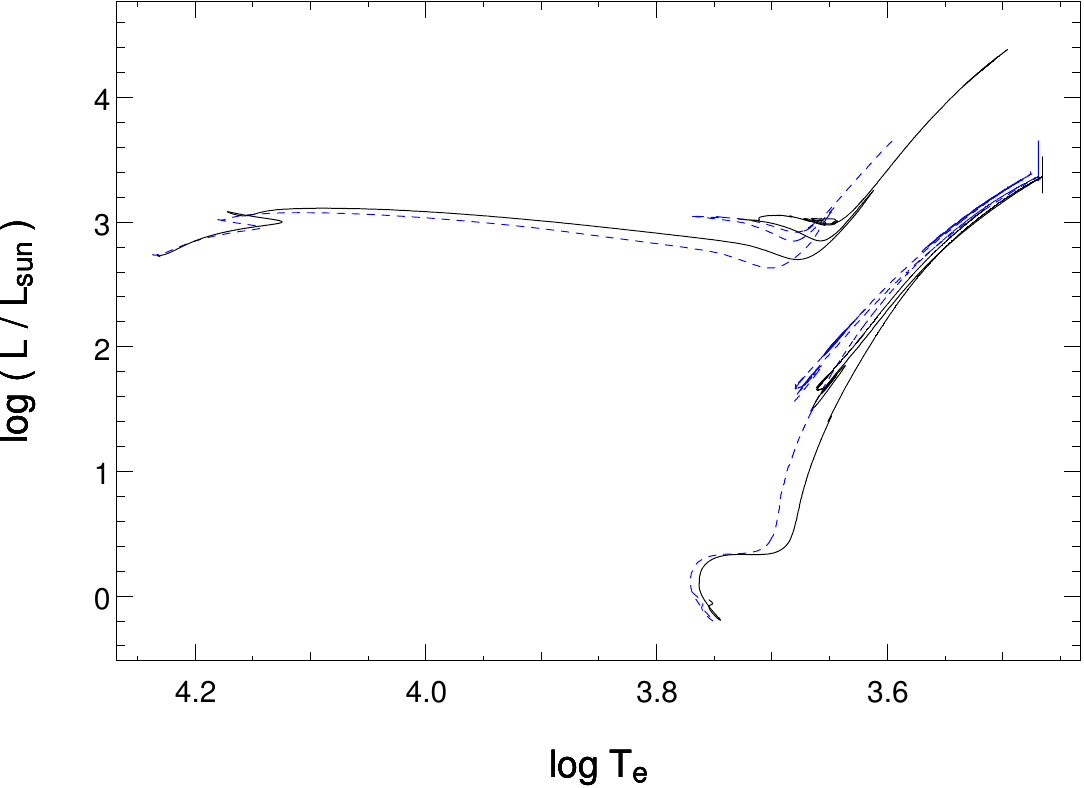}
\par\end{centering}
\caption{Comparison between the two pairs of stars in Figures \ref{fig-m5Compare}
and \ref{fig-m1compareHB}. The dashed lines are the models run with
the original SEV code whilst the solid are those from the SEV code
with diffusive mixing (with overshoot) and updated opacities. In the
5 M$_{\odot}$ case there are two differences -- the new code gives
slightly higher luminosities and the surface temperatures are slightly
lower. The first effect is due to the larger core (see Figure \ref{fig-m5Compare})
whilst the lower $T_{eff}$ is most likely due to the updated opacities
which include effects from molecules and grains at low temperatures.
In the 1 M$_{\odot}$ case the luminosity is practically identical
but the temperature is again shifted slightly cooler. Here there is
no convective core on the MS so the core masses are identical, leading
to identical luminosities. Again, the cooler $T_{eff}$ is due to
the updated opacities.\label{fig-hrds-compare1and5}}
\end{figure}

As a final comparison between the original SEV code and the modified
SEV code, we display in Figure \ref{fig-hrds-compare1and5} the same
evolutionary runs ($M=1$ M$_{\odot}$and 5 M$_{\odot}$) as discussed
above, in the HR diagram. The tracks are plotted for evolution up
to early AGB. In the 5 M$_{\odot}$ case there are two differences
-- the modified SEV code gives slightly higher luminosities and the
surface temperatures are slightly lower. The first effect is due to
the larger core (see Figure \ref{fig-m5Compare}) whilst the lower
$T_{eff}$ is most likely due to the updated opacities which include
effects from molecules and grains at low temperatures (see opacity
section). In the 1 M$_{\odot}$ case the luminosity is practically
identical but the temperature is again slightly cooler. Here there
is no convective core on the MS so the core masses are identical,
leading to identical luminosities. Again, the cooler $T_{eff}$ is
due to the updated opacities. We conclude that the diffusive mixing
and diffusive overshoot modifications have not drastically altered
the results of the code. The improved treatment of mixing will however
be of much use during certain difficult stages of evolution encountered
in zero- and low-metallicity stars, which is the focus of the current
study.

\subsubsection*{Diffusive Overshoot on the AGB}

The degree of overshoot from the convection zones during the AGB stage
of evolution is at least as uncertain as that during core helium burning.
During a shell helium flash there are two quite distinct types of
convection zones -- a He-flash-driven convective zone of small mass,
and a convective envelope of large mass (until mass loss reduces the
mass of the envelope). In some very recent work \citet{2006ApJ...642.1057H}
attempted to gauge the extent of overshoot from a He shell flash convective
zone during the AGB through multidimensional hydrodynamic modelling.
Their preliminary results suggest that some mixing across the formal
convective boundaries is likely to occur, but it is difficult to quantify
as yet. Interestingly, Herwig (2005, private communication via John
Lattanzio) finds that his `best guess' values of $f_{OS}$ on the
AGB for his code are 0.008 at the bottom of the convective intershell
and 0.130 at the bottom of the convective envelope (also see \citealt{1997AA...324L..81H}
for more information on the calibration of $f_{OS}$). The first value
compares well with our choice of $f_{OS}=0.01$ derived from core
helium burning comparisons. The second value seems very large to us.
It does however give a significant $^{13}$C pocket which leads to
s-processing in Herwig's code (\citealt{2004ApJ...605..425H}), which
an important observational constraint in low mass AGB stars. However,
Herwig also notes that a systematic calibration with observations
is still needed and is yet to be carried out by anyone. As this is
outside of the scope of this study we choose to retain $f_{OS}=0.01$
for all convective zones. The fact that it lies at the lower end of
Herwig's values suggests that our overshoot may be relatively conservative
in some cases. We note that there may be differences between the stellar
codes so the values may not be directly comparable, but we also reiterate
that the values are quite uncertain anyway.

\begin{figure}
\begin{centering}
\includegraphics[width=0.75\columnwidth,keepaspectratio]{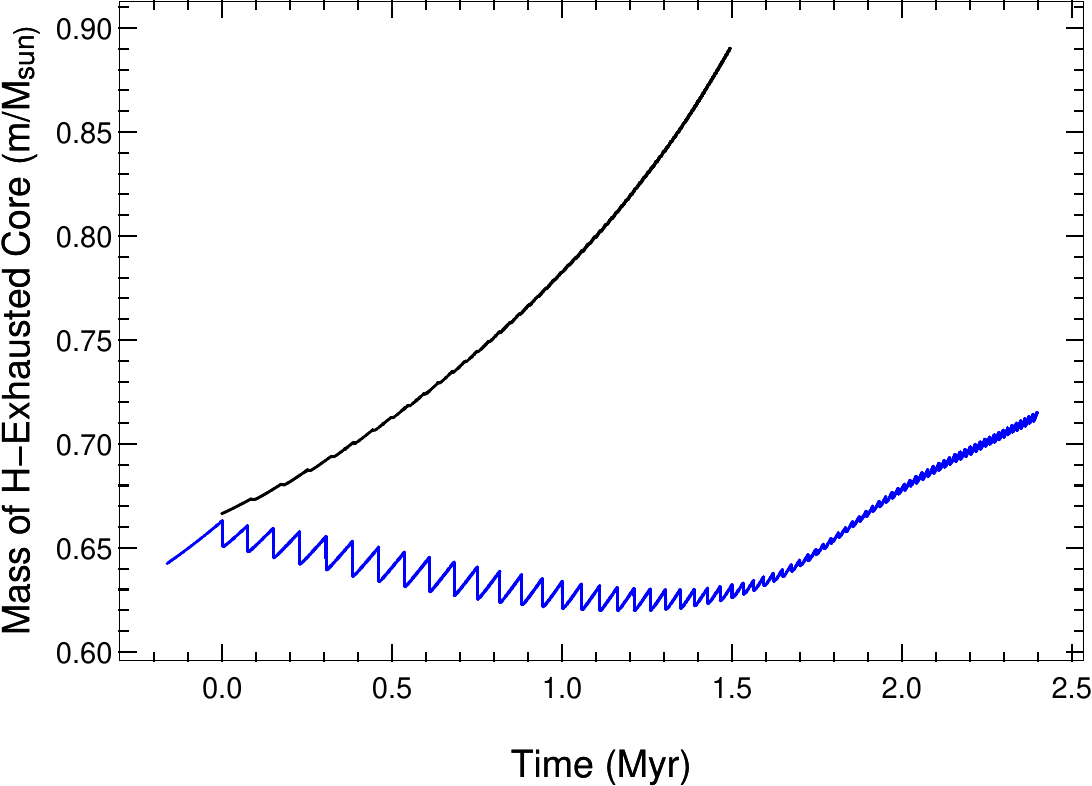}
\par\end{centering}
\caption{Highlighting the strong effect of including overshoot on the AGB.
The time evolution of the mass of the H-exhausted core is plotted.
Both stars, having a mass of 2 M$_{\odot}$ and a metallicity of $\textrm{[Fe/H]=-5.45}$,
started with the same initial conditions except for the inclusion
of overshoot in one (lower curve, $f_{OS}=0.01$). An enormous difference
in core mass evolution is clearly seen. The model with no overshoot
(upper curve) has virtually no 3dup whilst the model with overshoot
initially has $\lambda_{3dup}>1$. As the core mass is the primary
factor in AGB evolution, the vastly different core masses represent
a very large uncertainty in AGB evolution. \label{fig-AGB-OS-noOS-3dup}}
\end{figure}

As an example of the effect that overshoot can have during the AGB
phase of evolution, we show in Figure \ref{fig-AGB-OS-noOS-3dup}
the evolution of the core mass for two stars that had otherwise started
with identical initial conditions. The difference is striking --
there is virtually zero third dredge-up (3dup) in the model with no
overshoot, whilst the model with overshoot has a very large amount
of 3dup, with $\lambda_{3dup}>1$ for the first $\sim25$ thermal
pulses. This large amount of 3dup leads to an enormous pollution of
the envelope, especially in low metallicity stars. We note that the
original version of the SEV code never gives $\lambda_{3dup}>1$,
even when using (the original method of) overshoot. This has consequences
for envelope pollution in terms of the chemical makeup of the pollution,
as more oxygen is mixed up since the PDCZ overshoots \emph{down} into
the C-O core, altering the chemical composition of the intershell,
enriching it in $^{16}$O. Furthermore, as the core mass is the primary
factor in AGB evolution, the vastly different core masses resulting
from the inclusion or non-inclusion of overshoot represent a very
large uncertainty in AGB evolution. 

Finally we note that we have retained the method suggested by \citet{1981ApJ...248..311W}
for taking into account the change in entropy due to the mixing up
of heavy nuclei during the third dredge-up on the AGB. The main effect
of including this physics is to slightly increase the depth of the
third dredge-up (by about $20\%$ in mass). We have however added
a numerical switch that enables the operator to choose whether to
use this routine or not. The routine only operates during 3dup on
the AGB so does not effect other stages of evolution. The routine
was not used in the models plotted in Figure \ref{fig-AGB-OS-noOS-3dup}.

\subsection{Semiconvection\label{semimods}}

Another topic of perennial discussion with regards to stellar convection
is semiconvection. We have previously described the method used for
semiconvection in the original version of the SEV code in subsection
\ref{sub-semiconvection}. Here we briefly discuss the physical basis
for semiconvection, the relevance to stars and describe our method
for including it taking into account the new diffusive mixing paradigm.

It seems whenever it is deduced that partial mixing is required to
circumvent apparent contradictions in the determination of convective
boundaries of stars, the term semiconvection is used to describe that
mixing. Thus here we view the term as meaning `any slow mixing process'
-- ie. non-turbulent mixing. 

The problems in determining convective boundaries have been present
since the first studies of stellar evolution. \citet{1958ApJ...128..348S}
identified a problem in the MS evolution of massive stars (M $\sim$
30 M$_{\odot}$). In their models the convective core was found to
expand on the MS (this was later found to be incorrect at this mass,
but the argument still applies to other stars). The problem was that
a growing discontinuity in opacity, due to the continual burning of
H reducing the opacity in the core but not the envelope, leads to
a contradictory situation. They noticed that since the opacity is
higher just outside the convective core then this region is actually
\emph{more} unstable than the material just inside the core (since
the temperature gradient depends on opacity and the other physical
variables are continuous) -- thus we have a contradiction -- the
original structure is not possible. Their solution was to assume a
partial mixing, which they called semiconvection, of the region just
outside the convective core. The mixing was undertaken until convective
neutrality was achieved (ie. $\nabla_{rad}=\nabla_{ad}$). This means
that the convective core is still given by $\nabla_{rad}>\nabla_{ad}$
but there is a (growing) region of partially mixed material on top
of the core that is just stable against normal (turbulent) convection.
This leads to a different thermal structure and has significant consequences,
such as the predicted ratio of red to blue supergiants (M $\sim$
13 to 30 M$_{\odot}$; see eg. \citealt{1994ApJ...431..797S}).

A related problem occurs in low mass core helium burning stars, where
a local minimum in the ratio $\nabla_{rad}/\nabla_{ad}$ develops
in the core, again driven by opacity (see eg. \citealt{1971ApSS..10..355C}).
The \citet{1958ApJ...128..348S} solution (and variants thereof) has
been widely used for these semiconvection situations. However, the
physical foundation for the occurrence of stellar semiconvection is
not well defined. 

The main physical arguments for semiconvection have their origin in
fluid mechanics, and in particular oceanography. Early experimental
work by \citet{TS64} showed that a stably stratified fluid which
is heated from below develops an increasing number of discrete convective
layers above the initial convective layer. The convection within these
layers homogenises the composition, removing the stabilising gradient.
The convective layers are separated by stable layers with steep composition
profiles. Over time the convective layers slowly expand and merge,
probably due to convective `scouring' -- ie. extra mixing via eddies
slightly overshooting into the stable regions (see eg. \citealt{1989JFM...209....1F}).
The nett result of this process is that we effectively have a slow
mixing process just outside the formal convective boundary -- if
the material is heated from below \emph{and} there is a negative molecular
weight (or density) gradient. This is reminiscent of the semiconvection
situation in stellar model calculations. It is known as Double Diffusive
Mixing, as theory describes the process as the competition between
the diffusion of two properties -- heat and chemical composition.
In stars (and salt water) the timescale is far higher for heat diffusion
than chemical (microscopic) diffusion and it is this that leads to
the layering effect. Although this is the most cited physical analogy
to stellar semiconvection (eg. \citealt{1992AA...253..131S}; \citealt{1996MNRAS.283.1165G}),
it still remains uncertain whether it really applies to the vastly
different physical conditions found in stars (as compared to the ocean).
The avenue of detailed multidimensional hydrodynamic calculations
is still is in its infancy due to the complexity and computational
costs of the problem (for recent attempts at simulating semiconvection
see eg. \citealt{1995ApJ...444..318M}; \citealt{2001PhDT.........8B}
). If we do accept this mechanism for semiconvection, then we note
that this type of convection is not taken into account by the Schwarzschild
condition for onset of convection:

\begin{equation}
\nabla_{rad}\geq\nabla_{ad}\label{eqn-schwarzschild}
\end{equation}

as it ignores the possibility of a stabilising effect from a (negative)
composition gradient that gives rise to the double diffusive mixing.
The \citet{1947ApJ...105..305L} criterion does take this into account:

\begin{equation}
\nabla_{rad}\geq\nabla_{ad}+\frac{\beta}{4-3\beta}\nabla_{\mu}\label{eqn-ledoux}
\end{equation}

where $\beta$ is the ratio of gas pressure to total pressure and

\begin{equation}
\nabla_{\mu}=\frac{\partial\textrm{ln}\mu}{\partial\textrm{ln}P}\label{eqn-muGradient}
\end{equation}

is the molecular weight gradient. \citet{1966PASJ...18..374K} however
argues that overstable convection -- the amplifying oscillatory motion
of mass elements due to thermal dissipation within a zone with a negative
composition gradient -- is strong enough to remove a chemical gradient
just outside a convective zone. Thus they argue that the criterion
for convection reduces to the Schwarzschild one and it is this, with
a region just outside the core that is `convectively neutral' (ie.
$\nabla=\nabla_{ad}$) due to the mixing by overstable convection,
that should be used for stellar modelling. Thus they agree with the
\citet{1958ApJ...128..348S} method. Other authors have however argued
that there should be a slow mixing in the region between the convective
boundary given by the Ledoux criterion and that given by the Schwarzschild
criterion (see eg. \citealt{1985AA...145..179L}; \citealt{1996MNRAS.283.1165G}).
This is more analogous to double diffusive mixing. 

\begin{figure}
\begin{centering}
\includegraphics[width=0.8\columnwidth,keepaspectratio]{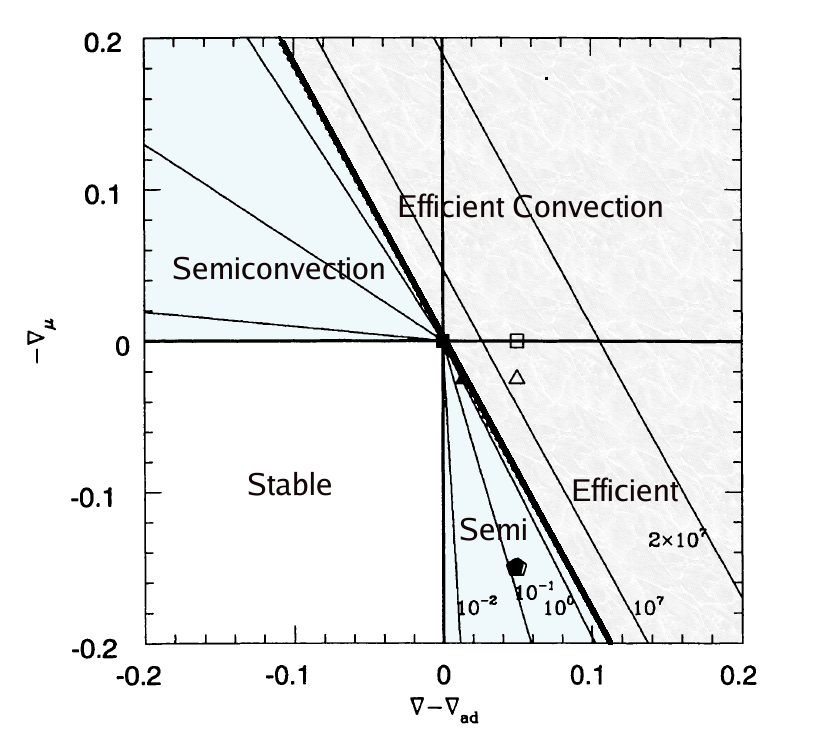}
\par\end{centering}
\caption{The stability diagram from \citet{1996MNRAS.283.1165G} (their Fig.
2). We have added some some extra labelling and shading. The heavy
solid line represents the Ledoux criterion, which \citet{1996MNRAS.283.1165G}
conclude is an excellent approximation for the stability criterion
given by their double-diffusive extended MLT. The thin solid lines
are contours of constant convective velocity. Note that the reduction/increase
in velocity is very sharp when crossing the Ledoux line. The plot
shows that 1) efficient convection can be reduced to semiconvection
in the presence of a $\mu$-gradient (quadrant 4), 2) Efficient convection
can occur even if $\nabla-\nabla_{ad}<0$, providing there is a positive
$\mu$-gradient (quadrant 1), and 3) slow ($\sim$ semiconvective)
mixing can occur well below $\nabla-\nabla_{ad}=0$ if there exists
a sufficiently steep and positive $\mu$-gradient (quadrant 1). {[}Figure
reproduced with permission from Prof. Ronald Taam{]}. \label{fig-grossmanTaammStability}}
\end{figure}

There are many types of instabilities that fluids can experience (eg.
Rayleigh-Taylor, Rayleigh-Bernard, Bernard-Marangoni, etc.). The manifestation
of one or the other depends on the prevailing physical conditions.
For example, \citet{TS64} list a range of physical scenarios in which
various types of instabilities arise, based on salt water experiments
in which the density, composition, and temperature profiles are varied.
However, in stellar applications we really only need to know the (rough)
timescale of mixing. \citet{1996MNRAS.283.1165G} have developed a
double-diffusive extended mixing length theory for stellar evolution
calculations. Their results show that there are various regimes of
convection and that their derived condition for stability is almost
exactly the Ledoux criterion. They present their results in a stability
plane which illustrates the situation concisely. Figure \ref{fig-grossmanTaammStability}
is a reproduction of their figure 2, with some minor alterations.
The main points which diverge from the strict Schwarzschild criterion
treatment are 1) efficient convection can be reduced to semiconvection
in the presence of a $\mu$-gradient, 2) efficient convection can
occur even if $\nabla-\nabla_{ad}<0$, providing there is a positive
$\mu$-gradient, and 3) semiconvection can occur well below $\nabla-\nabla_{ad}=0$
if there exists a sufficiently steep and positive $\mu$-gradient.
They also find that there is a dramatic change in characteristic convective
velocities when crossing the Ledoux line. We have thus interpreted
this in terms of three effective regimes: stable, semiconvective (ie.
any slow mixing) and convective (efficient/turbulent mixing). 

Given the discussion above, we have added the Ledoux criterion (Equation
\ref{eqn-ledoux}) to the SEV code. The criterion for deciding when
a region is semiconvective that we have adopted is:

\begin{equation}
0<(\nabla-\nabla_{ad})<\nabla_{L}\label{eqn-simsemiCriterion}
\end{equation}

where

\begin{equation}
\nabla_{L}=\frac{\beta}{4-3\beta}\nabla_{\mu}.\label{eqn-GradLedoux}
\end{equation}

As we now treat mixing time-dependently throughout the entire star,
it is an easy matter to also treat semiconvective regions in the same
manner. This is done by applying slow mixing velocities in the region
defined by Equation \ref{eqn-simsemiCriterion}. However, the problem
of not knowing how slow to mix remains. We have initially hard-wired
a fixed value velocity for semiconvection (a similar tactic to that
of \citealt{1972MNRAS.156..361E}). We note that this is insufficient
and must be improved upon, but it is outside the scope of this study.
We have however provided the basis for better treatment of semiconvection.
One option here is to accept the velocities given by the \citet{1996MNRAS.283.1165G}
double-diffusive MLT. We leave this for future work. Finally we note
that there is now a numerical switch in the SEV code that enables
one to use the Schwarzschild or Ledoux criteria at will.

\subsection{Summary, Practicalities \& Future Work}

In overview, the entire convective mixing paradigm in the SEV code
has been changed. The treatment is now a diffusion approximation to
turbulent mixing, allowing us to follow mixing time-dependently. Overshoot
and semiconvection are now also treated as diffusive mixing, through
the variation of convective velocities. 

We have however retained the instantaneous mixing routine (in \emph{abund.f}
-- the diffusion version is in \emph{abund-diffn.f}) and provide
a numerical switch so the operator can choose which to use (the switch
is called \emph{idiffusion}). The same switch, when given a value
of 2, lets the SEV code choose when it is necessary to use time-dependent
mixing or instantaneous mixing. This reduces calculation time as most
evolution only requires instant mixing which is faster to calculate.
It should be noted that doing this may lead to inconsistencies if
different overshoot and semiconvection methods are used when switching
between diffusive and instant mixing. Thus it is recommended to use
one or the other during a particular run.

We have also included a switch for diffusive overshoot, so the operator
can choose whether to include this or not. Related to this is another
control parameter in the input file called \emph{diffnosfac} which
is the $f_{OS}$ value for diffusive overshoot. This is a global value,
used for overshooting on all convective boundaries.

Finally, there is a switch for diffusive semiconvection called \emph{diffsemi}.
This allows the operator to choose whether to use the Ledoux criterion
to check for semiconvection or not (if this switch is set to `off'
then no form of semiconvection will be applied). 

Despite all these changes there is room for improvement and we suggest
the following as possible future work:
\begin{itemize}
\item The diffusion routine currently operates with the same timestep as
that used for the structural evolution. Whilst this is reasonable
(since we track the errors and the evolutionary timestep is taken
to keep changes in composition minimal) it may be more efficient to
use a sub-evolutionary timestep within the diffusion routine. Based
on the tests carried out above this would reduce errors in the diffusion
routine, removing the necessity to reduce the evolutionary timestep
for the diffusion routine. The facility for this is already available
in the code and would thus only require minor changes.
\item Although we argue that our method of `burn then mix' each iteration
should give self-consistent results, it would be instructive to implement
the other two methods to check, ie. 1) coupling the nuclear burning
and diffusive mixing and 2) coupling the nuclear burning, diffusive
mixing and structure.
\item The amount diffusive overshoot should be variable depending on the
physical conditions at each convective boundary. However we note to
reliably quantify this requires further multidimensional fluid dynamic
calculations in order to gauge the extent of overshoot under differing
conditions. When the (parameterised) results from these calculations
become available it will be an easy matter to implement them into
the existing code.
\item The diffusive semiconvection method could easily be improved to take
into account Rayleigh-Taylor instabilities that arise with positive
$\mu$-gradients. At the moment we have a traditional treatment of
semiconvection. The mixing velocities in the semiconvective regions
should also be treated more self-consistently. This could be done
by using a modified MLT (see semiconvection section above for more
details).
\item Given that a diffusion routine is now employed in the SEV code it
should be a fairly easy task to utilise it for diffusing angular momentum
also. This is the usual method for including the effects of rotation
in stars. Microscopic diffusion may also be a possibility.
\end{itemize}

\section{Opacity\label{opacmods}}

The opacity setup for the original version of the SEV code is described
in section \ref{sec-SEV-IncludedPhysics}. Here we describe the changes
made over the course of the current study.

\subsection{Low Temperature: $500-8000\,\textrm{K}$}

Low temperature opacities become important during the giant phases
of low and intermediate mass stellar evolution. The temperatures can
become so cool as to allow formation of molecules and grains. Therefore
these particles need to be taken into account when calculating opacities.
In the original version of SEV code molecular opacities were taken
into account using the formulae suggested by \citet{1989AAS...77....1B}
which are based on the molecular (and grain) opacity tables from \citet{1975ApJS...29..363A}
and \citet{1983ApJ...272..773A}. 

For the current version of the SEV code we have updated to the tables
calculated by \citet{2005ApJ...623..585F}. These tables already include
molecular and grain opacity sources so the \citet{1989AAS...77....1B}
formulae have been removed. The range of temperature that the tables
cover is $500<\textrm{T(K)}<31,000$. We have however chosen to use
them in the range $500<\textrm{T(K)}<8,000$. The main reason for
this is that, although \citet{2005ApJ...623..585F} do offer to compute
opacities with other compositions on request\footnote{See the low-T opacity website at \url{http://www.webs.wichita.edu/physics/opacity/}},
the tables are not yet computable by the user online. This facility
is available on the OPAL website\footnote{The OPAL website is url is \url{http://www-phys.llnl.gov/Research/OPAL/new.html} }
and is very useful for computing opacities for arbitrary compositions.
However the OPAL opacities (\citealt{1996ApJ...464..943I}) do not
include very low temperatures so we retain the \citet{2005ApJ...623..585F}
tables for the aforementioned low temperature range. The transition
between the two sets of tables at 8000 K was chosen as this is close
to the lower limit of the OPAL tables. The OPAL tables are variable
in carbon and oxygen, which is needed for the change in opacity caused
by increasing amounts of these elements through nuclear burning and
third dredge-up -- but the \citet{2005ApJ...623..585F} tables are
not. The portion of the stellar model that this affects is very small
however, as discussed in Appendix \vref{section-LowTOpacUncertainties}.

The subroutine \emph{lowtintz} in the file \emph{otable2005.f} now
reads in the needed low temperature opacity tables from a directory
below the working directory (called \emph{/optables/lowT}/) and interpolates
within them (in Z) when the SEV code starts. The Z value is specified
by the operator in the file \emph{infiles.list.} Thus a set of low
temperature opacity tables for the requested Z is stored in memory
and used throughout the evolution. The tables are stored in the array
variable \emph{OOP} and vary in X, T, and R (R is the OPAL variable:
$R=\rho/T_{6}^{3}$, where $T_{6}$ is the temperature in MK). Interpolation
in theses variables is done during the evolution as needed. In regard
to the initial interpolation in Z we have decided to use \emph{linear}
interpolation because we found that quadratic interpolation was inadequate
at handling the discontinuities in the tables, giving wildly inaccurate
values. Although linear interpolation can be a little inaccurate at
times, it does not suffer from the huge deviations possible with ill-behaved
quadratic interpolation. We reiterate that these tables are not variable
in C and O and are currently only available for particular compositions.
Some of the compositions for which the opacity tables are available
(on the website) are from compilations by:
\begin{itemize}
\item \citet{1992RMxAA..23...19S} (including an alpha-enhanced version)
\item \citet{1993oee..conf...14G}
\item \citet{1998SSRv...85..161G}
\item \citet{2003ApJ...591.1220L}
\item \citet{2005ASPC..336...25A}
\end{itemize}
of which we can now use any. All of the calculations in this study
have used the solar composition of \citet{1993oee..conf...14G} but
\citet{2003ApJ...591.1220L} is also available in the opacity directory
for the SEV code (these were the two available at the time of installation). 

\subsection{Mid Temperature: $8000\,\textrm{K}-500\,\textrm{MK}$}

A number of months work was spent updating the mid-range temperature
opacity routines in the SEV code. The primary aims were 1) to update
to the new OPAL table format, and 2) to relax all the scaled-solar
composition assumptions. The old OPAL subroutine \emph{xcotrin.f}
was substantially modified to take the new OPAL format. This routine
was originally written by Arnold Boothroyd (using some routines from
M. J. Seaton, circa 1991). Once this was done, and the scaled solar
assumptions removed, it was possible to calculate opacity tables for
any composition on the OPAL website and use them with the SEV code.
This version of the code was used successfully for a few months. 

However, it came to the attention of the Author that Arnold Boothroyd
had greatly expanded his original routine himself, and had made it
available on his website\footnote{Boothroyd's website url is: \url{http://www.cita.utoronto.ca/~boothroy/kappa.html}}.
As this very large, detailed subroutine could actually perform both
the tasks required -- and much more -- and it had been through much
debugging, we decided to replace our new routine and install Boothroyd's.
Boothroyd's routine is capable of accounting for opacity effects due
to CNO variations (due to nuclear burning) at a particular Z value.
To do this one needs an extra set of opacity tables (also available
on Boothroyd's website). Another feature is the ability to include
opacities for compositions with abnormal {[}O/Fe{]}. The key feature
for the current low-metallicity study is however the ability to use
opacity tables of arbitrary composition generated on the OPAL website.
This has been done many times now throughout the study.

The OPAL tables (\citealt{1996ApJ...464..943I}) have a temperature
range of $5600\,\textrm{K}$ to $501\,\textrm{MK}$ and a $log_{10}(\textrm{R})$
range of $-8.0$ to $+1.0$. The tables are not rectangular however.
The tables do not cover the higher density end at high temperatures.
At high densities and temperatures conductive opacity becomes important.
In these regions we retain the treatment for conductive opacities
described in section \ref{sec-SEV-IncludedPhysics}. We have chosen
the transition temperature from low temperature opacities to the OPAL
opacities as 8000 K, for reasons given in the previous subsection.
The transition to high temperature opacity tables is taken at the
limit of the OPAL tables (501 MK).

The OPAL tables come in two types: \emph{type 1} and \emph{type 2}
tables. Type 1 tables contain a set of 126 tables with a fixed relative
metal distribution but varying X and Z. We use one of these tables
for a given composition. Type 2 tables contain a set of tables (varying
in number) also with a fixed relative metal distribution but varying
in two chosen metals. The two chosen metals (usually C and O) increase
incrementally in each table (independent of the metal distribution).
Each table has constant Z and X. At each Z there are five tables varying
in X. An important feature here is the variability of the two chosen
metals. This allows one to take into account the change in opacity
that occurs as carbon and oxygen start to dominate the composition
via nuclear burning. This is important in the C-O core and when the
envelopes of AGB stars become polluted with C through third dredge-up. 

All the OPAL opacity tables are located in the directory \emph{/optables/opal/}
below the working directory. Tables generated on the OPAL website
must be put in this directory for use, and naming conventions taken
into account. 

\subsection{High Temperature: $500\,\textrm{MK}-2000\,\textrm{MK}$}

The original version of the SEV code did not include opacities for
temperatures above the OPAL limit of 500 MK. This limits the range
of operation of the code, particularly at low metallicities and higher
masses where temperatures can be very high. We have remedied this
by adding a high temperature opacity table from Alessandro Chieffi
(private communication). The table file name is \emph{xhighTopac.tab}
and is placed in the \emph{/optables/highT/} directory below the working
directory. This table has a temperature range of 400 MK to 12.5 GK.
The SEV code does not include oxygen burning as yet, so structures
with temperatures above $\sim$2 GK can not be followed and we thus
only use the range of 500 MK to 2 GK in \emph{xhighTopac.tab}. The
new subroutine \emph{readhight} reads in the high temperature table
and stores the values in memory for interpolation when needed. The
interpolation routine \emph{HIGHTOPA} was adapted from that of A.
Chieffi. The main difference is a change from cubic to quadratic interpolation.
At such high temperatures there is no H or He, so the metals we have
chosen to include in the opacity interpolation are C, O and Ne. Conductive
opacities are included in the table already.

\subsection{Conductive: High T and R}

The method for calculating conductive opacity has not been changed
(see section \ref{sec-SEV-IncludedPhysics} for a description). However
the routine for calculating them has been merged with the SEV code.
This means that it is unnecessary to produce the file \emph{con\_opac.dat}
needed for the original version of the SEV code. Everything is now
done at run time. Another change is the removal of the `hard-wired'
scaled-solar composition assumptions. The composition is now read
in from the OPAL table headers, so the conductive opacity composition
automatically matches the radiative opacity composition.

\subsection{Summary, Practicalities and Future Work}

In overview, the opacity setup in the SEV code has been updated, extended
and streamlined. 

There is however some room for more improvement (as always!). We note
a few points here:
\begin{enumerate}
\item There are slight discontinuities going from one set of tables/temperature
regimes to the next. This may cause convergence problems (although
we have not noticed a problem in practice). A remedy for this would
be to interpolate between the various tables to get smoother transitions
between temperature regimes.
\item It would be more consistent to have the low temperature opacity composition
match the OPAL composition. This is not a problem when using scaled-solar
composition (as both have tables with this composition) but it is
when we compute non-scaled-solar compositions on the OPAL website,
as we can't do this for the low temperature tables. However the effect
may be a small one.
\item Related to 2. above is the fact that the low temperature opacities
do not take into account the variability of C and O in the envelope,
as no tables are provided to enable interpolation (although they could
be requested). Again, the degree of effect is uncertain. (Also see
the discussion in Appendix \vref{section-LowTOpacUncertainties}).
\item The routine that generates the conductive opacities may need updating.
\end{enumerate}
Finally we note that the opacity setup is now more streamlined, as
there is no need to pre-compute various opacity tables (such as the
old \emph{opalfile.dat} and \emph{codataa}, etc.). If all that is
needed is scaled solar opacities, at any Z, the only requirement on
the part of the operator is to specify the Z value in the text file
\emph{infiles.list}. This will normally be the case. This requires
that all the opacity tables for all Z values are transported with
the source code. As the full set amounts to $\sim$8.5 MB (eg. the
file \emph{OPALkappa.tar.gz} from Boothroyd's website) this is not
a problem with current computing resources. 

If abnormal compositions are required then they can be computed on
the OPAL website at \url{http://www-phys.llnl.gov/Research/OPAL/new.html}.
The generated files must then be put in the \emph{/optables/opal/}
directory.

\section{CNO Equilibrium with He Burning at $Z=0$}

\subsection{Motivation}

A unique feature of zero metallicity stars comes about because of
the complete lack of CNO catalysts. At the beginning of the MS stars
of masses that would normally burn hydrogen via the CNO cycle aren't
able to do so due to this lack of catalysts. Instead they burn H through
the PP chains. This reaction sequence has a much lower temperature
dependence than the CNO cycles (T$^{4}$ as opposed to T$^{20}$)
which has a significant effect on the structure. Indeed, much of the
star's mass can be involved in hydrogen burning. To maintain hydrostatic
equilibrium the PP chains must produce enough energy and, again due
to the weak temperature dependence, the central temperatures become
very high. In fact, for stars of sufficient mass ($>\sim0.9$ M$_{\odot}$)
the core will be so hot towards the end of the MS (earlier for more
massive stars) that helium starts burning via the triple-alpha ($3\alpha$)
reactions -- in the same location as the PP chains are operating.
The resultant $^{12}$C is then available for the CNO cycle, which
starts immediately. The structure of the star then changes, becoming
more `normal' (ie. like higher metallicity stars), and a small excursion
in the HR diagram ensues. 

The reason we detail this here is because the original version of
the SEV code did not take into account the possibility of simultaneous
H and He burning in the particular situation when the CN(O) cycles
are in equilibrium. Thus, in order to model $\textrm{Z=0}$ stars
for this study, a small modification was needed in \emph{abund.f}
and \emph{abund-diffn.f} to take the simultaneous H and He burning
into account in the equilibrium calculations.

\subsection{Equilibrium CNO Burning and Timestepping}

A nuclear species is in equilibrium when its production rate is equal
to its destruction rate. When the CNO cycle is operating the sum of
the CNO nuclei is conserved over time. Taking both these facts into
account the abundance distribution is then given by simple algebraic
expressions involving the current abundances and the nuclear rates,
which depend on density and temperature (see eg. \citealt{1983psen.book.....C}).
As the evolution of CNO abundances is calculable via these relations,
it is possible to burn more of a particular nucleus than is actually
present (ie. each nuclei can react more than once in a single timestep),
as we know a priori how the abundances will be distributed. The SEV
code has a check to see if CN(O) equilibrium will be achieved during
the current timestep and if it is the CNO abundances are scaled to
the equilibrium values. In theory the abundance changes calculated
explicitly through all the reactions (ie. without assuming equilibrium
abundances) should give the equilibrium values. However in practice
perfect equilibrium is not achieved as one nucleus inevitably gets
burnt slightly faster than it is replaced (invariably $^{12}$C in
the CNO cycle). This effect is exacerbated on the \emph{approach}
to equilibrium as the $^{12}$C will definitely not be returned through
the CNO cycle fast enough, due to the $^{14}$N$(p,\gamma)$$^{15}$O
bottleneck. In both of these situations we can end up with a negative
abundance and thus the timestep needs to be decreased. This has the
interesting consequence that (if we don't include equilibrium assumptions)
the timestep is dependent on the metallicity of the star, since the
timestepping is governed by the $^{12}$C abundance during this stage
of evolution. Higher metallicity stars can have much larger timesteps
during CNO burning due to their increased equilibrium CNO abundances
-- whilst low metallicity stars need smaller timesteps. This effect
is taken to the extreme with $\textrm{Z}=0$ stars, which is one of
the main topics in the present study. Thus we need to retain an equilibrium
treatment in the SEV. 

As an empirical example we computed the MS evolution a 1 M$_{\odot}$
$Z=0$ star with and without the equilibrium checks and calculations.
With equilibrium the run took $\sim$20 min, with timesteps of $\sim10^{5}$
yr towards the end of the MS (when the $3\alpha$ reactions had started).
Without (CNO) equilibrium we were unable to evolve it very far into
the stage in which He burning started as the timesteps were so small,
being $\sim10^{-2}$ yr. This huge difference of $10^{7}$ in timestep
suggests that it would take $\sim19$ \emph{years} (!) to complete
the end of the MS evolution.

\subsection{Dual Burning Modification}

\begin{figure}
\begin{centering}
\includegraphics[width=0.7\columnwidth,keepaspectratio]{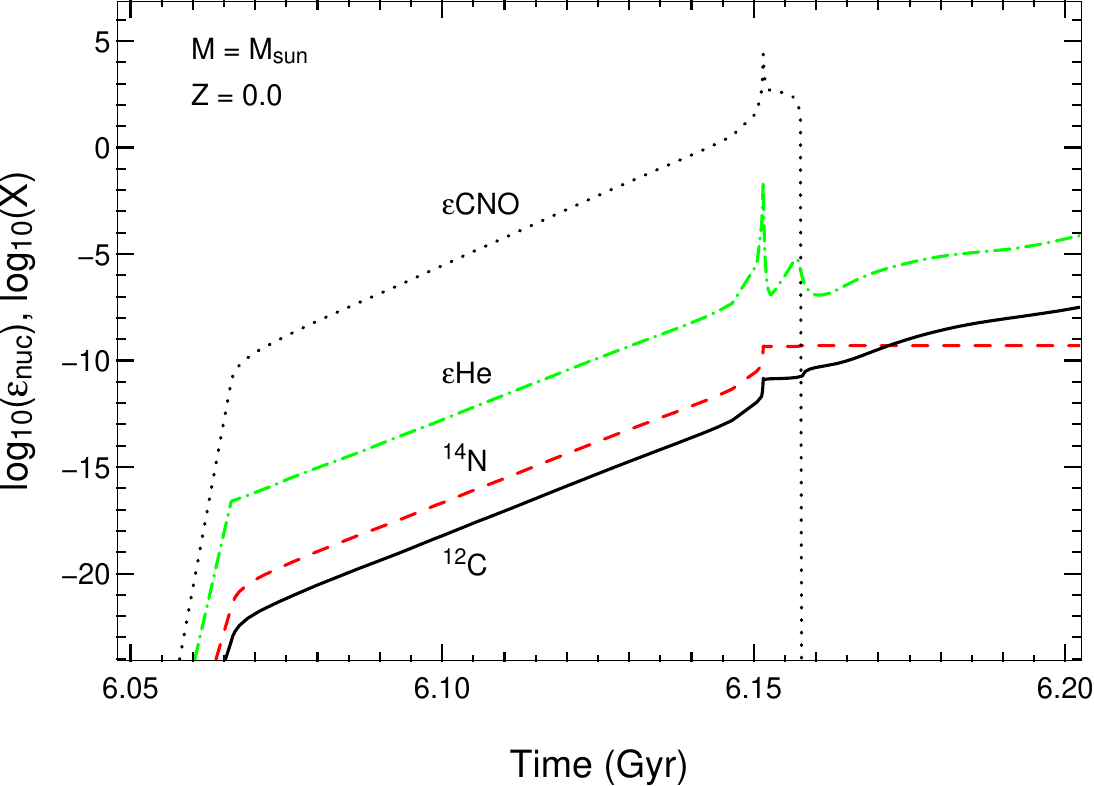}
\par\end{centering}
\caption{The evolution of the central $^{12}$C and $^{14}$N abundances, as
well as the central helium and CNO energy generation. The star has
$Z=0$ and a mass of 1 M$_{\odot}$. The evolution is shown from the
ignition of the $3\alpha$ reactions during the subgiant branch. The
consequent ignition of the CNO reactions is also evident. It can be
seen that the initiation of the CNO cycles occurs as soon as there
is a minute amount of $^{12}$C produced from the $3\alpha$ reactions
($X_{C}\sim10^{-23}$ in this case!), although the energy generation
is minimal when the $^{12}$C abundance very low. The $^{12}$C from
the $3\alpha$ reactions is mostly burnt to $^{14}$N by the CNO cycle.
At $t\sim6.16$ Gyr the CNO cycles cease due to the exhaustion of
H at the centre. \label{fig-m1Z0-CNO-3a-eqm}}
\end{figure}

We now return to the problem of having $3\alpha$ reactions happening
concomitantly with H burning. The $3\alpha$ reactions add $^{12}$C
to the plasma, which can then participate in the CNO cycle, increasing
the sum of CNO nuclei. If there were no CNO nuclei initially present
(ie. $\textrm{Z=0}$) then this fresh carbon initiates the CNO cycle.
In addition to the $^{12}$C there may also be some $^{16}$O produced
via $^{12}$C$(\alpha,\gamma)$. The method I have adopted to account
for this injection of extra nuclei into the CNO cycle is to add the
extra nuclei to the CNO cycle total -- if the cycle is found to be
in equilibrium. If it is not then all reactions are followed explicitly.
The original equilibrium tests remain, as they are sufficient to determine
when equilibrium is achieved, even if extra $^{12}$C is being produced.
As an example of the tests for equilibrium utilised in the SEV code
we show the one for CN equilibrium, such that if

\begin{equation}
12.12\,m_{p}\,X_{H}\,RC_{1}\,\Delta t>1.0\label{eqn-CNeqm}
\end{equation}

then the cycle is in equilibrium. Here $m_{p}$ is the proton mass,
$X_{H}$ the mass fraction of H and $RC_{1}$ the current rate of
the $^{12}$C$(p,\gamma)$$^{13}$N reaction. This essentially says
that if every $^{12}$C nucleus reacts more than once in the timestep
$\Delta t$ (via the reaction $RC_{1}$) then the whole CN cycle must
be in equilibrium, and thus we can use the simple expressions for
the equilibrium CNO abundances. In the original version of the SEV
code all the CNO nuclei were summed into the variable V4:

\begin{equation}
V4=X_{C}+\frac{6}{7}X_{N}+\frac{3}{4}X_{O}\label{eqn-old-CNO-eqm}
\end{equation}

which is a holding variable for all the CNO nuclei in terms of $C$.
All the $X_{i}$ are taken as the values from the previous timestep
(ie. at the beginning of the current timestep). This is then redistributed
amongst the CNO nuclei based on equilibrium assumptions. It can be
seen that the CNO nuclei are thus perfectly conserved from one timestep
to the next (as they should be). In order to account for the extra
nuclei from the reactions involving helium we have now included the
total production and destruction (from all reactions) from the current
timestep into the $V4$ sum:

\begin{equation}
V4=\left(X_{C}+\frac{\Delta t}{2}\frac{dX_{C}}{dt}\right)+\frac{6}{7}\left(X_{N}+\frac{\Delta t}{2}\frac{dX_{N}}{dt}\right)+\frac{3}{4}\left(X_{O}+\frac{\Delta t}{2}\frac{dX_{O}}{dt}\right)\label{eqn-CNO-3a-eqm}
\end{equation}

which is then redistributed in the usual way. In this way the increase
in CNO nuclei (mainly $^{12}$C in practice, which then gets burnt
to $^{14}$N) is now followed during equilibrium. This situation (combined
H and He burning) will occur whenever hydrogen burning happens at
very high temperatures (starting at $\sim80\,\textrm{MK}$), given
the availability of helium and a high enough density. The four main
sites we imagine this happening are:
\begin{enumerate}
\item During the MS of $Z=0$ stars
\item During hot $Z=0$ H-shell burning
\item During very hot HBB on the AGB at any metallicity
\item During proton ingestion episodes in very low or zero metallicity stars
\end{enumerate}
In the MS $Z=0$ case the ignition of the CNO cycle due to the local
production of primary CNO nuclei quickly causes the temperature to
drop, bringing the $3\alpha$ reactions to a virtual halt. This happens
with CNO abundances of the order $\sim10^{-10}$. As an example of
this case we present in Figure \ref{fig-m1Z0-CNO-3a-eqm} some properties
of a 1 M$_{\odot}$ model as given by the modified SEV code. It can
be seen that the magnitude of CNO nuclei production is indeed minute.
It is however enough to alter the burning regime, even giving rise
to a convective core due to the extra (localised) energy release from
the CNO reactions. Also of note is the negligible luminosity provided
by the $3\alpha$ reactions, and the rapid transmutation of the newly
created C (and O) nuclei to $^{14}$N.

The main aim of this modification was to enable computation of $Z=0$
stars. We leave the discussion of the other sites of coincident equilibrium
CNO and He burning till those parts of the study in which they arise. 

\section{Other Modifications}

Other modifications made to the SEV code during the course of the
current study include:
\begin{itemize}
\item Removal of all scaled-solar abundance assumptions. This allows the
calculation of models with arbitrary abundance distributions. This
is relevant at low metallicity as observations show that these stars
often have non-solar abundance ratios (eg. alpha-enhanced Galactic
globular cluster stars).
\item Alteration of logical constraints that did not allow a zero value
of metallicity Z.
\item The addition of a treatment to handle proton ingestion episodes (PIEs)
during low mass $\textrm{Z}=0$ evolution. These extreme events cause
convergence problems and necessitate very small timesteps. See the
$\textrm{Z}=0$ results section for details.
\item The addition of a treatment to estimate the increase in opacity due
to a $^{14}$N dominated envelope during the AGB of low metallicity
stars. Any dredge up of CNO nuclei is quickly burnt to $^{14}$N due
the high temperatures at the bottom of the convective envelope in
these low-Z stars. The ratio C:N:O can become $\sim$10:100:1 on the
AGB. Since the opacity tables only include alterations due to increased
C and O, the opacity in these regions would be wrong. See the $Z=0$
results section for more details.
\end{itemize}

\part{STELLAR MODEL RESULTS}

\chapter{Zero Metallicity Stars: Structural Evolution\label{chapter-Z0-StructEvoln}}
\begin{quote}
``Education is not preparation for life; education is life itself.''
\begin{flushright}
\vspace{-0.3cm}-- John Dewey
\par\end{flushright}
\end{quote}

\section{Background}

\subsection{Overview}

The modifications to the structural evolution code (SEV code) detailed
in Chapter \ref{sevmods} have enabled us to model metal-free primordial
stars. In particular there exists a number of evolutionary features
peculiar to $Z=0$ (and extremely low metallicity) stars that require
the use of time-dependent mixing. The two most important of these
are the proton ingestion episodes (PIEs) that are induced by helium
flash convection breaking through to hydrogen-rich regions. We refer
to these events as the dual core flash (DCF) and dual shell flash
(DSF). The DCF occurs at the time of the core He flash in low mass
models whilst the DSF occurs during the first major He shell flash
at the beginning of the TP-AGB in intermediate mass models (sometimes
more than one DSF occurs). Although these events have been modelled
before by other authors they are still relatively unexplored phenomena
as not many studies have evolved through these difficult phases. These
two events are pivotal in terms of the possible contribution of these
stars to polluting the interstellar medium as they both result in
considerable pollution of the envelope, which is later lost through
stellar winds. The DCF and DSF are notoriously difficult to evolve
though numerically and have provided a serious challenge for the author
(and indeed previous authors!). 

The real novelty in this study though is that we have also taken (most)
of our models through to the completion of the AGB. To the best of
our knowledge this is the first time this has been achieved. This
allows us to make a range of interesting predictions, such as the
expected white dwarf masses, the lifetimes of the various stages of
evolution, and the chemical yield of primordial low- and intermediate-mass
stars. We should however add a caveat here. The models suffer from
the usual uncertainties such as that which derives from the use of
the Mixing Length Theory of convection, the neglect of rotation, and
the choice of mass-loss formalism. We discuss some of these uncertainties
in the next subsection when we describe the input parameters used. 

To elucidate the evolutionary features of $Z=0$ low- and intermediate-mass
models we describe two of our models in substantial detail in this
Chapter. We have chosen the 0.85 M$_{\odot}$ and 2.0 M$_{\odot}$
models because between them they display all the peculiar evolutionary
features found in our models. In particular the 0.85 M$_{\odot}$
model experiences the DCF and the 2.0 M$_{\odot}$ model experiences
the DSF. At the end of each detailed description we provide a comparison
with previous studies at similar masses. Finally, in the last section,
we summarise the evolution of all our models. The models evolved are
of initial mass 0.85, 1.0, 2.0 and 3.0 M$_{\odot}$. The detailed
nucleosynthesis is reported in the next chapter.

\subsection{Physical Parameters Used \& Initial Composition\label{subsection-z0-Struct-InputPhysics}}

\subsubsection*{Initial Composition}

We have adopted a standard Big Bang nucleosynthesis (SBBN) composition
for our models. A summary of Big Bang nucleosynthesis and details
of our adopted initial abundances for the $Z=0$ and extremely metal-poor
models was given in Section \vref{sec-BBN}. For convenience we redisplay
the abundances here in Table \ref{table-Z0-InitialAbunds}. It can
be seen that all species heavier than $^{4}$He have been set to zero,
in line with SBBN. 

\begin{table}
\begin{centering}
\begin{tabular}{|c|c|}
\hline 
Nuclide & Mass Fraction\tabularnewline
\hline 
\hline 
$^{1}$H & $0.754992$\tabularnewline
\hline 
$^{3}$He & $7.85\times10^{-6}$\tabularnewline
\hline 
$^{4}$He & $0.24500$\tabularnewline
\hline 
$^{12}$C & $0.0$\tabularnewline
\hline 
$^{14}$N & $0.0$\tabularnewline
\hline 
$^{16}$O & $0.0$\tabularnewline
\hline 
\end{tabular}
\par\end{centering}
\caption{Initial abundances used in the zero-metallicity stellar structure
models, as given by Standard Big Bang Nucleosynthesis calculations
by from \citet{Coc0104}. \label{table-Z0-InitialAbunds}}
\end{table}

\subsubsection*{Mass Loss}

Mass loss is one of the most uncertain factors in stellar modelling.
We have discussed the methods employed in the SEV code in Subsection
\vref{sub-MassLossDescription} but provide a brief discussion here
also, in relation to the special case of zero metallicity.

It is generally thought that mass loss occurs via the driving force
of radiation pressure. In low- and intermediate-mass (LM and IM) stars
observations show that mass loss becomes important on the giant branches,
and in particular at the end of the AGB. During these phases two important
things occur -- the luminosity increases, increasing the (potential)
radiation pressure, and the envelope cools, increasing the opacity
and allowing grains to form. In addition to this AGB stars pulsate,
providing a further (mechanical) contribution to mass loss. The most
important phase of mass loss occurs during the AGB, as this is when
most of the mass of a star is lost. 

In normal (metal-rich) AGB stars the main mechanism for mass loss
is thought to be radiation pressure on grains. In the case of zero
metallicity grains cannot form due to the lack of metals to form oxides
(for example). This implies that the mass loss rate in $Z=0$ stars
should be much lower (if not zero). Although $Z=0$ stars are more
luminous than their higher metallicity counterparts, giving rise to
higher radiation pressures, they also have a lower opacity, which
again indicates a lower mass loss rate is warranted. 

We note however that mass loss is not fully understood -- there may
be other mechanisms that operate at low- or zero-metallicity. In addition
to this we have found that $Z=0$ LM and IM stars experience self-enriching
episodes (namely the DCF and DSF discussed above and also detailed
within this chapter). This raises the metallicity of the envelope
to values similar to that of the LMC (for example) in which AGB mass
loss has been observed. Grains are able to form. Thus it would seem
that a standard mass-loss formalism would be needed for stages of
evolution occurring after the self-enriching episodes. In low-mass
models ($M\lesssim1.2$ M$_{\odot}$ at $Z=0$) these episodes occur
at the tip of the RGB, so that all of the RGB evolution occurs when
the stars still have $Z=0$ at the surface. However, as demonstrated
in the next section which describes our 0.85 M$_{\odot}$ model, the
RGB phase is so short that even if some mass loss formalism is included
(we use the \citet{1975MSRSL...8..369R} formula on the RGB, see Section
\vref{sub-MassLossDescription}) there is negligible mass lost over
this phase. The rest of the star's evolution is characterised by a
relatively metal-rich envelope. In the intermediate mass case ($M\gtrsim1.2$
M$_{\odot}$) we find that the models do not go through the RGB phase,
so mass loss during this stage is not an issue. In addition to this
these stars experience the DSF self-enrichment event at the beginning
of the AGB, so that their envelopes are also (relatively) metal-rich
for the key AGB mass-losing phase. This suggests that using the \citet{1993ApJ...413..641V}
formula for the AGB is warranted. We also note that the metallicity
of a stellar model is indirectly taken into account in the Reimers'
and VW93 formulae as the bulk physical properties used in these formulae
are a function of metallicity. This is because the entire structure
of the models change with Z -- through opacity and different nuclear
burning energy sources (for example) -- and thus there is a feedback
on the mass loss rate. For these reasons we have decided to leave
the mass loss prescriptions \emph{as standard} in the SEV code. 

\subsubsection*{Convective Boundaries and Mixing}

We have employed the new diffusive mixing routine in all our models,
for all phases of evolution. For details on this routine see Section
\vref{timedepmix}. No overshoot or semiconvection was included, such
that the convective boundaries are all `hard' Schwarzschild ones.
This provides a good benchmark as the degree of overshoot and semiconvection
is unknown, particularly at zero metallicity.

\subsubsection*{Opacity}

Updated opacities, as described in Section \vref{opacmods}, were
used for all models. As mentioned in Section \ref{opacmods} the low
temperature tables are not variable in C and O and are currently only
available for particular compositions. This should not cause a large
uncertainty in the models as the amount of mass that requires these
opacity values is minuscule. We have however investigated the discrepancy
in detail, and provide a separate section on this topic in the Discussion
(Section \vref{section-LowTOpacUncertainties}). We note that most,
if not all, previous studies have used non-CO-variable low temperature
opacities because this is all that is presently available.

\section{Detailed Evolution at  $0.85\,\textrm{M}_{\odot}$\label{Section-DetailedEvoln-m0.85z0}}

\subsection{Main Sequence to RGB Tip\label{section-m0.85z0-MStoRGB}}

\begin{figure}
\begin{centering}
\includegraphics[width=0.85\columnwidth,keepaspectratio]{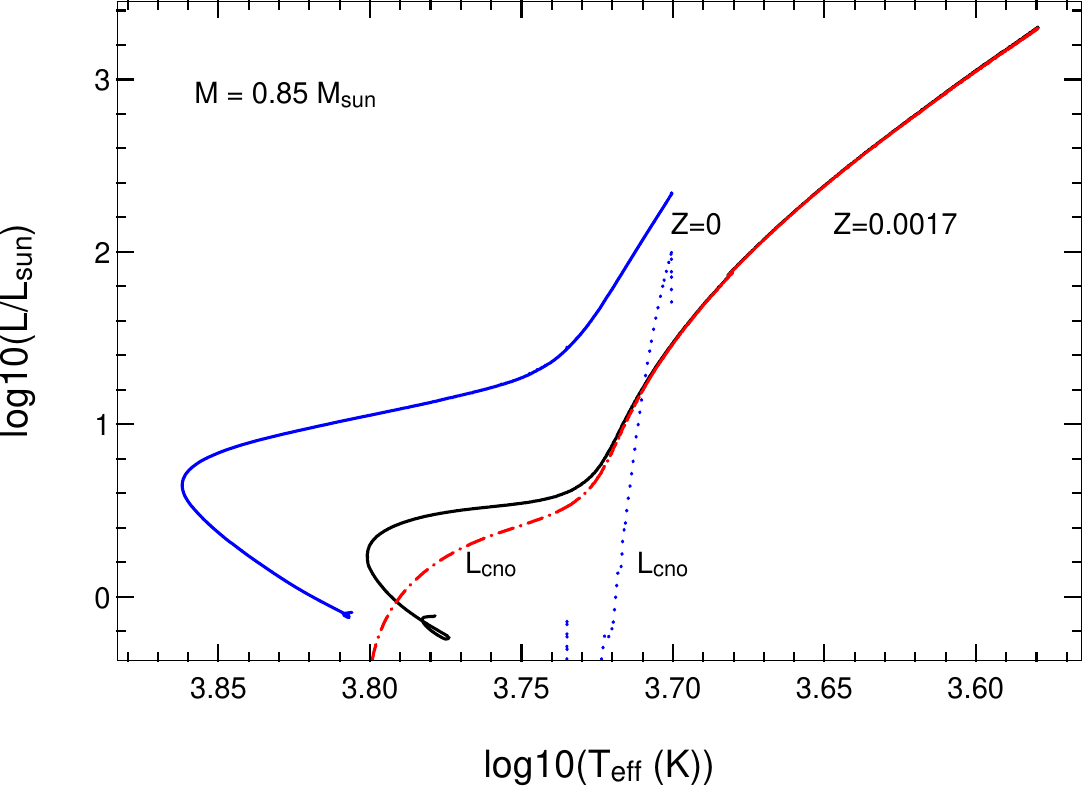}
\par\end{centering}
\caption{Comparison between the $M=0.85$ M$_{\odot}$ primordial star ($Z=0$)
and our GC model ($Z=0.0017$) in the HR diagram, from the MS to the
tip of the RGB. The main differences are i) the $Z=0$ star has a
higher surface temperature and luminosity and ii) the $Z=0$ star
is much less luminous at the RGB tip. We also plot the luminosity
from the CNO cycles for each star (dotted line for $Z=0$ and dash-dotted
for $Z=0.0017$). It can be seen that the ignition of a CNO H-shell
source is delayed in the $Z=0$ model in until the end of the RGB,
due to the lack of CNO catalysts. It thus relies on PP burning for
most of its pre-He-flash evolution. \label{fig-m0.85z0y245-HRD}}
\end{figure}

In order to give some indication of the difference between primordial
stars and more metal-rich stars we compare our $Z=0$, $M=0.85$ M$_{\odot}$
model with a star of the same mass but with $Z=0.0017$. This higher
metallicity model is taken from our globular cluster set of models
(see Chapter \ref{GC-chap}). Figure \ref{fig-m0.85z0y245-HRD} compares
the two stars in the HR diagram. It can be seen immediately that the
$Z=0$ star is significantly hotter and more luminous than the $Z=0.0017$
star during the MS and sub-giant branch (SGB) stages. The temperatures
become comparable from the start of the RGB, but the luminosity of
the $Z=0$ star continues to be higher than the more metal-rich star.
Also of note is that the $Z=0$ model has a much shorter RGB, in the
sense that the luminosity at the tip of the RGB is much lower. Thus
the conditions for He flash ignition are achieved `earlier' on the
RGB in the $Z=0$ model. In the same figure (Figure \ref{fig-m0.85z0y245-HRD})
we also display the CNO burning luminosities for each star. Due to
the lack of CNO catalysts in the $Z=0$ model the normal CNO-powered
H-shell does not ignite until the end of the RGB, whereas CNO burning
becomes significant during the SGB in the $Z=0.0017$ model. Thus
the $Z=0$ model is supported by PP chain burning for most of its
pre-He-core-flash evolution. 

\begin{figure}
\begin{centering}
\includegraphics[width=0.85\columnwidth,keepaspectratio]{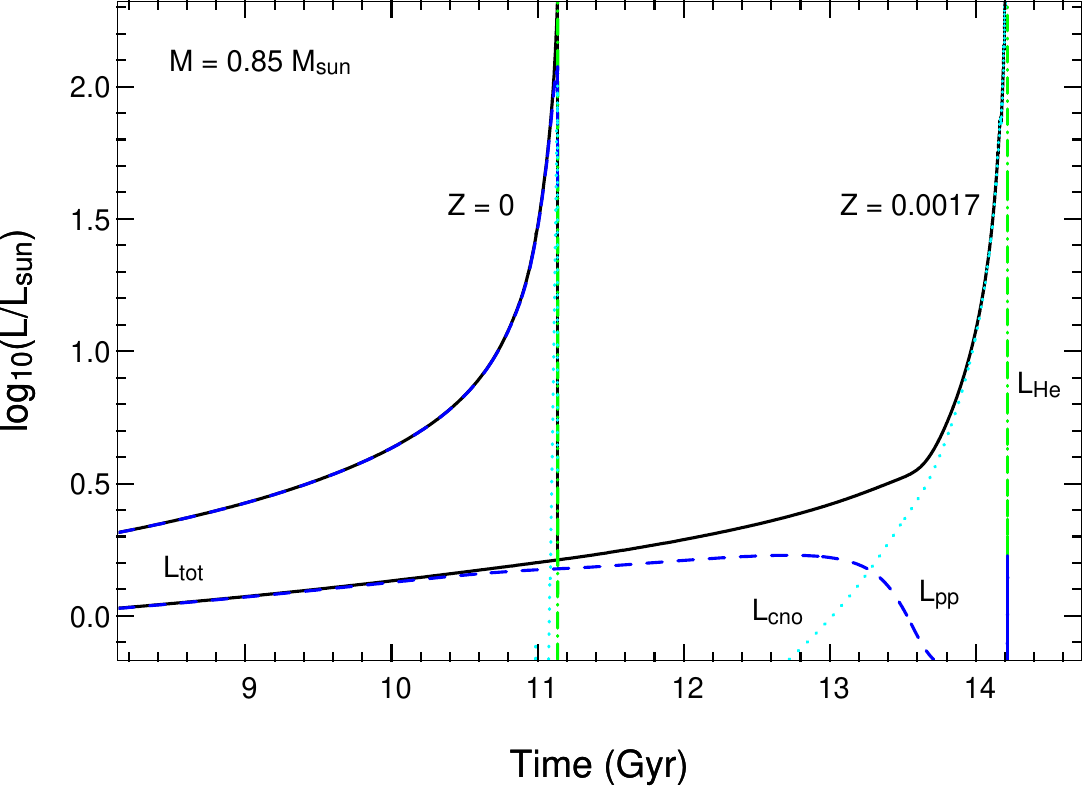}
\par\end{centering}
\caption{Evolution of the various luminosities over the last few Gyr of the
MS and up to the core He flash. The luminosity contributions from
the PP chains (dashed lines), CNO cycles (dotted lines) and He burning
(dash-dotted lines) are displayed, as well as the total luminosity.
Note that the PP chain luminosity overlies the total luminosity for
the $Z=0$ model, whilst the CNO luminosity only becomes significant
just before the core He flash. Whilst the RGB structure in the $Z=0.0017$
model is supported by shell CNO burning, it is supported by shell
PP burning in the $Z=0$ model. The shorter MS lifetime of the $Z=0$
star is also evident in this plot.\label{fig-m0.85z0y245-Lums-MS-RGB}}
\end{figure}

To further illustrate this phenomenon we present in Figure \ref{fig-m0.85z0y245-Lums-MS-RGB}
the time evolution of the luminosities generated by the PP, CNO, and
He burning reactions as well as the total luminosity for both stars.
An interesting feature in Figure \ref{fig-m0.85z0y245-Lums-MS-RGB}
is the similarity of the luminosity curves up until the end of the
MS. This is as expected because stars of such low mass do not have
hot enough cores to burn H via the CNO reactions on the MS (at least
until the very end of the MS). Thus the lack of CNO catalysts has
no effect on the MS nuclear burning regime for masses $\lesssim0.85$
M$_{\odot}$. The lack of metals does however have the effect of reducing
the opacity. This allows energy to escape the star more easily, leading
to the star requiring an increased burning rate in order to maintain
hydrostatic equilibrium. This increased burn rate gives rise to the
higher luminosity, higher surface temperature and shorter MS lifetime.

\begin{figure}
\begin{centering}
\includegraphics[width=0.85\columnwidth,keepaspectratio]{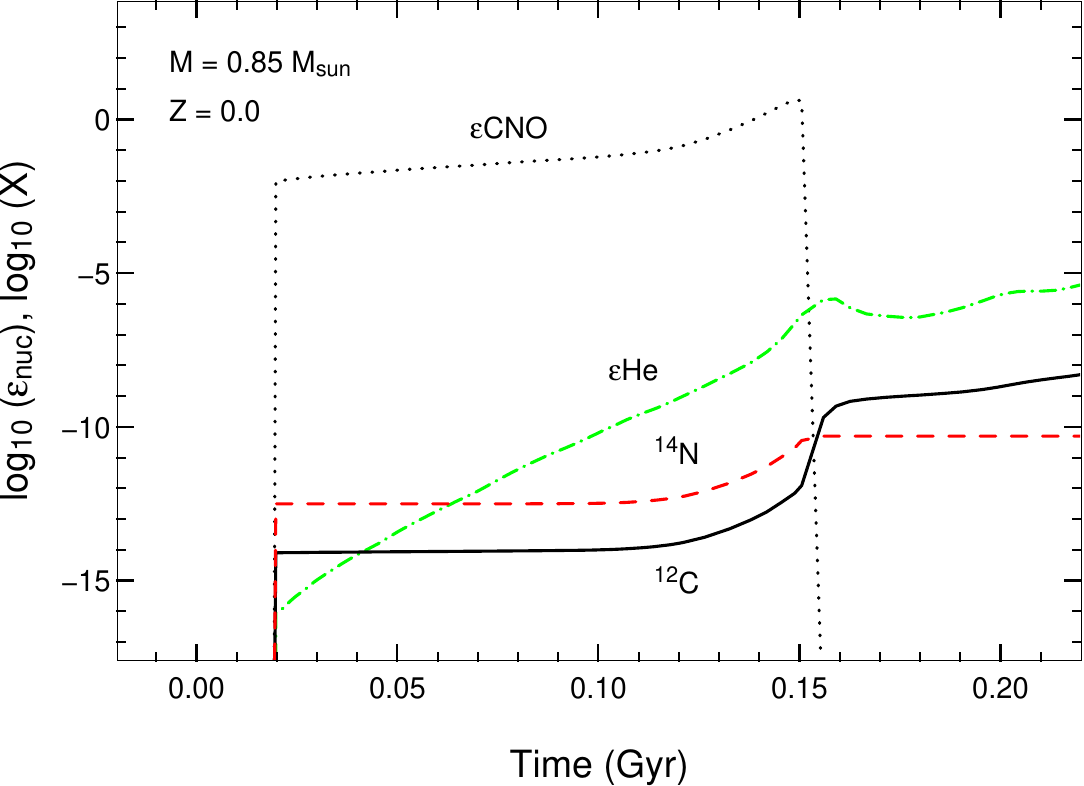}
\par\end{centering}
\caption{An example of CNO cycle ignition in material initially devoid of any
metals. In this case it is in the centre of our $Z=0$ model, during
the RGB. Time is arbitrarily offset for clarity (zero corresponds
to roughly 10.9 Gyr in Figure \ref{fig-m0.85z0y245-Lums-MS-RGB}).
It can be seen that a minute amount of He burning gives rise to a
tiny amount of $^{12}$C. This can then react with the protons (of
which not much are left), instigating the CNO cycles. Of note is just
how little $^{12}$C is required, with the mass fraction being of
the order $10^{-14}$ to $10^{-12}$ (note that the $^{12}$C is burnt
to $^{14}$N very quickly). Hydrogen exhaustion occurs at $t\sim0.16$
Gyr, at which time the CNO cycles cease and helium burning dominates.
$^{14}$N is static after this, but $^{12}$C continues to build up
due to the He fusion. \label{fig-m0.85z0y245-CNOignition-centralConditions}}
\end{figure}

\begin{figure}
\begin{centering}
\includegraphics[width=0.85\columnwidth,keepaspectratio]{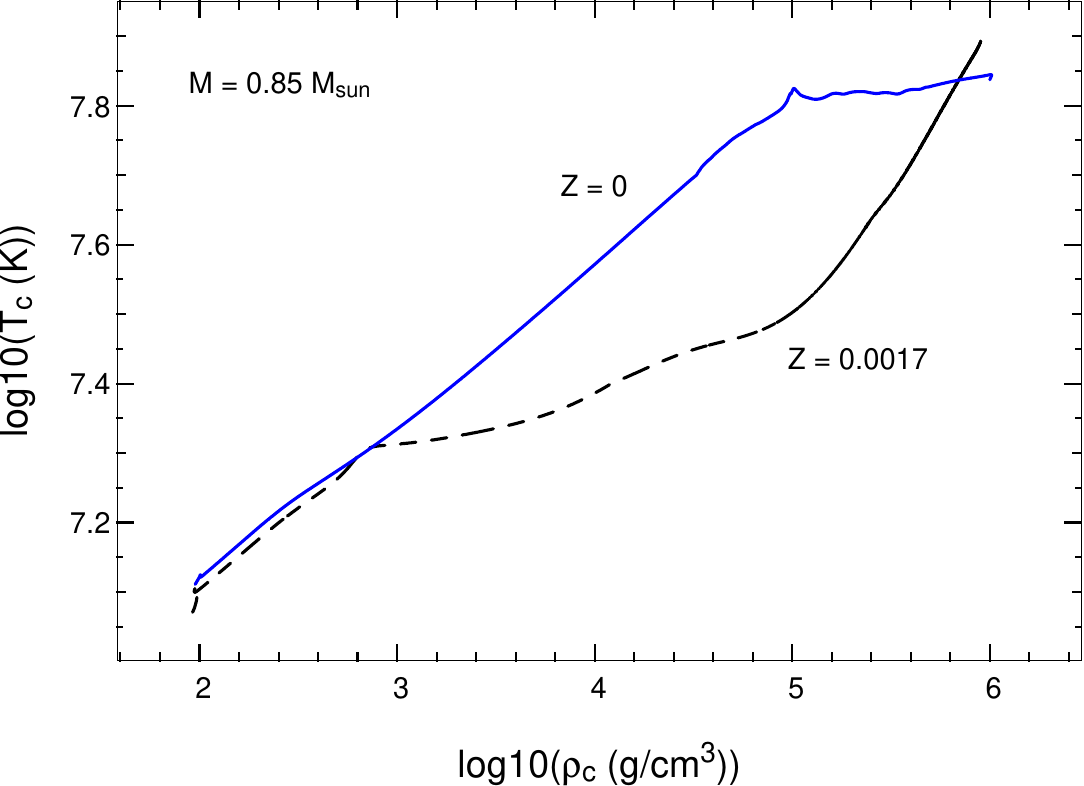}
\par\end{centering}
\caption{Central density versus central temperature for the $Z=0$ and $Z=0.0017$
stars, from the MS to the start of the core He flash. \label{fig-m0.85z0y245-TcRhoc}}
\end{figure}

The observation that CNO burning does not become important until late
on the RGB in the $Z=0$ model is explicable by the fact that these
reactions can only begin when some $^{12}$C is produced by the $3\alpha$
reactions -- which only happens as the temperature approaches $\sim60$
MK. In the current model this condition is reached only towards the
end of the RGB, as the density and temperature increase steadily till
then (see Fig. \ref{fig-m0.85z0y245-TcRhoc}). Indeed, the $Z=0$
RGB structure is never dominated by the CNO cycles, unlike the $Z=0.0017$
structure. In Figure \ref{fig-m0.85z0y245-CNOignition-centralConditions}
we show an example of CNO cycle ignition due to the local production
of $^{12}$C by $3\alpha$ reactions (in the core of the $Z=0$ model
in this case) . It can be seen that the amount of $^{12}$C needed
to initiate the CNO burning is very small, being of the order $X_{C}\sim10^{-12}$
or less. Note that the $^{12}$C is quickly burnt to $^{14}$N which
becomes the dominant metal species (albeit a very small amount!) until
H exhaustion. 

Returning to Figure \ref{fig-m0.85z0y245-Lums-MS-RGB} we also note
the difference in MS lifetimes between the two stars. The GC model
has a lifetime of $\sim11.5$ Gyr whilst the $Z=0$ model has a lifetime
of $\sim10$ Gyr. This has consequences in terms of the predicted
observability of low mass primordial stars as 10 Gyr is significantly
shorter than the age of the Universe ($\sim13$ Gyr, see Section \#
for a discussion). 

In Figure \ref{fig-m0.85z0y245-TcRhoc} we present the run of two
key central conditions -- density ($\rho_{c}$) and temperature.
The initial similarity of the models on the MS is also evident in
this graph. However a strong deviation from `normal' behaviour is
also apparent. The ignition of the CNO cycle in the $Z=0.0017$ model
causes a moderation of temperature in the core (and the burning shell).
This does not happen in the $Z=0$ star until the temperature is high
enough to ignite the $3\alpha$ reactions.

\begin{figure}
\begin{centering}
\includegraphics[width=0.85\columnwidth,keepaspectratio]{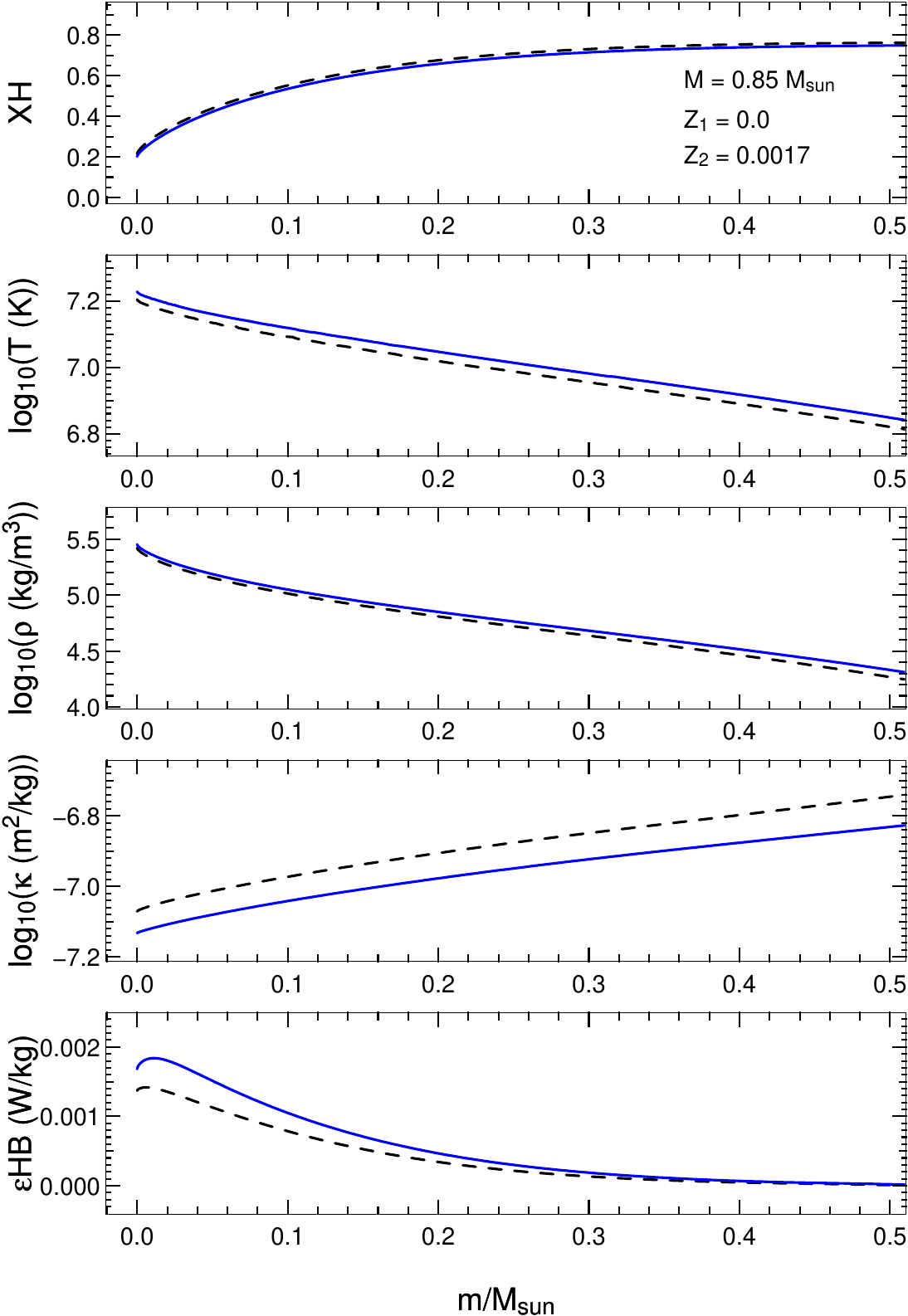}
\par\end{centering}
\caption{The run of physical parameters in both stars during the MS, from the
centre to $m_{r}=0.5$ M$_{\odot}$. The dashed line is the $Z=0.0017$
model whilst the solid line is the $Z=0$ model. From the top to bottom
we have: i) the mass fraction of hydrogen ($X_{H}$), ii) temperature,
iii) density, iv) opacity ($\kappa$) and v) the energy released by
H burning ($\epsilon_{HB}$, all sources). We have chosen models that
have a similar chemical profile ($X_{H}$) for comparison, where the
central $X_{H}$ is $\sim0.2$. \label{fig-m0.85z0y245-pltstar-MS-compare}}
\end{figure}

In Figure \ref{fig-m0.85z0y245-pltstar-MS-compare} we compare the
run of some physical parameters versus mass between the two stars
during the MS phase. We have chosen models in which both have $X_{H}=0.2$
in the centre for a direct comparison. It can be seen that the chemical
profiles are virtually identical, as are the density profiles. The
main points of difference are in the temperature and energy generation
rate -- both are higher in the $Z=0$ model. Also plotted is the
run of opacity, which is consistently lower in the $Z=0$ model, as
expected.

\begin{figure}
\begin{centering}
\includegraphics[width=0.85\columnwidth,keepaspectratio]{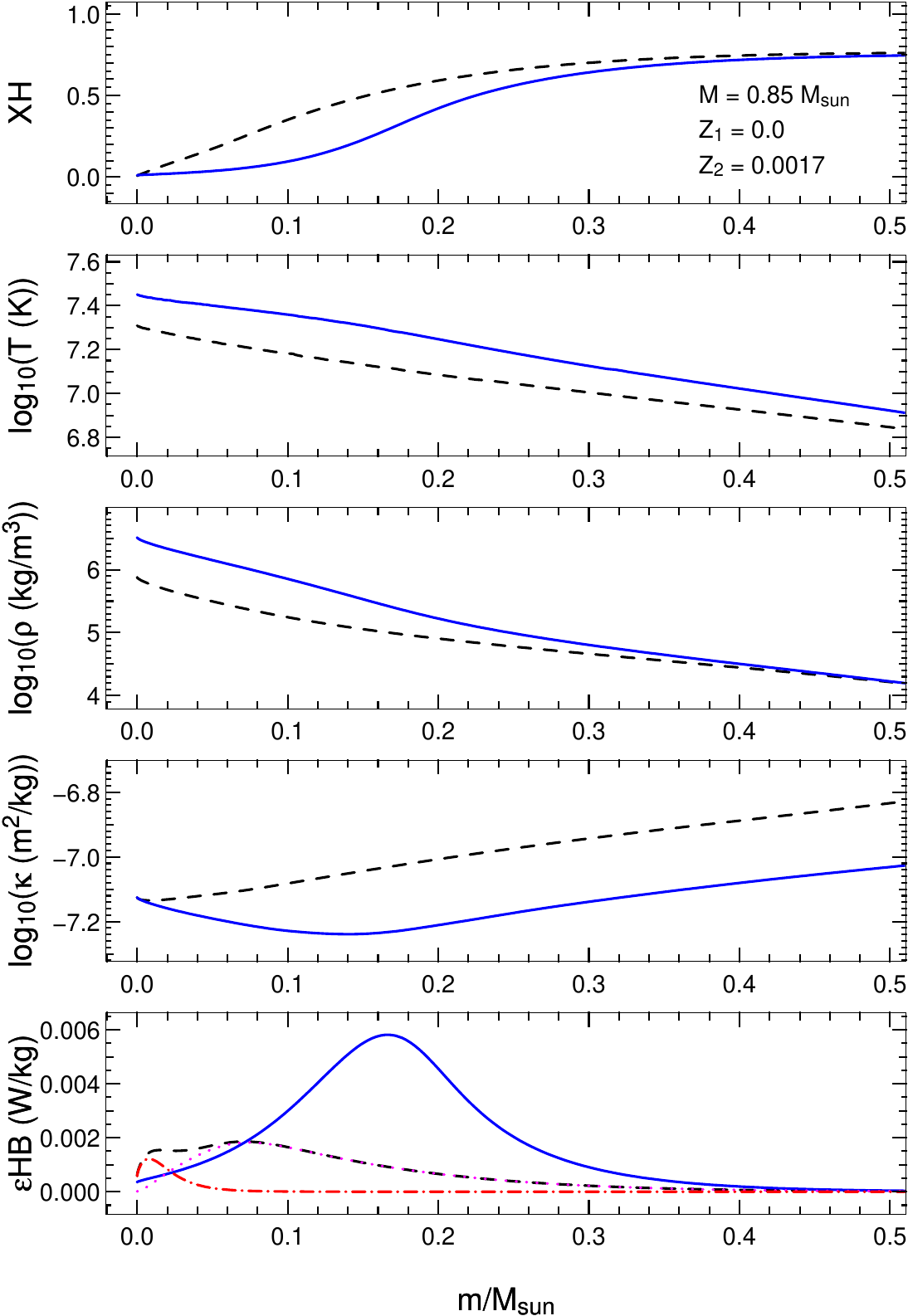}
\par\end{centering}
\caption{Same as Figure \ref{fig-m0.85z0y245-pltstar-MS-compare} but towards
the end of the MS (central $X_{H}\sim0.01$). Significant differences
in all physical quantities are apparent. Of particular interest is
the very different H burning regimes evident in the bottom panel.
Here we have over-plotted the contributions from the CNO cycles (dash-dotted
line, red) and the PP chains (dotted line, magenta) for the $Z=0.0017$
star (dashed line). The energy in the $Z=0$ model (solid line, blue)
is $100\%$ from the PP chains. Also of note is the broad spread of
energy generation in the $Z=0$ model, a consequence of the low temperature
dependence of these reactions. \label{fig-m0.85z0y245-pltstar-endMS-compareGC} }
\end{figure}

In Figure \ref{fig-m0.85z0y245-pltstar-endMS-compareGC} we make the
same comparison during the stage in which the models really begin
to diverge --- towards the end of the MS, when the $Z=0.0017$ model
switches to CNO burning. We compare models which both have central
H abundances of $X_{H}\sim0.01$. The chemical profile, temperature,
density and opacity are all now significantly different between the
models. The reason for this is clearly seen in the bottom panel, which
shows that the $Z=0.0017$ model has begun burning via the CNO cycles,
whilst the $Z=0$ star is supported only by the PP chains burning
in a very thick shell.

\begin{figure}
\begin{centering}
\includegraphics[width=0.85\columnwidth,keepaspectratio]{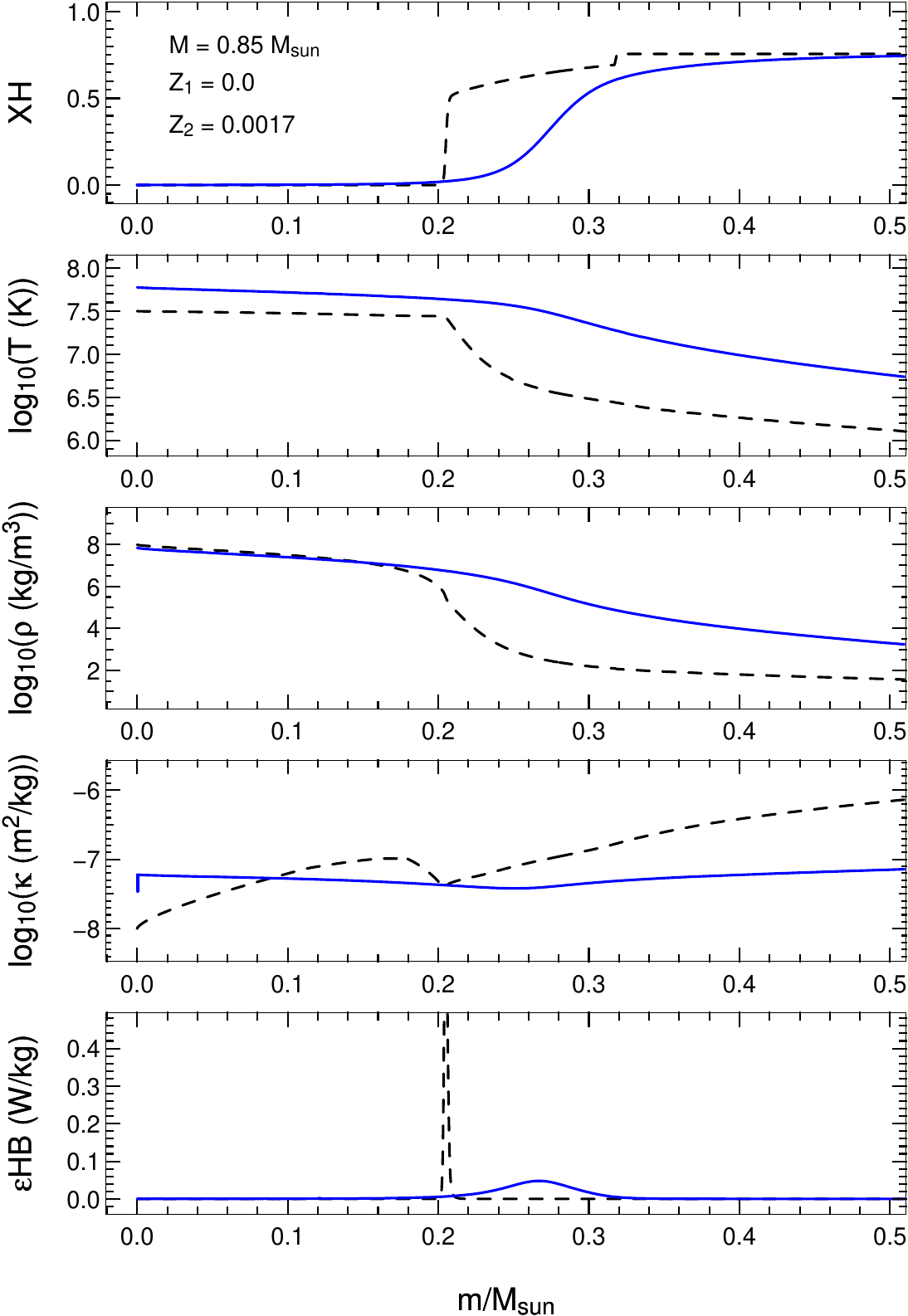}
\par\end{centering}
\caption{The same as Figure \ref{fig-m0.85z0y245-pltstar-MS-compare} except
for the stars when they are on the RGB, having a shell burning configuration.
Here the structures are quite different. The strong effect of having
CNO cycles dominating the H burn is evident in the bottom panel. The
very localised energy generation in the $Z=0.0017$ model (dashed
line) is a consequence of the strong temperature dependence of the
CNO cycles. The energy generation in the shell is almost entirely
from the CNO cycles. The $Z=0$ star on the other hand, which has
a PP-shell, has much smoother profiles throughout. \label{fig-m0.85z0y245-RGB-compareGC}}
\end{figure}

In Figure \ref{fig-m0.85z0y245-RGB-compareGC} we move further into
the evolution. Again we plot the run of physical quantities versus
mass, this time for models with a H-shell burning configuration (on
the RGB). The structural differences are more striking here. The $Z=0.0017$
model is now supported by a CNO burning shell. Due to the high temperature
dependence of these reactions, the burning is very localised, creating
a sharp composition change as it burns outwards in mass. In contrast,
the H burning shell in the $Z=0$ model is powered solely by the PP
chains, with the lower temperature dependence giving rise to a thick
burning shell that has a much lower peak luminosity (see bottom panel
of Figure \ref{fig-m0.85z0y245-RGB-compareGC}). This also gives rise
to the other smooth profiles such as the abundance and opacity profiles.

\begin{figure}
\begin{centering}
\includegraphics[width=0.85\columnwidth]{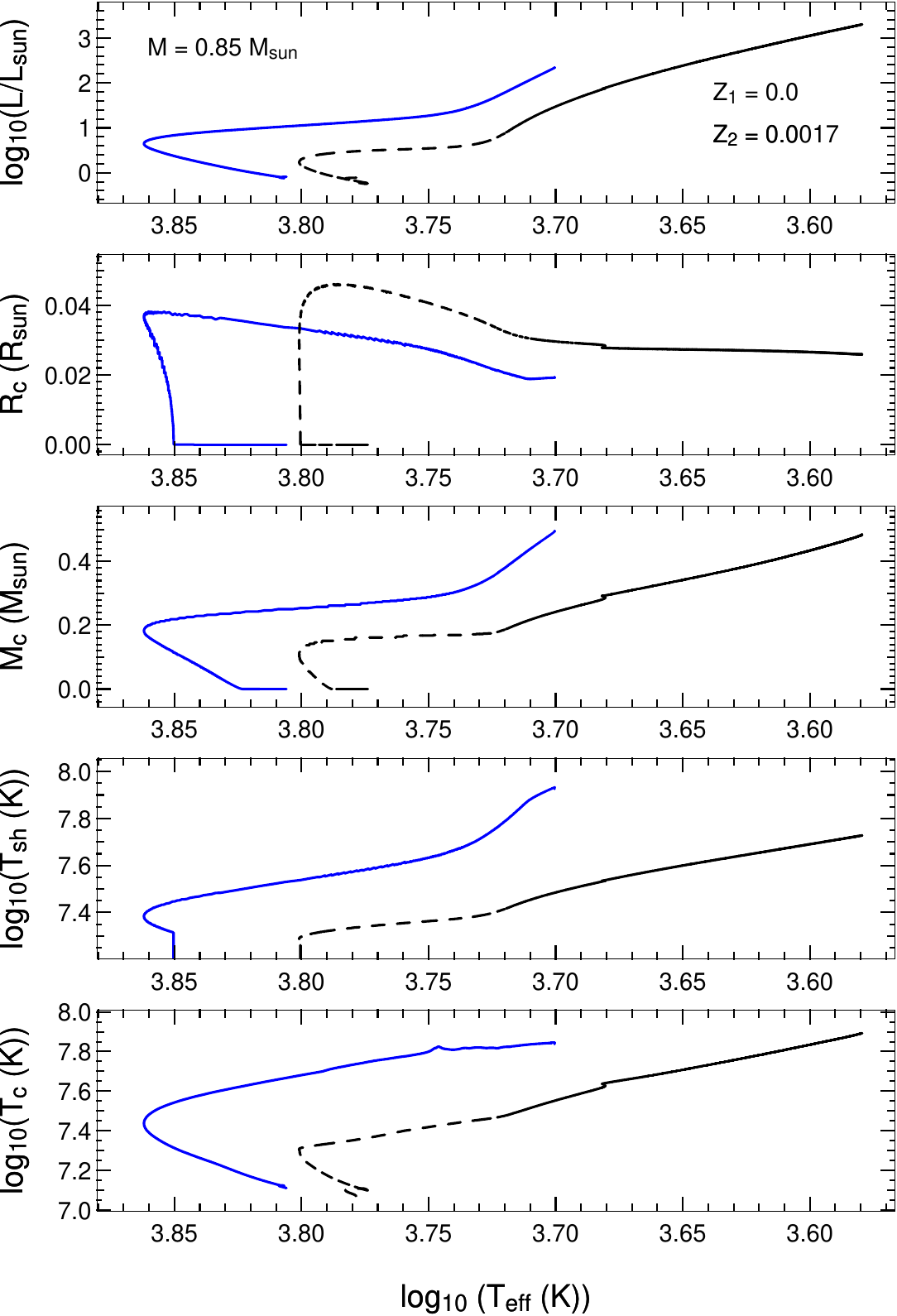}
\par\end{centering}
\caption{Comparison between the two stars, this time against surface temperature.
The top panel is the usual HR diagram, the second shows the evolution
of the H-exhausted core radius ($R_{c}$) and the third the H-exhausted
core mass. The gradual core collapse at the end of the MS onwards
can be seen, as can the ongoing growth of the core in terms of mass.
The bottom two panels show a) the temperature at the bottom of the
H-burning shell and b) the central temperature. Here it can be seen
that the shell temperature in the $Z=0$ star is very high, approaching
$10^{8}$ K. The core temperature of the $Z=0$ star is quite constant
during the RGB, and is very high. We note that both stars end up with
approximately the same core mass and core temperature at the tip of
the RGB.\label{fig-m0.85z0y245-coreRshellT-HRDPlus}}
\end{figure}

In Figure \ref{fig-m0.85z0y245-coreRshellT-HRDPlus} we show the evolution
of the core mass and radius (defined by $X_{H}<0.05$) of both stars,
as well as the evolution of the temperature of the H shell. The properties
are all plotted against surface temperature. In this way we can see
how the star evolves in relation to the familiar HR diagram, which
is shown in the top panel. The gradual increase of the core mass due
to the H-shell burning outwards in mass can be seen, as can the concurrent
collapse of the core in terms of radius. This collapse takes the H
shell into hotter and hotter regions, leading to increased energy
generation/luminosity. In the case of the $Z=0$ model the shell reaches
temperatures hot enough to ignite helium, as discussed earlier. The
core temperature of the $Z=0$ star is quite constant during the RGB,
and is also very high. We note that both stars end up with approximately
the same core mass and core temperature at the tip of the RGB, indicating
that these are key factors in determining the onset of the core He
flash that ensues.

\begin{figure}
\begin{centering}
\includegraphics[width=0.75\columnwidth]{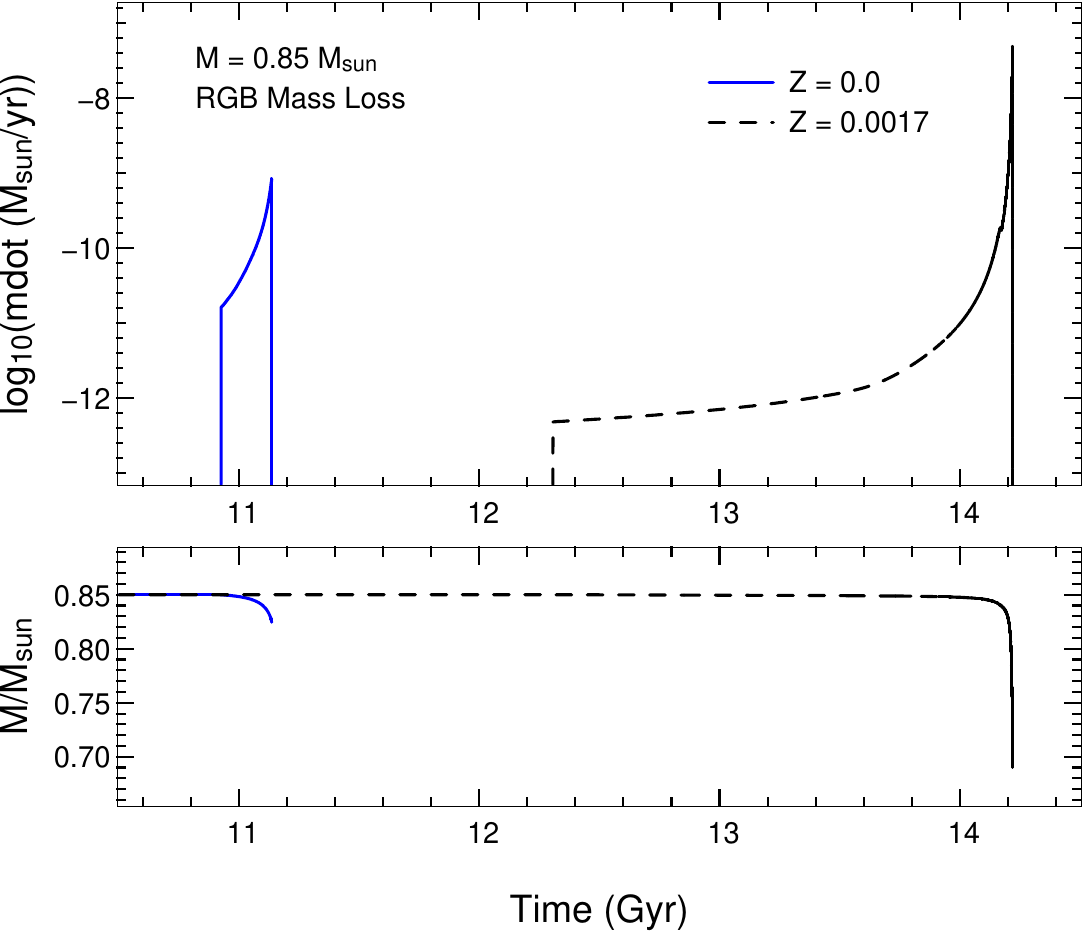}
\par\end{centering}
\caption{Time evolution of the mass loss rate (top) and total mass (bottom)
for both the $Z=0$ and $Z=0.0017$ models during the RGB. The difference
in time spent on the RGB is exaggerated in the top panel due to the
late onset of the Reimers mass loss in the $Z=0$ model. The $Z=0.0017$
model actually spends $\sim0.5$ Gyr on the RGB (mass loss is initiated
at the MS turn-off in this model) whilst the $Z=0$ model spends only
$\sim0.2$ Gyr on the RGB. This is due to the rapid evolution of the
core to the conditions in which He ignites. In the bottom panel we
see the direct effect of the shortened RGB -- the total mass of the
$Z=0$ star is considerably higher at the end of the RGB, due to a
shorter amount of time spent losing mass. We note that the late onset
of the Reimers mass loss in the $Z=0$ model has negligible effect
on the mass of the star at the end of the RGB, as it is such a low
rate initially (see bottom panel). \label{fig-m0.85z0y245-RGB-MassLoss} }
\end{figure}

Despite the similar \emph{core} masses at the end of the RGB we note
that the differing evolution of the $Z=0$ model leads to a higher
\emph{total} mass at the end of the RGB. Thus the envelope is more
massive in the $Z=0$ model at the end of the RGB. Figure \ref{fig-m0.85z0y245-RGB-MassLoss}
shows the time evolution of the mass loss rate and the total mass.
It is clear just how short the $Z=0$ RGB is in this diagram (more
so in the bottom panel). It is however exaggerated in the top panel
due to the late onset of the Reimers mass loss in this model. We note
that this late onset is a consequence of the algorithm that decides
when to switch on mass loss. In a `normal' star H is exhausted at
the end of the MS and this is when the SEV code initiates RGB mass
loss. In the $Z=0$ case however the H is \emph{not} exhausted at
the end of the MS (a little remains due to the inefficiency of the
pp-chains). The mass loss thus starts at the base of the RGB when
the last of the H is burnt. This delay actually has negligible effect
on the mass of the star at the end of the RGB, as the mass loss rate
is so low initially (see bottom panel). The difference in RGB lifetimes
is quite large, being a factor of $\sim2.7$ shorter in the $Z=0$
model. The $Z=0$ RGB lasts only 0.20 Gyr as compared to 0.54 Gyr
in the $Z=0.0017$ case. The reason for this is that the evolution
of the $Z=0$ model is cut short by the onset of the core He flash
thus it never loses much mass via winds. Indeed, the total mass at
the end of the RGB for the $Z=0$ star is $0.825$ M$_{\odot}$ whilst
that of the GC model is only $0.690$ M$_{\odot}$. 

\begin{figure}
\begin{centering}
\includegraphics[width=0.93\columnwidth]{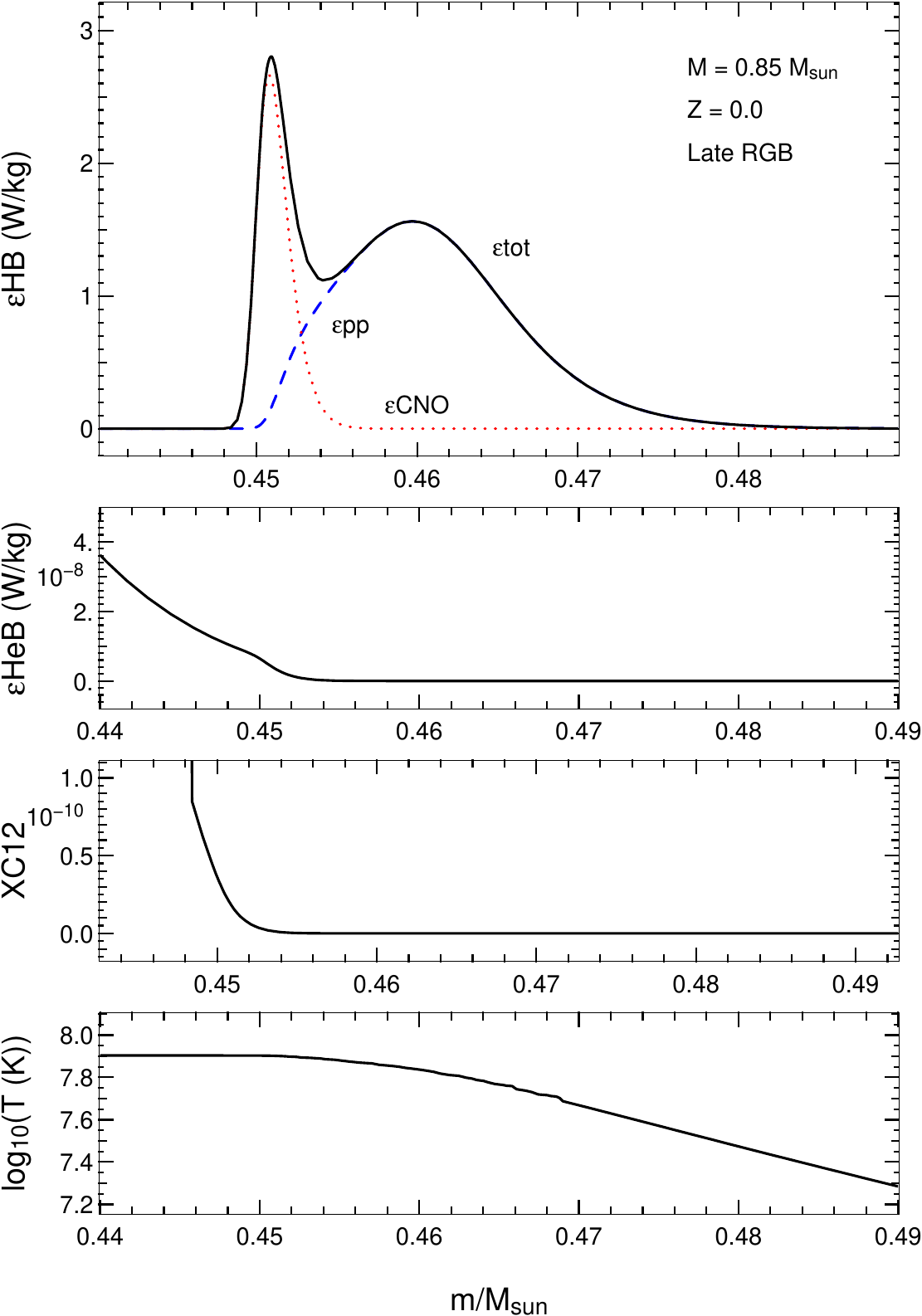}
\par\end{centering}
\caption{The hydrogen burning shell in the $Z=0$ model late on the RGB. The
two top panels show nuclear energy generation versus mass. The two
bottom panels show the run of $^{12}$C abundance and temperature,
also versus mass. It can be seen in the top panel that CNO burning
has started in the shell, but also that PP burning is still strong.
The reason CNO burning is occurring is because there has been some
He burning in the core/bottom of the H-shell (second panel), producing
some $^{12}$C. It is only a minute amount of He burning but produces
enough catalyst for the CNO cycles. The distribution of $^{12}$C
(panel 3) clearly indicates where the CNO cycles can operate. \label{fig-m0.85z0y245-LateRGBShell-lums-C12}}
\end{figure}

We now delve into the anatomy of the $Z=0$ model's H-shell towards
the end of the RGB, when some local $^{12}$C production has allowed
the CNO cycles to operate. Figure \ref{fig-m0.85z0y245-LateRGBShell-lums-C12}
shows the various nuclear energy generation profiles of the shell
against mass. It can be seen that the CNO cycle is indeed operating,
but only at the bottom of the H-shell. The top of the shell is still
burning via the PP chains, and is the dominant source of energy in
this model. To illustrate the cause of the CNO ignition we also plot
the He burning energy release, as there is a tiny amount of $3\alpha$
reactions occurring. The resultant $^{12}$C profile is also plotted
and it can be seen that the He burning at the top of the core/bottom
of the shell is producing the minute amount of $^{12}$C that is needed
for the CNO cycles. 

In terms of chemical pollution of the convective envelope on the RGB
the first dredge-up (FDUP) event that occurs in higher metallicity
stars is virtually non-existent at $Z=0$. Although the convective
envelope does make an incursion into the interior regions, the maximum
depth reached (in mass) is much less than higher metallicity stars.
For example, in our $Z=0.0017$ model the envelope reaches down to
a maximum depth of $m=0.63$ M$_{\odot}$ whilst in the $Z=0$ model
it only reaches to $m=0.30$ M$_{\odot}$. Although the H burning
shell/region is much larger in the $Z=0$ model, due to the low temperature
sensitivity of the p-p chains, the convection only just touches the
region of partial burning. In addition to this, as there is no CN
cycling (except right at the bottom of the H shell), the only product
that may be dredged up is He. The amount of He that is dredged up
is totally insignificant, such that the star continues to the tip
of the RGB with a pristine surface.

\begin{figure}
\begin{centering}
\includegraphics[width=0.85\columnwidth,keepaspectratio]{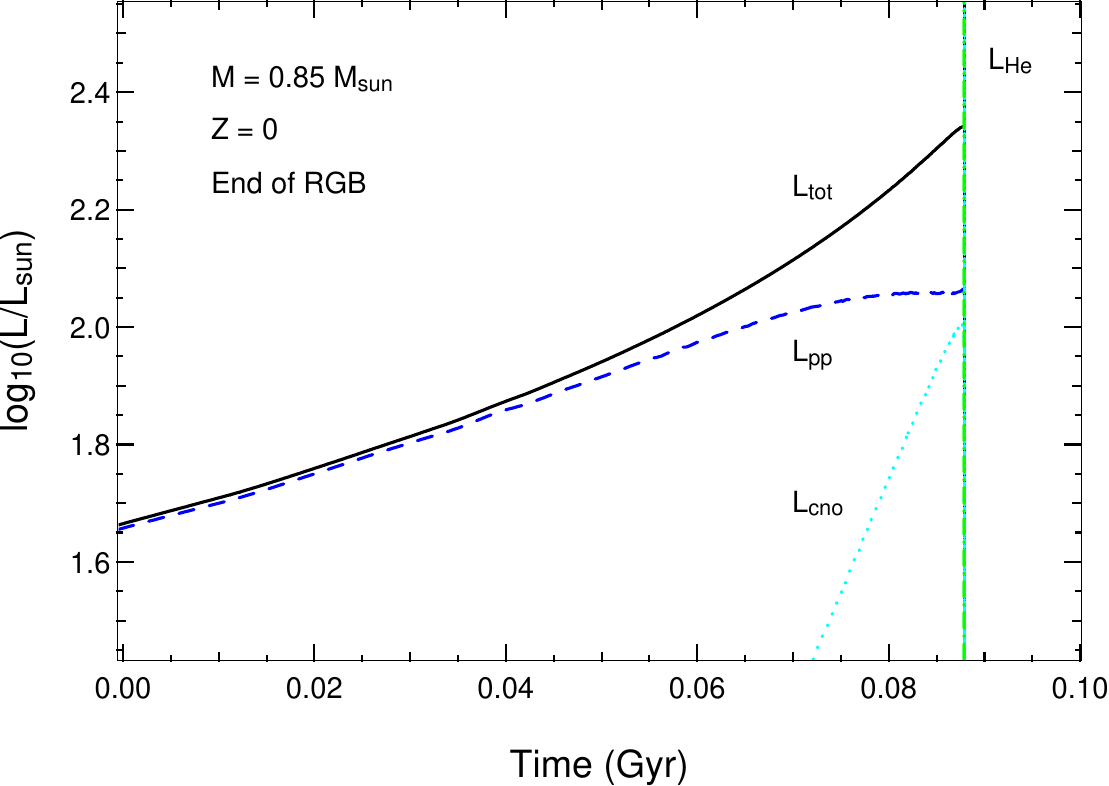}
\par\end{centering}
\caption{The final lead-up to the core He flash in the $Z=0$ model. Plotted
are the luminosities from the PP chains (dashed line), CNO cycles
(dotted line), He burning (dash-dotted, green) and also the total
luminosity of the star. Time is offset for clarity (zero corresponds
to roughly 11 Gyr in Figure \ref{fig-m0.85z0y245-Lums-MS-RGB}). It
can be seen that CNO burning never dominates the energy generation
in this star, as opposed to the $Z=0.0017$ model. The onset of the
core He flash is also evident (vertical line). \label{fig-m0.85z0y245-EndRGB-lums-time}}
\end{figure}

Finally we show in Figure \ref{fig-m0.85z0y245-EndRGB-lums-time}
a close-up of the penultimate evolution of the hydrogen burning luminosities.
It is only in the very last period before the core flash that the
total CNO luminosity finally becomes comparable to that of the PP
chains. The onset of the core He flash can also be seen. This interesting
phase of evolution is the subject of the next subsection.

\subsection{Dual Core Flash\label{section-m0.85z0y245-DualCoreFlash}}

It is well known that low mass stars ($M\lesssim2\,M_{\odot}$, dependent
on metallicity) ignite helium violently in their cores at the end
of the red giant branch (see eg.  \citet{1994sipp.book.....H} or
\citet{1992itsa.book.....B}  for an overview). The reason that the
ignition is `violent' is because a thermal runaway occurs due to the
matter being (electron-) degenerate. In degenerate matter the equation
of state is quite different to that of more normal plasma -- the
temperature and pressure are decoupled. This means that once the temperature
is high enough to ignite helium the gas will \emph{not} expand and
cool to compensate. This gives rise to a positive feedback loop in
which the energy release by the nuclear reactions increase the temperature
which in turn increases the rate of reactions, leading to the thermonuclear
runaway. Eventually the energy released will be so large so as to
remove the degeneracy (the velocity distribution of the electrons
becomes Maxwellian again), and the change in equation of state then
allows an expansion and cooling which ends the runaway. The reason
this happens in low mass stars only is that these stars do not reach
high enough temperatures in their cores to ignite helium before the
core becomes degenerate. The core becomes degenerate due to the density
increase during its gradual collapse, a collapse which is aided in
part by energy losses from neutrino production in the core. In this
subsection we look in detail at the core helium flash of our $0.85\,M_{\odot}$
model and find that a second flash -- a hydrogen flash -- is induced
by the helium flash, giving two large releases of energy within a
very short time. Hence the title `Dual Core Flash'.

\begin{figure}
\begin{centering}
\includegraphics[width=0.85\columnwidth,keepaspectratio]{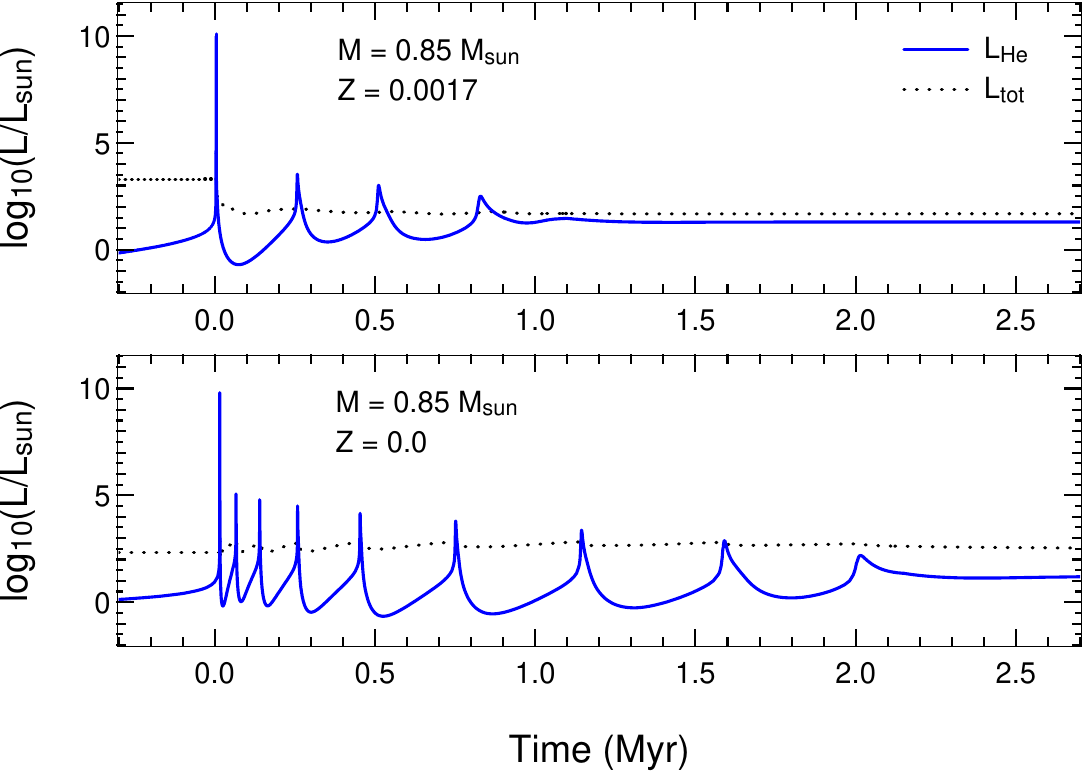}
\par\end{centering}
\caption{Time evolution of the helium and total luminosities in the $Z=0$
and $Z=0.0017$ models. Time is offset so zero coincides with the
main He flash and the x-y limits are identical. It can be seen that
the peak He luminosities of both flashes are approximately the same
($L_{He}\sim10^{10}$$L_{\odot}$). The subsequent flashes are however
stronger in the $Z=0$ model ($L_{He}\sim10^{5}$ as compared to $\sim10^{3.5}$
for the second flashes). The $Z=0$ flashes are also more numerous
and initially closer together in time. Also of note is the lower total
luminosity of the $Z=0$ model at the onset of the He flashes. \label{fig-m0.85z0y245-CoreHeFlash-HeLums-compareGC}}
\end{figure}

We begin by looking at the variation in helium burning luminosity
with time. Figure \ref{fig-m0.85z0y245-CoreHeFlash-HeLums-compareGC}
compares the luminosities of the $Z=0$ model with $Z=0.0017$ GC
model. Of note here is that the flash is not a single event, there
are a number of `mini-flashes' after the main flash. The main flash
is however of a much greater strength than the subsequent miniflashes.
It can be seen that the peak luminosity of the main He flashes are
approximately the same in both stars (marginally lower in the $Z=0$
model). Also apparent is the nature of the minipulses -- the $Z=0$
model flashes are closer together in time (initially) and many more
occur (8 versus 3 in the $Z=0.0017$ model). The miniflashes are also
initially stronger in the $Z=0$ model. This probably has little consequence
on the further evolution however. As mentioned in the previous subsection,
the total luminosity (also displayed in Figure \ref{fig-m0.85z0y245-CoreHeFlash-HeLums-compareGC})
at the time of He ignition (ie. the tip of the RGB) is $\sim1$ dex
lower in the $Z=0$ model. This has consequences for the expected
incidence of bright, low metallicity giants.

\begin{figure}
\begin{centering}
\includegraphics[width=0.85\columnwidth]{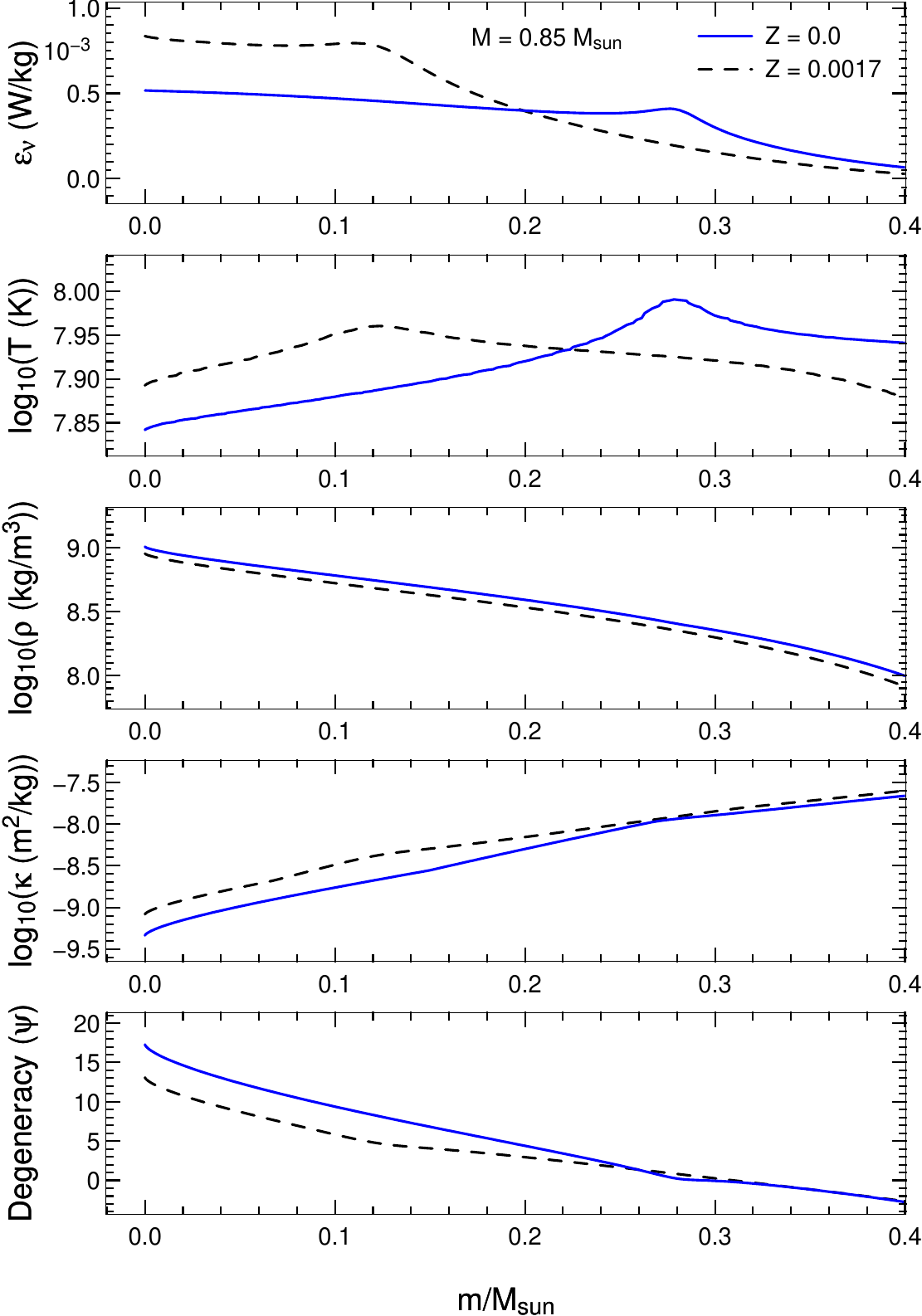}
\par\end{centering}
\caption{A selection of physical quantities against mass for the $Z=0$ and
$Z=0.0017$ models. Both models are taken just before the onset of
their core helium flashes (at a point where $L_{He}\sim10^{1.5}\,L_{\odot}$).
It can be seen that the location of the maximum temperatures for both
models are significantly off-centre. The reason for this is seen in
the top panel which shows the run of neutrino energy loss. The energy
loss cools the cores in the hot, dense regions. As the density drops
off so does the cooling efficiency, leading to the higher temperature
on the edge of the cooled region. This determines the location of
He ignition for the flash, which will be substantially further out
(in mass) in the $Z=0$ star. Also of note is that the opacity in
the $Z=0$ star is lower than the $Z=0.0017$ star but the degeneracy
is higher. \label{fig-M0.85-CHeF-Bef-plstar-compareGC}}
\end{figure}

We now look at the internal characteristics of the $Z=0$ and $Z=0.0017$
stars just before the onset of the core He flash. Figure \ref{fig-M0.85-CHeF-Bef-plstar-compareGC}
shows a variety of pertinent physical characteristics -- temperature,
density, opacity, degree of degeneracy and energy loss from neutrino
production. It can be seen that there is significant energy loss/cooling
of the inner core from neutrinos. Importantly, this leads to the temperature
profile (second panel) having a maximum that is offset from the centre.
The energy loss cools the cores in the hot, dense regions. As the
density drops off so does the cooling efficiency, leading to the higher
temperature on the edge of the cooled region. Comparing the GC star
with the $Z=0$ star we see that the $Z=0$ star has a maximum temperature
that is much more offset (located at $m\sim0.12$ and $m\sim0.28\,M_{\odot}$
respectively). Looking at the maximum temperature itself (which is
approaching $10^{8}$ K) it can be seen that this will be the location
of the He ignition for the flash. The fact that the ignition happens
so far out in the star will have drastic consequences for the $Z=0$
model. Also of note is that the opacity in the $Z=0$ star is lower
than the $Z=0.0017$ star but the degeneracy is higher.

\begin{figure}
\begin{centering}
\includegraphics[width=0.85\columnwidth,keepaspectratio]{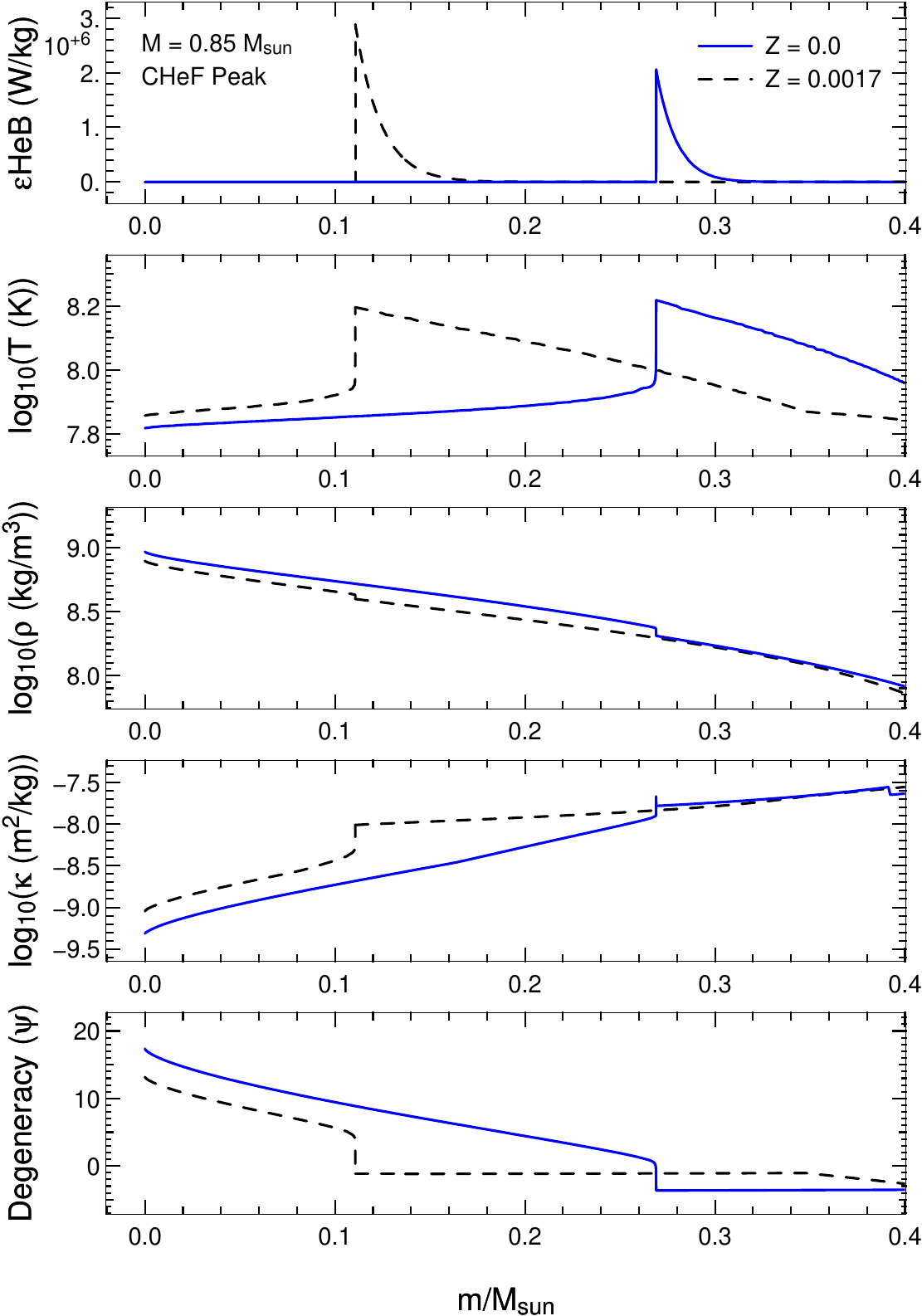}
\par\end{centering}
\caption{Same as Figure \ref{fig-M0.85-CHeF-Bef-plstar-compareGC} except the
models correspond to the peak of the He flash and the top panel displays
the He energy generation rather than the neutrino energy loss. It
can be seen that the ignition of helium has occurred at a mass location
of $0.11\,M_{\odot}$ in the $Z=0.0017$ model whereas it has occurred
at $0.27\,M_{\odot}$in the $Z=0$ model. The onset of the flash has
removed the degeneracy in the regions above the flash location. Also
of note is the fact that the ignition has occurred in the $Z=0$ model
in a region that is less dense and less degenerate than the $Z=0.0017$
model. The location of this model in time is marked `a' in Figure
\ref{fig-m0.85z0y245-HHeFlash-zoom-conv-lums}. \label{fig-m0.85z0y245-CHeF-Peak-pltstar-compareGC}}
\end{figure}

We now move to the core helium flash itself, showing the energy generation
from He burning, and the temperature, density, degeneracy and opacity
profiles for both stars in Figure \ref{fig-m0.85z0y245-CHeF-Peak-pltstar-compareGC}.
The enormous energy generation from the $3\alpha$ reactions is plainly
visible in the first panel, being $\sim10^{6}$ W/kg ($\sim10^{10}$
ergs/sec). Also of note are the very high temperatures at which the
burning is happening (panel b) -- much higher than is normal for
quiescent He burning. The effect of the temperature increase on the
degree of degeneracy can be seen in the bottom panel -- the degeneracy
has been lifted in all areas above the flash location. The material
below is however still degenerate, a fact that leads to the miniflashes
seen in Figure \ref{fig-m0.85z0y245-CoreHeFlash-HeLums-compareGC}.
Comparing the $Z=0$ model with the GC model it can be seen that the
flash ignition has occurred in a region which is less degenerate in
the $Z=0$ model (and less dense). This gives rise to the slightly
weaker flash in the $Z=0$ model also seen in Figure \ref{fig-m0.85z0y245-CoreHeFlash-HeLums-compareGC}.

\begin{figure}
\begin{centering}
\includegraphics[width=0.7\columnwidth,keepaspectratio]{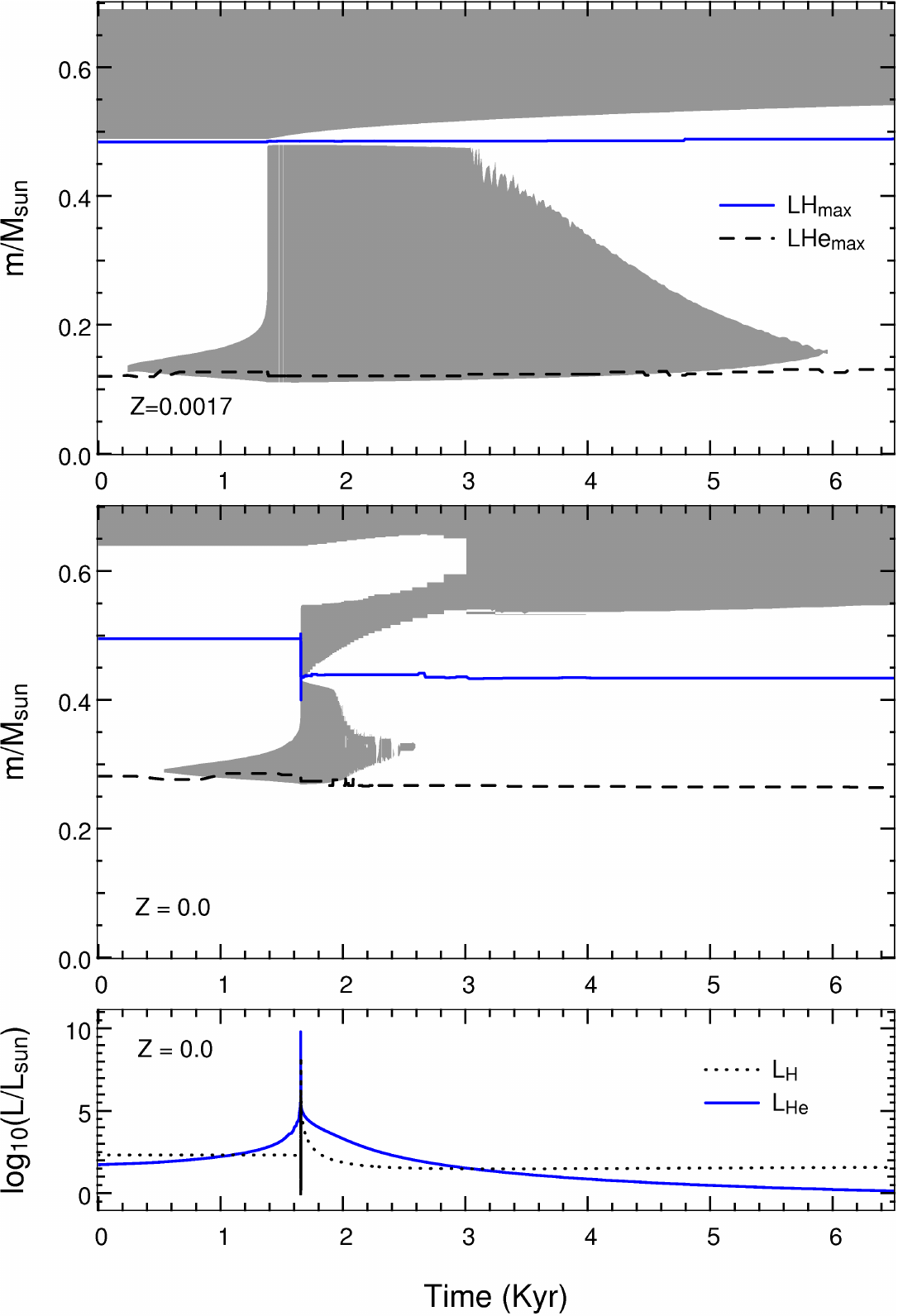}
\par\end{centering}
\caption{The time evolution of the convective zones (grey shading) in the $Z=0.0017$
model (top panel) and $Z=0$ model (second panel). The bottom panel
shows the evolution of the H and He burning luminosities in the $Z=0$
model. Note the short span of time ($\sim7000$ years). The axes for
the top two panels are identical, so a direct comparison can be made.
The duration of the He convective zone is much shorter in the $Z=0$
model ($\sim2000$ yr, as compared to $\sim6000$ yr in the $Z=0.0017$
model). This is due to the off-centre ignition which induces the PIE
(and consequent H-flash). The H-flash leads to a partial quenching
of the He flash (see also Fig. \ref{fig-m0.85z0y245-HHeFlash-zoom-conv-lums}).
In addition to this there is much less He to burn in the $Z=0$ HeCZ
as it is much smaller in mass ($\sim$$0.15$ $M_{\odot}$ compared
to $\sim0.4$ $M_{\odot}$). The location of the dual core flash is
clearly visible in the second panel, and in luminosity in the bottom
panel, at $t=1.6$ kyr. Finally we note the suddenness of the H flash
evident in the bottom panel -- on this timescale the H luminosity
appears to increase instantly at the start of the flash. This is in
contrast to the He flash which exhibits a gradual increase in strength.
\label{fig-m0.85z0y245-HeFlashConvection-lums-medWide}}
\end{figure}

\begin{figure}
\begin{centering}
\includegraphics[width=0.9\columnwidth]{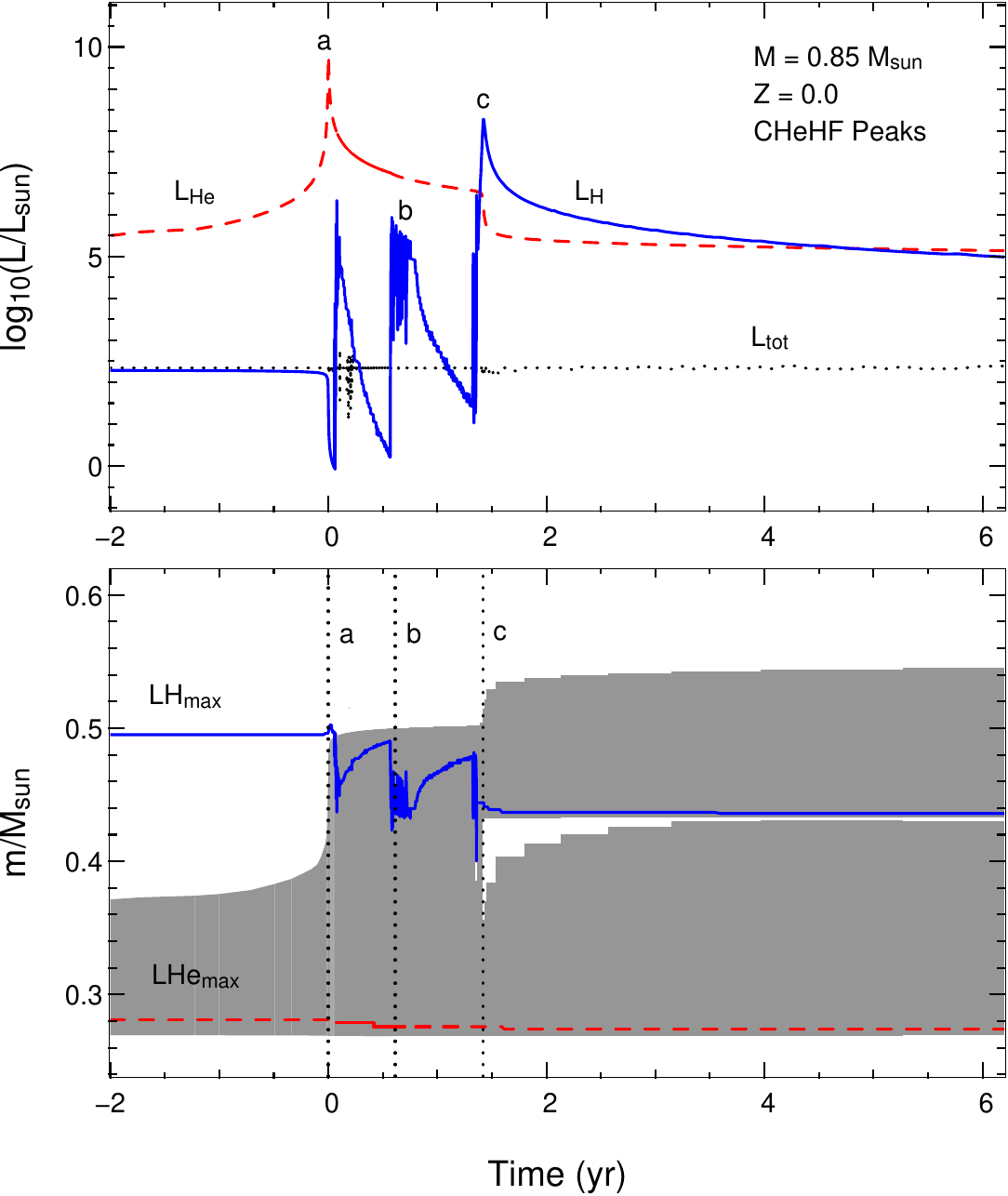}
\par\end{centering}
\caption{Zooming in on the dual core flash in the $Z=0$ model. Note the extremely
short timescale on the x-axis. In the top panel we show the evolution
of the H, He and total luminosities. In the bottom panel we show the
convective zones in shaded grey (except the convective envelope which
lies above this mass range) and the locations of maximum H and He
burning luminosities. A series of three proton ingestion episodes
(PIEs) can be seen, which give rise to the three peaks in H luminosity.
It is not until a large expansion of the He-H convection zone occurs,
bringing down a substantial amount of H, that the convection zone
splits into two -- a HCZ and a HeCZ. The H luminosity exceeds the
He luminosity at this stage. These two convection zones remain separate
for the rest of the evolution. The three vertical dotted lines marked
\emph{a}, \emph{b} and \emph{c} represent the time points at which
the three models in Figures \ref{fig-m0.85z0y245-CHeF-Peak-pltstar-compareGC},
\ref{fig-m0.85z0y245-miniHFlash-pltstar} and \ref{fig-m0.85z0y245-majorHFlash-pltstar}
were taken. They correspond to a) the peak of the He flash, b) the
second H mini-flash, and c) the main H flash. \label{fig-m0.85z0y245-HHeFlash-zoom-conv-lums}}
\end{figure}

In Figure \ref{fig-m0.85z0y245-HeFlashConvection-lums-medWide} we
display the convection zones that develop as a result of the He flashes
for both models. It is immediately apparent that the convection zone
is much larger (in mass) in the $Z=0.0017$ model than in the $Z=0$
model. Also apparent is the much larger offset from the centre. The
two models are quite similar in the early stages of He ignition in
terms of the extent of the He convection zone, location of the H-exhausted
core (approximately at the same location as the maximum in H burning
luminosity), and the timescale of convection zone expansion. However
the models diverge severely as the He-flash reaches its maximum. The
huge energy input drives an expansion of the He convection zone (a
consequence of the large temperature gradient now imposed). In the
$Z=0.0017$ model this convection reaches almost to the H-burning
shell, covering a mass of almost $0.4$ $M_{\odot}$ -- or practically
half the mass of the star. After the peak in luminosity the convection
recedes. In the $Z=0$ model the situation is quite different. As
the convection zone is already so far out (in mass) in the star, the
growth of the convection zone quickly reaches the H-burning shell
-- and indeed breaches the star's H-He discontinuity. This mixes
H down into the hot He convection zone very suddenly. Introducing
these protons into such a high temperature environment leads to rapid
H burning, releasing a large amount of energy (see Figure \ref{fig-m0.85z0y245-HHeFlash-zoom-conv-lums})
-- a second flash ensues, this time dominated by H burning. This
induced H-flash -- or Proton Ingestion Episode (PIE) -- is a very
short event. Indeed the increase in luminosity is much more sudden
than that of the He flash (see Figure \ref{fig-m0.85z0y245-HHeFlash-zoom-conv-lums})
although the luminosity diminishes at a similar, but still more rapid,
rate. In the bottom panel of Figure \ref{fig-m0.85z0y245-HeFlashConvection-lums-medWide}
it can be seen just how close together these two flashes are in time.

\begin{figure}
\begin{centering}
\includegraphics[width=0.85\columnwidth]{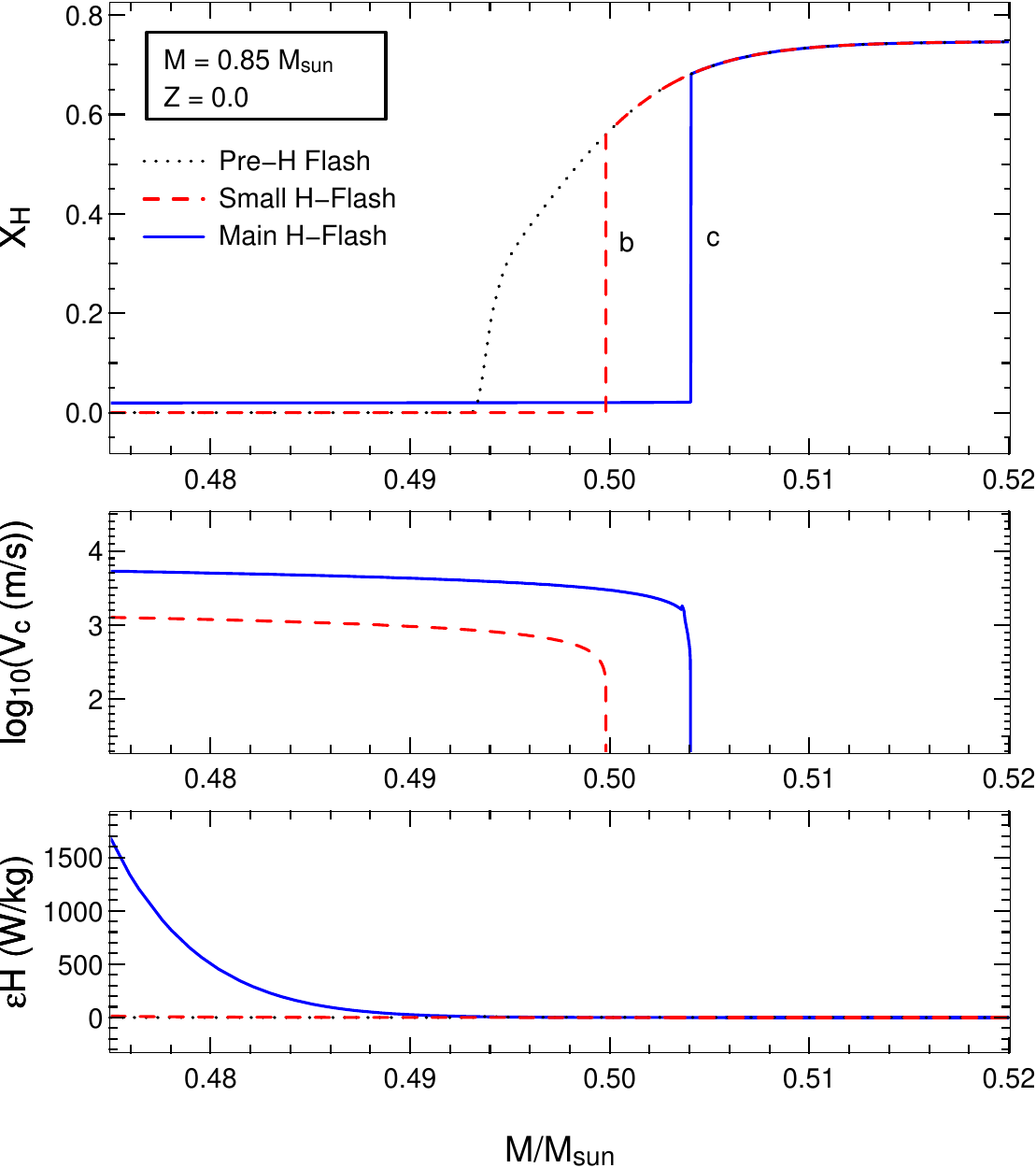}
\par\end{centering}
\caption{Eating the H tail. As the He-H convective zone grows it ingests more
and more protons per unit mass as the depleted tail of the H profile
(left behind by the H burning shell) is eroded. In the top panel we
show the H profile just outside the He-H convective zone (which is
identifiable through the velocity profiles in the second panel) for
three different stages of evolution during the core He flash. The
first model is taken just before the peak of the He flash (dotted
line), before the convective zone breaches the H-He boundary. The
second (dashed line) and third (solid line) models are taken during
the 2nd small H-flash and the major H-flash respectively. Their labelling
corresponds to points \emph{b} and \emph{c} in Figure \ref{fig-m0.85z0y245-HHeFlash-zoom-conv-lums}.
It can be seen in case \emph{c} there is a significant amount of H
in the convective shell. This is also evident in Figure \ref{fig-m0.85z0y245-majorHFlash-pltstar}.
\label{fig-m0.85z0y245-HFlashes-Hprofile-Conv}}
\end{figure}

\begin{figure}
\begin{centering}
\includegraphics[width=0.85\columnwidth]{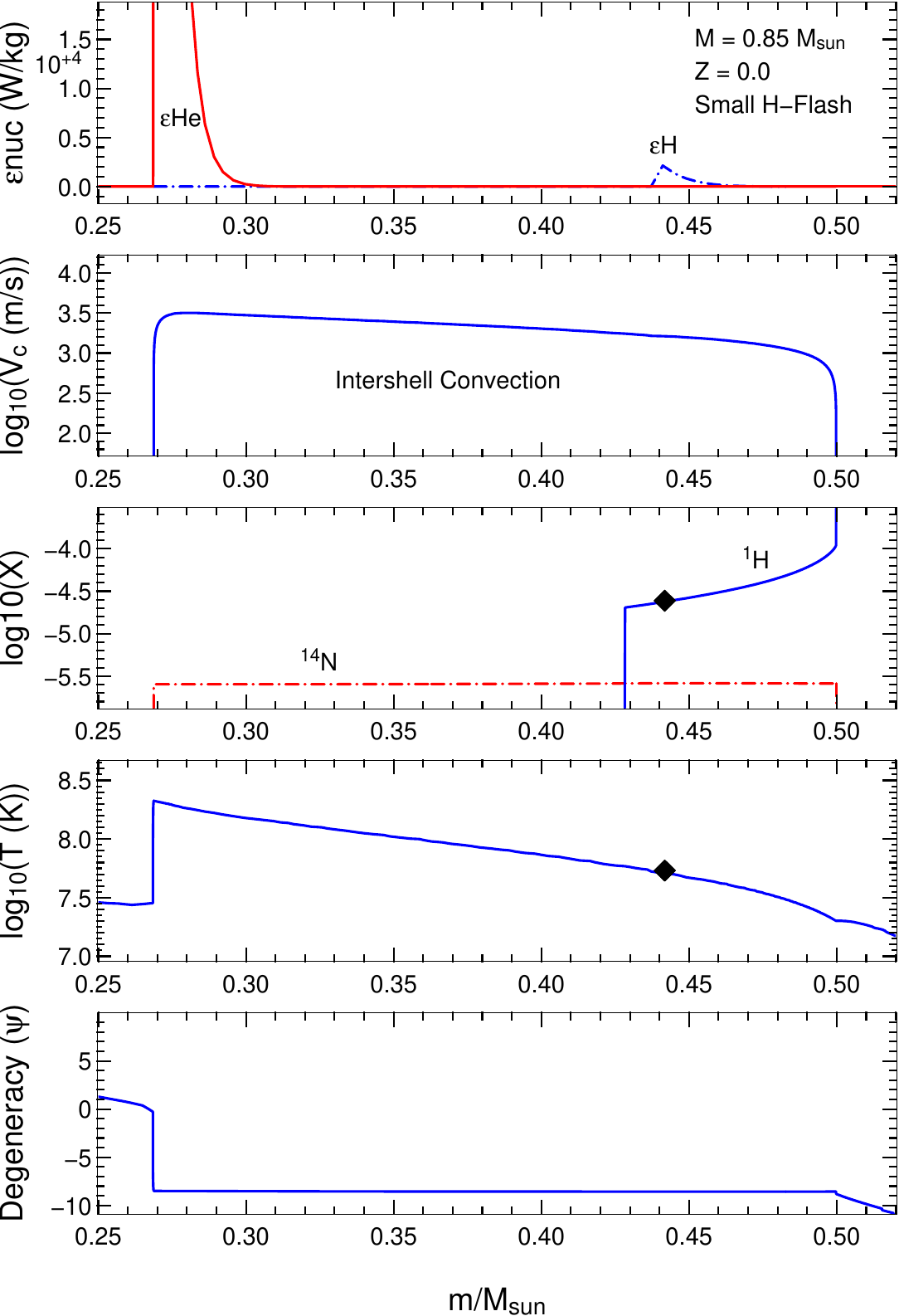}
\par\end{centering}
\caption{The run of some physical characteristics during the second H miniflash
(the location of this model in time is marked `b' in Figure \ref{fig-m0.85z0y245-HHeFlash-zoom-conv-lums}).
In the top panel we see that the He energy generation is still dominant.
There is only a single convection zone as yet but it contains some
H as well as He (panel two). The third panel shows the H abundance
profile. We note that, although the H is in a convective region the
profile has not been mixed homogeneously. This suggests that the evolutionary
timestep is smaller than the mixing timescale and is a good example
of the need for the new time dependent mixing routine. The black diamonds
give the mass location of the peak energy release from H burning,
which is occurring at a temperature of $10^{7.8}$K. Finally we show
the run of the degeneracy parameter to highlight the fact that the
H burning is happening in non-degenerate conditions. \label{fig-m0.85z0y245-miniHFlash-pltstar}}
\end{figure}

In Figure \ref{fig-m0.85z0y245-HHeFlash-zoom-conv-lums} we zoom in,
expanding the time axis to cover just a few years. At this time resolution
we see that the H-flash actually consists of three events. In the
bottom panel we plot the convection zones and the location of maximum
burning for H and He. It can be seen that the peaks in the H luminosity
correspond to the repetitive expansion of the He convection zone into
the H-rich layers. The differing amounts of protons ingested leads
to differing peak H luminosities. The differing amounts arise because
1) there is a H composition gradient left behind by the H burning
shell and 2) the degree to which the convection zone expands into
the H-rich layers varies. In Figure \ref{fig-m0.85z0y245-HFlashes-Hprofile-Conv}
we show the tail of the H profile, located just outside the convective
zone. It can be seen that the tail is progressively `eaten' by the
He-H convective zone, such that each PIE entrains material that is
more rich in protons than the previous PIE material. It is interesting
to note that the series of three flashes compares well with the $0.80$
$M_{\odot}$ $Z=0$ model by \citet{2004ApJ...609.1035P}, which also
exhibits three H-flashes (although the later evolution of their model
is different to our model). Of particular note in Figure \ref{fig-m0.85z0y245-HHeFlash-zoom-conv-lums}
is the fact that the protons only mix a certain distance down into
the convection zone. If the instantaneous mixing assumption were used
in this case the protons would have been evenly distributed over the
He convection zone after each timestep. Thus protons would have been
placed in regions of extremely high temperatures ($\sim10^{8.3}$
K, see Figure \ref{fig-m0.85z0y245-miniHFlash-pltstar}, panel 4),
leading to excessively violent H-burning. With our new time-dependent
mixing routine (described in Section \vref{timedepmix}) the mixing
velocity is taken into account, giving a more realistic situation
whereby the protons are burnt during their descent into the hot He
convection zone. The depth to which protons survive is dependent on
the interplay between the lifetime of the protons against capture
(H burning) and the timescale in which they are transported downwards
(ie. the mixing velocity). It can be seen in panel two of Figure \ref{fig-m0.85z0y245-HHeFlash-zoom-conv-lums}
that these factors lead to protons mixing down to a mass coordinate
of $m\sim0.44$ $M_{\odot}$. When the major entrainment occurs (just
before line \emph{c} in Figure \ref{fig-m0.85z0y245-HHeFlash-zoom-conv-lums})
the consequent energy generation exceeds that coming from He burning
(the He flash is receding by this stage). The He burning subsequently
reduces significantly, causing the He convection zone (HeCZ) to reduce
in size. We now have a distinct structure in which there are two burning
shells with two associated convective regions (as well as a convective
envelope). By 4 years after the peak of the He flash the energy generation
from each shell is approximately equal. 

To analyse this event in more detail we first look at the physical
characteristics of the star during the second proton ingestion episode
(PIE). The model in question is displayed in Figure \ref{fig-m0.85z0y245-miniHFlash-pltstar}
and corresponds to line \emph{b} in Figure \ref{fig-m0.85z0y245-HHeFlash-zoom-conv-lums}.
The dominance of the He burning source is clear in the first panel.
We note that the H burning is dominated by the CNO cycles as the He
shell is abundant in $^{12}$C due to the $3\alpha$ reactions. In
the second panel we see the extent of the convective zone and note
that the peak of H burning energy generation is well inside the HeCZ.
The H profile is plotted in the third panel. This is of interest in
relation to time-dependent mixing as it can be seen that, although
the H is in a highly convective region, it has not mixed to homogeneity,
indicating that the evolutionary timestep taken is shorter than the
mixing timescale in this instance. We mark the location of maximum
H burning energy generation on the H abundance profile as well as
the temperature profile. Hydrogen is present down to a location where
the temperature is $\sim10^{7.8}$ K. We also show the run of the
degeneracy parameter $\psi$ to highlight the fact that the H burning
happens in non-degenerate conditions.

\begin{figure}
\begin{centering}
\includegraphics[width=0.85\columnwidth]{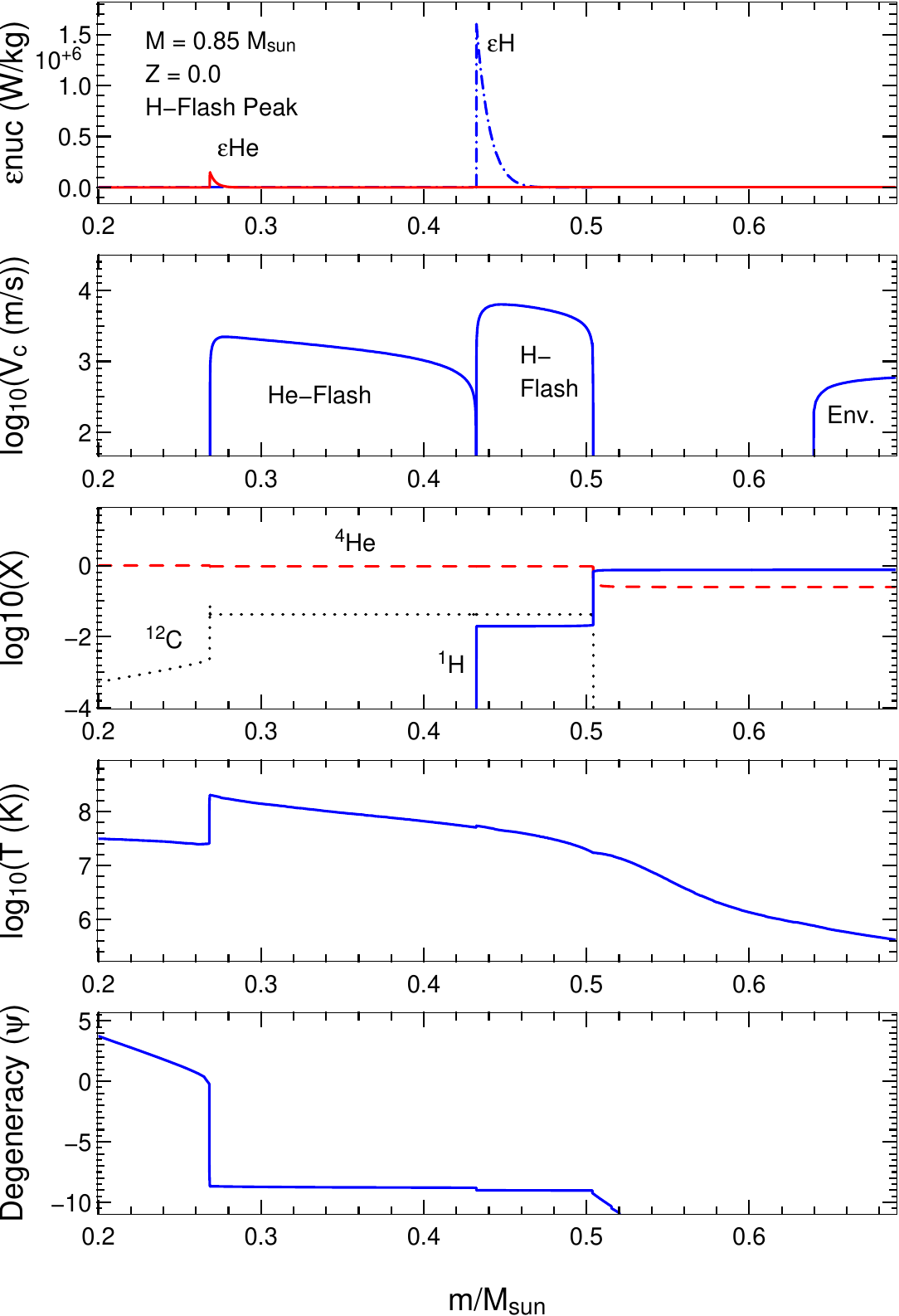}
\par\end{centering}
\caption{Same as Figure \ref{fig-m0.85z0y245-miniHFlash-pltstar} except for
the main H flash (the location of this model in time is marked `c'
in Figure \ref{fig-m0.85z0y245-HHeFlash-zoom-conv-lums}). We now
see the energy from H burning dominating. The development of a dual
convective zone structure has also occurred (panel two). In the third
panel we show the abundance profiles. It can be seen that there is
a substantial amount of H in the HCZ (although it is still dominated
by He). There is also a large fraction of $^{12}$C present, which
is allowing the CNO cycles to dominate the H energy production. The
peak in H energy generation is occurring at a temperature of $10^{7.9}$K.
\label{fig-m0.85z0y245-majorHFlash-pltstar}}
\end{figure}

In Figure \ref{fig-m0.85z0y245-majorHFlash-pltstar} we move to the
major H-flash/PIE. In terms of time the model displayed is approximately
indicated by Line \emph{c} in Figure \ref{fig-m0.85z0y245-HHeFlash-zoom-conv-lums}.
Evident in both figures is the phenomenon of the development of dual
convective zones. As mentioned earlier the PIE produces a huge amount
of energy generation ($\sim10^{6}$ W/kg). This can be seen in the
top panel of Figure \ref{fig-m0.85z0y245-majorHFlash-pltstar}. This
drives a new HCZ and leads to a reduction in He burning energy output
and hence a shrinking of the HeCZ. As can be seen in the second panel,
we now have a highly convective HCZ above a `less convective' HeCZ.
The third panel shows the abundance profiles. Here it can be seen
that there is abundant $^{12}$C in both convection zones (a product
of the previous complete mixing seen in Figure \ref{fig-m0.85z0y245-miniHFlash-pltstar})
and that the HCZ has a large amount of He as well as the newly-entrained
H. It can be seen in Figures \ref{fig-m0.85z0y245-HHeFlash-zoom-conv-lums}
and \ref{fig-m0.85z0y245-EndHflash-miniHeflashes-conv-lums-CNO} that
the HCZ continues to expand and ingest fresh H as time goes on, before
the whole convective region moves outwards in mass, leaving behind
a `polluted' composition profile. In the fourth panel we see that
the H burning is occurring at very high temperatures of $\sim10^{7.9}$K.
This is slightly lower than the temperature of $\sim10^{8.0}$ K found
in the $0.80$ $M_{\odot}$ model of \citet{2004ApJ...609.1035P}.

\begin{figure}
\begin{centering}
\includegraphics[width=0.8\columnwidth]{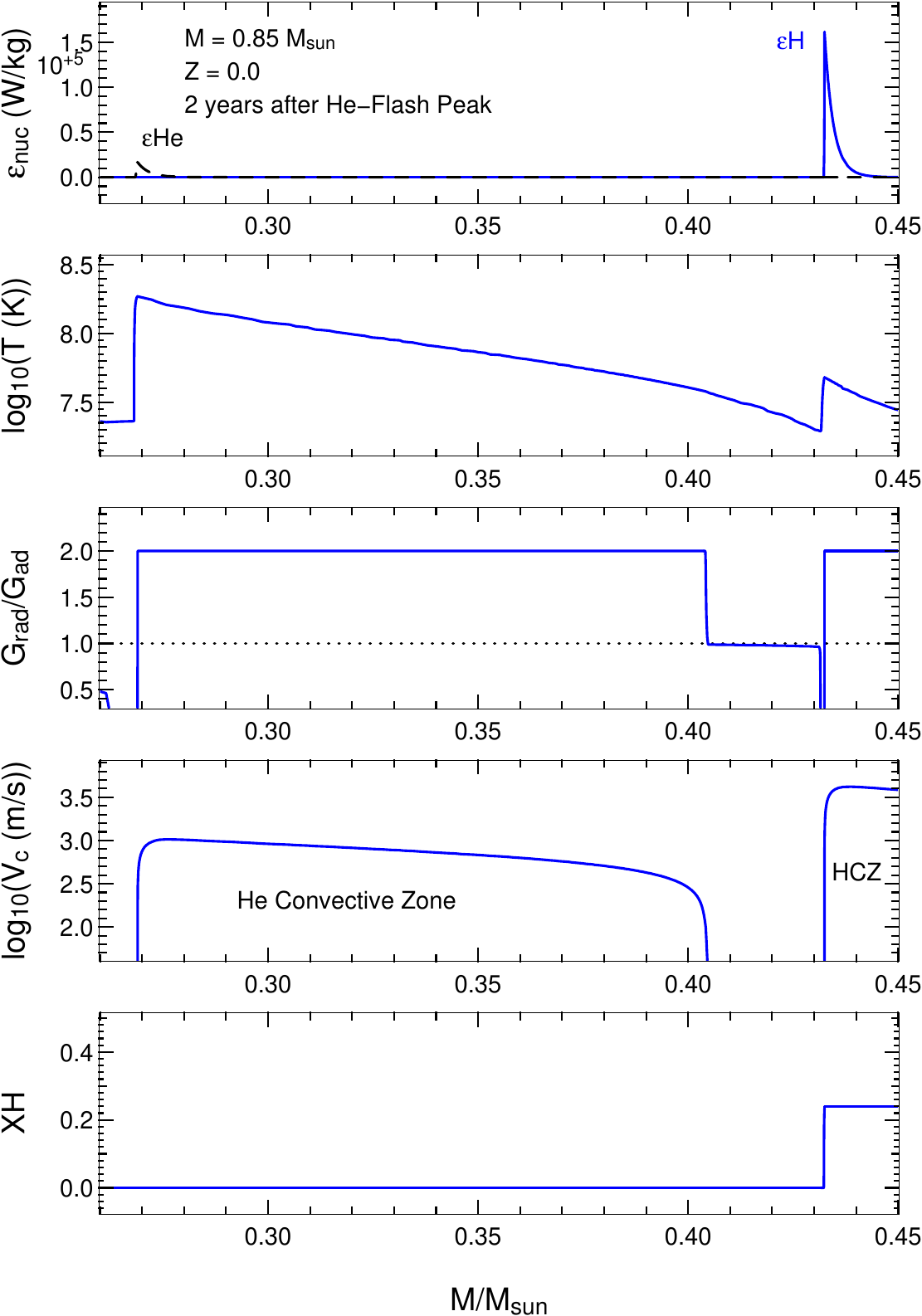}
\par\end{centering}
\caption{The run of various physical characteristics versus mass in the $Z=0$
star $\sim2$ years after the peak of the core He flash (which is
$\sim0.5$ years after the main H-flash peak -- see Figure \ref{fig-m0.85z0y245-HHeFlash-zoom-conv-lums}).
Panel 1 (top) shows the energy generation from H and He burning. It
can be seen that the H burning dominates at this stage. The consequent
temperature inversions are seen in panel 2. The effect of this temperature
profile can be seen in panel 3 where we show the ratio of the temperature
gradients ($\nabla_{rad}/\nabla_{ad}$; note that we have set all
values greater than 2 equal to 2 for clarity). The temperature inversion
at the bottom of the HCZ gives rise to a small region highly stable
against convection (ie. radiative), causing the continuing separation
of the two convective zones. Also apparent is a substantial region
between the two convective zones that is (roughly) convectively neutral.
This is short-lived however as the HeCZ expands soon after this.  The
dotted line in panel 3 at a ratio of 1 delineates the boundary between
convective and radiative regions as defined by the Schwarzschild criterion,
which we are using here. \label{fig-m0.85z0y245-HFlashLater-pltstar-RatioGradients}}
\end{figure}

\begin{figure}
\begin{centering}
\includegraphics[width=0.75\columnwidth]{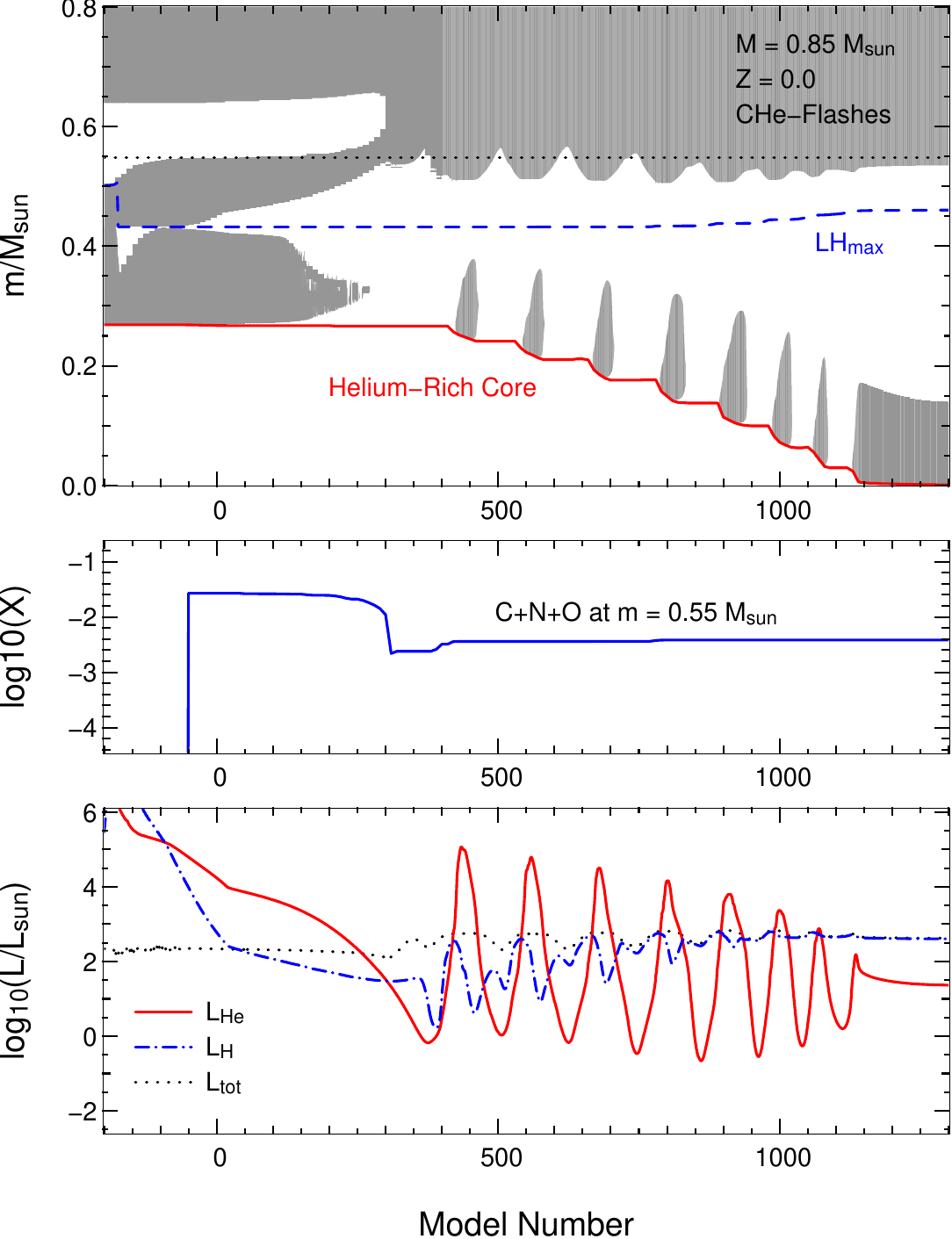}
\par\end{centering}
\caption{The evolution of the convective zones (top), CNO metallicity (middle)
and luminosities (bottom), all against \emph{model number}. We have
chosen to plot against model number because it is time-like but it
also spreads out the rapid changes occurring -- since the SEV code
timestep is mandated by the rate of change of physical properties.
In the top panel we can see the PIE (marked by the inward movement
of the location of maximum H burning) and the consequent splitting
of the convection zones. The horizontal line at $m=0.55$ $M_{\odot}$
represents the mass coordinate at which the metallicity evolution
in the middle panel was taken. This mass coordinate was chosen as
it initially samples the HCZ but also the convective envelope at later
times. The discontinuous jump from $Z_{CNO}=0.0$ to $Z_{CNO}=10^{-1.6}$
is due to the movement of the HCZ out to this mass coordinate whilst
the gradual reduction (starting around model 200) is due to dilution.
The sudden drop at model 300 is due to the dredge up by the envelope
-- the CNO nuclei are now diluted over the entire envelope. The final
metallicity is seen to be $Z_{CNO}=0.004$, or $\sim\frac{1}{5}$
solar. We note that this $Z=0$ star now has a similar surface $Z_{cno}$
metallicity to the 0.85 M$_{\odot}$ GC model. The oscillatory nature
of the step-wise He burning after the dual core flash can be also
be seen in the luminosities (bottom panel). This is the Secondary
RGB stage. The total luminosity is also oscillatory, leading to an
`up-and-down' motion in the HR diagram.\label{fig-m0.85z0y245-EndHflash-miniHeflashes-conv-lums-CNO}}
\end{figure}

We now move on to the later stages of the violent H-He burning episodes.
Figure \ref{fig-m0.85z0y245-EndHflash-miniHeflashes-conv-lums-CNO}
shows the evolution following (and including) that shown in Figure
\ref{fig-m0.85z0y245-HHeFlash-zoom-conv-lums}. The `time' axis (actually
model number for clarity) covers a much larger span in this case.
The behaviour of the H and He convection zones is clear in this plot,
showing that there is initially one HeCZ and that a second develops
as the PIE occurs. The subsequent evolution sees the two zones remaining
separate, such that further dredge-up of $^{12}$C and other species
from the HeCZ does not occur. We note that this is in contrast with
the $0.80$ $M_{\odot}$ model by \citet{2004ApJ...609.1035P} which
ends up dredging up the entire HeCZ at this stage. Our results are
however very similar to those of \citet{1990ApJ...351..245H} (see
their Figure 9). \citet{1990ApJ...351..245H} argue that the convection
splits into two separate zones due to the fact that the energy flux
from the H burning region is now higher than that coming from the
He burning region just below, so that a radiative zone necessarily
forms just below the H burning region. In Figure \ref{fig-m0.85z0y245-HFlashLater-pltstar-RatioGradients}
we show the run of some physical characteristics with mass for a model
taken during the dual convection zone phase ($\sim2$ years after
the He flash peak, see Figure \ref{fig-m0.85z0y245-HHeFlash-zoom-conv-lums}).
Like \citet{1990ApJ...351..245H} we show that there is a temperature
inversion created by the H burning region at the bottom of the HCZ.
To complement this we also show the ratio of the temperature gradients
($\nabla_{rad}/\nabla_{ad}$). Here we see the effect of the temperature
inversion clearly, whereby the positive gradient leads to a small
region that is highly stable against convection. We also find that
there is a substantial region between the HeCZ and HCZ that has virtually
a neutral gradient -- it appears that a semiconvection zone has developed.
We note however that this is soon wiped out by the subsequent re-expansion
of the HeCZ.

As both flashes recede their associated convection zones reduce in
mass. The convective envelope consequently cools and extends into
the regions affected by the H burning and HCZ mixing, thereby dredging
up some processed material and polluting the surface. We highlight
this process of pollution by following the abundance evolution of
total C+N+O mass fraction at one mass coordinate (second panel of
\ref{fig-m0.85z0y245-EndHflash-miniHeflashes-conv-lums-CNO}). We
chose a mass coordinate that reveals the composition of the HCZ as
well as the convective envelope post-dredge-up. It can be seen that
the material is initially devoid of any CNO but, as soon as the HCZ
extends out to this point the metallicity jumps to $\sim Z_{CNO}=0.025$
-- super Solar. As time progresses and the HCZ mixes in more H, the
CNO abundance is diluted somewhat. Finally the convective envelope
moves in and the mass point becomes part of the envelope, showing
the large dilution effect which leads to the envelope having an abundance
of $Z_{CNO}=0.004$. We thus have a star that was initially devoid
of any metals now having a surface abundance of $~\frac{1}{5}$ Solar,
at least in terms of the lower mass elements. Also shown in Figure
\ref{fig-m0.85z0y245-EndHflash-miniHeflashes-conv-lums-CNO} are the
He `mini-flashes' evident in the luminosity diagram in Figure \ref{fig-m0.85z0y245-CoreHeFlash-HeLums-compareGC}.
It can be seen that some of the helium left in the core due to the
off-centre nature of the (major) He flash is periodically burnt in
weak flashes. We discuss this in more detail in the next subsection
which deals with the core He burning stage and the early AGB. 

Another interesting result at the end of the core flashes is that
the GC model actually ends up with a slightly higher H-exhausted core
mass. While the GC model has a core of mass $\sim0.5$ M$_{\odot}$
the $Z=0$ model has a core mass of just $\sim0.45$ M$_{\odot}$.
This is due to the PIE whereby H has moved deeper into the star in
the $Z=0$ case. 

In summary of this Dual Core Flash (DCF) subsection we note that the
main reasons that the HeCZ breaches the H-He discontinuity are (1)
because the core He flash begins significantly off-centre -- so the
HeCZ is already relatively close to the H shell -- and (2), as noted
by \citet{1990ApJ...349..580F}, the entropy barrier in the hydrogen
shell is very low in models of extremely-low (or $Z=0$) models (we
note that we have not explored this feature here; see eg. \citet{1977ApJ...217..788I}
or \citet{1967ApJ...150..961S} for early discussions on entropy barriers
in H-shells). Finally we note that the nett effect of the DCF phenomenon
is a large pollution of the surface with CNO nuclei. This occurs $\sim3000$
years after the peak of the core He flash (see Figure \ref{fig-m0.85z0y245-HeFlashConvection-lums-medWide}).
We have quantified this pollution ($Z_{CNO}\sim\frac{1}{5}$ solar)
here but leave the discussion of the detailed nucleosynthesis for
the next chapter. 

\subsection{Secondary RGB and Core Helium Burning\label{section-m0.85z0y245-SRGBandCHeB}}

We now move into the post core He-flash evolution. Both stars begin
this stage with comparable core masses (the $Z=0$ model's is slightly
lower) but the GC model has a substantially less massive envelope
(see previous subsection for details). 

\begin{figure}
\begin{centering}
\includegraphics[width=0.85\columnwidth]{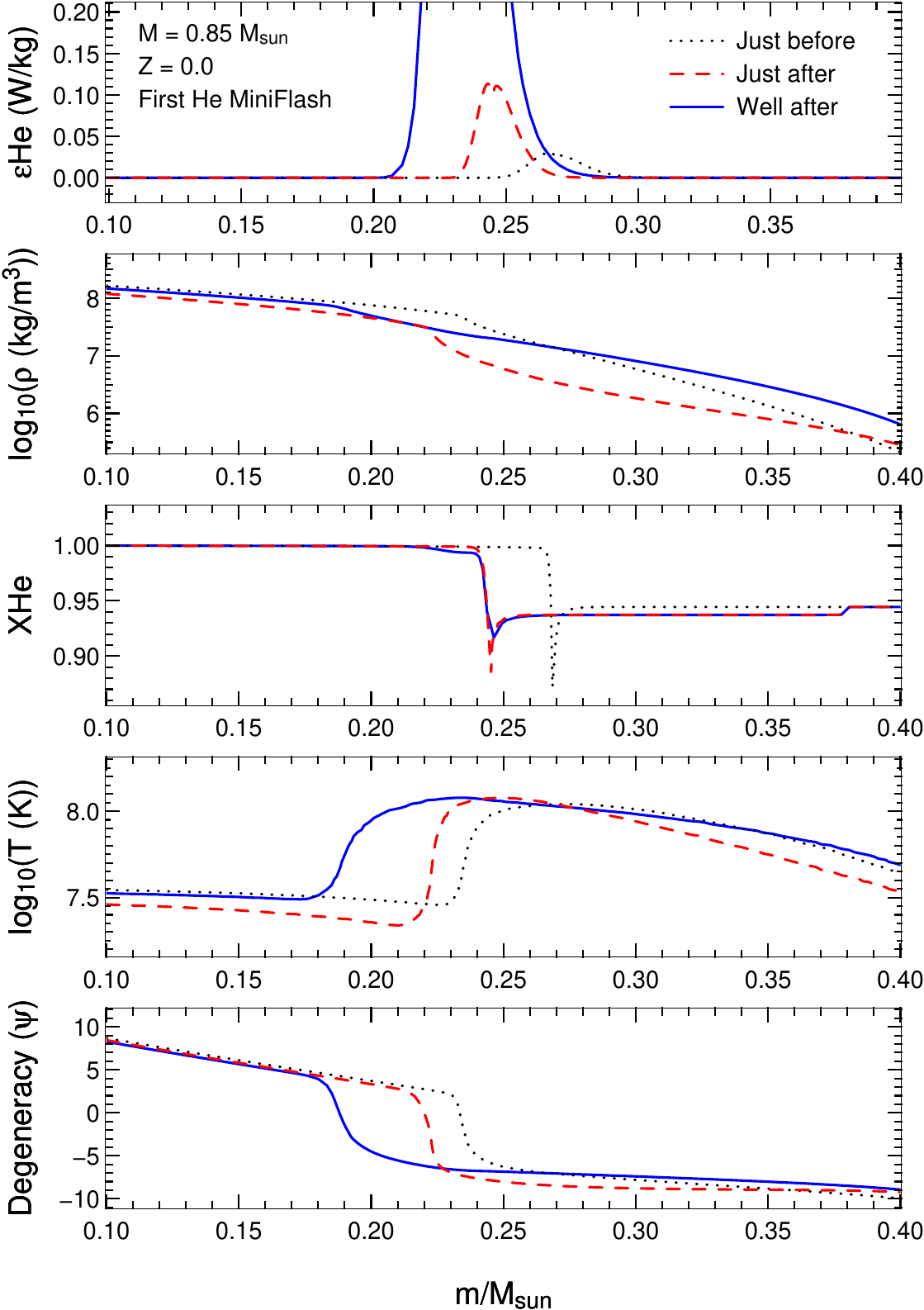}
\par\end{centering}
\caption{The run of some physical characteristics taken at three different
times during the Secondary RGB (SRGB) evolution of the $Z=0$ star.
The dotted line represents a structure just before the onset of a
He minipulse. It can be seen that the He burning is approaching regions
of higher degeneracy. Indeed, the inner core is quite degenerate.
Contrasting this model with the next one (taken just after the minipulse)
the effect of the minipulse can be seen. The temperature in most regions
has been reduced, as has the density. The degeneracy has also been
removed down to a lower mass coordinate and He has been burnt in a
small region. We note that the burning is incomplete - only $\sim5\%$
of the He has been burnt (the rest is mostly $^{12}$C). The next
model (blue line), taken just before the onset of another miniflash,
highlights the fact that the star has returned to a hotter structure,
ripe for the next flash. Finally we note that He burning occurs between
the minipulses, but at a low rate.  \label{fig-m0.85z0y245-HeMiniFlashes-pltstar}}
\end{figure}

As mentioned above it can be seen in Figure \ref{fig-m0.85z0y245-EndHflash-miniHeflashes-conv-lums-CNO}
that a series of He burning flashes ensue after the main He (and H)
flashes abate. These He `mini-flashes' are also evident in the luminosity
diagram in Figure \ref{fig-m0.85z0y245-CoreHeFlash-HeLums-compareGC}.
To elaborate on this phenomenon we show in Figure \ref{fig-m0.85z0y245-HeMiniFlashes-pltstar}
the run of some physical characteristics against mass in a selection
of three models for the $Z=0$ star. We have chosen the models to
illustrate what the effect of a minipulse has on the structure. It
can be seen that the flashes occur in partially degenerate conditions.
The flash has the effect of reducing the temperature and density throughout
the region (see dashed lines in Figure \ref{fig-m0.85z0y245-HeMiniFlashes-pltstar}).
This causes a reduction in the He burning -- but only temporarily.
As can be seen in the final model (solid lines), the star heats up
again (due to ongoing contraction) and He is again burnt at a high
rate. As the evolution continues this will produce a new convective
zone. This cycle repeats until the degeneracy in the entire core is
reduced. We note that each minipulse burns only a small proportion
of the local He ($\sim5\%$ in mass, see third panel). This incomplete
burning thus leaves a core that is still mainly composed of helium
($^{12}$C makes up most of the other $5\%$).

\begin{figure}
\begin{centering}
\includegraphics[width=0.9\columnwidth]{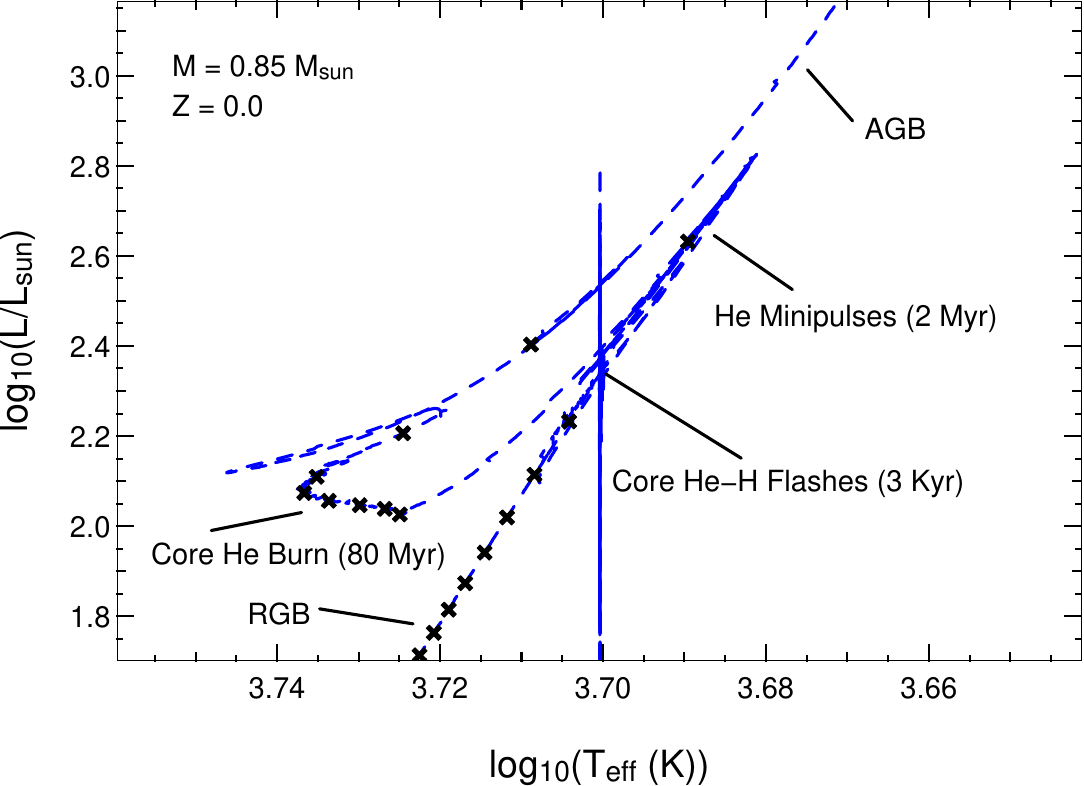}
\par\end{centering}
\caption{Zooming in on the HR diagram at the end of the RGB. The spike in luminosity
marks the (numerically difficult) phase of the dual core flash. This
occurs at a luminosity of $10^{2.4}$, quite low compared to the $Z=0.0017$
model (see also Figure \ref{fig-m0.85z0y245-CoreHeFlash-HeLums-compareGC}).
The star does become more luminous during the He minipulse/Secondary
RGB stage but this is short lived. The crosses indicate time and are
placed at intervals of 10 Myr. \label{fig-m0.85z0y245-CoreHeBurn-HRD}}
\end{figure}

Another important feature of this stage of evolution is apparent in
the HR diagram (see Figure \ref{fig-m0.85z0y245-CoreHeBurn-HRD})
where we can see a significant increase in luminosity \emph{after}
the core He-H flashes. As mentioned earlier the tip of the RGB in
the $Z=0$ model is much less luminous than that of the $Z=0.0017$
model. The fact that the luminosity increases again (by $0.4$ dex
above the location of the Core He-H flash) leads to a feature in the
HR diagram which we refer to as the \emph{Secondary RGB} (SRGB). This
excursion to higher luminosities may have observational consequences.
We note however that the time spent in this region of the HR diagram
is quite brief, being $\sim2$ Myr. In addition to this the star actually
oscillates between this higher luminosity and the lower luminosity
(due to the oscillatory nature of the He miniflashes), such that even
less time is spent at the high luminosity end. Thus we would predict
that not many $Z=0$ RGB stars would be observed above a luminosity
of $\sim10^{2.4}$ L$_{\odot}$ and below a temperature of $\sim10^{3.7}$
K (although there is some uncertainty in the temperature due to opacity,
see the discussion in Section \vref{section-LowTOpacUncertainties}).
We also suggest that this stage of evolution could be named the \emph{TP-RGB},
as it is characterised by a series of pulses.

\begin{figure}
\begin{centering}
\includegraphics[width=0.8\columnwidth]{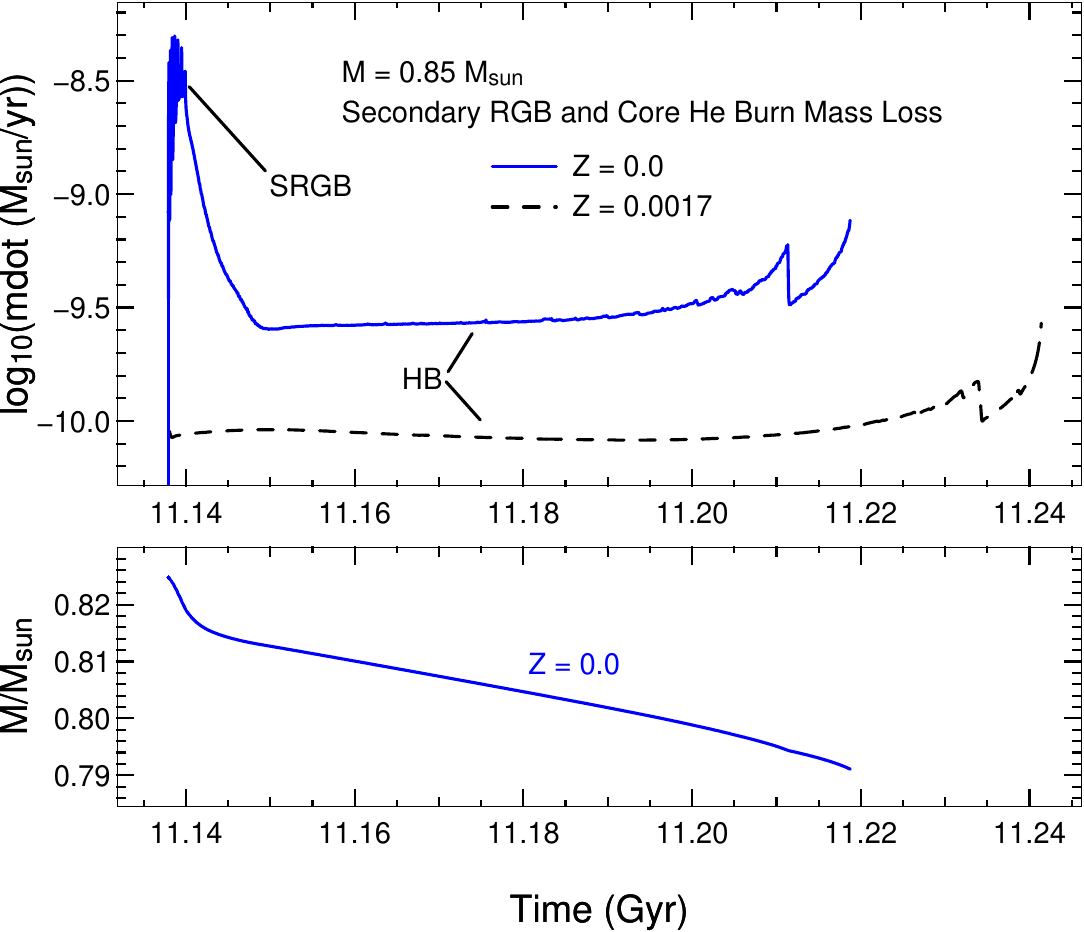}
\par\end{centering}
\caption{The mass loss history for the $Z=0$ and $Z=0.0017$ models (top panel)
and the evolution of total mass for the $Z=0$ model. In the top panel
we have offset the GC time scale so that the beginning of the HB coincides
with that of the $Z=0$ model. It can be seen that the mass loss rates
for the $Z=0$ model are higher in all phases of evolution shown.
Despite this the $Z=0$ star never attains a mass as low as the GC
model (which has a mass of $0.69$ at the end of the RGB), it begins
the AGB phase with a mass of $0.79$ M$_{\odot}$. \label{fig-m0.85z0y245-Mloss-SRGB-HB}}
\end{figure}

In Figure \ref{fig-m0.85z0y245-Mloss-SRGB-HB} we display the mass
loss history of the $Z=0$ and $Z=0.0017$ models during the SRGB/TP-RGB
and horizontal branch (HB) stages. It can be seen that the SRGB is
an important source of mass loss in the $Z=0$ model, due to the high
luminosity (here the mass loss rate is determined from the Reimers
formula). The star in effect `catches up' on some of the mass-losing
that it missed out on during its short RGB. We note however that since
this stage is also brief ($\sim2$ Myr) so the star still doesn't
end up losing as much mass as the GC star did on its RGB. The mass
loss rate is also much higher on the HB, where it loses about twice
as much mass as the GC model does during this stage (the HB lifetimes
are comparable). Despite these two sources of increased mass loss
the star still ends the HB substantially more massive than the GC
model ($0.79$ versus $0.68$ M$_{\odot}$). This has ramifications
for the AGB phase of evolution, which is discussed later. 

\begin{figure}
\begin{centering}
\includegraphics[width=0.9\columnwidth]{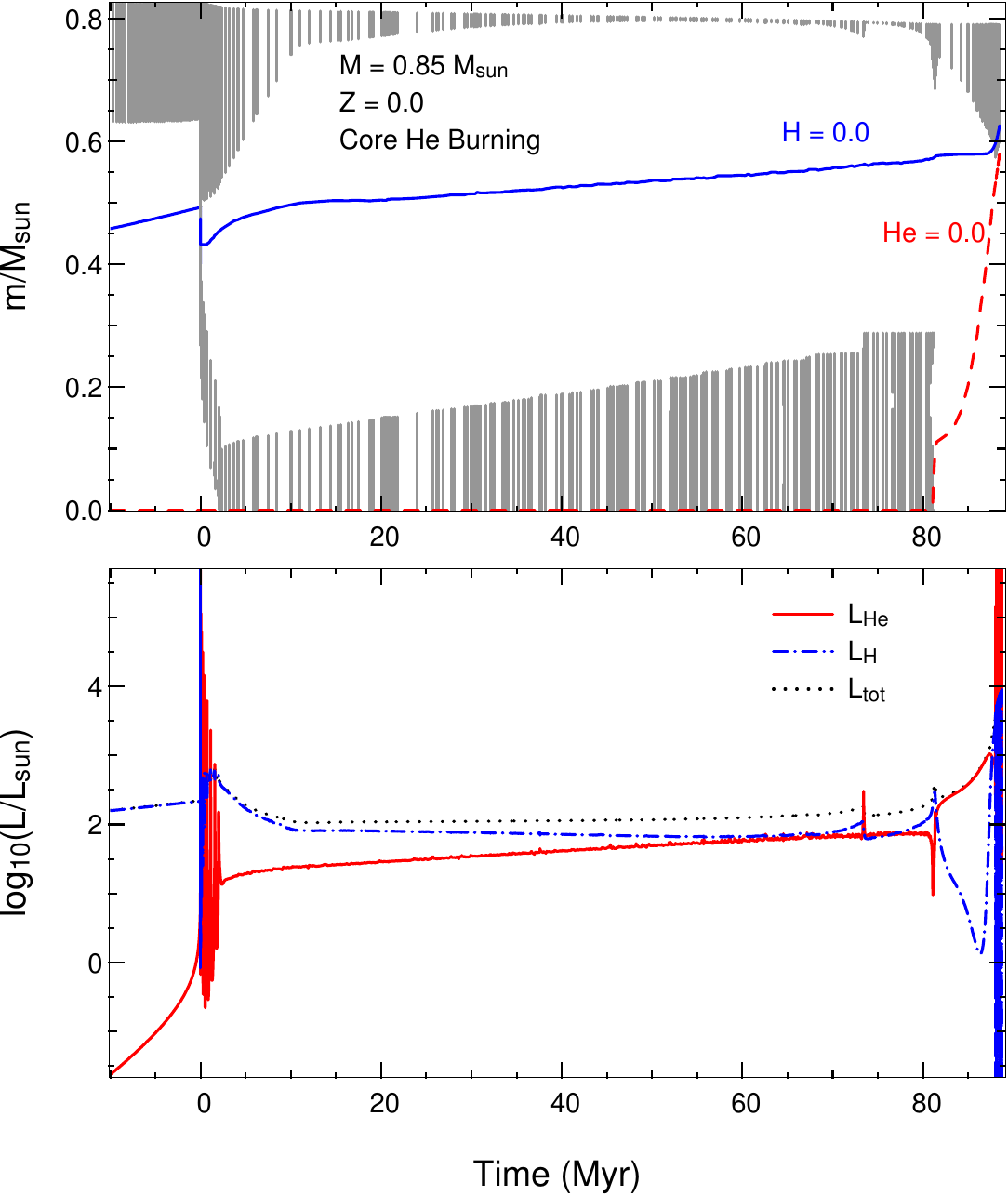}
\par\end{centering}
\caption{The time evolution of the convective regions (marked by individual
grey lines representing every 10th model) and luminosities for the
$M=0.85$ $M_{\odot}$ $Z=0$ model. The top panel also shows the
evolution of the H- and He-exhausted cores (as defined by regions
below the abundance = 0 lines). Time has been taken with a zero point
at the dual core flash. An interesting feature is the displacement
of the H burning shell during the core flash, which subsequently burns
outwards with a large luminosity release. The short duration of the
He minipulse/SRGB phase can be seen ($\sim2$ Myr in total), as well
as the long quiescent core He burning phase ($\sim80$ Myr). There
is a small core breathing pulse at $t\sim73$ Myr accompanied by a
modest peak in the He luminosity. Also of note is the fact that the
H burning shell is actually the main source of energy during most
of the core He burning phase. When the He is exhausted at the centre
the star moves to a double shell structure, whence the He shell takes
over as the main energy source. Just visible on the far right of the
plots is the beginning of the thermally pulsing AGB phase. \label{fig-m0.85z0y245-CoreHeBurn-conv-lums}}
\end{figure}

Quiescent central He burning begins two million years after the main
He-H flash. In Figure \ref{fig-m0.85z0y245-CoreHeBurn-conv-lums}
we display the time evolution of the He convective zones and luminosities.
The core He burning phase is a relatively long-lived stage of evolution,
with a duration of 80 Myr . During this time the convective core gradually
grows in mass, reaching a maximum extent of 0.3 M$_{\odot}$. Note
that we have not applied any overshoot or semiconvection to this model
(at any stage during its evolution). Despite this we find a `core
breathing pulse' (albeit minor) towards the end of the core burning
where the convective core suddenly expands and mixes in fresh He. 

During most of the quiescent core burning phase the dominant source
of energy is actually the H burning shell (see panel two of Figure
\ref{fig-m0.85z0y245-CoreHeBurn-conv-lums}). It is evident from the
higher luminosity in the $Z=0$ star that the H shell is burning out
in mass at a higher rate than in the $Z=0.0017$ model. This has the
important consequence that the $Z=0$ model ends up with a substantially
more massive core at the end of the HB, and hence begins the AGB with
a higher core mass. Despite the GC model beginning the HB stage with
a (slightly) higher core mass it ends the HB with a core mass of $0.53$
M$_{\odot}$ whilst the $Z=0$ star has a core mass of $0.60$ M$_{\odot}$.
This will also affect the the AGB evolution. 

Once He is exhausted at the centre the star moves to a double shell
structure, with the He burning shell then providing the bulk of the
energy. This is the early AGB (EAGB) stage. Also visible towards the
end of the timespan of Figure \ref{fig-m0.85z0y245-CoreHeBurn-conv-lums}
are large luminosity variations. These are the first thermal pulses
of the AGB, which are discussed in the next subsection. 

\subsection{AGB \label{section-m0.85z0-Struct-AGB}}

\begin{figure}
\begin{centering}
\includegraphics[width=0.7\columnwidth]{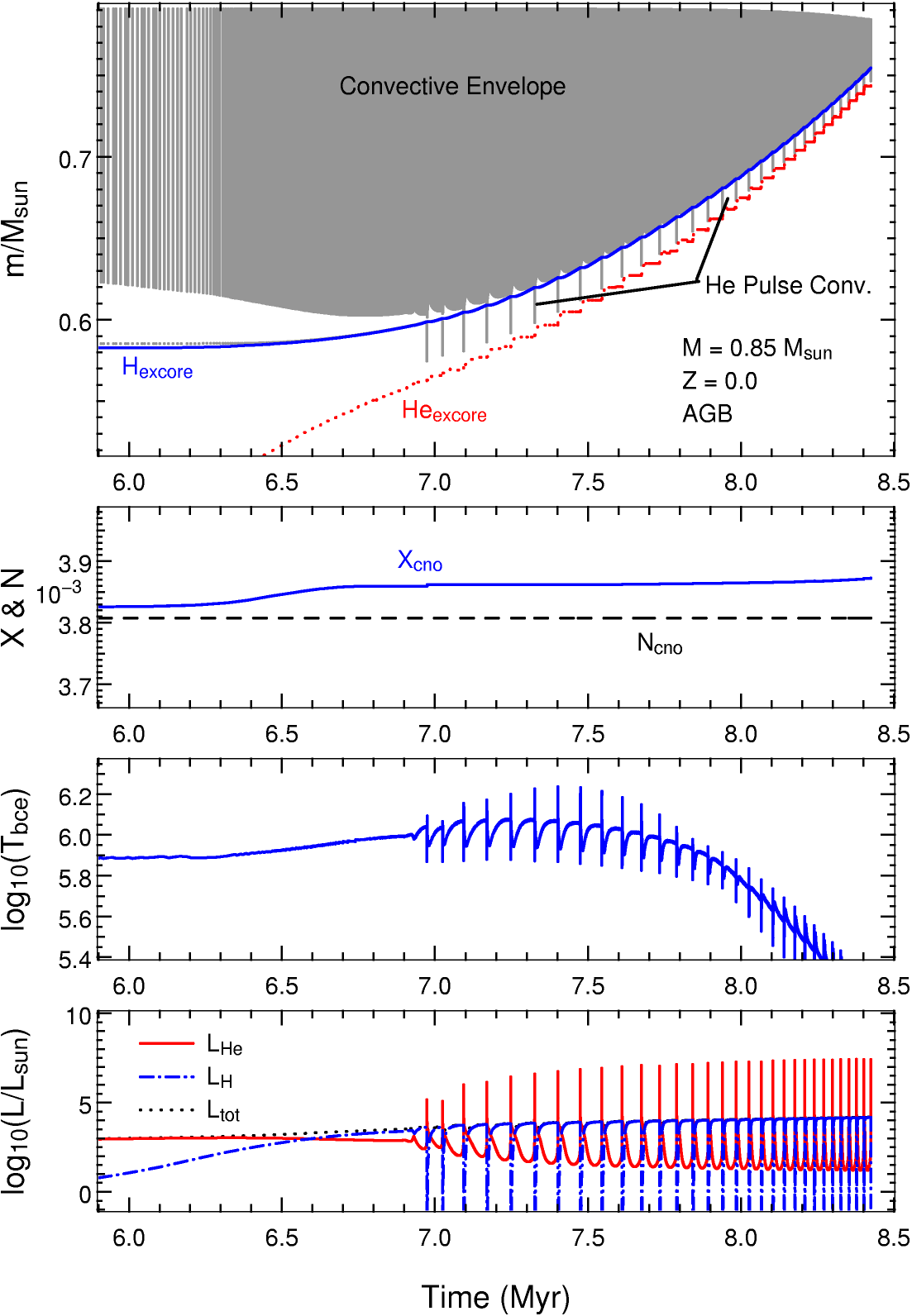}
\par\end{centering}
\caption{The entire EAGB and AGB evolution for the $Z=0$ model. In the first
(top) panel we show the evolution of the convective zones as well
as the H- and He-exhausted cores. The rapid core growth through the
AGB can be seen, as can the intershell convection zones created by
the periodic shell He flashes. We note that the evolution terminates
with an envelope mass of $\sim0.03$ M$_{\odot}$, due to numerical
difficulties. The (very short) remainder of the evolution is handled
with a synthetic code (see Nucleosynthesis section for details). In
the second panel we show the evolution of the surface CNO abundance
(using the same method as that in Fig. \ref{fig-m0.85z0y245-EndHflash-miniHeflashes-conv-lums-CNO})
by sampling at a mass coordinate of $m=0.77$ M$_{\odot}$. We plot
both the mass fraction ($X_{CNO}$) and the number fraction ($N_{CNO}$).
This allows us to see two things: 1) if primary CNO has been dredged
up (which would cause an increase in $N_{CNO}$ and $X_{CNO}$) and
2) if hot bottom burning has been active (which would cause an increase
in $X_{CNO}$ but not $N_{CNO}$, as the CNO cycles conserve the \emph{number}
of nuclei). It can be seen that there has been no 3DUP at all, but
there has been a (very) slight conversion of $^{12}$C to $^{14}$N.
The third panel shows the temperature at the site in which this occurs
-- the bottom of the convective envelope, which consistently reaches
temperatures of $\sim1$ million degrees. Finally, in the bottom panel
we show the variation of the luminosities. Of note here is the shortness
of the interpulse periods ($\sim10^{4}$ yr). Indeed, the pulse frequency
increases towards the end of the evolution, ending up with interpulse
periods of $\sim20,000$ yr. \label{fig-m0.85z0y245-AGB-all-ConvLums}}
\end{figure}

The complete evolution of the AGB for our $M=0.85$ M$_{\odot}$,
$Z=0$ star is shown in Figure \ref{fig-m0.85z0y245-AGB-all-ConvLums}.
Due to the lack of third dredge-up (3DUP) the evolution on the AGB
is somewhat uneventful. Indeed, the composition of the envelope only
changes very minimally, due to a tiny amount of hot bottom burning
(HBB) at the base of the convective envelope. It is thus the post-dual-core-flash
dredge-up that has the most significant effect on the chemical yield
from this star. That said, there are some interesting evolutionary
characteristics of the $Z=0$ model, which we now detail. 

\begin{figure}
\begin{centering}
\includegraphics[width=0.85\columnwidth]{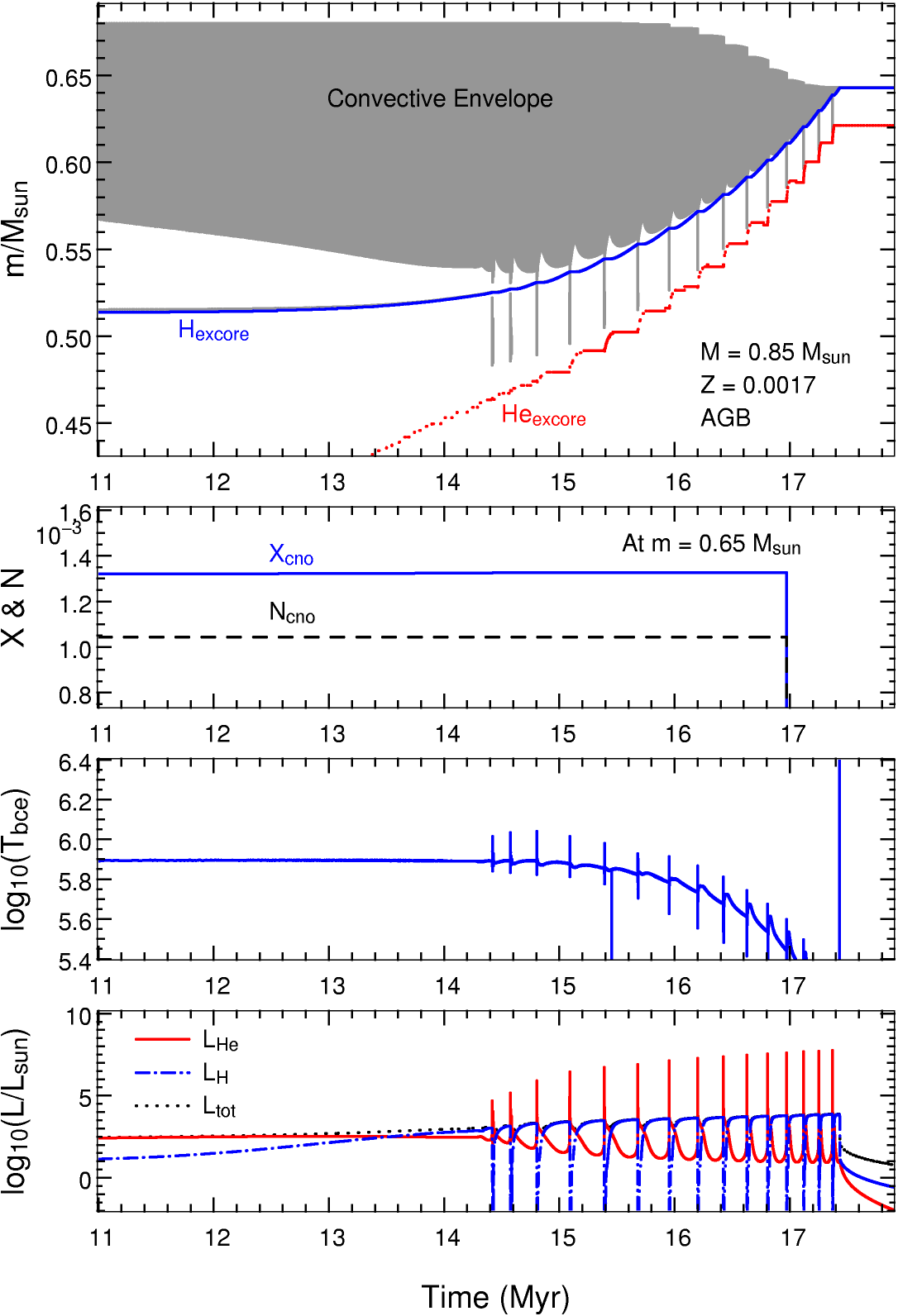}
\par\end{centering}
\caption{Same as Figure \ref{fig-m0.85z0y245-AGB-all-ConvLums} except plotted
for the $Z=0.0017$ GC model for comparison purposes. Of particular
note is the smaller core mass at the beginning and end of the TP-AGB
in this model and also the longer time spent on the AGB. The interpulse
periods are also longer in this case. Also note that the envelope
abundances don't change at all, as there is not 3DUP or HBB. \label{fig-m0.85gc-AGB-all-ConvLums}}
\end{figure}

In terms of the thermal pulses we can see in Figure \ref{fig-m0.85z0y245-AGB-all-ConvLums}
that the interpulse period is quite short, being $\sim10^{4}$ years.
We provide a corresponding Figure for our $Z=0.0017$ model for comparison
(Figure \ref{fig-m0.85gc-AGB-all-ConvLums}). In this star the interpulse
periods are of the order $10^{5}$ yr. The conditions at the beginning
of the AGB for each star are quite different also. Due to the short
RGB the $Z=0$ star did not lose much mass by the time it started
the AGB, at which time it has a mass of $0.79$ M$_{\odot}$. On the
contrary, the $Z=0.0017$ model had a significant amount of mass loss
during the RGB and starts the AGB with a mass of only $0.68$ M$_{\odot}$.
Also of note is the difference in \emph{core} mass between the two
models. The $Z=0$ model has a core mass of $\sim0.60$ M$_{\odot}$
at the first (convective) thermal pulse whilst the $Z=0.0017$ model
has a core mass of $\sim0.53$ M$_{\odot}$. 

Despite the $Z=0$ star having twice as many pulses as the GC model
(30 versus 15), the entire TP-AGB evolution is considerably shorter
than that of the GC model. In fact the $Z=0$ TP-AGB phase lasts half
as long -- $\sim1.6$ Myr as opposed to $\sim3$ Myr. Also apparent
in comparing Figures \ref{fig-m0.85z0y245-AGB-all-ConvLums} and \ref{fig-m0.85gc-AGB-all-ConvLums}
is that the temperature at the bottom of the convective envelope is
slightly higher in the $Z=0$ model, being $\sim10^{6}$ K for much
of the evolution, compared to $<10^{5.9}$ K in the GC model. This
is however still too low a temperature for significant HBB. 

\begin{figure}
\begin{centering}
\includegraphics[width=0.85\columnwidth]{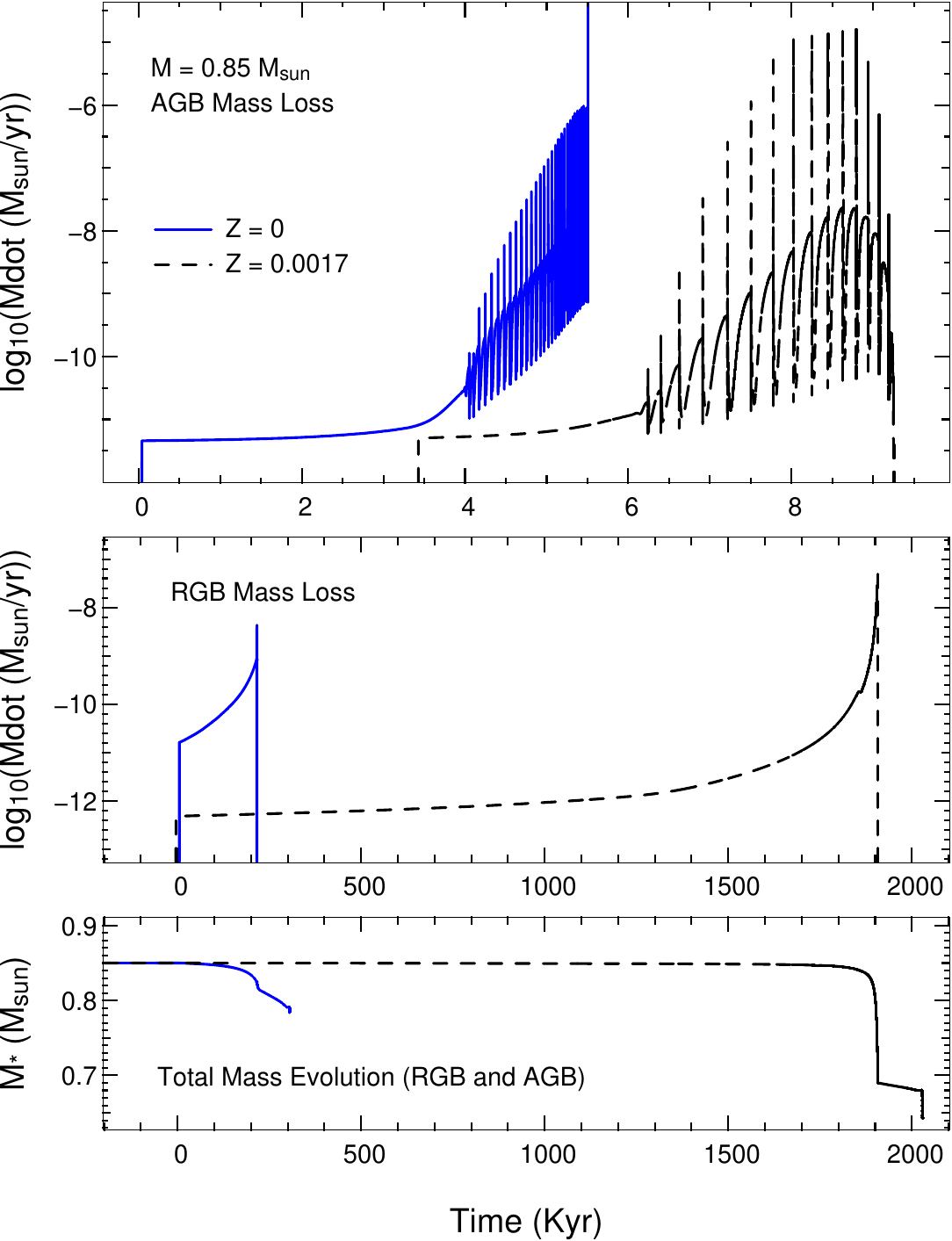}
\par\end{centering}
\caption{The mass loss history (top two panels) and evolution of the total
mass (bottom panel) for the $Z=0$ and $Z=0.0017$ stars. The models
have been shifted in time but the time scaling is identical, enabling
a direct comparison. Each panel has the same amount of time shift.
The top panel shows the mass loss rate during the AGB only. It is
apparent that the mass loss rates are actually quite similar on the
AGB, despite the much shorter overall evolutionary time. Thus it is
not the mass loss rate that is causing the shorter lifetime (see text
for a discussion). The middle panel shows the RGB mass loss history.
This illustrates where the difference in total mass at the beginning
of the AGB originated -- the $Z=0$ RGB evolution is extremely short
due to the early ignition of He (and H) in the core, truncating the
mass loss. The bottom panel shows the evolution of the total mass
of the stars. The end points represent the masses of the white dwarves
produced. \label{fig-m0.85z0y245-Mloss-AGB-GCcompare}}
\end{figure}

In order to shed some light on why the $Z=0$ star spends such a short
time on the AGB compared to the GC star we present the mass loss histories
for both stars in Figure \ref{fig-m0.85z0y245-Mloss-AGB-GCcompare}.
Firstly, the very large difference in time spent on the \emph{RGB}
between the two models is clear in these plots, as is the consequent
lower total mass loss during this stage of the $Z=0$ star evolution.
The low amount of mass loss occurs even though the $Z=0$ model begins
the RGB with a much higher mass loss rate, due to its higher luminosity.
The reason for this is that its evolution is severely truncated due
to the early ignition of He in the core (ie. the onset of the dual
core flash). Thus the $Z=0$ star enters the AGB phase with significantly
more mass, as mentioned earlier.

The difference in AGB evolution time is less severe than on the RGB,
but the factor of two difference is significant. Looking at panel
1 in Figure \ref{fig-m0.85z0y245-Mloss-AGB-GCcompare} it is apparent
that the mass loss rates are quite similar on the AGB -- despite
the much shorter interpulse periods in the $Z=0$ model noted earlier.
It therefore appears that the dominant factor determining the AGB
lifetime is actually the amount of mass each star `needs' to lose
over the AGB phase. This is easily verified by quick (approximate)
calculations of the overall \emph{mass loss rate} on the AGB for the
GC model:

\[
\frac{\Delta m_{env}}{\Delta t_{AGB}}\approx\frac{0.04}{3}=0.0133\,M_{\odot}/Myr
\]

and the $Z=0$ model:

\[
\frac{\Delta m_{env}}{\Delta t_{AGB}}\approx\frac{0.02}{1.6}=0.0125\,M_{\odot}/Myr
\]

which confirm that the average mass loss rates are practically identical.
So why is the AGB timespan so much shorter for the $Z=0$ model? A
further quick calculation of the \emph{rate of core growth} through
the AGB for the GC model:

\[
\frac{\Delta m_{core}}{\Delta t_{AGB}}\approx\frac{0.11}{3}=0.037\,M_{\odot}/Myr
\]

and the $Z=0$ model:

\[
\frac{\Delta m_{core}}{\Delta t_{AGB}}\approx\frac{0.17}{1.6}=0.106\,M_{\odot}/Myr
\]

shows that the core grows much more rapidly in the $Z=0$ model (by
a factor of $\sim3$). It is thus the competition between core mass
growth and mass loss that determines the AGB lifetime. 

\begin{figure}
\begin{centering}
\includegraphics[width=0.85\columnwidth]{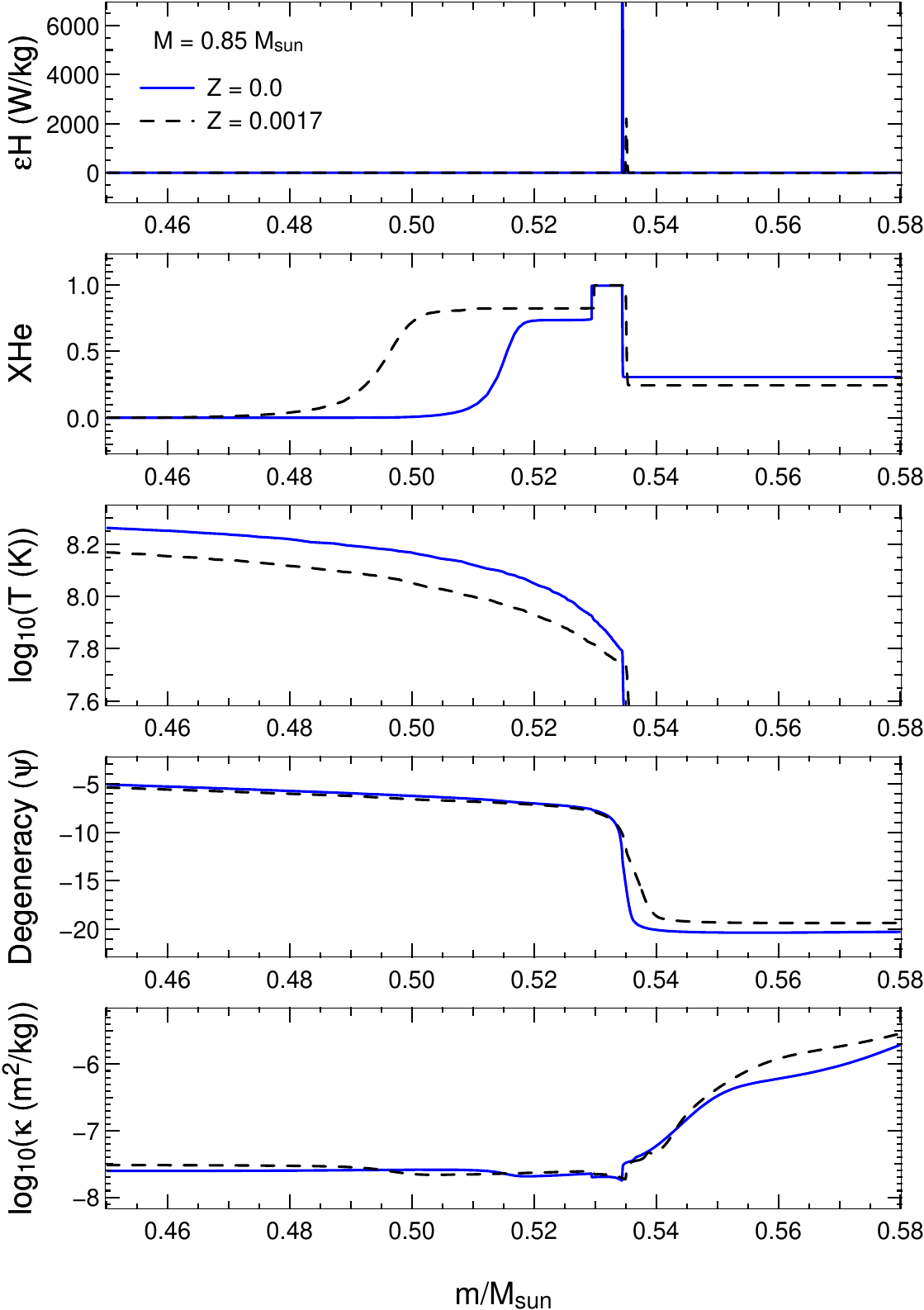}
\par\end{centering}
\caption{The run of various physical characteristics against mass for single
AGB interpulse models from the $Z=0$ and $Z=0.0017$ stars. In the
top panel we see that the energy release from the H burning shell
is much higher in the $Z=0$ case (the $Z=0$ peak is outside the
plotted region, at $2\times10^{4}$ W/kg). We suggest that this rapid
burning leads to the shorter AGB lifetime of the $Z=0$ star. Also
of note is the higher temperature in the $Z=0$ model and the much
larger extent of the He intershell in the $Z=0.0017$ model. \label{fig-m0.85z0y245-AGB-pltstar-IntPulseMod}}
\end{figure}

In order to ascertain \emph{why} the core grows so fast in the $Z=0$
case we present in Figure \ref{fig-m0.85z0y245-AGB-pltstar-IntPulseMod}
the run of some physical characteristics for structures taken during
the interpulse periods of the GC and $Z=0$ models. In particular
we show the region around and including the He intershell. It can
be seen that the $Z=0$ structure is hotter and slightly more dense
than the GC model. It is also supported by a H shell producing much
more energy than that in the GC model (this can also be seen in Figure
\ref{fig-m0.85z0y245-AGB-all-ConvLums} and Figure \ref{fig-m0.85gc-AGB-all-ConvLums}).
We suggest that this is the reason for the rapid core growth -- the
H shell is burning out at a faster rate, thereby leaving the He `ashes'
behind at a faster rate. 

Taking the next step down the rabbit hole we suggest that there are
two reasons that the H shell is so hot in the $Z=0$ model compared
to the GC model: 1) the core mass is initially higher and 2) the opacity
in the envelope in (slightly) lower. Both these facts lead to higher
temperatures and thus higher burning rates. 

Also of interest is that the extent of the He intershell is very different
between the models. The $Z=0$ model has an intershell with a mass
of $\sim0.02$ M$_{\odot}$ while the $Z=0.0017$ has one of twice
this mass. This can also be seen in Figures \ref{fig-m0.85z0y245-AGB-all-ConvLums}
and \ref{fig-m0.85gc-AGB-all-ConvLums} and is a characteristic feature
of the evolution. We suggest that the higher temperatures in the (partly)
degenerate regions in the intershell lead to an earlier ignition of
He each pulse cycle, thus maintaining a smaller mass of He that defines
the intershell. Thus the higher temperatures and consequent higher
burning rates also give rise to the short interpulse periods in the
$Z=0$ star. We note however that it is the rate of core growth that
determines the length of time the star spends on the AGB -- the short
interpulse period is a secondary effect. 

\begin{figure}
\begin{centering}
\includegraphics[width=0.9\columnwidth]{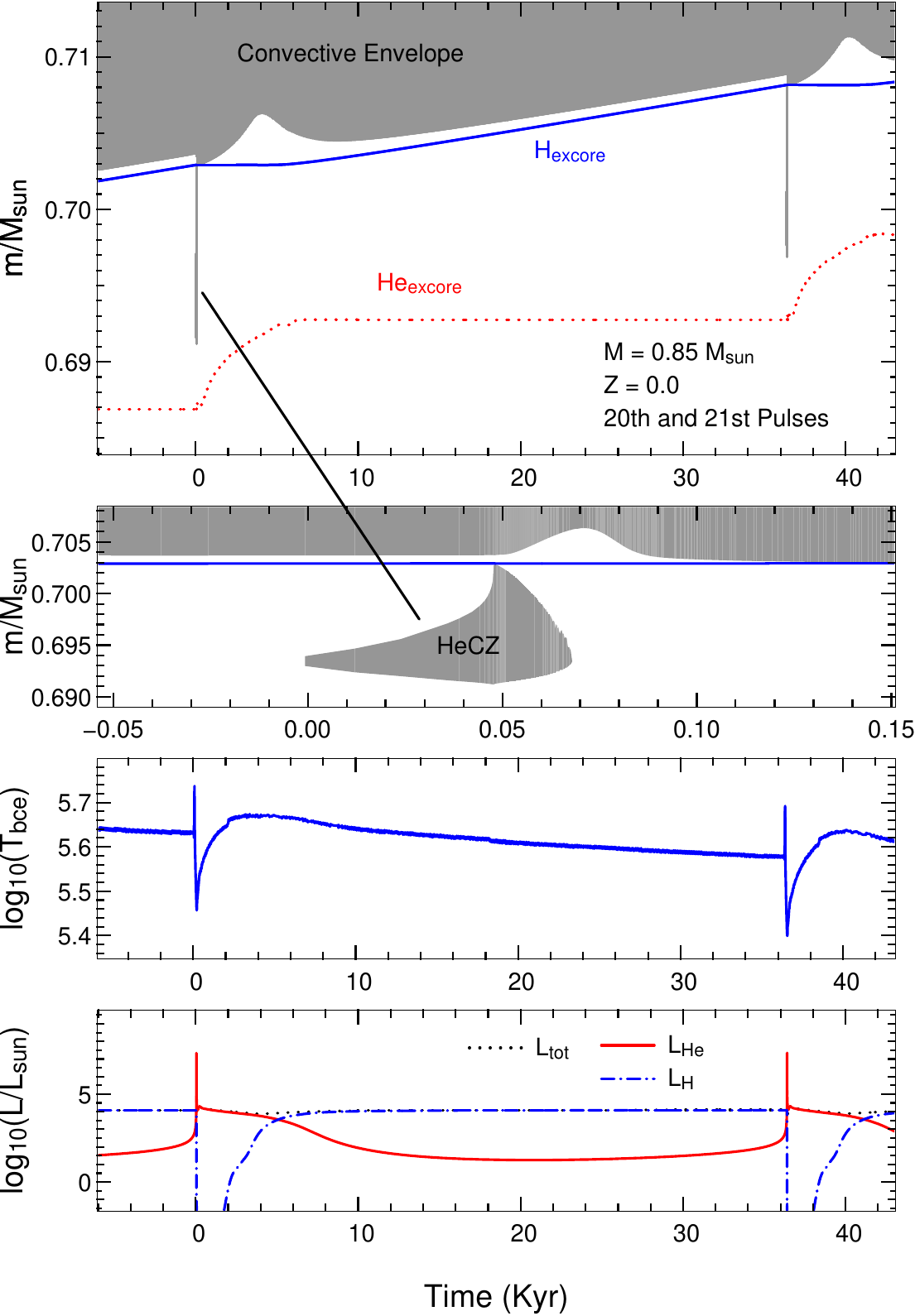}
\par\end{centering}
\caption{Zooming in on the evolution during one AGB pulse cycle in the $Z=0$
model. The thermal pulses shown occur about 2/3 of the way through
the TP-AGB evolution. It can be seen that the interpulse period is
36,000 years, which is short compared to the characteristic period
of $\sim10^{5}$ yr in the $Z=0.0017$ model. Importantly there is
no 3DUP occurring, which means that no extra heavy nuclei are added
to the envelope during the AGB. \label{fig-m0.85z0y245-AGB-singlePulse-ConvLums}}
\end{figure}

Moving beyond the issue of rapid evolution, we present in Figure \ref{fig-m0.85z0y245-AGB-singlePulse-ConvLums}
the time evolution of a single interpulse period, bounded by two thermal
pulses, from the $Z=0$ model. At this stage (the 20th pulse) the
interpulse period is 36,000 yr. It can be seen that the convective
zone does not reach down very close to the He-exhausted core. It can
also be seen that it does not breach the H-He discontinuity above,
remaining solely a self-contained He intershell convective zone (ISCZ).
About 50 years after the ISCZ recedes the convective envelope moves
inwards. It never breaks through to the He (and C) rich layers however
-- ie. there is no 3DUP, as also indicated by panel 2 in Figure \ref{fig-m0.85z0y245-AGB-all-ConvLums}.
In Figure \ref{fig-m0.85z0y245-AGB-singlePulse-ConvLums} we also
see that the peak He burning luminosity during the shell flashes reaches
$\sim10^{7.5}$ L$_{\odot}$, which is comparable to the peaks in
the GC model. 

An important result from AGB evolution is the final core mass of the
stars, which is the white dwarf (WD) mass. The $Z=0$ model produces
a much more massive WD, where $M_{WD}\sim0.77$ M$_{\odot}$ as opposed
to the $Z=0.0017$ model which has $M_{WD}\sim0.64$ M$_{\odot}$.
We note that the difference in core mass between these stars at the
beginning of the AGB was $0.07$ M$_{\odot}$, whilst the difference
at the end of the AGB is $0.13$ M$_{\odot}$. This highlights the
rapid core growth in the $Z=0$ model.

\begin{figure}
\begin{centering}
\includegraphics[width=0.95\columnwidth]{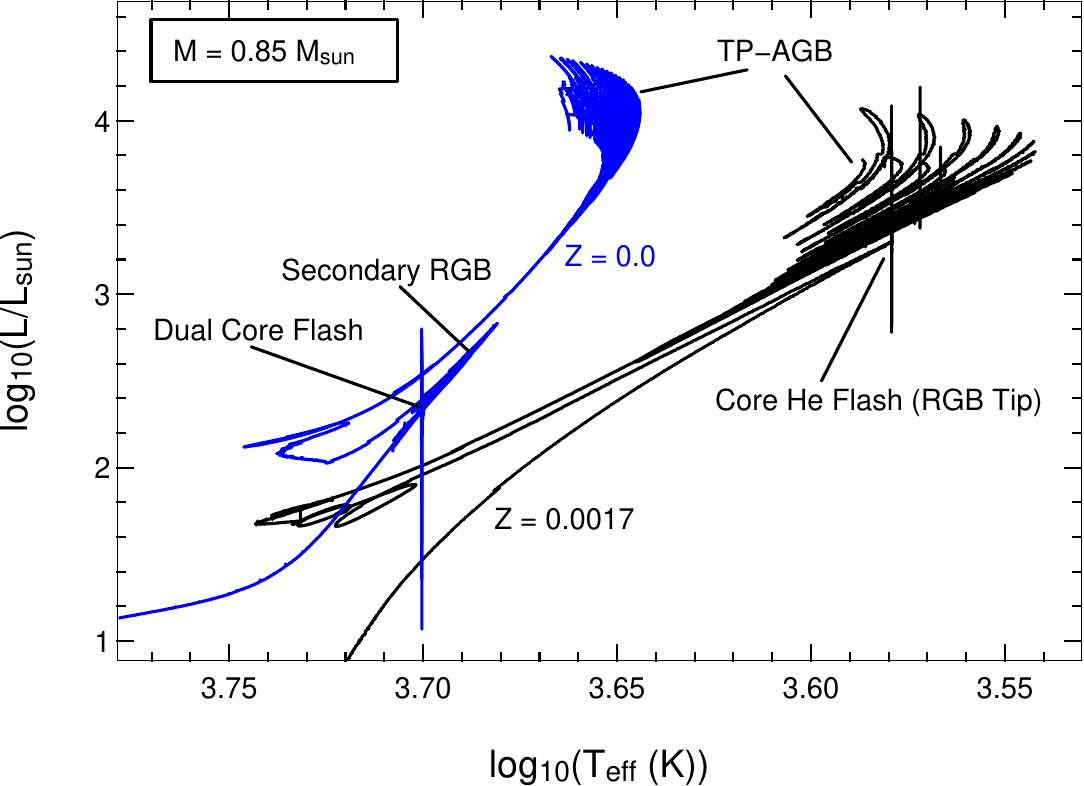}
\par\end{centering}
\caption{Evolution of the $Z=0$ and $Z=0.0017$ models in the HR diagram,
from the lower RGB till the TP-AGB. The peculiar evolutionary features
in the $Z=0$ model, namely the DCF and the SRGB are marked in the
diagram. The vertical lines (eg. at $log_{10}(T_{eff})=3.70$ for
the $Z=0$ model) are numerical artefacts located at the core flashes.
\label{fig-m0.85z0y245-HRD-AGB-GCcompare}}
\end{figure}

For completeness we show in Figure \ref{fig-m0.85z0y245-HRD-AGB-GCcompare}
the RGB, SRGB, HB and AGB evolution of both stars in the HR diagram.
We again note that there is some uncertainty in the surface temperature
of the $Z=0$ model after the post-dual-core-flash dredge-up (ie.
on the secondary RGB and the AGB) due to uncertainties in the low
temperature opacities (see discussion in Section \ref{section-LowTOpacUncertainties}
for more detail). 

\subsection{Executive Summary of Peculiar Evolution\label{subsection-m0.85z0-EvolutionSummary}}

Here we give a brief summary of the peculiarities at each stage of
evolution, in point form. The `peculiarities' are defined relative
to more metal-rich models, such as the GC model described throughout
the previous discussion. In order of evolutionary phase we have:
\begin{enumerate}
\item MS to RGB Tip:

\begin{enumerate}
\item The MS lifetime is substantially shorter.
\item The surface temperature and luminosity are higher.
\item The CNO cycle does not dominate energy production until the end of
the RGB, when the star produces its own primary $^{12}$C. 
\item The time spent on the RGB is very much shorter.
\item The luminosity at the tip of the RGB is substantially lower.
\item The mass of the envelope is significantly larger (but the core mass
is normal) at the end of the RGB.
\end{enumerate}
\item Dual Core Flash

\begin{enumerate}
\item The He core flash is much more off-centre.
\item The He core flash convective zone breaches the H-He discontinuity,
mixing protons down to high temperatures (and $^{12}$C upwards),
causing a strong H-flash. We refer to this as a proton ingestion episode
(PIE). The fact that it happens \emph{during} the core He flash gives
rise to the term Dual Core Flash (DCF)
\item An incursion of the convective envelope dredges up CNO-enriched and
CNO cycle processed material, polluting the pristine surface enough
to raise the $Z$ value to $\sim10^{-3}$.
\end{enumerate}
\item Secondary RGB and Core He Burning

\begin{enumerate}
\item A sustained series of weak core He flashes provide enough luminosity
to create a (thermally pulsing) Secondary RGB (SRGB), such that the
surface luminosity attains values higher than the RGB tip whilst the
star remains on the red side of the HR diagram.
\item Significant mass loss occurs on the SRGB despite its short duration
but the star remains comparatively massive.
\item Although the quiescent core He burning phase is approximately normal
in length, the H-exhausted core growth is very rapid, resulting in
a more massive core at the end of He burning.
\end{enumerate}
\item AGB

\begin{enumerate}
\item The time spent on the AGB is shorter by a factor of two.
\item The interpulse periods are much shorter, and more numerous.
\item The final core mass (ie. WD mass) is substantially higher, which also
means less matter is lost to the interstellar environment.
\item Although there is no 3DUP (which is normal) the chemical yield composition
will be very different from the formation material composition because
it will reflect the pollution from the dual core flash dredge-up episode.
\end{enumerate}
\end{enumerate}

\subsection{Comparison With Previous Work\label{subsec-m0.85z0-ComparePrevStudies}}

A broad review of the literature for $Z=0$ and very low Z stellar
modelling was given in Section \ref{Section-PreviousModels}. Here
we discuss and compare in detail our current model ($M=0.85$ M$_{\odot}$,
$Z=0$, $Y=0.245$) with similar models from the literature. 

In Table \ref{Table-m0.85z0y245-literatureComaprisons} we provide
a list of the relevant studies for comparison with our model. We find
no studies with exactly the same mass as our star but there are a
substantial number with masses close to 0.85 M$_{\odot}$.

\subsubsection*{MS to RGB Tip}

The study by \citet{1982AA...115L...1D} was the first to examine
the evolution of $Z=0$ models of mass $\lessapprox1$ M$_{\odot}$
past the MS. In their 0.9 M$_{\odot}$ model they find the MS turnoff
age to be 15 Gyr (11 Gyr for their 1 M$_{\odot}$ model). In our model
the MS turnoff age is 10.1 Gyr. \citet{1983AA...118..262G} report
an age of 14 Gyr for their 0.9 M$_{\odot}$ model. Thus there is a
substantial difference between the MS lifetimes when comparing with
these authors. We note however that these studies are quite old, and
many improvements to the input physics of stellar models has occurred
since then. Unfortunately the authors that calculated the model which
has the closest mass to ours (0.82 M$_{\odot}$, \citealt{2002AA...395...77S})
did not provide a MS lifetime (they were primarily interested in the
DCF). They do however provide an age on the RGB of 13.7 Gyr. We have
made an educated guess that the MS lifetime was approximately 12.5
Gyr. This is still some $25$\% greater than ours but we note that
\citet{1990ApJ...349..580F} find a MS lifetime of just 7.6 Gyr for
their 0.77 M$_{\odot}$ model, much shorter than our 0.85 M$_{\odot}$
model. Furthermore, if we interpolate in the MS lifetimes in the grid
of models by \citet{2001AA...371..152M} we find an age for a 0.85
M$_{\odot}$ star of $\sim11$ Gyr, about 10\% longer than our model.
Thus it appears, surprisingly, that there is a degree of uncertainty
in the MS lifetimes of these low-mass stars. We suggest that our value
is not inconsistent with the current spread. The uncertainty in this
area does however warrant an investigation, which we shall pursue
as a future study.

As far as we know all authors find that there is some $^{12}$C production
in the core near the bottom of the RGB (earlier in higher mass stars)
which leads to a minor amount of CNO burning (there is a small amount
of H left in the core at this stage, due to the inefficiency of the
p-p chains). We also find this occurs and it can essentially be seen
in our Figure \vref{fig-m0.85z0y245-HRD}. \citet{1982AA...115L...1D}
reported that, although there is a minor amount of locally-produced
$^{12}$C in the the H burning shell on the RGB, the shell energy
release only becomes dominated by the CNO cycles just before the core
He flash (also evident in our Figure \ref{fig-m0.85z0y245-HRD}).
\citeauthor{1982AA...115L...1D} also noted that the surface luminosity
at the tip of the RGB in their 1 M$_{\odot}$ model is about 1 dex
lower than that of a Pop II model. We also find this for our 0.85
M$_{\odot}$ model. Looking at Table \ref{Table-m0.85z0y245-literatureComaprisons}
(10th column, $L_{RGB}$) this is one feature that is reliably reproduced
across all the studies. 

\begin{sidewaystable}

\begin{center}

\begin{tabular}{|c|c|c|c|c|c|c|c|c|c|c|c|}
\hline 
Authors & $M_{*}$ & $\tau_{MS}$ & $\tau_{SGB}$ & $\tau_{RGB}$ & $\tau_{He}$ & $m_{ign}$ & $M_{core}$ & $\Delta m_{ign}$ & $L_{RGB}$ & DCF & $Z_{cno}$\tabularnewline
\hline 
\hline 
\citet{1990ApJ...349..580F}\footnote{Also see the companion paper by \citet{1990ApJ...351..245H}.} & 0.77 & 7.6 & -- & -- & -- & 0.41 & 0.53 & 0.12 & 2.49 & Yes & 0.004\tabularnewline
\hline 
\citet{2001AA...371..152M} & 0.80 & 13.8 & -- & -- & 0.12 & -- & 0.50 & -- & 2.37 & -- & --\tabularnewline
\hline 
\citet{2004ApJ...609.1035P} & 0.80 & -- & -- & 0.04 & 0.04 & 0.35 & 0.52 & 0.17 & 2.45 & Yes & 0.018\tabularnewline
\hline 
\citet{2002AA...395...77S} & 0.82 & 12.5? & -- & -- & -- & 0.31 & 0.61 & 0.30 & -- & Yes & --\tabularnewline
\hline 
\textbf{Current Study ($Z=0$)} & \textbf{0.85} & \textbf{10.1} & \textbf{0.84} & \textbf{0.20} & \textbf{0.07} & \textbf{0.27} & \textbf{0.49} & \textbf{0.22} & \textbf{2.30} & \textbf{Yes} & \textbf{0.004}\tabularnewline
\hline 
\textbf{Current Study (}\textbf{\emph{$Z=0.0017$}}\textbf{)} & \textbf{0.85} & \textbf{11.6} & \textbf{2.1} & \textbf{0.54} & \textbf{0.10} & \textbf{0.11} & \textbf{0.48} & \textbf{0.37} & \textbf{3.35} & -- & --\tabularnewline
\hline 
\citet{1982AA...115L...1D} & 0.90 & 15 & -- & 0.6? & -- & 0.27 & 0.52 & 0.25 & 1.80 & Yes & --\tabularnewline
\hline 
\citet{1983AA...118..262G} & 0.90 & 14 & -- & -- & -- & -- & -- & -- & -- & -- & --\tabularnewline
\hline 
\citet{1993ApJS...88..509C}\footnote{This model has $log(Z/Z_{\odot})=-8$.} & 1.0 & 6.5 & 0.5 & 0.1 & -- & 0.25 & 0.49 & -- & 2.34 & -- & --\tabularnewline
\hline 
\citet{2000ApJ...533..413W} & 1.0 & 6.3 & -- & -- & -- & -- & 0.50 & -- & 2.36 & No & --\tabularnewline
\hline 
\citet{2001ApJ...559.1082S} & 1.0 & -- & -- & -- & -- & 0.15 & 0.48 & 0.33 & 2.31 & Yes & 0.013\tabularnewline
\hline 
\citet{2002ApJ...570..329S} & 1.0 & 6.9 & -- & -- & 0.18 & 0.31 & 0.49 & 0.18 & 2.36 & No & --\tabularnewline
\hline 
\end{tabular}

\caption{Comparing our $M=0.85$ M$_{\odot}$ model with similar $Z=0$ models
from the literature. Also included for further comparison is our GC
model ($Z=0.0017)$ and the 1 M$_{\odot}$, $log(Z/Z_{\odot})=-8$
model by \citet{1993ApJS...88..509C}. The table is ordered by stellar
mass. Note that we were unable to obtain many of the comparison values,
as each study tends to focus on particular facets of the evolution.
The table columns show $M_{*}$(stellar mass), $\tau_{MS}$ (stellar
age at MS turn-off), $\tau_{SGB}$ (lifetime of shell burning stage
between MSTO and start of RGB), $\tau_{RGB}$ (RGB lifetime), $\tau_{He}$
(core He burning lifetime), $m_{ign}$ (ignition point, in mass, of
core He flash), $M_{core}$ (mass of H-exhausted core at time of core
He flash ignition), $L_{RGB}$ (luminosity at RGB tip, in L$_{\odot}$),
$\Delta m_{ign}$ (the distance, in mass, between the point of He
ignition and the H-shell), DCF (whether or not the He convective zone
penetrated the H-shell, or was about to, producing a dual core flash)
and $Z_{cno}$ (the C+N+O metallicity of the envelope after the DCF
dredge-up). All lifetimes are in Gyr and all masses are in M$_{\odot}$.\label{Table-m0.85z0y245-literatureComaprisons}}

\end{center}

\end{sidewaystable}

\begin{figure}
\begin{centering}
\includegraphics[width=0.8\columnwidth,keepaspectratio]{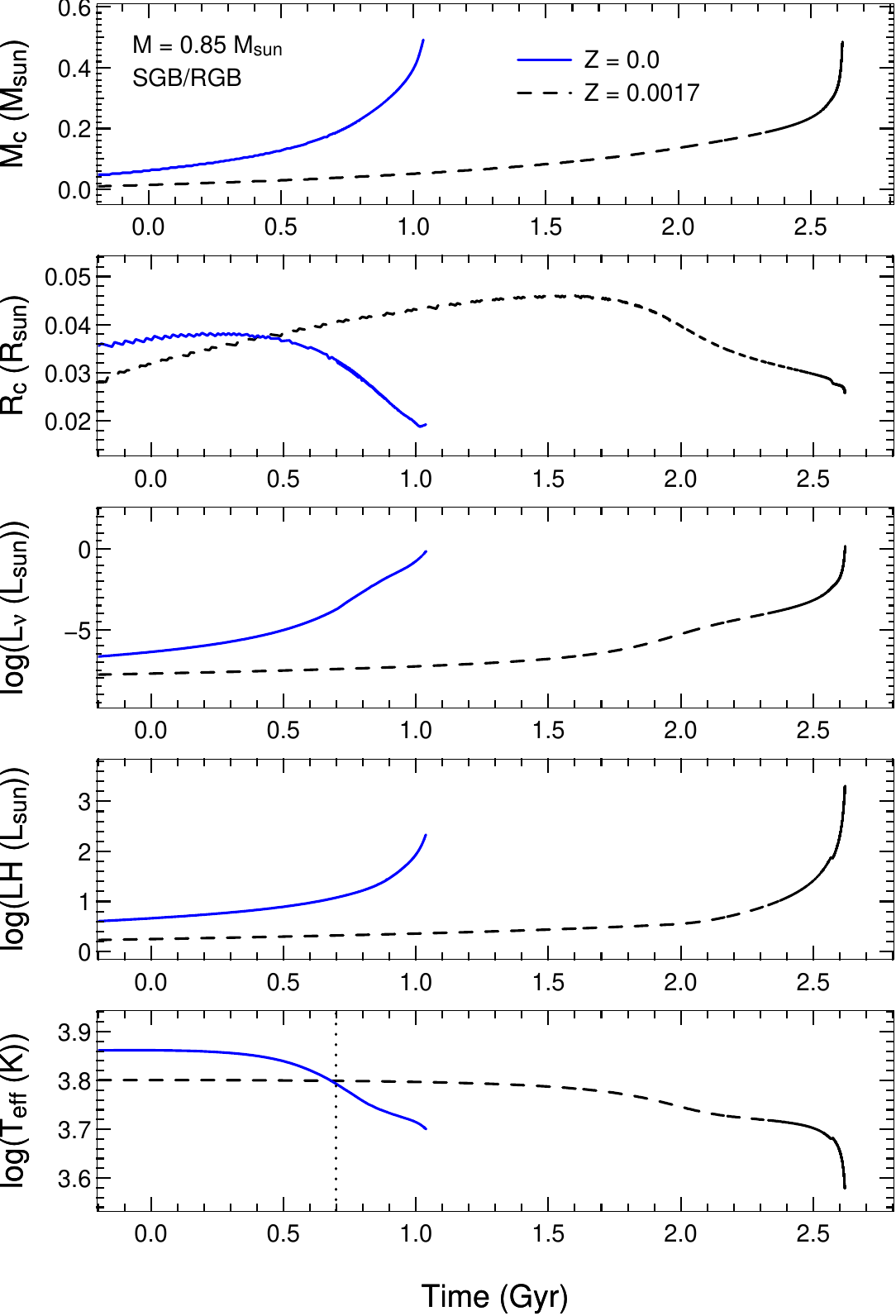}
\par\end{centering}
\caption{Time evolution of various physical characteristics for the $Z=0$
and $Z=0.0017$ models from the MS turn-off to the tip of the RGB.
Both models have been offset in time to place their MS turn-off times
at $t=0$. The models can thus be directly compared in time. The vertical
dotted line in the bottom panel roughly marks the time of H exhaustion
in the centre (this happens to almost coincide between the two models).
The rapid evolution of the $Z=0$ model to the end of the RGB is obvious
in this figure, the lifetime being a factor of $\sim2.6$ shorter
than the $Z=0.0017$ model. Note that the H-exhausted core mass and
radius are defined at $X_{H}=0.05$. \label{fig-m0.85z0y245-SGB-RGB-timeEv-compareGC}}
\end{figure}

Another finding by \citet{1982AA...115L...1D} was that the RGB lifetime
of their $Z=0$ model is a factor of $\sim1.7$ longer than a comparable
Pop II model. This is in stark contrast with the lifetimes of our
models -- we find that the RGB lifetime is a factor of $\sim2.8$
\emph{shorter} than our $Z=0.0017$ comparison model. As the \citeauthor{1982AA...115L...1D}
study is quite old now we suggest that it is wise not to make quantitative
comparisons in this case. The only other $Z=0$ RGB lifetime given
in Table \ref{Table-m0.85z0y245-literatureComaprisons} is that from
the study by \citet{2004ApJ...609.1035P}. They find an exceptionally
short-lived RGB in their 0.8 M$_{\odot}$ model (0.04 Gyr, a factor
of 5 shorter than ours). We can not determine why it is so extremely
short in their model. On the other hand we see that the $log(Z/Z_{\odot})=-8$
model by \citet{1993ApJS...88..509C} has a similar lifetime to our
model (taking the increased mass into account). It has an RGB lifetime
of $\sim0.1$ Gyr compared to $\sim0.2$ Gyr in our model. Although
this model does contain a small amount of metals, the analysis by
\citeauthor{1993ApJS...88..509C} shows that this star exhibits very
similar evolution to our $Z=0$ model. It is interesting to note that
this model deviates by a factor $\sim1.8$ or so in all the evolutionary
times listed but all other values are practically identical. This
suggests that the differences are mainly due to the larger mass. We
conclude that our model is consistent with this study but not others,
although we note that there is not much to compare with at this stage.
In order to expand on this topic of RGB lifetimes we briefly look
in more detail at the rapid evolution of our model. In Table \ref{Table-m0.85z0y245-literatureComaprisons}
we give the lifetime for the shell burning phase between the MS turnoff
and the base of the RGB for our models ($\tau_{SGB}$). In the $Z=0$
model $\tau_{SGB}$ is shorter than the same phase in the $Z=0.0017$
model by almost the same factor ($\sim2.6$) as it is for the RGB.
We note that this lifetime also compares well with the value of $\tau_{SGB}=0.5$
Gyr in the model by \citet{1993ApJS...88..509C}, especially when
factoring in the increased mass. As mentioned in Section \ref{section-m0.85z0-MStoRGB}
the main difference between our $Z=0.0017$ model and our $Z=0$ model
is that the $Z=0$ model has its H-shell powered by the p-p chains,
whilst in the $Z=0.0017$ model the CNO cycles are always operating
in the shell. This contributes to the modest difference in MS lifetime
but the effect becomes stronger as the models diverge in terms of
energy supply source as they turn off the MS. Whilst the $Z=0.0017$
model switches to shell CNO burning, the $Z=0$ model continues to
burn via the p-p chains. It rapidly evolves to an RGB structure and
then quickly evolves up the RGB, reaching the conditions in which
the He flash ignites in a relatively short time. Thus the RGB evolution
is essentially `cut short'. In Figure \ref{fig-m0.85z0y245-SGB-RGB-timeEv-compareGC}
we show the time evolution of some salient physical characteristics
of the models from the MS turn-off to the tip of the RGB (also see
Figure \vref{fig-m0.85z0y245-coreRshellT-HRDPlus}). The rapid evolution
of the $Z=0$ star is clear. We can see that the rate of core mass
growth is significantly higher, as is the subsequent collapse of the
core. This is most likely due to the rapid, high temperature burning
via the p-p chains, coupled with the increased rate of energy loss
due to the low opacity. If our results are correct then this would
imply that $Z=0$ RGB stars should be relatively rare (assuming $Z=0$
stars exist at all of course). Finally we note that this stage of
evolution also requires more investigation. This is quite important
in terms of potential observations, as bright (RGB) stars are more
readily observable and pinning down the lifetimes would provide predictions
for expected distributions. In terms of chemical pollution of the
convective envelope on the RGB the first dredge-up (FDUP) event found
in higher metallicity stars is virtually non-existent at $Z=0$. As
discussed in Section \ref{section-m0.85z0-MStoRGB} we found that
the star retains its original surface composition throughout the RGB.

\subsubsection*{Dual Core Flash\label{subsubsec-m0.85z0-comparePrevstudies-DCF-EMPHs}}

Another feature reported by \citet{1982AA...115L...1D} is that the
maximum temperature location just before the core He flash is very
off centre in their 1 M$_{\odot}$ model. Although they were unable
to calculate the further evolution they suggested that the He flash
occurs so off centre (at $m=0.27$ M$_{\odot}$, very similar to our
0.85 M$_{\odot}$ model -- see Figure \vref{fig-m0.85z0y245-CHeF-Peak-pltstar-compareGC})
that the resulting convection zone is likely to breach the H-He discontinuity,
causing a proton ingestion episode (the DCF). Indeed this is what
most later studies found (including the current one). We list in Table
\ref{Table-m0.85z0y245-literatureComaprisons} which studies have
found the DCF to occur, and which have not. We also list the mass
location of He-flash ignition ($m_{ign}$, roughly the location of
the bottom of the He convective zone), the mass of the H-exhausted
core at ignition ($M_{core}$), and also the distance (in mass) between
the ignition point and the H-shell ($\Delta m_{ign}$). Almost all
studies in this mass range find that the DCF occurs. Actually, as
the \citet{2000ApJ...533..413W} group have recently found that the
DCF does now occur in their models (\citealt{2001ApJ...559.1082S}),
\citet{2002ApJ...570..329S} are the only group that do not find it
occurring (at 1 M$_{\odot}$). Looking at Table \ref{Table-m0.85z0y245-literatureComaprisons}
we can see that the core mass at the end of the RGB ($M_{core}$)
is remarkably consistent between studies, the exception being the
model of \citet{2002AA...395...77S} which has a significantly higher
core mass. The location at which the He flash occurs is less consistent
between studies, varying from $0.15\rightarrow0.41$ M$_{\odot}$.
However, if we remove the `outliers' (\citealt{2001ApJ...559.1082S}
and \citealt{1990ApJ...349..580F} respectively), we find there is
a much narrower range, centred on $\sim0.3$ M$_{\odot}$. We should
not remove the outliers however. Thus the large spread is probably
indicative of uncertainties in the modelling. This also warrants further
investigation, but is outside the scope of the present study. We note
that our value of 0.27 M$_{\odot}$ lies almost exactly in the centre
of the range of values. In terms of the chemical pollution of the
envelope due the the DCF episode (CNO enriched material is dredged
up soon after the flashes recede), there appears to be a general consensus
that a huge amount of pollution occurs. The $Z_{cno}$ of the envelope
becomes so high that it exceeds the observations of the C- and N-rich
EMPHs. \citet{1990ApJ...351..245H} were the first to quantify this,
finding that the surface N abundance in their model was $\sim2$ dex
higher than observed for the star CD -38.245$^{\circ}$. \citet{2001ApJ...559.1082S}
find similar results, with their models exhibiting $\sim2$ to 3 dex
too much C and N compared to observations. They suggest that the C
and N production could be reduced by an improved modelling of the
strongly convective H burning region, as this is still modelled in
a very simple way (1D mixing length theory) that may not be sufficiently
accurate in this situation. Indeed, by reducing the mixing efficiency
by a factor of $\sim10^{4}$ \citet{2002AA...395...77S} found that
the C and N pollution results were closer to the observations. There
is no physical argument for altering the mixing efficiency so much.
However we note that there is no reason to believe that the MLT is
applicable in this situation of high temperature and strong nuclear
burning -- so the mixing efficiency may well be thought of as a free
parameter in this case. \citet{2004ApJ...609.1035P} also find that
C and N are over-produced. They highlight the fact that the models
always predict that the $^{12}$C/$^{13}$C ratio will be close to
equilibrium ($\approx4$) but the observations show values of $\sim50$.
In addition to this, they find the C/N ratio to be $\sim1$ in their
models but it is $\sim50$ in the CEMPHs. Another interesting result
that they report is that their model produces large amounts of Li.
They conclude that all the evidence points to these models \emph{not}
being the explanation for the CEMPs. \citet{2004AA...422..217W} came
to the same conclusion and also note that in an earlier investigation
of theirs (\citealt{2002AA...395...77S}) that explored the robustness
of the pollution predictions they found that the amount of C and N
produced is quite insensitive to the model parameters. The fact that
all studies have found essentially the same over-abundances despite
the wide variation of the location of He ignition further substantiates
this finding. In Section \ref{section-m0.85z0y245-DualCoreFlash}
we report that our model ends up with a surface composition having
$Z_{cno}=0.004$. This is in exact agreement of that found by \citet{1990ApJ...351..245H}.
The other two studies which report quantitative values for the DCF
pollution of pure $Z=0$ models find somewhat higher abundances of
C and N. The models by \citet{2001ApJ...559.1082S} and \citet{2004ApJ...609.1035P}
end up with $Z_{cno}\sim0.015$, a factor of about 4 greater than
ours. We note that this higher abundance only amounts to about 0.6
dex difference when comparing with observations, thus, although it
means that their models are further from fitting observations, the
overall discrepancy is about the same (ie. $\sim1\rightarrow3$ dex
too high). The discrepancy between models is probably a product of
the evolutionary details during the DCF. In the model of \citet{2004ApJ...609.1035P}
the two convective zones that are initially produced at the start
of the H-flash soon merge to become one again. This naturally mixes
up even more C-rich material that is later dredged up. \citet{1990ApJ...351..245H}
find that the two convection zones remain separated, as we also find
in our model (see eg. Figure \vref{fig-m0.85z0y245-HHeFlash-zoom-conv-lums}).
\citet{2001ApJ...559.1082S} also finds that they remain separated
but in another model by the same group (\citealt{2002AA...395...77S})
they appear to merge. As mentioned earlier the nett result of CNO
pollution is \emph{roughly} the same across the studies so it appears
that the details of the DCF evolution are not very significant. That
said, the fact that there is some variation between the studies --
and that one study does not even find that the DCF occurs -- suggests
that more work needs to be done on this phase of evolution as well.
It may also affect further evolutionary stages. Finally we note that,
in terms of the aims of the current study, we find that our results
are reasonably consistent with the previous studies in relation to
the DCF phase. 

\subsubsection*{Secondary RGB and Core He Burning}

\citet{2001ApJ...559.1082S} found that, after the DCF had abated
and the CNO-rich material has been dredged up, their star settled
back into an `new-born' RGB structure (Secondary RGB). It continued
evolving up the SRGB for 50 Myr, finally ending in a second (weaker)
core He flash. This second flash did not induce another H flash and
the star settled into quiescent core He burning. \citet{2004ApJ...609.1035P}
also find a secondary RGB in their 0.80 M$_{\odot}$ model. It is
however substantially different to that found by \citeauthor{2001ApJ...559.1082S}.
Most of the SRGB evolution in their model involves a series of weak
off-centre He flashes. Each flash occurs progressively less off-centre
and creates a short-lived convection zone, removing the degeneracy
in the core step by step. This is essentially identical to our 0.85
M$_{\odot}$ model (see Figures \vref{fig-m0.85z0y245-CoreHeFlash-HeLums-compareGC}
and \vref{fig-m0.85z0y245-EndHflash-miniHeflashes-conv-lums-CNO}).
They also report an SRGB lifetime of 2 Myr, which is identical to
what we find. We are unsure why the model by \citet{2001ApJ...559.1082S}
is so different, having an SRGB life time an order of magnitude greater
than ours.

There is only one study that we have found (see Table \ref{Table-m0.85z0y245-literatureComaprisons})
that is directly comparable (and provides quantitative results) to
our model in terms of the quiescent core He burning phase. Other studies
either did not evolve through the DCF phase or did not find the DCF
to occur. \citet{2004ApJ...609.1035P} find a core He burning lifetime
of 0.04 Gyr for their 0.80 M$_{\odot}$ model, which is somewhat shorter
than that of our 0.85 M$_{\odot}$ model which has a lifetime of 0.07
Gyr. Although we find our model increases its core mass substantially
over the core He burning stage (from $\sim0.45$ to 0.60 M$_{\odot},$
see Section \ref{section-m0.85z0y245-SRGBandCHeB}) the \citeauthor{2004ApJ...609.1035P}
model does so even more. They say the reason for this is that the
$Z_{cno}$ is very high in their model, which gives rise to very efficient
CNO shell burning, thereby adding He to the core at a great rate.
Their model increases its core mass from 0.49 to 0.75 M$_{\odot}$
during this phase. This is an extremely large growth when compared
to our $Z=0.0017$ model which only grows from 0.49 to 0.53 M$_{\odot}$.
Since the $Z_{cno}$ abundance is significantly smaller in our model
it is to be expected that our He burning lifetime is somewhat longer
and the core growth slower. We note that the \citeauthor{2004ApJ...609.1035P}
model is one of the models which has higher envelope $Z_{cno}$ values
after the DCF, thus we expect that our core He burning lifetime would
be more similar to the the study of \citet{1990ApJ...351..245H} who
find similar pollution levels to us. If the envelope metallicity is
indeed the primary reason for the difference in evolutionary timescales
then this reinforces the suggestion that more work needs to be done
on the DCF, as this is where the metallicity difference arises.

\subsubsection*{AGB}

With regards to the AGB we find that there have not been any studies
specifically dedicated to investigating this interesting phase of
evolution at $Z=0$ and $M\lesssim1$ M$_{\odot}$. This is probably
due to the large uncertainties that become important during this phase
-- mass loss and the treatment of 3DUP -- and the large computational
demand. The semi-analytic study by \citet{1984ApJ...287..749F} predicted
that at $Z=0$ thermal pulses will only occur when the core mass is
$\lesssim0.73$ M$_{\odot}$. The core mass in our model is well below
this limit (0.49 M$_{\odot}$) and indeed, we do find that the model
experiences thermal pulses (TPs). Despite the TPs we find that 3DUP
does not occur. There have been a couple of detailed modelling studies
that briefly touch on the AGB. \citet{1998IAUS..189P.150F} only give
qualitative descriptions of their models on the AGB but report that
thermal pulses do occur, and that they also find no 3DUP at these
masses. \citet{2001ApJ...559.1082S} evolve their 1 M$_{\odot}$ model
about 10 pulses into the AGB (they use the Reimers mass-loss formula).
Although the internal regions of this model have $Z=0$, the surface
of this model has been artificially polluted, thus the pulsation characteristics
are likely to be different to ours. Indeed, they find interpulse periods
of $\sim0.006$ Myr whereas ours are $\sim0.08$ Myr (at the start
of the AGB). They don't report any 3DUP. \citet{2002ApJ...570..329S}
also evolve their 1 M$_{\odot}$ model about 10 pulses into the AGB.
They find interpulse periods about a factor of two longer than ours
($\tau_{intp}\sim0.2$ Myr). We note however that this model does
not experience the DCF, whereas ours does. The DCF can effect the
TP-AGB characteristics since it dredges up so much CNO nuclei, altering
the opacity. Their model did however experience the DSF which also
pollutes the envelope but to a lesser extent. Interestingly \citeauthor{2002ApJ...570..329S}
find that their 1 M$_{\odot}$ model experiences \emph{intermittent}
3DUP. However they suggest that it is most likely a numerical resolution
problem. We note that \citeauthor{2002ApJ...570..329S} included some
overshoot, whereas we have not. To the best of our knowledge the studies
just mentioned constitute the extent of the research in to the TP-AGB
phase of these stars. Based on these few studies (and the present
one) it seems that thermal pulses do occur in these stars but 3DUP
does not. This has implications for the chemical composition of the
polluting material that will be released via winds. Without 3DUP the
yield will directly reflect the pollution brought about by the DCF
at the tip of the RGB. Hence the choice of mass loss prescription
will have little effect on the yield composition, but will effect
the amount of matter that is released and also the mass of the resulting
white dwarf.

We close this subsection by asserting that the evolutionary characteristics
of our models are comparable to the majority of previous work. The
main exception may be the AGB phase but we note that this has not
been studied in detail by anyone as yet. Indeed, our models appear
to be the first full TP-AGB calculations of these stars.

\section{Detailed Evolution at $2.0\,\textrm{M}_{\odot}$\label{section-m2z0-Structural}}

\subsection{MS and CNO Mini-Flash}

\begin{figure}
\begin{centering}
\includegraphics[width=0.8\columnwidth]{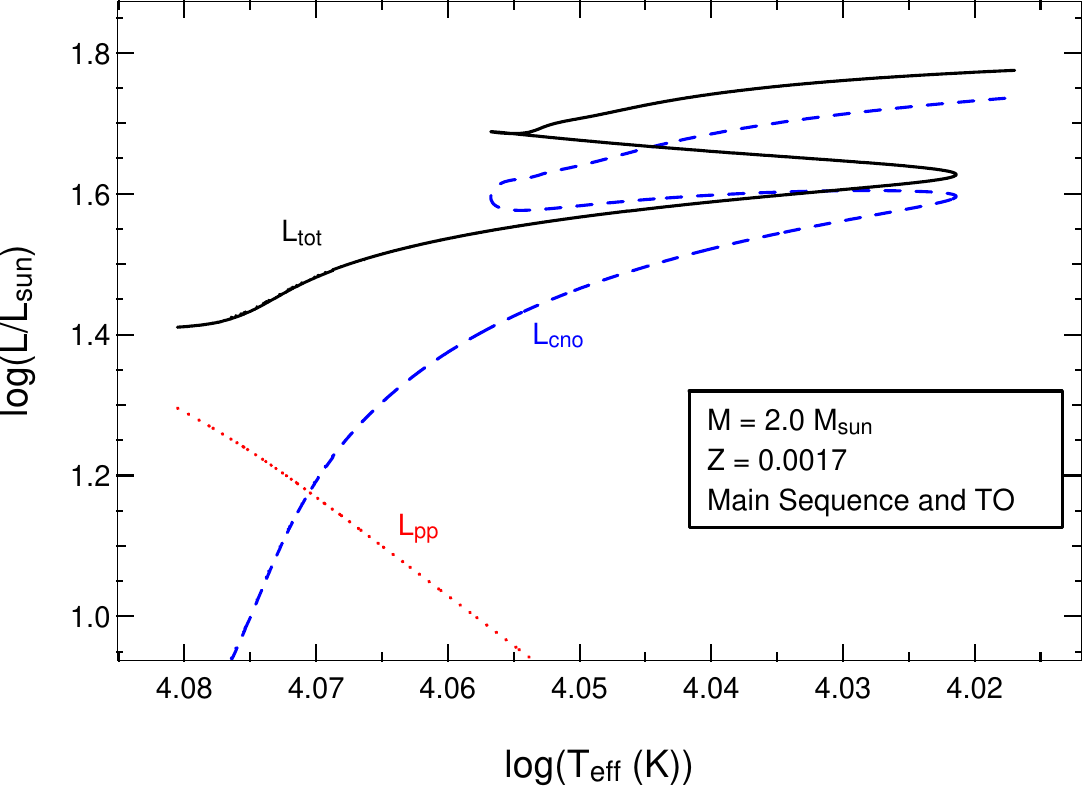}
\par\end{centering}
\caption{The main sequence HR diagram for our $Z=0.0017$ model. It can be
seen that the MS is dominated by CNO cycle energy generation in this
`normal' star. \label{fig-m2gc-HRD-LppLcno}}
\end{figure}

\begin{figure}
\begin{centering}
\includegraphics[width=0.8\columnwidth]{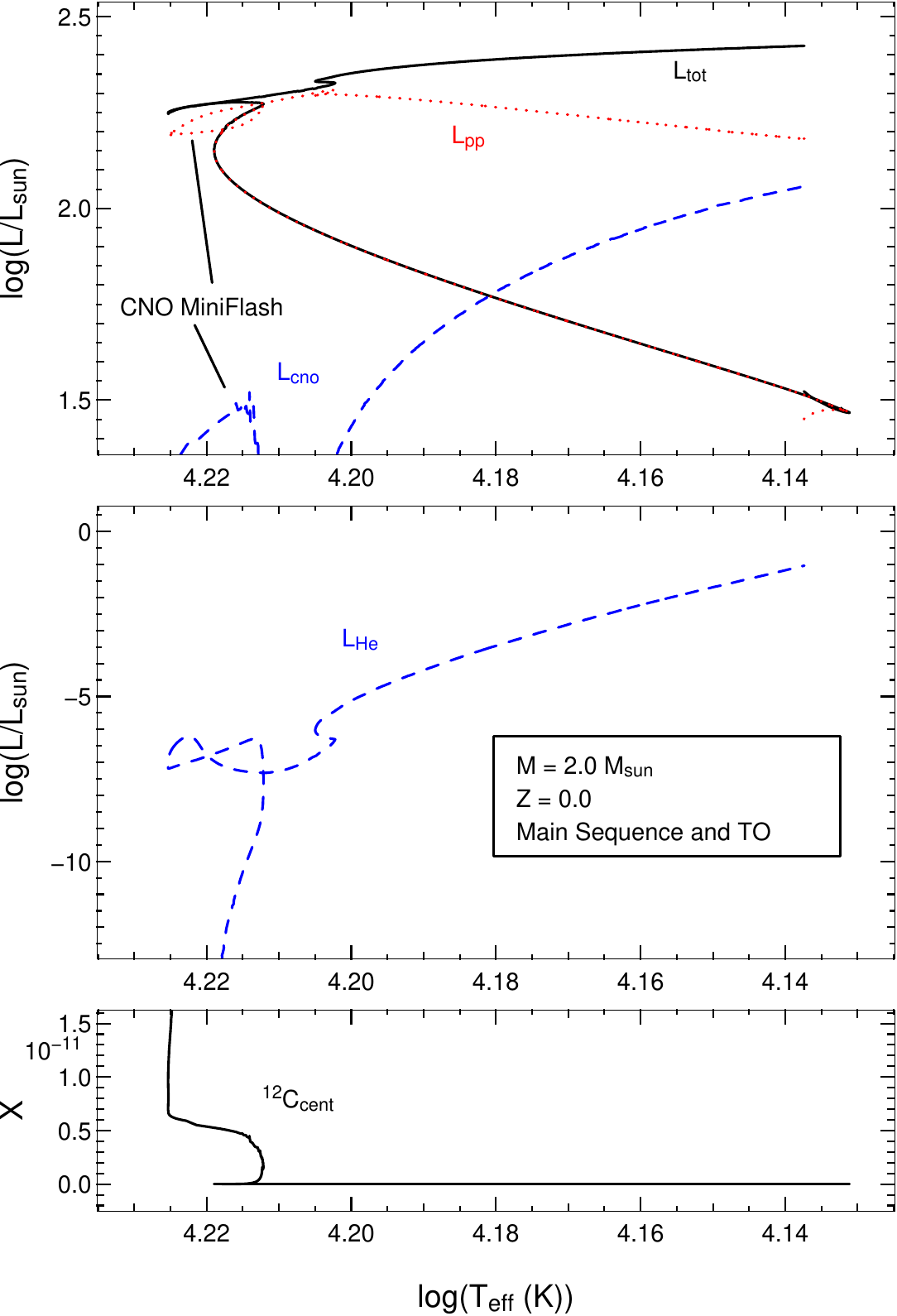}
\par\end{centering}
\caption{The MS HR diagram (top panel) for the 2 M$_{\odot}$ $Z=0$ model,
as well as the helium burning luminosity (middle) and resultant $^{12}$C
production (bottom). It can be seen that the p-p chains dominate for
the entire MS phase in this star, and that the star behaves more like
a low-mass model as its surface temperature increases with luminosity.
Also evident in the top panel is the brief CNO mini-flash, and the
increasing importance of the CNO cycles as the star moves to the shell
burning stage. The bottom panel indicates just how little carbon production
is needed to ignite the CNO cycles. \label{fig-m2z0y245-HRDs-LppLHeC12}}
\end{figure}

The main sequence evolution of the $Z=0$ model with $M=0.85$ M$_{\odot}$
described in the previous section was not substantially different
to that of a Pop II model, as Pop II stars of low mass also burn H
via the p-p chains on the MS. At higher masses ($M\gtrsim1$ M$_{\odot}$)
Pop II MS burning starts to become dominated by the CNO cycles. This
is demonstrated in Figure \ref{fig-m2gc-HRD-LppLcno} where we display
the various nuclear burning luminosities for a 2 M$_{\odot}$ Pop
II model (which we also refer to as our globular cluster (GC) model).
We have calculated this model specifically for comparison with our
2 M$_{\odot}$, $Z=0$ model. It has a metallicity of $Z=0.0017$
(scaled solar composition). All other input parameters were identical
to that of the $Z=0$ model. It can be seen in Figure \ref{fig-m2gc-HRD-LppLcno}
that p-p chain energy generation only dominates at the beginning of
the MS in the GC model, with the CNO cycles providing most of the
luminosity from then on. In fact, about half way through the MS (at
$\log(T_{eff})\sim4.05$) $80$\% of the star's total luminosity is
supplied by the CNO cycles. This contrasts sharply with the $Z=0$
model where, as can be seen in Figure \ref{fig-m2z0y245-HRDs-LppLHeC12},
$\sim100$\% of the MS luminosity is provided by the p-p chains throughout
the MS. As there are no CNO catalysts in the $Z=0$ model the CNO
cycles can not operate -- until the star produces its own $^{12}$C.

\begin{figure}
\begin{centering}
\includegraphics[width=0.8\columnwidth]{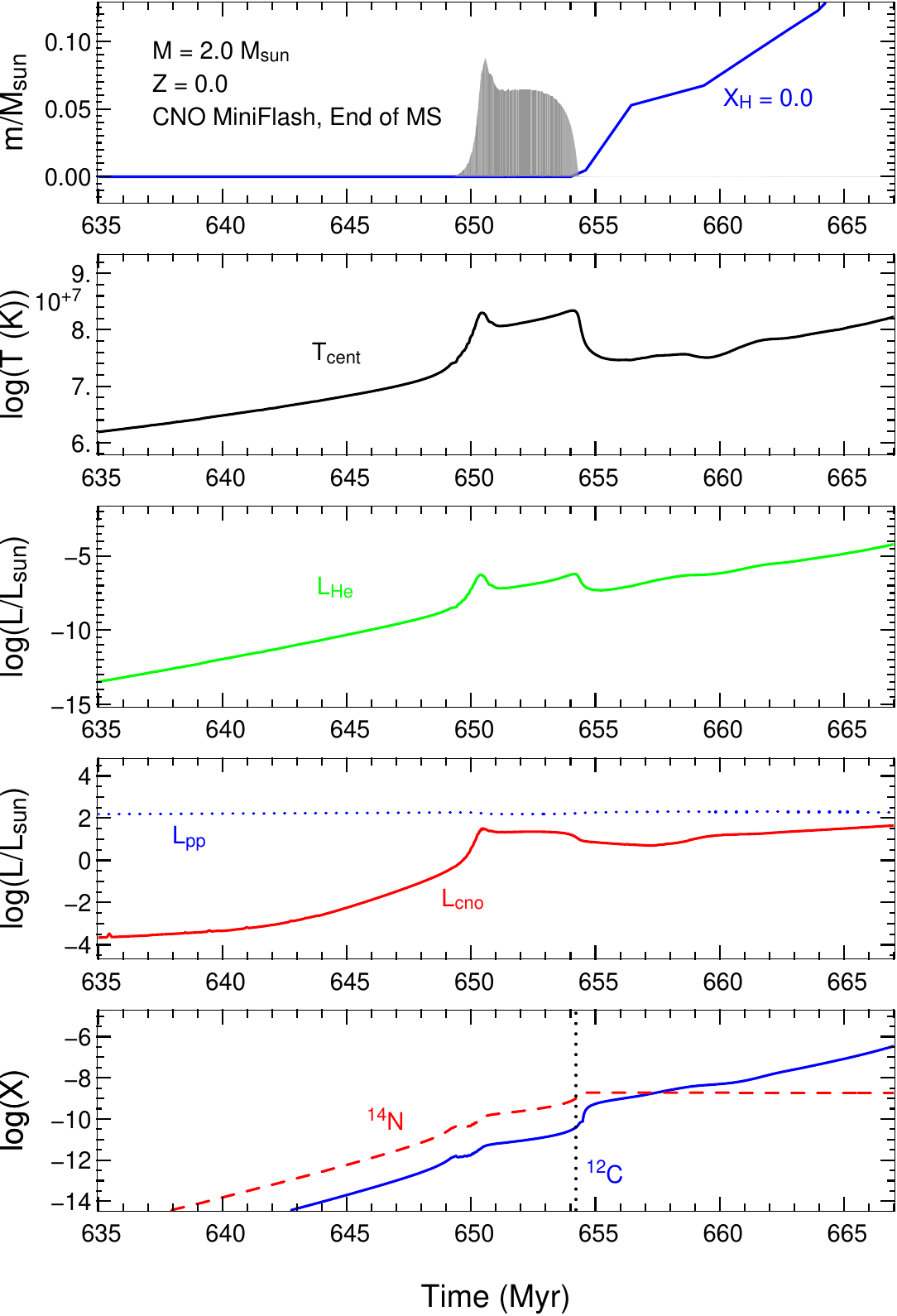}
\par\end{centering}
\caption{The anatomy of the CNO mini-flash in the 2 M$_{\odot}$ $Z=0$ model.
The small convective core is displayed in panel 1 (grey shading).
It can be seen that the miniflash lasts $\lesssim1$ Myr and that
the CNO burning core takes $\sim5$ Myr to burn the remaining hydrogen
(X$_{H}=0.013$). In the fourth panel we can see that the CNO burning
luminosity never dominates the energy production. The vertical dotted
line in the bottom panel indicates the time of core H exhaustion.
This event is also indicated by the fact that C stops getting cycled
to N, and the C abundance continues to rise due to the increasing
rate of $3\alpha$ burning. \label{fig-m2z0y245-plstarEv-CNOFlash-lums-conv}}
\end{figure}

\begin{figure}
\begin{centering}
\includegraphics[width=0.8\columnwidth]{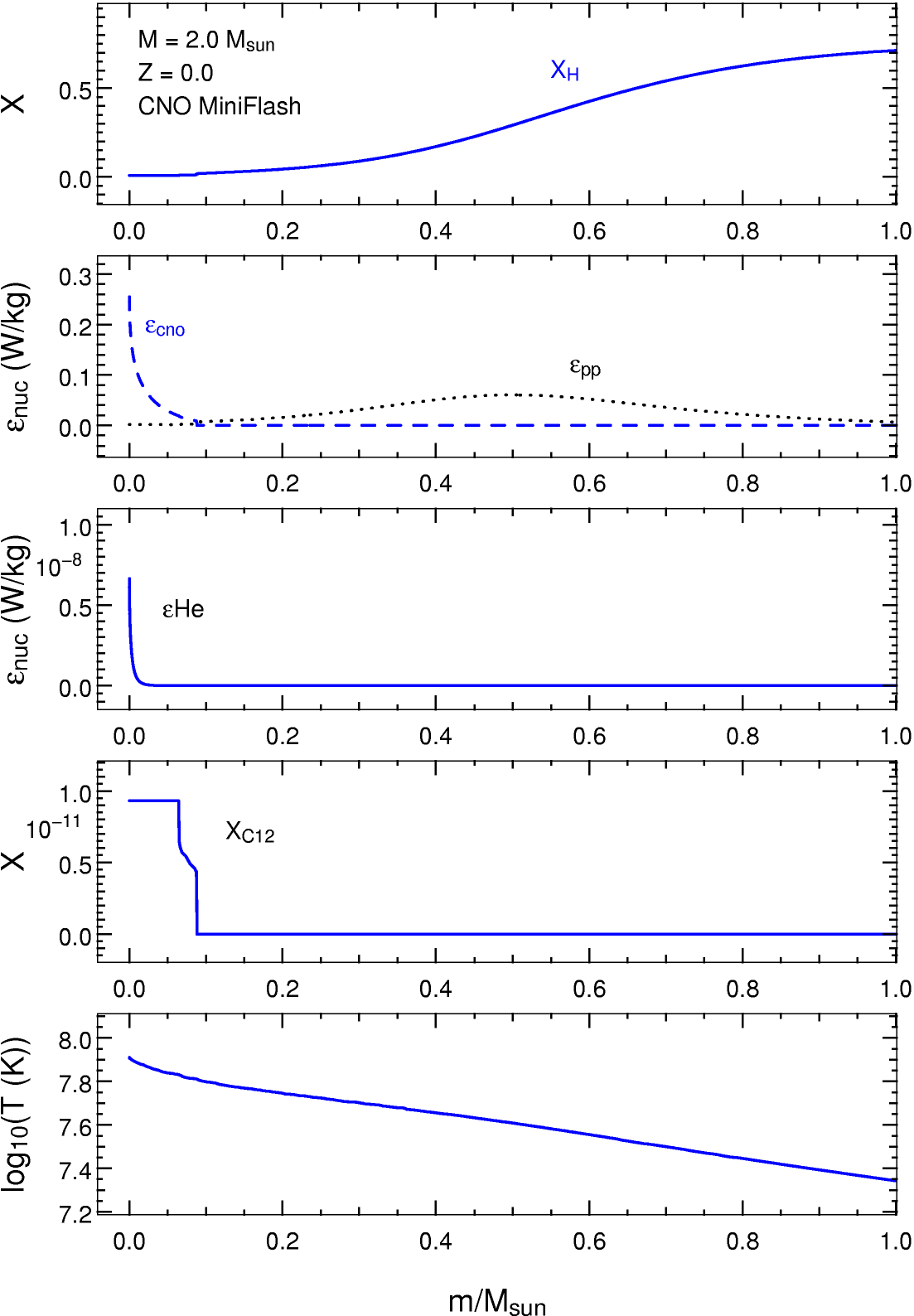}
\par\end{centering}
\caption{The run of some physical characteristics in a single model taken during
the CNO miniflash. The top panel shows just how little H is left in
the core. In the second panel we can see that the peak p-p burning
has already moved off centre, and that the CNO burning is very concentrated
right at the centre (due to its high temperature dependence). Also
evident is the small amount of He burning (third panel) that produces
the tiny amount of $^{12}$C (fourth panel). \label{fig-m2z0y245-pltsar-CNOminiflash}}
\end{figure}

Carbon production occurred in the $M=0.85$ M$_{\odot}$, $Z=0$ model
at the very bottom of the RGB, allowing the CNO cycles to burn the
tiny amount of remaining hydrogen. In the 2 M$_{\odot}$ model this
occurs just as the star turns off the MS, when there is a small but
significant amount of H still present ($X_{H}\sim0.010$). Once the
$^{12}$C abundance reaches $\sim10^{-12}$ there is a small thermal
runaway due to the sudden increase of CNO burning efficiency. The
burning is strong enough to give rise to a small convective core (see
top panel in Figure \ref{fig-m2z0y245-plstarEv-CNOFlash-lums-conv})
which mixes in some extra H, increasing the central H abundance a
little (up to $X_{H}\sim0.013$). This extra fuel slightly prolongs
the MS, giving rise to a CNO core burn that lasts $\sim5$ Myr before
the H is exhausted. We refer to the initial (minor) thermal runaway,
which lasts $\sim1$ Myr, as a `CNO Mini-Flash'. The rapid increase
in CNO burning luminosity can be seen in Figures \ref{fig-m2z0y245-HRDs-LppLHeC12}
and \ref{fig-m2z0y245-plstarEv-CNOFlash-lums-conv}. We show some
of the anatomy of the miniflash in Figure \ref{fig-m2z0y245-pltsar-CNOminiflash}.
This figure shows that the p-p burning has already moved off-centre,
forming a thick shell, whilst the CNO burning is concentrated at the
centre due to the high temperature dependence of these reaction cycles.
In the $Z=0$ HR diagram (Figure \ref{fig-m2z0y245-HRDs-LppLHeC12})
it can be seen that the star then assumes `normal' behaviour in terms
of surface temperature evolution, such that the temperature decreases
with luminosity. The previous evolution in the HR diagram is quite
the opposite to that of the $Z=0.0017$ model (see Figure \ref{fig-m2gc-HRD-LppLcno}),
as the temperature \emph{increases} with luminosity rather than decreasing.
This is reminiscent of the evolution of low-mass stars which are also
powered solely by the p-p chains (see Figure \vref{fig-m0.85z0y245-HRD}).
Another significant difference is that the $Z=0$ model increases
luminosity by a large amount during the MS. This is also similar to
low mass p-p powered models. This will have observational consequences
as the MS luminosity of the $Z=0$ model ranges from $\sim0.5$ to
1.5 dex higher than the $Z=0.0017$ model. The $Z=0$ model is also
much bluer throughout the MS. Interestingly these differences have
little effect on the MS lifetime -- the $Z=0$ lifetime is only $5\%$
\emph{longer} than that of the GC model (see Figure \ref{fig-m2z0y245-conv-lums-GCcompare}).

\subsection{Core He Burning and Second Dredge-Up}

\begin{figure}
\begin{centering}
\includegraphics[width=0.8\columnwidth]{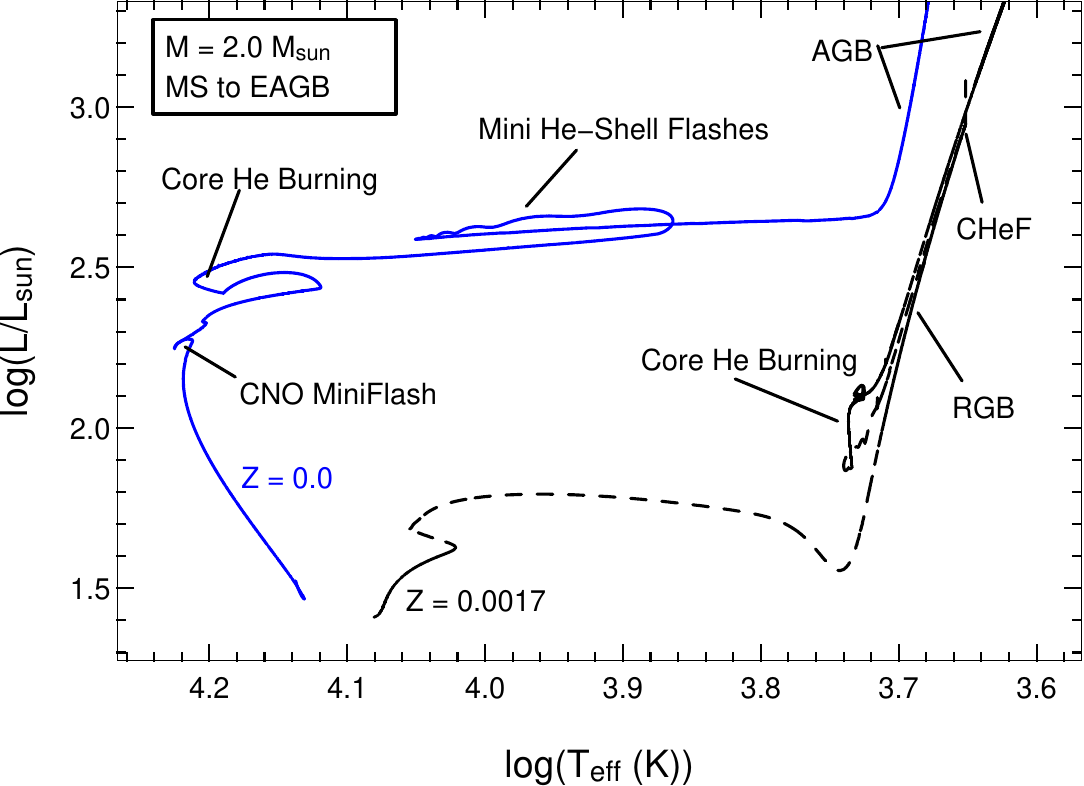}
\par\end{centering}
\caption{HR diagrams comparing the $Z=0$ and $Z=0.0017$ models. Significant
differences are evident in the $Z=0$ model, such as the large increase
in luminosity during the MS, the lack of an RGB and the lack of a
core He flash. Core He burning also occurs at a significantly higher
luminosity in the $Z=0$ model, and at a much higher surface temperature.
\label{fig-m2z0y245-HRDs-toEAGB-GCcompare}}
\end{figure}

\begin{figure}
\begin{centering}
\includegraphics[width=0.8\columnwidth]{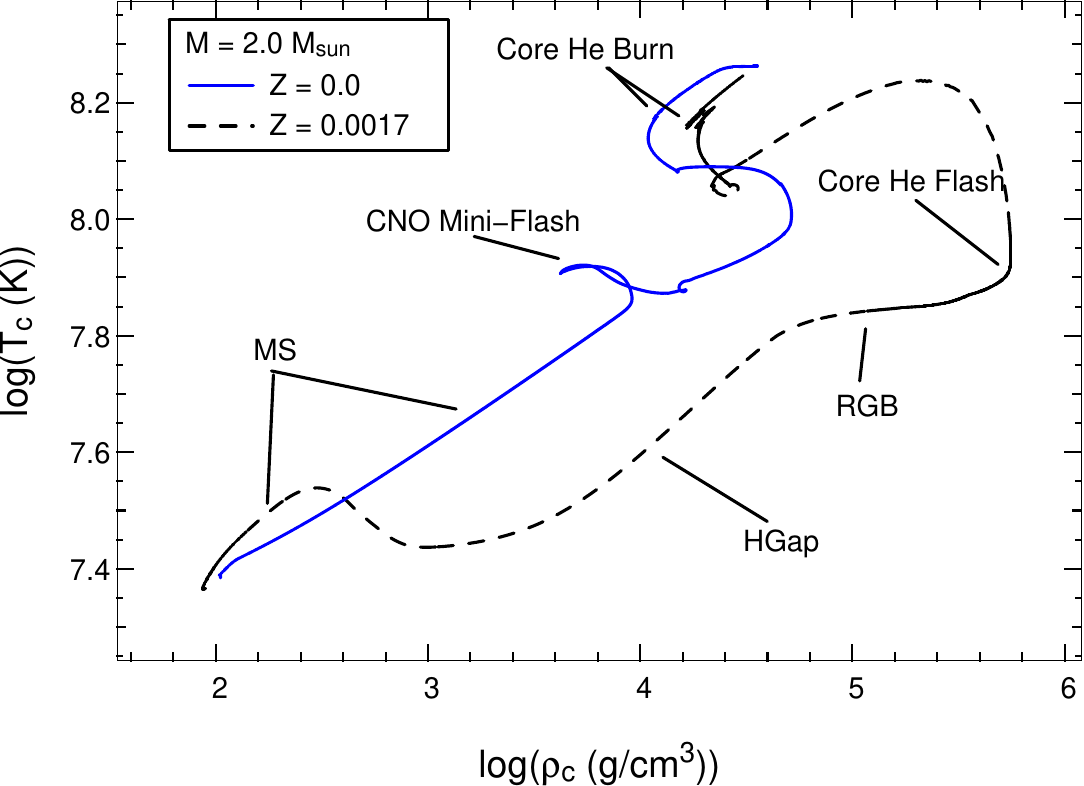}
\par\end{centering}
\caption{Central temperature versus central density for the $Z=0$ and $Z=0.0017$
2 M$_{\odot}$ models. The evolution to high temperatures and densities
on the MS in the $Z=0$ model is evident, as is the reduction in density
during the CNO miniflash. Unlike the $Z=0.0017$ model the $Z=0$
model reaches He ignition temperatures at a lower density, thereby
avoiding the core He flash due to the lower degeneracy. The $Z=0$
model also avoids the RGB phase, practically going straight from the
MS to quiescent core He burning. \label{fig-m2z0y245-RhoC-Tc-GCcompare}}
\end{figure}

\begin{figure}
\begin{centering}
\includegraphics[width=0.9\columnwidth]{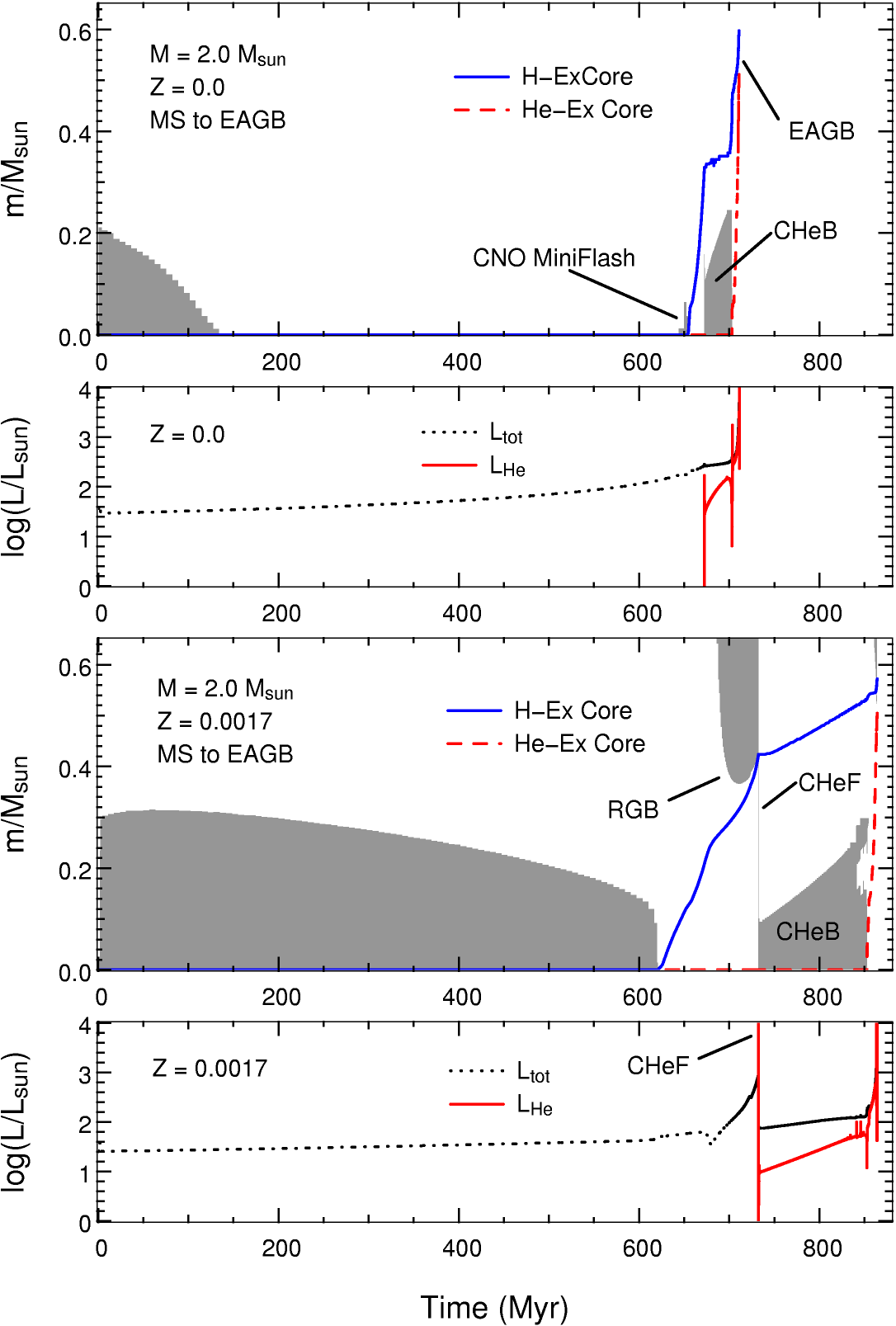}
\par\end{centering}
\caption{The MS and core He burning evolution in the $Z=0$ model (top two
panels) and $Z=0.0017$ model (bottom two panels). For each pair of
panels convection is shown in grey shading in the first panel and
the evolution of luminosities in the second. The striking evolutionary
differences in the $Z=0$ model can be seen here -- the occurrence
of the CNO miniflash, the non-occurrence of an RGB phase and core
He flash, and the very short-lived He burning lifetime. Interestingly
both stars have quite similar H burning lifetimes, with the $Z=0$
model lifetime (654 Myr) being slightly \emph{longer} ($\sim5\%$)
than the $Z=0.0017$ lifetime (623 Myr).\label{fig-m2z0y245-conv-lums-GCcompare}}
\end{figure}

In Figure \ref{fig-m2z0y245-HRDs-toEAGB-GCcompare} we present the
further evolution in the HR diagram for the $Z=0$ and GC models.
The evolution of the $Z=0$ model is strikingly different to that
of the Pop II model. The $Z=0$ star barely travels towards the red
(on the Hertzsprung Gap) before it ignites helium in the core. The
star thus avoids an RGB configuration entirely, quickly changing from
core H burning to core He burning. In addition to this He ignition
does not occur in a flash (as it does in the $Z=0.0017$ model), rather
the He ignites quiescently and the star settles on the HB straight
away. The reason for this is illustrated in Figure \ref{fig-m2z0y245-RhoC-Tc-GCcompare},
in which it can be seen that the $Z=0$ star reaches He burning temperatures
at much lower densities than the GC model. Thus the degeneracy in
the core is lower, and a thermal runaway is avoided, much like a higher
mass star at higher Z. Another interesting feature in Figure \ref{fig-m2z0y245-RhoC-Tc-GCcompare}
is the reaction of the star to the onset of the CNO miniflash. The
density reduces at (roughly) a constant temperature, due to an expansion
of the core. It is interesting to note that despite all the previous
differences in evolution the core He burning in both stars occurs
under approximately the same central conditions (although at a slightly
lower density in the $Z=0$ case). The core He burning lifetimes are
however very different. The $Z=0.0017$ model has a lifetime of 120
Myr whilst the $Z=0$ HB lifetime is only 30 Myr -- a factor of four
shorter. We note that other studies also find short-lived HBs at $Z=0$
(see next section for comparisons with other studies). 

\begin{figure}
\begin{centering}
\includegraphics[width=0.9\columnwidth]{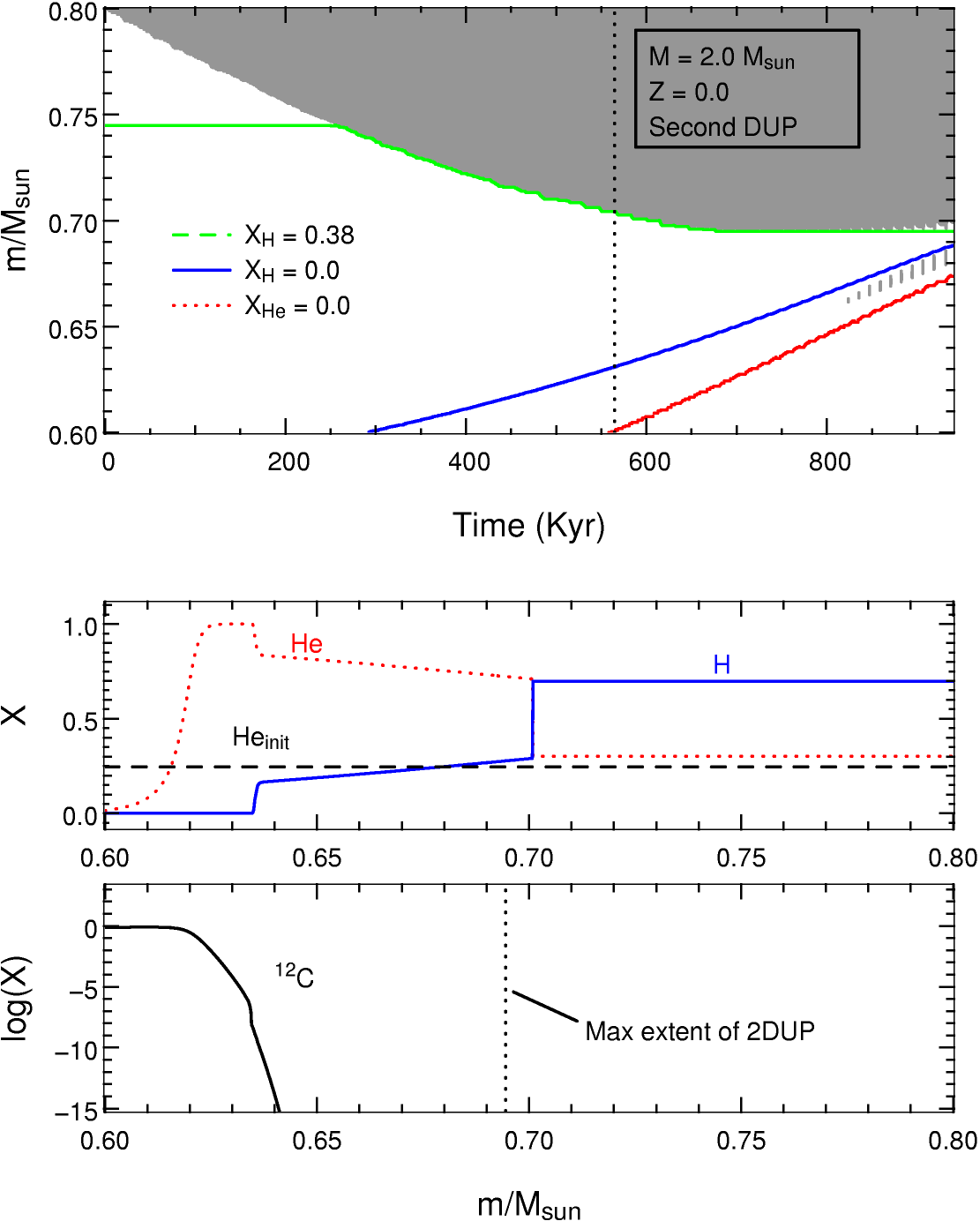}
\par\end{centering}
\caption{The second dredge-up episode in the $Z=0$ model (during the EAGB).
Note that the top panel shows evolution with time (which has been
offset), whilst the bottom two panels show the run of abundances against
\emph{mass} for a single model. The location in time at which the
single model was taken is represented by the vertical dotted line
in panel 1. It can be seen that the envelope reaches down into regions
that have been subjected to partial H burning. As these regions have
been burnt only via the p-p chains the only species enhanced by this
episode is $^{4}$He. The bottom panel shows the lack of carbon in
the region -- the $3\alpha$ reactions are only active right at the
bottom of the H burning shell. The dashed line in the middle panel
represents the initial He abundance ($Y=0.245$). The final He abundance
after 2DUP is $Y\sim0.30$. \label{fig-m2z0y245-2DUP-conv-abunds} }
\end{figure}

Due to the lack of an RGB first dredge-up does not occur in this model.
The `second' dredge-up (2DUP) event (actually the first dredge-up
event for this star) does occur in this model, but with limited consequences.
In Figure \ref{fig-m2z0y245-2DUP-conv-abunds} we show the evolution
of the convective envelope incursion along with the evolution of the
burning shells. We also show the abundance profiles of one model during
the 2DUP episode. It can be seen that the convective envelope does
reach down into regions that have been subjected to partial hydrogen
burning. However, as is evident from the carbon abundance profile,
the burning in this region has only been via the p-p chains. It is
only at the bottom of the H burning shell that carbon is being produced
in situ, allowing the CNO cycles to operate. Thus the usual surface
abundance increase in $^{14}$N cannot occur. The surface is however
polluted with a substantial amount of $^{4}$He, raising the envelope
abundance to $Y\sim0.30$ by the end of 2DUP (the model began with
$Y=0.245$). We note that the $^{3}$He abundance is also slightly
lowered, as it has been destroyed in the regions of partial H burning.
By contrast, in the $Z=0.0017$ model most of the surface abundance
changes happen at first dredge-up. The mass fraction of nitrogen increases
and that of carbon decreases. The helium mass fraction after FDUP
(and 2DUP) is $Y=0.26$, much lower than that in the $Z=0$ model
(this star had the same initial He abundance as the GC model). The
abundance changes during 2DUP are negligible in this star. 

\subsection{TP-AGB: Dual Shell Flashes and Third Dredge-Up\label{section-m2z0-DSF-TPAGB}}

\begin{figure}
\begin{centering}
\includegraphics[width=0.85\columnwidth]{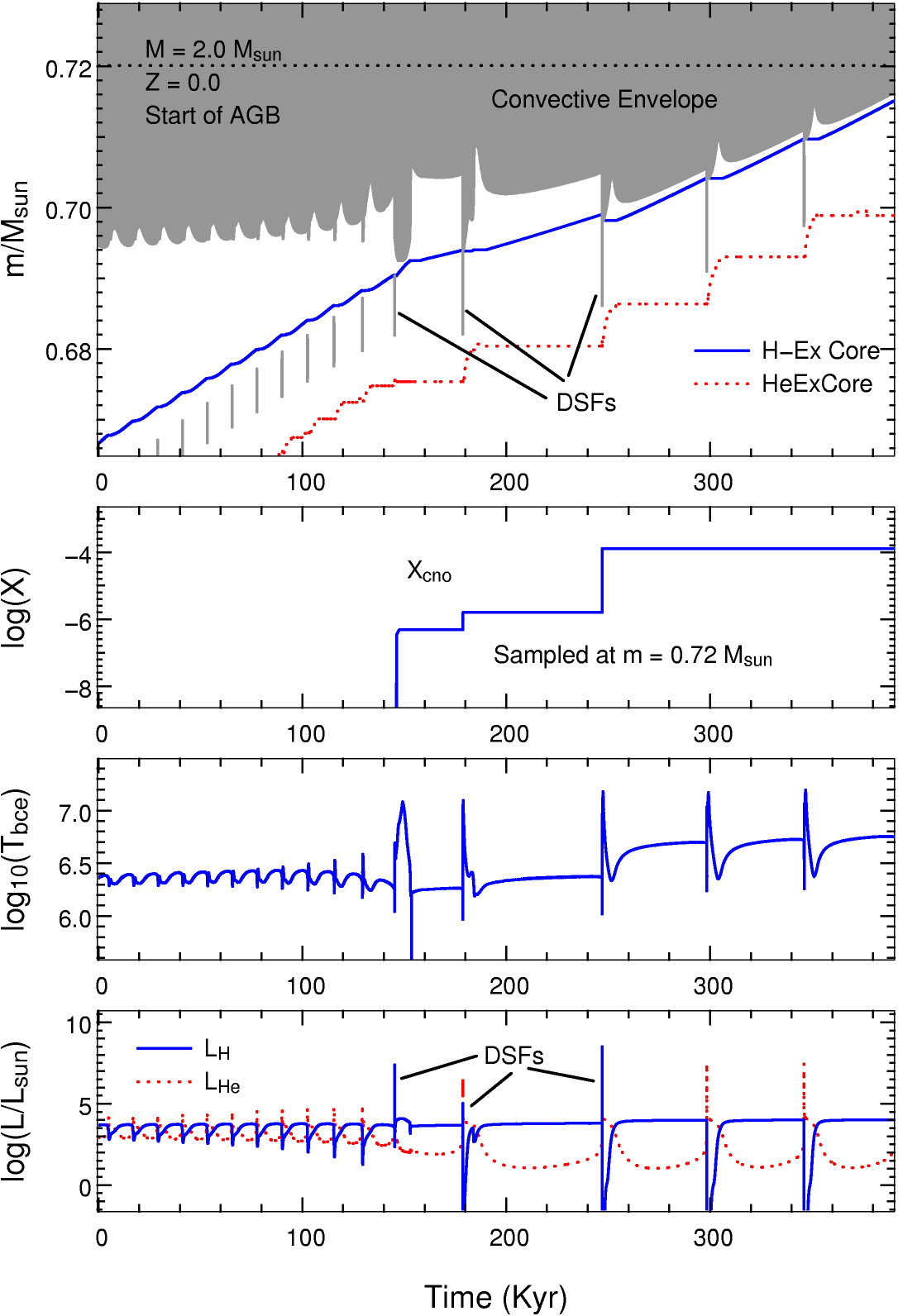}
\par\end{centering}
\caption{Evolution of the $Z=0$ model during the first few pulses on the TP-AGB.
A series of weak pulses can be seen leading up to a major H-flash
(the first flash marked `DSF'). This is the first of three consecutive
dual shell flashes (DSFs). The H shell flash is induced by the He
shell flash, whereby the convective intershell breaks through the
H-He discontinuity, consequently dredging down protons and dredging
up CNO nuclei -- much like the DCF in lower mass models. The CNO-processed
pocket that forms just above the H-He discontinuity is later dredged
up, polluting the envelope. This can be seen in the second panel which
shows that the CNO abundance in the envelope increases after each
DSF dredge-up episode. In the bottom panel it can be seen that the
magnitude of the H flashes ranges between $10^{5}$ and $10^{8.5}$
L$_{\odot}$. Once the envelope abundance reaches $Z_{cno}\sim10^{-4}$
the H-flashes cease and normal He-flashes continue (initially with
no 3DUP). \label{fig-m2z0y245-DSFs-cvn-lums-wide}}
\end{figure}

\begin{figure}
\begin{centering}
\includegraphics[width=0.9\columnwidth]{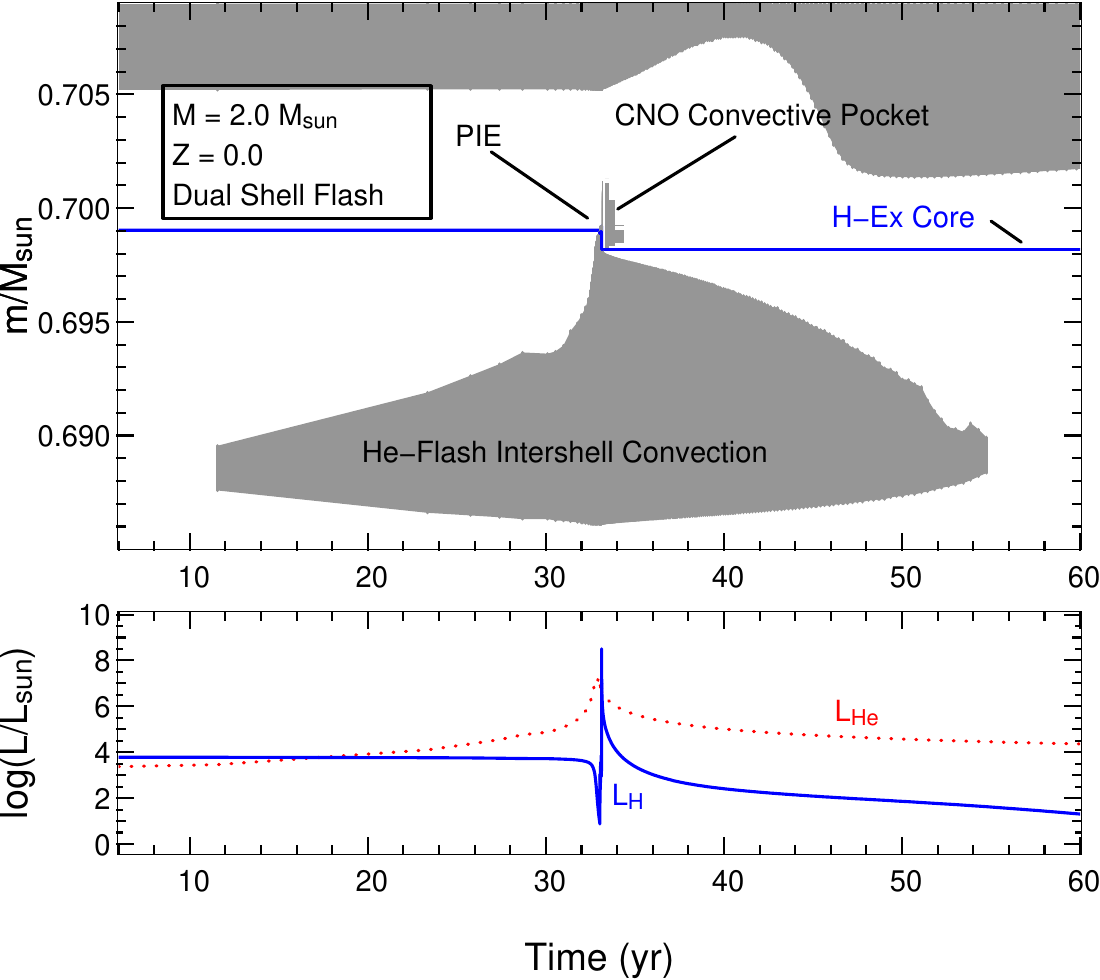}
\par\end{centering}
\caption{Zooming in on the evolution of the main DSF in the $Z=0$ model (the
third H-flash in Figure \ref{fig-m2z0y245-DSFs-cvn-lums-wide}). The
breaching of the H-He discontinuity can be seen in the top panel --
protons are mixed downwards and He intershell material mixed upwards.
This gives rise to an extra convection zone -- a small CNO burning
pocket. PIE stands for `proton ingestion episode'. In the bottom panel
it can be seen that the H burning luminosity temporarily exceeds that
of the He burning. Time has been offset for clarity. \label{fig-m2z0y245-DSF-conv-lums-zoomMedium}}
\end{figure}

In Figure \ref{fig-m2z0y245-DSFs-cvn-lums-wide} we present the evolution
of the $Z=0$ model during the first few pulses on the TP-AGB. A series
of weak pulses can be seen leading up to a major H-flash. This is
the first of three consecutive H flashes. Each H shell flash is induced
by a He shell flash, whereby the convective intershell breaks through
the H-He discontinuity (see Figure \ref{fig-m2z0y245-DSF-conv-lums-zoomMedium}
for a closer view of one of the DSFs). This event dredges down protons
and dredges up CNO nuclei -- much like the dual core flash (DCF)
in lower mass $Z=0$ models. We refer to these events as AGB dual
shell flashes (DSFs), as the luminosity peaks occur at roughly the
same time. A CNO-burning convective pocket forms just above the H-He
discontinuity (see Figure \ref{fig-m2z0y245-DSF-conv-lums-zoomMedium}).
This processed material is later dredged up, significantly polluting
the envelope. Once the envelope abundance reaches $Z_{cno}\sim10^{-4}$
the H-flashes cease and normal He-flashes continue. The star now appears
to be quite similar to a higher metallicity model -- it has a CO
core and a CNO-rich envelope. Indeed, it initially behaves in a more
`normal' (higher metallicity) fashion, with interpulse periods of
the order $10^{4}$ to $10^{5}$ years, with no 3DUP. 

\begin{figure}
\begin{centering}
\includegraphics[width=0.9\columnwidth]{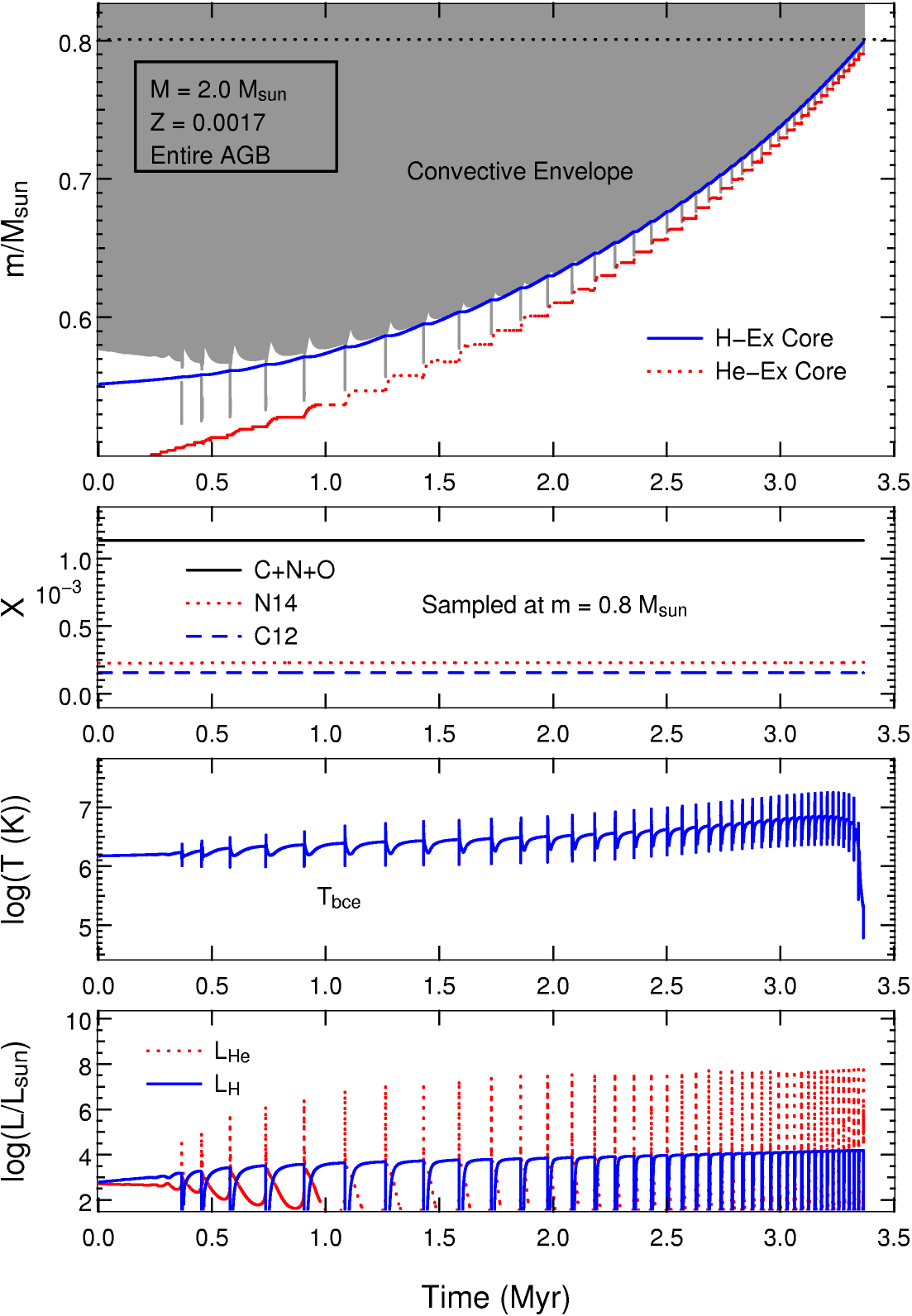}
\par\end{centering}
\caption{The entire AGB evolution of the $Z=0.0017$ model. The dotted line
in the top panel marks the sampling point for the abundance plot in
the second panel. The model experiences 41 thermal pulses (compared
to 282 in the $Z=0$ model). As can be seen by the lack of variation
in surface abundances (second panel) the star does not experience
3DUP or HBB. Time has been offset for clarity. \label{fig-m2gc-AGB-conv-lums-wide}}
\end{figure}

\begin{figure}
\begin{centering}
\includegraphics[width=0.9\columnwidth]{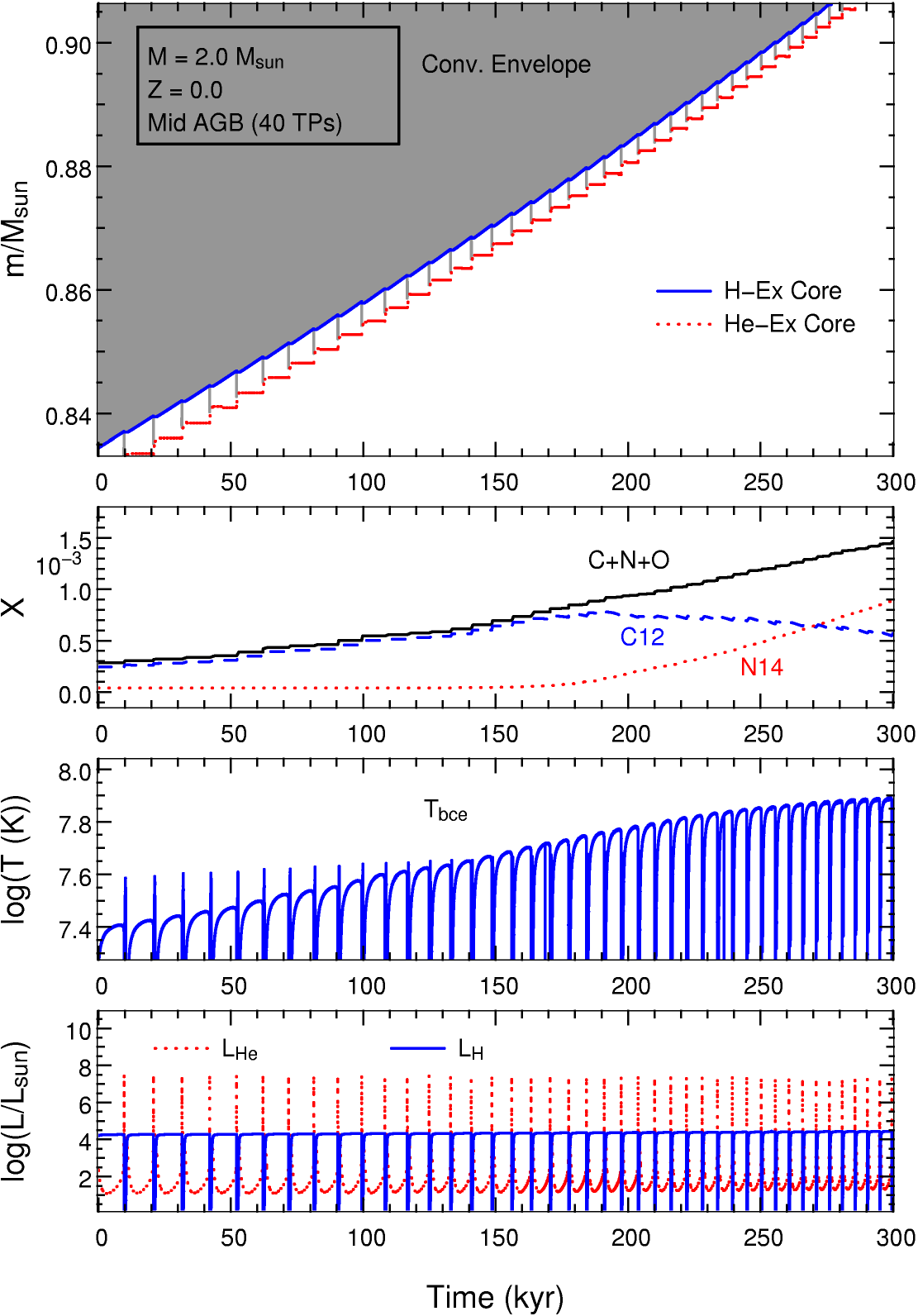}
\par\end{centering}
\caption{A section of the AGB evolution of the $Z=0$ model ($\sim40$ pulses).
The second panel shows the surface abundance evolution (sampled at
$m=1.0$ M$_{\odot}$). It can be seen that a small amount of 3DUP
occurs during each pulse cycle ($\lambda\sim0.01$), increasing the
abundances of the CNO isotopes. The conversion of $^{12}$C to $^{14}$N
due to HBB can also be seen towards the end of the timespan. This
HBB is due to the high temperature at the bottom of the convective
envelope, which is shown in panel 3. The interpulse periods are much
shorter than in the $Z=0.0017$ model. Time has been offset for clarity.
\label{fig-m2z0y245-AGB-conv-lums-wide}}
\end{figure}

However on closer inspection there are some significant differences
between the $Z=0$ and $Z=0.0017$ models. The key difference is that
the $Z=0$ model has a substantially higher core mass at the start
of the AGB. The core mass of the $Z=0$ model at the first major He
shell flash ($M_{c}=0.69$ M$_{\odot}$) is $\sim20\%$ larger than
that of the $Z=0.0017$ star ($M_{c}=0.56$ M$_{\odot}$). As mentioned
above the $Z=0$ model also has a higher He abundance in the envelope.
It also has no nuclei heavier than those produced by H and He burning.
These differences soon lead to a divergence in AGB evolution. As shown
in Figure \ref{fig-m2gc-AGB-conv-lums-wide} the $Z=0.0017$ model
experiences no 3DUP throughout its AGB evolution -- the surface abundances
remain the same. The $Z=0$ model on the other hand does have 3DUP
episodes. This is evident in Figure \ref{fig-m2z0y245-AGB-conv-lums-wide}
where it can be seen that the envelope is progressively enriched by
CNO nuclei, albeit in small steps. The dredge-up parameter $\lambda$
is very small throughout the evolution, reaching a maximum of only
$\sim0.01$. Also evident in Figure \ref{fig-m2z0y245-AGB-conv-lums-wide}
is that the temperature at the base of the convective envelope reaches
quite high temperatures ($T_{bce}\sim7.9$ K), such that hot bottom
burning (HBB) is active in this star. In the same figure it can be
seen that the $^{12}$C in the envelope is gradually cycled to $^{14}$N.
Indeed, the star's envelope becomes dominated by $^{14}$N (in terms
of heavy nuclei) towards the end of the AGB. We note that we have
included an approximation to the increase in opacity due to the excess
nitrogen in our $Z=0$ models (in most studies only changes in C and
O are taken into account, as the standard opacity tables only provide
this option). The GC model on the other hand never attains envelope
temperatures high enough for HBB (see Figure \ref{fig-m2gc-AGB-conv-lums-wide}).

\begin{figure}
\begin{centering}
\includegraphics[width=0.9\columnwidth]{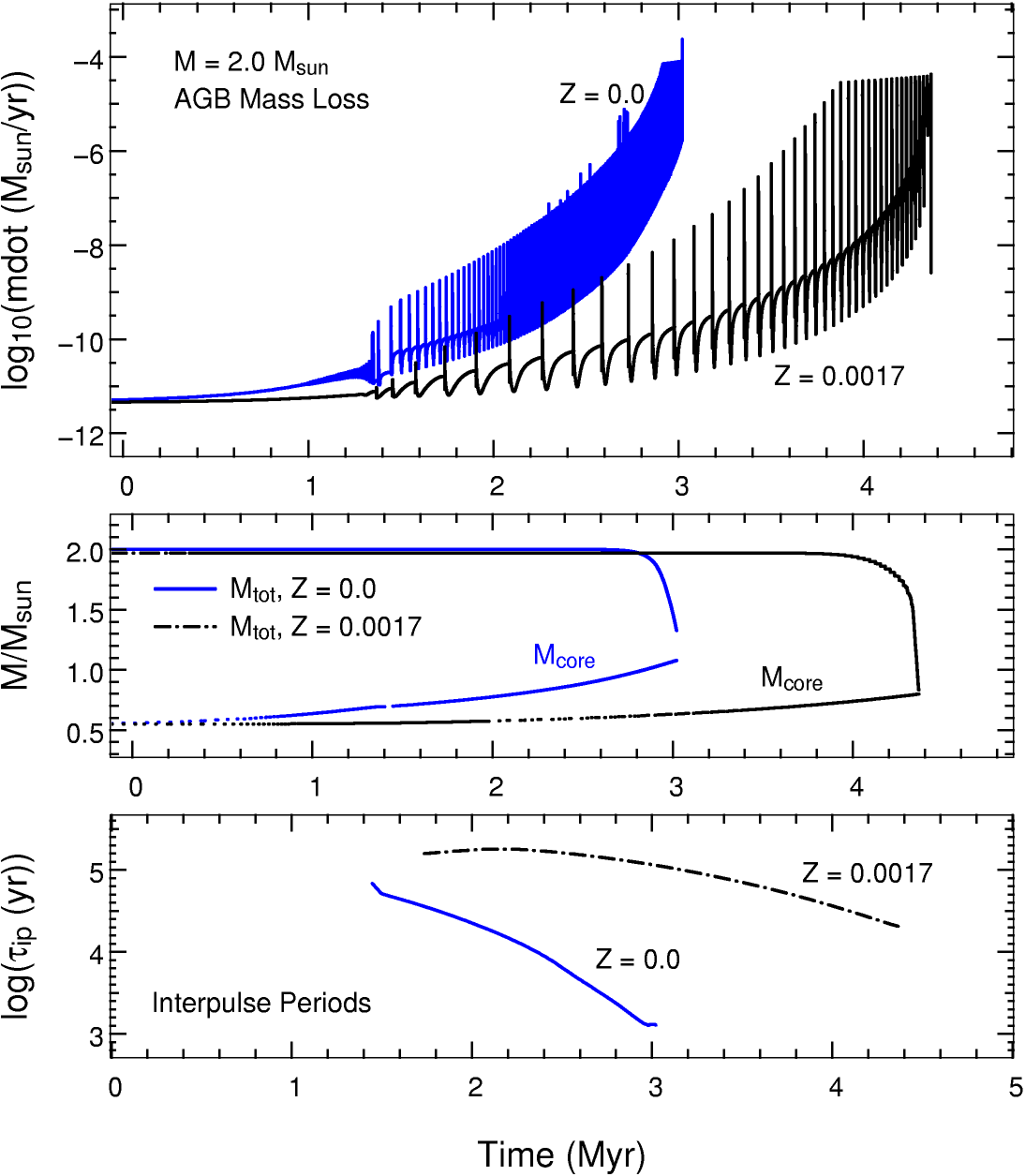}
\par\end{centering}
\caption{The AGB mass-loss history for both stars. It can be seen that the
$Z=0$ model quickly evolves to a higher mass loss rate than the $Z=0.0017$
model (the luminosity and core mass is higher in this model), and
that the AGB lifetime of the $Z=0$ model is roughly half that of
the GC model. In panel 2 we show the evolution of the total mass of
the stars and well as the evolution of the core masses. The gap between
the two $Z=0$ curves indicates that not all the envelope was lost
before the code failed to converge. The core mass in this case would
not increase much more as the star has entered the superwind stage
at the very end of the AGB. The bottom panel displays the evolution
of the interpulse periods. It can be seen that the $Z=0$ model has
very short interpulse periods by the end of the AGB. Time has been
offset for clarity. \label{fig-m2z0y245-AGB-mloss-mcore-IPperiod-GCcompare}}
\end{figure}

A feature of both models is that the interpulse period gradually decreases.
This occurs as the core mass of each star increases. The core mass
of the $Z=0$ model increases at a much faster rate than that of the
GC model. The bottom panel in Figure \ref{fig-m2z0y245-AGB-mloss-mcore-IPperiod-GCcompare}
demonstrates the increasing frequency of the pulses and also shows
just how short they become in the $Z=$ model, reducing from $\sim10^{5}$
yr to $\sim10^{3}$ yr. In the top panel of the same figure it can
be seen that this results in a great many pulses on the $Z=0$ AGB
-- 282 in total, as compared to only 41 in the $Z=0.0017$ model
-- even though the $Z=0$ AGB lifetime is a factor of two shorter
than the GC model's ($\sim$1.6 Myr compared to $\sim$ 3.5 Myr).
Finally we note that the core masses at the end of the AGB (ie. the
white dwarf masses) are substantially different -- the $Z=0$ WD
mass (1.1 M$_{\odot}$) is $\sim25\%$ larger than that of the $Z=0.0017$
(0.8 M$_{\odot}$).

\subsection{Comparison With Previous Work\label{subsec-m2z0-ComparePrevStudies}}

In Table \ref{Table-m0.85z0y245-literatureComparisons} we list, to
the best of our knowledge, all the $Z=0$ models of $M\sim2.0$ M$_{\odot}$
in the literature. We find that there are not many key values published
with which we can make comparisons. This is due to studies focusing
on particular phases of evolution or focusing on describing models
of other masses. 

\subsubsection*{MS, CNO Mini-Flash, Core He Burning and 2DUP}

Two studies give main sequence lifetimes. \citet{2001AA...371..152M}
find core H burning lasts 635 Myr in their model which compares very
well with our value of 654 Myr, being only $3\%$ shorter. \citet{2002ApJ...570..329S}
find a MS lifetime of 742 Myr, in reasonable agreement with ours,
being $12\%$ longer. As our MS lifetime lies between those found
by these two studies we suggest that it is in reasonable agreement
with the literature to date, but stress that the sample is still quite
small. In regards to the CNO miniflash we note that all the studies
find that this phenomena occurs. As it occurs right at the end of
the MS and thus involves such a small amount of H burning we suggest
that it is of little observational consequence, particularly since
core He burning proper begins soon after.

The same two studies that provided MS lifetimes also provide core
He burning lifetimes. Here there is more variation, with \citet{2001AA...371..152M}
finding a lifetime $\sim40\%$ longer than ours (43 Myr versus our
30 Myr) and \citet{2002ApJ...570..329S} finding a lifetime \emph{more
than a factor of 2} \emph{longer} than ours (68 Myr). We are unsure
why the lifetime of \citet{2002ApJ...570..329S} is so much longer
than ours and that found by \citet{2001AA...371..152M}. Such large
discrepancies between studies indicates that further work is really
needed in this area of $Z=0$ intermediate mass stellar evolution,
which we shall pursue as a future study. 

An interesting feature of our model is the extent to which helium
is enhanced in the envelope due to the second dredge-up event. We
find that the He mass fraction increases from $Y=0.245$ to 0.30.
\citet{2002ApJ...570..329S} find a similar increase, from $Y=0.23$
to 0.31. The surface He abundance in the 2.0 M$_{\odot}$ model of
\citet{2001AA...371..152M} barely increases at all (see Table \ref{Table-m0.85z0y245-literatureComparisons}
) but interestingly their 2.1 M$_{\odot}$ model ends up with exactly
the same abundance as our model ($Y=0.30$). This suggests that 2.0
M$_{\odot}$ is near the lower limit of significant 2DUP. 

\begin{sidewaystable}

\begin{center}

\begin{tabular}{|c|c|c|c|c|c|c|c|c|}
\hline 
Authors & $M_{*}$ & $\tau_{MS}$ & $\tau_{He}$ & $Y_{2DUP}$ & DSF & $M_{c,DSF}$ & $Z_{cno}(10^{-6})$ & $M_{WD}$\tabularnewline
\hline 
\hline 
\citet{1993ApJS...88..509C}\footnote{This model has $log(Z/Z_{\odot})=-8$.} & 2.3 & -- & -- & -- & -- & -- & -- & --\tabularnewline
\hline 
\citet{2000ApJ...529L..25F} & 2.0 & -- & -- & -- & Yes & -- & -- & --\tabularnewline
\hline 
\citet{2001AA...371..152M} & 2.0 & 635 & 43 & 0.23 (0.30)\footnote{Their 2.1 M$_{\odot}$ model has $Y_{2DUP}=0.30$} & -- & -- & -- & --\tabularnewline
\hline 
\citet{2002ApJ...570..329S} & 2.0 & 742 & 68 & 0.31 & Yes & 0.69 & 6 & --\tabularnewline
\hline 
\citet{2003ASPC..304..318H} & 2.0 & -- & -- & -- & Yes & -- & $\sim100$ & --\tabularnewline
\hline 
\citet{2004ApJ...611..476S} & 2.0 & -- & -- & -- & Yes & -- & -- & --\tabularnewline
\hline 
\textbf{Current Study ($Z=0$)} & \textbf{2.0} & \textbf{654} & \textbf{30} & \textbf{0.30} & \textbf{Yes} & \textbf{0.70} & \textbf{$0.5,2,160$}\footnote{This model experienced three DSFs.} & \textbf{1.1}\tabularnewline
\hline 
\textbf{Current Study ($Z=0.0017$)} & \textbf{2.0} & \textbf{623} & \textbf{120} & \textbf{0.26} & \textbf{No} & \textbf{--} & \textbf{--} & \textbf{0.8}\tabularnewline
\hline 
\end{tabular}

\caption{Comparing our $M=2.0$ M$_{\odot}$ model with similar $Z=0$ models
from the literature. Also included for further comparison is our GC
model ($Z=0.0017)$ and the 2.3 M$_{\odot}$, $log(Z/Z_{\odot})=-8$
model by \citet{1993ApJS...88..509C}. The table is ordered by date.
Note that we were unable to obtain many of the comparison values (indicated
by `--'), as each study tends to focus on particular facets of the
evolution. The table columns show $M_{*}$(initial stellar mass),
$\tau_{MS}$ (age at MS turn-off), $\tau_{He}$ (core He burning lifetime),
$M_{core}$ (mass of H-exhausted core at time of main dual shell flash),
DSF (whether or not the He convective zone penetrated the H-shell,
or was about to, producing a dual shell flash) and $Z_{cno}$ (the
C+N+O metallicity of the envelope after the DSF dredge-up). All lifetimes
are in Myr and all masses are in M$_{\odot}$.\label{Table-m0.85z0y245-literatureComparisons}}

\end{center}

\end{sidewaystable}

\subsubsection*{AGB: Dual Shell Flash and Third Dredge-Up}

Moving on to the AGB we find only one study that provides a quantitative
comparison for our model. This indicates that this phase of evolution
has not been well studied. Qualitatively all studies (that have evolved
their models this far) appear to find the same behaviour at the beginning
of the AGB. In particular they all find that a proton ingestion episode
(PIE) occurs at the start of the TP-AGB due to the penetration of
the shell He convective zone into the H-rich layers just above. This
is the dual shell flash (DSF) phenomenon that we detailed in the previous
subsection. It results in a significant pollution of the envelope,
thus allowing for the (more) normal evolution of the star on the AGB.
The fact that all studies have found this indicates that the phenomenon
is robust (at least in 1D stellar models) since the differing input
physics between stellar codes has not inhibited it. As a key indicator
of the conditions for the dual shell flash we have chosen the mass
of the H-exhausted core at the time of the first DSF. \citet{2002ApJ...570..329S}
find that the core mass at this stage is 0.69 M$_{\odot}$ -- in
almost perfect agreement with our value of 0.70 M$_{\odot}$. We note
that this is despite the large difference in He burning lifetime between
our model and theirs. \citet{2002ApJ...570..329S} also calculate
a number of further thermal pulses on the AGB. Examining their Figure
10 we see that the interpulse period in this early stage of the TP-AGB
is $\sim10^{4.8}$ yr. This compares well with our interpulse periods
for the same stage. The model of \citet{2002ApJ...570..329S} diverges
with ours soon after this though, as deep 3DUP is evident in their
graph. Whilst our model has a characteristic $\lambda\sim0.01$ their
model has $\lambda\sim0.6$ -- even after only 12 pulses. The reason
for this is most likely because they include a small amount of overshoot
on the AGB, whilst we have not. Another difference between these two
models is the number of DSFs that occur at the start of the AGB. Our
model experiences three, whilst theirs has only one. We note however
that their 3 M$_{\odot}$ model experiences \emph{four} DSFs. They
comment that the star only stops experiencing DSFs when the envelope
CNO abundance is $\sim10^{-7}$. Finally we note that the final $Z_{cno}$
abundance in the envelope of our model after the DSF episodes is $\sim10^{-4}$.
This compares well with that given by \citet{2003ASPC..304..318H}
(see Table \ref{Table-m0.85z0y245-literatureComparisons}) but is
2 dex lower than that found by \citet{2002ApJ...570..329S}. Although
this is a large discrepancy it appears that it may be insignificant
(in terms of nett envelope pollution and chemical yield), as the operation
of the 3DUP soon increases the $Z_{cno}$ abundance by more than an
order of magnitude above this value. However, if the pollution of
the envelope through this event is the reason that 3DUP occurs, which
is probably due to the increase in opacity it causes, then the resultant
pollution from the DSFs is important. We note however that it is the
model of \citet{2002ApJ...570..329S} that has a much lower pollution
from the DSF that has the deepest dredge-up, although we restate that
they have included some overshoot so it is not an equivalent comparison.
These differences, and the lack of quantitative studies of this phase
of evolution (at this mass) suggest that this is another area that
requires more investigation. We attempt to clarify the nature of the
DSF phenomenon in the broad discussion of all our $Z=0$ models in
Section \ref{sec-All-Z0-Structural}.

\section{Overview and Discussion of All $Z=0$ Models\label{sec-All-Z0-Structural}}

\subsection{M = 0.85 M$_{\odot}$ to 3.0 M$_{\odot}$ }

Apart from the two models discussed in detail in the preceding two
sections we have calculated two more models, with masses 1.0 M$_{\odot}$
and 3.0 M$_{\odot}$. Unfortunately we encountered convergence difficulties
with the the 3.0 M$_{\odot}$ model during the AGB so this model has
not been taken to completion. It has however been evolved through
the DSF phase and also 950 thermal pulses on the AGB so we are able
to present some results for this model and thus include it in the
following discussion. 

\begin{figure}
\begin{centering}
\includegraphics[width=0.8\columnwidth]{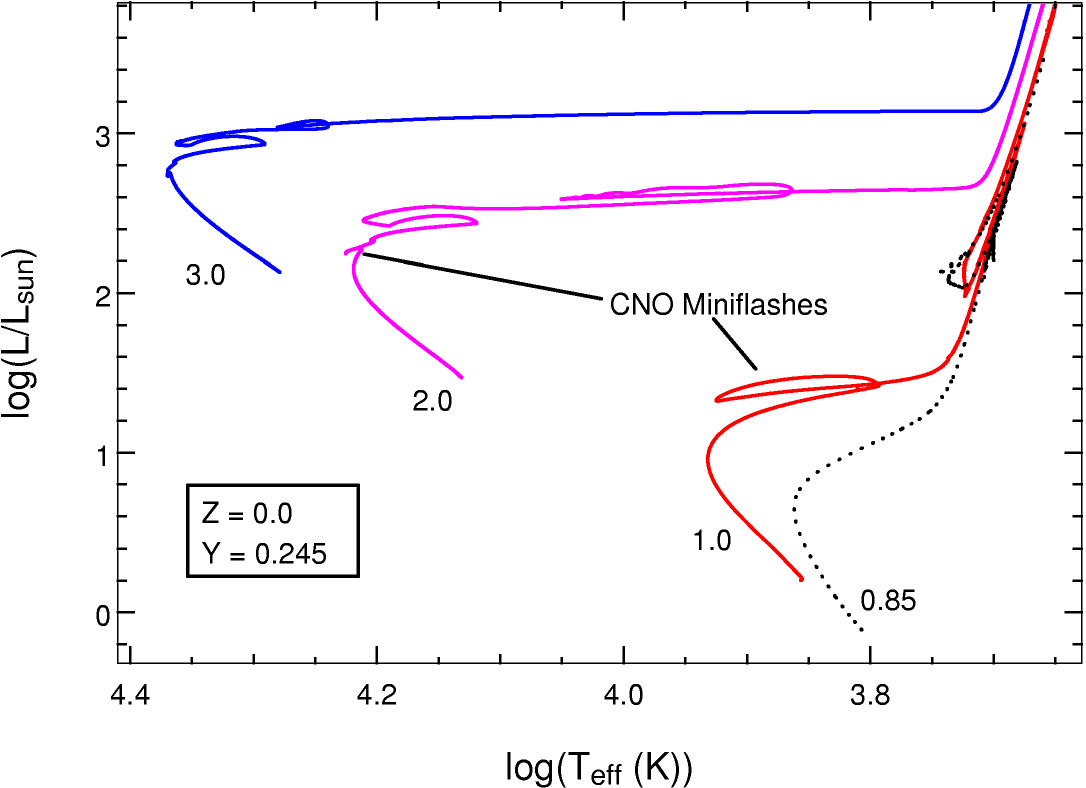}
\par\end{centering}
\caption{The HR diagrams for all our $Z=0$ models. The initial mass of each
is marked on the diagram (in units of M$_{\odot}$). Also indicated
are the CNO miniflashes in two of the models. All the models experience
the CNO miniflash to varying degrees. We note that the 2 M$_{\odot}$
model does not experience a core He flash, unlike more metal-rich
models. It is thus classed as an intermediate-mass star. \label{fig-HRDs-all-Z0}}
\end{figure}

In Figure \ref{fig-HRDs-all-Z0} we show the evolution of all our
$Z=0$ ($Y=0.245$) models in the HR diagram. It can be seen that
there is a transition between 1 and 2 M$_{\odot}$ such that stars
do not reach an RGB configuration due to the early onset of quiescent
core He burning. Like our 0.85 M$_{\odot}$ model we find that our
1 M$_{\odot}$ model also experiences a dual core flash (DCF) at the
tip of the RGB. However the 2 M$_{\odot}$ model does not undergo
a core He flash -- unlike the $Z=0.0017$ model of the same mass,
and indeed all more metal-rich models. Thus the transition between
low mass and intermediate mass stars, as defined by the occurrence
or non-occurrence of the core He flash, is much lower at $Z=0$. Our
work shows that the cut-off is below 2 M$_{\odot}$ but above 1 M$_{\odot}$,
which is consistent with other studies (eg. \citet{2001AA...371..152M}
find the cut-off to be at $\sim1.1$ M$_{\odot}$). We thus class
the 2 M$_{\odot}$ model as an intermediate-mass star.

In Table \ref{table-Z0-lifetimes-all} we present the lifespans of
various evolutionary stages for our models. In the previous two sections
we have compared our lifetimes with all other studies available (for
masses 0.85 M$_{\odot}$and 2.0 M$_{\odot}$). Here we shall compare
our results primarily with the work by \citet{2001AA...371..152M}
as they have calculated a comprehensive grid of models. We note however
that they did not calculate models for the AGB phase so not all comparisons
are possible. Along with our MS and core He burning lifetimes in Table
\ref{table-Z0-lifetimes-all} we display the percentage differences
between our lifetimes and those of \citet{2001AA...371..152M}. It
can be seen that the MS lifetimes are in reasonable agreement, never
being more than $\sim11\%$ different than each other. This is despite
the inclusion of a small amount of overshoot in the Marigo et al.
models (for $M>1.1$ M$_{\odot}$). We note however that overshoot
should not have much effect on the MS at $Z=0$ as the convective
cores are smaller and shorter-lived, due to the dominance of the p-p
chains. The core He burning lifetimes are consistently shorter in
our models, by values ranging from $\sim10\rightarrow64\%$. This
is explicable by the fact that \citet{2001AA...371..152M} have used
overshoot in all models during core He burning. Taking this into account
we suggest that all our models are, in general, in fair agreement
with those of Marigo et al., except for the $M=0.85$ M$_{\odot}$
model during core He burning. An important factor here is that Marigo
et al. did not evolve their low mass stars through the core He flash/DCF.
They constructed zero-age HB models to investigate the further evolution.
Thus it is probably not surprising that the lifetimes are different
in this case. We compared this model to other studies in an earlier
section (see Table \vref{Table-m0.85z0y245-literatureComaprisons}),
only finding one with which we could make a direct comparison. In
this case the 0.80 M$_{\odot}$ model of \citet{2004ApJ...609.1035P}
had a considerably \emph{shorter} core He burning lifetime than ours.
We concluded that more work needs to be done in this area. 

Comparisons between our other lifetimes given in Table \ref{table-Z0-lifetimes-all}
with the literature is not possible as studies tend not to provide
these. We provide them for future studies to make comparisons with.
These lifetimes are useful in terms of predicting the relative numbers
of stars in a particular stage of evolution at a given time. 

The details of an extra $Z=0$ model is also given in Table \ref{table-Z0-lifetimes-all}.
It has an initial mass of 3 M$_{\odot}$ but a different initial helium
abundance, being $Y=0.230$ rather than 0.245. We have added this
model because it did complete the AGB phase whilst the $Y=0.245$
model did not. We have discussed the nucleosynthesis of this 3 M$_{\odot}$
model in Section \vref{section-Yields-NS-all-Z0} and thus provide
some structural evolution details here (and in Table \ref{table-Z0-Mcore-Zcno-MWD-all}).

\begin{table}
\begin{center}\begin{threeparttable}\centering

\begin{tabular}{c|cccccc}
\hline 
M$_{*}$ & $\tau_{MS}$ & $\tau_{SGB}$ & $\tau_{RGB}$ & $\tau_{HeB}$ & $\tau_{EAGB}$ & $\tau_{AGB}$\tabularnewline
\hline 
0.85 & 10100 ($-11\%$) & 840 & 200 & 70 ($-64\%$) & 6.8 & 1.5\tabularnewline
1.0 & 5740 ($-6\%$) & 470 & 96 & 99 ($-10\%$) & 9 & 1.7\tabularnewline
2.0 & 654 ($+3\%$) & 18 & -- & 30 ($-30\%$) & 8 & 1.8\tabularnewline
3.0 & 200 ($-5\%$) & 13 & -- & 12 ($-25\%$) & 4.2 & $\sim2$\tnote{a}\tabularnewline
3.0\tnote{b} & 221 ($+5\%$) & 8 & -- & 12 ($-25\%$) & 5.0 & 1.2\tabularnewline
\hline 
\end{tabular}

\caption{Various lifetimes for all the $Z=0,$ $Y=0.245$ models, and one 3.0
M$_{\odot}$ model with $Y=0.230$. All ages are in Myr and all masses
are in M$_{\odot}$. M$_{*}$ is the initial mass of the star. Lifetimes
are: $\tau_{MS}$ (ZAMS to MS turn-off), $\tau_{SGB}$ (MS turn-off
to base of RGB, or to the start of core He burning in the more massive
models), $\tau_{RGB}$(red giant branch), $\tau_{HeB}$ (core helium
burning), $\tau_{EAGB}$ (from end of core He burning to first thermal
pulse on AGB), and $\tau_{AGB}$ (thermally-pulsing AGB). The percentages
in brackets are the differences between our lifetimes for the MS and
core He burning and those of \citet{2001AA...371..152M}. A negative
value implies that out lifetime is shorter. \label{table-Z0-lifetimes-all}}

\line(1,0){100}

\begin{tablenotes}\scriptsize

\item[a] This model did not fully complete the AGB. 

\item[b] This model had an initial composition for pure H and He, with $Y=0.230$ rather than 0.245.

\end{tablenotes}

\end{threeparttable} \end{center}
\end{table}

In Table \ref{table-Z0-Mcore-Zcno-MWD-all} we have provided some
more key values that characterise our $Z=0$ models. Interesting features
seen here are that the core masses of the two low mass models were
identical at the time of dual core flash onset (M$_{c}=0.49$ M$_{\odot}$)
and that none of the models experienced much third dredge-up at all.
A vital statistic with regards to $Z=0$ low- and intermediate-mass
stars is the degree of envelope pollution that occurs as a result
of the DCF or DSF events. Table \ref{table-Z0-Mcore-Zcno-MWD-all}
indicates that the level of pollution from this event in the 1 M$_{\odot}$
model is similar to that experienced by the 0.85 M$_{\odot}$ model,
such that the envelope abundance reached $Z_{cno}\sim10^{-3}$. Interestingly
the dual shell flash (DSF) event occurring in the intermediate mass
models ($M=2$ and 3 M$_{\odot}$) at the beginning of the AGB produced
1 to 2 dex less pollution in the envelope. As discussed earlier one
of the main problems of postulating $Z\sim0$ low mass stars as the
source of pollution for the carbon-enhanced extremely metal poor halo
stars (CEMPHS) is that models produce too much carbon and nitrogen,
usually by about 1 to 2 dex. The result that intermediate mass stars
produce less C and N suggests that these stars are better candidates
for the source of the C in CEMPHS. We shall return to this topic in
the main discussion of the thesis. 

\begin{table}
\begin{center}\begin{threeparttable}\centering

\begin{tabular}{c|c|ccccc}
\hline 
$M_{*}$ & M$_{c,DF}$ & $Z_{cno,DF}$ & $\lambda_{3dup}$ & N$_{TP}$ & $Z_{cno,f}$ & M$_{WD}$\tabularnewline
\hline 
0.85 & 0.49 & $4\times10^{-3}$ & 0.0 & 32 & $4\times10^{-3}$ & 0.77\tabularnewline
1.0 & 0.49 & $6\times10^{-3}$ & 0.0 & 60 & $6\times10^{-3}$ & 0.83\tabularnewline
2.0 & 0.70 & $2\times10^{-4}$ & 0.01 & 282 & $3\times10^{-3}$ & 1.08\tabularnewline
3.0\tnote{a} & 0.78 & $2\times10^{-5}$ & $<0.01$ & $>950$ & $3\times10^{-5}$ & $>1.17$\tabularnewline
3.0 \tnote{b} & 0.77 & $2\times10^{-5}$ & $\sim0.2$ & 390 & $1.2\times10^{-2}$ & 1.10\tabularnewline
\hline 
\end{tabular}

\caption{Various characteristic values for the $Z=0,$ $Y=0.245$ models, and
one 3.0 M$_{\odot}$ model with $Y=0.230$. The column definitions
are: $M_{*}$ (initial stellar mass), M$_{c,DF}$ (core mass at the
onset of the dual core flash or dual shell flash), $Z_{cno,DF}$ (the
envelope metallicity after the dredge-up events(s) associated with
the DCF/DSF episodes), $\lambda_{3dup}$ (the maximum value reached
for the dredge-up parameter on the AGB), N$_{TP}$ (number of thermal
pulses on the AGB), $Z_{cno,f}$ (final metallicity of the envelope),
and M$_{WD}$ (white dwarf mass). All masses are in M$_{\odot}$.
\label{table-Z0-Mcore-Zcno-MWD-all}}

\line(1,0){100}

\begin{tablenotes}\scriptsize

\item[a] This model did not fully complete the AGB. 

\item[b] This model had an initial composition for pure H and He, with $Y=0.230$ rather than 0.245.

\end{tablenotes}

\end{threeparttable} \end{center}
\end{table}

Further surface abundance enhancements can come about by the third
dredge-up (3DUP) on the TP-AGB. As we have included no overshoot at
all in these models we note that our 3DUP results represent somewhat
of a lower limit to the possibilities of AGB envelope enhancements.
Table \ref{table-Z0-Mcore-Zcno-MWD-all} highlights the zero and very
weak 3DUP occurring in our models. The 3DUP parameter $\lambda_{3dup}$
is only marginally above zero in the intermediate mass models (except
for the $Y=0.23,$ 3 M$_{\odot}$ model). Despite this the 2 M$_{\odot}$
model ends up with an extra order of magnitude increase in $Z$ (over
the earlier DSF pollution) because there are so many 3DUP episodes
(282 in total). Interestingly this brings the final envelope abundance
of this model roughly in line with the two low mass models -- all
these models end the AGB with $Z\sim10^{-3}$ (the 3 M$_{\odot}$
model did not complete the AGB). Interestingly the $Y=0.23$, 3 M$_{\odot}$
model had much deeper dredge-up, with $\lambda\sim0.2$. Although
evolutionary dependencies on helium abundance are outside the scope
of the present work, it is interesting to note that a lower He abundance
can have such a strong effect. The high degree of surface pollution
at the end of this model's evolution is due to this increased 3DUP
efficiency. 

As most mass loss occurs at the end of the AGB (the superwind phase)
the $Z_{cno}$ abundances at the end of the AGB (ie. column 6 in Table
\ref{table-Z0-Mcore-Zcno-MWD-all}) roughly represent the composition
of the material returned to the interstellar medium. Detailed nucleosynthesis
and associated yields from these models is the subject of the next
chapter.

Finally we note the interesting result that all the models calculated
here, which have started with a pristine H-He composition, experience
some form of major pollution event in their lifetimes. One of the
(similar) $Z=0$ phenomena -- a DCF or a DSF -- always occurs. If
a $Z=0$ star undergoes a core He flash then the DCF will occur. If
it doesn't then a DSF will occur on the early TP-AGB. Both events
involve some dredge-up of CN(O) nuclei from a He-rich convective shell,
consequent CNO burning and then a dredge-up of this processed material
by an incursion of the convective envelope. If the models are correct
then this suggests that much pollution would have been caused by a
population of $Z=0$ low- and intermediate-mass stars (if such a population
existed).

\chapter{Zero Metallicity Stars: Nucleosynthetic Evolution\label{Chapter-ZeroZ-Nucleo}}
\begin{quotation}
``The most incomprehensible thing about our universe is that it can
be comprehended.''
\begin{flushright}
\vspace{-0.4cm}-- Albert Einstein
\par\end{flushright}
\end{quotation}

\section{Background\label{section-Z0-NSevolution-NScode}}

In contrast to the structural evolution code only minor modifications
were needed to the nucleosynthesis code (NS code) in order to enable
the calculation of $Z=0$ (and very metal-poor) models. Removing the
scaled-solar composition assumptions to allow arbitrary stellar chemical
compositions was the main modification. Aside from this some computational
problems were encountered with the dual core flash phenomenon. In
particular it was found that the standard resolution of the NS code
was not sufficient to follow the small radiative zones that often
form after the splitting of the He convection zones in these models.
If these zones were not resolved then a large degree of extra mixing
would occur, leading to enrichment of the envelope inconsistent with
the structural code results. The solution to this was simply to increase
the resolution in the central regions during these events. This problem
is discussed in more detail later in this chapter. We note that the
NS code was described in detail in Section \vref{nscode}.

\subsection{Initial Composition}

We have discussed the rationale for choosing our primordial composition
earlier, in Section \vref{sec-BBN}. Here we re-display the abundances,
in Table \ref{table-bbnabunds}, for easy reference. The nucleosynthesis
code handles many more species than the SEV code (see Section \ref{nscode}
for a description of this code), so all the species predicted by BBN
are accounted for in the initial composition.

\begin{table}
\begin{centering}
\begin{tabular}{|c|c|}
\hline 
Nuclide & Primordial Mass Fraction\tabularnewline
\hline 
\hline 
$^{1}$H & $0.754796$\tabularnewline
\hline 
$^{2}$H & $1.96\times10^{-4}$\tabularnewline
\hline 
$^{3}$He & $7.85\times10^{-6}$\tabularnewline
\hline 
$^{4}$He & $0.24500$\tabularnewline
\hline 
$^{7}$Li & $3.13\times10^{-10}$\tabularnewline
\hline 
$^{12}$C & $0.0$\tabularnewline
\hline 
$^{14}$N & $0.0$\tabularnewline
\hline 
$^{16}$O & $0.0$\tabularnewline
\hline 
\end{tabular}
\par\end{centering}
\caption{Initial abundances used in the zero-metallicity stellar models, as
given by Standard Big Bang Nucleosynthesis calculations by \citet{Coc0104}.
\label{table-bbnabunds}}
\end{table}

\section{Detailed Nucleosynthesis at $M=0.85$ M$_{\odot}$}

In this section we detail the key nucleosynthetic events that occur
during the evolution of the 0.85 M$_{\odot}$ $Z=0$ model. In particular
we describe and quantify the events that lead to surface pollution
and hence affect the chemical yield of the star. In this sense the
defining event in this model is the dual core flash (DCF) that occurs
at the tip of the RGB. Processed material from this event is dredged-up
into the convective envelope, causing the surface metallicity (and
helium abundance) to rise dramatically. Before this event the surface
retains its primordial abundances (apart from a minor change at first
dredge-up). After the DCF dredge-up the star basically retains the
abundance pattern arising from this event for the rest of its evolution.
The AGB phase has little effect on the surface composition (apart
from some low-mass species such as $^{7}$Li and $^{3}$He), as there
is no third dredge-up (3DUP) and only very minor hot bottom burning
(HBB). We also present the chemical yield for the model.

\subsection{First Dredge-Up\label{subsection-m0.85z0-Nucleo-FDUP}}

\begin{figure}
\begin{centering}
\includegraphics[width=0.7\columnwidth]{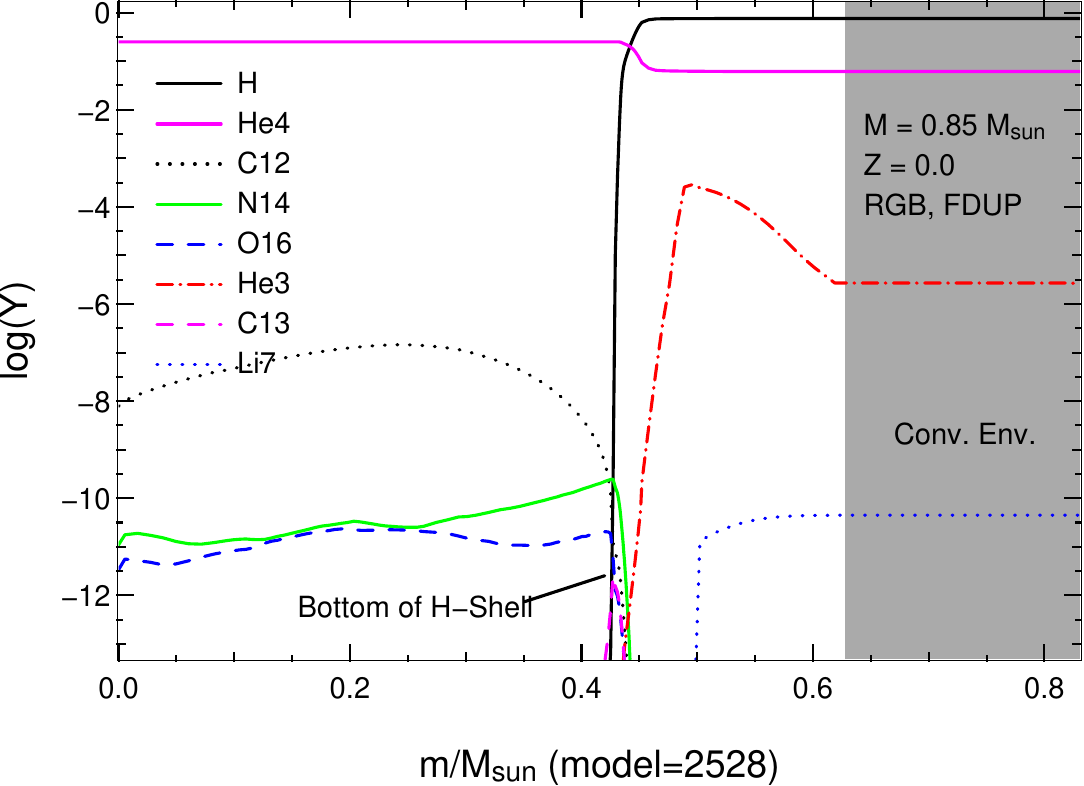}
\par\end{centering}
\caption{Chemical composition (log of mole fraction, Y) versus mass for the
0.85 M$_{\odot}$ model at the time of maximum incursion by the envelope
during FDUP. The grey shaded area represents convection.\label{fig-m0.85z0y245-cmp-FDUP}}
\end{figure}

Low mass $Z=0$ stars do evolve to an RGB configuration, although
their surface temperatures are somewhat higher than Pop II stars.
Their luminosities are also higher, due to hotter H-burning and lower
opacity. These factors lead to a weaker first dredge-up (FDUP) episode,
such that the envelope convection never reaches down into regions
in which any appreciable nuclear burning has previously occurred.
In Figure \ref{fig-m0.85z0y245-cmp-FDUP} we plot some of the abundance
profiles at the time of maximum dredge-up. It can be seen that the
only isotope that may have its surface abundance altered is $^{3}$He,
and only by a minute amount. The abundance did actually increase,
but by only $\sim0.02$ dex, such that chemical pollution from the
FDUP is essentially non-existent in this star.

An interesting feature in Figure \ref{fig-m0.85z0y245-cmp-FDUP} is
that there is a (relatively) substantial amount of $^{12}$C in the
core -- despite the star initially having a zero abundance of C and
not having reached a standard He burning phase as yet. As shown in
the previous chapter there is actually a small amount of He burning
just below the H burning shell (see Figure \vref{fig-m0.85z0y245-LateRGBShell-lums-C12}).
This feeds the bottom of the H shell with some primary $^{12}$C which
allows the CNO cycles to operate. This only occurs at the bottom of
the H shell -- the p-p chains still dominate most of the H-burning
shell. As the H shell burns its way outward (in mass) it thus leaves
a small trail of $^{12}$C (and associated burning products) in its
wake. Although this interesting abundance profile does not directly
affect the chemical yield of the star, it may be significant enough
to affect the nature of the He core flash burning that occurs in this
region at a later time. 

\subsection{Dual Core Flash \label{subsec-m0.85z0-DCF-NS}}

\begin{figure}
\begin{centering}
\includegraphics[width=0.95\columnwidth]{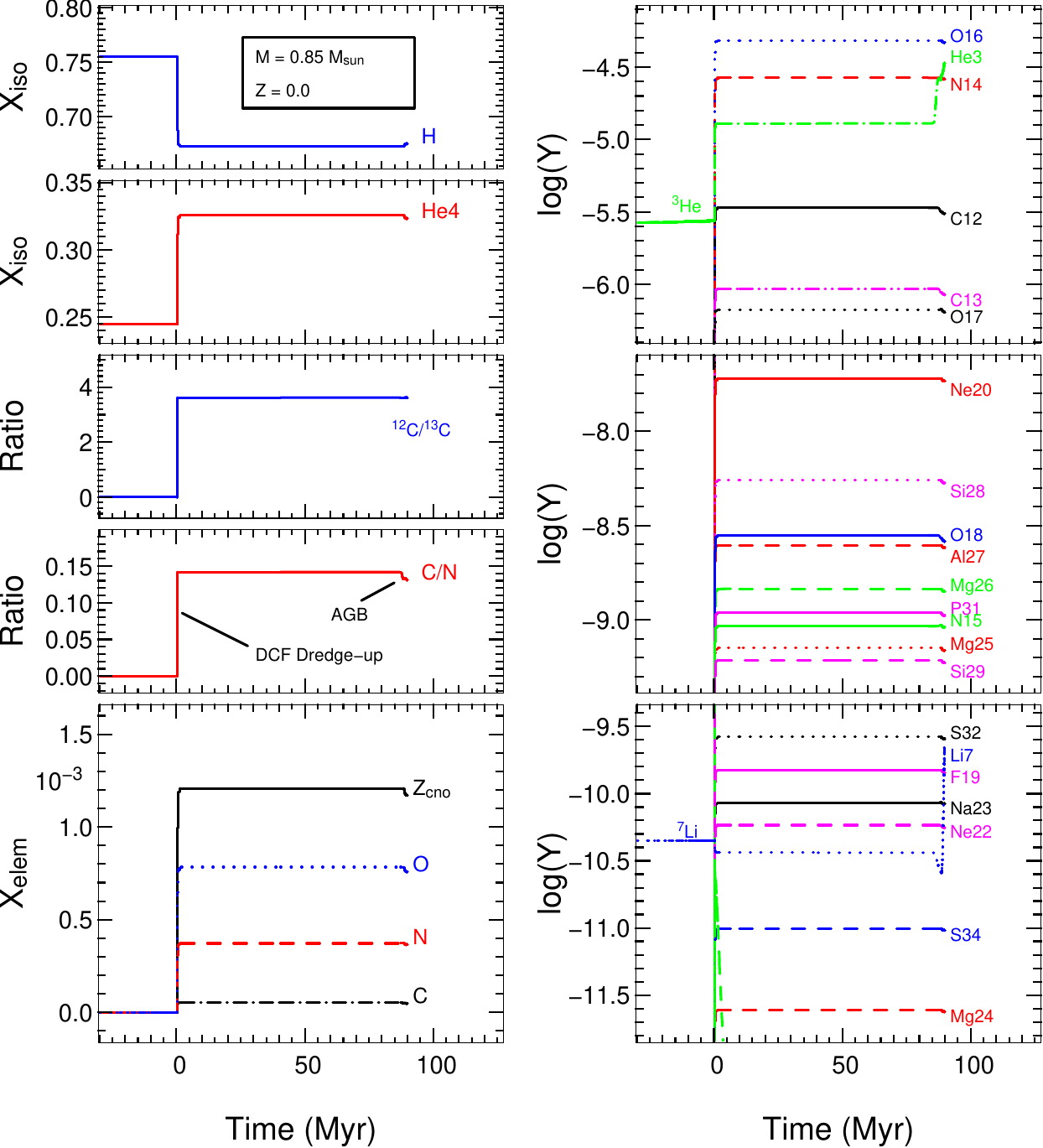}
\par\end{centering}
\caption{Time evolution of the surface abundances in the $M=0.85$ M$_{\odot}$,
$Z=0$ model. Time has been offset so that $t=0$ coincides with the
DCF dredge-up. It can be seen that the DCF dredge-up episode is by
far the main source of envelope pollution, causing a large increase
in He and CNO isotopes. The $^{12}$C/$^{13}$C ratio is quite low,
indicating that CN burning achieved equilibrium during the DCF episode.
The only species produced in significant amounts during the AGB are
$^{7}$Li and $^{3}$He. We note that this star initially had $Z=0$
but ends its evolution with $Z\sim\frac{1}{10}$ solar (although it
is still very metal-deficient in terms of the heavier elements).\label{fig-m0.85z0y245-srf-DCF-AGB}}
\end{figure}

\begin{figure}
\begin{centering}
\includegraphics[width=0.9\columnwidth]{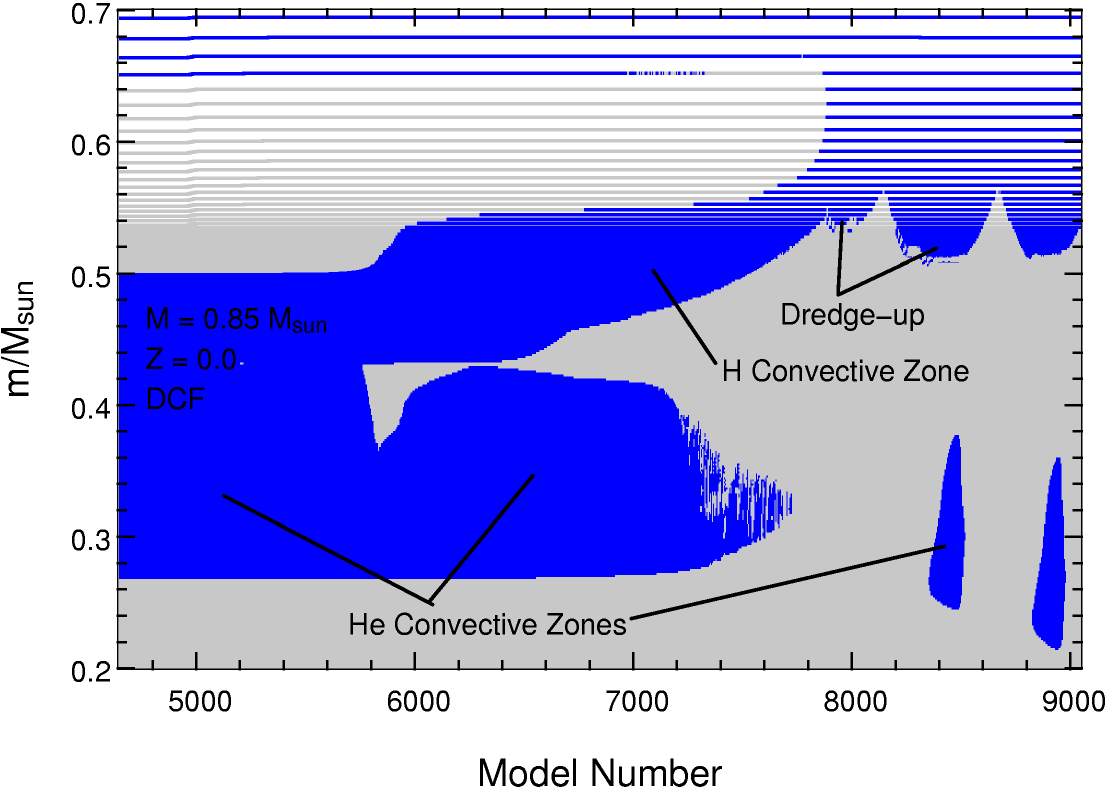}
\par\end{centering}
\caption{The mesh used in the nucleosynthesis code during the dual core flash
phase in the 0.85 M$_{\odot}$, $Z=0$ model. Each mesh point is represented
by a short horizontal line and on this wide timespan (actually model-span)
they appear as long horizontal lines. Dark blue represents convective
points and light grey radiative points. It can be seen that the mesh
in the outer regions is much less refined than in the central regions
(the resolution is so high in the core that the mesh appears as solid
colour). We found it necessary to increase the resolution in the core
due to the splitting convection zones in models such as this (this
can be seen here at model $\sim5800$). In particular the narrow radiative
zones between the convective zones need to be resolved (eg. at model
$\sim6200$ in this case). The resolution in the outer regions is
$\sim10^{-2}$ M$_{\odot}$, the inner regions $\sim10^{-3}$ M$_{\odot}$
and in the H burning shell it is as high as $\sim10^{-9}$ M$_{\odot}$.
The two small convective zones after model 8000 are He minipulses
occurring during the secondary RGB phase. \label{fig-m0.85z0y245-mesh-DCF}}
\end{figure}

As FDUP dredges up no processed material (practically), the star retains
its primordial surface composition till the end of the RGB. In fact
the surface remains unpolluted even through the dual core flash phase
itself. It is not until both flashes have abated and the convection
moves inwards (in mass, see eg. Figure \ref{fig-m0.85z0y245-mesh-DCF})
that the envelope is polluted. Figure \ref{fig-m0.85z0y245-srf-DCF-AGB}
demonstrates this pollution event clearly, showing that a large amount
of processed material is dredged up to the surface, taking the metallicity
from zero to $\sim10^{-3}$ in a very short time. Also evident in
Figure \ref{fig-m0.85z0y245-srf-DCF-AGB} is that, apart from some
$^{7}$Li and $^{3}$He production on the AGB, the DCFDUP is the defining
event in terms of the composition of the envelope for all future stages
of evolution. The DCFDUP thus has the greatest effect on the chemical
yield of this star. This figure gives a good overview of the bulk
nucleosynthetic evolution (in terms of surface pollution) for a large
range of chemical species and, due to the relative invariance of the
composition in the later mass-losing stages of evolution, it also
gives a good indication of the expected chemical yield. 

As we have identified the DCF (and associated dredge-up) as the defining
envelope pollution event for this star we now look into the nucleosynthesis
of this phenomenon in detail, as well as some related computational
aspects. We note that the structural evolution of this event has been
described in Section \vref{section-m0.85z0y245-DualCoreFlash}. 

Figure \ref{fig-m0.85z0y245-mesh-DCF} displays the numerical mesh
that was employed in the nucleosynthesis code (NS code) during the
DCF. Two key events are clear in this plot. The first is that the
He convective zone splits into two as its expansion dredges down protons
(this is the major H flash event). The second is that the convective
envelope later extends downwards into regions that have been polluted
by processed material, thereby altering the surface composition. The
splitting of the convection zone initially caused some problems in
the nucleosynthesis calculations due to the fact that the two resulting
convection zones are divided only by a small radiative region (for
some of the time). With our previous spatial resolution the two convective
zones would merge, resulting in the mixing of the upper H convective
zone (HCZ) with the lower He convective zone (HeCZ). Thus the HCZ
would be super-enriched and the later dredge-up by the envelope would
result in surface pollution inconsistent with the structural code
results. Thus this phase of evolution necessitated increasing the
resolution of the mesh. We found that an order of magnitude increase
in resolution (in the central regions) was necessary to reliably resolve
the DCF. It can be seen in Figure \ref{fig-m0.85z0y245-mesh-DCF}
that the radiative zone is now easily resolved. However, using such
high resolution slowed the calculations considerably. Thus, after
this demanding evolutionary phase was complete the resolution was
reduced back to the standard level. We used the same method in our
other models that experienced the DCF (or DSF). 

\begin{figure}
\begin{centering}
\includegraphics[width=0.8\columnwidth]{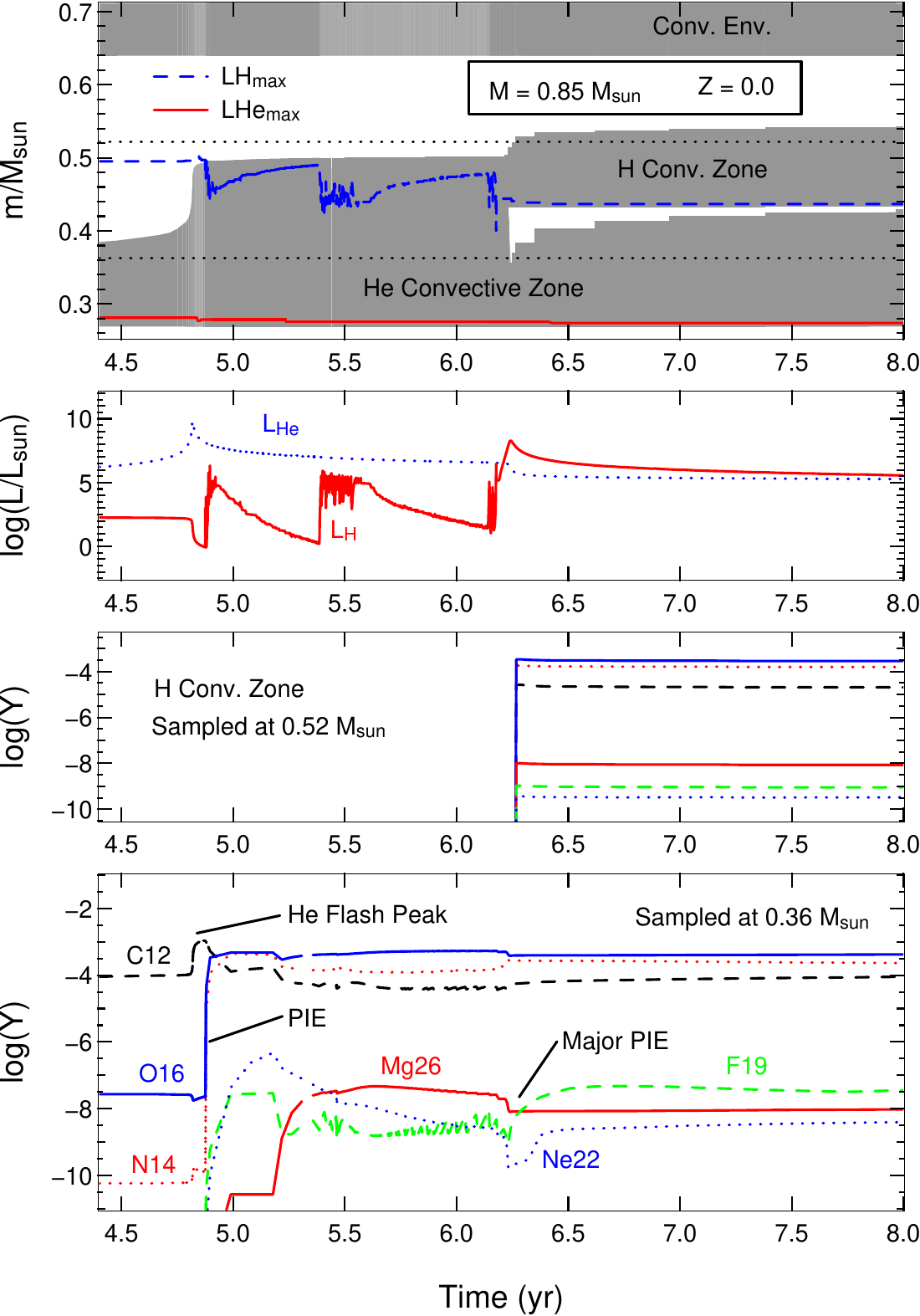}
\par\end{centering}
\caption{Zooming in on the evolution of convection, luminosities and abundances
during the DCF. Time has been offset for clarity. In the top panel
it can be seen that the convection zone does not split during the
first two proton ingestion episodes (PIEs). This is because the amount
of H ingested is relatively small, leading to lower H burning energy
release. The two dotted lines in the top panel indicate the mass sampling
points for the lower two panels which display the abundance evolution
in the HCZ and HeCZ. The abundances in the HCZ are a result of the
previous nucleosynthesis in the HeCZ before the split occurs -- it
can be seen that no significant changes are occurring in the abundance
pattern in the HCZ after the convection zone splits. In the HeCZ there
is a huge production of $^{16}$O and $^{14}$N during the first PIE
which occurs primarily via the $^{12}$C($p,\gamma$)$^{13}$N($\beta^{+}$)$^{13}$C($\alpha,n$)$^{16}$O
and $^{12}$C($p,\gamma$)$^{13}$N($\beta^{+}$)$^{13}$C($p,\gamma$)$^{14}$N
chains respectively. It is this nucleosynthesis which later gives
the envelope its abundance profile (via the post-DCF DUP). Note that
the line colours in the third panel are the same as those in the bottom
panel. \label{fig-m0.85z0y245-DCF-allPIEs-conv-nucleo}}
\end{figure}

In Figure \ref{fig-m0.85z0y245-DCF-allPIEs-conv-nucleo} we show the
evolution of the early stages of the DCF. The plot covers a timespan
of only 3.5 years but elucidates the important features of the event.
In particular we can see the three proton ingestion episodes (PIEs)
which lead to varying intensities of H burning inside the He convective
zone. The first two H flashes are not as strong as the final one because
the HeCZ is `eating' only into the tail of the H burning shell, where
the H abundance is fairly low (see Figure \vref{fig-m0.85z0y245-HFlashes-Hprofile-Conv}
in the structural evolution section). In these cases the energy release
from H burning is not large enough to cause a splitting of the convection
zone. In the third PIE a feedback effect is instigated whereby the
H burning is so strong due the influx of proton-rich material (this
is the major H-flash) that the convection zone expands upwards, causing
further proton ingestion. In this case there is a splitting of the
He convective zone as a radiative zone develops just below the region
of peak H burning energy release (there is a temperature inversion
here). The energy release from the He burning shell below also reduces
significantly, causing the lower convective zone to reduce in size.
These two convection zones -- the HCZ above and the HeCZ below --
remain separated for the remainder of the evolution, as both flashes
subsequently recede. 

Figure \ref{fig-m0.85z0y245-DCF-allPIEs-conv-nucleo} also displays
the chemical evolution during this phase, as calculated using the
NS code. We have sampled the composition over time at two mass coordinates,
one that takes in the HCZ and one that takes in the HeCZ. This gives
an insight into the nucleosynthesis that is occurring in both regions.
Of most interest is the composition of the HCZ, as it is this region
that is dredged up later, polluting the envelope. Looking at the chemical
evolution in the HCZ we see that the metal abundances discontinuously
jump from zero to $\sim10^{-4}$ (in mole fraction $Y$). From then
on they are quite stable (at least in this short timespan). Thus it
is apparent that, in terms of bulk metallicity (ie. $Y_{cno}=Y_{C}+Y_{N}+Y_{O}$),
there is no significant nucleosynthesis occurring in the HCZ (apart
from H being burnt to He). This is due to the fact that the bulk of
the nuclear  burning in this region is via the CNO cycles. As the
CN cycle is operating in equilibrium the relative abundances remain
approximately constant. The $^{12}$C/$^{13}$C ratio is also $\approx4$,
as can be seen in the surface abundance plot after the DCFDUP (Figure
\ref{fig-m0.85z0y245-srf-DCF-AGB}). This leads us to explore further
down the rabbit hole, as the HCZ is not the site of the nucleosynthesis
that has given it its chemical distribution. The final panel in Figure
\ref{fig-m0.85z0y245-DCF-allPIEs-conv-nucleo} displays the chemical
evolution in the HeCZ -- before and after the convection zone split
occurs. Here we can see that before the HeCZ first penetrates the
H-He discontinuity the chemical mix in this region is characteristic
of a standard core He flash. $^{12}$C is dominant due to the $3\alpha$
reactions, but $^{16}$O is also being produced via the $^{12}$C($\alpha,\gamma$)$^{16}$O
reaction. As the He burning luminosity starts to reach its peak, $^{12}$C
is produced at a much greater rate. Not long after the He flash peak
the HeCZ expands and causes a PIE. Although this is the first proton
ingestion episode and thus the material dredged down is relatively
proton-poor, the consequent nucleosynthesis causes a striking alteration
in the chemical composition of the HeCZ. In particular $^{16}$O is
increased by \emph{4 orders of magnitude} and $^{14}$N by more than
\emph{6 orders of magnitude}. Conversely $^{12}$C decreases by about
an order of magnitude. Furthermore, not long after, many other species
are produced in considerable amounts, such as $^{26}$Mg, $^{22}$Ne
and $^{19}$F. The huge variations in the CNO abundances is easily
understood, such that the addition of protons into the $^{12}$C-rich
HeCZ causes the CN(O) cycles to operate. $^{12}$C acts as an efficient
sink for the protons, producing $^{14}$N via the usual CN chain $^{12}$C($p,\gamma$)$^{13}$N($\beta^{+}$)$^{13}$C($p,\gamma$)$^{14}$N.
The oxygen production occurs primarily via the $^{12}$C($p,\gamma$)$^{13}$N($\beta^{+}$)$^{13}$C($\alpha,n$)$^{16}$O
chain. This chain of reactions is so prominent because there is a
large amount of He in the HeCZ. Usually the aforementioned CN chain
dominates, and the CN cycle can complete, mostly producing $^{14}$N
and resulting in little $^{16}$O production (or even $^{16}$O depletion).
It can also be seen in the last panel of Figure \ref{fig-m0.85z0y245-DCF-allPIEs-conv-nucleo}
that, as expected, the sum of C+N+O nuclei is approximately constant,
as the resulting abundances of $^{14}$N and $^{16}$O sum to the
$^{12}$C deficit. Thus $Z_{cno}$ is primarily dependent on the amount
of $^{12}$C produced in the HeCZ by the core He flash. So, even though
the structural evolution code (SEV code) does not include the $^{13}$C($\alpha,n$)$^{16}$O
reaction (ie. the $^{16}$O production is neglected) we note that
$Z_{cno}$ will be (practically) the same in both calculations. In
terms of possibly neglecting some important energy production (or
reduction) in the SEV code we also note that the energy budget in
the HeCZ is easily dominated by the He burning during this first PIE.

\begin{figure}
\begin{centering}
\includegraphics[width=0.95\columnwidth]{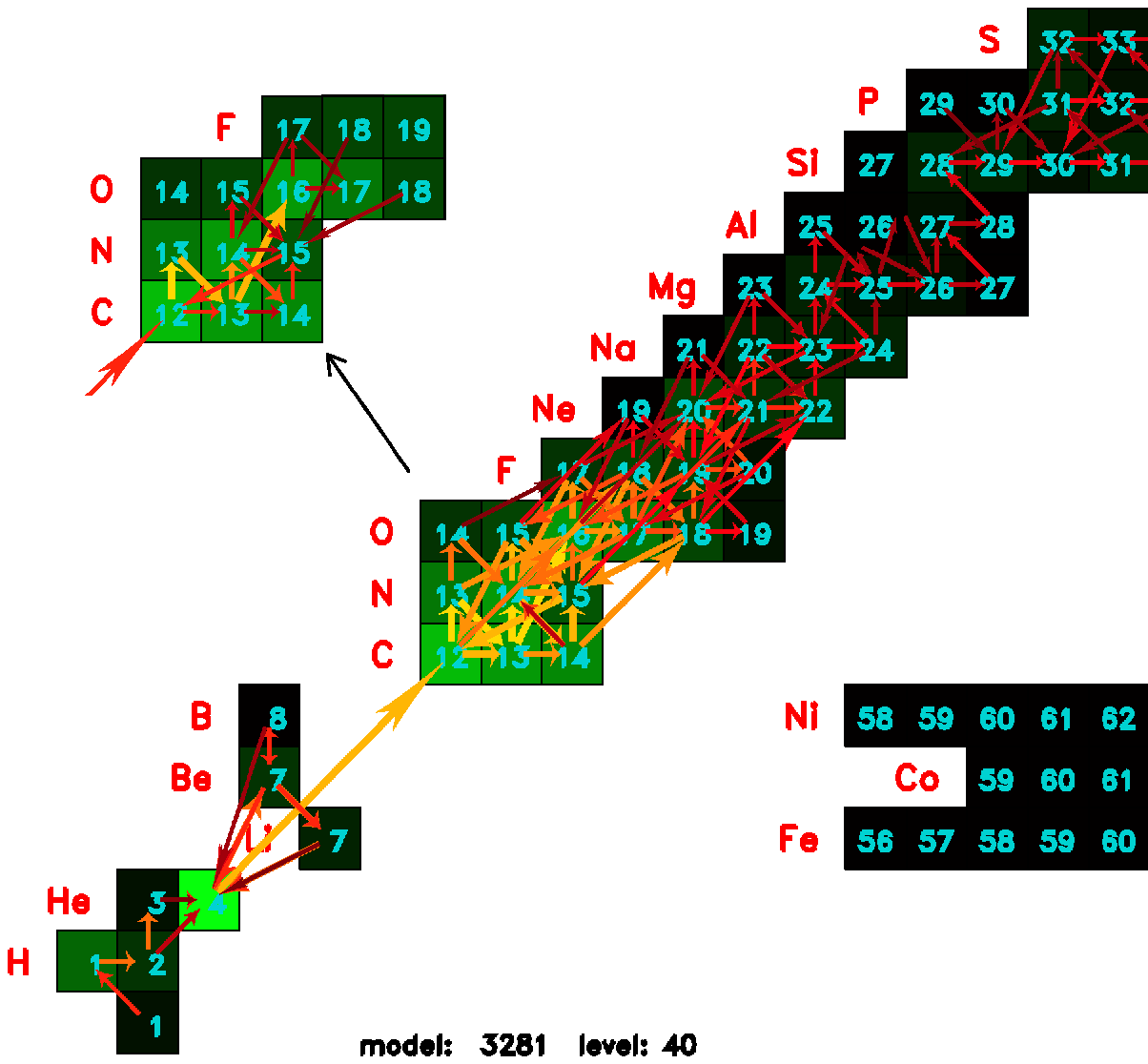}
\par\end{centering}
\caption{The full set of species included in our network. Arrows represent
reactions that are occurring at rates greater than a lower cutoff.
Yellow $\rightarrow$ red indicates faster $\rightarrow$ slower rates
of reactions. Thicker arrows also indicate reactions happening at
faster rates. The model plotted is taken during the first PIE and
the `level' corresponds to $m=0.40$ M$_{\odot}$ which is in the
middle of the HeCZ (see Figure \ref{fig-m0.85z0y245-DCF-allPIEs-conv-nucleo}).
A second CNO(F) group (top left) is given since there are too many
arrows in the main diagram to identify the reactions (due to the use
of a low rate cut-off in order to display the reactions occurring
amongst the heavier nuclides). It can be seen that the dominant reactions
in this region are the $3\alpha$, $^{13}$C($\alpha,n$ )$^{16}$O,
and also proton captures on the carbon nuclei. Neutron capture reactions
can also be seen (most easily in the heavier elements), giving rise
to heavy isotopes. Note that the Fe group species displayed (Ni, Fe,
Co) remain at zero abundance as we did not include the species between
these and $^{35}$S. \label{fig-m0.85z0y245-ARP-peak-neutrons-fullNetwork}}
\end{figure}

Panel 4 of Figure \ref{fig-m0.85z0y245-DCF-allPIEs-conv-nucleo} also
shows that further nucleosynthesis has occurred during the first PIE.
We show a few selected species to highlight this: $^{19}$F, $^{22}$Ne
and $^{26}$Mg, which are all produced in substantial amounts (considering
the star began with $Z=0$). In Figure \ref{fig-m0.85z0y245-ARP-peak-neutrons-fullNetwork}
we show the full nucleosynthesis network used for these calculations,
displayed as a nuclide chart. Overlaid on the nuclide chart are arrows
indicating the nucleosynthesis occurring in the middle of the HeCZ
during the initial stages of the first PIE. Although the bulk of the
nucleosynthesis involves the CNO cycles (and He burning) there is
much further processing via the NeNa cycle, MgAl chains, and even
neutron capture reactions, all the way up to the most massive species
included (note that we do not follow the production of the iron group
species, which remain at zero abundance). From this figure we can
elucidate the nucleosynthesis that leads to the production of the
species in Figure \ref{fig-m0.85z0y245-DCF-allPIEs-conv-nucleo}.
Indeed it can be seen that the dominant reactions include $^{13}$C($\alpha,n$)$^{16}$O
and the CN cycle. The bulk of the $^{19}$F is created through proton
capture on $^{18}$O (although $^{15}$N($\alpha,\gamma$)$^{19}$F
is also operating). The $^{18}$O required by this reaction is primarily
produced via neutron capture on $^{14}$N and subsequent $\alpha$-capture
on $^{14}$C: $^{14}$N($n,p$)$^{14}$C($\alpha,\gamma$)$^{18}$O.
The other main channel to $^{19}$F production also involves $^{18}$O,
via the $^{18}$O($n,\gamma$)$^{19}$O($p,n$)$^{19}$F reactions.
Thus it can be seen that the substantial nucleosynthesis in the proton-polluted
HeCZ arises from the combination of proton capture, $\alpha$ capture
and neutron capture reactions. The $^{22}$Ne also seen in Figure
\ref{fig-m0.85z0y245-DCF-allPIEs-conv-nucleo} arises through two
main channels. The first also relies on $^{18}$O: $^{18}$O($\alpha,\gamma$)$^{22}$Ne.
The second is via neutron captures: $^{20}$Ne($n,\gamma$)$^{21}$Ne($n,\gamma$)$^{22}$Ne.
We note that there is a substantial abundance of neutrons available
for these reactions due to the $^{13}$C($\alpha,n$)$^{16}$O reaction.
We shall discuss the potential for s-process nucleosynthesis in the
next subsection. The neutrons also lend themselves to $^{26}$Mg production
via $^{24}$Mg($n,\gamma$)$^{25}$Mg($n,\gamma$)$^{26}$Mg. This
isotope of Mg is also produced through successive proton captures
(the MgAl chains are operating): $^{23}$Na($p,\gamma$)$^{24}$Mg($p,\gamma$)$^{25}$Al($\beta^{+}$)$^{25}$Mg($p,\gamma$)$^{26}$Al($\beta^{+}$)$^{26}$Mg. 

It can be seen that a rich nucleosynthesis has occurred in the HeCZ
before its splitting, and in particular during the three proton ingestion
episodes. This abundance pattern is essentially frozen in the HCZ
after the split (see Panel 3 in Figure \ref{fig-m0.85z0y245-DCF-allPIEs-conv-nucleo}).
Consequently the surface abundances after the DCF dredge-up (see Figure
\ref{fig-m0.85z0y245-srf-DCF-AGB}) are also due to the nucleosynthesis
during this stage.

\begin{figure}
\begin{centering}
\includegraphics[width=0.8\columnwidth]{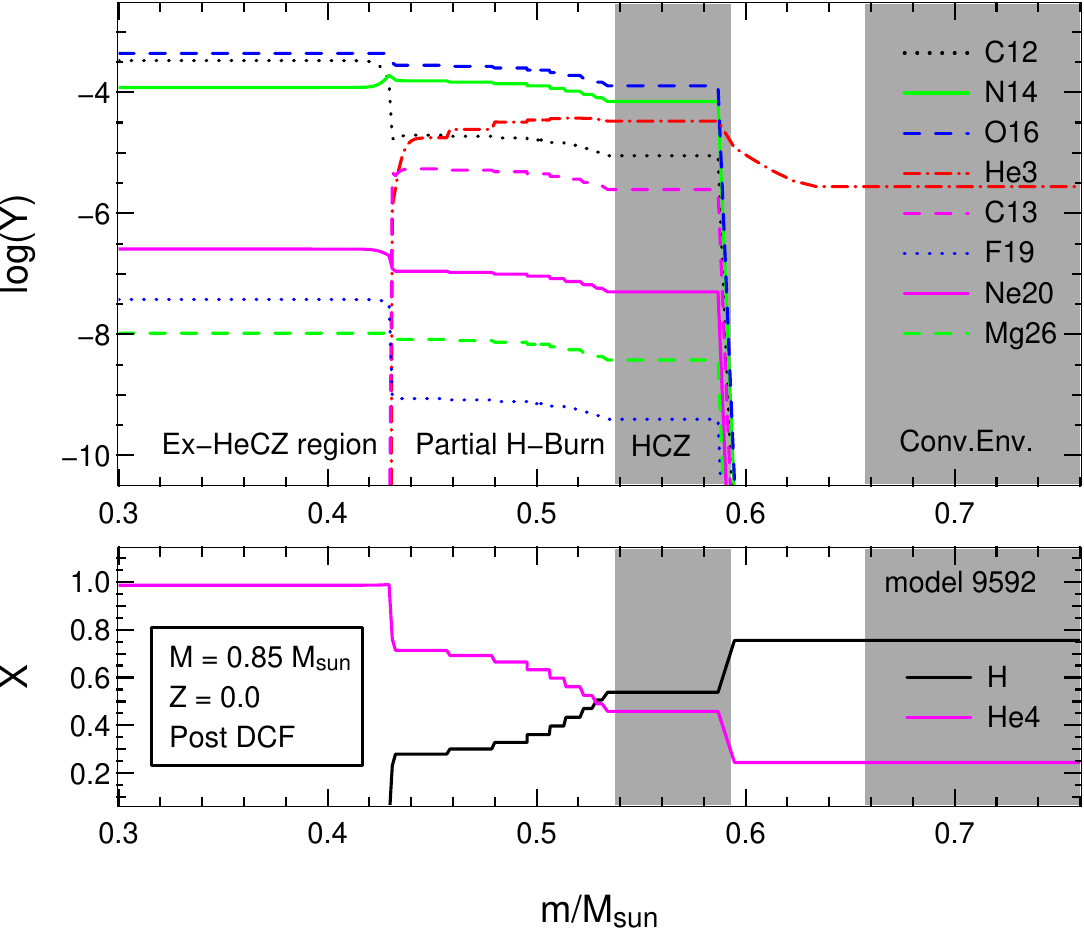}
\par\end{centering}
\caption{The abundance profiles after the HeCZ has subsided, just before the
post-DCF dredge-up. This gives an indication of the chemical profile
that the envelope will soon have. Of particular interest is the high
He abundance in the HCZ, $X_{He}\sim0.46$. A number of different
regions are identified (see text for details).\label{fig-m0.85z0y245-comp-PostDCF-HCZ-justBeforeDUP}}
\end{figure}

As the He convective zone recedes the HCZ initially expands and then
`moves' outwards (in mass, see Figure \ref{fig-m0.85z0y245-mesh-DCF}).
As both these changes involve mixing in pristine material from above,
the dredged-up material is progressively diluted and an interesting
abundance profile is left behind. This profile is displayed in Figure
\ref{fig-m0.85z0y245-comp-PostDCF-HCZ-justBeforeDUP}. The profile
consists of four parts -- the CNO rich zone that was the HeCZ, the
region between this and the remaining HCZ, which has a varying abundance
profile due to dilution, the HCZ itself, and the still unpolluted
convective envelope. Also of interest is that the HCZ, soon to be
dredged up, has quite a high He abundance of $X_{He}\sim0.46$. In
Figure \ref{fig-m0.85z0y245-comp-PostDCF-HCZ-justBeforeDUP} we also
display the profiles of some other species. Of particular interest
is the $^{13}$C abundance which is $\sim\frac{1}{4}$ of the $^{12}$C
abundance, indicating that the CN cycle reached equilibrium during
the earlier burn. $^{20}$Ne is also displayed. Two of the main paths
for its production under these conditions are: $^{16}$O($n,\gamma$)$^{17}$O($\alpha,p$)$^{20}$Ne
and $^{19}$F($n,\gamma$)$^{20}$F($\beta^{-}$)$^{20}$Ne. 

\begin{figure}
\begin{centering}
\includegraphics[width=0.9\columnwidth]{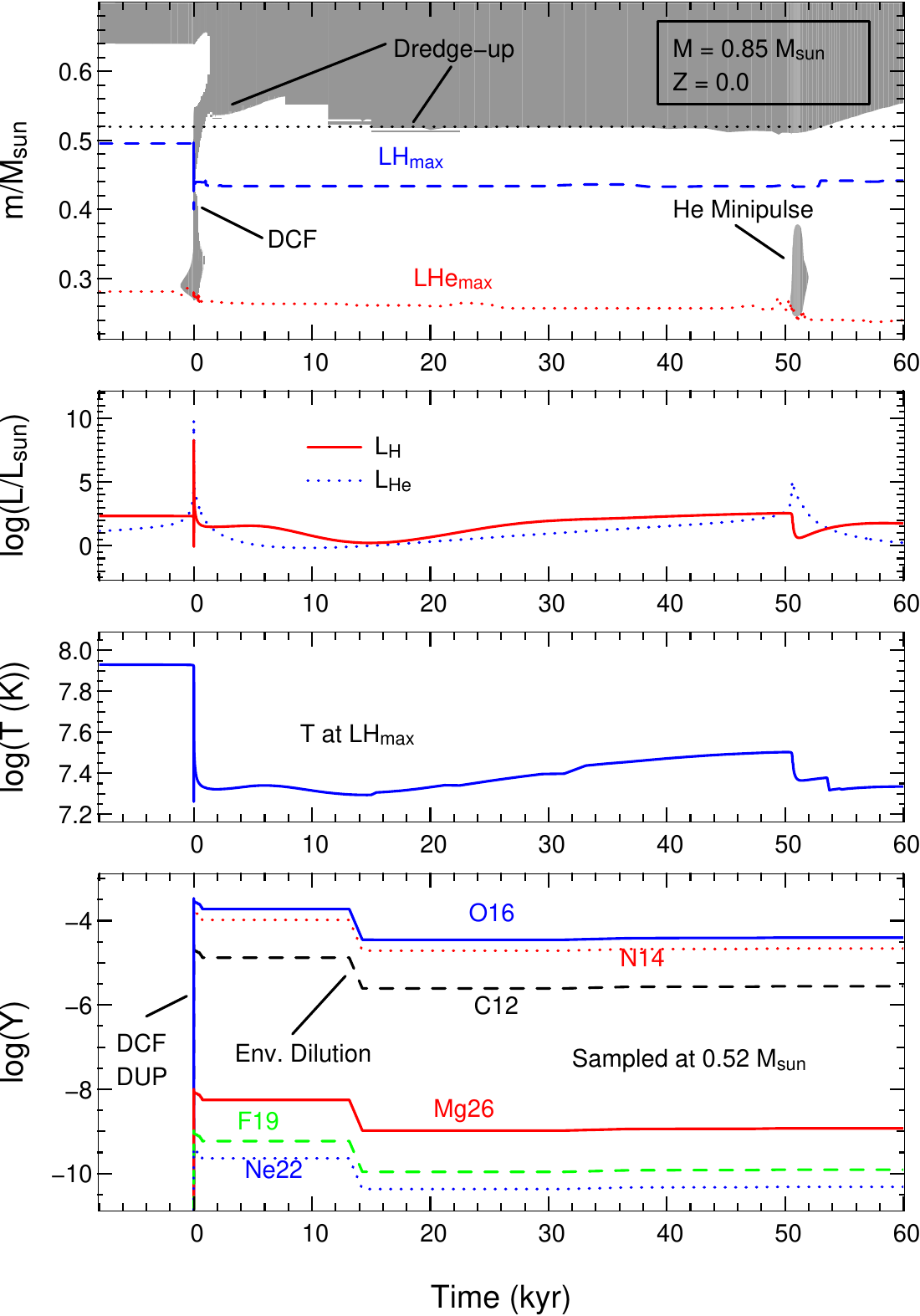}
\par\end{centering}
\caption{A broad view of the evolution around the DCF. The post-DCF dredge-up
can be seen in the top panel. The dotted line at $m=0.52$ M$_{\odot}$
in this panel indicates the sampling point for the abundance evolution
in the bottom panel. It takes in the HeCZ during the DCF as well as
the envelope during the DCF DUP. The dilution of the processed material
throughout the envelope is evident. \label{fig-m0.85z0y245-DCF-wide-nucleo-conv-temp}}
\end{figure}

Figure \ref{fig-m0.85z0y245-DCF-wide-nucleo-conv-temp} shows the
evolution of the convection zones, luminosities and temperature over
an extended time period. It also shows the evolution of the abundances
at a mass location that initially samples the HCZ but also the convective
envelope at later times. The timespan takes in the DCF and post-DCF
dredge-up (DCFDUP) so that the pollution of the convective envelope
is evident. It can be seen that the bulk changes in composition in
this region after the convection zone splits are due to dilution events
rather than nucleosynthesis. After DCFDUP the star retains this chemical
abundance pattern for the rest of its evolution -- apart from some
production of $^{7}$Li and $^{3}$He during the AGB, which is discussed
further below.

\subsection{Is the DCF a New s-Process Site?\label{subsec-m0.85z0-DCF-sProcess}}

In the previous subsection we discussed the nucleosynthesis occurring
during the dual core flash (DCF) and in particular during the first
(minor) proton ingestion episode (PIE). It was noted that one of the
most active reactions at this time is $^{13}$C($\alpha,n$)$^{16}$O.
This is evident in Figures \ref{fig-m0.85z0y245-DCF-allPIEs-conv-nucleo}
and \ref{fig-m0.85z0y245-ARP-peak-neutrons-fullNetwork}. An enormous
amount of $^{16}$O is produced in a very short time, primarily at
the expense of the already abundant $^{12}$C (via $^{12}$C($p,\gamma$)$^{13}$N($\beta^{+}$)$^{13}$C)
which itself is being produced by the core He flash $3\alpha$ reactions.
With such an active $^{13}$C($\alpha,n$)$^{16}$O reaction there
is a significant amount of neutrons being produced. Thus there is
the potential for s-process nucleosynthesis in the proton-polluted
HeCZ. 

\begin{figure}
\begin{centering}
\includegraphics[width=0.9\columnwidth]{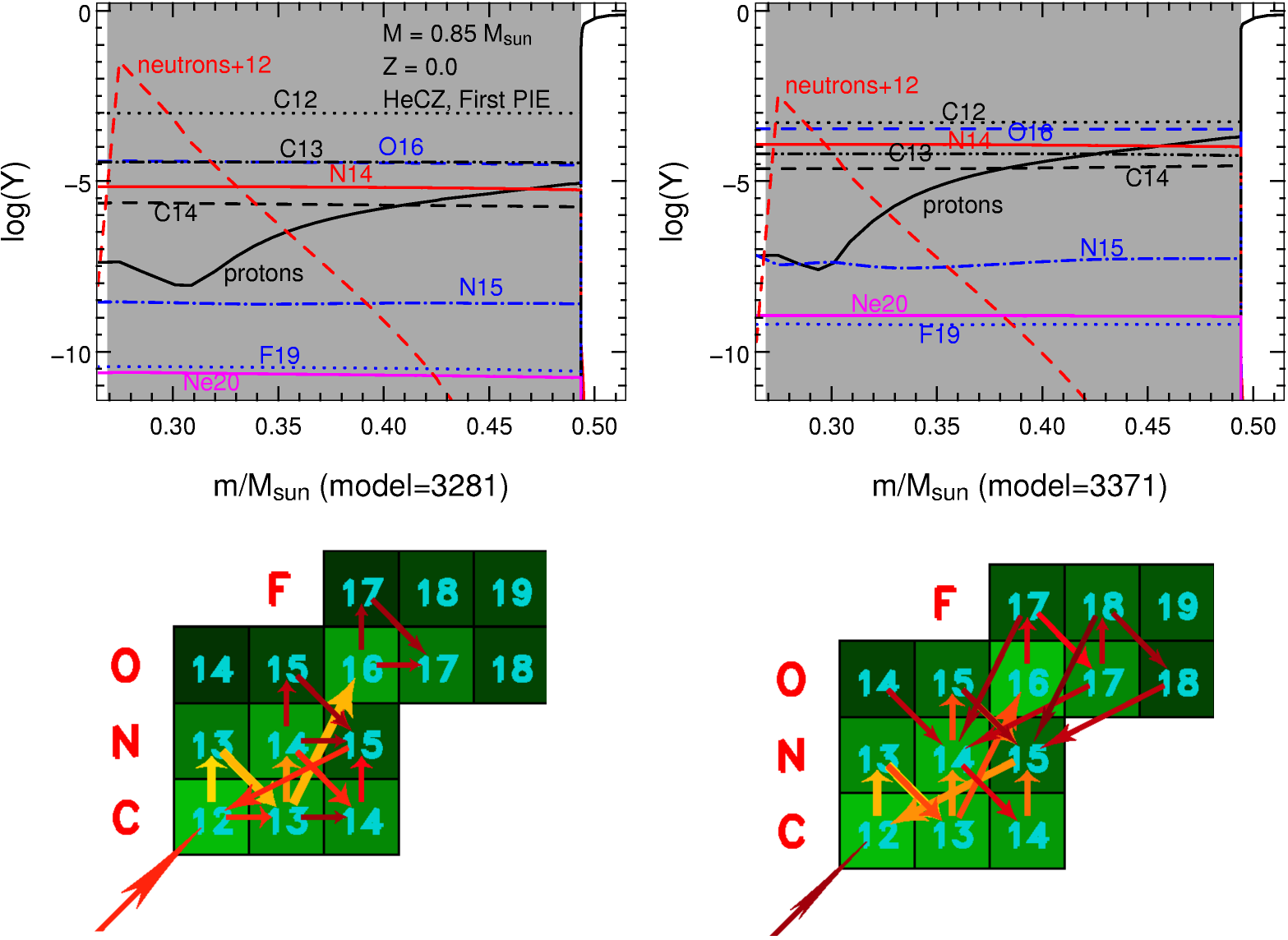}
\par\end{centering}
\caption{Investigating the potential for s-process nucleosynthesis during the
first DCF proton ingestion episode. Two models are displayed. The
first is taken at a time when the neutron abundance is near its peak
and the second when the proton abundance is near its peak (they are
separated in time by $\sim0.02$ years). Grey shading represents convection.
Below each abundance plot are corresponding rate plots (see Figure
\ref{fig-m0.85z0y245-ARP-peak-neutrons-fullNetwork} for an explanation
of this type of plot) which both sample the reactions occurring at
$m=0.28$ M$_{\odot}$, near the bottom of the HeCZ. The large arrow
at the bottom left of each arrow plot represents the $3\alpha$ reaction.
Comparing the two models it can be seen that $^{16}$O is produced
prodigiously, increasing by $\sim1$ dex, primarily via $^{13}$C($\alpha,n$)$^{16}$O.
$^{20}$Ne increases by $\sim1.5$ dex, mainly via a reaction that
also releases neutrons: $^{17}$O($\alpha,n$)$^{20}$Ne. The resulting
neutron abundance has been offset by $+12$ dex. This shows that the
main neutron production is towards the bottom of the HeCZ. The carbon
isotopes are initially much more abundant than $^{14}$N, lending
themselves to strong neutron production. However $^{14}$N also increases
by about 1 dex over this timespan due to proton capture reactions.
It then starts to become important as a neutron sink (via $^{14}$N($n,p$)$^{14}$C),
dampening the (possible) s-process nucleosynthesis. This reaction
also releases protons, causing the inversion in this abundance profile
at the bottom of the HeCZ. \label{fig-m0.85z0y245-CMPandARP-sprocess}}
\end{figure}

In Figure \ref{fig-m0.85z0y245-CMPandARP-sprocess} we show the abundance
profiles of two models taken during the first PIE. From these plots
we can see that the HeCZ is initially dominated by $^{12}$C (apart
from H and He). A well known neutron sink is $^{14}$N. If it is present
in large amounts then the $^{14}$N($n,p$)$^{14}$C reaction will
dominate over other neutron capture reactions, essentially muting
the s-process. In the present case however $^{14}$N is initially
present only at the level of $log(Y)\sim-10$, as compared to the
$^{12}$C abundance of $\sim-3$. Thus other neutron capture reactions
are expected to proceed. Indeed we see this in Figure \ref{fig-m0.85z0y245-ARP-peak-neutrons-fullNetwork}
where, for example, S and P are undergoing neutron captures to produce
the heavier isotopes of these elements. Unfortunately our network
did not include the further reactions (or species) that lead to the
production of the iron group and beyond. We note that Ni, Co and Fe
are included in the network but that they remain at zero abundance
because only reactions \emph{within} this group were included. Thus
we were unable to follow any s-processing that may have occurred.

Since these proton ingestion episodes are a common feature of $Z=0$
and extremely metal-poor low-mass models, this evolutionary phase
may be an important source of s-process nuclei in the early Universe.
As far as we are aware this phase has not been explored as a possible
s-process site before (whereas the AGB dual shell flash has; see eg.
\citealt{2004AA...421L..25G}). We shall definitely pursue this very
interesting possibility in a future study, when we have extended the
nuclear network of the NS code. 

\subsection{TP-AGB}

\begin{figure}
\begin{centering}
\includegraphics[width=0.8\columnwidth]{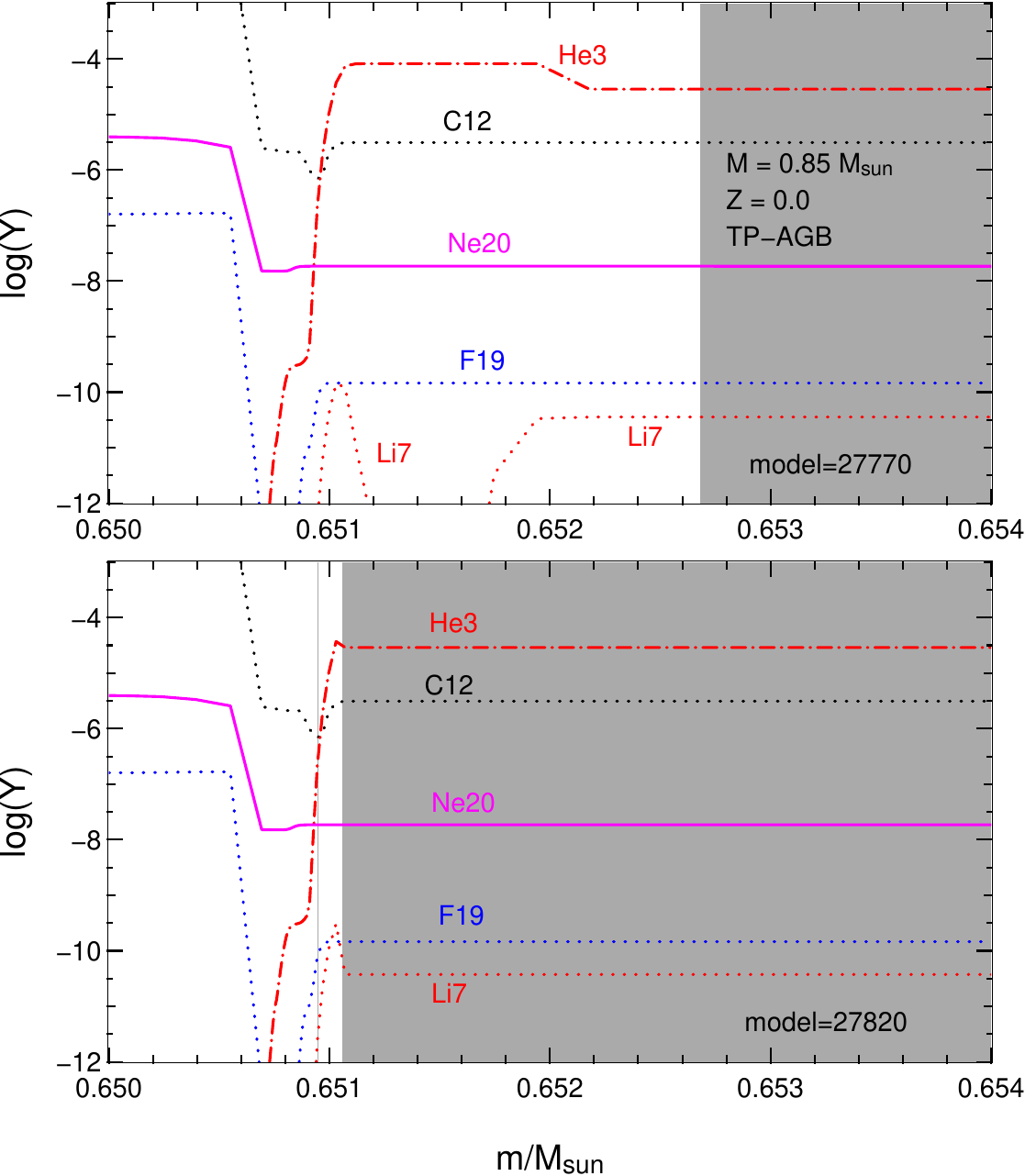}
\par\end{centering}
\caption{The composition profiles of two models during the AGB. The first (top)
shows the abundance profiles during the interpulse period. The second
(bottom) shows them just after the thermal pulse when the envelope
becomes more deeply convective. This convection mixes up some material
enhanced in $^{7}$Li and $^{3}$He, which, after many pulses, gives
rise to surface enrichment at the end of the AGB of $\sim0.7$ and
$\sim1.0$ dex respectively (over the initial abundances). \label{fig-m0.85z0y245-CMP-AGB-Li7He3-dup}}
\end{figure}

As can be seen in Figure \ref{fig-m0.85z0y245-srf-DCF-AGB} the AGB
has little effect on most of the species present in the envelope.
This is because the third dredge-up does not operate in this model,
and because the temperature at the bottom of the convective envelope
is too low for any significant hot bottom burning. The only species
affected are $^{7}$Li and $^{3}$He. We display the abundance profiles
of these species in Figure \ref{fig-m0.85z0y245-CMP-AGB-Li7He3-dup}
where it can be seen that both are produced just above the H shell
(during the interpulse period) and subsequently mixed upwards when
the envelope convection moves inwards each TP cycle. By the end of
the AGB $^{7}$Li is enhanced over the initial abundance by $\sim0.7$
dex and $^{3}$He by $\sim1.0$ dex. 

\subsection{Chemical Yield}

\begin{figure}
\begin{centering}
\includegraphics[width=0.7\columnwidth,keepaspectratio]{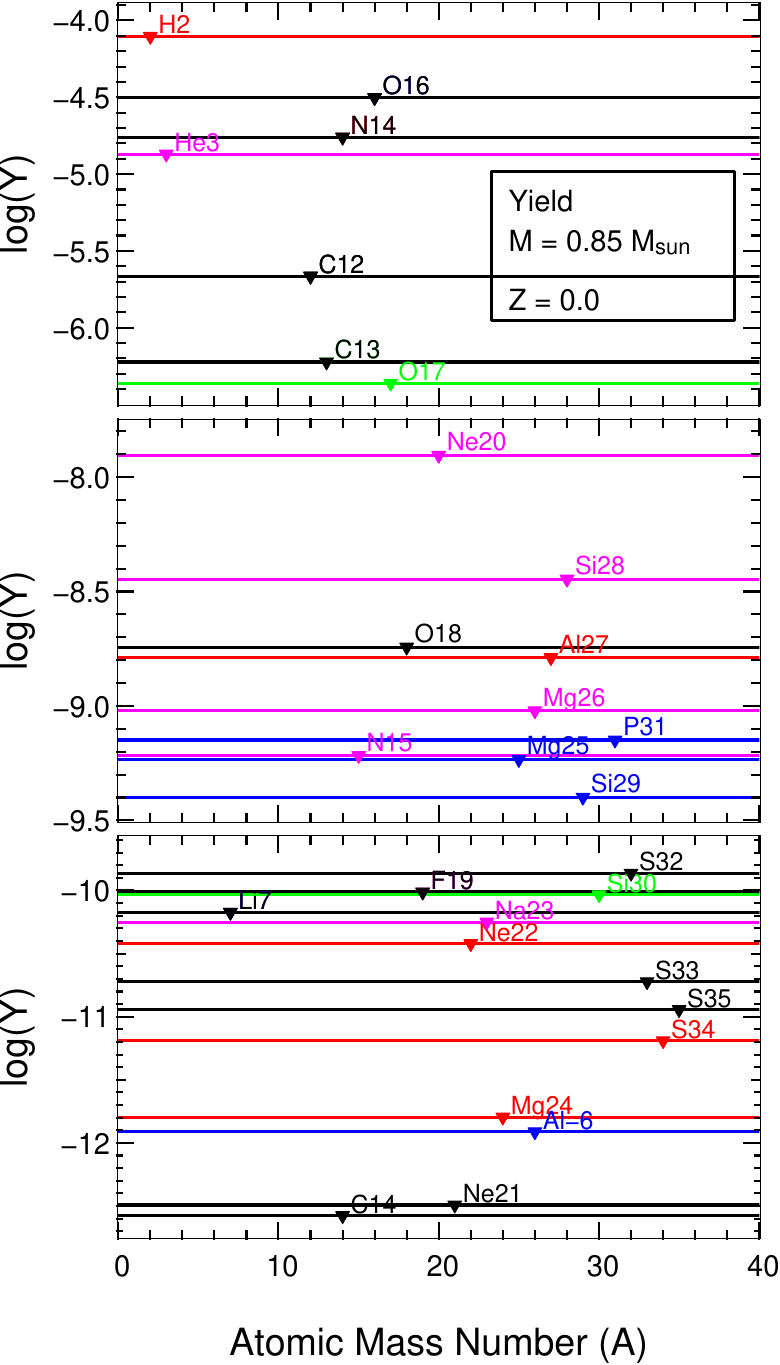}
\par\end{centering}
\caption{The chemical yield for the 0.85 M$_{\odot}$, $Z=0$ model. All isotopes
included in the network (74 species) with an abundance greater than
$10^{-13}$ are plotted. \label{fig-m0.85z0y245-YIELD-LogY-All}}
\end{figure}

The integrated output of material via stellar winds over the lifetime
of this model (the chemical yield), is plotted in Figure \ref{fig-m0.85z0y245-YIELD-LogY-All}.
In total 0.074 M$_{\odot}$ of material was lost via winds, leaving
a WD remnant of 0.776 M$_{\odot}$. 

In terms of metals $^{16}$O is the dominant species ejected into
the ISM, with $^{14}$N not far behind. $^{12}$C was output at a
level $\sim1$ dex lower. The $^{12}$C/$^{13}$C ratio is low, being
$\sim3.6$, whilst $^{4}$He is significantly enhanced, up from the
primordial value of $X_{He4}=0.245$ to 0.30. $^{7}$Li is marginally
enhanced, with an increase of $\sim0.2$ dex. This is despite a large
increase in Li on the AGB of $\sim0.7$ dex. The reason for this difference
is twofold. Firstly the $^{7}$Li is periodically enhanced on the
AGB, such that the surface abundance only reaches its peak towards
the end of the AGB. As mass loss is occurring throughout the AGB the
earlier winds do not have such a high Li abundance. Secondly, as seen
in the structural evolution section some mass loss occurs on the RGB
-- when the surface of the star still has zero metallicity. Thus
some of the mass lost to the ISM has a zero Li abundance, diluting
the overall yield of Li. The RGB mass loss has a small but noticeable
effect on the yields of all species, such that all are slightly lower
than their values on the AGB (see Figure \ref{fig-m0.85z0y245-srf-DCF-AGB}
for a comparison). The mass lost on the RGB represents $\sim20\%$
of the mass loss via winds, whilst the other $\sim80\%$ was lost
on the AGB. 

\begin{figure}
\begin{centering}
\includegraphics[width=0.85\columnwidth]{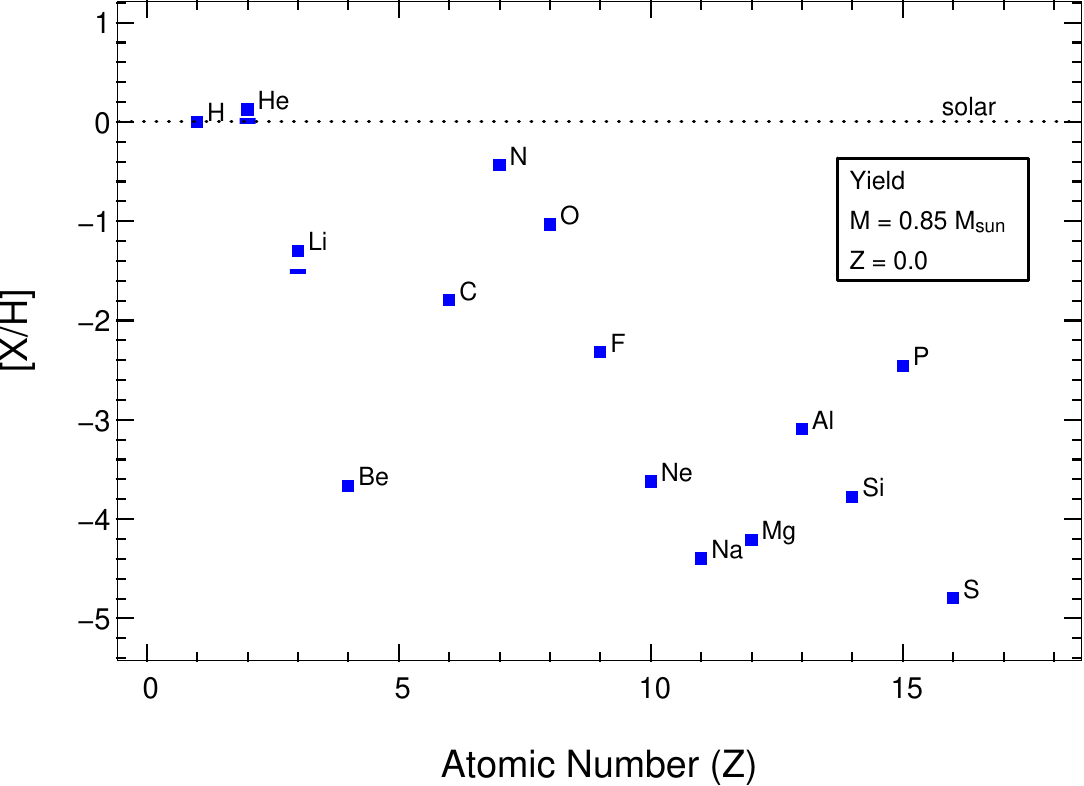}
\par\end{centering}
\caption{The yield for all the elements included in the network (that have
abundances $[X/H]>-5$), relative to solar, for the 0.85 M$_{\odot}$
model. Solar abundances are from \citet{2003ApJ...591.1220L}. Small
horizontal lines indicate the initial abundance for for H, He and
Li (all others were zero initially). Despite oxygen being the most
abundant metallic species in absolute terms (see previous figure)
it is second to nitrogen when taken relative to solar. We note that
Li is only marginally enhanced over the primordial abundance even
though it reached relatively high abundances on the AGB (see text
for a discussion). \label{fig-m0.85z0y245-YIELD-Elems-XonH}}
\end{figure}

In order to place the present yield in perspective we provide a second
figure (Figure \ref{fig-m0.85z0y245-YIELD-Elems-XonH}) that presents
the yield in relation to the Solar composition. We have used the Solar
abundances from \citet{2003ApJ...591.1220L}. It can be seen that
none of the metals are present in the ejecta at super-solar abundances.
Nitrogen comes the closest, being only $\sim0.4$ dex lower than the
solar {[}N/H{]} value. Interestingly oxygen is the second most abundant
metal in relative terms, even though in absolute terms it was the
most abundant. Lithium is only marginally up from the primordial value,
despite its large increase on the AGB. All the other elements are
$\lesssim-2$ dex lower than solar. We note that the Fe group elements
are not displayed as they all remain at zero abundance due to the
limitations of the network used, as mentioned in the previous subsection.

We shall discuss and compare this yield with observations in Section
\ref{Section-HaloStarModels-CompareObs}, after we have presented
the yields for the extremely metal-poor models.

\section{Detailed Nucleosynthesis at $M=2.0$ M$_{\odot}$}

In this section we detail the key nucleosynthetic events that occur
during the evolution of the $Z=0$, 2.0 M$_{\odot}$ model. In particular
we describe and quantify the events that lead to surface pollution
and hence affect the chemical yield of the star. 

Prior to the TP-AGB the only surface pollution event that this model
experiences is that of second dredge-up (2DUP). The main effect of
this is to mix up considerable amounts of helium, raising the He abundance
from the primordial $Y=0.245$ to $Y\sim0.31$. 

This model experiences three dual shell flashes (DSFs) at the start
of the TP-AGB. These significantly pollute the envelope, raising the
$Z_{cno}$ metallicity from zero to $\sim10^{-4}$. The DSF events
are similar to the DCF event found in the 0.85 M$_{\odot}$ model
described in the previous section. In this case it is the AGB He flash
convection zone that breaches the H-He discontinuity, dredging down
protons and dredging up CNO nuclei. Convection zone splitting also
occurs, due to the induced H-flash. The subsequent dredge-up by the
convective envelope transports this processed material to the surface.

Unlike the 0.85 M$_{\odot}$ model this model does experience third
dredge-up (3DUP) that periodically increases $Z_{cno}$. Although
the magnitude of 3DUP is quite small ($\lambda\sim0.01$) there are
so many pulses that $Z_{cno}$ increases by an order of magnitude
by the end of the AGB. This results in a chemical yield characterised
by $Z_{cno}\sim0.004$. Thus, in terms of surface pollution, 3DUP
has a greater effect than the prior DSFs. We note however that the
occurrence of 3DUP itself may be dependent on the occurrence of the
envelope pollution arising from the DSFs, due to the increase in envelope
opacity that this causes. 

Also unlike the 0.85 M$_{\odot}$ model this model experiences strong
hot bottom burning (HBB). This envelope burning couples with the periodic
dredge-up of $^{12}$C, $^{16}$O and $^{22}$Ne (for example), and
the residual pollution from the DSF episodes, to provide a rich nucleosynthetic
yield.

\subsection{Second Dredge-Up}

\begin{figure}
\begin{centering}
\includegraphics[width=0.75\columnwidth,keepaspectratio]{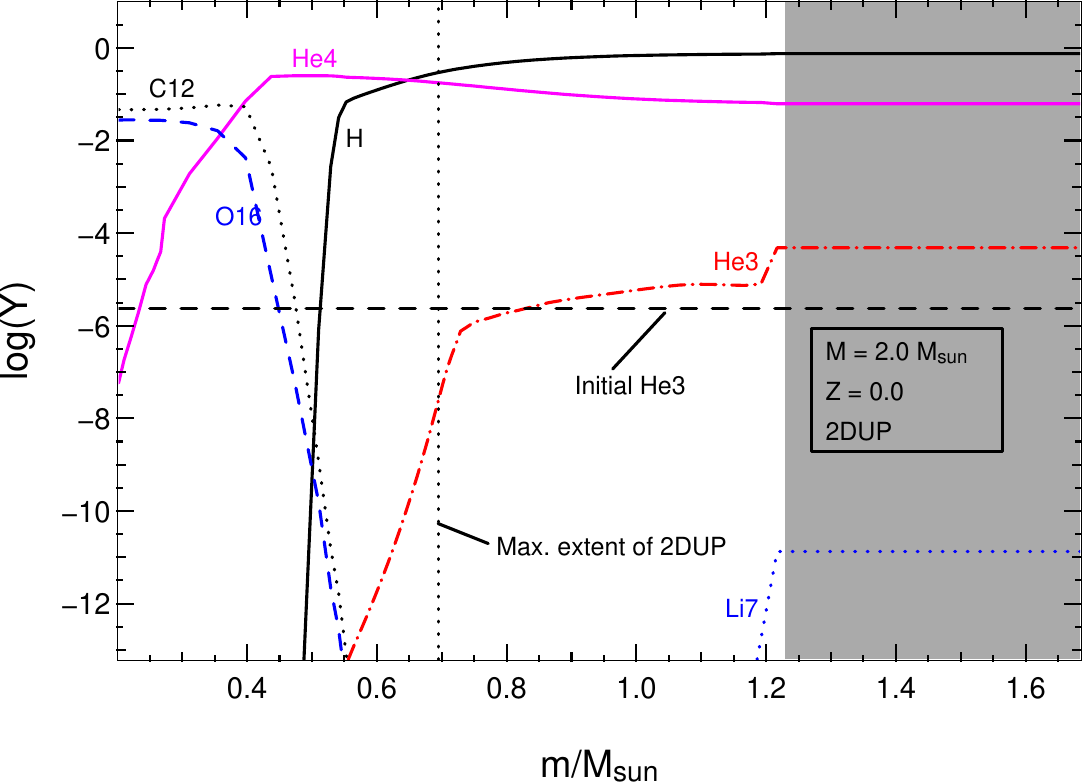}
\par\end{centering}
\caption{The chemical profiles of selected species in a model during second
dredge-up. The deepest incursion of the convective envelope is marked
by a vertical dotted line (convection is denoted by grey shading).
It can be seen that $^{4}$He will be enriched in the envelope, due
to previous $Z=0$ p-p chain burning in these regions. $^{3}$He will
also be enriched, and then slightly depleted (the original abundance
is marked by the horizontal dashed line), whilst $^{7}$Li will be
depleted. \label{fig-m2z0y245-2DUP-CMP}}
\end{figure}

This model does not reach an RGB configuration so first dredge-up
does not occur (see structural evolution discussion in Section \vref{section-m2z0-Structural}).
However second dredge-up (2DUP) does occur, although with limited
consequences when compared to metal-rich models. The maximum incursion
of the convective envelope on the EAGB is marked in Figure \ref{fig-m2z0y245-2DUP-CMP}.
Convection only reaches down into regions that have been (partially)
burnt via the p-p chains, as the material here still has $Z=0$. Nonetheless,
due to the low temperature dependence of the p-p chains there has
been significant production of $^{4}$He quite far out (in mass) from
the H-exhausted core. In fact much of the star has undergone some
H burning (out to at least $m\sim$1.5 M$_{\odot}$). Figure \ref{fig-m2z0y245-2DUP-CMP}
also shows the abundance profiles of selected species. It can be seen
that $^{7}$Li is depleted in the interior regions above the core
and that $^{3}$He is enriched (apart from a drop-off just below the
point of maximum envelope incursion). Thus, when the envelope moves
inwards it is substantially polluted with $^{4}$He and $^{3}$He
and depleted in $^{7}$Li.

\begin{figure}
\begin{centering}
\includegraphics[width=0.8\columnwidth,keepaspectratio]{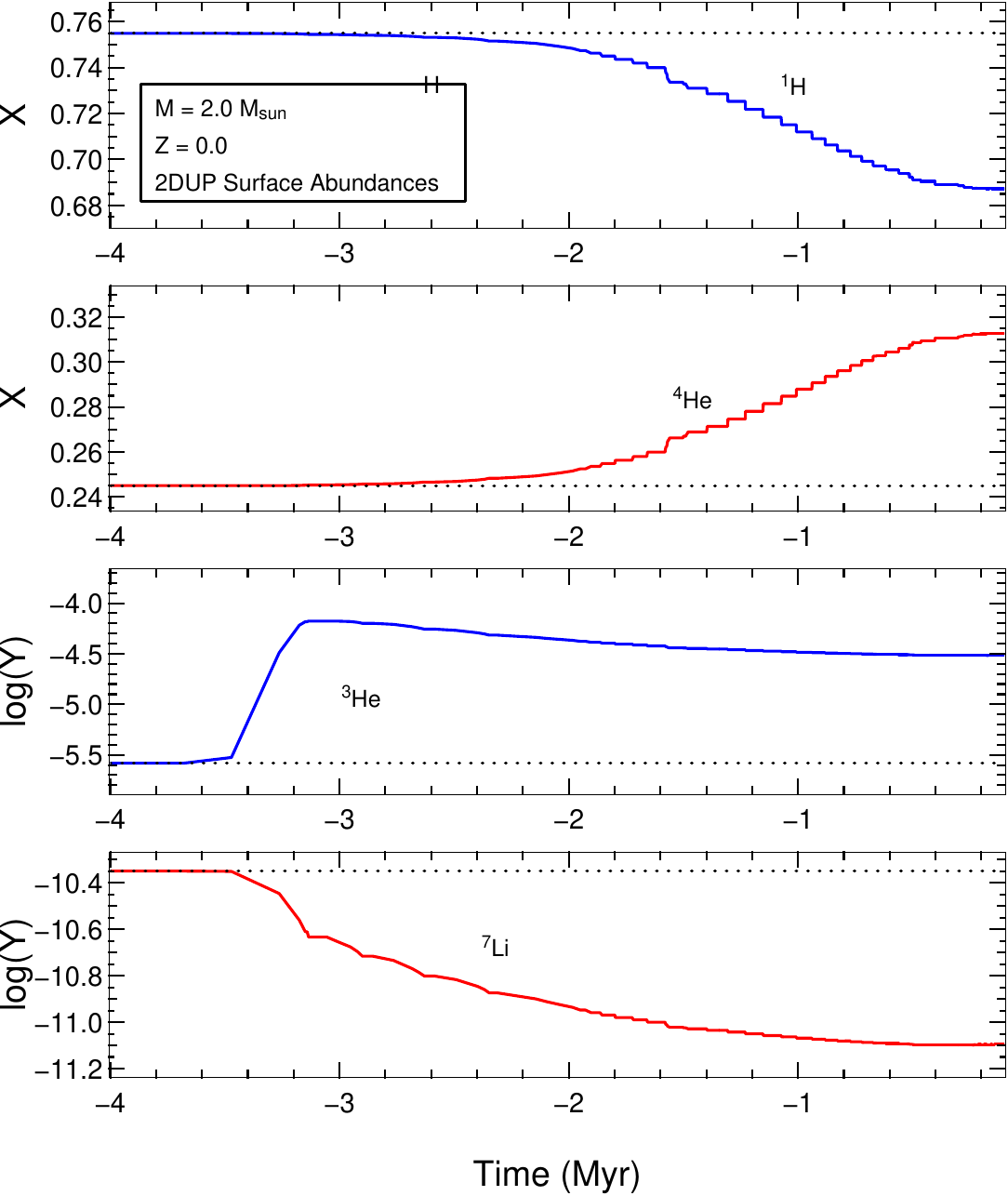}
\par\end{centering}
\caption{The time evolution of the surface abundances during the 2DUP event.
Time has been offset for clarity. Horizontal dotted lines mark the
initial abundance of each species. The main result from this event
is that $^{4}$He increases substantially, from the primordial value
of $Y=0.245$ to $\sim0.31$. \label{fig-m2z0y245-2DUP-AbundsSurfaceEvoln}}
\end{figure}

We show the time evolution of these envelope abundances in Figure
\ref{fig-m2z0y245-2DUP-AbundsSurfaceEvoln}. It can be seen that the
$^{4}$He abundance rises by a large amount, from the primordial value
of $Y=0.245$ to $Y\sim0.31$. We note that this 2DUP event has the
largest effect on the yield of $^{4}$He, as the increase on the TP-AGB
is less than this. It can also be seen that $^{3}$He increases by
$\sim$1 dex and $^{7}$Li decreases by $\sim$0.8 dex. In terms of
the chemical yield from this star this is inconsequential as both
these species are almost totally destroyed by HBB on the TP-AGB.

\subsection{Dual Shell Flashes/CNO-PIEs}

A key characteristic of $Z=0$ intermediate mass models is the occurrence
of one or more dual shell flashes (DSFs) at the start of the TP-AGB.
We suggest that these events may be pivotal for the further evolution
of the star because they lead to pollution of the convective envelope.
This pollution causes an increase in opacity consequently making third
dredge-up more likely to occur. If 3DUP does occur, as it does in
this model, it has a profound impact on the chemical yield of the
star. 

\begin{figure}
\begin{centering}
\includegraphics[width=0.85\columnwidth,keepaspectratio]{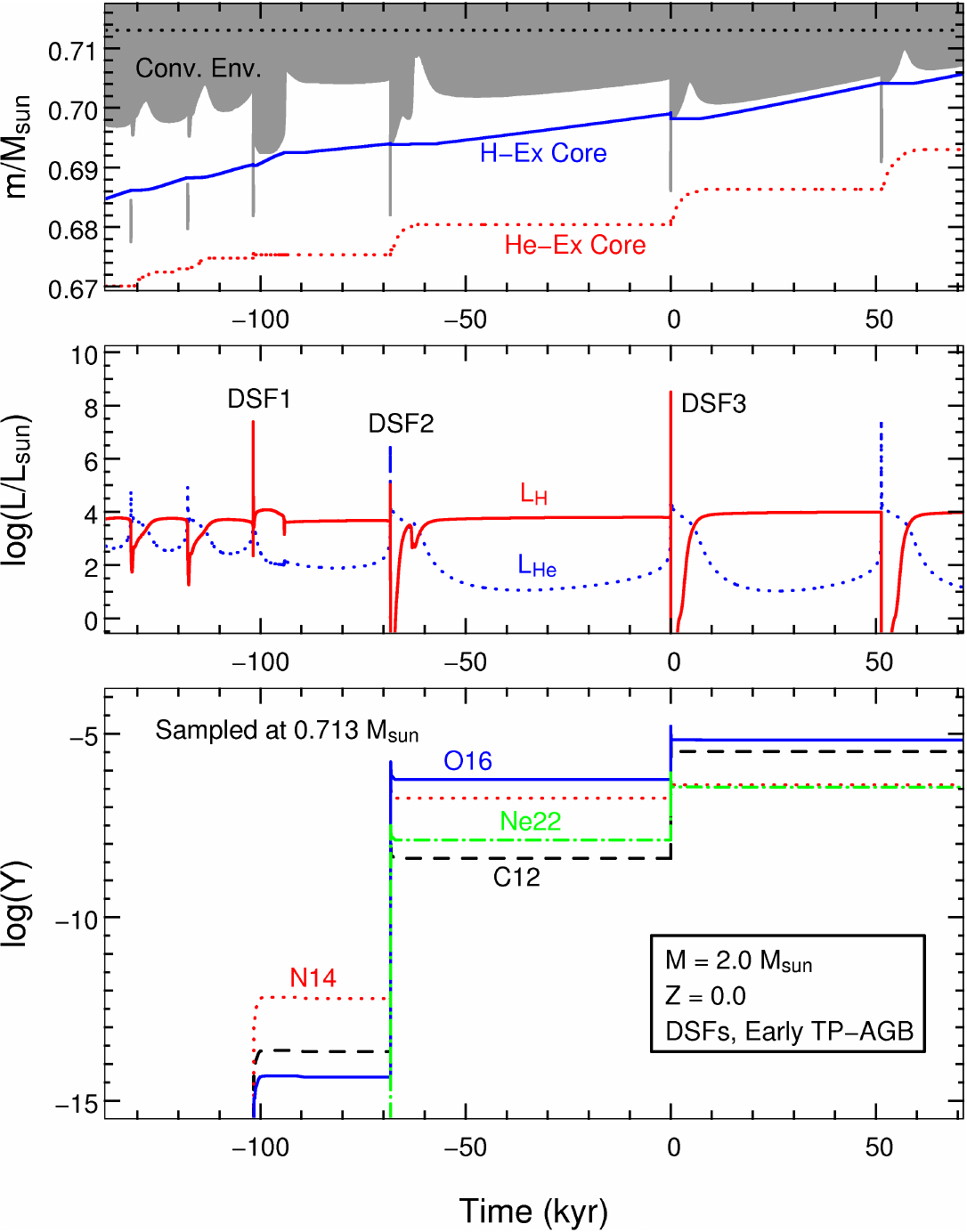}
\par\end{centering}
\caption{An overview of the three dual shell flashes and the resultant pollution
of the envelope. Time has been offset so zero corresponds to DSF3.
The DSFs are clearly marked by the peak in H luminosities which occur
at the same time as the peaks of the He shell flashes. The H flashes
only start to occur as the TP-AGB He flash luminosity rises above
$\sim$10$^{5}$ L$_{\odot}$. In the bottom panel the envelope pollution
that arises due to post-DSF dredge-up of the erstwhile HCZ is shown.
The dotted line in the top panel indicates the sample point for composition.
DSF1 has little effect on the envelope composition whilst DSF2 raises
$Z_{cno}$ to $\sim10^{-5}$ and DSF to $\sim10^{-4}$. It can also
be seen that DSF2 gives rise to a $^{14}$N-rich pollution whilst
DSF3 does not. Interestingly both give rise to pollution in which
$^{16}$O is the dominant metal component. \label{fig-m2z0y245-DSFs-wide-conv-lums-SRFabunds}}
\end{figure}

We find three DSFs to occur in this model. Each event consists of
the He intershell convection zone (associated with a thermal pulse)
penetrating the H-He discontinuity. This occurs near the start of
the TP-AGB, when the He flash luminosity first starts to exceed $\sim$10$^{5}$
L$_{\odot}$. We refer to these three DSFs as DSF1, DSF2 and DSF3.
The three events are seen in context in Figure \ref{fig-m2z0y245-DSFs-wide-conv-lums-SRFabunds}.
They are easily identified by the peaks in H burning luminosity occurring
at the same time as the peaks of the He shell flash luminosities.
Associated with each DSF is a dredge-up event, which we refer to as
post-DSF dredge-up (DSFDUP). This brings up the material processed
in the H convection zone (HCZ) that forms above the HeCZ during the
DSF. The envelope pollution resulting from these DSFDUPs can be seen
in panel 3 of Figure \ref{fig-m2z0y245-DSFs-wide-conv-lums-SRFabunds}.
The first DSF does not result in a significant pollution of the envelope
($Z_{cno}\sim10^{-11}$) so we shall concentrate on the details of
DSFs 2 and 3. DSF2 does result in substantial pollution, raising the
envelope metal abundance to $Z_{cno}\sim10^{-5}$, primarily in the
form of oxygen and nitrogen. At the next thermal pulse DSF3 raises
the metallicity even further, so that the envelope has a metallicity
of $Z_{cno}\sim10^{-4}$ after all the DSF episodes have completed.
Interestingly the post-DSF3 DUP results in an envelope composition
dominated by $^{16}$O and $^{12}$C, where DSF2 brought up more $^{14}$N. 

We now explore the details of the DSF episodes to ascertain what gives
rise to the composition of the polluting material from each DSF.

\subsubsection*{DSF2}

\begin{figure}
\begin{centering}
\includegraphics[width=0.75\columnwidth,keepaspectratio]{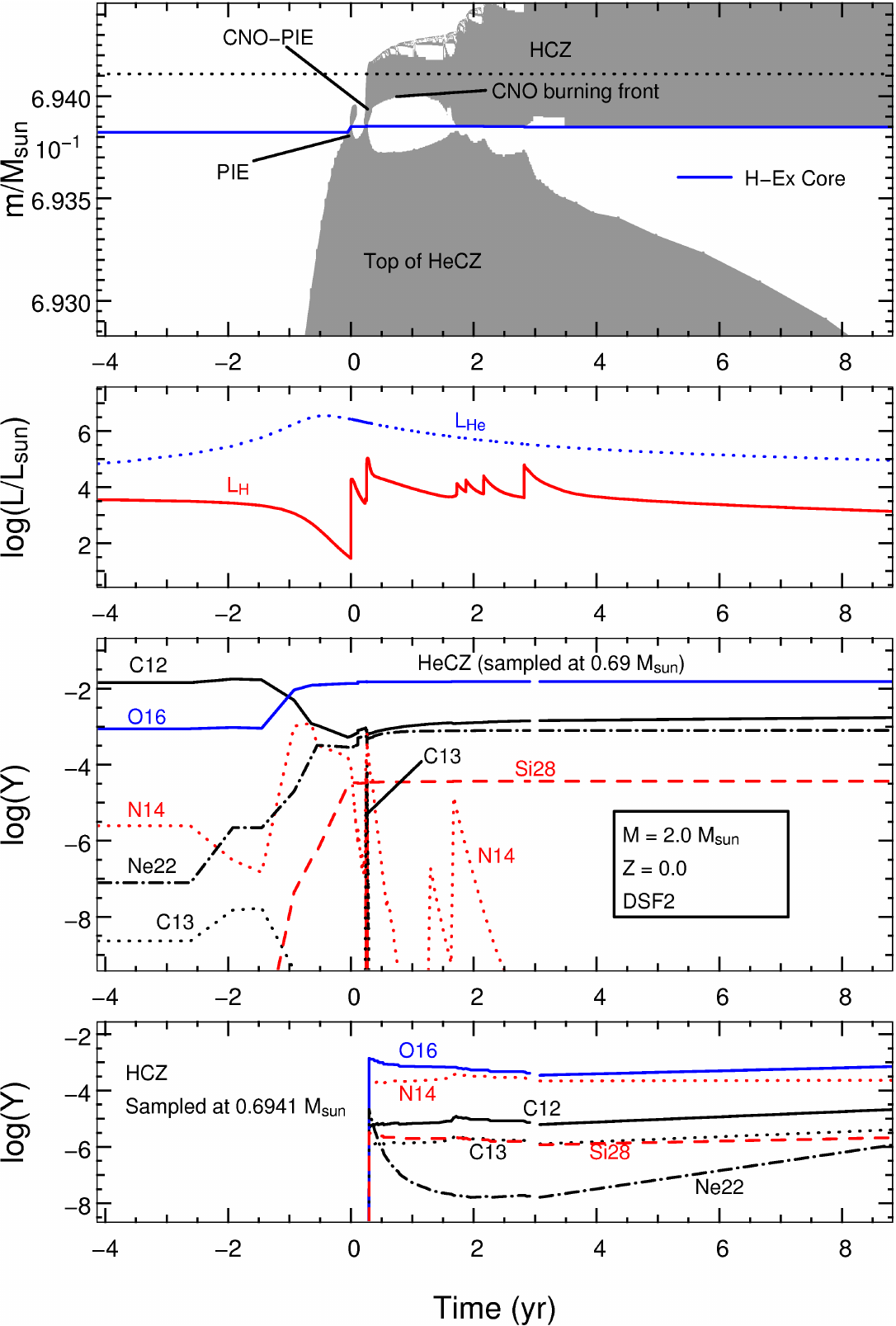}
\par\end{centering}
\caption{Zooming in on the evolution of DSF2. Time has been offset. The CNO
H-flash is instigated by the mixing up of CNO nuclei during the `CNO-PIE'
(CNO proton ingestion episode). The composition of the HeCZ is shown
in panel 3 and that of the HCZ in panel 4. The horizontal dotted line
in the top panel indicates the sampling point for the HCZ composition.
The HeCZ is sampled below the mass range of the plot, towards the
bottom of the convective zone where it is more actively burning. It
can be seen that, at the time of the CNO-PIE, the HeCZ is (comparatively)
$^{14}$N-rich. The presence of $^{28}$Si indicates that advanced
proton capture nucleosynthesis has occurred in the HeCZ. Some of this
material is mixed up into the newly formed HCZ. As the HCZ expands
the polluting material is diluted and also undergoes strong CN cycling,
so that the resultant chemical profile has $^{14}$N more abundant
than $^{12}$C. The $^{12}$C/$^{13}$C is also low, being $\sim5$.
The dominance of $^{16}$O is the result of the $^{13}$C($\alpha,n$)$^{16}$O
reaction. This nucleosynthesis of $^{16}$O has a lasting effect as
$^{16}$O remains the dominant species in the AGB envelope until 3DUP
of $^{12}$C eclipses it (see text for details). In the bottom panel
proton capture nucleosynthesis can be seen occurring in the HCZ. In
particular $^{22}$Ne is first destroyed then produced. \label{fig-m2z0y245-DSF2-medium-conv-lums-NS}}
\end{figure}

\begin{figure}
\begin{centering}
\includegraphics[width=0.7\columnwidth,keepaspectratio]{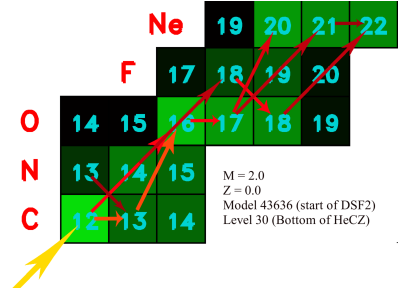}
\par\end{centering}
\caption{A rates plot showing the dominant reactions occurring at the bottom
of the HeCZ near the start of dual shell flash 2. Lighter coloured
(yellow/orange) thicker arrows indicate reactions occurring at higher
rates than those represented by darker coloured (red) thinner arrows.
The large yellow arrow at the bottom left represents the $3\alpha$
reaction. Apparent are the two active channels cycling $^{12}$C to
$^{16}$O, $^{12}$C($n,\gamma$)$^{13}$C($\alpha,\gamma$)$^{16}$O
and $^{12}$C($\alpha,\gamma$)$^{16}$O, as well as a second neutron
source, $^{16}$O($n,\gamma$)$^{17}$O($\alpha,n$)$^{20}$Ne. \label{fig-m2z0y245-ARP-DSF2-HeCZ-O16production}}
\end{figure}

Figure \ref{fig-m2z0y245-DSF2-medium-conv-lums-NS} shows the evolution
of various characteristics of dual shell flash 2. An important nucleosynthetic
event occurs at the beginning of the DSF -- an addition of protons
into the HeCZ leads to rapid CN cycling, producing $^{14}$N in an
amount comparable to the abundance of $^{12}$C. However, before the
$^{14}$N abundance rises to high levels a large production of $^{16}$O
occurs. This happens through the activation of the $^{13}$C($\alpha,n$)$^{16}$O
reaction. Neutrons from this reaction and the $^{17}$O($\alpha,n$)$^{20}$Ne
reaction are captured by $^{12}$C to produce $^{13}$C ($^{13}$C
is also produced by proton captures: $^{12}$C($p,\gamma$)$^{13}$N($\beta^{+}$)$^{13}$C
but the neutron capture channel is dominant). Thus there is a recycling
of neutrons that converts $^{12}$C to $^{16}$O: a C$\rightarrow$O
channel is set up. In addition to this the $^{12}$C($\alpha,\gamma$)$^{16}$O
reaction is still very active, so there are now two strong channels
to $^{16}$O production (see the rates plot for the bottom of the
HeCZ in Figure \ref{fig-m2z0y245-ARP-DSF2-HeCZ-O16production}). The
resultant high abundance from this nucleosynthesis is important in
one sense as this material is dredged up after the DSF to pollute
the envelope and $^{16}$O remains the dominant species in the AGB
envelope until 3DUP of $^{12}$C eclipses it. However, in terms of
the chemical yield of this model it is relatively unimportant because
most of the mass loss occurs during the HBB phase of the AGB. During
HBB the $^{16}$O is largely destroyed via ON cycling and other CNO
nuclei become dominant. 

\begin{figure}
\begin{centering}
\includegraphics[width=0.8\columnwidth]{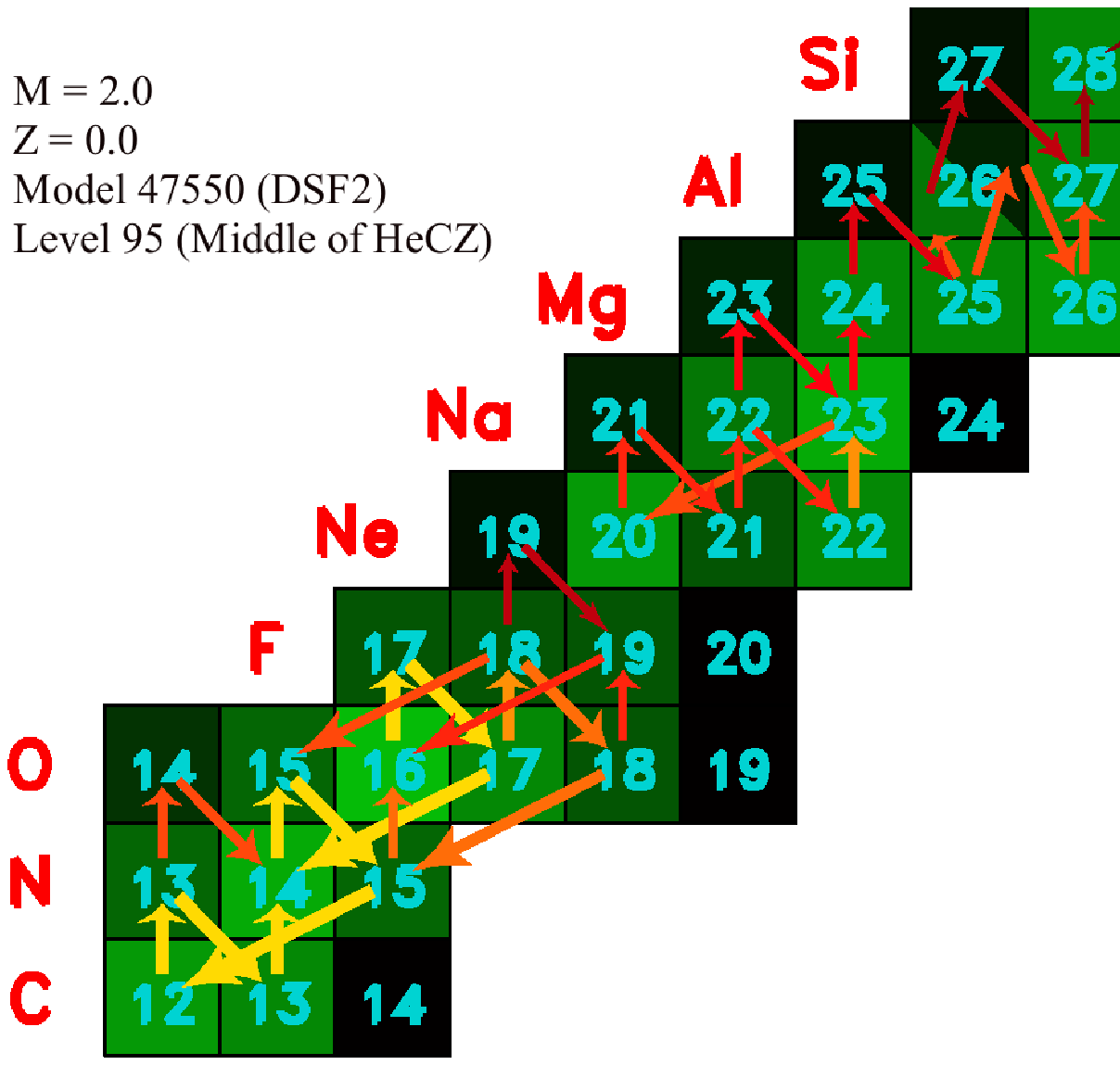}
\par\end{centering}
\caption{Same as Figure \ref{fig-m2z0y245-ARP-DSF2-HeCZ-O16production} except
sampled further out in mass in the HeCZ, during the proton ingestion
episode. Many proton capture reactions are occurring -- the full
CNO cycle is very active and the NeNa and MgAl chains can be seen
operating. $^{28}$Si is also produced. \label{fig-m2z0y245-ARP-DSF2-HeCZ}}
\end{figure}

Despite strong CN(O) cycling during the main H-flash (at $t\sim0.25$
yr in Figure \ref{fig-m2z0y245-DSF2-medium-conv-lums-NS}, panel 3)
$^{16}$O remains the dominant species in the HeCZ as there is not
enough time for the ON cycle to reach equilibrium. We note that the
$^{14}$N from this CN burning is later converted to neon by the continuing
He shell flash via $^{14}$N($\alpha,\gamma$)$^{18}$F($\beta^{+}\nu$)$^{18}$O($\alpha,\gamma$)$^{22}$Ne,
which is the usual fate of $^{14}$N in He shell flashes. However
in terms of envelope pollution it is the composition of the material
dredged up into the H-rich layers just above the HeCZ that is of most
interest. At the time of the PIE (and indeed, because of the PIE)
the abundance of $^{14}$N is still quite high in the HeCZ, and the
composition of the newly formed HCZ consequently reflects this. The
HeCZ also has a substantial amount of $^{28}$Si which is also dredged
up into the HCZ. This indicates that advanced proton capture nucleosynthesis
has occurred in the HeCZ. This is also evident in Figure \ref{fig-m2z0y245-ARP-DSF2-HeCZ}
where we show the dominant reactions occurring in the middle of the
HeCZ. 

It is important to note here that, from the perspective of the H-rich
layers, the PIE may also be seen as a CNO ingestion episode (or `CNO-PIE'),
as CNO nuclei are dredged upwards at the same time as protons are
mixed downwards. The addition of CNO nuclei into the H-rich layers
instantly causes an H-flash that peaks at $\sim10^{5}$ L$_{\odot}$.
During the H-flash the HeCZ (temporarily) recedes. Interestingly,
the burning shell \emph{moves outwards in mass} (like a flame front)
and creates a convection zone that mixes CNO nuclei further out and
also mixes in fresh H fuel. This behaviour is most likely caused by
the fact that the material just above the H-He discontinuity is still
extremely metal-poor, such that the H shell was mainly burning via
the p-p chains. The introduction of significant amounts of CNO material
initiates a runaway `CNO flash' as the burning switches from p-p chains
to CNO cycling. After the CNO flash recedes the HeCZ reestablishes
itself below. 

Some of the nucleosynthesis that occurs in the HCZ is evident in panel
4 of Figure \ref{fig-m2z0y245-DSF2-medium-conv-lums-NS}. The material
dredged up from the HeCZ during the CNO-PIE is first diluted throughout
the HCZ, reducing the local metallicity by $\sim1$ dex (as compared
to that of the HeCZ). Taking this into account it can be seen that
the abundance of $^{14}$N is enhanced by about 1 dex over the HeCZ
value, whilst $^{12}$C is depleted by $\sim1$ dex. This indicates
that strong CN cycling has occurred and further contributes to making
the HCZ relatively $^{14}$N-rich. Further proton-capture reactions
occur throughout the lifetime of the HCZ, as indicated by the gradual
destruction of $^{22}$Ne seen in Figure \ref{fig-m2z0y245-DSF2-medium-conv-lums-NS},
which occurs via the Ne-Na chains. Finally we note that the mass involved
in this DSF is quite small (as compared to that of the next DSF),
with the HCZ containing $\sim10^{-4}$ M$_{\odot}$ and the mass exchange
between the two convective zones only amounting to only $\sim10^{-5}$
M$_{\odot}$, as evidenced by the size of the HCZ during the CNO-PIE
(see top panel in Figure \ref{fig-m2z0y245-DSF2-medium-conv-lums-NS}).

\subsubsection*{An Error During DSF2}

\begin{figure}
\begin{centering}
\includegraphics[width=0.75\columnwidth,keepaspectratio]{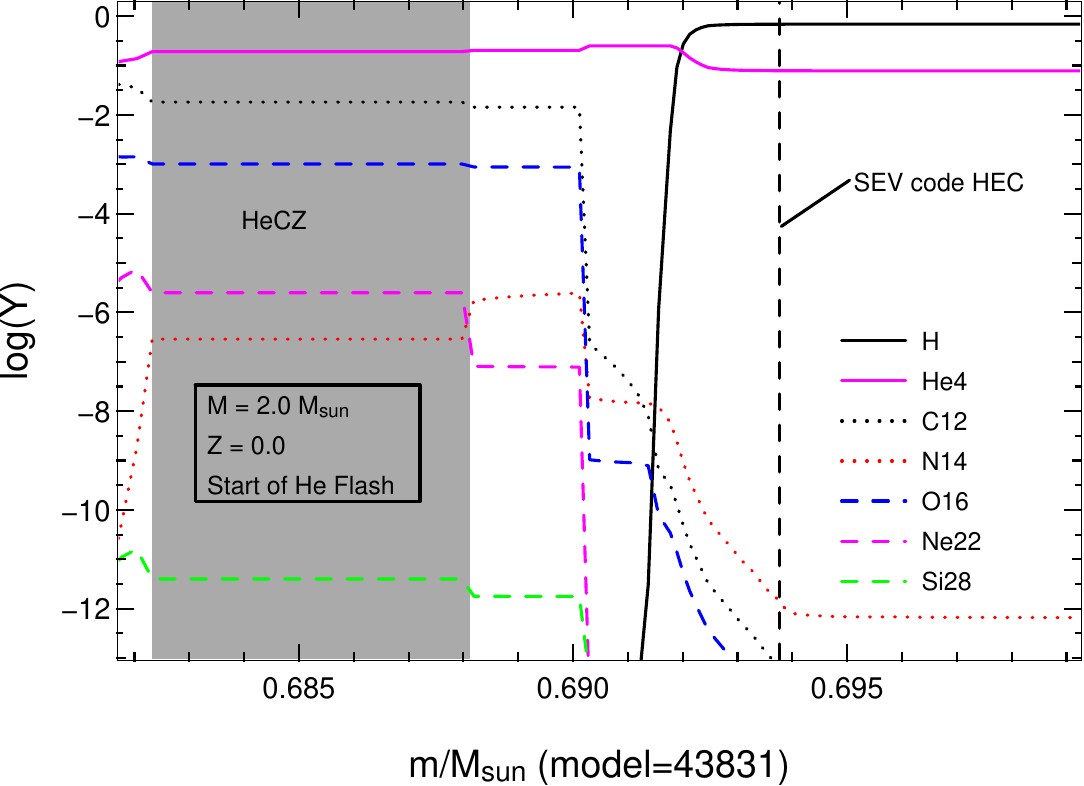}
\par\end{centering}
\caption{The chemical profile of some key species taken just as the He flash
convection starts at the beginning of DSF2. Grey shading marks the
He convective zone. The vertical dashed line marks the position of
the H-exhausted core as given by the structural evolution code. As
can be seen the H profile given by the NS code has placed the H-shell
at a lower mass coordinate (see text for discussion). The region with
high $^{12}$C abundance was created by He burning in the previous
He-flash. Regions with $^{14}$N $>$ $^{12}$C indicate advanced
CN cycling by the H-shell moving outwards. \label{fig-m2z0y245-CMP-DSF2start-HshellProblem}}
\end{figure}

A peculiar feature seen in the HeCZ chemical evolution plot (panel
4 of Figure \ref{fig-m2z0y245-DSF2-medium-conv-lums-NS}) is that
the C$\rightarrow$O conversion, and some CN cycling, \emph{occurs}
\emph{before the H-flash begins}. This implies that protons have been
mixed into the HeCZ \emph{before} it breaches the H-He discontinuity
marked in the top panel of Figure \ref{fig-m2z0y245-DSF2-medium-conv-lums-NS}.
In order to ascertain how this could be possible we present in Figure
\ref{fig-m2z0y245-CMP-DSF2start-HshellProblem} the chemical profiles
of a few species at a time near the start of the He flash. It appears
that the NS code has arrived at a different solution for the placement
of the H shell at this point in time, as compared to the structural
evolution code (SEV code). The difference between the two locations
is $\sim10^{-3}$ M$_{\odot}$. What this means is that, when the
He convection zone expands as the He flash reaches its peak, there
is some H ingested into it earlier than predicted with the SEV code.
This leads to the early production of $^{16}$O and $^{14}$N seen
in Figure \ref{fig-m2z0y245-DSF2-medium-conv-lums-NS}. We note that
we have never seen such a discrepancy between the two codes before.
In the 0.85 M$_{\odot}$ model's dual core flash this did not occur
(see Subsection \vref{subsec-m0.85z0-DCF-NS}), and it does not occur
in the next DSF of this model (DSF3). Unfortunately we have not been
able to trace the error due to the time constraints on the present
work. This discrepancy will however be addressed before we publish
the results of this model as a journal article. It will require a
close examination to find the source of the error and re-run the model.
Despite this uncertainty we believe that the yield results will not
be significantly affected. There are two reasons for this. Firstly,
the nucleosynthesis that occurs as a result of the extra and premature
proton ingestion calculated by the NS code is likely to be very similar
to that of the PIE that occurs directly after (the `real' PIE, as
given by the SEV code). Looking at the dual core flash nucleosynthesis
of the 0.85 M$_{\odot}$ model (see Figure \vref{fig-m0.85z0y245-DCF-allPIEs-conv-nucleo}),
which is a very similar phenomenon, we see that the same C$\rightarrow$O
processing occurs at the very first proton ingestion episode, such
that $^{16}$O becomes the dominant species at the expense of $^{12}$C.
We suggest that this is a robust outcome of the PIE phenomenon for
DCFs and DSFs. Moreover, when the $^{16}$O abundance has been raised
to such high levels it is difficult for any process to reduce it,
such that $^{16}$O remains dominant. This brings us to our second
reason. The present model soon undergoes HBB and 3DUP. 3DUP quickly
increases the surface $^{12}$C abundance to values much higher than
$^{16}$O. Meanwhile strong HBB causes ON cycling, depleting the $^{16}$O
dredged up in the DSF -- and thus essentially erases this nucleosynthetic
signature. It is 3DUP and HBB that dominate the yield of this model.
Thus we believe that the yield will not be significantly altered by
the error in the calculation but note again that it will nonetheless
need to be addressed in future work in the interests of scientific
robustness. 

\subsubsection*{DSF3}

\begin{figure}
\begin{centering}
\includegraphics[width=0.8\columnwidth,keepaspectratio]{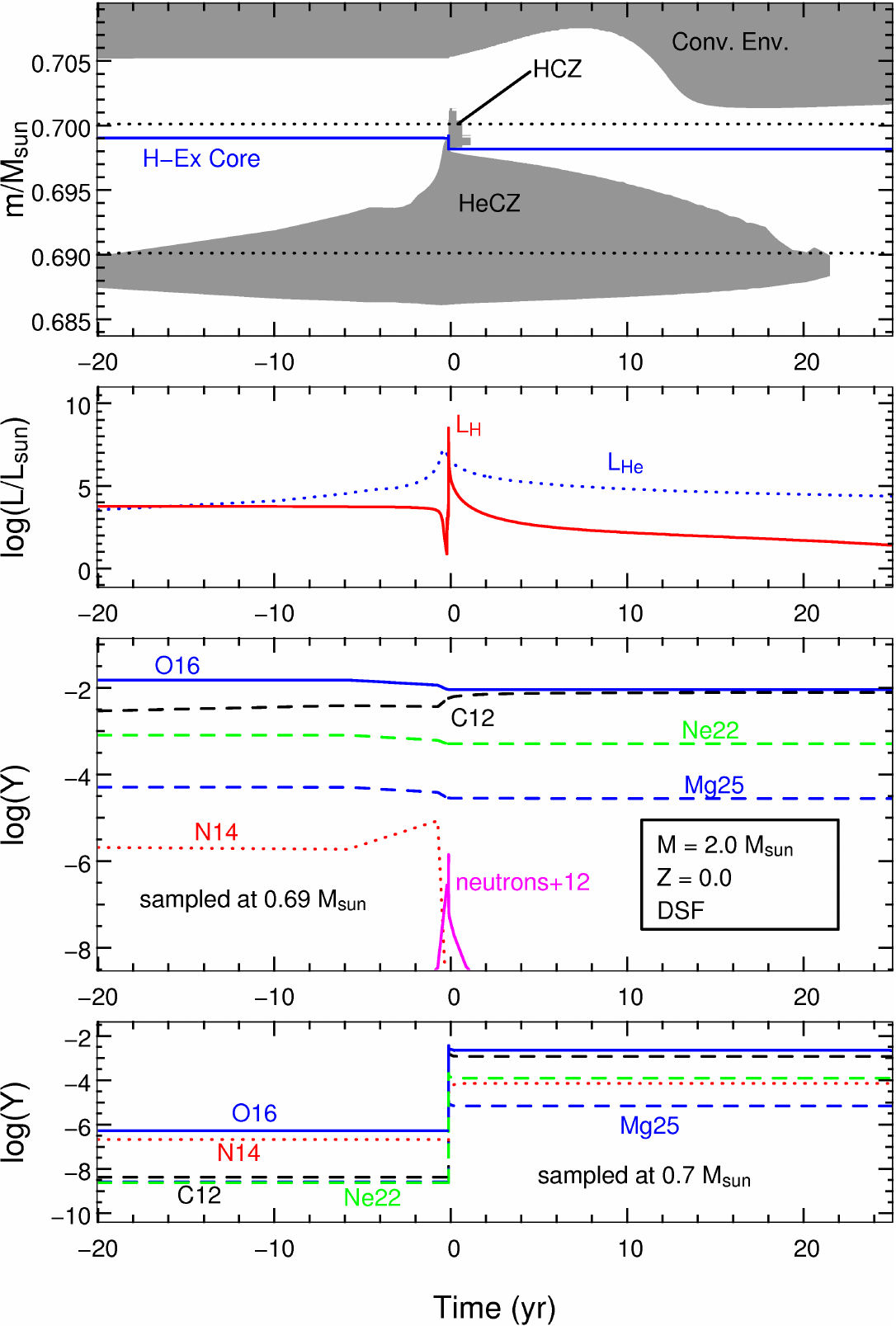}
\par\end{centering}
\caption{An overview of the third dual shell flash (DSF3). Time has been offset
to coincide with the peak of the H flash. The two horizontal dotted
lines in the top panel show the sample points for the composition
of the HCZ and HeCZ shown in the bottom two panels. The extent in
mass of the HCZ, and the significant movement inwards of the H burning
shell, is much larger than that in DSF2. This leads to a large envelope
pollution when the erstwhile HCZ is later dredged up. It can be seen
that the post-PIE HCZ composition basically reflects the composition
of the HeCZ apart from some CN cycling ($^{14}$N is enhanced and
$^{12}$C depleted). Also evident is a short neutron burst during
the peak of the flashes, which may give rise to some s-processing.
\label{fig-m2z0y245-DSF-conv-nucleo}}
\end{figure}

\begin{figure}
\begin{centering}
\includegraphics[width=0.8\columnwidth,keepaspectratio]{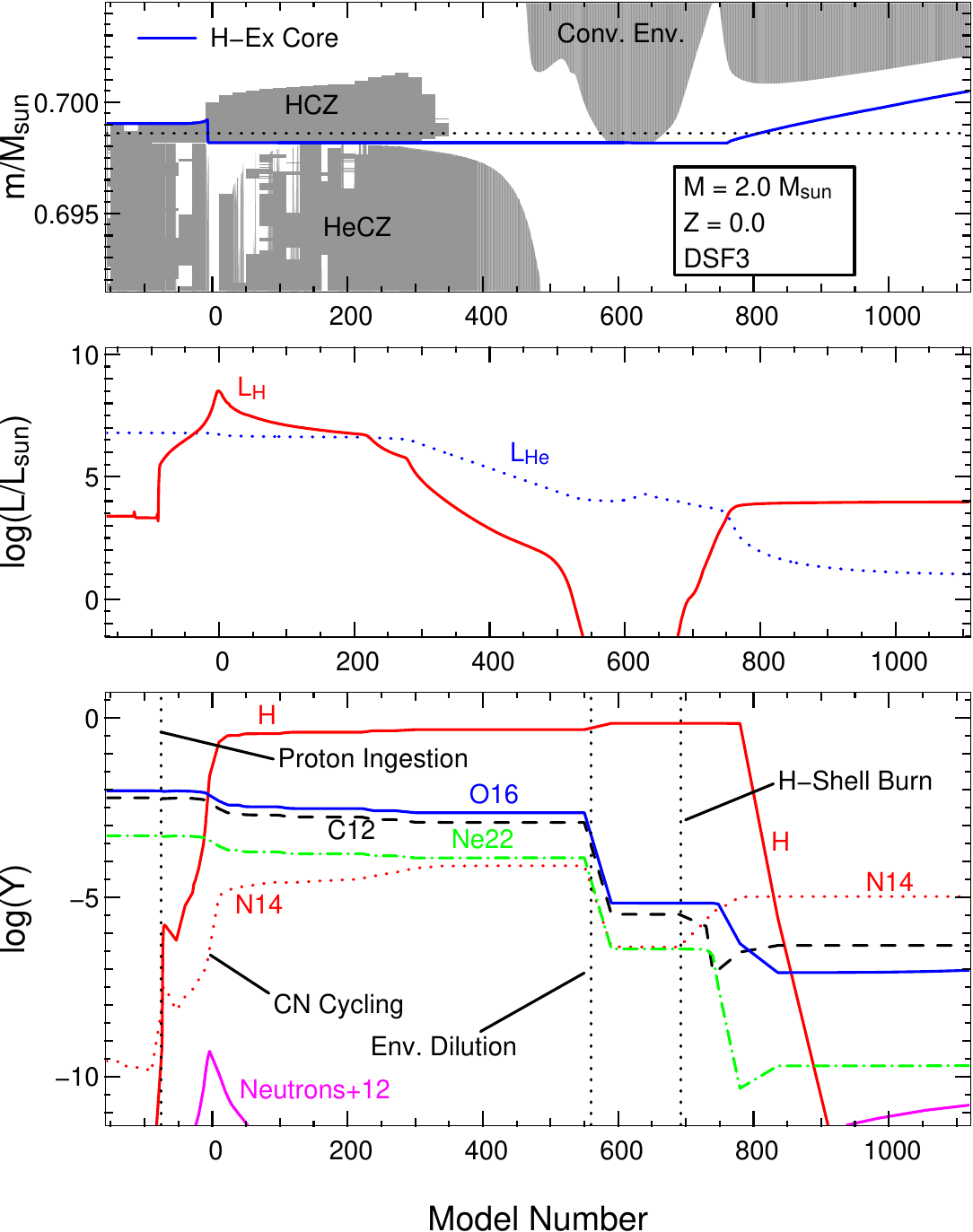}
\par\end{centering}
\caption{Zooming in on DSF3, the strongest and most polluting DSF. Note that
the evolution is against model number rather than time. This allows
us to see the rapid evolution more clearly, as the code timestepping
is governed by the changes in physical quantities. The dotted line
in panel 1 shows the sampling point for the composition evolution
in the bottom panel. The messy part of the HeCZ (top panel, convection)
occurs during the peak of the H flash and is due to the HeCZ convection
essentially dying off at this time. The splitting of the convection
zones is evident, as is the later dredge-up of the ex-HCZ by the convective
envelope. Many regimes are marked in the bottom panel. Of particular
note is the partial CN cycling during the very beginning of the PIE
(before the convective zones split) and also in the HCZ after the
convective zones split. \label{fig-m2z0y245-DSF3-detail-conv-lums-HCZabunds}}
\end{figure}

Out of all the DSFs the third gives rise to the largest pollution
of the envelope and thus it is the most important in terms of the
chemical yield for the model. Figure \ref{fig-m2z0y245-DSF-conv-nucleo}
gives an overview of the DSF, showing the He and H convective zones
and the evolution of the composition within these zones. As with DSF2
the composition of the HCZ after the PIE reflects the composition
of the HeCZ at the time of the ingestion, aside from some modifications
that occur during the PIE itself. As also seen in DSF2 the modifications
are due to CN cycling: $^{14}$N is strongly enhanced over the HeCZ
value, and $^{12}$C is slightly depleted. However, at odds with the
HCZ composition in DSF2, this HCZ is still dominated by $^{16}$O
and $^{12}$C, rather than $^{16}$O and $^{14}$N. The reason for
this appears to be the difference in starting abundances of $^{14}$N
during the PIEs. In the HCZ of DSF2 we saw that there was a small
PIE just before the H flash that produced a large amount of $^{14}$N
in the HeCZ. This high abundance was then added to via CN cycling
during the H flash (in the HCZ). In the DSF3 case the $^{14}$N abundance
has dropped to very low levels ($Y_{N14}\sim10^{-7}$) by the time
this PIE occurs. Figure \ref{fig-m2z0y245-DSF3-detail-conv-lums-HCZabunds}
shows that, like DSF2, there is some CN cycling during the PIE and
during the rest of the H flash, but it not complete. The CN cycle
does not reach equilibrium within this time and $^{12}$C remains
dominant. Thus the composition of the material dredged up to the envelope
later is relatively $^{14}$N-poor. 

We suggest that the difference between the two DSFs that is expressed
in the $^{14}$N abundance may be due to the different environments
in which each occurs. In the case of DSF2 the CNO-rich material from
the HeCZ is mixed up into regions of extremely low metallicity, causing
a CNO-flash as the previous burning could only occur via the p-p chains.
In DSF3 the material just above the H-He discontinuity is already
relatively metal-rich, so that CNO cycling is already occurring and
the H flash does not occur in the same manner. In addition to this
-- and possibly more importantly -- the strength of the He flashes
are different in the two episodes. The peak He luminosity in DSF2
is $\sim6.4$ L$_{\odot}$ whilst in DSF3 it is $\sim7.3$ L$_{\odot}$.
This leads to stronger convection in the second case, and hence to
a greater breaching of the H-He discontinuity. We suggest that it
is the weaker He flash that gives rise to the small proton ingestion
episode just before the main H flash in the DSF2 case (this can be
seen in luminosity in panel 2 of Figure \ref{fig-m2z0y245-DSF2-medium-conv-lums-NS}),
allowing some CN processing in the HeCZ prior to the main CNO-PIE.
In DSF3 the PIE is more `assertive' due to the stronger He flash --
the HeCZ quickly breaks through the H-He discontinuity to mix down
substantial amounts of H, giving rise to an immediate strong H flash
and a splitting of convection zones (although there is a very small
amount of proton ingestion just before the convection zone splitting).
It is interesting to note here that we have not used any form of overshoot
in these models. If we were to include overshoot then we would expect
even more `assertive' PIEs. This may remove the small PIEs and thus
any prior $^{14}$N production in the HeCZs. Then $^{12}$C would
be more abundant than $^{14}$N in the HCZ. It may also reduce the
effectiveness of the C$\rightarrow$O conversion (due to $^{13}$C($\alpha,n$)$^{16}$O
reactions operating in tandem with $^{12}$C($\alpha,\gamma$)$^{16}$O),
and thus reduce the oxygen content of the HCZ. Advanced proton capture
nucleosynthesis may also be reduced. Thus, due to the uncertainty
of the convective boundaries, there is some uncertainty in the composition
of the polluting material from these events. However, as discussed
in the next subsection, the $^{14}$N yield from this star is primarily
from hot bottom burning of $^{12}$C on the TP-AGB. 

\begin{figure}
\begin{centering}
\includegraphics[width=0.8\columnwidth]{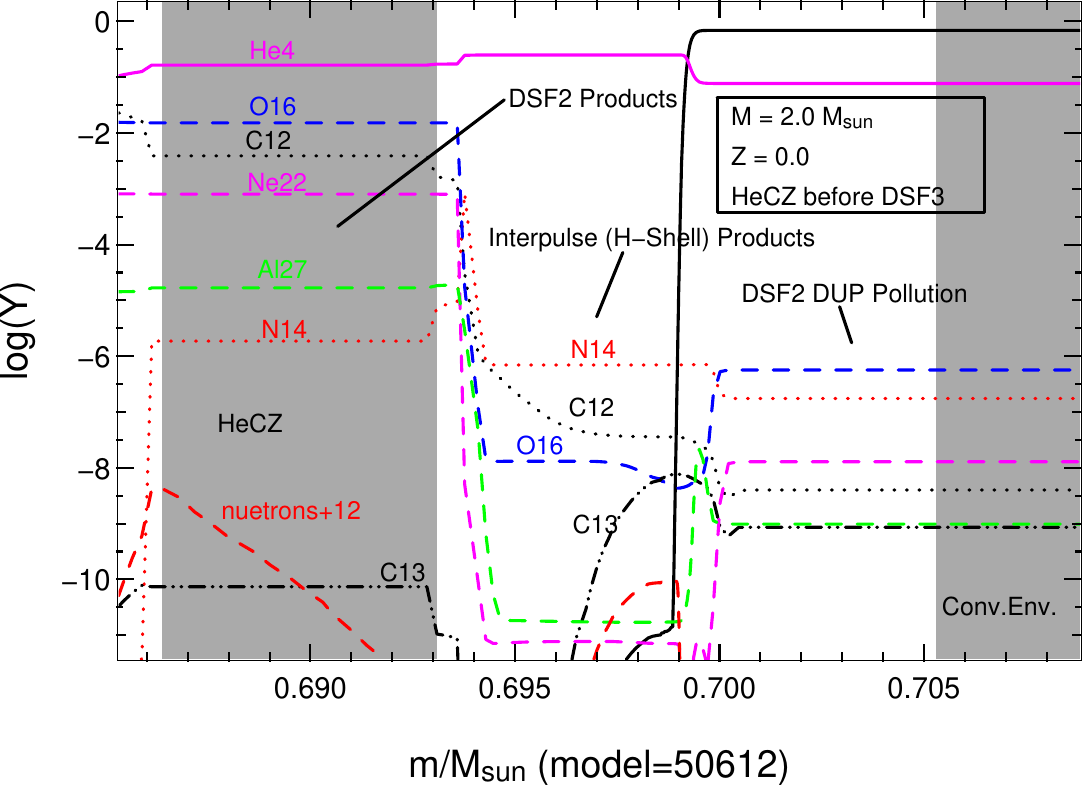}
\par\end{centering}
\caption{The run of abundances against mass just before the first proton ingestion
episode of DSF3. As indicated in the plot the HeCZ (which is still
expanding) is primarily composed of the material produced by the previous
thermal pulse (DSF2). In terms of metals this material is dominated
by $^{16}$O (see text for details). This chemical signature is also
visible in the envelope (although diluted). Between the envelope and
the HeCZ we see the products of the H shell burning that occurred
during the previous interpulse period, which has cycled the $^{16}$O
and $^{12}$C to $^{14}$N. Interestingly there is a $^{13}$C pocket
at the bottom of the H shell, where the proton abundance is $\sim10^{-8}$.
The neutron abundance is however very low ($Y_{n}\sim10^{-22}$).
\label{fig-m2z0y245-CMP-DSF3-HeCZ-beforeHFlash}}
\end{figure}

In Figure \ref{fig-m2z0y245-DSF3-detail-conv-lums-HCZabunds} we zoom
in closer on the evolution of DSF3. Here we can see that the He convective
zone splits in to two, similar to the case of the dual core flash
of the 0.85 M$_{\odot}$ model. The mass involved in the exchange
between the HeCZ and HCZ is $\sim10^{-3}$ M$_{\odot}$, much more
than that involved in DSF2 ($\sim10^{-5}$ M$_{\odot}$). Also similar
to the 0.85 M$_{\odot}$ model these two convection zones remain separated
from then on. The bottom panel in Figure \ref{fig-m2z0y245-DSF3-detail-conv-lums-HCZabunds}
shows the chemical evolution at a particular mass coordinate that
initially takes in the HeCZ and then the HCZ. $^{16}$O is the dominant
species in the HeCZ and thus becomes the dominant species in the HCZ
after the PIE mass exchange. Apparent in Figure \ref{fig-m2z0y245-CMP-DSF3-HeCZ-beforeHFlash}
is that the HeCZ primarily consists of material from the previous
thermal pulse nucleosynthesis (DSF2). As detailed in the previous
subsection this material underwent significant C$\rightarrow$O processing
which gave rise to the dominance of $^{16}$O. 

After the third DSF abates the entire (erstwhile) HCZ is dredged up
into the convective envelope (see Figure \ref{fig-m2z0y245-DSF3-detail-conv-lums-HCZabunds}).
This results in a significant increase in surface metallicity, raising
it to $Z_{cno}\sim10^{-4}$. Interestingly this means $log(Z/Z_{\odot})\sim-2$
such that it is comparable to the metal-poor Galactic globular cluster
stars. However, in terms of the heavier elements it remains ultra
metal-poor. 

\subsection{Dual Shell Flashes -- Potential s-Process Sites?\label{SubSec-DSFs-PotentialS-process}}

We suggested in Section \vref{subsec-m0.85z0-DCF-sProcess} that the
dual \emph{core} flash of the 0.85 M$_{\odot}$ model may be a good
site for s-process nucleosynthesis due to the $^{13}$C($\alpha,n$)$^{16}$O
reaction being so active during the flash. The DSF event is quite
similar to the DCF as they both involve protons being introduced into
a HeCZ, giving rise to interesting nucleosynthesis. 

\begin{figure}
\begin{centering}
\includegraphics[width=0.7\columnwidth,keepaspectratio]{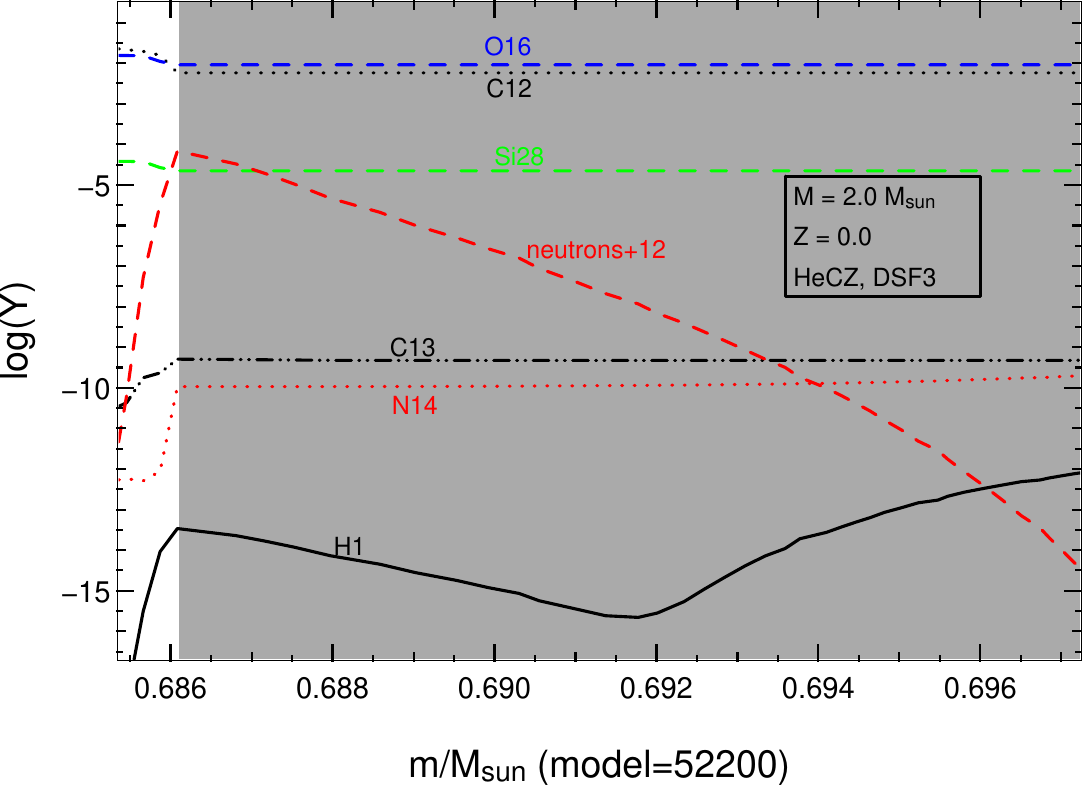}
\par\end{centering}
\caption{The chemical composition of the HeCZ near the peak of the core He-H
flash (DSF3). Grey shading represents convection. The (relatively)
large abundance of neutrons ($Y_{n}\sim10^{-16}$) may give rise to
some s-process nucleosynthesis. \label{fig-m2z0y245-CMP-sProcess-DSF3}}
\end{figure}

\begin{figure}
\begin{centering}
\includegraphics[width=0.8\columnwidth,keepaspectratio]{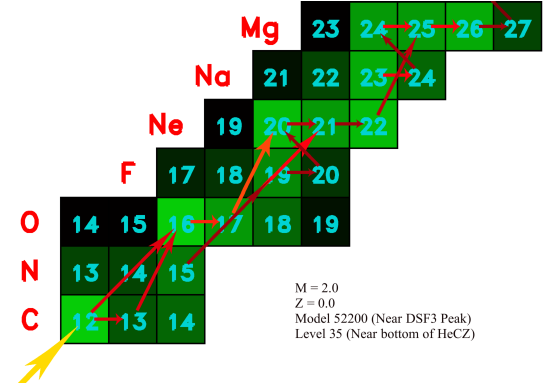}
\par\end{centering}
\caption{A rates plot taken during the peak of the He-H flash (DSF3). See the
caption of Figure \ref{fig-m2z0y245-ARP-DSF2-HeCZ-O16production}
for an explanation of this type of plot. The mass coordinate sampled
is towards the bottom of the HeCZ. Interestingly there are three competing
($\alpha,n$) reactions producing neutrons; the usual $^{22}$Ne and
$^{13}$C sources as well as the $^{17}$O($\alpha,n$)$^{20}$Ne
reaction. Neutron captures can also be seen occurring, leading to
the production the heavy Mg isotopes, for example. \label{fig-m2z0y245-ARP-sProcess-HeCZ-DSF3}}
\end{figure}

We suggest that the DSF may also be a good s-process site. In Figure
\ref{fig-m2z0y245-DSF-conv-nucleo} we do see a relatively large spike
in the neutron abundance at the peak of the H flash ($Y_{n}\sim10^{-18}$),
at the mass level sampled. In Figure \ref{fig-m2z0y245-CMP-sProcess-DSF3}
we show the chemical composition of the HeCZ near the peak of the
DSF. The neutron abundance is highest at the base of the HeCZ and
reaches values up to $Y_{n}\sim10^{-16}$. Although this is a much
lower abundance than found in the 0.85 M$_{\odot}$ DCF (where $Y_{n}$
peaked at $\sim10^{-13}$) we note again that, as we have not included
enough species to follow the s-process, we cannot be certain how significant
the resulting s-process nucleosynthesis would be. $^{13}$C is only
present in trace amounts ($Y_{C13}\sim10^{-9}$) but it is more abundant
than the neutron sink $^{14}$N. $^{16}$O is already the most abundant
metallic species. The reason for this is that both main C$\rightarrow$O
channels are operational ($^{12}$C($\alpha,\gamma$)$^{16}$O and
$^{13}$C($\alpha,n$)$^{16}$O), as can be seen in the rates plot
in Figure \ref{fig-m2z0y245-ARP-sProcess-HeCZ-DSF3}. An interesting
feature in this plot is that there are three competing ($\alpha,n$)
reactions producing neutrons; the usual $^{22}$Ne and $^{13}$C sources
as well as the $^{17}$O($\alpha,n$)$^{20}$Ne reaction. 

An interesting possibility from this potential nucleosynthesis is
that some primary Fe may be produced via neutron captures in the DSFs.
After the DSF dredge-up the surface chemical composition of the star
may then appear like that of an extremely metal poor Halo star. However
we suggest that the amount of Fe that would be produced would be negligible.
Nonetheless this possibility should be explored properly just incase.
We leave this and the other potential s-process nucleosynthesis for
future work, when we have extended the reaction network. 

\subsection{TP-AGB\label{subsection-m2z0-Nucleo-AGB}}

\begin{figure}
\begin{centering}
\includegraphics[width=0.95\columnwidth,keepaspectratio]{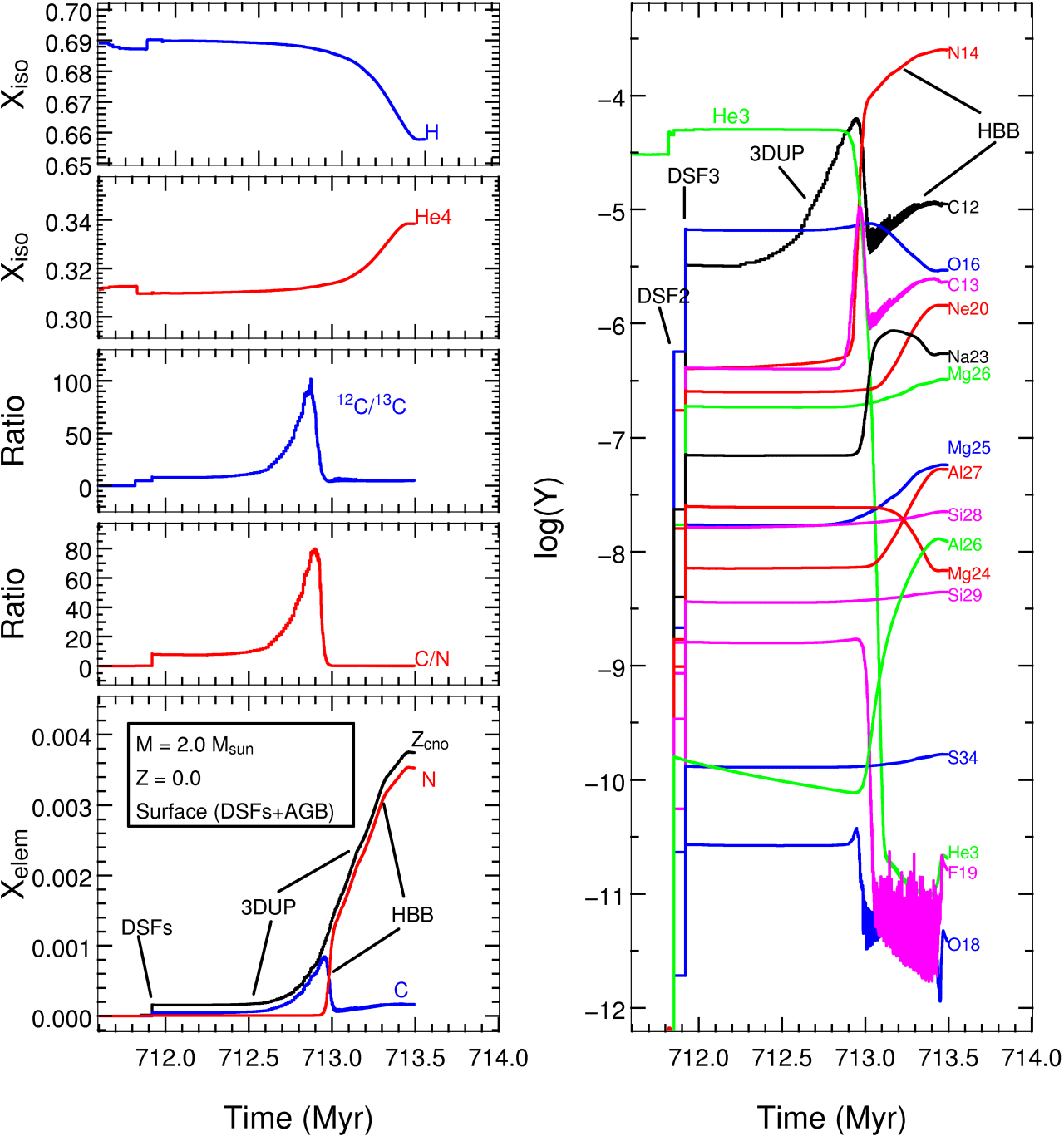}
\par\end{centering}
\caption{The surface chemical evolution during the DSFs and the AGB. Only selected
species are shown, for clarity. In the bottom panel on the left it
can be seen that the envelope pollution from the DSF episodes is swamped
by the pollution from ongoing 3DUP. The onset of (significant) HBB
is evident in the same panel by the strong C$\rightarrow$N conversion.
This star becomes a `nitrogen giant' at this stage, as $^{14}$N by
far dominates the metal content of the envelope. It also dominates
the yield as the majority of the mass loss occurs during this stage.
\label{fig-m2z0y245-SRF-AGB-all}}
\end{figure}

After the three DSF episodes have completed the star settles into
a relatively normal thermally-pulsing AGB evolution. Figure \ref{fig-m2z0y245-SRF-AGB-all}
displays the surface chemical evolution of the star on the AGB, including
the DSFs at the start of the AGB. Unlike the 0.85 M$_{\odot}$ model
this star does undergo third dredge-up (3DUP) and hot bottom burning
(HBB), both of which severely modify the composition of the envelope.
One thing immediately apparent in Figure \ref{fig-m2z0y245-SRF-AGB-all}
is that, in terms of bulk ($Z_{cno}$) metallicity, the surface pollution
the star experiences during the AGB far outweighs the pollution caused
by the DSFs. Thus, on the surface, it would seem that the computational
effort and detailed analysis of the previous subsection was unwarranted.
However we note that the DSFs are important for a couple of reasons.
Firstly, it may be argued that the 3DUP that pollutes the surface
on the AGB may not have occurred if the envelope had not been first
polluted by the DSFs. The increase in opacity that these pollution
events create allows deeper envelope convection, making 3DUP more
likely to occur. Secondly, although 3DUP drastically increases the
abundance of the CNO isotopes (which are redistributed by HBB) in
the latter part of the AGB, the star retains the DSF pollution composition
for about $2/3$ of the time spent on the AGB. Related to this is
the fact that the DSFs also provide a starting chemical distribution
for the AGB, which affects the final composition of the star due to
HBB acting on this material. 

It is important to note here that \emph{any} form of dredge-up from
the HeCZ early on the AGB would lead to a pollution of the envelope,
possibly to a similar level as that provided by the DSFs. However
the nature of the DSFs gives rise to a distinct chemical distribution.
This is mainly due to the proton ingestion episodes that occur at
the start of the DSFs. The main chemical signature that the DSFs provide
is that of $^{16}$O being very prominent at the start of the AGB.
This would also be the case in a model with scaled-solar composition
that experienced just a few 3DUP episodes, because the initial composition
is O-rich, but it would not be expected from a 3DUP event in a star
with a metal-free envelope. We would expect the composition to reflect
that of the HeCZ which would result in $^{12}$C dominating the metals
in the envelope. As shown in the previous subsection the DSF $^{16}$O
comes from the $^{12}$C($p,\gamma$)$^{13}$N($\beta^{+}\nu$)$^{13}$C($\alpha,n$)$^{16}$O
chain of reactions (and also the ongoing $^{12}$C($\alpha,\gamma$)$^{16}$O
reaction) when protons are ingested into the HeCZ . That this chain
of reactions also results in a substantial neutron production may
indicate that there may be an s-process signature accompanying the
$^{16}$O one. As mentioned earlier we did not follow the s-process
with the NS code but it will be an interesting avenue for further
work. 

Although the current work predicts an $^{16}$O signature we should
reiterate a caveat here: the DSF episode is dependent on the treatment
of convective boundaries as this affects the nature of the DSFs. For
example, the inclusion of overshoot would cause a larger ingestion
of protons very quickly, as the HeCZ expands and breaches the H-He
discontinuity earlier and more abruptly. This would alter the timescale
of the DSF (the extra H fuel would lead to a higher H luminosity sooner
and also a quicker splitting of the convection zone). It would also
alter the nucleosynthesis in the HeCZ. Thus the resultant composition
of the DSF pollution is somewhat uncertain due to the uncertainty
of the location of the convective boundaries. We stress that this
uncertainty should be kept in mind when examining the nucleosynthetic
products coming from DSFs (and DCFs). 

Returning to Figure \ref{fig-m2z0y245-SRF-AGB-all} it can be seen
that the $^{4}$He abundance rises significantly, primarily through
strong HBB towards the end of the AGB. As shown in the structural
evolution section for this star (Section \vref{section-m2z0-Structural})
the temperature at the bottom of the convective envelope reaches $\sim80$
MK. Under these conditions we expect significant ON cycling and also
further proton capture nucleosynthesis such as that via the Ne-Na
and Mg-Al chains. Looking at the last third of the AGB in Figure \ref{fig-m2z0y245-SRF-AGB-all}
we indeed see a rich nucleosynthesis has occurred. $^{16}$O is depleted
by $\sim0.5$ dex and proton chain products such as $^{23}$Na and
$^{27}$Al have been enhanced by about 1 dex (over the DSF pollution
levels). Before HBB sets in the $^{12}$C/$^{13}$C ratio rises to
very high values due to 3DUP of $^{12}$C. Once HBB is operating the
ratio then drops down to equilibrium values ($\sim3$). The most striking
nucleosynthetic feature is however the enormous production of $^{14}$N.
It is so abundant that, to first order, this star can be considered
a \emph{nitrogen giant} such that N totally dominates the metal content
of the envelope. The next most abundant species is $^{12}$C, but
this is 1.5 dex less prevalent.

\begin{figure}
\begin{centering}
\includegraphics[width=0.8\columnwidth]{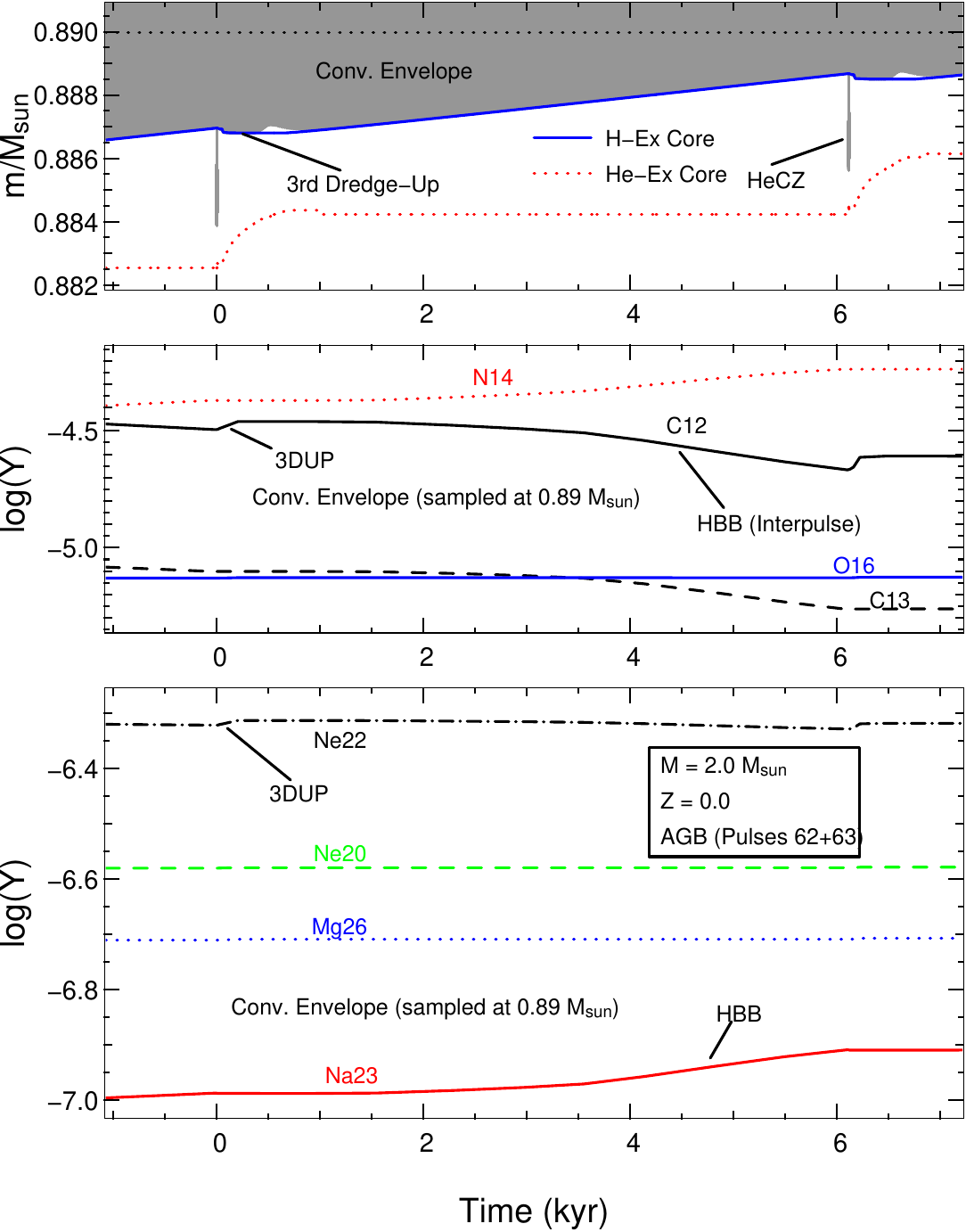}
\par\end{centering}
\caption{The evolution of an interpulse, just as $^{14}$N is beginning to
dominate the metal content of the envelope. The dredge-up of $^{12}$C
just after the He flashes can be seen in the second panel, as can
the decrease of this species during the interpulse period. The consequent
increase in $^{14}$N is also evident. In the bottom panel (sampled
at the same mass coordinate) a small dredge-up of $^{22}$Ne can also
be seen, as can its partial destruction later on. It is burnt to $^{23}$Na,
which can be seen increasing during the interpulse period. \label{fig-m2z0y245-HBB-Intpulse-conv-abunds}}
\end{figure}

$^{14}$N production comes about via the combination of 3DUP and HBB.
In Figure \ref{fig-m2z0y245-HBB-Intpulse-conv-abunds} we show this
occurring between two thermal pulses. First, fresh $^{12}$C is dredged
up into the convective envelope just after the pulse, then CN cycling
at the bottom of the convective envelope during the interpulse period
converts it to $^{14}$N. Also visible is the same combination of
events occurring for $^{22}$Ne, which is primarily burnt to $^{23}$Na
(note that this is before HBB is strong enough to destroy $^{23}$Na).
It is evident in this Figure that the amount of mass dredged up is
very small ($\lambda\sim0.01$) but, as noted in the structural evolution
section, the star undergoes 282 thermal pulses and thus achieves a
high level of pollution despite this. 

\begin{figure}
\begin{centering}
\includegraphics[width=0.85\columnwidth,keepaspectratio]{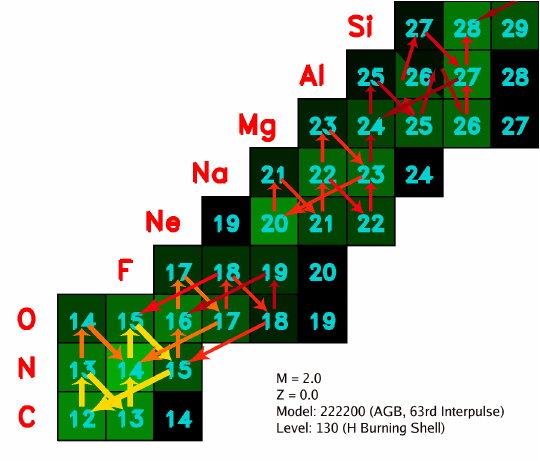}
\par\end{centering}
\caption{A rates plot showing the dominant reactions occurring at the base
of the convective envelope during the same interpulse period as that
displayed in Figure \ref{fig-m2z0y245-HBB-Intpulse-conv-abunds}.
See the caption of Figure \ref{fig-m2z0y245-ARP-DSF2-HeCZ-O16production}
for an explanation of this type of plot. This is where HBB is occurring
and, naturally, proton capture reactions are dominant. It can be seen
that, in addition to the CN(O) cycling, the NeNa and MgAl chains/cycles
are also operating. It is this nucleosynthesis that leads to the large
movement of abundances during the last third of the AGB seen in Figure
\ref{fig-m2z0y245-SRF-AGB-all}. As most of the mass loss also occurs
during this stage it is this nucleosynthesis (combined with ongoing
3DUP) that dominates the yield of the star. \label{fig-m2z0y245-ARP-AGBintpulse-HburningShell}}
\end{figure}

Figure \ref{fig-m2z0y245-ARP-AGBintpulse-HburningShell} displays
the dominant reactions occurring at the base of the convective envelope
during the interpulse. It can clearly be seen that the C-N, O-N, Ne-Na
and Mg-Al chains/cycles are all operating. In addition to this there
are proton captures on $^{26}$Mg and $^{27}$Al causing a leakage
out of the Mg-Al chains, producing $^{28}$Si. 

\begin{figure}
\begin{centering}
\includegraphics[width=0.8\columnwidth]{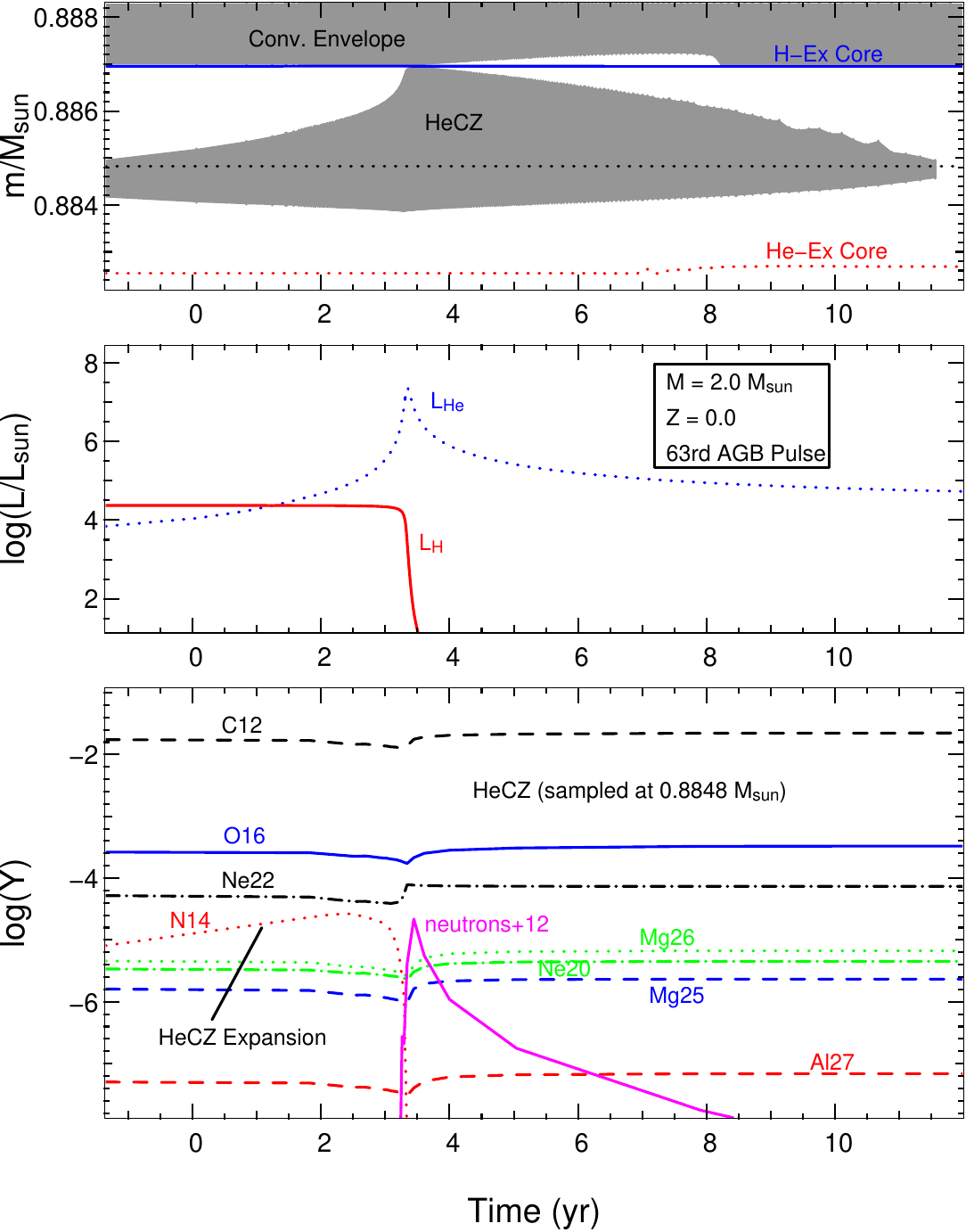}
\par\end{centering}
\caption{The evolution of the 63rd thermal pulse on the TP-AGB. The horizontal
dotted line in the top panel shows the sample point for the composition
evolution in the bottom panel. As can be seen in panel 2 there is
no associated H flash -- this is a normal thermal pulse. The evolution
of the abundances of some selected species is shown in the bottom
panel. The main nucleosynthesis happening here is the conversion of
$^{14}$N to $^{22}$Ne via alpha captures, and the production of
$^{12}$C via the usual He flash reactions. Interestingly there is
a (relatively) large spike in neutron abundance at the peak of the
flash (see text for further details). \label{fig-m2z0y245-AGB-HeCZ-TP63-conv-abunds}}
\end{figure}

We now delve into the evolution of a single thermal pulse (the 63rd
for this model) and associated HeCZ, to ascertain what nucleosynthesis
is occurring inside the HeCZ. This is important because it is this
material that is dredged up into the envelope at each 3DUP episode.
Figure \ref{fig-m2z0y245-AGB-HeCZ-TP63-conv-abunds} shows the evolution
of the composition in the HeCZ as well as that of the luminosities
and convection zones. The only species (of those displayed) which
has its abundance altered significantly is $^{14}$N. It is quickly
destroyed as the He flash gets close to its peak, ending up as $^{22}$Ne
via $^{14}$N($\alpha,\gamma$)$^{18}$F($\beta^{+}\nu$)$^{18}$O($\alpha,\gamma$)$^{22}$Ne.
Thus the abundance of $^{22}$Ne is seen to increase somewhat. We
note that the $^{14}$N initially in the HeCZ was produced before
the He flash by CN cycling in the H burning shell when it moved through
this region during the previous interpulse. 

\begin{figure}
\begin{centering}
\includegraphics[width=0.8\columnwidth]{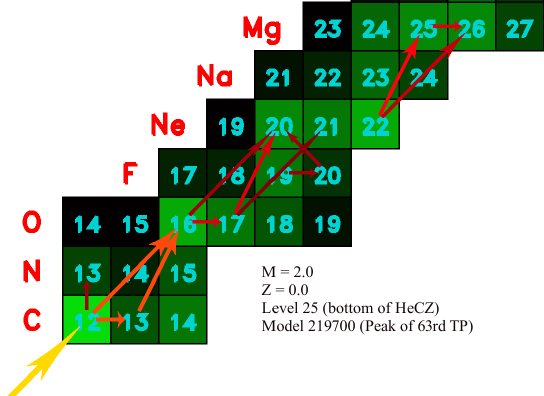}
\par\end{centering}
\caption{A rates plot showing the dominant reactions at the bottom of the HeCZ
at the peak of the He flash of the 63rd thermal pulse. See the caption
of Figure \ref{fig-m2z0y245-ARP-DSF2-HeCZ-O16production} for an explanation
of this type of plot. Three key neutron producing reactions are occurring:
$^{22}$Ne($\alpha,n$ )$^{25}$Mg, $^{13}$C($\alpha,n$)$^{16}$O
and $^{17}$O($\alpha,n$)$^{20}$Ne. This is allowing further production
of $^{13}$C through neutron capture on the abundant $^{12}$C so
that it appears that neutrons are being recycled between these two
reactions: $^{12}$C($n,\gamma$)$^{13}$C($\alpha,n$)$^{16}$O.
This neutron processing may lead to some s-process nucleosynthesis
during each thermal pulse. \label{fig-m2z0y245-ARP-TPAGB-HeCZ-sProcess}}
\end{figure}

Also slightly increasing in abundance during this phase are the heavy
Mg isotopes $^{25}$Mg and $^{26}$Mg, and $^{20}$Ne. $^{25}$Mg
is primarily produced via the $^{22}$Ne($\alpha,n$)$^{25}$Mg reaction.
This reaction liberates neutrons, and a neutron abundance spike can
be seen in Figure \ref{fig-m2z0y245-AGB-HeCZ-TP63-conv-abunds} during
the He flash. The neutrons are then available for capture by $^{25}$Mg,
which produces the $^{26}$Mg. The neutrons also open up another channel
for $^{20}$Ne production through $^{19}$F($n,\gamma$)$^{20}$F($\beta^{-}\nu$)$^{20}$Ne.
The relatively high neutron abundance during the flash may lead to
significant s-process nucleosynthesis. Again, as we have not included
enough species (and reactions) to follow the s-process we can only
speculate on the degree of its importance here. In Figure \ref{fig-m2z0y245-ARP-TPAGB-HeCZ-sProcess}
we show the main reactions occurring at the bottom of the HeCZ during
the peak of the He flash (after $^{14}$N has been depleted). The
$^{22}$Ne($\alpha,n$)$^{25}$Mg reaction can be seen operating and,
interestingly, so can the $^{13}$C($\alpha,n$)$^{16}$O neutron-producing
reaction. This reaction is able to operate because of the production
of $^{13}$C via neutron capture on the abundant $^{12}$C. It appears
that $^{12}$C is acting as an efficient neutron sink. It also appears
that neutrons are being recycled between these two reactions: $^{12}$C($n,\gamma$)$^{13}$C($\alpha,n$)$^{16}$O.
We suggest that this neutron cycle may have been triggered by the
$^{22}$Ne source, but note that this is difficult to ascertain. With
so many free neutrons available, and with a lack of the $^{14}$N
neutron sink, further neutron-capture reactions are likely. We shall
investigate the s-process in detail in future work with an extended
nucleosynthetic network. Finally we note that s-processing here on
the AGB may provide a useful nucleosynthetic signature for these stars.

\subsection{Chemical Yield}

\begin{figure}
\begin{centering}
\includegraphics[width=0.8\columnwidth,keepaspectratio]{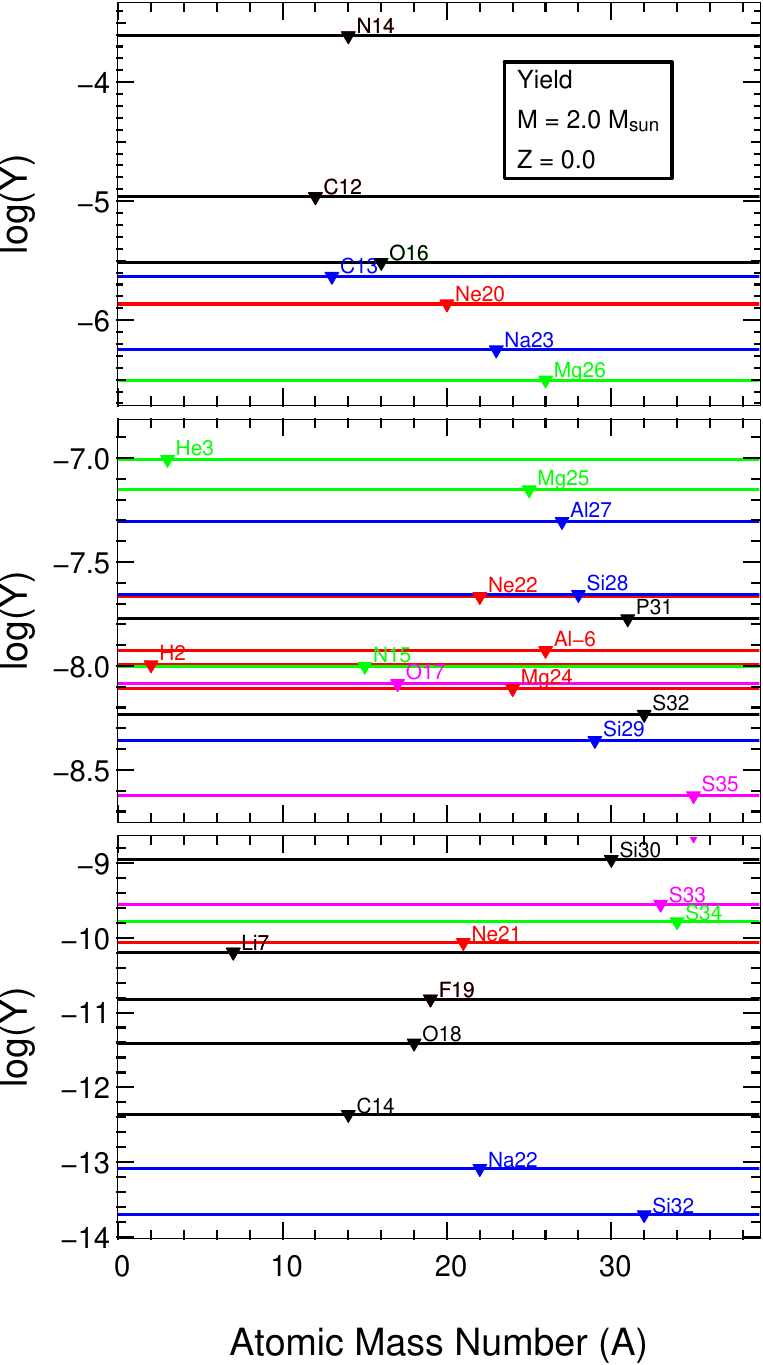}
\par\end{centering}
\caption{The yield of every species except $^{1}$H and $^{4}$He that was
included in the network (which have abundances of $log(Y)>-14$).
$^{14}$N is the dominant product, followed by $^{12}$C, $^{16}$O,
$^{13}$C and $^{20}$Ne. \label{fig-m2z0y245-YIELD-logY-All}}
\end{figure}

We display the entire nucleosynthetic yield (of metals) for this model
in Figure \ref{fig-m2z0y245-YIELD-logY-All}. $^{14}$N is the dominant
product, exceeding $^{12}$C by 1.5 dex. It was primarily produced
on the AGB via HBB of $^{12}$C which was dredged up periodically.
As seen in the previous subsection $^{12}$C was the dominant metal
species on the AGB just before HBB CN cycling set in ($^{16}$O from
the DSFs was dominant before this). However, as most of the mass loss
occurs towards the end of the AGB and as this model did not have any
RGB mass loss, the yield reflects the composition during the late
(HBB) phase of the AGB. Likewise, the $^{12}$C/$^{13}$C ratio in
the ejected material is low, being at a C-N cycle equilibrium level
of $\sim4$. $^{20}$Ne is the next most prominent species in the
yield. It was produced in the HeCZ (intershell) and also through HBB.
In the HeCZ the dominant reactions were alpha captures on $^{16}$O
and $^{17}$O. It was also produced via neutron capture through the
$^{19}$F($n,\gamma$)$^{20}$F($\beta^{-}$)$^{20}$Ne chain (see
Figure \ref{fig-m2z0y245-ARP-TPAGB-HeCZ-sProcess}). This material
was mixed to the surface via 3DUP. The bulk of the $^{20}$Ne was
however produced through HBB where proton captures destroyed much
of the $^{23}$Na via $^{23}$Na($p,\alpha$)$^{20}$Ne. $^{23}$Na
was initially in nett production on the AGB and was more abundant
than $^{20}$Ne before HBB set in. It was produced by the destruction
of $^{22}$Ne in the Ne-Na cycle. $^{22}$Ne is always abundant in
the HeCZ due to the destruction of the $^{14}$N left by the previous
H shell burning: $^{14}$N($\alpha,\gamma$)$^{18}$F($\beta^{+}\nu$)$^{18}$O($\alpha,\gamma$)$^{22}$Ne.
When HBB became hot enough for $^{23}$Na destruction it became the
main contribution to the increase in $^{20}$Ne. Some leakage out
of the Ne-Na cycle also occurred during HBB (see Figure \ref{fig-m2z0y245-ARP-AGBintpulse-HburningShell}),
primarily through $^{23}$Na($p,\gamma$)$^{24}$Mg. Thus the Mg-Al
cycle was able to operate, resulting in the modest yield of $^{25}$Mg,
$^{26}$Mg and $^{27}$Al seen in Figure \ref{fig-m2z0y245-YIELD-logY-All}.
Furthermore, HBB proton captures on these species allowed the production
of more massive species. $^{28}$Si was produced through $^{27}$Al($p,\gamma$)$^{28}$Si.

$^{3}$He was almost totally destroyed during HBB, although a small
amount was released in the yield as some of the mass loss occurred
before it reached an extremely low abundance. $^{31}$P and $^{32}$S
were released in small amounts ($log(Y)\sim-8$). The main source
for these two species was the DSFDUP after DSF3. Their abundances
were only slightly increased during the AGB. $^{19}$F was only released
in trace amounts ($log(Y)\sim-10$) due to its almost total destruction
via HBB (it reached a peak abundance of $log(Y)\sim-8$ after the
DSFs). $^{7}$Li was also mostly destroyed by HBB but it was initially
produced, before the HBB temperature reached high temperatures. Its
abundance reached a peak of $log(Y)\sim-8$ before plummeting to $\sim-15$.
The final yield was between these two extremes, being $\sim-10.2$
which, coincidentally, is very close to the initial (primordial) abundance
of $-10.4$.

The final yield of $^{4}$He was $X_{\textrm{He4}}=0.34$, an increase
of $\sim40\%$ over the initial (primordial) value of 0.245. About
$70\%$ of this increase was a product of second DUP whilst the rest
occurred via HBB of H on the AGB. This is quite a large increase (especially
for a 2 M$_{\odot}$ star) and represents another significant nucleosynthetic
signature of this model. The He increase was at the expense of $^{1}$H
which is present in the yield at the level of $X_{\textrm{H}}=0.66$,
down from an initial value of 0.75, roughly a $10\%$ decrease. 

\begin{figure}
\begin{centering}
\includegraphics[width=0.85\columnwidth,keepaspectratio]{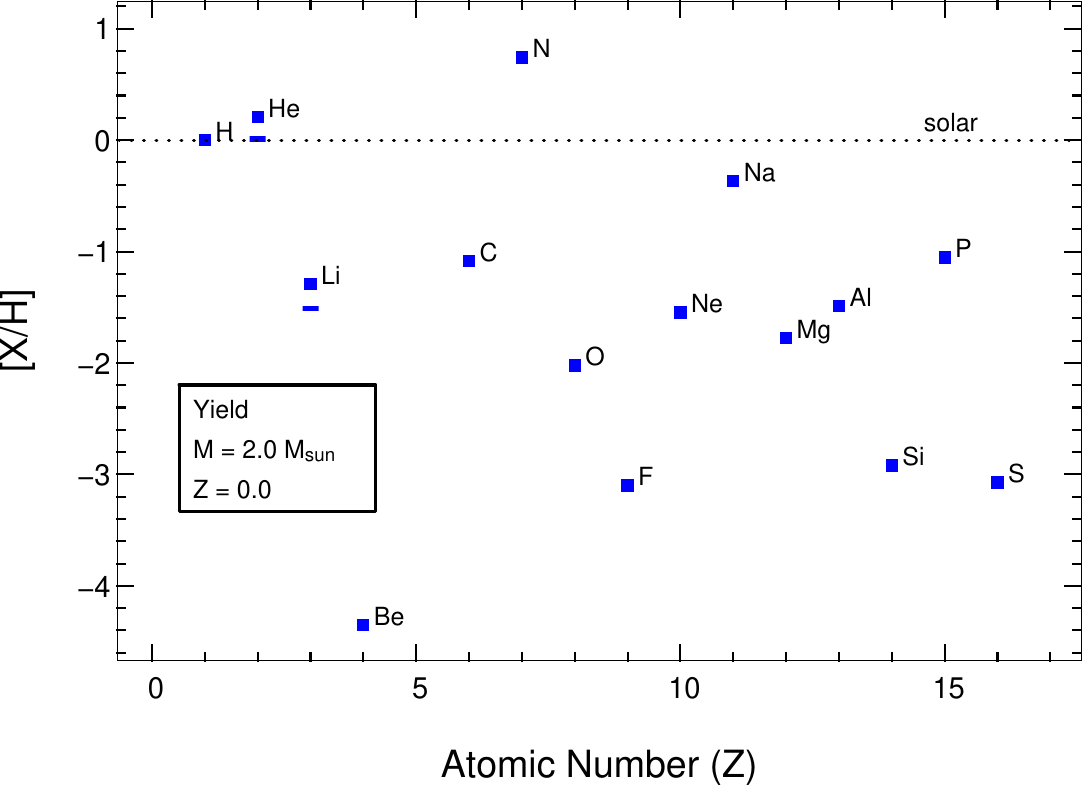}
\par\end{centering}
\caption{The yield for all the elements included in the network (that have
abundances $[X/H]>-5$), relative to solar. Solar abundances are from
\citet{2003ApJ...591.1220L}. Small horizontal lines indicate the
initial abundances for H, He and Li (all others were zero initially).
It can be seen that He has increased to a super-solar value. $^{14}$N
is very abundant compared to the solar N/H, being $\sim$0.8 dex higher.
The rest of the elements remain substantially sub-solar, except for
Na which is only about 0.2 dex lower than the solar value. Interestingly
Li remains close to its initial (primordial) value. \label{fig-m2z0y245-YIELD-XoH-elemental-all}}
\end{figure}

In Figure \ref{fig-m2z0y245-YIELD-XoH-elemental-all} we display the
chemical yield relative to the solar composition (in elemental form
rather than isotopic). This gives some perspective on the chemical
makeup of the yield for this star. $^{14}$N still remains dominant
when scaled in this manner, being $\sim$ 0.8 dex higher than the
solar ratio. Helium is substantially super-solar and Na is just below
solar (only $\sim0.2$ dex lower). As mentioned earlier Li remains
close to its initial primordial value. Carbon and P are the next most
abundant relative to solar, being about 1 dex sub-solar. Oxygen is
quite low -- despite its large production during the DSFs -- as
it was cycled to N during the HBB phase of the AGB. 

Finally we note that, if we had included reactions and species pertaining
to the s-process, there may have been an additional s-process signature
in the yield. This processing would have occurred during the DSFs
and normal He shell flashes on the AGB (see previous subsections for
more detail).

\section{Summary of All $Z=0$ Models and Yields \label{section-Yields-NS-all-Z0}}

Here we \emph{briefly} discuss the nucleosynthetic evolution of the
remaining two $Z=0$ models, which have masses of 1 and 3 M$_{\odot}$.
We then go on to present all the yields for the four stars, in graphical
and tabular form. 

Some structural properties of the 1 M$_{\odot}$ and 3 M$_{\odot}$
models are given in Section \vref{sec-All-Z0-Structural}. The nucleosynthetic
evolution of the models is discussed here in context with comparable
models presented in the previous Sections. The 1 M$_{\odot}$ experiences
a dual core flash, much like the 0.85 M$_{\odot}$ model. We thus
class it as a low mass star. The 3.0 M$_{\odot}$ does not experience
a DCF but it does go through dual shell flashes and 3DUP (and HBB),
much like the 2.0 M$_{\odot}$ model just discussed in detail above.
Thus we class this model as an intermediate mass star. We note that
the 3.0 M$_{\odot}$ model had a slightly different starting composition
to that of the other models. It was composed of a pure H-He mix, with
$Y=0.230$ (rather than 0.245). The reason we present this model instead
of the $Y=0.245$ one is because the $Y=0.245$ model did not complete
the AGB. Thus we cannot present the yield for that model. We note
that the evolution of this model is quite similar to the $Y=0.245$
model. Furthermore, as shall be discussed below, its yield closely
resembles that of the 2 M$_{\odot}$ model with $Y=0.245$. This is
despite the fact that it has significantly deeper third dredge-up.
Similarities between the yields of the models such as this, and differences
in the yields, are discussed when we present all the yields together
at the end of this section.

\subsection{Low Mass: The 1 M$_{\odot}$ Model}

\begin{figure}
\begin{centering}
\includegraphics[width=0.95\columnwidth,keepaspectratio]{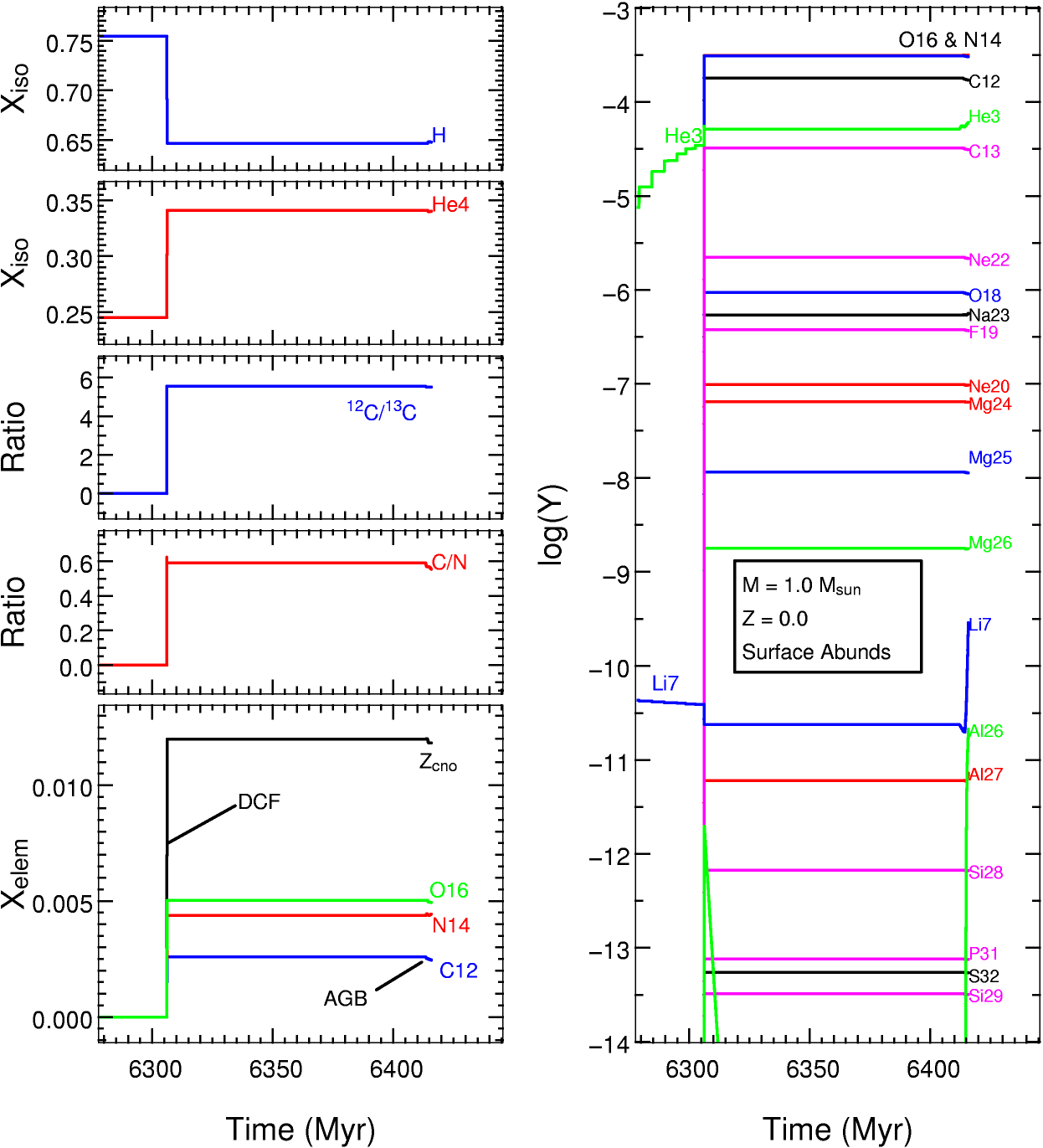}
\par\end{centering}
\caption{Time evolution of the surface abundances in the $1.0$ M$_{\odot}$,
$Z=0$ model (for selected species). It can be seen that the DCF dredge-up
episode is by far the main source of envelope pollution, causing a
large increase in He and CNO isotopes. The $^{12}$C/$^{13}$C ratio
is quite low, indicating that CN burning (almost) achieved equilibrium
during the DCF episode. The only species produced in a significant
amount during the AGB is $^{7}$Li. We note that this star initially
had $Z=0$ but ends its evolution with $Z\sim$ solar (although it
is still very metal-deficient in terms of the heavier elements).\label{fig-m1z0y245-SRF-all}}
\end{figure}

In Figure \ref{fig-m1z0y245-SRF-all} we show the surface evolution
of the 1 M$_{\odot}$ model, from just before the dual core flash
to the end of the TP-AGB. It can be seen that the overall behaviour
of this model is the same as that of the 0.85 M$_{\odot}$ model (see
Figure \vref{fig-m0.85z0y245-srf-DCF-AGB}). The surface retains its
primordial abundance up until the DCF dredge-up, when $Z_{cno}$ jumps
to $\sim0.012$. We note that this is close to the solar metallicity.
From then on the surface composition basically stays constant, reflecting
the DCF DUP pollution. Only slight changes occur on the AGB as there
is no 3DUP and only very minor HBB. In terms of the degree of pollution
coming from the DCF episode, it is much higher in this model than
in the 0.85 M$_{\odot}$ model. In fact it is almost exactly one order
of magnitude higher, as the 0.85 M$_{\odot}$ model ended up with
a surface pollution characterised by $Z_{cno}\sim0.0012$. Regrettably
we do not discuss the reasons behind this difference here, due to
time and space constraints. Suffice to say that it must be due to
a larger transfer of mass during the DCF. The degree of He enrichment
is similar also, although slightly higher in the 1 M$_{\odot}$ case
($Y=0.33$ versus $Y=0.30$ in the final yield). The distribution
of the CNO elements is also similar, with $^{16}$O and $^{14}$N
being the main products, followed by $^{12}$C. These three species
are however closer together in abundance in the 1 M$_{\odot}$ model,
with $^{16}$O and $^{14}$N being present in almost equal amounts.
This also means that the C/N ratio is higher in this model than in
the 0.85 M$_{\odot}$ model. The $^{12}$C/$^{13}$C ratio is also
slightly higher in the 1 M$_{\odot}$ model. Like the 0.85 M$_{\odot}$
model $^{13}$C and $^{17}$O are the next most abundant species,
having abundances of $\textrm{log(Y)}\sim-4.5$ and -5.5 respectively.
Interestingly the compositions diverge with regards to the less abundant
species. The 1 M$_{\odot}$ shows the signs of strong Ne-Na burning,
manifested in the large abundances of $^{22}$Ne and $^{23}$Na, whilst
the 0.85 M$_{\odot}$ model has very low abundances of these species.
This compositional difference appears to have arisen during the DCF.
A further difference is seen in the Mg isotopes. In the 0.85 M$_{\odot}$
model the isotopes follow the pattern: $^{26}$Mg > $^{25}$Mg > $^{24}$Mg
whilst in the 1 M$_{\odot}$ model this pattern is reversed, such
that $^{24}$Mg is the most abundant isotope. Finally we note that
the $^{28}$Si abundance is much greater in the 0.85 M$_{\odot}$
case (it is virtually non-existent in the 1 M$_{\odot}$ model). Again,
all these differences arise from the time of the DCF. That such different
nucleosynthesis has come about in the two models, at least for the
less abundant species, is very interesting. We shall explore these
differences in future work, in particular exploring the dependence
on overshoot.

\begin{figure}
\begin{centering}
\includegraphics[width=0.8\columnwidth,keepaspectratio]{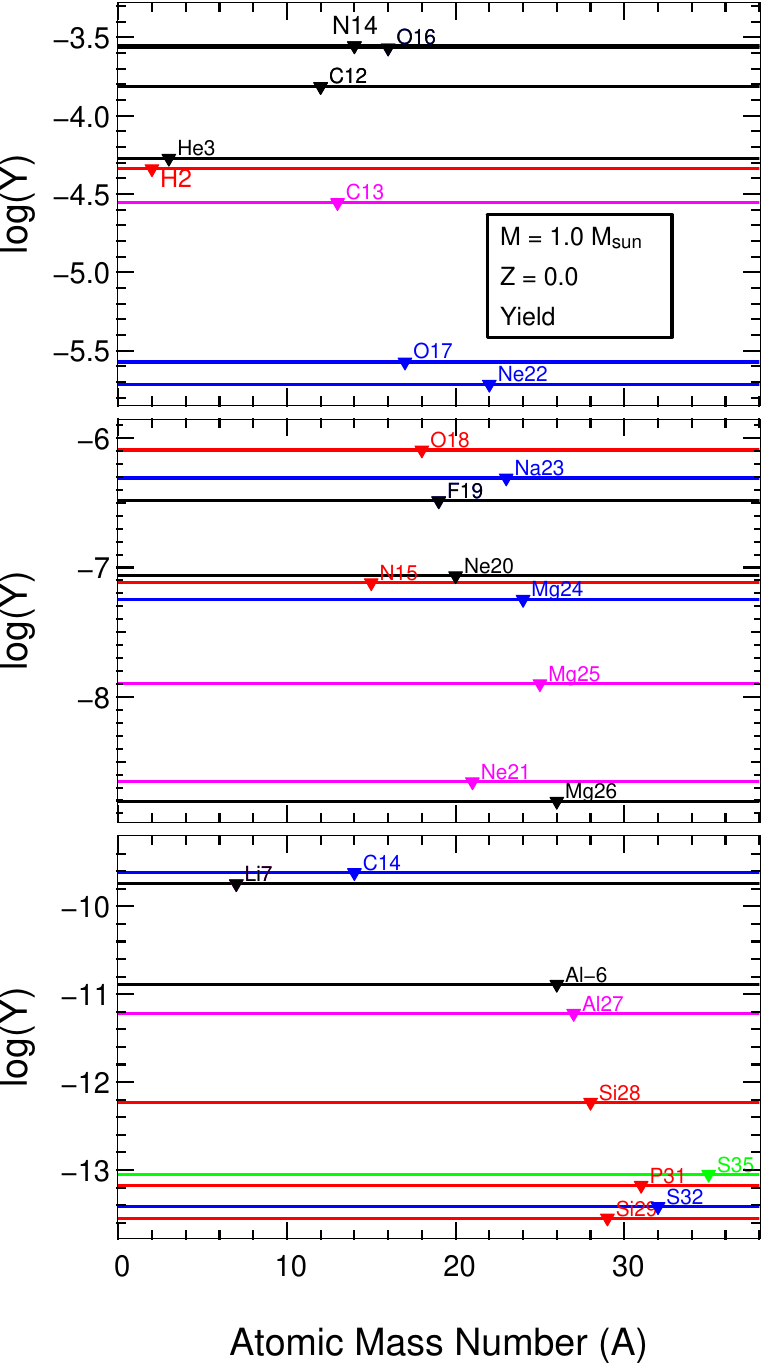}
\par\end{centering}
\caption{The yield from the 1 M$_{\odot}$ model, in log$_{10}$ of mole fraction
$Y$. All species used in the network are displayed (if their abundance
is $>10^{-14}$). \label{fig-m1z0y245-Yield-logY-all}}
\end{figure}

Figure \ref{fig-m1z0y245-Yield-logY-all} displays the yield obtained
from the 1 M$_{\odot}$ model, for all species. As also seen in Figure
\ref{fig-m1z0y245-SRF-all} the yield naturally reflects the composition
of the pollution from the DCF event, as no 3DUP and virtually no HBB
occurred on the AGB. 

\subsection{Intermediate Mass: The 3 M$_{\odot}$ Model}

\begin{figure}
\begin{centering}
\includegraphics[width=0.95\columnwidth,keepaspectratio]{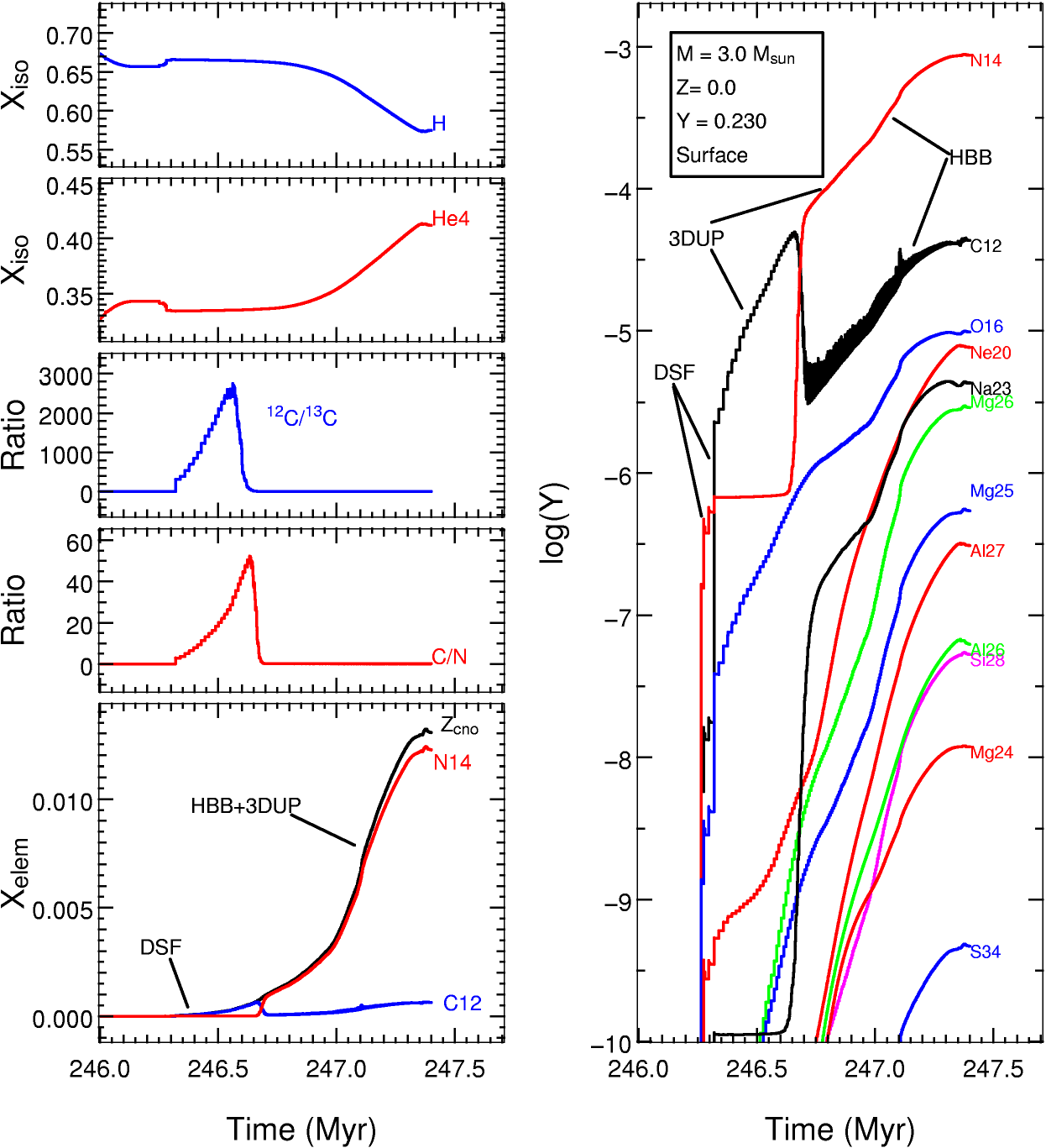}
\par\end{centering}
\caption{Time evolution of the surface abundances in the $3.0$ M$_{\odot}$,
$Z=0$ model (for selected species). The dominant contribution to
surface pollution comes from the combined effects of HBB and 3DUP.
As can be seen in the left-hand bottom panel the enrichment coming
from the dual shell flash is insignificant compared to this. The onset
of (significant) HBB is evident in the same panel by the strong C$\rightarrow$N
conversion. This star becomes a `nitrogen giant' at this stage, as
$^{14}$N by far dominates the metal content of the envelope, just
like in the 2 M$_{\odot}$ model. It also dominates the yield as the
majority of the mass loss occurs during this stage. Another consequence
of the strong HBB is that the $^{12}$C/$^{13}$C ratio is quite low.
We note that this star initially had $Z=0$ but ends its evolution
with $Z_{cno}\sim$ solar (although it is still very metal-deficient
in terms of the heavier elements). \label{fig-m3z0y2300-SRF-all-DSF-AGB}}
\end{figure}

Figure \ref{fig-m3z0y2300-SRF-all-DSF-AGB} shows the evolution of
the surface abundances for the 3.0 M$_{\odot}$ model during the entire
AGB phase. We note that this model had an initial composition of pure
H and He, with $Y=0.230$ rather than 0.245. This mainly affects the
$^{7}$Li abundance evolution but we note that $^{7}$Li is easily
destroyed by strong HBB on the AGB anyway. The chemical evolution
of the surface is very similar to that of the 2 M$_{\odot}$model
(see Figure \vref{fig-m2z0y245-SRF-AGB-all}). The envelope is pristine
up until the start of the TP-AGB, when a series of dual shell flashes
occur. In the case of the 3 M$_{\odot}$ model the pollution arising
from its three DSFs is substantially smaller that that experienced
by the 2 M$_{\odot}$ model. In the 2 M$_{\odot}$ model the surface
$Z_{cno}$ after the all the DSF episodes was $\sim10^{-4}$, whereas
in the 3 M$_{\odot}$ model it was only $\sim10^{-6}$. This difference
appears to be immaterial however, as pollution from 3DUP soon overwhelms
the DSF pollution in both cases. Furthermore, the high temperatures
at the base of the convective envelope cause strong HBB, erasing most
chemical signatures of the DSF episodes. In both models the dominant
product is $^{14}$N, by a long margin (the next most abundant species
is $^{12}$C, with an abundance $\sim1.5$ dex lower). They are both
`$^{14}$N giants' for the HBB portion of the AGB. As discussed in
the 2.0 M$_{\odot}$ nucleosynthesis section $^{22}$Ne is also periodically
dredged up with the $^{12}$C. This then participates in the Ne-Na
cycle/chain, producing $^{20}$Ne and $^{23}$Na. Leakage out of this
group of reactions then leads to further proton capture reactions
in the the Mg-Al cycle, which gradually increases the Mg and Al abundances.
$^{28}$Si is also moderately produced. 

The $^{4}$He abundance is significantly increased in both models
at second dredge-up and on the AGB. The increase in the abundance
in the 3 M$_{\odot}$ model is greater for each phase. The resultant
3 M$_{\odot}$ yield has $Y=0.41$ whilst that from the 2 M$_{\odot}$
has $Y=0.33$. In general the chemical evolution is very similar between
the two models, the main difference being that the 3 M$_{\odot}$
model has, in general, abundances $\sim0.5$ dex lower than the 2
M$_{\odot}$. This is due to the increased depth of 3DUP in this model
-- with $\lambda$ up to 0.2 as compared to 0.01 in the 2 M$_{\odot}$
model -- as well as the increased number of thermal pulses. It is
interesting to note that the $Y=0.245$ 3 M$_{\odot}$ model only
had $\lambda\sim0.01$, similar to the 2 M$_{\odot}$ model. It thus
appears that deeper 3DUP occurs with a lower helium abundance, although
we note that the evolution of the 3 M$_{\odot}$ $Y=0.230$ model
was not fully completed ($\lambda$ may have increased later in the
evolution). Another effect of the deeper 3DUP was to allow the $^{12}$C/$^{13}$C
ratio in the 3 M$_{\odot}$ model to (temporarily) attain much higher
values than in the 2 M$_{\odot}$ model. 

\begin{figure}
\begin{centering}
\includegraphics[width=0.8\columnwidth,keepaspectratio]{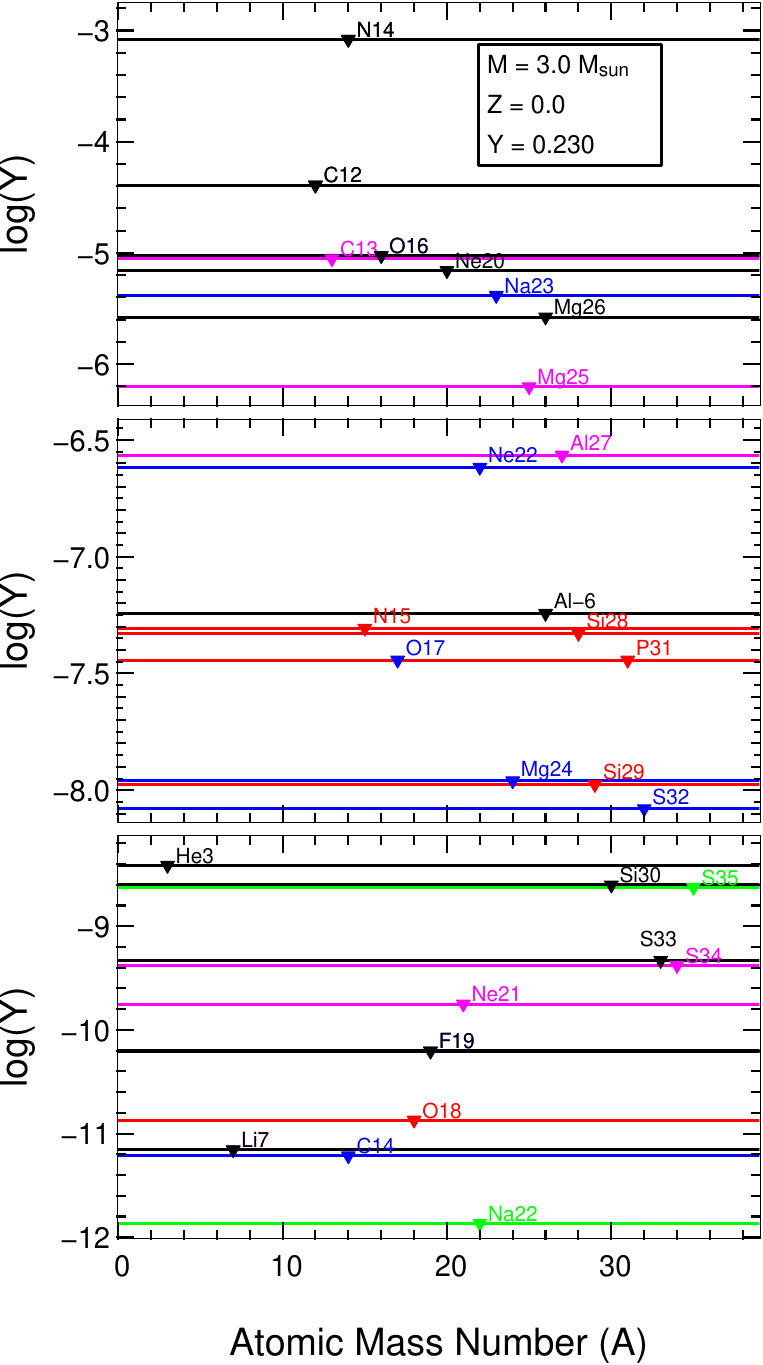}
\par\end{centering}
\caption{The yield from the 3 M$_{\odot}$ model, in log$_{10}$ of mole fraction
$Y$. All species used in the network are displayed (if their abundance
is $>10^{-14}$). \label{fig-m3z0y2300-Yield-logY-all}}
\end{figure}

Figure \ref{fig-m3z0y2300-Yield-logY-all} displays the yield obtained
from the 3 M$_{\odot}$ model, for all species. The yield reflects
the processing that occurred during the TP-AGB, ie. the combination
of 3DUP and HBB. 

\subsection{All $Z=0$ Yields}

\begin{figure}
\begin{centering}
\includegraphics[width=0.8\columnwidth]{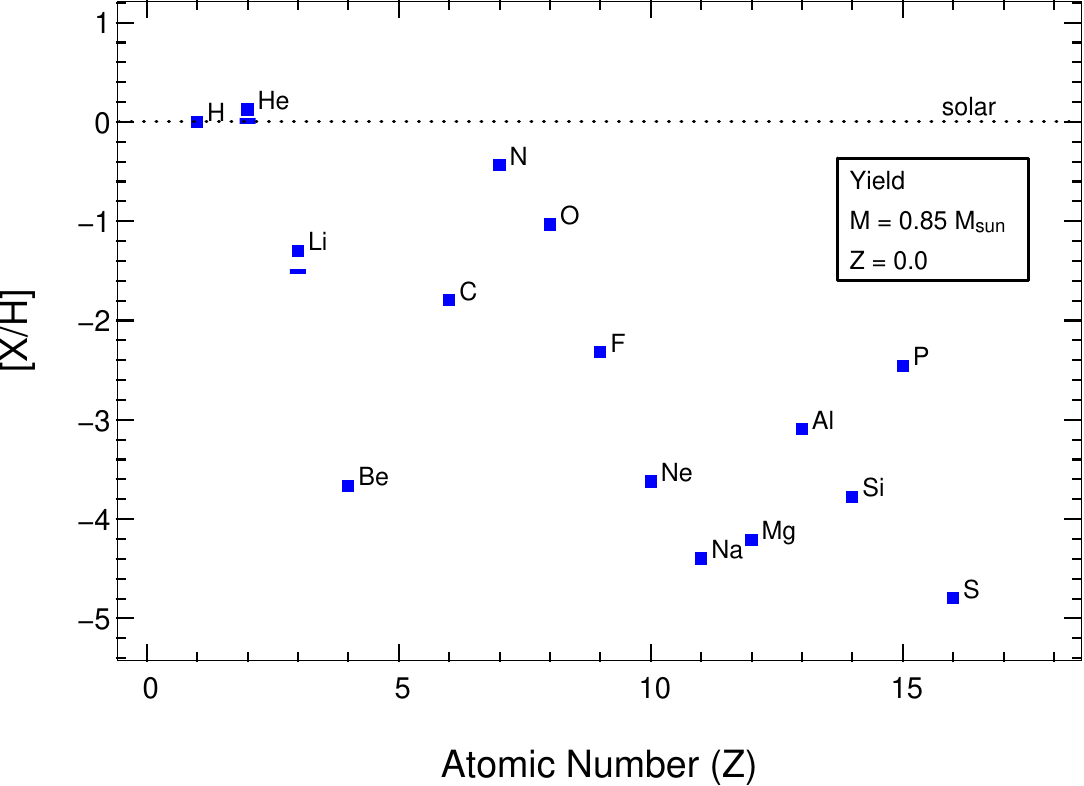}
\par\end{centering}
\caption{The yield for all the elements included in the network (that have
abundances $[X/H]>-5$), relative to solar, for the 0.85 M$_{\odot}$
model. Solar abundances are from \citet{2003ApJ...591.1220L}. Small
horizontal lines indicate the initial abundance for for H, He and
Li (all others were zero initially). Note that the Fe group is not
shown as these elements have zero yields (due to the limitations of
the nuclear network -- we shall expand the network in future studies).
\label{fig-m0.85z0y245-YIELD-Elems-XonH-2}}
\end{figure}

\begin{figure}
\begin{centering}
\includegraphics[width=0.8\columnwidth,keepaspectratio]{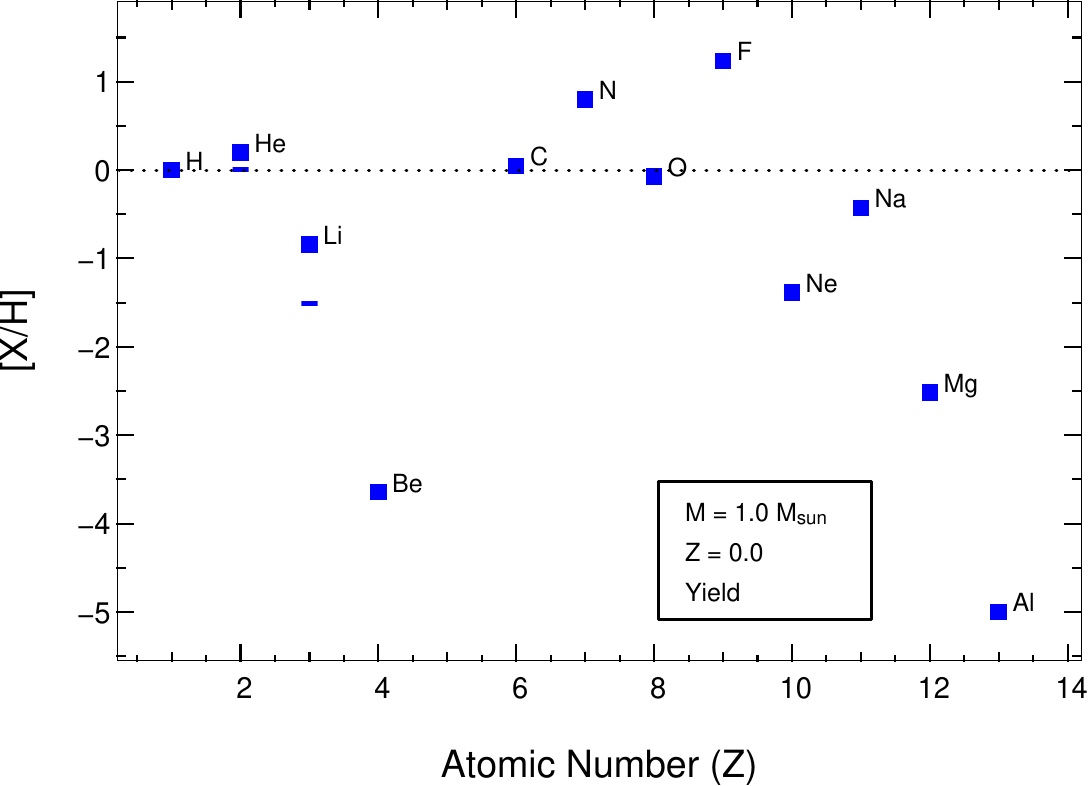}
\par\end{centering}
\caption{Same as Figure \ref{fig-m0.85z0y245-YIELD-Elems-XonH-2} but for the
1 M$_{\odot}$ model. \label{fig-m1z0y245-Yield-all-XoH}}
\end{figure}

Figures \ref{fig-m0.85z0y245-YIELD-Elems-XonH-2} to \ref{fig-m3z0y2300-Yield-XoH-all}
display the chemical yield for our four models ($M=0.85$, 1.0, 2.0
and 3.0 M$_{\odot}$) in terms of elemental abundance relative to
solar. We have redisplayed the figures from the 0.85 and 2.0 M$_{\odot}$
sections for ease of comparison. The yields naturally split into two
groups -- the low mass stars (0.85 and 1.0 M$_{\odot}$) and the
intermediate mass stars (2.0 and 3.0 M$_{\odot}$). 

\subsubsection*{Low Mass Models}

Comparing the two low mass stars similarities can be seen in the light
elements. Helium is increased in both models (solely due to the DCF
DUP, see eg. Figure \ref{fig-m1z0y245-SRF-all}), as is lithium. As
mentioned earlier the degree of pollution from the DCF was greater
in the 1 M$_{\odot}$ model. This is reflected in the increased abundances
of the CNO elements, such that N and O are both $\sim$1 dex higher
in this model. Thus the pattern of these elements is the same in both
low mass stars. Carbon is substantially different however. Even accounting
for the increased pollution in the 1 M$_{\odot}$ model it is much
more abundant in this model than the 0.85 M$_{\odot}$model. A larger
divergence occurs in the heavier elements, starting from fluorine.
In the 1 M$_{\odot}$ model F, Ne, Na and Mg are all present in (relatively)
large amounts, being a few orders of magnitude more abundant than
in the 0.85 M$_{\odot}$ model. Fluorine in particular is very abundant.
In fact it is the most abundant species in these terms (relative to
solar), being $\sim1$ dex super-solar. Fluorine in the 0.85 M$_{\odot}$
model is $\sim2$ dex sub-solar by comparison. We note that it would
be interesting to try and observe fluorine in the extremely metal-poor
halo stars to give come constraints on this nucleosynthesis. Interestingly
the heaviest elements (that are included in our network) are actually
much more abundant in the 0.85 M$_{\odot}$ model. Aluminium is $\sim3$
dex higher. Silicon, P and S are present in modest amounts but virtually
non-existent in the 1 M$_{\odot}$ model ($>5$ dex below solar).
These differences are very interesting however, as mentioned earlier,
we shall have to explore the detailed evolution of the 1 M$_{\odot}$
model in a future study due to time and space constraints. In summary
it appears that the yields for the low-mass models are similar in
abundance pattern for elements lighter than F but diverge significantly
for heavier elements. This must reflect the nucleosynthesis occurring
in the DCF episodes, as it is this pollution that drives the surface
composition and therefore yield.

\subsubsection*{Intermediate Mass Models}

\begin{figure}
\begin{centering}
\includegraphics[width=0.8\columnwidth,keepaspectratio]{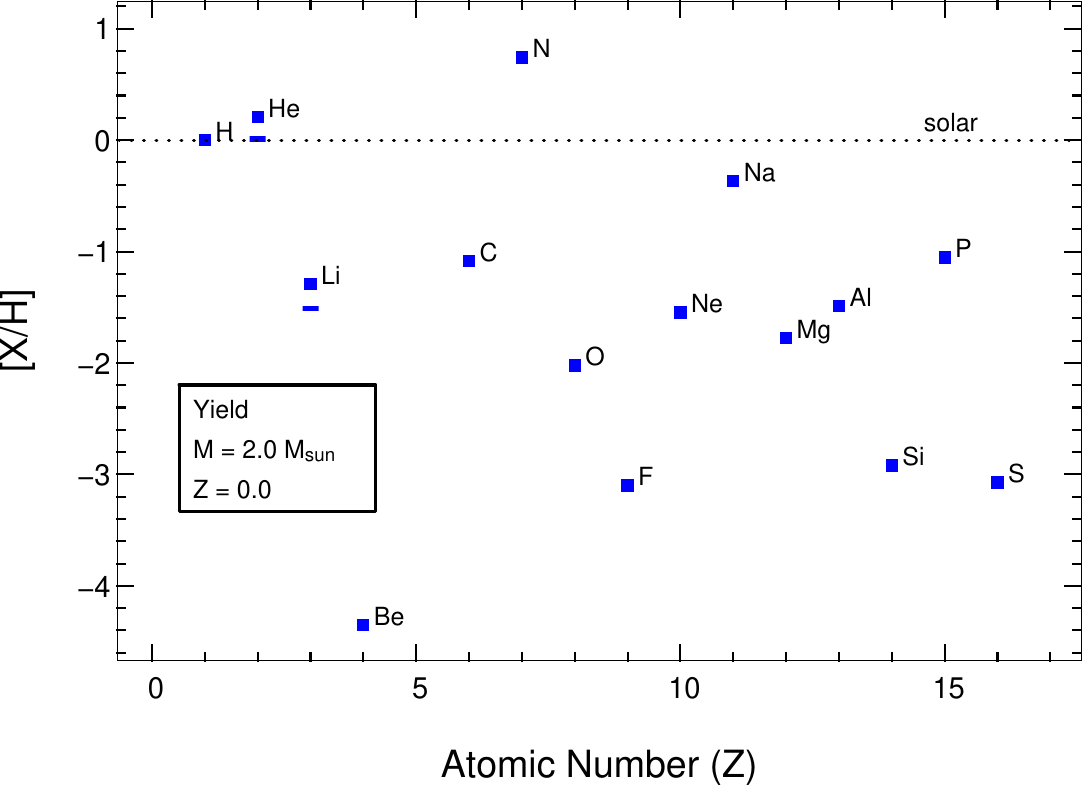}
\par\end{centering}
\caption{Same as Figure \ref{fig-m0.85z0y245-YIELD-Elems-XonH-2} but for the
2 M$_{\odot}$ model. \label{fig-m2z0y245-YIELD-XoH-elemental-all-2}}
\end{figure}

\begin{figure}
\begin{centering}
\includegraphics[width=0.8\columnwidth,keepaspectratio]{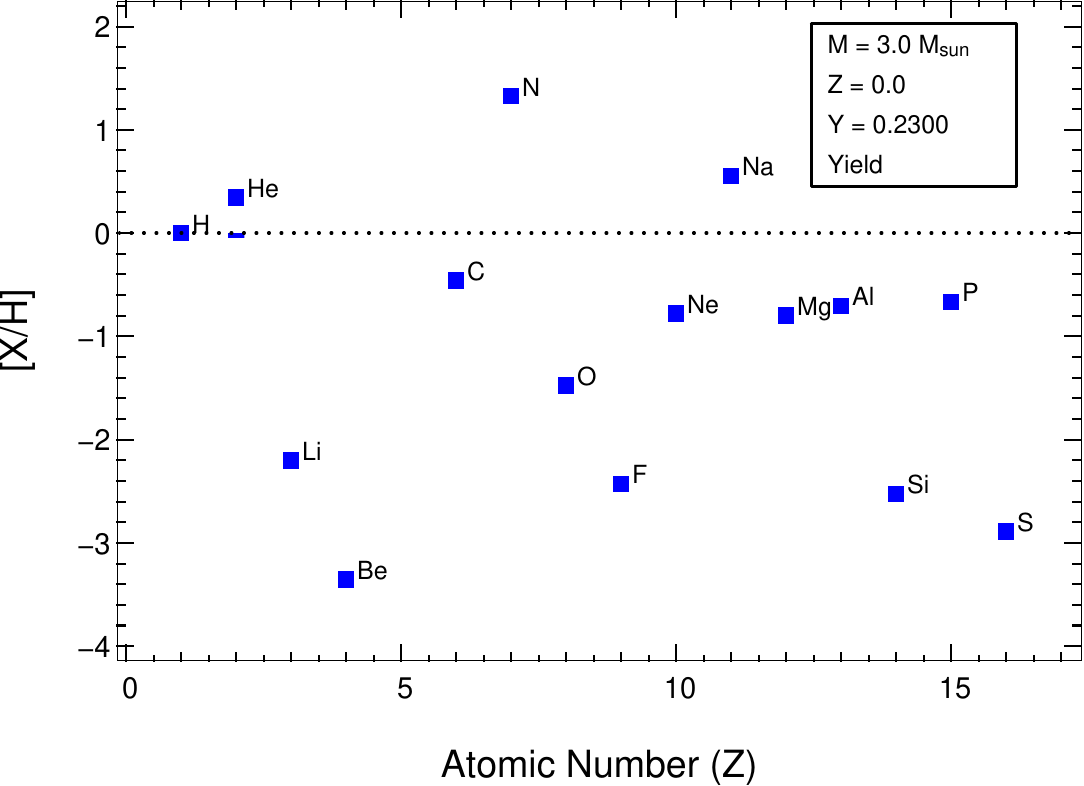}
\par\end{centering}
\caption{Same as Figure \ref{fig-m0.85z0y245-YIELD-Elems-XonH-2} but for the
3 M$_{\odot}$ model. Note that this model started with a pure H-He
composition, and with $Y=0.230$ instead of 0.245. \label{fig-m3z0y2300-Yield-XoH-all}}
\end{figure}

Moving on to the intermediate-mass model yields (Figures \ref{fig-m2z0y245-YIELD-XoH-elemental-all-2}
and \ref{fig-m3z0y2300-Yield-XoH-all}) we see that there is a striking
similarity between the yield patterns of the 2.0 and 3.0 M$_{\odot}$
models. All the elements expected from proton capture reactions are
produced substantially: N, Ne, Na, Mg, Al, etc. There is however a
difference in the degree of pollution. The 3 M$_{\odot}$ model suffered
somewhat more pollution than the 2 M$_{\odot}$ model due to the deeper
3DUP. This is reflected in the $\sim0.5$ dex increase in the yield
of most species. The \emph{lower} abundances of C and O in the 3 M$_{\odot}$
model are readily understood in terms of the increased efficiency
of HBB due to the higher temperatures at the base of the convective
envelope in this higher mass model. The similarity between the yields
of these models indicates that the yields of $Z=0$ intermediate mass
models are less uncertain than that of the low-mass models -- if
3DUP and HBB occur. If 3DUP and HBB did not occur then the yields
would probably reflect the pollution brought up by the dual shell
flash episodes, much like the yields of the low-mass models reflect
the DCF pollution. A more likely scenario is that HBB would operate
on the DCF pollution, redistributing the nuclei an hence resulting
in a different abundance pattern. However the occurrence of 3DUP and
HBB in our models virtually erases the DSF pollution pattern, giving
rise to the consistent abundance pattern between the models. 

\subsubsection*{Comparing All Models}

\begin{figure}
\begin{centering}
\includegraphics[width=0.9\columnwidth,keepaspectratio]{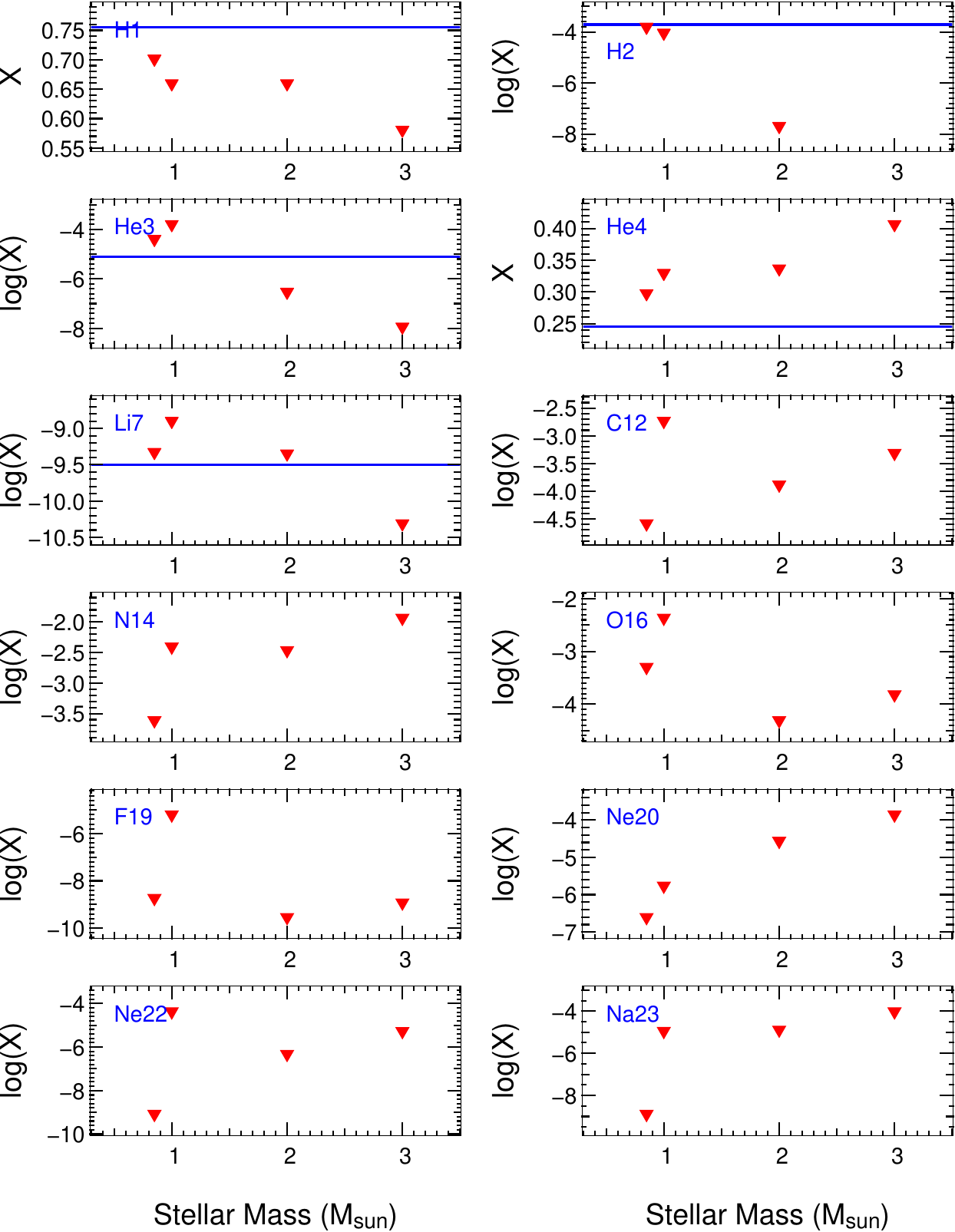}
\par\end{centering}
\caption{Yields of selected species against initial stellar mass, for all the
$Z=0$ models. More species are given in the next figure. Yields are
given in log of mass fraction X, except for $^{1}$H and $^{4}$He
which are given as X. The solid horizontal lines (blue) represent
the initial abundances for the H, He and Li nuclides (all others were
zero initially). We note that the 3 M$_{\odot}$ model started with
a pure H-He composition, with X$_{\textrm{He4}}=0.230$ rather than
0.245. \label{fig-z0-Yields-allmasses-logX-HHeCNOF}}
\end{figure}

\begin{figure}
\begin{centering}
\includegraphics[width=0.9\columnwidth,keepaspectratio]{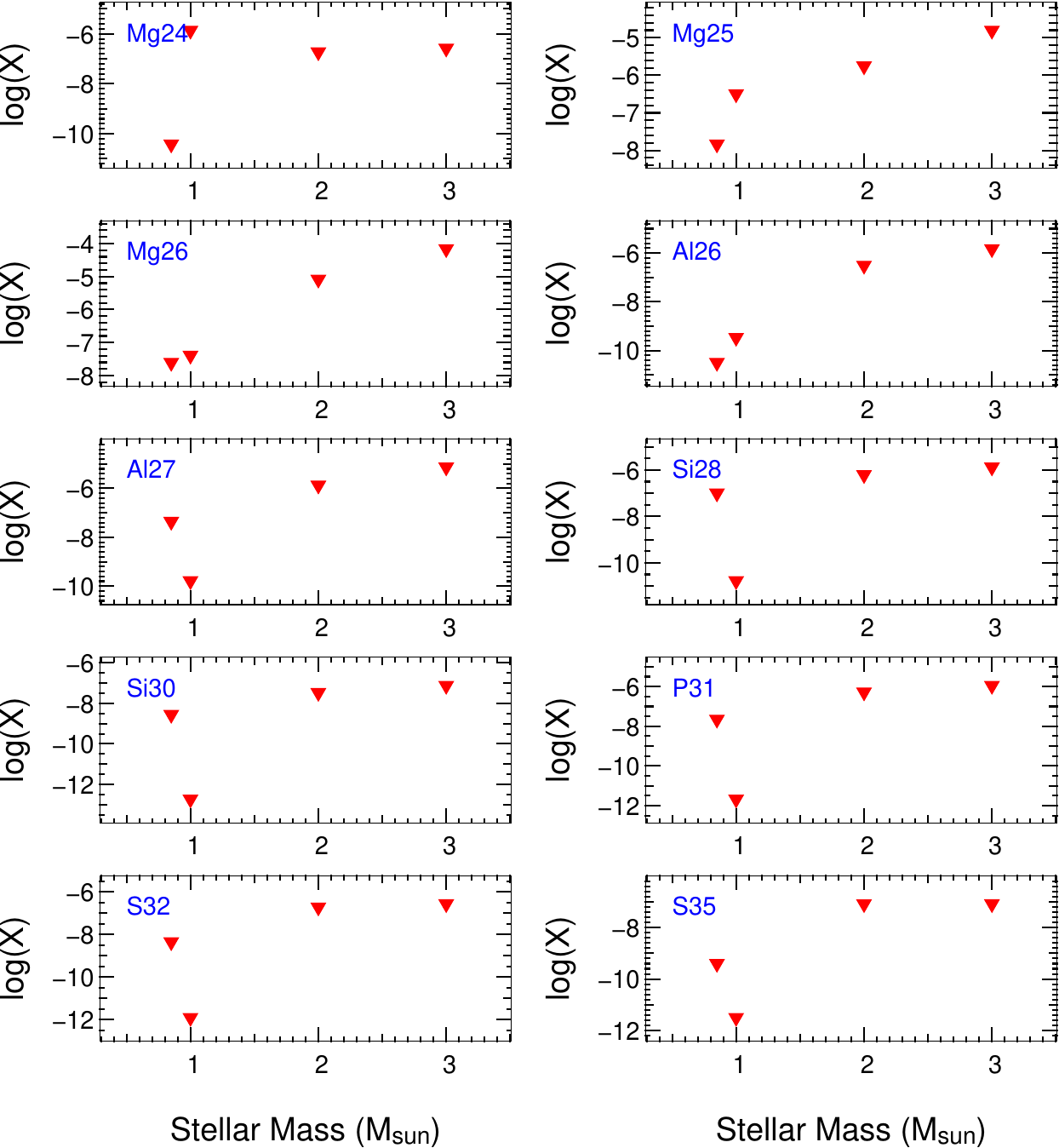}
\par\end{centering}
\caption{Same as Figure \ref{fig-z0-Yields-allmasses-logX-HHeCNOF} except
for the heavier species. Note that the Fe group is not shown as these
elements have zero yields (due to the limitations of the nuclear network
-- we shall expand the network in future studies). \label{fig-z0-Yields-allmasses-logX-MgToS}}
\end{figure}

We now compare the yields from all our $Z=0$ models, to gain an overview
of how the ejecta composition varies with mass. Figures \ref{fig-z0-Yields-allmasses-logX-HHeCNOF}
and \ref{fig-z0-Yields-allmasses-logX-MgToS} display the yields of
all stars in mass fraction X. Some of the lighter, more easily burnt
species follow a trend of decreasing in abundance with mass (eg. $^{3}$He,
$^{7}$Li, $^{2}$H). However the abundances of most species follow
a pattern of increasing with stellar mass. This is particularly evident
in Figure \ref{fig-z0-Yields-allmasses-logX-MgToS} which displays
the heavier nuclides. This is unexpected as the evolution leading
to the surface pollution was quite different between the two mass
ranges -- the LM stars are (almost) solely polluted by the DCF event,
whilst the IM stars undergo the (similar) DSF event but their surface
pollution is dominated by 3DUP and HBB. The exception to this general
trend is the 1 M$_{\odot}$ model in which, as mentioned earlier,
the heavier species are only present in very low abundances but the
lower mass species are quite abundant. For example $^{12}$C and $^{22}$Ne
are very abundant in the this model but $^{28}$Si is almost non-existent.
As mentioned earlier we shall investigate the details of the nucleosynthetic
evolution of the 1 M$_{\odot}$ model as a future study. 

An interesting piece of information to glean from comparing these
yields would be to find a chemical signature (or signatures) that
could be used to separate the LM models from the IM models. There
appear to be a few possibilities. It can be seen that the LM models
do not destroy the very light elements $^{2}$H and $^{3}$He, whilst
the IM stars do. In fact there is an increase above the primordial
value of $^{3}$He in the low mass stars. $^{16}$O may be another
tracer, as it is higher in the LM stars, but we note that there is
only 0.5 dex difference between the 0.85 M$_{\odot}$ model and the
3 M$_{\odot}$ model. $^{26}$Mg looks like the best candidate, as
its abundance is $\sim$3 dex lower in both the LM stars. $^{20}$Ne
is similar, but with only a $\sim$1 dex separation. 

\begin{figure}
\begin{centering}
\includegraphics[width=0.95\columnwidth,keepaspectratio]{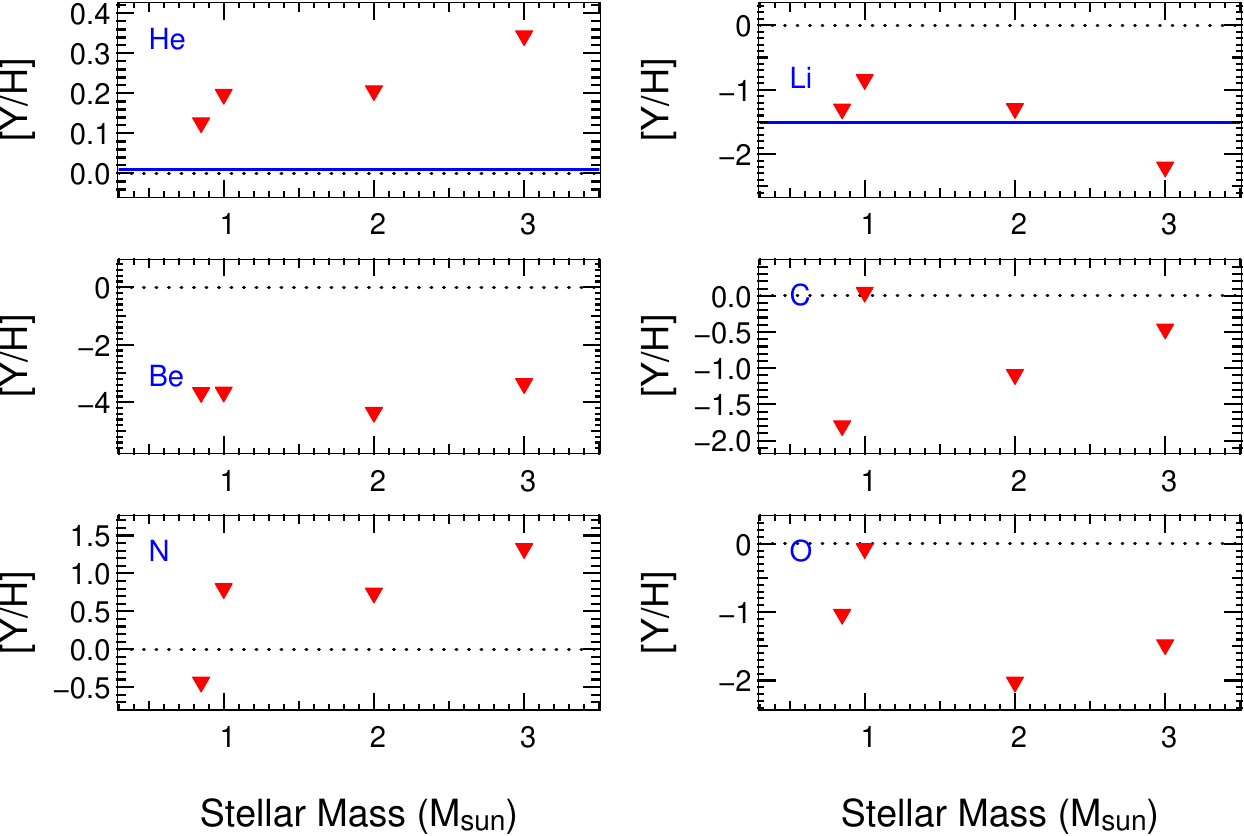}
\par\end{centering}
\caption{Elemental yields against initial stellar mass, for all the $Z=0$
models. The rest of the elements are given in the next figure. Yields
are given relative to solar (see Equation \ref{eqn-YoH} for the definition
of {[}Y/H{]}). Solar abundances are from \citet{2003ApJ...591.1220L}.
The solid horizontal lines (blue) represent the initial abundances
for He and Li (all others were zero initially). Dotted horizontal
lines at {[}Y/H$]=0$ indicate solar composition. We note that the
3 M$_{\odot}$ model started with a pure H-He composition, with X$_{\textrm{He4}}=0.230$
rather than 0.245. \label{fig-z0-Yields-allmasses-YoH-HeCNO}}
\end{figure}

\begin{figure}
\begin{centering}
\includegraphics[width=0.95\columnwidth,keepaspectratio]{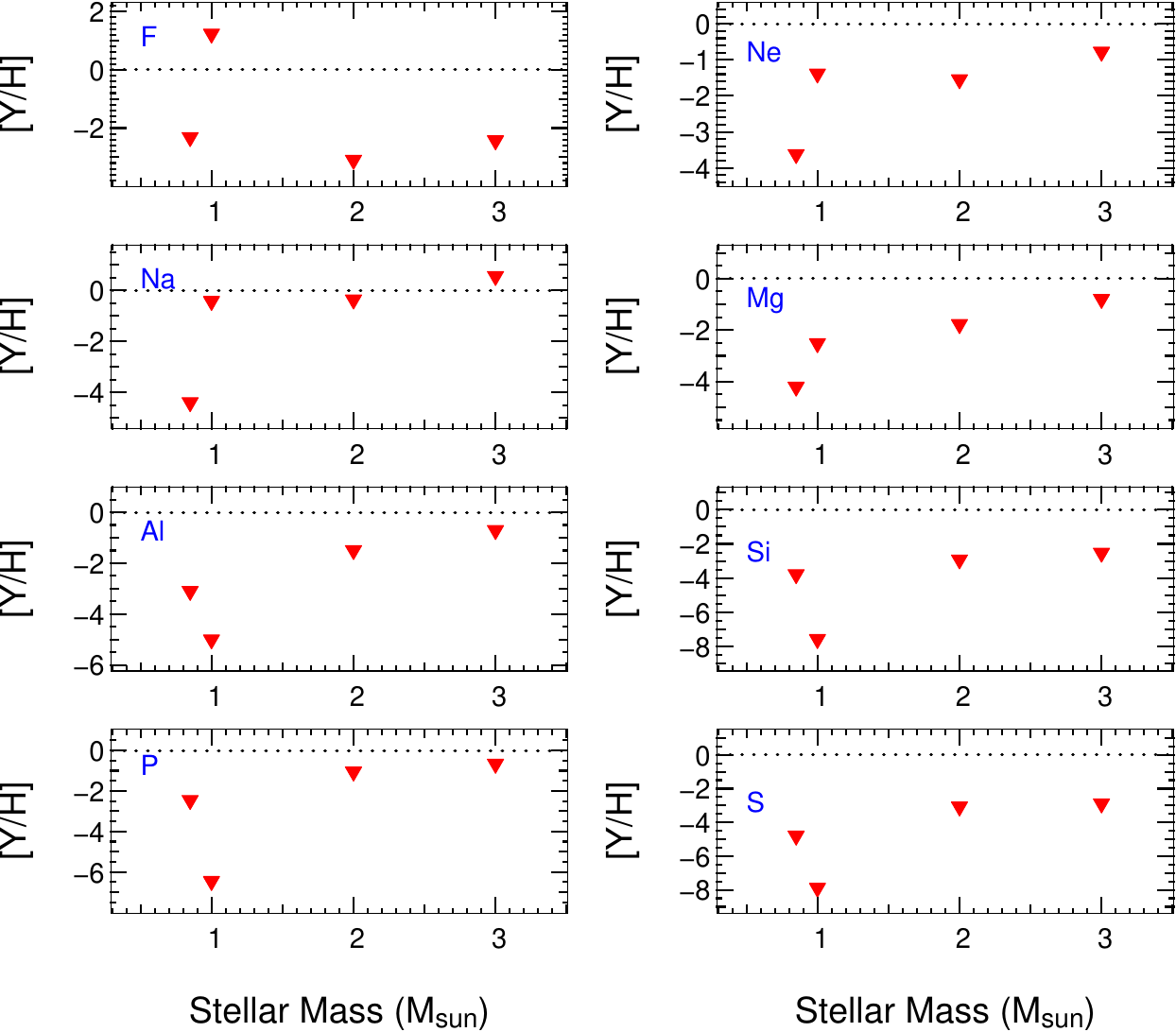}
\par\end{centering}
\caption{Same as Figure \ref{fig-z0-Yields-allmasses-YoH-HeCNO} except for
the heavier elements. Note that the Fe group is not shown as these
elements have zero yields (due to the limitations of the nuclear network
-- we shall expand the network in future studies). \label{fig-z0-Yields-allmasses-YoH-FtoS}}
\end{figure}

In order to give some perspective on the yields we also present them
relative to the solar abundances in Figures \ref{fig-z0-Yields-allmasses-YoH-HeCNO}
and \ref{fig-z0-Yields-allmasses-YoH-FtoS}. Here we have used the
usual notation:

\begin{equation}
[\textrm{Y/H}]=\log_{10}\left(\frac{N_{Y}}{N_{H}}\right)_{Star}-\log_{10}\left(\frac{N_{Y}}{N_{H}}\right)_{\odot}\label{eqn-YoH}
\end{equation}

where Y represents the particular species and N is the mole fraction.
We present the yields relative to H as there is no Fe in these models.
In these terms carbon is below solar in all cases except the 1 M$_{\odot}$
model, in which the abundance is almost exactly solar. Nitrogen is
generally super-solar, with the exception of the 0.85 M$_{\odot}$
model, which is only 0.5 dex lower than solar. Oxygen is generally
lower than solar, except again for the 1 M$_{\odot}$ model which
is $\sim$ solar. Interestingly Na is roughly solar in all the models
except for the 0.85 M$_{\odot}$ model. The 1 M$_{\odot}$ model has
an anomalous F abundance, being $\sim1$ dex super-solar, whilst in
the other models' yields it is well below solar ($\sim2$ dex subsolar).
As noted earlier observations of F in EMP halo stars could useful
for discriminating between these two cases. 

Finally, in Table \ref{table-allZ0yields-massFrac}  we present the
yields for all the $Z=0$ models and all the species, in tabular form.
The yields are given in mass fraction. We compare the yields to observations
in Chapter \ref{chapter-Discussion}.

\scriptsize

\begin{longtable}{|>{\centering}m{1.3cm}|>{\centering}m{1.5cm}||>{\centering}m{1.8cm}|>{\centering}m{1.8cm}|>{\centering}m{1.8cm}|>{\centering}m{1.8cm}|}
\caption[All $Z=0$ Yields]{Yields and initial composition for all the $Z=0$ models. All species in the network are listed. Abundances are in mass fraction, normalised to 1.0. The remnant masses (white dwarf masses) are in brackets below the initial stellar masses in the table header.}
\label{table-allZ0yields-massFrac}\\ 

\hline

\textbf{Nuclide}& 

\textbf{Initial \linebreak Abund.}\footnotemark&

\textbf{0.85 M$_{\odot}$\linebreak (0.776)}& 

\textbf{1.0 M$_{\odot}$\linebreak (0.843)}& 

\textbf{2.0 M$_{\odot}$\linebreak (1.080)}&

\textbf{3.0 M$_{\odot}$\linebreak (1.101)}\tabularnewline 

\hline \hline

\endfirsthead

\multicolumn{6}{c}{{\bfseries \tablename\ \thetable{} -- continued from previous page.}} \tabularnewline 

\hline  

\textbf{Nuclide}& 

\textbf{Initial} \linebreak \textbf{Abund.}& 

\textbf{0.85 M$_{\odot}$\linebreak (0.776)}& 

\textbf{1.0 M$_{\odot}$\linebreak (0.843)}& 

\textbf{2.0 M$_{\odot}$\linebreak (1.080)}&

\textbf{3.0 M$_{\odot}$\linebreak (1.101)}\tabularnewline 

\hline \hline

\endhead

\hline \multicolumn{6}{|r|}{{Continued on next page...}} \\ \hline 

\endfoot

\hline \hline 

\endlastfoot

\setcounter{footnote}{0}\stepcounter{footnote}\footnotetext{Except for the 3.0 M$_{\odot}$ model which started with a pure $^1$H + $^4$He composition ($Y=0.230$).}\stepcounter{footnote}\footnotetext{Note that there is a gap in the network between $^{35}$S and $^{56}$Fe, such that further nucleosynthesis is not possible with $Z=0$ initial composition.}\stepcounter{footnote}\footnotetext{g is a synthetic species that ends the nuclear network. It gives an approximate indication of the expected yield of species heavier than $^{56}$Ni. Due to the gap in the network at $^{35}$S this yield will always be zero with $Z=0$ initial composition.}
h1&
0.7548&
0.701&
0.660&
0.660&
0.581\tabularnewline
\hline 
h2&
1.960E-04&
1.57E-04&
9.17E-05&
2.02E-08&
3.34E-18\tabularnewline
\hline 
he3&
7.851E-06&
4.04E-05&
1.60E-04&
2.95E-07&
1.16E-08\tabularnewline
\hline 
he4&
0.2450&
0.298&
0.330&
0.337&
0.407\tabularnewline
\hline 
li7&
3.130E-10&
4.70E-10&
1.26E-09&
4.49E-10&
4.88E-11\tabularnewline
\hline 
be7&
0.0&
2.73E-14&
2.68E-14&
5.27E-15&
4.61E-14\tabularnewline
\hline 
b8&
0.0&
1.65E-27&
7.54E-27&
6.34E-27&
1.50E-27\tabularnewline
\hline 
c12&
0.0&
2.60E-05&
1.84E-03&
1.31E-04&
4.88E-04\tabularnewline
\hline 
c13&
0.0&
7.78E-06&
3.62E-04&
3.03E-05&
1.15E-04\tabularnewline
\hline 
c14&
0.0&
3.72E-12&
3.37E-09&
6.09E-12&
8.53E-11\tabularnewline
\hline 
n13&
0.0&
0.0&
0.0&
0.0&
0.0\tabularnewline
\hline 
n14&
0.0&
2.44E-04&
3.92E-03&
3.43E-03&
1.17E-02\tabularnewline
\hline 
n15&
0.0&
9.10E-09&
1.14E-06&
1.49E-07&
7.41E-07\tabularnewline
\hline 
o14&
0.0&
0.0&
0.0&
0.0&
0.0\tabularnewline
\hline 
o15&
0.0&
0.0&
0.0&
0.0&
0.0\tabularnewline
\hline 
o16&
0.0&
5.03E-04&
4.33E-03&
4.88E-05&
1.52E-04\tabularnewline
\hline 
o17&
0.0&
7.40E-06&
4.55E-05&
1.40E-07&
6.13E-07\tabularnewline
\hline 
o18&
0.0&
3.25E-08&
1.46E-05&
7.03E-11&
2.42E-10\tabularnewline
\hline 
o19&
0.0&
0.0&
0.0&
0.0&
0.0\tabularnewline
\hline 
f17&
0.0&
0.0&
0.0&
0.0&
0.0\tabularnewline
\hline 
f18&
0.0&
0.0&
0.0&
0.0&
0.0\tabularnewline
\hline 
f19&
0.0&
1.85E-09&
6.22E-06&
2.88E-10&
1.19E-09\tabularnewline
\hline 
f20&
0.0&
0.0&
0.0&
0.0&
0.0\tabularnewline
\hline 
ne19&
0.0&
0.0&
0.0&
0.0&
0.0\tabularnewline
\hline 
ne20&
0.0&
2.48E-07&
1.73E-06&
2.74E-05&
1.39E-04\tabularnewline
\hline 
ne21&
0.0&
6.77E-12&
4.64E-08&
1.81E-09&
3.73E-09\tabularnewline
\hline 
ne22&
0.0&
8.33E-10&
4.24E-05&
4.74E-07&
5.32E-06\tabularnewline
\hline 
na21&
0.0&
0.0&
0.0&
0.0&
0.0\tabularnewline
\hline 
na22&
0.0&
2.30E-26&
4.86E-17&
1.81E-12&
2.99E-11\tabularnewline
\hline 
na23&
0.0&
1.29E-09&
1.13E-05&
1.29E-05&
9.54E-05\tabularnewline
\hline 
na24&
0.0&
0.0&
0.0&
0.0&
0.0\tabularnewline
\hline 
mg23&
0.0&
0.0&
0.0&
0.0&
0.0\tabularnewline
\hline 
mg24&
0.0&
3.84E-11&
1.36E-06&
1.86E-07&
2.63E-07\tabularnewline
\hline 
mg25&
0.0&
1.46E-08&
3.17E-07&
1.76E-06&
1.56E-05\tabularnewline
\hline 
mg26&
0.0&
2.47E-08&
4.06E-08&
8.16E-06&
6.89E-05\tabularnewline
\hline 
mg27&
0.0&
0.0&
0.0&
0.0&
0.0\tabularnewline
\hline 
al25&
0.0&
0.0&
0.0&
0.0&
0.0\tabularnewline
\hline 
al-6&
0.0&
3.18E-11&
3.36E-10&
3.08E-07&
1.49E-06\tabularnewline
\hline 
al{*}6&
0.0&
0.0&
0.0&
0.0&
0.0\tabularnewline
\hline 
al27&
0.0&
4.39E-08&
1.63E-10&
1.34E-06&
7.36E-06\tabularnewline
\hline 
al28&
0.0&
0.0&
0.0&
0.0&
0.0\tabularnewline
\hline 
si27&
0.0&
0.0&
0.0&
0.0&
0.0\tabularnewline
\hline 
si28&
0.0&
1.00E-07&
1.65E-11&
6.17E-07&
1.31E-06\tabularnewline
\hline 
si29&
0.0&
1.16E-08&
8.23E-13&
1.27E-07&
3.06E-07\tabularnewline
\hline 
si30&
0.0&
2.77E-09&
1.83E-13&
3.34E-08&
7.49E-08\tabularnewline
\hline 
si31&
0.0&
0.0&
0.0&
0.0&
0.0\tabularnewline
\hline 
si32&
0.0&
8.70E-21&
7.99E-25&
6.37E-13&
2.20E-12\tabularnewline
\hline 
si33&
0.0&
0.0&
0.0&
0.0&
0.0\tabularnewline
\hline 
p29&
0.0&
0.0&
0.0&
0.0&
0.0\tabularnewline
\hline 
p30&
0.0&
0.0&
0.0&
0.0&
0.0\tabularnewline
\hline 
p31&
0.0&
2.20E-08&
2.07E-12&
5.22E-07&
1.12E-06\tabularnewline
\hline 
p32&
0.0&
3.38E-24&
3.10E-28&
2.47E-16&
8.55E-16\tabularnewline
\hline 
p33&
0.0&
0.0&
0.0&
4.75E-24&
4.23E-22\tabularnewline
\hline 
p34&
0.0&
0.0&
0.0&
0.0&
0.0\tabularnewline
\hline 
s32&
0.0&
4.38E-09&
1.22E-12&
1.87E-07&
2.66E-07\tabularnewline
\hline 
s33&
0.0&
6.24E-10&
3.36E-14&
9.23E-09&
1.54E-08\tabularnewline
\hline 
s34&
0.0&
2.19E-10&
1.32E-13&
5.56E-09&
1.42E-08\tabularnewline
\hline 
s35\addtocounter{footnote}{-2}\footnotemark&
0.0&
3.98E-10&
3.13E-12&
8.31E-08&
8.31E-08\tabularnewline
\hline 
fe56&
0.0&
0.0&
0.0&
0.0&
0.0\tabularnewline
\hline 
fe57&
0.0&
0.0&
0.0&
0.0&
0.0\tabularnewline
\hline 
fe58&
0.0&
0.0&
0.0&
0.0&
0.0\tabularnewline
\hline 
fe59&
0.0&
0.0&
0.0&
0.0&
0.0\tabularnewline
\hline 
fe60&
0.0&
0.0&
0.0&
0.0&
0.0\tabularnewline
\hline 
fe61&
0.0&
0.0&
0.0&
0.0&
0.0\tabularnewline
\hline 
co59&
0.0&
0.0&
0.0&
0.0&
0.0\tabularnewline
\hline 
co60&
0.0&
0.0&
0.0&
0.0&
0.0\tabularnewline
\hline 
co61&
0.0&
0.0&
0.0&
0.0&
0.0\tabularnewline
\hline 
ni58&
0.0&
0.0&
0.0&
0.0&
0.0\tabularnewline
\hline 
ni59&
0.0&
0.0&
0.0&
0.0&
0.0\tabularnewline
\hline 
ni60&
0.0&
0.0&
0.0&
0.0&
0.0\tabularnewline
\hline 
ni61&
0.0&
0.0&
0.0&
0.0&
0.0\tabularnewline
\hline 
ni62&
0.0&
0.0&
0.0&
0.0&
0.0\tabularnewline
\hline 
g\addtocounter{footnote}{-0}\footnotemark&
0.0&
0.0&
0.0&
0.0&
0.0\tabularnewline
\hline
\end{longtable}
\normalsize

\chapter{Metal Poor Halo Star Models\label{Chapter-HaloStarModels}}
\begin{quote}
``If we knew what it was we were doing, it would not be called research,
would it?''
\begin{flushright}
\vspace{-0.6cm}--Albert Einstein 
\par\end{flushright}
\end{quote}

\section{Background}

\subsection{Introduction}

One of the theories put forth to explain the extremely low metallicities
and strange abundance patterns observed in many Galactic Halo stars
(see \citet{BC05} for a recent review of the observations) is that
some of them may have formed from a primordial gas cloud that had
only been polluted by a few -- or even one -- Pop III supernovae.
Such a dilution of supernova ejecta allows the iron content of a star
forming from this material to be very low, and, depending on the supernova
mass, may explain some of the observed abundance patterns (see eg.
\citealt{1998ApJ...507L.135S}; \citealt{2002ApJ...565..385U}; \citealt{2003ApJ...594L.123L};
\citealt{2004MNRAS.348L..23Z}). It has also often been suggested
that the shocks from the first SNe may have triggered the formation
of these low-mass stars in primordial gas clouds.

Another scenario utilises supernovae to provide the heavy element
distribution, whilst the lighter elements are provided by the low-mass
stars themselves, via some form of auto-pollution (such as the dual
shell flash or dual core flash, see eg. \citealt{2000ApJ...529L..25F};
\citealt{2002AA...395...77S}; \citealt{2004ApJ...609.1035P}; \citealt{2004AA...422..217W}),
or perhaps via mass transfer from a low-Z binary companion (see eg.
\citealt{2004ApJ...611..476S}).

In this chapter we seek to explore the expected chemical signatures
of low- and intermediate-mass stars that may have formed from a gas
cloud polluted by a single supernova. To this end we have calculated
a grid of stellar models of very low and extremely low metallicity
which have an initial abundance pattern arising from the mixing of
a $Z=0$ supernova yield (as given by a theoretical calculation, see
below for details) with Big Bang material.

Our modifications to the structural evolution code (SEV code) detailed
in Chapter \ref{sevmods} have enabled us to model extremely metal-deficient
and metal-free stars such as these. As described in the $Z=0$ stellar
evolution chapters there exists a number of evolutionary features
peculiar to $Z=0$ and extremely low metallicity stars that require
the use of time-dependent mixing. The two most important of these
are the two proton ingestion episodes (PIEs) that are induced by helium
flash convection breaking through to hydrogen-rich regions. We refer
to these events as the dual core flash (DCF) and dual shell flash
(DSF). The DCF occurs at the time of the core He flash in low mass
models whilst the DSF occurs during the first major He shell flash
at the beginning of the TP-AGB in intermediate mass models (sometimes
more than one DSF occurs). Although these events have been modelled
before by other authors they are still relatively unexplored phenomena
as not many studies have evolved through these difficult phases. These
two events are pivotal as they provide considerable pollution of the
envelope and could provide the source of the large amounts of C, N
and O observed in the C-enhanced metal poor halo stars (CEMPHs, see
eg. \citet{2007ApJ...655..492A} for current observations and eg.
\citealt{2004ApJ...611..476S}; \citealt{2000ApJ...529L..25F}; \citealt{2001ApJ...559.1082S}
for theoretical calculations). We note that the DCF and DSF are notoriously
difficult to evolve though numerically and have provided a serious
challenge for the author (and indeed previous authors!). 

As mentioned in the $Z=0$ chapters the novelty in this study is that
we have also taken (most) of our models through to the completion
of the AGB. To the best of our knowledge this is the first time this
has been achieved for stars of such low metallicity. This allows us
to make a range of interesting predictions, such as the expected white
dwarf masses, the lifetimes of the various stages of evolution, and
the chemical yields of extremely metal poor low- and intermediate-mass
stars. As also mentioned in the $Z=0$ section we should however add
a caveat here. The models suffer from the usual uncertainties such
as that which derives from the use of the Mixing Length Theory of
convection, the neglect of rotation, and the choice of mass-loss formalism.
With this in mind we shall compare our models with observations of
the Galactic Halo stars in a later chapter.

\subsection{Brief Overview of Code Inputs\label{subsection-HaloStarCodeInputs}}

Apart from the initial composition all of the physical parameters
used for the structural evolution of these models were exactly the
same as those used for the $Z=0$ models presented in the previous
two chapters. We thus refer the reader to the introduction of Chapter
\vref{chapter-Z0-StructEvoln} for a detailed description of the included
physics (mass loss, opacity, etc.). Some key physical inputs are however
important enough to warrant a brief reiteration here.

\subsubsection*{Convective Boundaries and Mixing}

We have employed the new diffusive mixing routine in all our metal-deficient
models, for all phases of evolution. For details on this routine see
Section \vref{timedepmix}. As with the $Z=0$ models we do not include
any overshoot beyond the classical Schwarzschild boundary. Thus our
models are \emph{conservative} in terms of the amount of mixing that
occurs during episodes such as the DCF, DSF and third dredge-up. 

\subsubsection*{Mass Loss}

Mass loss is one of the most uncertain factors in stellar modelling.
We have discussed the methods employed in the SEV code in Subsection
\vref{sub-MassLossDescription} but provide a brief discussion here
also, in relation to the special case of very- and extremely-low metallicity.

First we note that the interested reader will find a more detailed
discussion on mass loss in the introduction to the chapter dealing
with the structural evolution of the $Z=0$ models (in particular
Section \vref{subsection-z0-Struct-InputPhysics}). That section gives
some extra background on mass loss in general, and in relation to
metal-free stars. 

As with $Z=0$ stars one may expect very metal-deficient stars to
be less susceptible to mass loss due to the lack of metals from which
grains could form. This expectation is based on the assumption that
mass loss occurs (only) through radiation pressure acting on grains.
As noted in Section \vref{subsection-z0-Struct-InputPhysics} this
may or may not be a valid assumption since, in reality, the mechanisms
for mass loss are not yet properly understood. Nevertheless we shall
continue our discussion with this assumption, but keep this caveat
in mind.

Similar evolutionary traits are found in our metal deficient models
as those found in the $Z=0$ models. In particular almost all of our
models experience self-enriching episodes (namely the DCF and DSF
discussed in the previous chapters and also within this chapter).
This raises the metallicity of the envelope to values similar to that
of the LMC (and in some cases much higher). As AGB mass loss has been
observed in the LMC it follows that grains are able to form at these
metallicities (assuming this is necessary for mass loss). Thus it
would seem that a standard mass-loss formalism would be needed for
stages of evolution occurring after the self-enriching episodes. In
the low-mass models ($M\leq1$ M$_{\odot}$) these episodes occur
at the tip of the RGB, so that in the subsequent evolutionary phases
(HB, AGB) the star has a `metal-rich' surface. Thus for these stages
we suggest that it is reasonable to use the standard mass-loss formulae.
In the IM mass models the main polluting event, the DSF, occurs at
the start of the AGB. In these cases the envelope is enriched by a
more modest amount (lower than the LMC metallicity), such that the
normal mass loss prescriptions may not apply. However in all of the
IM models 3DUP occurs, which, by the time the mass loss rate is significantly
high, also enriches the surface by a large amount (to $\sim$LMC or
super-LMC levels). Thus in this case also we suggest it is reasonable
to use the standard mass-loss formulae. As the IM models do not in
general go through the RGB phase (if they do then it is for a very
brief period) it appears that the standard procedure used in the SEV
code suffices for all the intermediate mass models (the \citet{1993ApJ...413..641V}
(VW93) formula for the AGB is the only one that will actually be utilised
in this case).

In our low mass and lower metallicity models we find that the RGB
phase is very short, as also reported for the 0.85 M$_{\odot},$ $Z=0$
model. In these cases the mass loss on the RGB is relatively small
when compared to the total mass lost over the whole evolution, at
least for our 1 M$_{\odot}$ models (the 0.85 M$_{\odot}$ models
spend longer in the RGB phase). Thus, in these cases we suggest that
using the standard mass-loss formulae introduces a small amount of
uncertainty. 

There is however one set of cases where the choice of mass loss prescriptions
introduces a significant uncertainty. Unlike the $Z=0$ and extremely
metal-poor cases the RGB phases in our more metal-'rich' low-mass
models are quite long. So long in fact that up to $90\%$ of the mass
lost from the star by the end of its evolution can be lost on the
RGB (we note that we use the \citet{1975MSRSL...8..369R} formula
on the RGB, see Section \vref{sub-MassLossDescription} for details).
If this degree of mass loss is wrong then our results will be wrong.
For example, since the mass lost on the RGB has the initial composition
of the models, it is thus very metal-poor. This naturally has a strong
impact on the yields of these models, effectively diluting the yield
from the AGB. If, in reality, the RGB mass loss should be much lower
then our yields will be wrong -- they should be more metal-rich.
The RGB evolutionary paths of the models are also rendered uncertain
as the rate of core growth would be different with a lower mass loss
rate. This then has implications for the HB and RGB evolution. Thus
many uncertainties arise through the unknown nature of mass-loss rates
at low metallicity.

We note however that this uncertainty in the low-mass models is based
on the assumption that the mass loss rate should reduce severely with
metallicity, i.e. that mass loss occurs only via radiation pressure
acting on grains. As mentioned above this may not be the case. Assuming
for now (for the sake of argument) that this \emph{is} the main mechanism
for mass loss then one path that we could have taken, and that other
studies have taken, would be to scale the mass loss rate with metallicity
-- such that at low metallicity the mass loss rate is lower. The
question that then arises is: By what factor or function do we scale
the rate? As mass loss is not yet fully understood theoretically,
and there are no observations of mass loss at the very low metallicities
in question, any scaling function would necessarily introduce uncertainties
itself. So it appears that neither of the options -- scaling the
rate with Z or using the existing formulae -- can be argued to minimise
uncertainty in the modelling. With this in mind we have chosen the
simplest option -- to use the existing mass loss formalisms in the
SEV code.

Finally we also note that the metallicity of a stellar model is, to
some degree, indirectly taken into account in the Reimers' and VW93
formulas. This occurs because the entire structure of the models change
with Z, through opacity and different nuclear burning energy sources
(for example). Since the mass loss formulae utilise bulk physical
properties there is thus a feedback from the composition of the star
onto the mass loss rates. Some evidence for this arises during the
presentation of results below, with regards to the horizontal branch
and RGB mass evolution in the low mass models (Section \ref{subsection-HaloStarStruct-MassLoss}).

\subsubsection*{Opacity Tables}

We note that we have calculated opacity tables specifically for these
metal-deficient models. This was done because the initial composition
is not scaled-solar, whereas opacity tables are usually calculated
for solar composition. Thus our models are self consistent insomuch
as the opacity composition matches the composition of the star. Most
studies use scaled solar opacities as the differences arising in opacity
are not thought to significantly affect the evolution of the models.
In light of experience gained form the calculation of models for this
study the Author now agrees with this approximation, at least for
models with abundance profiles not too different from solar. We note
however that it is now an easy task to calculate custom opacity tables
for use with the SEV code by using the OPAL web service. See Section
\vpageref{opacmods} for the web address and more details on opacity
in the SEV code. 

\subsubsection*{Nucleosynthesis Code Inputs}

In regards to the nucleosynthesis code (NS code) only minor modifications
were needed to be able to calculate the nucleosynthetic evolution
of our metal-deficient models. Removing the scaled-solar composition
assumptions to allow arbitrary stellar chemical compositions was the
main modification. Also, as we found many of the models experience
Dual Core Flashes and Dual Shell Flashes, it was necessary to increase
the resolution during these phases. This was mentioned in the $Z=0$
chapter (see Section \vref{section-Z0-NSevolution-NScode} and references
therein). 

\subsection{The Grid of Models}

The range of mass for our small grid is:

\[
\textrm{M }=0.85,1.0,2.0,3.0\,\textrm{M}_{\odot}
\]

and the range of metallicity:

\[
\textrm{[Fe/H]}=-\infty,-6.5,-5.45,-4.0,-3.0
\]

where we have included our $Z=0$ models (as $-\infty$) since we
shall be referring to these models in the present chapter for various
comparisons. The peculiar value of $\textrm{[Fe/H]}=-5.45$ was chosen
to match the metallicity (at least in terms of {[}Fe/H{]}) of the
most metal poor star currently known -- the Galactic Halo star HE1327-2326,
(\citealt{2005Natur.434..871F}). 

In mass fraction of CNO nuclides the metallicities are:

\begin{center}
$Z_{cno}=4\times10^{-9},4\times10^{-8},1\times10^{-6},1\times10^{-5}$
\par\end{center}

which are very near the total $Z$ value as the CNO nuclides are the
dominant `metal' species in the initial composition.

\subsection{Initial Composition\label{subSec-HaloStar-InitialComp}}

The composition for these stars was arrived at by mixing the ejecta
of a 20 M$_{\odot}$, Z=0 supernova model with pristine Big Bang material.
The Big Bang abundances are from the calculations by \citet{Coc0104},
displayed in Table \ref{table-bbnabunds-EMPHs}, whilst the supernova
yield is from a calculation by Limongi et al. (2002, private communication)
and is displayed in Table \ref{table-20msunSNyield}. 

\begin{table}
\begin{centering}
\begin{tabular}{|c|c|}
\hline 
Nuclide & Primordial Mass Fraction\tabularnewline
\hline 
\hline 
$^{1}$H & $0.754796$\tabularnewline
$^{2}$H & $1.96\times10^{-4}$\tabularnewline
$^{3}$He & $7.85\times10^{-6}$\tabularnewline
$^{4}$He & $0.24500$\tabularnewline
$^{7}$Li & $3.13\times10^{-10}$\tabularnewline
$^{12}$C & $0.0$\tabularnewline
$^{14}$N & $0.0$\tabularnewline
$^{16}$O & $0.0$\tabularnewline
\hline 
\end{tabular}
\par\end{centering}
\caption{The chemical composition of the primordial gas cloud that the Pop
III supernova pollutes. Taken from the Standard Big Bang Nucleosynthesis
calculations by \citet{Coc0104}. The 20 M$_{\odot}$ supernova yield
was diluted using this material. \label{table-bbnabunds-EMPHs}}
\end{table}

\begin{table}
\begin{centering}
\begin{tabular}{|cc||cc||cc|}
\hline 
Nuclide & Mass Frac. & Nuclide & Mass Frac. & Nuclide & Mass Frac.\tabularnewline
\hline 
\hline 
H1 & 0.5453 & Si29 & 6.84E-05 & Ti50 & 2.98E-12\tabularnewline
H2 & 3.63E-18 & Si30 & 4.83E-05 & V50 & 1.06E-11\tabularnewline
He3 & 1.60E-06 & P31 & 1.80E-05 & V51 & 2.79E-07\tabularnewline
He4 & 0.3506 & S32 & 2.85E-03 & Cr50 & 1.09E-06\tabularnewline
Li6 & 1.40E-19 & S33 & 1.35E-05 & Cr52 & 8.55E-05\tabularnewline
Li7 & 1.11E-11 & S34 & 8.58E-05 & Cr53 & 5.16E-06\tabularnewline
Be9 & 1.99E-38 & S36 & 5.93E-09 & Cr54 & 1.69E-10\tabularnewline
B10 & 2.06E-18 & Cl35 & 4.27E-06 & Mn55 & 1.55E-05\tabularnewline
B11 & 1.23E-16 & Cl37 & 1.40E-06 & Fe54 & 1.11E-04\tabularnewline
C12 & 2.26E-02 & Ar36 & 4.64E-04 & Fe56 & 7.86E-03\tabularnewline
C13 & 2.30E-09 & Ar38 & 2.31E-05 & Fe57 & 2.38E-04\tabularnewline
N14 & 3.93E-09 & Ar40 & 1.29E-10 & Fe58 & 4.84E-11\tabularnewline
N15 & 2.25E-11 & K39 & 1.12E-06 & Co59 & 1.25E-04\tabularnewline
O16 & 3.87E-02 & K40 & 1.11E-09 & Ni58 & 1.11E-04\tabularnewline
O17 & 5.88E-10 & K41 & 9.39E-08 & Ni60 & 1.29E-07\tabularnewline
O18 & 3.92E-09 & Ca40 & 3.92E-04 & Ni61 & 6.93E-10\tabularnewline
F19 & 6.79E-11 & Ca42 & 3.58E-07 & Ni62 & 4.13E-13\tabularnewline
Ne20 & 1.60E-02 & Ca43 & 7.54E-10 & Ni64 & 6.37E-20\tabularnewline
Ne21 & 3.57E-06 & Ca44 & 8.44E-06 & Cu63 & 1.18E-14\tabularnewline
Ne22 & 8.05E-07 & Ca46 & 2.80E-13 & Cu65 & 5.36E-14\tabularnewline
Na23 & 2.13E-04 & Ca48 & 1.10E-18 & Zn64 & 8.10E-09\tabularnewline
Mg24 & 7.16E-03 & Sc45 & 3.52E-07 & Zn66 & 7.31E-17\tabularnewline
Mg25 & 7.78E-05 & Ti46 & 1.77E-07 & Zn67 & 1.73E-22\tabularnewline
Mg26 & 8.59E-05 & Ti47 & 4.16E-09 & Zn68 & 1.02E-23\tabularnewline
Al27 & 1.73E-04 & Ti48 & 1.60E-05 &  & \tabularnewline
Si28 & 6.49E-03 & Ti49 & 2.94E-07 & \textbf{Z$_{total}$} & \textbf{0.104}\tabularnewline
\hline 
\end{tabular}
\par\end{centering}
\caption{The chemical composition of the $Z=0$, 20 M$_{\odot}$ supernova
yield from Limongi (2002, private communication). Abundances are given
in mass fraction, normalised to 1.0. \label{table-20msunSNyield}}
\end{table}

The variation in initial metallicity between the four sets of metal-deficient
models was achieved by varying the amount of Big Bang material with
which the supernova ejecta was diluted. For example, the {[}Fe/H{]}$=-4.0$
models required the supernova yield to be diluted into $1.0\times10^{6}$
M$_{\odot}$ of Big Bang material, whilst the {[}Fe/H{]}$=-3.0$ models
required $1.0\times10^{7}$ M$_{\odot}$. We display the initial composition
used in the SEV code calculations of the {[}Fe/H{]}$=-4.0$ set of
models as an example in Table \ref{table-Z0-InitialAbundsEMPHs}.
The full composition, as used in the nucleosynthesis code (NS code),
is given in Table \ref{table-NS-initialComp-EMPHs}. The abundance
profiles for the other sets of models are basically scaled versions
of this. Naturally the species provided by the Big Bang gas are roughly
invariant between the sets of models. In particular this means that
$^{7}$Li, $^{3}$He and $^{2}$H are always present in primordial
amounts, as their abundance in the $Z=0$ gas cloud far outweighs
that produced by the supernova. All the heavier species are present
in progressively lower fractional abundances as the metallicity is
lowered. As we use Fe to define the metallicity in this case the SN
yield has a direct effect on the abundance pattern. The resulting
composition is thus different to that given by just scaling the solar
composition. The most notable difference is the lack of nitrogen,
which, for example, is only present in the {[}Fe/H{]}$=-4.0$ models
at the level of $\sim10^{-14}$ (by mass), as compared to $\sim10^{-7}$
for carbon and oxygen. This is due to the fact that it is not present
in a significant amount in the yield of the $Z=0$, 20 M$_{\odot}$
supernova, nor is it synthesised in (standard) BBN. It must be noted
that this lack of Nitrogen is model-dependent --- some $Z=0$ supernova
models do in fact produce significant amounts of N. However we do
not think that this uncertainty is of much importance, in most respects.
In terms of structural evolution N is quickly produced through the
CNO cycle due to the presence of C and O, and thus the (energetically
important) CNO cycle operates as normal. In terms of chemical yield
we have found that the initial N abundance is swamped by the pollution
events occurring in these stars (eg. the DCF, DSF and/or the third
dredge-up). Of course it does however affect the surface composition
for early phases of evolution, before any surface pollution occurs.
This may be of importance in comparing our MS models with observations
of (unpolluted) dwarf Halo stars. 

\begin{table}
\begin{centering}
\begin{tabular}{|c|c|}
\hline 
Nuclide & Initial Mass Fraction\tabularnewline
\hline 
\hline 
$^{1}$H & 0.75479\tabularnewline
$^{3}$He & $7.85\times10^{-6}$\tabularnewline
$^{4}$He & 0.24500\tabularnewline
$^{12}$C & $4.04\times10^{-7}$\tabularnewline
$^{14}$N & $7.03\times10^{-14}$\tabularnewline
$^{16}$O & $6.91\times10^{-7}$\tabularnewline
Z$_{other}$ & $7.65\times10^{-7}$\tabularnewline
\hline 
Z$_{total}$ & $1.86\times10^{-6}$\tabularnewline
\hline 
\end{tabular}
\par\end{centering}
\caption{An example of the initial abundances used in the stellar structure
calculations of the low and intermediate mass metal poor stars. In
this case it is the composition of the {[}Fe/H{]}$=-4.0$ set of models.
This composition was arrived at by diluting the 20 M$_{\odot}$ yield
with $1.0\times10^{6}$ M$_{\odot}$ of Big Bang material. Z$_{other}$
is a synthetic species that keeps track of all nuclides heavier than
$^{16}$O. Mass fraction is normalised to 1.0. \label{table-Z0-InitialAbundsEMPHs}}
\end{table}

\begin{table}
\begin{centering}
\begin{tabular}{|cc||cc||cc|}
\hline 
Nuclide & Mass Frac. & Nuclide & Mass Frac. & Nuclide & Mass Frac.\tabularnewline
\hline 
\hline 
h1 & 0.7548 & ne21 & 6.39E-11 & p31 & 3.21E-10\tabularnewline
h2 & 0.00E+00 & ne22 & 1.44E-11 & p32 & 0.00E+00\tabularnewline
he3 & 7.85E-06 & na21 & 0.00E+00 & p33 & 0.00E+00\tabularnewline
he4 & 0.2450 & na22 & 0.00E+00 & p34 & 0.00E+00\tabularnewline
li7 & 3.13E-10 & na23 & 3.81E-09 & s32 & 5.10E-08\tabularnewline
be7 & 0.00E+00 & na24 & 0.00E+00 & s33 & 2.41E-10\tabularnewline
b8 & 0.00E+00 & mg23 & 0.00E+00 & s34 & 1.64E-08\tabularnewline
c12 & 4.04E-07 & mg24 & 1.28E-07 & s35 & 0.00E+00\tabularnewline
c13 & 4.12E-14 & mg25 & 1.39E-09 & fe56 & 1.41E-07\tabularnewline
c14 & 0.00E+00 & mg26 & 1.54E-09 & fe57 & 4.25E-09\tabularnewline
n13 & 0.00E+00 & mg27 & 0.00E+00 & fe58 & 8.65E-16\tabularnewline
n14 & 7.03E-14 & al25 & 0.00E+00 & fe59 & 0.00E+00\tabularnewline
n15 & 4.03E-16 & al-6 & 0.00E+00 & fe60 & 0.00E+00\tabularnewline
o14 & 0.00E+00 & al{*}6 & 0.00E+00 & fe61 & 0.00E+00\tabularnewline
o15 & 0.00E+00 & al27 & 3.10E-09 & co59 & 2.24E-09\tabularnewline
o16 & 6.91E-07 & al28 & 0.00E+00 & co60 & 0.00E+00\tabularnewline
o17 & 1.05E-14 & si27 & 0.00E+00 & co61 & 0.00E+00\tabularnewline
o18 & 7.02E-14 & si28 & 1.16E-07 & ni58 & 1.99E-09\tabularnewline
o19 & 0.00E+00 & si29 & 1.22E-09 & ni59 & 0.00E+00\tabularnewline
f17 & 0.00E+00 & si30 & 8.63E-10 & ni60 & 2.31E-12\tabularnewline
f18 & 0.00E+00 & si31 & 0.00E+00 & ni61 & 1.24E-14\tabularnewline
f19 & 1.21E-15 & si32 & 0.00E+00 & ni62 & 7.38E-18\tabularnewline
f20 & 0.00E+00 & si33 & 0.00E+00 & g & 1.45E-13\tabularnewline
ne19 & 0.00E+00 & p29 & 0.00E+00 &  & \tabularnewline
ne20 & 2.87E-07 & p30 & 0.00E+00 & \textbf{Z$_{total}$} & \textbf{1.86E-06}\tabularnewline
\hline 
\end{tabular}
\par\end{centering}
\caption{An example of the initial abundances used in the nucleosynthesis calculations
of the low and intermediate mass metal-deficient stars. In this case
it is the composition of the {[}Fe/H{]}$=-4.0$ set of models. This
composition was arrived at by diluting the 20 M$_{\odot}$ yield with
$1.0\times10^{6}$ M$_{\odot}$ of Big Bang material. `$g$' is a
synthetic species that ends the network, representing all species
heavier than $^{62}$Ni (see the NS code description in Section \ref{nscode}
for more details on the network). Abundances are given in mass fraction,
normalised to 1.0. Starting compositions for the rest the models are
given with the yields towards the end of the chapter. \label{table-NS-initialComp-EMPHs}}
\end{table}

The other main difference that the SN+BB composition exhibits, as
compared to Solar, is a modest overabundance of the alpha elements.
We display some of these abundances in Table \ref{table-SN-alphas}.

\begin{table}
\begin{centering}
\begin{tabular}{|c|c|}
\hline 
Element Ratio & Abundance\tabularnewline
\hline 
\hline 
{[}O/Fe{]} & $0.00$\tabularnewline
{[}Ne/Fe{]} & $0.36$\tabularnewline
{[}Mg/Fe{]} & $0.25$\tabularnewline
{[}Si/Fe{]} & 0.15\tabularnewline
{[}S/Fe{]} & 0.20\tabularnewline
\hline 
\end{tabular}
\par\end{centering}
\caption{Abundances of some of the alpha elements in the Pop III supernova
yield, relative to Solar. Solar abundances are from \citet{2003ApJ...591.1220L}.
\label{table-SN-alphas}}
\end{table}

\section{Structural Evolution}

Here we give an overview of the evolution of all our metal-deficient
models. We have also included the $Z=0$ models of the previous chapters
in most of the tables and plots, for comparison. The $Z=0$ models
provide a useful handle on the lower limit of metal-deficient stellar
evolution. In particular, gradual changes in evolution can be seen
as the abundance of heavy elements is increased from zero. 

We follow the definition often used to distinguish between low mass
(LM) and intermediate mass (IM) models, such that low mass models
ignite He under degenerate conditions (in a core He flash -- or possibly
a dual core flash) whilst IM models ignite He quiescently. In the
extremely low metallicity regime of this study this means that our
2 and 3 M$_{\odot}$ models are IM stars. 

We give an overall summary of our results, and compare them to previous
studies, at the end of this chapter (Section \vref{section-SummaryAndCompare-HaloMods}).

\subsection{Main Sequence to RGB Tip/Core He Ignition\label{section-HaloStars-Struct-MSandRGB}}

\subsubsection*{Low Mass Models: 0.85 \& 1.0 M$_{\odot}$}

\begin{figure}
\begin{centering}
\includegraphics[width=0.8\columnwidth]{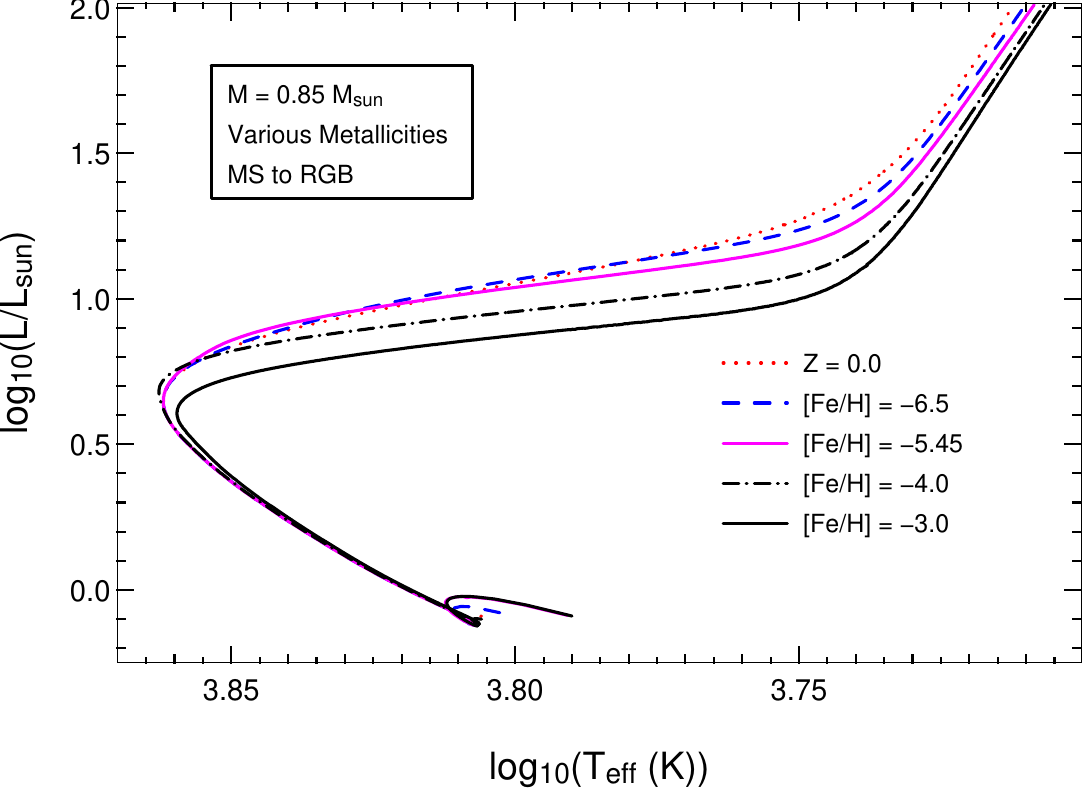}
\par\end{centering}
\caption{The HR diagram for all the 0.85 M$_{\odot}$ low metallicity models,
from ZAMS to the start of the RGB. Our $Z=0$ model has been included
for comparison. \label{fig-hrds-m0.85all-MS.eps}}
\end{figure}

\begin{figure}
\begin{centering}
\includegraphics[width=0.8\columnwidth,keepaspectratio]{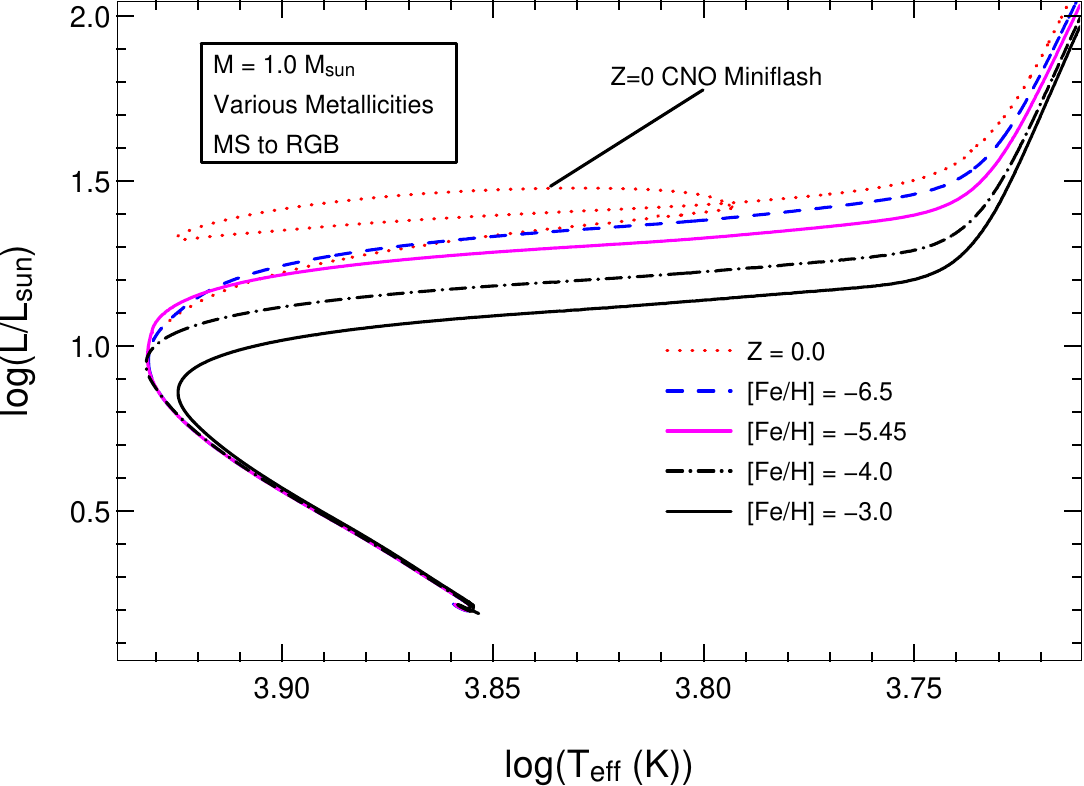}
\par\end{centering}
\caption{Same as Figure \ref{fig-hrds-m0.85all-MS.eps} but for the 1 M$_{\odot}$models.
\label{fig-hrds.m1all.MS.eps}}
\end{figure}

In Figures \ref{fig-hrds-m0.85all-MS.eps} and \ref{fig-hrds.m1all.MS.eps}
we show the evolution of the five 0.85 M$_{\odot}$ models and five
1.0 M$_{\odot}$ models in the HR diagram, from ZAMS to the start
of the RGB. The models have metallicities of {[}Fe/H{]}$=-6.5,-5.45,-4.0,-3.0$
and $Z=0$. We have magnified this section of the HR diagrams to highlight
the differing luminosities and surface temperatures between the models.
However, also apparent is the similarity between the models during
most of the MS phase. This is due to the fact that stars of higher
metallicity are not hot enough to burn hydrogen efficiently via the
CNO cycles, and are powered mainly by the p-p chains. Thus the paucity
(or absence) of CNO catalysts in the metal-deficient models has no
effect during this phase of evolution. It also appears that the opacity
difference in this range of metallicity has little effect on the surface
temperature during this phase either. 

Differences begin to appear as the stars reach the end of the MS.
The more metal-rich models begin to burn H more efficiently via the
CNO cycles due to the increasing core temperatures and decreasing
H abundance. The higher temperature dependence of energy generation
in the CNO cycles ($\sim$T$^{20}$ compared to $\sim$T$^{4}$ in
the p-p chains) leads to lower burning temperatures and thus lower
luminosities in the more metal-rich models. Thus the MS turn-offs
of the lower metallicity stars are at slightly higher temperatures
and luminosities. These qualities persist as the stars change to shell
burning energy sources and move off the MS, crossing the HR diagram
to become red giants. The shell burning in the lower metallicity models
is again dominated by the p-p chains whilst the higher metallicity
models have more of a contribution from the CNO cycles (Figure \vref{fig-m0.85z0y245-HRD}
in the $Z=0$ chapter shows this clearly). The differences in surface
temperature and luminosities is more pronounced in the 1 M$_{\odot}$
models than the 0.85 M$_{\odot}$ models, leading to a modest variation
of MS turn-off luminosities of $\sim0.15$ dex. An interesting feature
in the $Z=0$, 1 M$_{\odot}$ model (Figure \ref{fig-hrds.m1all.MS.eps})
not seen in any of the other low-mass models is the occurrence of
a `CNO Miniflash' during the Hertzsprung Gap evolution that leads
to a loop in the HR diagram. This is similar to the CNO miniflash
reported for the $Z=0$, 2 M$_{\odot}$ model in Section \vref{section-m2z0-Structural}
but it occurs earlier in this case. It indicates that a small amount
of He burning has produced enough $^{12}$C to allow the CNO cycles
to operate, burning the small amount of H left in the core. As this
carbon production does not affect the shell the further evolution
is unaffected, so that the star returns to a high luminosity and temperature
position after the episode.

\begin{figure}
\begin{centering}
\includegraphics[width=0.8\columnwidth,keepaspectratio]{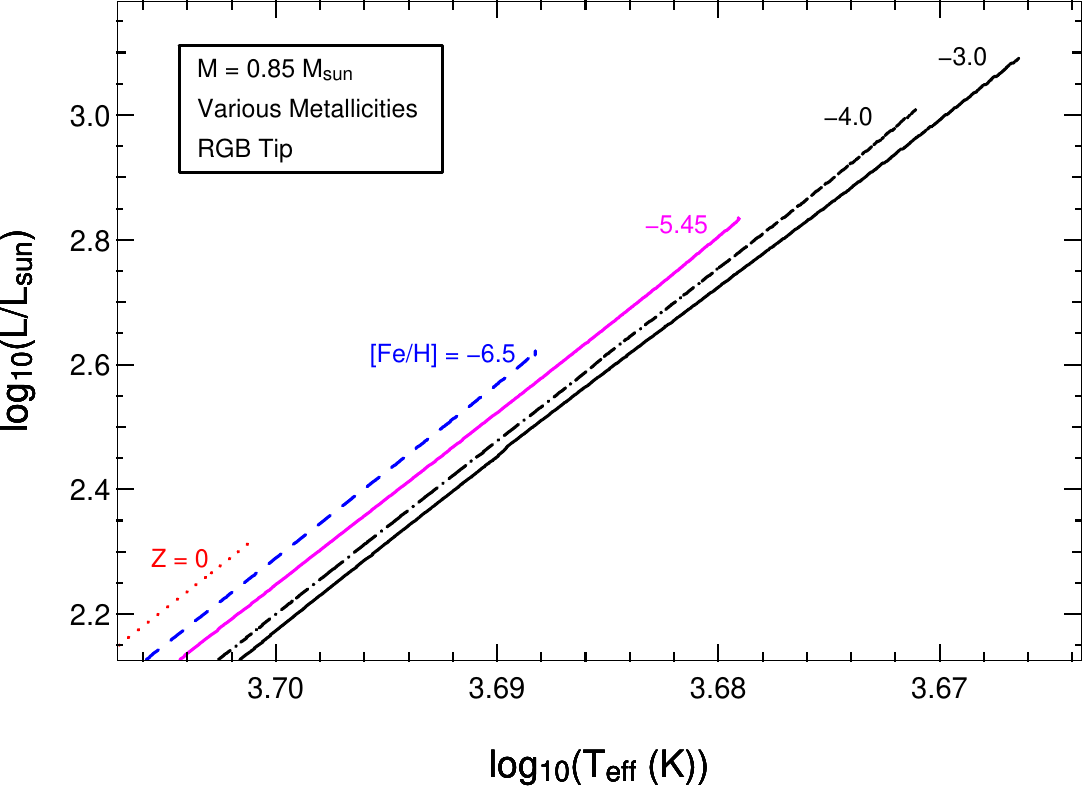}
\par\end{centering}
\caption{The last portion of the RGB evolution in the HR diagram for all the
0.85 M$_{\odot}$ models. The curves end when He ignites degenerately
in the cores. It can be seen that the luminosity at the tip of the
RGB varies by $\sim0.8$ dex over this metallicity range. \label{fig-hrds-m0.85all-RGBtip}}
\end{figure}

\begin{figure}
\begin{centering}
\includegraphics[width=0.8\columnwidth,keepaspectratio]{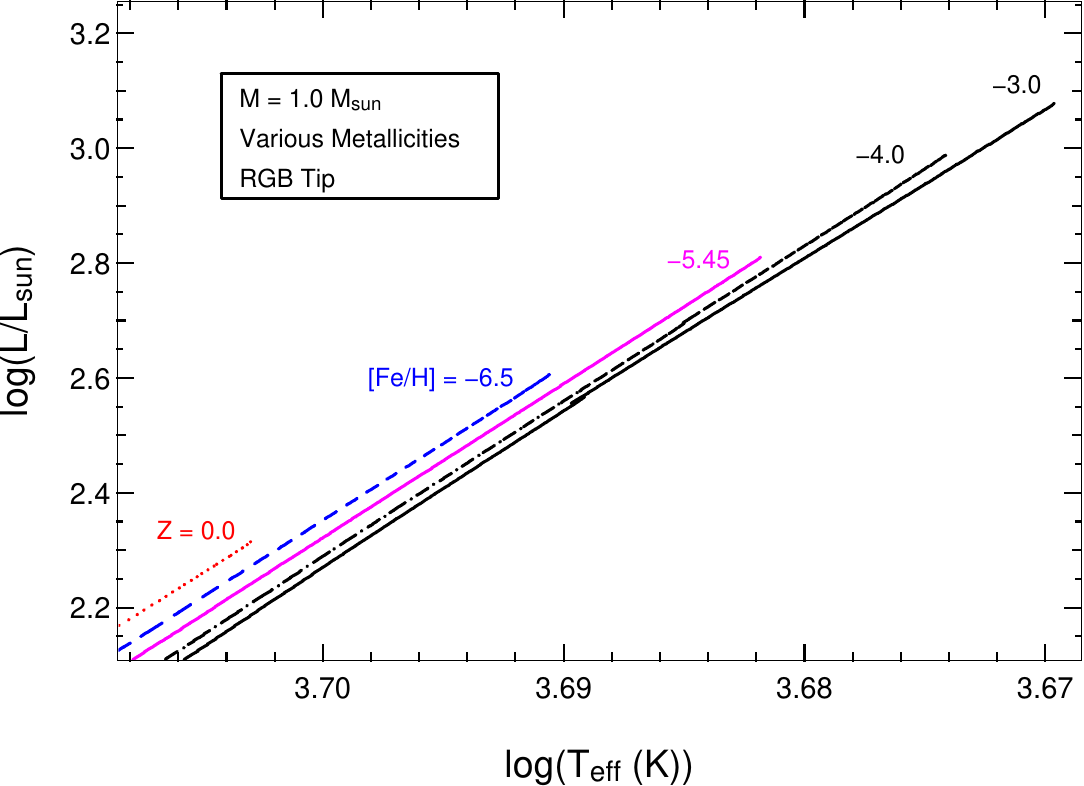}
\par\end{centering}
\caption{Same as Figure \ref{fig-hrds-m0.85all-RGBtip} but for the 1.0 M$_{\odot}$
models. \label{fig-hrds.m1all.RGBtip}}
\end{figure}

As noted in \ref{section-m0.85z0-MStoRGB} the luminosity of the $Z=0$,
0.85 M$_{\odot}$ model at the time of core He ignition at the top
of the RGB is significantly lower than in higher metallicity models.
In Figures \ref{fig-hrds-m0.85all-RGBtip} and \ref{fig-hrds.m1all.RGBtip}
we show `close-ups' of the end of the RGB in the HR diagrams for all
the 0.85 and 1.0 M$_{\odot}$ models. We again include the $Z=0$
models for comparison. It can be seen that as the metallicity increases
He is ignited when the stars are at higher and higher luminosities.
Between the $Z=0$ models and our most metal-rich models, with {[}Fe/H{]}$=-3.0$,
there is a difference of 0.8 dex in luminosity. Interestingly the
range is almost identical in the 1.0 M$_{\odot}$ models as in the
0.85 M$_{\odot}$ models. 

\subsubsection*{Intermediate Mass Models: 2.0 \& 3.0 M$_{\odot}$}

\begin{figure}
\begin{centering}
\includegraphics[width=0.8\columnwidth,keepaspectratio]{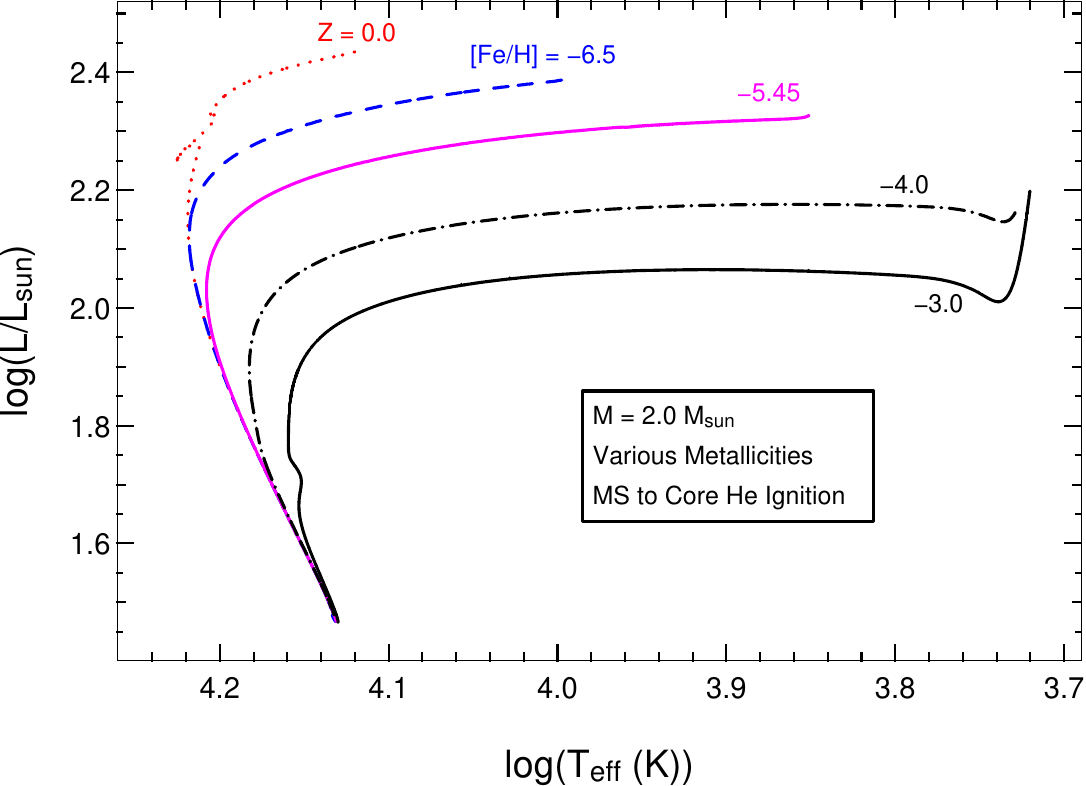}
\par\end{centering}
\caption{The HR diagrams for all our low metallicity 2 M$_{\odot}$ models,
from ZAMS to the ignition of helium in the centre. As metallicity
decreases the stars ignite He earlier and earlier. Only the most metal-rich
model reaches the RGB, and then only spends a brief time there. \label{fig-hrds.m2.all.MStoCHeIgnition}}
\end{figure}

\begin{figure}
\begin{centering}
\includegraphics[width=0.8\columnwidth,keepaspectratio]{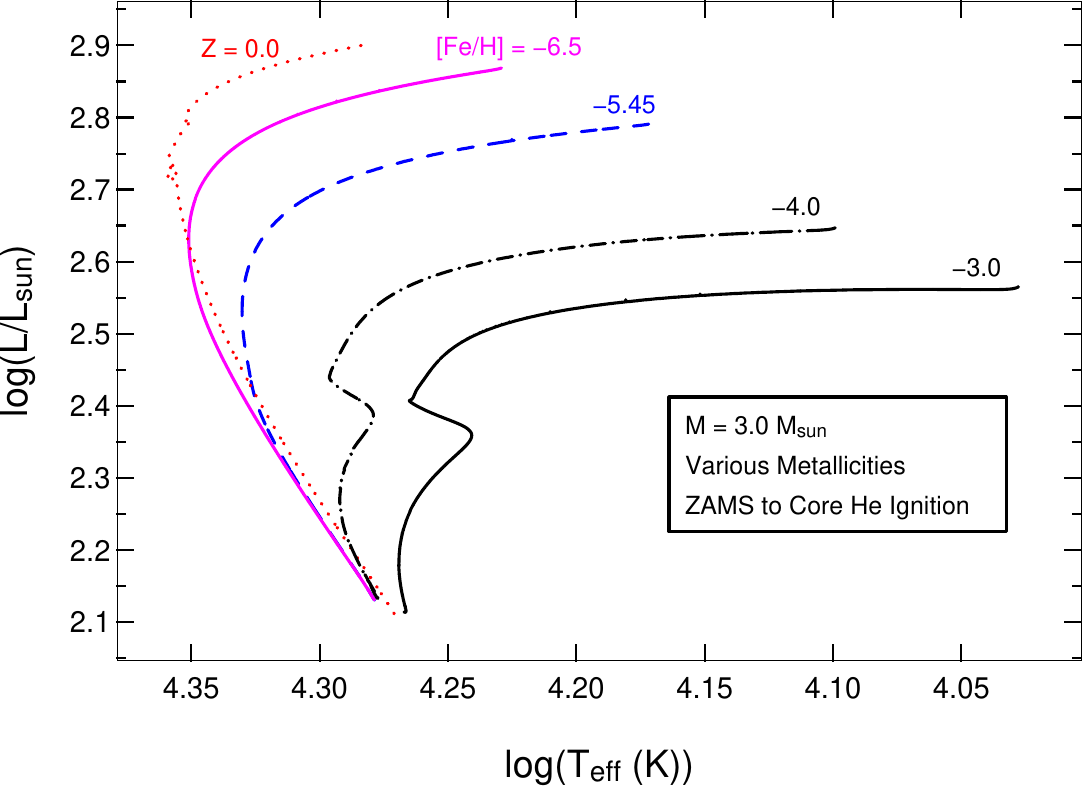}
\par\end{centering}
\caption{Same as Figure \ref{fig-hrds.m2-z0andEMP.MStoEAGB} but for the 3
M$_{\odot}$ models. In this case none of the models become red giants
as they ignite helium early and thus move back to the blue. \label{fig-hrds.m3.all.MStoCHeIgnition}}
\end{figure}

Figure \ref{fig-hrds.m2.all.MStoCHeIgnition} displays the evolution
of our five 2 M$_{\odot}$ models in the HR diagram, from ZAMS to
core He ignition. By definition these intermediate mass stars do not
ignite He degenerately, and thus do not experience a core He flash/DCF.
Differences in evolution are seen to occur earlier in these models,
as compared to the low mass models. In solar metallicity models the
CNO cycles operate efficiently at 2 M$_{\odot}$. At $Z=0$ the CNO
cycles do not contribute until the end of the MS. This CNO ignition
can be seen in the HR diagram (it is labeled in Figure \ref{fig-hrds.m2-z0andEMP.MStoEAGB}).
A significant difference in MSTO luminosities ($\sim0.4$ dex) between
the models can be seen in Figure \ref{fig-hrds.m2.all.MStoCHeIgnition},
again due to the varying degrees of paucity of CNO catalysts. A more
moderate variation in MSTO temperature is also present, being $\sim0.05$
dex over this range of metallicity. The stepwise reduction in MSTO
luminosity mirrors the reduction in metallicity. All these traits
are also present in the 3 M$_{\odot}$ models, as displayed in Figure
\ref{fig-hrds.m3.all.MStoCHeIgnition}. Our models thus predict a
bluer and more luminous MS at lower and lower metallicities. 

An additional feature in the 3 M$_{\odot}$ models is the appearance
of the CNO convective core `kink' in the HR diagram. This is a characteristic
seen at quite low masses (starting at $\sim$1.2 M$_{\odot}$) at
solar metallicity. It indicates that a significant convective core
has formed -- due to the efficient operation of the CNO cycles. The
core becomes convective because the high temperature of the CNO cycle
results in a very steep temperature gradient. The kink is thus a useful
indicator for our current models. Indeed, in Figure \ref{fig-hrds.m2.all.MStoCHeIgnition}
it can be seen that a small kink is seen in the most metal-rich of
our 2 M$_{\odot}$ models, but not the others. This highlights the
fact that the CNO cycle is not operating efficiently in the lower-Z
models. In the 3 M$_{\odot}$ models the two most metal-rich models
show well developed kinks, indicating that, due to the higher temperatures
in the cores of these more massive models, the CNO cycles are able
to operate efficiently at such low metallicities. Interestingly the
{[}Fe/H{]}$=-6.5$ and $-5.45$ models, of both masses, show perfectly
smooth evolution in the HR diagram -- they do not experience the
$Z=0$ CNO miniflash or the CNO convective core kink. Thus they behave
more like a 1 M$_{\odot}$ star at solar metallicity, although they
are much hotter and much more luminous. 

Also of note are the differing surface temperatures at which He is
ignited in the cores. The 2 M$_{\odot}$ models show a large range
of core He ignition temperatures -- from just after the MSTO to the
(short lived) RGB in the {[}Fe/H{]}$=-3.0$ model. Out of the ten
2.0 and 3.0 M$_{\odot}$ models this is the only model that reaches
the RGB. The differing ignition points (in surface temperature) for
He reflect the pace at which the conditions for He burning is achieved
at different metallicities (ie. temperature and density). Thus all
these low-Z intermediate mass models predict a lack of an RGB in colour
magnitude diagrams at very low metallicity and quite low masses, as
compared to solar. 

\subsection{On the Occurrence of Dual \emph{Core} Flashes}

By definition Dual Core Flashes (DCFs) can only occur in low-mass
models, as the secondary (hydrogen) flash is induced by the normal
degenerate core He flash. We have found that the DCF event also occurs
in some of our metal-deficient models -- but not all of them. 

\begin{figure}
\begin{centering}
\includegraphics[width=0.8\columnwidth,keepaspectratio]{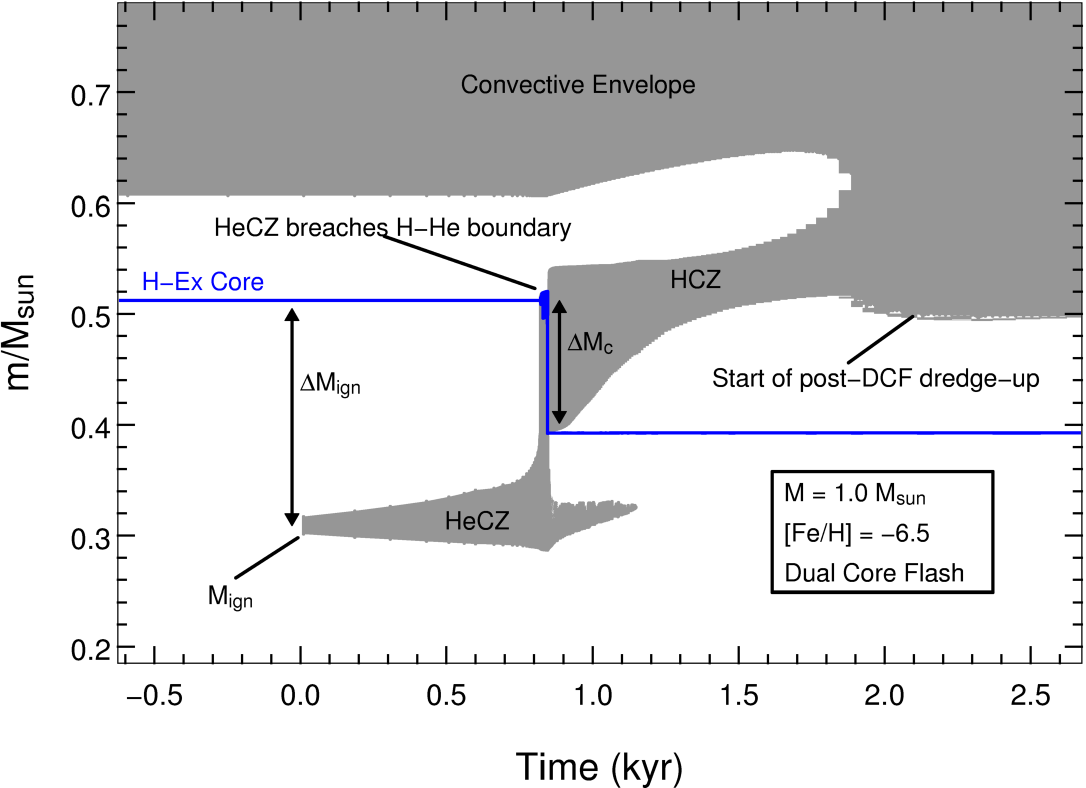}
\par\end{centering}
\begin{centering}
\includegraphics[width=0.8\columnwidth]{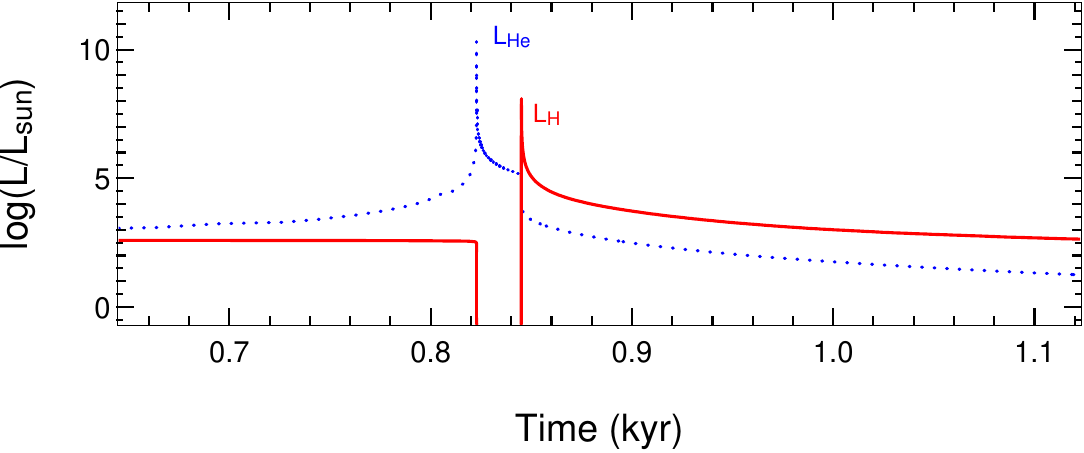}
\par\end{centering}
\begin{centering}
\includegraphics[width=0.8\columnwidth,keepaspectratio]{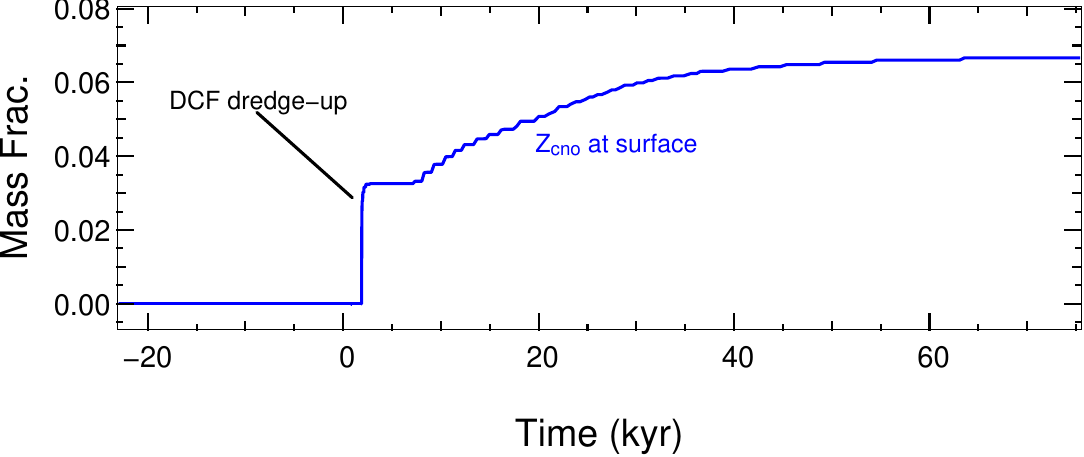}
\par\end{centering}
\caption{An example of one of the Dual Core Flashes that occurred in some of
our low metallicity models. In this case it is the 1 M$_{\odot}$
model with {[}Fe/H{]}$=-6.5$. Note that the three panels have the
same time scale and have been offset by the same amount (ie. zero
corresponds to the start of the He flash in all panels) -- but they
all cover different time-spans. The top panel shows the convective
regions (grey shading) and the mass of the hydrogen-exhausted core.
It also shows the definitions of some of the quantities referred to
in Table \ref{table-AllHaloModelDualFlashProperties}. The large inward
movement of the H-He boundary (H-Ex. core) due to the ingestion of
protons into the HeCZ can be seen ($\Delta$M$_{c}$). In the second
panel we zoom in on the luminosity evolution. The bottom panel shows
the evolution of the surface pollution arising from the DCF event,
in terms of $Z_{cno}$, the sum the mass fractions of all the CNO
isotopes. \label{fig-DCF-example-1mmp}}
\end{figure}

In Section \vref{section-m0.85z0y245-DualCoreFlash} we presented
an account of the DCF that occurred in our $Z=0$, 0.85 M$_{\odot}$
model. We refer the reader to that section for a detailed description
of this type of evolutionary event but reiterate the main properties
here. As a guide we display some details of the DCF that occurred
in one of our metal-poor models in Figure \ref{fig-DCF-example-1mmp}
(the 1 M$_{\odot}$, {[}Fe/H{]}$=-6.5$ model). As occurred in our
$Z=0$ models He ignited a long way off centre in the metal-poor models
which experienced the DCF. Helium convection zones (HeCZs) developed,
as normal, due to the large energy release from the thermonuclear
runaways. The extent of the convection zones increased over time,
mainly outwards, again as normal. However in these cases the convection
zones penetrated the H-He discontinuity. This penetration occurs for
two reasons: firstly, as just mentioned, He ignites significantly
off-centre, so the HeCZ is already relatively close to the H shell,
and secondly, as noted by \citet{1990ApJ...349..580F}, the entropy
barrier in the hydrogen shell is very low in models of extremely-low
(or $Z=0$) models. The extension of the He convection zones into
H-rich layers mixed protons downwards into the He convection zones,
so that H was now present in regions of very high temperature. The
large energy release from the furious burning of this H then caused
a creation of a secondary convection zone, the H convection zone (HCZ).
Thus two (burning) convection zones now existed in the stars -- a
HCZ and a HeCZ. The base of the HCZ now marked the top of the H-exhausted
cores, the boundaries of which had moved inwards due to the dredging
down of protons. After both flashes and their associated convection
zones abated the envelopes became more deeply convective, dredging
up the erstwhile HCZs. As the HCZs were polluted by He burning products,
which were present due to the previous He-flash burning at the locations
that the HCZs formed, these post-DCF dredge-up events led to a strong
pollution of the envelopes, raising their metallicities from zero
to $Z_{cno}\sim10^{-2}$. Since there was no further pollution of
the envelope via third dredge-up (3DUP) on the AGB in any of these
models, the pollution from these DCF events defined the chemical yield
of the models. 

As mentioned above we have found that the DCF event occurs in some
of our metal-deficient models but not others. In column five of Table
\ref{table-AllHaloModelDualFlashProperties} we show which stars experienced
the DCF and which did not, for all our metal deficient models (the
$Z=0$ models are also included for reference). It can be seen in
this table that those models which did not experience a DCF did however
go on to have a Dual Shell Flash (DSF, the topic of the next subsection).
It can also be seen that the occurrence of the DCF is a strong function
of the metallicity and mass of the stars. The DCF only occurs for
extremely metal-deficient models, and occurs more at low masses. At
M$=0.85$ M$_{\odot}$ we find that only our three most metal poor
models experience the DCF: the $Z=0$, {[}Fe/H{]}$=-6.5$ and {[}Fe/H{]}$=-5.45$
ones. At M$=1.0$ M$_{\odot}$ only the two most metal-poor models
do: the $Z=0$ and {[}Fe/H{]}$=-6.5$ models. The main reason for
this is that the most metal-poor models ignite He a long way off centre.
As metallicity increases the ignition point moves closer to the centre,
making the breaching of the H-He boundary less likely (although we
note that the $Z=0$ models ignite slightly further out than the {[}Fe/H{]}$=-6.5$
models). In Table \ref{table-AllHaloModelDualFlashProperties} we
give the location of the ignition points for all the models. We also
show the mass of the core at the time of core He ignition. Taking
the difference of these two quantities gives $\Delta$M$_{ign}$,
the distance (in mass) that separates the ignition point from the
H-He boundary (also see Figure \ref{fig-DCF-example-1mmp} for a definition
of the diagnostic quantities mentioned here). This value increases
with metallicity, as expected, ranging from $\sim0.2$ M$_{\odot}$
to $\sim0.4$ M$_{\odot}$ in these models. It appears that a value
less than 0.25 M$_{\odot}$ leads to a DCF. Another interesting diagnostic
is $\Delta$M$_{c}$, the amount that the mass of the H-exhausted
core decreases by, due to the DCF. In other words it is the distance
(in mass) that protons are mixed down before the secondary (hydrogen)
convection zone forms. The envelope later moves inwards, usually all
the way down to the new location of the H burning shell, dredging
up the ex-HCZ. Thus this value $\Delta$M$_{c}$ gives a (rough) indication
of the amount of pollution that occurs as a result of the DCF. Interestingly
the magnitude of surface pollution from the DCFs is similar in all
the models. They all end up with $Z_{cno}$ a few times $10^{-2}$,
with the exception of the 0.85 M$_{\odot}$, $Z=0$ model which ends
up less polluted, having $Z_{cno}\sim10^{-3}$ at the surface. Looking
more closely there is however some variation between the models, such
that the resultant pollution ranges from $Z_{cno}=1\times10^{-3}$
to $7\times10^{-2}$. This variation mostly correlates with the value
of $\Delta$M$_{c}$ but in one case, the 0.85 M$_{\odot}$, {[}Fe/H{]}$=-5.45$
model, the pollution is lower than expected. 

\begin{table}
\begin{center}\begin{threeparttable}\centering

\begin{tabular}{cccccccc}
\multicolumn{7}{c}{Mass $=0.85$ M$_{\odot}$} & \tabularnewline
\hline 
\hline 
Metallicity & M$_{ign}$ & M$_{c}$ & $\Delta$M$_{ign}$ & DCF/DSF & $\Delta$M$_{c}$ & LH$_{max}$ & Z$_{cno}$\tabularnewline
\hline 
-3.00 & 0.14 & 0.50 & 0.36 & \textcolor{blue}{Shell} & 0.014 & 8.1 & 1E-02\tabularnewline
-4.00 & 0.21 & 0.52 & 0.31 & \textcolor{blue}{Shell} & 0.004 & 7.7 & 1E-03\tabularnewline
-5.45 & 0.30 & 0.53 & 0.23 & \textcolor{red}{Core} & 0.140 & 8.1 & 3E-02\tabularnewline
-6.50 & 0.33 & 0.52 & 0.19 & \textcolor{red}{Core} & 0.095 & 7.6 & 6E-02\tabularnewline
$Z=0$ & 0.28 & 0.49 & 0.21 & \textcolor{red}{Core} & 0.056 & 8.3 & 1E-03\tabularnewline
\hline 
 &  &  &  &  &  &  & \tabularnewline
\multicolumn{7}{c}{Mass $=1.0$ M$_{\odot}$} & \tabularnewline
\hline 
\hline 
Metallicity & M$_{ign}$ & M$_{c}$ & $\Delta$M$_{ign}$ & DCF/DSF & $\Delta$M$_{c}$ & LH$_{max}$ & Z$_{cno}$\tabularnewline
\hline 
-3.00 & 0.09 & 0.49 & 0.40 & \textcolor{blue}{Shell} & 0.010 & 7.3 & 5E-03\tabularnewline
-4.00 & 0.15 & 0.51 & 0.36 & \textcolor{blue}{Shell} & 0.019 & 10.5 & 1E-02\tabularnewline
-5.45 & 0.27 & 0.52 & 0.25 & \textcolor{blue}{Shell} & 0.010 & 10.6 & 4E-03\tabularnewline
-6.50 & 0.31 & 0.52 & 0.21 & \textcolor{red}{Core} & 0.120 & 8.2 & 7E-02\tabularnewline
$Z=0$ & 0.25 & 0.49 & 0.23 & \textcolor{red}{Core} & 0.083 & 8.7 & 1E-02\tabularnewline
\hline 
 &  &  &  &  &  &  & \tabularnewline
\multicolumn{7}{c}{Mass $=2.0$ M$_{\odot}$} & \tabularnewline
\hline 
\hline 
Metallicity & M$_{ign}$ & M$_{c}$ & $\Delta$M$_{ign}$ & DCF/DSF & $\Delta$M$_{c}$ & LH$_{max}$ & Z$_{cno}$\tabularnewline
\hline 
-3.00 & 0.62 & 0.64 & 0.018 & \textcolor{darkgreen}{None}

 & -- & -- & --\tabularnewline
-4.00 & 0.64 & 0.66 & 0.016 & \textcolor{blue}{Shell} & 4E-03 & 10.1 & 6E-04\tabularnewline
-5.45 & 0.65 & 0.67 & 0.014 & \textcolor{blue}{Shell} & 6E-04 & 8.3 & 2E-05\tabularnewline
-6.50 & 0.67 & 0.69 & 0.012 & \textcolor{blue}{Shell} & 4E-03 & 10.5 & 7E-04\tabularnewline
$Z=0$ & 0.69 & 0.70 & 0.011 & \textcolor{blue}{Shell} & 1E-03 & 8.6 & 2E-04\tabularnewline
\hline 
 &  &  &  &  &  &  & \tabularnewline
\multicolumn{7}{c}{Mass $=3.0$ M$_{\odot}$} & \tabularnewline
\hline 
\hline 
Metallicity & M$_{ign}$ & M$_{c}$ & $\Delta$M$_{ign}$ & DCF/DSF & $\Delta$M$_{c}$ & LH$_{max}$ & Z$_{cno}$\tabularnewline
\hline 
-3.00 & 0.798 & 0.804 & 0.006 & \textcolor{darkgreen}{None}

 & -- & -- & --\tabularnewline
-4.00 & 0.820 & 0.825 & 0.005 & \textcolor{darkgreen}{None}

 & -- & -- & --\tabularnewline
-5.45 & 0.834 & 0.837 & 0.003 & \textcolor{blue}{Shell} & 9E-05 & 7.6 & 4E-08\tabularnewline
-6.50 & 0.817 & 0.821 & 0.003 & \textcolor{blue}{Shell} & 1E-04 & 7.1 & 8E-06\tabularnewline
$Z=0$\tnote{a} & 0.770 & 0.775 & 0.005 & \textcolor{blue}{Shell} & 3E-05 & 8.4 & 8E-06\tabularnewline
\hline 
\end{tabular}

\caption{Various properties pertaining to the dual core flashes (DCFs) and
dual shell flashes (DSFs) for all the low metallicity models, grouped
in terms of initial mass. We also redisplay the properties of the
$Z=0$ models for ease of comparison. All masses are in M$_{\odot}$.
Metallcity is given as {[}Fe/H{]}, except for $Z=0$. Properties displayed
are: M$_{ign}$ (the mass coordinate at which the He flash ignites,
be it a core He flash or an AGB shell flash), M$_{c}$ (core mass
at the onset of the flash), $\Delta$M$_{ign}$ (the distance, in
mass, between the He flash ignition point and the H shell), $\Delta$M$_{c}$
(the distance, in mass, that the H shell moves inwards during the
dual flash; this gives an indication of the degree of mass transfer
between the HeCZ and the H-rich zone above), LH$_{max}$ (log of the
maximum H burning luminosity during the dual flash), and $Z_{cno}$
(the envelope metallicity after the dredge-up events(s) associated
with the DCF/DSF episodes; this excludes pollution from other episodes
that occur later, such as the 3DUP). \label{table-AllHaloModelDualFlashProperties}}

\line(1,0){100}

\begin{tablenotes}\scriptsize

\item[a] This model started with a pure H-He composition, with $Y=0.230$ rather than 0.245. 

\end{tablenotes}

\end{threeparttable} \end{center}
\end{table}

\subsection{On the Occurrence of Dual \emph{Shell} Flashes\label{subsection-HaloStars-Struct-DSFs}}

\begin{figure}
\begin{centering}
\includegraphics[width=0.9\columnwidth,keepaspectratio]{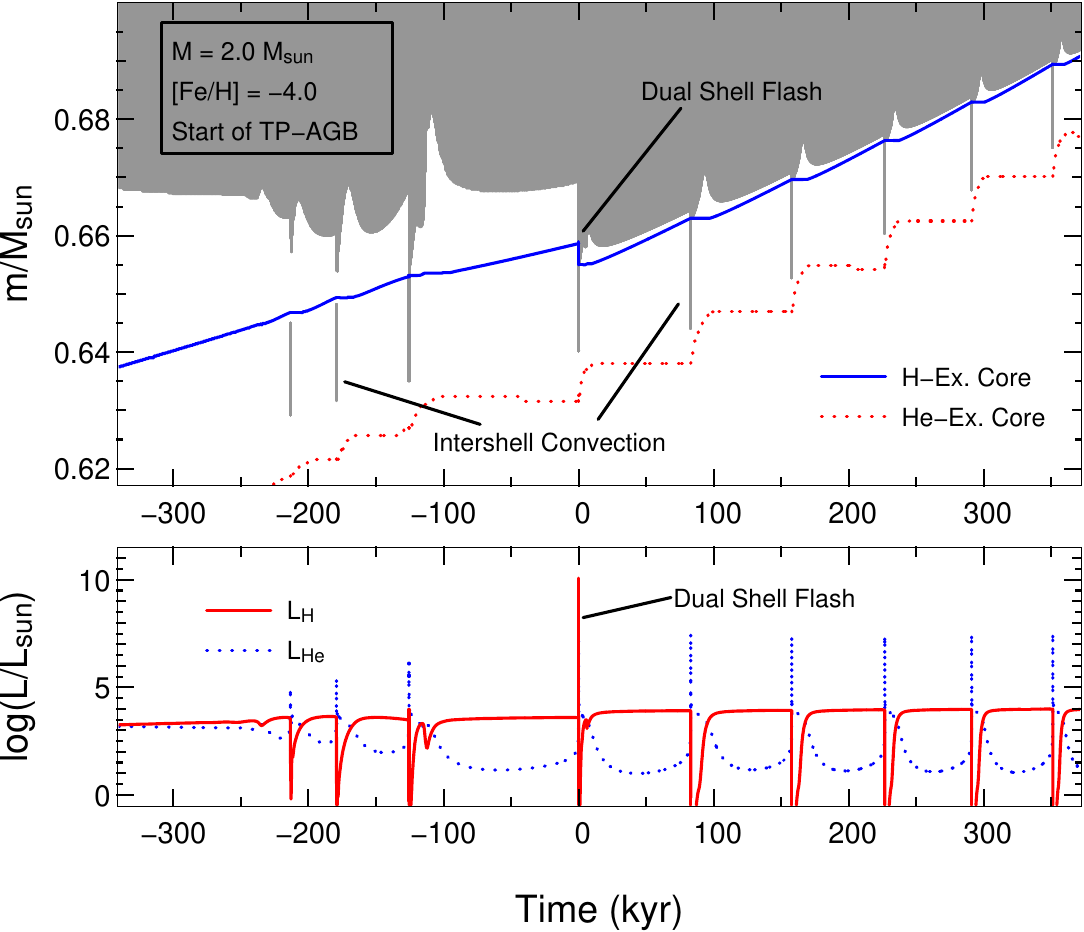}
\par\end{centering}
\caption{A wide-angle view of the start of the TP-AGB in the 2 M$_{\odot}$
model with {[}Fe/H{]}$=-4.0$. Time is offset for clarity. This model
experienced a Dual Shell Flash (DSF) at the 4th thermal pulse, as
indicated in both panels. It can be seen that the hydrogen burning
luminosity exceeds that of the He flash luminosity during the DSF.
Figure \ref{fig-m2ump-DSF-Example} provides a close-up view of this
event. \label{fig-m2ump-DSF-wide-cvn-lums}}
\end{figure}

\begin{figure}
\begin{centering}
\includegraphics[width=0.75\columnwidth,keepaspectratio]{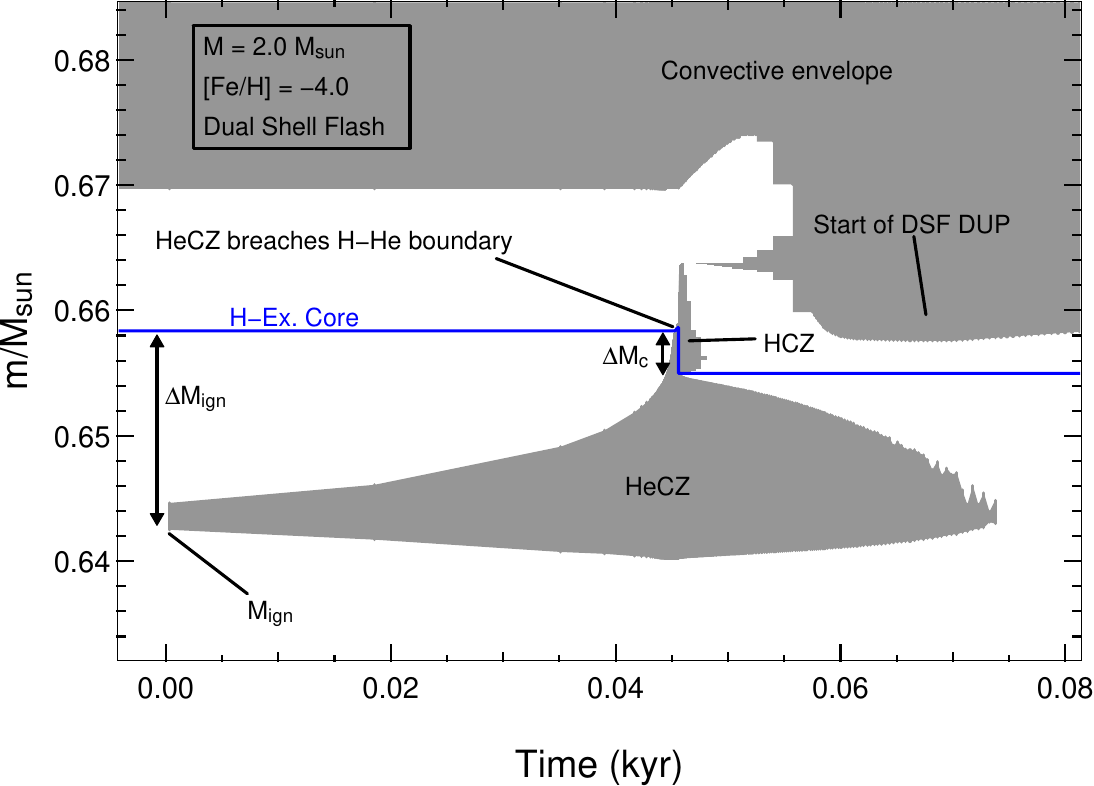}
\par\end{centering}
\begin{centering}
\includegraphics[width=0.75\columnwidth]{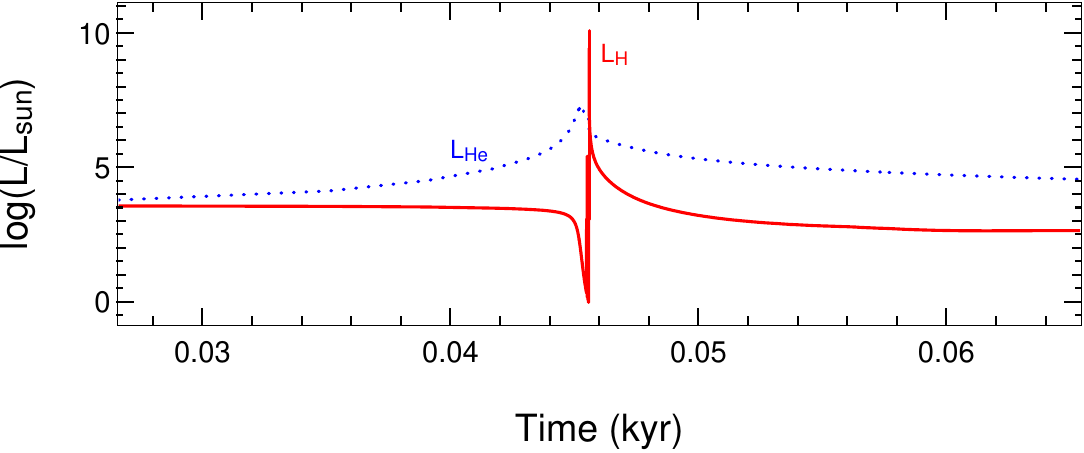}
\par\end{centering}
\begin{centering}
\includegraphics[width=0.75\columnwidth,keepaspectratio]{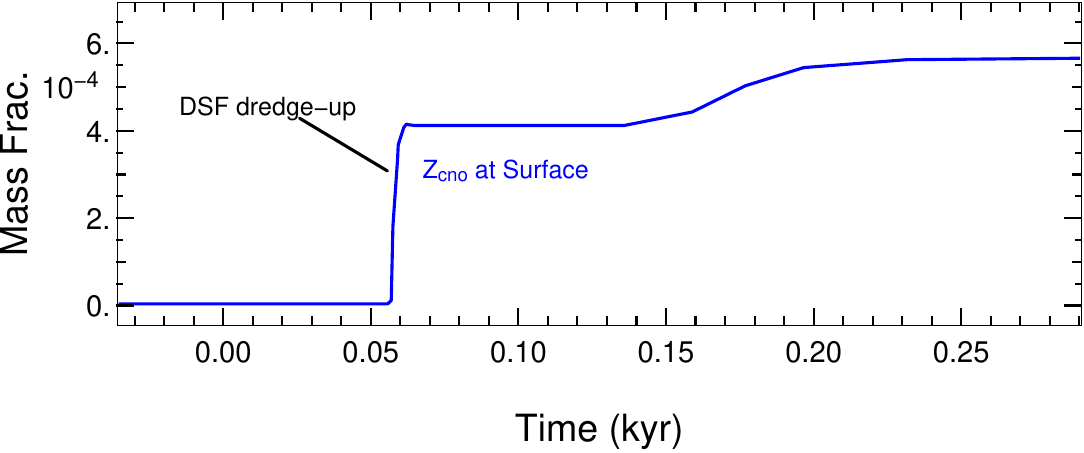}
\par\end{centering}
\caption{An example of one of the Dual Shell Flashes (DSFs) that occurred in
some of our low metallicity models. In this case it is the 1 M$_{\odot}$
model with {[}Fe/H{]}$=-4.0$. Note that the three panels have the
same time scale and have been offset by the same amount (ie. zero
corresponds to the start of the He flash in all panels) -- but they
all cover different time-spans. The top panel shows the convective
regions (grey shading) and the mass of the hydrogen-exhausted core.
It also shows the definitions of some of the quantities referred to
in Table \ref{table-AllHaloModelDualFlashProperties}. The inward
movement of the H-He boundary (H-Ex. core) due to the ingestion of
protons into the HeCZ can be seen ($\Delta$M$_{c}$). In the second
panel we zoom in on the luminosity evolution. The bottom panel shows
the evolution of the surface pollution arising from the DCF event,
in terms of $Z_{cno}$, the sum of the mass fractions of all the CNO
isotopes. The degree of pollution is much less than in the DCF case
(Figure \ref{fig-DCF-example-1mmp}), such that $Z_{cno}$ is 2 dex
lower. \label{fig-m2ump-DSF-Example}}
\end{figure}

In \vref{section-m2z0-DSF-TPAGB} we described in detail the Dual
Shell Flash (DSF) event occurring near the beginning of the TP-AGB
in our 2 M$_{\odot}$, $Z=0$ model. We refer the reader to that section
for a detailed description of this evolutionary event but reiterate
the main properties here. As a guide we display some details of the
DSF that occurred in one of our metal-poor models in Figures \ref{fig-m2ump-DSF-wide-cvn-lums}
and \ref{fig-m2ump-DSF-Example} (the 2 M$_{\odot}$, {[}Fe/H{]}$=-4.0$
model). The DSF is a similar event to the DCF insomuch as it involves
a helium convection zone expanding upwards and breaching the H-He
interface. This proton ingestion event (PIE) also results in the creation
of a second convection zone, the HCZ. Like the DCF HCZ this convection
zone is polluted with He (and H) burning products, as its extent includes
some of the mass that was previously part of the HeCZ (see panel 1
in Figure \ref{fig-m2ump-DSF-Example}). Importantly this HCZ material
is dredged up at a later time (when the convection has stopped), polluting
the envelope. In the case of the low mass stars (M$=0.85$ and 1 M$_{\odot}$)
this pollution is the main contributor to the chemical yield, as these
models experience no 3DUP. Conversely, in the case of the intermediate
mass models (M$=2$ and 3 M$_{\odot}$) the chemical pollution from
this event is erased/swamped by the 3DUP pollution occurring later
on the AGB. In Table \ref{table-AllHaloModelDualFlashProperties}
we display the pollution resulting from the DSF events in all the
models. We also list which models do and do not experience the DSF.
From this list we can see that the DSF is also very metallicity and
mass dependent, like the DCF. As mentioned earlier the low mass models
that do not go through the DCF always experience the DSF. At intermediate
mass we see that only the most metal-poor models go through the DSF.
Also, the metallicity cut-off for the DSF decreases with increasing
mass -- only one of our 2 M$_{\odot}$ models does not experience
the DSF, the {[}Fe/H{]}$=-3.0$ model -- but the two most metal-rich
3 M$_{\odot}$ models avoid it (the {[}Fe/H{]}$=-3.0$ and {[}Fe/H{]}$=-4.0$
models). 

\subsection{Further Evolution in the HR Diagram}

\begin{figure}
\begin{centering}
\includegraphics[width=0.8\columnwidth,keepaspectratio]{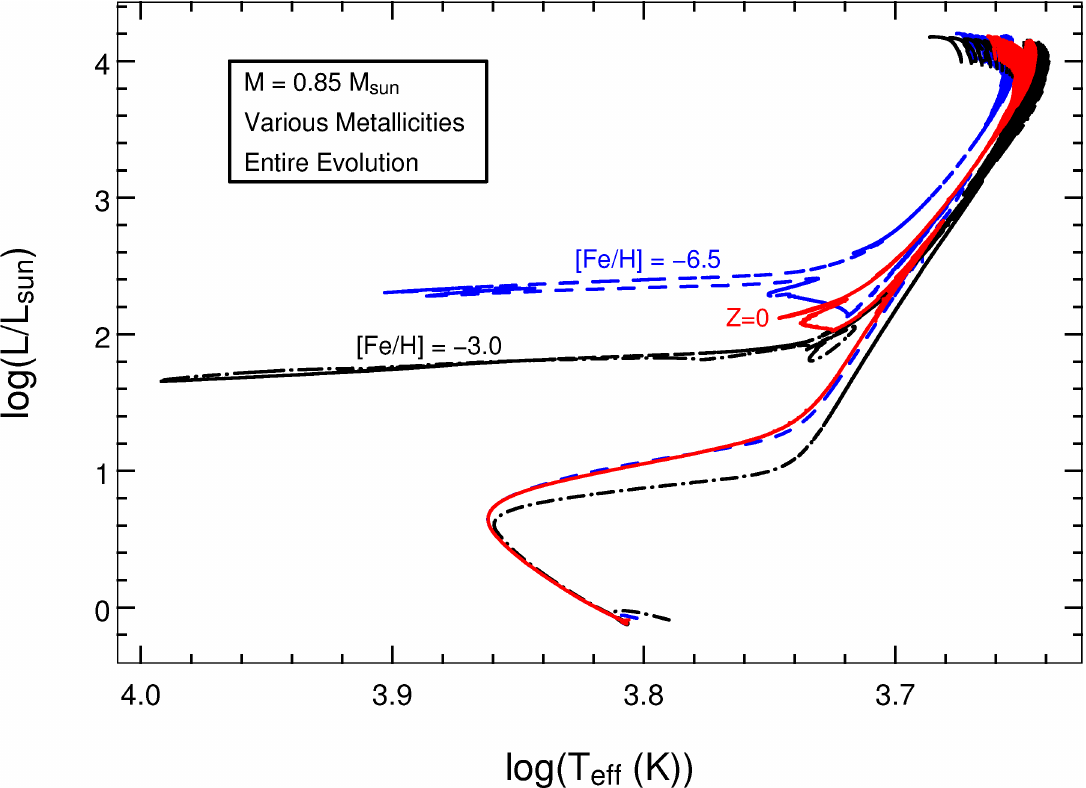}
\par\end{centering}
\caption{The full evolution in the HR diagram for some of the 0.85 M$_{\odot}$
models. \label{fig-hrds-m0.85all-HBs-AGB}}
\end{figure}

In Figure \ref{fig-hrds-m0.85all-HBs-AGB} we show the full evolution
of the some of the 0.85 M$_{\odot}$ models in the HR diagram. The
MS and RGB features were discussed earlier. Of note in this diagram
is the variation in the blueward excursions during the horizontal
branch (HB) stage. It appears that the HB becomes redder as metallicity
\emph{decreases} -- the {[}Fe/H{]}$=-3.0$ model still has a blue
HB, like the (mildly) metal-poor stars of globular clusters, but the
{[}Fe/H{]}$=-6.5$ model HB is less blue. Indeed, the $Z=0$ HB is
almost like a metal-rich `red clump' star, insomuch as it does not
become much bluer than its RGB. 

\begin{figure}
\begin{centering}
\includegraphics[width=0.8\columnwidth,keepaspectratio]{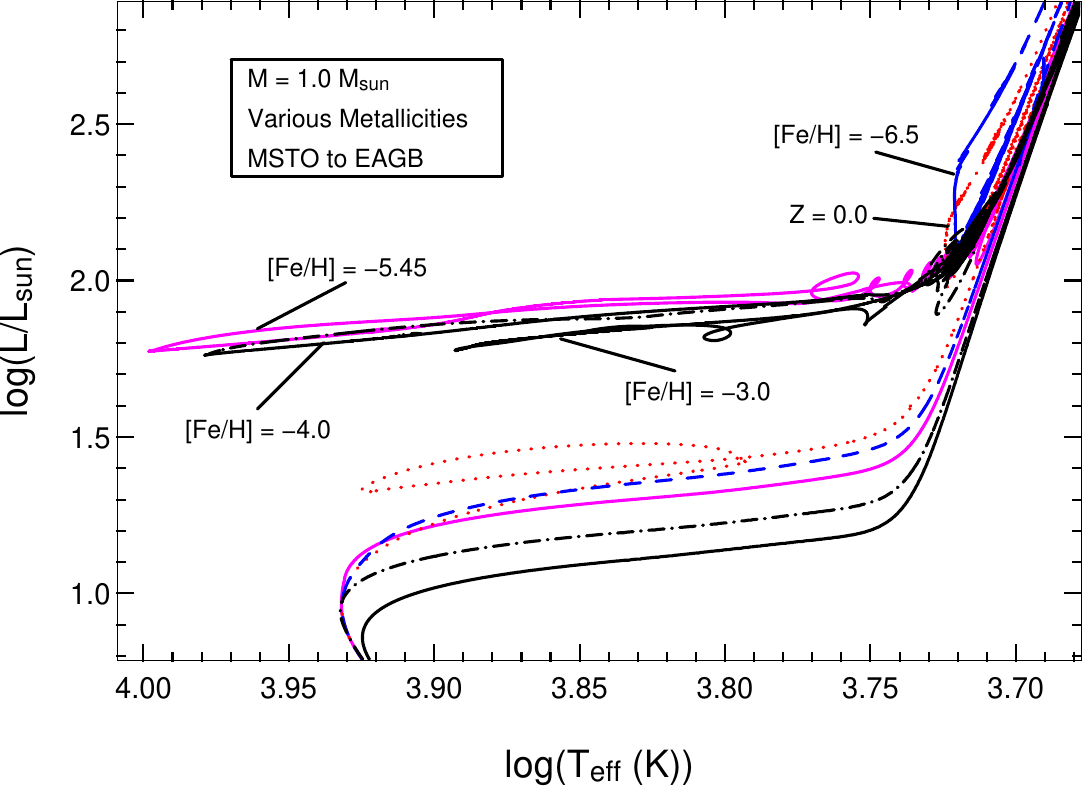}
\par\end{centering}
\caption{Some of the evolution in the HR diagram for all of the 1.0 M$_{\odot}$
models. We note that the two most metal-deficient models have quite
red horizontal branches. \label{fig-hrds.m1all.HB}}
\end{figure}

In Figure \ref{fig-hrds.m1all.HB} we show the horizontal branches
of all the 1 M$_{\odot}$ models in the HR diagram. Here again it
can be seen that the blueward excursions decrease with metallicity.
However this only occurs at extremely low metallicity -- the blueward
excursions first increase, maximising at {[}Fe/H{]}$=-5.45$ before
rapidly getting redder at {[}Fe/H{]}$=-6.5$ and $Z=0$. 

\begin{figure}
\begin{centering}
\includegraphics[width=0.8\columnwidth,keepaspectratio]{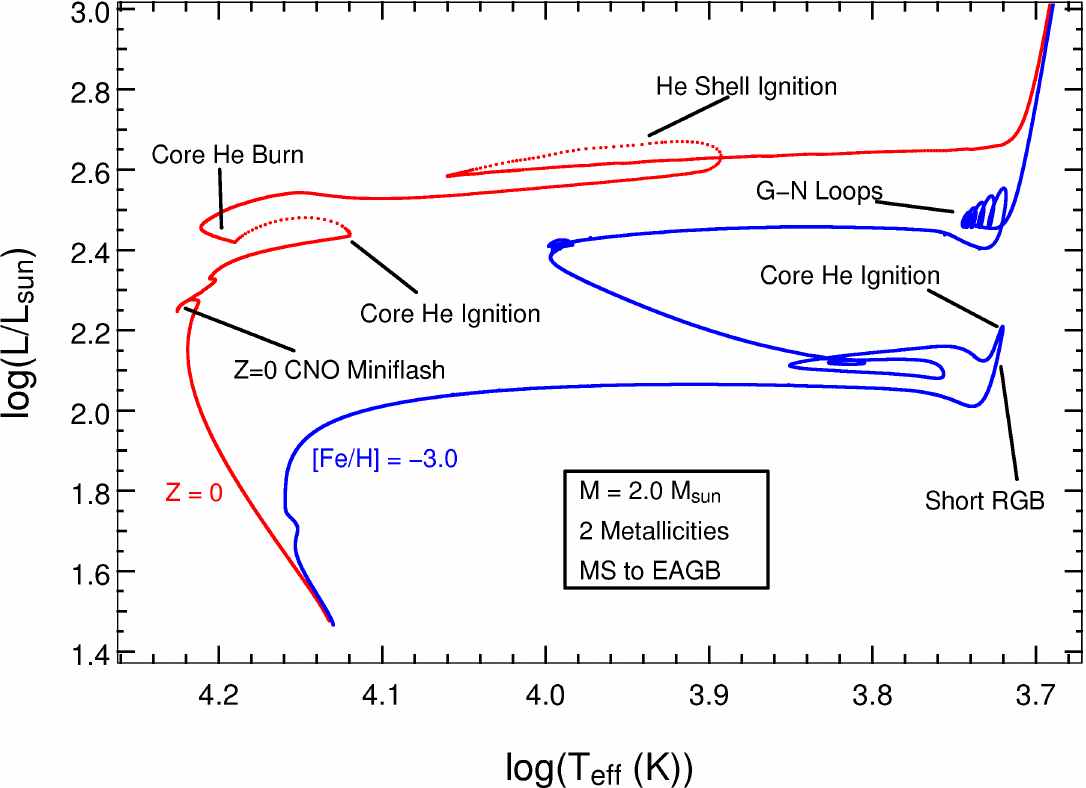}
\par\end{centering}
\caption{The evolution in the HR diagram for two of the 2 M$_{\odot}$ models.
We show only our most metal-rich and most metal-poor models for clarity.
The rest of the models display intermediary evolution, gradually turning
to the blue earlier and earlier (due to core He ignition) as the metallicity
decreases. Also indicated is the location of the CNO `miniflash' (see
the $Z=0$, 2 M$_{\odot}$ structural evolution chapter for more information
on this event), and some `gravo-nuclear loops' in the {[}Fe/H{]}$=-3.0$
model (see Section \vref{overshootMods} for details on this feature).
\label{fig-hrds.m2-z0andEMP.MStoEAGB}}
\end{figure}

In Figure \ref{fig-hrds.m2-z0andEMP.MStoEAGB} we present the further
evolution in the HR diagram of a couple of the 2.0 M$_{\odot}$ models.
For clarity we have plotted only the most metal-poor (the $Z=0$ model)
and our most metal-rich model ({[}Fe/H{]}$=-3.0$). All the others
show intermediary behaviour, so it suffices to discuss the extreme
cases here. In the {[}Fe/H{]}$=-3.0$ case it can be seen that it
reaches the start of the RGB. However, before it increases its luminosity
very much it ignites He in the core and consequently moves back to
the blue. After completing core He burning it then returns to the
red, moving up the AGB. The $Z=0$ model on the other hand never reaches
the red side of the HR diagram. It ignites He very soon after the
MS turnoff and spends its core He burning lifetime as a very blue
star. In fact it is as blue as its main sequence. Like the more metal-rich
models it then moves to the red to become an AGB star. 

\begin{figure}
\begin{centering}
\includegraphics[width=0.8\columnwidth,keepaspectratio]{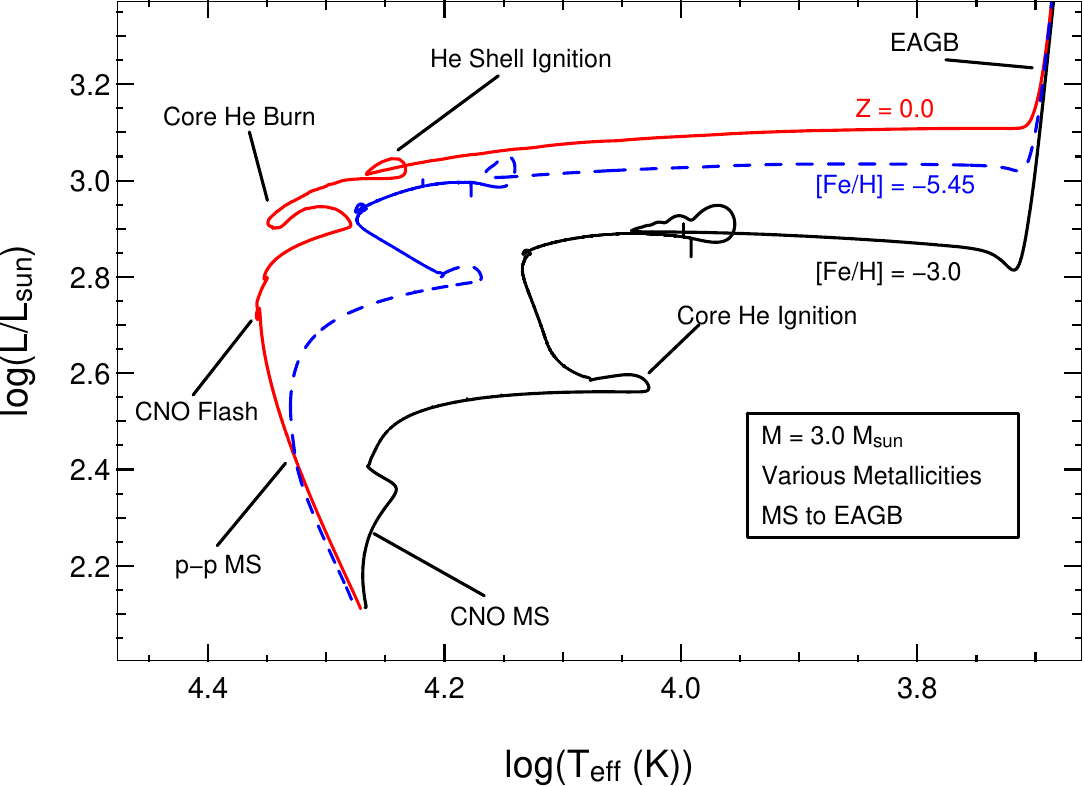}
\par\end{centering}
\caption{The evolution in the HR diagram of some of the 3 M$_{\odot}$ models.
We show only some of the models for clarity. The rest of the models
display intermediary evolution, gradually turning to the blue earlier
and earlier (due to core He ignition) as metallicity decreases. We
note that none of the 3 M$_{\odot}$ models make it to the RGB. The
difference in the MS due to the competition between p-p and CNO burning
is indicated (this is also discussed in Section \ref{section-HaloStars-Struct-MSandRGB}),
as is the location of the CNO `miniflash' (see the $Z=0$, 2 M$_{\odot}$
structural evolution chapter for more information on this event).
\label{fig-hrds.m3.some.MStoEAGB}}
\end{figure}

Figure \ref{fig-hrds.m3.some.MStoEAGB} displays the HR diagrams of
some of the 3 M$_{\odot}$ models, from ZAMS to the EAGB. As mentioned
earlier none of these models reach the RGB at all. Like the more metal-poor
2 M$_{\odot}$ models they ignite He in their cores whilst moving
across the Hertzsprung gap (HG), thus turning back towards the blue
where they spend their core He burning lifetimes. The colour of the
stars during the core He burn stage is progressively bluer with metallicity.
We note that this would effect the integrated light from a population
of extremely metal-deficient stars. 

\subsection{TP-AGB\label{subsection-AGB-Struct-HaloStars}}

All stars in our grid of models go on to become thermally-pulsing
AGB stars. This is the stage where they lose the most mass, and thus
the composition of their surfaces during this stage has the most impact
on the chemical yields. Below we begin by delineating three groups
that we have identified from our grid, based on the nature of the
pollution that defines the chemical composition of their AGBs. We
then go on to present some details of the models. As there are so
many models we have chosen two as examples, for which we give detailed
figures. However we also provide selected salient quantities for the
whole grid of models -- in Table \ref{table-AllHaloModelAGBProperties}
(which complements Table \ref{table-AllHaloModelDualFlashProperties}). 

\subsubsection*{DCF Dominated AGB: Low-Mass \& Extremely Metal-Deficient Models}

In the low mass models that experienced the dual core flash ($\textrm{[Fe/H]}\leqslant-5.45$
at $\textrm{M}=0.85$ M$_{\odot}$ and $\textrm{[Fe/H]}\leqslant-6.5$
at 1.0 M$_{\odot}$ -- see Table \ref{table-AllHaloModelDualFlashProperties}
for a list) we found that no further pollution events occurred. This
had the consequence that the AGB surface composition reflected the
dual core flash (DCF) pollution for the rest of the evolution. In
Table \ref{table-AllHaloModelAGBProperties} we show the final composition
of the envelope (in terms of $Z_{cno}$) for all the models. Indeed,
it can be seen that the pollution levels in these low-mass models
has not changed since that given by the DCFs. Thus the chemical yields
for these models are defined by the DCF pollution event. This result
was also mentioned in the 0.85 M$_{\odot}$, $Z=0$ sections.

\subsubsection*{DSF Dominated AGB: Low-Mass \& Very Metal-Deficient Models}

Those low-mass models that did \emph{not} experience the DCF all went
on to experience a dual shell flash at the start of the TP-AGB (see
Table \ref{table-AllHaloModelDualFlashProperties} for a list). The
pollution from this event raised the $Z_{cno}$ value to similar levels
as that achieved by the DCF event (although somewhat lower in some
cases). As the further evolution on the TP-AGB did not include any
periodic 3DUP, the chemical composition of the AGB envelopes remained
imprinted by the DSF event till the end of their evolution. Thus the
chemical yields from these models are essentially dominated by the
DSF pollution event, much like how those of the lower metallicity
models are dominated by the DCF pollution.

\subsubsection*{3DUP Dominated AGB: Intermediate-Mass -- All Metallicities}

In the case of the intermediate mass models we found that the most
metal-deficient ones did experience a DSF event at the start of the
TP-AGB but the more metal-rich ones did not. The pollution that resulted
from the DSF events in the IM models was however orders of magnitude
less than that arising in the low mass models ($Z_{cno}\sim10^{-6}\rightarrow10^{-4}$
versus $\sim10^{-3}\rightarrow10^{-2}$). Also unlike the low mass
models the IM DSF models all went on to experience many 3DUP events
through their further AGB evolution. Thus the surface pollution from
the DSF events was quickly `swamped', such that the periodically increasing
pollution from 3DUP dominated the surface composition for the majority
of the AGB. Hence the chemical yield was defined by 3DUP (and HBB)
rather than the DSF in these models. 

In the models that did not experience a DSF event (the more metal-rich
models) it was found that a similar amount of 3DUP occurred, polluting
the AGB envelopes to a similar degree. Thus all of the IM mass models
ended up with similar degrees of pollution, due to 3DUP. 

\subsubsection*{AGB Properties}

Table \ref{table-AllHaloModelAGBProperties} gives a range of AGB
properties for the entire grid of models, namely the number of thermal
pulses, the average of the dredge-up parameter $\lambda$, the average
length of the interpulse period, and the remnant mass (WD mass). The
number of thermal pulses is seen to increase with mass, ranging from
$\sim30$ at 0.85 M$_{\odot}$ to $\sim370$ at 3.0 M$_{\odot}$.
This is a function of a few factors, such as the mass of the envelope
that needed to be expelled, the characteristic interpulse period of
each model, and the mass loss rate, which itself is dependent on luminosity
and radius. The average interpulse periods for the intermediate mass
models were generally of the order of $10^{3}$ years, whilst they
were longer in the low mass models, being generally around $10^{4}$
years. 

As mentioned earlier, 3DUP was absent in all our low mass models but
a significant amount was found to occur in the IM models. Note that
we have not included any form of overshoot so these results represent
a lower limit on 3DUP. In the 2 M$_{\odot}$ models the average of
the dredge-up parameter $\lambda$ was around 0.1, but was as low
as 0.05 in a couple of models. In the 3 M$_{\odot}$ models $\lambda$
was higher, the average values ranging from 0.20 to 0.46. This substantial
degree of 3DUP combined with the large number of thermal pulses led
the envelopes of all the 3 M$_{\odot}$ models to become heavily polluted,
reaching $Z_{cno}\sim10^{-2}$ (super-solar in some cases). The 2
M$_{\odot}$ models reached a lower level of pollution, characterised
by $Z_{cno}\sim10^{-3}$. As this pollution peaks towards the end
of the AGB when the mass-loss rate is the highest, the chemical yield
is also heavily polluted. 

As examples of our metal deficient models we now detail the AGB evolution
of our 1 M$_{\odot}$, $\textrm{[Fe/H]}=-5.45$ model and 3 M$_{\odot}$,
$\textrm{[Fe/H]}=-5.45$ model.

\begin{table}
\begin{center}\begin{threeparttable}\centering

\small

\begin{tabular}{ccccccc}
\multicolumn{7}{c}{Mass $=0.85$ M$_{\odot}$}\tabularnewline
\hline 
\hline 
Metallicity & N$_{TP}$ & $\lambda_{ave}$ & IPP$_{ave}$ & Z$_{cno,f}$ & M$_{WD}$ & N$_{mods}$\tabularnewline
\hline 
-3.00 & 40 (1) & zero & 1E+05 & 1E-02 & 0.75 & 0.4\tabularnewline
-4.00 & 35 (4) & zero & 1E+05 & 1E-03 & 0.76 & 0.4\tabularnewline
-5.45 & 18 (8) & zero & 9E+03 & 3E-02 & 0.74 & 0.6\tabularnewline
-6.50 & 27 (3) & zero & 3E+04 & 6E-02 & 0.76 & 0.3\tabularnewline
$Z=0$ & 37 (6) & zero & 5E+04 & 1E-03 & 0.78 & 0.1\tabularnewline
\hline 
 &  &  &  &  &  & \tabularnewline
\multicolumn{7}{c}{Mass $=1.0$ M$_{\odot}$}\tabularnewline
\hline 
\hline 
Metallicity & N$_{TP}$ & $\lambda_{ave}$ & IPP$_{ave}$ & Z$_{cno,f}$ & M$_{WD}$ & N$_{mods}$\tabularnewline
\hline 
-3.00 & 60 (3) & zero & 7E+04 & 5E-03 & 0.85 & 0.8\tabularnewline
-4.00 & 69 (3) & zero & 7E+04 & 1E-02 & 0.85 & 0.8\tabularnewline
-5.45 & 63 (3) & zero & 8E+04 & 3E-03 & 0.87 & 0.5\tabularnewline
-6.50 & 63 (4) & zero & 2E+04 & 6E-02 & 0.85 & 0.7\tabularnewline
$Z=0$ & 65 (4) & zero & 3E+04 & 1E-02 & 0.84 & 0.7\tabularnewline
\hline 
 &  &  &  &  &  & \tabularnewline
\multicolumn{7}{c}{Mass $=2.0$ M$_{\odot}$}\tabularnewline
\hline 
\hline 
Metallicity & N$_{TP}$ & $\lambda_{ave}$ & IPP$_{ave}$ & Z$_{cno,f}$ & M$_{WD}$ & N$_{mods}$\tabularnewline
\hline 
-3.00 & 218 (2) & 0.12 & 1E+04 & 1E-02 & 1.04 & 2.1\tabularnewline
-4.00 & 306 (4) & 0.04 & 7E+03 & 5E-03 & 1.08 & 1.2\tabularnewline
-5.45 & 257 (3) & 0.12 & 8E+03 & 1E-02 & 1.06 & 1.9\tabularnewline
-6.50 & 269 (3) & 0.09 & 7E+03 & 8E-03 & 1.07 & 1.3\tabularnewline
$Z=0$ & 286 (4) & 0.05 & 6E+03 & 4E-03 & 1.08 & 1.1\tabularnewline
\hline 
 &  &  &  &  &  & \tabularnewline
\multicolumn{7}{c}{Mass $=3.0$ M$_{\odot}$}\tabularnewline
\hline 
\hline 
Metallicity & N$_{TP}$ & $\lambda_{ave}$ & IPP$_{ave}$ & Z$_{cno,f}$ & M$_{WD}$ & N$_{mods}$\tabularnewline
\hline 
-3.00 & 314 (4) & 0.33 & 4E+03 & 3E-02 & 1.07 & 2.5\tabularnewline
-4.00\tnote{a} & 343 & 0.46 & 4E+03 & 4E-02 & 1.06 & 4.0\tabularnewline
-5.45 & 369 (6) & 0.26 & 3E+03 & 2E-02 & 1.10 & 2.6\tabularnewline
-6.50 & 403 (7) & 0.21 & 3E+03 & 1E-02 & 1.11 & 2.1\tabularnewline
$Z=0$\tnote{b} & 387 (6) & 0.20 & 3E+03 & 1E-02 & 1.10 & 3.7\tabularnewline
\hline 
\end{tabular}

\caption{ Various properties pertaining to the AGB phase of evolution for all
the low metallicity models, grouped in terms of initial mass. We also
redisplay the properties of the $Z=0$ models. Metallcity is given
as {[}Fe/H{]}, except for $Z=0$. Properties displayed are: N$_{TP}$
(number of thermal pulses; the number in brackets is how many synthetic
pulses were calculated to finish the evolution once the SEV code failed
to converge), $\lambda_{ave}$ (the average of the dredge-up parameter
over all the pulses), IPP$_{ave}$ (average of the interpulse period
over all the pulses, in years), Z$_{cno,f}$ (the final surface metallicity
in terms of the sum of the CNO nuclides), M$_{WD}$ (mass of the white
dwarf remnant, in M$_{\odot}$), and N$_{mods}$ (the total number
of models calculated for each star over its entire evolution, in units
of 10$^{6}$ models). \label{table-AllHaloModelAGBProperties}}

\line(1,0){100}

\begin{tablenotes}\scriptsize

\item[a] Due to a loss of data we were unable to calculate the final few  thermal pulses of this model (or the yield). Here we give the number of pulses and the mass of the core at the end of the SEV code calculation.

\item[b] This model started with a pure H-He composition, with $Y=0.230$ rather than 0.245. 

\end{tablenotes}

\end{threeparttable} \end{center}

\normalsize
\end{table}

\subsubsection*{AGB Evolution of the 1 M$_{\odot}$, $\textrm{[Fe/H]}=-5.45$ Model}

\begin{figure}
\begin{centering}
\includegraphics[width=0.8\columnwidth,keepaspectratio]{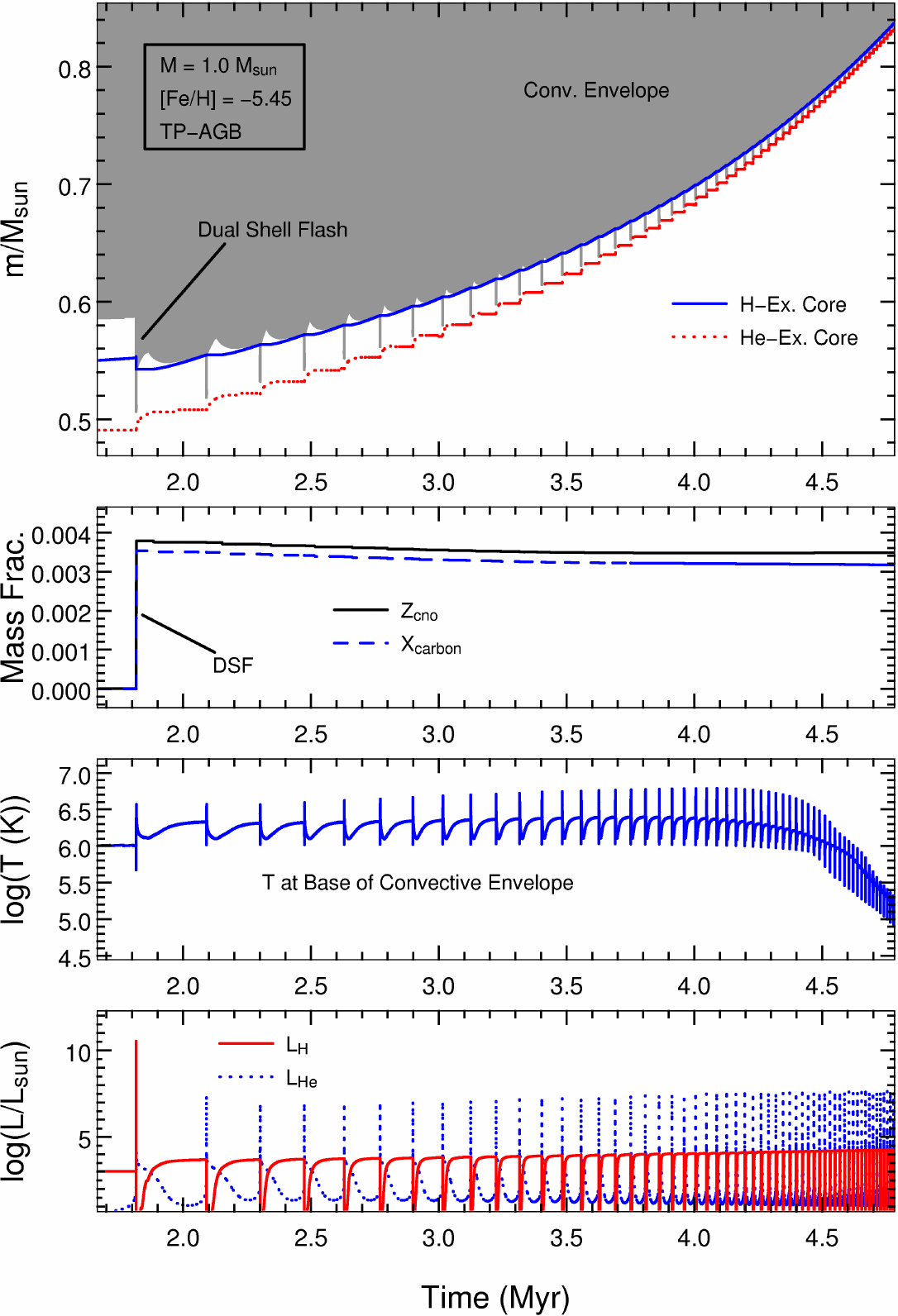}
\par\end{centering}
\caption{Most of the AGB evolution of the 1 M$_{\odot}$, {[}Fe/H{]}$=-5.45$
model. Time has been offset. We present this model as representative
of the models that have their AGB envelopes dominated by the pollution
arising from the DSF event at the start of the AGB, which thus defines
the chemical yield (see text for details). In the top panel the many
intershell convection zones can be seen (convection is in grey shading).
Indicated in the top panel is the dual shell flash that occurs near
the start of the TP-AGB. This feature can also be seen as a large
spike in H burning luminosity in the bottom panel. In the second panel
we show the evolution of the surface abundances of carbon and $Z_{cno}$.
$Z_{cno}$ is seen not to increase, indicating that no 3DUP is occurring.
More details of this model are presented in Figure \ref{fig-m1hmp-AGB-mdot}.
\label{fig-agb.m1hmp}}
\end{figure}

In Figure \ref{fig-agb.m1hmp} we show most of the AGB evolution of
this model. This model is representative of the low mass models that
have their AGB envelope compositions dominated by the pollution from
the DSF event occurring at the start of the AGB. The DSF event thus
defines the chemical makeup of the yields in these cases. For an example
of a case where the AGB composition is dominated by the DCF we refer
the reader to the detailed description of the 0.85 M$_{\odot}$, $Z=0$
model in Section \vref{section-m0.85z0-Struct-AGB}. 

\begin{figure}
\begin{centering}
\includegraphics[width=0.8\columnwidth,keepaspectratio]{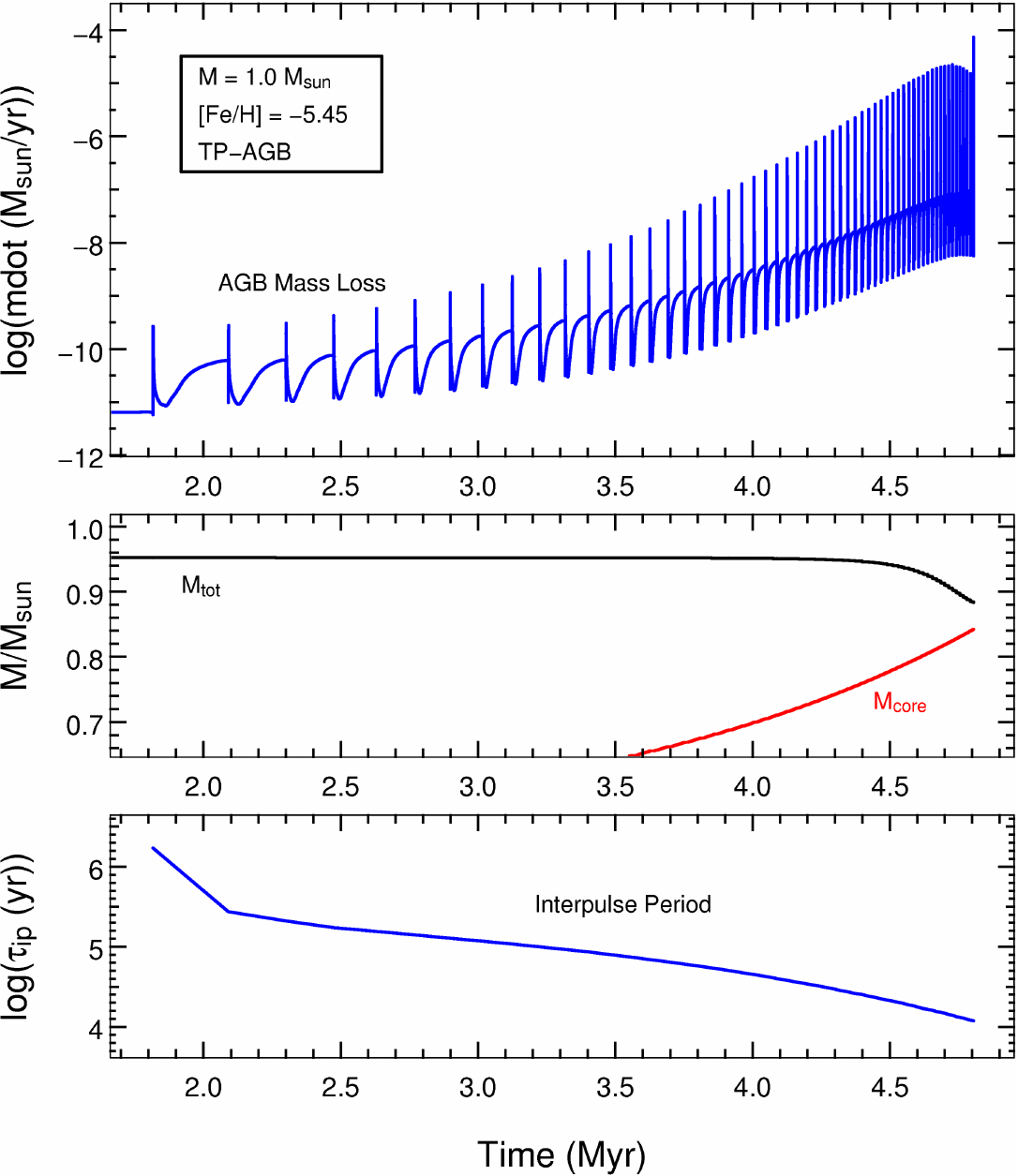}
\par\end{centering}
\caption{Some more properties of the AGB evolution of the 1 M$_{\odot}$, {[}Fe/H{]}$=-5.45$
model. It can be seen that most of the envelope mass is lost towards
the end of the AGB, and that the interpulse period decreases with
the increase of the core mass. The core mass increases at a high rate
due to the lack of 3DUP in this model. \label{fig-m1hmp-AGB-mdot}}
\end{figure}

The 1 M$_{\odot}$, $\textrm{[Fe/H] }=-5.45$ model experienced a
total of 63 thermal pulses on the AGB. However, as can be seen in
panel two of Figure \ref{fig-agb.m1hmp}, only the one associated
with the dual shell flash affected the composition of the envelope
in any substantial way, since no 3DUP episodes occurred. The temperature
at the base of the convective envelope was quite high for a star of
this mass, being $\sim10^{6.2}$ K ($1.6$ MK) through much of the
AGB. This is high enough for minor HBB, the nucleosynthesis of which
will be detailed in the next Section. The very high H burning luminosity
attained during the DSF can be seen in panel 4. Figure \ref{fig-m1hmp-AGB-mdot}
shows some more AGB properties of this model. The mass loss rate does
not acquire very high values until the end of the AGB, when the majority
of mass is lost via winds. Interestingly the yield from this star
is not affected by the mass loss rate since the envelope composition
is essentially set at the beginning of the AGB by the DSF event. The
mass loss rate does however affect the mass of the remnant, as it
puts a limit on the amount of core growth that can occur before the
entire envelope is lost. The interpulse period is seen to decrease
over the AGB, as normal. 

\subsubsection*{AGB Evolution of the 3 M$_{\odot}$, $\textrm{[Fe/H]}=-5.45$ Model}

\begin{figure}
\begin{centering}
\includegraphics[width=0.75\columnwidth,keepaspectratio]{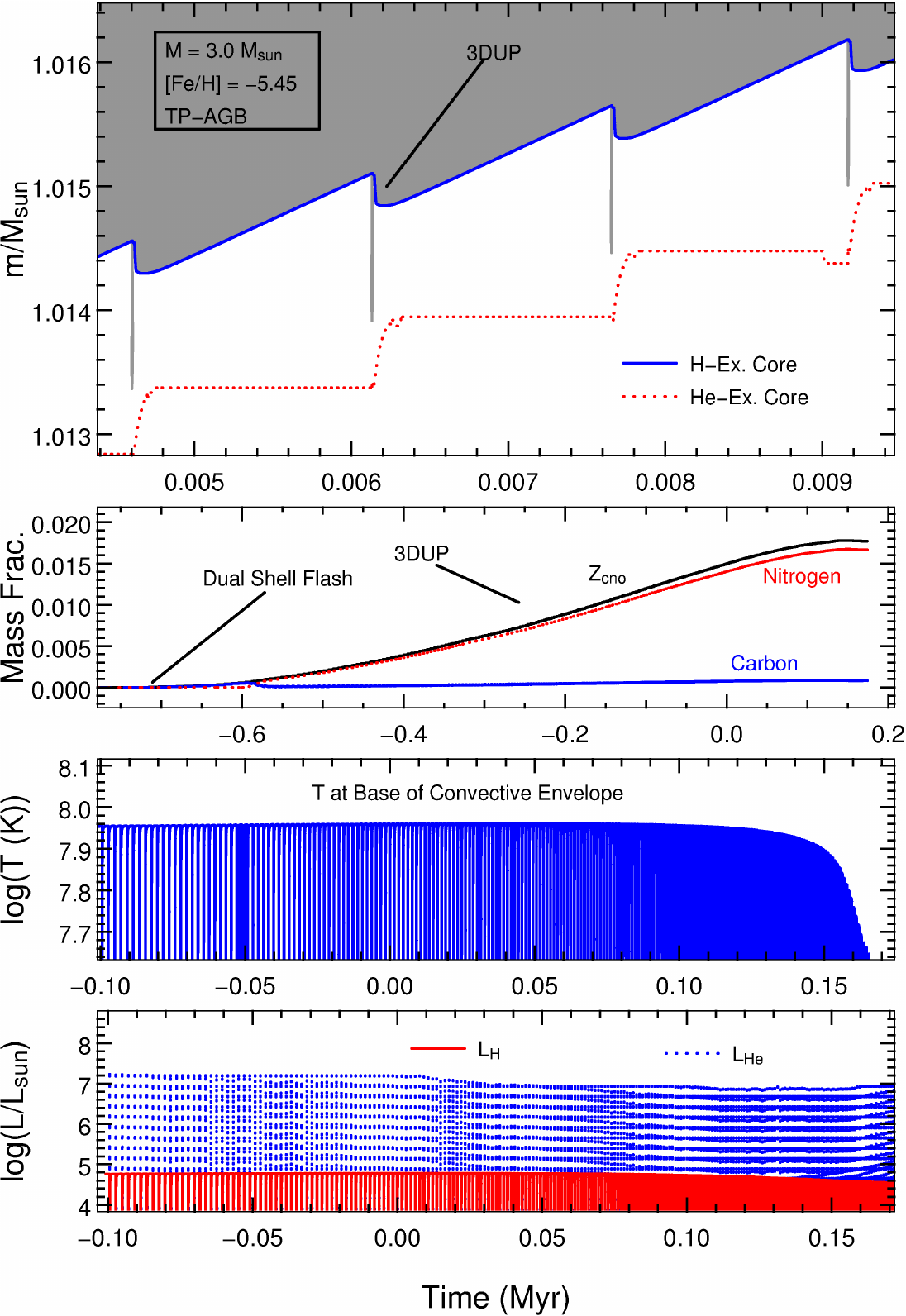}
\par\end{centering}
\caption{Some properties of the AGB evolution of the 3 M$_{\odot}$, {[}Fe/H{]}$=-5.45$
model. Time has been offset by the same amount in all panels but note
that the time span covered in each panel is different. We present
this model as representative of the models that have their AGB envelopes
dominated by the pollution arising from many 3DUP events during the
AGB, which thus defines the chemical yield (see text for details).
In the top panel a few thermal pulses can be seen (grey shading represents
convection). The degree of 3DUP can be seen here. In the second panel
we show the evolution of the surface abundances of carbon, nitrogen
and $Z_{cno}$ for the entire AGB. $Z_{cno}$ is seen to increase
by a very large amount due to 3DUP. The location of the DSF pollution
is marked, highlighting the fact that it is dwarfed by the 3DUP pollution
and thus has little effect on the yield. The composition of the AGB
envelope is always dominated by nitrogen, due to the strong HBB caused
by the high temperature at the base of the convective envelope (T$_{bce}$$\sim10^{7.95}$),
which is shown in the third panel. Note that there are so many pulses
shown in the bottom two panels ($\sim230$) that they tend to blend
together in the plot. More details of this model are presented in
Figure \ref{fig-m3hmp-AGB-mdot}. \label{fig-agb.m3hmp}}
\end{figure}

\begin{figure}
\begin{centering}
\includegraphics[width=0.8\columnwidth,keepaspectratio]{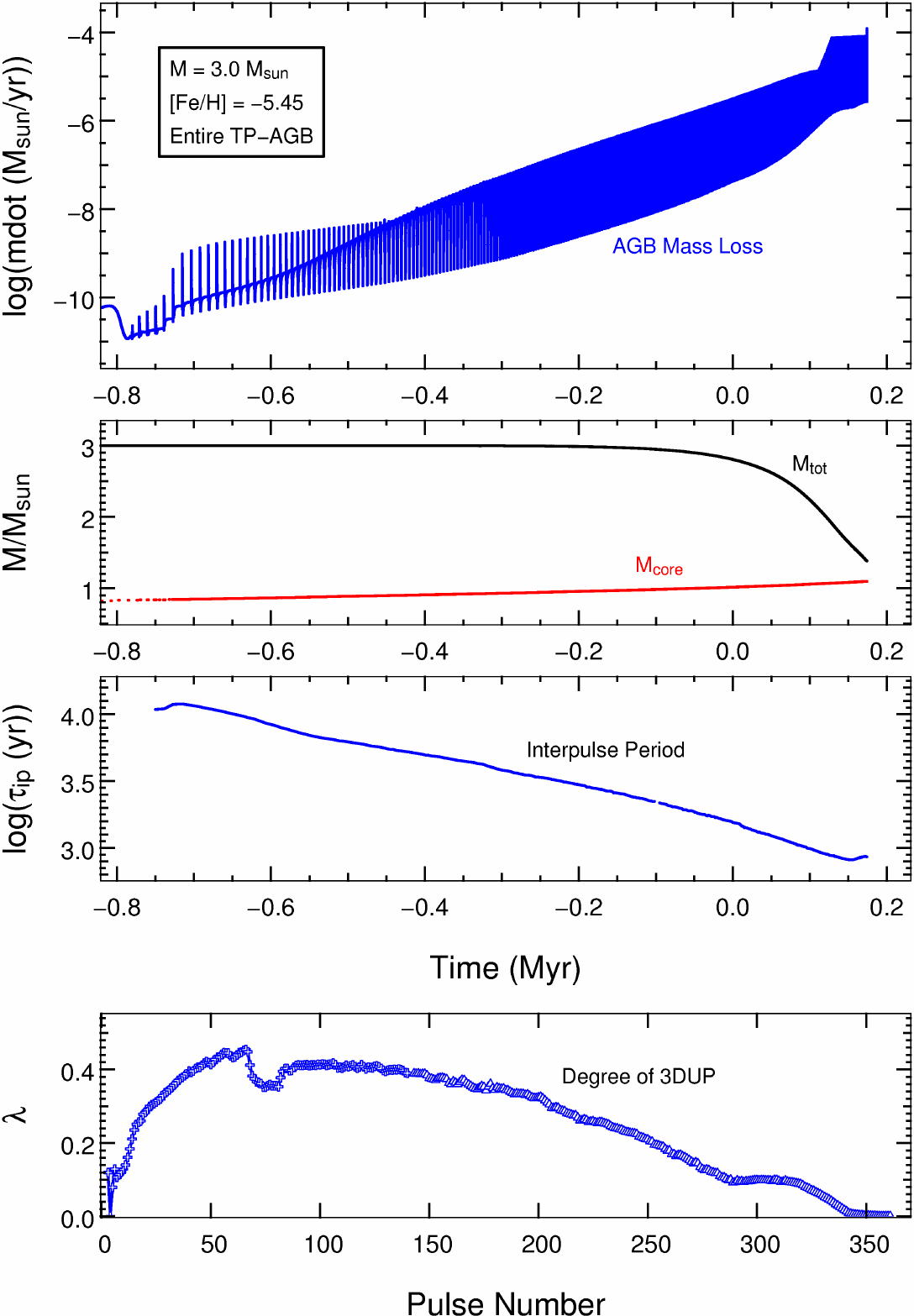}
\par\end{centering}
\caption{Some more properties of the AGB evolution of the 3 M$_{\odot}$, {[}Fe/H{]}$=-5.45$
model. Time has been offset in the top three panels. All panels cover
the entire AGB evolution. It can be seen that the mass of the star
does not change significantly until half way through the AGB. Thus
the chemical yield is dominated by the products of 3DUP and HBB. In
the bottom panel we show the variation of the dredge-up parameter
$\lambda$ with pulse number. It can be seen that the depth of dredge
up rises to a maximum early on the TP-AGB before slowly declining
to zero by the end of the evolution. \label{fig-m3hmp-AGB-mdot}}
\end{figure}

In Figures \ref{fig-agb.m3hmp} and \ref{fig-m3hmp-AGB-mdot} we show
some properties of the AGB evolution of our 3 M$_{\odot}$, $\textrm{[Fe/H] }=-5.45$
model. This model is representative of all the intermediate mass models
in our grid ($\textrm{M}=2$ and 3 M$_{\odot}$, all metallicities)
since it experiences many 3DUP episodes which define the degree of
chemical pollution on the AGB. The depth of 3DUP in this model can
be seen for a few pulses in panel 1 of Figure \ref{fig-agb.m3hmp},
having a $\lambda$ value of $\sim0.25$. This model goes through
369 pulses, and thus ends up with a highly polluted envelope, which
is characterised by $Z_{cno}\sim10^{-2}$ by the end of the evolution.
We also present the evolution of the surface abundances of C, N and
$Z_{cno}$. It can be seen that nitrogen dominates the composition
of the envelope for practically all of the AGB. This is due to strong
HBB driven by temperatures as high as $10^{7.95}$ K ($\sim90$ MK)
at the base of the convective envelope. Again, as most of the mass
loss occurs towards the end of the AGB (see Figure \ref{fig-m3hmp-AGB-mdot}),
the yield for this model will be dominated by nitrogen (after H and
He of course). The 3DUP moderates the core growth in this star, as
can be seen in Figure \ref{fig-m3hmp-AGB-mdot}. It ends up as a WD
with a mass of 1.1 M$_{\odot}$. The evolution of the depth of 3DUP
($\lambda$) is plotted in the same figure. It rises to a maximum
early on the TP-AGB before slowly declining to zero by the end of
the evolution. The average value of $\lambda$ for all the models
are given in Table \ref{table-AllHaloModelAGBProperties}. For this
model the average is 0.26. 

Finally we note that we shall discuss the nucleosynthesis of these
models in the next Section, after the subsection below which details
the various lifetimes of the models.

\subsection{Mass Loss\label{subsection-HaloStarStruct-MassLoss}}

\begin{figure}
\begin{centering}
\includegraphics[width=0.8\columnwidth,keepaspectratio]{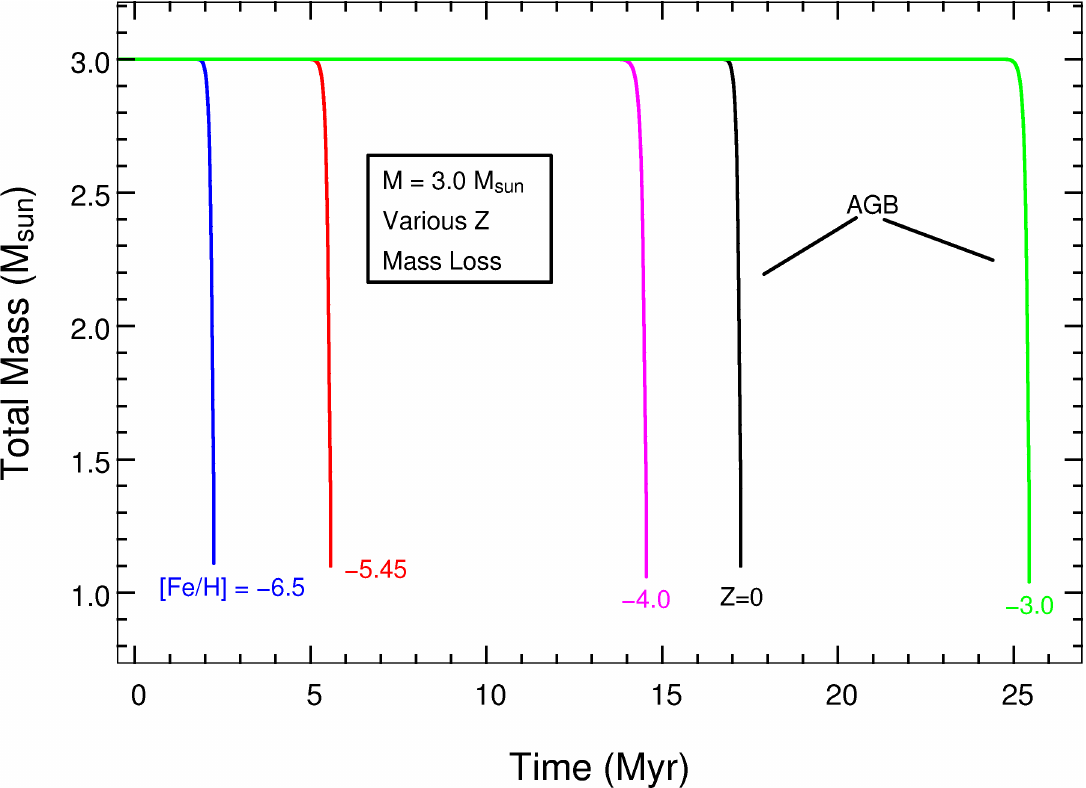}
\par\end{centering}
\caption{The evolution of total mass with time for all our 3 M$_{\odot}$ models.
Time has been offset for clarity (but by the same amount in all models).
It can be seen that all the mass loss in these models occurs on the
AGB. This is mainly due to the fact that these IM models do not go
through the RGB phase. \label{fig-massLoss-allm3}}
\end{figure}

The mass loss formalisms used in the SEV code for these models was
briefly discussed in the introduction to this chapter (Section \vref{subsection-HaloStarCodeInputs}).
The reduction of the total stellar mass of a couple of the models
was touched on in the preceding subsections. In particular it was
noted that the most rapid mass loss occurs at the end of the AGB phase,
when the luminosities are highest. Indeed, in panel 2 of Figure \ref{fig-m3hmp-AGB-mdot}
it can be seen that essentially all of the mass loss in the $\textrm{[Fe/H]}=-5.45,$
3 M$_{\odot}$ occurs at the end of the AGB. This is mainly due to
the fact that it does not make it to the RGB phase. In Figure \ref{fig-massLoss-allm3}
we show the evolution of the total mass for all of the 3 M$_{\odot}$
models. It can be seen that all these IM mass models behave in the
same way with regards to mass loss, i.e. all the mass lost from the
stars occurs during the AGB phase. Thus in these models the composition
of the AGB envelope wholly determines the yields. As mentioned in
the introduction to the current chapter the mass loss formula used
for the AGB (\citealt{1993ApJ...413..641V}) is expected to be valid
in these (initially) low-Z models since their envelopes are heavily
enriched by the DSF and 3DUP episodes, especially as the AGB evolution
progresses and the mass loss increases. These models are really quite
similar to metal-rich models once on the AGB, as they have metal-rich
envelopes and CO cores.

\begin{figure}
\begin{centering}
\includegraphics[width=0.8\columnwidth,keepaspectratio]{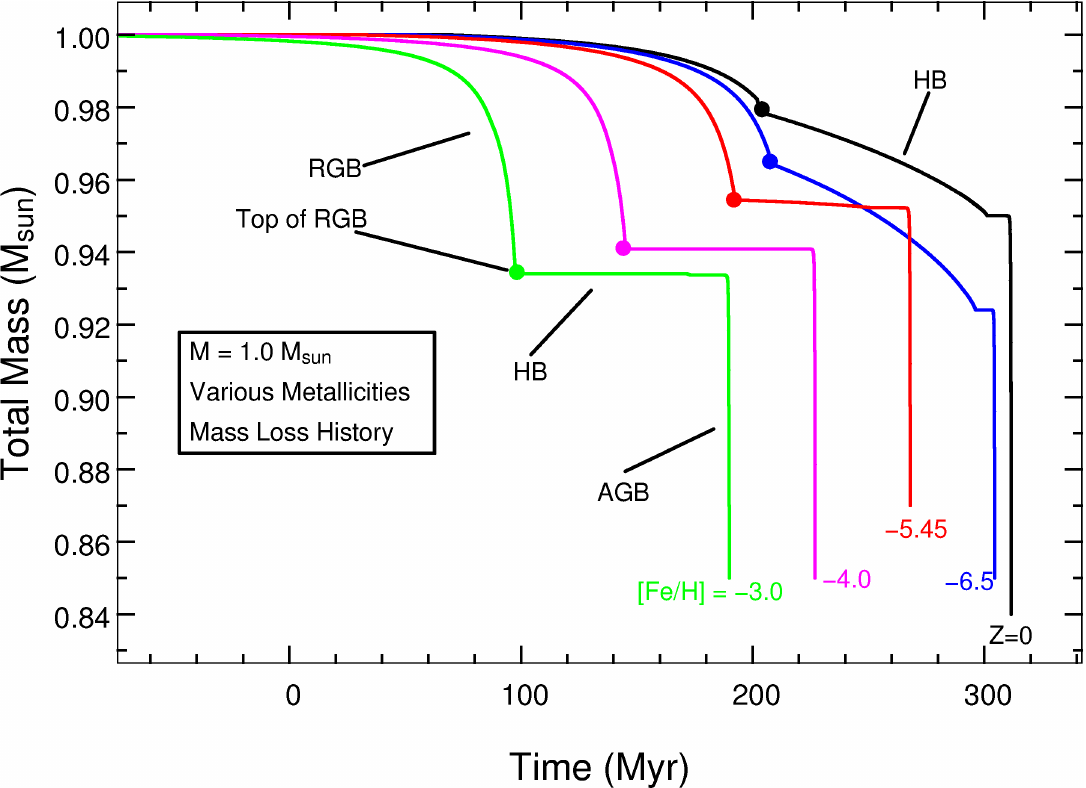}
\par\end{centering}
\caption{The evolution of total mass with time for all our 1 M$_{\odot}$ models.
Time has been offset for clarity (but by the same amount in all models).
It can be seen that substantial mass loss occurs during a few different
stages of evolution. In the higher metallicity models the mass loss
is split 40:60 on the RGB and AGB respectively, whilst in the most
metal-poor models it is split roughly 20:30:50 between the RGB, HB
and AGB. The increased mass loss on the HB in the lowest metallicity
models is due to the pollution arising from the DCF event that terminates
the RGB. After this event their envelopes have a metallicity at least
$\frac{1}{10}$ solar, whereas the (initially) higher metallicity
models (with no DCF) still have very metal poor envelopes on the HB.\label{fig-massLoss-allm1}}
\end{figure}

The situation is more complex in the low mass models. In Figure \ref{fig-massLoss-allm1}
we plot the evolution of total mass for all our 1 M$_{\odot}$ models.
In the figure we point out the various stages of evolution during
which mass is lost. A significant fraction of the mass is lost during
the RGB phase in these models. The situation is complicated by the
fact that the stars turn off the RGB at earlier and earlier stages
as the metallicity decreases. This was discussed above, in Section
\ref{section-HaloStars-Struct-MSandRGB}, where this low-Z characteristic
was shown in the HR diagram. As the most rapid stage of mass loss
is at the top of the RGB this has a strong effect on the mass of the
star at the beginning of core He burning. This is clearly seen in
Figure \ref{fig-massLoss-allm1}, where, for example, the $\textrm{[Fe/H]}=-3.0$
model loses $\sim0.06$ M$_{\odot}$ of envelope material on the RGB
whilst the $Z=0$ model only loses $\sim0.02$ M$_{\odot}$. In the
more metal-rich models the mass loss on the RGB represents $\sim40\%$
of the total mass loss from the star (the rest is lost on the AGB).
This lowers to $\sim10\rightarrow20\%$ at extremely low metallicity.
The more mass loss there is on the RGB the more uncertain the further
evolution and yields are. This is because the RGB composition is unpolluted,
and thus of very low metallicity, so the mass loss rate is uncertain
(see the discussion in the introduction: Section \vref{subsection-HaloStarCodeInputs}).
Since the RGB mass loss is greater in the more metal-rich low-mass
models, this is where the most uncertainty lies. 

At the lowest metallicities a second effect comes into play -- there
is a significant amount of mass loss on the horizontal branch (HB,
core He burning phase). In fact, in the $Z=0$ and $\textrm{[Fe/H]}=-6.5$
models (1 M$_{\odot}$) there is more mass lost during this phase
than is lost on the RGB ($50\%$ more in the $Z=0$ case). However
the dominant source of mass loss is the AGB in these models. In the
$Z=0$ model, $\sim70\%$ of the mass is lost during the AGB ($\sim50\%$
in the $\textrm{[Fe/H]}=-6.5$ model). In the more metal-rich models
the AGB mass loss represents $\sim60\%$ of the total. 

\begin{figure}
\begin{centering}
\includegraphics[width=0.8\columnwidth,keepaspectratio]{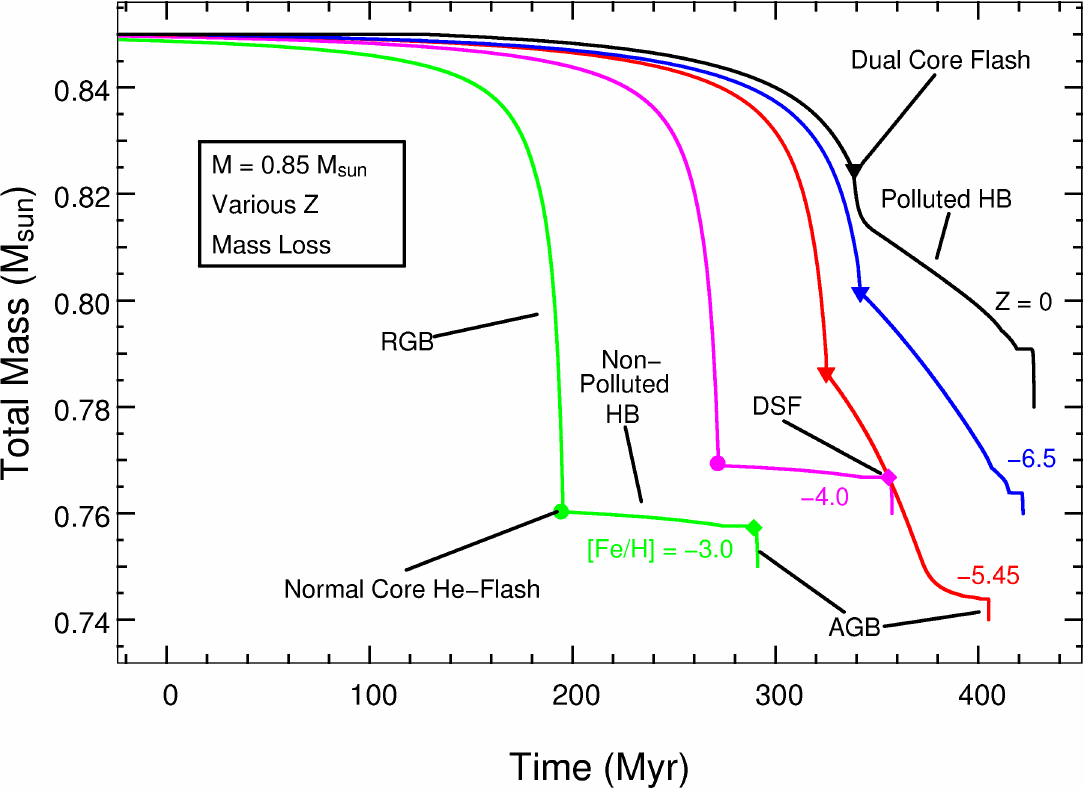}
\par\end{centering}
\caption{The evolution of total mass with time for all our 0.85 M$_{\odot}$
models. Time has been offset for clarity (but by the same amount in
all models). It can be seen that substantial mass loss occurs during
a few different stages of evolution. In the higher metallicity models
the mass loss is split about 90:10 on the RGB and AGB respectively,
whilst in the most metal-poor models it is split roughly 55:40:5 between
the RGB, HB and AGB. The increased mass loss on the HB in the lowest
metallicity models is due to the pollution arising from the DCF event
that terminates the RGB (marked by triangles). After this event their
envelopes have a metallicity at least $\frac{1}{10}$ solar, whereas
the (initially) higher metallicity models (with no DCF) still have
very metal poor envelopes on the HB. Thus the initially most metal-poor
models are actually the metal-rich models during this phase. The DSFs
in the higher metallicity models are marked by diamonds, and normal
core He flashes by small filled circles. \label{fig-massLoss-allm8}}
\end{figure}

In summary, in the higher metallicity 1 M$_{\odot}$ models the mass
loss is generally split 40:60 on the RGB and AGB respectively, whilst
in the most metal-poor models it is split 20:30:50 between the RGB,
HB and AGB (roughly). The splits are more severe at lower mass, where,
for example, $90\%$ of the total mass loss in the 0.85 M$_{\odot},$
$\textrm{[Fe/H]}=-3.0$ model occurs on the RGB (see Figure \ref{fig-massLoss-allm8}).
As the mass loss occurs at different stages of evolution the yields
will reflect a mix of the surface compositions from these stages.
In particular, before the dual core flashes occur in the low metallicity
models, and before the dual shell flashes occur in the higher metallicity
models, the surface compositions reflect the initial, metal-poor compositions.
Thus the mass lost on the RGB will have a diluting effect on the metal-rich
yield of the AGB. 

The mass loss on the HB in the extremely metal-deficient models is
an interesting feature. As shown above in Figure \vref{fig-hrds.m1all.HB},
the HBs of the $Z=0$ and $\textrm{[Fe/H]}=-6.5$ models are substantially
more luminous and very much redder than those of the other models.
We suggest that this is an effect arising from the DCF pollution that
occurred in these models just before their descent onto the HB. This
event severely pollutes the envelopes, such that they are, from then
on, much more like metal-rich models. Indeed, the change in HB morphology
is much like that seen in Galactic globular clusters, where the more
metal-rich populations show red HBs and the more metal-poor populations
show blue HBs. It appears that this, combined with their higher luminosities,
leads to a high mass-loss rate, as can be seen in the most metal-poor
HB phases in Figures \ref{fig-massLoss-allm1} and \ref{fig-massLoss-allm8}.
Indeed, the mass lost on the HB in the more metal-`rich' models is
much lower. Ironically this is due to the fact that by this stage
the more metal-`rich' models are actually metal-poor compared to the
models which have experienced DCF pollution (the initially metal-poor
models). For example, on the HB $Z_{cno}$ is $10^{-2}$ in the $\textrm{[Fe/H]}=-6.5$
model, whilst it remains at the initial value of $10^{-6}$ in the
$\textrm{[Fe/H]}=-4.0$ model. It is interesting to see this metallicity
effect on mass loss despite the fact that we have not explicitly included
any metallicity dependence in the mass loss formulae.

To end this subsection we reiterate/clarify the main finding in regards
to uncertainties in the modelling arising from mass-loss, as it is
an important caveat to keep in mind. The main finding was that the
mass loss from the RGB in some of the low-mass models -- in particular
the lowest mass and highest metallicity models -- makes up a significant
proportion of the total mass loss. This causes a problem because the
envelope compositions of these models are still very metal-poor during
the RGB phase, and, as the mass-loss rate is uncertain at these metallicities,
an uncertainty arises in the evolution. This naturally has a strong
impact on the yields of these models, effectively diluting the yield
from the AGB. If, in reality, the RGB mass loss should be much lower
then our yields will be wrong -- they should be more metal-rich.
The RGB evolutionary paths of the models are also rendered uncertain
as the rate of core growth would be different with a lower mass loss
rate. This then has implications for the HB and RGB evolution. Thus
many uncertainties arise through the unknown nature of mass-loss rates
at low metallicity. However we note that the metallicity of a stellar
model is, to some degree, indirectly taken into account in the mass
loss formulae -- as evidenced by the varying amounts of mass lost
during the HB phases of the models discussed above.

\subsection{Table of Lifetimes}

In Table \ref{table-AllHaloModelLifetimes} we present the lifetimes
of various stages of evolution for the whole grid of models. The MS
lifetimes of the low-mass stars is essentially constant across this
metallicity range, being $\sim10.1$ Gyr. In the intermediate mass
models ($\textrm{M}=2$ and 3 M$_{\odot}$) we find a variation of
$\sim20\%$ at each mass. In general the time spent on the MS increases
with the decrease in metallicity. This is due to the increased dominance
of p-p chain energy generation in stars that would normally be powered
primarily by the CNO cycles. The lifetimes of these stars are much
shorter than those of the low-mass stars (as normal), being $\sim600$
Myr at 2 M$_{\odot}$ and $\sim180$ Myr at 3 M$_{\odot}$. On the
subgiant branch (SGB), which we define as the time between the MSTO
and the bottom of the RGB (or He burning ignition in the more massive
models), there is a moderate variation in lifetime in the low mass
stars, generally seen as a decrease with an increase in metallicity.
The intermediate mass models however show a very strong variation
with metallicity. This is due to the fact that stars at these masses
turn off the HG earlier and earlier (most do not reach the RGB at
all, as mentioned earlier), as the metallicity decreases. The most
extreme of these is the 3 M$_{\odot}$, $Z=0$ model which only spends
8 Myr evolving towards the red before igniting He in its core. This
early turning off the HG also manifests itself in the RGB lifetimes
which, naturally, are zero for most of the IM models. only the two
most metal-rich 2 M$_{\odot}$ models actually spend time on the RGB
-- but only a few Myr. On the other hand all of the low mass models
complete the full evolution of the RGB, igniting He at the top of
the RGB. In an opposite trend to their SGB lifetimes the low-mass
RGB lifetimes \emph{decrease} with metallicity. At 0.85 M$_{\odot}$
the lifetime ranges from 200 Myr at $Z=0$ to 250 Myr at $\textrm{[Fe/H]}=-3$.
At 1 M$_{\odot}$ the lifetime ranges from 110 Myr at $Z=0$ to 155
Myr at $\textrm{[Fe/H]}=-3$. This reflects the early termination
of the RGB in the low-Z models due to earlier and earlier He ignition,
as reported in an earlier section. Horizontal branch (HB) lifetimes
in the low mass models appear to first decrease then increase with
metallicity. The picture is clearer in the IM models, where a clear
decrease of HB lifetime with metallicity can be seen. In the 2 M$_{\odot}$
models it ranges from 95 Myr at $\textrm{[Fe/H]}=-3$ down to a short
30 Myr at $Z=0$. In the 3 M$_{\odot}$ models it ranges from 47 Myr
at $\textrm{[Fe/H]}=-3$ down to a very short 12 Myr at $Z=0$. AGB
lifetimes are generally seen to decrease with metallicity. There is
not much variation at in the IM models, with AGB lifetimes being $\sim2$
Myr at 2 M$_{\odot}$ and $\sim1$ Myr at 3 M$_{\odot}$, but a strong
variation is seen at low mass. At 1 M$_{\odot}$ the lifetimes range
from 4.4 Myr at $\textrm{[Fe/H]}=-3$ down to 1.7 Myr at $Z=0$ (although
the $\textrm{[Fe/H]}=-6.5$ model lifetime is actually shorter). The
picture is similar at 0.85 M$_{\odot}$ except that there is a drastic
drop from 3.8 Myr at $\textrm{[Fe/H]}=-4$ to 0.05 Myr at $\textrm{[Fe/H]}=-5.45$,
before an increase at lower Z. We are unsure what are the driving
factors in this case, and leave the investigation of this interesting
variation in lifetimes for future work. 

\begin{table}
\begin{center}\begin{threeparttable}\centering

\begin{tabular}{cccccccc}
\multicolumn{8}{c}{Mass $=0.85$ M$_{\odot}$}\tabularnewline
\hline 
\hline 
{[}Fe/H{]} & $L_{tip}$ & MS (Gyr) & SGB & RGB & CHeB & AGB & Total (Gyr)\tabularnewline
\hline 
-3.00 & 3.1 & 10.0 & 740 & 250 & 85 & 4.00 & 11.09\tabularnewline
-4.00 & 3.0 & 10.1 & 750 & 220 & 75 & 3.80 & 11.16\tabularnewline
-5.45 & 2.8 & 10.1 & 820 & 205 & 75 & 0.05 & 11.21\tabularnewline
-6.50 & 2.6 & 10.1 & 840 & 200 & 75 & 0.60 & 11.22\tabularnewline
$Z=0$ & 2.3 & 10.1 & 840 & 200 & 80 & 1.50 & 11.23\tabularnewline
\hline 
 &  &  &  &  &  &  & \tabularnewline
\multicolumn{8}{c}{Mass $=1.0$ M$_{\odot}$}\tabularnewline
\hline 
\hline 
{[}Fe/H{]} & $L_{tip}$ & MS (Gyr) & SGB & RGB & CHeB & AGB & Total (Gyr)\tabularnewline
\hline 
-3.00 & 3.1 & 5.58 & 470 & 155 & 74 & 4.35 & 6.29\tabularnewline
-4.00 & 3.0 & 5.73 & 360 & 150 & 61 & 4.34 & 6.33\tabularnewline
-5.45 & 2.8 & 5.73 & 410 & 131 & 57 & 3.03 & 6.37\tabularnewline
-6.50 & 2.6 & 5.73 & 450 & 127 & 86 & 1.05 & 6.40\tabularnewline
$Z=0$ & 2.3 & 5.70 & 480 & 110 & 96 & 1.74 & 6.41\tabularnewline
\hline 
 &  &  &  &  &  &  & \tabularnewline
\multicolumn{8}{c}{Mass $=2.0$ M$_{\odot}$}\tabularnewline
\hline 
\hline 
{[}Fe/H{]} & $L_{tip}$ & MS & SGB & RGB & CHeB & AGB & Total\tabularnewline
\hline 
-3.00 & 2.2 & 492 & 140 & 5.7 & 95 & 2.50 & 745\tabularnewline
-4.00 & 2.1 & 562 & 84 & 1.5 & 72 & 2.30 & 733\tabularnewline
-5.45 & -- & 605 & 58 & zero & 50 & 2.27 & 722\tabularnewline
-6.50 & -- & 625 & 43 & zero & 40 & 2.10 & 718\tabularnewline
$Z=0$ & -- & 660 & 12 & zero & 30 & 2.02 & 713\tabularnewline
\hline 
 &  &  &  &  &  &  & \tabularnewline
\multicolumn{8}{c}{Mass $=3.0$ M$_{\odot}$}\tabularnewline
\hline 
\hline 
{[}Fe/H{]} & $L_{tip}$ & MS & SGB & RGB & CHeB & AGB & Total\tabularnewline
\hline 
-3.00 & -- & 185 & 18 & zero & 47 & 1.24 & 255\tabularnewline
-4.00 & -- & 174 & 27 & zero & 38 & 1.19 & 245\tabularnewline
-5.45 & -- & 177 & 30 & zero & 24 & 0.96 & 236\tabularnewline
-6.50 & -- & 189 & 21 & zero & 17 & 1.04 & 232\tabularnewline
$Z=0$\tnote{a} & -- & 222 & 8 & zero & 12 & 1.16 & 247\tabularnewline
\hline 
\end{tabular}

\caption{Various lifetimes for all the low metallicity models, grouped in terms
of initial mass. We redisplay the lifetimes for the $Z=0$ models
for ease of comparison. Also included, in the second column, are the
luminosities at the tip of the RGB ($L_{tip}$) -- if the model made
it to the RGB. All ages are in Myr unless otherwise stated. Metallcity
is given as {[}Fe/H{]}, except for $Z=0$. Lifetimes are given for:
MS (ZAMS to MS turn-off), SGB (MS turn-off to base of RGB, or to the
start of core He burning in the more massive models), RGB (red giant
branch), CHeB (core helium burning), AGB (thermally-pulsing AGB) and
Total (the entire lifetime of the star, from ZAMS to the end of the
TP-AGB). \label{table-AllHaloModelLifetimes}}

\line(1,0){100}

\begin{tablenotes}\scriptsize

\item[a] This model started with a pure H-He composition, with $Y=0.230$ rather than 0.245. 

\end{tablenotes}

\end{threeparttable} \end{center}
\end{table}

\section{Nucleosynthetic Evolution\label{section-HaloStars-Nucleo}}

\subsection{Introduction}

In this section we give an overview of the nucleosynthetic evolution
of all our metal-deficient models. We have also included the $Z=0$
models of the previous chapters in the discussion, as well as in the
tables, for comparison. We discuss all of the key evolutionary events
in separate subsections. Within those subsections we limit the discussion
to some key nucleosynthetic signatures, due to time and space constraints.
However, for the interested reader, we present detailed yield plots
and tables in the next section and also in the appendices, from which
much detailed information can be gleaned. We also give an overall
summary of our results, and compare them to previous studies, at the
end of this chapter (Section \vref{section-SummaryAndCompare-HaloMods}).

A finding from the structural evolution section is relevant for the
discussion of the nucleosynthesis in these models. In subsection \vref{subsection-AGB-Struct-HaloStars}
we showed that our models can be divided into three distinct groups
based on the source of the main contribution to AGB envelope pollution.
These groups are:
\begin{enumerate}
\item DCF Dominated: Low-Mass \& \emph{Extremely} Metal-Deficient Models
\item DSF Dominated: Low-Mass \& Very Metal-Deficient Models
\item 3DUP Dominated: Intermediate-Mass -- All Metallicities
\end{enumerate}
This grouping gives a good overview of the competing processes that
pollute the envelopes of these metal-deficient models. Below we detail
the nucleosynthetic evolution arising from these major events, as
well as from other more minor events. Although we discuss all the
models in the grid (20 in total) we give detailed figures of only
two, in order to illustrate the various evolutionary events. The models
we display both have a metallicity of $\textrm{[Fe/H]}=-5.45$, and
masses of 1.0 and 3.0 M$_{\odot}$. These cover two of the groups
outlined above (groups 2 and 3). We note that this metallicity corresponds
to the most metal poor star known to date (HE1327-2326, \citealt{2005Natur.434..871F})
-- at least in terms of {[}Fe/H{]}. For an example of a group 1 model
we refer the reader to the detailed chapters on the 0.85 M$_{\odot}$,
$Z=0$ model (Chapters \ref{chapter-Z0-StructEvoln} and \ref{Chapter-ZeroZ-Nucleo}).
We do however provide tables containing key surface composition quantities
for the whole grid of models -- Tables \ref{table-HaloStars-Nucleo-YandZ}
and \ref{table-HaloStars-Nucleo-CandN}. These complement Tables \ref{table-AllHaloModelDualFlashProperties},
\ref{table-AllHaloModelAGBProperties} and \ref{table-AllHaloModelLifetimes}
in the previous (structural evolution) section. As mentioned above
we present the yields in the next section (on page \pageref{section-Yields-HaloStars}).

\subsection{First Dredge-Up}

Apart from the two most metal-rich 2 M$_{\odot}$ models, it is only
the low mass models that make it to the RGB to experience first dredge-up
(FDUP). Thus most of the IM models retain their initial surface composition
until second dredge-up. The models that do experience FDUP all show
only minor enhancements in helium. For example, in the 1 M$_{\odot}$,
$\textrm{[Fe/H]}=-5.45$ model the He abundance at the surface only
changes from $Y=0.245$ to 0.249. This universal trait is seen in
column 3 of Table \ref{table-HaloStars-Nucleo-YandZ}. 

In Table \ref{table-HaloStars-Nucleo-CandN} we give the $^{12}$C/$^{13}$C
ratio after FDUP. We note that the initial value of this ratio is
very high in all the models, due to the non-scaled-solar composition.
For example, in the $\textrm{[Fe/H]}=-5.45$ models it is $\approx10^{7}$.
In some of the models the effect of FDUP on this ratio is very strong,
reducing it to $\sim10$. In some models, particularly the lowest
metallicity models, FDUP is less efficient, and the ratio remains
super-solar after FDUP. The $Z=0$ models are a distinct case, as
there is no CN cycling above the main H burning region due to a lack
of catalysts. As mentioned in Section \vref{subsection-m0.85z0-Nucleo-FDUP}
(in regards to the 0.85 M$_{\odot},$ $Z=0$ model) the main species
to be affected by the FDUP at $Z=0$ is $^{3}$He, as it is produced
far above the main H burning shell in p-p reactions. Also due to the
lack of CN-cycling the post-FDUP C/N ratio (given in Table \ref{table-HaloStars-Nucleo-CandN})
remains at zero (or $\infty$) in the $Z=0$ models. In the metal-deficient
models the C/N ratio shows a similar variation as the $^{12}$C/$^{13}$C
ratio does, again reflecting the degree of FDUP. The C/N ratio of
the initial composition of these models is also very high, being $\sim10^{6}$
in the $\textrm{[Fe/H]}=-5.45$ models. This is due to the paucity
of N in the 20 M$_{\odot},$ $Z=0$ supernova yield. Thus it can be
seen in Table \ref{table-HaloStars-Nucleo-CandN} that the FDUP operates
very effectively in the higher metallicity models, reducing the ratio
to between $\sim5$ and $\sim180$ by mixing up N-enhanced material
left over from CN cycling in the outer reaches of the H-burning shell.
However in the more metal-poor models it can be seen that the ratio
remains very high. This is mainly a reflection of the depth of FDUP.
For instance, in the 0.85 M$_{\odot},$ $\textrm{[Fe/H]}=-3$ model
the maximum depth of FDUP was 0.4 M$_{\odot}$ whilst it was only
0.52 M$_{\odot}$ in the 0.85 M$_{\odot},$ $\textrm{[Fe/H]}=-5.45$
model. It also reflects the reduced efficiency of the CN cycle at
very low Z. Indeed, in column 7 of Table \ref{table-HaloStars-Nucleo-CandN}
a strong correlation between FDUP efficiency and metallicity can be
seen. 

The metallicity $Z$ of the models does not change due to FDUP as
the `polluted' material dredged up has only undergone CN and/or p-p
burning, so the catalysts have just been rearranged (ie. C$\rightarrow$N)
without producing any (significant) additional number of heavy nuclei.
This is why we do not display the $Z_{cno}$ value for FDUP in Table
\ref{table-HaloStars-Nucleo-YandZ}.

\begin{table}
\begin{center}\begin{threeparttable}\centering

\begin{tabular}{ccccccccc}
\multicolumn{9}{c}{Mass $=0.85$ M$_{\odot}$}\tabularnewline
\hline 
\hline 
{[}Fe/H{]} & DF & Y$_{fdu}$ & Y$_{sdu}$ & Y$_{DF}$ & Y$_{agb}$ & Z$_{i}$ & Z$_{DF}$ & Z$_{f}$\tabularnewline
\hline 
-3.00 & \textcolor{blue}{Shell} & 0.25 & 0.26 & 0.28 & 0.28 & 1E-05 & 1E-02 & 1E-02\tabularnewline
-4.00 & \textcolor{blue}{Shell} & 0.25 & 0.25 & 0.27 & 0.27 & 1E-06 & 1E-03 & 1E-03\tabularnewline
-5.45 & \textcolor{red}{Core} & 0.25 & 0.44 & 0.44 & 0.44 & 4E-08 & 3E-02 & 3E-02\tabularnewline
-6.50 & \textcolor{red}{Core} & 0.25 & 0.31 & 0.31 & 0.31 & 4E-09 & 6E-02 & 6E-02\tabularnewline
$Z=0$ & \textcolor{red}{Core} & 0.25 & 0.33 & 0.33 & 0.33 & zero & 1E-03 & 1E-03\tabularnewline
\hline 
 &  &  &  &  &  &  &  & \tabularnewline
\multicolumn{9}{c}{Mass $=1.0$ M$_{\odot}$}\tabularnewline
\hline 
\hline 
{[}Fe/H{]} & DF & Y$_{fdu}$ & Y$_{sdu}$ & Y$_{DF}$ & Y$_{agb}$ & Z$_{i}$ & Z$_{DF}$ & Z$_{f}$\tabularnewline
\hline 
-3.00 & \textcolor{blue}{Shell} & 0.26 & 0.27 & 0.28 & 0.28 & 1E-05 & 5E-03 & 5E-03\tabularnewline
-4.00 & \textcolor{blue}{Shell} & 0.25 & 0.27 & 0.28 & 0.28 & 1E-06 & 1E-02 & 1E-02\tabularnewline
-5.45 & \textcolor{blue}{Shell} & 0.25 & 0.26 & 0.28 & 0.28 & 4E-08 & 4E-03 & 3E-03\tabularnewline
-6.50 & \textcolor{red}{Core} & 0.25 & 0.33 & 0.33 & 0.33 & 4E-09 & 7E-02 & 6E-02\tabularnewline
$Z=0$ & \textcolor{red}{Core} & 0.25 & 0.34 & 0.34 & 0.34 & zero & 1E-02 & 1E-02\tabularnewline
\hline 
 &  &  &  &  &  &  &  & \tabularnewline
\multicolumn{9}{c}{Mass $=2.0$ M$_{\odot}$}\tabularnewline
\hline 
\hline 
{[}Fe/H{]} & DF & Y$_{fdu}$ & Y$_{sdu}$ & Y$_{DF}$ & Y$_{agb}$ & Z$_{i}$ & Z$_{DF}$ & Z$_{f}$\tabularnewline
\hline 
-3.00 & \textcolor{darkgreen}{None}

 & 0.25 & 0.26 & -- & 0.31 & 1E-05 & -- & 1E-02\tabularnewline
-4.00 & \textcolor{blue}{Shell} & 0.25 & 0.27 & 0.27 & 0.30 & 1E-06 & 6E-04 & 5E-03\tabularnewline
-5.45 & \textcolor{blue}{Shell} & -- & 0.29 & 0.29 & 0.34 & 4E-08 & 2E-05 & 1E-02\tabularnewline
-6.50 & \textcolor{blue}{Shell} & -- & 0.30 & 0.30 & 0.34 & 4E-09 & 7E-04 & 8E-03\tabularnewline
$Z=0$ & \textcolor{blue}{Shell} & -- & 0.31 & 0.31 & 0.34 & zero & 2E-04 & 4E-03\tabularnewline
\hline 
 &  &  &  &  &  &  &  & \tabularnewline
\multicolumn{9}{c}{Mass $=3.0$ M$_{\odot}$}\tabularnewline
\hline 
\hline 
{[}Fe/H{]} & DF & Y$_{fdu}$ & Y$_{sdu}$ & Y$_{DF}$ & Y$_{agb}$ & Z$_{i}$ & Z$_{DF}$ & Z$_{f}$\tabularnewline
\hline 
-3.00 & \textcolor{darkgreen}{None}

 & -- & 0.26 & -- & 0.36 & 1E-05 & -- & 3E-02\tabularnewline
-4.00\tnote{a} & \textcolor{darkgreen}{None}

 & -- & n/a & -- & 0.40 & 1E-06 & -- & 4E-02\tabularnewline
-5.45 & \textcolor{blue}{Shell} & -- & 0.30 & 0.30 & 0.39 & 4E-08 & 4E-08 & 2E-02\tabularnewline
-6.50 & \textcolor{blue}{Shell} & -- & 0.32 & 0.32 & 0.39 & 4E-09 & 8E-06 & 1E-02\tabularnewline
$Z=0$\tnote{b} & \textcolor{blue}{Shell} & -- & 0.33 & 0.33 & 0.41 & zero & 8E-06 & 1E-02\tabularnewline
\hline 
\end{tabular}

\caption{Various surface abundance quantities at different stages of evolution.
Values for all the low metallicity models are displayed, grouped in
terms of initial mass. We also show the $Z=0$ models for comparison.
Initial metallcity (first column) is given as {[}Fe/H{]}, except for
$Z=0$. The second column specifies whether the model experienced
a dual flash (DF) or not, be it a DCF or DSF. The abundances given
are: Y$_{fdu}$ (the mass fraction of He after the first dredge up
episode), Y$_{sdu}$ (after 2DUP), Y$_{DF}$ (after a dual flash),
Y$_{agb}$(at the end of the AGB), Z$_{i}$ (the mass fraction of
CNO nuclei initially present), Z$_{DF}$ (after a dual flash), Z$_{f}$
(at the end of the evolution/AGB). Most of the intermediate mass models
did not reach the RGB so Y$_{fdu}$ is blank in these cases. \label{table-HaloStars-Nucleo-YandZ}}

\line(1,0){100}

\begin{tablenotes}\scriptsize

\item[a] Due to a loss of data we are unable to provide the Y$_{sdu}$ value.

\item[b] This model started with a pure H-He composition, with $Y=0.230$ rather than 0.245. 

\end{tablenotes}

\end{threeparttable} \end{center}
\end{table}

\begin{table}
\begin{center}\begin{threeparttable}\centering

\footnotesize

\begin{tabular}{cccccccccc}
\multicolumn{10}{c}{Mass $=0.85$ M$_{\odot}$}\tabularnewline
\hline 
\hline 
{[}Fe/H{]} & DF & $\frac{^{12}\textrm{C}}{^{13}\textrm{C}}$$_{fd}$ & $\frac{^{12}\textrm{C}}{^{13}\textrm{C}}$$_{sd}$ & $\frac{^{12}\textrm{C}}{^{13}\textrm{C}}$$_{DF}$ & $\frac{^{12}\textrm{C}}{^{13}\textrm{C}}$$_{agb}$ & $\frac{\textrm{C}}{\textrm{N}}$$_{fd}$ & $\frac{\textrm{C}}{\textrm{N}}$$_{sd}$ & $\frac{\textrm{C}}{\textrm{N}}$$_{DF}$ & $\frac{\textrm{C}}{\textrm{N}}$$_{agb}$\tabularnewline
\hline 
-3.00 & \textcolor{blue}{S} & 71 & 34 & 21 & 21 & 34 & 1 & 107 & 44\tabularnewline
-4.00 & \textcolor{blue}{S} & 152 & 41 & 6 & 6 & 181 & 1 & 4 & 3\tabularnewline
-5.45 & \textcolor{red}{C} & 2E4 & 6 & 7 & 7 & 2E6 & 2 & 7 & 4\tabularnewline
-6.50 & \textcolor{red}{C} & 4E6 & 8 & 8 & 8 & 6E6 & 3 & 4 & 3\tabularnewline
$Z=0$ & \textcolor{red}{C} & 0 & 4 & 4 & 4 & 0 & 0.14 & 0.14 & 0.14\tabularnewline
\hline 
 &  &  &  &  &  &  &  &  & \tabularnewline
\multicolumn{10}{c}{Mass $=1.0$ M$_{\odot}$}\tabularnewline
\hline 
\hline 
{[}Fe/H{]} & DF & $\frac{^{12}\textrm{C}}{^{13}\textrm{C}}$$_{fd}$ & $\frac{^{12}\textrm{C}}{^{13}\textrm{C}}$$_{sd}$ & $\frac{^{12}\textrm{C}}{^{13}\textrm{C}}$$_{DF}$ & $\frac{^{12}\textrm{C}}{^{13}\textrm{C}}$$_{agb}$ & $\frac{\textrm{C}}{\textrm{N}}$$_{fd}$ & $\frac{\textrm{C}}{\textrm{N}}$$_{sd}$ & $\frac{\textrm{C}}{\textrm{N}}$$_{DF}$ & $\frac{\textrm{C}}{\textrm{N}}$$_{agb}$\tabularnewline
\hline 
-3.00 & \textcolor{blue}{S} & 43 & 29 & 20 & 20 & 5 & 1 & 117 & 33\tabularnewline
-4.00 & \textcolor{blue}{S} & 41 & 25 & 14 & 14 & 8 & 1 & 47 & 25\tabularnewline
-5.45 & \textcolor{blue}{S} & 10 & 5 & 14 & 14 & 27 & 0.03 & 55 & 23\tabularnewline
-6.50 & \textcolor{red}{C} & 316 & 10 & 10 & 10 & 963 & 4 & 5 & 4\tabularnewline
$Z=0$ & \textcolor{red}{C} & 0 & 6 & 6 & 6 & 0 & 0.60 & 0.60 & 0.60\tabularnewline
\hline 
 &  &  &  &  &  &  &  &  & \tabularnewline
\multicolumn{10}{c}{Mass $=2.0$ M$_{\odot}$}\tabularnewline
\hline 
\hline 
{[}Fe/H{]} & DF & $\frac{^{12}\textrm{C}}{^{13}\textrm{C}}$$_{fd}$ & $\frac{^{12}\textrm{C}}{^{13}\textrm{C}}$$_{sd}$ & $\frac{^{12}\textrm{C}}{^{13}\textrm{C}}$$_{DF}$ & $\frac{^{12}\textrm{C}}{^{13}\textrm{C}}$$_{agb}$ & $\frac{\textrm{C}}{\textrm{N}}$$_{fd}$ & $\frac{\textrm{C}}{\textrm{N}}$$_{sd}$ & $\frac{\textrm{C}}{\textrm{N}}$$_{DF}$ & $\frac{\textrm{C}}{\textrm{N}}$$_{agb}$\tabularnewline
\hline 
-3.00 & \textcolor{darkgreen}{N}

 & 29 & 26 & -- & 5 & 2 & 1 & -- & 0.05\tabularnewline
-4.00 & \textcolor{blue}{S} & 420 & 26 & 12 & 5 & 8E3 & 0.20 & 31 & 0.05\tabularnewline
-5.45 & \textcolor{blue}{S} & -- & 29 & 5 & 5 & -- & 0.50 & 0.04 & 0.05\tabularnewline
-6.50 & \textcolor{blue}{S} & -- & 30 & 15 & 5 & -- & 0.44 & 30 & 0.05\tabularnewline
$Z=0$ & \textcolor{blue}{S} & -- & 5 & 8 & 5 & -- & 0.01 & 9 & 0.05\tabularnewline
\hline 
 &  &  &  &  &  &  &  &  & \tabularnewline
\multicolumn{10}{c}{Mass $=3.0$ M$_{\odot}$}\tabularnewline
\hline 
\hline 
{[}Fe/H{]} & DF & $\frac{^{12}\textrm{C}}{^{13}\textrm{C}}$$_{fd}$ & $\frac{^{12}\textrm{C}}{^{13}\textrm{C}}$$_{sd}$ & $\frac{^{12}\textrm{C}}{^{13}\textrm{C}}$$_{DF}$ & $\frac{^{12}\textrm{C}}{^{13}\textrm{C}}$$_{agb}$ & $\frac{\textrm{C}}{\textrm{N}}$$_{fd}$ & $\frac{\textrm{C}}{\textrm{N}}$$_{sd}$ & $\frac{\textrm{C}}{\textrm{N}}$$_{DF}$ & $\frac{\textrm{C}}{\textrm{N}}$$_{agb}$\tabularnewline
\hline 
-3.00 & \textcolor{darkgreen}{N}

 & -- & 21 & -- & 5 & -- & 0.46 & -- & 0.05\tabularnewline
-4.00\tnote{a} & \textcolor{darkgreen}{N}

 & -- & n/a & -- & 5 & -- & n/a & -- & 0.05\tabularnewline
-5.45 & \textcolor{blue}{S} & -- & 21 & 21 & 5 & -- & 0.24 & 0.26 & 0.05\tabularnewline
-6.50 & \textcolor{blue}{S} & -- & 26 & 81 & 5 & -- & 0.23 & 0.51 & 0.05\tabularnewline
$Z=0$\tnote{b} & \textcolor{blue}{S} & -- & 5 & 5 & 5 & -- & 0.01 & 0.04 & 0.05\tabularnewline
\hline 
\end{tabular}

\caption{The $^{12}$C/$^{13}$C and C/N ratios at different stages of evolution.
Values for all the low metallicity models are displayed, grouped in
terms of initial mass. We also show the $Z=0$ models for comparison.
Initial metallcity (first column) is given as {[}Fe/H{]}, except for
$Z=0$. The second column specifies whether the model experienced
a dual flash (DF) or not, be it a DCF (C), DSF (S) or neither (N).
The subscripts in the table headers indicate the stage of evolution
at which the values are taken: $fd$ (just after the first dredge
up episode), $sd$ (after 2DUP), $DF$ (after a dual flash), $agb$
(at the end of the AGB). Most of the intermediate mass models did
not reach the RGB so the FDUP entries are blank in these cases. \label{table-HaloStars-Nucleo-CandN}}

\line(1,0){100}

\begin{tablenotes}\scriptsize

\item[a] Due to a loss of data we are unable to provide the 2DUP values.

\item[b] This model started with a pure H-He composition, with $Y=0.230$ rather than 0.245. 

\end{tablenotes}

\end{threeparttable} \end{center}
\end{table}

\begin{figure}
\begin{centering}
\includegraphics[width=0.8\columnwidth,keepaspectratio]{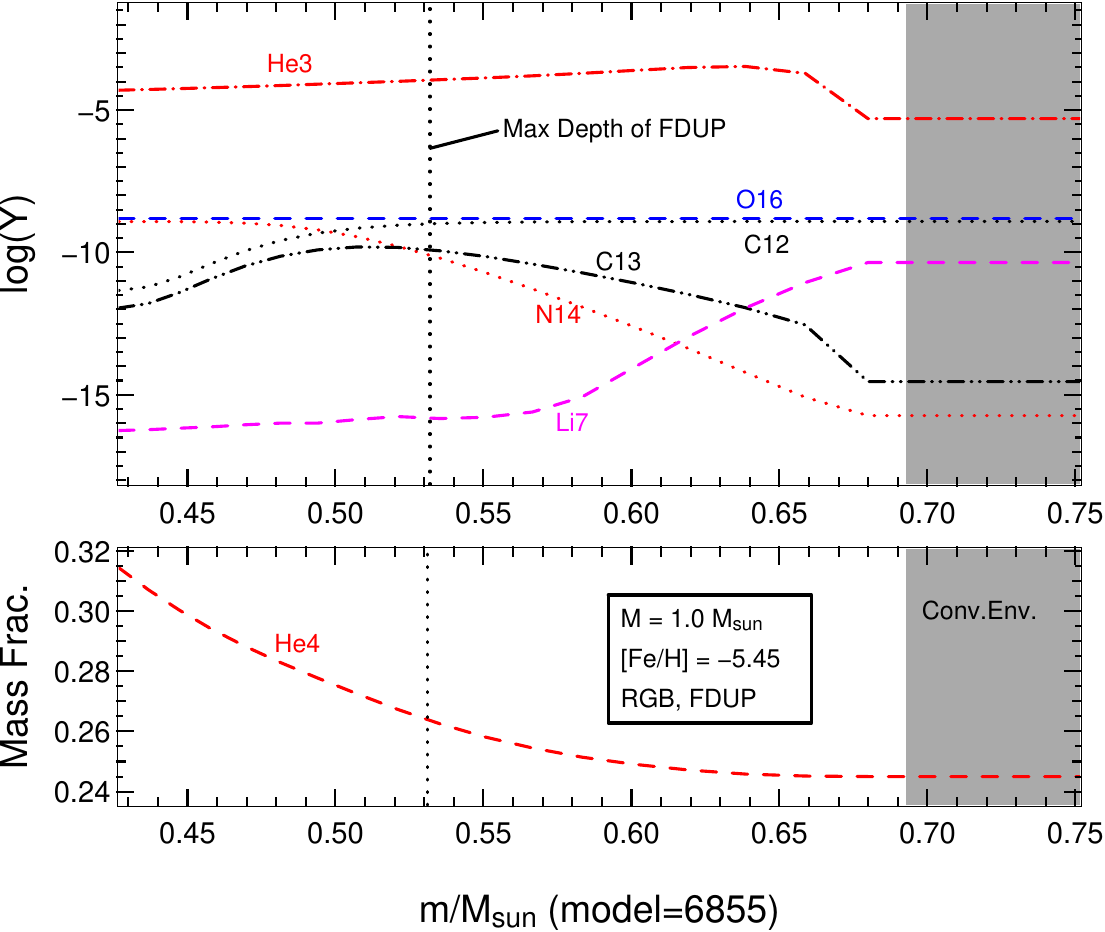}
\par\end{centering}
\caption{The abundance profiles of some selected species just as the envelope
convection begins to deepen at FDUP (at the start of the RGB). Due
to the CN cycle operating over a wide mass range it can be seen that
the abundances of the CN(O) nuclides have been altered. This causes
strong changes in the $^{12}$C/$^{13}$C and C/N ratios. We note
however that the changes occur amongst extremely small abundances
of these species. Helium is slightly increased by the end of FDUP,
as expected from panel 2. \label{fig-m1hmp-FDUP-cmp}}
\end{figure}

In Figure \ref{fig-m1hmp-FDUP-cmp} we give an example of FDUP in
one of our models (the 1 M$_{\odot},$ $\textrm{[Fe/H]}=-5.45$ model).
We show the profiles of some species just as the envelope convection
is extending inwards. It can be seen that a moderate amount of H burning
has occurred in the region that is subsequently dredged up. The $^{14}$N
abundance has increased by up to 5 orders of magnitude (though we
note that is is still present in very small amounts in an absolute
sense), as has $^{13}$C. $^{12}$C and $^{16}$O are scarcely affected
but $^{3}$He has increased and $^{7}$Li has decreased. Thus, as
the envelope moves inwards, it can be seen that this is the source
of the decrease in the $^{12}$C/$^{13}$C and C/N ratios detailed
above. In the bottom panel we display the $^{4}$He abundance, which
has also increased slightly in the region of interest, leading to
the slight increase in surface $Y$ by the end of FDUP. 

\subsection{Dual Core Flashes}

In our models the dual core flash (DCF) event is found to occur only
in the low mass models with the lowest metallicities. In Table \ref{table-HaloStars-Nucleo-YandZ}
we show which models go through this event, along with the surface
helium and $Z_{cno}$ abundances after the event. The DCF occurs at
the time of the core He flash, at the top of the RGB. Thus the only
previous envelope pollution has come from FDUP. In terms of $Y$ and
$Z_{cno}$ FDUP has little effect in these models. Thus the values
for these quantities post-DCF represent the pollution resulting from
this event alone. It can be seen that the abundance of He increases
by a large amount. In all these models the initial $Y$ was $\sim0.25$.
After the DCFs the average value is $\sim0.35$ (averaged over all
the DCF models) -- a $30\%$ increase. This large increase reflects
the large amount of He-rich material dredged up after the DCF (see
$\Delta$M$_{c}$ in Table \vref{table-AllHaloModelDualFlashProperties},
in the structural evolution section). As there is no 3DUP and no significant
HBB in these models this surface He abundance is essentially retained
till the end of the AGB evolution. Thus the chemical yield from these
models will have a He-enriched contribution (just how strongly this
affects the yields depends on the mass-loss history of each model). 

In all the models which experience the DCF the $^{12}$C/$^{13}$C
ratio remains very high after FDUP, or, in the case of the $Z=0$
models, it remains undefined (see Table \ref{table-HaloStars-Nucleo-CandN}).
The story after the DCF is much different though -- in all the models
the ratio is then quite low, being between 4 and 10. This reflects
the CN processing that has occurred in the DCF hydrogen convective
zone (HCZ) and HeCZ. The CN cycle equilibrium value of the $^{12}$C/$^{13}$C
ratio is usually $\sim4$, so these low values show that considerable
CN cycling has occurred, as expected. Unlike the FDUP event the absolute
abundances of these species increases enormously, which is seen in
the $Z_{cno,DF}$ values of Table \ref{table-HaloStars-Nucleo-YandZ}. 

The C/N ratios also remained high after the FDUP in these models.
In the low-metallicity models the ratio dropped to much lower levels
after the DCF. In both of the $Z=0$ models the ratio, which was previously
undefined due to the lack of metals, was $<1$ after the DCF. Thus
the surface composition contained more N than C. This may give an
indication that the polluting material from the DCF event is N rich
in the metal-poor models too, but it was diluted in those cases. The
dominance of N is another sign of advanced CN cycling during the DCF. 

As mentioned above all the models that experienced the DCF had a very
large increase in surface metallicity, as defined by $Z$ (naturally
{[}Fe/H{]} remains essentially unchanged). In Table \ref{table-HaloStars-Nucleo-YandZ}
we give the initial $Z_{cno}$ values (the sum of the mass fractions
of the CNO nuclides) and also the $Z_{cno}$ resulting from the DCFs.
It can be seen that the metal abundance jumps, almost uniformly amongst
the models, to $Z_{cno}\sim10^{-2}$ (the 0.85 M$_{\odot},$ $Z=0$
model `only' reaches 10$^{-3}$). Thus most of the DCF stars have
a super-solar Z after the DCF. We note that the $Z$ values do not
change significantly between this stage of evolution and the end of
the AGB, as can be seen from the final $Z_{cno}$ values given in
the same table 

As second dredge-up has little effect on the surface abundances in
these models (apart from a modest decrease in the C/N ratio in some
of the models), and as 3DUP and HBB do not occur to any significant
degree, the models essentially retain the chemical signatures from
this event till the end of the AGB. Thus the metallicity and chemical
profile of the yields are, to varying extents, set by the DCF events
(the extent depends on the mass-loss history of each model). This
also means that this contribution to the yields is directly dependent
on the treatment of the DCF, and thus on the uncertainties associated
with modelling this event (eg. the unknown degree of overshoot, which
we have assumed to be zero). We note that this group of stars constitute
our Group 1 specified in the introduction to this section -- the
group that we defined due to their AGB surface composition being dominated
by the DCF. For a more detailed description of the nucleosynthesis
of the DCF event we refer the reader to the 0.85 M$_{\odot}$, $Z=0$
model in Chapter \ref{Chapter-ZeroZ-Nucleo}. 

\subsection{Second Dredge-Up}

In the models that experience the dual core flash (low mass and lower
metallicities) the 2DUP has little effect on the surface He abundance.
As mentioned above the He abundance arising from the DCF was quite
high in all these models, and thus the 2DUP would have to bring up
an enormous amount of He to alter this. The surface He abundance in
these (`Group 1') models is set by the DCF for the rest of the evolution
as the AGB has a negligible effect. The metallicity of these models
is also unaltered by the 2DUP. Likewise the $^{12}$C/$^{13}$C and
C/N ratios do not change significantly (the 0.85 M$_{\odot}$, $\textrm{[Fe/H]}=-5.45$
model is an exception, where the C/N ratio reduces from 7 to 2 after
2DUP), as they were already quite low due to the preceding DCF. 

In the low-mass and intermediate-mass models that experienced the
dual shell flash, 2DUP increased the surface He abundance by varying
amounts. The degree of increase is mainly a function of metallicity
but also of mass, with the lowest metallicity and highest mass models
having the greatest increases. In the low mass models, only the most
metal-rich of which went through DSFs, a typical increase was from
$Y\sim0.25$ to $Y\sim0.27$. As mentioned below the DSF that comes
after the 2DUP further increased $Y$, usually by a comparable amount
to that gained from 2DUP. The surface abundance on the AGB in these
low mass models then remained at this level of He enrichment. Thus,
in these `Group 2' models, the 2DUP is a significant contributor to
the He yields, but not the only one. In the lowest metallicity intermediate-mass
models the abundance reached as high as $Y\sim0.33$ as a result of
2DUP (no FDUP occurred in these models so this increase was from the
primordial value of $\sim0.25$). In this group (Group 3) HBB on the
AGB also increases $Y$ in the yields. Interestingly, just like the
DSF in the low-mass models (which has little effect on $Y$ in these
IM models), the HBB increases the surface He abundance by a similar
amount as that gained from 2DUP. For example, in the $\textrm{[Fe/H]}=-5.45,$
2 M$_{\odot}$ model 2DUP raised $Y$ from 0.25 to 0.29, then HBB
and 3DUP raised it to 0.34 (see Table \ref{table-HaloStars-Nucleo-YandZ}).
Thus, like in the low-mass models, the 2DUP has a significant effect
on the final yield of He, but is not the only contributor. 

In regards to the $^{12}$C/$^{13}$C ratio in the low- and intermediate-mass
models (that do not go through the DCF -- ie. Groups 1 and 2), it
is seen to decrease in all the models to a similar level. In the intermediate
mass models the ratio ends up in the range $20\rightarrow30$ whilst
it is slightly higher in the low mass models, being in the range $25\rightarrow40$.
In most of the IM models there was no FDUP, thus these values were
down from the (very high) primordial ones. In the LM models FDUP had
already reduced the $^{12}$C/$^{13}$C ratio substantially so the
change was not as great. An exception here is the 1 M$_{\odot}$,
$\textrm{[Fe/H]}=-5.45$ model, which had quite a low value after
FDUP ($=10$) and thus a very low value after 2DUP ($=5$). The $Z=0$
IM models are a distinct case, as their $^{12}$C/$^{13}$C ratios
were initially zero. In this case 2DUP \emph{increases} the ratio,
universally bringing it up to 5 (the CN equilibrium value). 

The C/N ratio is reduced by 2DUP in all the low- and intermediate-mass
models (that do not go through the DCF -- ie. Groups 1 and 2). At
low mass the ratio ends up equal to 1 in all models except for the
1 M$_{\odot}$, $\textrm{[Fe/H]}=-5.45$ model, in which it ends up
very low (0.03). We note that it is this model that also has an anomalously
low $^{12}$C/$^{13}$C ratio after 2DUP (and FDUP). In the intermediate
mass models the reduction is much stronger, always resulting in ratios
less than 1. Thus all the IM models are N-rich after 2DUP (except
for the 2 M$_{\odot}$, $\textrm{[Fe/H]}=-3$ model, which ends up
with C/N$=1$). The ratio in the majority of the models lies in the
range $0.2\rightarrow0.5$. We note that this reduction is from the
(very large) primordial value as most of these models did not experience
FDUP. Again, the $Z=0$ models represent a special case, since there
was initially no C or N, so the C/N ratio \emph{increased} to 0.01.
Thus these stars are very N-rich as they begin the TP-AGB stage. 

\begin{figure}
\begin{centering}
\includegraphics[width=0.8\columnwidth,keepaspectratio]{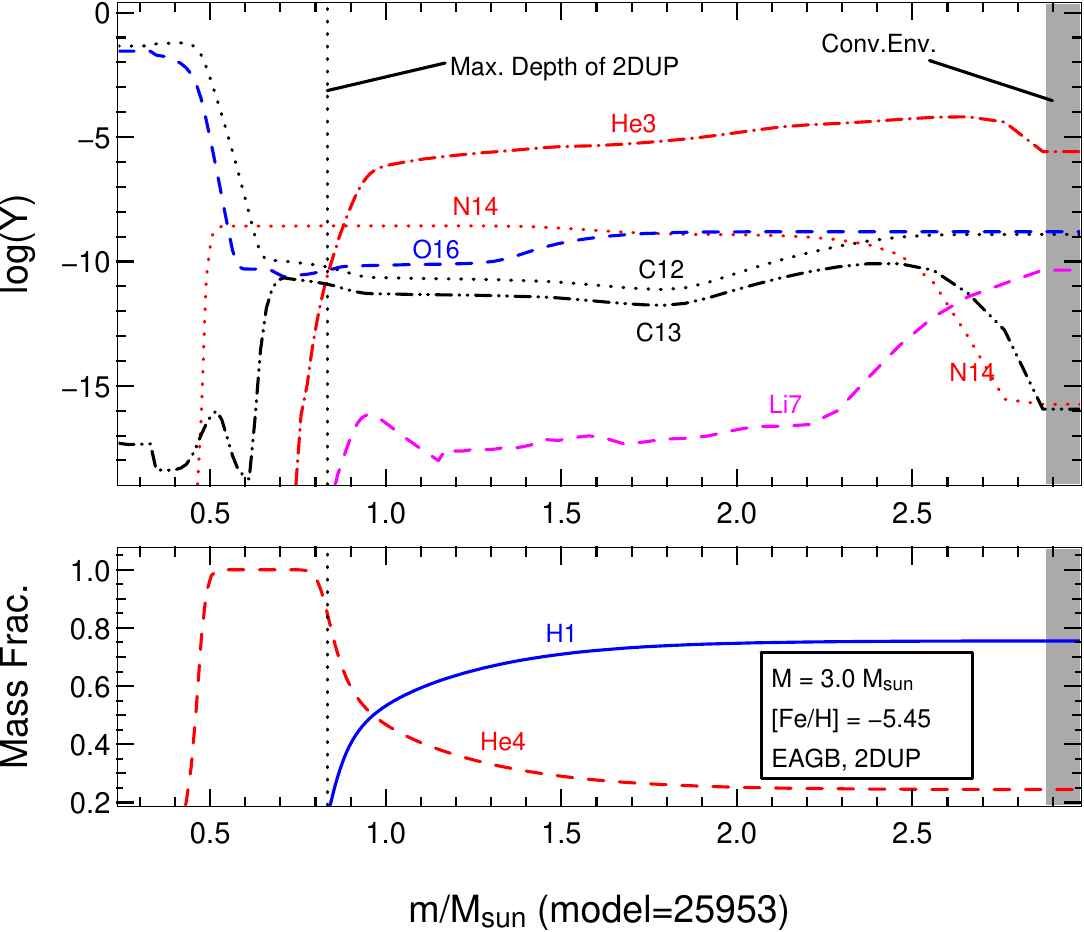}
\par\end{centering}
\caption{The abundance profiles of selected species against mass just as 2DUP
is starting in the 3 M$_{\odot}$, $\textrm{[Fe/H]}=-5.45$ model.
The convective envelope can be seen as grey shading on the right-hand
side. Interestingly a very large mass fraction of the entire star
has been affected by CN(O) cycling, such that N and $^{13}$C is enhanced.
Indeed a large mass fraction has reached CN equilibrium. This alters
the $^{12}$C/$^{13}$C and C/N ratios as the material is mixed to
the surface. In the bottom panel it can be seen that the He abundance
will also increase, as seen in Table \ref{table-HaloStars-Nucleo-YandZ}.
\label{fig-m3hmp-CMP-2DUP}}
\end{figure}

As an example of 2DUP in action we present in Figure \ref{fig-m3hmp-CMP-2DUP}
some composition profiles of our 3 M$_{\odot}$, $\textrm{[Fe/H]}=-5.45$
model, just as the convective envelope starts to move in. The reason
for the reduction in the $^{12}$C/$^{13}$C and C/N ratios, and the
increase in He abundance at the surface, can clearly be seen. Substantial
CN(O) cycling has occurred above the core, in a very thick H-burning
shell. Indeed, over much of the mass of the star the CN cycle has
reached equilibrium (out to $\sim2.2$ M$_{\odot}$, where the mass
of the core is only $\sim0.8$ M$_{\odot}$), such that the material
is in these regions is N-rich and has a $^{12}$C/$^{13}$C ratio
$\approx5$. Thus, as the envelope convection deepens, the surface
composition will be altered in the ways detailed above. We also note
that from this diagram we expect the surface abundance of $^{7}$Li
to be significantly reduced by 2DUP as well. 

\subsection{Dual Shell Flashes}

The dual shell flash (DSF) event occurs at the beginning of the AGB
and is similar to the DCF insomuch as it consists of a He-flash convection
zone breaching the H-He discontinuity above, thus mixing protons down
and He burning products up (see Section \ref{subsection-HaloStars-Struct-DSFs}
for more information on this event). More than half of the models
in our grid (including the $Z=0$ models) experience the DSF event.
At low mass it appears that if the model does not go through a DCF
it will always go through a DSF, such that all low-mass, very low-metallicity
models experience one type of dual flash. At intermediate mass most
of the models go through a DSF. The only exceptions are the most metal-rich
models (the $\textrm{[Fe/H]}=-3$ model at 2 M$_{\odot}$ and the
$\textrm{[Fe/H]}=-3$ and $-4$ models at 3 M$_{\odot}$, which experience
no dual flashes at all (see Table \ref{table-HaloStars-Nucleo-YandZ}). 

\begin{figure}
\begin{centering}
\includegraphics[width=0.95\columnwidth,keepaspectratio]{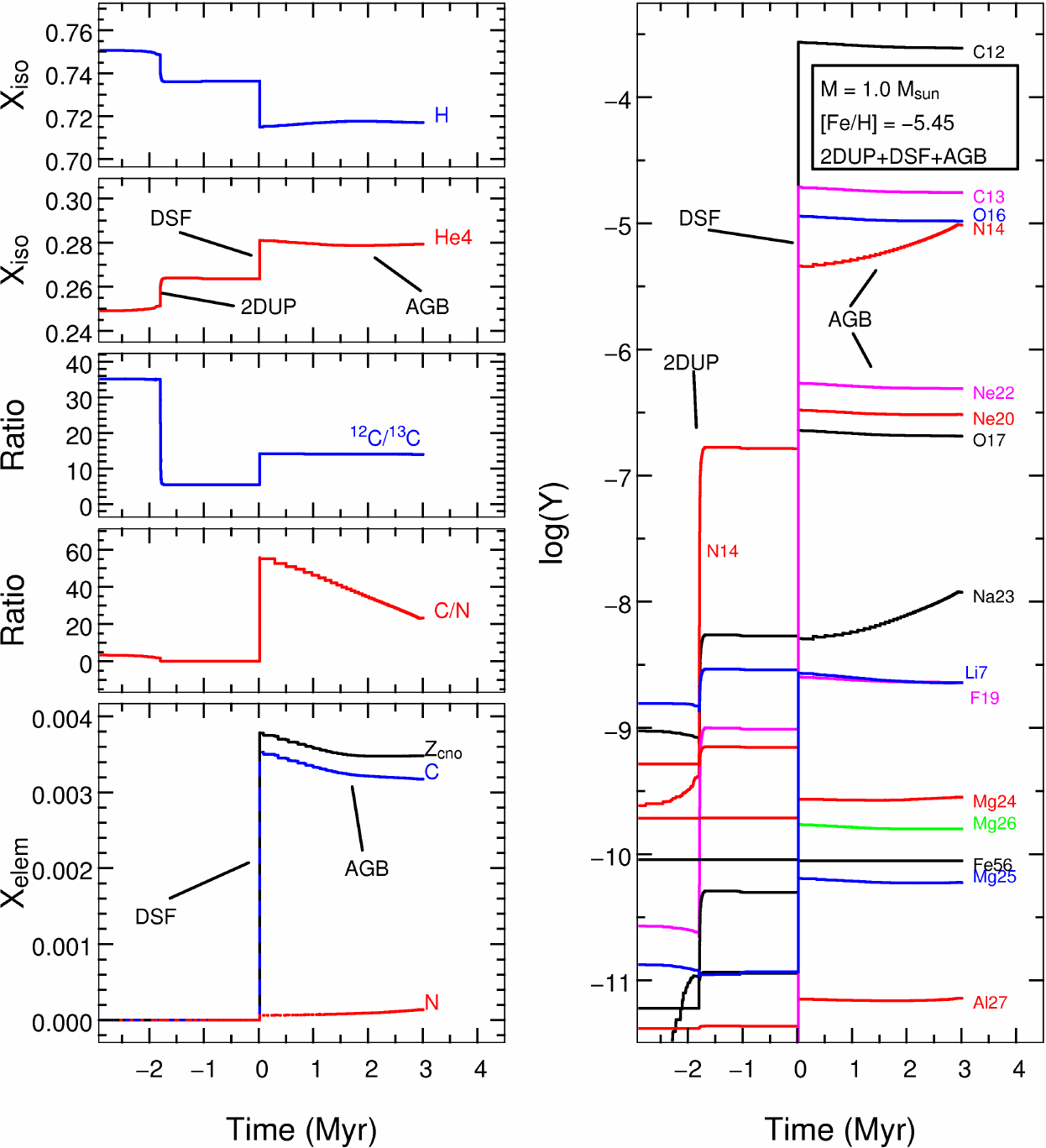}
\par\end{centering}
\caption{The time evolution of the surface composition of the $\textrm{[Fe/H]}=-5.45,$
1 M$_{\odot}$ model, for selected species. This model fits into our
Group 2, as its surface composition on the AGB is dominated by the
DSF event. This event is clearly seen in the abrupt change in surface
composition at $t=0$ (time has been offset for clarity). In particular
the metallicity of the model increases by a huge amount, from $Z_{cno}\sim10^{-8}$
to $10^{-3}$. Another interesting feature is that there is some HBB
occurring on the AGB, as evidenced by the increase in $^{14}$N and
$^{23}$Na. It is however not strong enough to significantly increase
the surface He abundance (as it does in the intermediate mas models).
\label{fig-m1hmp-SRF-2DUP-DSF-AGB}}
\end{figure}

In terms of helium the DSF has essentially no effect in the IM models.
There are two reasons for this. The first is that the mass involved
in the DSF is quite small, whilst the envelope mass is relatively
large (see $\Delta$M$_{c}$ in Table \vref{table-AllHaloModelDualFlashProperties},
in the structural evolution section). The second is that the second
dredge-up in these models significantly enhances the He abundance
in the envelope before the DSF, so that the DSF pollution is negligible
next to this. In the low-mass models the increase in He is significant
but still rather small. In general the envelope is enhanced by a similar
amount as that given by 2DUP ($\Delta Y\sim0.1\rightarrow0.2$, see
Table \ref{table-HaloStars-Nucleo-YandZ}). Of significance here is
that the He abundance in the LM models remains at this level (there
is no substantial He production on the AGB), such that the AGB contributions
to the He yields are (roughly) affected equally by the DSF and 2DUP
events. Thus the He yields are not highly sensitive to the treatment
of the DSF at low mass, but they are affected to a moderate degree
(the degree also depends on the mass-loss history of the models, as
in some models a substantial amount of mass is lost during the unpolluted
RGB phase). In Figure \ref{fig-m1hmp-SRF-2DUP-DSF-AGB} we show an
example of a low-mass model that goes through the DSF episode. The
similar level of surface He pollution given by the DSF and 2DUP is
clear in panel 2 of this figure. In the IM models it is HBB on the
AGB that dominates the He yields. Thus the yields in these cases are
essentially unaffected by the treatment of the DSF. 

The surface metallicity arising from the DSF events is also more mass-dependent
than metallicity-dependent. At 3 M$_{\odot}$ the pollution from the
DSFs only raises the envelope $Z_{cno}$ to $\sim10^{-6}$ whereas
at 2 M$_{\odot}$ it is usually raised to $\sim10^{-4}$. At both
masses 3DUP on the AGB increases $Z_{cno}$ to $\sim10^{-3}\rightarrow10^{-2}$
(the average is higher at 3 M$_{\odot}$), such that this pollution
effect is more dominant than the DSF. At low mass we find that the
reverse is true -- the DSF pollution far outweighs any pollution
on the AGB (which is virtually nil due to the lack of 3DUP). Interestingly
the DSF pollution in the LM models raises the surface metallicity
to similar levels as that given by the hundreds of 3DUP episodes in
the IM models, such that $Z_{cno}$ ends up being $\sim10^{-3}\rightarrow10^{-2}$
in these models as well (see Table \ref{table-HaloStars-Nucleo-YandZ}
and an example of the time evolution of one model in Figure \ref{fig-m1hmp-SRF-2DUP-DSF-AGB}).
Thus the AGB surface composition, and therefore a component of the
yields, is set by the DSF event in the low-mass models. It is this
group of models that we designated as Group 2 at the beginning of
this section -- low-mass models that experience the DSF and have
their AGB envelope compositions dominated by this pollution event.
This means that the yields from these models are partly dependent
on the treatment of the DSF itself (the degree of dependence depends
on the mass loss history of each model). 

In Table \ref{table-HaloStars-Nucleo-CandN} we also provide the $^{12}$C/$^{13}$C
and C/N ratios after the DSF event. The $^{12}$C/$^{13}$C ratios
in the low mass models had been reduced to $\sim30$ in most (DSF)
cases by the 2DUP. The DSF episode further reduced this ratio in most
of the models, indicating that the DSF pollution material had undergone
CN cycling. There was however one exception -- the $\textrm{[Fe/H]}=-5.45,$
1 M$_{\odot}$ model had a very low $^{12}$C/$^{13}$C ratio after
2DUP and the DSF \emph{increased} it to 14. Interestingly this is
a similar ratio to that in the other models after their DSFs, suggesting
that this may be a characteristic signature of the DSF (ie. the ratio
in the DSF pollution material is moderately higher than the equilibrium
value of $\sim4$, being between 6 and 21). We show the time evolution
of the surface$^{12}$C/$^{13}$C ratio for this model in panel 3
of Figure \ref{fig-m1hmp-SRF-2DUP-DSF-AGB}. In the intermediate-mass
models the DF had a stronger effect on the $^{12}$C/$^{13}$C ratio.
In general the ratio reduced from a post-2DUP value of $\sim30$ to
a post-DSF value of $\sim5$. Thus the polluting material appears
to have attained CN equilibrium in these stars. The one exception
here is that the $\textrm{[Fe/H]}=-5.45,$ 3 M$_{\odot}$ model suffered
no change in the ratio. This model appears to have had an extremely
mild DSF, as the $Z_{cno}$ value did not change significantly from
the initial value either. We also note that in the $Z=0$ models the
$^{12}$C/$^{13}$C ratio was already $\sim5$ after the 2DUP episode.
Thus the DSF kept the status quo in these cases. 

The C/N ratio in the low-mass models was universally very low after
2DUP, being $\leq1$. The DSF was also universal in its effect, raising
the ratio in all the models. Despite this the final ratio was however
somewhat varied between the models. In most of the models the ratio
ended up between 47 and 107. The exception was the $\textrm{[Fe/H]}=-4,$
0.85 M$_{\odot}$ model. It ended up with a ratio of only 4 (up from
1). In the intermediate mass models the C/N ratios were all initially
low, being $<1$. However at these masses the DSF had the opposite
effect -- the ratio was \emph{decreased} further. In fact the ratio
dropped to $\sim0.04$ which indicates that the DSF pollution material
had reached complete CN equilibrium (this concurs with the $^{12}$C/$^{13}$C
ratios). Indeed, it is similar to the final AGB value attained in
all the IM models after strong HBB. The one exception is again the
$\textrm{[Fe/H]}=-5.45,$ 3 M$_{\odot}$ which, as mentioned above,
appears to have had an extremely mild DSF, so no change in the ratio
occurred.

\subsection{TP-AGB}

All of our models become thermally-pulsing AGB stars. However it is
only the more massive ones that experience third dredge-up (3DUP)
and significant hot bottom burning (HBB). The lack of 3DUP in the
low-mass models is evident in the fact that the $Z_{cno}$ metallicities
at the end of the AGB are the same as those just after the DCF or
DSF episodes. It is also evident in the unchanging $^{12}$C/$^{13}$C
ratios. Thus, as mentioned earlier, these models have their AGB surface
composition dominated by the DCF/DSF pollution events. Interestingly
there is a slight exception to this general rule. It is seen in the
C/N ratios of all the low-mass, low-Z models -- they all reduce (we
note that at $Z=0$ they remain constant, since they are already very
low). The reason for this is seen in our example model in Figure \ref{fig-m1hmp-SRF-2DUP-DSF-AGB}
-- $^{14}$N is actually being produced on the AGB. Thus it appears
that a small amount of HBB is occurring even in the low mass models
(the 0.85 M$_{\odot}$ models also show a reduction in the C/N ratio).
The $^{14}$N is produced via proton captures on the very abundant
$^{12}$C. $^{23}$Na is also produced, via proton captures on the
abundant $^{22}$Ne. Interestingly both these species increase by
$\sim0.3$ dex on the AGB. $^{7}$Li is also (very) slightly depleted.
We note however that the abundance changes occurring from this minor
HBB only marginally alters the pollution given by the DCF or DSF events.

\begin{figure}
\begin{centering}
\includegraphics[width=0.95\columnwidth,keepaspectratio]{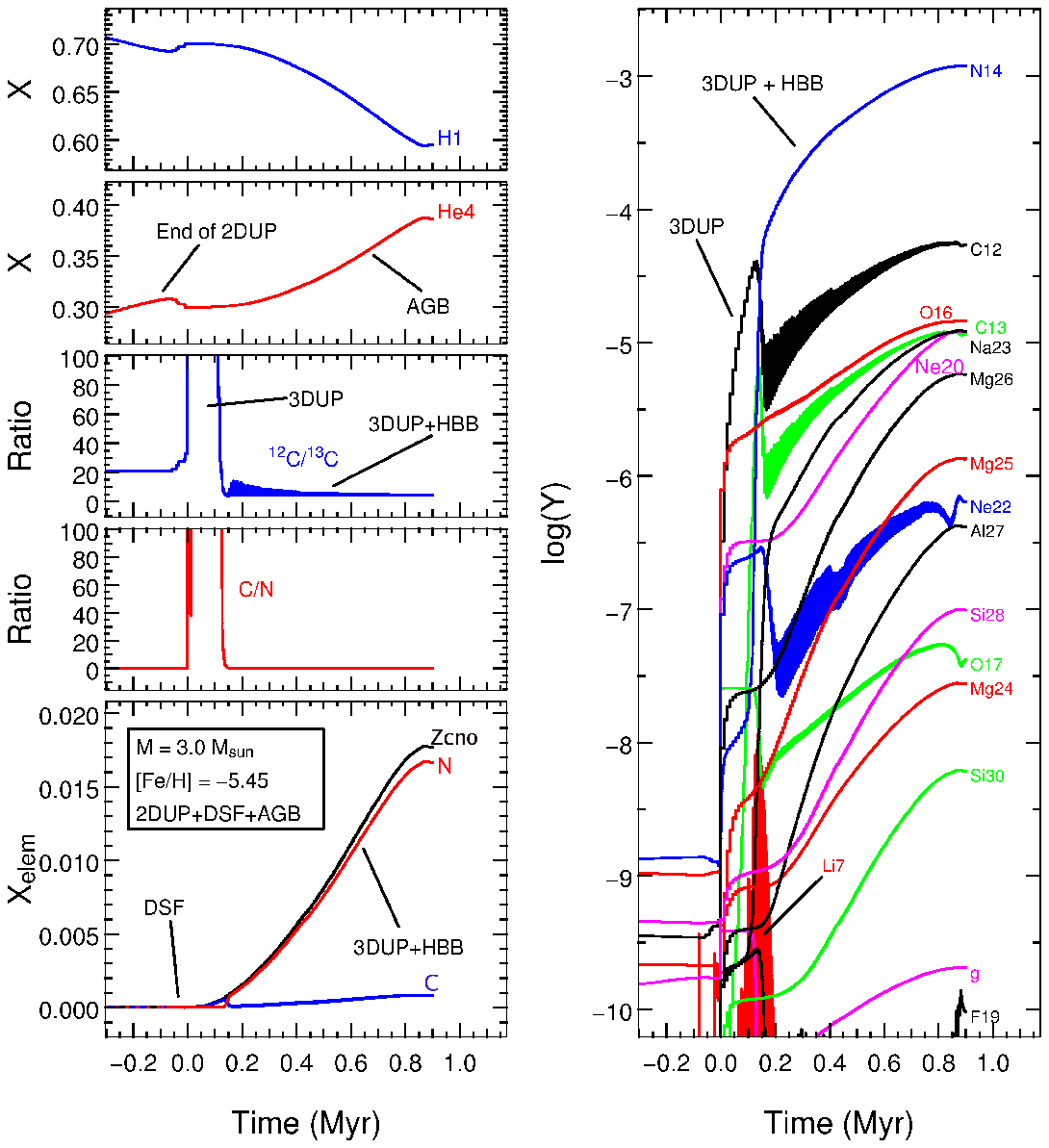}
\par\end{centering}
\caption{The time evolution of the surface composition of the $\textrm{[Fe/H]}=-5.45,$
3 M$_{\odot}$ model, for selected species. This model fits into our
Group 3, as its surface composition on the AGB (and thus the yield,
as practically all of the mass loss occurs during the AGB in this
model) is dominated by the combination of 3DUP and HBB. The relative
importance of this form of pollution compared to that from the DSF
event is highlighted in the bottom left-hand panel, where the metallicity
is seen to increase enormously on the AGB. The metallicity of the
model increases from $Z_{cno}\sim10^{-8}$ to $10^{-2}$. The rich
nucleosynthesis arising from 3DUP and HBB is seen in the right-hand
panel. In particular the CN cycling product $^{14}$N dominates most
of the AGB. The $^{12}$C/$^{13}$C and C/N ratios quickly tend to
equilibrium values once the (strong) HBB starts. \label{fig-m3hmp.SRF-AGB-DSF-2DUP}}
\end{figure}

The situation in the IM models is very different. 3DUP is active in
all the models, as is strong HBB. The mass fraction of He increases
by a large amount, mainly due to H burning at the base of the convective
envelope. Between the end of the DSF events and the end of the AGB
the difference in He abundance is $\Delta Y\sim0.03\rightarrow0.05$
in the 2 M$_{\odot}$ models and $\sim0.07\rightarrow0.09$ in the
3 M$_{\odot}$models, reflecting the higher temperatures in the more
massive models. Thus the final He abundance in the most metal-poor
models is $\sim0.34$ at 2 M$_{\odot}$ and $\sim0.39$ at 3 M$_{\odot}$
(see Table \ref{table-HaloStars-Nucleo-YandZ}). Due to the many 3DUP
episodes in the IM models ($\sim300$ on average) large amounts of
He burning products get mixed into the AGB envelopes. This raises
the surface metallicity in all of the models. Interestingly all the
models end up with a similar level of pollution, characterised by
$Z_{cno}\sim10^{-3}\rightarrow10^{-2}$ (the 3 M$_{\odot}$ models
end up slightly more metal-rich on average, at $Z_{cno}\sim10^{2}$).
Thus the bulk level of pollution coming from the DSF events is eclipsed
by the 3DUP pollution. It is these IM models, combined with the IM
models that do not experience any DSF episode at all, that we designated
as Group 2 at the start of this section. They are defined by the fact
that their AGB surfaces are dominated by the effects of 3DUP and HBB.
As an example of a member of this group we show the time evolution
of the surface abundances for our $\textrm{[Fe/H]}=-5.45,$ 3 M$_{\odot}$
model in Figure \ref{fig-m3hmp.SRF-AGB-DSF-2DUP}. In panel 5 of this
panel the relative insignificance of the DSF pollution can be clearly
seen. The rich nucleosynthesis arising from the combination of 3DUP
and HBB can be seen in the right-hand panel of the same figure. For
most of the AGB evolution the dominant (metal) species is $^{14}$N,
as it is continually increased through the conversion of dredged-up
$^{12}$C via proton captures (ie. the CN(O) cycles). Thus it can
be seen that strong HBB starts early on the TP-AGB in this model.
The CN cycling naturally affects the $^{12}$C/$^{13}$C and C/N ratios
(see Table  \ref{table-HaloStars-Nucleo-CandN}), causing both of
them to drop to equilibrium values well before the end of the AGB.
In fact \emph{every} IM model ends the AGB with a $^{12}$C/$^{13}$C
ratio of 5 and a C/N ratio of 0.05, regardless of its previous pollution
history (some of the models did not experience a DSF). The C/N ratio
implies that N is 20 times more abundant than C, ie. $\sim1$dex.
This is a robust chemical signature of these metal-deficient (and
metal-free) IM models that could be compared with observations. The
chemical composition of the surface during the AGB is similar in all
the IM models. We thus refer the reader to the section describing
the nucleosynthesis in the $Z=0,$ 2 M$_{\odot}$ model in Chapter
\vref{Chapter-ZeroZ-Nucleo} for more details of this type of model.

\section{Yields\label{section-Yields-HaloStars}}

In this section the yields for all the models in the grid are presented,
in graphical and tabular form (although some of these are actually
located in the appendices for practicality). For details on the definition
of Yield that we use throughout the current study we refer the reader
to Section \vref{subsection-YieldCodeCalcsAndSynthPulses}. The sources
of the yields for some selected species are discussed, as well as
a suggested classification scheme which we use to split the yields
into three chemically similar groups. We have however chosen to begin
the section with a discussion on one of the major uncertainties affecting
the yields of a subgroup of our models, as we believe it is important
to keep this key uncertainty in mind when examining the yields given
below. 

\subsection{Uncertainty in the Yields of the LM, Higher-Z Models\label{subsec-YieldUncert-UnpollutedRGBmloss}}

By comparing the surface He abundances at the end of the AGB phase
($Y_{agb}$) in Table \ref{table-HaloStars-Nucleo-YandZ} with the
helium yields plotted in Figure \ref{fig-YandZ-AllYields} it can
be seen that the low-mass yields display lower helium abundances than
their respective AGB surfaces. This is due to the fact that these
models all go through the RGB stage, where they lose some mass via
winds. In all cases the surface composition during the RGB is the
initial one, such that $Y$ is still equal to 0.245 (although in some
cases there is a secondary RGB at which time the surface is polluted
-- but this is short-lived). Thus, even though the matter lost on
the AGB is high in He, it is partially diluted by the He-poor material
on the RGB. The degree of this dilution effect is naturally dependent
on the amount of mass lost on the AGB as compared to that lost on
the RGB. The relative importance of each phase of mass loss was discussed
in the previous section. In particular Figures \ref{fig-massLoss-allm3}
to \ref{fig-massLoss-allm8} display the evolution of total stellar
mass with time for all of the 3, 1 and 0.85 M$_{\odot}$ models. These
figures make it clear when and to what degree mass loss occurs throughout
the various stages of evolution. As stated in that section it is a
complex function of mass, metallicity and the occurrence or non-occurrence
of polluting events such as the DCF -- not to mention the mass loss
formulae themselves. However a clear conclusion was reached at the
end of that discussion -- a significant uncertainty arises in the
low mass, higher metallicity models. The uncertainty is due to the
unknown rates of mass loss at very low metallicity whilst the reason
it affects this particular group of models is because they spend a
significant amount of time at high luminosities on the RGB (see eg.
Figure \vref{fig-hrds-m0.85all-RGBtip}), such that much mass is lost
while the surface is still very metal-poor. Thus, in all the following
discussion on yields, we need to remember that the yields from the
low-mass, higher metallicity models come with a caveat. In particular
we note that, if mass loss at very low metallicity is in reality much
less efficient, then our yields (for these models) will be over-diluted
by the material lost on the RGB. Hence we would expect that our yields
would be much more metal-rich (and more helium-rich) if most of the
mass was lost during the AGB phase. 

With regards to the more massive and more metal-poor models, we note
that they did not go through the RGB stage (or did so only for a brief
time, at low luminosity), so their yields are totally dominated by
the AGB mass loss, as can be seen by the close correspondence between
$Y_{agb}$ and the yields. Thus they are not affected by the uncertainty
in mass-loss rates at low metallicity -- the low-mass higher-metallicity
caveat does not apply. 

\subsection{Grouping the Models Into Yield Types}

At the beginning of Section \ref{section-HaloStars-Nucleo} we defined
three groups into which all the models fit. The groups were delineated
by the dominant source of pollution of the AGB envelope -- the DCF,
the DSF or 3DUP + HBB. We referred to them as groups 1, 2 and 3, respectively.
We were of the belief that this classification scheme should translate
directly to a natural delineation within the yields, as much of the
mass loss occurs on the AGB. However the unpolluted mass-loss on the
RGB in some of the models partly undermines this categorisation because
it causes a dilution of the AGB yields. Nevertheless it is apparent
from the presentation of the yields below that the yields are still
quite distinguishable into the three categories despite this. In the
following we also refer to Groups 1, 2, and 3 as the `DCF group',
the `DSF group' and the `IM group'. The IM (intermediate mass) group
is so called because all the 2 and 3 M$_{\odot}$ models experience
considerable 3DUP and HBB on the AGB, the resulting composition of
which dominates the yields.

\subsection{Some Key Properties of The Yields}

To continue on from the nucleosynthesis discussion in the previous
section we first present the He, $Z_{cno}$, C/N and $^{12}$C/$^{13}$C
compositions of the yields for the entire grid of models in Figures
\ref{fig-YandZ-AllYields} and \ref{fig-CNratios-AllYields}. 

\subsubsection*{Helium}

As mentioned just above, the He mass fractions in the low mass yields
are quite low when compared to the composition of the envelopes of
these models just after the DCF and DSF pollution episodes (see Table
\ref{table-HaloStars-Nucleo-YandZ}). The extreme case is the 0.85
M$_{\odot},$ $\textrm{[Fe/H]}=-5.45$ model. It had $Y=0.44$ on
the surface after its DCF but only $Y=0.33$ in the yield (see Section
\vref{subsection-YieldCodeCalcsAndSynthPulses} for the definition
of Yield used in the current study). This is due to the fact that
it suffered considerable mass loss on the RGB, when its envelope still
had the initial He-poor composition, thus the AGB (and HB) mass loss
was diluted. This effect reduces with metallicity and increasing mass,
as the RGB phases in the models get progressively shorter. In the
most metal-rich low-mass models (the ones that did not experience
a DCF, i.e. Group 2) the mass fraction of He is virtually the same
as the initial one. This is partly due to the DSF and 2DUP not increasing
He very much (up to $Y\sim0.28$) but mostly due to the large amount
of unpolluted mass loss on the AGB (see eg. Figure \vref{fig-massLoss-allm8}). 

The helium situation in the intermediate-mass yields is much more
uniform. They all reflect the composition at the end of the AGB phase.
This is due to the fact that practically all of the mass loss occurs
towards the end of the AGB in these Group 3 models. In most of the
IM models the He comes from two main sources -- 2DUP and HBB. As
noted in the nucleosynthesis section these two sources increase the
helium content in the envelope by similar amounts. Interestingly the
DSF, which occurs in the most metal-poor of the IM models, has little
effect on the He abundance.

\subsubsection*{Metallicity}

\begin{figure}
\begin{centering}
\includegraphics[width=0.75\columnwidth,keepaspectratio]{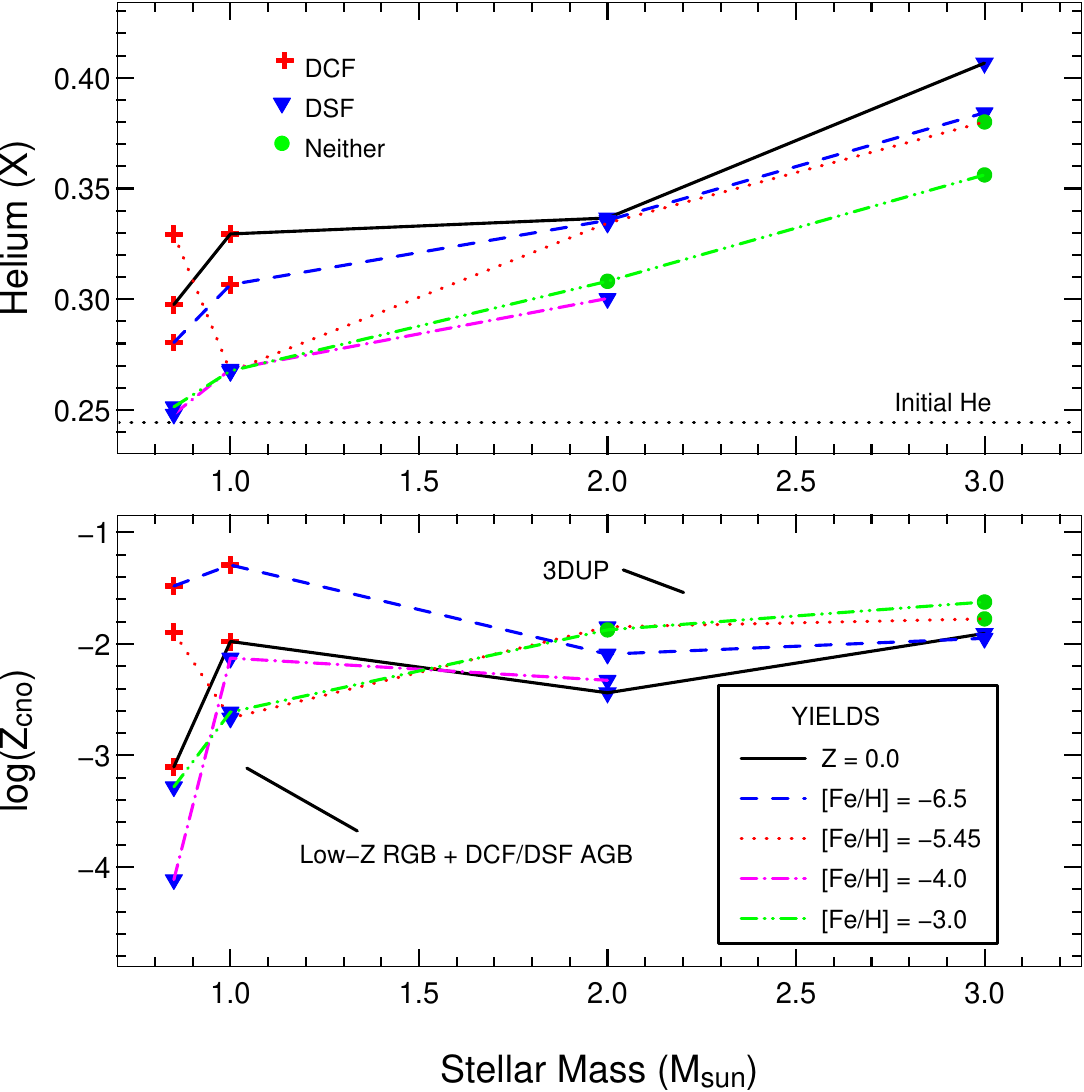}
\par\end{centering}
\caption{The helium yields (in mass fraction $X$) and metallicity yields (in
$Z_{cno}$, the mass fraction of all of the CNO nuclides) against
initial stellar mass. All models in the grid are shown, including
the $Z=0$ models. The different pollution events that occur in the
models are indicated by different shaped markers, as per the key in
the top panel. The metallicity key in the bottom panel applies to
both panels. The main points are that He is produced more in the IM
models as compared to the LM models and that all of the models have
very metal-rich yields considering their extremely low initial metallicities.
Note that there are no points included for the 3 M$_{\odot},$ $\textrm{[Fe/H] =}-4.0$
model due to a loss of data. \label{fig-YandZ-AllYields}}
\end{figure}

In Panel 2 of Figure \ref{fig-YandZ-AllYields} we show the metallicity
in the yields, as $\log(Z_{cno})$, where $Z_{cno}$ is the mass fraction
of all the CNO nuclides combined. This gives a good overview of the
degree of pollution that has made it into the yields. At intermediate
masses the pollution is quite uniform -- the yields are all characterised
by $\log(Z_{cno})\sim-2\pm0.4$. We suggest that this is due to the
fact that all these models experienced a similar amount of 3DUP over
the hundreds of thermal pulses. As can be seen in Table \vref{table-HaloStars-Nucleo-YandZ}
3DUP is the main contributor to the metal pollution of the AGB envelope
(the DSF episode only raises the metallicity to a more moderate level).
In the low-mass models the situation is more complex. Again the RGB
dilution effect comes into play. At the lowest metallicities $Z_{cno}$
in the yields is not far from that reported for the end of the AGB.
This is due to their short-lived RGB phases. In the more metal-rich
models there is a decrease of up to an order of magnitude in $Z_{cno}$
(from the AGB values), due to these models losing up to $90\%$ of
their mass on the unpolluted RGB. 

The general result for overall metallicity is one of a huge increase
over the initial abundances in all the models. There is however some
variation in the degree of pollution. Ironically the (initially) highest
metallicity low-mass models end up with the lowest metallicity yields,
though they are still well above the initial composition. The main
reason for this is that they have much more mass-loss during the RGB
phase, since they spend longer at higher luminosities on the RGB.
As discussed above it is these models that suffer from the most uncertainty
arising from the unknown mass-loss rates at very low Z. It is interesting
to note that if the mass loss rate were reduced on the RGB, then these
models would have yields as metal-rich as the more metal-poor models,
as the severe pollution from the DSF events at the start of the AGB
raise the metallicity of their envelopes to similar levels as the
DCF events do in the more metal-poor models. Thus the lines in panel
2 of Figure \ref{fig-YandZ-AllYields} would be flatter in the low
mass regime. With regards to the IM yields the spread is much smaller.
Unlike the low-mass yields a general trend of increasing metallicity
with increasing initial metallicity can be seen, but it is not a strict
rule since there are some exceptions. 

\subsubsection*{$^{12}$C/$^{13}$C and C/N Ratios}

\begin{figure}
\begin{centering}
\includegraphics[width=0.75\columnwidth,keepaspectratio]{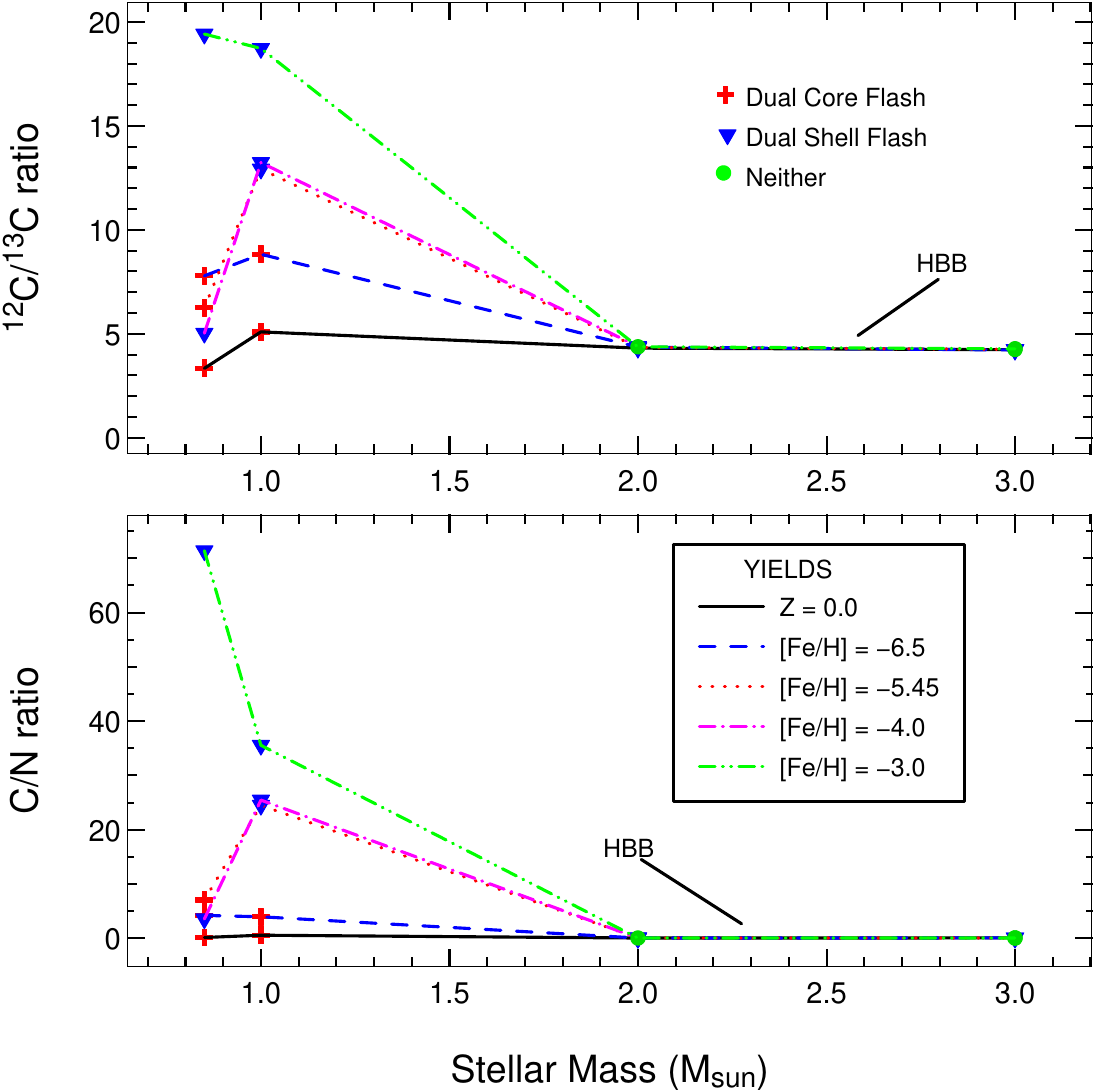}
\par\end{centering}
\caption{The same as Figure \ref{fig-YandZ-AllYields} but for the $^{12}$C/$^{13}$C
and C/N ratios in the yields. We note that the ratios in IM model
yields are all at CN-equilibrium levels (due to strong HBB) whilst
only the most metal-poor LM models are close to equilibrium, which
is due to dual core flash pollution. \label{fig-CNratios-AllYields}}
\end{figure}

In Figure \ref{fig-CNratios-AllYields} it can be seen that the $^{12}$C/$^{13}$C
and C/N ratios follow the same pattern as $Z$ and $Y$ insomuch as
they are very uniform in the IM yields but have a large spread in
the LM yields. The $^{12}$C/$^{13}$C ratios in the IM model yields
are uniformly $\approx5$, which reflects the fact that the AGB envelopes
all reached CN equilibrium due to strong HBB. The C/N ratios are also
perfectly uniform, being $\approx0.05$, for the same reason. Interestingly
this means that the yields are very N-rich. Indeed, N makes up the
largest part of Z in these yields. 

In the low mass models we see that the (initially) lower-Z models
have quite low $^{12}$C/$^{13}$C ratios, being $\lesssim10$, which
is approaching equilibrium values. In these models it is the N production
by the DCF event that causes the drop to such low values (the ratio
in the initial composition was extremely high). Interestingly in one
of the models that experienced the dual shell flash the ratio is also
very low (the 0.85 M$_{\odot},$ $\textrm{[Fe/H]}=-4$ model). This
low value was a result of the DSF pollution itself. This didn't occur
in the $\textrm{[Fe/H]}=-3$ model however, which also experienced
a DSF, but which has a relatively high ratio in the yield. In fact
the 0.85 M$_{\odot},$ $\textrm{[Fe/H]}=-4$ model is an exception
amongst all those that experienced DSFs, as the rest of them have
moderately high $^{12}$C/$^{13}$C ratios. It appears that more advanced
CN processing occurred during the DSF in this model, making the chemical
signature appear more like that of a DCF. 

In regards to the C/N ratios in the low-mass models, they follow the
same general pattern as the $^{12}$C/$^{13}$C ratios -- moderately
high in the DSF models (i.e. more the metal-rich, Group 2 models),
and quite low in the DCF models (metal-poor, Group 1 models).

\subsection{Elemental Yields: Helium to Phosphorus}

Figures \ref{fig-mmp.yields.XoHvsMass-elems-most} and \ref{fig-hmp.yields.XoHvsMass-elems-most}
show the yields of most of the elements in our network for all the
$\textrm{[Fe/H]}=-6.5$ and $\textrm{[Fe/H]}=-5.45$ models, respectively.
We have displayed the yields of each element versus initial stellar
mass in these groups of plots. This gives a quick visual overview
of how the yield of each element is dependent on mass at a constant
metallicity. The rest of the figures of this type are given in Appendix
\vref{Appx-MetalPoorAndZ0Yields-ElemVsMass}.

\subsubsection*{Helium}

\begin{figure}
\begin{centering}
\includegraphics[width=0.95\columnwidth,keepaspectratio]{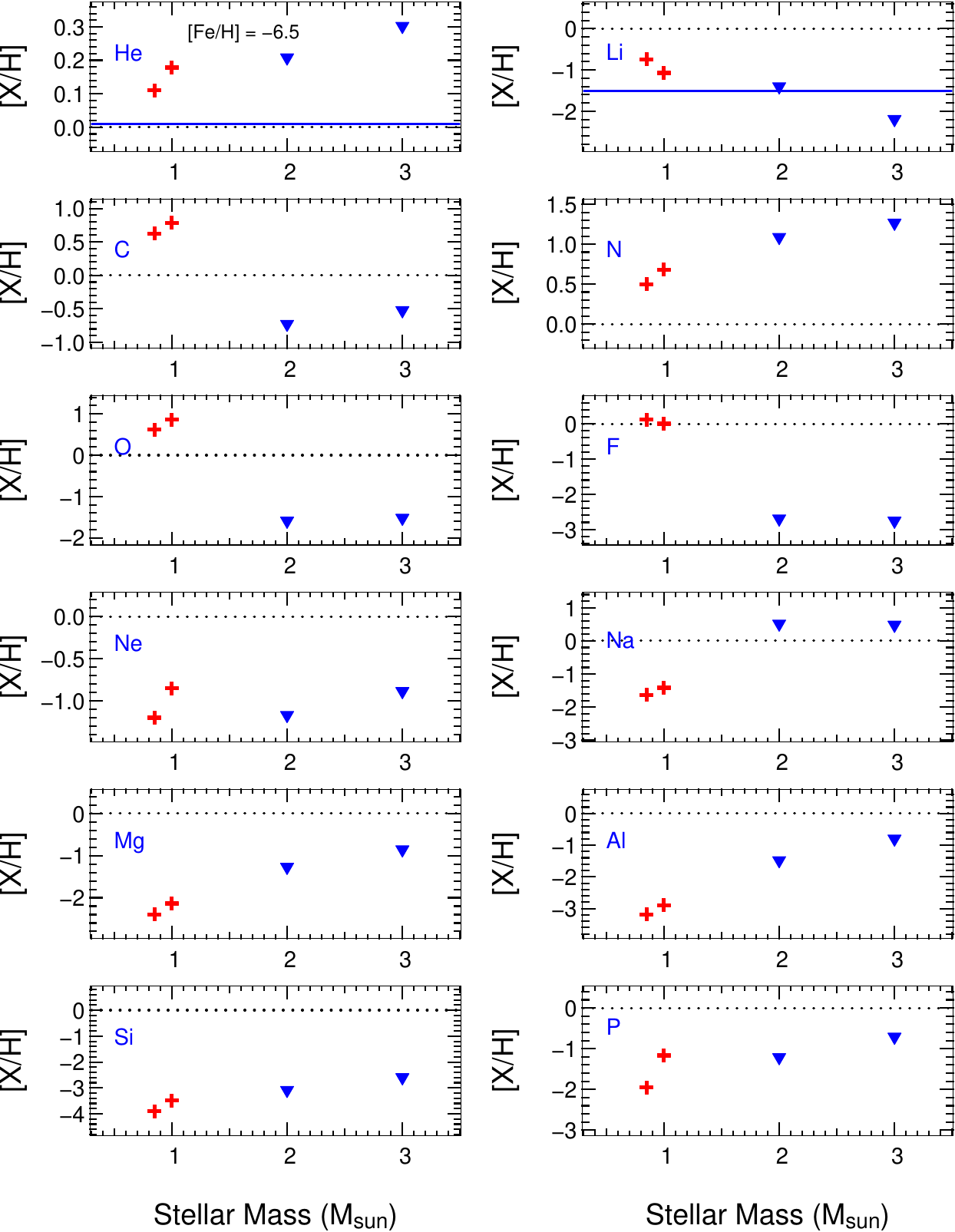}
\par\end{centering}
\caption{Selected elemental yields for all the $\textrm{[Fe/H]}=-6.5$ models,
given as relative to solar against initial stellar mass. Red crosses
indicate models that experienced a dual core flash (at the top of
the RGB), whilst blue triangles indicate models that experienced a
dual shell flash (at the beginning of the AGB). Solid horizontal lines
(blue) indicate initial abundances, but are often not visible due
to their relatively low values. The horizontal dotted (black) lines
at $\textrm{[X/H]}=0$ indicate the solar abundances (Solar abundances
are from \citealt{2003ApJ...591.1220L}). For the rest of these types
of figures see Appendix \vref{Appx-MetalPoorAndZ0Yields-ElemVsMass}.
\label{fig-mmp.yields.XoHvsMass-elems-most}}
\end{figure}

The helium yields were discussed in detail above but we give a quick
summary here, using the $\textrm{[Fe/H]}=-6.5$ and $\textrm{[Fe/H]}=-5.45$
models as examples. In the low mass models which experience the DCF
(the 0.85 and 1 M$_{\odot}$ models in Figure \ref{fig-mmp.yields.XoHvsMass-elems-most},
and the 0.85 M$_{\odot}$ model in Figure \ref{fig-hmp.yields.XoHvsMass-elems-most})
the He mainly comes from the DCF pollution. It is however diluted
by the unpolluted mass loss on the RGB in these models. In the low-mass
model which does not experience the DCF (1 M$_{\odot}$, $\textrm{[Fe/H]}=-5.45$)
the He comes from the combination of the 2DUP and DSF episodes, and
is also partially diluted by RGB mass loss. In regards to the IM mass
models the He yields are consistently high due to the combination
of 2DUP and HBB. This is true whether the model goes through a DSF
episode or not (the 3 M$_{\odot}$ in Figure \ref{fig-hmp.yields.XoHvsMass-elems-most}
did not experience a DSF). 

\subsubsection*{Lithium}

\begin{figure}
\begin{centering}
\includegraphics[width=0.95\columnwidth,keepaspectratio]{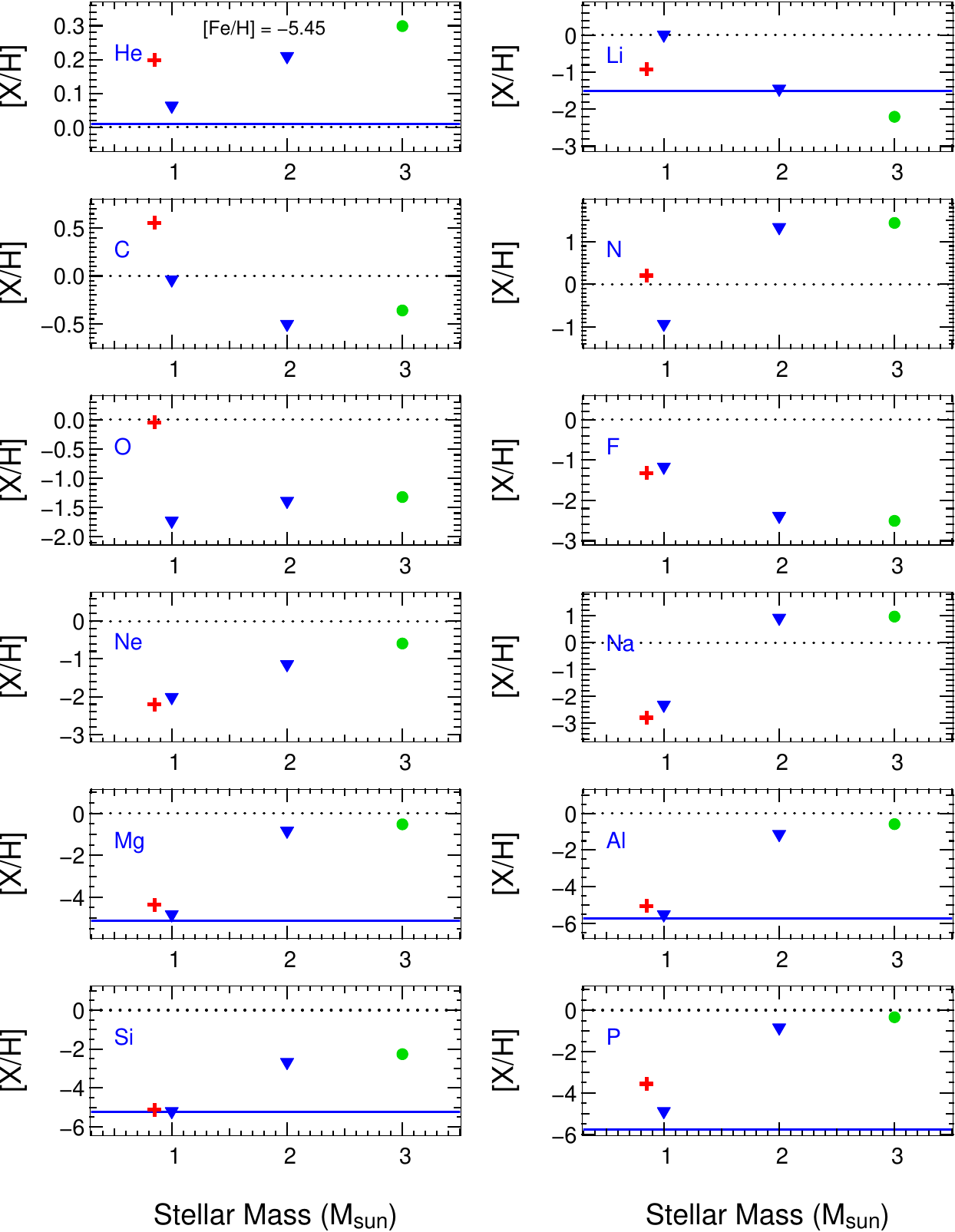}
\par\end{centering}
\caption{Same as Figure \ref{fig-mmp.yields.XoHvsMass-elems-most} but for
the $\textrm{[Fe/H]}=-5.45$ models. Green dots indicate the model
that did not experience a DCF or a DSF episode. For the rest of these
types of figures see Appendix \vref{Appx-MetalPoorAndZ0Yields-ElemVsMass}.
\label{fig-hmp.yields.XoHvsMass-elems-most}}
\end{figure}

Lithium is enhanced above the primordial value in all the low-mass
models of Figures \ref{fig-mmp.yields.XoHvsMass-elems-most} and \ref{fig-hmp.yields.XoHvsMass-elems-most},
usually by $\sim0.5$ dex. It is still below the solar value though.
However in one case (1 M$_{\odot}$, $\textrm{[Fe/H]}=-5.45$) it
has increased by 1.5 dex, to become equal to the solar abundance.
Looking at Figure \ref{fig-Lithium.yields.XoHvsFeH-Appx} in the Appendix
it can be seen that this is true of most of the 1 M$_{\odot}$ models
that experience the DSF rather than the DCF (ie. Group 2 models).
By contrast the Group 2 models of 0.85 M$_{\odot}$ have similar Li
enrichment levels as that of the DCF (Group 1) models. The source
of Li in all these low-mass models are the DCF and DSF episodes themselves.
The Li production from these events is however moderated in the yields
due to destruction on the AGB via (weak) HBB and also due to dilution
by the unpolluted RGB mass loss. In the IM mass models Li is essentially
unchanged from the primordial abundance in the 2 M$_{\odot}$ model
yields -- which is an interesting result -- but depleted in the
3 M$_{\odot}$ yields. Being such a `fragile' nuclide, in these relatively
hot models it has a complicated history of production and destruction
through the various stages of evolution. However the dominant factor
is HBB on the AGB. Here it is repetitively destroyed and created.
It is created via electron captures on $^{7}$Be, which itself is
produced by HBB (the \citealt{1971ApJ...164..111C} mechanism, also
see \citealt{1992ApJ...392L..71S}). All the models go through Li-rich
phases on the AGB, for varying lengths of time, but as evidenced by
the nett destruction in the yields, the 3 M$_{\odot}$ models are
Li-poor for more of the time. The Li-poor portion of the 2 M$_{\odot}$
winds coincidentally combine with the Li-rich portion to average to
(roughly) the primordial value. 

\subsubsection*{Carbon}

\begin{figure}
\begin{centering}
\includegraphics[width=0.95\columnwidth,keepaspectratio]{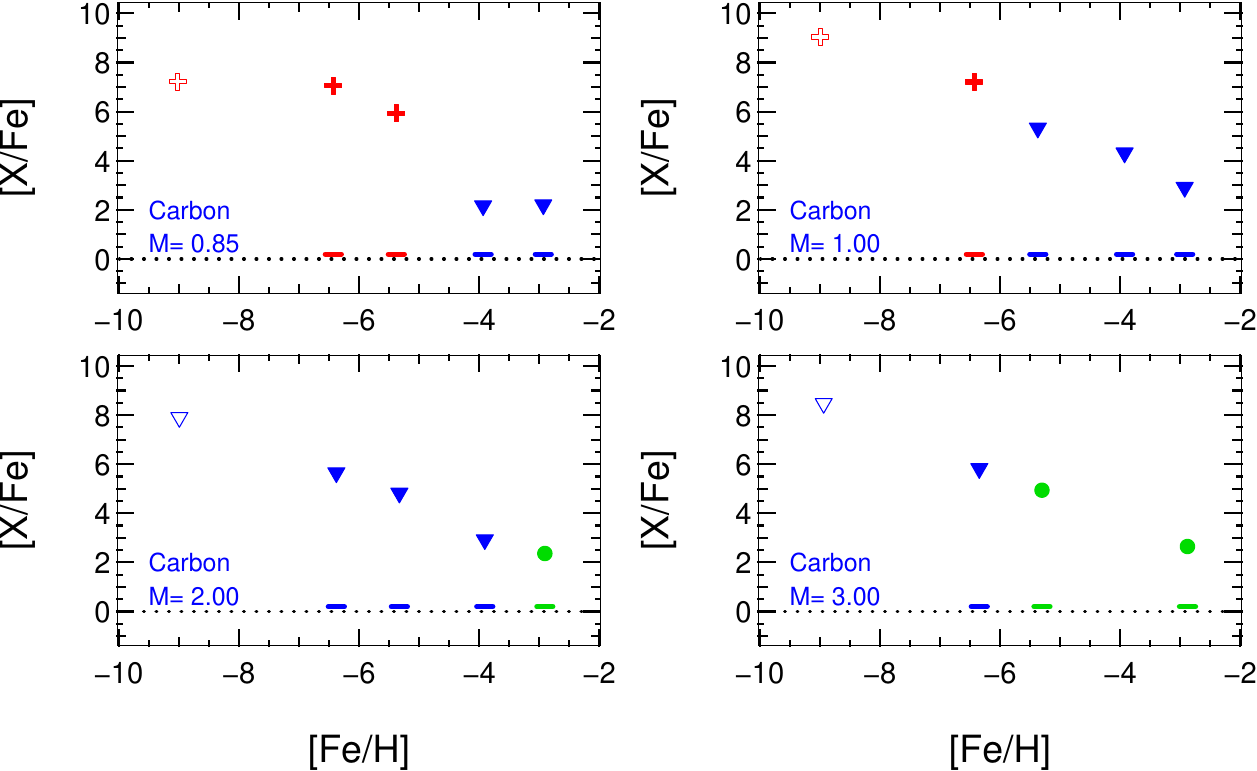}
\par\end{centering}
\begin{centering}
\includegraphics[width=0.95\columnwidth,keepaspectratio]{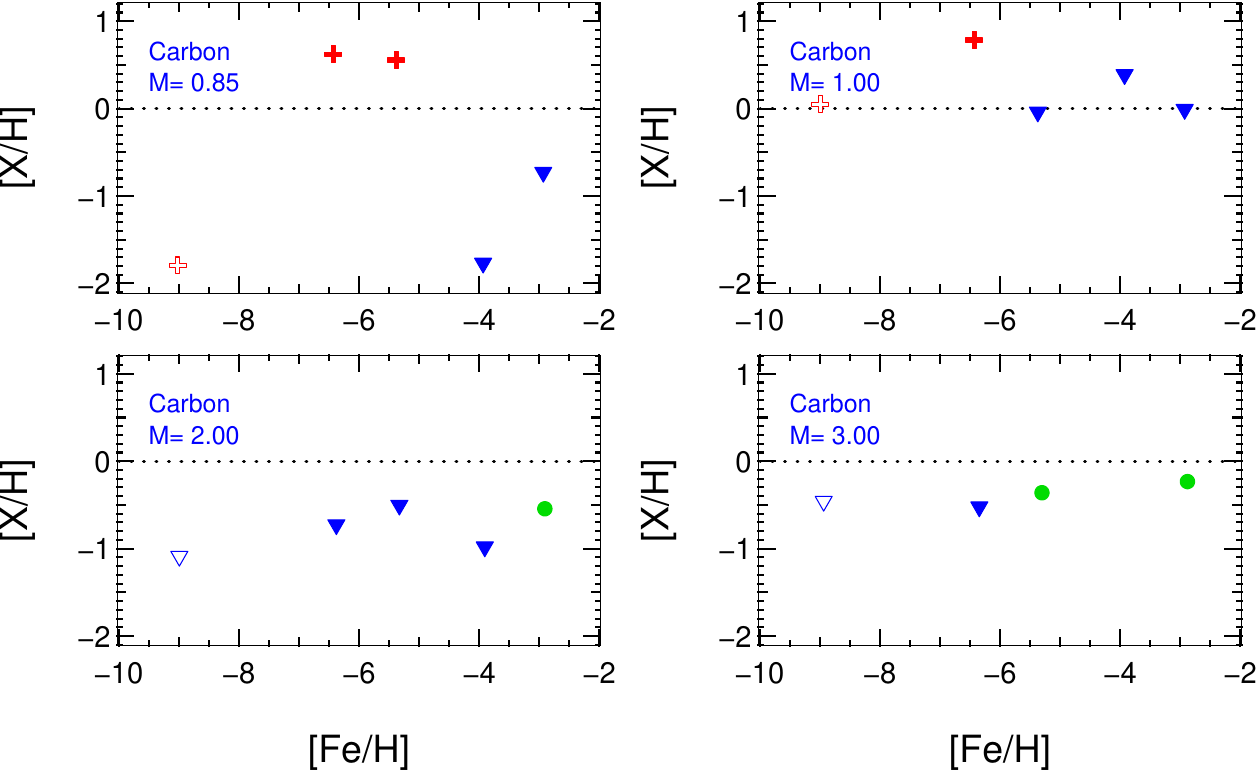}
\par\end{centering}
\caption{Carbon yields versus initial {[}Fe/H{]} for all the models in the
grid, grouped by initial mass. The $Z=0$ models have also been included
for comparison (they have been given an artificial Fe abundance corresponding
to $\textrm{[Fe/H]}=-9$ for display purposes). The top set of four
plots displays the yields as {[}X/Fe{]}, where X is the species (i.e.
carbon) whilst the bottom set displays them as {[}X/H{]}. Comparing
the two sets highlights the fact that the extremely high {[}C/Fe{]}
abundances are mainly driven by the very low Fe abundances in the
yields. Red crosses indicate models that experienced a dual core flash
(at the top of the RGB), blue triangles models that experienced a
dual shell flash (at the beginning of the AGB) and green dots indicate
models that experienced neither. Short horizontal lines indicate the
initial abundance for each model (which are well below the scale of
the bottom panel). Note that within each group of four plots the vertical
axes are identical to allow direct comparison. \label{fig-carbon.yields.XoHvsFeH}}
\end{figure}

In Figure \ref{fig-carbon.yields.XoHvsFeH} we show all the Carbon
yields as {[}C/Fe{]} versus {[}Fe/H{]} and also as {[}C/H{]} versus
{[}Fe/H{]}. We do the same for nitrogen and oxygen in Figures \ref{fig-nitrogen.yields.XoHvsFeH}
and \ref{fig-oxygen.yields.XoHvsFeH} but only for {[}X/H{]} (where
X represents the elemental species). In Appendix \vref{secAppx:Elemental-Yields-VersusFeH}
we give the same type of plots for most of the elements in the network
up to sulphur.

It can be seen in the {[}C/Fe{]} plots of Figure \ref{fig-carbon.yields.XoHvsFeH}
that carbon is extremely enhanced -- reaching values of up to $\sim9.5$
dex above solar! However, looking at the {[}C/H{]} plots it is apparent
that this is primarily driven by the very small abundances of Fe in
our metal-poor models. This highlights the fact that the abundance
levels of light elements reached in the yields of these models are,
in absolute terms, approaching solar or are super-solar. It also highlights
the problem with using {[}Fe/H{]} as a metallicity indicator. These
stars still retain their original very low values of {[}Fe/H{]} but
their Z values are now very high. So in one sense the yields are metal-poor
and in another they are metal-rich. 

In terms of {[}C/H{]}, C is overabundant (i.e. super-solar) in the
yields of all the low-mass models that experienced the DCF, except
in the case of the $Z=0$ model (which we have included artificially
at $\textrm{[Fe/H]}=-9$). The source of the C enrichment in these
models is the DCF event. In the low-mass models that don't go through
the DCF (the more metal-rich models, Group 2) carbon is less enhanced,
being significantly sub-solar in the 0.85 M$_{\odot}$ yields and
at, or slightly above, solar in the 1 M$_{\odot}$ yields. The C enhancement
in these cases is a result of the DSF at the start of the AGB. We
note that the C yields are reduced by any time the models spend losing
unpolluted matter on the RGB. Thus, in some of the models, there is
an uncertainty in the C yield due to the uncertainty of the mass loss
rates at low Z. In particular one would expect that the yield of C
would increase with a lower mass loss rate on the RGB. This is of
course true for all the elements as it is just a dilution effect.
In the IM models, which do not spend much time, if any, on the RGB,
the C yield is wholly determined by the combination of 3DUP and HBB
on the AGB. As the temperature at the base of the convective envelope
is very high (usually slightly less than $10^{8}$ K), and due to
the length of time spent on the AGB, all the IM model envelopes reach
CN equilibrium. Thus the relative abundances of C and N are fixed,
as mentioned above, with N/C $\sim20$. The underlying factors in
the degree of C (and N) pollution are thus the depth of 3DUP, the
number of 3DUP episodes (i.e. pulses) and the mass of the envelope
that is polluted. Interestingly, in Figure \ref{fig-carbon.yields.XoHvsFeH}
it can be seen that the yields are fairly constant with metallicity
but appear to increase with mass. The 2 M$_{\odot}$ yields are all
$\sim0.8\pm0.3$ dex below solar whilst the 3 M$_{\odot}$ yields
are only $\sim0.4\pm0.2$ dex below solar. It appears that a strong
C enrichment is a robust feature of all our metal-poor models. Even
though not all of them reach solar {[}C/H{]} we note that they initially
had very minimal abundances of this element. 

\subsubsection*{Nitrogen}

\begin{figure}
\begin{centering}
\includegraphics[width=0.95\columnwidth,keepaspectratio]{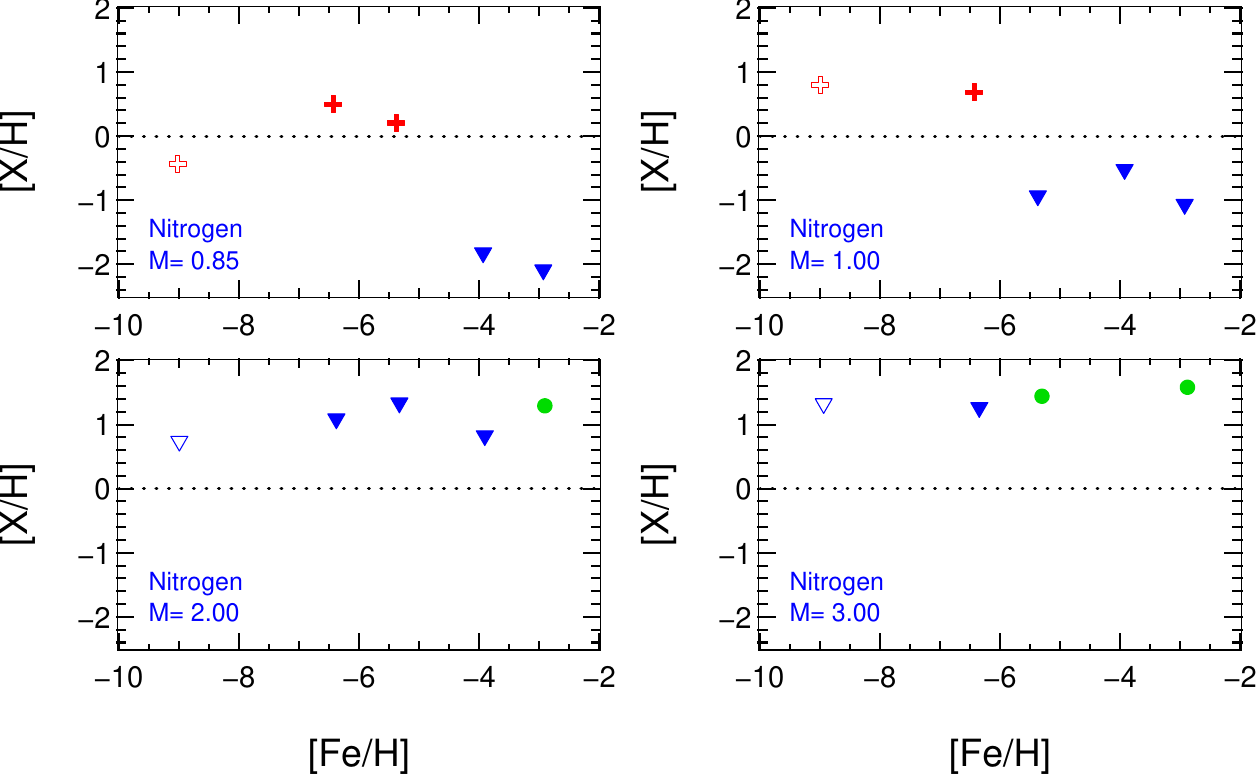}
\par\end{centering}
\caption{Same as Figure \ref{fig-carbon.yields.XoHvsFeH} except for nitrogen
and only for {[}N/H{]}. The large amount of N in the yields of the
IM models (Group 3 as described in the text) is due to the combination
of 3DUP and HBB during the AGB phase. In the low mass DCF yields (red
crosses, Group 1) the N comes from the DCF event whilst in the low
mass DSF yields (Group 2) it comes from the DSF event. \label{fig-nitrogen.yields.XoHvsFeH}}
\end{figure}

In Figure \ref{fig-nitrogen.yields.XoHvsFeH} it can be seen that
nitrogen has also been raised from very low initial levels to abundances
often super-solar. In all the low-mass models that experience the
DCF the nitrogen yields are super-solar (although slightly below in
the 0.85 M$_{\odot},$ $Z=0$ case). Again this is mainly due to the
DCF, but 2DUP also plays a part -- even in some of the 0.85 M$_{\odot}$
models. Slight HBB in the 1 M$_{\odot}$ models also increases N slightly,
at the expense of C. The situation is uniformly different in the low-mass,
higher-Z models that do not go through the DCF (i.e. Group 2). All
these models show sub-solar {[}N/H{]}. The source of N in these models
is mainly the DSF but, particularly in the 0.85 M$_{\odot}$ models,
it is heavily diluted in the yields due to unpolluted mass loss on
the RGB. Again we note that these yields would be higher if a low
mass-loss rate were used on the RGB. In regards to the IM models their
N yields, like their C yields, are quite constant with metallicity,
again with the 3 M$_{\odot}$ yields being slightly more enhanced
than the 2 M$_{\odot}$ yields. The yields are all very much super-solar,
with $\textrm{[N/H]}\sim+0.8\rightarrow+1.6$. The source of the N
here is HBB of the carbon brought up by the many 3DUP episodes. Assuming
the occurrence of 3DUP and HBB are correct in our models, this is
a very robust result for our IM yields. We note that we have used
no form of overshoot in these models, so this may represent a lower
limit to the N pollution from the low metallicity models. 

\subsubsection*{Oxygen}

\begin{figure}
\begin{centering}
\includegraphics[width=0.95\columnwidth,keepaspectratio]{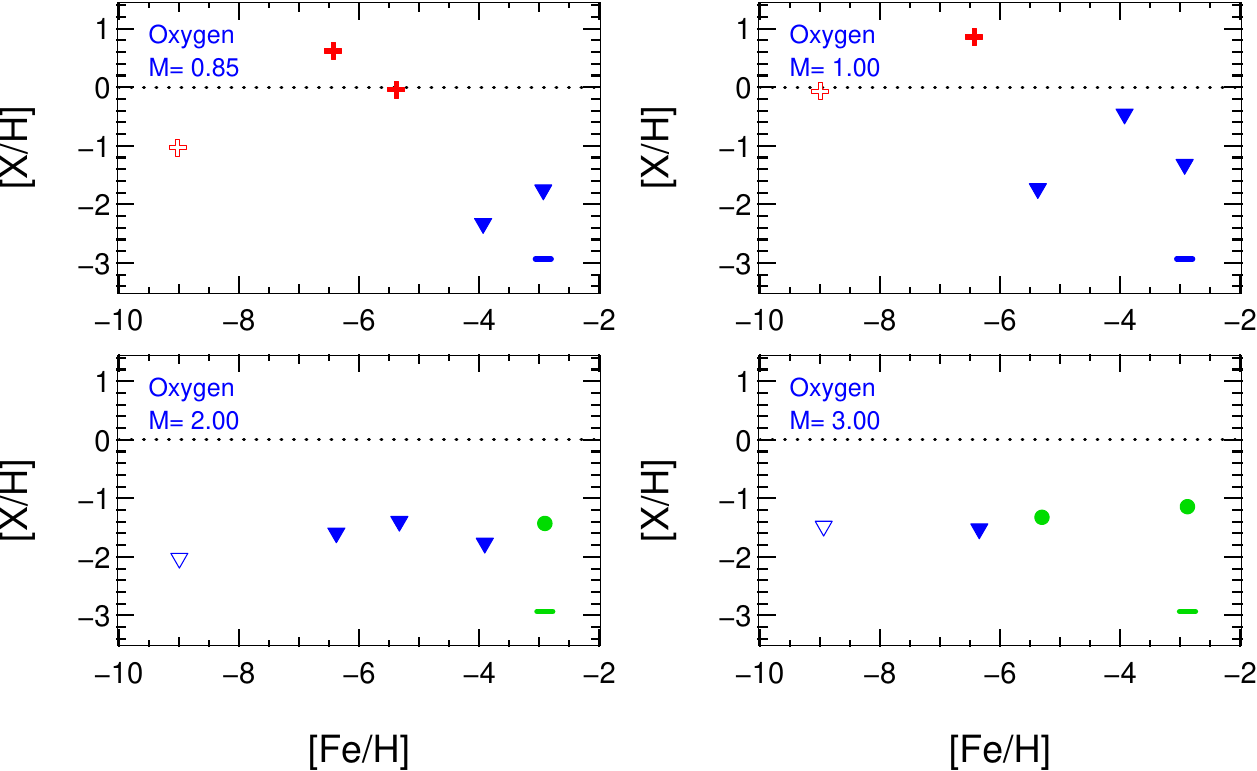}
\par\end{centering}
\caption{Same as Figure \ref{fig-carbon.yields.XoHvsFeH} except for oxygen
and only for {[}O/H{]}. The large amount of O in in the low mass DCF
yields (red crosses, Group 1) comes mainly from the DCF events. In
the low mass DSF models (Group 2) it comes from the DSF events whilst
the O in the yields of the IM models (Group 3) arises from the combination
of 3DUP and HBB during the AGB phase -- 3DUP brings up the O and
HBB destroys it. \label{fig-oxygen.yields.XoHvsFeH}}
\end{figure}

Figure \ref{fig-oxygen.yields.XoHvsFeH} shows the oxygen yields in
terms of {[}O/H{]}. Again there is a strong enrichment in all the
models. In the DCF models' yields {[}O/H{]} is between solar and 1
dex super-solar, except again for the 0.85 M$_{\odot},$ $Z=0$ model
which is 1 dex sub-solar. The main sources of the oxygen in these
models are the DCF events. Due to this single event being the main
source we note that this means that the yields are dependent on the
treatment of this event in the structural evolution code, and thus
the uncertainties therein (eg. the unknown degree of overshoot). In
the DSF low-mass models (Group 2) we see that all the O yields are
sub-solar, to varying degrees. The main sources of O here are the
DSF events but we note that the O pollution is highly diluted in some
of the models, again due to the unpolluted RGB mass-loss. Thus the
O yields in these models may be more of a lower limit. In the IM yields
the situation is more uniform. All the {[}O/H{]} yields are substantially
sub-solar, being on average $\sim1.5$ dex lower than solar in all
the models. The source of the O enrichment in these models is primarily
the AGB. The 3DUP episodes also mix up O with the C, thus repetitively
enriching the envelope. The O yields from the IM models would have
been higher if it were not for HBB cycling some of this O to N (even
at 2 M$_{\odot}$). 

\subsubsection*{Fluorine}

Moving back to Figures \ref{fig-mmp.yields.XoHvsMass-elems-most}
and \ref{fig-hmp.yields.XoHvsMass-elems-most} it can be seen that
the yields of the $\textrm{[Fe/H]}=-6.5$ and -5.45 DCF models are
enriched in fluorine also (a plot of all the fluorine yields is available
in Figure \ref{fig-fluorine.yields.XoHvsFeH} of Appendix \ref{secAppx:Elemental-Yields-VersusFeH}).
In fact the DCF group of models is the most F-enhanced group in the
grid. This is due to the production of F by the DCF event and the
fact that the F is not destroyed by HBB (to a significant degree)
at these low masses. In the low-mass DSF group, which does not experience
3DUP or significant HBB, the F yields are substantially lower. This
is due to two factors. The first is that less F pollution occurs as
a result of the DSF event, and the second is that these yields are
diluted by unpolluted RGB mass loss. Again, due to HBB, the situation
in the IM yields is quite uniform. The F yields in this group are
all substantially sub-solar, with {[}F/H{]} being $\sim-3\rightarrow-2$.
In these models F follows a similar fate to that of carbon. Its abundance
first increases as it is dredged up from the intershell after each
pulse (for a description of the F production process in the intershell
of AGB stars see eg. \citealt{2003agbs.conf.....H}). However, when
the temperature at the base of the convective envelope increases,
it is destroyed by HBB (via $^{19}$F($p,\alpha$)$^{16}$O). Hence
the yield of this element is relatively low at these masses because
of the extensive HBB. 

\subsubsection*{Neon to Phosphorus}

The main channel for neon production is $^{14}$N($\alpha,\gamma$)$^{18}$F($\beta^{+},\nu$)$^{18}$O($\alpha,\gamma$)$^{22}$Ne,
which generally occurs in the intershell of AGB stars. Indeed, in
our IM models the source of Ne in the AGB envelope is 3DUP of this
intershell $^{22}$Ne. With scaled-solar abundances $^{20}$Ne is
normally the dominant isotope of neon but in these extremely metal-poor
models $^{22}$Ne easily dominates as soon as some 3DUP occurs. The
neon is however quickly burnt in the envelope via HBB. This is reflected
in the yields, as can be seen in Figure \ref{fig-mmp.yields.XoHvsMass-elems-most}
for example, where Ne is always sub-solar (see Appendix \ref{APPX-YieldsLowZandZ0}
for more plots). We note that if 3DUP occurred without HBB then the
Ne yield would be quite high in these Group 3 models. At low mass
and metallicity (the DCF group) the Ne yield is also always sub-solar
in terms of {[}Ne/H{]}, ranging from $\sim-1\rightarrow-4$. The source
of the Ne is the DCF itself. In the low-mass models which do not experience
the DCF (Group 2) the yields are similar, but in these cases the source
is the DSF. As the base of the convective envelope in the low mass
models is not hot enough for significant HBB the Ne in the AGB envelope
survives. Thus it is apparent that the Ne production from the DCF
and DSF events is relatively small. Since the main sources of the
Ne from these models are the DCF and DSF, we again note that this
means that the yields are dependent on the treatment of these events
in the structural evolution code, and thus the uncertainties therein
(eg. the unknown degree of overshoot). 

\begin{figure}
\begin{centering}
\includegraphics[width=0.85\columnwidth,keepaspectratio]{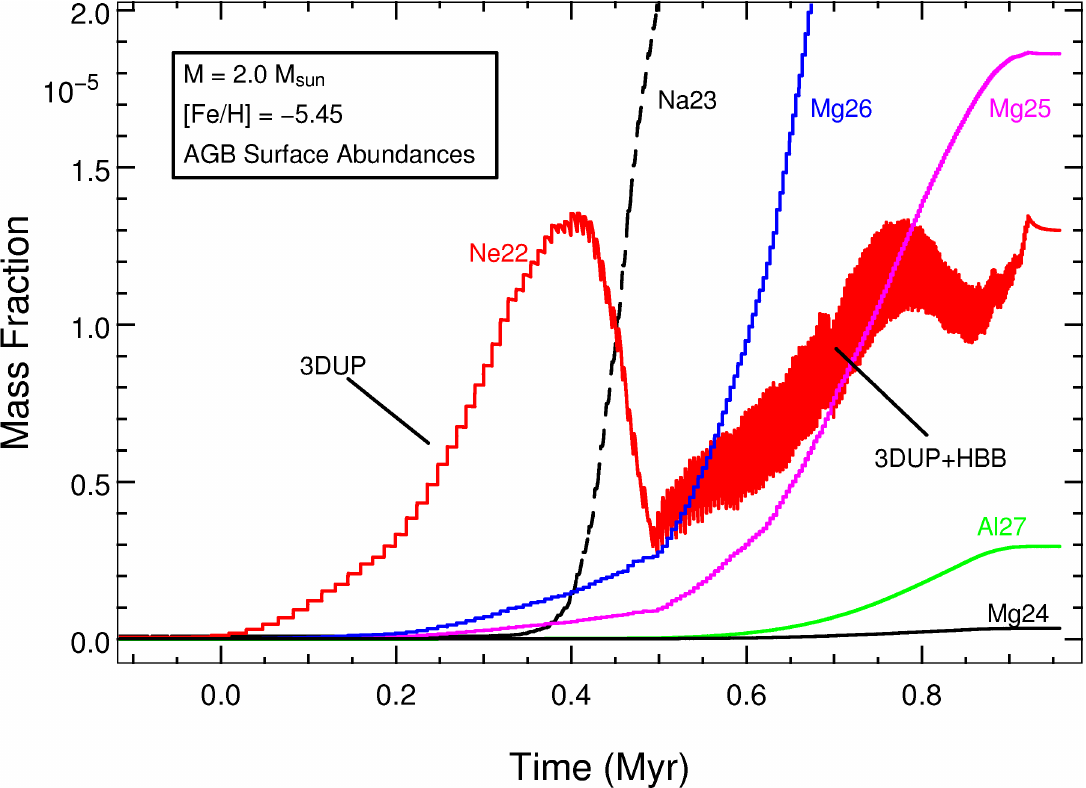}
\par\end{centering}
\caption{The evolution of the surface abundances for a selection of nuclides
involved in the Ne-Na and Mg-Al chains/cycles during the AGB phase
of the 2 M$_{\odot},$ $\textrm{[Fe/H}]=-5.45$ model. Time has been
offset. The initial enrichment of the surface with Ne due to 3DUP
can be seen, as can the activation of strong HBB which transforms
the repeatedly dredged-up Ne to Na. The increase in $^{26}$Mg and
$^{25}$Mg occurs because they are dredged up at the same time as
the $^{22}$Ne. HBB is hot enough in this model for the Mg-Al proton
capture reactions to operate, which produce $^{27}$Al from the $^{25}$Mg.
\label{fig-m2hmp-NeNaMgAl-AGBsurf}}
\end{figure}

As mentioned above the periodically dredged-up $^{22}$Ne in the IM
models is burnt to sodium in the Ne-Na cycle/chain. The main product
from these reactions is $^{23}$Na. Indeed Na is present in large
amounts in the IM yields, always being between $\sim$ solar and $1.5$
dex super-solar. As an example of the Ne-Na cycle in action we show
in Figure \ref{fig-m2hmp-NeNaMgAl-AGBsurf} the surface evolution
of some salient nuclides during the AGB phase of the 2 M$_{\odot},$
$\textrm{[Fe/H]}=-5.45$ model. The initial enrichment of the envelope
with $^{22}$Ne from 3DUP is seen, as is the subsequent destruction
via HBB as the temperature increases. We also plot some Mg isotopes
and $^{27}$Al in the same figure. The yields in the IM models are
enriched in Mg and Al, although not to the same extent as Na. In general
the Mg yield is between $\sim$ solar and $\sim2$ dex below solar.
The majority of this Mg is in the form of $^{26}$Mg and comes from
3DUP of AGB intershell material. It is produced in the He intershell
via alpha captures on $^{22}$Ne. It can also be seen that Al is enhanced
in the yields, but at an even lower level. This is apparent in all
the IM yields, where {[}Al/H{]} is always sub-solar, by about $0.5$
to 2 dex. The Al in these yields is mainly a product of HBB destruction
of $^{25}$Mg via $^{25}$Mg($p,\gamma$)$^{26}$Al($p,\gamma$)$^{27}$Si($\beta^{+}\nu$)$^{27}$Al.
The source of the $^{25}$Mg seeds is the AGB intershell (via $^{22}$Ne($\alpha,n$)$^{25}$Mg).
Thus it can be seen that the Ne$\rightarrow$Na and Mg$\rightarrow$Al
chains operate independently, insomuch as the dominant source of the
initial seeds, $^{22}$Ne and $^{25}$Mg, are both fed in from 3DUP
rather than via leakage between the chains (or from other nucleosynthetic
channels). Since the yields of these species all ultimately depend
on the degree of 3DUP they are thus dependent on the amount of overshoot
used in the modelling. Here we have used none, so the the results
of our 3DUP pollution is probably a lower limit, all else being constant.

\begin{figure}
\begin{centering}
\includegraphics[width=0.95\columnwidth,keepaspectratio]{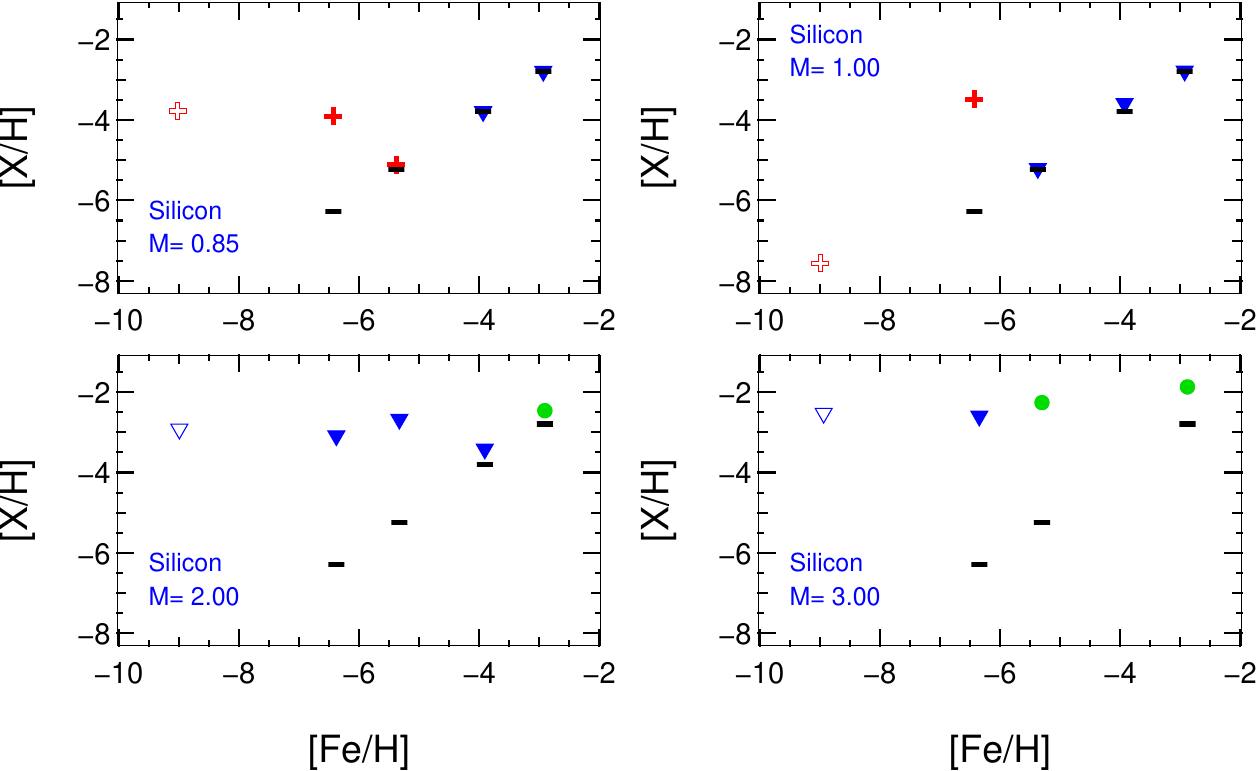}
\par\end{centering}
\caption{Same as Figure \ref{fig-carbon.yields.XoHvsFeH} except for silicon
and only for {[}X/H{]}. It can be seen that , for most of the LM models
the Si yield is practically the same as the initial abundances (shown
as short horizontal lines) -- except in some of the DCF models. In
the IM models there is always a nett production but it is still $\sim2$
to 3 dex below solar. \label{fig-silicon.yields.XoHvsFeH}}
\end{figure}

Returning to the low-mass models we see that the Na yields are similar
to the Ne yields (relative to solar). As no significant HBB occurred
in these models the source of this Na was also the DCF and DSF events.
The lack of significant amounts of advanced proton- or $\alpha$-capture
nucleosynthesis in these models becomes increasingly apparent in the
Mg and Al yields (see Appendix \ref{APPX-YieldsLowZandZ0} for the
figures). The yields for both of these elements are significantly
lower than the HBB/3DUP yields of the IM models, being at least 2
dex below solar. Indeed, in roughly half of the LM models the abundances
in the yields have hardly changed from the initial values. Looking
at the silicon yields (Figure \ref{fig-silicon.yields.XoHvsFeH})
this trend is continued. It is only the most metal-poor models that
have any significant Si production, and this only occurs during DCFs
(and not all of them). In contrast to this the IM models do all produce
Si (via 3DUP), although the yields are only between $\sim3$ and 2
dex sub-solar. Phosphorous is produced in large quantities relative
to the lighter Si in the IM models (at least relative to solar --
in absolute terms the yield is lower). In most of the yields {[}P/H{]}
is only $\sim1$ dex sub-solar. It is mainly present in the form of
$^{31}$P, which is dredged up from the AGB intershell where it is
produced via $^{27}$Al($\alpha,\gamma$)$^{31}$P. At low mass it
is produced to a moderate extent in the Group 1 (DCF) models (at least
1 dex below solar in the yields, usually lower) but hardly enhanced
above the initial values in the Group 2 (DSF) yields. 

\subsection{Nuclidic Yield Tables by Mass Fraction}

We supply the yields and initial compositions for the entire grid
of models in tabular form in Appendix \ref{AppxSec-YieldTablesNuclidicMassFrac}.
All species in the network are listed (except for neutrons). Abundances
are given in mass fraction. The stellar remnant masses are also supplied
for easy reference. Here we present just one of the tables as a sample,
the one for the $\textrm{[Fe/H]}=-6.5$ models (Table \ref{table-allMMPyields-massFrac2}).

\newpage

\scriptsize

\begin{longtable}[c]{|>{\centering}m{1.1cm}|>{\centering}m{1.8cm}||>{\centering}m{1.5cm}|>{\centering}m{1.5cm}|>{\centering}m{1.5cm}|>{\centering}m{1.5cm}|}
\caption{Yields and initial composition for all the [Fe/H]$=-6.5$
 models.
 All species in the network are listed except for neutrons.
 Abundances are in mass fraction, normalised to 1.0.
 The remnant masses (white dwarf masses) are in brackets below the initial
 stellar masses in the table header.}
\label{table-allMMPyields-massFrac2}\\ 

\hline

\textbf{Nuclide}& 

\textbf{Initial} \linebreak \textbf{\tiny{[Fe/H]$=-6.5$}}&
\textbf{0.85 M$_{\odot}$\linebreak (0.758)}& 

\textbf{1.0 M$_{\odot}$\linebreak (0.846)}& 

\textbf{2.0 M$_{\odot}$\linebreak (1.066)}&

\textbf{3.0 M$_{\odot}$\linebreak (1.107)}\tabularnewline 

\hline \hline

\endfirsthead

\multicolumn{6}{c}{{\bfseries \tablename\ \thetable{} -- continued from previous page}} \tabularnewline 

\hline  

\textbf{Nuclide}& 

\textbf{Initial} \linebreak \textbf{\tiny{[Fe/H]$=-6.5$}}& 

\textbf{0.85 M$_{\odot}$\linebreak (0.758)}& 

\textbf{1.0 M$_{\odot}$\linebreak (0.846)}& 

\textbf{2.0 M$_{\odot}$\linebreak (1.066)}&

\textbf{3.0 M$_{\odot}$\linebreak (1.107)}\tabularnewline 

\hline \hline

\endhead

\hline \multicolumn{6}{|r|}{{Continued on next page...}} \\ \hline 

\endfoot

\hline \hline 

\endlastfoot

\hline
h1 &   0.7548 & 0.686 & 0.642   & 0.656   & 0.604\tabularnewline
\hline 
h2 & 0.00E+00 & 3.44E-19 & 2.08E-18 & 3.99E-18 & 3.53E-18\tabularnewline
\hline 
he3 & 7.85E-06 & 1.33E-05 & 1.62E-04 & 1.61E-07 & 1.72E-08\tabularnewline
\hline 
he4 & 0.2450 & 0.280 & 0.307 & 0.336 & 0.384\tabularnewline
\hline 
li7 & 3.13E-10 & 1.65E-09 & 7.22E-10 & 3.46E-10 & 5.14E-11\tabularnewline
\hline 
be7 & 0.00E+00 & 5.26E-16 & 7.96E-15 & 6.63E-14 & 3.71E-14\tabularnewline
\hline 
b8 & 0.00E+00 & 5.04E-26 & 4.00E-27 & 1.04E-26 & 5.32E-27\tabularnewline
\hline 
c12 & 1.30E-09 & 7.59E-03 & 1.04E-02 & 2.99E-04 & 4.38E-04\tabularnewline
\hline 
c13 & 1.33E-16 & 9.74E-04 & 1.18E-03 & 6.86E-05 & 1.04E-04\tabularnewline
\hline 
c14 & 0.00E+00 & 8.58E-07 & 2.07E-07 & 4.84E-11 & 6.76E-11\tabularnewline
\hline 
n13 & 0.00E+00 & 0.00E+00 & 0.00E+00 & 0.00E+00 & 0.00E+00\tabularnewline
\hline 
n14 & 2.27E-16 & 2.03E-03 & 2.92E-03 & 7.60E-03 & 1.06E-02\tabularnewline
\hline 
n15 & 1.30E-18 & 7.36E-06 & 8.92E-06 & 3.09E-07 & 7.02E-07\tabularnewline
\hline 
o14 & 0.00E+00 & 0.00E+00 & 0.00E+00 & 0.00E+00 & 0.00E+00\tabularnewline
\hline 
o15 & 0.00E+00 & 0.00E+00 & 0.00E+00 & 7.86E-37 & 0.00E+00\tabularnewline
\hline 
o16 & 2.23E-09 & 2.10E-02 & 3.03E-02 & 1.32E-04 & 1.42E-04\tabularnewline
\hline 
o17 & 3.39E-17 & 1.37E-03 & 6.21E-03 & 3.45E-07 & 5.70E-07\tabularnewline
\hline 
o18 & 2.26E-16 & 8.12E-05 & 1.64E-05 & 1.47E-10 & 2.30E-10\tabularnewline
\hline 
o19 & 0.00E+00 & 0.00E+00 & 0.00E+00 & 0.00E+00 & 0.00E+00\tabularnewline
\hline 
f17 & 0.00E+00 & 0.00E+00 & 0.00E+00 & 1.41E-42 & 0.00E+00\tabularnewline
\hline 
f18 & 0.00E+00 & 0.00E+00 & 0.00E+00 & 2.79E-34 & 0.00E+00\tabularnewline
\hline 
f19 & 3.92E-18 & 4.90E-07 & 3.55E-07 & 7.24E-10 & 5.80E-10\tabularnewline
\hline 
f20 & 0.00E+00 & 0.00E+00 & 0.00E+00 & 0.00E+00 & 0.00E+00\tabularnewline
\hline 
ne19 & 0.00E+00 & 0.00E+00 & 0.00E+00 & 0.00E+00 & 0.00E+00\tabularnewline
\hline 
ne20 & 9.24E-10 & 3.34E-05 & 6.73E-05 & 6.12E-05 & 1.11E-04\tabularnewline
\hline 
ne21 & 2.06E-13 & 2.48E-06 & 6.69E-06 & 1.04E-08 & 2.60E-09\tabularnewline
\hline 
ne22 & 4.64E-14 & 3.14E-05 & 6.66E-05 & 5.05E-06 & 5.36E-06\tabularnewline
\hline 
na21 & 0.00E+00 & 0.00E+00 & 0.00E+00 & 0.00E+00 & 0.00E+00\tabularnewline
\hline 
na22 & 0.00E+00 & 3.24E-30 & 1.92E-17 & 4.83E-12 & 2.30E-11\tabularnewline
\hline 
na23 & 1.23E-11 & 7.33E-07 & 1.14E-06 & 9.86E-05 & 8.39E-05\tabularnewline
\hline 
na24 & 0.00E+00 & 0.00E+00 & 0.00E+00 & 6.93E-37 & 0.00E+00\tabularnewline
\hline 
mg23 & 0.00E+00 & 0.00E+00 & 0.00E+00 & 0.00E+00 & 0.00E+00\tabularnewline
\hline 
mg24 & 4.13E-10 & 4.81E-07 & 4.82E-07 & 1.89E-07 & 2.08E-07\tabularnewline
\hline 
mg25 & 4.49E-12 & 3.22E-07 & 2.44E-07 & 5.74E-06 & 1.39E-05\tabularnewline
\hline 
mg26 & 4.95E-12 & 1.68E-06 & 3.53E-06 & 2.62E-05 & 6.25E-05\tabularnewline
\hline 
mg27 & 0.00E+00 & 0.00E+00 & 0.00E+00 & 0.00E+00 & 0.00E+00\tabularnewline
\hline 
al25 & 0.00E+00 & 0.00E+00 & 0.00E+00 & 0.00E+00 & 0.00E+00\tabularnewline
\hline 
al-6 & 0.00E+00 & 1.66E-12 & 2.48E-11 & 4.16E-07 & 1.17E-06\tabularnewline
\hline 
al{*}6 & 0.00E+00 & 0.00E+00 & 0.00E+00 & 0.00E+00 & 0.00E+00\tabularnewline
\hline 
al27 & 1.00E-11 & 3.47E-08 & 6.31E-08 & 1.25E-06 & 6.22E-06\tabularnewline
\hline 
al28 & 0.00E+00 & 0.00E+00 & 0.00E+00 & 0.00E+00 & 0.00E+00\tabularnewline
\hline 
si27 & 0.00E+00 & 0.00E+00 & 0.00E+00 & 0.00E+00 & 0.00E+00\tabularnewline
\hline 
si28 & 3.75E-10 & 6.64E-08 & 1.62E-07 & 4.00E-07 & 1.16E-06\tabularnewline
\hline 
si29 & 3.95E-12 & 1.17E-08 & 2.77E-08 & 1.01E-07 & 2.74E-07\tabularnewline
\hline 
si30 & 2.78E-12 & 5.61E-09 & 1.45E-08 & 2.55E-08 & 6.90E-08\tabularnewline
\hline 
si31 & 0.00E+00 & 0.00E+00 & 0.00E+00 & 1.70E-41 & 0.00E+00\tabularnewline
\hline 
si32 & 0.00E+00 & 3.02E-16 & 2.30E-17 & 6.52E-12 & 4.98E-12\tabularnewline
\hline 
si33 & 0.00E+00 & 0.00E+00 & 0.00E+00 & 0.00E+00 & 0.00E+00\tabularnewline
\hline 
p29 & 0.00E+00 & 0.00E+00 & 0.00E+00 & 0.00E+00 & 0.00E+00\tabularnewline
\hline 
p30 & 0.00E+00 & 0.00E+00 & 0.00E+00 & 0.00E+00 & 0.00E+00\tabularnewline
\hline 
p31 & 1.04E-12 & 6.82E-08 & 3.91E-07 & 3.59E-07 & 1.05E-06\tabularnewline
\hline 
p32 & 0.00E+00 & 1.17E-19 & 8.91E-21 & 2.53E-15 & 1.93E-15\tabularnewline
\hline 
p33 & 0.00E+00 & -3.08E-44 & 0.00E+00 & 1.17E-23 & 4.58E-22\tabularnewline
\hline 
p34 & 0.00E+00 & 0.00E+00 & 0.00E+00 & 0.00E+00 & 0.00E+00\tabularnewline
\hline 
s32 & 1.65E-10 & 2.61E-08 & 2.05E-07 & 7.95E-08 & 2.47E-07\tabularnewline
\hline 
s33 & 7.78E-13 & 1.95E-09 & 2.86E-09 & 9.74E-09 & 1.63E-08\tabularnewline
\hline 
s34 & 5.30E-11 & 3.19E-09 & 6.39E-09 & 6.35E-09 & 1.56E-08\tabularnewline
\hline 
s35 & 0.00E+00 & 1.16E-07 & 6.94E-07 & 3.22E-08 & 9.27E-08\tabularnewline
\hline 
fe56 & 4.54E-10 & 4.13E-10 & 3.86E-10 & 4.42E-10 & 4.39E-10\tabularnewline
\hline 
fe57 & 1.37E-11 & 1.25E-11 & 1.17E-11 & 1.35E-11 & 1.34E-11\tabularnewline
\hline 
fe58 & 2.79E-18 & 8.97E-15 & 1.33E-15 & 4.87E-13 & 5.27E-13\tabularnewline
\hline 
fe59 & 0.00E+00 & 0.00E+00 & 0.00E+00 & 0.00E+00 & 5.86E-31\tabularnewline
\hline 
fe60 & 0.00E+00 & 5.60E-16 & 1.19E-16 & 3.35E-13 & 4.94E-13\tabularnewline
\hline 
fe61 & 0.00E+00 & 0.00E+00 & 0.00E+00 & 0.00E+00 & 0.00E+00\tabularnewline
\hline 
co59 & 7.22E-12 & 6.57E-12 & 6.14E-12 & 7.29E-12 & 7.24E-12\tabularnewline
\hline 
co60 & 0.00E+00 & 4.87E-20 & 1.04E-20 & 2.92E-17 & 4.30E-17\tabularnewline
\hline 
co61 & 0.00E+00 & 0.00E+00 & 0.00E+00 & 0.00E+00 & 0.00E+00\tabularnewline
\hline 
ni58 & 6.41E-12 & 5.83E-12 & 5.46E-12 & 6.24E-12 & 6.20E-12\tabularnewline
\hline 
ni59 & 0.00E+00 & 1.01E-18 & 1.93E-24 & 4.61E-19 & 1.06E-18\tabularnewline
\hline 
ni60 & 7.43E-15 & 4.18E-14 & 1.66E-14 & 1.44E-12 & 1.53E-12\tabularnewline
\hline 
ni61 & 4.00E-17 & 4.69E-11 & 7.68E-11 & 1.06E-11 & 1.40E-11\tabularnewline
\hline 
ni62 & 2.38E-20 & 2.04E-20 & 1.28E-20 & 1.99E-24 & 2.05E-25\tabularnewline
\hline 
g & 4.67E-16 & 6.35E-08 & 3.16E-07 & 1.41E-09 & 6.38E-10\tabularnewline
\hline 
\end{longtable}

\normalsize

\section{Comparison with Previous Studies\label{section-SummaryAndCompare-HaloMods}}

First we note that we have made detailed comparisons of our $0.85$
and 2 M$_{\odot},$ $Z=0$ models with with those in the literature
in Sections \ref{subsec-m0.85z0-ComparePrevStudies} (on page \pageref{subsec-m0.85z0-ComparePrevStudies})
and \ref{subsec-m2z0-ComparePrevStudies} (page \pageref{subsec-m2z0-ComparePrevStudies}).
Given the small sets of models we have found to compare with below,
and the fact that these models suffer the same `peculiar' evolutionary
traits as the $Z=0$ models, we suggest that reading the aforementioned
sections in conjunction with this one would be advantageous.

\subsubsection*{Background}

In the $\sim20$ years after \citet{1974ApJ...191..173W} calculated
his grid of metal-poor models there were only two studies that explored
this low-mass, low-metallicity regime again -- \citet{1986MNRAS.220..529T}
and \citet{1993ApJS...88..509C}. This was mainly due to the paucity
of observational evidence for such low metallicity stars (although
we do note that much work was done on Pop III, $Z=0$ modelling in
the same time period). Indeed, in 1993 \citeauthor{1993ApJS...88..509C}
noted that there was ``... a lack of metal-deficient stars found
in the Galactic halo ...''. They did however note that this may be
due to the restricted observational sample at the time. Observations
of lower and lower metallicity (i.e. $\textrm{[Fe/H]}<-3$) halo stars
started to be reported in the 80s and 90s (eg. \citealt{1984ApJ...285..622B};
\citealt{1990AA...228..426M}). However it was the HK survey (\citealt{BPS85},
\citeyear{BPS92}) that began to discover significant numbers of halo
stars with $\textrm{[Fe/H]}<-3.0$, fleshing out the low metallicity
tail of the halo metallicity distribution function. More detailed
follow-up observations started to reveal that there is a subclass
of extremely metal-poor halo stars (EMPHs) that have large carbon
(and nitrogen) overabundances (i.e. higher than solar, relative to
Fe) -- the CEMPs. These observations sparked a flurry of theoretical
modelling in the late 90s and early 2000s. This increase in modelling
can be seen in Table \ref{table-LowZ-LitReview-HaloSubset} where
we give a summary of the literature on this particular topic of low
metallicity stellar evolution at low- and intermediate-mass.

In this section we first compare our structural evolution results
with those in the literature. Then we compare results from the primary
low-metallicity polluting events (the dual core flash and dual shell
flash) and finally provide a summary/overview of the expected polluting
affects of the yields from these models. Note that we have chosen
a subset of our models with which to make comparisons, partly for
brevity and partly because of the fact that there are not very many
models to compare with (we chose subsets that had the most overlap
with the models available in the literature). We compare our results
with observations of the CEMPs in the next section.

\begin{sidewaystable}

\begin{center}

\normalsize

\begin{tabular}{|c|c|c|c|c|c|}
\hline 
 Year &  Author & Mass & Metallicity & Max. Evolution & Helium\tabularnewline
\hline 
\hline 
1974 & \citeauthor{1974ApJ...191..173W} & $0.65\rightarrow2.5$ & $-6,-4,-2$ & RGB & 0.26\tabularnewline
\hline 
1986 & \citeauthor{1986MNRAS.220..529T} & $2.5\rightarrow8.0$ & $-6,-4,-2$ & EAGB  & 0.2\tabularnewline
\hline 
1993 & \citeauthor{1993ApJS...88..509C} & $0.7\rightarrow15$ & $-8,-4,-2$ & RGB/EAGB & 0.23\tabularnewline
\hline 
1996 & \citeauthor{1996ApJ...459..298C} & $0.7\rightarrow1.1$ & $-8,-4,-3$ & $\sim$DSF (M=0.8) & 0.23\tabularnewline
\hline 
2000 & \citeauthor{2000ApJ...529L..25F} & $0.8\rightarrow4$ & zero,$-4,-2$ & DCF/DSF & 0.23\tabularnewline
\hline 
2004 & \citeauthor{2004ApJ...602..377I} & $1\rightarrow3$ & -2.7 & DSF, $\sim$AGB & 0.24\tabularnewline
\hline 
2004 & \citeauthor{2004ApJ...609.1035P} & $0.8\rightarrow1.5$ & zero,$-6,-5,-4$ & DCF, SRGB, $\sim$EAGB & 0.23 \& 0.27\tabularnewline
\hline 
2004 & \citeauthor{2004AA...422..217W} & 0.82 & zero, -5 & DCF, SRGB & 0.23?\tabularnewline
\hline 
2006 & Campbell (This study) & $0.8\rightarrow3+$\footnote{Some models with $M=4$ \& 5 M$_{\odot}$ were also calculated but
only part of the way through the AGB.} & zero, $-6,-5,-4,-3$ & DCF, DSF, SRGB, AGB \& Yields & 0.245\tabularnewline
\hline 
\end{tabular}

\end{center}

\caption{A summary of the literature (to the best of our knowledge) for theoretical
studies of low- and intermediate-mass stars of very low metallicity.
This is basically a subset of the literature review table in Section
\ref{Section-PreviousModels} (Table \vref{table-LowZ-LitReview})
. Mass is given in units of M$_{\odot}$ (an arrow indicates a range
of masses). Metallicity is given as $\textrm{log}_{10}(Z/Z_{\odot})$
except when $Z=0$ where we write `zero'. Using $\textrm{log}_{10}(Z/Z_{\odot})$
gives a rough approximation to {[}Fe/H{]}, which is useful for comparisons
with observational studies. In column 5 we show the maximum stage
of evolution that the models were evolved \emph{through} (so it is
an inclusive designation). The `$\sim$' symbol is used where the
models were evolved only part of the way through the indicated phase
of evolution. In the last column we show the helium mass fraction
used. {[}Abbrevations for the evolutionary stages are: CHeB (core
helium burning), DCF (dual core flash, the H-He core flash in low
mass low-Z stars, also known as HEFM in the literature), SRGB (secondary
RGB), DSF (dual shell flash, the H-He shell flash near the start of
the AGB in low- and intermediate-mass stars, also known as HCE).{]}\label{table-LowZ-LitReview-HaloSubset}}

\end{sidewaystable}

\normalsize

\subsection{Low Mass Structural Evolution}

We define low mass models as models that go through the degenerate
core He flash (or DCF) at the top of the RGB. At $Z=0$ the boundary
is at roughly 1.2 M$_{\odot}$ (\citealt{2001AA...371..152M}). At
very low metallicity ($\textrm{[Fe/H]}\sim-4$) it appears to be slightly
higher, at $\sim1.4$ M$_{\odot}$ (\citealt{1993ApJS...88..509C}).
Our models are consistent with these values, as all of our 1 M$_{\odot}$
models experience the core He flash whilst none of our 2 M$_{\odot}$
models do. A finer grid of models would be required to pinpoint the
transition at each metallicity.

\subsubsection*{Extremely Low Metallicity}

Here we compare our 0.85 and 1.0 M$_{\odot},$ $\textrm{[Fe/H]}=-5.45$
models with models of similar metallicity and mass from the literature.
In the top panel of Table \ref{Table-CompareStudies-LowZ-LowMass}
we display a selection of evolutionary characteristics for these models.
It can be seen that the intersection of our chosen comparison models
and those of comparable mass and metallicity (and which give the required
comparison values) from the literature is quite small. Indeed, for
the 1 M$_{\odot}$ case we find only one study (\citealt{1974ApJ...191..173W})
to compare with. In this case the luminosity at the top of the RGB
and the MS lifetime agree reasonably well, considering there is more
that 30 years between the studies. The RGB lifetime is however very
different between the models. Our lifetime is 131 Myr whilst Wagner's
is 76 Myr -- almost a factor of two shorter. This may be a result
of the updated physics in our model but it may also be a result of
RGB lifetime definition. We define it as the length of time from the
end of the Hertzsprung gap to the time of core He ignition. We note
that our model experienced a dual shell flash (DSF) at the beginning
of the AGB. \citealt{1974ApJ...191..173W} do not report a DSF as
they were probably unable to evolve through the He flash at that stage. 

In the 0.85 M$_{\odot},$ $\textrm{[Fe/H]}=-5.45$ case there are
more studies to compare to. The key finding is that \citet{2004ApJ...609.1035P}
report that a dual core flash (DCF) occurs in their 0.80 M$_{\odot},$
$\textrm{[Fe/H]}=-6$ model, as it does in our 0.85 M$_{\odot}$ model.
\citet{2004AA...422..217W} did not find this event to occur in their
0.81 M$_{\odot},$ $\textrm{[Fe/H]}=-5.3$ model but this is expected
because the initial composition of their model was heavily enriched
in carbon and oxygen (eg. $\textrm{[C/Fe]}=4$), giving it an increased
metallicity as defined by $Z_{cno}$. The luminosity at the top of
the RGB in the \citet{2004ApJ...609.1035P} model is exactly the same
as ours. However their RGB lifetime is very different -- it is a
factor of four shorter than ours. We are unsure of the reason for
this but note that the \citealt{1974ApJ...191..173W} lies between
our lifetime and that of \citeauthor{2004ApJ...609.1035P}. Despite
this the core masses at core He ignition are all quite similar between
the studies.

\begin{table}
\begin{center}\begin{threeparttable}

\scriptsize

\begin{tabular}{ccccccccc}
\multicolumn{9}{c}{Metallicity$=-5.45$}\tabularnewline
\hline 
\hline 
Study & M$_{*}$ & {[}Fe/H{]} & $\tau_{MS}$ & $\tau_{RGB}$ & $L_{tip}$ & M$_{c}$ & DF & $Z_{DF}$\tabularnewline
\hline 
\citet{1974ApJ...191..173W} & 0.85 & $-6$ & 8.2 & 140 & 2.6 & -- & -- & --\tabularnewline
\citet{2004ApJ...609.1035P} & 0.80 & $-6$ & -- & 47 & 2.8 & 0.54 & \textcolor{red}{C} & 2E-2\tabularnewline
\citet{2004AA...422..217W} & 0.81 & $-5.3$ & -- & -- & -- & -- & \textcolor{darkgreen}{N}\tnote{a} & --\tabularnewline
\textit{This study} & 0.85 & $-5.5$ & 10 & 205 & 2.8 & 0.53 & \textcolor{red}{C} & 3E-2\tabularnewline
\hline 
\citet{1974ApJ...191..173W} & 1.0 & $-6$ & 4.7 & 76 & 2.6 & -- & -- & --\tabularnewline
\textit{This study} & 1.0 & $-5.5$ & 5.7 & 131 & 2.8 & 0.52 & \textcolor{blue}{S} & 4E-3\tabularnewline
\hline 
 &  &  &  &  &  &  &  & \tabularnewline
\multicolumn{9}{c}{Metallicity$=-4.0$}\tabularnewline
\hline 
Study & M$_{*}$ & {[}Fe/H{]} & $\tau_{MS}$ & $\tau_{RGB}$ & $L_{tip}$ & M$_{c}$ & DF & $Z_{DF}$\tabularnewline
\hline 
\citet{1974ApJ...191..173W} & 0.85 & $-4$ & 7.9 & 160 & 2.9 & -- & -- & --\tabularnewline
\citet{1996ApJ...459..298C} & 0.80 & $-4$ & -- & -- & 3.1 & 0.53 & -- & --\tabularnewline
\citet{2000ApJ...529L..25F} & 0.80 & $-4$ & -- & -- & -- & -- & \textcolor{blue}{S} & 6E-3\tabularnewline
\textit{This study} & 0.85 & $-4$ & 10 & 220 & 3.0 & 0.52 & \textcolor{blue}{S} & 1E-3\tabularnewline
\hline 
\citet{1974ApJ...191..173W} & 1.0 & $-4$ & 4.5 & 93 & 2.8 & -- & -- & --\tabularnewline
\citet{1993ApJS...88..509C} & 1.0 & $-4$ & 6.5 & -- & 3.0 & 0.52 & -- & --\tabularnewline
\citet{1996ApJ...459..298C} & 1.0 & $-4$ & -- & -- & 3.0 & 0.52 & -- & --\tabularnewline
\citet{2000ApJ...529L..25F} & 1.0 & $-4$ & -- & -- & -- & -- & \textcolor{blue}{S} & --\tabularnewline
\citet{2004ApJ...602..377I} & 1.0 & $-2.7$ & -- & -- & 3.2 & -- & \textcolor{blue}{S} & 1E-2\tabularnewline
\textit{This study} & 1.0 & $-4$ & 5.7 & 150 & 3.0 & 0.51 & \textcolor{blue}{S} & 1E-2\tabularnewline
\hline 
\end{tabular}

\caption{Comparing some of our low-mass models with those in the literature.
The top panel displays characteristic values for our low-mass $\textrm{[Fe/H]}=-5.45$
models and those with comparable mass and metallicity from the literature.
The second panel does the same for our $\textrm{[Fe/H]}=-4$ low-mass
models. Column values are: M$_{*}$ (initial mass of the model), $\tau_{MS}$
(main sequence lifetime, in Gyr), $\tau_{RGB}$ (RGB lifetime, in
Myr), $L_{tip}$ (the maximum luminosity attained on the RGB, in log($L/L_{\odot}$),
M$_{c}$ (mass of the H-exhausted core at the time of the core He
flash), $DF$ (which type of flash the model experienced, if any --
Core, Shell or Neither) and $Z_{DF}$ (the $Z$ metallicity of the
surface after the dual flash episode, whether it be a DCF or DSF).
A dash means that the information was either not supplied in the paper
or not relevant to the model. \label{Table-CompareStudies-LowZ-LowMass}}

\line(1,0){100}

\begin{tablenotes}\scriptsize

\item[a]{This model had an initial composition with [C/Fe]$=4$ and [O/Fe]$=4$ so the Z-defined metallicty was actually relatively high.}

\end{tablenotes}

\end{threeparttable} \end{center}
\end{table}

\normalsize

\subsubsection*{Very Low Metallicity}

In panel 2 of Table \ref{Table-CompareStudies-LowZ-LowMass} we show
comparisons for our 0.85 and 1.0 M$_{\odot},$ $\textrm{[Fe/H]}=-4.0$
models. At this higher metallicity we have been more fruitful in finding
comparison models. We begin with the 0.85 M$_{\odot}$ comparisons.
The key finding here is that \citet{2000ApJ...529L..25F} report a
DSF to occur in their 0.80 M$_{\odot}$ model, as we do in our model.
\citet{1996ApJ...459..298C} and \citet{1974ApJ...191..173W} do not
report DSFs but we note that they do not evolve through the core He
flash (\citet{1996ApJ...459..298C} create a zero-age horizontal branch
model for the further evolution of their 0.80 M$_{\odot}$ star).
The RGB tip luminosities all agree reasonably well between these models
but we note that the RGB lifetime of \citet{1974ApJ...191..173W}
is again substantially shorter than ours (we have found no other RGB
lifetimes to compare with here). 

At 1 M$_{\odot}$ we find five models to compare with (although not
many studies give the comparison values we use). Again, a key finding
is the occurrence of DSFs in these models. This time there are two
other studies that find DSFs to occur -- \citet{2000ApJ...529L..25F}
and \citet{2004ApJ...602..377I}. Our model also experiences a DSF.
The fact that independent modellers find the same (general) result
is reassuring. The differences in surface pollution from the DCF events
is discussed below. We note that our 1 M$_{\odot},$ $\textrm{[Fe/H]}=-3.0$
model (which is not listed in the table) is actually closer in metallicity
to the model of \citet{2004ApJ...602..377I}. This model also experiences
a DSF, concurring with their findings. Finally we note that the RGB
tip luminosities show a small variation between the studies ($L_{tip}/L_{\odot}=2.8\rightarrow3.2$)
but this is probably not significant.

\subsection{Intermediate Mass Structural Evolution}

\subsubsection*{Low and Extremely Low Metallicity}

In panel 1 of Table \ref{Table-CompareStudies-LowZ-LowMass} we show
comparisons for our 2.0 and 3.0 M$_{\odot},$ $\textrm{[Fe/H]}=-5.45$
models. Unfortunately we have not found many studies to compare with
at these masses and metallicities. However an important result is
that \citet{2000ApJ...529L..25F} also find DSFs to occur in their
$\textrm{[Fe/H]}=-5$ intermediate mass models. 

At $\textrm{[Fe/H]}=-4.0$ both the \citet{2000ApJ...529L..25F} and
\citet{2004ApJ...602..377I} studies find a DSF to occur in their
2 M$_{\odot}$ models, which is consistent with our model. There is
however some disagreement between the 3 M$_{\odot}$ models. Our model
and the model of \citeauthor{2004ApJ...602..377I} do \emph{not} experience
DSFs, whereas the model of \citeauthor{2000ApJ...529L..25F} does.
It is important to mention that the \citeauthor{2004ApJ...602..377I}
model is significantly more metal-rich than than the other two models,
having $\textrm{[Fe/H]}=-2.7$. Looking at our $\textrm{[Fe/H]}=-3.0$
model we see that it does not experience a DSF either, so the results
are consistent between these two studies. The discrepancy between
our $\textrm{[Fe/H]}=-4.0$ model and that of \citeauthor{2000ApJ...529L..25F}
remains however. Thus it appears that different modellers find (at
least slightly) different metallicity limits on the occurrence of
the DSF. This is not surprising as these events are dependent on the
details of the treatment of convective boundaries. \citeauthor{2004ApJ...602..377I}
have used the same minimalist approach as we have, insomuch as they
do not add any overshoot to the standard Schwarzschild boundary. As
far as we aware \citeauthor{2000ApJ...529L..25F} do the same (from
the code description in \citealt{1990ApJ...351..245H}). Of course
this does not necessarily mean that all the modellers treat the numerical
details of meshpoint placing in the same way. Thus there may be some
form of `numerical diffusion' across the Schwarzschild boundaries
in all the models -- which could cause the DSF to occur at slightly
higher metallicities. The different results may also be due to a difference
in input physics.

\begin{table}
\begin{center}\begin{threeparttable}

\normalsize

\begin{tabular}{ccc>{\centering}p{1.3cm}>{\centering}p{1.4cm}c}
\multicolumn{6}{c}{Metallicity$=-5.45$}\tabularnewline
\hline 
\hline 
Study & M$_{*}$ & {[}Fe/H{]} & $\tau_{MS}$ & DF & $Z_{DF}$\tabularnewline
\hline 
\citet{1974ApJ...191..173W} & 2.0 & $-6$ & 545 & -- & --\tabularnewline
\citet{2000ApJ...529L..25F} & 2.0 & $-5$ & -- & \textcolor{blue}{Shell} & --\tabularnewline
\textit{This study} & 2.0 & $-5.5$ & 605 & \textcolor{blue}{Shell} & 2E-5\tabularnewline
\hline 
\citet{1974ApJ...191..173W} & 2.5 & $-6$ & 289 & -- & --\tabularnewline
\citet{2000ApJ...529L..25F} & 3.0 & $-5$ & -- & \textcolor{blue}{Shell} & --\tabularnewline
\textit{This study} & 3.0 & $-5.5$ & 177 & \textcolor{blue}{Shell} & 4E-8\tnote{a}\tabularnewline
\hline 
 &  &  &  &  & \tabularnewline
\multicolumn{6}{c}{Metallicity$=-4.0$}\tabularnewline
\hline 
Study & M$_{*}$ & {[}Fe/H{]} & $\tau_{MS}$ & DF & $Z_{DF}$\tabularnewline
\hline 
\citet{1974ApJ...191..173W} & 2.0 & $-4$ & 478 & -- & --\tabularnewline
\citet{2000ApJ...529L..25F} & 2.0 & $-4$ & -- & \textcolor{blue}{Shell} & --\tabularnewline
\citet{2004ApJ...602..377I} & 2.0 & $-2.7$ & -- & \textcolor{blue}{Shell} & 2E-3\tabularnewline
\textit{This study} & 2.0 & $-4$ & 562 & \textcolor{blue}{Shell} & 6E-4\tabularnewline
\hline 
\citet{1974ApJ...191..173W} & 2.5 & $-4$ & 242 & -- & --\tabularnewline
\citet{2000ApJ...529L..25F} & 3.0 & $-4$ & -- & \textcolor{blue}{Shell} & --\tabularnewline
\citet{2004ApJ...602..377I} & 3.0 & $-2.7$ & -- & \textcolor{darkgreen}{None} & --\tabularnewline
\textit{This study} & 3.0 & $-4$ & 174 & \textcolor{darkgreen}{None} & --\tabularnewline
\hline 
\end{tabular}

\caption{Comparing some of our intermediate-mass (IM) models with those in
the literature. The top panel displays characteristic values for our
IM $\textrm{[Fe/H]}=-5.45$ models and those with comparable mass
and metallicity from the literature. The second panel does the same
for our $\textrm{[Fe/H]}=-4$ IM models. Column values are: M$_{*}$
(initial mass of the model), $\tau_{MS}$ (main sequence lifetime,
in Gyr), $DF$ (whether the model experienced a dual shell flash or
not), and $Z_{DF}$ (the $Z$ metallicity of the surface after the
DSF episode). A dash means that the information was either not supplied
in the paper or not relevant to the model. \label{Table-CompareStudies-LowZ-IntMass}}

\line(1,0){100}

\begin{tablenotes}\scriptsize

\item[a]{Note that this is the same as the initial composition -- the DSF has practically no effect on the surface $Z$ in this case.}

\end{tablenotes}

\end{threeparttable} \end{center}
\end{table}

\normalsize

\subsection{The Polluting Episodes}

As this is the first study to calculate the entire AGB evolution and
detailed yields for stars of such low metallicity there are no yields
with which we can directly compare. However, inspired by the CEMP
stars, there has been much quantitative work on the envelope pollution
resulting from the dual core flash (DCF) and dual shell flash (DSF)
events. We thus compare the chemical pollution results of our models
at these stages of evolution with those in the literature. Some studies
have evolved models a short way in to the AGB evolution, so we also
discuss the occurrence (or non-occurrence) of 3DUP and HBB. We note
that the comparisons of our $Z=0$ models with those in the literature
can be found in Sections \ref{subsec-m0.85z0-ComparePrevStudies}
(0.85 M$_{\odot}$) and \ref{subsec-m2z0-ComparePrevStudies} (2.0
M$_{\odot}$). 

\subsubsection*{Dual Core Flashes}

In the 0.85 M$_{\odot},$ $Z=0$ section we showed that there was
a variation in surface $Z_{cno}$ resulting from the DCF between studies
($Z_{cno}\sim0.015\rightarrow0.004$). However, despite this factor
of four variation, the general result from all the studies was that
the models produce $\sim1\rightarrow3$ dex too much C and N, as compared
to CEMP observations. Here we make further comparisons with the models
in the literature, this time with regards to our low metallicity models.

In panel 1 of Table \ref{Table-CompareStudies-AllDualFlashes} we
show the resulting pollution from some models in the literature that
experience the DCF. We include $Z=0$ models as there are not many
studies at very low metallicity that give quantitative results for
the pollution. All the models are naturally of low mass as the DCF
only occurs in stars that go through the core He flash. Our $\textrm{[Fe/H]}=-5.45,$
0.85 M$_{\odot}$ model agrees very well with the 0.80 M$_{\odot}$
models at $\textrm{[Fe/H]}=-5$ and $-6$ by \citet{2004ApJ...609.1035P}.
In fact, our value of surface $Z$ after the DCF ($Z_{DCF}$) is exactly
the same as that in their $\textrm{[Fe/H]}=-5$ model ($3\times10^{-2}$).
Looking at the $Z=0$ models we can see that there is not a very large
amount of variation in pollution at $0.80\rightarrow0.85$ M$_{\odot}$,
even when comparing to the higher metallicity models. They are all
in the range $0.9\rightarrow3\times10^{-2}$, which is reassuring
to see considering the results come from three independent stellar
structure codes. It would seem that the `ballpark' bulk pollution
expected from modelling the DCF at these masses is now reasonably
well constrained. The details of the pollution may be another matter
-- eg. degree of s-process, ratios of nuclides. We shall discuss
some chemical details of the models in the next section when we compare
our (and other studies') results to observations. 

In the case of our $\textrm{[Fe/H]}=-5.45,$ 1.0 M$_{\odot}$ model
there is an order of magnitude less pollution arising from the DCF
event than in the 0.85 M$_{\odot}$ model. We have found no low-Z
models to compare with (that experience the DCF) so we compare with
two $Z=0,$ 1 M$_{\odot}$ models. The model by \citet{2001ApJ...559.1082S}
has a factor of $\sim3$ more pollution from the DCF than ours, which,
in the context of this complex evolutionary phase, and the difference
in metallicity, we consider as a reasonable agreement. The situation
is better when comparing to the model of \citet{1990ApJ...349..580F},
which has exactly the same level of pollution as our 1 M$_{\odot}$
model ($Z_{DCF}=4\times10^{-3}$). We note that if it were not for
the larger amount of pollution in the \citet{2001ApJ...559.1082S}
model one might say that the DCF produces more pollution at lower
masses, but the number of studies is too small to make any serious
conclusions on this. 

In summary it appears that the pollution arising from the DCF at low
mass and extremely low Z is reasonably consistent between models,
being $\sim2\times10^{-2}$ at 0.8 M$_{\odot}$, and ranging between
$4\rightarrow10\times10^{-3}$ at 1 M$_{\odot}$. However we add the
caveat that this conclusion is not based on a great many studies.

\subsubsection*{Dual Shell Flashes and the TP-AGB}

In Table \ref{Table-CompareStudies-AllDualFlashes} it is apparent
that there is, in general, more pollution arising from the DSF at
low mass ($M=0.8\rightarrow1.0$ M$_{\odot}$) than at intermediate
mass. The $Z_{DSF}$ value (which is defined as the surface metallicity
after the DSF) ranges from $2\times10^{-4}\rightarrow1\times10^{-2}$
in the LM models and $6\times10^{-6}\rightarrow2\times10^{-3}$ in
the IM (2 M$_{\odot}$) models. We note however that our literature
sample covers a wide range of initial metallicity which explains much
of the variation, as will become clear in the model-to-model comparisons
below. 

We reported in Section \ref{subsec-m2z0-ComparePrevStudies} that
the amount of pollution resulting from DSF episodes in the $Z=0$
intermediate mass models is relatively small compared to that arising
in low-mass models. Moreover, since these models experience 3DUP immediately
after the DSF, which also brings up processed material, the polluting
effect of the DSF is quickly swamped. Thus the detailed results in
terms of the bulk pollution from these events are rendered insignificant.
This was also recognised by \citet{2002ApJ...570..329S}. This phenomenon
is also true of all the metal-deficient 2 M$_{\odot}$ models in Table
\ref{Table-CompareStudies-AllDualFlashes} -- as all studies find
3DUP to occur. We note that there may however be more subtle nucleosynthetic
signatures from this event, such as that from s-processing. Putting
3DUP aside, we see that the resultant $Z_{DSF}$ from the 2 M$_{\odot},$
$\textrm{[Fe/H]}=-2.7$ model of \citet{2004ApJ...602..377I} and
that of our closest comparison model of 2 M$_{\odot}$ and $\textrm{[Fe/H]}=-4$,
differ by a factor of $\sim3$. We suggest that this is a reasonably
close result, given the differences in initial composition. Note that
our 2 M$_{\odot},$ $\textrm{[Fe/H]}=-3$ model did \emph{not} experience
a DSF. This is an interesting difference in itself, as it shows that
the line between the DSF occurring and not occurring (at a particular
mass and/or metallicity) is uncertain. The only other 2 M$_{\odot}$
DSF models that we have found in the literature are of zero metallicity.
We compared these with our $Z=0$ models in the $Z=0$ chapters. 

Returning to DSFs in low mass models it can be seen in Table \ref{Table-CompareStudies-AllDualFlashes}
that the $Z_{DSF}$ from our 0.85 M$_{\odot},$ $\textrm{[Fe/H]}=-4$
model and that from the 0.80 M$_{\odot}$ model of the same metallicity
by \citet{2000ApJ...529L..25F} differ by a factor of $\sim6$. A
difference of this size is quite significant when comparing to observations,
as it translates to $\sim0.8$ dex. Comparing our 1 M$_{\odot},$
$\textrm{[Fe/H]}=-3$ model with the 1 M$_{\odot},$ $\textrm{[Fe/H]}=-2.7$
model of \citet{2004ApJ...602..377I} there is a much closer match,
such that $Z_{DSF}$ only differs by a factor of $\sim2$. A discrepancy
of this size is roughly the size of the error bars in observational
studies. Interestingly \citet{2004ApJ...602..377I} report that 3DUP
occurs in their model, which is absent from all our 1 M$_{\odot}$
models (and the model of \citealt{2000ApJ...529L..25F}). However,
on inspection of their Figure 5 it appears that it is very minor,
at least during the beginning of the TP-AGB. 

In summary we have shown that there is much variation in the results
given in the literature for pollution from the DSF events. At IM this
is of little importance due to the dominating effect of 3DUP on the
AGB surface composition. We note however that if the DSF event leads
to some form of envelope ejection (that may pollute a binary companion)
then the details of the DSF are important and thus this polluting
event would need to be studied more rigorously. At low mass the DSF
does remain the dominant source of pollution for the rest of the AGB
evolution, due to the lack of 3DUP. Although this material is usually
diluted (sometimes by a very large amount) by unpolluted RGB mass
loss, the details of the DCF pollution would also be important in
a binary star mass-transfer scenario. At any rate it does have an
impact on the yields, unlike in the IM models. Our results have been
found to differ by factors of $\sim3\rightarrow6$ (in $Z_{DSF}$)
with similar models in the literature. Although the discrepancies
are reasonably large they are not catastrophic when comparing to the
error bars in observational studies but, in the worst cases, they
are quite significant. Thus we suggest that the current state of the
field is that we have a rough prediction of DSF pollution from the
models, but more work needs to be done. We again add the caveat that
these conclusions are based on a small number of comparison models
available in the literature. Finally we note that a discussion and
summary of the pollution effects of these models is given in Section
\vref{section-HaloStarModelsPollutionSummary}.

\begin{table}
\begin{center}\begin{threeparttable}

\normalsize

\begin{tabular}{ccccc}
\multicolumn{5}{c}{Dual Core Flashes}\tabularnewline
\hline 
\hline 
Study & M$_{*}$ & {[}Fe/H{]} & $Z_{DCF}$ & 3DUP\tabularnewline
\hline 
\citet{2000ApJ...529L..25F} & 0.80 & $Z=0$ & $9\times10^{-3}$ & No\tabularnewline
\citet{2004ApJ...609.1035P} & 0.80 & $Z=0$ & $2\times10^{-2}$ & --\tabularnewline
'' & 0.80 & $-6$ & $2\times10^{-2}$ & --\tabularnewline
'' & 0.80 & $-5$ & $3\times10^{-2}$ & --\tabularnewline
\citet{1990ApJ...349..580F} & 1.0 & $Z=0$ & $4\times10^{-3}$ & --\tabularnewline
\citet{2001ApJ...559.1082S} & 1.0 & $Z=0$ & $1\times10^{-2}$ & --\tabularnewline
\hline 
\textit{This study} & 0.85 & $-5.5$ & $3\times10^{-2}$ & No\tabularnewline
'' & 1.0 & $-5.5$ & $4\times10^{-3}$ & No\tabularnewline
\hline 
 &  &  &  & \tabularnewline
\multicolumn{5}{c}{Dual Shell Flashes}\tabularnewline
\hline 
\hline 
Study & M$_{*}$ & {[}Fe/H{]} & $Z_{DSF}$ & 3DUP\tabularnewline
\hline 
\citet{2000ApJ...529L..25F} & 0.8 & $-4$ & $6\times10^{-3}$ & No\tabularnewline
\citet{2002ApJ...570..329S} & 1.0 & $Z=0$ & $2\times10^{-4}$ & --\tabularnewline
\citet{2004ApJ...602..377I} & 1.0 & $-2.7$ & $1\times10^{-2}$ & Yes\tabularnewline
\citet{2002ApJ...570..329S} & 2.0 & $Z=0$ & $6\times10^{-6}$ & Yes\tabularnewline
\citet{2003ASPC..304..318H} & 2.0 & $Z=0$ & $1\times10^{-4}$ & Yes\tabularnewline
\citet{2004ApJ...602..377I} & 2.0 & $-2.7$ & $2\times10^{-3}$ & Yes\tabularnewline
\hline 
\textit{This study} & 0.85 & $-4$ & $1\times10^{-3}$ & No\tabularnewline
'' & 1.0 & $-3$ & $5\times10^{-3}$ & No\tabularnewline
'' & 2.0 & $-4$ & $6\times10^{-4}$ & Yes\tabularnewline
\hline 
\end{tabular}

\caption{Most of the dual core flashes (DCFs, top panel) and dual shell flashes
(DSFs, bottom panel) reported in the literature for masses in the
range $0.8\rightarrow2.0$ M$_{\odot}$. Only studies that have given
a quantitative value for the resultant pollution (i.e. the surface
metallicity after the DSF, $Z_{DSF}$) are included. In column two
we give the initial mass of the star (M$_{*}$), in M$_{\odot}$ whilst
in the last column we indicate if the model experienced 3DUP during
the early stages of the AGB or not. A dash indicates that the information
is not given in the study (or is ambiguous). \label{Table-CompareStudies-AllDualFlashes}}

\line(1,0){100}

\begin{tablenotes}\scriptsize

\item[a]{This model did not experience a DSF.}

\end{tablenotes}

\end{threeparttable} \end{center}
\end{table}

\normalsize

\section{Comparison with Halo Star Observations\label{Section-HaloStarModels-CompareObs}}

\subsection{Overview}

We gave a brief overview of Galactic Halo star observations in Section
\ref{subsection-EMPHs}. It was noted there that the tail of the (currently
known) Halo metallicity distribution function (MDF) reaches down to
$\textrm{[Fe/H]}\sim-5.5$. Thus our grid of models, which begins
at $\textrm{[Fe/H]}=-3.0$ and ends at $Z=0$ has a significant overlap
with the observed metallicity range but also extends well below this.
This enables us to make comparisons with the current observations
and also to make predictions about future, lower-metallicity discoveries
-- assuming stars of such low metallicity exist. 

\begin{figure}
\begin{centering}
\includegraphics[width=0.95\columnwidth,keepaspectratio]{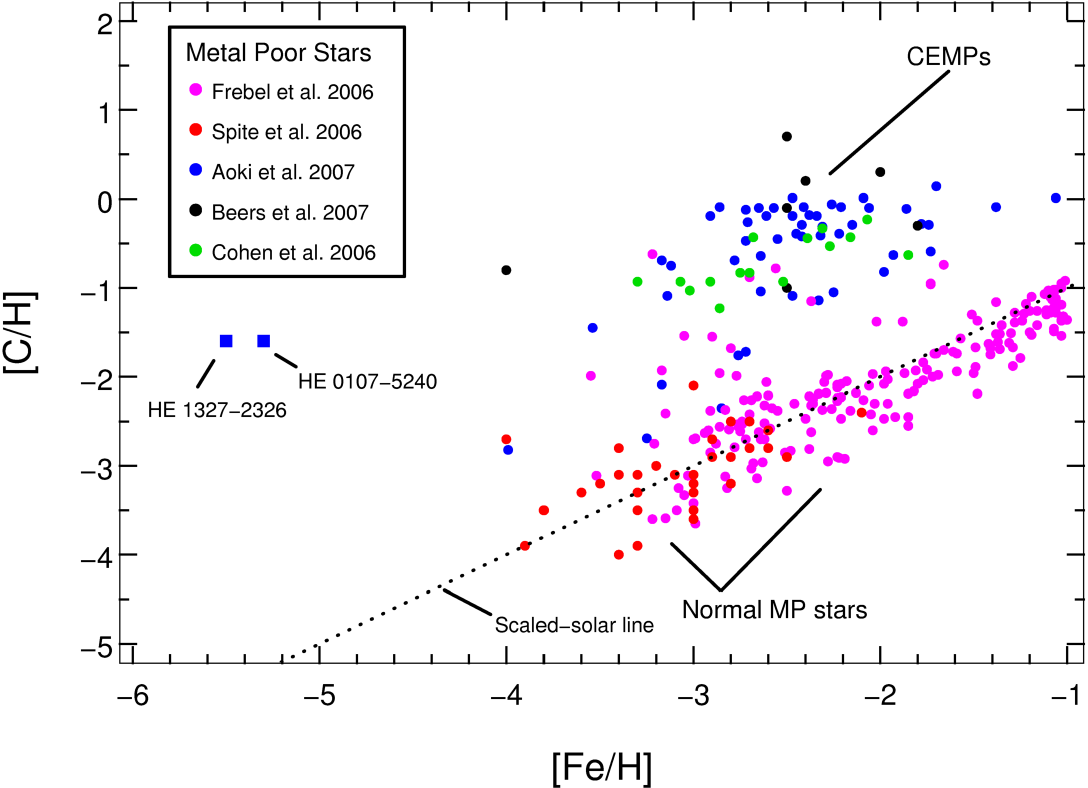}
\par\end{centering}
\caption{Carbon abundance observations versus {[}Fe/H{]} for metal-poor stars
compiled from the literature. The dotted line marks $\textrm{[C/Fe]}=0.0$.
The normal EMPs lie on this Galactic chemical evolution line, whilst
the C-EMPs (which we define as having $\textrm{[C/Fe]}>+0.7$) lie
above this. The two most Fe-poor stars known, HE 0107-5240 ($\textrm{[Fe/H]}=-5.3,$
\citealt{2004ApJ...603..708C}) and HE 1327-2326 ($\textrm{[Fe/H]}=-5.4,$
\citealt{2005IAUS..228..207F}) are labelled. We note that the most
C-rich stars have similar amounts of carbon as the Sun, giving them
a $Z$-defined metallicity that is roughly solar. Data sets are from
\citet{2006ApJ...652.1585F}, \citet{2006AA...455..291S}, \citet{2007ApJ...655..492A},
\citet{2007AJ....133.1193B}, and \citet{2006AJ....132..137C}. \label{fig-Observations-EMPs-CHvsFeH}}
\end{figure}

As also mentioned in Section \ref{subsection-EMPHs} there has been
two major low-metallicity halo field star surveys: the HK survey (\citealt{BPS85};
\citealt{BPS92}) and the Hamburg/ESO (HES) survey (\citealt{WKG+96};
\citealt{CWR+99}). These are the studies that have fleshed out the
low-metallicity tail of the Halo MDF (see \citet{BC05} for a review
of this area). There has been a huge amount of observational work
in recent years, spurred on by the discoveries of lower and lower
metallicity stars such as HE 0107-5240 ($\textrm{[Fe/H]}=-5.3,$ \citealt{2004ApJ...603..708C})
and HE 1327-2326 ($\textrm{[Fe/H]}=-5.4,$ \citealt{2005IAUS..228..207F}).
Indeed, even in just the last few months there has been quite a few
publications of large data sets of abundance observations for extremely
metal-poor stars (EMPs). These have been very useful for the current
study. In Figure \ref{fig-Observations-EMPs-CHvsFeH} we display the
{[}C/H{]} observations for a selection of these studies. The sample
starts at $\textrm{[Fe/H]}=-1$ and ends at the aforementioned two
most metal-poor stars. It is important to note that the combined sample
is biased towards carbon-enhanced MP stars (CEMPs), as much interest
has arisen in these objects due to their apparently very high numbers.
They are often reported to represent $\sim20\%$ of the low metallicity
halo star population (eg. \citealt{BC05}; \citealt{2006ApJ...652L..37L}),
although \citet{2005ApJ...633L.109C} report that it is somewhat lower
($\sim9\rightarrow14\%$). Regardless of whether the proportion is
$10\%$ or $20\%$ they are much more abundant than CH stars at higher
metallicity ($\sim1\%$, eg. \citealt{1991ApJS...77..515L}) and thus
represent quite a mystery. A few of the studies from which we have
collected observational data focus on CEMPs in particular, so this
group is over-represented in our plots. Since our model results give
the chemical pollution expected from evolved stars (of low- and intermediate-mass)
a large observational set that may reflect the composition of the
ejecta of that generation of stars is actually very useful for comparisons.
The `normal' EMPs (ie. having $\sim$scaled-solar abundances of C)
lie on the Galactic chemical evolution line (which we define as $\textrm{[C/Fe]}=0$).
For example, in Figure \ref{fig-Observations-EMPs-CHvsFeH} carbon
is seen to evolve with Fe in the normal EMPs. The `polluted' EMPs
mostly lie above this line (hence the name CEMPs). In the figures
below, where we compare our models with observations of other elements,
we will follow convention by identifying the CEMPs as a separate population.
This will show if there are any general trends common to this group
over the larger elemental parameter space. We quantitatively define
extremely metal-poor carbon stars (C-EMPs = CEMPs) as those MP stars
having $\textrm{[C/Fe]}>+0.7$. It is interesting to note that the
most C-rich CEMPs have as much carbon as the Sun, meaning that they
are not metal-poor in the strictest sense. Indeed their $Z$ metallicity
is roughly solar! A better term may be heavy-metal-poor, or Fe-poor.
We shall however continue to follow tradition and use MP to refer
to Fe-poor stars. Another interesting feature in Figure \ref{fig-Observations-EMPs-CHvsFeH}
is that there is a spread of stars between the scaled-solar {[}C/Fe{]}
line and the $\textrm{[C/H]}=0$ line. Indeed, it looks as if there
is an upper envelope at $\textrm{[C/H]}\sim0$. This must reflect
the maximum amount of C pollution that the low-metallicity stars can
produce and will be an interesting observational feature to compare
with our models. 

The first comparison we make against the observations is with the
surface compositions of our models just after the dual core flash
(DCF) and dual shell flash (DSF) events. We have two reasons for doing
this. Firstly, since these are potentially violent events, there may
be an enhanced probability of mass transfer within a binary system
during (or just after) these events. Secondly, if mass loss \emph{is}
a function of metallicity (as defined by $Z$) then these stars would
lose more mass after the dual flash (DF) events as their surfaces
are heavily polluted from then on. These winds could also pollute
a binary companion. It has already been shown that many of the C-EMP
stars are in binary systems (in fact possibly \emph{all} s-process-rich
CEMP stars are binaries: \citealt{2005ApJ...625..825L}). Thus we
believe it is important to confront the surface compositions at these
stages of evolution with the observations. We note that others (eg.
\citealt{2004ApJ...611..476S}; \citealt{2004ApJ...603..708C}) also
suggest a binary scenario for the formation of CEMPs. After these
DF composition comparisons we then compare our \emph{yields} with
the observations.

\subsection{Comparisons With Dual Flash Abundances}

\begin{figure}
\begin{centering}
\includegraphics[width=0.95\columnwidth,keepaspectratio]{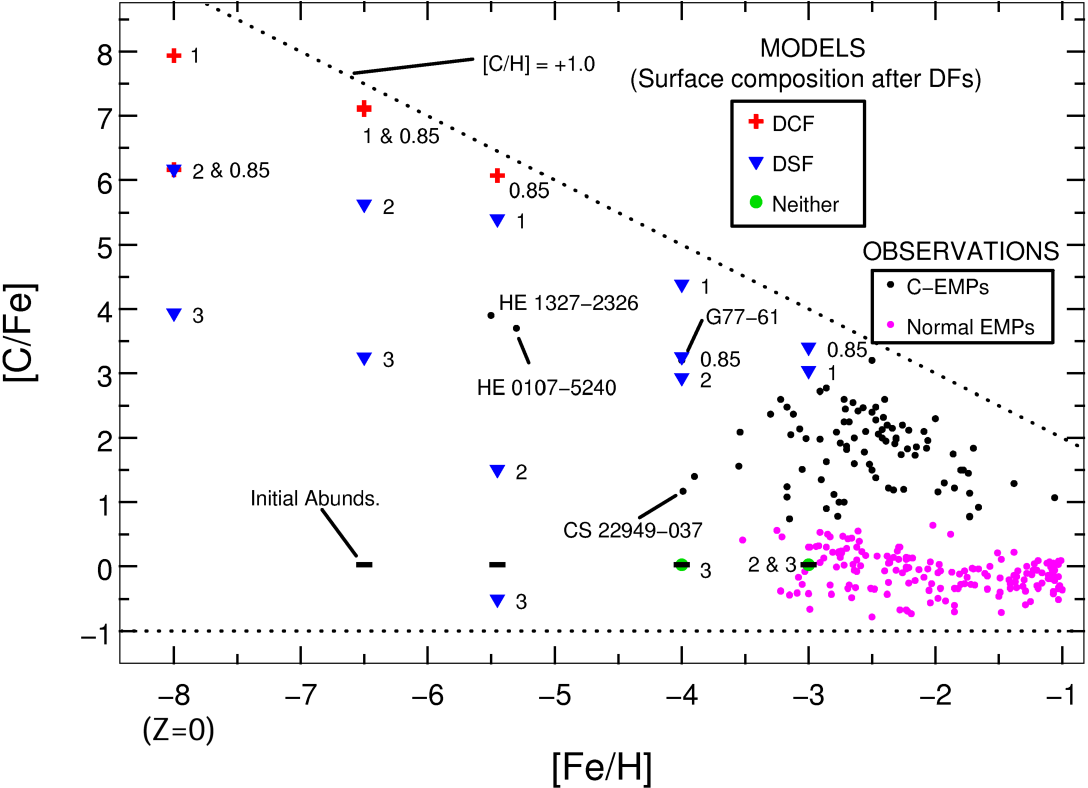}
\par\end{centering}
\caption{Surface {[}C/Fe{]} versus {[}Fe/H{]} of all our models -- just after
the dual flash events. We show the composition at this stage since
these violent events may induce extra mass transfer in binary systems.
Also shown for comparison are the observations (see Figure \ref{fig-Observations-EMPs-CHvsFeH}
for data sources). Here we have colour-coded the data into {[}C/Fe{]}-rich
and {[}C/Fe{]}-normal, where {[}C/Fe{]}-rich is defined by $\textrm{[C/Fe]}>+0.7$.
We keep this colour coding in all the figures below. The short horizontal
lines indicate the starting composition of the models (in this case
they are all at $\textrm{[C/Fe]}=0$, except for the $Z=0$ models).
We have plotted the $Z=0$ models at $\textrm{[Fe/H]}=-8$ for comparison.
The two most metal-poor stars can be seen at $\textrm{[Fe/H]}\sim-5.5$
(they are both C-rich). We also point out two $\textrm{[Fe/H]}\sim-4$
stars (G77-61 and CS 22949-037) which we discuss later. An upper envelope
to the pollution in the models -- and the observations -- is marked
by the dotted line at $\textrm{[C/H]}=+1.0$. A very interesting feature
of the models is that the mass range that suffers surface pollution
for these DF events increases with decreasing metallicity. See text
for a discussion.\label{fig-compare-CFe-DFcomp-Obs-all}}
\end{figure}

We begin with Figure \ref{fig-compare-CFe-DFcomp-Obs-all} by comparing
the dual flash surface abundances of carbon with the observations.
In this figure our definition of CEMP stars can be seen clearly. Also
indicated in the figure are the initial abundances used in the models,
which are all scaled-solar for C (except the $Z=0$ models of course).
It can be seen that practically all models that experience a DF, whether
it be a DCF or a DSF, have large surface C enhancements over the initial
composition. An interesting bulk feature of the grid of models is
that there is an upper envelope of C enhancement -- just like the
observations (although we don't have many observations below $\textrm{[Fe/H]}<-3.5$).
We have marked this envelope with a line at $\textrm{[C/H]}=+1.0$.
We have also marked a lower envelope at $\textrm{[C/Fe]}=-1.0$ which
probably indicates the lower limit to C and Fe pollution coming from
low metallicity SNe. In terms of our models we can say little about
the lower limit except that the DFs do not reduce the amount of C
in the envelope so they never violate this constraint. 

Looking more closely it can be seen that our higher metallicity models
($\textrm{[Fe/H]}=-3,-4$) of lower mass all undergo dual shell flashes.
At $\textrm{[Fe/H]}=-4$ the 2 M$_{\odot}$ model does also. This
has raised the C abundances in these models to high levels ($\textrm{[C/Fe]}\sim3\rightarrow4.5$).
At $\textrm{[Fe/H]}=-3$ this is (roughly) consistent with the most
extreme observations. The same can not be said at $\textrm{[Fe/H]}=-4$
but we have very little observational data to compare with here. If
we extrapolated the more metal-rich observations to $-4$ (or interpolated
up to the two most metal-poor stars) the 0.85 M$_{\odot}$ and 2 M$_{\odot}$
models would fit this line well. An important point to make here is
that the compositions from our models really represent an \emph{upper
limit} to the level of pollution expected in a binary scenario as
the material transferred would be mixed into the envelope of the secondary
(possibly even if that envelope was radiative, see eg. \citealt{2007astro.ph..2138S}).
The material would also be diluted by any previous mass loss on the
RGB (which varies with mass and metallicity in our models -- see
section \vref{subsection-HaloStarStruct-MassLoss}). In the case of
the two $\textrm{[Fe/H]}\sim-5.5$ stars we see that the lowest mass
models are much more C-rich (by $\sim2$ dex) than the observations
whilst the intermediate mass models (2 and 3 M$_{\odot}$) are $\sim2\rightarrow4$
dex too low. Since it is easier to dilute stellar material, the low
mass models can be considered a better `fit' than the IM models. 

In terms of the extremely low metallicity end of the spectrum ($\textrm{[Fe/H]}<-6$),
for which no stars have been discovered, we note that our models predict
the trend of very high {[}C/Fe{]} to increase. Indeed, one would expect
from these results that an $\textrm{[Fe/H]}=-6.5$ star discovered
tomorrow could have $\textrm{[C/Fe]}\sim5$ or 6 -- although it could
be anywhere below this. It is also very interesting that the two $-5.5$
stars are both CEMPs. This may indicate that \emph{all} (or most)
EMP stars are C-rich. This is however speculation based on a very
small sample. We would really expect to see a population at $\textrm{[C/Fe]}\sim0$
also. 

Another very interesting result from the models is that the mass range
in which C enhancement events occur increases with decreasing metallicity.
This is clearly seen when contrasting the $\textrm{[Fe/H]}=-3$ models
with the $-6.5$ models. At $\textrm{[Fe/H]}=-3$ only the two lowest
mass models (0.85 and 1 M$_{\odot}$) are C-enhanced due to the DSFs,
whilst at $-6.5$ \emph{all} the models are. Since a significantly
larger portion of the IMF is able to pollute at extremely low metallicity
(assuming an IMF peaked at low mass) then we would predict that larger
and larger proportions of EMP stars would be C-rich at lower and lower
metallicities. This certainly seems to be the case at $\textrm{[Fe/H]}=-5.5$
but again there are only two observations so far. The caveat here
is that this pollution arises from dual flashes -- the story may
be different when considering the yields. We shall return to this
result in the Discussion as it is a key finding.

\begin{figure}
\begin{centering}
\includegraphics[width=0.95\columnwidth,keepaspectratio]{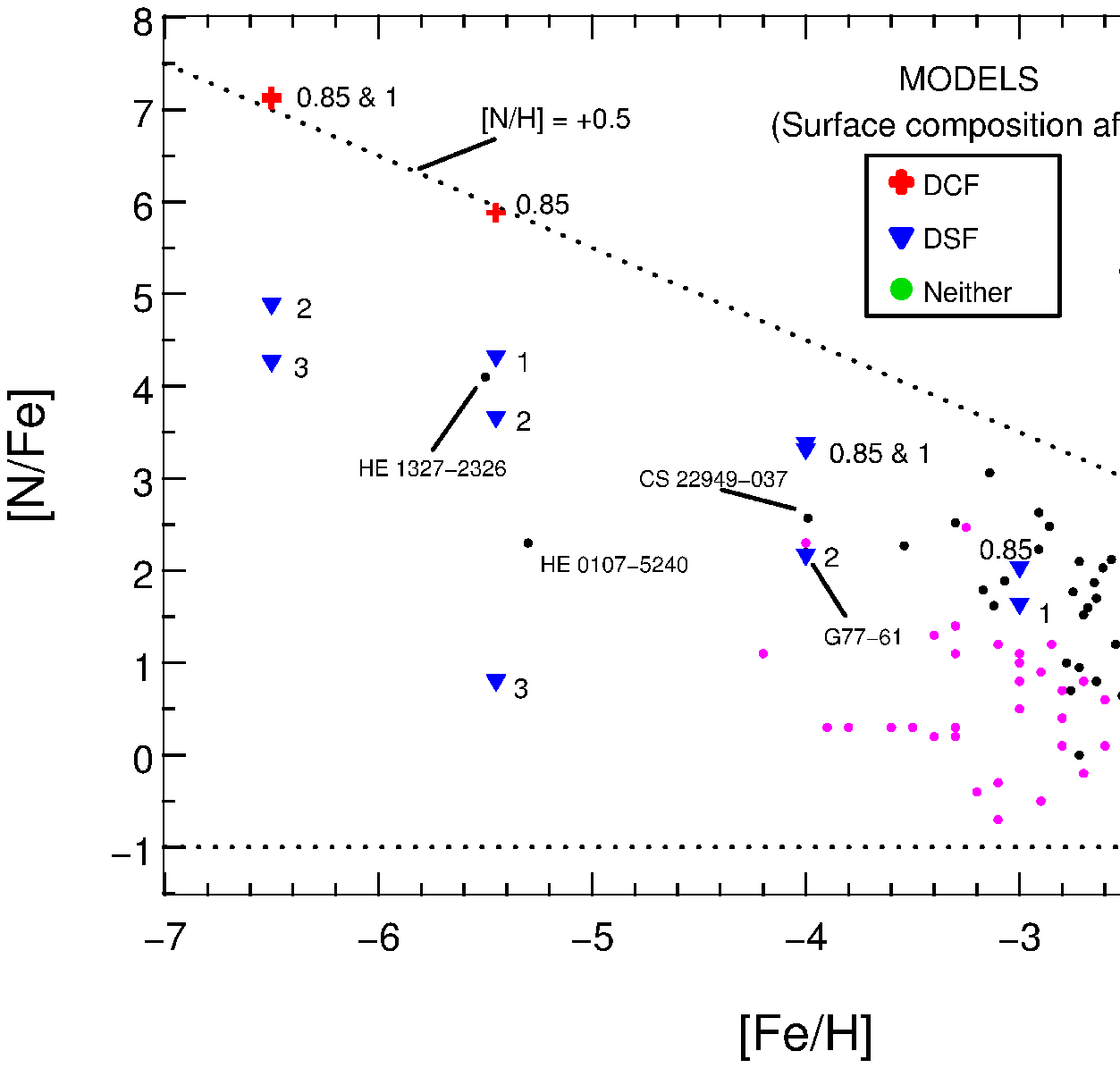}
\par\end{centering}
\caption{Same as Figure \ref{fig-compare-CFe-DFcomp-Obs-all} but for nitrogen.
Upper and lower envelopes are sketched with dotted lines. The two
most metal-poor stars observed to date are indicated, as are two $\textrm{[Fe/H]}=-4$
stars (G77-G1 and CS22949-037). There appears to be a fairly uniform
spread in N enhancement in the observations (although we note that
our sample is not complete). The C-rich EMPs are, in general, more
N-rich than the C-normal EMPs. Both groups show a spread. Our models
are in reasonable agreement with the CEMP observations. The missing
higher-mass models still have their initial N abundances at this stage,
which are well below the scale of the plot. Note that we have not
plotted the $Z=0$ models in this case. \label{fig-compare-NFe-DFcomp-Obs}}
\end{figure}

In Figure \ref{fig-compare-NFe-DFcomp-Obs} we compare the surface
nitrogen abundances of our models taken just after the DF events with
those of the observations. First we note that the distribution of
the observations is best described as a spread between an upper boundary
of $\textrm{[N/H]}\sim+0.5$ and a lower boundary of $\textrm{[N/Fe]}\sim-1.0$.
Ignoring the distinction between the CEMPs and normal EMPs the observations
do not break into two groups as with carbon. That said we have to
note that our sample is far from perfect -- we have displayed all
the stars for which we have N abundances, from the full CEMP-biased
sample. Despite these limitations we can safely say that most of the
C-EMPs are also N-rich. Most of them lie above $\textrm{[N/Fe]}=1.0$
(and even higher at lower metallicities). Conversely most of the C-normal
EMPs lie below $\textrm{[N/Fe]}\sim1.5$. We note that \citet{2005AA...434.1117P}
also mention these trends in N and/or C enhancement. 

In terms of our models it can be seen that the higher metallicity
low-mass models fall nicely into the spread. The higher mass models
did not go through the DSF event so their N abundances are still extremely
low due to the low initial abundances used. At $\textrm{[Fe/H]}=-4$
our DSF-polluted models appear to be in reasonable agreement with
the observations. The low mass models have quite high N but dilution
effects could lower this. We note that the observation sample is quite
small here. Looking at the two most metal-poor objects we see that
our low-mass models match the observations well. However we note that
they weren't well matched by the same models in terms of carbon (we
discuss abundance patterns in detail below). As seen with carbon the
mass range in which our models show N overabundances from these DF
events increases with decreasing metallicity. Our models thus predict
that the frequency (and degree) of N-enhancements arising from these
events (which may be a source of polluted material in a binary mass
transfer) will increase at lower metallicities.

\begin{figure}
\begin{centering}
\includegraphics[width=0.95\columnwidth,keepaspectratio]{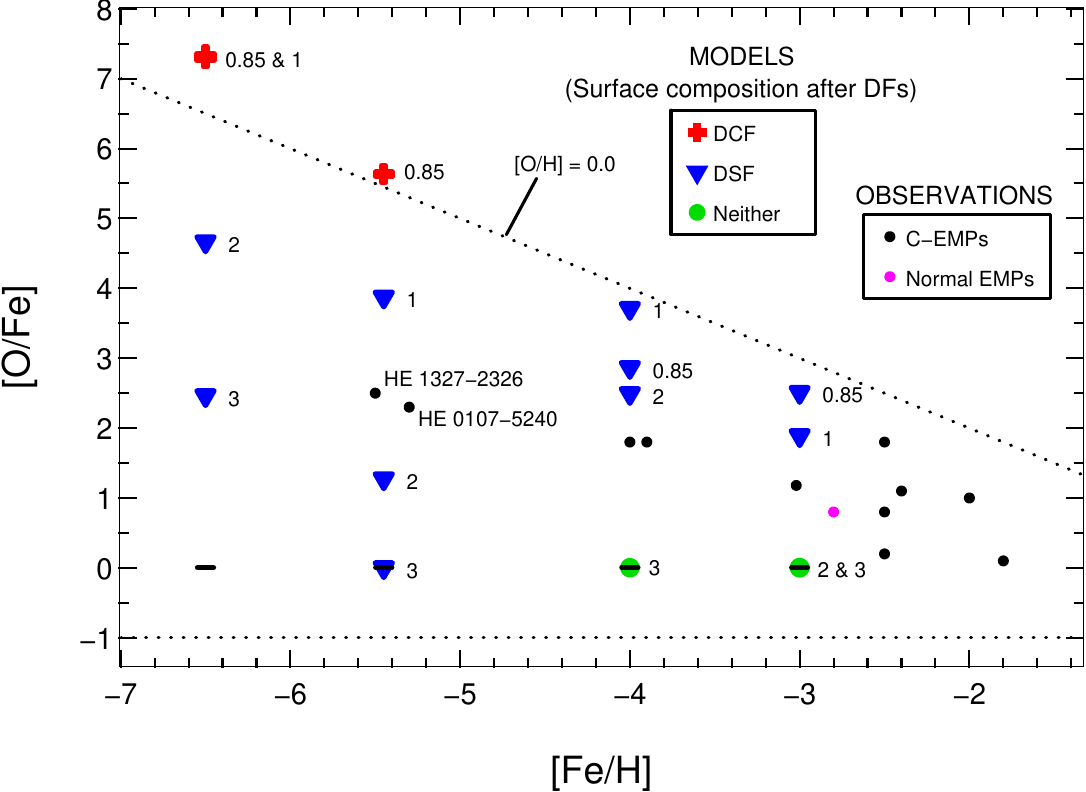}
\par\end{centering}
\caption{Same as Figure \ref{fig-compare-CFe-DFcomp-Obs-all} but for oxygen.
Very sketchy upper and lower envelopes are marked with dotted lines
(there not not many observations to compare with as oxygen is difficult
to measure). The two most metal-poor stars observed to date are indicated.
Their oxygen abundances are from \citet{2004ApJ...612L..61B} (HE
0107-5240) and \citet{2006ApJ...638L..17F} (HE 1327-2326). Note that
we have not plotted the $Z=0$ models in this case. \label{fig-compare-OFe-DFcomp-Obs-all}}
\end{figure}

We make comparisons with the stellar oxygen abundances in Figure \ref{fig-compare-OFe-DFcomp-Obs-all}.
Oxygen is notoriously difficult to measure, thus there are very few
observations to compare with -- and these are uncertain (eg. \citealt{2004AA...419.1095I}
give two values for each of their stars, depending on the line measured).
At higher metallicities we see that there is a spread of oxygen abundances,
with no discernible trend with C-richness. The DF pollution from our
models is generally on the high side compared with the observations
but again, these could be diluted. There is a trend in the lower metallicity
stars -- the O overabundances increase with decreasing metallicity
(although the sample is very small). Our models also follow the same
trend. The two most metal-poor stars fall between our 1 and 2 M$_{\odot}$
models but we note that there is a vary large gap of $\sim2.5$ dex
between the O abundances of these models (for the oxygen abundance
determinations for these two stars see \citealt{2004ApJ...612L..61B}
and \citealt{2006ApJ...638L..17F}). In summary, our models do produce
the large amounts of oxygen observed but often produce much more --
but our observational sample from the literature is very small. 

\begin{figure}
\begin{centering}
\includegraphics[width=0.95\columnwidth,keepaspectratio]{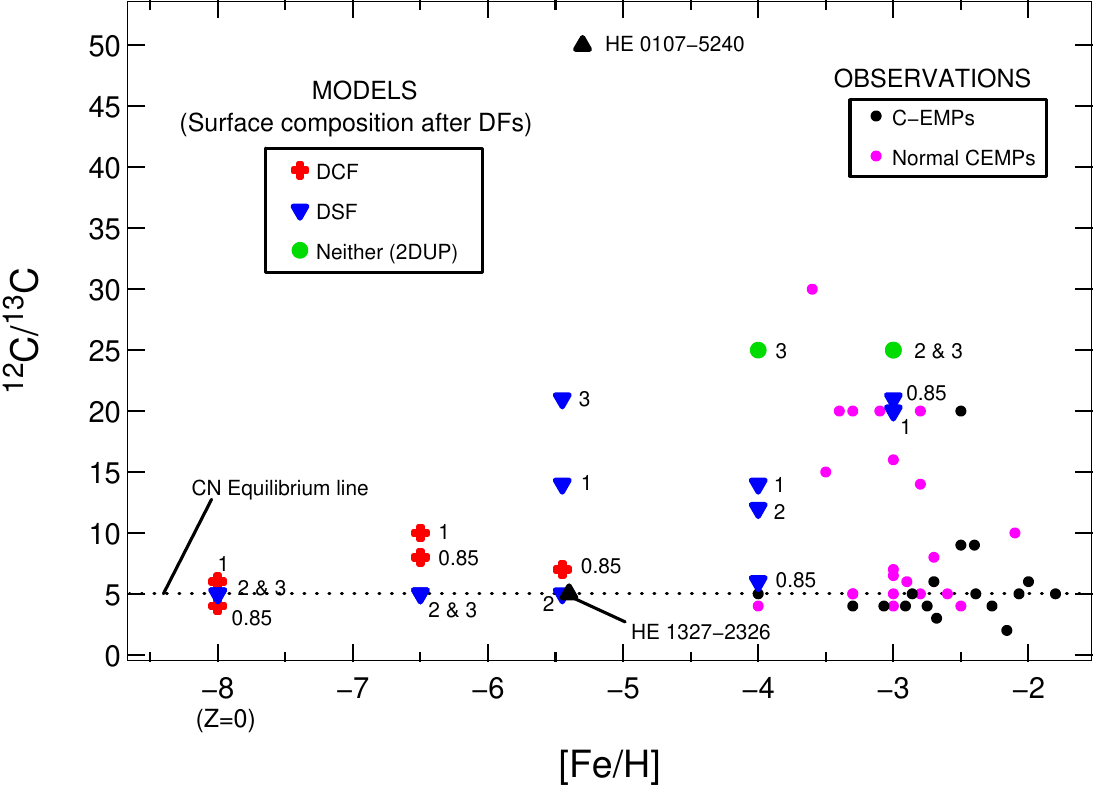}
\par\end{centering}
\caption{Same as Figure \ref{fig-compare-CFe-DFcomp-Obs-all} except for $^{12}$C/$^{13}$C.
We have plotted the $Z=0$ models at $\textrm{[Fe/H]}=-8$ for comparison.
Again our observational sample does not contain many data points.
Interestingly most of the C-enhanced EMPs have low $^{12}$C/$^{13}$C
ratios. The upper envelope of our models reduces with metallicity,
such that the $^{12}$C/$^{13}$C ratios of the most metal-poor models
are all very low. The two most metal-poor stars are represented by
upright triangles as the measurements for both these stars are lower
limits. The ratios in these two stars are very different -- in HE
1327-2326 it is low and HE 0107-5240 it is quite high ($>50$, \citealt{2004ApJ...603..708C}).
Note that the green symbols (no DF episodes) give the ratios in the
surface composition of these models after second dredge-up. \label{fig-compare-c12c13-DFcomp-Obs-all}}
\end{figure}

Some of the studies in our observational sample also report the $^{12}$C/$^{13}$C
ratio for (some of) their objects. We plot these in Figure \ref{fig-compare-c12c13-DFcomp-Obs-all}
along with the surface abundances from our models just after the DF
events. The values of the ratio are also given for three of our stars
that did not experience any DF events (the 2 and 3 M$_{\odot}$ models
with $\textrm{[Fe/H]}=-3$ and the 3 M$_{\odot}$ model with $-4$).
These values are taken from the surface at the start of the AGB (where
the DSF would normally occur) so they reflect the ratios after second
dredge-up. An interesting feature of the observations is that most
of the C-EMPs have low $^{12}$C/$^{13}$C ratios, whereas the normal
EMPs have a range of values. We again note that our dataset is small
so our categorisations may not hold with a set of observations that
is more complete (we shall compare again as the data becomes available).
There is one notable exception to the categorisation that most C-EMPs
have low ratios -- the second most metal poor star, HE 0107-5240,
has a high ratio ($>50$, \citealt{2004ApJ...603..708C}). This contrasts
strongly with the $^{12}$C/$^{13}$C ratio of the most metal-poor
star HE 1327-2326, which is estimated at $\sim5$ (\citealt{2006ApJ...639..897A}).
Note that these values are both lower limits. \citet{2004ApJ...609.1035P}
cite the high value in HE 0107-5240 as one of the good reasons to
reject the DCF scenario for explaining this star's abundances as they
found that the ratio resulting from this event was low ($\sim5$).
Indeed, as can be seen in Figure \ref{fig-compare-c12c13-DFcomp-Obs-all},
the ratio in our 0.85 M$_{\odot},$ $\textrm{[Fe/H]}=-5.45$ model
is also close to the equilibrium value. We note that the value of
$>50$ is more reminiscent of a first dredge-up value. However, looking
at our Table \vref{table-HaloStars-Nucleo-CandN} we see that it is
only (slightly) higher metallicity models that have FDUP $^{12}$C/$^{13}$C
ratios of this order (although the lower metallicity models have very
high values, which may also be consistent with the lower limit). It
is interesting that these two most metal poor stars (of very similar
metallicity) both have similar C enhancements but have such different
isotopic contributions to the enhancements. This seems to imply that
different C-enhancement mechanisms lead to similar levels of pollution. 

It can also be seen in Figure \ref{fig-compare-c12c13-DFcomp-Obs-all}
that the DSF events produce a range of $^{12}$C/$^{13}$C ratios,
although they are all low or moderately low, being $\lesssim25$.
At higher metallicities ($\textrm{[Fe/H]}=-3$) the ratios in our
models all sit around 20. This is consistent with some of the observations
but not all. In particular there is a notable lack of low $^{12}$C/$^{13}$C
ratios in these models. Interestingly it is the C-enhanced EMPs, which
mostly have very low ratios, that are \emph{not} matched by our models
at this metallicity (at least from DSF pollution). At $\textrm{[Fe/H]}=-4$
our models show a spread in $^{12}$C/$^{13}$C ratio, much more like
the observations. There are however very few observational data points
to compare with at this metallicity. 

Finally, two general points can be made about the $^{12}$C/$^{13}$C
ratios arising from the DF pollution in our models. The first is that
we find that the bulk of our models at higher metallicity ($\textrm{[Fe/H]}=-3,-4$)
show a similar spread as the observations (although the lack of low
ratios at $\textrm{[Fe/H]}=-3$ is a notable exception). The second
is that this spread is seen to reduce with metallicity -- the lower
the metallicity the lower the upper envelope of the $^{12}$C/$^{13}$C
ratio. Indeed, at $Z=0$ all the models show equilibrium values of
$\sim5$. 

\subsection{Comparisons With Yields}

Here we compare the average composition of the yields from the stellar
winds with observations. 

\subsubsection*{Carbon}

\begin{figure}
\begin{centering}
\includegraphics[width=0.95\columnwidth,keepaspectratio]{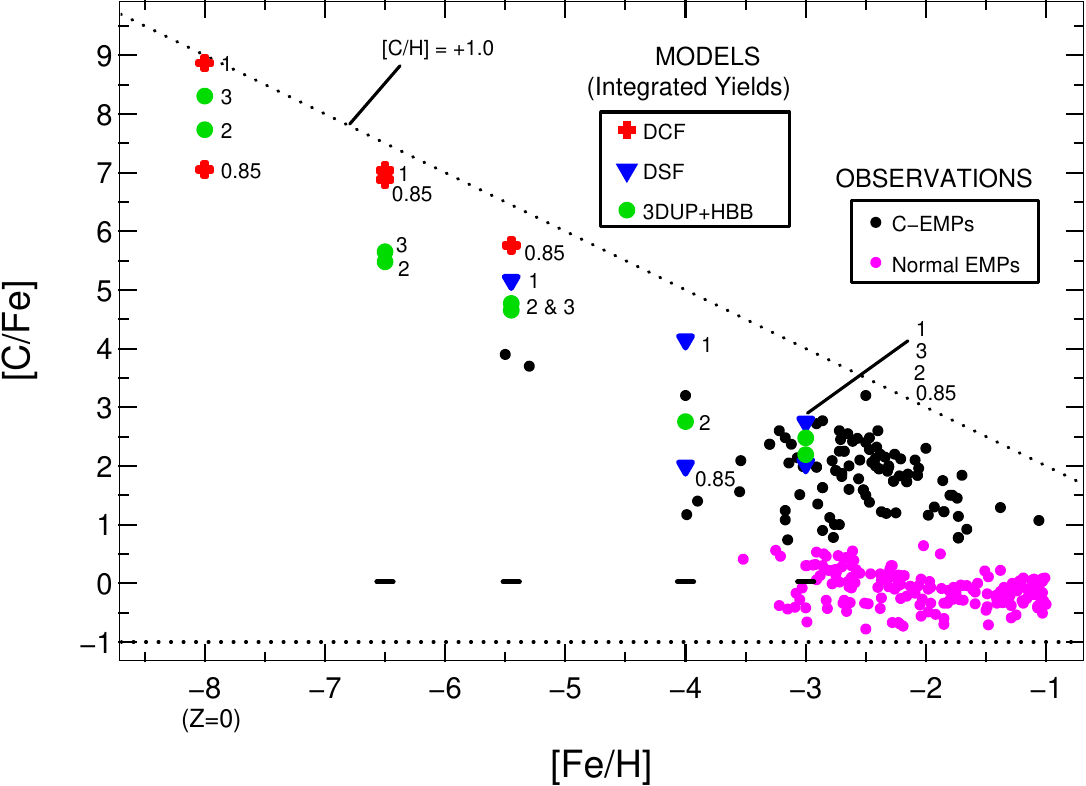}
\par\end{centering}
\caption{Comparing the yields from all our models with observations of EMP
stars. See Figure \ref{fig-Observations-EMPs-CHvsFeH} for the observational
data sources. Here we have colour-coded the observations into {[}C/Fe{]}-rich
and {[}C/Fe{]}-normal, where {[}C/Fe{]}-rich is defined by $\textrm{[C/Fe]}>+0.7$.
We keep this colour coding in all the figures below. The short horizontal
lines indicate the starting composition of the models (in this case
they are all at $\textrm{[C/Fe]}\sim0$, except for the $Z=0$ models).
We have plotted the $Z=0$ models at $\textrm{[Fe/H]}=-8$ for comparison.
The two most metal-poor stars can be seen at $\textrm{[Fe/H]}\sim-5.5$
(they are both C-rich). An upper envelope to the pollution of the
models -- and the observations -- is marked by the dotted line at
$\textrm{[C/H]}=+1.0$. The yields from our models are colour- and
shape-coded to highlight the different episodes that produced the
bulk of the pollution in each yield. \label{fig-compare-CFe-Yields-Obs-all}}
\end{figure}

We begin with {[}C/Fe{]} in Figure \ref{fig-compare-CFe-Yields-Obs-all}.
As mentioned above we have defined C-enhanced EMPs as those having
$\textrm{[C/Fe]}>+0.7$, as shown in the figure by different coloured
dots. The models are represented by coloured shapes in the same way
as in the DF comparison figures above, but this time we have categorised
the yields in terms of the evolutionary episodes that play key roles
in the resultant yield compositions (see eg. Figure \vref{fig-PollutionSummary}
for an overview). In practice this just means that the IM model yields
are represented by green dots (= AGB pollution) rather than blue triangles
(= DSF pollution) since the 3DUP and HBB on the AGB in these models
far outweigh the DSF pollution in the yields. 

It can be seen in Figure \ref{fig-compare-CFe-Yields-Obs-all} that
\emph{all} of our models contribute very C-enhanced material to the
ISM. We have marked in an upper envelope line at $\textrm{[C/H]}=+1.0$
that is consistent with our models and with the observations. The
trend in the models is for increasing {[}C/Fe{]} with decreasing metallicity,
which reflects the fact that they all lie within $2\rightarrow3$
dex below the $\textrm{[C/H]}=+1.0$ line. It is interesting to note
that strong C-enhancement occurs via all of the types of polluting
episodes. This is expected as the DSF, DCF and 3DUP all bring up carbon
from the He burning regions. Indeed, it is the composition and the
size of these He burning regions that sets an upper limit to the amount
of surface pollution that can occur (dilution through the envelope
also plays a part of course). It is this that sets the upper envelope
in our figures. It appears that nature may also do the same!

Looking at the observations we see that \emph{none} of the yields
from our models are low enough in C to account for the normal EMPs.
Thus our models predict that none of the C-normal EMPs are on the
AGB -- or have been polluted by stars in this phase of evolution
-- in the mass range of our study ($0.85\rightarrow3.0$ M$_{\odot}$).
This extends to the HB in the DCF models, as in these models the surfaces
become polluted at the top of the RGB. Indeed, due to the various
C-enriching episodes in all our models, the lowest C abundance of
\emph{all} our yields is $\textrm{[C/Fe]}\sim+2$ dex (in the 0.85
M$_{\odot},$ $\textrm{[Fe/H]}=-4$ model). This profusion of carbon
matches well with the most of the observed abundances in the C-EMPs,
although some dilution is needed to cover the lower {[}C/Fe{]} observations.
We note that our 5 M$_{\odot}$ models of $\textrm{[Fe/H]}=-3$ and
$-4$ also experience 3DUP (see Figure \vref{fig-PollutionSummary})
which means that their yields would also be C-enhanced (although possibly
to a lesser extent due to their larger envelopes and smaller intershells).
The lower metallicity 5 M$_{\odot}$ models did not experience 3DUP
so their yields would be unpolluted (SNe type 1.5 events withstanding).
Thus the prediction is strengthened on widening the mass range --
our models predict all yields coming from low- and intermediate-mass
stars of {[}Fe/H{]} at least $<-3$ to be C-enhanced. We note that
the intermediate mass models of \citet{2002PASA...19..515K} (which
have $\textrm{[Fe/H]}\sim-1$) and those of \citet{2004ApJS..155..651H}
($\textrm{[Fe/H]}=-2.3$) also show significant 3DUP. Thus the 3DUP+HBB
pollution from our extremely metal-poor IM models can be seen an extension
of that given by the moderately metal-poor IM models.

The situation is more intriguing at low mass. The 1 M$_{\odot}$ model
of \citet{2002PASA...19..515K} ($\textrm{[Fe/H]}\sim-1$), like our
1 M$_{\odot}$ EMP models, does not show significant 3DUP -- even
though that study utilises a form of overshoot (we note that the \citeauthor{2002PASA...19..515K}
study uses an earlier version of same code as the current work). However,
in a recent study \citet{2005MNRAS.356L...1S} do find 3DUP to occur
in models of mass 1 M$_{\odot}$ at a metallicity of $\textrm{[Fe/H]}\sim-1$.
The fact that they find 3DUP at such low masses (the lower limit is
usually $\sim1.5$ M$_{\odot}$ at solar metallicity, see eg. \citealt{1996MmSAI..67..651S})
allows them to match the low luminosity end of the carbon star luminosity
function for the LMC. They also report that 3DUP is metallicity-dependent,
such that their lower metallicity models experience deeper 3DUP (\citealt{2002PASA...19..515K}
also report a metallicity dependence, as do earlier studies). We note
that they did not include any form of overshoot. Thus their code may
be expected to produce 3DUP in a 1 M$_{\odot}$ model at extremely
low metallicity. If this were the case then this would be another
avenue/mass range for CEMP pollution. Despite the fact that our models
do not have 3DUP they do have very C-enhanced yields due to the DCF
and DSF events. Thus we have two possible channels for C production
in low mass EMP models. One way to distinguish between the pollution
from these two channels is through their different chemical signatures.
Low mass stars that have 3DUP are not predicted to experience HBB
so N remains low whilst C is periodically enhanced. By contrast our
DCF and DSF models show \emph{C and N enhanced at the same time}.
Interestingly this dual enhancement appears to be required by the
observations, which we discuss below (see Figure \ref{fig-compare-CN-Yields-Obs-all}).
Related to this are the $^{12}$C/$^{13}$C ratios -- they would
be high in the 3DUP case due to repetitive dredge-up of $^{12}$C
but are low in the DCF and DSF models because of CN cycling (they
are lowest in the DCF models). 

Another interesting possibility is that DFs \emph{and} 3DUP could
occur at low mass (which may be possible in our models by using overshoot).
In this case the models would again be C and N rich. This is because
the N from the DSF and DCF events has no way of being depleted in
the H-rich envelopes. Since 3DUP appears to enrich the yields with
C to a similar degree as the DF events, the C enhancement would probably
not increase significantly in this scenario over our present DF-only
results. The $^{12}$C/$^{13}$C ratios may however increase but we
are uncertain how significant this would be. Clearly this scenario
requires further modelling -- the combination of DFs with 3DUP may
well provide a better match to the observations. We shall pursue this
avenue at a later date and will return to this important topic in
the Discussion chapter.

\subsubsection*{$^{12}$C/$^{13}$C Ratio}

\begin{figure}
\begin{centering}
\includegraphics[width=0.95\columnwidth,keepaspectratio]{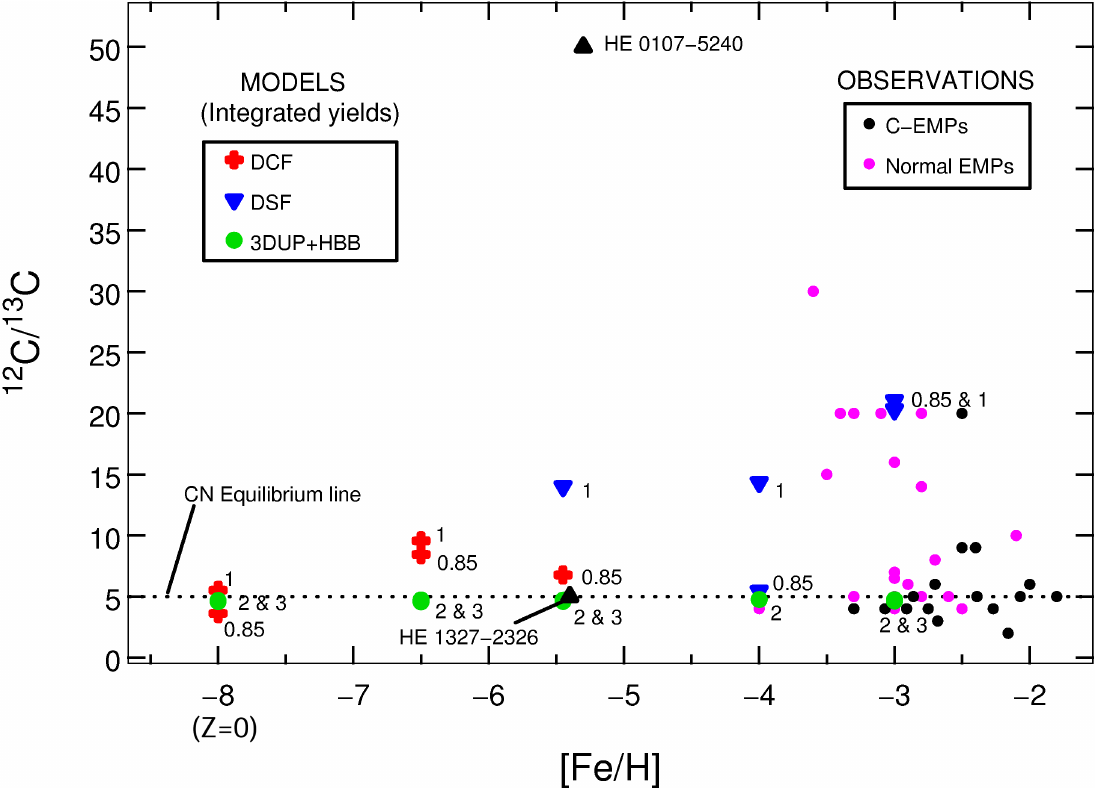}
\par\end{centering}
\caption{Comparing the $^{12}$C/$^{13}$C ratios in our models with those
from observations. We have plotted the $Z=0$ models at $\textrm{[Fe/H]}=-8$
for comparison. See Figure \ref{fig-Observations-EMPs-CHvsFeH} for
the observational data sources (note that only some have reported
$^{12}$C/$^{13}$C ratios). C-EMPs and normal EMPs are defined in
Figure \ref{fig-compare-CFe-Yields-Obs-all}, as are the symbols used
for the yields. The dotted line at $^{12}$C/$^{13}$C$=5$ indicates
(roughly) the equilibrium value from the CN cycle. All the HBB models
end up on this line. An interesting feature here is that most of the
C-EMPs have low ratios, whilst the normal EMPs show a spread. \label{fig-compare-c12c13-Yields-Obs-all}}
\end{figure}

Moving back to the yield-observation comparisons we present in Figure
\ref{fig-compare-c12c13-Yields-Obs-all} the $^{12}$C/$^{13}$C ratios
in the yields of all our models. In most cases the ratios are basically
the same as those reported earlier for the DF surface compositions.
The only changes have been brought about by HBB in the IM models.
All the 2 and 3 M$_{\odot}$ models now have low ratios, reflecting
the substantial CN cycling that occurred in their AGB envelopes. The
pattern of reducing spread in the $^{12}$C/$^{13}$C ratio with reducing
metallicity remains in the low-mass yields. As mentioned above \emph{all}
the models have CN equilibrium values at $Z=0$ (the $Z=0$ models
are included at $\textrm{[Fe/H]}=-8$ in the figure). Most of the
carbon-enhanced EMPs have low ratios. From the perspective of our
models this means that they could have been polluted by AGB HBB material
or DCF material (or possibly DSF material in some cases). The observations
of the C-normal EMPs reveal a spread in $^{12}$C/$^{13}$C ratio.
This may reflect stars at different stages of evolution (eg. deep
mixing in RGB stars can reduce the ratio -- see \citealt{2006AA...455..291S}
for some observational evidence), or inhomogeneous formation material.
HE 0107-5240 is again an outlier. It has a lower limit to its ratio
of 50, well outside the range of our models. In terms of an AGB mass
transfer scenario we suggest that it could have such a high ratio
if the matter was transferred from an IM star during its early TP-AGB
phase -- when some 3DUP of $^{12}$C had occurred but HBB had not
set in (see \citealt{2004ApJ...603..708C} for a detailed discussion
on possible sources of the abundance pattern of this star).

\subsubsection*{Nitrogen}

\begin{figure}
\begin{centering}
\includegraphics[width=0.95\columnwidth,keepaspectratio]{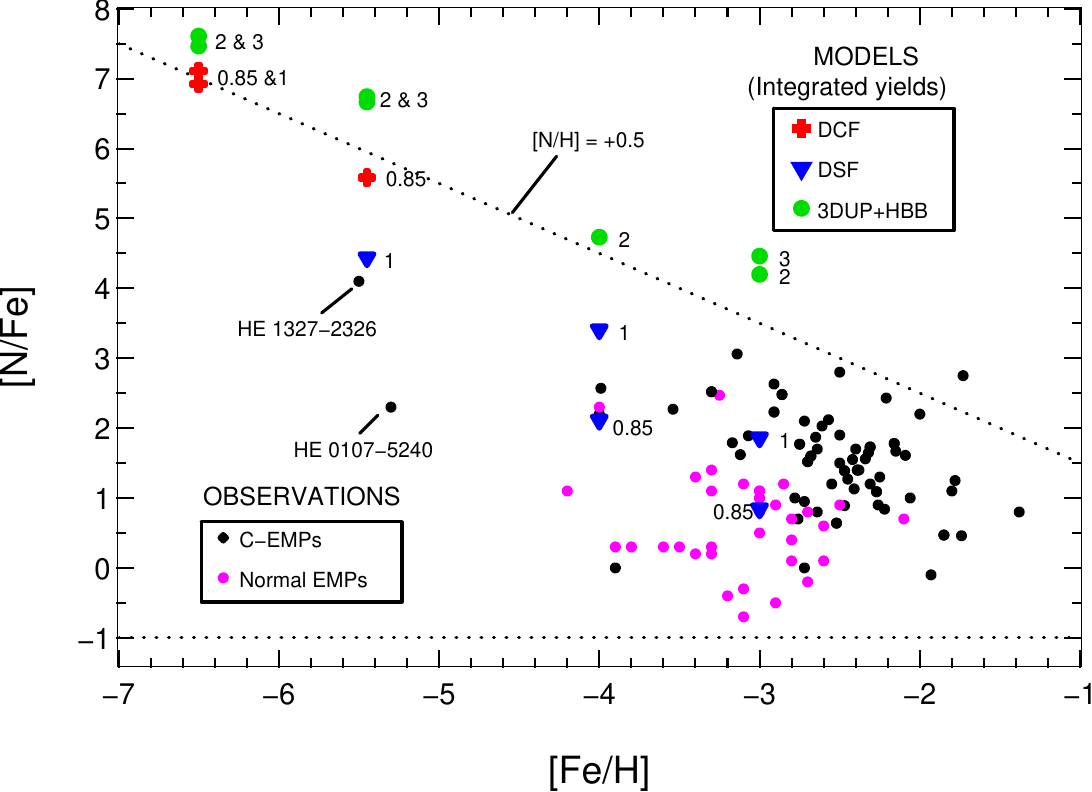}
\par\end{centering}
\caption{Comparing the {[}N/Fe{]} ratios in the yields of our models with those
of the observations. See Figure \ref{fig-Observations-EMPs-CHvsFeH}
for the observational data sources. C-EMPs and normal EMPs are defined
in Figure \ref{fig-compare-CFe-Yields-Obs-all}, as are the symbols
used for the yields. All the HBB models (2 and 3 M$_{\odot}$) produce
enormous amounts of nitrogen, whilst the DSF models produce less (although
still substantial amounts). The observations show that C-EMPs are
generally more N-rich than normal CEMPs, but there is an overlap in
the distributions. We have sketched in upper and lower envelopes for
the observations (dotted lines). Note that we have not plotted the
$Z=0$ models in this case. \label{fig-compare-NFe-Yields-Obs-all} }
\end{figure}

In Figure \ref{fig-compare-NFe-Yields-Obs-all} we display {[}N/Fe{]}
against {[}Fe/H{]} for the observations and our model yields. First
we note that our N sample is small and also biased. That said, it
can be seen that the C-enhanced EMPs are also N-enhanced (in general).
Some of the C-normal stars are also N-enhanced but, in general, not
as much as the CEMPs. The two populations appear to each have a large
spread in N, which overlaps between the populations. Looking at our
model yields we see that they are all enhanced in N. The lowest abundance
is $\textrm{[N/Fe]}\sim+1.0,$ in the 0.85 M$_{\odot},$ $\textrm{[Fe/H]}=-3$
model. Thus \emph{all} our yields are enhanced in C \emph{and} N.
The yields that are dominated by 3DUP+HBB contain very large amounts
of N. In fact all of these yields lie above the rough upper envelope
we have inscribed for the observations in Figure \ref{fig-compare-NFe-Yields-Obs-all}
at $\textrm{[N/H]}=+0.5$. The large amount of N production is a robust
prediction of the models that have 3DUP and HBB. This N excess may
indicate that dredge-up is not as deep in real stars, since this would
reduce the C seeds for the N production. We note that this may not
actually be a problem in a binary scenario due to (probable) dilution
effects. Another interesting feature of the model yields is that the
DSF models all produce much less nitrogen (whilst still having large
amounts of C). These yields lie right in amongst the observations,
at least at higher metallicities. Looking at the two most metal-poor
stars we see that HE 1327-2326 is not far from the lower envelope
of our models ($\sim0.2$ dex from our 1 M$_{\odot}$ model). However
HE 0107-5240 is well below the envelope. This star has very high C
but not so high N (although it is enhanced in both). Our EMP models
produce too much N to explain this star. 

\subsubsection*{C/N Ratio}

\begin{figure}
\begin{centering}
\includegraphics[width=0.95\columnwidth,keepaspectratio]{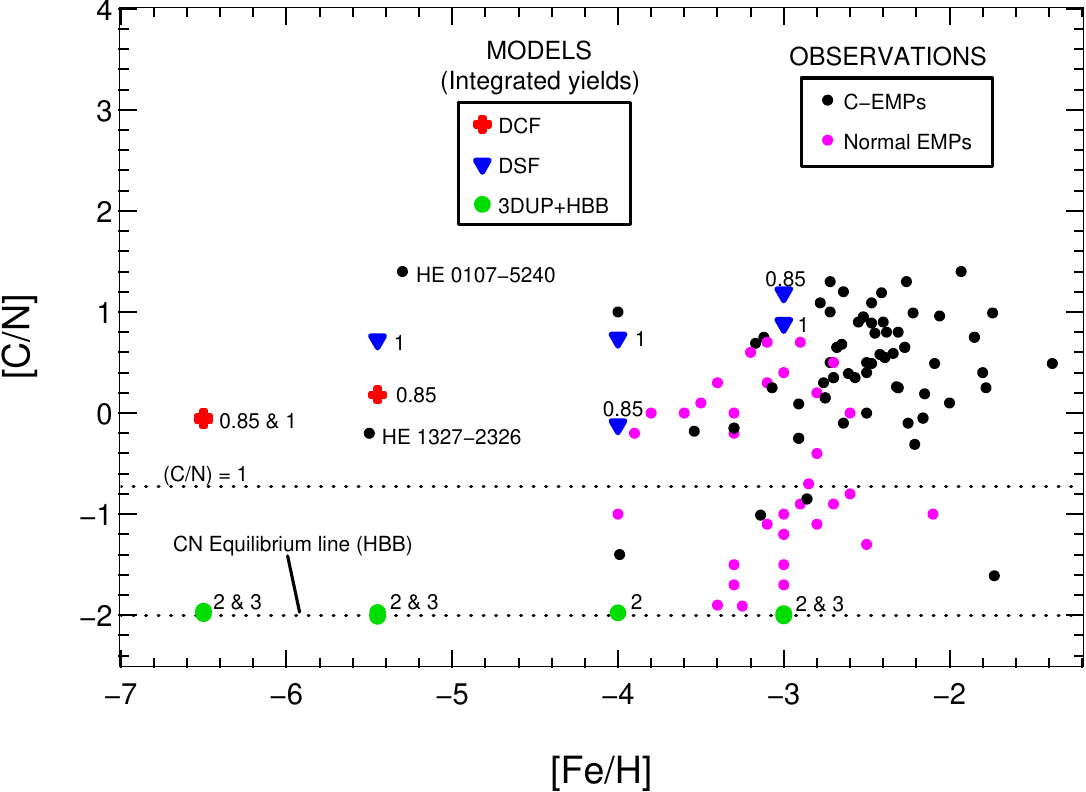}
\par\end{centering}
\caption{Comparing the {[}C/N{]} ratios in the yields of our models with those
of the observations. See Figure \ref{fig-Observations-EMPs-CHvsFeH}
for the observational data sources. C-EMPs and normal EMPs are defined
in Figure \ref{fig-compare-CFe-Yields-Obs-all}, as are the symbols
used for the yields. We have marked in the CN equilibrium line (for
HBB) at $\textrm{[C/N]}=-2$ (ie. $\textrm{C/N}=0.05$). All the HBB
models fall on this line. We have also marked in a dotted line at
$\textrm{[C/N]}\sim-0.8$ which is where $\textrm{C/N}\sim1$, so
that C dominates N above this line. An interesting feature of the
observations is that most of the CEMPs lie above the $\textrm{C/N}=1$
line whilst the normal CEMPs show a spread. Thus the CEMPs are rich
in C \emph{and} N but have more C. There are however a few stars that
show strong C enhancements and have $\textrm{N}>\textrm{C}$. The
yields of our models are consistent with the spread of the observations.
However HE 0107-5240 lies significantly above out models of comparable
metallicity. Note that we have not plotted the $Z=0$ models in this
case. \label{fig-compare-CN-Yields-Obs-all}}
\end{figure}

This brings us to our next figure where we show the C:N ratios for
model yields and observations, relative to solar (Figure \ref{fig-compare-CN-Yields-Obs-all}).
HE 0107-5240 is seen to be on the upper envelope of the observations
in this parameter space. The normal EMPs appear to occupy the whole
range between $\textrm{[C/N]}=-2.0\rightarrow+1.0$. This is in contrast
to the C-EMPs which are mainly found in the range $-0.5\rightarrow+1.5$.
This shows that C dominates over N in the CEMPs, even though they
are rich in both (there are exceptions though). The spread probably
reflects differing degrees of C$\rightarrow$N processing in the polluting
material, whether the material be supplied internally or externally.
Looking at the models the most robust prediction is the very low {[}C/N{]}
values in the 3DUP+HBB yields -- they all have $\textrm{[C/N]}\sim-2.0$.
This reflects the fact that CN equilibrium was achieved in the AGB
envelopes, such that C/N $\approx0.05$ (ie. N/C $\sim20$). Interestingly
this also appears to be the lower envelope for the observations. All
of the yields for our other models fall above the $\textrm{C/N}=1$
line. Thus -- like the observations -- C is always dominant over
N in the yields. There is a bit of a trend with metallicity in the
yields, insomuch as the lower the metallicity of the model, the lower
the value of the upper envelope. All the yields are however either
at or above the solar ratio, again similar to the observations. 

\subsubsection*{Oxygen}

\begin{figure}
\begin{centering}
\includegraphics[width=0.95\columnwidth,keepaspectratio]{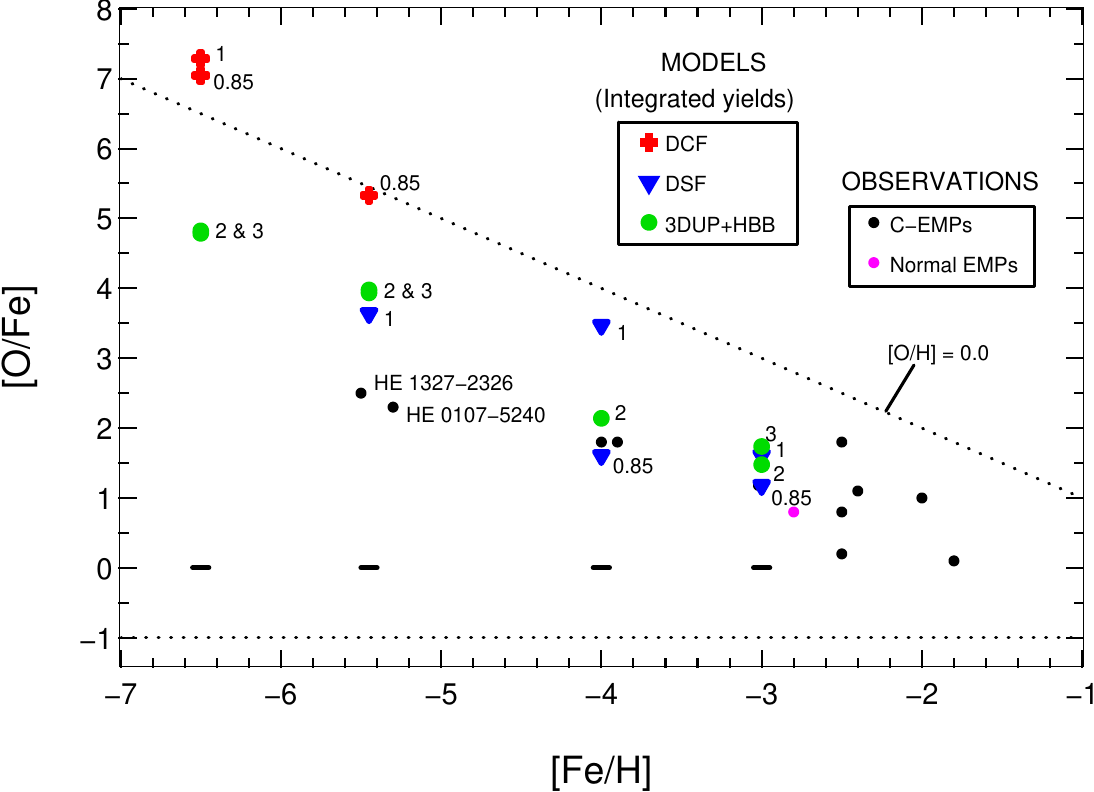}
\par\end{centering}
\caption{Comparing the {[}O/Fe{]} ratios in the yields of our models with those
of the observations. See Figure \ref{fig-Observations-EMPs-CHvsFeH}
for the observational data sources. C-EMPs and normal EMPs are defined
in Figure \ref{fig-compare-CFe-Yields-Obs-all}, as are the symbols
used for the yields. Very sketchy upper and lower envelopes are marked
with dotted lines (there not not many observations to compare with
as oxygen is difficult to measure). The two most metal-poor stars
observed to date are indicated. Their oxygen abundances are from \citet{2004ApJ...612L..61B}
(HE 0107-5240) and \citet{2006ApJ...638L..17F} (HE 1327-2326). It
can be seen that all models produce large oxygen enhancements, but
the DCF events produce the most. The yields form HBB and DSF models
match the observations well at the higher metallicities but are $\sim1$
dex higher than those of HE 0107-5240 and HE 1327-2326 at $\textrm{[Fe/H]}\sim-5.5$.
Note that we have not plotted the $Z=0$ models in this case. \label{fig-compare-OFe-Yields-Obs-all}}
\end{figure}

In Figure \ref{fig-compare-OFe-Yields-Obs-all} we compare the observations
of oxygen with our model yields. As mentioned earlier oxygen is difficult
to measure so there are not many observations to compare with. This
is evident in our literature sample. The trend in the observations
is one of increasing O enhancement with decreasing metallicity (although
the sample is small). There is however a spread at the higher metallicities.
The spread is partially covered by our yields at $\textrm{[Fe/H]}=-3$
but the lowest values are not matched. Interestingly there are a couple
of C-EMPs that have approximately solar {[}O/Fe{]}, where we might
expect the C-normal stars to lie (we only have one C-normal star with
an O abundance unfortunately). At $\textrm{[Fe/H]}\sim-4$ our 0.85
M$_{\odot}$ DSF model and our 2 M$_{\odot}$ 3DUP+HBB model both
have similar O abundances to the two stars for which we have measurements.
On the other hand our models at $\textrm{[Fe/H]}=-5.5$ all overproduce
O relative to the two most metal-poor stars, by at least $\sim1$
dex. Again this could be remedied by invoking dilution of the yields.
In the figure we also sketch a rough upper envelope for O enhancement
from out models (note that the $Z=0$ models have exaggerated O since
we have artificially chosen an Fe abundance for plotting purposes).
This limit is also consistent with the observations. Finally we note
that more observations of oxygen are really needed to constrain the
models. 

\subsubsection*{Sodium}

\begin{figure}
\begin{centering}
\includegraphics[width=0.95\columnwidth,keepaspectratio]{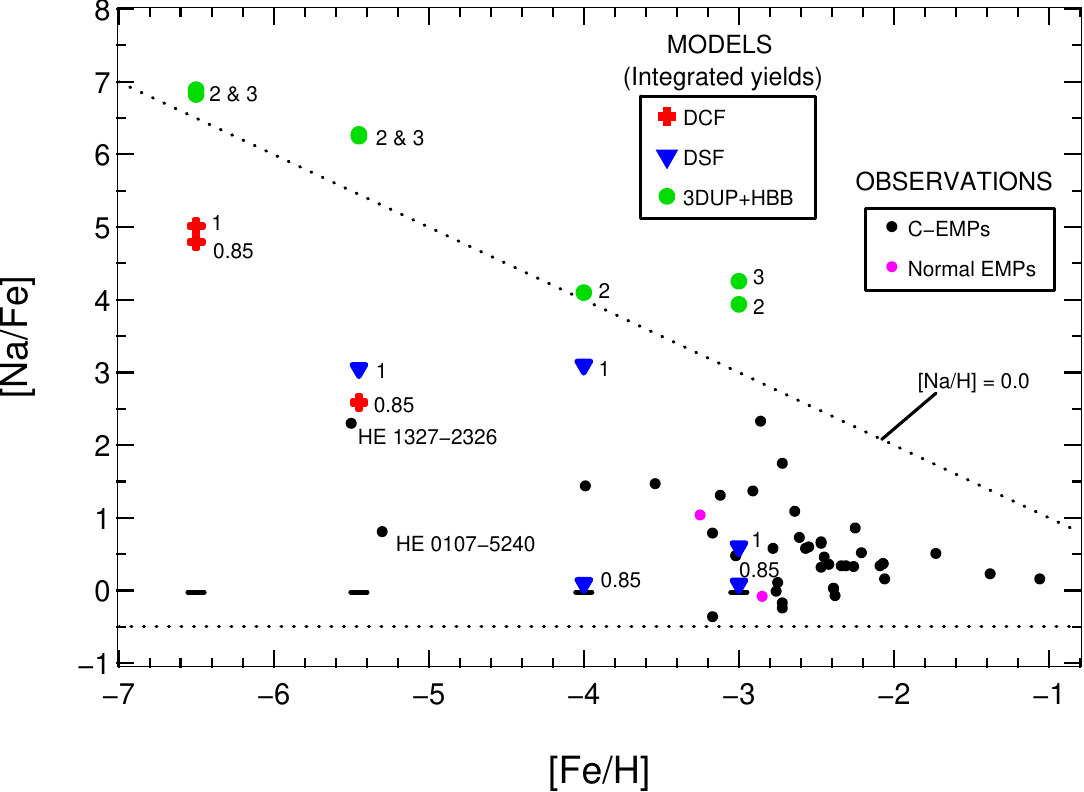}
\par\end{centering}
\caption{Comparing the {[}Na/Fe{]} ratios in the yields of our models with
those of the observations. See Figure \ref{fig-Observations-EMPs-CHvsFeH}
for the observational data sources. C-EMPs and normal EMPs are defined
in Figure \ref{fig-compare-CFe-Yields-Obs-all}, as are the symbols
used for the yields. Upper and lower envelopes are sketched with dotted
lines at $\textrm{[Na/H]}=0.0$ and $\textrm{[Na/Fe]}=-0.5$. The
observations show a wide spread in Na amongst the C-EMPs and it can
be seen that our HBB models produce enormous mounts of Na. Note that
we have not plotted the $Z=0$ models in this case. \label{fig-compare-NaFe-Yields-Obs-all}}
\end{figure}

We now move onwards to sodium, which we show in Figure \ref{fig-compare-NaFe-Yields-Obs-all}.
Most of the observational data we have for Na is for the C-EMPs. The
observations show a large spread at $\textrm{[Fe/H]}\sim-3$, ranging
from $\textrm{[Na/Fe]}\sim0.5$ dex below solar to $\sim2.5$ dex
super-solar. Our models at this metallicity show an even larger spread,
from solar to $\sim4$ dex super-solar. This is mainly driven by the
huge amounts of Na from the 2 and 3 M$_{\odot}$ models, which is
a product of HBB of Ne dredged up from the intershells. These very
large overabundances continue at lower metallicities and are reminiscent
of the huge N overabundances -- which have the same source (with
C as the seed). Thus Na is another argument for less 3DUP in the models
. This may be consistent with C as well, but this is less clear. If
not, then these results may imply that HBB is too efficient in our
models. Looking at the two most metal-poor stars we see that HE 1327-2326
is very close to our 0.85 M$_{\odot}$ yield at $\textrm{[Na/Fe]}\sim2.5$,
but HE 0107-5240 again defies the models, having a relatively low
Na abundance. Interestingly all of our models produce Na enhancements
at $\textrm{[Fe/H]}=-6.5$ and $Z=0$. 

\subsection{Comparisons with Individual EMP Halo Stars\label{SubSection-HaloStars-Obs-Compare-IndividStars}}

\subsubsection*{Stars With $\textrm{[Fe/H]}\approx-4.0$}

\begin{figure}
\begin{centering}
\includegraphics[width=0.8\columnwidth,keepaspectratio]{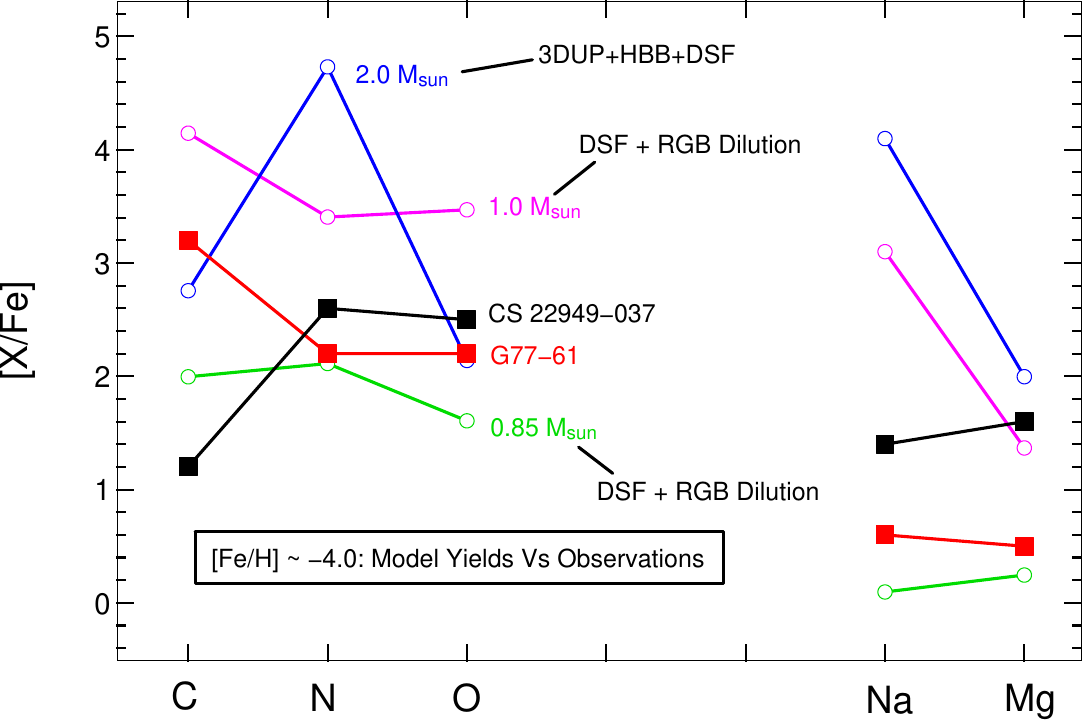}
\par\end{centering}
\caption{Comparing the abundance pattern in our $\textrm{[Fe/H]}=-4.0$ yields
with observations of stars with the same metallicity. See text for
sources of the abundance determinations in these stars. The clear
3DUP+HBB signature of high C and O combined with very enhanced N can
be seen in the 2 M$_{\odot}$ model (note that we do not display the
3 M$_{\odot}$ yield due to a loss of data, but it would be very similar
to the 2 M$_{\odot}$ one). The AGB yields of both the 0.85 and 1
M$_{\odot}$ have been diluted by varying amounts of unpolluted mass
loss on the RGB. See text for a discussion. \label{fig-compare-FeH-4.0-Yields-Obs-2stars}}
\end{figure}

In our sample from the literature we have about 10 stars which have
$\textrm{[Fe/H]}\sim-4.$ However there are only a couple for which
we have C, N, O, Na and Mg abundances. In Figure \ref{fig-compare-FeH-4.0-Yields-Obs-2stars}
we plot the abundances of these two stars along with the abundances
of the yields from all our $\textrm{[Fe/H]}=-4.0$ models. It is thus
a raw, direct comparison (ie. no scaling of Fe). The abundances for
the C-EMP star G77-61 are from \citet{2005AA...434.1117P} and those
for CS 22949-037 (also a C-EMP) are from \citet{2004AA...419.1095I}
and \citet{2006AA...455..291S}. In the figure we indicate which were
the dominant sources of pollution in the yields for each model. At
this metallicity we have representatives from only two of the three
pollution groups -- the DSF and 3DUP+HBB groups. This is because
the DCF only occurs in models of lower metallicity. The clear signature
of HBB can be seen in the 2 M$_{\odot}$ yield -- N is very abundant,
being $\sim2$ dex above C in the same model. Oxygen is overabundant
but still less abundant than C and N. The large Na production from
the combination of HBB and 3DUP can be seen, as can a $\sim2$ dex
overabundance of Mg (from 3DUP). The DSF yields show variation. The
0.85 M$_{\odot}$ model shows roughly a 2 dex lower enhancement in
C, N and O than the 1 M$_{\odot}$ model. Nitrogen is more dominant
in the 0.85 M$_{\odot}$ yield (relative to its C and O). With regards
to Na and Mg the two yields diverge -- Na is much more prevalent
in the 1 M$_{\odot}$ model ($\sim3$ dex higher) whilst Mg is only
$\sim1$ dex higher. The $\sim2$ dex gap in the CNO group yield between
these models is partly due to the dilution of the yields by unpolluted
RGB mass loss (see Figures \ref{fig-massLoss-allm8} and \vref{fig-massLoss-allm1}
in the structural evolution section), and partly due to the greater
surface enrichment from the DSF event in the 1 M$_{\odot}$ model.
The difference in abundance \emph{patterns} is due to the differences
in the dual shell flashes. The gas in the 1 M$_{\odot}$ model achieved
more advanced proton burning on top of the dredged up He intershell
material, whilst the 0.85 M$_{\odot}$ yield shows more signs of CN
cycling. 

Looking at the abundance patterns in the two $\textrm{[Fe/H]}=-4$
stars in Figure \ref{fig-compare-FeH-4.0-Yields-Obs-2stars} we see
that they have similar N and O abundances. Their C abundances are
however very different. G77-61 has $\sim2$ dex more C than CS 22949-037.
On the other hand CS 22949-037 is more enhanced in Na and Mg, with
both being about 1 dex higher in this star. The abundance pattern
of CS 22949-037 is closest to that of the HBB model (2 M$_{\odot}$)
except for the fact that its oxygen abundance is similar to its N
abundance, whereas in the HBB model O is much lower than N. Thus it
appears that the CS 22949-037 material has undergone CN cycling but
not ON cycling. This would be the case in a star with less efficient
HBB. Deep mixing could also do this but the large CNO abundances would
need to be explained as no dredge-up occurs on the RGB. This may be
a candidate for a scenario in which a low mass star with DSF+3DUP
and only weak HBB gets its CNO enhancements from the DSF and then
cycles C to N whilst O remains high from the DSF. The moderate enhancements
in Mg and Na would come from the 3DUP+HBB also. Again, it will be
interesting to calculate some models at these metallicities with 3DUP
via the inclusion of overshoot, which we leave as a future project.
Moving now to G77-61 we see that there is a very nice match to its
CNO abundance pattern in the 1 M$_{\odot}$ yield, albeit offset upwards
by $\sim1.5$ dex. The model yield and the star both have high C overabundances
while N and O about 1 dex lower than C. Magnesium is similarly offset
upwards. Sodium is however $\sim2.5$ dex higher in the model. Interestingly
the Na-Mg pattern is very similar to the HBB Na-Mg pattern. We discuss
this interesting (partial) match in terms of the DSF abundance patterns
below.

\subsubsection*{Comparing With the Dual Flash Abundances: The G77-61 Case}

\begin{figure}
\begin{centering}
\includegraphics[width=0.8\columnwidth,keepaspectratio]{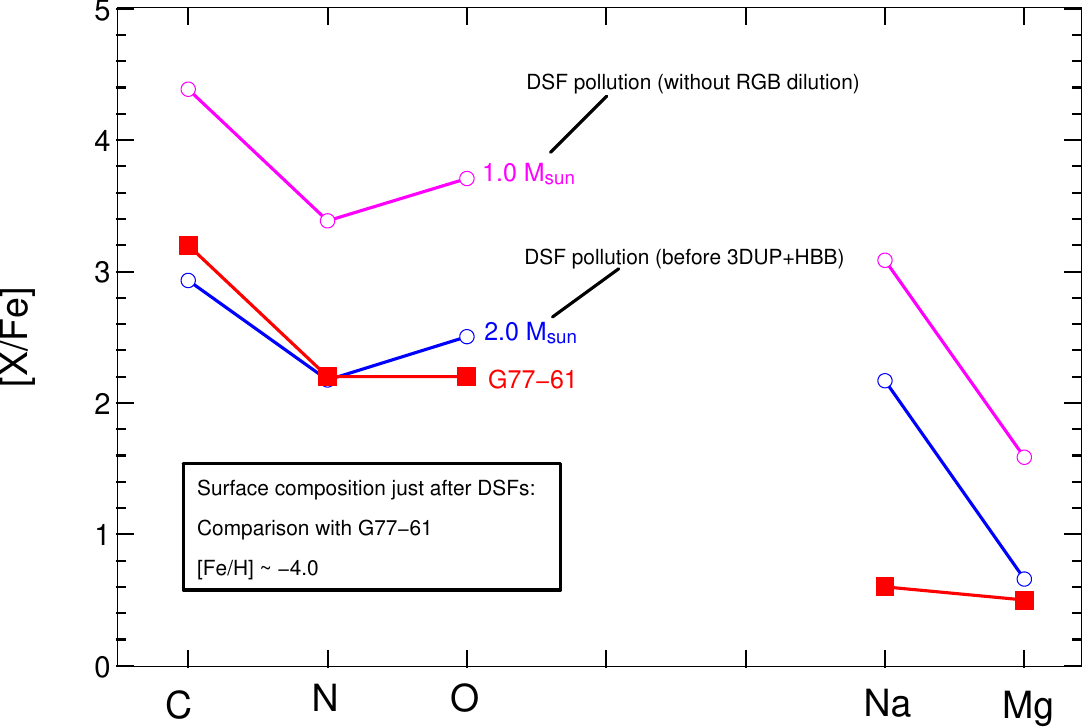}
\par\end{centering}
\caption{Comparing the abundance patterns in our $\textrm{[Fe/H]}=-4.0$ models
taken just after the DSF events with the C-EMP dwarf star G77-61.
See text for sources of the abundance determinations of this star.
By contrasting the chemical signature of this event with that of the
integrated yields (Figure \ref{fig-compare-FeH-4.0-Yields-Obs-2stars})
it can be seen how strongly the 3DUP+HBB affects the yields (eg. in
the 2 M$_{\odot}$ model). Interestingly the DSF CNO signature matches
fairly well with G77-61, particularly if observational uncertainties
are considered (usually $\sim\pm0.3$ dex or so). The only problem
with the match is that the models produce by far too much Na. \label{fig-compare-FeH-4.0-DSFs-Obs-G77-61}}
\end{figure}

In Figure \ref{fig-compare-FeH-4.0-DSFs-Obs-G77-61} we make a comparison
between the surface compositions of our 1 and 2 M$_{\odot}$ models
taken just after the dual shell flash events and the observed composition
of the star G77-61. This star is a cool carbon-rich dwarf (also identified
as a carbon star, due to the presence of strong C-based molecular
bands). Since it is a dwarf it can not have gained its surface abundances
via autopollution. The binary mass transfer hypothesis (from a low
or IM companion) has always been the most favoured scenario for explaining
its abundance pattern (eg. \citealt{1977ApJ...216..757D}; \citealt{2005AA...434.1117P}).
\citet{1986ApJ...300..314D} found that it indeed had a binary companion.
By comparing the abundance pattern of this star with our DSF surface
compositions we are sampling the early TP-AGB stage as a candidate
for the mass transfer hypothesis. As mentioned earlier we suggest
that the DSF (and DCF) events may lead to enhanced mass loss since
they are (at least) mildly violent events. In addition to this the
surfaces of these models are not enhanced in carbon until this stage,
so any mass transfer before this would not have a significant effect
on the surface composition of the secondary star. Moreover, the pollution
may also enhance the stellar winds due to grain formation, again increasing
the mass transfer rate. 

Looking now at our models in Figure \ref{fig-compare-FeH-4.0-DSFs-Obs-G77-61}
we see that both models (1 and 2 M$_{\odot},$ $\textrm{[Fe/H]}=-4.0$)
show a very similar abundance pattern to that of G77-61 at this stage
of their evolution. In the case of the 1 M$_{\odot}$ model this abundance
pattern is retained for the rest of its evolution, as it experiences
no 3DUP. This is evident in the yields shown in Figure \ref{fig-compare-FeH-4.0-Yields-Obs-2stars}.
In this case the yield is diluted by unpolluted RGB mass loss, so
the abundances are scaled down. This could be an explanation for the
G77-61 abundance pattern, especially if there were more dilution.
However, looking at Na and Mg we see that the ratio of these two elements
is very different to that in G77-61. These two elements have similar
overabundances in G77-61 ($\sim0.5$ dex above solar) but in the models
Na is $\sim1.5$ dex higher than Mg. This excess Na problem is reminiscent
of the globular cluster abundance problem, where standard AGB models
produce too much Na (see next chapter). We note that this may be a
product of poorly constrained reaction rates (Illiadis and Lattanzio
2007, private communication) and we shall pursue this uncertainty
as a future study. Looking now at the 2.0 M$_{\odot}$ DSF surface
composition it can be seen that there is a good match with the CNO
pattern -- and absolute abundances -- of G77-61. Taking into account
abundance determination uncertainties (which can vary between studies
by $\sim\pm0.3$ dex or more), it is a very good match. We reiterate
that no scaling has been performed on these compositions. There is
however the same problem with excess sodium (the Mg matches very well
though). An important point here is that the 2 M$_{\odot}$ model
goes on to experience 3DUP soon after the DSF event. This means that
this abundance pattern is not maintained for very long, as C will
quickly increase. Once HBB sets in this C is burnt to N and the classic
HBB abundance pattern will emerge (see the yield pattern for this
model in Figure \ref{fig-compare-FeH-4.0-Yields-Obs-2stars}), which
is very different to that of G77-61. One way to circumvent this problem
would be to postulate that the DSF gives rise to an envelope ejection
(or partial envelope ejection). In this way the companion star, G77-61,
would accrete matter with the DSF pattern. This is however very speculative
-- fluid dynamics simulations are really needed to test if the DSF
event can do this.

\subsubsection*{Stars With $\textrm{[Fe/H]}\approx-5.5$: Comparisons with Yields
\& DF Abundances}

This metallicity represents the lower boundary of all observations
made of the Universe by humanity to date. No objects with metallicities
lower than the halo stars HE 0107-5240 ($\textrm{[Fe/H]}=-5.3$, \citealt{2002Natur.419..904C})
and HE 1327-2326 ($\textrm{[Fe/H]}=-5.4$, \citealt{2005IAUS..228..207F})
have been discovered -- yet. There are of course projects underway
to better sample the Halo (eg. GAIA, SEGUE), which may reveal stars
as metal poor as our models :) 

A fact that has raised considerable interest is that both of these
stars are very C-rich. This implies that the \emph{majority} of stars
at such low metallicities should be C-rich. However the sample is
of course very small. 

HE 1327-2326 has been identified as either a dwarf or a subgiant (\citealt{2006ApJ...639..897A}),
whilst HE 0107-5240 is probably an RGB star (\citealt{2004ApJ...603..708C}).
Thus HE 1327-2326 should not have altered its own abundance pattern,
but HE 0107-5240 may have (eg. via deep mixing or FDUP). 

\begin{figure}
\begin{centering}
\includegraphics[width=0.8\columnwidth,keepaspectratio]{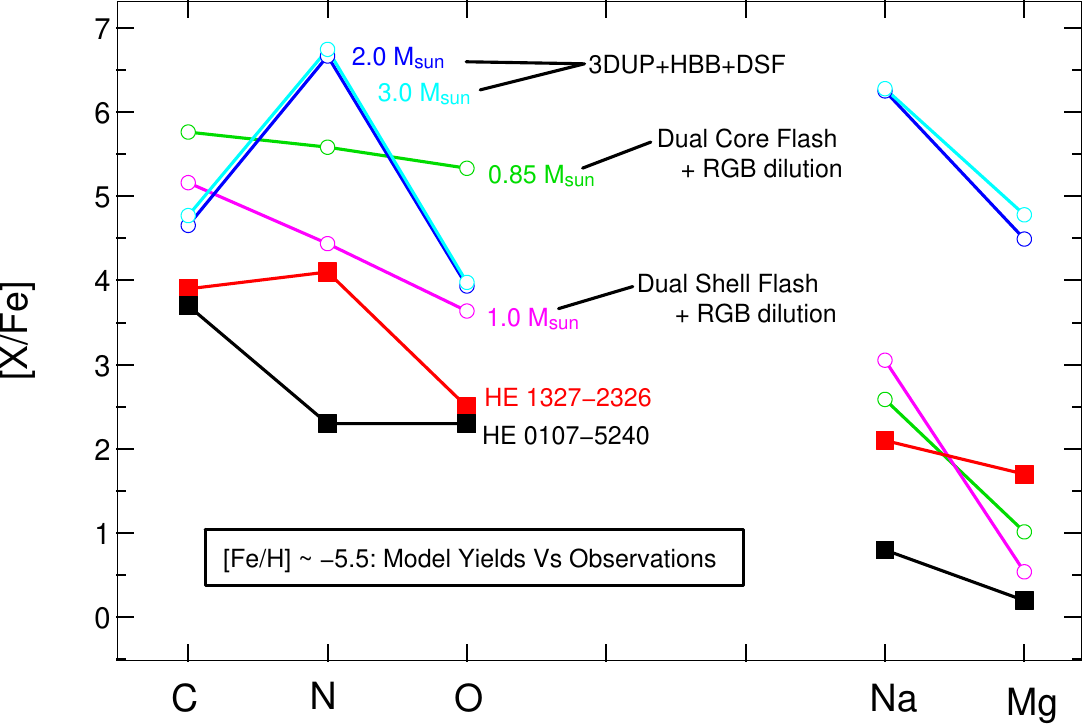}
\par\end{centering}
\caption{Comparing the abundance pattern in our $\textrm{[Fe/H]}=-5.45$ yields
with observations of stars with the same metallicity. The abundance
determinations in these stars are from \citet{2006ApJ...639..897A}
(HE 1327-2326) and \citet{2004ApJ...603..708C} (HE 0107-5240). The
clear 3DUP+HBB signature of high C and O combined with very enhanced
N can be seen in the 2 and 3 M$_{\odot}$ models. The AGB yields of
both the 0.85 and 1 M$_{\odot}$ have been diluted by varying amounts
of unpolluted mass loss on the RGB. It can be seen that the HBB models
are the closet match to the pattern of HE 1327-2326, although there
is not so much N in this star. These models produce far too much Na
and Mg. See text for a discussion. \label{fig-compare-FeH-5.5-Yields-Obs-2stars-new}}
\end{figure}

\begin{figure}
\begin{centering}
\includegraphics[width=0.8\columnwidth,keepaspectratio]{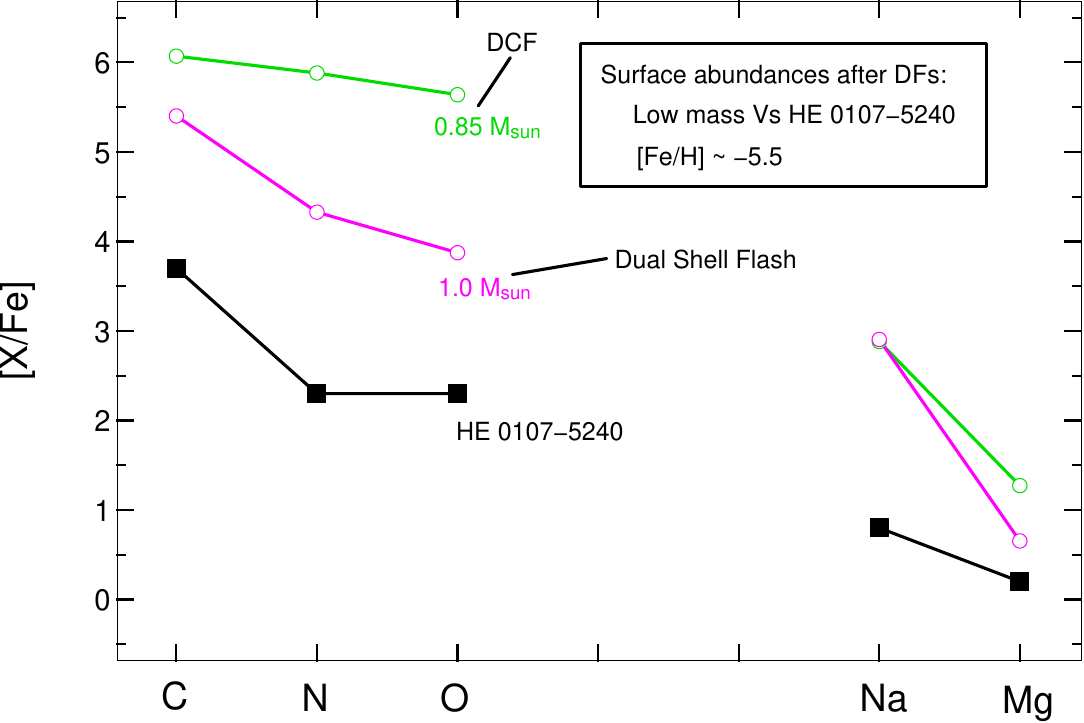}
\par\end{centering}
\caption{Comparing the abundance patterns in our low-mass $\textrm{[Fe/H]}=-5.45$
models at times just after the DSF and DCF events with the RGB CEMP
HE 0107-5240. This star has a similar abundance pattern to that of
the more metal-rich star G77-61. Contrasting these abundances with
those of the yields (Figure \ref{fig-compare-FeH-5.5-Yields-Obs-2stars-new})
it can be seen that this material was slightly diluted by unpolluted
RGB mass loss, and that slight HBB CN processing occurred in the 1
M$_{\odot}$model. The pattern of the 1 M$_{\odot}$ matches well
with the observations -- apart from the fact that a scaled down version
(through dilution) would have too much Mg. \label{fig-compare-FeH-5.5-DSFs-Obs-5240}}
\end{figure}

In Figure \ref{fig-compare-FeH-5.5-Yields-Obs-2stars-new} we show
the abundances of C, N, O, Na and Mg in the yields of all our $\textrm{[Fe/H]}=-5.45$
models. We also do the same for the observations of the two stars.
The abundance pattern of HE 1327-2326 is different to all the stars
we have looked at so far. It has high C and N but low(er) oxygen,
unlike either of the $\textrm{[Fe/H]}=-4$ stars and HE 0107-5240.
It is interesting to note that amongst our four comparison stars we
have three different patterns in CNO! This suggests that there are
many paths to becoming a C-rich EMP, possibly combined with many paths
to altering the abundance patterns post-pollution. This naturally
makes attempts to decipher pollution scenarios quite difficult. On
the other hand the abundance pattern of HE 0107-5240 is very similar
to that of the $\textrm{[Fe/H]}=-4$ star G77-61. We noted in the
previous subsection that this pattern was matched by the DSF patterns
of our 1 and 2 M$_{\odot}$ models -- except for Na. Looking at the
patterns in our $\textrm{[Fe/H]}=-5.45$ yields we see that none of
them match very well with the HE 0107-5240 pattern, although the 1
M$_{\odot}$ model comes reasonably close. These yields are however
averaged over the lifetime of the stars. In Figure \ref{fig-compare-FeH-5.5-DSFs-Obs-5240}
we plot the patterns of the 0.85 and 1 M$_{\odot}$ surface compositions
just after the DCF (0.85 M$_{\odot}$) and DSF (1 M$_{\odot}$) events.
Here it can be seen that, at this stage of evolution, the 1 M$_{\odot}$
does provide a reasonable match to the pattern of HE 0107-5240. The
reason for this improved match is due to the fact that the 1 M$_{\odot}$
model undergoes (slight) HBB on the AGB that converts some of the
C to N, thus altering the pattern. It is not a prefect match though.
If we scale the yield down (via some form of dilution with unpolluted
material) then we see that C, N, O and Na would all match fairly well
-- but Mg would be underabundant. We suggest that the extra Mg needed
could be supplied by some 3DUP. Since N is relatively low compared
to C, it would need to be a low mass AGB star that supplies the pollution,
without HBB. 

\begin{figure}
\begin{centering}
\includegraphics[width=0.8\columnwidth,keepaspectratio]{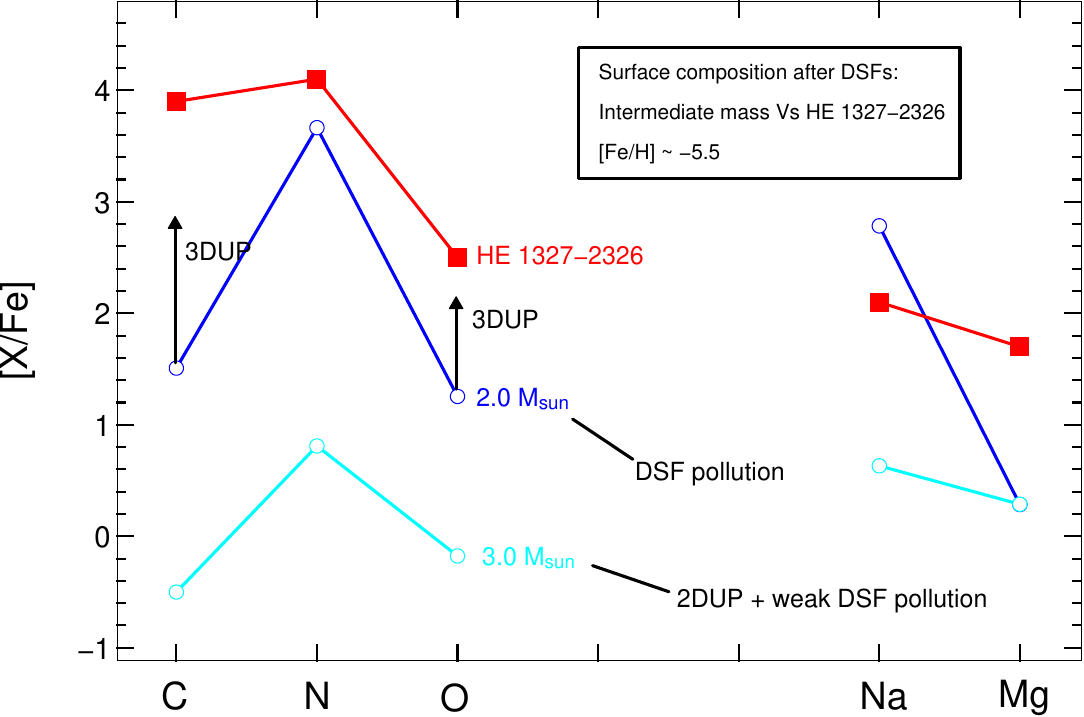}
\par\end{centering}
\caption{Comparing the abundance patterns in our intermediate-mass $\textrm{[Fe/H]}=-5.45$
models at times just after the DSF events with the CEMP dwarf (or
subgiant) star HE 1327-2326. By contrasting these abundances with
those of the yields (Figure \ref{fig-compare-FeH-5.5-Yields-Obs-2stars-new})
it can be seen that 3DUP+HBB increased the abundances of all these
elements by a few orders of magnitude. Interestingly the pattern from
the DSFs in these intermediate-mass models is similar to that from
HBB (and different from the DF patterns of the low-mass models). The
2 M$_{\odot}$ model has a similar amount of N as the star, but has
lower C and O. We suggest that a few 3DUP episodes would increase
the C and O whilst keeping N constant (before HBB sets in), as indicated
by the vertical arrows, making a better fit. Na and Mg may still be
a problem though. We note that the 3 M$_{\odot}$ model abundance
pattern has mainly come about via 2DUP, as the DSF pollution was very
weak. \label{fig-compare-FeH-5.5-IMDSFs-Obs-2326}}
\end{figure}

Moving back to Figure \ref{fig-compare-FeH-5.5-Yields-Obs-2stars-new}
we can see the strong HBB pattern in the 2 and 3 M$_{\odot}$ yields.
This pattern appears to be the best match for the pattern of HE 1327-2326.
However carbon is a bit high relative to N in this star for CN equilibrium.
This may indicate that the pollution material underwent only partial
CN cycling. This would be the case before HBB fully sets in near the
start of the AGB in an IM star, or for a lower mass star that does
not undergo very hot HBB. Since this star is a dwarf it should not
have altered its CNO composition itself (as deep-mixing RGB stars
can). In Figure \ref{fig-compare-FeH-5.5-IMDSFs-Obs-2326} we compare
the abundance pattern of HE 1327-2326 with the surface composition
of our IM models just after their DSF events. In the 3 M$_{\odot}$
case the DSF had very little effect on the surface composition (the
abundance ratios are all $\sim$solar which are the initial abundances,
except for N which was initially very sub-solar). The increased N
and decreased C at the surface in this model is mainly due 2DUP. The
2 M$_{\odot}$ model on the other hand has a HBB-like surface composition.
The magnitude of the surface pollution is also similar to that of
HE 1327-2326. However C and O in this model are much less abundant
than in HE 1327-2326. The operation of 3DUP after the DSF could remedy
this by dredging up these two elements whilst N remains constant.
We have indicated this in the diagram. This raises a significant issue
in relation to these sorts of comparisons. The time \emph{evolution}
of the surface abundances may be a better way to make comparisons.
Weighting would need to be applied due to the different timescales
of the different evolutionary stages. We shall attempt these types
of comparisons in our future publications, due to time and space constraints
here, although we note that we have partly included this dimension
by comparing with the DSF surface abundances as well as the yields.
Returning to the 2 M$_{\odot}$ model we see that the abundances of
Na and Mg are poor matches for those of HE 1327-2326. Na is again
too high whereas Mg is too low. Although Mg increases through 3DUP
in this model, so does Na, so the problem remains.

\subsection{Summary\label{subsection-Summary-HaloStarObservationCompare}}

\begin{figure}
\begin{centering}
\includegraphics[width=0.8\columnwidth,keepaspectratio]{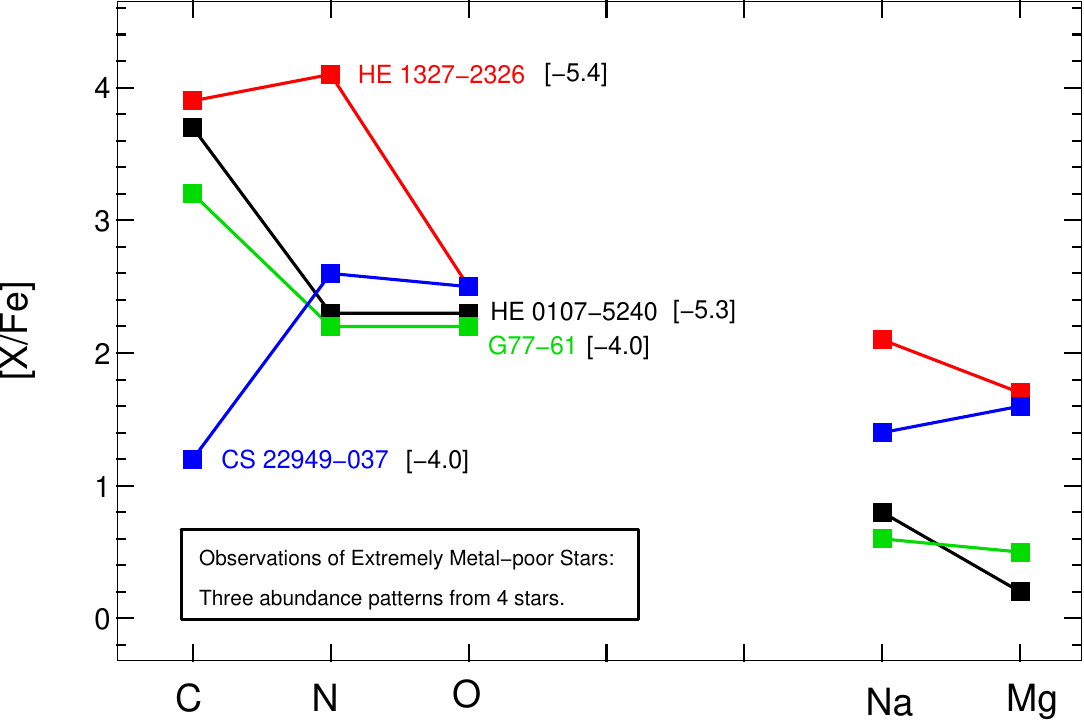}
\par\end{centering}
\caption{Highlighting the complex nature of the abundance patterns at extremely
low metallicity. It can be seen that in our sample of four stars there
are no less than three distinct abundance patterns. The (similar)
patterns of G77-61 and HE 0107-5240 are reasonably well matched by
the abundance patterns of some of our DSF models (at the time of the
DSFs) -- except for Na, which is overproduced. We note that CS 22949-037
is an odd CEMP since it has $\textrm{C/N}<1$, whereas most CEMPs
do not. \label{fig-EMP-Obs-5elems-4stars-3abundPatterns}}
\end{figure}

In summary we find that our yields only offer partial fits to the
abundance patterns at extremely low metallicities -- none of the
models match any of the stars' patterns across the five elements considered.
We note however that the time evolution of the surface composition
of the models is not taken into account in these types of comparisons
-- these stars may have been polluted by binary companions at any
stage of the companion's evolution. To further complicate matters
their gas may have been polluted by more than one star. The fact that
our sample of four stars has three distinct abundance patterns shows
that the formation of the patterns in Nature is indeed complex. We
highlight this by comparing all of the observed patterns (from our
four comparison stars) in Figure \ref{fig-EMP-Obs-5elems-4stars-3abundPatterns}.
It can be seen that the patterns of G77-61 and HE 0107-5240 are almost
identical (especially so when considering the uncertainties in the
observations) -- despite the order of magnitude difference in Fe
abundance between them. We note however that one of the stars, CS
22949-037, is an odd one. Unlike almost all of the other CEMPs in
our large sample it has a C/N ratio less than 1 -- such that it is
C-rich but N-richer (see Figure \ref{fig-compare-CN-Yields-Obs-all}
for C/N ratios in the observations). The other three stars in Figure
\ref{fig-EMP-Obs-5elems-4stars-3abundPatterns} are `normal CEMPs',
having $\textrm{C/N}>1$. 

On the bright side of our comparisons we recall a few key indicators
that our models do reproduce:
\begin{enumerate}
\item All our models produce large C overabundances in their yields, as
observed in the CEMPs.
\item All models produce large N overabundances, as observed in most of
the CEMPs.
\item Many of our models show low $^{12}$C/$^{13}$C ratios, as seen in
most CEMPs.
\item All our models, except for the HBB ones, have $\textrm{C/N}>1$ --
as seen in most CEMPs.
\item All our models produce oxygen overabundances, as seen in most CEMPs.
\item The yields show a spread in Na, as seen amongst the CEMPs.
\item Some of the DSF CNO abundance patterns match those of some of the
most metal-poor CEMPs. 
\end{enumerate}
Problems arise when we attempt to match all these -- and the extended
abundance patterns -- simultaneously (and quantitatively). Clearly
there is much work to be done here in attempting to decipher the pollution
scenarios. Some major uncertainties that need to be explored include
the nuclear reaction rates and the degree of overshoot in the models.
We shall continue to pursue this complex task in future studies --
it is quite a challenge! One first step will be to compile all the
observational data to determine if there are some common abundance
patterns that dominate the CEMPs.

Finally we note that there are other pollution scenarios in the literature
that attempt to explain the abundance patterns of these stars apart
from the binary mass transfer and autopollution scenarios mentioned
in the current study. In particular massive stars ($\textrm{M}\gtrsim10$
M$_{\odot}$) are often invoked to explain the abundance patterns
of the heavier elements (eg. \citealt{2002ApJ...577..281C}). Indeed
these stars are needed in the low- and intermediate-mass scenarios
as the LM and IM stars cannot produce many of the heavy elements (s-process
elements withstanding). \citet{2002ApJ...577..281C} showed that standard
SNe models in the mass range $15\rightarrow80$ M$_{\odot}$ can not
explain \emph{all} the patterns in the observations, but they do match
a significant number of them. However \citet{2003Natur.422..871U}
showed it may be possible for all the elements to come from SNe --
including the large CNO excesses -- by assuming the low-energy SN
`mixing and fallback' scenario in their models. They match the abundance
pattern of one of the stars considered here, HE 0107-4240, but note
that by varying the nature of the explosion they should be able to
explain stars with other abundance patterns. In this scenario low-
and intermediate-mass stars are not required. 

\chapter{\label{GC-chap}Galactic Globular Cluster Stars}
\begin{quote}
``The activist movements of the past 40 years have had a significant
civilizing effect.''
\begin{flushright}
\vspace{-0.6cm}-- Noam Chomsky
\par\end{flushright}

\end{quote}

\section{Introduction}

\subsection{The GC Abundance Anomalies Mystery\label{subsection-GC-Anomalies-Intro}}

Most Galactic globular clusters (GCs) have a very uniform distribution
of heavy elements. This indicates that the cluster gas was well mixed
when the stars formed. However, in contrast to the Fe group, it has
been known since the early 1970s that there is a large spread in Carbon
and Nitrogen in many GCs (eg. \citealt{1974MNRAS.166...89B}; \citealt{1980ApJ...236L..83D}).

The first `anticorrelation' was found 25 years ago -- C is low when
N is high. This observation is explicable in terms of the CN cycle,
where $^{12}$C is burnt to $^{14}$N. It has also been found that
the C abundance decreases with luminosity on the red giant branch
(RGB) in many GCs. This is known as the C-L anticorrelation and is
also observed in halo field stars (see eg. figures 4 and 5 in \citealt{2002PASP..114.1215S}).
This is thought to be the result of Deep Mixing -- extra mixing of
the convective envelope into the top of the H-shell, allowing C-N
cycling to alter the composition of the surface. Another effect of
C-N cycling is the reduction of the $^{12}$C/$^{13}$C ratio (it
approaches an equilibrium value of $\sim4$). This has been observed
in GCs and in the field (see eg. figure 2 in \citealt{2003ApJ...585L..45S}).

GCs also show other abundance features. The most famous of these is
the O-Na anticorrelation: O decreases with increased Na. This is readily
explained by hot(ter) hydrogen burning, where the O-N and Ne-Na chains
are operating -- the ON cycling reduces O, whilst the Ne-Na chain
increases Na (at T $\sim45$ million K). \emph{Where} this occurs
is still a mystery. The notable thing about this abundance trend is
that it only occurs in GCs \emph{-- it is not seen in field halo
stars} (see \citealt{2000AA...354..169G} for a comparison of M13
stars and fields stars of similar metallicity).

There also appears to be a Mg-Al anticorrelation in some GCs (eg.
NGC 6752, \citealt{2002AA...385L..14G}). This can also be explained
through high-temperature (T $\sim65$ million K) proton capture nucleosynthesis,
via the MgAl chain (Mg depleted, Al enhanced). Again the site for
this is uncertain and it \emph{is not seen in field stars}. Furthermore,
the light elements show various (anti)correlations amongst themselves
(see eg. Figure 9 in \citealt{1997AJ....113..279K} for O, Na, Mg,
and Al in M13). All these (anti)correlations point to hydrogen burning
-- the CN, ON, MgAl, NeNa cycles/chains -- at various temperatures.
The most popular theory for the site of this burning is at the base
of the convective envelope in AGB stars - where hot bottom burning
(HBB, first proposed as a possible site by \citealt{1981ApJ...245L..79C})
occurs. HBB provides the proton-capture nucleosynthesis needed, and
at low metallicities (like those of GCs) HBB occurs at higher temperatures.
Qualitatively it seems to fit the observations-- although some deep
mixing is required also, to explain the C-L anticorrelation. We note
that another possible site for the abundance anomalies has recently
gained increased popularity -- that of rotating massive stars (see
eg. \citealt{2006AA...448L..37M}).

\subsection{GC Stellar Models: The Grids and Input Physics}

\subsubsection*{The Grids of Models}

During the course of the current study we have investigated whether
the AGB GC pollution scenario also fits \emph{quantitatively}. To
this end we have calculated a few small grids of models to emulate
a generation of stars in NGC 6752 ($\textrm{[Fe/H]}=-1.4$) which
may have given rise to the abundance patterns -- either via supplying
the material for star formation of the current generation, or by supplying
the material for accretion by the current generation. The reason we
have calculated more than one grid of models is so we can investigate
the uncertainties in the model predictions. We have calculated four
sets of models, although not all sets contain the full mass range:
\begin{enumerate}
\item The Standard set ($M=1.25$, 2.5, 3.5, 5.0, 6.5 M$_{\odot}$) --
using our standard mass loss formalism on the AGB (VW93) and our standard
reaction rates.
\item Mass loss variation set ($M=2.5$ \& 5.0 M$_{\odot}$) -- using Reimers'
mass loss on the AGB and our standard reaction rates.
\item Reaction rate variation set ($M=2.5$ \& 5.0 M$_{\odot}$) -- using
the NACRE compilation for some important reactions, whilst using the
standard mass-loss rate on the AGB (VW93).
\item No 3DUP set ($M=2.5$, 3.5, 5.0, 6.5 M$_{\odot}$) -- standard mass
loss and rates but 3DUP inhibited.
\end{enumerate}

\subsubsection*{Initial Composition}

All the sets of models we calculated using the same initial composition.
This composition was arrived at through a chemical evolution model
in which the gas was enriched by a population of $Z=0$ stars. The
IMF used was top-heavy, so most of the pollution was from massive
Pop. III stars (see Section \ref{Section-GCChemEvoln} for more details).
In order to match the metallicity of NGC 6752 ($\textrm{[Fe/H]}=-1.4$)
the mostly supernova material was mixed with pristine Big Bang material.
We list the initial composition in each yield table for each set of
models in Appendix \ref{Appx-GCYields}. The key differences between
this composition and the Solar composition is that N is very underabundant
and the $\alpha$ elements are enhanced. This is a direct result of
the $Z=0$ supernovae yields used in the chemical evolution model. 

\subsubsection*{Input Physics}

Since the models were computed using the non-scaled-solar composition
mentioned above, it was necessary to remove all scaled-solar assumptions
from the codes. In addition to this we calculated opacity tables specifically
for this composition on the OPAL website -- so that the opacity would
be consistent with the stellar composition (rather than scaled-solar). 

The version of the SEV code that we used for the first three sets
of models listed above did not have time-dependent mixing (as this
had not been completed yet). It did however have all the opacity updates
(see Chapter \ref{sevmods} for details). The version of the SEV code
used for the fourth set of models, the ones with no 3DUP, was the
same as that used for our $Z=0$ and halo star models. Thus it included
all the modifications reported in Chapter \ref{sevmods}. In particular
it contained time-dependent mixing and the option of using the Ledoux
criterion for convective boundaries. For our standard mass-loss rates
and reaction rates see Chapter \ref{Chapter-NumericalCodes}. 

In the nucleosynthesis code (NS code) most of our reaction rates come
from the Reaclib library (\citealt{Thi87}). However many of the important
rates involving the Mg and Al isotopes have been updated using various
sources. These include: $^{24}$Mg($p,\gamma$)$^{25}$Al (\citealt{DCP+99});
$^{25}$Mg($p,\gamma$)$^{26}$Al (\citealt{1996PhRvC..53..475I});
$^{26}$Mg($p,\gamma$)$^{27}$Al (\citealt{1990NuPhA.512..509I});
$^{22}$Ne($\alpha,n$)$^{25}$Mg and $^{22}$Ne($\alpha,\gamma$)$^{26}$Mg
(\citealt{1994ApJ...437..396K}). We refer to this combination of
rates as our `standard' set, and it is this set that we use for all
of our models, unless otherwise stated. See Section \ref{nscode}
for further details on the NS code.

\section{Models of NGC 6752 Stars\label{section-GC-Models-Standard}}

We have calculated three different sets of models for this part of
our globular cluster study. The first is a `standard' set whilst the
other two were designed to test the sensitivity of the results to
two of the major uncertainties in the model calculations -- reaction
rates and mass-loss.

\subsection{The Standard Set}

\subsubsection*{Input Physics}

Some of the key input physics used in the structural evolution code
(SEV code) for this `Standard' set of models are:
\begin{itemize}
\item \citet{1975MSRSL...8..369R} mass loss formula on the RGB.
\item \citet{1993ApJ...413..641V} mass loss formula on the AGB.
\item Instantaneous mixing (see Chapter \ref{Chapter-NumericalCodes}).
\item Updated opacities which also match the non-scaled solar composition
(see Section \ref{opacmods}).
\end{itemize}
Convective boundaries were treated using the Schwarzschild criterion
plus the standard SEV code overshoot method described in Section \vref{sub-evoln-overshoot}. 

In the NS code we have used the standard set of reaction rates, as
described in the introduction to this chapter (also see Section \ref{nscode}
for further details on the NS code). 

\subsubsection*{Results}

\begin{figure}
\begin{centering}
\includegraphics[width=0.9\columnwidth,keepaspectratio]{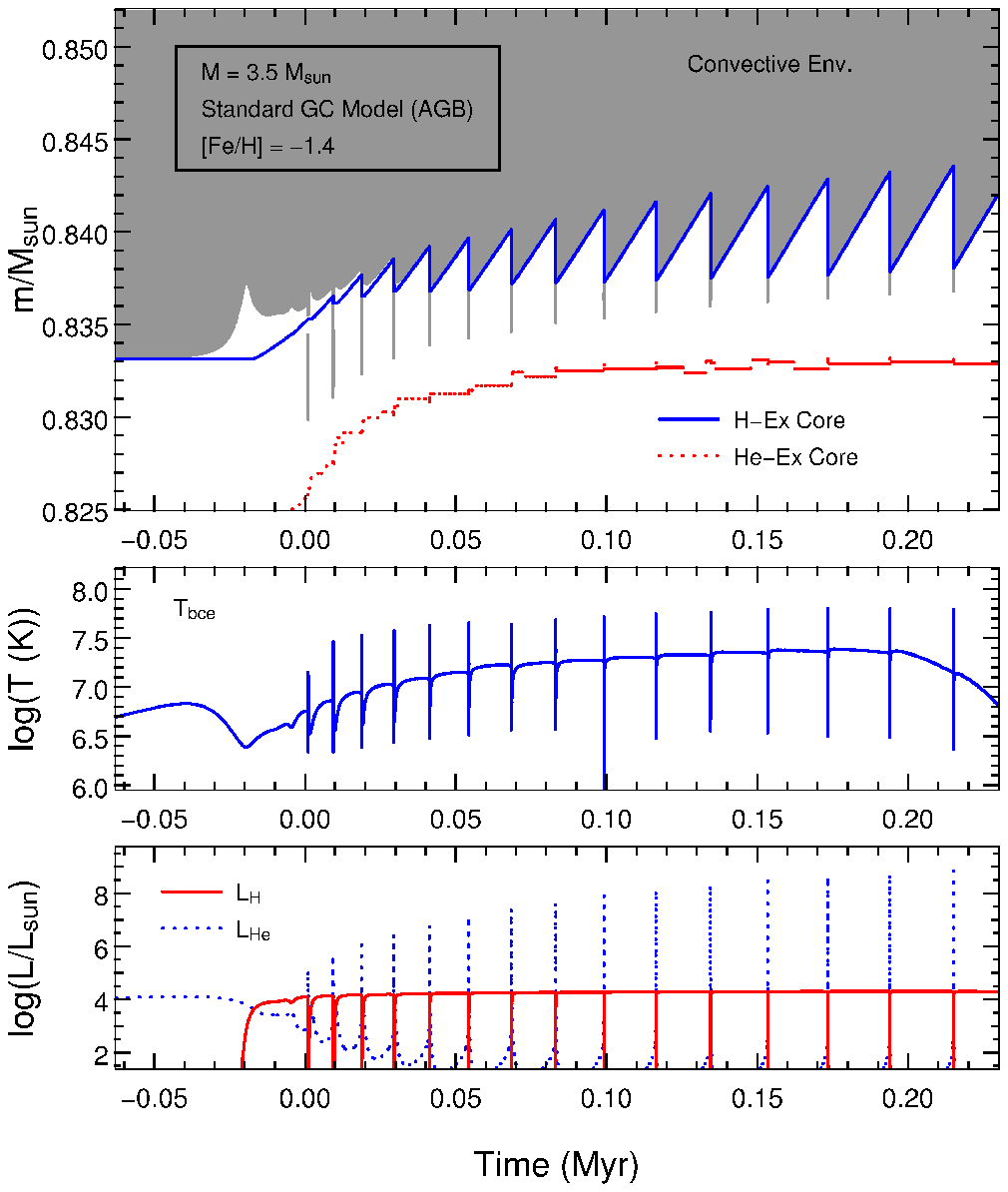}
\par\end{centering}
\caption{Most of the AGB evolution of our standard GC model of 3.5 M$_{\odot}$.
Time has been offset. It can be seen in the top panel that this model
experiences deep 3DUP, with $\lambda\sim0.98$. The temperature at
the base of the convective envelope (second panel) is only high enough
for minor HBB, so this model becomes a carbon star during the AGB.
\label{fig-m3.5gc-Standard-AGB-conv}}
\end{figure}

\begin{figure}
\begin{centering}
\includegraphics[width=0.7\columnwidth,keepaspectratio]{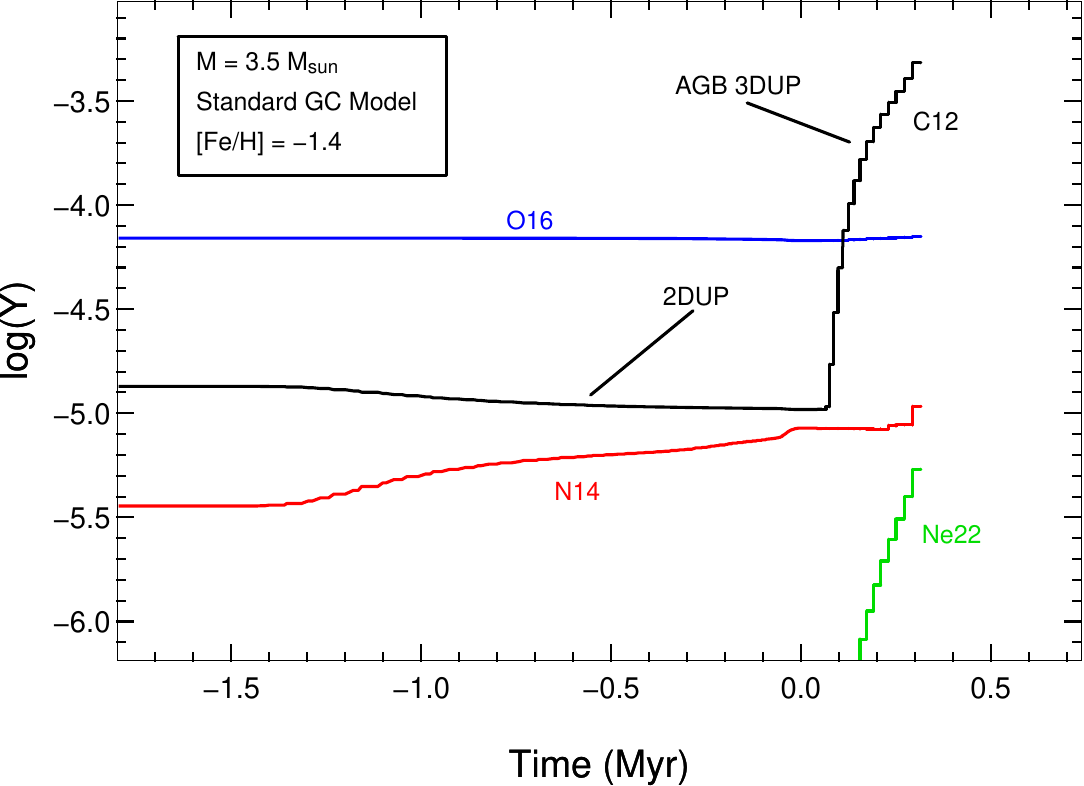}
\par\end{centering}
\caption{The surface abundance evolution of some of the CNO isotopes and $^{22}$Ne
in our standard GC model of 3.5 M$_{\odot}$. Time has been offset.
Included in the timeframe is the 2DUP phase where $^{14}$N-rich material
is dredged up into the envelope. The stepwise increase in $^{12}$C
and $^{22}$Ne is due to 3DUP of intershell material. The temperature
at the base of the convective envelope is only high enough for minor
HBB, as can be seen in the constancy of $^{16}$O and $^{14}$N during
the AGB. \label{fig-m3.5gc-Standard-SRF-AGB-CNO}}
\end{figure}

\begin{figure}
\begin{centering}
\includegraphics[width=0.7\columnwidth,keepaspectratio]{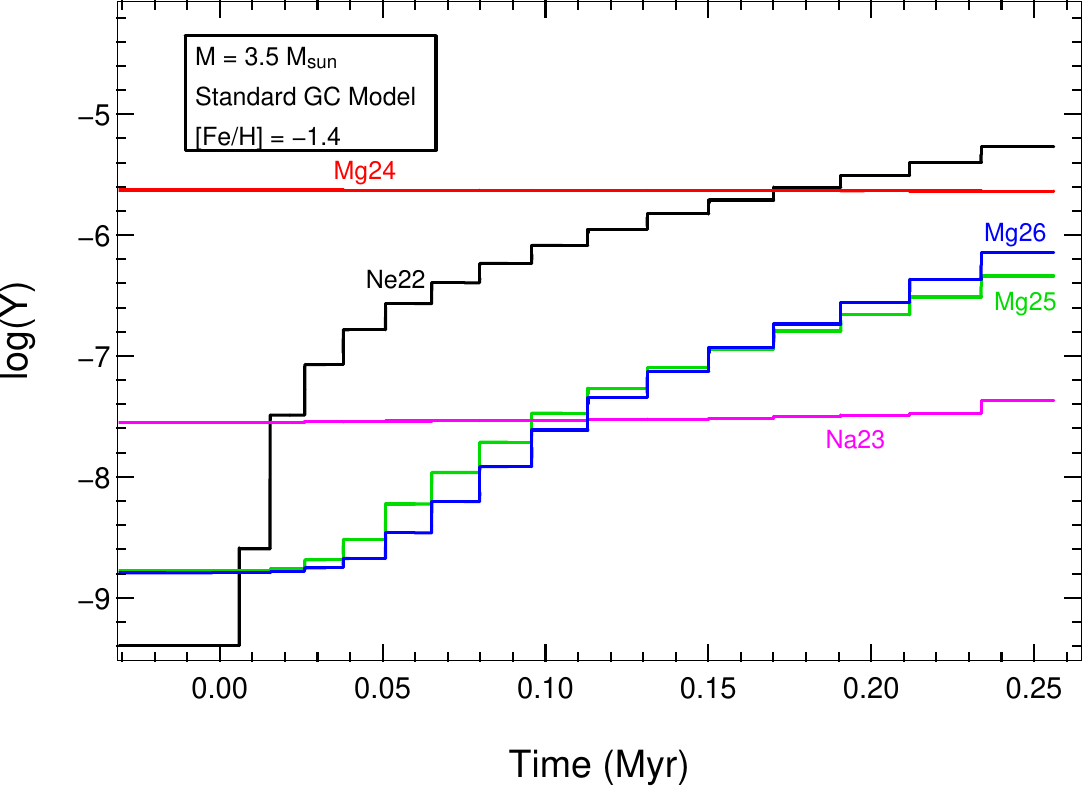}
\par\end{centering}
\caption{The surface abundance evolution of some of the Mg isotopes, $^{22}$Ne
and $^{23}$Na in our standard GC model of 3.5 M$_{\odot}$. Time
has been offset. The stepwise increase in the Mg isotopes is due to
3DUP of intershell material. It can be seen that $^{23}$Na only increases
very marginally. No significant Na-O anticorrelation is present in
this model. \label{fig-m3.5gc-Standard-SRF-AGB-MgAlNeNa}}
\end{figure}

Here we briefly describe a couple of representative models from our
standard set. We shall provide more detail of the evolution of all
our models in a forthcoming paper. These models follow quite normal
evolutionary paths -- unlike the $Z=0$ and EMP models presented
in the previous chapters (they experience no dual flash events). For
reference we note that the version of the code used for these models
is similar to that used in the studies by \citet{2003PASA...20..279K}
and \citet{2002PASA...19..515K} -- apart from some opacity updates
in our version. The third dredge-up (3DUP) results are similar to
that found by \citet{2002PASA...19..515K}. In particular our $Z=0.0017$
($\textrm{[Fe/H]}=-1.4$) GC models have very similar $\lambda$ values
to their $Z=0.004$ models. To highlight this we show in Figure \ref{fig-m3.5gc-Standard-AGB-conv}
some of the AGB evolution in our 3.5 M$_{\odot}$ model. It can be
seen that $\lambda\approx0.98$, which is the same as that found by
\citet{2002PASA...19..515K} in their 3.5 M$_{\odot}$ model. This
model experiences $\sim18$ thermal pulses before leaving the AGB.
The temperature at the base of the convective envelope does not become
high enough for significant HBB ($T_{bce}\lesssim25$ MK). In Figures
\ref{fig-m3.5gc-Standard-SRF-AGB-CNO} and \ref{fig-m3.5gc-Standard-SRF-AGB-MgAlNeNa}
we show some of the surface abundance evolution from the NS code for
this model, during the AGB phase. The effect of 3DUP of intershell
material on the surface abundances can be seen in the periodic jumps
in $^{12}$C and $^{22}$Ne. The heavy Mg isotopes are also enhanced
by 3DUP, increasing by $\sim3$ dex at the surface over the lifetime
of the AGB (Figure \ref{fig-m3.5gc-Standard-SRF-AGB-MgAlNeNa}). A
slight increase in $^{14}$N can be seen but this is mainly due to
2DUP. Oxygen, $^{24}$Mg and $^{23}$Na are all barely affected on
the AGB in this model. 

\begin{figure}
\begin{centering}
\includegraphics[width=0.9\columnwidth,keepaspectratio]{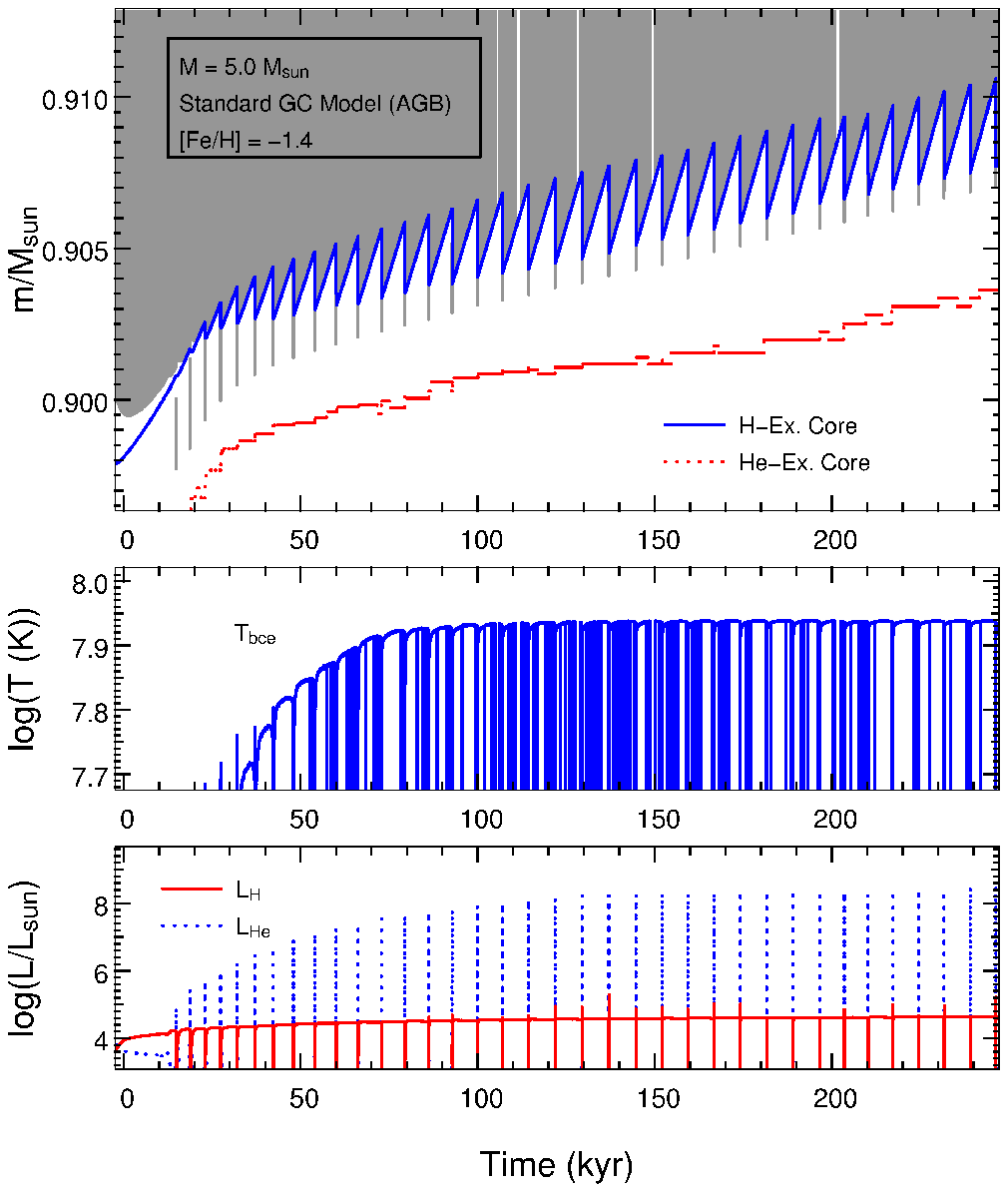}
\par\end{centering}
\caption{About half of the AGB evolution of our standard GC model of 5.0 M$_{\odot}$
(this model experiences $\sim80$ thermal pulses in total). Time has
been offset. It can be seen in the top panel that this model experiences
deep 3DUP, with $\lambda\sim0.9$. The temperature at the base of
the convective envelope (second panel) approaches $10^{8}$ K --
so strong HBB occurs. \label{fig-m5.0gc-Standard-AGB-conv}}
\end{figure}

\begin{figure}
\begin{centering}
\includegraphics[width=0.8\columnwidth,keepaspectratio]{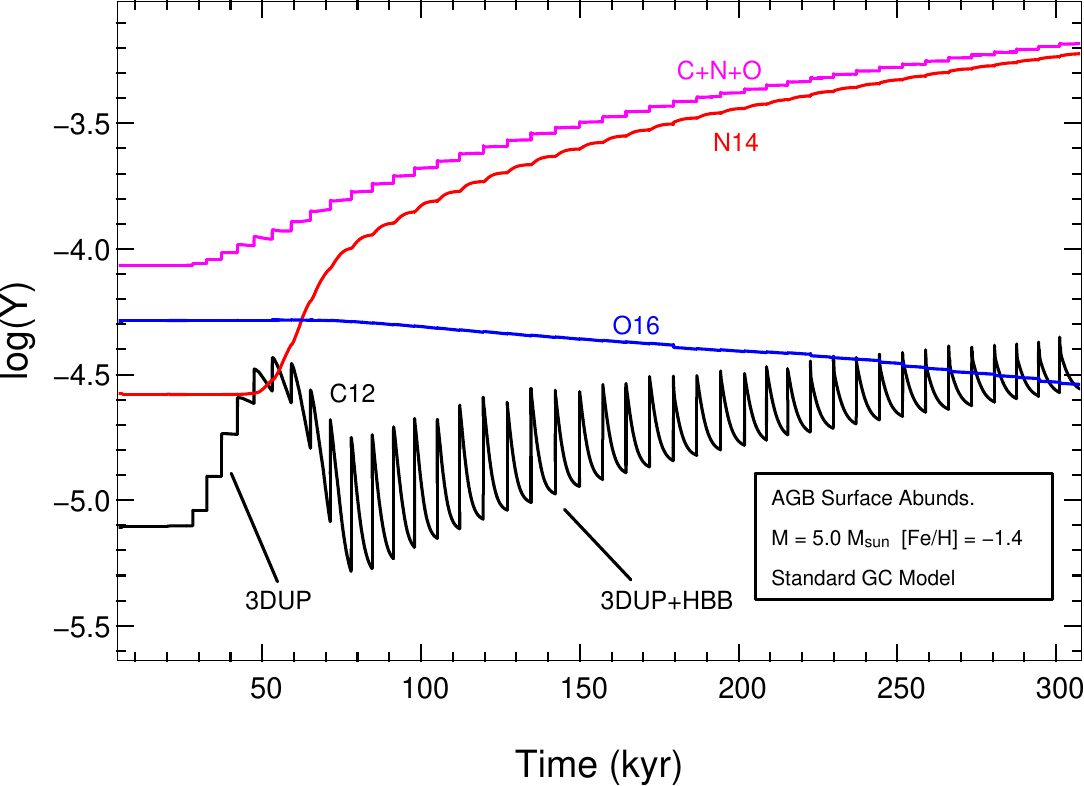}
\par\end{centering}
\caption{The surface abundance evolution of some of the CNO isotopes and the
sum of C+N+O in our standard GC model of 5.0 M$_{\odot}$. Time has
been offset. An initial increase in $^{12}$C and can be seen at the
start of the AGB before the temperature at the bottom of the convective
envelope increases and HBB sets in. Then the $^{12}$C is converted
to $^{14}$N which increases substantially. $^{16}$O is also seen
to reduce, indicating that O$\rightarrow$N cycling is operating as
well. \label{fig-m5.0gc-Standard-SRF-AGB-CNO}}
\end{figure}

\begin{figure}
\begin{centering}
\includegraphics[width=0.8\columnwidth,keepaspectratio]{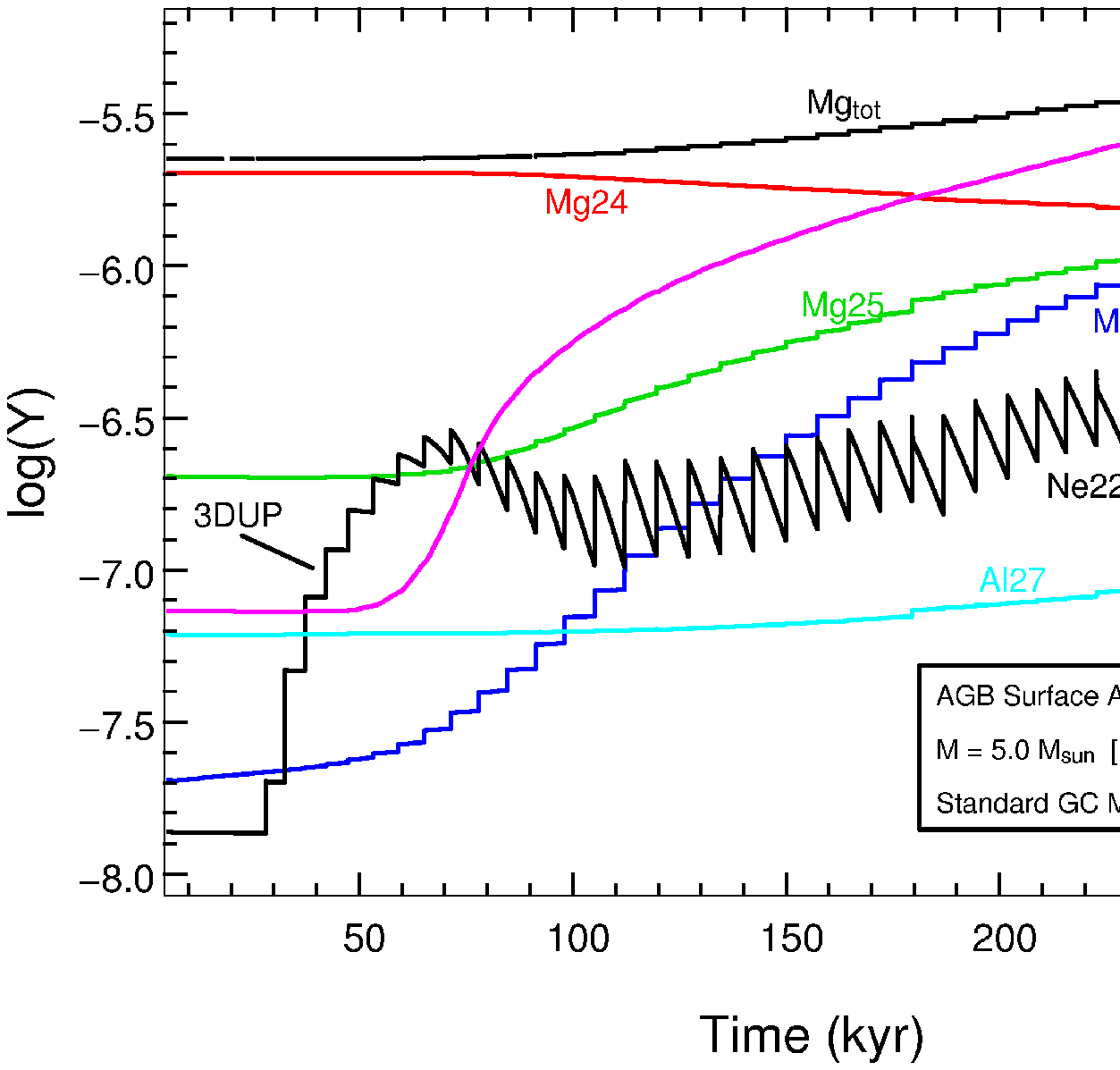}
\par\end{centering}
\caption{The surface abundance evolution of some of the Mg isotopes, $^{22}$Ne
and $^{23}$Na in our standard GC model of 5.0 M$_{\odot}$. Also
plotted is the sum of the Mg isotopes (Mg$_{tot}$). Time has been
offset. Here we can see the HBB of $^{22}$Ne to $^{23}$Na (when
the temperature increases at $t\sim70$ kyr), as well as the dredge-up
of the heavy Mg isotopes and the HBB destruction of $^{24}$Mg. This
model shows a nett increase in Mg and Al.\label{fig-m5.0gc-Standard-SRF-AGB-MgAlNeNa}}
\end{figure}

Moving to a higher mass model we see that the temperature at the bottom
of the convective envelope is much higher -- reaching $\sim90$ MK
in the 5 M$_{\odot}$ model (see panel 2 of Figure \ref{fig-m5.0gc-Standard-AGB-conv}).
This gives rise to strong HBB after only $\sim5$ thermal pulses into
the AGB (this model experiences $\sim80$ TPs in total). This can
be seen clearly in the $^{12}$C and $^{14}$N surface abundance evolution
in Figure \ref{fig-m5.0gc-Standard-SRF-AGB-CNO}. Once the temperature
is high enough C$\rightarrow$N cycling converts the dredged-up $^{12}$C
to $^{14}$N very quickly. It can also be seen that some O$\rightarrow$N
cycling is occurring, as the $^{16}$O abundance decreases towards
the end of the AGB -- despite the fact that it is being dredged up
with the carbon. Thus there is full CNO cycling occurring in the envelope
of this model. In the context of the GC abundance anomalies this is
good, since oxygen is seen to be depleted in some stars (by up to
$\sim1$ dex in NGC 6752). Na is seen to anticorrelate with oxygen
in many GCs. In Figure \ref{fig-m5.0gc-Standard-SRF-AGB-MgAlNeNa}
we can see that the abundance of $^{23}$Na indeed increases with
evolution along the AGB. Thus an anticorrelation is present in our
(higher-mass) models. The increase in Na mainly comes from the HBB
destruction of $^{22}$Ne, which is dredged up from the intershell.
It can be seen in Figure \ref{fig-m5.0gc-Standard-SRF-AGB-MgAlNeNa}
that $^{22}$Ne and $^{23}$Na play similar roles to $^{12}$C and
$^{14}$N, respectively. $^{22}$Ne increases through 3DUP initially
because $T_{bce}$ is too low for Ne-Na cycling, but as the temperature
increases most of it ends up in the form of $^{23}$Na. Hence the
abundance of $^{23}$Na increases enormously, just like that of $^{14}$N.
In terms of the Mg-Al anticorrelation in NGC 6752 we can see in the
same figure that $^{27}$Al is increased by $\sim0.5$ dex over the
AGB phase. However the heavy Mg isotopes are also produced in large
amounts ($^{24}$Mg is depleted slightly). Thus our models show a
\emph{positive} correlation between Mg and Al -- which is not present
in the observations.

\begin{figure}
\begin{centering}
\includegraphics[width=0.9\columnwidth,keepaspectratio]{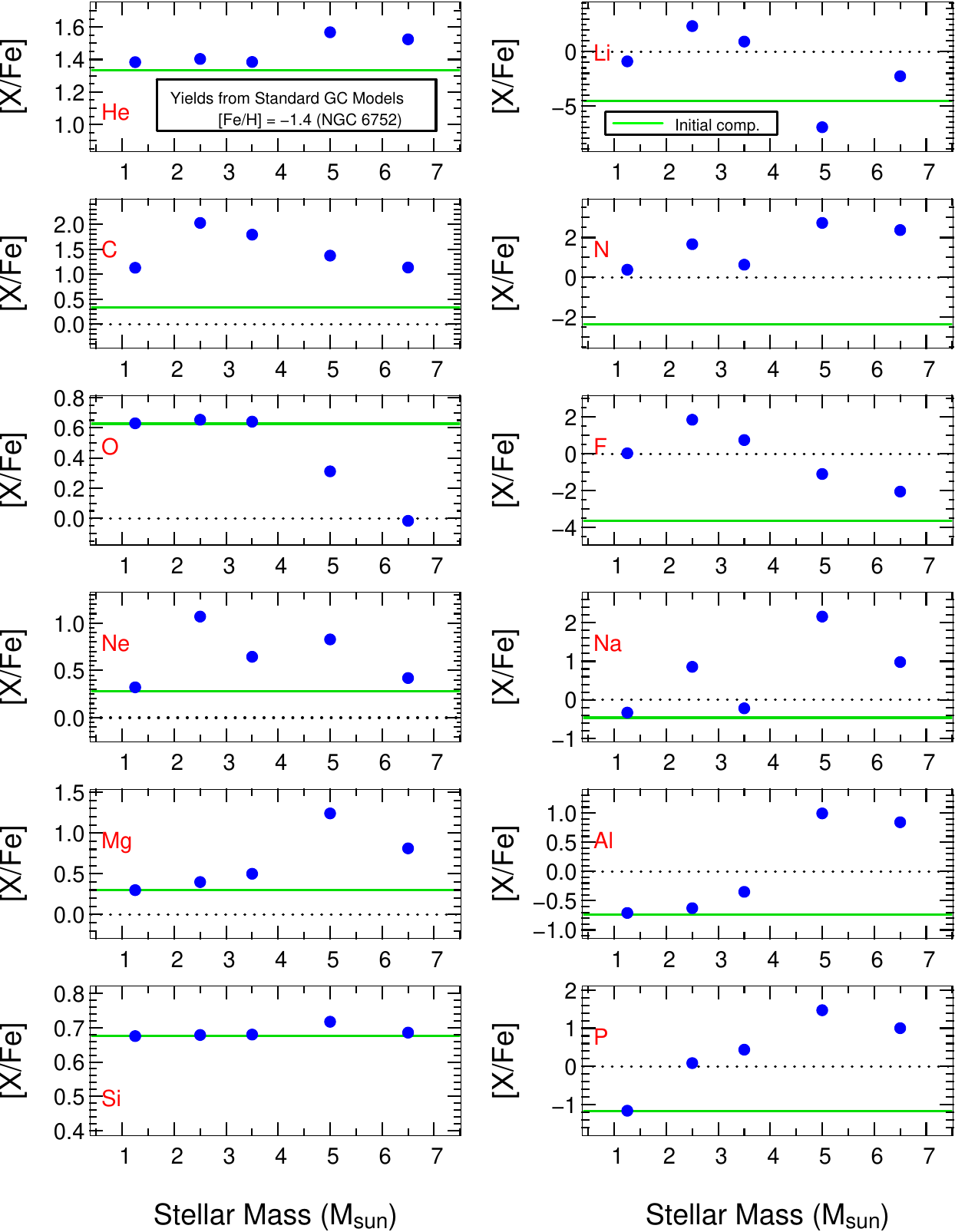}
\par\end{centering}
\caption{The yields of all the standard GC models, relative to solar, for a
selection of elements. The green horizontal lines show the initial
composition of the models. \label{fig-yields-All-GCstandard-mostElems-XFe}}
\end{figure}

Figure \ref{fig-yields-All-GCstandard-mostElems-XFe} shows the yields
for all the stars in our standard-model grid. We show the yields for
each element (relative to solar) against the initial mass of each
model. This gives a clear idea of the mass dependence of the yields.
Starting at the lower masses it can be seen that C is between $\sim1$
and 2 dex super-solar in the yields. In general it is more prevalent
in the lower mass yields (M$\le3.5$ M$_{\odot}$), due to the fact
that deep dredge-up occurs in these models and there is no HBB to
destroy it. The higher-mass model yields have on average less C but
they are still enhanced by $\sim1$ dex above solar (and $\sim0.7$
dex above the initial composition). N is enhanced in all the models.
In the low mass models it is first and 2DUP that does this. N is more
abundant at higher masses due to HBB CN(O) cycling. Oxygen follows
a clear pattern with initial stellar mass. In the lower mass models
it is unchanged, whilst the higher mass models burn it via the ON
cycle during HBB. It can be seen that the higher the mass the more
O is depleted, due to the increasing $T_{bce}$. Interestingly oxygen
ends up at the solar abundance (relative to Fe) in the 6.5 M$_{\odot}$
model. Na is produced mainly in the higher mass models due to HBB
of dredged-up $^{22}$Ne (the 2.5 M$_{\odot}$ is an exception at
low mass, it has produced a significant amount of Na, but less than
the higher mass models). Magnesium and Al do not start increasing
over the initial abundance until $M\sim3.5$ M$_{\odot}$. In the
5 and 6.5 M$_{\odot}$ model yields it is enhanced by $\sim1$ dex.
As mentioned above this tandem increase is the opposite to that seen
in the NGC 6752 stars. Finally we note that phosphorous is very enhanced
in most of the yields, peaking at $\sim$3 dex above the initial composition
in the 5 M$_{\odot}$ model. 

\subsection{Models Varying Mass-Loss Rates and Reaction Rates}

\subsubsection*{Overview}

In order to estimate the sensitivity of our results to two of the
major uncertainties in the model calculations, we calculated two sets
of comparison models in which we altered:
\begin{enumerate}
\item Some of the reaction rates used in the NS code.
\item The mass-loss formula used during the AGB phase in the SEV code.
\end{enumerate}
Two representative models were calculated for each set, with masses
of 2.5 and 5.0 M$_{\odot}.$ In the first comparison set we altered
the rates used for the Mg-Al, Ne-Na chains/cycles and that for the
$^{22}$Ne+$\alpha$ reactions. The standard rates were replaced with
those from the NACRE compilation (\citealt{1999NuPhA.656....3A}).
In the second comparison set we altered the mass loss formula on the
AGB from that of \citet{1993ApJ...413..641V} to that from \citet{1975MSRSL...8..369R}
(with $\eta_{AGB}=3.5$).

\subsubsection*{Results}

\begin{figure}
\begin{centering}
\includegraphics[width=0.85\columnwidth,keepaspectratio]{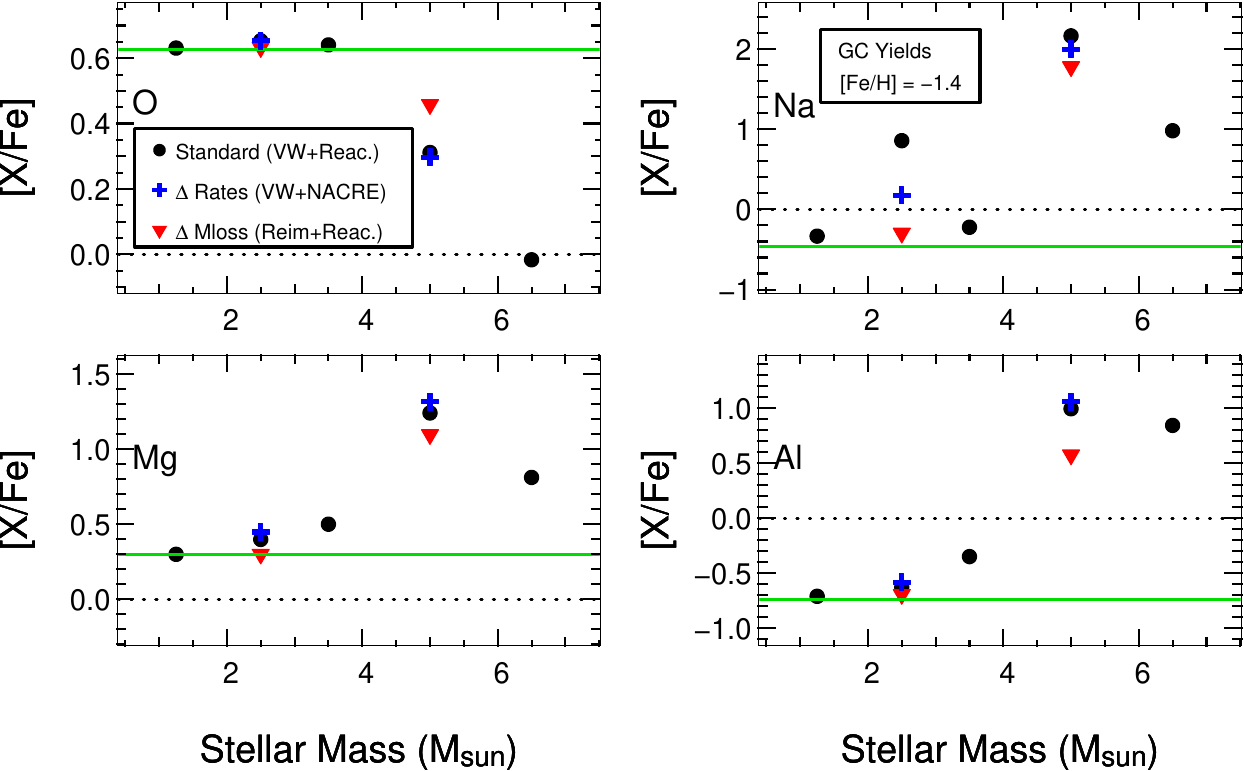}
\par\end{centering}
\caption{Comparing the standard yields (black circles) with those from the
models that used the Reimers' mass loss formula on the AGB instead
of VW93 (red triangles) and those from the models that used the NACRE
rates for the Ne-Na and Mg-Al cycles/chains (blue crosses). It can
be seen that altering the mass loss rate had the most effect on the
yields. \label{fig-yields-compare-Standard-Reim-Nacre-NaOMgAl}}
\end{figure}

In Figure \ref{fig-yields-compare-Standard-Reim-Nacre-NaOMgAl} we
show the yields from the three sets of models -- the two sets just
described above, plus the standard set. The elements pertaining to
the Na-O and Mg-Al anticorrelations in NGC 6752 are displayed. It
can be seen that altering the reaction rates had very little effect
on the final yields. The only exception is the Na production in the
2.5 M$_{\odot}$ model, which was lowered. However, even in this case,
it is the mass-loss formula for the AGB that had the greatest impact
on the yields. Sodium yields are significantly higher with the \citet{1993ApJ...413..641V}
(VW93) mass-loss formula, owing to the increased number of third dredge-up
episodes that progressively increase the Na abundance at the surface,
and to the fact that much of the convective envelope is lost during
the final few thermal pulses (the `superwind' phase) -- when the
surface abundance of Na is at its highest. Since mass-loss is more
evenly spread through the AGB phase when using the \citet{1975MSRSL...8..369R}
formula, more material is lost earlier on in the AGB phase, prior
to the high envelope abundance of Na. Looking at oxygen we see that
the use of Reimers' mass-loss reduces the oxygen destruction. This
is because depletion of $^{16}$O in the envelopes is due to HBB,
which operates for a shorter time in models with Reimers mass-loss
(due to the faster initial mass-loss rate as compared to VW93). 

In summary it appears that the mass loss uncertainties are more important
than the reaction rate uncertainties -- at least in the case of these
particular elements. We however add a caveat that further alterations
of reaction rates may alter this conclusion. For example the large
uncertainty in the $^{26}$Mg(p,$\gamma$)$^{27}$Al
 reaction rate may allow more production of $^{27}$Al, at the expense
of $^{26}$Mg (\citealt{1999AA...347..572A}). This would probably
not effect the broad outcome of the models -- that Mg production
is correlated with Al and that Na is anticorrelated with O -- but
it is worth testing. We will pursue this in a future study. 

\section{GC Chemical Evolution with AGB Yields\label{Section-GCChemEvoln}}

\subsection{Abstract}

Here we make a brief report on our test of the popular theory that
the observed abundance anomalies in the Galactic globular cluster
NGC 6752 are due to `internal pollution' from intermediate mass asymptotic
giant branch stars. For a more detailed report we refer the reader
to the main article by \citet{2004MNRAS.353..789F}, of which I am
the second author. In the test we used a chemical evolution model
to track the composition of the intracluster medium over time. Custom-made
stellar evolution models were calculated using an earlier version
of the SEV code (with updated opacities but no time-dependent mixing)
and the nucleosynthesis was calculated with the current version of
the NS code (these models were just presented in Section \ref{section-GC-Models-Standard}).
Yields from these calculations were used as feedback in the chemical
evolution model. A novelty of this study is that the stellar evolution
of the second generation stars was calculated using the appropriate
composition (not scaled-solar), as given by the chemical evolution
model itself. By tracing the chemical evolution of the intracluster
gas we were able to test the internal pollution scenario, in which
the Na- and Al-enhanced ejecta from intermediate mass AGB stars is
either accreted onto the surfaces of other stars, or goes toward forming
new stars. 

We found that our model \emph{cannot} account for the Na-O \emph{or}
Mg-Al anticorrelations. In addition, we found that the sum of $C+N+O$
should vary by up to an order of magnitude (between stars with varying
levels of anomalies) if the pollution is coming from AGB stars. However,
the sum is observed to be (roughly) constant in globular clusters
for which this has been measured. Thus the results of our model do
not compare well with observational data --- the qualitative theory
is not supported by this quantitative study.

\subsection{The Pollution Model}

As heavy elements are primarily produced by massive stars, the observations
of GCs suggest that there was a generation of massive stars which
polluted the pristine protocluster gas, followed by a later release
of lighter elements, presumably from lower mass stars. In our model
we assume a two stage star formation/pollution history, similar to
that used in the dynamical evolution study by \citet{1999AA...352..138P}.
We now describe those two stages.

\subsubsection*{Stage 1: Initial Pollution by `Generation A' (Pop. III) Stars}

The initial mass distribution used to model the first stars that pollute
the primordial gas was based on the work of \citet{2001ApJ...548...19N}.
They predict a bimodal initial mass function (IMF) for a Z=0 population.
In addition to being bimodal, it is also `top-heavy', favouring massive
star formation.

Stellar yields were taken as input for the chemical evolution (CE)
model. The yields of \citet{2002ApJ...565..385U} were used for the
$150\rightarrow270$ M$_{\odot}$ range, \citet{2002ApJ...577..281C}
for the $13\rightarrow80$ M$_{\odot}$ range, and Karakas (2003,
private communication) for the intermediate mass stars (nb: these
were $Z=0.0001$ models rather than $Z=0$ models but they played
a negligible role in polluting the early cluster due to the top-heavy
IMF). 

Star formation occurred in a single burst, with newly synthesised
elements returned on timescales prescribed by mass dependent lifetimes
(\citealt{1982VA.....26..159G}). In this way the cluster gas was
brought up to $\textrm{[Fe/H]}=-1.4$ (the yields were diluted with
pristine Big Bang material), which is the current value (see e.g.
\citealt{2001AA...369...87G}). The next generation of stars formed
from this polluted gas.

\subsubsection*{Stage 2: Pollution by `Generation B' Stars}

Stage 2 in the model sees a population of stars forming from a mix
of the ejecta of the $Z=0$ stars (from stage 1) and Big Bang material.
A standard IMF (\citealt{1993MNRAS.262..545K}) was adopted for Generation
B. However, it was assumed that the GC only retained the ejecta from
stars with mass $<7$ M$_{\odot}$ -- the winds and ejecta from SNe
are assumed to have escaped the system due to their high velocities.
Thus only the yields from intermediate mass stars impact upon the
chemical evolution from then on. Yields from a specifically calculated
grid of models were used as self-consistent feedback in the CE model
(the models from Section \ref{section-GC-Models-Standard} of the
current study). 

\subsection{Generation B AGB Stars}

\subsubsection*{The Fiducial Set of Models}

This generation of stars has a unique chemical composition given by
the first stage of pollution - they lack nitrogen, are $\alpha$-enhanced,
and are of low metallicity. The models were computed using the exact
(non-scaled-solar) composition that resulted from the evolution of
the chemical pollution model described above. This required removing
all scaled-solar assumptions from the codes and computing new opacity
tables specifically for these stars. For this `fiducial/standard'
set of models, the \citet{1975MSRSL...8..369R} mass-loss law was
used during the RGB and the \citet{1993ApJ...413..641V} law during
the AGB (see Section \ref{section-GC-Models-Standard} for more details
on these models). 

\subsubsection*{The Effects of Switching Reaction Rate Compilations \& Mass Loss
Rates}

To estimate the sensitivity of the GCCE model to the various prescriptions
used in the stellar model calculations, some experiments were run
in which the following were altered: 1) the reaction rates for the
NeNa chain, MgAl chain and Ne22 + $\alpha$ reactions and 2) the mass-loss
formula used for the AGB evolution. Reaction rates and mass-loss are
two of the key uncertainties in the stellar models (these models were
presented in Section \ref{section-GC-Models-Standard}). It was found
that altering the mass-loss formula on the AGB had a greater impact
on the stellar yields than changing reaction rate compilations. The
effects of these changes are further diluted when the yields are integrated
in the CE model. 

\subsection{Results}

\begin{figure}
\begin{centering}
\includegraphics[width=0.7\columnwidth,keepaspectratio]{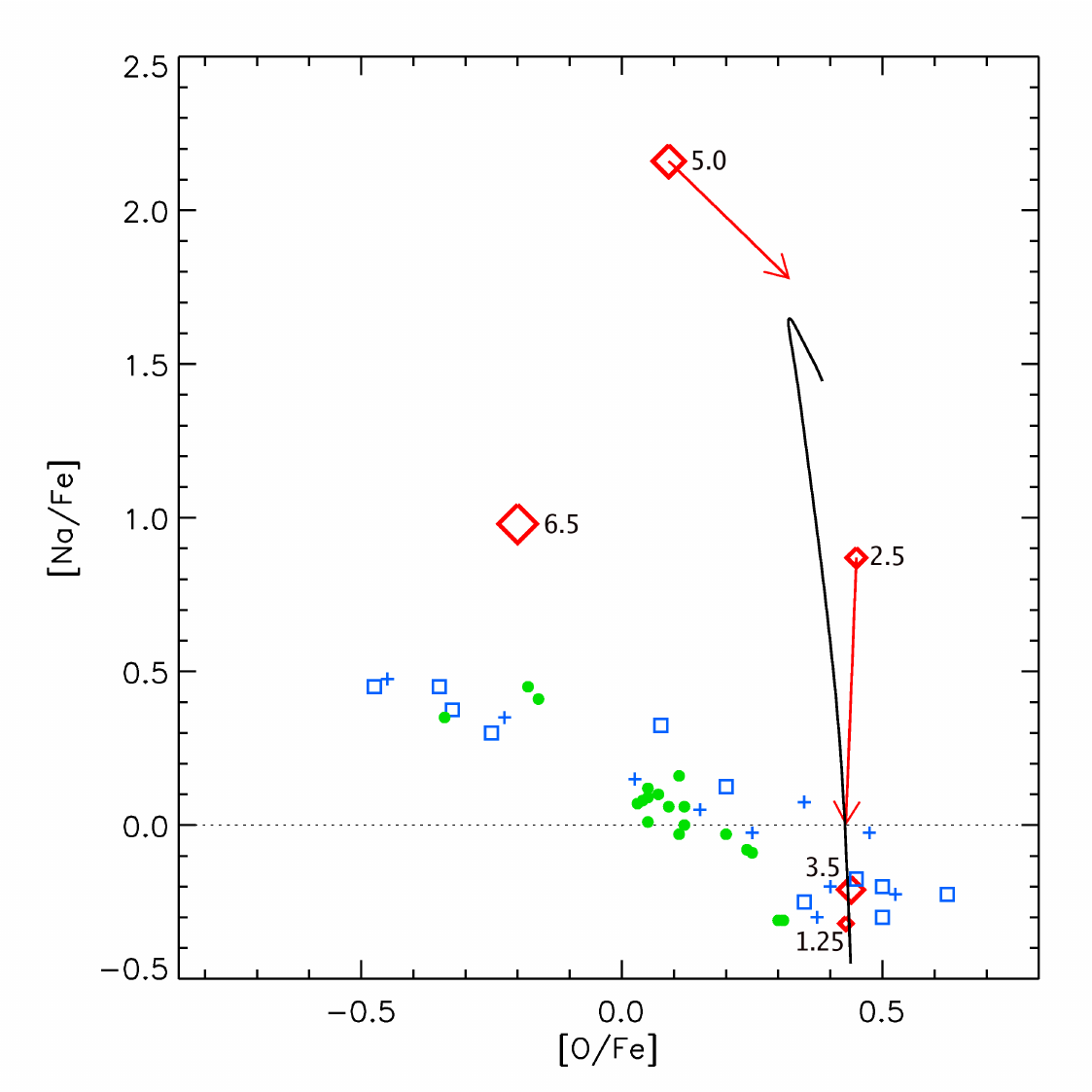}
\par\end{centering}
\caption{The O-Na anticorrelation in NGC 6752. The solid line is the predicted
trend given by tracking the chemical evolution of the intracluster
medium in the CE model. Blue and green symbols are the observational
data (from \citealt{2002AA...385L..14G} and \citealt{2003AA...402..985Y}
respectively), while the diamonds are the AGB yields (initial stellar
masses are marked, in M$_{\odot}$). The red arrows indicate the change
due to the use of Reimers' mass-loss law on the AGB (instead of VW93).
Note that oxygen is shifted $\sim0.2$ dex compared to our yields
plots in Section \ref{section-GC-Models-Standard} due to the use
of a different Solar oxygen abundance (to keep in line with that used
for plotting the observations). {[}Figure courtesy of Yeshe Fenner,
Harvard University, USA. Also see \citet{2004MNRAS.353..789F}.{]}
\label{fig-NaO-6752paper}}
\end{figure}

\begin{figure}
\begin{centering}
\includegraphics[width=0.7\columnwidth,keepaspectratio]{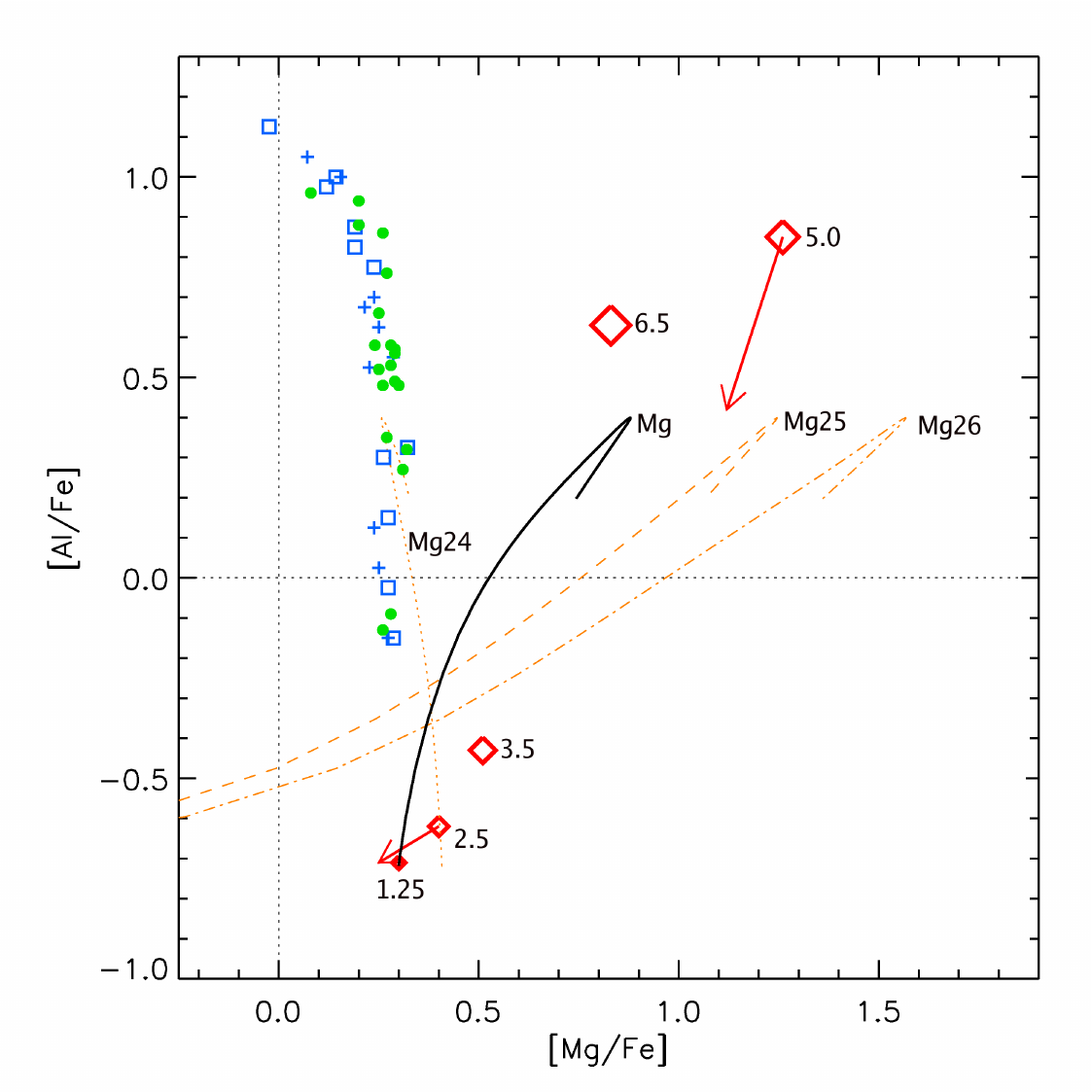}
\par\end{centering}
\caption{Same as Figure \ref{fig-NaO-6752paper} except for the Mg-Al anticorrelation
in NGC 6752. Chemical evolution tracks are shown for the Mg isotopes
as well as total Mg. It can be seen that the CE model predicts a \emph{positive}
correlation rather than an anticorrelation between these elements.
{[}Figure courtesy of Yeshe Fenner, Harvard University, USA. Also
see \citet{2004MNRAS.353..789F}.{]} \label{fig-MgAl-6752paper}}
\end{figure}

Figure \ref{fig-NaO-6752paper} compares the GCCE model results with
observational data for the O-Na anticorrelation. As can be seen, the
predicted variations in Na and O do \emph{not} match the observations.
The spread in Na is easily achieved (actually too much is produced),
but the oxygen is not depleted enough. Although O is significantly
depleted in the 5.0 and 6.5 M$_{\odot}$ stellar models, through HBB
on the AGB, these stars are not very numerous in a standard IMF. Biasing
the IMF towards higher mass AGB stars may allow the matching of the
Na-O anticorrelation but the Mg-Al anticorrelation would still be
a problem. This anticorrelation is not matched by the model either
(see Figure \ref{fig-MgAl-6752paper}). The increase in Al is achieved
(although offset), but the Mg \emph{also} increases -- which the
opposite to the observations (which show Mg decreasing as Al increases).

\begin{figure}
\begin{centering}
\includegraphics[width=0.7\columnwidth]{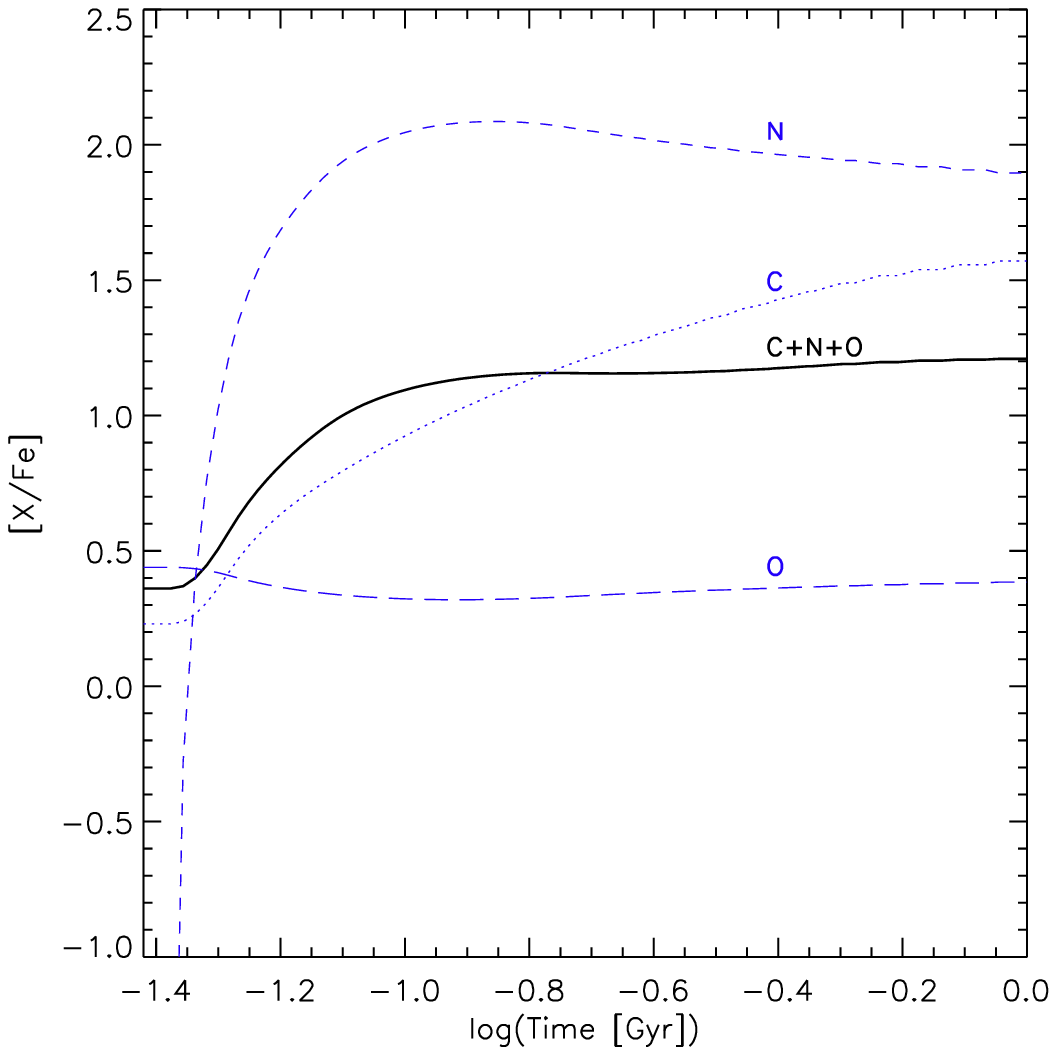}
\par\end{centering}
\caption{The time evolution of C, N, O and the sum \textrm{C+N+O}
 in the chemical evolution model. It can be seen that the model predicts
a $\sim1$ dex variation in C+N+O -- which is not observed. {[}Figure
courtesy of Yeshe Fenner, Harvard University, USA. Also see \citet{2004MNRAS.353..789F}.{]}
\label{fig-CNO-6752paper}}
\end{figure}

In addition to these problems, we note that HBB AGB stars are known
to produce primary C (which arrives at the surface via 3DUP). Indeed,
AGB stars are predicted to enhance the abundance of C by about an
order of magnitude. This alters the sum of $\textrm{C+N+O}$ in the
polluting material (which CNO cycling alone does not do). Therefore,
if it is these stars supplying the material for the GC pollution,
we would expect a variation in the sum of $\textrm{C+N+O}$ from star
to star, depending on the amount of polluting material each has received.
The CE model results in Figure \ref{fig-CNO-6752paper} support this
prediction. 

However, $\textrm{C+N+O}$ has been observed to be (at least roughly)
constant in all clusters measured so far (eg. in M4, \citealt{1999AJ....118.1273I};
M13, \citealt{1996AJ....112.1511S}; and M92, \citealt{1988ApJ...326L..57P}).
Thus, unless GC AGB stars do not undergo 3DUP, this is a strong argument
against AGB pollution causing the anomalies.

\subsection{Summary}

Using detailed nucleosynthetic yields we have computed the chemical
evolution of the intracluster medium in the globular cluster NGC 6752.
Abundance spreads in Na and Al were found, however too much Na was
produced, while the Al was offset from the observations. The O-Na
and Mg-Al anticorrelations were not matched. In particular, neither
O nor Mg are sufficiently depleted to account for the observations.
Altering the IMF to favour more massive AGB stars may allow the matching
of the Na-O anticorrelation but the Mg-Al anticorrelation would still
be a problem. Furthermore, stars bearing the imprint of AGB ejecta
are predicted to be strongly enhanced in CNO nuclei (ie. the sum of
C+N+O), also in conflict with the observational data. Thus, although
the second generation intermediate-mass AGB stars do show the hot
hydrogen burning (via HBB) that is required to explain the observations,
this quantitative study suggests that HBB in AGB stars may \emph{not}
be the site. Altering the mass-loss prescriptions and reaction rate
compilations in the stellar models does not alter this conclusion.
We note however that the single stellar models of \citet{2005ApJ...635L.149V}
do come close to matching the abundance anomalies. This is due to
their use and development of an alternative theory of convection,
which alters the evolution of their models significantly (see eg.
\citealt{2005AA...431..279V} for details). It would be interesting
to see a chemical evolution study using their models, to see if the
(partial) match is maintained when convolving with an IMF.

\section{Models of NGC 6752 Stars \emph{Without 3DUP\label{Section-GC-No3DUP}}}

\subsection{Motivation}

Current standard models of intermediate-mass AGB stars with third
dredge-up (3DUP) do not reproduce the observed C-N, O-Na or Mg-Al
inverse-correlations observed in most GCs. This is supported by the
models presented earlier in this chapter (also see \citealt{2004MNRAS.353..789F}).
We note however that the models of \citet{2005ApJ...635L.149V} do
come close, due to their use and development of an alternative theory
of convection, which alters the evolution of their models significantly
(see eg. \citealt{2005AA...431..279V} for details). In the standard
models the combination of hot bottom burning (HBB) and 3DUP during
AGB evolution does predict enhancements of nitrogen, sodium and aluminium,
and a reduction in oxygen -- all of which are needed to explain the
observations -- however they also predict an increase of carbon and
magnesium, giving positive C-N and Mg-Al correlations which is in
direct violation of the observations. In addition to this the models
do not \emph{quantitatively} account for the large degree of O depletion
and tend to over-produce Na, thus falling short of the mark with the
O-Na inverse-correlation as well (e.g., \citealt{1997AA...320..115D};
\citealt{2003ApJ...590L..99D}; \citealt{2004MmSAI..75..335V}). Furthermore,
the standard AGB models produce ejecta with an overall increase in
CNO nuclei of $\sim$ 1
 to 3 dex (mainly due to the dredge-up of carbon), whilst observations
show that the sum of CNO nuclei is virtually constant, varying by
a factor of only two or so at most (e.g., \citealt{1988ApJ...326L..57P};
\citealt{1996AJ....112.1511S}; \citealt{1999AJ....118.1273I}). These
discrepancies are based on confronting single stellar models with
observations. Taking the evolution of an entire cluster of stars into
account -- in particular convolving with an IMF -- presents further
problems as the stars with the prerequisite hot hydrogen burning are
few in number when using a standard IMF (e.g., \citealt{2004MNRAS.353..789F}),
whilst these stars need to contribute an amount of polluting material
that is a large fraction of the GC mass. Despite all these discrepancies
it should be stressed that the AGB pollution scenario should not be
dismissed outright just yet, as the models are well known to have
many serious uncertainties (eg. the treatment of convection, overshoot,
mass loss and the uncertainties in reaction rates). For example the
non-standard models of \citet{2005ApJ...635L.149V}, which use a different
convection model, can reproduce most of the abundance anomalies, although
some overshooting and the adoption of a high mass-loss rate is required.
In addition to the uncertainties in the input physics of all stellar
models, numerical details can also significantly affect the results.
For example, time-stepping during diffusive mixing can cause changes
in the stellar structure, hence altering the temperature at the bottom
of the convective envelope, and thus the resulting nucleosynthesis
and yields (Siess, private communication). Attempts have been made
to reconcile the (standard) models with the observations but have
universally ended with a need to `tweak' the relative effects of HBB
and 3DUP to an unpalatable degree (e.g., \citealt{2003ApJ...590L..99D};
also suggested by \citealt{2004MNRAS.353..789F}). 

With the results of these stellar modelling studies in mind we have
chosen to investigate the limiting case -- by calculating yields
for AGB models that do not experience any 3DUP. Thus, in models that
are hot enough (ie. massive enough), we will see pure HBB as the primary
source of the abundance patterns in the yields. By decoupling HBB
from 3DUP we hope to shed light on the relative importance of these
two processes in shaping the yields, particularly in relation to the
Na-O and Mg-Al anticorrelations. In addition to this most of the evidence
points to (possibly pure) hot hydrogen burning as the source of the
abundance anomalies -- which HBB in AGB stars provides.

\subsection{Inhibiting 3DUP}

Although `turning off' 3DUP in our models is ad-hoc, it was surprisingly
easy to achieve. By adding an extra physical condition for the determination
of convective boundaries -- the molecular weight gradient -- via
implementing the \citet{1947ApJ...105..305L} criterion rather than
the Schwarzschild criterion we found that no dredge-up at all occurred
in our models. 

In Section \vref{semimods} we discussed the SEV code modifications
with regards to semiconvection and the addition of the Ledoux criterion.
For ease of reference we repeat some details of the changes here.

Given the discussion in Section \ref{semimods}, we decided to add
the Ledoux criterion (Equation \vref{eqn-ledoux}) to the SEV code
as an alternative option for simulating semiconvection. Our adopted
criterion for deciding when a region is semiconvective was:

\begin{equation}
0<(\nabla-\nabla_{ad})<\nabla_{L}\label{eqn-simsemiCriterion}
\end{equation}

where

\begin{equation}
\nabla_{L}=\frac{\beta}{4-3\beta}\nabla_{\mu}.\label{eqn-GradLedoux}
\end{equation}

The definitions of the rest of the variables are available in Section
\vref{semimods}. Since we now treat mixing time-dependently throughout
the entire star, it was an easy matter to also treat semiconvective
regions in the same manner. This was done by applying slow mixing
velocities in the region defined by Equation \ref{eqn-simsemiCriterion}.
In Section \ref{semimods} we pointed out that there is the problem
of not knowing how slow/fast to mix. We initially hard-wired a fixed
value velocity for semiconvection (a similar tactic to that of \citealt{1972MNRAS.156..361E}).
In order to inhibit 3DUP in this case we have chosen a mixing velocity
of zero -- such that the convective boundary is now a `hard' Ledoux
one (we have used no overshoot either). As mentioned above, this was
sufficient to inhibit 3DUP totally -- in models of all masses considered
here.

\subsection{The Grid of Models }

We have calculated four no-3DUP models, with masses 2.5, 3.5, 5.0,
6.5 M$_{\odot}$. This covers most of the masses of our standard models
presented in Section \ref{section-GC-Models-Standard}, so direct
comparisons can be made. The initial composition is identical to the
standard models.

\subsubsection*{Input Physics}

The version of the SEV code used for these models was the same one
used for our $Z=0$ and EMP models in chapters \ref{chapter-Z0-StructEvoln}
and \ref{Chapter-HaloStarModels}. In particular it had the updates
for time dependent mixing, opacities, diffusive overshoot and choice
of convective boundary criteria. As mentioned above 3DUP was inhibited
by using the Ledoux criterion for convective boundaries in these models.
A couple of other key inputs of note are:
\begin{enumerate}
\item Time dependent mixing was always used.
\item No overshoot was used.
\item Scaled-solar opacities were used.
\end{enumerate}
We believe that using time-dependent mixing rather than instantaneous
mixing (as used in the standard models) should not have effected the
results significantly. Particularly since no very rapid events like
the dual flashes of the $Z=0$ models occur in these models. The use
of scaled-solar opacities rather than opacities with the exact composition
(as used in the standard models) also has little effect. The suppression
of 3DUP is by far the major factor affecting the results of these
models. The non-use of overshoot helps with this suppression.

In regards to the nucleosynthesis we used the same nuclear reaction
rates as those used in the standard models in Section \ref{section-GC-Models-Standard}.
As mentioned there altering the rates had little effect on the yields. 

\subsection{Results}

\subsubsection*{AGB Surface Abundance Evolution}

\begin{figure}
\begin{centering}
\includegraphics[width=0.85\columnwidth,keepaspectratio]{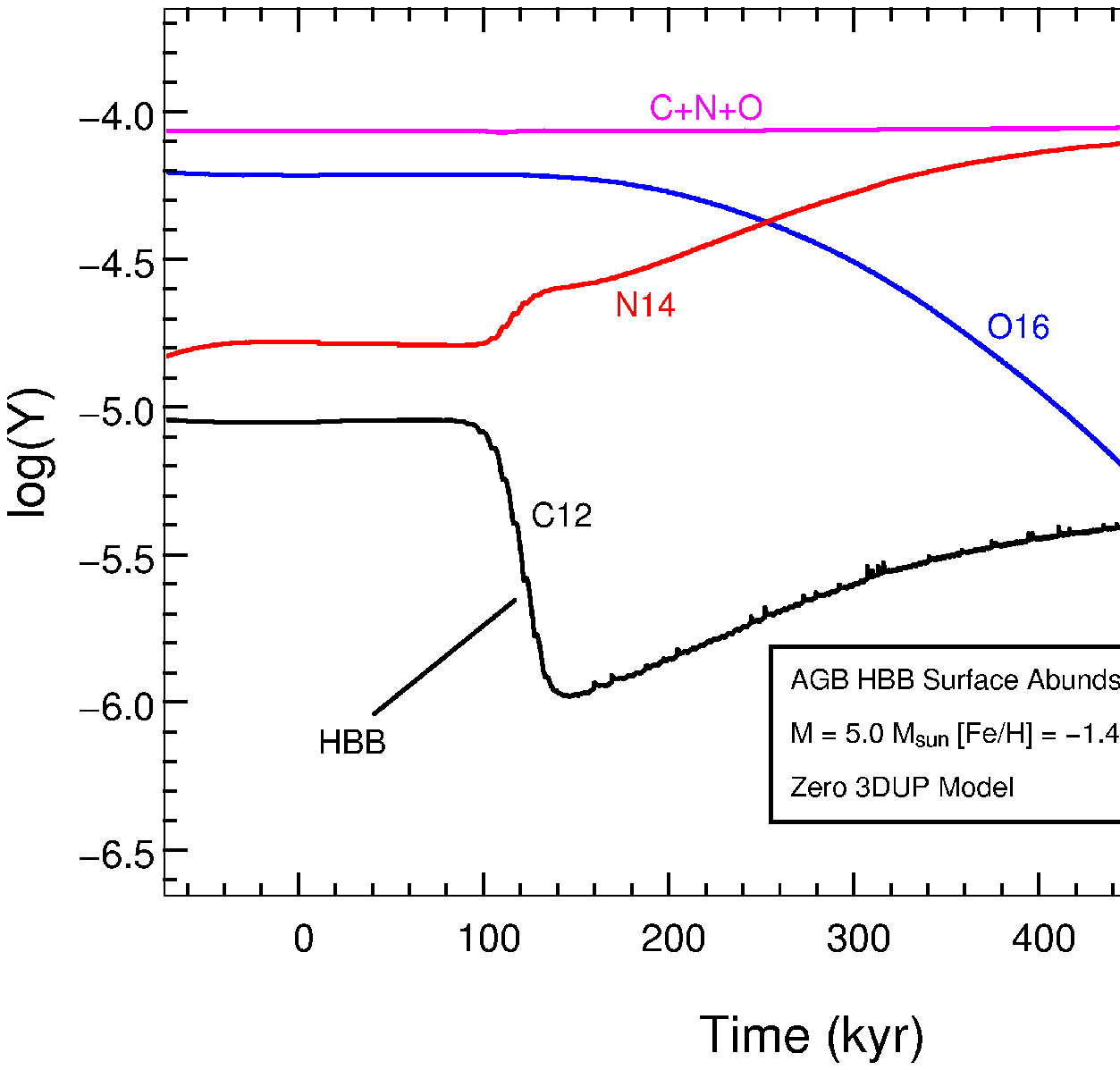}
\par\end{centering}
\caption{Surface abundance evolution of some of the CNO nuclides in the 5.0
M$_{\odot}$ no-3DUP model during the AGB. Time has been offset. Also
plotted is the sum of C+N+O which can be seen to stay constant despite
large changes in its constitutes. This is due to the fact that the
CNO cycles conserve the number of CNO nuclei. \label{fig-m5gc-Zero3Dup-SRF-CNO-AGB}}
\end{figure}

In Figure \ref{fig-m5gc-Zero3Dup-SRF-CNO-AGB} we show the surface
abundance evolution of the main CNO isotopes in the 5 M$_{\odot}$
model. In contrast to the standard 5 M$_{\odot}$ model (see Figure
\vref{fig-m5.0gc-Standard-SRF-AGB-CNO}) it can be seen that there
is no initial enrichment of the surface in $^{12}$C. This is of course
due to the total lack of 3DUP. However the decrease in $^{12}$C is
seen once $T_{bce}$ increases and HBB sets in. The corresponding
increase in $^{14}$N is also seen, just as in the standard case.
The increase in the no-3DUP case is however much less than in the
standard case. Interestingly $^{14}$N continues to increase throughout
the AGB in the no-3DUP case also, despite the lack of fresh C for
CN processing. This is due to significant O$\rightarrow$N cycling,
as can be seen in Figure \ref{fig-m5gc-Zero3Dup-SRF-CNO-AGB}. The
difference between the final N abundances is very different though.
In the standard model $^{14}$N increases by $\sim1.5$ dex over the
abundance at the start of the AGB phase, whilst in the no-3DUP model
it only increases by $\sim0.7$ dex. Naturally the reason for this
is the lack of fresh C and O nuclei to be cycled to N in the no-3DUP
model. Indeed, it can be seen that the sum of C+N+O is constant in
the no-3DUP case, whilst it increases by $\sim1$ dex in the standard
case. Thus this model satisfies this key observational constraint
where the standard model did not. Furthermore, it can be seen that
$^{16}$O is heavily depleted (by $\sim1.7$ dex) in the no-3DUP model.
In the standard model is is only depleted by $\sim0.3$ dex. The reason
for this difference is also the lack of 3DUP. In the standard case
HBB does deplete oxygen but it is also periodically replenished by
3DUP, so the decrease is moderated. Thus the no-3DUP model satisfies
another constraint in the observations of NGC 6572 which show the
most extreme stars having O depleted by $\sim$1 dex. Moreover, C
is seen to be strongly reduced whilst N is enhanced, also matching
the observations (the standard models show N \emph{and} C enhancement).

\begin{figure}
\begin{centering}
\includegraphics[width=0.85\columnwidth,keepaspectratio]{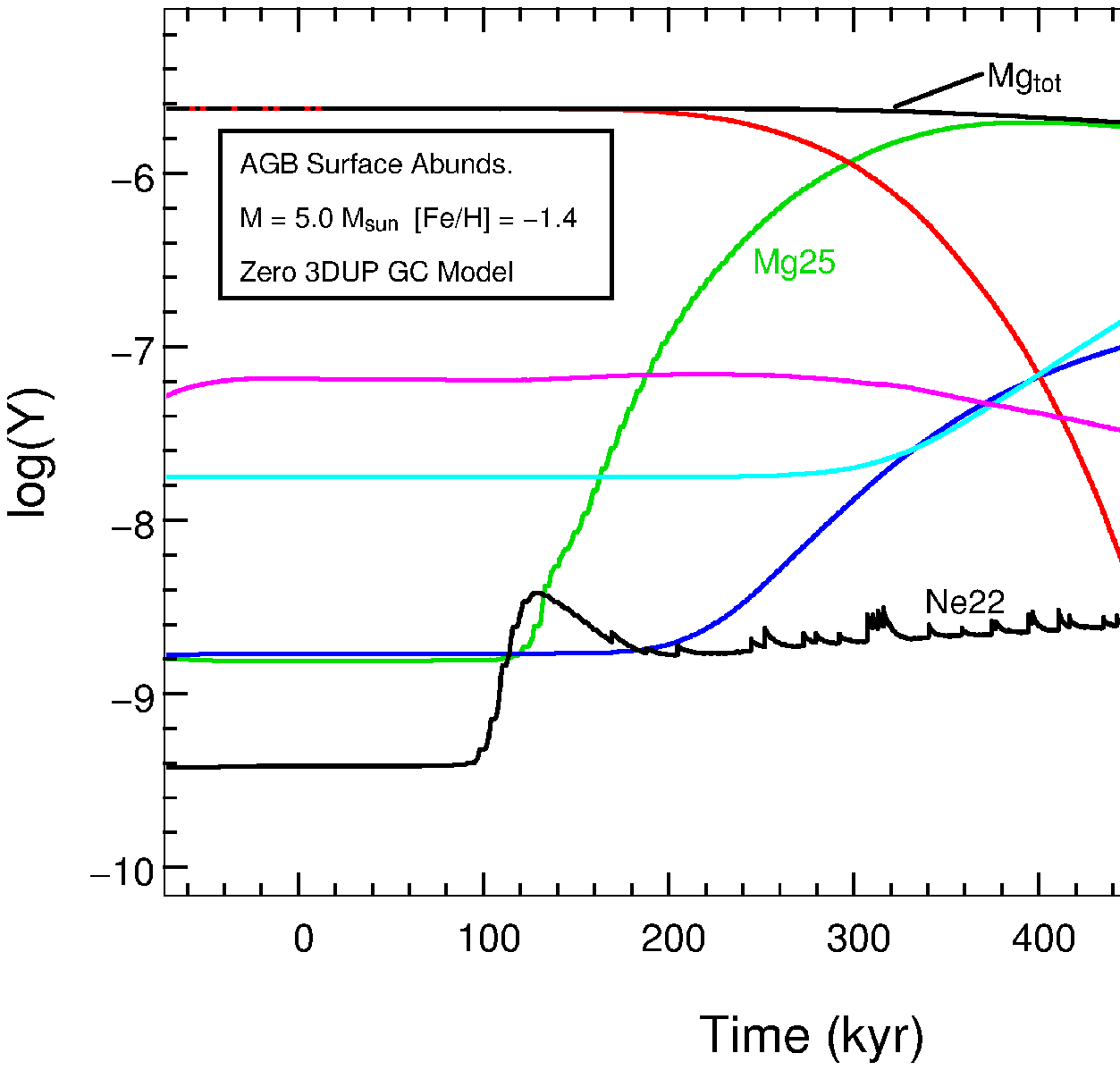}
\par\end{centering}
\caption{Surface abundance evolution of some of the Mg, Al, Ne and Na nuclides
in the 5.0 M$_{\odot}$ no-3DUP model during the AGB. Time has been
offset. Also plotted is Mg$_{tot}$, the sum of the Mg isotopes. This
is seen to reduce slightly towards the end of the AGB. It can be seen
that the Mg isotope ratios are very non-solar by the end of the AGB.
\label{fig-m5gc-Zero3Dup-SRF-MgAlNeNa-AGB}}
\end{figure}

In Figure \ref{fig-m5gc-Zero3Dup-SRF-MgAlNeNa-AGB} we show the surface
abundance evolution of some key isotopes involved in the Ne-Na and
Mg-Al cycles/chains. Starting with Ne we see that $^{22}$Ne is initially
enhanced at the beginning of the AGB. This comes about through the
destruction of $^{21}$Ne. As the temperature increases $^{22}$Ne
is itself destroyed, cycling through to $^{23}$Na. Likewise, as the
temperature increases further, $^{23}$Na is cycled through to $^{20}$Ne.
$^{20}$Ne is very abundant ($\textrm{log}(Y)\sim-5.3$) compared
to the other Ne isotopes so its abundance is scarcely enhanced by
this addition. Some leakage occurs out of the Ne-Na cycle through
$^{23}$Na($p,\gamma$)$^{24}$Mg but this has little effect as $^{24}$Mg
is (initially) so abundant. The rearrangement of the Mg isotopes is
clearly seen in Figure \ref{fig-m5gc-Zero3Dup-SRF-MgAlNeNa-AGB}.
$^{24}$Mg is almost totally destroyed, mainly ending up as $^{25}$Mg
via $^{24}$Mg(p,$\gamma$)$^{25}$Al($\beta^{+}$)$^{25}$Mg. Also
enhanced by proton captures is $^{26}$Mg, at the expense of $^{25}$Mg
and $^{27}$Al at the expense of $^{26}$Mg. Leakage out of the Mg-Al
cycle causes a slight increase in $^{28}$Si but this is minor due
to its (relatively) large initial abundance. It does however cause
a small but significant decrease in the total abundance of the Mg
isotopes. Thus the elemental Mg abundance reduces, as seen in NGC
6752 stars. Add to this the increase of $^{27}$Al and we see that
this model shows a Mg-Al anticorrelation, which is again good news
for comparing with GCs. This contrasts with the situation in the standard
models (see Figure \vref{fig-m5.0gc-Standard-SRF-AGB-MgAlNeNa} for
the standard 5.0 M$_{\odot}$ surface abundance evolution). In that
case we saw that the total Mg abundance \emph{increased} by about
$\sim0.5$ dex, due to the fresh injection of heavy Mg isotopes from
each 3DUP episode. Also present in that model is a very large increase
in $^{23}$Na, due to HBB of dredged-up $^{22}$Ne. In the no-3DUP
case we see that $^{23}$Na actually suffers a nett \emph{decrease}
(although small) over the AGB lifetime. Given the decrease in oxygen
this means Na and O are positively correlated -- in contradiction
to the observations. 

\subsubsection*{Yields}

\begin{figure}
\begin{centering}
\includegraphics[width=0.9\columnwidth,keepaspectratio]{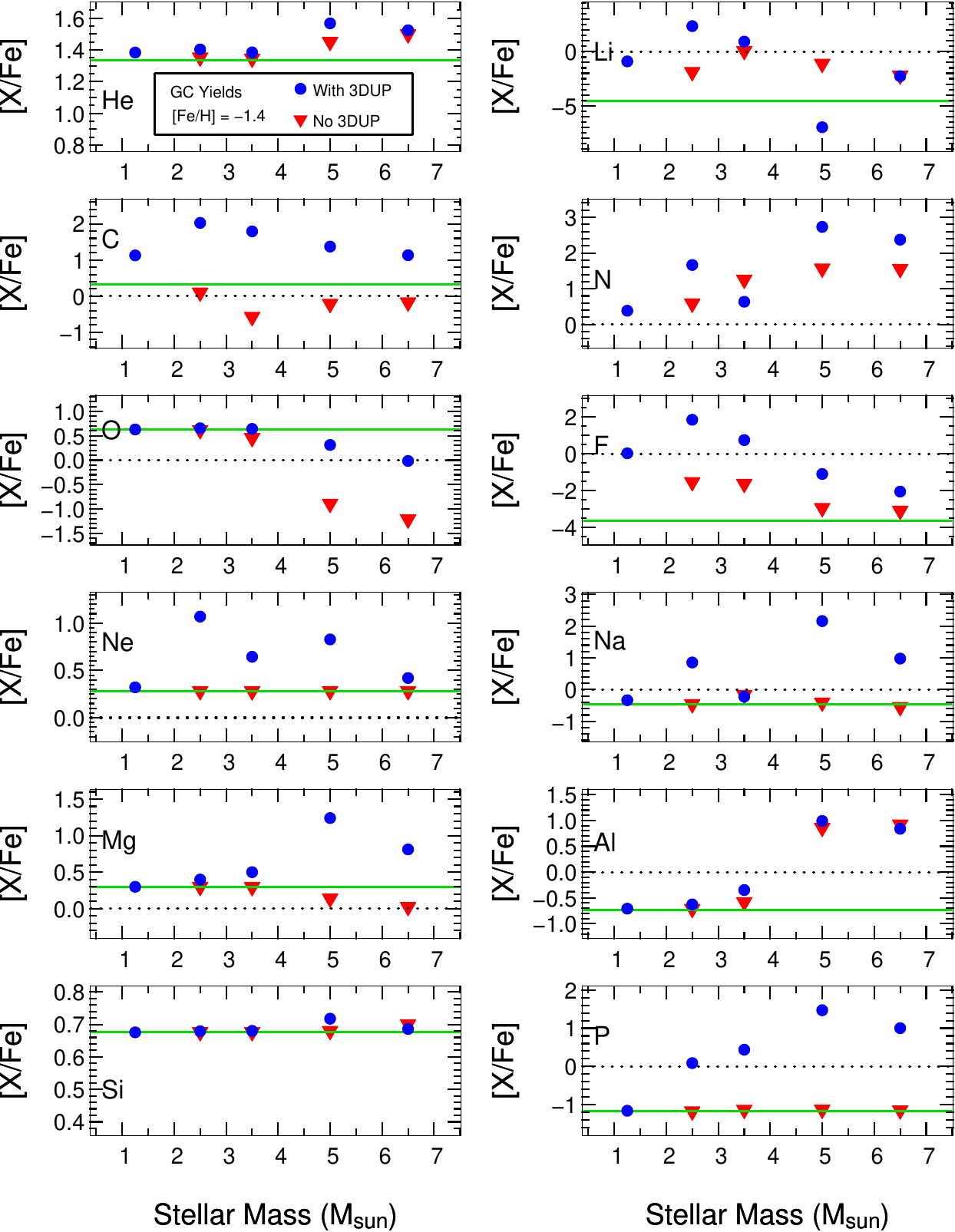}
\par\end{centering}
\caption{Yields of most of the elements in the nuclear network. The yields
from the no-3DUP models are shown as red triangles whilst those from
the standard models are shown as blue circles. The initial composition
of each element is marked by a green horizontal line in each graph.
\label{fig-yields-GC-3DUPvsNo3DUP-mostElems-XFe}}
\end{figure}

In Figure \ref{fig-yields-GC-3DUPvsNo3DUP-mostElems-XFe} we present
the yields of most of the elements in the network for all our no-3DUP
models. Also plotted in this figure are all the yields for the standard
set of GC models. This enables a quick comparison that reveals the
key role 3DUP plays in the production of many elements. Conversely
it also reveals which elements are primarily a product of HBB. 

It can be seen that carbon is drastically reduced in the yields of
all the no-3DUP models, as compared to the standard models. In fact
all the no-3DUP yields show a nett decrease in C from the initial
abundance. This is due to first dredge-up and 2DUP in the lower mass
models. An added factor in the higher mass models is HBB which cycles
much of the C to N. Due to the lack of 3DUP in all the models C can
never increase after these events. Nitrogen on the other hand is increased
in all the models, as it is in the standard models. In the higher
mass models it is however much less enhanced, since there is no 3DUP
to provide the C seeds for C$\rightarrow$N conversion. Indeed, its
abundance is limited by the initial C (and O) abundance since the
only way it can increase is via CN(O) cycling, which conserves the
sum of C+N+O nuclei. 

Oxygen exhibits the same pattern as that in the standard models --
it decreases with initial stellar mass. The difference here is that
O is more severely depleted in the no-3DUP models. As mentioned earlier
this is due to the lack of O replenishment by 3DUP that moderates
the O reduction. Interestingly fluorine is seen to have a strong dependence
on 3DUP -- it is consistently lower in the no-3DUP models. The same
applies to Ne (as expected). In fact Ne does not change from its initial
abundance at all. As mentioned above Na comes from the destruction
of $^{22}$Ne from 3DUP, so this element is also not enhanced in the
no-3DUP yields. Mg shows the opposite trend to that in the standard
models -- it decreases slightly with initial stellar mass. Interestingly
aluminium is present in the yields of the two sets of models in practically
the same amounts. Thus it is, in this case, primarily produced by
HBB rather than 3DUP. The yields are similar because there is more
than enough $^{24}$Mg and $^{25}$Mg available (Mg is much more abundant
than Al initially) for its production via $^{24}$Mg(p,$\gamma$)$^{25}$Al($\beta^{+}$)$^{25}$Mg(p,$\gamma$)$^{26}$Al($\beta^{+}$)$^{26}$Mg(p,$\gamma$)$^{27}$Al
or via $^{24}$Mg(p,$\gamma$)$^{25}$Al($\beta^{+}$)$^{25}$Mg(p,$\gamma$)$^{26}$Al(p,$\gamma$)$^{27}$Si($\beta^{+}$)$^{27}$Al
in both sets of models. 

\subsection{Discussion}

\begin{figure}
\begin{centering}
\includegraphics[width=0.85\columnwidth,keepaspectratio]{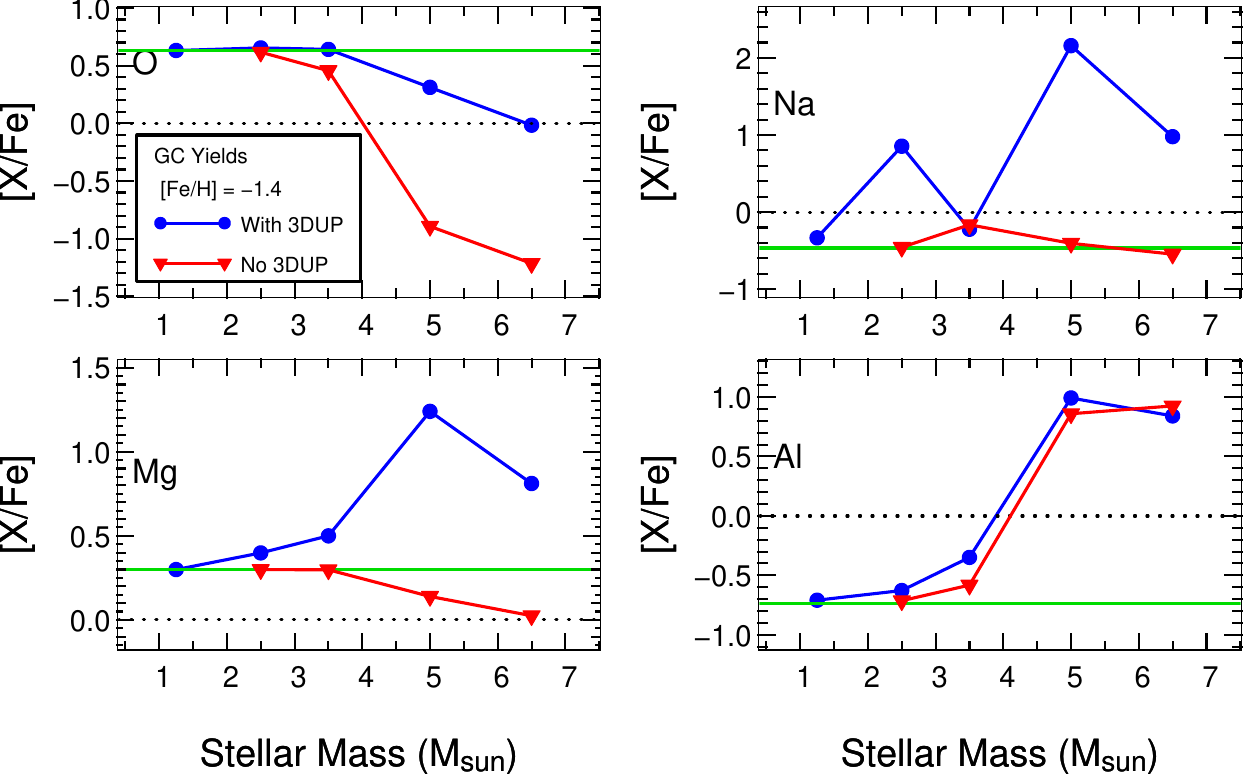}
\par\end{centering}
\caption{Comparing the Na, O, Mg and Al yields from the standard models (blue
circles) and the no-3DUP models (red triangles). In terms of the Na-O
and Mg-Al anticorrelations in NGC 6752 it can be seen that the lack
of 3DUP (1) increases the O destruction, (2) causes a Mg destruction
(3) retains the Al production but (4) removes the Na production. Thus
these models are a better fit to the observations but problems remain.
\label{fig-yields-compare-3DUP-Zero3DUP-NaOMgAl}}
\end{figure}

A salient result from our new stellar models is that oxygen is now
heavily depleted in the more massive models. Also of importance is
the the fact that there is now an opposite trend for Mg (in the same
stars) -- it is depleted instead of produced, whilst Al is still
enhanced. Furthermore, the C is depleted whilst the N yield remains
quite high. All of these features (except the increase in Al) derive
from the lack of 3DUP, as O is now not periodically replenished, C
cannot increase and fresh fuel for the MgAl cycle is not available.
The lack of 3DUP also allows the sum of CNO nuclei to remain constant,
as fresh CNO is not added to the envelope and the CNO burning cycles
(occurring at the base of the convective envelope) conserve the number
of CNO nuclei. All these features indicate that a better fit to the
observations is expected (IMF withstanding). To clearly demonstrate
the situation with respect to O-Na and Mg-Al we re-plot these elements
together in Figure \ref{fig-yields-compare-3DUP-Zero3DUP-NaOMgAl}.
The effect of removing 3DUP is beneficial (in relation to matching
the anticorrelations) in all cases -- except for sodium. Sodium is
observed to be enhanced in some NGC 6752 stars by $\sim1$ dex, however
in the no-3DUP yields it is barely enhanced at all ($\lesssim0.2$
dex). In the 6.5 M$_{\odot}$ case it is even depleted slightly. Thus,
although we can reproduce a large O reduction needed, our no-3DUP
models cannot reproduce the Na-O anticorrelation!

\begin{figure}
\begin{centering}
\includegraphics[width=0.85\columnwidth,keepaspectratio]{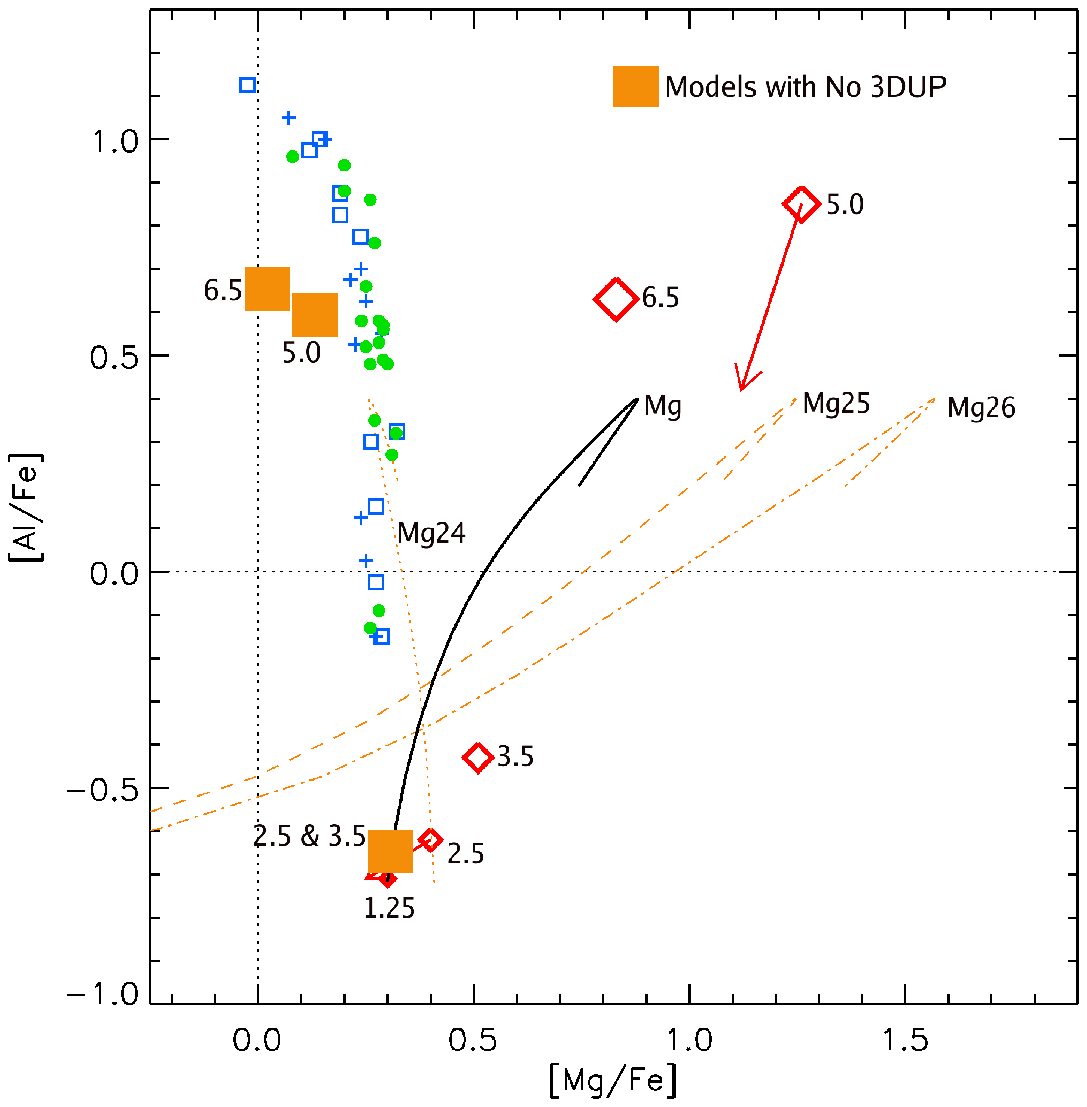}
\par\end{centering}
\caption{The Mg-Al anticorrelation in NGC 6752. Same as Figure \vref{fig-MgAl-6752paper}
but with our no-3DUP model yields over-plotted (large orange filled
squares). Blue and green symbols are the observational data (from
\citealt{2002AA...385L..14G} and \citealt{2003AA...402..985Y} respectively),
while the diamonds are the AGB yields (initial stellar masses are
marked, in M$_{\odot}$). The red arrows indicate the change due to
the use of Reimers' mass-loss law on the AGB (instead of VW93). The
solid line is the predicted trend given by tracking the chemical evolution
of the intracluster medium in the CE model. Chemical evolution tracks
are shown for the Mg isotopes as well as total Mg (broken orange lines).
It can be seen that, apart from an offset downwards in Al, the no-3DUP
yields match well with this anticorrelation. The Al offset would be
easily remedied by adopting the Al abundance of the unpolluted stars
as the initial composition in the stellar models. We note however
that the Mg isotope results from the no-3DUP models are even a worse
match than that of the standard models (see text for a discussion).
\label{fig-MgAl-6752paper-Zero3DUP}}
\end{figure}

\begin{figure}
\begin{centering}
\includegraphics[width=0.85\columnwidth,keepaspectratio]{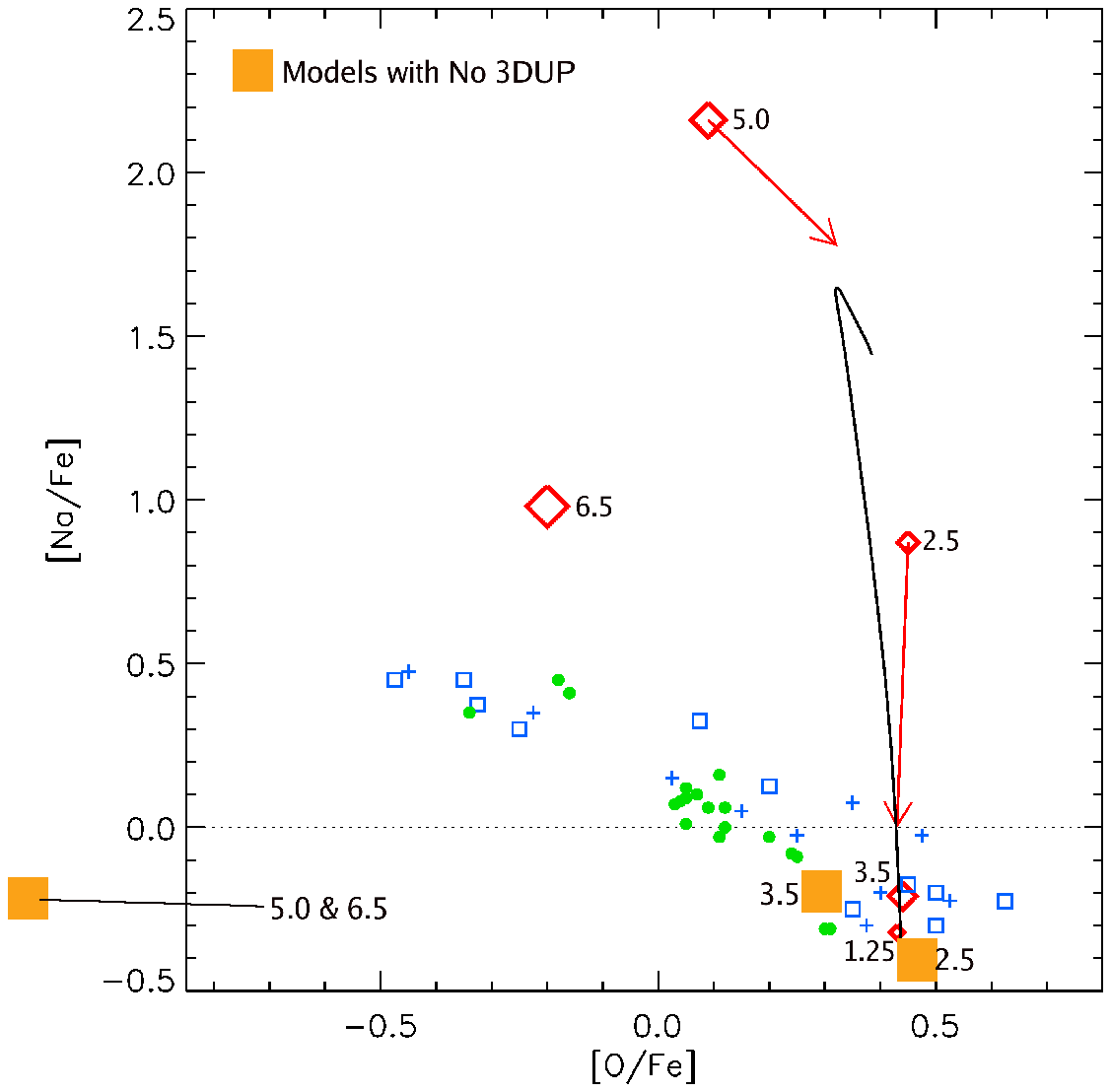}
\par\end{centering}
\caption{Same as Figure \ref{fig-MgAl-6752paper-Zero3DUP} except for the Na-O
anticorrelation in NGC 6752. It can be seen that not enough Na is
present in the yields of the higher mass models to match the observations.
The dilemma is that it is these models that provide the strong O depletion
needed in the `polluted' stars. Note that the oxygen yields of the
5 and 6.5 M$_{\odot}$ models are off the scale of this figure. They
should be located at $\textrm{[O/Fe]}\approx-1.2$ and $-1.4$ respectively.
Thus they show more O depletion than the most polluted GC stars. The
{[}O/Fe{]} values of the yields are shifted $\sim-0.2$ dex compared
to our yield plots due to the use of a different Solar oxygen abundance
(to keep in line with that used for plotting the observations). \label{fig-NaO-6752paper-Zero3DUP}}
\end{figure}

At the high masses (ie. temperatures) required to deplete O and Mg
via the MgAl and ON cycles, Na is not produced enough, or rather it
is destroyed too much by proton captures, resulting in the low yields.
Moving to lower masses (temperatures) improves the Na situation (slightly)
but prevents O and Mg from being depleted. Interestingly this is the
identical problem that \citet{2005ApJ...635L.149V} have with their
best-fit (for the Na-O inverse-correlation) model. \citet{2003ApJ...590L..99D}
also report this problem. We note that the models of \citet{2005ApJ...635L.149V}
are quite independent as they use a different treatment for convection.
In a recent and very timely study \citet{2006AA...457..995V} have
investigated this problem in detail, finding that their stellar model
requires either the addition of some overshooting or a reduction of
the $^{23}$Na(p,$\alpha$)$^{20}$Ne reaction rate in order to produce
the required Na yield whilst maintaining the O destruction. The Ventura
and D'Antona (\citeyear{2005ApJ...635L.149V} , \citeyear{2006AA...457..995V})
models, like our more ad-hoc models, otherwise match the abundance
anomalies fairly well. We highlight this by plotting some of these
no-3DUP yields against the observations in Figures \ref{fig-MgAl-6752paper-Zero3DUP}
and \ref{fig-NaO-6752paper-Zero3DUP}. It can be seen in Figure \ref{fig-MgAl-6752paper-Zero3DUP}
that the magnitude of Mg destruction is very similar to that seen
in the NGC 6752 stars. The magnitude of increase in Al is also a very
good match (the absolute values of Al are easily shifted upwards by
using an initial composition that matches the lower-Al observations).
Thus it appears that the no-3DUP yields provide a good match to the
Mg-Al anticorrelation. However a serious problem actually remains.
\citet{2003AA...402..985Y} have been able to deduce the ratios of
the Mg isotopes in the stars of this cluster. They find that the `unpolluted'
stars (ie. low Al, low Na stars) show a roughly solar ratio of $^{24}$Mg:$^{25}$Mg:$^{26}$Mg
($\sim$80:10:10; see eg. \citealt{1971SoPh...19..330B} for the solar
ratio), whilst the polluted stars show a ratio of $\sim$60:10:30.
Thus there appears to have been an increase in $^{26}$Mg, probably
at the expense of $^{24}$Mg, whilst $^{25}$Mg is constant (in a
relative sense -- the sum of the absolute abundances has reduced
in these stars). Turning to the surface abundance evolution of our
models (Figures \ref{fig-m5.0gc-Standard-SRF-AGB-MgAlNeNa} and \ref{fig-m5gc-Zero3Dup-SRF-MgAlNeNa-AGB})
we see at a glance that the Mg isotopic ratios are \emph{extremely}
different to those observed! $^{24}$Mg is heavily depleted in all
the no-3DUP models that produce Al (higher masses), whilst $^{25}$Mg
is dominant. The ratio in the 5 M$_{\odot}$ model \emph{yield} is
$\sim$0.1:10:1, which compares very poorly with the observed ratio.
Interestingly the situation is better in the yield from our standard
model (of 5 M$_{\odot}$). It has a ratio of $\sim$2:10:26. This
model does however still show a huge depletion of $^{24}$Mg which
is not seen in the observations -- not to mention the fact that the
standard models have other problems in explaining the anticorrelations.
We are thus in a dire situation with respect to the Mg isotopes, despite
having a good fit to the Mg-Al (elemental) anticorrelation. How this
can be remedied is a difficult question. A partial solution may be
found in using the observed ratios in the unpolluted stars in the
initial composition of models. We note however that this would be
scaled-solar, which is what most studies usually use anyway. Reaction
rate uncertainties are also another avenue to explore. We shall pursue
this tantalising mystery in our future work. 

In Figure \ref{fig-NaO-6752paper-Zero3DUP} we compare the no-3DUP
yields with the Na-O anticorrelation observations in NGC 6752. As
mentioned earlier oxygen is efficiently depleted in the higher mass
models, and much more than in the standard models. In fact our models
display O deficiencies much greater than that observed in the stars
(up to $\sim-1.7$ dex in the models versus $\sim-1$ dex in the observations).
This effect may be reduced by dilution effects but we note that the
discrepancy is significant. The most serious problem lies however
in Na. As discussed above it is barely produced at all -- especially
in the O-depleted models where it is most needed to match the observations.

\subsection{Conclusion}

In summary we have used these ad-hoc models to investigate the interplay
between 3DUP and HBB and the effects they have on the predictions
of (anti)correlations from stellar model yields, in the context of
the GC abundance anomalies problem. We have found that the yields
from the no-3DUP models match the Mg-Al anticorrelation in NGC 6752
well, and also have oxygen significantly depleted, as needed (although
too much in some cases). However Na is not produced enough (even depleted
in some cases). This is in direct contradiction with the observations.
Furthermore, the Mg isotope ratios are very different to those of
the observations. In particular $^{24}$Mg is always heavily depleted
by HBB whilst it is still relatively abundant in the polluted star
observations. This occurs at the same temperatures that are needed
to produce the Al enhancements and oxygen depletions, making the solution
to this problem seemingly impossible to resolve in the AGB HBB scenario
(or possibly any pure hydrogen-burning site). Thus, although we appear
to have a better fit to the observations with the no-3DUP models,
there remains some serious discrepancies. We shall continue with this
fascinating (and frustrating) problem in our future studies -- one
of which is a globular cluster observing project described in the
next section.

Finally we note that these models may actually simulate the effects
of a previously ignored population of stars, the Super-AGB stars (8
M$_{\odot}$ $\lesssim M\lesssim11$ M$_{\odot}$). We may expect
the sort of pollution as produced by our no-3DUP models to come from
these stars since they are expected to either have either zero or
very little 3DUP. Even if they do have 3DUP characterised by a large
$\lambda$ value the pollution visited on their envelopes would be
quite small due to their small intershells and massive envelopes (Siess
and Doherty, 2007, private communication). They would also probably
suffer strong HBB -- much like our no-3DUP models (for more information
on SAGB stars see eg. \citealt{1996MmSAI..67..675R}; \citealt{2006MmSAI..77..846P};
\citealt{2006MmSAI..77..834G})

\section{Cyanogen in GC AGB Stars: Observing Project\label{Section-GC-CN-AGB-ObsProject}}

\subsection{Abstract}

On reading an old paper about galactic globular cluster abundance
observations (of NGC 6752) we came across an intriguing result. \citet{1981ApJ...244..205N}
found that there was a distinct lack of cyanogen-strong (CN-strong)
stars in their sample of AGB stars, as compared to their sample of
RGB stars (which had roughly equal numbers of CN-normal and CN-strong
stars). Further reading revealed that similar features have been discovered
in the AGB populations of other clusters. Recently, \citet{2000MmSAI..71..657S}
followed up on this possibility (and considered other proton-capture
products) by compiling the existing data at the time and came to a
similar conclusion for two more clusters. Unfortunately all of these
studies suffer from low AGB star counts so the conclusions are not
necessarily robust --- larger, statistically significant, sample
sizes are needed. 

In this section we outline the results of a literature search for
relevant CN observations and describe our observing proposal to test
the suggestion that there are substantial abundance pattern differences
between the AGB and RGB in galactic globular clusters. The literature
search revealed that the AGB star counts for all studies (which are
not, in general, studies about AGB stars in particular) are low, usually
being $\lesssim10$. The search also revealed that the picture may
not be consistent between clusters. Although most clusters appear
to have CN-weak AGBs, at least two seem to have CN-strong AGBs (M5
and 47 Tuc). To further complicate the picture, clusters often appear
to have a combination of both CN-strong and CN-weak stars on their
AGBs -- although one population tends to dominate. Again, all these
assertions are however based on small sample sizes. We aim to increase
the sample sizes by \emph{an order of magnitude} using existing high
quality photometry in which the AGB and RGB can be reliably separated.
For the observations we will use a wide-field, low- to mid-resolution
multi-object spectroscope to obtain data not only on the AGB but also
on the horizontal branches and first giant branches of a sample of
clusters. With the new information we hope to ascertain whether significant
abundance differences really exist. 

\subsection{Introduction}

We are attempting to perform a conclusive test of the suggestion put
forward by \citet{1981ApJ...244..205N}, which has been touched upon
by many authors since and recently explored by \citet{2000MmSAI..71..657S},
that there are differences in cyanogen abundance distributions between
the first and second giant branches in galactic globular clusters. 

Although galactic globular clusters (GCs) are chemically homogeneous
with respect to Fe and most other heavy elements (see eg. \citealt{1992AJ....104..645K}),
it has long been known that they show inhomogeneities in many lighter
elements (eg. C, N, O, Mg, Al). These inhomogeneities are considered
anomalous because they are not seen in halo field stars of similar
metallicity (see eg. \citealt{2000AA...354..169G}). 

One of the first inhomogeneities discovered was that of the molecule
Cyanogen (CN, often used as a proxy for nitrogen). A picture of `CN-bimodality'
emerged in the early 1980s whereby there appears to be two distinct
chemical populations of stars in most, if not all, GCs. One population
is known as `CN-strong', the other `CN-weak' (the CN-weak population
might be more informatively called `CN-normal' -- as these stars
show CN abundances similar to the Halo field stars). Originally, observations
of CN were mainly made in stars on the giant branches but more recently
there have been observations on the main sequence (MS) and sub-giant
branch (SGB) of some clusters (eg. \citealt{1998MNRAS.298..601C}).
These observations show that there is little difference in the bimodal
CN pattern on the MS and SGB as compared with the giants --- indicating
a primordial origin for the differing populations. Figure 6 in \citet{1998MNRAS.298..601C}
exemplifies this situation.

Due to the paucity of asymptotic giant branch (AGB) stars in GCs (a
result of their short lifetimes) there have been very few systematic
observational studies of the CN anomaly on the AGB in globular clusters
(\citealt{1978AA....70..115M} is one that the Authors are aware of).
What little that has been done has been an aside in more general papers
(eg. \citealt{1981ApJ...244..205N}; \citealt{1993AJ....106..142B};
\citealt{1999AJ....118.1273I}). However these studies have hinted
at a tantalising characteristic: most (observed) GCs show a lack of
CN-strong stars on the AGB. If this is true then it is in stark contrast
to the red giant branch (RGB) and earlier phases of evolution, where
the ratio of CN-Strong to CN-Weak stars is roughly unity in many clusters. 

This \emph{possible} discrepancy was noted by \citealt{1981ApJ...244..205N}
in their paper about abundances in giant stars in NGC 6752. They state
that ``The behaviour of the CN bands in the AGB stars is... quite
difficult to understand... not one of the stars studied here has enhanced
CN... yet on the {[}first{]} giant branch there are more CN strong
stars than CN weak ones.'' (also see Figure 3 in that paper). More
recently \citet{2000MmSAI..71..657S} presented a conference paper
on this exact topic. Compiling the contemporaneous preexisting data
in the literature they discussed the relative amounts of CN in AGB
and RGB stars in the GCs NGC 6752 (data from \citealt{1981ApJ...244..205N}),
M13 (data from \citealt{1981ApJS...47....1S}) and M4 (data from \citealt{1981ApJ...244..205N};
\citealt{1991ApJ...381..160S}). They also discuss Na abundance variations
in M13 (data from \citealt{1996AJ....112..545P}; \citealt{1996AJ....111.1689P}).
Their conclusion for the CN variations was that the clusters in question
all showed significantly less CN on the AGB as compared to the RGB.
However the data compiled only contained about 10 AGB stars per cluster.
In their closing remarks they suggest observations with larger sample
sizes are needed --- which may be done using wide-field multi-object
spectroscopes. This is exactly the conclusion the present authors
also came to, inspiring a presentation at the Eighth Torino Workshop
on Nucleosynthesis AGB Stars held at the Universidad de Granada, Spain,
in 2006.

\subsection{Literature Search Results and the Observing Proposal}

We conducted a literature search (which may not be complete) to ascertain
what work had already been done in terms on CN on the AGB in galactic
globular clusters. The results are displayed in Table \ref{Table-GC-CN-AGB-LitSearch}.
The main result from this search was that the available number of
AGB star observations are not statistically significant enough to
come to any real conclusion about the nature of the CN abundance distributions.
This has mainly been due to technological constraints. However, the
data that does exist shows that there appears to be a strong trend
towards CN-weak asymptotic giant branches. The picture is not so simple
though, as two clusters in Table \ref{Table-GC-CN-AGB-LitSearch}
actually have CN-strong AGBs. In addition to this, most clusters have
some stars of the opposite class on their AGBs -- the classifications
given in Table \ref{Table-GC-CN-AGB-LitSearch} (usually) refer to
strong majorities in each cluster, rather than totally homogeneous
populations.

\begin{table}
\begin{centering}
\begin{tabular}{|c|c|c|c|}
\hline 
Cluster &  No. AGB Stars &  AGB CN &  Reference\tabularnewline
\hline 
\hline 
M3 &  8 &  weak &  \citet{1981ApJS...47....1S}\tabularnewline
M4 &  11 &  weak &  \citet{1999AJ....118.1273I}\tabularnewline
M5 &  8 &  strong &  \citet{1993AJ....105..173S}\tabularnewline
M13 &  12 &  weak & \citet{1981ApJS...47....1S}\tabularnewline
M15 &  2 &  weak & \citet{2000JKAS...33..137L}\tabularnewline
M55 &  10 &  weak & \citet{1993AJ....106..142B}\tabularnewline
NGC 6752 &  12 &  weak & \citet{1993AJ....105..173S}\tabularnewline
47 Tuc &  14 &  strong &  \citet{1978AA....70..115M}\tabularnewline
\hline 
\end{tabular}
\par\end{centering}
\caption{Results of the literature search for CN abundances in GC AGB stars.
Note that `weak' or `strong' means that there is a very significant
majority of that class of star in each case.\label{Table-GC-CN-AGB-LitSearch}}
\end{table}

\begin{figure}
\begin{centering}
\includegraphics[width=0.7\columnwidth,keepaspectratio]{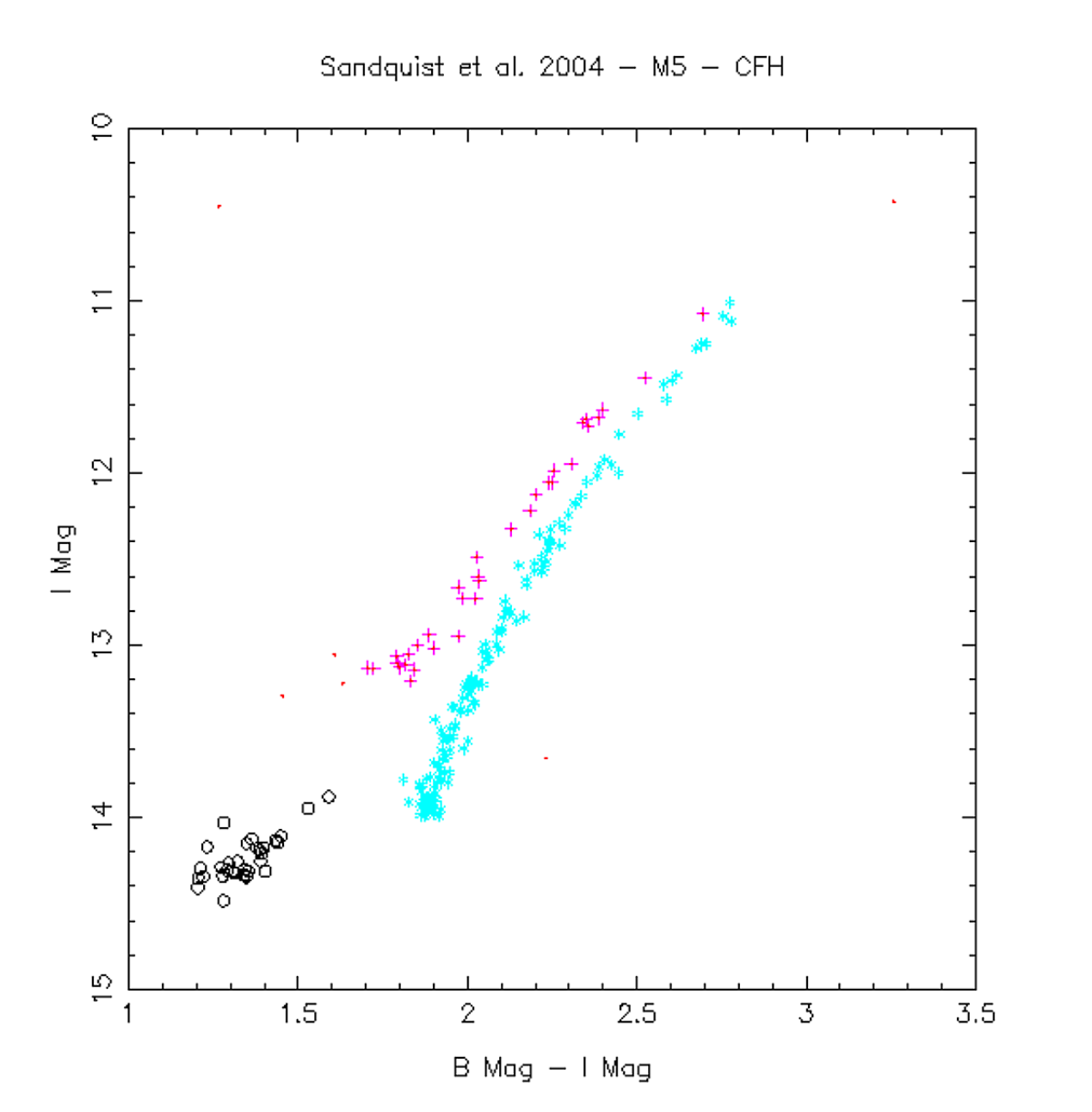}
\par\end{centering}
\caption{An example of AGB star identification in a CMD. Data is from \citet{2004ApJ...611..323S}.
Crosses, stars and open circles represent AGB, RGB and HB stars respectively.
\label{fig-Sandquist-M5-CMD}}
\end{figure}

A vital part in being able to observe significant numbers of AGB stars
is having photometry good enough to separate the AGB from the RGB.
Photometric observations have now reached such high accuracy that
it is becoming feasible to separate the AGB and RGB populations reliably.
In addition to this, wide-field multi-object spectroscopes are now
available. During our literature search we came across some very high-quality
photometric studies. For example, the study of M5 done by \citet{2004ApJ...611..323S}.
Their set of observations is complete out to 8-10 arc min. They also
tabulate all their stars according to evolutionary status -- and
find 105 AGB stars! This represents a sample size increase of \emph{one
order of magnitude}. In Figure \ref{fig-Sandquist-M5-CMD} we plot
some of the observations from \citet{2004ApJ...611..323S} in a colour-magnitude
diagram. A clear separation between the AGB and RGB can be seen. Further
to this we found colour magnitude diagrams for two more GC candidates
that have the required accuracy (and high AGB star counts). Thus our
current study involves three GCs, one of which appears to have a majority
of CN-strong stars on its AGB (M5) which makes it an important outlier
that may cause problems for some explanations of the (possible) phenomenon.

\begin{figure}
\begin{centering}
\includegraphics[width=0.65\columnwidth,keepaspectratio]{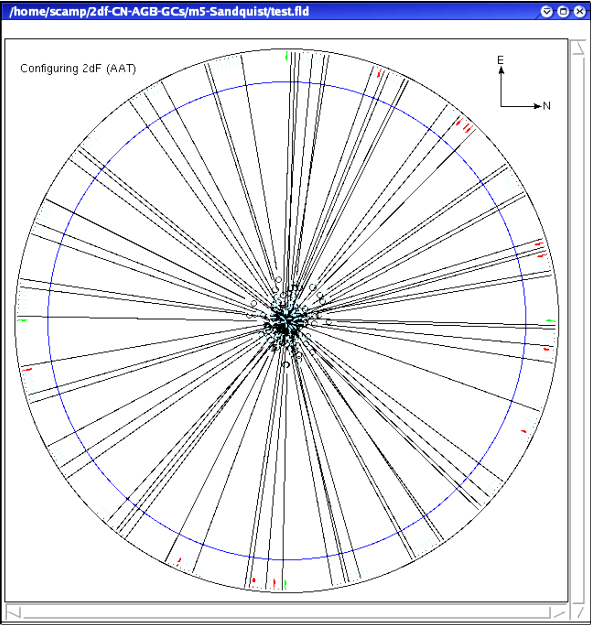}
\par\end{centering}
\caption{One of our 2dF plate setups for the M5 sample from \citet{2004ApJ...611..323S}.
Lines represent optical fibres whilst open circles represent the mini-prism
placements for the stars. Note that most of the objects are concentrated
in the centre. This is due to the fact that we need to sample the
dense regions of the cluster to get reasonable numbers of AGB stars,
and that the 2dF instrument has such a large field of view. \label{fig-2df-plateSetup}}
\end{figure}

Fortuitously, observations of CN bands require only low- to mid-resolution
spectroscopes. This combines well with the fact that a large sample
is required, which is easily achievable with multi-object spectroscopes
which also tend to have lower resolutions. We show a sample fibre
setup for the 2dF multi-object spectrograph on the Anglo Australian
Telescope in Figure \ref{fig-2df-plateSetup} (note that this instrument
has now been superceded, but similar instruments are available). Our
proposal also includes some (red) HB stars, as this may let us know
which stars reach the HB only, and which stars proceed to the second
giant branch. RGB stars will be used as control stars as they are
very well studied already -- and have similar temperatures and luminosities
to the AGB stars. Depending on the quality of the final data we will
also attempt to derive abundances for aluminium and CH (a proxy for
carbon). 

\subsection{Discussion}

Assuming for the purpose of discussion that the lack of CN-normal
AGB stars is real, then this is actually the opposite to what we would
expect based on observations at the tip of the RGB. These stars, which
are the precursors to the AGB stars (via the HB), actually tend to
become \emph{more} N-dominated due to `extra mixing' (the results
of extra mixing are routinely observed in Halo field RGB stars as
well GC RGB stars -- see eg. \citealt{2003ApJ...585L..45S}). Thus
we would predict an \emph{increase} in the number of CN-strong AGB
stars over the RGB mean -- rather than a decrease.

\citet{1981ApJ...244..205N} proposed two possible explanations to
explain the (apparent) lack of CN-strong stars on the AGB:
\begin{enumerate}
\item The two populations in NGC 6752 have different He abundances (they
suggest $\Delta Y\sim0.05$). This may have come about through a merger
of two proto-cluster clouds with differing chemical histories or through
successive generations of stars (ie. self-pollution). The He-rich
material would also be N-rich. The He-rich stars would then evolve
to populate the blue end of the HB -- and not ascend the AGB --
leaving only CN-normal stars to evolve to the AGB.
\item Mixing in about half of the RGB stars pollutes their surfaces (increasing
N) and also increases mass-loss rates, again leading to two separate
mass groups on the HB. As before, the CN-strong, low mass group does
not ascend the AGB.
\end{enumerate}
A constraint on the first explanation (for NGC 6752) is that about
\emph{half the mass of the cluster} must be polluted, as the number
of CN-strong and CN-normal stars is roughly equal. As Norris et al.
state, this is not a serious problem for the merger scenario, as the
merging clouds/proto-clusters may very well have had similar masses.
However, due to the constancy of Fe group elements, the chemical histories
of the two clouds/proto-clusters would have to have been identical
with respect to these heavy elements. This is more difficult to explain
since we require a differing chemical histories for the light elements.

The self-pollution scenario, whereby a second generation of stars
pollutes the cluster at an early epoch, also needs to satisfy these
two constraints. \citet{2004MNRAS.353..789F} have recently explored
this scenario. To maintain the heavy element abundances whilst increasing
N (and other elements) they assume that the cluster does not retain
the ejecta from the second generation supernovae but does retain the
material from the less energetic winds from intermediate mass AGB
stars. Qualitatively AGB stars have a perfect site for the hydrogen
burning needed to produce many of the abundance anomalies in GCs --
the bottom of the convective envelope (so-called `hot-bottom burning').
However, the theoretical study of \citet{2004MNRAS.353..789F} suggests
that there are actually serious problems for the scenario as the AGB
stars not only produce the N needed but also produce primary carbon
(which is dredged up to the surface). This also alters the sum of
C+N+O significantly which is observed to be (roughly) constant in
GCs. Constraints from other hydrogen burning products also cause this
model to fail.

In light of recent observations on the MS and SGBs of some clusters,
the second explanation by Norris et al. may require some clarification.
As N appears to have a preformation source (as evidenced my MS observations),
the extra mixing is not required (although it does still exist). However,
the general suggestion that the differing compositions may affect
mass-loss rates and lead to different mass populations on the HB may
be a valid one. 

An important point visible in Table \ref{Table-GC-CN-AGB-LitSearch}
is that it appears that there may be variation between the clusters
themselves -- some asymptotic giant branches seem to be CN-strong
as opposed to the majority which appear to be CN-normal. In addition,
the fact that most clusters have a \emph{mix} of CN-strong and CN-normal
AGB stars (although usually strongly dominated by one population),
rather than a homogeneous set, suggests that there may be a continuum
of CN-strong to CN-normal ratios. Theories such as those of \citet{1981ApJ...244..205N}
will have to account for these points also if the conclusions from
observations to date are proven correct. Of course, the low sample
sizes may be artificially complicating the issue.

If there really are substantial abundance differences between the
RGB and AGB then this may also reveal other clues to the GC abundance
anomaly problems (ie. those of the heavier p-capture products - see
\citealt{2000MmSAI..71..657S} for a discussion), and the second parameter
problem.

Our study seeks to clarify the understanding of abundance differences
between the various stages of evolution by very significantly increasing
the amount of information available about the asymptotic branch. 

Finally we note that the AGB stars in question are generally \emph{early}
AGB stars -- they are not thermally pulsing. However, this should
have no impact on the testing for abundance differences as they are
not expected to reduce their surface abundance of nitrogen. Indeed,
third dredge-up on top of preformation pollution and deep mixing would
make the issue even more complex.

\part{CONCLUSION}

\chapter{Summary of $Z=0$ and EMP Models\label{chapter-Discussion}}
\begin{quote}
``Science is facts; just as houses are made of stone, so is science
made of facts; but a pile of stones is not a house, and a collection
of facts is not necessarily science.'' 
\begin{flushright}
\vspace{-0.5cm}-- Jules Henri Poincar\'{e} (1854-1912) 
\par\end{flushright}
\end{quote}

\section{Peculiar Evolution}

A large part of this study has centred on the peculiar evolutionary
traits found in $Z=0$ and extremely metal-poor (EMP) stellar models.
In Section \ref{Section-DetailedEvoln-m0.85z0} we gave a detailed
report on the evolutionary properties of our 0.85 M$_{\odot},$ $Z=0$
model, whilst in Section \ref{section-m2z0-Structural} we did the
same for our 2.0 M$_{\odot},$ $Z=0$ model. We note that in Section
\vref{subsection-m0.85z0-EvolutionSummary} we gave a concise summary
(`executive summary') of the interesting evolution of our 0.85 M$_{\odot},$
$Z=0$ model that may be useful as a quick reference. Two of the key
peculiar evolutionary events found in these models are present in
many of our $Z=0$ and EMP stars. We refer to these events as the
`Dual Core Flash' (DCF) and `Dual Shell Flash' (DSF). The term dual
is used because these events are characterised by a twin peak in luminosity.
They occur during -- and are induced by -- the normally occurring
helium flashes (which also occur in higher-metallicity models). We
note that other authors have used (many) different names for these
events. The He flashes lead to a breaching of the H-He interface above
the He-burning region. This brings down protons into the hot He burning
regions, causing a secondary (but often stronger) hydrogen flash.
The difference between a DCF and a DSF is that the DCF happens during
the core He flash in low-mass models of $Z=0$ and extremely low metallicity,
whilst the DSF happens during the first few thermal pulses (i.e. He
shell flashes) in intermediate-mass models of $Z=0$ and extremely
low metallicity. As just mentioned both events involve the same phenomenon
-- a He flash inducing a mixing between the H-rich zones and the
He-burning zones. Thus we have given a general term to the H-flash
portion of these events -- proton ingestion episodes (PIEs). We note
that, whilst protons are dredged down and lead to the secondary H-flash,
He burning products are also mixed upwards. This has significant consequences
for the $Z=0$ models as it leads to an ignition of the CNO cycles
where there was only p-p chain burning occurring (due to a lack of
CNO catalysts). However the main consequence of both the upward and
downward mixing is that polluted material is then located above the
He burning region, and this is dredged up to pollute the surface at
a later stage (the pollution arising from this is summarised in the
next section).

Although the DF events appear to be ubiquitous in the stellar models
of all studies there is some variation in the details of exactly which
masses and metallicities they occur at, and also how much pollution
arises from them. This is most likely due to the different input physics
and numerical schemes used between the studies. A key factor is the
degree of overshoot used. We used none in our models but the inclusion
of overshoot would most likely increase the frequency of occurrence
of these events, widening the metallicity range in which they occur.
If the increase in the proportion of CEMP stars at low {[}Fe/H{]}
is due to these DF events (as conjectured below) then the observations
may be able to constrain the models. Indeed, the upper metallicity
limit may already be visible in the observations. We shall investigate
this in the near future.

\section{Chemical Pollution\label{section-HaloStarModelsPollutionSummary}}

\subsection{Mass-Metallicity Pollution Diagram}

\begin{figure}
\begin{centering}
\includegraphics[width=0.9\columnwidth,keepaspectratio]{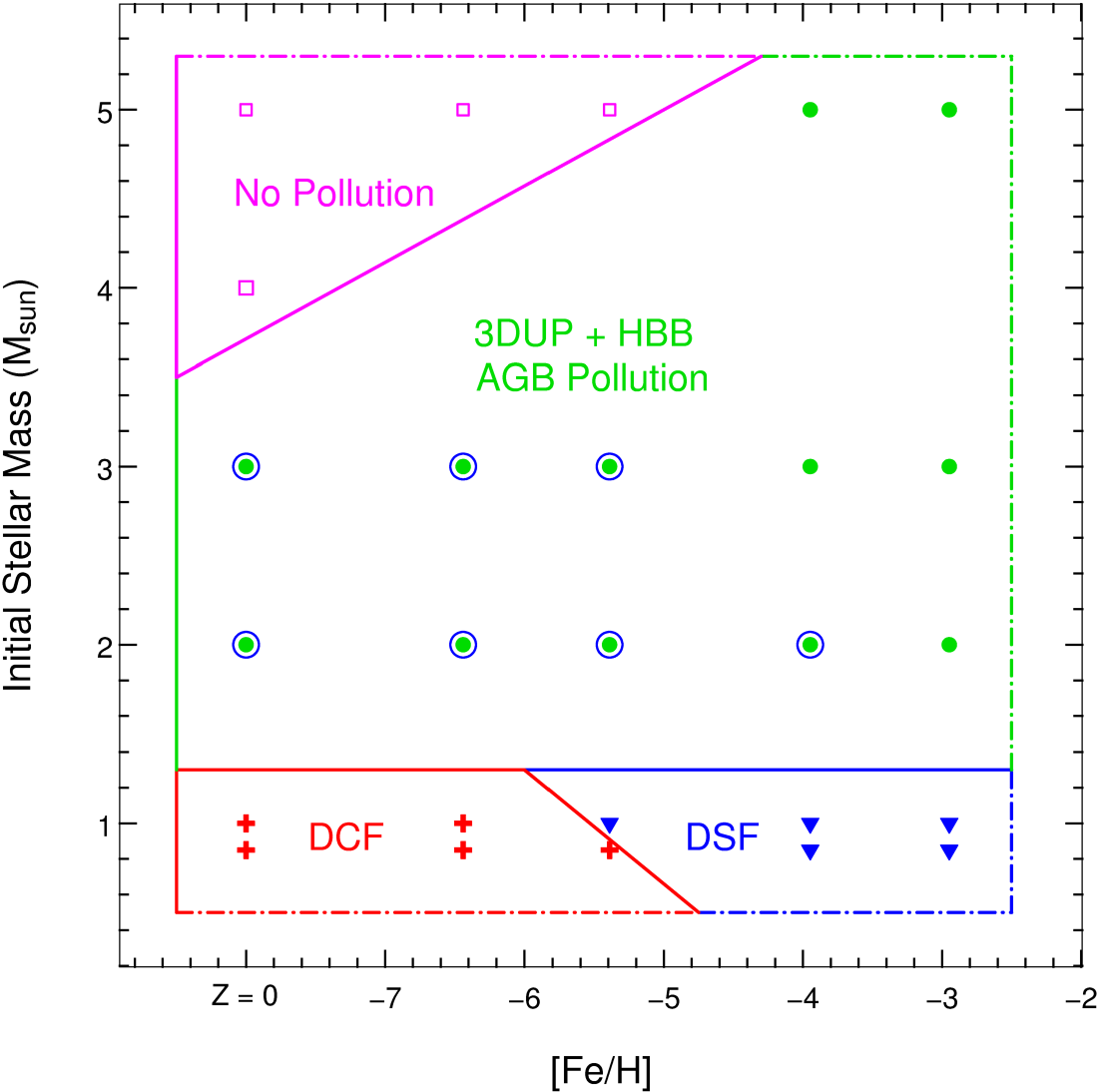}
\par\end{centering}
\caption{Mass-metallicity diagram summarising the dominant sources of pollution
in the yields. Each symbol represents one of our detailed stellar
models. Note that we have only calculated detailed nucleosynthesis
for the models in the mass range $0.85\rightarrow3.0$ M$_{\odot}$,
but we have calculated a sufficient amount of the AGB evolution of
the 4 and 5 M$_{\odot}$ models to be quite certain of the nature
of the yields (type 1.5 supernovae events withstanding). We identify
four categories into which we group the chemical outputs from our
models: $DCF$ (pollution primarily from the dual core flash event
= Group 1), $DSF$ (primarily from the dual shell flash event = Group
2), $AGB$ (from third dredge-up and hot-bottom burning on the AGB
= Group 3) and $No\,\,Pollution$ (no major events pollute the AGB
envelope, thus no enhancements in the yields = Group 4). The open
circles (blue) around the filled circles (green) indicate intermediate
mass models that experienced DSFs. Pollution from 3DUP (and HBB) easily
dominates the pollution from the DSF events at IM mass so the yields
of these models fall into the AGB group. The dash-dotted lines represent
the fact that these categorisations may extend out to higher metallicities
and/or higher/lower masses. As we have included our $Z=0$ models
(artificially at $\textrm{[Fe/H]}=-8$, as marked on the x-axis) the
low metallicity boundary is known, and therefore represented by solid
lines. We note, for the interested reader, that \citet{2000ApJ...529L..25F}
also provide a mass-metallicity sketch for their low-metallicity study.
\label{fig-PollutionSummary}}
\end{figure}

\citet{2000ApJ...529L..25F} appear to have been the first to provide
a clear overview of the expected polluting effects from low- and intermediate-mass
very metal-deficient models. In their Figure 2 they sketch the various
regimes of pollution episodes in the mass-metallicity plane. The current
study is the first to evolve stars of such low metallicity through
to the end of the AGB, and thus to produce yields from detailed calculations.
In Section \ref{section-HaloStars-Nucleo} we categorised our models
into three groups, based on the source of pollution for their AGB
envelopes. Using these three groups we now also provide an overview
of expected pollution in the mass-metallicity plane, which we summarise
graphically in Figure \ref{fig-PollutionSummary}. Note that we describe
all our categories in the caption of this figure. In the figure we
have added a fourth group, based on some higher-mass models for which
we have calculated a large portion of the structural evolution (hundreds
of pulses on the AGB). Although we have not fully completed these
calculations we are confident that the yields from these models will
not be affected by any significant pollution events (unlike the lower-mass
cases). These models, which have the same metallicity range as our
other models, but are of mass 5 M$_{\odot}$ (there is also one of
mass 4 M$_{\odot}$ with $Z=0$), experience no third dredge-up at
all. As they do not experience the DSF event either their yields will
be unpolluted -- i.e. practically the same as their initial compositions.
The only pollution event that slightly alters the yield compositions
is 2DUP, but this is (relatively) minor. We note however that the
core masses in these models are all approaching the Chandrasekhar
mass and thus they may end up as type I$\frac{1}{2}$ supernovae,
as described by \citet{1983ARAA..21..271I}. This would then cause
the yields to be polluted with SN products and our `no pollution'
(Group 4) categorisation would need to be revised. The evolution to
supernovae type I$\frac{1}{2}$ is however strongly dependent on the
mass loss rates used. We shall pursue this interesting avenue of low
metallicity stellar evolution as a future study (we note that some
recent work has been done on this topic, see eg. \citealt{2006astroph12267G}).

Comparing our `pollution diagram' (Figure \ref{fig-PollutionSummary})
with the diagram of \citet{2000ApJ...529L..25F} (Figure 2 of that
paper) we note that our results are very similar. They have similar
boundaries for the different dominant polluting events, and they also
delineate four groups that match with ours (their extra group, Case
IV, is outside our metallicity range). The main point of difference
is that our boundary for the AGB-DSF models (green filled circles
with blue haloes in Figure \ref{fig-PollutionSummary}) is at a lower
metallicity than theirs. Our diagram also shows a mass dependency
in addition to the metallicity dependency for the pollution events.
Finally we note that one dimension is not elucidated by these types
of pollution diagrams -- the degree of dilution that some of the
yields suffer through unpolluted RGB mass-loss. Naturally this effect
is only relevant at low masses, but we have found it to be very significant
in some of the models (see Section \ref{subsec-YieldUncert-UnpollutedRGBmloss}).
This adds some (further) uncertainty to the pollution diagram. The
RGB mass-loss dilution is however not important in the case of a binary
mass-transfer event just after (or during) a DCF or DSF event. 

\subsection{Extra Channels for Producing CEMPs\label{SubSec-CEMPchannels-Conclusion}}

As noted above, the dual flash (DF) events have strong consequences
for the surface composition of many of our models. The surfaces of
the models are polluted after the flashes recede -- the convective
envelope extends inwards in mass and dredges up polluted material
that was left behind just above the He shell. This material has been
through He burning \emph{and} H burning. Thus it is dominated by the
three main products of these types of burning -- carbon, oxygen and
nitrogen (in terms of metals). The magnitude of the surface pollution
is usually very large: $Z=0$ and EMP models often end up with super-solar
metallicity in terms of $Z$ (they do of course remain EMP in terms
of {[}Fe/H{]}).

\begin{figure}
\begin{centering}
\includegraphics[width=1\columnwidth,keepaspectratio]{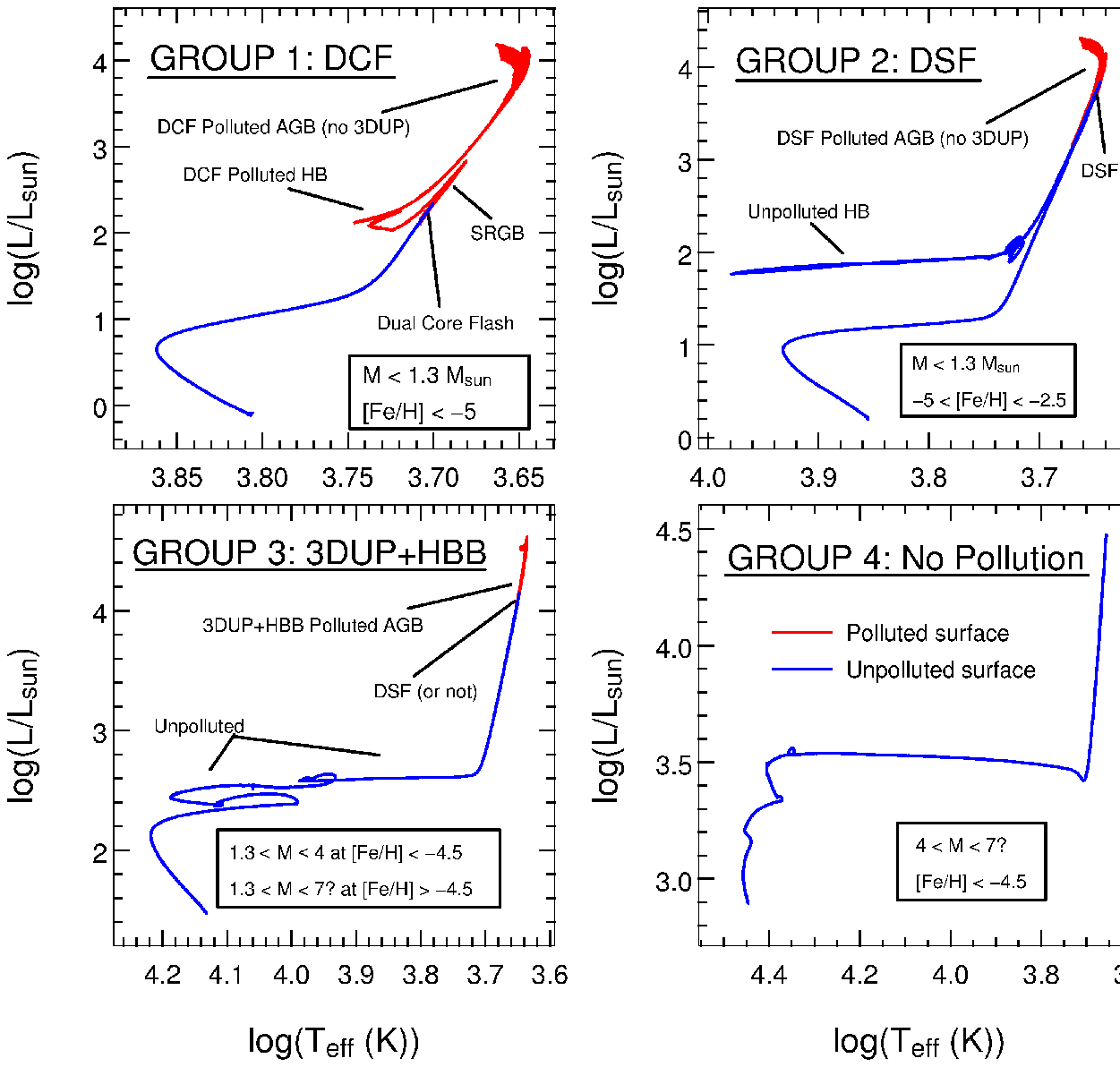}
\par\end{centering}
\caption{An evolutionary perspective on the four pollution groups. Displayed
in each HR diagram is a representative example from our grid of models
for each pollution group (see Figure \ref{fig-PollutionSummary} for
the definitions of the groups). Red lines indicate phases of the evolution
in which the surface is strongly polluted with CNO nuclides (from
the DCF, DSF or 3DUP events). Blue lines indicate that the surface
still retains the initial metal-poor composition. Evolutionary stages
and pollution sources are marked, as are the mass and metallicity
ranges of each group. Question marks indicate unknown upper boundaries
(due to the limited mass range of the current study). Two key features
are (1) the DCF group has a polluted HB and (2) both the DCF and DSF
groups, which are of low mass, have polluted surfaces during the AGB
despite the lack of 3DUP. These features mean that our models predict
a greater proportion of C-rich stars at extremely low metallicity,
as these events do not occur at higher metallicities. \label{fig-pollutionSummary-HRD}}
\end{figure}

In Figure \ref{fig-pollutionSummary-HRD} we give an overview of the
four pollution groups discussed in the previous subsection, in terms
of evolutionary status. We do this by indicating in HR diagrams at
which stages of evolution the models have polluted surfaces. The most
notable result visible from this perspective is that the low mass
models (M $\lesssim1.3$ M$_{\odot}$) have their AGB surfaces polluted
by the DF events only. Our models do not show 3DUP at these masses
and metallicities so this remains the dominant source of pollution
throughout the AGB evolution. If it is correct that 3DUP does not
occur in stars of this mass and metallicity then \textbf{\emph{these
DF events represent a serious increase in C production at extremely
low metallicities}}, since models without the DF events would release
no polluted matter (and DFs do not occur at higher metallicities).
\textbf{\emph{Thus the model results predict that the proportion of
stars that are C-enhanced should be higher at extremely low metallicity.}}
Moreover, our models predict that \textbf{\emph{the proportion of
CEMPs should continue to increase at lower and lower metallicities.}}
This is due to the fact the mass range in which the DF events occur
increases with decreasing metallicity \emph{--} i.e. the carbon-enriching
DF events become more common with decreasing metallicity. The increase
in the proportion of CEMPs would be compounded through the binary
star population (via mass transfer events) since some of the primary
stars would have undergone DFs. Furthermore, as these events are quite
violent, one might expect that enhanced mass-loss would occur during
these phases -- possibly even envelope ejection. This is something
that needs to be explored properly with fluid dynamics simulations.
If this were the case then this would \emph{again} increase the proportion
of CEMPs in EMP populations, since the secondary components in binaries
would be preferentially polluted by material from these events. 

Another salient result visible in Figure \ref{fig-pollutionSummary-HRD}
is that the models which experience the DCF at the top of the RGB
have polluted surfaces for a significant portion of their lifetimes.
This is because their surfaces are already polluted during the HB
phase (and secondary RGB phase, but this is quite short-lived like
the AGB). Interestingly our models predict that, at ultra low metallicities
($\textrm{[Fe/H]}\lesssim-5$), \textbf{\emph{all low-mass stars in
the HB phase should have polluted surfaces}}. In terms of the binary
mass transfer scenario this is of particular importance. Since these
stars are polluted for a longer portion of their evolution it is more
likely that companions would accrete polluted material. Thus we have
\emph{another} CEMP formation channel.

It will be very interesting to see if future observations of ultra-metal-poor
halo stars show such high proportions of CEMPs, as forecast by the
models. 

Finally, as mentioned above, our higher mass models ($\textrm{M}\gtrsim4$
M$_{\odot}$) of extremely low metallicity do not undergo DF or 3DUP
events, so their surfaces remain unpolluted. We note that this should
have little (direct) observational consequence in regards to halo
observations as these stars have short lifetimes. Interestingly it
also means that secondary stars in binary systems with these stars
would not be (significantly) polluted upon mass transfer. Thus we
have one \emph{less} way of creating a CEMP star at extremely low
metallicities (since higher metallicity stars in this mass range \emph{do}
experience 3DUP). 

\subsubsection*{Simultaneous Enhancement of C and N}

As mentioned above our models have very C-enhanced yields due to the
DCF and DSF events. Another way to produce C enrichment is via 3DUP.
Most studies find that 3DUP efficiency increases at low metallicity
(eg. \citealt{2002PASA...19..515K}; \citealt{2005MNRAS.356L...1S}).
\citet{2005MNRAS.356L...1S} even find 3DUP to occur in 1 M$_{\odot}$
models of LMC metallicity. If 3DUP were to occur in low-mass ($M\lesssim1$
M$_{\odot}$) EMP stars then this would be another avenue for CEMP
pollution (we note however that, like our models, the $\textrm{[Fe/H]}=-2.7,$
1 M$_{\odot}$ model of \citet{2004ApJ...602..377I} does \emph{not}
experience 3DUP). This would again lead to an increase in the proportion
of C-rich stars at extremely low metallicities since models of higher
metallicity generally do not experience 3DUP. Thus it appears that
there may be two possible channels for C production in low mass EMP
models. One way to distinguish between these two channels is through
their different chemical signatures. Since low mass stars that have
3DUP are not predicted to experience HBB their surface abundance of
N remains low whilst C is periodically enhanced. By contrast our \textbf{\emph{DCF
and DSF models show C and N enhanced at the same time}}\textbf{.}
Interestingly this dual enhancement appears to be required by the
observations, which we discussed in Section \vref{Section-HaloStarModels-CompareObs}. 

Another interesting possibility discussed in Section \ref{Section-HaloStarModels-CompareObs}
is that DFs \emph{and} 3DUP could occur in low mass EMP stars (which
may be possible in our models by using overshoot). In this case the
models \textbf{\emph{would again be C and N rich}}. This is because
the N from the DSF and DCF events has no way of being depleted in
the H-rich envelopes at such low masses. Since 3DUP appears to enrich
the yields with C to a similar degree as the DF events, the overall
C enhancement would probably not increase significantly in this scenario
over our present DF-only results. The $^{12}$C/$^{13}$C ratios may
however increase but we are uncertain how significant this would be.
Clearly this scenario requires further modelling -- the combination
of DFs with 3DUP may well provide a better match to the observations.
We shall pursue this avenue in future work.

\subsubsection*{CEMP Abundance Patterns}

Finally we note that in Section \vref{subsection-Summary-HaloStarObservationCompare}
we gave a summary of all the comparisons we made between our models
and the EMP halo star observations. It was found there that many features
of the CEMPs could be explained by our models that experience the
DCF and DSF episodes (some of which have just been discussed above).
However our comparison with the extended abundance \emph{patterns}
of a few of the most metal-poor objects known (which are, interestingly,
all CEMPs) showed that \textbf{\emph{our models fail to fully explain
these abundance patterns}}. There were however some abundance patterns
amongst the (small sample of) CEMPs that showed strong similarities
with our models. We also noted that there are uncertainties in our
models that may account for the remaining differences in some cases.
A useful continuation for this line of study would be to collate all
the observed abundance patterns of CEMPs to see if there are any consistent
groups of patterns. We shall pursue this in our future work as well.

\chapter{Concluding Remarks\label{chapter-ConcludingRemarks}}
\begin{quote}
``Doing a PhD is a strange state of being.\textquotedbl{}
\begin{flushright}
\vspace{-0.5cm}-- Lisa J. Pinter
\par\end{flushright}
\end{quote}

\section{Reflecting on the Journey}

In the current study we have investigated a wide range of stellar
modelling at low metallicities. We began by modifying the structural
evolution code (SEV code) to enable it to compute the evolution of
$Z=0$ and extremely metal-poor models. The main change we made to
the code was to alter the mixing paradigm. We removed the assumption
of instantaneous mixing and replaced it with a time-dependent formalism
(Chapter \ref{sevmods}). The formalism chosen was that of `diffusive
mixing'. This change was necessary due to the peculiar evolutionary
events that these model stars experience. During the `Dual Flash'
events the evolution proceeds so rapidly that the evolutionary timesteps
become comparable to the convective turnover timescales -- so the
assumption of instantaneous mixing no longer holds. Other changes
were also made to the SEV code to facilitate the modelling of these
types of stars. These included increasing the temperature range in
which the SEV code could operate (eg. via adding high-temperature
opacity tables) and allowing arbitrary compositions to be used (eg.
by removing scaled-solar assumptions). We note that these changes
have also also brought the code closer to being able to model other
types of stars as well -- such as `Super AGB' stars (eg. \citealt{2006MmSAI..77..828D}).

With the SEV code armed with the new capabilities we embarked on a
few different investigations within the fascinating world of low-metallicity
stellar evolution.

The first challenge (chronologically) was that given by the Galactic
globular cluster abundance anomalies mystery (Chapter \ref{GC-chap}).
This notorious problem has been around for at least 30 years -- and
has defied explanation to the current day. In our early work on the
subject we decided to quantitatively investigate the oft-cited qualitative
explanation of the abundance anomalies -- that the source of pollution
is AGB stars. We calculated a small grid of models (Chapter \ref{GC-chap})
of which the initial composition was taken from a chemical evolution
model by \citet{2004MNRAS.353..789F}. The yields from our (second
generation) stellar models were then fed back into the chemical evolution
model. We found that the qualitative theory was not supported by our
quantitative study. The reasons for this turned out to be fundamental
problems, such as the strong increases in C+N+O and magnesium, which
are both due to third dredge-up (3DUP). Only very significant changes
in the AGB models could remedy the situation. One such change is the
alteration of the convective mixing model (\citealt{2005ApJ...635L.149V}).
As a diagnostic tool we then calculated some models without 3DUP.
The inhibition of 3DUP had been made possible due to the inclusion
of the Ledoux criterion for convection in our time-dependent mixing
scheme. These models (Section \ref{Section-GC-No3DUP}) proved to
be a much better match to the observations, but still fell short in
some comparisons. We note that these models have recently been utilised
in a chemical evolution model that uses a different paradigm for the
source of the polluting gas that formed the abundance anomalies. In
this case the gas is theorised to come from an external source, namely
the original host of the GC -- a low-mass dwarf galaxy (\citealt{2007astro.ph..2289B}).
To complete our foray into the GC problem we began an investigation
into an interesting GC observation first noted by \citet{1981ApJ...244..205N}
-- that there appears to be differences in the numbers of cyanogen-strong
stars on the AGB as compared to the RGB in NGC 6752. Our literature
search (Section \ref{Section-GC-CN-AGB-ObsProject}) appeared to support
this conjecture for most GCs, but the samples were all very small
(including that for NGC 6752). We therefore put together an observing
proposal -- that should increase the sample sizes by an order of
magnitude -- to investigate the situation. This work is still underway.

The next challenge was to return to the originally proposed topic
of the thesis -- modelling $Z=0$ stars. With the time-dependent
mixing routine in place this was now possible. These models did however
present some difficult problems due to the severe nature of the dual
flash events. Code convergence was often an issue. Much time and effort
was spent in getting these models to work. Nevertheless it was a worthwhile
experience as it forced the author to \emph{really} get to know the
SEV code. It is often the case that we learn the most when things
do \emph{not} go smoothly. When we did finally get the code to handle
the $Z=0$ stars, we were inspired, in a momentary lapse of reason,
to explore the whole parameter space of low- and intermediate-mass
extremely metal-poor stellar evolution. Spurred on by the fact that
there were some very interesting observations at these extreme metallicities,
and by the fact that no one had quantitatively explored the pollution
arising from these models through full computation of AGB structure,
nucleosynthesis and yield calculations, we embarked on a large modelling
crusade. The grid of models we calculated amounted to 30 stars, of
which about 20 were successfully completed and presented in Chapter
\ref{Chapter-HaloStarModels}. Here we need to acknowledge the support
of the Australian Partnership of Advanced Computing (APAC) who were
very helpful, particularly in increasing our quota of CPU hours, of
which we used $\sim15000$ for this grid of models (see Figures \ref{fig-APAC-Usage}
and \ref{fig-apacmodels-ShellScreenGrab} on page \pageref{fig-APAC-Usage}).
These models, although certainly full of uncertainties, represent
the most comprehensive investigation into EMP evolution and nucleosynthesis
to date. In comparing our results to observations we found some broad
agreements but note that some of the details do not match well (see
Section \ref{Section-HaloStarModels-CompareObs}). Although naturally
also full of uncertainties, we have provided the yields from all of
our models in the appendices. These will be of use to chemical evolution
studies of the early Universe. Indeed, they are the only yields at
these masses and metallicities of which the author is aware.

\section{Future Work}

Throughout the current study we have noted many areas that require
more work. We list some of them here, with references to the parts
of the study in which we noted them:
\begin{itemize}
\item Main sequence lifetimes of low-mass $Z=0$ models vary between studies
(Section \vref{subsec-m0.85z0-ComparePrevStudies}).
\item Core helium burning in intermediate-mass $Z=0$ models vary widely
between studies (Section \vref{subsec-m2z0-ComparePrevStudies}). 
\item Investigate the effects of mass loss on the yields of EMP and $Z=0$
models (Section \vref{subsection-HaloStarStruct-MassLoss}).
\item Possible s-processing during the Dual Core Flash of low-mass $Z=0$
and EMP models (Section \vref{subsec-m0.85z0-DCF-sProcess}).
\item Possible s-processing during the Dual Shell Flash events (\vref{SubSec-DSFs-PotentialS-process}).
\item Determine the reason that the 1 M$_{\odot}$ and 0.85 M$_{\odot},$
$Z=0$ models have a different abundance pattern amongst the heavier
proton-capture elements (Section \vref{section-Yields-NS-all-Z0}).
\item Investigate the potential for supernova type 1.5 events in the 4 and
5 M$_{\odot}$ models (\vref{section-HaloStarModelsPollutionSummary}).
\item Investigate the effect of the combination of 3DUP + dual flash polluting
episodes in low-mass models via the inclusion of overshoot (Section
\vref{SubSection-HaloStars-Obs-Compare-IndividStars}).
\item Make comparisons with observations including the time evolution of
the abundances (Section \vref{SubSection-HaloStars-Obs-Compare-IndividStars}).
\item Collate abundance patterns of CEMPs to determine whether there are
any consistent pollution patterns amongst them (Section \vref{subsection-Summary-HaloStarObservationCompare}).
\item Try (more) alternative rates in relation to the Na-O and Mg-Al anticorrelations
in GCs (Section \vref{section-GC-Models-Standard}).
\item Investigate how the Mg isotopic ratios could be so different between
the models and the observations of GCs (Section \vref{Section-GC-No3DUP}).
\item Investigate the implications of the surface opacity uncertainty described
in Section \vref{section-LowTOpacUncertainties}.
\end{itemize}
It can be seen from this (non-exhaustive) list that many uncertainties
and unknowns remain in this field of low- and intermediate-mass metal-poor
stellar evolution -- and thus that there is much work that still
needs to be done. We shall certainly not be lost for things to do!

\begin{spacing}{1.2}

\part{APPENDICES}

\appendix

\chapter{Yields: Metal-Deficient and $Z=0$ Models\label{APPX-YieldsLowZandZ0}}

This appendix holds all the yields for all the EMP and $Z=0$ models.
It is divided into three sections. 

The first section gives the yields in elemental form (i.e. the sum
of all nuclides of each element) relative to solar, against initial
stellar mass. One plot is given for each metallicity (5 plots in total). 

The second section gives the yields in elemental form relative to
solar, against the initial {[}Fe/H{]} of the models. This gives an
idea of how the yields vary with metallicity. They are presented in
graphical form (13 plots). 

The third section gives the yields for every species in the nuclear
network (except neutrons), in mass fraction. These are presented in
tabular form (5 tables). 

\newpage \begin{center}

\vskip 2in~\vskip 1in~

\section{Plots: Elemental Yields Versus Initial Mass\label{Appx-MetalPoorAndZ0Yields-ElemVsMass}}

\newpage

\begin{figure}[H]
\begin{centering}
\vspace{2cm}\includegraphics[width=0.9\columnwidth,keepaspectratio]{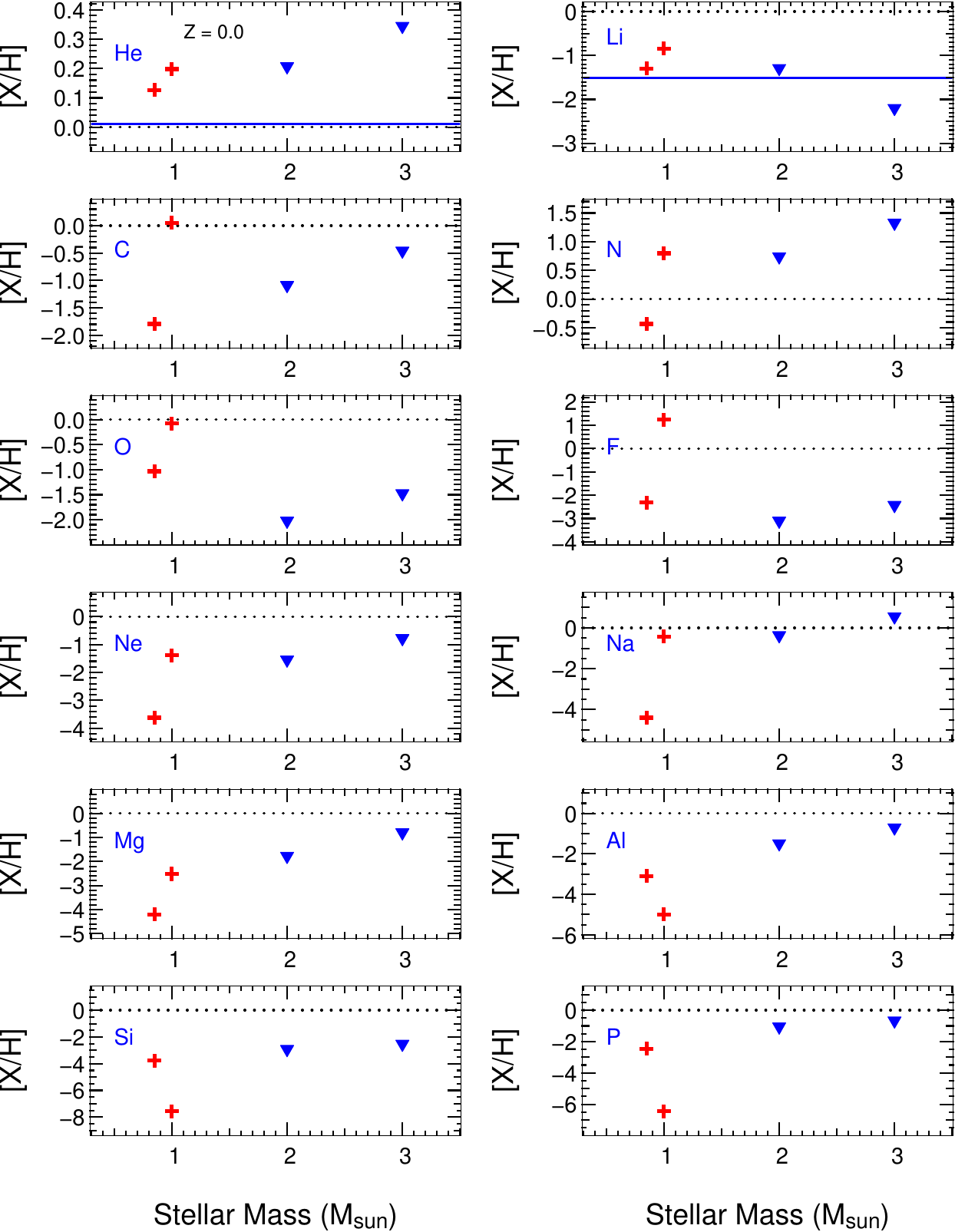}
\par\end{centering}
\caption{Selected elemental yields for all the $Z=0$ models, given as relative
to solar against initial stellar mass. Red crosses indicate models
that experienced a dual core flash (at the top of the RGB), whilst
blue triangles indicate models that experienced a dual shell flash
(at the beginning of the AGB). Solid horizontal lines (blue) indicate
initial abundances, but are often not visible due to their relative
paucity. The horizontal dotted (black) lines at $\textrm{[X/H]}=0$
indicate the solar abundances (Solar abundances are from \citealt{2003ApJ...591.1220L}).
\label{fig-Appx-z0-Yields1}}
\end{figure}

\begin{figure}[H]
\begin{centering}
\vspace{2.5cm}\includegraphics[width=0.9\columnwidth,keepaspectratio]{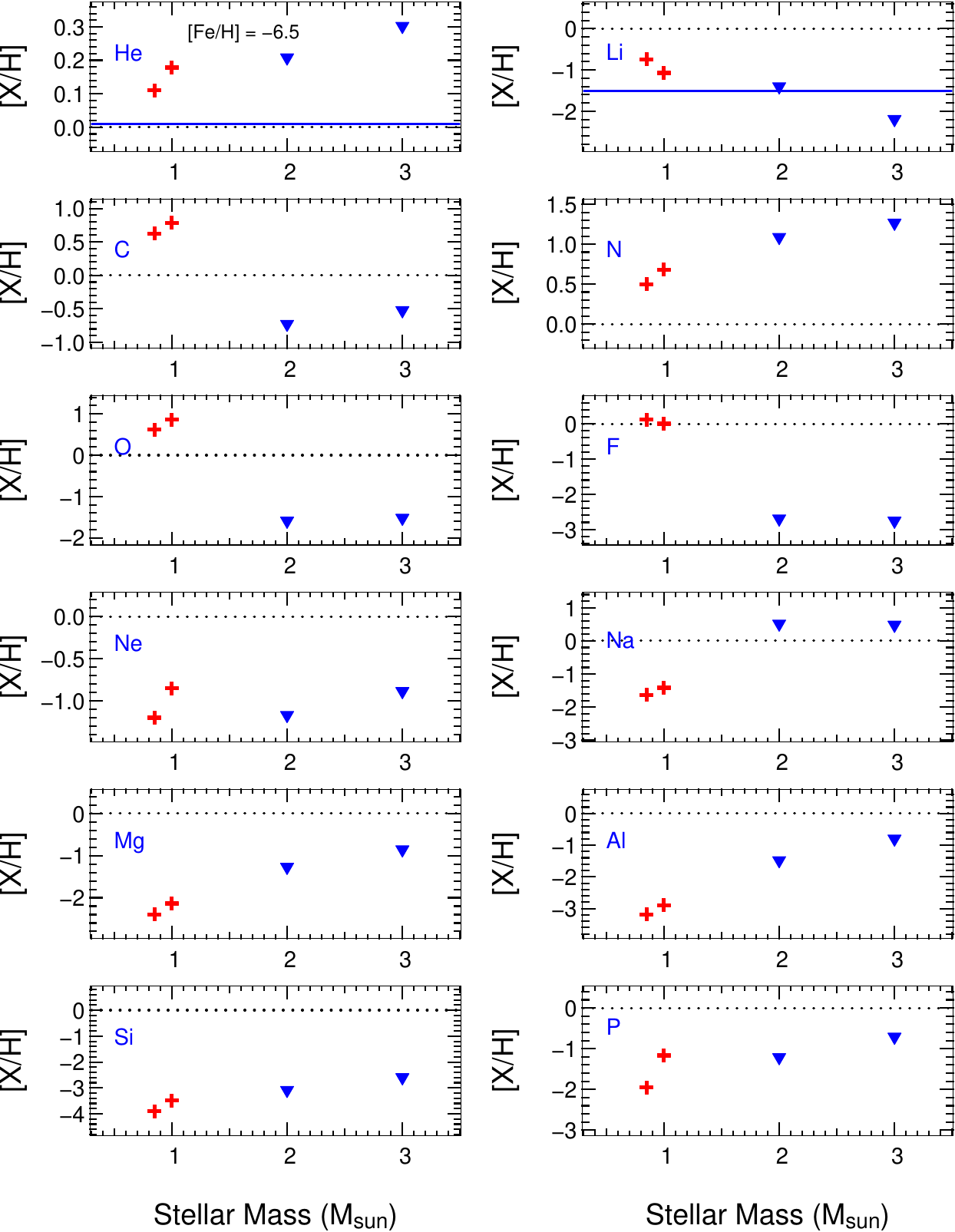}
\par\end{centering}
\caption{Same as Figure \ref{fig-Appx-z0-Yields1}, but for the $\textrm{[Fe/H]}=-6.5$
models.}
\end{figure}

\begin{figure}[H]
\begin{centering}
\vspace{2.5cm}\includegraphics[width=0.9\columnwidth,keepaspectratio]{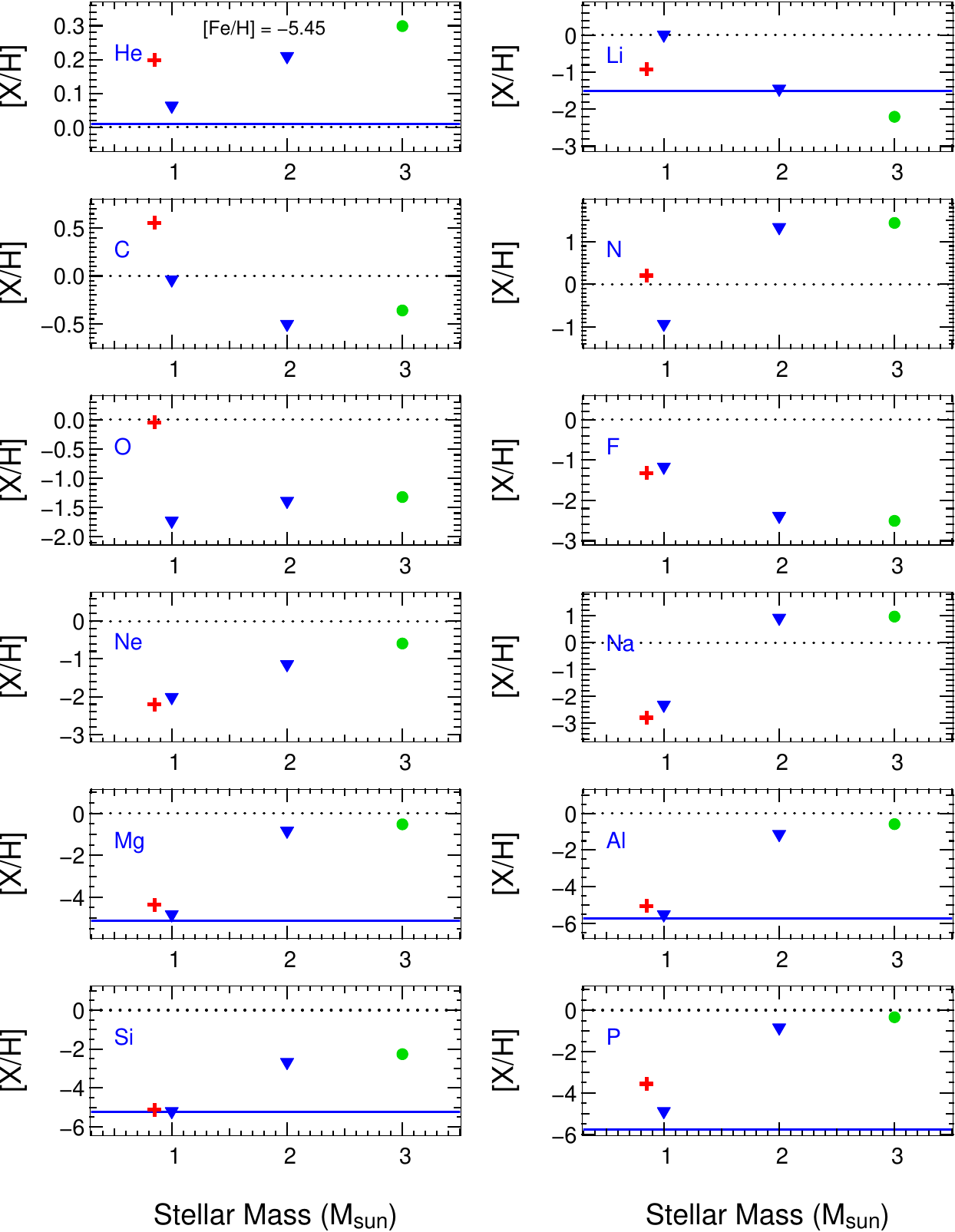}
\par\end{centering}
\caption{Same as Figure \ref{fig-Appx-z0-Yields1}, but for the $\textrm{[Fe/H]}=-5.45$
models. Green dots indicate that the model did not experience a DCF
or a DSF.}
\end{figure}

\begin{figure}
\begin{centering}
\includegraphics[width=0.9\columnwidth,keepaspectratio]{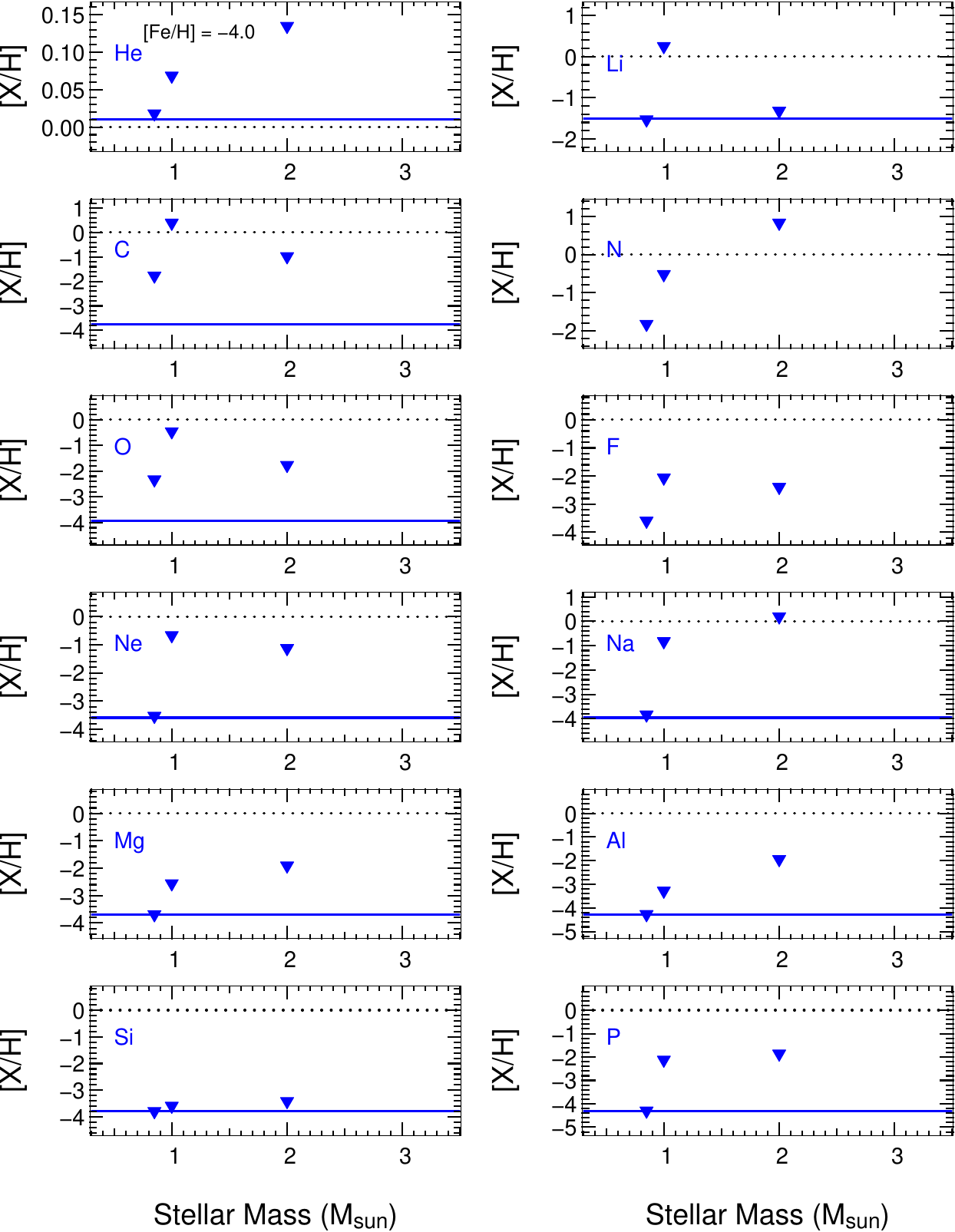}
\par\end{centering}
\caption{Same as Figure \ref{fig-Appx-z0-Yields1}, but for the $\textrm{[Fe/H]}=-4.0$
models.}
\end{figure}

\begin{figure}
\begin{centering}
\includegraphics[width=0.9\columnwidth,keepaspectratio]{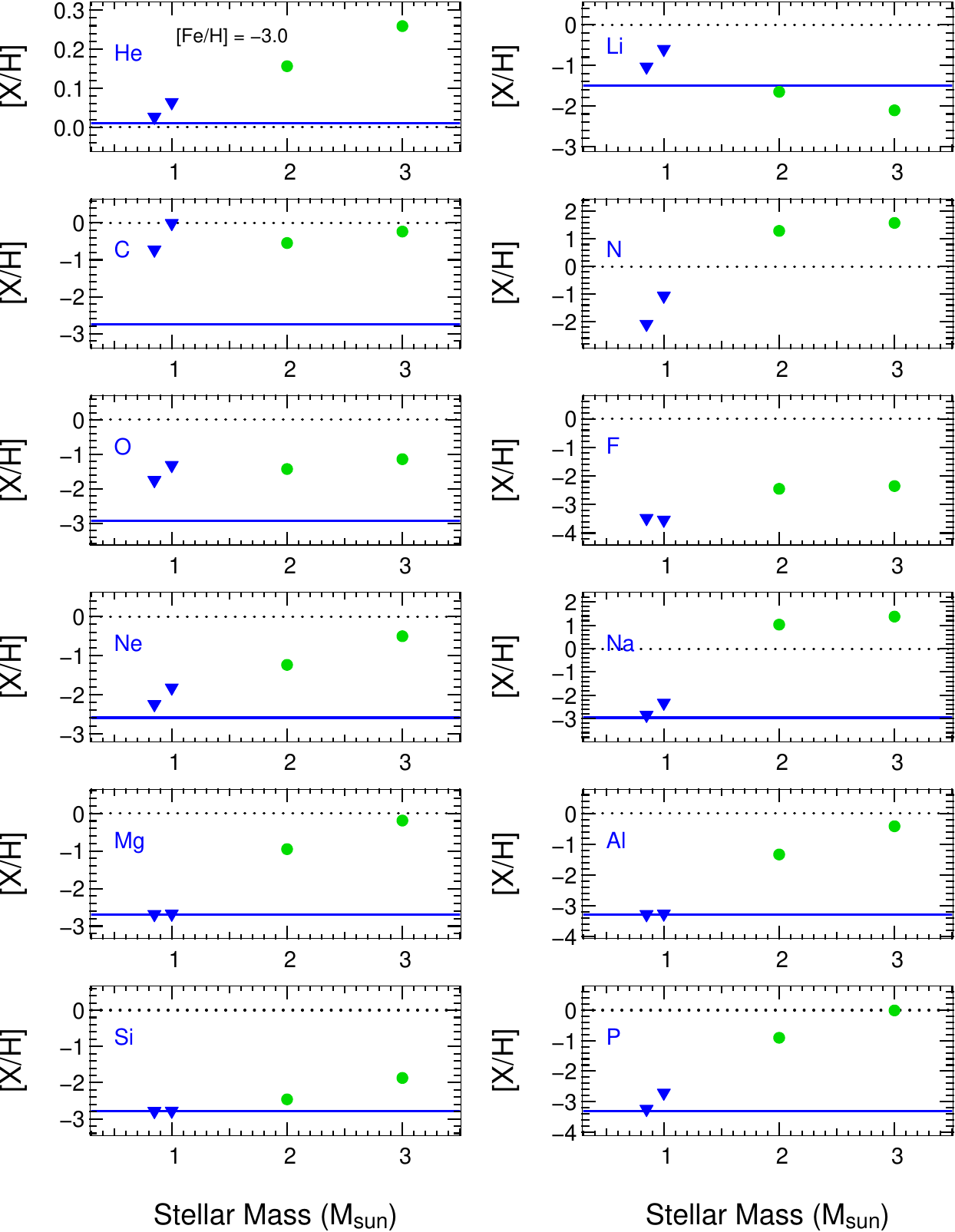}
\par\end{centering}
\caption{Same as Figure \ref{fig-Appx-z0-Yields1}, but for the $\textrm{[Fe/H]}=-3.0$
models. Green dots indicate that the model did not experience a DCF
or a DSF.}
\end{figure}

\newpage

\vskip 2in~\vskip 1in~

\section{Plots: Elemental Yields Versus {[}Fe/H{]}\label{secAppx:Elemental-Yields-VersusFeH}}

\end{center} \newpage

\begin{figure}[H]
\begin{centering}
\vspace{5cm}\includegraphics[width=0.9\columnwidth,keepaspectratio]{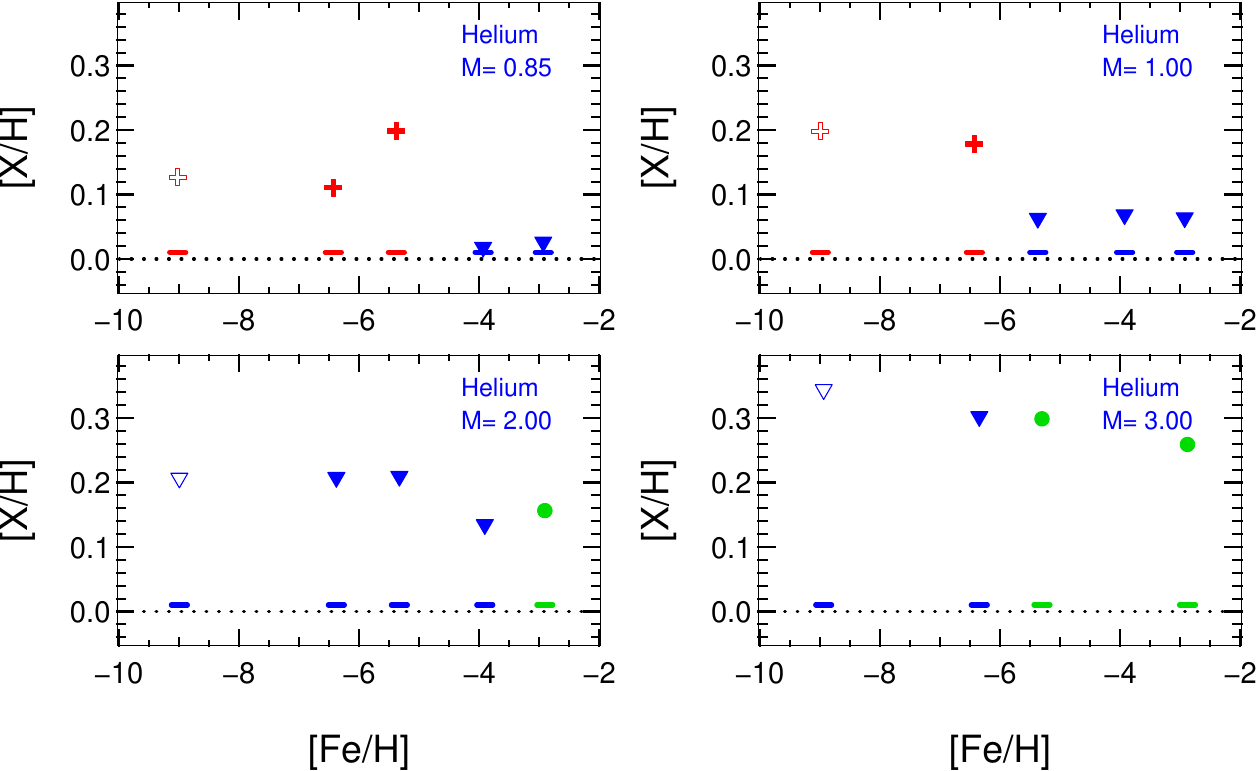}
\par\end{centering}
\caption{Helium yields versus initial {[}Fe/H{]} for all the models in the
grid, grouped by initial mass. The $Z=0$ models have also been included
for comparison (they have been given an artificial Fe abundance corresponding
to $\textrm{[Fe/H]}=-9$ for display purposes). Red crosses indicate
models that experienced a dual core flash (at the top of the RGB),
blue triangles models that experienced a dual shell flash (at the
beginning of the AGB) and green dots indicate models that experienced
neither. Short horizontal lines indicate the initial abundance for
each model (which are well below the scale of the bottom panel). Note
that within each group of four plots the vertical axes are identical
to allow direct comparison. \label{fig-Appx-HeliumVsFeH}}
\end{figure}

\newpage

\begin{figure}
\begin{centering}
\vspace{1cm}\includegraphics[width=0.9\columnwidth,keepaspectratio]{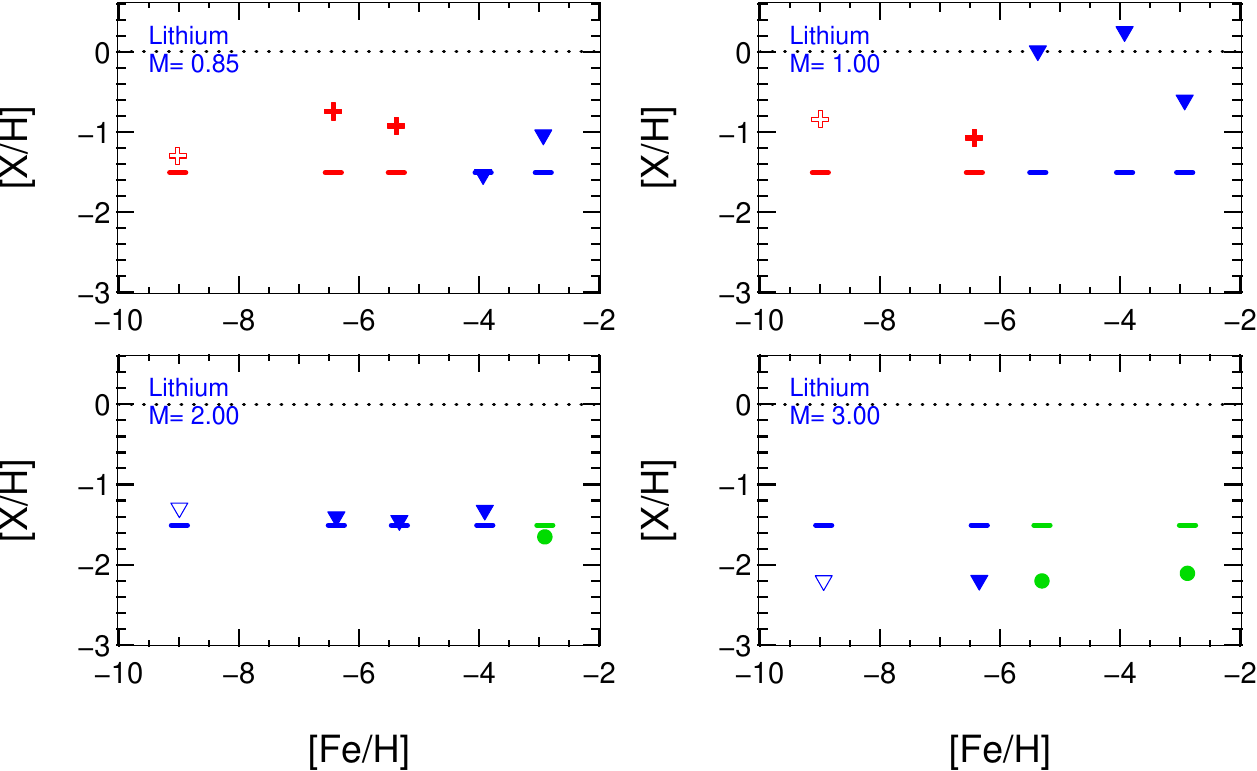}
\par\end{centering}
\caption{Same as Figure \ref{fig-Appx-HeliumVsFeH} except for lithium. \label{fig-Lithium.yields.XoHvsFeH-Appx}}
\end{figure}

\begin{figure}[H]
\begin{centering}
\includegraphics[width=0.9\columnwidth,keepaspectratio]{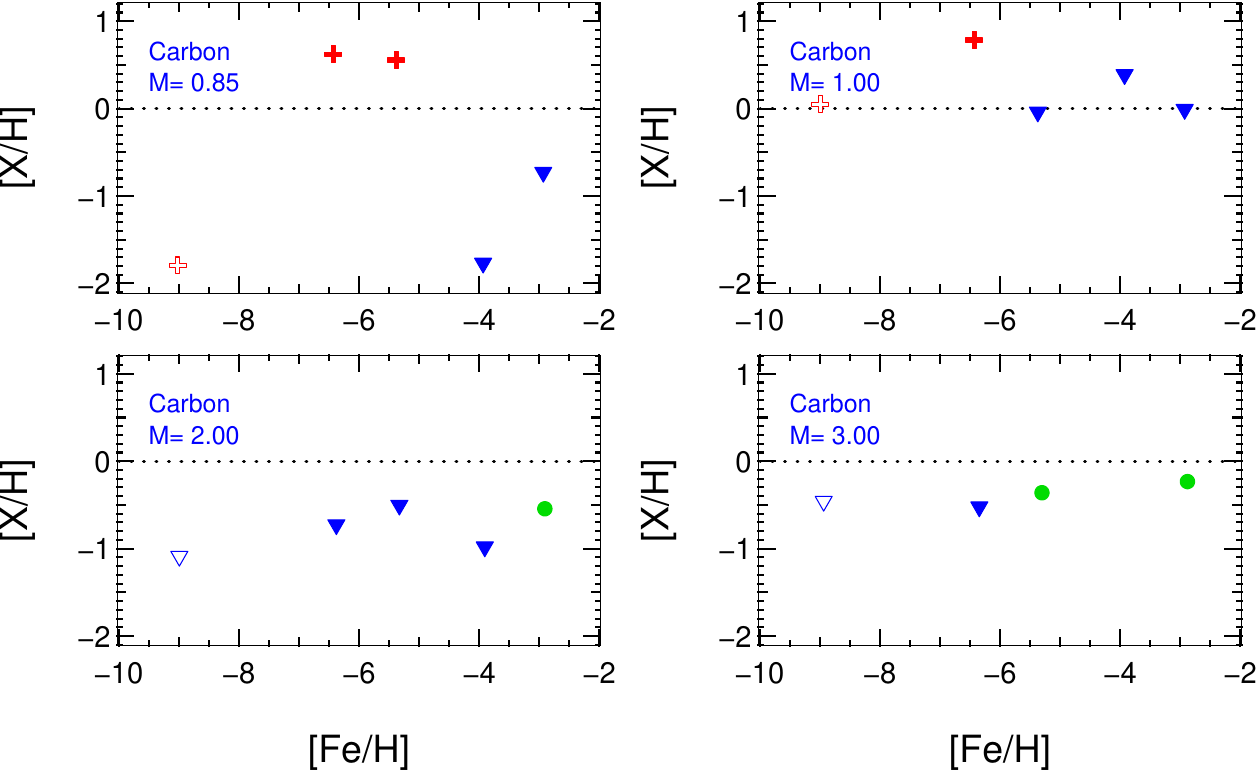}
\par\end{centering}
\caption{Same as Figure \ref{fig-Appx-HeliumVsFeH} except for carbon.}
\end{figure}

\begin{figure}
\begin{centering}
\includegraphics[width=0.9\columnwidth,keepaspectratio]{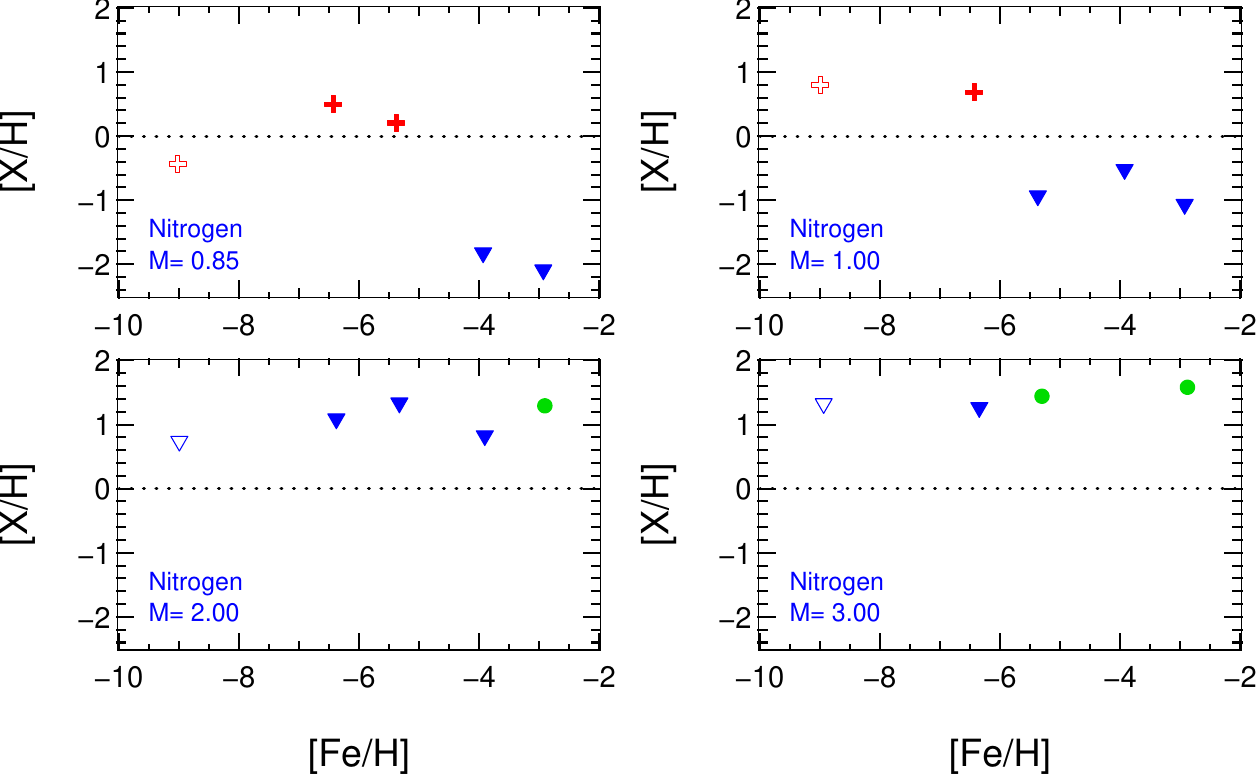}
\par\end{centering}
\caption{Same as Figure \ref{fig-Appx-HeliumVsFeH} except for nitrogen.}
\end{figure}

\begin{figure}
\begin{centering}
\includegraphics[width=0.9\columnwidth,keepaspectratio]{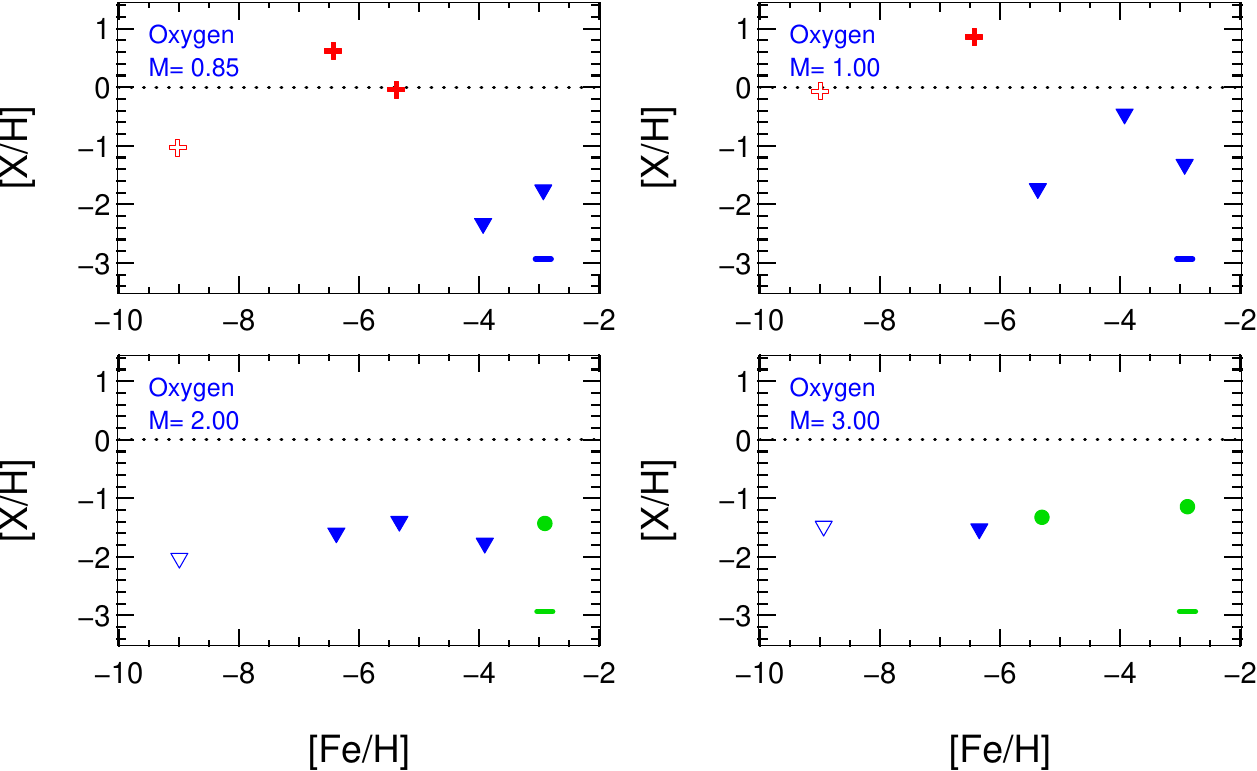}
\par\end{centering}
\caption{Same as Figure \ref{fig-Appx-HeliumVsFeH} except for oxygen.}
\end{figure}

\begin{figure}
\begin{centering}
\includegraphics[width=0.9\columnwidth,keepaspectratio]{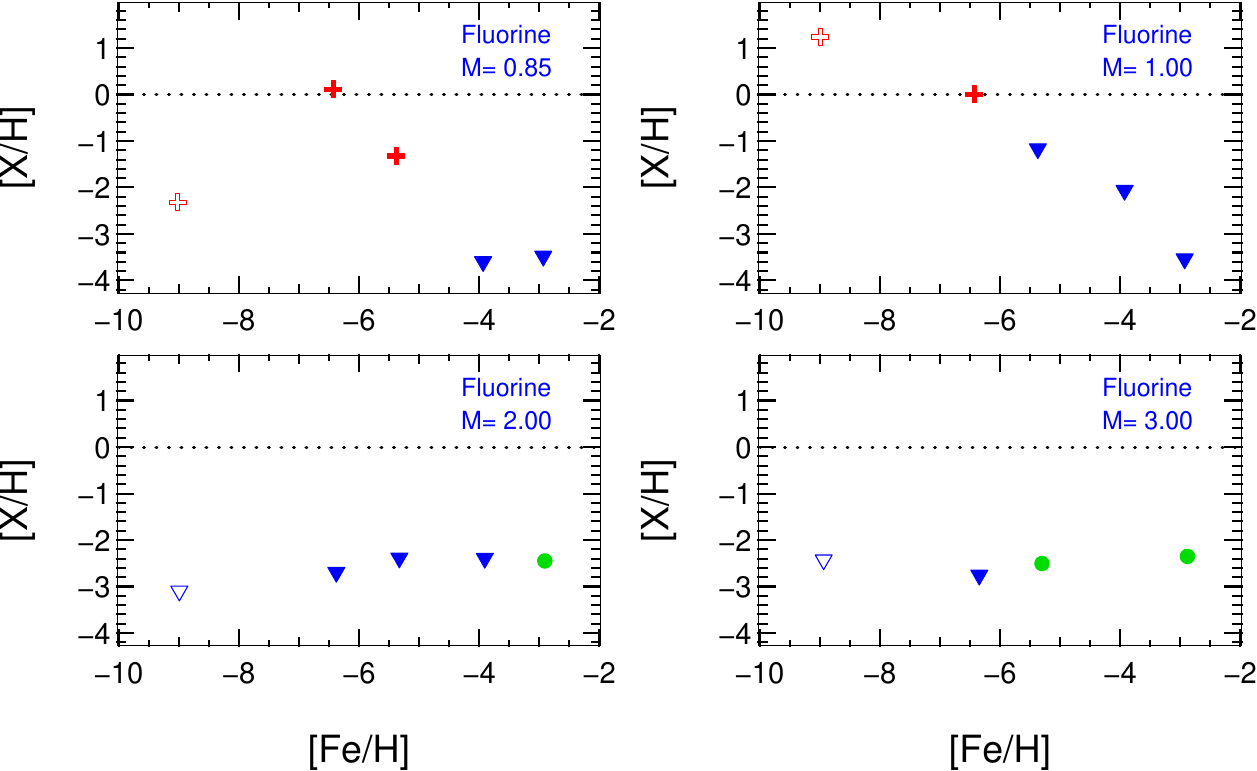}
\par\end{centering}
\caption{Same as Figure \ref{fig-Appx-HeliumVsFeH} except for fluorine. \label{fig-fluorine.yields.XoHvsFeH}}
\end{figure}

\begin{figure}
\begin{centering}
\includegraphics[width=0.9\columnwidth,keepaspectratio]{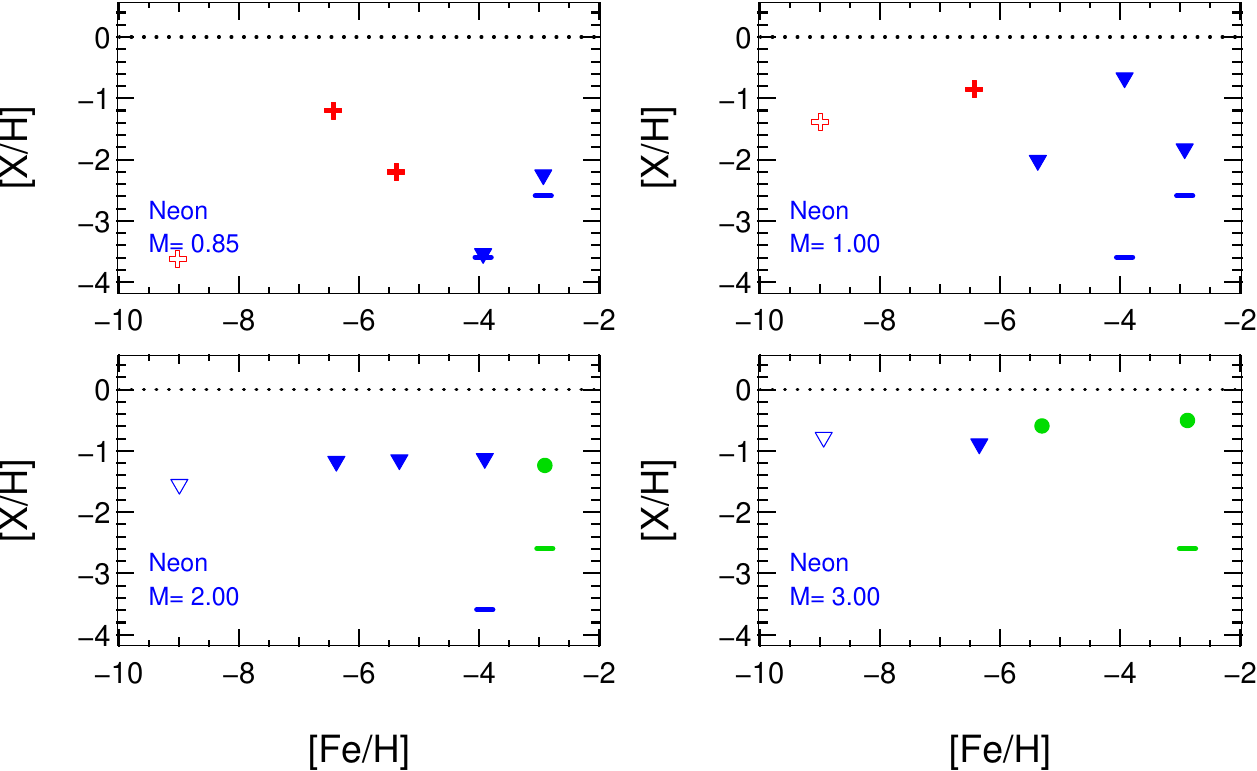}
\par\end{centering}
\caption{Same as Figure \ref{fig-Appx-HeliumVsFeH} except for neon.}
\end{figure}

\begin{figure}
\begin{centering}
\includegraphics[width=0.9\columnwidth,keepaspectratio]{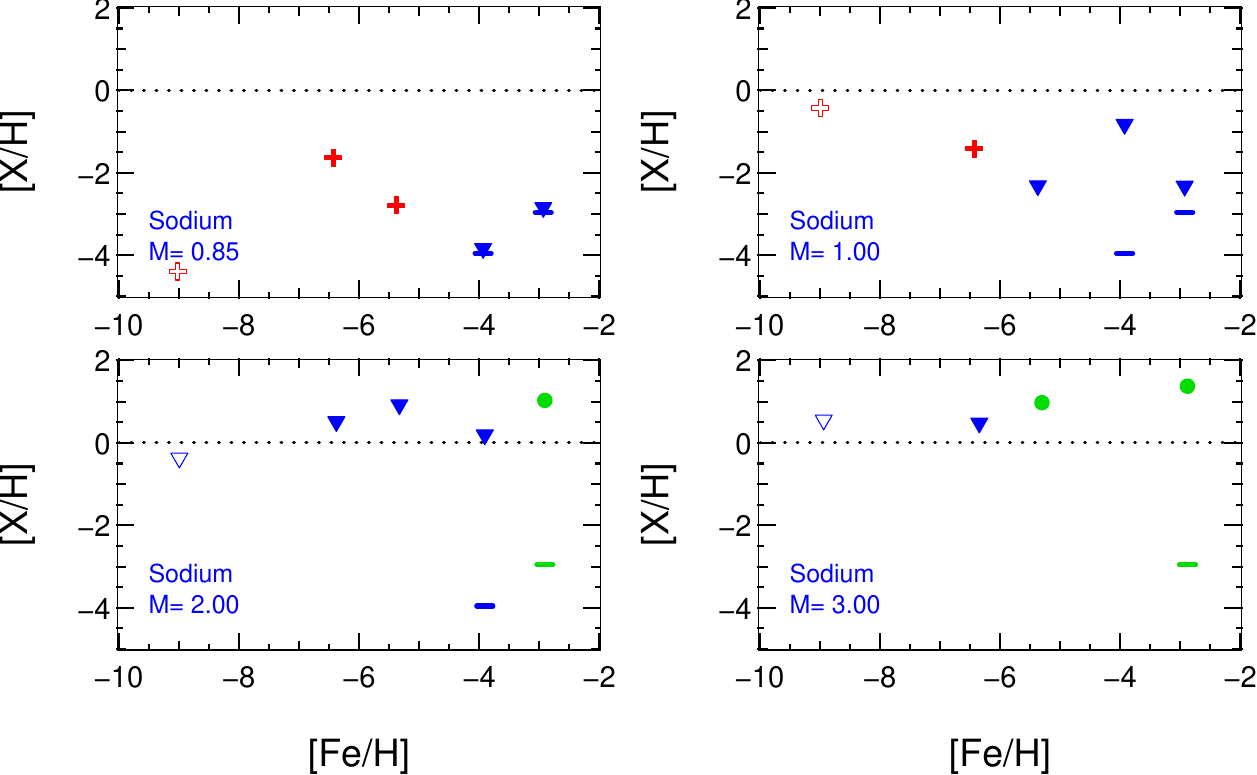}
\par\end{centering}
\caption{Same as Figure \ref{fig-Appx-HeliumVsFeH} except for sodium.}
\end{figure}

\begin{figure}
\begin{centering}
\includegraphics[width=0.9\columnwidth,keepaspectratio]{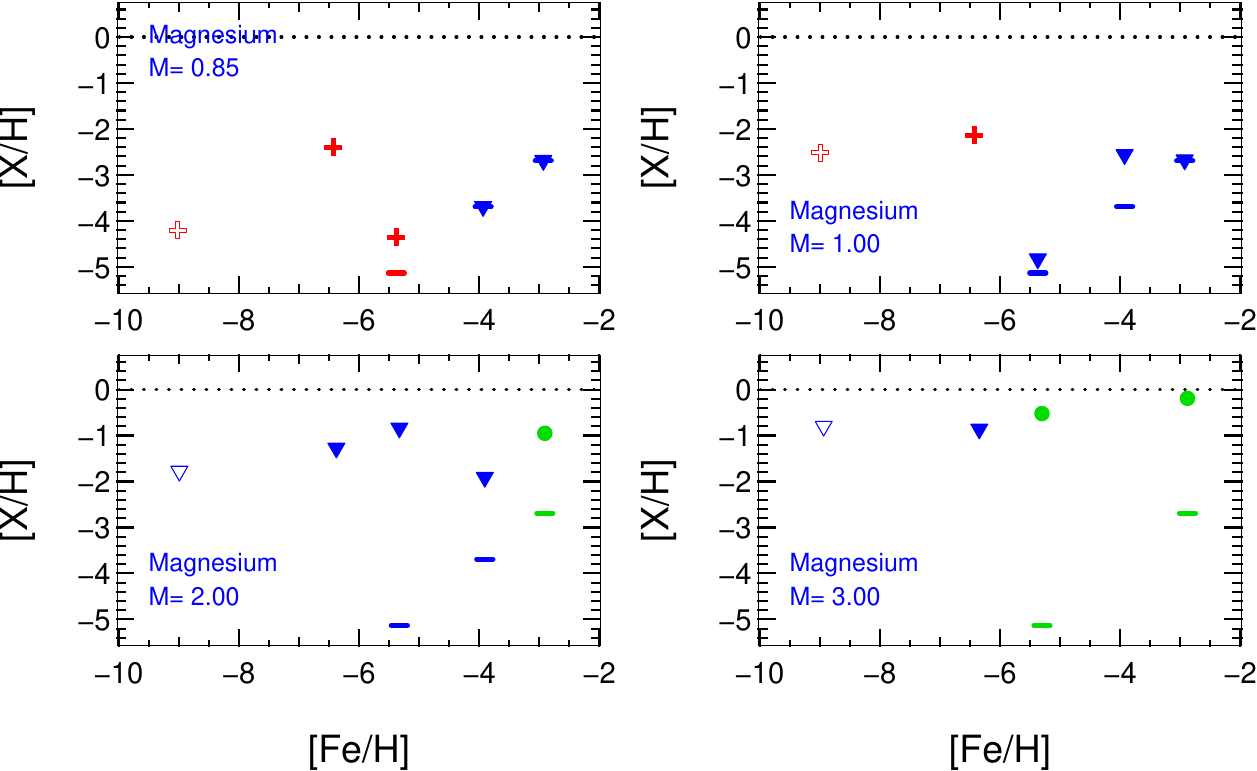}
\par\end{centering}
\caption{Same as Figure \ref{fig-Appx-HeliumVsFeH} except for magnesium.}
\end{figure}

\begin{figure}
\begin{centering}
\includegraphics[width=0.9\columnwidth,keepaspectratio]{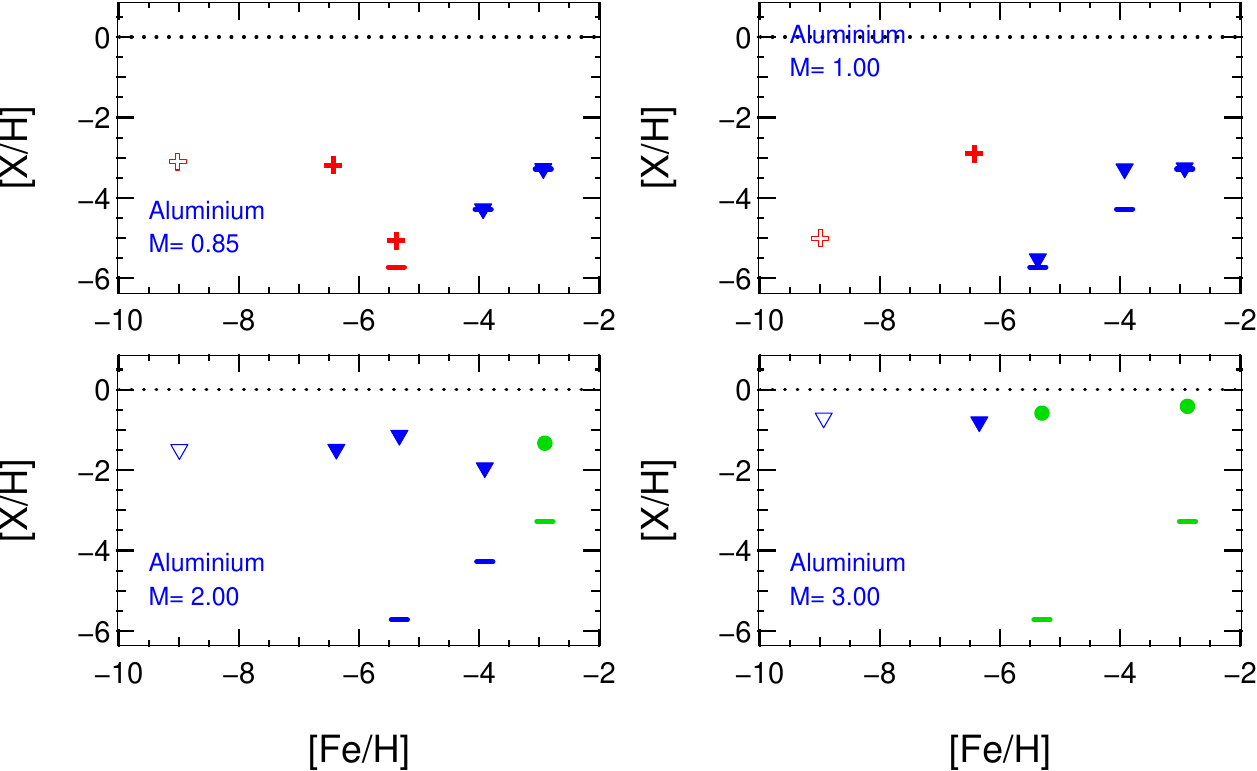}
\par\end{centering}
\caption{Same as Figure \ref{fig-Appx-HeliumVsFeH} except for aluminium.}
\end{figure}

\begin{figure}
\begin{centering}
\includegraphics[width=0.9\columnwidth,keepaspectratio]{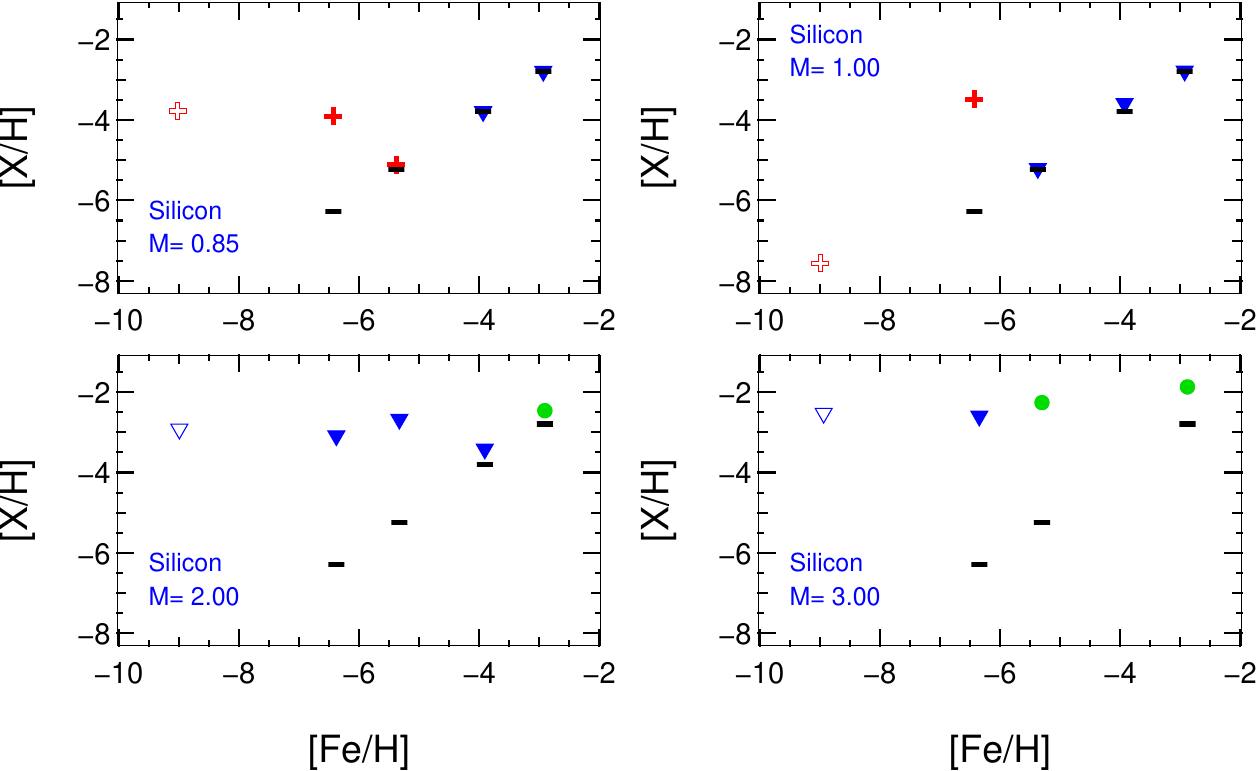}
\par\end{centering}
\caption{Same as Figure \ref{fig-Appx-HeliumVsFeH} except for silicon.}
\end{figure}

\begin{figure}
\begin{centering}
\includegraphics[width=0.9\columnwidth,keepaspectratio]{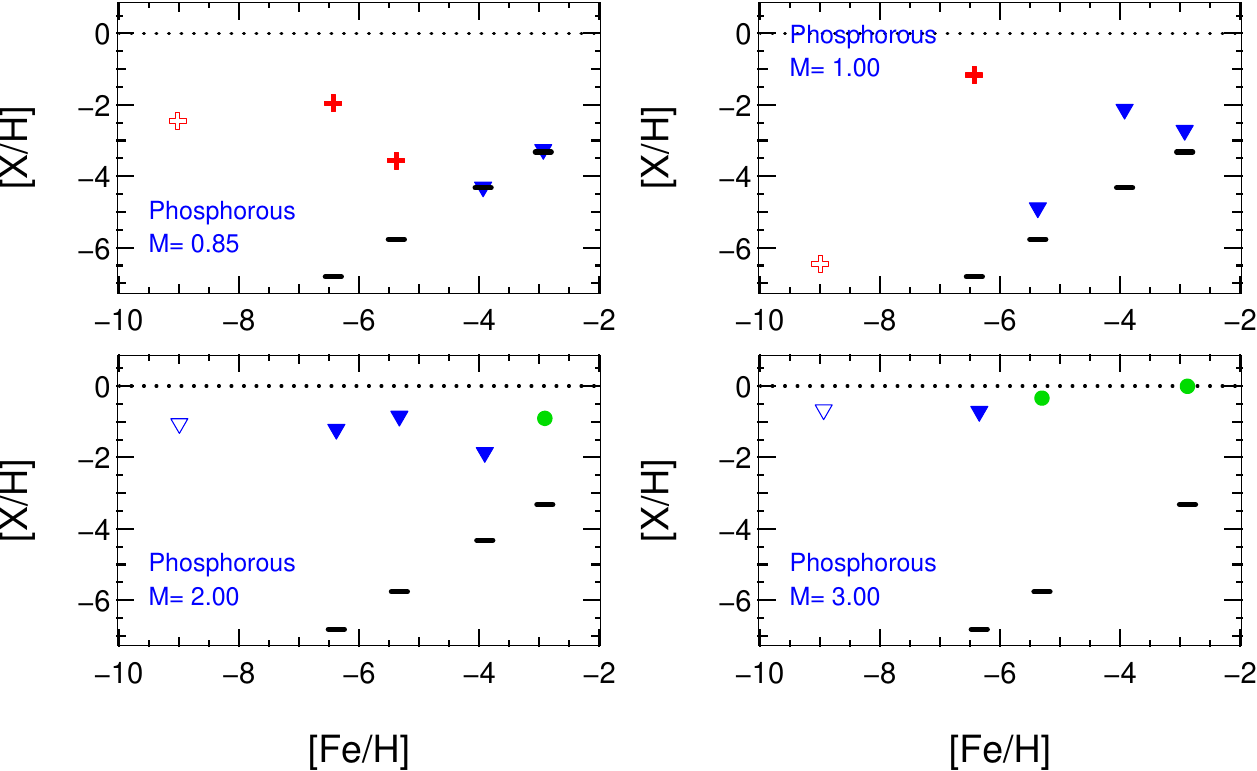}
\par\end{centering}
\caption{Same as Figure \ref{fig-Appx-HeliumVsFeH} except for phosphorous.}
\end{figure}

\begin{figure}
\begin{centering}
\includegraphics[width=0.9\columnwidth,keepaspectratio]{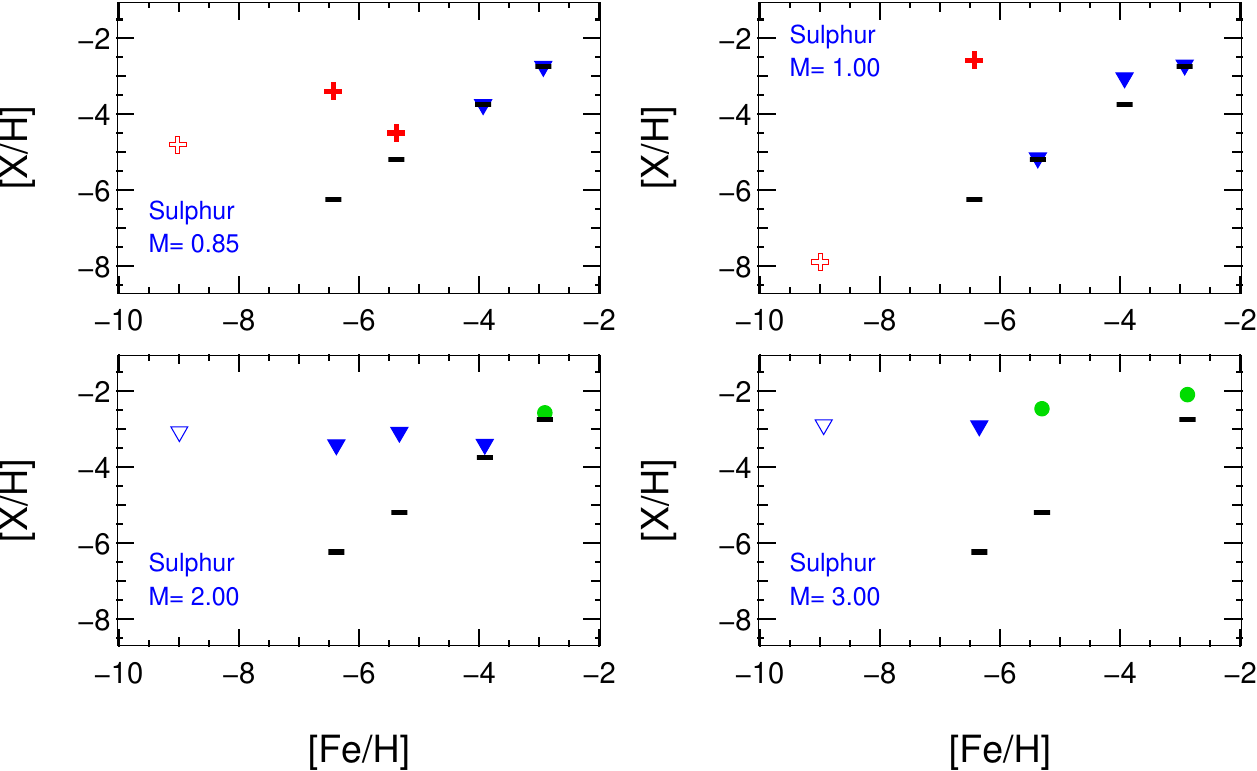}
\par\end{centering}
\caption{Same as Figure \ref{fig-Appx-HeliumVsFeH} except for sulphur.}
\end{figure}

\newpage

\vskip 2in~\vskip 1in~

\section{Tables: Nuclidic Yields by Mass Fraction\label{AppxSec-YieldTablesNuclidicMassFrac}}

\newpage

\newpage

\scriptsize

\begin{longtable}[c]{|>{\centering}m{1.3cm}|>{\centering}m{1.5cm}||>{\centering}m{1.8cm}|>{\centering}m{1.8cm}|>{\centering}m{1.8cm}|>{\centering}m{1.8cm}|}
\caption{Yields and initial composition for all the $Z=0$
 models.
 All species in the network are listed.
 Abundances are in mass fraction, normalised to 1.0.
 The remnant masses (white dwarf masses) are in brackets below the initial
 stellar masses in the table header.}\\

\hline

\textbf{Nuclide}& 

\textbf{Initial \linebreak Abund.}\footnotemark&

\textbf{0.85 M$_{\odot}$\linebreak (0.776)}& 

\textbf{1.0 M$_{\odot}$\linebreak (0.843)}& 

\textbf{2.0 M$_{\odot}$\linebreak (1.080)}&

\textbf{3.0 M$_{\odot}$\linebreak (1.101)}\tabularnewline 

\hline \hline

\endfirsthead

\multicolumn{6}{c}{{\bfseries \tablename\ \thetable{} -- continued from previous page.}} \tabularnewline 

\hline  

\textbf{Nuclide}& 

\textbf{Initial} \linebreak \textbf{Abund.}& 

\textbf{0.85 M$_{\odot}$\linebreak (0.776)}& 

\textbf{1.0 M$_{\odot}$\linebreak (0.843)}& 

\textbf{2.0 M$_{\odot}$\linebreak (1.080)}&

\textbf{3.0 M$_{\odot}$\linebreak (1.101)}\tabularnewline 

\hline \hline

\endhead

\hline \multicolumn{6}{|r|}{{Continued on next page...}} \\ \hline 

\endfoot

\hline \hline 

\endlastfoot
\hline
h1 & 0.7548  & 0.701 & 0.660  & 0.660  & 0.581\tabularnewline
\hline 
h2 & 1.960E-04 & 1.57E-04 & 9.17E-05 & 2.02E-08 & 3.34E-18\tabularnewline
\hline 
he3 & 7.851E-06 & 4.04E-05 & 1.60E-04 & 2.95E-07 & 1.16E-08\tabularnewline
\hline 
he4 & 0.2450 & 0.298 & 0.330 & 0.337 & 0.407\tabularnewline
\hline 
li7 & 3.130E-10 & 4.70E-10 & 1.26E-09 & 4.49E-10 & 4.88E-11\tabularnewline
\hline 
be7 & 0.0 & 2.73E-14 & 2.68E-14 & 5.27E-15 & 4.61E-14\tabularnewline
\hline 
b8 & 0.0 & 1.65E-27 & 7.54E-27 & 6.34E-27 & 1.50E-27\tabularnewline
\hline 
c12 & 0.0 & 2.60E-05 & 1.84E-03 & 1.31E-04 & 4.88E-04\tabularnewline
\hline 
c13 & 0.0 & 7.78E-06 & 3.62E-04 & 3.03E-05 & 1.15E-04\tabularnewline
\hline 
c14 & 0.0 & 3.72E-12 & 3.37E-09 & 6.09E-12 & 8.53E-11\tabularnewline
\hline 
n13 & 0.0 & 0.0 & 0.0 & 0.0 & 0.0\tabularnewline
\hline 
n14 & 0.0 & 2.44E-04 & 3.92E-03 & 3.43E-03 & 1.17E-02\tabularnewline
\hline 
n15 & 0.0 & 9.10E-09 & 1.14E-06 & 1.49E-07 & 7.41E-07\tabularnewline
\hline 
o14 & 0.0 & 0.0 & 0.0 & 0.0 & 0.0\tabularnewline
\hline 
o15 & 0.0 & 0.0 & 0.0 & 0.0 & 0.0\tabularnewline
\hline 
o16 & 0.0 & 5.03E-04 & 4.33E-03 & 4.88E-05 & 1.52E-04\tabularnewline
\hline 
o17 & 0.0 & 7.40E-06 & 4.55E-05 & 1.40E-07 & 6.13E-07\tabularnewline
\hline 
o18 & 0.0 & 3.25E-08 & 1.46E-05 & 7.03E-11 & 2.42E-10\tabularnewline
\hline 
o19 & 0.0 & 0.0 & 0.0 & 0.0 & 0.0\tabularnewline
\hline 
f17 & 0.0 & 0.0 & 0.0 & 0.0 & 0.0\tabularnewline
\hline 
f18 & 0.0 & 0.0 & 0.0 & 0.0 & 0.0\tabularnewline
\hline 
f19 & 0.0 & 1.85E-09 & 6.22E-06 & 2.88E-10 & 1.19E-09\tabularnewline
\hline 
f20 & 0.0 & 0.0 & 0.0 & 0.0 & 0.0\tabularnewline
\hline 
ne19 & 0.0 & 0.0 & 0.0 & 0.0 & 0.0\tabularnewline
\hline 
ne20 & 0.0 & 2.48E-07 & 1.73E-06 & 2.74E-05 & 1.39E-04\tabularnewline
\hline 
ne21 & 0.0 & 6.77E-12 & 4.64E-08 & 1.81E-09 & 3.73E-09\tabularnewline
\hline 
ne22 & 0.0 & 8.33E-10 & 4.24E-05 & 4.74E-07 & 5.32E-06\tabularnewline
\hline 
na21 & 0.0 & 0.0 & 0.0 & 0.0 & 0.0\tabularnewline
\hline 
na22 & 0.0 & 2.30E-26 & 4.86E-17 & 1.81E-12 & 2.99E-11\tabularnewline
\hline 
na23 & 0.0 & 1.29E-09 & 1.13E-05 & 1.29E-05 & 9.54E-05\tabularnewline
\hline 
na24 & 0.0 & 0.0 & 0.0 & 0.0 & 0.0\tabularnewline
\hline 
mg23 & 0.0 & 0.0 & 0.0 & 0.0 & 0.0\tabularnewline
\hline 
mg24 & 0.0 & 3.84E-11 & 1.36E-06 & 1.86E-07 & 2.63E-07\tabularnewline
\hline 
mg25 & 0.0 & 1.46E-08 & 3.17E-07 & 1.76E-06 & 1.56E-05\tabularnewline
\hline 
mg26 & 0.0 & 2.47E-08 & 4.06E-08 & 8.16E-06 & 6.89E-05\tabularnewline
\hline 
mg27 & 0.0 & 0.0 & 0.0 & 0.0 & 0.0\tabularnewline
\hline 
al25 & 0.0 & 0.0 & 0.0 & 0.0 & 0.0\tabularnewline
\hline 
al-6 & 0.0 & 3.18E-11 & 3.36E-10 & 3.08E-07 & 1.49E-06\tabularnewline
\hline 
al{*}6 & 0.0 & 0.0 & 0.0 & 0.0 & 0.0\tabularnewline
\hline 
al27 & 0.0 & 4.39E-08 & 1.63E-10 & 1.34E-06 & 7.36E-06\tabularnewline
\hline 
al28 & 0.0 & 0.0 & 0.0 & 0.0 & 0.0\tabularnewline
\hline 
si27 & 0.0 & 0.0 & 0.0 & 0.0 & 0.0\tabularnewline
\hline 
si28 & 0.0 & 1.00E-07 & 1.65E-11 & 6.17E-07 & 1.31E-06\tabularnewline
\hline 
si29 & 0.0 & 1.16E-08 & 8.23E-13 & 1.27E-07 & 3.06E-07\tabularnewline
\hline 
si30 & 0.0 & 2.77E-09 & 1.83E-13 & 3.34E-08 & 7.49E-08\tabularnewline
\hline 
si31 & 0.0 & 0.0 & 0.0 & 0.0 & 0.0\tabularnewline
\hline 
si32 & 0.0 & 8.70E-21 & 7.99E-25 & 6.37E-13 & 2.20E-12\tabularnewline
\hline 
si33 & 0.0 & 0.0 & 0.0 & 0.0 & 0.0\tabularnewline
\hline 
p29 & 0.0 & 0.0 & 0.0 & 0.0 & 0.0\tabularnewline
\hline 
p30 & 0.0 & 0.0 & 0.0 & 0.0 & 0.0\tabularnewline
\hline 
p31 & 0.0 & 2.20E-08 & 2.07E-12 & 5.22E-07 & 1.12E-06\tabularnewline
\hline 
p32 & 0.0 & 3.38E-24 & 3.10E-28 & 2.47E-16 & 8.55E-16\tabularnewline
\hline 
p33 & 0.0 & 0.0 & 0.0 & 4.75E-24 & 4.23E-22\tabularnewline
\hline 
p34 & 0.0 & 0.0 & 0.0 & 0.0 & 0.0\tabularnewline
\hline 
s32 & 0.0 & 4.38E-09 & 1.22E-12 & 1.87E-07 & 2.66E-07\tabularnewline
\hline 
s33 & 0.0 & 6.24E-10 & 3.36E-14 & 9.23E-09 & 1.54E-08\tabularnewline
\hline 
s34 & 0.0 & 2.19E-10 & 1.32E-13 & 5.56E-09 & 1.42E-08\tabularnewline
\hline 
s35 & 0.0 & 3.98E-10 & 3.13E-12 & 8.31E-08 & 8.31E-08\tabularnewline
\hline 
fe56 & 0.0 & 0.0 & 0.0 & 0.0 & 0.0\tabularnewline
\hline 
fe57 & 0.0 & 0.0 & 0.0 & 0.0 & 0.0\tabularnewline
\hline 
fe58 & 0.0 & 0.0 & 0.0 & 0.0 & 0.0\tabularnewline
\hline 
fe59 & 0.0 & 0.0 & 0.0 & 0.0 & 0.0\tabularnewline
\hline 
fe60 & 0.0 & 0.0 & 0.0 & 0.0 & 0.0\tabularnewline
\hline 
fe61 & 0.0 & 0.0 & 0.0 & 0.0 & 0.0\tabularnewline
\hline 
co59 & 0.0 & 0.0 & 0.0 & 0.0 & 0.0\tabularnewline
\hline 
co60 & 0.0 & 0.0 & 0.0 & 0.0 & 0.0\tabularnewline
\hline 
co61 & 0.0 & 0.0 & 0.0 & 0.0 & 0.0\tabularnewline
\hline 
ni58 & 0.0 & 0.0 & 0.0 & 0.0 & 0.0\tabularnewline
\hline 
ni59 & 0.0 & 0.0 & 0.0 & 0.0 & 0.0\tabularnewline
\hline 
ni60 & 0.0 & 0.0 & 0.0 & 0.0 & 0.0\tabularnewline
\hline 
ni61 & 0.0 & 0.0 & 0.0 & 0.0 & 0.0\tabularnewline
\hline 
ni62 & 0.0 & 0.0 & 0.0 & 0.0 & 0.0\tabularnewline
\hline 
g & 0.0 & 0.0 & 0.0 & 0.0 & 0.0\tabularnewline
\hline 
\end{longtable}

\newpage

\scriptsize

\begin{longtable}[c]{|>{\centering}m{1.1cm}|>{\centering}m{1.8cm}||>{\centering}m{1.5cm}|>{\centering}m{1.5cm}|>{\centering}m{1.5cm}|>{\centering}m{1.5cm}|}

\caption{Yields and initial composition for all the [Fe/H]$=-6.5$ models. 
All species in the network are listed except for neutrons. 
Abundances are in mass fraction, normalised to 1.0. The remnant masses (white dwarf masses) are in brackets below the initial stellar masses in the table header.}
\label{table-allMMPyields-massFrac}\\ 

\hline

\textbf{Nuclide}& 

\textbf{Initial} \linebreak \textbf{\tiny{[Fe/H]$=-6.5$}}&

\textbf{0.85 M$_{\odot}$\linebreak (0.758)}& 

\textbf{1.0 M$_{\odot}$\linebreak (0.846)}& 

\textbf{2.0 M$_{\odot}$\linebreak (1.066)}&

\textbf{3.0 M$_{\odot}$\linebreak (1.107)}\tabularnewline 

\hline \hline

\endfirsthead

\multicolumn{6}{c}{{\bfseries \tablename\ \thetable{} -- continued from previous page}} \tabularnewline 

\hline  

\textbf{Nuclide}& 

\textbf{Initial} \linebreak \textbf{\tiny{[Fe/H]$=-6.5$}}& 

\textbf{0.85 M$_{\odot}$\linebreak (0.758)}& 

\textbf{1.0 M$_{\odot}$\linebreak (0.846)}& 

\textbf{2.0 M$_{\odot}$\linebreak (1.066)}&

\textbf{3.0 M$_{\odot}$\linebreak (1.107)}\tabularnewline 

\hline \hline

\endhead

\hline \multicolumn{6}{|r|}{{Continued on next page...}} \\ \hline 

\endfoot

\hline \hline 

\endlastfoot


\hline
h1  &  0.7548  & 0.686  &  0.642  &  0.656  &  0.604\tabularnewline
\hline 
h2 & 0.00E+00 & 3.44E-19 & 2.08E-18 & 3.99E-18 & 3.53E-18\tabularnewline
\hline 
he3 & 7.85E-06 & 1.33E-05 & 1.62E-04 & 1.61E-07 & 1.72E-08\tabularnewline
\hline 
he4 & 0.2450 & 0.280 & 0.307 & 0.336 & 0.384\tabularnewline
\hline 
li7 & 3.13E-10 & 1.65E-09 & 7.22E-10 & 3.46E-10 & 5.14E-11\tabularnewline
\hline 
be7 & 0.00E+00 & 5.26E-16 & 7.96E-15 & 6.63E-14 & 3.71E-14\tabularnewline
\hline 
b8 & 0.00E+00 & 5.04E-26 & 4.00E-27 & 1.04E-26 & 5.32E-27\tabularnewline
\hline 
c12 & 1.30E-09 & 7.59E-03 & 1.04E-02 & 2.99E-04 & 4.38E-04\tabularnewline
\hline 
c13 & 1.33E-16 & 9.74E-04 & 1.18E-03 & 6.86E-05 & 1.04E-04\tabularnewline
\hline 
c14 & 0.00E+00 & 8.58E-07 & 2.07E-07 & 4.84E-11 & 6.76E-11\tabularnewline
\hline 
n13 & 0.00E+00 & 0.00E+00 & 0.00E+00 & 0.00E+00 & 0.00E+00\tabularnewline
\hline 
n14 & 2.27E-16 & 2.03E-03 & 2.92E-03 & 7.60E-03 & 1.06E-02\tabularnewline
\hline 
n15 & 1.30E-18 & 7.36E-06 & 8.92E-06 & 3.09E-07 & 7.02E-07\tabularnewline
\hline 
o14 & 0.00E+00 & 0.00E+00 & 0.00E+00 & 0.00E+00 & 0.00E+00\tabularnewline
\hline 
o15 & 0.00E+00 & 0.00E+00 & 0.00E+00 & 7.86E-37 & 0.00E+00\tabularnewline
\hline 
o16 & 2.23E-09 & 2.10E-02 & 3.03E-02 & 1.32E-04 & 1.42E-04\tabularnewline
\hline 
o17 & 3.39E-17 & 1.37E-03 & 6.21E-03 & 3.45E-07 & 5.70E-07\tabularnewline
\hline 
o18 & 2.26E-16 & 8.12E-05 & 1.64E-05 & 1.47E-10 & 2.30E-10\tabularnewline
\hline 
o19 & 0.00E+00 & 0.00E+00 & 0.00E+00 & 0.00E+00 & 0.00E+00\tabularnewline
\hline 
f17 & 0.00E+00 & 0.00E+00 & 0.00E+00 & 1.41E-42 & 0.00E+00\tabularnewline
\hline 
f18 & 0.00E+00 & 0.00E+00 & 0.00E+00 & 2.79E-34 & 0.00E+00\tabularnewline
\hline 
f19 & 3.92E-18 & 4.90E-07 & 3.55E-07 & 7.24E-10 & 5.80E-10\tabularnewline
\hline 
f20 & 0.00E+00 & 0.00E+00 & 0.00E+00 & 0.00E+00 & 0.00E+00\tabularnewline
\hline 
ne19 & 0.00E+00 & 0.00E+00 & 0.00E+00 & 0.00E+00 & 0.00E+00\tabularnewline
\hline 
ne20 & 9.24E-10 & 3.34E-05 & 6.73E-05 & 6.12E-05 & 1.11E-04\tabularnewline
\hline 
ne21 & 2.06E-13 & 2.48E-06 & 6.69E-06 & 1.04E-08 & 2.60E-09\tabularnewline
\hline 
ne22 & 4.64E-14 & 3.14E-05 & 6.66E-05 & 5.05E-06 & 5.36E-06\tabularnewline
\hline 
na21 & 0.00E+00 & 0.00E+00 & 0.00E+00 & 0.00E+00 & 0.00E+00\tabularnewline
\hline 
na22 & 0.00E+00 & 3.24E-30 & 1.92E-17 & 4.83E-12 & 2.30E-11\tabularnewline
\hline 
na23 & 1.23E-11 & 7.33E-07 & 1.14E-06 & 9.86E-05 & 8.39E-05\tabularnewline
\hline 
na24 & 0.00E+00 & 0.00E+00 & 0.00E+00 & 6.93E-37 & 0.00E+00\tabularnewline
\hline 
mg23 & 0.00E+00 & 0.00E+00 & 0.00E+00 & 0.00E+00 & 0.00E+00\tabularnewline
\hline 
mg24 & 4.13E-10 & 4.81E-07 & 4.82E-07 & 1.89E-07 & 2.08E-07\tabularnewline
\hline 
mg25 & 4.49E-12 & 3.22E-07 & 2.44E-07 & 5.74E-06 & 1.39E-05\tabularnewline
\hline 
mg26 & 4.95E-12 & 1.68E-06 & 3.53E-06 & 2.62E-05 & 6.25E-05\tabularnewline
\hline 
mg27 & 0.00E+00 & 0.00E+00 & 0.00E+00 & 0.00E+00 & 0.00E+00\tabularnewline
\hline 
al25 & 0.00E+00 & 0.00E+00 & 0.00E+00 & 0.00E+00 & 0.00E+00\tabularnewline
\hline 
al-6 & 0.00E+00 & 1.66E-12 & 2.48E-11 & 4.16E-07 & 1.17E-06\tabularnewline
\hline 
al{*}6 & 0.00E+00 & 0.00E+00 & 0.00E+00 & 0.00E+00 & 0.00E+00\tabularnewline
\hline 
al27 & 1.00E-11 & 3.47E-08 & 6.31E-08 & 1.25E-06 & 6.22E-06\tabularnewline
\hline 
al28 & 0.00E+00 & 0.00E+00 & 0.00E+00 & 0.00E+00 & 0.00E+00\tabularnewline
\hline 
si27 & 0.00E+00 & 0.00E+00 & 0.00E+00 & 0.00E+00 & 0.00E+00\tabularnewline
\hline 
si28 & 3.75E-10 & 6.64E-08 & 1.62E-07 & 4.00E-07 & 1.16E-06\tabularnewline
\hline 
si29 & 3.95E-12 & 1.17E-08 & 2.77E-08 & 1.01E-07 & 2.74E-07\tabularnewline
\hline 
si30 & 2.78E-12 & 5.61E-09 & 1.45E-08 & 2.55E-08 & 6.90E-08\tabularnewline
\hline 
si31 & 0.00E+00 & 0.00E+00 & 0.00E+00 & 1.70E-41 & 0.00E+00\tabularnewline
\hline 
si32 & 0.00E+00 & 3.02E-16 & 2.30E-17 & 6.52E-12 & 4.98E-12\tabularnewline
\hline 
si33 & 0.00E+00 & 0.00E+00 & 0.00E+00 & 0.00E+00 & 0.00E+00\tabularnewline
\hline 
p29 & 0.00E+00 & 0.00E+00 & 0.00E+00 & 0.00E+00 & 0.00E+00\tabularnewline
\hline 
p30 & 0.00E+00 & 0.00E+00 & 0.00E+00 & 0.00E+00 & 0.00E+00\tabularnewline
\hline 
p31 & 1.04E-12 & 6.82E-08 & 3.91E-07 & 3.59E-07 & 1.05E-06\tabularnewline
\hline 
p32 & 0.00E+00 & 1.17E-19 & 8.91E-21 & 2.53E-15 & 1.93E-15\tabularnewline
\hline 
p33 & 0.00E+00 & -3.08E-44 & 0.00E+00 & 1.17E-23 & 4.58E-22\tabularnewline
\hline 
p34 & 0.00E+00 & 0.00E+00 & 0.00E+00 & 0.00E+00 & 0.00E+00\tabularnewline
\hline 
s32 & 1.65E-10 & 2.61E-08 & 2.05E-07 & 7.95E-08 & 2.47E-07\tabularnewline
\hline 
s33 & 7.78E-13 & 1.95E-09 & 2.86E-09 & 9.74E-09 & 1.63E-08\tabularnewline
\hline 
s34 & 5.30E-11 & 3.19E-09 & 6.39E-09 & 6.35E-09 & 1.56E-08\tabularnewline
\hline 
s35 & 0.00E+00 & 1.16E-07 & 6.94E-07 & 3.22E-08 & 9.27E-08\tabularnewline
\hline 
fe56 & 4.54E-10 & 4.13E-10 & 3.86E-10 & 4.42E-10 & 4.39E-10\tabularnewline
\hline 
fe57 & 1.37E-11 & 1.25E-11 & 1.17E-11 & 1.35E-11 & 1.34E-11\tabularnewline
\hline 
fe58 & 2.79E-18 & 8.97E-15 & 1.33E-15 & 4.87E-13 & 5.27E-13\tabularnewline
\hline 
fe59 & 0.00E+00 & 0.00E+00 & 0.00E+00 & 0.00E+00 & 5.86E-31\tabularnewline
\hline 
fe60 & 0.00E+00 & 5.60E-16 & 1.19E-16 & 3.35E-13 & 4.94E-13\tabularnewline
\hline 
fe61 & 0.00E+00 & 0.00E+00 & 0.00E+00 & 0.00E+00 & 0.00E+00\tabularnewline
\hline 
co59 & 7.22E-12 & 6.57E-12 & 6.14E-12 & 7.29E-12 & 7.24E-12\tabularnewline
\hline 
co60 & 0.00E+00 & 4.87E-20 & 1.04E-20 & 2.92E-17 & 4.30E-17\tabularnewline
\hline 
co61 & 0.00E+00 & 0.00E+00 & 0.00E+00 & 0.00E+00 & 0.00E+00\tabularnewline
\hline 
ni58 & 6.41E-12 & 5.83E-12 & 5.46E-12 & 6.24E-12 & 6.20E-12\tabularnewline
\hline 
ni59 & 0.00E+00 & 1.01E-18 & 1.93E-24 & 4.61E-19 & 1.06E-18\tabularnewline
\hline 
ni60 & 7.43E-15 & 4.18E-14 & 1.66E-14 & 1.44E-12 & 1.53E-12\tabularnewline
\hline 
ni61 & 4.00E-17 & 4.69E-11 & 7.68E-11 & 1.06E-11 & 1.40E-11\tabularnewline
\hline 
ni62 & 2.38E-20 & 2.04E-20 & 1.28E-20 & 1.99E-24 & 2.05E-25\tabularnewline
\hline 
g\addtocounter{footnote}{-1}\footnotemark & 4.67E-16 & 6.35E-08 & 3.16E-07 & 1.41E-09 & 6.38E-10\tabularnewline
\hline 
\end{longtable}

\newpage

\scriptsize

\begin{longtable}[c]{|>{\centering}m{1.1cm}|>{\centering}m{1.8cm}||>{\centering}m{1.5cm}|>{\centering}m{1.5cm}|>{\centering}m{1.5cm}|>{\centering}m{1.5cm}|}
\caption[All \[Fe/H\]$=-5.45$ Yields]{Yields and initial composition for all the [Fe/H]$=-5.45$ models. All species in the network are listed except for neutrons. Abundances are in mass fraction, normalised to 1.0. The remnant masses (white dwarf masses) are in brackets below the initial stellar masses in the table header.}
\label{table-allHMPyields-massFrac}\\ 

\hline

\textbf{Nuclide}& 

\textbf{Initial} \linebreak \textbf{\tiny{[Fe/H]$=-5.45$}}&

\textbf{0.85 M$_{\odot}$\linebreak (0.744)}& 

\textbf{1.0 M$_{\odot}$\linebreak (0.873)}& 

\textbf{2.0 M$_{\odot}$\linebreak (1.058)}&

\textbf{3.0 M$_{\odot}$\linebreak (1.097)}\tabularnewline 

\hline \hline

\endfirsthead

\multicolumn{6}{c}{{\bfseries \tablename\ \thetable{} -- continued from previous page}} \tabularnewline 

\hline  

\textbf{Nuclide}& 

\textbf{Initial} \linebreak \textbf{\tiny{[Fe/H]$=-5.45$}}& 

\textbf{0.85 M$_{\odot}$\linebreak (0.744)}& 

\textbf{1.0 M$_{\odot}$\linebreak (0.873)}& 

\textbf{2.0 M$_{\odot}$\linebreak (1.058)}&

\textbf{3.0 M$_{\odot}$\linebreak (1.097)}\tabularnewline 

\hline \hline

\endhead

\hline \multicolumn{6}{|r|}{{Continued on next page...}} \\ \hline 

\endfoot

\hline \hline 

\endlastfoot

\hline
h1  &  0.7548  &  0.658  &  0.730  &  0.651  & 0.603\tabularnewline
\hline 
h2 & 0.00E+00 & 4.47E-19 & 3.78E-18 & 4.04E-18 & 3.54E-18\tabularnewline
\hline 
he3 & 7.85E-06 & 6.21E-05 & 1.68E-04 & 1.71E-07 & 1.89E-08\tabularnewline
\hline 
he4 & 0.2450 & 0.329 & 0.267 & 0.334 & 0.380\tabularnewline
\hline 
li7 & 3.13E-10 & 1.05E-09 & 1.00E-08 & 3.09E-10 & 5.03E-11\tabularnewline
\hline 
be7 & 0.00E+00 & 0.00E+00 & 2.17E-14 & 5.37E-14 & 1.40E-14\tabularnewline
\hline 
b8 & 0.00E+00 & 7.29E-20 & 2.35E-25 & 3.13E-26 & 2.28E-25\tabularnewline
\hline 
c12 & 1.46E-08 & 6.05E-03 & 1.84E-03 & 4.96E-04 & 6.35E-04\tabularnewline
\hline 
c13 & 1.49E-15 & 9.68E-04 & 1.42E-04 & 1.14E-04 & 1.49E-04\tabularnewline
\hline 
c14 & 0.00E+00 & 3.26E-08 & 5.12E-12 & 1.14E-10 & 1.30E-10\tabularnewline
\hline 
n13 & 0.00E+00 & 0.00E+00 & 0.00E+00 & 0.00E+00 & 0.00E+00\tabularnewline
\hline 
n14 & 2.54E-15 & 9.96E-04 & 8.08E-05 & 1.34E-02 & 1.58E-02\tabularnewline
\hline 
n15 & 1.45E-17 & 2.91E-07 & 2.81E-09 & 5.27E-07 & 9.22E-07\tabularnewline
\hline 
o14 & 0.00E+00 & 0.00E+00 & 0.00E+00 & 0.00E+00 & 0.00E+00\tabularnewline
\hline 
o15 & 0.00E+00 & 0.00E+00 & 0.00E+00 & 0.00E+00 & 0.00E+00\tabularnewline
\hline 
o16 & 2.50E-08 & 3.89E-03 & 1.04E-04 & 2.06E-04 & 2.23E-04\tabularnewline
\hline 
o17 & 3.80E-16 & 7.93E-04 & 2.18E-06 & 5.04E-07 & 8.00E-07\tabularnewline
\hline 
o18 & 2.53E-15 & 1.22E-06 & 2.29E-09 & 2.10E-10 & 2.43E-10\tabularnewline
\hline 
o19 & 0.00E+00 & 0.00E+00 & 0.00E+00 & 0.00E+00 & 0.00E+00\tabularnewline
\hline 
f17 & 0.00E+00 & 0.00E+00 & 0.00E+00 & 0.00E+00 & 0.00E+00\tabularnewline
\hline 
f18 & 0.00E+00 & 0.00E+00 & 0.00E+00 & 0.00E+00 & 0.00E+00\tabularnewline
\hline 
f19 & 4.39E-17 & 1.72E-08 & 2.72E-08 & 1.46E-09 & 1.03E-09\tabularnewline
\hline 
f20 & 0.00E+00 & 0.00E+00 & 0.00E+00 & 0.00E+00 & 0.00E+00\tabularnewline
\hline 
ne19 & 0.00E+00 & 0.00E+00 & 0.00E+00 & 0.00E+00 & 0.00E+00\tabularnewline
\hline 
ne20 & 1.03E-08 & 2.95E-06 & 3.80E-06 & 5.82E-05 & 2.17E-04\tabularnewline
\hline 
ne21 & 2.31E-12 & 4.80E-07 & 6.12E-07 & 4.40E-09 & 5.21E-09\tabularnewline
\hline 
ne22 & 5.19E-13 & 2.97E-06 & 6.72E-06 & 1.20E-05 & 1.23E-05\tabularnewline
\hline 
na21 & 0.00E+00 & 0.00E+00 & 0.00E+00 & 0.00E+00 & 0.00E+00\tabularnewline
\hline 
na22 & 0.00E+00 & 0.00E+00 & 7.61E-16 & 3.97E-12 & 2.88E-11\tabularnewline
\hline 
na23 & 1.37E-10 & 4.90E-08 & 1.62E-07 & 2.50E-04 & 2.61E-04\tabularnewline
\hline 
na24 & 0.00E+00 & 0.00E+00 & 0.00E+00 & 0.00E+00 & 0.00E+00\tabularnewline
\hline 
mg23 & 0.00E+00 & 0.00E+00 & 0.00E+00 & 0.00E+00 & 0.00E+00\tabularnewline
\hline 
mg24 & 4.62E-09 & 5.13E-09 & 5.94E-09 & 3.10E-07 & 6.03E-07\tabularnewline
\hline 
mg25 & 5.02E-11 & 7.54E-10 & 9.44E-10 & 1.72E-05 & 3.02E-05\tabularnewline
\hline 
mg26 & 5.54E-11 & 2.00E-08 & 2.58E-09 & 6.96E-05 & 1.35E-04\tabularnewline
\hline 
mg27 & 0.00E+00 & 0.00E+00 & 0.00E+00 & 0.00E+00 & 0.00E+00\tabularnewline
\hline 
al25 & 0.00E+00 & 0.00E+00 & 0.00E+00 & 0.00E+00 & 0.00E+00\tabularnewline
\hline 
al-6 & 0.00E+00 & 1.50E-14 & 1.16E-11 & 1.04E-06 & 2.32E-06\tabularnewline
\hline 
al{*}6 & 0.00E+00 & 0.00E+00 & 0.00E+00 & 0.00E+00 & 0.00E+00\tabularnewline
\hline 
al27 & 1.12E-10 & 4.48E-10 & 1.62E-10 & 2.63E-06 & 9.79E-06\tabularnewline
\hline 
al28 & 0.00E+00 & 0.00E+00 & 0.00E+00 & 0.00E+00 & 0.00E+00\tabularnewline
\hline 
si27 & 0.00E+00 & 0.00E+00 & 0.00E+00 & 0.00E+00 & 0.00E+00\tabularnewline
\hline 
si28 & 4.19E-09 & 4.60E-09 & 4.32E-09 & 1.02E-06 & 2.43E-06\tabularnewline
\hline 
si29 & 4.42E-11 & 1.95E-10 & 7.43E-11 & 2.67E-07 & 6.44E-07\tabularnewline
\hline 
si30 & 3.12E-11 & 9.94E-11 & 3.71E-11 & 6.29E-08 & 1.62E-07\tabularnewline
\hline 
si31 & 0.00E+00 & 0.00E+00 & 0.00E+00 & 0.00E+00 & 0.00E+00\tabularnewline
\hline 
si32 & 0.00E+00 & 9.95E-18 & 1.21E-15 & 1.14E-11 & 7.41E-12\tabularnewline
\hline 
si33 & 0.00E+00 & 0.00E+00 & 0.00E+00 & 0.00E+00 & 0.00E+00\tabularnewline
\hline 
p29 & 0.00E+00 & 0.00E+00 & 0.00E+00 & 0.00E+00 & 0.00E+00\tabularnewline
\hline 
p30 & 0.00E+00 & 0.00E+00 & 0.00E+00 & 0.00E+00 & 0.00E+00\tabularnewline
\hline 
p31 & 1.16E-11 & 1.61E-09 & 8.49E-11 & 8.33E-07 & 2.47E-06\tabularnewline
\hline 
p32 & 0.00E+00 & 3.86E-21 & 4.71E-19 & 4.41E-15 & 2.88E-15\tabularnewline
\hline 
p33 & 0.00E+00 & 0.00E+00 & 0.00E+00 & 1.41E-23 & 3.14E-22\tabularnewline
\hline 
p34 & 0.00E+00 & 0.00E+00 & 0.00E+00 & 0.00E+00 & 0.00E+00\tabularnewline
\hline 
s32 & 1.84E-09 & 2.35E-09 & 1.84E-09 & 1.82E-07 & 5.96E-07\tabularnewline
\hline 
s33 & 8.71E-12 & 8.49E-11 & 2.19E-11 & 1.58E-08 & 3.32E-08\tabularnewline
\hline 
s34 & 5.93E-10 & 5.67E-10 & 5.87E-10 & 1.31E-08 & 3.62E-08\tabularnewline
\hline 
s35 & 0.00E+00 & 8.21E-09 & 1.81E-10 & 5.97E-08 & 3.89E-07\tabularnewline
\hline 
fe56 & 5.08E-09 & 4.43E-09 & 5.00E-09 & 4.90E-09 & 4.82E-09\tabularnewline
\hline 
fe57 & 1.54E-10 & 1.34E-10 & 1.52E-10 & 1.51E-10 & 1.48E-10\tabularnewline
\hline 
fe58 & 3.12E-17 & 2.75E-17 & 3.98E-12 & 1.30E-11 & 8.79E-12\tabularnewline
\hline 
fe59 & 0.00E+00 & 0.00E+00 & 0.00E+00 & 6.65E-30 & 5.35E-31\tabularnewline
\hline 
fe60 & 0.00E+00 & 2.01E-18 & 1.06E-14 & 5.61E-12 & 6.16E-12\tabularnewline
\hline 
fe61 & 0.00E+00 & 0.00E+00 & 0.00E+00 & 0.00E+00 & 0.00E+00\tabularnewline
\hline 
co59 & 8.08E-11 & 7.05E-11 & 8.21E-11 & 8.41E-11 & 8.09E-11\tabularnewline
\hline 
co60 & 0.00E+00 & 1.75E-22 & 9.19E-19 & 4.90E-16 & 5.36E-16\tabularnewline
\hline 
co61 & 0.00E+00 & 0.00E+00 & 0.00E+00 & 0.00E+00 & 0.00E+00\tabularnewline
\hline 
ni58 & 7.17E-11 & 6.25E-11 & 7.07E-11 & 6.92E-11 & 6.80E-11\tabularnewline
\hline 
ni59 & 0.00E+00 & 1.79E-25 & 1.91E-19 & 4.08E-17 & 1.03E-17\tabularnewline
\hline 
ni60 & 8.32E-14 & 7.26E-14 & 1.48E-11 & 2.46E-11 & 2.97E-11\tabularnewline
\hline 
ni61 & 4.47E-16 & 7.48E-10 & 6.14E-11 & 1.44E-10 & 2.43E-10\tabularnewline
\hline 
ni62 & 2.67E-19 & 2.15E-19 & 1.08E-19 & 1.70E-23 & 6.59E-24\tabularnewline
\hline 
g\addtocounter{footnote}{-1}\footnotemark & 5.23E-15 & 1.95E-06 & 4.73E-08 & 5.28E-09 & 1.25E-08\tabularnewline
\hline 
\end{longtable}

\newpage

\scriptsize

\begin{longtable}[c]{|>{\centering}m{1.1cm}|>{\centering}m{1.8cm}||>{\centering}m{1.5cm}|>{\centering}m{1.5cm}|>{\centering}m{1.5cm}|}
\caption[All \[Fe/H\]$=-4.0$ Yields]{Yields and initial composition for all the [Fe/H]$=-4.0$ models. All species in the network are listed except for neutrons. Abundances are in mass fraction, normalised to 1.0. The remnant masses (white dwarf masses) are in brackets below the initial stellar masses in the table header. Due to loss of data we were unable to calculate the yield for the 3 M$_\odot$ model in this case.}
\label{table-allUMPyields-massFrac}\\ 

\hline

\textbf{Nuclide}& 

\textbf{Initial} \linebreak \textbf{\tiny{[Fe/H]$=-4.0$}}&

\textbf{0.85 M$_{\odot}$\linebreak (0.761)}& 

\textbf{1.0 M$_{\odot}$\linebreak (0.852)}& 

\textbf{2.0 M$_{\odot}$\linebreak (1.085)}\tabularnewline 

\hline \hline

\endfirsthead

\multicolumn{5}{c}{{\bfseries \tablename\ \thetable{} -- continued from previous page}} \tabularnewline 

\hline  

\textbf{Nuclide}& 

\textbf{Initial} \linebreak \textbf{\tiny{[Fe/H]$=-4.0$}}& 

\textbf{0.85 M$_{\odot}$\linebreak (0.761)}& 

\textbf{1.0 M$_{\odot}$\linebreak (0.852)}& 

\textbf{2.0 M$_{\odot}$\linebreak (1.085)}\tabularnewline 

\hline \hline

\endhead

\hline \multicolumn{5}{|r|}{{Continued on next page...}} \\ \hline 

\endfoot

\hline \hline 

\endlastfoot

\hline
h1 &  0.7548  &  0.752  &  0.724  &  0.695\tabularnewline

\hline 
h2 & 0.00E+00 & 1.70E-18 & 3.93E-18 & 4.51E-18\tabularnewline
\hline 
he3 & 7.85E-06 & 2.10E-04 & 2.00E-04 & 2.74E-06\tabularnewline
\hline 
he4 & 0.2450 & 0.248 & 0.268 & 0.300\tabularnewline
\hline 
li7 & 3.13E-10 & 2.94E-10 & 1.73E-08 & 4.43E-10\tabularnewline
\hline 
be7 & 0.00E+00 & 4.53E-15 & 2.53E-15 & 2.09E-13\tabularnewline
\hline 
b8 & 0.00E+00 & 2.59E-27 & 1.61E-25 & 1.49E-26\tabularnewline
\hline 
c12 & 4.04E-07 & 3.19E-05 & 4.88E-03 & 1.77E-04\tabularnewline
\hline 
c13 & 4.12E-14 & 6.34E-06 & 3.69E-04 & 4.03E-05\tabularnewline
\hline 
c14 & 0.00E+00 & 3.02E-10 & 9.87E-09 & 2.08E-11\tabularnewline
\hline 
n13 & 0.00E+00 & 0.00E+00 & 0.00E+00 & 0.00E+00\tabularnewline
\hline 
n14 & 7.03E-14 & 1.07E-05 & 2.06E-04 & 4.41E-03\tabularnewline
\hline 
n15 & 4.03E-16 & 3.35E-09 & 3.76E-08 & 1.78E-07\tabularnewline
\hline 
o14 & 0.00E+00 & 0.00E+00 & 0.00E+00 & 0.00E+00\tabularnewline
\hline 
o15 & 0.00E+00 & 0.00E+00 & 0.00E+00 & 0.00E+00\tabularnewline
\hline 
o16 & 6.91E-07 & 2.48E-05 & 1.73E-03 & 9.31E-05\tabularnewline
\hline 
o17 & 1.05E-14 & 3.13E-06 & 2.66E-04 & 2.71E-07\tabularnewline
\hline 
o18 & 7.02E-14 & 7.84E-09 & 1.90E-07 & 1.65E-10\tabularnewline
\hline 
o19 & 0.00E+00 & 0.00E+00 & 0.00E+00 & 0.00E+00\tabularnewline
\hline 
f17 & 0.00E+00 & 0.00E+00 & 0.00E+00 & 0.00E+00\tabularnewline
\hline 
f18 & 0.00E+00 & 0.00E+00 & 0.00E+00 & 0.00E+00\tabularnewline
\hline 
f19 & 1.21E-15 & 1.05E-10 & 3.43E-09 & 1.54E-09\tabularnewline
\hline 
f20 & 0.00E+00 & 0.00E+00 & 3.26E-38 & 0.00E+00\tabularnewline
\hline 
ne19 & 0.00E+00 & 0.00E+00 & 0.00E+00 & 0.00E+00\tabularnewline
\hline 
ne20 & 2.87E-07 & 3.01E-07 & 2.06E-05 & 7.59E-05\tabularnewline
\hline 
ne21 & 6.39E-11 & 1.38E-09 & 2.69E-06 & 1.68E-08\tabularnewline
\hline 
ne22 & 1.44E-11 & 2.61E-08 & 2.29E-04 & 1.64E-06\tabularnewline
\hline 
na21 & 0.00E+00 & 0.00E+00 & 0.00E+00 & 0.00E+00\tabularnewline
\hline 
na22 & 0.00E+00 & -2.22E-29 & 7.82E-15 & 3.99E-12\tabularnewline
\hline 
na23 & 3.81E-09 & 5.01E-09 & 4.95E-06 & 4.97E-05\tabularnewline
\hline 
na24 & 0.00E+00 & 0.00E+00 & 0.00E+00 & 0.00E+00\tabularnewline
\hline 
mg23 & 0.00E+00 & 0.00E+00 & 0.00E+00 & 0.00E+00\tabularnewline
\hline 
mg24 & 1.28E-07 & 1.28E-07 & 3.16E-07 & 2.26E-07\tabularnewline
\hline 
mg25 & 1.39E-09 & 1.40E-09 & 7.57E-08 & 1.59E-06\tabularnewline
\hline 
mg26 & 1.54E-09 & 1.94E-09 & 1.42E-06 & 6.00E-06\tabularnewline
\hline 
mg27 & 0.00E+00 & 0.00E+00 & 0.00E+00 & 0.00E+00\tabularnewline
\hline 
al25 & 0.00E+00 & 0.00E+00 & 0.00E+00 & 0.00E+00\tabularnewline
\hline 
al-6 & 0.00E+00 & 4.17E-13 & 1.00E-10 & 1.75E-07\tabularnewline
\hline 
al{*}6 & 0.00E+00 & 0.00E+00 & 0.00E+00 & 0.00E+00\tabularnewline
\hline 
al27 & 3.10E-09 & 3.13E-09 & 2.97E-08 & 4.37E-07\tabularnewline
\hline 
al28 & 0.00E+00 & 0.00E+00 & 0.00E+00 & 0.00E+00\tabularnewline
\hline 
si27 & 0.00E+00 & 0.00E+00 & 0.00E+00 & 0.00E+00\tabularnewline
\hline 
si28 & 1.16E-07 & 1.16E-07 & 1.66E-07 & 2.30E-07\tabularnewline
\hline 
si29 & 1.22E-09 & 1.23E-09 & 1.28E-08 & 2.45E-08\tabularnewline
\hline 
si30 & 8.63E-10 & 8.63E-10 & 3.06E-09 & 6.68E-09\tabularnewline
\hline 
si31 & 0.00E+00 & 0.00E+00 & 0.00E+00 & 0.00E+00\tabularnewline
\hline 
si32 & 0.00E+00 & 2.93E-17 & 1.36E-11 & 9.56E-13\tabularnewline
\hline 
si33 & 0.00E+00 & 0.00E+00 & 0.00E+00 & 0.00E+00\tabularnewline
\hline 
p29 & 0.00E+00 & 0.00E+00 & 0.00E+00 & 0.00E+00\tabularnewline
\hline 
p30 & 0.00E+00 & 0.00E+00 & 0.00E+00 & 0.00E+00\tabularnewline
\hline 
p31 & 3.21E-10 & 3.37E-10 & 4.88E-08 & 8.55E-08\tabularnewline
\hline 
p32 & 0.00E+00 & 1.14E-20 & 5.29E-15 & 3.71E-16\tabularnewline
\hline 
p33 & 0.00E+00 & 0.00E+00 & 0.00E+00 & 1.97E-24\tabularnewline
\hline 
p34 & 0.00E+00 & 0.00E+00 & 0.00E+00 & 0.00E+00\tabularnewline
\hline 
s32 & 5.10E-08 & 5.10E-08 & 9.01E-08 & 7.05E-08\tabularnewline
\hline 
s33 & 2.41E-10 & 2.51E-10 & 2.84E-08 & 7.52E-09\tabularnewline
\hline 
s34 & 1.64E-08 & 1.64E-08 & 2.46E-08 & 1.87E-08\tabularnewline
\hline 
s35 & 0.00E+00 & 5.02E-10 & 2.06E-07 & 4.44E-08\tabularnewline
\hline 
fe56 & 1.41E-07 & 1.41E-07 & 1.38E-07 & 1.38E-07\tabularnewline
\hline 
fe57 & 4.25E-09 & 4.25E-09 & 4.16E-09 & 4.22E-09\tabularnewline
\hline 
fe58 & 8.65E-16 & 2.41E-14 & 4.30E-14 & 1.87E-10\tabularnewline
\hline 
fe59 & 0.00E+00 & 0.00E+00 & 0.00E+00 & 8.39E-27\tabularnewline
\hline 
fe60 & 0.00E+00 & 4.56E-17 & 1.98E-16 & 4.27E-11\tabularnewline
\hline 
fe61 & 0.00E+00 & 0.00E+00 & 0.00E+00 & 0.00E+00\tabularnewline
\hline 
co59 & 2.24E-09 & 2.24E-09 & 2.19E-09 & 2.29E-09\tabularnewline
\hline 
co60 & 0.00E+00 & 3.97E-21 & 1.82E-20 & 3.73E-15\tabularnewline
\hline 
co61 & 0.00E+00 & 0.00E+00 & 0.00E+00 & 0.00E+00\tabularnewline
\hline 
ni58 & 1.99E-09 & 1.98E-09 & 1.94E-09 & 1.95E-09\tabularnewline
\hline 
ni59 & 0.00E+00 & 1.51E-19 & 1.07E-20 & 1.71E-15\tabularnewline
\hline 
ni60 & 2.31E-12 & 2.33E-12 & 2.43E-12 & 3.49E-10\tabularnewline
\hline 
ni61 & 1.24E-14 & 8.03E-11 & 3.52E-09 & 1.86E-09\tabularnewline
\hline 
ni62 & 7.38E-18 & 5.48E-18 & 2.75E-18 & 1.68E-20\tabularnewline
\hline 
g\addtocounter{footnote}{-1}\footnotemark & 1.45E-13 & 1.59E-07 & 3.82E-06 & 9.96E-07\tabularnewline
\hline 
\end{longtable}

\newpage

\scriptsize

\begin{longtable}[c]{|>{\centering}m{1.1cm}|>{\centering}m{1.8cm}||>{\centering}m{1.5cm}|>{\centering}m{1.5cm}|>{\centering}m{1.5cm}|>{\centering}m{1.5cm}|}
\caption[All \[Fe/H\]$=-3.0$ Yields]{Yields and initial composition for all the [Fe/H]$=-3.0$ models. All species in the network are listed except for neutrons. Abundances are in mass fraction, normalised to 1.0. The remnant masses (white dwarf masses) are in brackets below the initial stellar masses in the table header.}
\label{table-allHMPyields-massFrac}\\ 

\hline

\textbf{Nuclide}& 

\textbf{Initial} \linebreak \textbf{\tiny{[Fe/H]$=-3.0$}}&

\textbf{0.85 M$_{\odot}$\linebreak (0.753)}& 

\textbf{1.0 M$_{\odot}$\linebreak (0.855)}& 

\textbf{2.0 M$_{\odot}$\linebreak (1.038)}&

\textbf{3.0 M$_{\odot}$\linebreak (1.069)}\tabularnewline 

\hline \hline

\endfirsthead

\multicolumn{6}{c}{{\bfseries \tablename\ \thetable{} -- continued from previous page}} \tabularnewline 

\hline  

\textbf{Nuclide}& 

\textbf{Initial} \linebreak \textbf{\tiny{[Fe/H]$=-3.0$}}& 

\textbf{0.85 M$_{\odot}$\linebreak (0.753)}& 

\textbf{1.0 M$_{\odot}$\linebreak (0.855)}& 

\textbf{2.0 M$_{\odot}$\linebreak (1.038)}&

\textbf{3.0 M$_{\odot}$\linebreak (1.069)}\tabularnewline 

\hline \hline

\endhead

\hline \multicolumn{6}{|r|}{{Continued on next page...}} \\ \hline 

\endfoot

\hline \hline 

\endlastfoot

\hline
h1  &  0.7548  &  0.748  &  0.730  &  0.678  &  0.619\tabularnewline
\hline 
h2 & 0.00E+00 & 2.20E-18 & 3.87E-18 & 4.24E-18 & 3.70E-18\tabularnewline
\hline 
he3 & 7.85E-06 & 2.68E-04 & 2.29E-04 & 2.56E-06 & 3.90E-08\tabularnewline
\hline 
he4 & 0.2450 & 0.251 & 0.268 & 0.308 & 0.356\tabularnewline
\hline 
li7 & 3.13E-10 & 9.15E-10 & 2.44E-09 & 2.01E-10 & 6.43E-11\tabularnewline
\hline 
be7 & 0.00E+00 & 0.00E+00 & 1.43E-14 & 9.57E-14 & 1.34E-13\tabularnewline
\hline 
b8 & 0.00E+00 & 9.75E-27 & 2.64E-26 & 1.04E-26 & 1.43E-27\tabularnewline
\hline 
c12 & 4.04E-06 & 3.93E-04 & 2.01E-03 & 4.71E-04 & 8.76E-04\tabularnewline
\hline 
c13 & 4.12E-13 & 2.02E-05 & 1.07E-04 & 1.08E-04 & 2.05E-04\tabularnewline
\hline 
c14 & 0.00E+00 & 1.92E-09 & 1.93E-12 & 6.84E-11 & 1.96E-10\tabularnewline
\hline 
n13 & 0.00E+00 & 0.00E+00 & 0.00E+00 & 0.00E+00 & 0.00E+00\tabularnewline
\hline 
n14 & 7.03E-13 & 5.79E-06 & 5.94E-05 & 1.26E-02 & 2.23E-02\tabularnewline
\hline 
n15 & 4.03E-15 & 7.83E-10 & 2.68E-09 & 4.70E-07 & 1.13E-06\tabularnewline
\hline 
o14 & 0.00E+00 & 0.00E+00 & 0.00E+00 & 0.00E+00 & 0.00E+00\tabularnewline
\hline 
o15 & 0.00E+00 & 0.00E+00 & 0.00E+00 & 0.00E+00 & 0.00E+00\tabularnewline
\hline 
o16 & 6.91E-06 & 9.33E-05 & 2.60E-04 & 1.98E-04 & 3.48E-04\tabularnewline
\hline 
o17 & 1.05E-13 & 1.05E-05 & 1.61E-05 & 4.38E-07 & 1.08E-06\tabularnewline
\hline 
o18 & 7.01E-13 & 8.02E-09 & 1.73E-08 & 1.90E-10 & 2.72E-10\tabularnewline
\hline 
o19 & 0.00E+00 & 0.00E+00 & 0.00E+00 & 0.00E+00 & 0.00E+00\tabularnewline
\hline 
f17 & 0.00E+00 & 0.00E+00 & 0.00E+00 & 0.00E+00 & 0.00E+00\tabularnewline
\hline 
f18 & 0.00E+00 & 0.00E+00 & 0.00E+00 & 0.00E+00 & 0.00E+00\tabularnewline
\hline 
f19 & 1.21E-14 & 1.36E-10 & 1.16E-10 & 1.32E-09 & 1.50E-09\tabularnewline
\hline 
f20 & 0.00E+00 & 0.00E+00 & 0.00E+00 & 0.00E+00 & 0.00E+00\tabularnewline
\hline 
ne19 & 0.00E+00 & 0.00E+00 & 0.00E+00 & 0.00E+00 & 0.00E+00\tabularnewline
\hline 
ne20 & 2.87E-06 & 4.20E-06 & 8.43E-06 & 4.59E-05 & 2.58E-04\tabularnewline
\hline 
ne21 & 6.39E-10 & 2.50E-07 & 1.09E-06 & 6.98E-09 & 9.33E-09\tabularnewline
\hline 
ne22 & 1.44E-10 & 1.99E-06 & 7.41E-06 & 1.35E-05 & 3.23E-05\tabularnewline
\hline 
na21 & 0.00E+00 & 0.00E+00 & 0.00E+00 & 0.00E+00 & 0.00E+00\tabularnewline
\hline 
na22 & 0.00E+00 & 1.43E-17 & 4.12E-16 & 2.49E-12 & 2.94E-11\tabularnewline
\hline 
na23 & 3.81E-08 & 4.91E-08 & 1.58E-07 & 3.34E-04 & 6.74E-04\tabularnewline
\hline 
na24 & 0.00E+00 & 0.00E+00 & 0.00E+00 & 0.00E+00 & 0.00E+00\tabularnewline
\hline 
mg23 & 0.00E+00 & 0.00E+00 & 0.00E+00 & 0.00E+00 & 0.00E+00\tabularnewline
\hline 
mg24 & 1.28E-06 & 1.28E-06 & 1.27E-06 & 1.43E-06 & 1.69E-06\tabularnewline
\hline 
mg25 & 1.39E-08 & 1.42E-08 & 1.42E-08 & 1.31E-05 & 6.76E-05\tabularnewline
\hline 
mg26 & 1.54E-08 & 2.24E-08 & 3.52E-08 & 5.49E-05 & 2.96E-04\tabularnewline
\hline 
mg27 & 0.00E+00 & 0.00E+00 & 0.00E+00 & 0.00E+00 & 0.00E+00\tabularnewline
\hline 
al25 & 0.00E+00 & 0.00E+00 & 0.00E+00 & 0.00E+00 & 0.00E+00\tabularnewline
\hline 
al-6 & 0.00E+00 & 8.84E-13 & 4.35E-11 & 7.02E-07 & 4.03E-06\tabularnewline
\hline 
al{*}6 & 0.00E+00 & 0.00E+00 & 0.00E+00 & 0.00E+00 & 0.00E+00\tabularnewline
\hline 
al27 & 3.10E-08 & 3.11E-08 & 3.16E-08 & 1.73E-06 & 1.43E-05\tabularnewline
\hline 
al28 & 0.00E+00 & 0.00E+00 & 0.00E+00 & 0.00E+00 & 0.00E+00\tabularnewline
\hline 
si27 & 0.00E+00 & 0.00E+00 & 0.00E+00 & 0.00E+00 & 0.00E+00\tabularnewline
\hline 
si28 & 1.16E-06 & 1.16E-06 & 1.15E-06 & 1.98E-06 & 6.30E-06\tabularnewline
\hline 
si29 & 1.22E-08 & 1.23E-08 & 1.23E-08 & 2.40E-07 & 1.49E-06\tabularnewline
\hline 
si30 & 8.63E-09 & 8.63E-09 & 9.07E-09 & 6.33E-08 & 3.70E-07\tabularnewline
\hline 
si31 & 0.00E+00 & 0.00E+00 & 0.00E+00 & 0.00E+00 & 0.00E+00\tabularnewline
\hline 
si32 & 0.00E+00 & 3.63E-12 & 3.61E-13 & 9.87E-12 & 6.57E-12\tabularnewline
\hline 
si33 & 0.00E+00 & 0.00E+00 & 0.00E+00 & 0.00E+00 & 0.00E+00\tabularnewline
\hline 
p29 & 0.00E+00 & 0.00E+00 & 0.00E+00 & 0.00E+00 & 0.00E+00\tabularnewline
\hline 
p30 & 0.00E+00 & 0.00E+00 & 0.00E+00 & 0.00E+00 & 0.00E+00\tabularnewline
\hline 
p31 & 3.21E-09 & 3.77E-09 & 1.26E-08 & 7.54E-07 & 5.44E-06\tabularnewline
\hline 
p32 & 0.00E+00 & 1.41E-15 & 1.40E-16 & 3.83E-15 & 2.55E-15\tabularnewline
\hline 
p33 & 0.00E+00 & 0.00E+00 & 0.00E+00 & 2.14E-24 & 2.14E-22\tabularnewline
\hline 
p34 & 0.00E+00 & 0.00E+00 & 0.00E+00 & 0.00E+00 & 0.00E+00\tabularnewline
\hline 
s32 & 5.10E-07 & 5.09E-07 & 5.09E-07 & 6.57E-07 & 1.70E-06\tabularnewline
\hline 
s33 & 2.41E-09 & 2.57E-09 & 6.10E-09 & 1.71E-08 & 5.88E-08\tabularnewline
\hline 
s34 & 1.64E-07 & 1.64E-07 & 1.64E-07 & 1.69E-07 & 2.38E-07\tabularnewline
\hline 
s35 & 0.00E+00 & 1.35E-08 & 5.37E-08 & 6.75E-08 & 5.21E-07\tabularnewline
\hline 
fe56 & 1.41E-06 & 1.40E-06 & 1.39E-06 & 1.34E-06 & 1.31E-06\tabularnewline
\hline 
fe57 & 4.25E-08 & 4.25E-08 & 4.21E-08 & 4.17E-08 & 3.99E-08\tabularnewline
\hline 
fe58 & 8.64E-15 & 9.64E-14 & 3.37E-13 & 4.11E-09 & 2.41E-09\tabularnewline
\hline 
fe59 & 0.00E+00 & 0.00E+00 & 0.00E+00 & 8.44E-29 & 1.01E-29\tabularnewline
\hline 
fe60 & 0.00E+00 & 1.24E-15 & -2.56E-22 & 9.75E-10 & 1.95E-09\tabularnewline
\hline 
fe61 & 0.00E+00 & 0.00E+00 & 0.00E+00 & 0.00E+00 & 0.00E+00\tabularnewline
\hline 
co59 & 2.24E-08 & 2.23E-08 & 2.21E-08 & 2.32E-08 & 2.20E-08\tabularnewline
\hline 
co60 & 0.00E+00 & 2.37E-19 & -2.23E-26 & 8.52E-14 & 1.70E-13\tabularnewline
\hline 
co61 & 0.00E+00 & 0.00E+00 & 0.00E+00 & 0.00E+00 & 0.00E+00\tabularnewline
\hline 
ni58 & 1.99E-08 & 1.98E-08 & 1.96E-08 & 1.90E-08 & 1.84E-08\tabularnewline
\hline 
ni59 & 0.00E+00 & 1.13E-18 & -5.49E-25 & 6.71E-16 & 8.40E-16\tabularnewline
\hline 
ni60 & 2.31E-11 & 2.32E-11 & 2.43E-11 & 8.88E-09 & 1.10E-08\tabularnewline
\hline 
ni61 & 1.24E-13 & 3.10E-09 & 1.75E-08 & 5.27E-08 & 9.81E-08\tabularnewline
\hline 
ni62 & 7.38E-17 & 4.89E-17 & 2.88E-17 & 2.64E-19 & 1.09E-21\tabularnewline
\hline 
g\addtocounter{footnote}{-1}\footnotemark & 1.45E-12 & 3.13E-06 & 9.88E-06 & 2.27E-06 & 4.28E-06\tabularnewline
\hline 
\end{longtable}

\normalsize

\chapter{Yields: Globular Cluster Models\label{Appx-GCYields}}

This appendix holds the yields for the four sets of globular cluster
models ($\textrm{[Fe/H]}=-1.4$) -- the standard models, the models
calculated using the NACRE rates (2 masses only), the models calculated
using Reimers' mass-loss formula on the AGB (also only 2 masses),
and the models with no 3DUP. The yields for every species in the nuclear
network (except neutrons) are given, in mass fraction. There are 3
tables in total. 

\newpage \begin{center}

\vskip 2in~\vskip 1in~

\section{Tables: Nuclidic Yields by Mass Fraction}

\newpage

\newpage

\scriptsize

\begin{longtable}[c]{>{\centering}m{1.1cm}>{\centering}m{1.5cm}|>{\centering}m{1.3cm}>{\centering}m{1.2cm}>{\centering}m{1.2cm}>{\centering}m{1.2cm}>{\centering}m{1.2cm}}

\caption[All Standard GC Yields]{Yields and initial composition for all the standard GC models (for NGC 6752, [Fe/H]$=-1.4$). By standard models we mean that 3DUP has not been inhibited and we have used the standard mass loss rates and nuclear reaction rates. All species in the network are listed except for neutrons. Abundances are in mass fraction, normalised to 1.0. The remnant masses (white dwarf masses) are in brackets below the initial stellar masses in the table header.}
\label{table-allGCyields-Standard}\\ 

\hline

\textbf{Nuclide}& 

\textbf{Initial}&

\textbf{1.25 M$_{\odot}$\linebreak (0.645)}& 

\textbf{2.5 M$_{\odot}$\linebreak (0.676)}& 

\textbf{3.5 M$_{\odot}$\linebreak (0.838)}&

\textbf{5.0 M$_{\odot}$\linebreak (0.915)}&

\textbf{6.5 M$_{\odot}$\linebreak (1.047)}\tabularnewline 

\hline \hline

\endfirsthead

\multicolumn{7}{c}{{\bfseries \tablename\ \thetable{} -- continued from previous page}} \tabularnewline 

\hline  

\textbf{Nuclide}& 

\textbf{Initial}& 

\textbf{1.25 M$_{\odot}$\linebreak (0.645)}& 

\textbf{2.5 M$_{\odot}$\linebreak (0.676)}& 

\textbf{3.5 M$_{\odot}$\linebreak (0.838)}&

\textbf{5.0 M$_{\odot}$\linebreak (0.915)}&

\textbf{6.5 M$_{\odot}$\linebreak (1.047)}\tabularnewline 

\hline \hline

\endhead

\hline \multicolumn{7}{r}{{Continued on next page...}} \\

\endfoot

\hline \hline 

\endlastfoot
h1  &  0.76860  &  0.74035  &  0.71976  &  0.73714  &  0.60457  &  0.64137\tabularnewline
h2 & 2.26E-20 & 3.79E-18 & 4.23E-18 & 4.45E-18 & 3.66E-18 & 3.50E-18\tabularnewline
he3 & 1.27E-09 & 3.02E-04 & 1.07E-04 & 8.97E-05 & 2.91E-07 & 1.49E-07\tabularnewline
he4 & 0.22970 & 0.25644 & 0.26610 & 0.25487 & 0.37499 & 0.34909\tabularnewline
li7 & 1.15E-14 & 5.75E-11 & 1.06E-07 & 3.90E-09 & 4.27E-17 & 2.37E-12\tabularnewline
be7 & 0.00E+00 & 6.61E-18 & 2.34E-39 & 7.38E-14 & 1.19E-13 & 6.46E-15\tabularnewline
b8 & 0.00E+00 & 1.08E-27 & 1.11E-18 & 1.73E-24 & 1.25E-21 & 9.93E-22\tabularnewline
c12 & 2.10E-04 & 1.32E-03 & 9.97E-03 & 6.04E-03 & 2.02E-03 & 1.21E-03\tabularnewline
c13 & 4.68E-10 & 4.62E-06 & 5.06E-04 & 3.32E-05 & 2.15E-04 & 1.25E-04\tabularnewline
c14 & 0.00E+00 & 3.03E-11 & 1.19E-08 & 5.83E-10 & 4.82E-11 & 4.67E-13\tabularnewline
n13 & 0.00E+00 & 0.00E+00 & 0.00E+00 & 0.00E+00 & 0.00E+00 & 0.00E+00\tabularnewline
n14 & 1.45E-07 & 7.70E-05 & 1.45E-03 & 1.36E-04 & 1.63E-02 & 7.36E-03\tabularnewline
n15 & 4.74E-12 & 4.39E-09 & 5.08E-08 & 4.86E-09 & 7.44E-07 & 6.71E-07\tabularnewline
o14 & 0.00E+00 & 0.00E+00 & 0.00E+00 & 0.00E+00 & 0.00E+00 & 0.00E+00\tabularnewline
o15 & 0.00E+00 & 0.00E+00 & 0.00E+00 & 0.00E+00 & 0.00E+00 & 0.00E+00\tabularnewline
o16 & 1.11E-03 & 1.12E-03 & 1.17E-03 & 1.13E-03 & 5.13E-04 & 2.48E-04\tabularnewline
o17 & 5.73E-11 & 7.51E-07 & 5.38E-06 & 2.73E-06 & 1.22E-06 & 1.05E-06\tabularnewline
o18 & 1.05E-10 & 1.52E-09 & 1.54E-08 & 2.24E-09 & 3.84E-10 & 8.82E-10\tabularnewline
o19 & 0.00E+00 & 0.00E+00 & 0.00E+00 & 0.00E+00 & 0.00E+00 & 0.00E+00\tabularnewline
f17 & 0.00E+00 & 0.00E+00 & 0.00E+00 & 0.00E+00 & 0.00E+00 & 0.00E+00\tabularnewline
f18 & 0.00E+00 & 0.00E+00 & 0.00E+00 & 4.08E-39 & 3.12E-35 & 4.51E-30\tabularnewline
f19 & 4.22E-12 & 1.98E-08 & 1.29E-06 & 1.02E-07 & 1.41E-09 & 1.61E-10\tabularnewline
f20 & 0.00E+00 & 0.00E+00 & 0.00E+00 & 0.00E+00 & 0.00E+00 & 0.00E+00\tabularnewline
ne19 & 0.00E+00 & 0.00E+00 & 0.00E+00 & 0.00E+00 & 0.00E+00 & 0.00E+00\tabularnewline
ne20 & 9.49E-05 & 9.48E-05 & 9.82E-05 & 9.68E-05 & 2.24E-04 & 1.23E-04\tabularnewline
ne21 & 1.12E-08 & 2.20E-08 & 3.20E-07 & 3.05E-07 & 3.48E-07 & 9.51E-08\tabularnewline
ne22 & 9.85E-09 & 1.03E-05 & 5.27E-04 & 1.32E-04 & 1.04E-04 & 5.52E-06\tabularnewline
na21 & 0.00E+00 & 0.00E+00 & 0.00E+00 & 0.00E+00 & 0.00E+00 & 0.00E+00\tabularnewline
na22 & 0.00E+00 & 2.21E-31 & 3.33E-32 & 4.64E-15 & 3.41E-11 & 5.84E-11\tabularnewline
na23 & 5.25E-07 & 7.12E-07 & 1.09E-05 & 9.11E-07 & 2.12E-04 & 1.44E-05\tabularnewline
na24 & 0.00E+00 & 0.00E+00 & 0.00E+00 & 1.44E-42 & 6.28E-36 & 1.43E-30\tabularnewline
mg23 & 0.00E+00 & 0.00E+00 & 0.00E+00 & 0.00E+00 & 0.00E+00 & 0.00E+00\tabularnewline
mg24 & 5.67E-05 & 5.67E-05 & 5.66E-05 & 5.54E-05 & 2.17E-05 & 9.79E-07\tabularnewline
mg25 & 4.25E-08 & 8.31E-08 & 8.68E-06 & 1.32E-05 & 1.36E-04 & 8.03E-05\tabularnewline
mg26 & 4.10E-08 & 6.65E-08 & 6.23E-06 & 2.30E-05 & 3.49E-04 & 1.12E-04\tabularnewline
mg27 & 0.00E+00 & 0.00E+00 & 0.00E+00 & 0.00E+00 & 0.00E+00 & 0.00E+00\tabularnewline
al25 & 0.00E+00 & 0.00E+00 & 0.00E+00 & 0.00E+00 & 0.00E+00 & 0.00E+00\tabularnewline
al-6 & 0.00E+00 & 1.28E-08 & 7.09E-09 & 2.25E-09 & 8.59E-06 & 6.85E-06\tabularnewline
al{*}6 & 0.00E+00 & 0.00E+00 & 0.00E+00 & 0.00E+00 & 0.00E+00 & 0.00E+00\tabularnewline
al27 & 4.78E-07 & 4.96E-07 & 6.02E-07 & 1.15E-06 & 1.55E-05 & 1.07E-05\tabularnewline
al28 & 0.00E+00 & 0.00E+00 & 0.00E+00 & 0.00E+00 & 0.00E+00 & 0.00E+00\tabularnewline
si27 & 0.00E+00 & 0.00E+00 & 0.00E+00 & 0.00E+00 & 0.00E+00 & 0.00E+00\tabularnewline
si28 & 1.54E-04 & 1.54E-04 & 1.53E-04 & 1.53E-04 & 1.59E-04 & 1.54E-04\tabularnewline
si29 & 7.36E-08 & 8.72E-08 & 9.22E-07 & 7.95E-07 & 2.74E-06 & 9.56E-07\tabularnewline
si30 & 4.14E-08 & 4.27E-08 & 1.78E-07 & 1.88E-07 & 7.38E-07 & 3.40E-07\tabularnewline
si31 & 0.00E+00 & 0.00E+00 & 0.00E+00 & 0.00E+00 & 9.81E-45 & 4.83E-37\tabularnewline
si32 & 0.00E+00 & 1.51E-18 & 6.36E-12 & 6.12E-12 & 2.28E-10 & 3.69E-10\tabularnewline
si33 & 0.00E+00 & 0.00E+00 & 0.00E+00 & 0.00E+00 & 0.00E+00 & 0.00E+00\tabularnewline
p29 & 0.00E+00 & 0.00E+00 & 0.00E+00 & 0.00E+00 & 0.00E+00 & 0.00E+00\tabularnewline
p30 & 0.00E+00 & 0.00E+00 & 0.00E+00 & 0.00E+00 & 0.00E+00 & 0.00E+00\tabularnewline
p31 & 2.01E-08 & 2.08E-08 & 3.63E-07 & 8.13E-07 & 8.46E-06 & 2.96E-06\tabularnewline
p32 & 0.00E+00 & 5.84E-22 & 2.47E-15 & 2.37E-15 & 8.85E-14 & 1.43E-13\tabularnewline
p33 & 0.00E+00 & 0.00E+00 & 0.00E+00 & 0.00E+00 & 1.05E-18 & 7.20E-17\tabularnewline
p34 & 0.00E+00 & 0.00E+00 & 0.00E+00 & 0.00E+00 & 0.00E+00 & 0.00E+00\tabularnewline
s32 & 7.83E-06 & 7.83E-06 & 7.76E-06 & 7.86E-06 & 9.22E-06 & 8.29E-06\tabularnewline
s33 & 1.68E-08 & 1.72E-08 & 2.29E-08 & 2.16E-08 & 9.59E-08 & 5.05E-08\tabularnewline
s34 & 7.74E-08 & 2.38E-06 & 2.33E-06 & 2.35E-06 & 2.42E-06 & 2.41E-06\tabularnewline
s35 & 0.00E+00 & 1.21E-09 & 5.44E-08 & 4.59E-08 & 6.73E-07 & 2.61E-07\tabularnewline
fe56 & 5.49E-05 & 5.48E-05 & 5.34E-05 & 5.40E-05 & 5.21E-05 & 5.38E-05\tabularnewline
fe57 & 4.82E-07 & 5.08E-07 & 8.66E-07 & 5.48E-07 & 5.19E-07 & 5.54E-07\tabularnewline
fe58 & 3.27E-14 & 1.02E-08 & 6.14E-07 & 2.13E-07 & 1.66E-07 & 6.15E-08\tabularnewline
fe59 & 0.00E+00 & 4.63E-45 & 0.00E+00 & 5.79E-30 & 2.80E-45 & 1.50E-31\tabularnewline
fe60 & 0.00E+00 & 2.28E-12 & 3.02E-08 & 1.20E-07 & 1.20E-07 & 9.30E-08\tabularnewline
fe61 & 0.00E+00 & 0.00E+00 & 0.00E+00 & 0.00E+00 & 0.00E+00 & 0.00E+00\tabularnewline
co59 & 1.18E-07 & 1.19E-07 & 2.80E-07 & 1.98E-07 & 1.84E-07 & 1.43E-07\tabularnewline
co60 & 0.00E+00 & 1.99E-16 & 2.63E-12 & 1.05E-11 & 1.05E-11 & 8.10E-12\tabularnewline
co61 & 0.00E+00 & 0.00E+00 & 0.00E+00 & 0.00E+00 & 0.00E+00 & 1.83E-42\tabularnewline
ni58 & 2.40E-07 & 2.39E-07 & 2.30E-07 & 2.35E-07 & 2.27E-07 & 2.34E-07\tabularnewline
ni59 & 0.00E+00 & 1.89E-11 & 4.21E-11 & 1.61E-11 & 3.49E-13 & 5.64E-12\tabularnewline
ni60 & 2.06E-10 & 7.86E-10 & 1.51E-07 & 1.18E-07 & 3.65E-07 & 1.01E-07\tabularnewline
ni61 & 3.28E-13 & 1.51E-10 & 1.78E-07 & 3.53E-07 & 2.20E-06 & 8.47E-07\tabularnewline
ni62 & 2.87E-16 & 0.00E+00 & 9.83E-19 & 5.95E-19 & 4.71E-19 & 3.13E-19\tabularnewline
g\addtocounter{footnote}{-1}\footnotemark & 7.29E-14 & 1.16E-10 & 1.06E-06 & 6.57E-06 & 6.28E-05 & 2.47E-05\tabularnewline
\end{longtable}

\normalsize

\newpage

\scriptsize

\begin{longtable}[c]{>{\centering}m{1.1cm}>{\centering}m{1.6cm}|>{\centering}m{1.5cm}>{\centering}m{1.5cm}|>{\centering}m{1.5cm}>{\centering}m{1.5cm}}
\caption[GC Yields with alterations to mass loss and reaction rates]{Yields and initial composition for all the comparison GC models ([Fe/H]$=-1.4$). There are two groups of comparison models. One group has been calculated with Reimers' mass loss formula on the AGB instead of that of VW93, whilst the other has been calculated using some updated rates from NACRE. All species in the network are listed except for neutrons. Abundances are in mass fraction, normalised to 1.0. The remnant masses (white dwarf masses) are in brackets below the initial stellar masses in the table header.}
\label{table-allGCyields-Reim-Nacre}\\ 

\hline

\multicolumn{2}{c|}{ }&

\multicolumn{2}{c|}{{\bfseries Reimers' on AGB}}&

\multicolumn{2}{c}{{\bfseries NACRE Rates}} \tabularnewline

\textbf{Nuclide}& 

\textbf{Initial}&

\textbf{2.5 M$_{\odot}$\linebreak (0.681)}& 

\textbf{5.0 M$_{\odot}$\linebreak (0.914)}& 

\textbf{2.5 M$_{\odot}$\linebreak (0.676)}&

\textbf{5.0 M$_{\odot}$\linebreak (0.915)}\tabularnewline 

\hline \hline

\endfirsthead

\multicolumn{6}{c}{{\bfseries \tablename\ \thetable{} -- continued from previous page}} \tabularnewline 

\hline  

\multicolumn{2}{c|}{ }&

\multicolumn{2}{c|}{{\bfseries Reimers' on AGB}}&

\multicolumn{2}{c}{{\bfseries NACRE Rates}} \tabularnewline

\textbf{Nuclide}& 

\textbf{Initial}& 

\textbf{2.5 M$_{\odot}$\linebreak (0.681)}& 

\textbf{5.0 M$_{\odot}$\linebreak (0.914)}& 

\textbf{2.5 M$_{\odot}$\linebreak (0.676)}&

\textbf{5.0 M$_{\odot}$\linebreak (0.915)}\tabularnewline 

\hline \hline

\endhead

\hline \multicolumn{6}{r}{{Continued on next page...}} \\

\endfoot

\hline \hline 

\endlastfoot

h1  &  0.76860  &  0.75389  &  0.66114  &  0.72009  &  0.61272\tabularnewline
h2 & 2.26E-20 & 5.99E-18 & 4.35E-18 & 5.51E-18 & 3.71E-18\tabularnewline
he3 & 1.27E-09 & 1.57E-04 & 3.73E-06 & 1.48E-04 & 3.31E-07\tabularnewline
he4 & 0.22970 & 0.24173 & 0.32500 & 0.26588 & 0.36692\tabularnewline
li7 & 1.15E-14 & 2.50E-11 & 3.24E-09 & 2.46E-10 & 0.00E+00\tabularnewline
be7 & 0.00E+00 & 5.51E-18 & 6.56E-12 & 3.97E-19 & 1.31E-13\tabularnewline
b8 & 0.00E+00 & 6.92E-28 & 3.62E-22 & 5.75E-24 & 0.00E+00\tabularnewline
c12 & 2.10E-04 & 2.58E-03 & 3.49E-03 & 1.16E-02 & 2.03E-03\tabularnewline
c13 & 4.68E-10 & 4.67E-06 & 2.27E-04 & 5.16E-06 & 2.35E-04\tabularnewline
c14 & 0.00E+00 & 5.92E-11 & 3.51E-11 & 2.84E-11 & 9.27E-14\tabularnewline
n13 & 0.00E+00 & 0.00E+00 & 1.54E-44 & 0.00E+00 & 0.00E+00\tabularnewline
n14 & 1.45E-07 & 1.09E-04 & 8.40E-03 & 1.30E-04 & 1.62E-02\tabularnewline
n15 & 4.74E-12 & 5.07E-09 & 3.75E-07 & 7.55E-09 & 7.53E-07\tabularnewline
o14 & 0.00E+00 & 0.00E+00 & 0.00E+00 & 0.00E+00 & 0.00E+00\tabularnewline
o15 & 0.00E+00 & 0.00E+00 & 0.00E+00 & 0.00E+00 & 0.00E+00\tabularnewline
o16 & 1.11E-03 & 1.12E-03 & 7.32E-04 & 1.17E-03 & 4.96E-04\tabularnewline
o17 & 5.73E-11 & 6.66E-06 & 1.60E-06 & 6.27E-06 & 1.17E-06\tabularnewline
o18 & 1.05E-10 & 3.75E-09 & 9.14E-10 & 2.86E-09 & 3.61E-10\tabularnewline
o19 & 0.00E+00 & 0.00E+00 & 0.00E+00 & 0.00E+00 & 0.00E+00\tabularnewline
f17 & 0.00E+00 & 0.00E+00 & 0.00E+00 & 0.00E+00 & 0.00E+00\tabularnewline
f18 & 0.00E+00 & 0.00E+00 & 1.93E-35 & 0.00E+00 & 3.94E-35\tabularnewline
f19 & 4.22E-12 & 6.40E-08 & 2.10E-09 & 1.58E-06 & 1.63E-09\tabularnewline
f20 & 0.00E+00 & 0.00E+00 & 0.00E+00 & 0.00E+00 & 0.00E+00\tabularnewline
ne19 & 0.00E+00 & 0.00E+00 & 0.00E+00 & 0.00E+00 & 0.00E+00\tabularnewline
ne20 & 9.49E-05 & 9.48E-05 & 1.31E-04 & 9.69E-05 & 1.98E-04\tabularnewline
ne21 & 1.12E-08 & 2.20E-08 & 7.33E-07 & 3.08E-07 & 2.78E-07\tabularnewline
ne22 & 9.85E-09 & 2.89E-05 & 1.35E-04 & 5.83E-04 & 1.01E-04\tabularnewline
na21 & 0.00E+00 & 0.00E+00 & 0.00E+00 & 0.00E+00 & 0.00E+00\tabularnewline
na22 & 0.00E+00 & 2.42E-26 & 2.68E-11 & 0.00E+00 & 2.96E-11\tabularnewline
na23 & 5.25E-07 & 7.76E-07 & 8.93E-05 & 2.28E-06 & 1.45E-04\tabularnewline
na24 & 0.00E+00 & 0.00E+00 & 8.72E-36 & 0.00E+00 & 7.51E-36\tabularnewline
mg23 & 0.00E+00 & 0.00E+00 & 0.00E+00 & 0.00E+00 & 0.00E+00\tabularnewline
mg24 & 5.67E-05 & 5.66E-05 & 3.52E-05 & 6.11E-05 & 7.05E-05\tabularnewline
mg25 & 4.25E-08 & 2.59E-07 & 9.04E-05 & 1.06E-05 & 1.55E-04\tabularnewline
mg26 & 4.10E-08 & 1.38E-07 & 2.43E-04 & 8.88E-06 & 3.73E-04\tabularnewline
mg27 & 0.00E+00 & 0.00E+00 & 0.00E+00 & 0.00E+00 & 0.00E+00\tabularnewline
al25 & 0.00E+00 & 0.00E+00 & 0.00E+00 & 0.00E+00 & 0.00E+00\tabularnewline
al-6 & 0.00E+00 & 7.19E-09 & 3.53E-06 & 1.27E-08 & 1.20E-05\tabularnewline
al{*}6 & 0.00E+00 & 0.00E+00 & 0.00E+00 & 0.00E+00 & 0.00E+00\tabularnewline
al27 & 4.78E-07 & 5.22E-07 & 5.83E-06 & 6.66E-07 & 1.59E-05\tabularnewline
al28 & 0.00E+00 & 0.00E+00 & 0.00E+00 & 0.00E+00 & 0.00E+00\tabularnewline
si27 & 0.00E+00 & 0.00E+00 & 0.00E+00 & 0.00E+00 & 0.00E+00\tabularnewline
si28 & 1.54E-04 & 1.54E-04 & 1.54E-04 & 1.53E-04 & 1.56E-04\tabularnewline
si29 & 7.36E-08 & 1.25E-07 & 1.80E-06 & 1.06E-06 & 3.51E-06\tabularnewline
si30 & 4.14E-08 & 4.72E-08 & 4.95E-07 & 4.26E-07 & 4.90E-06\tabularnewline
si31 & 0.00E+00 & 0.00E+00 & 1.20E-44 & 0.00E+00 & 1.34E-44\tabularnewline
si32 & 0.00E+00 & 4.71E-17 & 7.24E-10 & 1.09E-12 & 1.79E-10\tabularnewline
si33 & 0.00E+00 & 0.00E+00 & 0.00E+00 & 0.00E+00 & 0.00E+00\tabularnewline
p29 & 0.00E+00 & 0.00E+00 & 0.00E+00 & 0.00E+00 & 0.00E+00\tabularnewline
p30 & 0.00E+00 & 0.00E+00 & 0.00E+00 & 0.00E+00 & 0.00E+00\tabularnewline
p31 & 2.01E-08 & 2.46E-08 & 4.85E-06 & 1.17E-07 & 6.98E-06\tabularnewline
p32 & 0.00E+00 & 1.83E-20 & 2.81E-13 & 4.22E-16 & 6.94E-14\tabularnewline
p33 & 0.00E+00 & 0.00E+00 & 6.53E-19 & 3.04E-39 & 5.76E-19\tabularnewline
p34 & 0.00E+00 & 0.00E+00 & 0.00E+00 & 0.00E+00 & 0.00E+00\tabularnewline
s32 & 7.83E-06 & 7.83E-06 & 8.60E-06 & 7.75E-06 & 8.83E-06\tabularnewline
s33 & 1.68E-08 & 1.79E-08 & 5.18E-08 & 2.35E-08 & 8.55E-08\tabularnewline
s34 & 7.74E-08 & 2.38E-06 & 2.40E-06 & 2.34E-06 & 2.38E-06\tabularnewline
s35 & 0.00E+00 & 3.97E-09 & 4.03E-07 & 4.72E-08 & 5.32E-07\tabularnewline
fe56 & 5.49E-05 & 5.47E-05 & 5.31E-05 & 5.36E-05 & 5.22E-05\tabularnewline
fe57 & 4.82E-07 & 5.54E-07 & 5.17E-07 & 8.88E-07 & 5.25E-07\tabularnewline
fe58 & 3.27E-14 & 4.14E-08 & 1.20E-07 & 5.42E-07 & 1.72E-07\tabularnewline
fe59 & 0.00E+00 & 0.00E+00 & 0.00E+00 & 1.49E-37 & 1.13E-23\tabularnewline
fe60 & 0.00E+00 & 1.80E-10 & 1.26E-07 & 1.81E-08 & 1.25E-07\tabularnewline
fe61 & 0.00E+00 & 0.00E+00 & 0.00E+00 & 0.00E+00 & 0.00E+00\tabularnewline
co59 & 1.18E-07 & 1.23E-07 & 1.65E-07 & 2.47E-07 & 1.87E-07\tabularnewline
co60 & 0.00E+00 & 1.56E-14 & 1.10E-11 & 1.58E-12 & 1.09E-11\tabularnewline
co61 & 0.00E+00 & 0.00E+00 & 0.00E+00 & 0.00E+00 & 0.00E+00\tabularnewline
ni58 & 2.40E-07 & 2.38E-07 & 2.32E-07 & 2.30E-07 & 2.27E-07\tabularnewline
ni59 & 0.00E+00 & 7.03E-11 & 5.98E-12 & 7.97E-11 & 6.17E-13\tabularnewline
ni60 & 2.06E-10 & 3.07E-09 & 1.93E-07 & 1.06E-07 & 3.73E-07\tabularnewline
ni61 & 3.28E-13 & 1.01E-09 & 1.36E-06 & 1.05E-07 & 2.13E-06\tabularnewline
ni62 & 2.87E-16 & 0.00E+00 & 0.00E+00 & 0.00E+00 & 0.00E+00\tabularnewline
g\addtocounter{footnote}{-1}\footnotemark & 7.29E-14 & 1.08E-09 & 4.13E-05 & 3.72E-07 & 4.43E-05\tabularnewline
\end{longtable}

\normalsize

\newpage

\scriptsize

\begin{longtable}[c]{>{\centering}m{1.1cm}>{\centering}m{1.5cm}|>{\centering}m{1.3cm}>{\centering}m{1.2cm}>{\centering}m{1.2cm}>{\centering}m{1.2cm}}
\caption[All Zero-3DUP GC Yields]{Yields and initial composition for all the  GC models without 3DUP ([Fe/H]$=-1.4$). All species in the network are listed except for neutrons. Abundances are in mass fraction, normalised to 1.0. The remnant masses (white dwarf masses) are in brackets below the initial stellar masses in the table header.}
\label{table-allGCyields-Zero3DUP}\\ 

\hline

\textbf{Nuclide}& 

\textbf{Initial}&

\textbf{2.5 M$_{\odot}$\linebreak (0.869)}& 

\textbf{3.5 M$_{\odot}$\linebreak (0.996)}&

\textbf{5.0 M$_{\odot}$\linebreak (1.078)}&

\textbf{6.5 M$_{\odot}$\linebreak (1.132)}\tabularnewline 

\hline \hline

\endfirsthead

\multicolumn{6}{c}{{\bfseries \tablename\ \thetable{} -- continued from previous page}} \tabularnewline 

\hline  

\textbf{Nuclide}& 

\textbf{Initial}& 

\textbf{2.5 M$_{\odot}$\linebreak (0.869)}& 

\textbf{3.5 M$_{\odot}$\linebreak (0.996)}&

\textbf{5.0 M$_{\odot}$\linebreak (1.078)}&

\textbf{6.5 M$_{\odot}$\linebreak (1.132)}\tabularnewline 

\hline \hline

\endhead

\hline \multicolumn{6}{r}{{Continued on next page...}} \\

\endfoot

\hline \hline 

\endlastfoot

h1  &  0.76860  &  0.75924  &  0.76286  &  0.69852  &  0.66393\tabularnewline
h2 & 2.26E-20 & 5.88E-18 & 5.29E-18 & 4.20E-18 & 3.84E-18\tabularnewline
he3 & 1.27E-09 & 1.62E-04 & 1.15E-06 & 3.72E-07 & 1.93E-07\tabularnewline
he4 & 0.22970 & 0.23889 & 0.23541 & 0.29985 & 0.33447\tabularnewline
li7 & 1.15E-14 & 6.02E-12 & 5.41E-10 & 3.44E-11 & 2.84E-12\tabularnewline
be7 & 0.00E+00 & 4.69E-17 & 2.71E-14 & 2.08E-14 & 7.23E-14\tabularnewline
b8 & 0.00E+00 & 1.92E-29 & 2.75E-18 & 2.42E-27 & 3.28E-26\tabularnewline
c12 & 2.10E-04 & 1.20E-04 & 2.18E-05 & 4.89E-05 & 5.32E-05\tabularnewline
c13 & 4.68E-10 & 4.33E-06 & 4.82E-06 & 1.22E-05 & 1.44E-05\tabularnewline
c14 & 0.00E+00 & 8.17E-25 & 7.48E-14 & 4.36E-14 & 4.89E-13\tabularnewline
n13 & 0.00E+00 & 0.00E+00 & 0.00E+00 & 0.00E+00 & 0.00E+00\tabularnewline
n14 & 1.45E-07 & 1.23E-04 & 5.73E-04 & 1.16E-03 & 1.14E-03\tabularnewline
n15 & 4.74E-12 & 4.80E-09 & 2.10E-08 & 8.32E-08 & 1.52E-07\tabularnewline
o14 & 0.00E+00 & 0.00E+00 & 0.00E+00 & 0.00E+00 & 0.00E+00\tabularnewline
o15 & 0.00E+00 & 0.00E+00 & 0.00E+00 & 0.00E+00 & 0.00E+00\tabularnewline
o16 & 1.11E-03 & 1.08E-03 & 7.48E-04 & 3.34E-05 & 1.59E-05\tabularnewline
o17 & 5.73E-11 & 6.89E-06 & 1.86E-06 & 1.47E-07 & 8.15E-08\tabularnewline
o18 & 1.05E-10 & 2.79E-09 & 1.01E-09 & 2.34E-10 & 2.80E-10\tabularnewline
o19 & 0.00E+00 & 0.00E+00 & 0.00E+00 & 0.00E+00 & 0.00E+00\tabularnewline
f17 & 0.00E+00 & 0.00E+00 & 0.00E+00 & 0.00E+00 & 0.00E+00\tabularnewline
f18 & 0.00E+00 & 0.00E+00 & 0.00E+00 & 0.00E+00 & 0.00E+00\tabularnewline
f19 & 4.22E-12 & 5.41E-10 & 4.33E-10 & 2.12E-11 & 1.50E-11\tabularnewline
f20 & 0.00E+00 & 0.00E+00 & 0.00E+00 & 0.00E+00 & 0.00E+00\tabularnewline
ne19 & 0.00E+00 & 0.00E+00 & 0.00E+00 & 0.00E+00 & 0.00E+00\tabularnewline
ne20 & 9.49E-05 & 9.49E-05 & 9.47E-05 & 9.50E-05 & 9.49E-05\tabularnewline
ne21 & 1.12E-08 & 1.70E-08 & 1.90E-09 & 1.35E-09 & 1.38E-09\tabularnewline
ne22 & 9.85E-09 & 1.04E-08 & 3.56E-08 & 5.19E-08 & 7.99E-08\tabularnewline
na21 & 0.00E+00 & 0.00E+00 & 0.00E+00 & 0.00E+00 & 0.00E+00\tabularnewline
na22 & 0.00E+00 & 4.39E-15 & 1.43E-12 & 4.87E-11 & 1.06E-10\tabularnewline
na23 & 5.25E-07 & 5.36E-07 & 1.05E-06 & 6.04E-07 & 4.37E-07\tabularnewline
na24 & 0.00E+00 & 0.00E+00 & 0.00E+00 & 0.00E+00 & 9.39E-44\tabularnewline
mg23 & 0.00E+00 & 0.00E+00 & 0.00E+00 & 0.00E+00 & 0.00E+00\tabularnewline
mg24 & 5.67E-05 & 5.67E-05 & 5.01E-05 & 3.72E-07 & 2.81E-07\tabularnewline
mg25 & 4.25E-08 & 4.48E-08 & 6.64E-06 & 3.66E-05 & 2.79E-05\tabularnewline
mg26 & 4.10E-08 & 4.73E-08 & 1.12E-07 & 4.12E-06 & 3.43E-06\tabularnewline
mg27 & 0.00E+00 & 0.00E+00 & 0.00E+00 & 0.00E+00 & 0.00E+00\tabularnewline
al25 & 0.00E+00 & 0.00E+00 & 0.00E+00 & 0.00E+00 & 0.00E+00\tabularnewline
al-6 & 0.00E+00 & 5.40E-09 & 1.82E-07 & 7.17E-06 & 3.26E-06\tabularnewline
al{*}6 & 0.00E+00 & 0.00E+00 & 0.00E+00 & 0.00E+00 & 0.00E+00\tabularnewline
al27 & 4.78E-07 & 4.99E-07 & 4.94E-07 & 1.14E-05 & 1.85E-05\tabularnewline
al28 & 0.00E+00 & 0.00E+00 & 0.00E+00 & 0.00E+00 & 0.00E+00\tabularnewline
si27 & 0.00E+00 & 0.00E+00 & 0.00E+00 & 0.00E+00 & 0.00E+00\tabularnewline
si28 & 1.54E-04 & 1.54E-04 & 1.54E-04 & 1.56E-04 & 1.63E-04\tabularnewline
si29 & 7.36E-08 & 7.36E-08 & 7.74E-08 & 7.80E-08 & 8.05E-08\tabularnewline
si30 & 4.14E-08 & 4.14E-08 & 4.20E-08 & 4.21E-08 & 4.20E-08\tabularnewline
si31 & 0.00E+00 & 0.00E+00 & 0.00E+00 & 0.00E+00 & 0.00E+00\tabularnewline
si32 & 0.00E+00 & 7.26E-37 & 1.13E-15 & 1.40E-16 & 2.23E-14\tabularnewline
si33 & 0.00E+00 & 0.00E+00 & 0.00E+00 & 0.00E+00 & 0.00E+00\tabularnewline
p29 & 0.00E+00 & 0.00E+00 & 0.00E+00 & 0.00E+00 & 0.00E+00\tabularnewline
p30 & 0.00E+00 & 0.00E+00 & 0.00E+00 & 0.00E+00 & 0.00E+00\tabularnewline
p31 & 2.01E-08 & 2.01E-08 & 2.21E-08 & 2.22E-08 & 2.14E-08\tabularnewline
p32 & 0.00E+00 & 2.82E-40 & 4.40E-19 & 5.44E-20 & 8.65E-18\tabularnewline
p33 & 0.00E+00 & 2.09E-43 & 2.10E-31 & 1.53E-26 & 4.96E-24\tabularnewline
p34 & 0.00E+00 & 0.00E+00 & 0.00E+00 & 0.00E+00 & 0.00E+00\tabularnewline
s32 & 7.83E-06 & 7.83E-06 & 7.83E-06 & 7.83E-06 & 7.83E-06\tabularnewline
s33 & 1.68E-08 & 1.68E-08 & 1.68E-08 & 1.68E-08 & 1.71E-08\tabularnewline
s34 & 7.74E-08 & 2.38E-06 & 2.38E-06 & 2.38E-06 & 2.38E-06\tabularnewline
s35 & 0.00E+00 & 3.80E-19 & 1.83E-10 & 2.57E-10 & 1.77E-09\tabularnewline
fe56 & 5.49E-05 & 5.49E-05 & 5.49E-05 & 5.49E-05 & 5.48E-05\tabularnewline
fe57 & 4.82E-07 & 4.82E-07 & 4.83E-07 & 4.85E-07 & 5.32E-07\tabularnewline
fe58 & 3.27E-14 & 3.27E-14 & 1.88E-09 & 1.95E-09 & 4.03E-09\tabularnewline
fe59 & 0.00E+00 & 0.00E+00 & 1.51E-31 & 1.00E-28 & 1.68E-27\tabularnewline
fe60 & 0.00E+00 & 6.03E-38 & 7.92E-11 & 1.61E-10 & 3.18E-10\tabularnewline
fe61 & 0.00E+00 & 0.00E+00 & 0.00E+00 & 0.00E+00 & 0.00E+00\tabularnewline
co59 & 1.18E-07 & 1.18E-07 & 1.19E-07 & 1.19E-07 & 1.19E-07\tabularnewline
co60 & 0.00E+00 & 1.12E-25 & 6.92E-15 & 1.40E-14 & 2.77E-14\tabularnewline
co61 & 0.00E+00 & 0.00E+00 & 0.00E+00 & 0.00E+00 & 0.00E+00\tabularnewline
ni58 & 2.40E-07 & 2.40E-07 & 2.40E-07 & 2.40E-07 & 2.39E-07\tabularnewline
ni59 & 0.00E+00 & 2.32E-20 & 1.60E-13 & 3.66E-13 & 3.24E-13\tabularnewline
ni60 & 2.06E-10 & 2.06E-10 & 7.41E-10 & 7.38E-10 & 7.34E-10\tabularnewline
ni61 & 3.28E-13 & 3.28E-13 & 8.10E-10 & 1.04E-09 & 7.11E-10\tabularnewline
ni62 & 2.87E-16 & 5.02E-19 & 4.75E-19 & 3.98E-19 & 2.83E-19\tabularnewline
g\addtocounter{footnote}{-1}\footnotemark & 7.29E-14 & 7.01E-11 & 5.82E-09 & 9.19E-09 & 8.89E-09\tabularnewline
\end{longtable}

\normalsize

\end{center}

\chapter{Miscellaneous Material}

\section{Surface Opacity Uncertainty in $Z=0$ AGB Stars\label{section-LowTOpacUncertainties}}

\begin{figure}
\begin{centering}
\includegraphics[width=0.85\columnwidth]{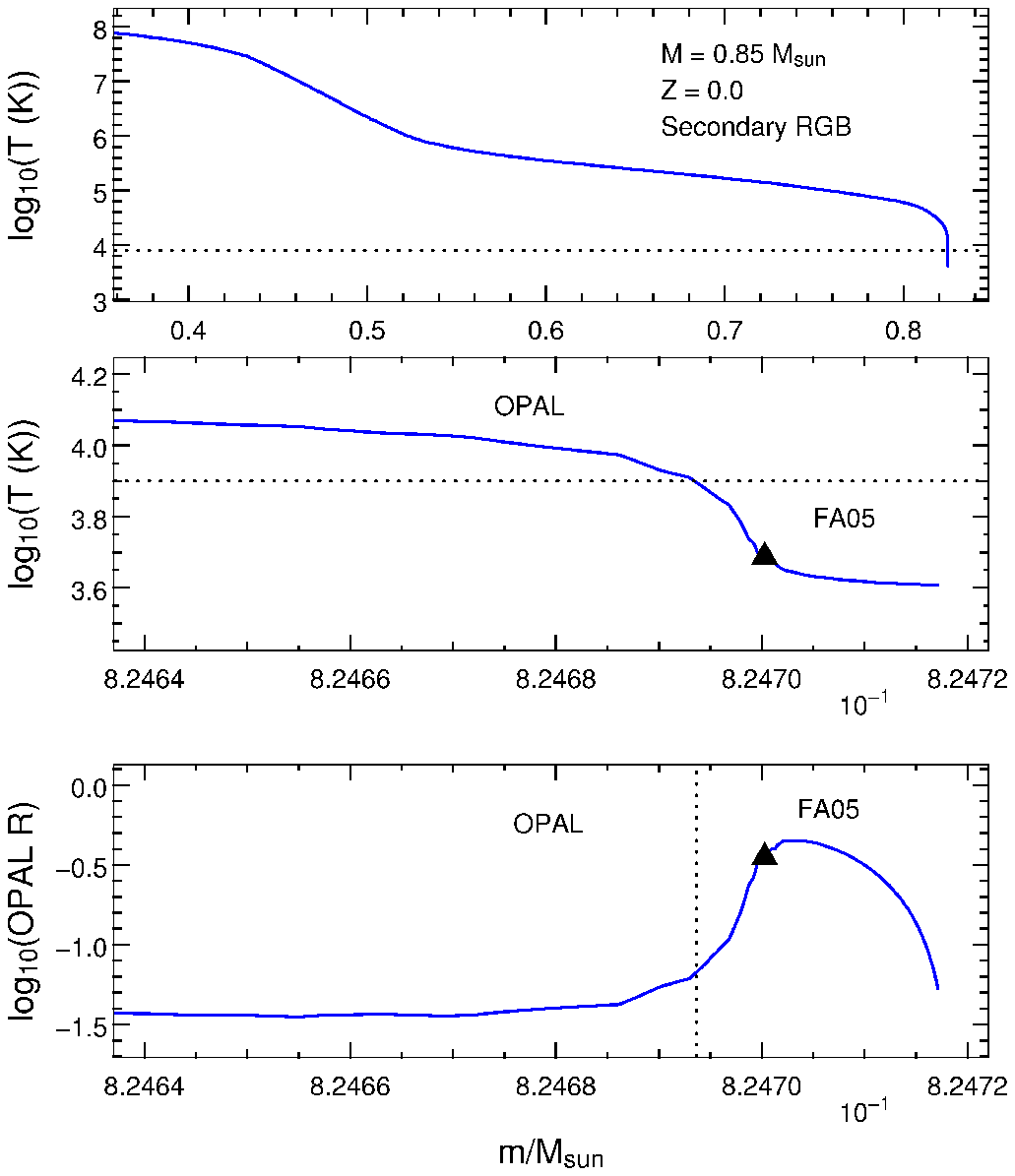}
\par\end{centering}
\caption{The run of temperature (top two panels) and density (bottom panel)
against mass for our $M=0.85$ $M_{\odot}$, $Z=0$ star during the
secondary red giant branch. The top panel is a low resolution plot,
showing the temperature for the outer $50\%$ of the star, whilst
the bottom two panels zoom in on the very outermost layers. The dotted
lines mark where we switch between the OPAL tables (\citealt{1996ApJ...464..943I})
and the low temperature opacity tables from \citet{2005ApJ...623..585F}
(FA05). This occurs at $\log(T)=3.9$, ie. 8000 K. As can be seen
the low temperature tables only come into effect in the extreme outer
layers of the model -- covering a mass range of only $\sim10^{-5}$
M$_{\odot}$. The triangles denote the point at which the surface
is defined (using optical depth arguments). It is the temperature
at these points that we plot in our HR diagrams. \label{fig-m0.85z0y245-OpacPltStar-SecondaryRGB}}
\end{figure}

We noted in Section \vref{opacmods} that the updated low temperature
opacities do not contain tables with varying amounts of C and O enhancements,
as the OPAL opacities do. This means that any enhancement over the
initial abundances (via dredge-up for example) is not taken into account.
For this reason we chose to use the low temperature tables over a
minimum temperature range only, making use of the OPAL tables down
to a temperature of 8000 K. Thus the non-inclusion of the enhancements
of CNO elements (for example) only affects the very outermost layers
of cool envelopes. To illustrate this we take our $\textrm{M}=0.85$
M$_{\odot}$, $Z=0$ model as an example. In Figure \ref{fig-m0.85z0y245-OpacPltStar-SecondaryRGB}
we show the run of temperature and density in the outer layers of
a model on the secondary RGB (see Section \vref{section-m0.85z0y245-SRGBandCHeB}
for the evolutionary background). It can be seen that the low temperature
opacities (from \citealt{2005ApJ...623..585F}) are only used in the
extreme outer layers of the model -- over a mass range of $\sim10^{-5}$
$M_{\odot}$. 

In an attempt to gauge the effect of neglecting the CNO enrichment
through post-dual-flash dredge-up we have run a few tests on this
star. The tests involved re-reading the low temperature opacity tables
and interpolating in them using a new $Z$, during run-time. This
was done if the $Z$ value of the tables stored in memory differed
by more than a factor of two from the $Z$ value at the mesh point
currently requiring an opacity value. In this case the surface metallicity
increases from $Z=0$ to $Z=10^{-3}$ very quickly, so the routine
was only called a few times. The new $Z$ value was approximated in
three different ways. For case 1 we used $Z=1-X-Y$, in case 2 we
assumed $Z$ was composed of only C, N and O, whilst in case 3 we
assumed $Z$ was entirely $^{16}$O (all mass fraction abundances
were naturally from the model itself). The reason for this variety
of choices is because it is quite uncertain as to which is the dominant
factor/species in determining the opacity at such low temperatures.
Indeed, the dominant molecular species (in terms of opacity contribution
at least) are all oxides (H$_{2}$O and TiO are the main ones, assuming
scaled-solar composition -- see Fig. 4 in \citealt{2005ApJ...623..585F}).
This may imply that an oxygen-poor composition may not be as opaque
as an oxygen-rich one, hence the $Z=Z_{^{16}\textrm{O}}$ test. We
note however that since opacity is very `non-linear' in terms of varying
chemical compositions it is perilous to assume that any of the assumptions
for $Z$ that are used to interpolate in the \emph{scaled-solar} mixes
provided will necessarily be close to reality. Indeed, the composition
of the envelope is very non-solar.

\begin{figure}
\begin{centering}
\includegraphics[width=0.9\columnwidth]{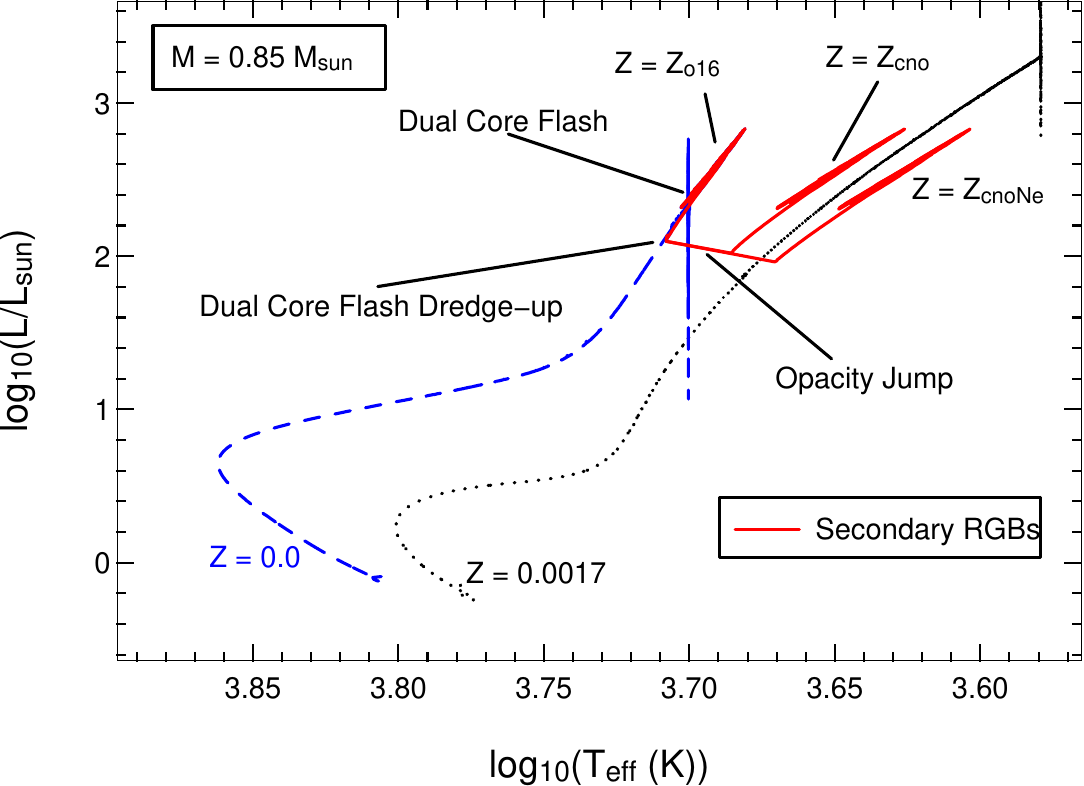}
\par\end{centering}
\caption{Evolution of the 0.85 M$_{\odot},$ $Z=0$ model in the HR diagram,
from the main sequence to the end of the red giant branch. Also shown
is the $Z=0.0017$ model for reference. The vertical lines (a numerical
artefact of this difficult phase of evolution) mark the core He/He-H
flashes. Note that the $Z=0$ model exhibits an extra/secondary RGB
above the location of the dual core flash, a feature peculiar to very
low metallicity, low mass stars. As mentioned in Section \vref{section-m0.85z0y245-DualCoreFlash}
the $Z=0$ model experiences some dredge-up of CNO-enriched material
just after the dual core flash. This means that the surface is enriched
(from $Z=0$ to $Z\sim10^{-3}$ in this case) on the secondary RGB.
The solid lines (red) represent three different evolutionary paths
of the SRGB, each using a different makeup of $Z$ for the purposes
of the low-temperature opacity calculation (see text for details).
As can be seen, the different approximations -- and consequently
the different scaled-solar low-temperature opacities used -- lead
to differing colours of the SRGB (and also the AGB). See text for
more details. \label{fig-m0.8z0y245-HRD-opacityIssue-SecondaryRGBs}}
\end{figure}

With this in mind we show the results of our tests on the secondary
RGB in Figure \ref{fig-m0.8z0y245-HRD-opacityIssue-SecondaryRGBs}.
It can be seen that an increase in scaled-solar Z used to obtain new
opacities for the surface has a marked effect as $Z$ increases. As
soon as the surface is polluted the model jumps discontinuously to
cooler temperatures, due to the increase in surface opacity. If we
use $Z=Z_{^{16}\textrm{O}}$ (under the assumption that the opacity
is dependent on oxide formation) then the temperature/opacity jump
becomes negligible. We note here however that the $^{16}$O abundance
obtained from the SEV code may be too low as some (energetically negligible)
reactions which create $^{16}$O are not included. We also note that
the models with cooler surfaces continue to be cooler on the AGB.
This shift in surface temperature is most likely the main effect of
neglecting the CNO enhancements, as the mass range in question is
of order $10^{-5}$ M$_{\odot}$, as mentioned above. 

\begin{figure}
\begin{centering}
\includegraphics[width=0.8\columnwidth]{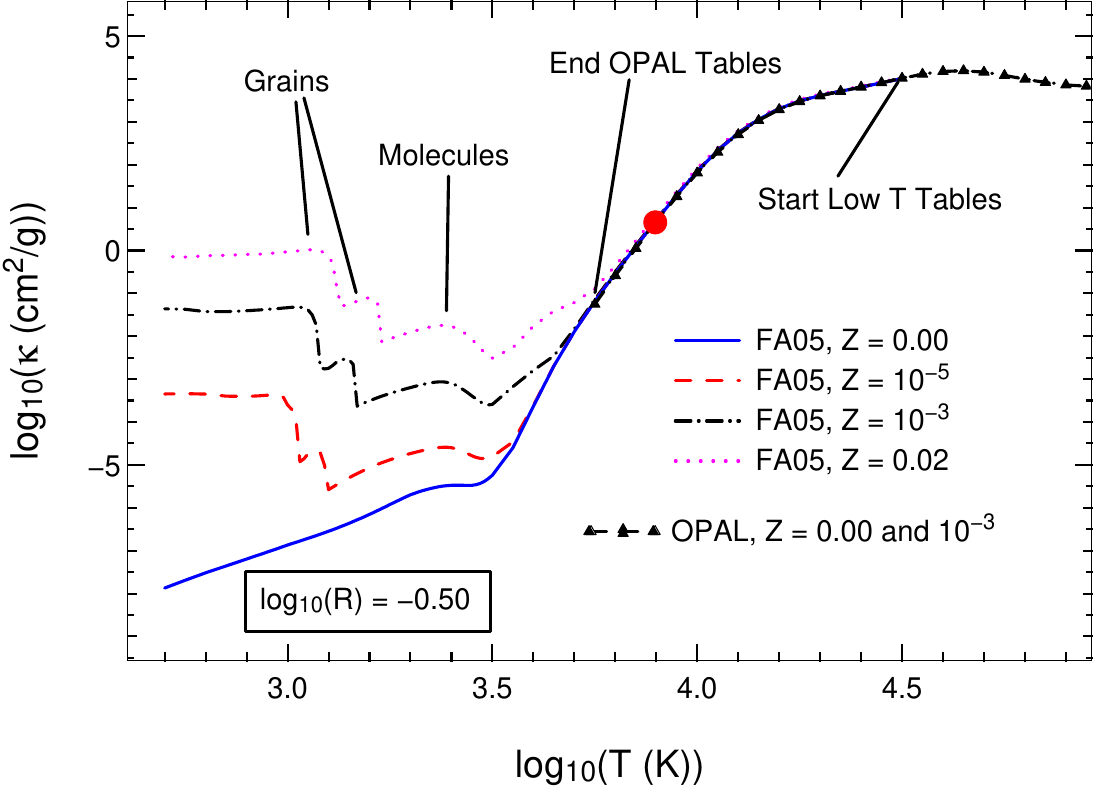}
\par\end{centering}
\caption{The run of opacity versus temperature at a log(R) value of -0.5, representative
of the outermost layers of the model in Figure \ref{fig-m0.85z0y245-OpacPltStar-SecondaryRGB}.
We show the low temperature opacities (from \citet{2005ApJ...623..585F},
FA05) for a range of metallicities and also the low temperature end
of the OPAL opacities (\citealt{1996ApJ...464..943I}). It can be
seen that the two sets overlap. The large dot (red) marks the point
at which we change from one set to the next in the SEV code. The tables
match up very well in this region. An interesting observation is that
the the run of opacities at all the metallicities shown are practically
identical to that of the $Z=0$ case for temperatures above 8000 K
-- the metals have no tangible effect at these temperatures and densities.
The story is different at low temperatures though, when molecules
and grains begin to form and dominate the opacity. See text for further
discussion. \label{fig-LowTOpacs-wideAngle}}
\end{figure}

We now delve into the issue a bit more deeply. Shown in Figure \ref{fig-LowTOpacs-wideAngle}
is the run of opacities around the region of interest, corresponding
to the outermost layers of the SRGB model in Figure \ref{fig-m0.85z0y245-OpacPltStar-SecondaryRGB}.
It is interesting to note that the opacity at these temperatures and
densities is wholly dominated by H, such that material with up to
$Z=0.02$ has essentially identical opacity to $Z=0$ material. The
situation does however change below $\sim8000$ K, when molecules
and then grains begin to form and dominate the opacity. 

\begin{figure}
\begin{centering}
\includegraphics[width=0.8\columnwidth]{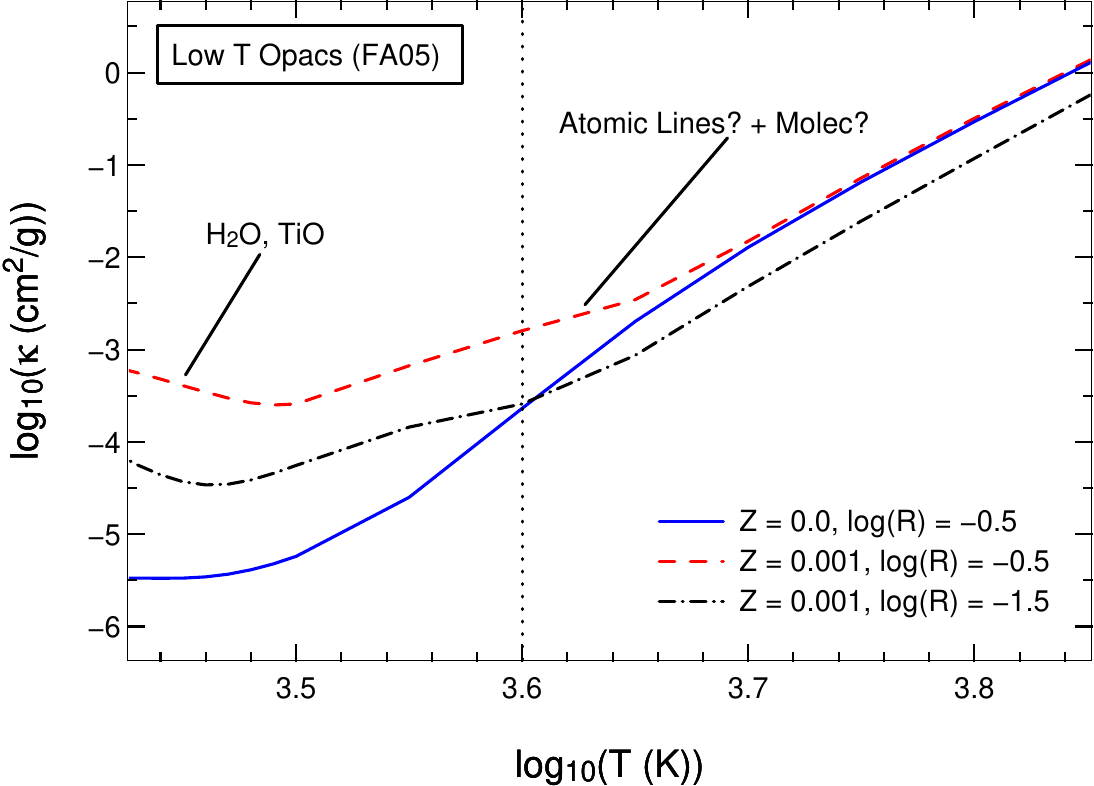}
\par\end{centering}
\caption{Zooming in on Figure \ref{fig-LowTOpacs-wideAngle} but with a reduced
number of cases plotted for clarity. We choose to plot $Z=0$ and
$Z=0.001$ because these are the metallicities of the surface before
and after the post-dual-core-flash dredge-up in the model of Figure
\ref{fig-m0.8z0y245-HRD-opacityIssue-SecondaryRGBs}. We also plot
the opacity at $Z=0.001$ for a log(R) of -1.5 which is representative
of the very outer point of the model (as opposed to the point defined
as the surface, see Figure \ref{fig-m0.85z0y245-OpacPltStar-SecondaryRGB}).
The temperature of this point is represented by the vertical dotted
line. Note that this is the coolest temperature reached in the envelope.
It can be seen that the regime of interest is just at the turn-off
from the $Z=0$ curve. As noted in the text we are uncertain whether
this turn-off is due mainly to atomic lines or molecular lines, or
how sensitive it is to deviations from scaled-solar composition. \label{fig-LowTOpacs-SRGB-zoom}}
\end{figure}

In Figure \ref{fig-LowTOpacs-SRGB-zoom} we zoom into the region of
interest. It is apparent that the region of opacity of most interest
is at the intersection of two opacity regimes -- the opacity from
H is declining rapidly whilst the molecules are beginning to form.
However we note that the main molecular contribution (from water vapour
and TiO) is yet to peak. It is difficult to ascertain from the \citet{2005ApJ...623..585F}
or \citet{1994ApJ...437..879A} papers whether the dominant contribution
is from \emph{atomic lines} or molecules in this very particular region
(where the opacity just diverges from the $Z=0$ curve). It seems
that our region of interest lies between the figures of monochromatic
opacity given in the aforementioned papers. We suggest that our region
of interest is the equivalent of 3500 K in Figure 4 of \citet{2005ApJ...623..585F},
which is a selection of plots for solar metallicity and a lower value
of log(R). Our 3500 K value lies between the their plots of 2000 K
and 5000 K. Whether the opacity is dominated by atomic or molecular
lines may be important, as different elements are important in each
source of opacity. For instance, if iron contributes a significant
amount of opacity then it may not be warranted to use $Z=Z_{CNO}$
as an approximation, as the Fe abundance in this star is still zero.
Conversely, if the CNO group (or individual elements therein) are
important contributors under these conditions then the opacity may
be \emph{higher} than that given by scaled-solar mixture opacities.
Similar arguments can be made for molecular contributions, particularly
noting that oxygen appears to be a key ingredient in the molecular
makeup of the gas, in which oxides dominate the opacity (although
this may be a consequence of the large amount of oxygen in the scaled-solar
mixture). 

In summary we suggest that additional sets of opacity tables are needed
at low temperatures in order to follow the evolution of surface temperature
in stars such as these. In particular tables with variable C, N, O
and possibly Ne are needed. This may also affect the surface temperature
evolution of other classes of stars that also increase their CNO abundances
due to dredge-up of processed material, such as thermally pulsing
AGB stars. We shall contact the Wichita State University group and
request these tables which we believe will be adopted widely by the
stellar modelling community. We finally note that a more thorough
investigation should be made into the feedback effects on the overall
evolution of these stars, although it is likely to be small since
the mass involved is so small. We shall pursue this in future work. 

\section{Plots from the Journey}

Here we display just a small, random selection of figures pertaining
to the the past year or so of the Author's candidature.

\begin{figure}[H]
\begin{centering}
\vspace{1.5cm}\includegraphics[width=1\columnwidth,keepaspectratio]{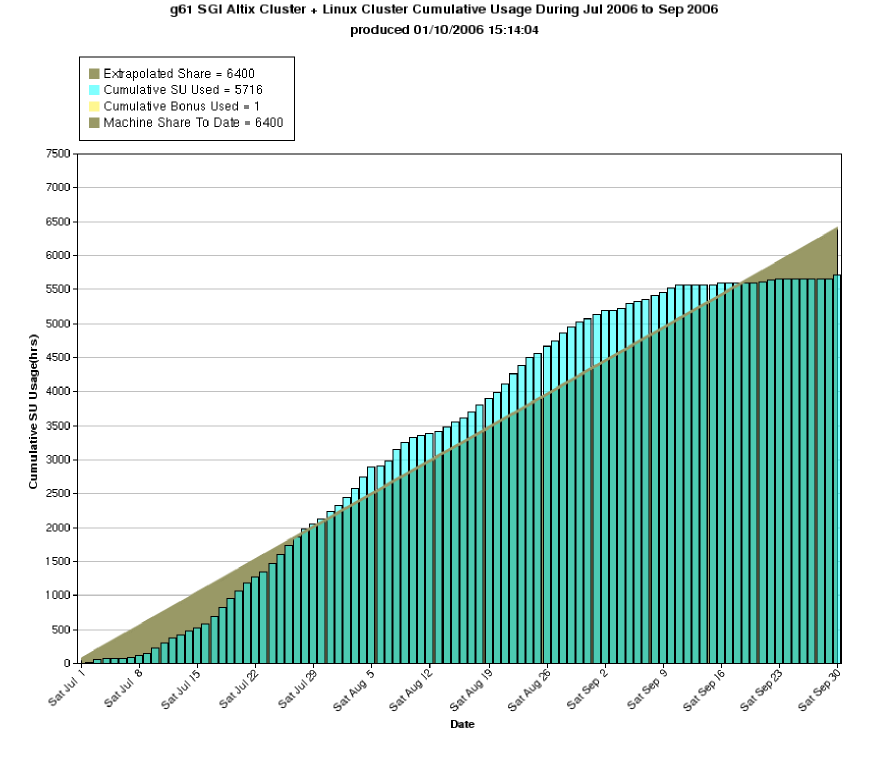}
\par\end{centering}
\caption{Cumulative CPU hour usage for the $Z=0$ and EMP models computed at
the Australian Partnership for Advanced Computing (APAC, Canberra,
Australia). About 15000 CPU hours were used in total (this graph represents
one quarter year, on the x-axis). Note that the y-axis units are in
SU, where $1\textrm{SU}=2$ hours. \label{fig-APAC-Usage}}
\end{figure}

\begin{figure}
\begin{centering}
\includegraphics[width=0.9\columnwidth,keepaspectratio]{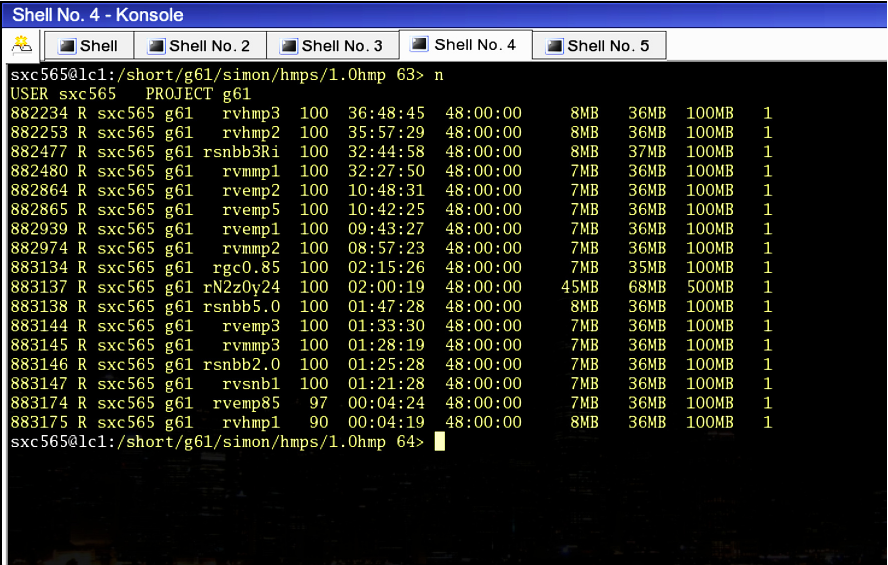}
\par\end{centering}
\caption{A screengrab of an APAC session showing all the $Z=0$ and EMP models
the Author had running at one time. Both nucleosynthesis and structural
evolution calculations can be seen.\label{fig-apacmodels-ShellScreenGrab}}
\end{figure}

\begin{figure}
\begin{centering}
\includegraphics[width=0.9\columnwidth,keepaspectratio]{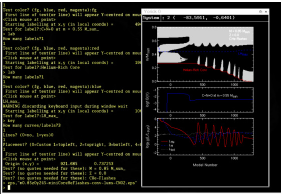}
\par\end{centering}
\caption{A screengrab of the Author's laptop computer showing a Yorick plotting
session. It can be seen that the plot is of a DCF model, with He minipulses
on the SRGB.}
\end{figure}

\begin{figure}
\begin{centering}
\includegraphics[width=0.8\columnwidth,keepaspectratio]{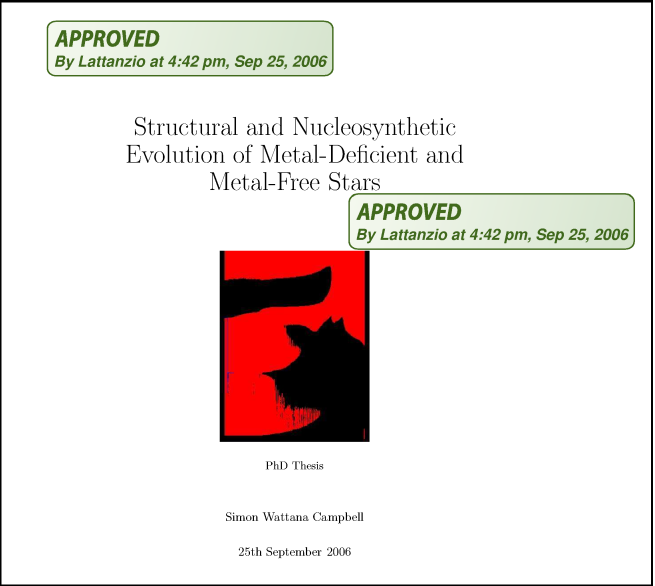}
\par\end{centering}
\caption{One of the working titles and front covers for the current thesis.
The picture is actually a convection plot for one of the $Z=0$ DCF
models (model number versus mass). As can be seen the Author's supervisor
has stamped it with his approval.}
\end{figure}

\begin{figure}
\begin{centering}
\includegraphics[width=0.8\columnwidth,keepaspectratio]{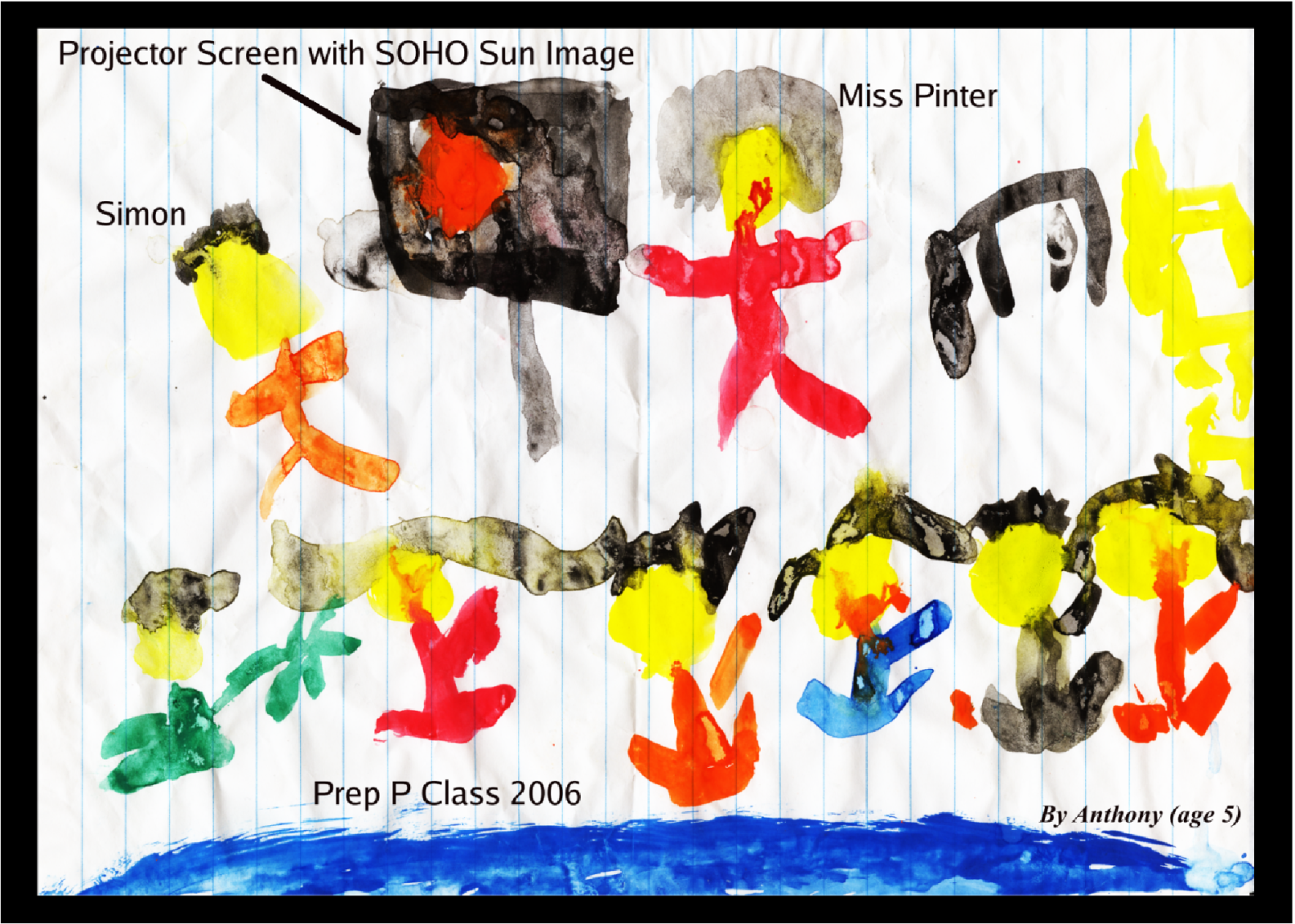}
\par\end{centering}
\caption{A painting by a student (Anthony) at a Primary School
in Melbourne, Australia. Pictured are Lisa Pinter (teacher) and the
Author who was giving an astronomy presentation to the Prep class
(5-6 year-olds).}
\end{figure}

\end{spacing}

\begin{spacing}{1.0}
\footnotesize\bibliographystyle{apalike}
\addcontentsline{toc}{chapter}{\bibname}\bibliography{thesis}

\begin{thebibliography}{}

\bibitem[Adelberger et~al., 2005]{ASS+05}
Adelberger, K.~L., Shapley, A.~E., Steidel, C.~C., Pettini, M., Erb, D.~K., and
  Reddy, N.~A. (2005).
\newblock The {C}onnection between {G}alaxies and {I}ntergalactic {A}bsorption
  {L}ines at {R}edshift $2<z<3$.
\newblock {\em ApJ}, 629(2):636--653.

\bibitem[{Alexander}, 1975]{1975ApJS...29..363A}
{Alexander}, D.~R. (1975).
\newblock {Low-Temperature Rosseland Opacity Tables}.
\newblock {\em ApJS}, 29:363.

\bibitem[{Alexander} and {Ferguson}, 1994]{1994ApJ...437..879A}
{Alexander}, D.~R. and {Ferguson}, J.~W. (1994).
\newblock {Low-temperature Rosseland opacities}.
\newblock {\em ApJ}, 437:879--891.

\bibitem[{Alexander} et~al., 1983]{1983ApJ...272..773A}
{Alexander}, D.~R., {Rypma}, R.~L., and {Johnson}, H.~R. (1983).
\newblock {Effect of molecules and grains on Rosseland mean opacities}.
\newblock {\em ApJ}, 272:773.

\bibitem[{Allende Prieto} et~al., 2006]{2006ApJ...636..804A}
{Allende Prieto}, C., {Beers}, T.~C., {Wilhelm}, R., {Newberg}, H.~J.,
  {Rockosi}, C.~M., {Yanny}, B., and {Lee}, Y.~S. (2006).
\newblock {A Spectroscopic Study of the Ancient Milky Way: F- and G-Type Stars
  in the Third Data Release of the Sloan Digital Sky Survey}.
\newblock {\em ApJ}, 636:804--820.

\bibitem[{Angulo} et~al., 1999]{1999NuPhA.656....3A}
{Angulo}, C., {Arnould}, M., {Rayet}, M., {Descouvemont}, P., {Baye}, D.,
  {Leclercq-Willain}, C., {Coc}, A., {Barhoumi}, S., {Aguer}, P., {Rolfs}, C.,
  {Kunz}, R., {Hammer}, J.~W., {Mayer}, A., {Paradellis}, T., {Kossionides},
  S., {Chronidou}, C., {Spyrou}, K., {degl'Innocenti}, S., {Fiorentini}, G.,
  {Ricci}, B., {Zavatarelli}, S., {Providencia}, C., {Wolters}, H., {Soares},
  J., {Grama}, C., {Rahighi}, J., {Shotter}, A., and {Lamehi Rachti}, M.
  (1999).
\newblock {A compilation of charged-particle induced thermonuclear reaction
  rates}.
\newblock {\em Nuclear Physics A}, 656:3--183.

\bibitem[{Aoki} et~al., 2007]{2007ApJ...655..492A}
{Aoki}, W., {Beers}, T.~C., {Christlieb}, N., {Norris}, J.~E., {Ryan}, S.~G.,
  and {Tsangarides}, S. (2007).
\newblock {Carbon-enhanced Metal-poor Stars. I. Chemical Compositions of 26
  Stars}.
\newblock {\em ApJ}, 655:492--521.

\bibitem[{Aoki} et~al., 2006]{2006ApJ...639..897A}
{Aoki}, W., {Frebel}, A., {Christlieb}, N., {Norris}, J.~E., {Beers}, T.~C.,
  {Minezaki}, T., {Barklem}, P.~S., {Honda}, S., {Takada-Hidai}, M., {Asplund},
  M., {Ryan}, S.~G., {Tsangarides}, S., {Eriksson}, K., {Steinhauer}, A.,
  {Deliyannis}, C.~P., {Nomoto}, K., {Fujimoto}, M.~Y., {Ando}, H., {Yoshii},
  Y., and {Kajino}, T. (2006).
\newblock {HE 1327-2326, an Unevolved Star with [Fe/H] $< -5.0$. I. A
  Comprehensive Abundance Analysis}.
\newblock {\em ApJ}, 639:897--917.

\bibitem[{Arnould} et~al., 1999]{1999AA...347..572A}
{Arnould}, M., {Goriely}, S., and {Jorissen}, A. (1999).
\newblock {Non-explosive hydrogen and helium burnings: abundance predictions
  from the NACRE reaction rate compilation}.
\newblock {\em A\&A}, 347:572--582.

\bibitem[{Asplund} et~al., 2005]{2005ASPC..336...25A}
{Asplund}, M., {Grevesse}, N., and {Sauval}, A.~J. (2005).
\newblock {The Solar Chemical Composition}.
\newblock In {Barnes}, III, T.~G. and {Bash}, F.~N., editors, {\em ASP Conf.
  Ser. 336: Cosmic Abundances as Records of Stellar Evolution and
  Nucleosynthesis}, page~25.

\bibitem[{Baerentzen}, 1965]{1965ZA.....62..221B}
{Baerentzen}, J. (1965).
\newblock {The Adiabatic Temperature-gradient and the Specific Heat $c_{p}$ in
  Outer Hydrogen-Helium Convection Zones}.
\newblock {\em Zeitschrift fur Astrophysik}, 62:221.

\bibitem[{Beaudet} et~al., 1967]{1967ApJ...150..979B}
{Beaudet}, G., {Petrosian}, V., and {Salpeter}, E.~E. (1967).
\newblock {Energy Losses due to Neutrino Processes}.
\newblock {\em ApJ}, 150:979.

\bibitem[Beers and Christlieb, 2005]{BC05}
Beers, T. and Christlieb, N. (2005).
\newblock The {D}iscovery and {A}nalysis of {V}ery {M}etal-{P}oor {S}tars in
  the {G}alaxy.
\newblock {\em ARA\&A}, 43(1):531--580.

\bibitem[{Beers} et~al., 2005]{2005AAS...20714704B}
{Beers}, T.~C., {Allende Prieto}, C., {Wilhelm}, R., {Norris}, J.~E., {Yanny},
  B., {Newberg}, H.~J., {Rockosi}, C., {Thirupathi}, S., and {Lee}, Y. (2005).
\newblock {Low Metallicity Stars in SDSS and SEGUE}.
\newblock {\em American Astronomical Society Meeting Abstracts}, 207:147.04.

\bibitem[Beers et~al., 2005]{BCN+05}
Beers, T.~C., Christlieb, N., Norris, J.~E., Bessell, M.~S., Wilhelm, R.,
  {Allende P. C.}, and {et al.} (2005).
\newblock The {M}etallicity {D}istribution {F}unction of the {H}alo of the
  {M}ilky {W}ay.
\newblock {\em IAUS}, 228:175--183.

\bibitem[Beers et~al., 1985]{BPS85}
Beers, T.~C., Preston, G.~W., and Shectman, S.~A. (1985).
\newblock A search for stars of very low metal abundance. {I}.
\newblock {\em AJ}, 90:2089--2102.

\bibitem[Beers et~al., 1992]{BPS92}
Beers, T.~C., Preston, G.~W., and Shectman, S.~A. (1992).
\newblock A search for stars of very low metal abundance. {II}.
\newblock {\em AJ}, 103(6):1987--2034.

\bibitem[{Beers} et~al., 2007]{2007AJ....133.1193B}
{Beers}, T.~C., {Sivarani}, T., {Marsteller}, B., {Lee}, Y., {Rossi}, S., and
  {Plez}, B. (2007).
\newblock {Near-Infrared Spectroscopy of Carbon-Enhanced Metal-Poor Stars. I. A
  SOAR/OSIRIS Pilot Study}.
\newblock {\em AJ}, 133:1193--1203.

\bibitem[{Bekki} et~al., 2007]{2007astro.ph..2289B}
{Bekki}, K., {Campbell}, S.~W., {Lattanzio}, J.~C., and {Norris}, J.~E. (2007).
\newblock {Origin of Abundance Inhomogeneity in Globular Clusters}.
\newblock {\em MNRAS (accepted Jan. 2007)}.

\bibitem[{Bell} and {Dickens}, 1974]{1974MNRAS.166...89B}
{Bell}, R.~A. and {Dickens}, R.~J. (1974).
\newblock {The CNO abundances of the CH stars in omega Cen.}
\newblock {\em MNRAS}, 166:89--99.

\bibitem[{Bessell} et~al., 1989]{1989AAS...77....1B}
{Bessell}, M.~S., {Brett}, J.~M., {Wood}, P.~R., and {Scholz}, M. (1989).
\newblock {Colors of extended static model photospheres of M giants}.
\newblock {\em A\&AS}, 77:1--30.

\bibitem[{Bessell} et~al., 2004]{2004ApJ...612L..61B}
{Bessell}, M.~S., {Christlieb}, N., and {Gustafsson}, B. (2004).
\newblock {On the Oxygen Abundance of HE 0107-5240}.
\newblock {\em ApJL}, 612:L61--L63.

\bibitem[{Bessell} and {Norris}, 1984]{1984ApJ...285..622B}
{Bessell}, M.~S. and {Norris}, J. (1984).
\newblock {The ultra-metal-deficient (Population III?) red giant CD 38.245
  deg}.
\newblock {\em ApJ}, 285:622--636.

\bibitem[{Bethe}, 1939]{1939PhRv...55..434B}
{Bethe}, H.~A. (1939).
\newblock {Energy Production in Stars}.
\newblock {\em Physical Review}, 55:434--456.

\bibitem[{Bica} et~al., 2006]{2006AA...450..105B}
{Bica}, E., {Bonatto}, C., {Barbuy}, B., and {Ortolani}, S. (2006).
\newblock {Globular cluster system and Milky Way properties revisited}.
\newblock {\em A\&A}, 450:105--115.

\bibitem[Biello, 2001]{2001PhDT.........8B}
Biello, J.~A. (2001).
\newblock Layer formation in semiconvection.
\newblock {\em Ph.D. Thesis}.

\bibitem[{Biermann}, 1951]{1951ZA.....28..304B}
{Biermann}, L. (1951).
\newblock {Bemerkungen {\"u}ber das Rotationsgesetz in irdischen und stellaren
  Instabilit{\"a}tszonen. Mit 1 Textabbildung}.
\newblock {\em Zeitschrift fur Astrophysik}, 28:304.

\bibitem[Boesgaard and King, 1993]{BK93}
Boesgaard, A.~M. and King, J.~R. (1993).
\newblock Galactic evolution of {B}eryllium.
\newblock {\em AJ}, 106(6):2309--2323.

\bibitem[Boesgaard and Steigman, 1985]{Boes0985}
Boesgaard, A.~M. and Steigman, G. (1985).
\newblock Big {B}ang {N}ucleosynthesis: {T}heories and {O}bservations.
\newblock {\em Annual Review of Astronomy and Astrophysics}, 23:319--378.

\bibitem[{B{\"o}hm-Vitense}, 1958]{1958ZA.....46..108B}
{B{\"o}hm-Vitense}, E. (1958).
\newblock {{\"U}ber die Wasserstoffkonvektionszone in Sternen verschiedener
  Effektivtemperaturen und Leuchtkr{\"a}fte. Mit 5 Textabbildungen}.
\newblock {\em Zeitschrift fur Astrophysik}, 46:108.

\bibitem[{B{\"o}hm-Vitense}, 1992]{1992itsa.book.....B}
{B{\"o}hm-Vitense}, E. (1992).
\newblock {\em {Introduction to stellar astrophysics. Vol. 3 - Stellar
  structure and evolution}}.
\newblock Cambridge, England and New York, Cambridge University Press, 1992,
  300 p.

\bibitem[Bond, 1981]{Bon81}
Bond, H.~E. (1981).
\newblock Where is {P}opulation {III}.
\newblock {\em ApJ}, 248:606--611.

\bibitem[Bono et~al., 1997]{1997ApJ...489..822B}
Bono, G., Caputo, F., Cassisi, S., Castellani, V., and Marconi, M. (1997).
\newblock Evolutionary and {P}ulsational {C}onstraints for
  {S}uper--{M}etal-rich {S}tars with {Z} = 0.04.
\newblock {\em ApJ}, 489:822.

\bibitem[{Boyer} et~al., 1971]{1971SoPh...19..330B}
{Boyer}, R., {Henoux}, J.~C., and {Sotirovski}, P. (1971).
\newblock {Isotopes of Magnesium in the Solar Atmosphere}.
\newblock {\em SoPh}, 19:330.

\bibitem[{Briley} et~al., 1993]{1993AJ....106..142B}
{Briley}, M.~M., {Smith}, G.~H., {Hesser}, J.~E., and {Bell}, R.~A. (1993).
\newblock {CN and CH variations in the globular cluster M55}.
\newblock {\em AJ}, 106:142--153.

\bibitem[Brodie and Strader, 2006]{BS06}
Brodie, J.~P. and Strader, J. (2006).
\newblock Extragalactic {G}lobular {C}lusters and {G}alaxy {F}ormation.
\newblock {\em Annual Review of Astronomy and Astrophysics}, 44:193--267.

\bibitem[{Burbidge} et~al., 1957]{1957RvMP...29..547B}
{Burbidge}, E.~M., {Burbidge}, G.~R., {Fowler}, W.~A., and {Hoyle}, F. (1957).
\newblock {Synthesis of the Elements in Stars}.
\newblock {\em Reviews of Modern Physics}, 29:547--650.

\bibitem[{Cameron} and {Fowler}, 1971]{1971ApJ...164..111C}
{Cameron}, A.~G.~W. and {Fowler}, W.~A. (1971).
\newblock {Lithium and the s-PROCESS in Red-Giant Stars}.
\newblock {\em ApJ}, 164:111.

\bibitem[{Cannon}, 1993]{1993MNRAS.263..817C}
{Cannon}, R.~C. (1993).
\newblock {Massive Thorne-Zytkow Objects - Structure and Nucleosynthesis}.
\newblock {\em MNRAS}, 263:817.

\bibitem[{Cannon} et~al., 1998]{1998MNRAS.298..601C}
{Cannon}, R.~D., {Croke}, B.~F.~W., {Bell}, R.~A., {Hesser}, J.~E., and
  {Stathakis}, R.~A. (1998).
\newblock {Carbon and nitrogen abundance variations on the main sequence of 47
  Tucanae}.
\newblock {\em MNRAS}, 298:601--624.

\bibitem[{Canuto} et~al., 1996]{1996ApJ...473..550C}
{Canuto}, V.~M., {Goldman}, I., and {Mazzitelli}, I. (1996).
\newblock {Stellar Turbulent Convection: A Self-consistent Model}.
\newblock {\em ApJ}, 473:550.

\bibitem[{Cary}, 1974]{1974ApSS..31....3C}
{Cary}, N. (1974).
\newblock {Main sequence models for massive zero-metal stars}.
\newblock {\em Ap\&SS}, 31:3--10.

\bibitem[{Cassisi} and {Castellani}, 1993]{1993ApJS...88..509C}
{Cassisi}, S. and {Castellani}, V. (1993).
\newblock {An evolutionary scenario for primeval stellar populations}.
\newblock {\em ApJS}, 88:509--527.

\bibitem[{Cassisi} et~al., 1996]{1996ApJ...459..298C}
{Cassisi}, S., {Castellani}, V., and {Tornambe}, A. (1996).
\newblock {The Evolutionary Properties and Peculiar Thermal Pulses of
  Metal-deficient Low-Mass Stars}.
\newblock {\em ApJ}, 459:298.

\bibitem[Castellani et~al., 1985]{1985ApJ...296..204C}
Castellani, V., Chieffi, A., Tornambe, A., and Pulone, L. (1985).
\newblock Helium-burning evolutionary phases in population {II} stars. {I}
  {B}reathing pulses in horizontal branch stars.
\newblock {\em ApJ}, 296:204--212.

\bibitem[Castellani et~al., 1971]{1971ApSS..10..355C}
Castellani, V., Giannone, P., and Renzini, A. (1971).
\newblock Induced {S}emi-{C}onvection in {H}elium-{B}urning
  {H}orizontal-{B}ranch {S}tars {II}.
\newblock {\em AP\&SS}, 10:355.

\bibitem[{Castellani} and {Paolicchi}, 1975]{1975ApSS..35..185C}
{Castellani}, V. and {Paolicchi}, P. (1975).
\newblock {Zero-population stars}.
\newblock {\em Ap\&SS}, 35:185--196.

\bibitem[{Caughlan} and {Fowler}, 1988]{CF88}
{Caughlan}, G.~R. and {Fowler}, W.~A. (1988).
\newblock {Thermonuclear Reaction Rates V}.
\newblock {\em Atomic Data and Nuclear Data Tables}, 40:283.

\bibitem[{Chieffi} et~al., 2001]{2001ApJ...554.1159C}
{Chieffi}, A., {Dom{\'{\i}}nguez}, I., {Limongi}, M., and {Straniero}, O.
  (2001).
\newblock {Evolution and Nucleosynthesis of Zero-Metal Intermediate-Mass
  Stars}.
\newblock {\em ApJ}, 554:1159--1174.

\bibitem[{Chieffi} and {Limongi}, 2002]{2002ApJ...577..281C}
{Chieffi}, A. and {Limongi}, M. (2002).
\newblock {The Explosive Yields Produced by the First Generation of Core
  Collapse Supernovae and the Chemical Composition of Extremely Metal Poor
  Stars}.
\newblock {\em ApJ}, 577:281--294.

\bibitem[{Chieffi} and {Tornambe}, 1984]{1984ApJ...287..745C}
{Chieffi}, A. and {Tornambe}, A. (1984).
\newblock {On the evolution of an intermediate-mass zero-metal star which does
  not experience thermal instabilities during the double shell burning phase}.
\newblock {\em ApJ}, 287:745--748.

\bibitem[{Chiosi} et~al., 1993]{1993ApJS...86..541C}
{Chiosi}, C., {Wood}, P.~R., and {Capitanio}, N. (1993).
\newblock {Theoretical models of Cepheid variables and their BVI(c) colors and
  magnitudes}.
\newblock {\em ApJS}, 86:541.

\bibitem[{Christlieb} et~al., 2002]{2002Natur.419..904C}
{Christlieb}, N., {Bessell}, M.~S., {Beers}, T.~C., {Gustafsson}, B., {Korn},
  A., {Barklem}, P.~S., {Karlsson}, T., {Mizuno-Wiedner}, M., and {Rossi}, S.
  (2002).
\newblock {A stellar relic from the early Milky Way}.
\newblock {\em Nature}, 419:904--906.

\bibitem[{Christlieb} et~al., 2004]{2004ApJ...603..708C}
{Christlieb}, N., {Gustafsson}, B., {Korn}, A.~J., {Barklem}, P.~S., {Beers},
  T.~C., {Bessell}, M.~S., {Karlsson}, T., and {Mizuno-Wiedner}, M. (2004).
\newblock {HE 0107-5240, a Chemically Ancient Star. I. A Detailed Abundance
  Analysis}.
\newblock {\em ApJ}, 603:708--728.

\bibitem[Christlieb et~al., 1999]{CWR+99}
Christlieb, N., Wisotzki, L., Reimers, D., Gehren, T., Reetz, J., and Beers,
  T.~C. (1999).
\newblock An {A}utomated {S}earch for {M}etal-{P}oor {H}alo {S}tars in the
  {H}amburg/{ESO} {O}bjective-{P}rism {S}urvey.
\newblock {\em ASPC}, 165:259.

\bibitem[{Clayton}, 1983]{1983psen.book.....C}
{Clayton}, D.~D. (1983).
\newblock {\em {Principles of stellar evolution and nucleosynthesis}}.
\newblock Chicago: University of Chicago Press, 1983.

\bibitem[Coc et~al., 2004]{Coc0104}
Coc, A., Vangioni-Flam, E., Descouvemont, P., Adahchour, A., and Angulo, C.
  (2004).
\newblock Updated {B}ig {B}ang {N}ucleosynthesis {C}ompared with {W}ilkinson
  {M}icrowave {A}nisotropy {P}robe {O}bservations and the {A}bundance of
  {L}ight {E}lements.
\newblock {\em ApJ}, 600:544--552.

\bibitem[{Cohen} et~al., 2006]{2006AJ....132..137C}
{Cohen}, J.~G., {McWilliam}, A., {Shectman}, S., {Thompson}, I., {Christlieb},
  N., {Melendez}, J., {Ramirez}, S., {Swensson}, A., and {Zickgraf}, F.-J.
  (2006).
\newblock {Carbon Stars in the Hamburg/ESO Survey: Abundances}.
\newblock {\em AJ}, 132:137--160.

\bibitem[{Cohen} et~al., 2005]{2005ApJ...633L.109C}
{Cohen}, J.~G., {Shectman}, S., {Thompson}, I., {McWilliam}, A., {Christlieb},
  N., {Melendez}, J., {Zickgraf}, F.-J., {Ram{\'{\i}}rez}, S., and {Swenson},
  A. (2005).
\newblock {The Frequency of Carbon Stars among Extremely Metal-poor Stars}.
\newblock {\em ApJL}, 633:L109--L112.

\bibitem[{Cottrell} and {Da Costa}, 1981]{1981ApJ...245L..79C}
{Cottrell}, P.~L. and {Da Costa}, G.~S. (1981).
\newblock {Correlated cyanogen and sodium anomalies in the globular clusters 47
  TUC and NGC 6752}.
\newblock {\em ApJL}, 245:L79--L82.

\bibitem[{Cox} and {Stewart}, 1970]{1970ApJS...19..261C}
{Cox}, A.~N. and {Stewart}, J.~N. (1970).
\newblock {Rosseland Opacity Tables for Population II Compositions}.
\newblock {\em ApJS}, 19:261.

\bibitem[Cyburt et~al., 2002]{Cyb0402}
Cyburt, R.~H., Fields, B.~D., and Olive, K.~A. (2002).
\newblock Primordial nucleosynthesis with {CMB} inputs: probing the early
  universe and light element astrophysics.
\newblock {\em APh}, 17(1):87--100.

\bibitem[{Da Costa} and {Cottrell}, 1980]{1980ApJ...236L..83D}
{Da Costa}, G.~S. and {Cottrell}, P.~L. (1980).
\newblock {Carbon and nitrogen abundances on the first giant branch of NGC
  6752}.
\newblock {\em ApJL}, 236:L83--L86.

\bibitem[{Dahn} et~al., 1977]{1977ApJ...216..757D}
{Dahn}, C.~C., {Liebert}, J., {Kron}, R.~G., {Spinrad}, H., and {Hintzen},
  P.~M. (1977).
\newblock {G77-61 - A dwarf carbon star}.
\newblock {\em ApJ}, 216:757--766.
\newblock Provided by the Smithsonian/NASA Astrophysics Data System.

\bibitem[D'Antona and Mazzetelli, 1982]{1982AA...115L...1D}
D'Antona, F. and Mazzetelli, I. (1982).
\newblock {Evolution of low mass zero metal giants up to the helium flash}.
\newblock {\em A\&A}, 115:L1--L3.

\bibitem[{Dearborn} et~al., 2001]{2001AAS...198.6513D}
{Dearborn}, D.~S.~P., {Bazan}, G., {Castor}, J., {Cavallo}, R., {Cohl}, H.,
  {Cook}, K., {Dossa}, D., {Eastman}, R., {Eggleton}, P.~P., {Eltgroth}, P.,
  {Keller}, S., {Murray}, S., {Taylor}, A., {Turcotte}, S., and {Djehuty Team}
  (2001).
\newblock {Djehuty: A 3D Hydrodynamic Stellar Evolution Code}.
\newblock {\em Bulletin of the American Astronomical Society}, 33:886.

\bibitem[{Dearborn} et~al., 1986]{1986ApJ...300..314D}
{Dearborn}, D.~S.~P., {Liebert}, J., {Aaronson}, M., {Dahn}, C.~C.,
  {Harrington}, R., {Mould}, J., and {Greenstein}, J.~L. (1986).
\newblock {On the nature of the dwarf carbon star G77-61}.
\newblock {\em ApJ}, 300:314--324.

\bibitem[Denissenkov and Herwig, 2003]{2003ApJ...590L..99D}
Denissenkov, P.~A. and Herwig, F. (2003).
\newblock The {A}bundance {E}volution of {O}xygen, {S}odium, and {M}agnesium in
  {E}xtremely {M}etal {P}oor {I}ntermediate-{M}ass {S}tars: {I}mplications for
  the {S}elf-{P}ollution {S}cenario in {G}lobular {C}lusters.
\newblock {\em ApJ}, 590:L99--L102.

\bibitem[Denissenkov et~al., 1997]{1997AA...320..115D}
Denissenkov, P.~A., Weiss, A., and Wagenhuber, J. (1997).
\newblock Could intermediate-mass {AGB} stars produce star-to-star abundance
  variations in globular-cluster red giants?
\newblock {\em A\&A}, 320:115--124.

\bibitem[{Denker} et~al., 1995]{1995AIPC..327..255D}
{Denker}, A., {Drotleff}, H.~W., {Grosse}, M., {Knee}, H., {Kunz}, R., {Mayer},
  A., {Seidel}, R., {Soin{\'e}}, M., {W{\"o}ohr}, A., {Wolf}, G., and {Hammer},
  J.~W. (1995).
\newblock {Neutron Producing Reactions in Stars}.
\newblock In {Busso}, M., {Raiteri}, C.~M., and {Gallino}, R., editors, {\em
  AIP Conf. Proc. 327: Nuclei in the Cosmos III}, page 255.

\bibitem[{Dicus} et~al., 1976]{1976ApJ...210..481D}
{Dicus}, D.~A., {Kolb}, E.~W., {Schramm}, D.~N., and {Tubbs}, D.~L. (1976).
\newblock {Neutrino pair bremsstrahlung including neutral current effects}.
\newblock {\em ApJ}, 210:481.

\bibitem[{Doherty} and {Lattanzio}, 2006]{2006MmSAI..77..828D}
{Doherty}, C.~L. and {Lattanzio}, J.~C. (2006).
\newblock {Evolution and nucleosynthesis in super-AGB stars.}
\newblock {\em Memorie della Societa Astronomica Italiana}, 77:828.

\bibitem[{Dominguez} et~al., 2000]{2000MmSAI..71..781D}
{Dominguez}, I., {Straniero}, O., {Limongi}, M., and {Chieffi}, A. (2000).
\newblock {Do zero metals intermediate mass stars experience thermal pulses?}
\newblock {\em Memorie della Societa Astronomica Italiana}, 71:781.

\bibitem[{Eddington}, 1919]{1919Obs....42..371E}
{Eddington}, A.~S. (1919).
\newblock {The sources of stellar energy}.
\newblock {\em The Observatory}, 42:371--376.

\bibitem[{Eddington}, 1920]{1920Obs....43..341E}
{Eddington}, A.~S. (1920).
\newblock {The internal constitution of the stars}.
\newblock {\em The Observatory}, 43:341--358.

\bibitem[Eggleton, 1972]{1972MNRAS.156..361E}
Eggleton, P.~P. (1972).
\newblock Composition changes during stellar evolution.
\newblock {\em MNRAS}, 156:361.

\bibitem[{El Eid} and {Champagne}, 1995]{1995ApJ...451..298E}
{El Eid}, M.~F. and {Champagne}, A.~E. (1995).
\newblock {Sodium Enrichment in A--F Type Supergiants}.
\newblock {\em ApJ}, 451:298.

\bibitem[{Eryurt-Ezer}, 1981]{1981ApSS..79..265E}
{Eryurt-Ezer}, D. (1981).
\newblock {Advanced evolutionary phase of a first-generation star}.
\newblock {\em Ap\&SS}, 79:265--287.

\bibitem[{Eryurt-Ezer} and {Kiziloglu}, 1985]{1985ApSS.117...95E}
{Eryurt-Ezer}, D. and {Kiziloglu}, N. (1985).
\newblock {The evolutionary behaviour of Population III intermediate mass
  stars}.
\newblock {\em Ap\&SS}, 117:95--109.

\bibitem[{Ezer}, 1961]{1961ApJ...133..159E}
{Ezer}, D. (1961).
\newblock {Models of Massive Pure Hydrogen Stars.}
\newblock {\em ApJ}, 133:159.

\bibitem[{Ezer}, 1972]{1972ApSS..18..226E}
{Ezer}, D. (1972).
\newblock {Theoretical Evolution of a Hydrogen-Helium Star of 3 M$_\odot$ from
  the Pre-Main Sequence to the Core Helium-Exhaustion Phase}.
\newblock {\em Ap\&SS}, 18:226.

\bibitem[{Ezer} and {Cameron}, 1971]{1971ApSS..14..399E}
{Ezer}, D. and {Cameron}, A.~G.~W. (1971).
\newblock {The Evolution of Hydrogen-Helium Stars}.
\newblock {\em Ap\&SS}, 14:399.

\bibitem[Fan et~al., 2003]{Fea03}
Fan, X., Strauss, M.~A., Schneider, D.~P., Becker, R.~H., White, R.~L., and
  Haiman, Z. (2003).
\newblock A {S}urvey of z>5.7 {Q}uasars in the {S}loan {D}igital {S}ky
  {S}urvey. {II}. {D}iscovery of {T}hree {A}dditional {Q}uasars at z>6.
\newblock {\em AJ}, 125(4):1649--1659.

\bibitem[Faulkner, 1968]{Fau68}
Faulkner, D.~J. (1968).
\newblock The evolution of helium shell-burning stars.
\newblock {\em MNRAS}, 140:223.
\newblock Original Evoln paper!

\bibitem[{Faulkner} and {Wood}, 1972]{FW72}
{Faulkner}, D.~J. and {Wood}, P.~R. (1972).
\newblock {Thermal pulses in He shell-burning stars.}
\newblock {\em ApJ}, 178:207.

\bibitem[{Fenner} et~al., 2004]{2004MNRAS.353..789F}
{Fenner}, Y., {Campbell}, S., {Karakas}, A.~I., {Lattanzio}, J.~C., and
  {Gibson}, B.~K. (2004).
\newblock {Modelling self-pollution of globular clusters from asymptotic giant
  branch stars}.
\newblock {\em MNRAS}, 353:789--795.

\bibitem[{Ferguson} et~al., 2005]{2005ApJ...623..585F}
{Ferguson}, J.~W., {Alexander}, D.~R., {Allard}, F., {Barman}, T., {Bodnarik},
  J.~G., {Hauschildt}, P.~H., {Heffner-Wong}, A., and {Tamanai}, A. (2005).
\newblock {Low-Temperature Opacities}.
\newblock {\em ApJ}, 623:585--596.

\bibitem[Fernando, 1989]{1989JFM...209....1F}
Fernando, H. J.~S. (1989).
\newblock Buoyancy transfer across a diffusive interface.
\newblock {\em Journal of Fluid Mechanics}, 209:1--34.

\bibitem[{Festa} and {Ruderman}, 1969]{1969PhRv..180.1227F}
{Festa}, G.~G. and {Ruderman}, M.~A. (1969).
\newblock {Neutrino-Pair Bremsstrahlung from a Degenerate Electron Gas}.
\newblock {\em Physical Review}, 180:1227--1231.

\bibitem[Fick, 1855]{fick}
Fick, A. (1855).
\newblock {\em Phil. Mag.}, 10:30.

\bibitem[{Frebel} et~al., 2005a]{2005Natur.434..871F}
{Frebel}, A., {Aoki}, W., {Christlieb}, N., {Ando}, H., {Asplund}, M.,
  {Barklem}, P.~S., {Beers}, T.~C., {Eriksson}, K., {Fechner}, C., {Fujimoto},
  M.~Y., {Honda}, S., {Kajino}, T., {Minezaki}, T., {Nomoto}, K., {Norris},
  J.~E., {Ryan}, S.~G., {Takada-Hidai}, M., {Tsangarides}, S., and {Yoshii}, Y.
  (2005a).
\newblock {Nucleosynthetic signatures of the first stars}.
\newblock {\em Nature}, 434:871--873.

\bibitem[{Frebel} et~al., 2005b]{2005IAUS..228..207F}
{Frebel}, A., {Aoki}, W., {Christlieb}, N., {Ando}, H., {Asplund}, M.,
  {Barklem}, P.~S., {Beers}, T.~C., {Eriksson}, K., {Fechner}, C., {Fujimoto},
  M.~Y., {Honda}, S., {Kajino}, T., {Minezaki}, T., {Nomoto}, K., {Norris},
  J.~E., {Ryan}, S.~G., {Takada-Hidai}, M., {Tsangarides}, S., and {Yoshii}, Y.
  (2005b).
\newblock {The new record holder for the most iron-poor star: HE 1327 2326, a
  dwarf or subgiant with [Fe/H] $=-5.4$}.
\newblock In {Hill}, V., {Fran{\c c}ois}, P., and {Primas}, F., editors, {\em
  IAU Symposium}, pages 207--212.

\bibitem[{Frebel} et~al., 2006a]{2006ApJ...638L..17F}
{Frebel}, A., {Christlieb}, N., {Norris}, J.~E., {Aoki}, W., and {Asplund}, M.
  (2006a).
\newblock {The Oxygen Abundance of HE 1327-2326}.
\newblock {\em ApJL}, 638:L17--L20.

\bibitem[{Frebel} et~al., 2006b]{2006ApJ...652.1585F}
{Frebel}, A., {Christlieb}, N., {Norris}, J.~E., {Beers}, T.~C., {Bessell},
  M.~S., {Rhee}, J., {Fechner}, C., {Marsteller}, B., {Rossi}, S., {Thom}, C.,
  {Wisotzki}, L., and {Reimers}, D. (2006b).
\newblock {Bright Metal-poor Stars from the Hamburg/ESO Survey. I. Selection
  and Follow-up Observations from 329 Fields}.
\newblock {\em ApJ}, 652:1585--1603.

\bibitem[{Freytag} et~al., 1996]{1996AA...313..497F}
{Freytag}, B., {Ludwig}, H.-G., and {Steffen}, M. (1996).
\newblock {Hydrodynamical models of stellar convection. The role of overshoot
  in DA white dwarfs, A-type stars, and the Sun.}
\newblock {\em A\&A}, 313:497--516.

\bibitem[{Frost} et~al., 1998]{Frost98}
{Frost}, C.~A., {Cannon}, R.~C., {Lattanzio}, J.~C., {Wood}, P.~R., and
  {Forestini}, M. (1998).
\newblock {The brightest carbon stars}.
\newblock {\em AAP}, 332:L17.

\bibitem[{Frost} and {Lattanzio}, 1996]{1996ApJ...473..383F}
{Frost}, C.~A. and {Lattanzio}, J.~C. (1996).
\newblock {On the Numerical Treatment and Dependence of the Third Dredge-up
  Phenomenon}.
\newblock {\em ApJ}, 473:383--+.

\bibitem[{Fujimoto} and {Iben}, 1998]{1998IAUS..189P.150F}
{Fujimoto}, M.~Y. and {Iben}, Jr., I. (1998).
\newblock {The peculiar evolution of metal-free stars and the search for
  Population III stars}.
\newblock In {Bedding}, T.~R., {Booth}, A.~J., and {Davis}, J., editors, {\em
  IAU Symp. 189: Fundamental Stellar Properties}, page 150P.

\bibitem[{Fujimoto} et~al., 1984]{1984ApJ...287..749F}
{Fujimoto}, M.~Y., {Iben}, Jr., I., {Chieffi}, A., and {Tornambe}, A. (1984).
\newblock {Hydrogen and helium burning in zero-metal asymptotic giant branch
  stars and the existence of thresholds (in core mass and CNO abundance) for
  the occurrence of helium shell flashes}.
\newblock {\em ApJ}, 287:749--760.

\bibitem[{Fujimoto} et~al., 1990]{1990ApJ...349..580F}
{Fujimoto}, M.~Y., {Iben}, I.~J., and {Hollowell}, D. (1990).
\newblock {Helium flashes and hydrogen mixing in low-mass population III
  stars}.
\newblock {\em ApJ}, 349:580--592.

\bibitem[{Fujimoto} et~al., 2000]{2000ApJ...529L..25F}
{Fujimoto}, M.~Y., {Ikeda}, Y., and {Iben}, I.~J. (2000).
\newblock {The Origin of Extremely Metal-poor Carbon Stars and the Search for
  Population III}.
\newblock {\em ApJ}, 529:L25--L28.

\bibitem[Fynbo et~al., 2006]{FSL+06}
Fynbo, J. P.~U., Starling, R. L.~C., Ledoux, C., Wiersema, K., Th\"one, C.~C.,
  and Sollerman, J. (2006).
\newblock Probing cosmic chemical evolution with gamma-ray bursts: {GRB} 060206
  at z = 4.048.
\newblock {\em A\&A}, 451(3):L47--L50.

\bibitem[{Garc{\'{\i}}a-Berro} et~al., 2006]{2006MmSAI..77..834G}
{Garc{\'{\i}}a-Berro}, E., {Gil-Pons}, P., and {Guti{\'e}rrez}, J. (2006).
\newblock {The evolution of primordial Super-AGB stars.}
\newblock {\em Memorie della Societa Astronomica Italiana}, 77:834.

\bibitem[{Gass} et~al., 1988]{1988AA...189..194G}
{Gass}, H., {Wehrse}, R., and {Liebert}, J. (1988).
\newblock {Spectrum analysis of the extremely metal-poor carbon dwarf star G
  77-61}.
\newblock {\em A\&A}, 189:194--198.

\bibitem[Giancoli, 1988]{Gian88}
Giancoli, D.~C. (1988).
\newblock {\em Physics for scientists and engineers}.
\newblock Prentice-Hall, Englewood Cliffs, N.J., 2nd edition.

\bibitem[{Gil-Pons} et~al., 2006]{2006astroph12267G}
{Gil-Pons}, P., {Gutierrez}, J., and {Garcia-Berro}, E. (2006).
\newblock {The late stages of the evolution of intermediate-mass primordial
  stars: the effects of overshooting}.
\newblock {\em ArXiv Astrophysics eprints}.

\bibitem[Gingold, 1974]{Gin74}
Gingold, R.~A. (1974).
\newblock Asymptotic {G}iant-{B}ranch {E}volution of a 0.6 {M}$_\odot$ {S}tar.
\newblock {\em ApJ}, 193:177.

\bibitem[Gnedin, 1998]{Gne98}
Gnedin, N.~Y. (1998).
\newblock Metal enrichment of the intergalactic medium.
\newblock {\em MNRAS}, 294:407.

\bibitem[{Goriely} and {Siess}, 2001]{2001AA...378L..25G}
{Goriely}, S. and {Siess}, L. (2001).
\newblock {Nucleosynthesis of s-elements in zero-metal AGB stars}.
\newblock {\em A\&A}, 378:L25--L28.

\bibitem[Goriely and Siess, 2004]{2004AA...421L..25G}
Goriely, S. and Siess, L. (2004).
\newblock S-process in hot {AGB} stars: {A} complex interplay between diffusive
  mixing and nuclear burning.
\newblock {\em A\&A}, 421:L25--L28.

\bibitem[{Gratton} et~al., 2001]{2001AA...369...87G}
{Gratton}, R.~G., {Bonifacio}, P., {Bragaglia}, A., {Carretta}, E.,
  {Castellani}, V., {Centurion}, M., {Chieffi}, A., {Claudi}, R., {Clementini},
  G., {D'Antona}, F., {Desidera}, S., {Fran{\c c}ois}, P., {Grundahl}, F.,
  {Lucatello}, S., {Molaro}, P., {Pasquini}, L., {Sneden}, C., {Spite}, F., and
  {Straniero}, O. (2001).
\newblock {The O-Na and Mg-Al anticorrelations in turn-off and early subgiants
  in globular clusters}.
\newblock {\em A\&A}, 369:87--98.

\bibitem[{Gratton} et~al., 2000]{2000AA...354..169G}
{Gratton}, R.~G., {Sneden}, C., {Carretta}, E., and {Bragaglia}, A. (2000).
\newblock {Mixing along the red giant branch in metal-poor field stars}.
\newblock {\em A\&A}, 354:169--187.

\bibitem[{Grevesse} and {Noels}, 1993]{1993oee..conf...14G}
{Grevesse}, N. and {Noels}, A. (1993).
\newblock {Cosmic Abundances of the Elements}.
\newblock In {Prantzos}, N., {Vangioni-Flam}, E., and {Casse}, M., editors,
  {\em Origin and evolution of the elements: {C}onference proceedings},
  page~14.

\bibitem[{Grevesse} and {Sauval}, 1998]{1998SSRv...85..161G}
{Grevesse}, N. and {Sauval}, A.~J. (1998).
\newblock {Standard Solar Composition}.
\newblock {\em Space Science Reviews}, 85:161--174.

\bibitem[Grossman and Taam, 1996]{1996MNRAS.283.1165G}
Grossman, S.~A. and Taam, R.~E. (1996).
\newblock Double-{D}iffusive {M}ixing-{L}ength {T}heory, {S}emiconvection and
  {M}assive {S}tar {E}volution.
\newblock {\em MNRAS}, 283:1165--1178.

\bibitem[{Grundahl} et~al., 2002]{2002AA...385L..14G}
{Grundahl}, F., {Briley}, M., {Nissen}, P.~E., and {Feltzing}, S. (2002).
\newblock {Abundances of RGB stars in NGC 6752}.
\newblock {\em A\&A}, 385:L14--L17.

\bibitem[{Guenther} and {Demarque}, 1983]{1983AA...118..262G}
{Guenther}, D.~B. and {Demarque}, P. (1983).
\newblock {Evolution of a population III star of low mass}.
\newblock {\em A\&A}, 118:262--266.

\bibitem[{Guesten} and {Mezger}, 1982]{1982VA.....26..159G}
{Guesten}, R. and {Mezger}, P.~G. (1982).
\newblock {Star formation and abundance gradients in the galaxy}.
\newblock {\em Vistas in Astronomy}, 26:159--224.

\bibitem[{Habing} and {Olofsson}, 2003]{2003agbs.conf.....H}
{Habing}, H.~J. and {Olofsson}, H., editors (2003).
\newblock {\em {Asymptotic giant branch stars}}.

\bibitem[{Hansen} and {Kawaler}, 1994]{1994sipp.book.....H}
{Hansen}, C.~J. and {Kawaler}, S.~D. (1994).
\newblock {\em {Stellar Interiors. Physical Principles, Structure, and
  Evolution.}}
\newblock Stellar Interiors. Physical Principles, Structure, and Evolution,
  XIII. Springer-Verlag Berlin Heidelberg New York.

\bibitem[{Harris} et~al., 1983]{Har83}
{Harris}, M.~J., {Fowler}, W.~A., {Caughlan}, G.~R., and {Zimmerman}, B.~A.
  (1983).
\newblock {Thermonuclear reaction rates. III}.
\newblock {\em ARAA}, 21:165.

\bibitem[{Harris}, 1996]{1996AJ....112.1487H}
{Harris}, W.~E. (1996).
\newblock {A Catalog of Parameters for Globular Clusters in the Milky Way}.
\newblock {\em AJ}, 112:1487.

\bibitem[{Haselgrove} and {Hoyle}, 1956]{1956MNRAS.116..515H}
{Haselgrove}, C.~B. and {Hoyle}, F. (1956).
\newblock {A mathematical discussion of the problem of stellar evolution, with
  reference to the use of an automatic digital computer}.
\newblock {\em MNRAS}, 116:515.

\bibitem[Henyey et~al., 1964]{Hen64}
Henyey, L.~G., Forbes, J.~E., and Gould, N.~L. (1964).
\newblock A {N}ew {M}ethod of {A}utomatic {C}omputation of {S}tellar
  {E}volution.
\newblock {\em ApJ}, 139:306.
\newblock Paper 2.

\bibitem[Henyey et~al., 1959]{Hen59}
Henyey, L.~G., Wilets, L., B\"{o}hm, K.~H., Lelevier, R., and Levee, R.~D.
  (1959).
\newblock A {M}ethod for {A}tomic {C}omputation of {S}tellar {E}volution.
\newblock {\em ApJ}, 129:628.
\newblock Paper I.

\bibitem[{Herwig}, 2003]{2003ASPC..304..318H}
{Herwig}, F. (2003).
\newblock {CNO in Low- and Zero-Metallicity AGB Stars}.
\newblock In {Charbonnel}, C., {Schaerer}, D., and {Meynet}, G., editors, {\em
  Astronomical Society of the Pacific Conference Series}, page 318.

\bibitem[Herwig, 2004]{2004ApJ...605..425H}
Herwig, F. (2004).
\newblock Dredge-up and {E}nvelope {B}urning in {I}ntermediate-{M}ass {G}iants
  of {V}ery {L}ow {M}etallicity.
\newblock {\em ApJ}, 605:425--435.

\bibitem[{Herwig}, 2004]{2004ApJS..155..651H}
{Herwig}, F. (2004).
\newblock {Evolution and Yields of Extremely Metal-poor Intermediate-Mass
  Stars}.
\newblock {\em ApJS}, 155:651--666.

\bibitem[Herwig et~al., 1997]{1997AA...324L..81H}
Herwig, F., Bloecker, T., Schoenberner, D., and El~Eid, M. (1997).
\newblock Stellar evolution of low and intermediate-mass stars. {IV}.
  {H}ydrodynamically-based overshoot and nucleosynthesis in {AGB} stars.
\newblock {\em A\&A}, 324:L81--L84.

\bibitem[{Herwig} et~al., 2006]{2006ApJ...642.1057H}
{Herwig}, F., {Freytag}, B., {Hueckstaedt}, R.~M., and {Timmes}, F.~X. (2006).
\newblock {Hydrodynamic Simulations of He Shell Flash Convection}.
\newblock {\em ApJ}, 642:1057--1074.

\bibitem[{Hollowell} et~al., 1990]{1990ApJ...351..245H}
{Hollowell}, D., {Iben}, I.~J., and {Fujimoto}, M.~Y. (1990).
\newblock {Hydrogen burning and dredge-up during the major core helium flash in
  a Z = 0 model star}.
\newblock {\em ApJ}, 351:245--257.

\bibitem[{Hoppe} et~al., 1995]{1995GeCoA..59.4029H}
{Hoppe}, P., {Amari}, S., {Zinner}, E., and {Lewis}, R.~S. (1995).
\newblock {Isotopic compositions of C, N, O, Mg, and Si, trace element
  abundances, and morphologies of single circumstellar graphite grains in four
  density fractions from the Murchison meteorite}.
\newblock {\em GeCoA}, 59:4029--4056.

\bibitem[{Hubbard} and {Lampe}, 1969]{1969ApJS...18..297H}
{Hubbard}, W.~B. and {Lampe}, M. (1969).
\newblock {Thermal Conduction by Electrons in Stellar Matter}.
\newblock {\em ApJS}, 18:297.

\bibitem[Huebner et~al., 1977]{HMMJ+77}
Huebner, W., Merts, A., Magee~Jr, N., Argo, M., and 1977 (1977).
\newblock Astrophysical {O}pacity {L}ibrary.
\newblock {\em Los Alamos Scientific Laboratory, LA-6760-M}.

\bibitem[{Iben}, 1975]{1975ApJ...196..525I}
{Iben}, Jr., I. (1975).
\newblock {Thermal pulses; p-capture, alpha-capture, s-process nucleosynthesis;
  and convective mixing in a star of intermediate mass}.
\newblock {\em ApJ}, 196:525.

\bibitem[{Iben}, 1977]{1977ApJ...217..788I}
{Iben}, Jr., I. (1977).
\newblock {Thermal pulse and interpulse properties of intermediate-mass stellar
  models with carbon-oxygen cores of mass 0.96, 1.16, and 1.36 solar masses}.
\newblock {\em ApJ}, 217:788--798.

\bibitem[{Iben} and {Renzini}, 1983]{1983ARAA..21..271I}
{Iben}, Jr., I. and {Renzini}, A. (1983).
\newblock {Asymptotic giant branch evolution and beyond}.
\newblock {\em ARA\&A}, 21:271--342.

\bibitem[{Iglesias} and {Rogers}, 1996]{1996ApJ...464..943I}
{Iglesias}, C.~A. and {Rogers}, F.~J. (1996).
\newblock {Updated Opal Opacities}.
\newblock {\em ApJ}, 464:943.

\bibitem[{Iliadis} et~al., 1996]{1996PhRvC..53..475I}
{Iliadis}, C., {Buchmann}, L., {Endt}, P.~M., {Herndl}, H., and {Wiescher}, M.
  (1996).
\newblock {New stellar reaction rates for $^{25}${M}g(p,{$\gamma$})$^{26}${A}l
  and $^{25}${A}l(p,{$\gamma$})$^{26}${S}i}.
\newblock {\em PhRvC}, 53:475--496.

\bibitem[{Iliadis} et~al., 1990]{1990NuPhA.512..509I}
{Iliadis}, C., {Schange}, T., {Rolfs}, C., {Schr{\"o}der}, U., {Somorjai}, E.,
  {Trautvetter}, H.~P., {Wolke}, K., {Endt}, P.~M., {Kikstra}, S.~W.,
  {Champagne}, A.~E., {Arnould}, M., and {Paulus}, G. (1990).
\newblock Low-energy resonances in $^{25}${M}g(p,$\gamma$)$^{26}${A}l,
  $^{26}${M}g(p,$\gamma$)$^{27}${A}l and $^{27}${A}l(p,$\gamma$)$^{28}${S}i.
\newblock {\em Nuclear Physics A}, 512:509--530.

\bibitem[{Israelian} et~al., 2004]{2004AA...419.1095I}
{Israelian}, G., {Shchukina}, N., {Rebolo}, R., {Basri}, G., {Gonz{\'a}lez
  Hern{\'a}ndez}, J.~I., and {Kajino}, T. (2004).
\newblock {Oxygen and magnesium abundance in the ultra-metal-poor giants CS
  22949-037 and CS 29498-043: Challenges in models of atmospheres}.
\newblock {\em A\&A}, 419:1095--1109.

\bibitem[{Itoh} et~al., 1983]{1983ApJ...273..774I}
{Itoh}, N., {Mitake}, S., {Iyetomi}, H., and {Ichimaru}, S. (1983).
\newblock {Electrical and thermal conductivities of dense matter in the liquid
  metal phase. I - High-temperature results}.
\newblock {\em ApJ}, 273:774.

\bibitem[{Ivans} et~al., 1999]{1999AJ....118.1273I}
{Ivans}, I.~I., {Sneden}, C., {Kraft}, R.~P., {Suntzeff}, N.~B., {Smith},
  V.~V., {Langer}, G.~E., and {Fulbright}, J.~P. (1999).
\newblock {Star-to-Star Abundance Variations among Bright Giants in the Mildly
  Metal-poor Globular Cluster M4}.
\newblock {\em AJ}, 118:1273--1300.

\bibitem[{Iwamoto} et~al., 2004]{2004ApJ...602..377I}
{Iwamoto}, N., {Kajino}, T., {Mathews}, G.~J., {Fujimoto}, M.~Y., and {Aoki},
  W. (2004).
\newblock {Flash-Driven Convective Mixing in Low-Mass, Metal-deficient
  Asymptotic Giant Branch Stars: A New Paradigm for Lithium Enrichment and a
  Possible s-Process}.
\newblock {\em ApJ}, 602:377--388.

\bibitem[Izotov and Thuan, 1999]{IT99}
Izotov, Y.~I. and Thuan, T.~X. (1999).
\newblock Heavy-{E}lement {A}bundances in {B}lue {C}ompact {G}alaxies.
\newblock {\em ApJ}, 511(2):639--659.

\bibitem[{Jorissen} and {Arnould}, 1989]{1989AA...221..161J}
{Jorissen}, A. and {Arnould}, M. (1989).
\newblock {Proton mixing in He-rich layers - The C-13(alpha,n)O-16 neutron
  source and associated nucleosynthesis}.
\newblock {\em A\&A}, 221:161--179.

\bibitem[{Kaeppeler} et~al., 1994]{1994ApJ...437..396K}
{Kaeppeler}, F., {Wiescher}, M., {Giesen}, U., {Goerres}, J., {Baraffe}, I.,
  {El Eid}, M., {Raiteri}, C.~M., {Busso}, M., {Gallino}, R., {Limongi}, M.,
  and {Chieffi}, A. (1994).
\newblock {Reaction rates for O-18(alpha, gamma)Ne-22, Ne-22(alpha,
  gamma)Mg-26, and Ne-22(alpha, n)Mg-25 in stellar helium burning and s-process
  nucleosynthesis in massive stars}.
\newblock {\em ApJ}, 437:396--409.

\bibitem[Karakas, 2003]{entry-5}
Karakas, A.~I. (2003).
\newblock {\em Asymptotic {G}iant {B}ranch {S}tars: {T}heir {I}nfluence on
  {B}inary {S}ystems and the {I}nterstellar {M}edium}.
\newblock PhD thesis, Monash University, Australia.

\bibitem[{Karakas} and {Lattanzio}, 2003]{2003PASA...20..279K}
{Karakas}, A.~I. and {Lattanzio}, J.~C. (2003).
\newblock {Production of Aluminium and the Heavy Magnesium Isotopes in
  Asymptotic Giant Branch Stars}.
\newblock {\em Publications of the Astronomical Society of Australia},
  20:279--293.

\bibitem[{Karakas} et~al., 2002]{2002PASA...19..515K}
{Karakas}, A.~I., {Lattanzio}, J.~C., and {Pols}, O.~R. (2002).
\newblock {Parameterising the Third Dredge-up in Asymptotic Giant Branch
  Stars}.
\newblock {\em Publications of the Astronomical Society of Australia},
  19:515--526.

\bibitem[Kato, 1966]{1966PASJ...18..374K}
Kato, S. (1966).
\newblock Overstable {C}onvection in a {M}edium {S}tratified in {M}ean
  {M}olecular {W}eight.
\newblock {\em PASJ}, 18:374.

\bibitem[{Kirchhoff} and {Bunsen}, 1860]{1860AnP...186..161K}
{Kirchhoff}, G. and {Bunsen}, R. (1860).
\newblock {Chemische Analyse durch Spectralbeobachtungen}.
\newblock {\em Annalen der Physik}, 186:161--189.

\bibitem[{Kiziloglu} and {Eryurt-Ezer}, 1987]{1987ApSS.136...83K}
{Kiziloglu}, N. and {Eryurt-Ezer}, D. (1987).
\newblock {Pregalactic-primordial low-mass stars}.
\newblock {\em Ap\&SS}, 136:83--90.

\bibitem[{Kraft} et~al., 1992]{1992AJ....104..645K}
{Kraft}, R.~P., {Sneden}, C., {Langer}, G.~E., and {Prosser}, C.~F. (1992).
\newblock {Oxygen abundances in halo giants. II - Giants in the globular
  clusters M13 and M3 and the intermediately metal-poor halo field}.
\newblock {\em AJ}, 104:645--668.

\bibitem[{Kraft} et~al., 1997]{1997AJ....113..279K}
{Kraft}, R.~P., {Sneden}, C., {Smith}, G.~H., {Shetrone}, M.~D., {Langer},
  G.~E., and {Pilachowski}, C.~A. (1997).
\newblock {Proton Capture Chains in Globular Cluster Stars.II.Oxygen, Sodium,
  Magnesium, and Aluminum Abundances in M13 Giants Brighter Than the Horizontal
  Branch}.
\newblock {\em AJ}, 113:279.

\bibitem[{Kroupa} et~al., 1993]{1993MNRAS.262..545K}
{Kroupa}, P., {Tout}, C.~A., and {Gilmore}, G. (1993).
\newblock {The distribution of low-mass stars in the Galactic disc}.
\newblock {\em MNRAS}, 262:545--587.

\bibitem[Kunth and \"Ostlin, 2000]{KO00}
Kunth, D. and \"Ostlin, G. (2000).
\newblock The most metal-poor galaxies.
\newblock {\em A\&ARv}, 10(1):1--79.

\bibitem[Langer et~al., 1985]{1985AA...145..179L}
Langer, N., El~Eid, M.~F., and Fricke, K.~J. (1985).
\newblock Evolution of massive stars with semiconvective diffusion.
\newblock {\em A\&A}, 145:179--191.

\bibitem[Laor et~al., 1995]{LBJ+95}
Laor, A., Bahcall, J.~N., Jannuzi, B.~T., Schneider, D.~P., and Green, R.~F.
  (1995).
\newblock The {U}ltraviolet {E}mission {P}roperties of 13 {Q}uasars.
\newblock {\em ApJS}, 99:1.

\bibitem[Lara et~al., 2006]{LKM06}
Lara, J.~F., Kajino, T., and Mathews, G.~J. (2006).
\newblock Inhomogeneous big bang nucleosynthesis revisited.
\newblock {\em PhRvD}, 78(8).

\bibitem[Lattanzio, 1984]{LATT84}
Lattanzio, J.~C. (1984).
\newblock {\em The {E}volution of {I}nitially {I}nhomogeneous {S}tars and {L}ow
  {M}ass {AGB} {S}tars}.
\newblock PhD thesis, Monash University, Australia, School of Mathematical
  Sciences.
\newblock Lattanzio's PhD Thesis.

\bibitem[{Lattanzio}, 1986]{1986ApJ...311..708L}
{Lattanzio}, J.~C. (1986).
\newblock {The asymptotic giant branch evolution of 1.0-3.0 solar mass stars as
  a function of mass and composition}.
\newblock {\em ApJ}, 311:708--730.
\newblock JL's thesis paper.

\bibitem[{Lattanzio}, 1989]{1989ApJ...344L..25L}
{Lattanzio}, J.~C. (1989).
\newblock {Carbon dredge-up in low-mass stars and solar metallicity stars}.
\newblock {\em ApJL}, 344:L25.

\bibitem[{Lattanzio}, 1992]{1992PASAu..10..120L}
{Lattanzio}, J.~C. (1992).
\newblock {Hot bottom burning in a 5 solar mass model}.
\newblock {\em Proceedings of the Astronomical Society of Australia}, 10:120.

\bibitem[Ledoux, 1947]{1947ApJ...105..305L}
Ledoux, W.~P. (1947).
\newblock Stellar {M}odels with {C}onvection and with {D}iscontinuity of the
  {M}ean {M}olecular.
\newblock {\em ApJ}, 105:305.

\bibitem[{Lee}, 2000]{2000JKAS...33..137L}
{Lee}, S.~G. (2000).
\newblock {CN and CH Band Strengths of Bright Giants in the Globular Cluster
  M15}.
\newblock {\em Journal of Korean Astronomical Society}, 33:137--142.

\bibitem[{Levshakov} et~al., 2001]{LKA01}
{Levshakov}, S.~A., {Kegel}, W.~H., and {Agafonova}, I.~I. (2001).
\newblock {Argon, silicon and iron abundances in the damped Ly-alpha system I
  Zw 18}.
\newblock {\em A\&A}, 373:836--842.

\bibitem[{Limongi} et~al., 2003]{2003ApJ...594L.123L}
{Limongi}, M., {Chieffi}, A., and {Bonifacio}, P. (2003).
\newblock {On the Origin of HE 0107-5240, the Most Iron-deficient Star
  Presently Known}.
\newblock {\em ApJL}, 594:L123--L126.

\bibitem[{Lodders}, 2003]{2003ApJ...591.1220L}
{Lodders}, K. (2003).
\newblock {Solar System Abundances and Condensation Temperatures of the
  Elements}.
\newblock {\em ApJ}, 591:1220--1247.

\bibitem[{Lucatello} et~al., 2006]{2006ApJ...652L..37L}
{Lucatello}, S., {Beers}, T.~C., {Christlieb}, N., {Barklem}, P.~S., {Rossi},
  S., {Marsteller}, B., {Sivarani}, T., and {Lee}, Y.~S. (2006).
\newblock {The Frequency of Carbon-enhanced Metal-poor Stars in the Galaxy from
  the HERES Sample}.
\newblock {\em ApJL}, 652:L37--L40.

\bibitem[{Lucatello} et~al., 2005]{2005ApJ...625..825L}
{Lucatello}, S., {Tsangarides}, S., {Beers}, T.~C., {Carretta}, E., {Gratton},
  R.~G., and {Ryan}, S.~G. (2005).
\newblock {The Binary Frequency Among Carbon-enhanced, s-Process-rich,
  Metal-poor Stars}.
\newblock {\em ApJ}, 625:825--832.

\bibitem[{Luck} and {Bond}, 1991]{1991ApJS...77..515L}
{Luck}, R.~E. and {Bond}, H.~E. (1991).
\newblock {Subgiant CH stars. II - Chemical compositions and the evolutionary
  connection with barium stars}.
\newblock {\em ApJS}, 77:515--540.

\bibitem[Lugaro, 2001]{Lug01}
Lugaro, M. (2001).
\newblock {\em Nucleosynthesis in {AGB} {S}tars}.
\newblock PhD thesis, Monash University, Australia.

\bibitem[{Maeder}, 1975]{1975AA....40..303M}
{Maeder}, A. (1975).
\newblock {Stellar evolution. III - The overshooting from convective cores}.
\newblock {\em A\&A}, 40:303--310.

\bibitem[{Maeder}, 1976]{Mad76}
{Maeder}, A. (1976).
\newblock {Stellar evolution. V - Evolutionary models of population I stars
  with or without overshooting from convective cores}.
\newblock {\em A\&A}, 47:389--400.

\bibitem[{Maeder} and {Meynet}, 2006]{2006AA...448L..37M}
{Maeder}, A. and {Meynet}, G. (2006).
\newblock {On the origin of the high helium sequence in {$\omega$} Centauri}.
\newblock {\em A\&A}, 448:L37--L41.

\bibitem[{Mallia}, 1978]{1978AA....70..115M}
{Mallia}, E.~A. (1978).
\newblock {Spectra of asymbiotic giant branch stars in four southern globular
  clusters.}
\newblock {\em A\&A}, 70:115--123.

\bibitem[{Marigo} et~al., 2001]{2001AA...371..152M}
{Marigo}, P., {Girardi}, L., {Chiosi}, C., and {Wood}, P.~R. (2001).
\newblock {Zero-metallicity stars. I. Evolution at constant mass}.
\newblock {\em A\&A}, 371:152--173.

\bibitem[Matsuura et~al., 2005]{Mats1205}
Matsuura, S., Fujimoto, S., Nishimura, S., Hashimoto, M., and Sato, K. (2005).
\newblock Heavy element production in inhomogeneous big bang nucleosynthesis.
\newblock {\em PhsRevD}, 72(12).

\bibitem[Merryfield, 1995]{1995ApJ...444..318M}
Merryfield, W.~J. (1995).
\newblock Hydrodynamics of semiconvection.
\newblock {\em ApJ}, 444:318--337.

\bibitem[{Messenger}, 2000]{2000PASA...17..284M}
{Messenger}, B. (2000).
\newblock {Abundance anomalies in globular cluster red giant stars. (Thesis
  abstract).}
\newblock {\em Publications of the Astronomical Society of Australia},
  17:284--284.

\bibitem[Meynet et~al., 2004]{2004AA...416.1023M}
Meynet, G., Maeder, A., and Mowlavi, N. (2004).
\newblock Diffusion in stellar interiors: {C}ritical tests of three numerical
  methods.
\newblock {\em A\&A}, 416:1023--1036.

\bibitem[{Mitake} et~al., 1984]{1984ApJ...277..375M}
{Mitake}, S., {Ichimaru}, S., and {Itoh}, N. (1984).
\newblock {Electrical and thermal conductivities of dense matter in the liquid
  metal phase. II - Low-temperature quantum corrections}.
\newblock {\em ApJ}, 277:375.

\bibitem[{Molaro} and {Castelli}, 1990]{1990AA...228..426M}
{Molaro}, P. and {Castelli}, F. (1990).
\newblock {A new ultra metal-deficient star - CS 22876}.
\newblock {\em A\&A}, 228:426--442.

\bibitem[{Nakamura} and {Umemura}, 2001]{2001ApJ...548...19N}
{Nakamura}, F. and {Umemura}, M. (2001).
\newblock {On the Initial Mass Function of Population III Stars}.
\newblock {\em ApJ}, 548:19--32.

\bibitem[{Nordlund} and {Dravins}, 1990]{1990AA...228..155N}
{Nordlund}, A. and {Dravins}, D. (1990).
\newblock {Stellar granulation. III - Hydrodynamic model atmospheres. IV - Line
  formation in inhomogeneous stellar photospheres. V - Synthetic spectral lines
  in disk-integrated starlight}.
\newblock {\em A\&A}, 228:155--217.

\bibitem[{Norris} et~al., 1981]{1981ApJ...244..205N}
{Norris}, J., {Cottrell}, P.~L., {Freeman}, K.~C., and {Da Costa}, G.~S.
  (1981).
\newblock {The abundance spread in the giants of NGC 6752}.
\newblock {\em ApJ}, 244:205--220.

\bibitem[\"Odman and Izzard, 2004]{2004MmSAI..75..631O}
\"Odman, C.~J. and Izzard, R.~G. (2004).
\newblock Stellardb: {A} {S}tellar {A}bundances {D}atabase.
\newblock {\em Memorie della Societa Astronomica Italiana}, 75:631.

\bibitem[Olive and Skillman, 2004]{Olive1204}
Olive, K.~A. and Skillman, E.~D. (2004).
\newblock A {R}ealistic {D}etermination of the {E}rror on the {P}rimordial
  {H}elium {A}bundance: {S}teps toward {N}onparametric {N}ebular {H}elium
  {A}bundances.
\newblock {\em ApJ}, 617(1):29--49.

\bibitem[{Olszewski}, 1993]{1993ASPC...48..351O}
{Olszewski}, E.~W. (1993).
\newblock {The Age and Metallicity Distributions Among the Magellanic Cloud
  Clusters}.
\newblock In {Smith}, G.~H. and {Brodie}, J.~P., editors, {\em ASP Conf. Ser.
  48: The Globular Cluster-Galaxy Connection}, page 351.

\bibitem[Olszewski et~al., 1996]{OSM96}
Olszewski, E.~W., Suntzeff, N.~B., and Mateo, M. (1996).
\newblock Old and {I}ntermediate-{A}ge {S}tellar {P}opulations in the
  {M}agellanic {C}louds.
\newblock {\em ARA\&A}, 34:511--550.

\bibitem[{Parmentier} et~al., 1999]{1999AA...352..138P}
{Parmentier}, G., {Jehin}, E., {Magain}, P., {Neuforge}, C., {Noels}, A., and
  {Thoul}, A.~A. (1999).
\newblock {The self-enrichment of galactic halo globular clusters. A clue to
  their formation?}
\newblock {\em A\&A}, 352:138--148.

\bibitem[{Peebles}, 1966]{1966ApJ...146..542P}
{Peebles}, P.~J.~E. (1966).
\newblock {Primordial Helium Abundance and the Primordial Fireball. II}.
\newblock {\em ApJ}, 146:542.

\bibitem[Pettini, 2004]{Pet04}
Pettini, M. (2004).
\newblock Element abundances through the cosmic ages.
\newblock {\em In: Cosmochemistry. The melting pot of the elements. XIII Canary
  Islands Winter School of Astrophysics}, page 275.

\bibitem[{Picardi} et~al., 2004]{2004ApJ...609.1035P}
{Picardi}, I., {Chieffi}, A., {Limongi}, M., {Pisanti}, O., {Miele}, G.,
  {Mangano}, G., and {Imbriani}, G. (2004).
\newblock {Evolution and Nucleosynthesis of Primordial Low-Mass Stars}.
\newblock {\em ApJ}, 609:1035--1044.

\bibitem[{Pilachowski}, 1988]{1988ApJ...326L..57P}
{Pilachowski}, C.~A. (1988).
\newblock {The abundance of oxygen in M92 giant stars}.
\newblock {\em ApJL}, 326:L57--L60.

\bibitem[{Pilachowski} et~al., 1996a]{1996AJ....111.1689P}
{Pilachowski}, C.~A., {Sneden}, C., and {Kraft}, R.~P. (1996a).
\newblock {Sodium Abundances in Field Metal-Poor Stars}.
\newblock {\em AJ}, 111:1689.

\bibitem[{Pilachowski} et~al., 1996b]{1996AJ....112..545P}
{Pilachowski}, C.~A., {Sneden}, C., {Kraft}, R.~P., and {Langer}, G.~E.
  (1996b).
\newblock {Proton Capture Chains in Glubular Cluster Stars. I. Evidence for
  Deep Mixing Based on Sodium and Magnesium Abundances in M13 Giants}.
\newblock {\em AJ}, 112:545.

\bibitem[{Plez} and {Cohen}, 2005]{2005AA...434.1117P}
{Plez}, B. and {Cohen}, J.~G. (2005).
\newblock {Analysis of the carbon-rich very metal-poor dwarf G77-61}.
\newblock {\em A\&A}, 434:1117--1124.

\bibitem[{Poelarends} et~al., 2006]{2006MmSAI..77..846P}
{Poelarends}, A.~J.~T., {Izzard}, R.~G., {Herwig}, F., {Langer}, N., and
  {Heger}, A. (2006).
\newblock Supernovae from massive {AGB} stars.
\newblock {\em Memorie della Societa Astronomica Italiana}, 77:846.

\bibitem[Pols and Tout, 2001]{2001MmSAI..72..299P}
Pols, O.~R. and Tout, C.~A. (2001).
\newblock Thermal pulses and dredge-up in {AGB} stars with a consistent
  solution for stellar structure and composition: dependence on convection
  prescriptions.
\newblock {\em Memorie della Societa Astronomica Italiana}, 72:299--308.

\bibitem[Powell et~al., 1999]{DCP+99}
Powell, D.~C., Iliadis, C., Champagne, A.~E., Grossmann, C.~A., Hale, S.~E.,
  Hansper, V.~Y., and McLean, L.~K. (1999).
\newblock Reaction rate of $^{24}${M}g($p,\gamma$)$^{25}${A}l.
\newblock {\em Nuclear Physics A}, 660(3):349.

\bibitem[Prandtl, 1925]{Pr25}
Prandtl, L. (1925).
\newblock {\em Zs. Angew. Math. Mech.}, 5:136.

\bibitem[Press et~al., 1992]{1992nrfa.book.....P}
Press, W.~H., Teukolsky, S.~A., Vetterling, W.~T., and Flannery, B.~P. (1992).
\newblock {\em Numerical recipes in {FORTRAN}. {T}he art of scientific
  computing}.
\newblock Cambridge: University Press, 2nd ed.

\bibitem[{Prochaska} et~al., 2003]{2003ApJS..147..227P}
{Prochaska}, J.~X., {Gawiser}, E., {Wolfe}, A.~M., {Cooke}, J., and {Gelino},
  D. (2003).
\newblock {The ESI/Keck II Damped Ly{$\alpha$} Abundance Database}.
\newblock {\em ApJS}, 147:227--264.

\bibitem[{Raikh} and {Iakovlev}, 1982]{1982ApSS..87..193R}
{Raikh}, M.~E. and {Iakovlev}, D.~G. (1982).
\newblock {Thermal and electrical conductivities of crystals in neutron stars
  and degenerate dwarfs}.
\newblock {\em Ap\&SS}, 87:193--203.

\bibitem[{Ramadurai}, 1976]{1976MNRAS.176....9R}
{Ramadurai}, S. (1976).
\newblock {Neutral current and the onset of the helium flash}.
\newblock {\em MNRAS}, 176:9.

\bibitem[{Reimers}, 1975]{1975MSRSL...8..369R}
{Reimers}, D. (1975).
\newblock {Circumstellar absorption lines and mass loss from red giants}.
\newblock {\em Memoires of the Societe Royale des Sciences de Liege},
  8:369--382.

\bibitem[{Renzini}, 1987]{1987AA...188...49R}
{Renzini}, A. (1987).
\newblock {Some embarrassments in current treatments of convective
  overshooting}.
\newblock {\em A\&A}, 188:49--54.

\bibitem[{Ritossa} et~al., 1996]{1996MmSAI..67..675R}
{Ritossa}, C., {Garcia-Berror}, E., and {Iben}, Jr., I. (1996).
\newblock {On the evolution of super AGB stars with ONe degenerate cores: the
  case of a 10 M$_{\odot}$ model}.
\newblock {\em Memorie della Societa Astronomica Italiana}, 67:675.

\bibitem[{Robertson} and {Faulkner}, 1972]{1972ApJ...171..309R}
{Robertson}, J.~W. and {Faulkner}, D.~J. (1972).
\newblock {Semiconvection in the Core-Helium Phase of Stellar Evolution}.
\newblock {\em ApJ}, 171:309.

\bibitem[{Rogers} et~al., 1996]{1996ApJ...456..902R}
{Rogers}, F.~J., {Swenson}, F.~J., and {Iglesias}, C.~A. (1996).
\newblock {OPAL Equation-of-State Tables for Astrophysical Applications}.
\newblock {\em ApJ}, 456:902.

\bibitem[{Sackmann} and {Boothroyd}, 1992]{1992ApJ...392L..71S}
{Sackmann}, I.-J. and {Boothroyd}, A.~I. (1992).
\newblock {The creation of superrich lithium giants}.
\newblock {\em ApJL}, 392:L71--L74.

\bibitem[Sackmann et~al., 1974]{1974ApJ...187..555S}
Sackmann, I.~J., Smith, R.~L., and Despain, K.~H. (1974).
\newblock Carbon and eruptive stars : surface enrichment of lithium, carbon,
  nitrogen, and {C}13 by deep mixing.
\newblock {\em ApJ}, 187:555--574.

\bibitem[{Sandquist} and {Bolte}, 2004]{2004ApJ...611..323S}
{Sandquist}, E.~L. and {Bolte}, M. (2004).
\newblock {Exploring the Upper Red Giant and Asymptotic Giant Branches: The
  Globular Cluster M5}.
\newblock {\em ApJ}, 611:323--337.

\bibitem[{Schlattl} et~al., 2001]{2001ApJ...559.1082S}
{Schlattl}, H., {Cassisi}, S., {Salaris}, M., and {Weiss}, A. (2001).
\newblock {On the Helium Flash in Low-Mass Population III Red Giant Stars}.
\newblock {\em ApJ}, 559:1082--1093.

\bibitem[{Schlattl} et~al., 2002]{2002AA...395...77S}
{Schlattl}, H., {Salaris}, M., {Cassisi}, S., and {Weiss}, A. (2002).
\newblock {The surface carbon and nitrogen abundances in models of ultra
  metal-poor stars}.
\newblock {\em A\&A}, 395:77--83.

\bibitem[{Schwarzschild} and {Selberg}, 1962]{1962ApJ...136..150S}
{Schwarzschild}, I.~M. and {Selberg}, H. (1962).
\newblock {Red Giants of Population II.}
\newblock {\em ApJ}, 136:150.

\bibitem[Schwarzschild, 1906]{SWCH1906}
Schwarzschild, K. (1906).
\newblock {\em Gott. Nach.}, 1:41.

\bibitem[{Schwarzschild}, 1958]{1958ses..book.....S}
{Schwarzschild}, M. (1958).
\newblock {\em {Structure and evolution of the stars.}}
\newblock Princeton, Princeton University Press, 1958.

\bibitem[Schwarzschild and H\"arm, 1958]{1958ApJ...128..348S}
Schwarzschild, M. and H\"arm, R. (1958).
\newblock Evolution of {V}ery {M}assive {S}tars.
\newblock {\em ApJ}, 128:348.

\bibitem[Schwarzschild and H\"arm, 1965]{1965ApJ...142..855S}
Schwarzschild, M. and H\"arm, R. (1965).
\newblock Therma {I}nstability in {N}on-{D}egenerate {S}tars.
\newblock {\em ApJ}, 142:855.

\bibitem[{Schwarzschild} and {H{\"a}rm}, 1967]{1967ApJ...150..961S}
{Schwarzschild}, M. and {H{\"a}rm}, R. (1967).
\newblock {Hydrogen Mixing by Helium-Shell Flashes}.
\newblock {\em ApJ}, 150:961.

\bibitem[{Searle} and {Sargent}, 1972]{1972ApJ...173...25S}
{Searle}, L. and {Sargent}, W.~L.~W. (1972).
\newblock {Inferences from the Composition of Two Dwarf Blue Galaxies}.
\newblock {\em ApJ}, 173:25.

\bibitem[{Sears}, 1959]{1959ApJ...129..489S}
{Sears}, R.~L. (1959).
\newblock {An Evolutionary Sequence of Solar Models.}
\newblock {\em ApJ}, 129:489.

\bibitem[{Seaton} et~al., 1992]{1992RMxAA..23...19S}
{Seaton}, M.~J., {Zeippen}, C.~J., {Tully}, J.~A., {Pradhan}, A.~K., {Mendoza},
  C., {Hibbert}, A., and {Berrington}, K.~A. (1992).
\newblock {The Opacity Project - Computation of Atomic Data}.
\newblock {\em Revista Mexicana de Astronomia y Astrofisica, vol.~23}, 23:19.

\bibitem[{Shaviv} and {Salpeter}, 1973]{1973ApJ...184..191S}
{Shaviv}, G. and {Salpeter}, E.~E. (1973).
\newblock {Convective Overshooting in Stellar Interior Models}.
\newblock {\em ApJ}, 184:191--200.

\bibitem[{Shetrone}, 2003]{2003ApJ...585L..45S}
{Shetrone}, M.~D. (2003).
\newblock {Carbon Isotopes in Globular Clusters Down to the Bump in the
  Luminosity Function}.
\newblock {\em ApJL}, 585:L45--L48.

\bibitem[{Shigeyama} and {Tsujimoto}, 1998]{1998ApJ...507L.135S}
{Shigeyama}, T. and {Tsujimoto}, T. (1998).
\newblock {Fossil Imprints of the First-Generation Supernova Ejecta in
  Extremely Metal-deficient Stars}.
\newblock {\em ApJl}, 507:L135--L139.

\bibitem[{Siess} et~al., 2002]{2002ApJ...570..329S}
{Siess}, L., {Livio}, M., and {Lattanzio}, J. (2002).
\newblock {Structure, Evolution, and Nucleosynthesis of Primordial Stars}.
\newblock {\em ApJ}, 570:329--343.

\bibitem[Simpson, 1971]{1971ApJ...165..295S}
Simpson, E.~E. (1971).
\newblock Evolutionary {M}odels of {S}tars of 15 and 30 {M}$_\odot$.
\newblock {\em ApJ}, 165:295.

\bibitem[{Singh} et~al., 1995]{1995AA...295..703S}
{Singh}, H.~P., {Roxburgh}, I.~W., and {Chan}, K.~L. (1995).
\newblock {Three-dimensional simulation of penetrative convection: penetration
  below a convection zone.}
\newblock {\em A\&A}, 295:703.

\bibitem[{Smith}, 2002]{2002PASP..114.1215S}
{Smith}, G.~H. (2002).
\newblock {The Incidence of CN-strong Giants in Globular Clusters}.
\newblock {\em PASP}, 114:1215--1221.

\bibitem[{Smith} and {Norris}, 1993]{1993AJ....105..173S}
{Smith}, G.~H. and {Norris}, J.~E. (1993).
\newblock {CN variations among asymptotic giant branch and horizontal branch
  stars in the intermediate metallicity globular clusters M5, M4, and NGC
  6752}.
\newblock {\em AJ}, 105:173--183.

\bibitem[{Smith} et~al., 1996]{1996AJ....112.1511S}
{Smith}, G.~H., {Shetrone}, M.~D., {Bell}, R.~A., {Churchill}, C.~W., and
  {Briley}, M.~M. (1996).
\newblock {CNO Abundances of Bright Giants in the Globular Clusters M3 and
  M13}.
\newblock {\em AJ}, 112:1511.

\bibitem[{Sneden} et~al., 2000]{2000MmSAI..71..657S}
{Sneden}, C., {Ivans}, I.~I., and {Kraft}, R.~P. (2000).
\newblock {Do AGB stars differ chemically from RGB stars in globular clusters?}
\newblock {\em Memorie della Societa Astronomica Italiana}, 71:657--665.

\bibitem[Songaila, 2001]{Son01}
Songaila, A. (2001).
\newblock The {M}inimum {U}niversal {M}etal {D}ensity between {R}edshifts of
  1.5 and 5.5.
\newblock {\em ApJ}, 561(2):L153--L156.

\bibitem[Songaila and Cowie, 1996]{SC96}
Songaila, A. and Cowie, L.~L. (1996).
\newblock Metal enrichment and {I}onization {B}alance in the {L}yman {A}lpha
  {F}orest at {Z} = 3.
\newblock {\em AJ}, 112:335.

\bibitem[{Spiegel}, 1971]{1971ARAA...9..323S}
{Spiegel}, E.~A. (1971).
\newblock {Convection in Stars: I. Basic Boussinesq Convection}.
\newblock {\em ARA\&A}, 9:323.

\bibitem[Spite and Spite, 1982]{SS82}
Spite, F. and Spite, M. (1982).
\newblock Abundance of lithium in unevolved halo stars and old disk stars -
  {I}nterpretation and consequences.
\newblock {\em A\&A}, 115(2):357--366.

\bibitem[{Spite} et~al., 2006]{2006AA...455..291S}
{Spite}, M., {Cayrel}, R., {Hill}, V., {Spite}, F., {Fran{\c c}ois}, P.,
  {Plez}, B., {Bonifacio}, P., {Molaro}, P., {Depagne}, E., {Andersen}, J.,
  {Barbuy}, B., {Beers}, T.~C., {Nordstr{\"o}m}, B., and {Primas}, F. (2006).
\newblock {First stars IX - Mixing in extremely metal-poor giants. Variation of
  the $^{12}$C/$^{13}$C, [Na/Mg] and [Al/Mg] ratios}.
\newblock {\em A\&A}, 455:291--301.

\bibitem[Spruit, 1992]{1992AA...253..131S}
Spruit, H.~C. (1992).
\newblock The rate of mixing in semiconvective zones.
\newblock {\em A\&A}, 253:131--138.

\bibitem[Sreenivasan and Ziebarth, 1971]{1971BAAS....3R.394S}
Sreenivasan, S.~R. and Ziebarth, K.~E. (1971).
\newblock The {E}volution of a 20 {M}$_\odot$ {S}tar.
\newblock {\em BAAS}, 3:394.

\bibitem[{Stancliffe} et~al., 2007]{2007astro.ph..2138S}
{Stancliffe}, R.~J., {Glebbeek}, E., {Izzard}, R.~G., and {Pols}, O.~R. (2007).
\newblock {Carbon-enhanced metal-poor stars and thermohaline mixing}.
\newblock {\em ArXiv Astrophysics e-prints}.

\bibitem[{Stancliffe} et~al., 2005]{2005MNRAS.356L...1S}
{Stancliffe}, R.~J., {Izzard}, R.~G., and {Tout}, C.~A. (2005).
\newblock {Third dredge-up in low-mass stars: solving the Large Magellanic
  Cloud carbon star mystery}.
\newblock {\em MNRAS}, 356:L1--L5.

\bibitem[Stothers and Chin, 1994]{1994ApJ...431..797S}
Stothers, R.~B. and Chin, C.-W. (1994).
\newblock Galactic stars applied to tests of the criterion for convection and
  semiconvection in an inhomogeneous star.
\newblock {\em ApJ}, 431:797--805.

\bibitem[{Straniero} et~al., 1996]{1996MmSAI..67..651S}
{Straniero}, O., {Chieffi}, A., {Limongi}, M., {Busso}, M., {Gallino}, R., and
  {Arlandini}, C. (1996).
\newblock {The third dredge-up and carbon star formation in population I low
  mass stars}.
\newblock {\em Memorie della Societa Astronomica Italiana}, 67:651.

\bibitem[Straniero et~al., 2003]{2003ApJ...583..878S}
Straniero, O., Dom\'{\i}nguez, I., Imbriani, G., and Piersanti, L. (2003).
\newblock The {C}hemical {C}omposition of {W}hite {D}warfs as a {T}est of
  {C}onvective {E}fficiency during {C}ore {H}elium {B}urning.
\newblock {\em ApJ}, 583:878--884.

\bibitem[{Suda} et~al., 2004]{2004ApJ...611..476S}
{Suda}, T., {Aikawa}, M., {Machida}, M.~N., {Fujimoto}, M.~Y., and {Iben},
  I.~J. (2004).
\newblock {Is HE 0107-5240 A Primordial Star? The Characteristics of Extremely
  Metal-Poor Carbon-Rich Stars}.
\newblock {\em ApJ}, 611:476--493.

\bibitem[{Suntzeff}, 1981]{1981ApJS...47....1S}
{Suntzeff}, N.~B. (1981).
\newblock {Carbon and nitrogen abundances in the giant stars of the globular
  clusters M3 and M13}.
\newblock {\em ApJS}, 47:1--32.

\bibitem[{Suntzeff} and {Smith}, 1991]{1991ApJ...381..160S}
{Suntzeff}, N.~B. and {Smith}, V.~V. (1991).
\newblock {Carbon isotopic abundances in giant stars in the CN-bimodal globular
  clusters NGC 6752 and M4}.
\newblock {\em ApJ}, 381:160--172.

\bibitem[Sweigart, 1991]{1991ASPC...13..299S}
Sweigart, A.~V. (1991).
\newblock Horizontal-branch evolution with time-dependent convective
  overshooting.
\newblock In Janes, K., editor, {\em Asp conf. ser. 13: the formation and
  evolution of star clusters}, pages 299--301.

\bibitem[Sweigart et~al., 2000]{2000LIACo..35..529S}
Sweigart, A.~V., Lattanzio, J.~C., Gray, J.~P., and Tout, C.~A. (2000).
\newblock Gravonuclear instabilities in post-horizontal-branch stars.
\newblock In Noels, A., Magain, P., Caro, D., Jehin, E., Parmentier, G., and
  Thoul, A.~A., editors, {\em Liege international astrophysical colloquia},
  page 529.

\bibitem[Thielemann et~al., 1987]{Thi87}
Thielemann, F., Arnould, M., and Truran, J.~W. (1987).
\newblock {\em in Vangioni-Flam E., Audouze J., Casse M., Chieze J. P., Tran
  Thanh Van J., (eds.), Advances of Nuclear Astrophysics. Editions Frontieres,
  France.}, page 525.

\bibitem[{Tornambe} and {Chieffi}, 1986]{1986MNRAS.220..529T}
{Tornambe}, A. and {Chieffi}, A. (1986).
\newblock {Extremely metal-deficient stars. II - Evolution of intermediate-mass
  stars up to carbon ignition or core degeneracy}.
\newblock {\em MNRAS}, 220:529--547.

\bibitem[Turner and Stommel, 1964]{TS64}
Turner, J. and Stommel, H. (1964).
\newblock A new case of convection in the presence of vertical salinity and
  temperature gradients.
\newblock {\em Proc. Natl Acad. Sci.}, 52:49--53.

\bibitem[{Umeda} and {Nomoto}, 2002]{2002ApJ...565..385U}
{Umeda}, H. and {Nomoto}, K. (2002).
\newblock {Nucleosynthesis of Zinc and Iron Peak Elements in Population III
  Type II Supernovae: Comparison with Abundances of Very Metal Poor Halo
  Stars}.
\newblock {\em ApJ}, 565:385--404.

\bibitem[{Umeda} and {Nomoto}, 2003]{2003Natur.422..871U}
{Umeda}, H. and {Nomoto}, K. (2003).
\newblock {First-generation black-hole-forming supernovae and the metal
  abundance pattern of a very iron-poor star}.
\newblock {\em Nature}, 422:871--873.

\bibitem[Vangioni-Flam et~al., 2000]{VCC00}
Vangioni-Flam, E., Coc, A., and Cass\'e, M. (2000).
\newblock Big bang nucleosynthesis updated with the {NACRE} compilation.
\newblock {\em A\&A}, 360:15--23.

\bibitem[{Vassiliadis} and {Wood}, 1993]{1993ApJ...413..641V}
{Vassiliadis}, E. and {Wood}, P.~R. (1993).
\newblock {Evolution of low- and intermediate-mass stars to the end of the
  asymptotic giant branch with mass loss}.
\newblock {\em ApJ}, 413:641--657.

\bibitem[{Ventura} et~al., 2005]{Ven05}
{Ventura}, P., {Castellani}, M., and {Straka}, C.~W. (2005).
\newblock {Diffusive convective overshoot in core He-burning intermediate mass
  stars. I. The LMC metallicity}.
\newblock {\em A\&A}, 440:623--636.

\bibitem[{Ventura} and {D'Antona}, 2005a]{2005AA...431..279V}
{Ventura}, P. and {D'Antona}, F. (2005a).
\newblock {Full computation of massive AGB evolution. I. The large impact of
  convection on nucleosynthesis}.
\newblock {\em A\&A}, 431:279--288.

\bibitem[{Ventura} and {D'Antona}, 2005b]{2005ApJ...635L.149V}
{Ventura}, P. and {D'Antona}, F. (2005b).
\newblock {Toward a Working Model for the Abundance Variations in Stars within
  Globular Clusters}.
\newblock {\em ApJL}, 635:L149--L152.

\bibitem[{Ventura} and {D'Antona}, 2006]{2006AA...457..995V}
{Ventura}, P. and {D'Antona}, F. (2006).
\newblock {Does the oxygen-sodium anticorrelation in globular clusters require
  a lowering of the $^{23}$Na(p,{$\alpha$})$^{20}$Ne reaction rate?}
\newblock {\em A\&A}, 457:995--1001.

\bibitem[Ventura et~al., 2004]{2004MmSAI..75..335V}
Ventura, P., D'Antona, F., and Mazzitelli, I. (2004).
\newblock New {AGB} models to explore the spread of abundances in {G}lobular
  {C}lusters.
\newblock {\em Memorie della Societa Astronomica Italiana}, 75:335.

\bibitem[{Wagner}, 1974]{1974ApJ...191..173W}
{Wagner}, R.~L. (1974).
\newblock {Theoretical Evolution of Extremely Metal-Poor Stars}.
\newblock {\em ApJ}, 191:173--182.

\bibitem[{Wagoner}, 1973]{1973ApJ...179..343W}
{Wagoner}, R.~V. (1973).
\newblock {Big-Bang Nucleosynthesis Revisited}.
\newblock {\em ApJ}, 179:343--360.

\bibitem[{Weiss} et~al., 2000]{2000ApJ...533..413W}
{Weiss}, A., {Cassisi}, S., {Schlattl}, H., and {Salaris}, M. (2000).
\newblock {Evolution of Low-Mass Metal-Free Stars including Effects of
  Diffusion and External Pollution}.
\newblock {\em ApJ}, 533:413--423.

\bibitem[{Weiss} et~al., 2004]{2004AA...422..217W}
{Weiss}, A., {Schlattl}, H., {Salaris}, M., and {Cassisi}, S. (2004).
\newblock {Models for extremely metal-poor halo stars}.
\newblock {\em A\&A}, 422:217--223.

\bibitem[Wisotzki et~al., 1996]{WKG+96}
Wisotzki, L., Koehler, T., Groote, D., and Reimers, D. (1996).
\newblock The {H}amburg/{ESO} survey for bright {QSO}s. {I}. {S}urvey design
  and candidate selection procedure.
\newblock {\em A\&AS}, 115:227.

\bibitem[Wolfe et~al., 1986]{WTS+86}
Wolfe, A.~M., Turnshek, D.~A., Smith, H.~E., and Cohen, R.~D. (1986).
\newblock Damped {L}yman-alpha absorption by disk galaxies with large
  redshifts. {I} - {T}he {L}ick survey.
\newblock {\em ApJS}, 61:249--304.

\bibitem[{Wood}, 1981]{1981ApJ...248..311W}
{Wood}, P.~R. (1981).
\newblock {On the entropy of mixing, with particular reference to its effect on
  dredge-up during helium shell flashes}.
\newblock {\em ApJ}, 248:311--314.

\bibitem[{Wood} and {Zarro}, 1981]{1981ApJ...247..247W}
{Wood}, P.~R. and {Zarro}, D.~M. (1981).
\newblock {Helium-shell flashing in low-mass stars and period changes in mira
  variables}.
\newblock {\em ApJ}, 247:247--256.

\bibitem[{Xiong}, 1985]{Xiong85}
{Xiong}, D.~R. (1985).
\newblock {Convective overshooting in stellar internal models}.
\newblock {\em A\&A}, 150:133--138.

\bibitem[{Yong} et~al., 2003]{2003AA...402..985Y}
{Yong}, D., {Grundahl}, F., {Lambert}, D.~L., {Nissen}, P.~E., and {Shetrone},
  M.~D. (2003).
\newblock {Mg isotopic ratios in giant stars of the globular cluster NGC 6752}.
\newblock {\em A\&A}, 402:985--1001.

\bibitem[York et~al., 2000]{YAA+00}
York, D.~G., Adelman, J., Anderson, J. E.~J., Anderson, S.~F., Annis, J., and
  Bahcall, N.~A. (2000).
\newblock The {S}loan {D}igital {S}ky {S}urvey: {T}echnical {S}ummary.
\newblock {\em AJ}, 120(3):1579--1587.

\bibitem[{Zhao}, 2005]{2005AIPC..752..155Z}
{Zhao}, G. (2005).
\newblock {Search for Metal-poor Stars with LAMOST}.
\newblock In {Mikolajewska}, J. and {Olech}, A., editors, {\em AIP Conf. Proc.
  752: Stellar Astrophysics with the World's Largest Telescopes}, pages
  155--158.

\bibitem[{Zijlstra}, 2004]{2004MNRAS.348L..23Z}
{Zijlstra}, A.~A. (2004).
\newblock {Low-mass supernovae in the early Galactic halo: source of the double
  r/s-process enriched halo stars?}
\newblock {\em MNRAS}, 348:L23--L27.

\end{thebibliography}
\end{spacing}
\end{document}